\documentclass[12pt,twoside,titlepage]{book}
\usepackage{latexsym,amsmath,amssymb,amsfonts,amsthm}
\usepackage{makeidx}
\usepackage{graphicx}
\usepackage{indentfirst}
\usepackage{chapterbib}
\def\nabla{\bigtriangledown}
\newtheorem{result}{Result}[chapter]

\newcounter{resultnum}[section]
\newtheorem{conclusion}{Conclusion}[chapter]

\newcounter{conclusionnum}[section]

\newcounter{conditionnum}[section]

\newcounter{conjecturenum}[section]
\newtheorem{example}{Example}[chapter]

\newcounter{examplenum}[section]

\newcounter{exercisenum}[section]
\newtheorem{lemma}{Lemma}[chapter]

\newcounter{lemmanum}[section]

\newcounter{notationnum}[section]
\newtheorem{theorem}{Theorem}[chapter]

\newcounter{theoremnum}[section]
\newtheorem{definition}{Definition}[chapter]

\newcounter{definitionnum}[section]
\newtheorem{corollary}{Corollary}[chapter]

\newcounter{corollarynum}[section]
\newtheorem{remark}{Remark}[chapter]

\newcounter{remarknum}[section]
\newtheorem{proposition}{Proposition}[chapter]

\newcounter{propositionnum}[section]

\newcounter{acknowledgementnum}[section]

\newcounter{algorithmnum}[section]

\newcounter{axiomnum}[section]

\newcounter{casenum}[section]

\newcounter{claimnum}[section]

\newcounter{summarynum}[section]

\newcounter{problemnum}[section]
\newcommand{ \R} {\mbox{\rm I$\!$R}}
\newcommand{ \C} {\mbox{\rm I$\!$C}}
\newcommand{ \Z} {\mbox{\rm I$\!$Z}}
\newcommand{ \I} {\mbox{\rm I$\!$I}}
\newcommand{ \D} {\mbox{\rm I$\!$\bf{D}}}

\begin{document}

\frontmatter

%%%%%%%%%%%
\begin{titlepage}
 -
\vskip100pt
\vbox{\begin{center} \Huge \bf Clifford and Riemann--Finsler
\\ Structures in Geometric \\
 Mechanics and Gravity
\end{center}}

\vskip65pt
\vbox{\begin{center}
\Large \sf
 Selected Works by   \end{center}}
%\vskip10pt
\vbox{\begin{center} {\Large S.  Vacaru, P. Stavrinos, E. Gaburov
and D. Gon\c ta}
\end{center}}
%\vspace*{2mm}

\vskip130pt \vbox{\begin{center} \Large \it  Geometry Balkan Press,\
2005
\end{center}}
 \vskip01pt

%---------------------------------

%{\copyright \quad S. Vacaru}
\end{titlepage}

\newpage

\chapter*{Preface}

The researches resulting in this massive book have been initiated by S. Vacaru
fifteen years ago when he prepared a second Ph. Thesis in Mathematical Physics.
 Studying Finsler--Lagrange geometries he became aware of the potential applications
 of these geometries in exploring nonlinear aspects and nontrivial symmetries arising
 in various models of gravity, classical and quantum field theory and geometric mechanics.

Along years he convinced many people to enroll in solving some of his open
problems and especially he attracted young students to specialize in this field.
 Some of his collaborators are among the co--authors of this book.

The book contains a collection of works on Riemann--Cartan and metric-affine manifolds
 provided with nonlinear connection structure and on generalized Finsler--Lagrange
 and Cartan--Hamilton geometries and Clifford structures modelled on such manifolds.

The authors develop and use the method of anholonomic frames with associated nonlinear
connection structure and apply it to a great number of concrete problems: constructing
 of generic off--diagonal exact solution, in general, with nontrivial torsion and
 nonmetricity, possessing noncommutative symmetries and describing black ellipsoid/ torus
  configurations, locally anisotropic wormholes, gravitational solitons and warped
  factors and investigation of stability of such solutions; classification of
  Lagrange/Finsler affine spaces; definition of nonholonomic Dirac operators and
  their applications in commutative and noncommutative Finsler geometry.

This collection of works enriches very much the literature on generalized Finsler
 spaces and opens new ways toward applications by proposing new geometric approaches
  in gravity, string theory, quantum deformations and noncommutative models.

The book is extremely useful for the researchers in Differential Geometry and
Mathematical Physics.

\vskip2pt

February, 2006

\vskip3pt

Prof. Dr. Mihai Anastasiei

\vskip2pt

Faculty of Mathematics, University "Al.I.Cuza" Iasi ,

Iasi, 700506, Romania

\newpage

\tableofcontents

\newpage
% - \newpage

%%%%%%%%%%%%%

\chapter{Foreword}

The general aim of this Selection of Works  is to outline the
methods of Riemann--Finsler geometry and generalizations as an aid
in exploring certain less known nonlinear aspects and nontrivial
symmetries of field equations defined by nonholonomic and
noncommutative structures arising in various models of gravity,
classical and quantum field theory and geometric mechanics.
Accordingly, we move primarily in the realm of the geometry of
nonholonomic manifolds for which the tangent bundles are provided
with nonintegrable (anholonomic) distributions defining nonlinear
connection (in brief, N--connection) structures. Such
N--connections may be naturally associated to certain general
off--diagonal metric terms and distinguish some preferred classes
of adapted local frames and linear connections. This amounts to a
program of unification when  the Riemann--Cartan,
Finsler--Lagrange spaces and  various generalizations are commonly
described by the corresponding geometric  objects on
N--anholonomic manifolds.

Our purposes and main concern are to illuminate common aspects in spinor
differential geometry, gravity and geometric mechanics from the viewpoint of
N--connection geometry and methods elaborated in investigating
Finsler--Lagrange and related metric--affine spaces (in general, with
nontrivial torsion and nonmetricity), to elaborate a corresponding language
and techniques of nonholonomic deformations of geometric structures with
various types of commutative and noncommutative symmetries and to benefit
physicists interested in more application of advanced commutative and
noncommutative geometric methods.

The guiding principle of the selected here works has been to show
that the concept of N--anholono\-mic space seeks in roots when
different type of geometries can be modelled by certain
parametrizations of the N--connection structure and
correspondingly adapted linear connection and metric structures.
For instance, a class of such objects results in (pseudo) Riemann
spaces but with preferred systems of reference, other classes of
objects give rise into models of Finsler (Lagrange) geometries
with metric compatible, or noncompatible, linear connections, all
defined by the fundamental  Finsler (regular Lagrange) functions
and corresponding parametrizations  and prescribed symmetries.

\newpage

Despite a number of last decade works on research and applications
of Finsler--Lagrange geometry by leading schools and prominent
scholars in Romania (R. Miron, M. Anastasiei, A. Bejancu, ...),
Japan (K.  Matsumoto, S. Ikeda, H. Shimada,...), USA (S. S. Chern,
S. Bao, Z. Shen, J.  Vargas, R. G. Beil,...), Russia (G. Asanov,
G. Yu. Bogoslovsky, ...),  Germany (H. F. Goenner, K. Buchner, H.
B. Rademacher,...), Canada  (P. Antonelli, D. Hrimiuc,...),
Hungary (L. Tamassy, S. Basco,...)  and other Countries, there
have not been yet obtained explicit results  related to the
phenomenology of Standard Model of particle physics, string
theory, standard cosmological scenaria and astrophysics. The
problem is that the main approaches and constructions in Finsler
geometry and generalizations were elaborated in the bulk on
tangent/ vector bundles and their higher generalizations. In this
case all type of such locally anisotropic models are related to
violations of the local Lorentz symmetry which is a fashion, for
instance,  in brane physics but, nevertheless, is subjected to
substantial  theoretical and experimental restrictions (J. D.
Beckenstein and C. Will).

Our idea\footnote{%
hereafter, in this Preface, we shall briefly outline the  main ideas,
concepts and results obtained during the last decade  by a team of young
researches in the Republic of Moldova  (S. Vacaru, S. Ostaf, Yu.
Goncharenko, E. Gaburov, D. Gon\c ta, N. Vicol, I. Chiosa ...)
 in collaboration (or
having certain support) with some scientific groups  and scholars
in Romania, USA, Germany, Greece, Portugal and Spain  (D.
Singleton, H. Dehnen, P. Stavrinos, F. Etayo, B. Fauser, O. \c
Tint\v areanu--Mircea,  F. C. Popa, J. F. Gonzales--Hernandez, R.
R. Santamar\'{\i}a and hosting by R. Miron, M. de Leon, M. Vi\c
sinescu,  M. Anastasiei, T. Wolf, S. Anco, I. Gottlieb, C. Mociu\c tchi, B. Fauser,
 J. P. S. Lemos, L. Boya, P. Almeida, R. Picken, M. E. Gomez,
 M. Piso,  M. Mars, L. Al\'{\i}as, C. Udri\c ste, D.
Balan, V. Blanu\c t\v a, G. Zet, ...)} was to define and work with
Finsler (Lagrange) like geometric objects and structures not only
on the tangent/vector bundles (and their higher order
generalizations) but to model them on usual manifolds enabled with
certain classes of nonholonomic distributions defined by exact
sequences of subspaces of the tangent space to such manifolds.
This way, for instance, we can model a Finsler geometry as a
Riemann--Cartan manifold provided with certain types of
N--connection and adapted linear connection and metric structures.
The constructions have to be generalized for the metric--affine
spaces provided with N--connection structure if there are
considered the so--called Berwald--Moor or Chern connections  for
Finsler geometry, or, in an alternative way, one can be imposed
such  nonholonomic constrains on the frame structure when some
subclasses  of Finsler metrics are equivalently modelled on
(pseudo) Riemann  spaces provided with corresponding preferred
systems of reference.  This is possible for such configurations
when the Ricci tensor for the so--called canonical distinguished
connection  in generalized Lagrange (or Finsler) space is
constrained to be equal to the Ricci tensor for the Levi--Civita
connection even the curvature  tensors are different.

%\newpage

It was a very surprising result when a number of exact solutions  modelling
Finsler like structures were constructed in the Einstein  and string
gravity. Such solutions are defined by generic  off--diagonal metrics,
nonholonomic frames and linear connections  (in general, with nontrivial
torsion; examples of solutions with nontrivial  nonmetricity were also
constructed), when a subset of variables are  holonomic and the subset of
the rest ones are nonholonomic.  That was an explicit proof that effective
local anisotropies can be induced  by off--diagonal metric terms and/or from
extra dimensions.  In a particular case, the locally anisotropic
configurations can be  modelled as exact solutions of the vacuum or
nonvacuum Einstein equations.  Sure, such Einstein--Finsler/ generalized
Lagrange metrics and related  nonholnomic frame structures are not subjected
to the existing experimental  restrictions and theoretical considerations
formulated for the Finsler models  on tangent/vector bundles.

The new classes of exact solutions describe three, four, or five dimensional
space--times (there are possibilities for extensions to higher dimensions)
with generalized symmetries when the metric, connection and frame
coefficients depend on certain integration functions on  two/ three / four
variables. They may possess noncommutative  symmetries even for commutative
gravity models, or any  generalizations to Lie/ Clifford algebroid
structures, and can be  extended to stable configurations in complex
gravity. \footnote{%
The bibliography presented for the Introduction  and at the end of
Chapters contains exact citations of our works on  anholonomic
black ellipsoid/torus and disk solutions,  locally anisotropic
wormholes and Taub NUT spaces, nonholonomic  Einstein--Dirac wave
solitons, locally anisotropic cosmological  solutions, warped
configurations or with Lie/Clifford algebroid and/or
noncommutative symmetries, ...} \  Here it is appropriate to
emphasize that the proposed 'anholnomic  frame method' of
constructing exact solutions was derived by  using explicit
methods from the  Finsler--Lagrange geometry. Perhaps,  this is
the most general method of constructing exact solutions  in
gravity: it was elaborated as a geometric method
by using the N--connection formalism.

The above mentioned results derived by using moving frames and
nonholonomic structures feature several fundamental constructions:
1) Any Finsler--Lagrange geometry can be equivalently realized as
an effective Riemann--Cartan nonholonomic manifold and, inversely,
2) any space-time with generic off--diagonal metric and
nonholonomic frame and affine  connection structures can be
equivalently nonholonomically deformed  into various types of
Finsler/ Lagrange geometries. 3) As a matter of principle,
realizing a Finsler configuration as a Riemannian nonholonomic
manifold (with nonholonomically induced  torsion), we may assemble
this construction from the ingredients  of noncommutative spin
geometry, in the A. Connes approach, or  we can formulate a
noncommutative gauge--Finsler geometry via the Seiberg--Witten
transform.

The differential geometry of N--anholonomic spinors and related Clifford
structures provided with N--connections predated the results on
'nonholonomic' gravity and related classes of exact solutions. The first and
second  important results were, respectively, the possibility to give a
rigorous definition for spinors in Finsler spaces and  elaboration of the
concept of Finsler superspaces. The third  result was the classification of
such spaces in terms of  nearly autoparallel maps (generalizing the classes
of geodesic maps  and conformal transforms) and their basic equations and
invariants suggesting variants of definition of conservation  lows for such
(super) spaces. The forth such a fundamental result was a nontrivial proof
that Finsler like  structures can be derived in low energy limits of (super)
string theory if the (super) frames with associated  N--connection structure
are introduced into consideration.  There were obtained a set of results in
the theory of locally  anisotropic stochastic, kinetic and thermodynamic
processes in  generalized curved spacetimes. Finally, we mention here the
constructions  when from a regular Lagrange (Finsler) fundamental functions
one derived  canonically a corresponding Clifford/spinor structure which in
its turn induces canonical noncommutative Lagrange (Finsler)  geometries,
nonholonomic Fedosov manifolds and generalized  Lagrange (Finsler)
Lie/Clifford algebroid structures.

The ideas that we can deal in a unified form, by applying the  N--connection
formalism, with various types of nonholnomic  Riemann--Cartan--Weyl and
generalized Finsler--Lagrange or  Cartan--Hamilton spaces scan several new
directions in modern  geometry and physics: We hope that they will appeal
researches  (we also try to contribute explicitly in our works) in
investigating  nonholonomic Hopf strutures, N--anholonomic gerbes and
noncommutative/ algebroid extensions, Atyiah--Singer theorems  for
Clifford--Lagrange spaces and in constructing exact  solutions with
nontrivial topological structure modelling  Finsler gerbes, Ricci and
Finsler--Lagrange fluxes, and applications  in gravity and string theory,
analogous modelling of gravity and  gauge interactions and geometric
mechanics.

\subsection*{Acknowledgments}

The selection of works reflects by 15 years of authors' researches
 on generalized Finsler geometry and applications in modern
 physics. The most pleasant aspect in finishing this Foreword is
 having the opportunity to thank the many people (their names are given
 in footnote 1) who helped to perform this work, provided substantial support,
 collaborated and improved the results and suggested new very important
 ideas.

 \vskip6pt

 Editor:\ Sergiu I. Vacaru,
 \vskip3pt

 {\qquad} IMAFF, CSIC, Madrid, Spain, and

  {\qquad} Brock  University, St. Catharines, Ontario, Canada

  \vskip3pt

 {\qquad} vacaru@imaff.cfmac.csic.es,\

{\qquad} svacaru@brocku.ca

  \newpage

\chapter{Introduction}

This collection of works grew out from explicit constructions  proving
that the Finsler and Lagrange  geometries can be modelled as certain
type nonintegrable distributions on Riemann--Cartan manifolds
if the metric and connection structures on such spaces are
compatible\footnote{one has to consider metric--affine spaces
with nontrivial  torsion and  nonmetricity fields defined by the
N--connection and adapted linear connection and  metric structures
if we work, for instance, with the metric noncompatible Berwald and
Chern connections (they can be defined both in Finsler and Lagrange
geometries)}. This is a rather surprising fact because the standard
 approaches were based on the idea that the Finsler geometry is
 more {\bf general} then the Riemannian one when, roughly speaking,
the metric anisotropically depends on "velocity" and the
geometrical and physical models are elaborated on the tangent
bundle.  Much confusing may be  made from such a generalization if
one does not pay a  due attention to the second fundamental
geometric structure for the Finsler spaces called the  nonlinear
connection (N--connection), being defined by a nonholonomic
distribution on the tangent bundle and related to a corresponding
class of preferred systems of reference. There is the third
fundamental geometric object, the linear connection, which for the
Finsler like geometries is usually adapted to the N--connection
structure.

In the former (let us say standard) approach, the Finsler and Lagrange spaces
(the second class of spaces are derived similarly to the Finsler
ones but for regular Lagrangians) are assembled from the mentioned
three fundamental objects (the metric, N--connection and  linear
connection) defined in certain  adapted forms on the tangent bundles
provided with a canonical nonholonomic splitting into horizontal and
vertical subspaces (stated by the exact sequence just defining
the N--connection).

\newpage

As a matter of principle, we may consider that certain exact
sequences and related nonintegrable distributions, also defining a
N--connection structure, are prescribed, for instance, for a class
of Riemann--Cartan manifolds. In this case, we work  with sets of
mixed holonomic coordinates (corresponding to the horizontal
coordinates on the tangent bundle) and anholonomic\footnote{in
literature, one introduced two equivalent terms: nonholonomic or
anholonomic; we shall use both terms} coordinates (corresponding
to the vertical coordinates). The splitting into
holonomic--anholonomic local coordinates and the corresponding
conventional horizontal--vertical decomposition are globally
stated by the N--connection structure as in the standard
approaches to Finsler geometry. We are free to consider that a
fibered structure is some way established as a generalized
symmetry by a prescribed nonholonomic distribution defined for a
usual manifold and not for a tangent or vector bundle.

There is a proof that for any vector  bundles over paracompact
manifolds the N--connection structure always exists, see Ref. \cite{00ma}.
On general manifolds this does not hold true but we can restrict our
considerations to such Riemann--Cartan (or metric--affine) spaces when
the metric structure\footnote{the corresponding metric tensor can not
be diagonalized by any coordinate transforms} is someway related via
nontrivial off--diagonal metric coefficients to the coefficients of
the N--connection and associated nonholonomic frame structure. This way
we can model Finsler like geometries on nonholonomic manifolds when the
anisotropies depend on anholonomic coordinates (playing the role of
"velocities" if to compare with the standard approaches to the Finsler
geometry and generalizations). It is  possible the case when such generic
off--diagonal metric and N--connection and the linear connection structures
are subjected to the condition to  satisfy a variant of gravitational filed
equations in Einstein--Cartan or string gravity. Any such solution is
described by a  Finsler like gravitational configuration which for corresponding
constraints on the frame structure and distributions of matter
defines a nonholonomic Einstein space.

\section{For Experts in Differential Geometry and Gravity}
 The aim of this section is to present a self--contained treatment
 of generalized Finsler strucures modelled on Riemann--Cartan spaces
 provided with N--connections (we follow Chapters 2
 and 3 in Ref. \cite{00vjfg}, see also  details on modelling Lagrange and Finsler
  geometries on metric--affine spaces presented in Part I of this  book).

\subsection{N--anholonomic manifolds}

We formulate a coordinate free introduction into the geometry of
nonholonomic manifolds. The reader may consult details in Refs. \cite%
{00vhep2,00vs,00vggmaf,00esv}. Some important component/coordinate formulas are
given in the next section.

\subsubsection{Nonlinear connection structures}

Let $\mathbf{V}$ be a smooth manifold of dimension $(n+m)$ with a local
splitting in any point $u\in \mathbf{V}$ of type \ $\mathbf{V}%
_{u}=M_{u}\oplus V_{u},$ where $M$ is a $n$--dimensional subspace and $V$ is
a $m$--dimensional subspace. It is supposed that one exists such a local
decomposition when $\mathbf{V}\rightarrow M$ is a surjective submersion. Two
important particular cases are that of a vector bundle, when we shall write $%
\mathbf{V=E}$ (with $\mathbf{E}$ being the total space of a vector bundle $%
\pi :$ $\mathbf{E}\rightarrow M$ with the base space $M)$ and that of
tangent bundle when we shall consider $\mathbf{V=TM.}$\ The differential of
a map $\pi :\mathbf{V}\rightarrow M$ defined by fiber preserving morphisms
of the tangent bundles $T\mathbf{V}$ and $TM$ is denoted by $\pi ^{\top }:T%
\mathbf{V}\rightarrow TM.$ The kernel of $\pi ^{\top }$ defines the vertical
subspace $v\mathbf{V}$ with a related inclusion mapping $i:v\mathbf{V}%
\rightarrow T\mathbf{V}.$

\begin{definition}
\label{dnc}A nonlinear connection (N--connection) $\mathbf{N}$ on a manifold
$\mathbf{V}$ is defined by the splitting on the left of an exact sequence%
\begin{equation}
0\rightarrow v\mathbf{V}\overset{i}{\rightarrow }T\mathbf{V}\rightarrow T%
\mathbf{V}/v\mathbf{V}\rightarrow 0,  \label{exseq}
\end{equation}%
i. e. by a morphism of submanifolds $\mathbf{N:\ \ }T\mathbf{V}\rightarrow v%
\mathbf{V}$ such that $\mathbf{N\circ i}$ is the unity in $v\mathbf{V}.$
\end{definition}

The exact sequence (\ref{exseq}) states a nonintegrable (nonholonomic,
equivalently, anholonomic) distribution on $\mathbf{V},$ i.e. this manifold
is nonholonomic. We can say that a N--connection is defined by a global
splitting into conventional horizontal (h) subspace, $\left( h\mathbf{V}%
\right) ,$ and vertical (v) subspace, $\left( v\mathbf{V}\right) ,$
corresponding to the Whitney sum
\begin{equation}
T\mathbf{V}=h\mathbf{V}\oplus _{N}v\mathbf{V}  \label{00whitney}
\end{equation}%
where $h\mathbf{V}$ is isomorphic to $M.$ We put the label $N$ to the symbol
$\oplus $ in order to emphasize that such a splitting is associated to a
N--connection structure.

For convenience, in the next Section, we give some important local
formulas (see, for instance, the local representation for a
N--connection (\ref{nclf})) for the basic geometric objects and
formulas on spaces provided with N--connection structure. Here, we
note that the concept of N--connection came from E. Cartan's works
on Finsler geometry \cite{00cart} (see a detailed historical study
in Refs. \cite{00ma,00esv,00vncl} and alternative approaches developed
by using the Ehressmann connection \cite{00ehr,00dl}). Any manifold
admitting an exact sequence of type (\ref{exseq}) admits a
N--connection
structure. If $\mathbf{V=E,}$ a N--connection exists for any vector bundle $%
\mathbf{E}$ over a paracompact manifold $M,$ see proof in Ref. \cite{00ma}.

The geometric objects on spaces provided with N--connection structure are
denoted by ''bolfaced'' symbols. Such objects may be defined in
''N--adapted'' form by considering h-- and v--decompositions (\ref{00whitney}%
). Following the conventions from \cite{00ma,00vfs,00vstav,00vncl}, one
call such objects to be d--objects (i. e. they are distinguished
by the N--connection; one considers d--vectors, d--forms,
d--tensors, d--spinors, d--connections, ....). For instance, a
d--vector is an element $\mathbf{X}$ of the module of the vector
fields $\chi (\mathbf{V)}$ on $\mathbf{V},$ which in N--adapted
form may be written
\begin{equation*}
\mathbf{X=}\ h\mathbf{X}+v\mathbf{X}\mbox{ or }\mathbf{X=\ }X\oplus _{N}\
^{\bullet }X,
\end{equation*}%
where $h\mathbf{X}$ (equivalently, $X$) is the h--component and $v\mathbf{X}$
(equivalently, $\ ^{\bullet }X)$ is the v--component of $\mathbf{X.}$

A N--connection is characterized by its \textbf{N--connection curvature }%
(the Nijenhuis tensor)%
\begin{equation}
\Omega (\mathbf{X,Y})\doteqdot \lbrack \ ^{\bullet }X,\ ^{\bullet }Y]+\
^{\bullet }[\mathbf{X,Y}]-\ ^{\bullet }[\ ^{\bullet }X\mathbf{,Y}]-\
^{\bullet }[\mathbf{X,}\ ^{\bullet }Y]  \label{njht}
\end{equation}%
for any $\mathbf{X,Y\in }\chi (\mathbf{V),}$ where $[\mathbf{X,Y]\doteqdot
XY-}$ $\mathbf{YX}$ and $\ ^{\bullet }[\mathbf{,]}$ is the v--projection of $%
[\mathbf{,],}$ see also the coordinate formula (\ref{00ncurv}) in
section \ref{brfg}. This d--object $\Omega $ was introduced in
Ref. \cite{00grifone} in order to define the curvature of a
nonlinear connection in the tangent bundle over a smooth manifold.
But this can be extended for any nonholonomic manifold,
nonholnomic Clifford structure and any noncommutative /
supersymmetric versions of bundle spaces provided with
N--connection structure, i. e. with nonintegrable distributions of
type (\ref{00whitney}), see \cite{00esv,00vncl,00vv}.

\begin{proposition}
\label{pddv}A N--connection structure on $\mathbf{V}$ defines a nonholonomic
N--adapted frame (vielbein) structure $\mathbf{e}=(e,^{\bullet }e)$ and its
dual $\widetilde{\mathbf{e}}=\left( \widetilde{e},\ ^{\bullet }\widetilde{e}%
\right) $ with $e$ and $\ ^{\bullet }\widetilde{e}$ linearly depending on
N--connection coefficients.\
\end{proposition}

\begin{proof}
It follows from explicit local constructions, see formulas (\ref{00ddif}), (%
\ref{00dder}) and (\ref{00anhrel}).\
\end{proof}

\begin{definition}
A manifold \ $\mathbf{V}$ is called N--anholonomic if it is defined a local
(in general, nonintegrable) distribution (\ref{00whitney}) on its tangent
space $T\mathbf{V,}$ i.e. $\mathbf{V}$ is N--anholonomic if it is enabled
with a N--connection structure (\ref{exseq}).
\end{definition}

\subsubsection{Curvatures and torsions of N--anholonomic manifolds}

One can be defined N--adapted linear connection and metric structures on $%
\mathbf{V:}$

\begin{definition}
\label{00ddc}A distinguished connection (d--connection) $\mathbf{D}$ on a
N--anho\-lo\-no\-mic manifold $\mathbf{V}$ is a linear connection conserving
under parallelism the Whitney sum (\ref{00whitney}). For any $\mathbf{X\in }%
\chi (\mathbf{V),}$ one have a decomposition into h-- and v--covariant
derivatives,%
\begin{equation}
\mathbf{D}_{\mathbf{X}}\mathbf{\doteqdot X}\rfloor \mathbf{D=\ }X\rfloor
\mathbf{D+}\ ^{\bullet }X\rfloor \mathbf{D=}D_{X}+\ ^{\bullet }D_{X}.
\label{dconcov}
\end{equation}
\end{definition}

The symbol ''$\rfloor "$ in (\ref{dconcov}) denotes the interior product. We
shall write conventionally that $\mathbf{D=}(D,\ ^{\bullet }D).$

For any d--connection $\mathbf{D}$ on a N--anholonomic manifold $\mathbf{V},$
it is possible to define the curvature and torsion tensor in usual form but
adapted to the Whitney sum (\ref{00whitney}):
\begin{definition}
The torsion
\begin{equation}
\mathbf{T(X,Y)\doteqdot D}_{\mathbf{X}}\mathbf{Y-D}_{\mathbf{Y}}\mathbf{%
X-[X,Y]}  \label{tors1}
\end{equation}%
of a d--connection $\mathbf{D=}(D,\ ^{\bullet }D)\mathbf{,}$ for any $%
\mathbf{X,Y\in }\chi (\mathbf{V),}$ has a N--adapted decomposition
\begin{equation}
\mathbf{T(X,Y)=T(}X,Y\mathbf{)+T(}X,\ ^{\bullet }Y\mathbf{)+T(\ }^{\bullet
}X,Y\mathbf{)+T(\ }^{\bullet }X,\ ^{\bullet }Y\mathbf{).}  \label{tors2}
\end{equation}
\end{definition}

By further h- and v--projections of (\ref{tors2}), denoting $h\mathbf{%
T\doteqdot }T$ and $v\mathbf{T\doteqdot \ }^{\bullet }T,$ taking in the
account that $h\mathbf{[\ }^{\bullet }X,\ ^{\bullet }Y\mathbf{]=}0\mathbf{,}$
one proves

\begin{theorem}
\label{tht}The torsion of a d--connection $\mathbf{D=}(D,\ ^{\bullet }D)$ is
defined by five nontrivial d--torsion fields adapted to the h-- and
v--splitting by the N--connection structure%
\begin{eqnarray*}
T(X,Y) &\mathbf{\doteqdot }&D_{X}Y-D_{Y}X-h[X,Y],\ \mathbf{\ }^{\bullet
}T(X,Y)\mathbf{\doteqdot \ }^{\bullet }[Y,X], \\
T(X,\mathbf{\ }^{\bullet }Y) &\mathbf{\doteqdot }&\mathbf{-\ }^{\bullet
}D_{Y}X-h[X,\mathbf{\ }^{\bullet }Y],\mathbf{\ }^{\bullet }T(X,\mathbf{\ }%
^{\bullet }Y)\mathbf{\doteqdot \ }^{\bullet }D_{X}Y-\mathbf{\ }^{\bullet }[X,%
\mathbf{\ }^{\bullet }Y], \\
\mathbf{\ }^{\bullet }T(\mathbf{\ }^{\bullet }X,\mathbf{\ }^{\bullet }Y) &%
\mathbf{\doteqdot }&\mathbf{\ }^{\bullet }D_{X}\mathbf{\ }^{\bullet }Y-%
\mathbf{\ }^{\bullet }D_{Y}\mathbf{\ }^{\bullet }X-\mathbf{\ }^{\bullet }[%
\mathbf{\ }^{\bullet }X,\mathbf{\ }^{\bullet }Y].
\end{eqnarray*}%
\end{theorem}

The d--torsions $T(X,Y),\mathbf{\ }^{\bullet }T(\mathbf{\ }^{\bullet }X,%
\mathbf{\ }^{\bullet }Y)$ are called respectively the $h(hh)$--torsion,\\
 $v(vv)$--torsion and so on. The formulas (\ref{00dtors})  present a
local proof of this Theorem.

\begin{definition}
The curvature of a d--connection $\mathbf{D=}(D,\ ^{\bullet }D)$ is defined%
\textbf{\ }
\begin{equation}
\mathbf{R(X,Y)\doteqdot D}_{\mathbf{X}}\mathbf{D}_{\mathbf{Y}}\mathbf{-D}_{%
\mathbf{Y}}\mathbf{D}_{\mathbf{X}}\mathbf{-D}_{\mathbf{[X,Y]}}  \label{curv1}
\end{equation}
for any $\mathbf{X,Y\in }\chi (\mathbf{V).}$
\end{definition}

Denoting $h\mathbf{R}=R$ and $v\mathbf{R}=\ ^{\bullet }R,$ by
straightforward calculations, one check the properties%
\begin{eqnarray*}
R(\mathbf{X,Y})\mathbf{\ }^{\bullet }Z &=&0,\mathbf{\ }^{\bullet }R(\mathbf{%
X,Y})Z=0, \\
\mathbf{R(X,Y)Z} &\mathbf{=}&R\mathbf{(X,Y)}Z+\mathbf{\ }^{\bullet }R\mathbf{%
(X,Y)\ }^{\bullet }Z
\end{eqnarray*}%
for any for any $\mathbf{X,Y,Z\in }\chi (\mathbf{V).}$

\begin{theorem}
\label{thr}The curvature $\mathbf{R}$ of a d--connection $\mathbf{D=}(D,\
^{\bullet }D)$ is completely defined by six d--curvatures
\begin{eqnarray*}
\mathbf{R(}X\mathbf{,}Y\mathbf{)}Z &\mathbf{=}&\left(
D_{X}D_{Y}-D_{Y}D_{X}-D_{[X,Y]}-\mathbf{\ }^{\bullet }D_{[X,Y]}\right) Z, \\
\mathbf{R(}X\mathbf{,}Y\mathbf{)\ }^{\bullet }Z &\mathbf{=}&\left(
D_{X}D_{Y}-D_{Y}D_{X}-D_{[X,Y]}-\mathbf{\ }^{\bullet }D_{[X,Y]}\right) \
^{\bullet }Z, \\
\mathbf{R(\ }^{\bullet }X\mathbf{,}Y\mathbf{)}Z &\mathbf{=}&\left( \mathbf{\
}^{\bullet }D_{X}D_{Y}-D_{Y}\mathbf{\ }^{\bullet }D_{X}-D_{[\mathbf{\ }%
^{\bullet }X,Y]}-\mathbf{\ }^{\bullet }D_{[\mathbf{\ }^{\bullet
}X,Y]}\right) Z, \\
\mathbf{R(\ }^{\bullet }X\mathbf{,}Y\mathbf{)\mathbf{\ }}^{\bullet }Z &%
\mathbf{=}&\left( \mathbf{\ }^{\bullet }D_{X}\mathbf{\ }^{\bullet }D_{Y}-%
\mathbf{\ }^{\bullet }D_{Y}\mathbf{\ }^{\bullet }D_{X}-D_{[\mathbf{\ }%
^{\bullet }X,Y]}-\mathbf{\ }^{\bullet }D_{[\mathbf{\ }^{\bullet
}X,Y]}\right) \ ^{\bullet }Z, \\
\mathbf{R(\ }^{\bullet }X\mathbf{,\ }^{\bullet }Y\mathbf{)}Z &\mathbf{=}%
&\left( \mathbf{\ }^{\bullet }D_{X}D_{Y}-D_{Y}\mathbf{\ }^{\bullet }D_{X}-%
\mathbf{\ }^{\bullet }D_{[\mathbf{\ }^{\bullet }X,\mathbf{\ }^{\bullet
}Y]}\right) Z, \\
\mathbf{R(\ }^{\bullet }X\mathbf{,\ }^{\bullet }Y\mathbf{)\mathbf{\mathbf{\ }%
}}^{\bullet }Z &\mathbf{=}&\left( \mathbf{\ }^{\bullet }D_{X}D_{Y}-D_{Y}%
\mathbf{\ }^{\bullet }D_{X}-\mathbf{\ }^{\bullet }D_{[\mathbf{\ }^{\bullet
}X,\mathbf{\ }^{\bullet }Y]}\right) \mathbf{\mathbf{\ }}^{\bullet }Z.
\end{eqnarray*}
\end{theorem}

The proof of Theorems \ref{tht} and \ref{thr} is given for vector bundles
provided with N--connection structure in Ref. \cite{00ma}. Similar Theorems
and respective proofs hold true for superbundles \cite{00vncsup}, for
noncommutative projective modules \cite{00vncl} and for N--anholonomic
metric--affine spaces \cite{00vggmaf}, where there are also give the main
formulas in abstract coordinate form. The formulas (\ref{dcurv})
consist a coordinate proof of Theorem \ref{thr}.

\begin{definition}
A metric structure $\ \breve{g}$ on a N--anholonomic space $\mathbf{V}$ is a
symmetric covariant second rank tensor field which is not degenerated and of
constant signature in any point $\mathbf{u\in V.}$
\end{definition}

In general, a metric structure is not adapted to a N--connection structure.

\begin{definition}
\label{ddm}A d--metric $\mathbf{g}=g\oplus _{N}\ ^{\bullet }g$ is a usual
metric tensor which contracted to a d--vector results in a dual d--vector,
d--covector (the duality being defined by the inverse of this metric tensor).
\end{definition}

The relation between arbitrary metric structures and d--metrics is
established by

\begin{theorem}
\label{tdm}Any metric $\ \breve{g}$ can be equivalently transformed into a
d--metric
\begin{equation}
\mathbf{g}=g(X,Y)+\mathbf{\ }^{\bullet }g(\mathbf{\ }^{\bullet }X,\mathbf{\ }%
^{\bullet }Y)  \label{dmetra}
\end{equation}%
for a corresponding N--connection structure.
\end{theorem}

\begin{proof}
We introduce denotations $h\breve{g}(X,Y)\doteqdot g(X,Y)$ and $\mathbf{\ }v%
\breve{g}(\mathbf{\ }^{\bullet }X,\mathbf{\ }^{\bullet }Y)$ $=\mathbf{\ }%
^{\bullet }g(\mathbf{\ }^{\bullet }X,\mathbf{\ }^{\bullet }Y)$ and try to
find a N--connection when
\begin{equation}
\breve{g}(X,\mathbf{\ }^{\bullet }Y)=0  \label{algn01}
\end{equation}%
for any $\mathbf{X,Y\in }\chi (\mathbf{V).}$ In local form, \ the equation (%
\ref{algn01}) is just an algebraic equation for $\mathbf{N}=\{N_{i}^{a}\},$
see formulas (\ref{00metr}), (\ref{00ansatz}) and (\ref{dmetr}) and related
explanations in section \ref{brfg}.% $\square $
\end{proof}

\begin{definition}
A d--connection $\mathbf{D}$ on $\mathbf{V}$ is said to be metric, i.e. it
satisfies the metric compatibility (equivalently, metricity) conditions with
a metric $\ \breve{g}$ and its equivalent d--metric $\mathbf{g},$ if there
are satisfied the conditions
\begin{equation}
\mathbf{D}_{\mathbf{X}}\mathbf{g=0.}  \label{mcc}
\end{equation}
\end{definition}

Considering explicit h-- and v--projecting of (\ref{mcc}), one proves

\begin{proposition}
\label{pmcc}A d--connection $\mathbf{D}$ on $\mathbf{V}$ is metric if and
only if
\begin{equation*}
D_{X}g=0,\ D_{X}\mathbf{\ }^{\bullet }g=0,\ ^{\bullet }D_{X}g=0,\ ^{\bullet
}D_{X}\mathbf{\ }^{\bullet }g=0.
\end{equation*}
\end{proposition}

One holds this important
\begin{conclusion}
Following Propositions \ref{pddv} and \ref{pmcc} and Theorem \ref{tdm}, we
can elaborate the geometric constructions on a N--anholonomic manifold $%
\mathbf{V}$ $\ $\ in N--adapted form by considering N--adapted frames $%
\mathbf{e}=(e,^{\bullet }e)$ and co--frames $\widetilde{\mathbf{e}}=\left(
\widetilde{e},\ ^{\bullet }\widetilde{e}\right) ,$ d--connection $\mathbf{D}$
and d--metric $\mathbf{g}=[g,\mathbf{\ }^{\bullet }g]$ fields.
\end{conclusion}

In Riemannian geometry, there is a preferred linear Levi--Civita connection $%
\bigtriangledown $ which is metric compatible and torsionless, i.e.
\begin{equation*}
\ ^{\bigtriangledown }\mathbf{T(X,Y)\doteqdot }\bigtriangledown _{\mathbf{X}}%
\mathbf{Y-}\bigtriangledown _{\mathbf{Y}}\mathbf{X-[X,Y]=}0,
\end{equation*}%
and defined by the metric structure. On a general N--anholonomic manifold $%
\mathbf{V}$ provided with a d--metric structure $\mathbf{g}=[g,\mathbf{\ }%
^{\bullet }g],$ the Levi--Civita connection defined by this metric is not
adapted to the N--connection, i. e. to the splitting (\ref{00whitney}). The
h-- and v--distributions are nonintegrable ones and any d--connection
adapted to a such splitting contains nontrivial d--torsion coefficients.
Nevertheless, one exists a minimal extension of the Levi--Civita connection
to a canonical d--connection which is defined only by a metric $\breve{g}.$

\begin{theorem}
\label{thcdc}For any d--metric $\mathbf{g}=[g,\mathbf{\ }^{\bullet }g]$ on a
N--anholonomic manifold $\mathbf{V,}$ there is a unique metric canonical
d--connection $\widehat{\mathbf{D}}$ satisfying the conditions $\widehat{%
\mathbf{D}}\mathbf{g=}0$ and with vanishing $h(hh)$--torsion, $v(vv)$%
--torsion, i. e. $\widehat{T}(X,Y)=0$ and $\mathbf{\ }^{\bullet }\widehat{T}(%
\mathbf{\ }^{\bullet }X,\mathbf{\ }^{\bullet }Y)=0.$
\end{theorem}

\begin{proof}
The formulas (\ref{00cdc}) and (\ref{candcon}) and related discussions state
 a proof, in component form, of this Theorem. %$\square $
\end{proof}

The following Corollary gathers some basic information about
N--anho\-lo\-no\-mic manifolds.

\begin{corollary}
\label{cncs}A N--connection structure defines three important geometric
objects:

\begin{enumerate}
\item a (pseudo) Euclidean N--metric structure $\ ^{\eta }\mathbf{g}=\eta
\oplus _{N}\ ^{\bullet }\eta ,$ i.e. a d--metric with (pseudo) Euclidean
metric coefficients with respect to $\widetilde{\mathbf{e}}$ defined only by
$\mathbf{N;}$

\item a N--metric canonical d--connection $\widehat{\mathbf{D}}^{N}$ defined
only by $^{\eta }\mathbf{g}$ and $\mathbf{N;}$

\item a nonmetric Berwald type linear connection $\mathbf{D}^{B}.$
\end{enumerate}
\end{corollary}

\begin{proof}
Fixing a signature for the metric, $sign^{\ \eta }\mathbf{g=(\pm ,\pm
,...,\pm ),}$ we introduce these values in (\ref{dmetr}) and get $^{\eta }%
\mathbf{g}=\eta \oplus _{N}\ ^{\bullet }\eta $ of type (\ref{dmetra}), i.e.
we prove the point 1. The point 2 is to be proved by an explicit
construction by considering the coefficients of $^{\eta }\mathbf{g}$ into (%
\ref{candcon}). This way, we get a canonical d--connection induced by the
N--connection coefficients and satisfying the metricity conditions (\ref{mcc}%
). In an approach to Finsler geometry \cite{00cbs}, one emphasizes the
constructions derived for the so--called Berwald type d--connection $\mathbf{%
D}^{B},$ considered to be the ''most'' minimal (linear on $\Omega )$
extension of the Levi--Civita connection, see formulas (\ref{dbc}). Such
d--connections can be defined for an arbitrary d--metric $\mathbf{g}=[g,%
\mathbf{\ }^{\bullet }g],$ or for any $^{\eta }\mathbf{g}=\eta \oplus _{N}\
^{\bullet }\eta .$ They are only ''partially'' metric because, for instance,
$D^{B}g=0$ and $\mathbf{\ }^{\bullet }D^{B}\mathbf{\ }^{\bullet }g=0$ but,
in general, $D^{B}\mathbf{\ }^{\bullet }g\neq 0$ and $\mathbf{\ }^{\bullet
}D^{B}g\neq 0,$ i. e. $\mathbf{D}^{B}\mathbf{g}\neq 0,$ see Proposition \ref%
{pmcc}. It is a more sophisticate problem to define spinors and
supersymmetric physically valued models for such Finsler spaces, see
discussions in \cite{00vncl,00vclalg,00vggmaf}. %$\square $
\end{proof}

\begin{remark}
The d--connection $\widehat{\mathbf{D}}^{N}$ or $\mathbf{D}^{B},$ for $%
^{\eta }\mathbf{g,}$ nonholonomic bases $\mathbf{e}=(e,^{\bullet }e)$ and $%
\widetilde{\mathbf{e}}=\left( \widetilde{e},\ ^{\bullet }\widetilde{e}%
\right) ,$ see Proposition \ref{pddv} and the N--connection curvature $%
\mathbf{\Omega }$ (\ref{njht}), define completely the main properties of a
N--anholonomic manifold $\mathbf{V.}$
\end{remark}

It is possible to extend the constructions for any additional d--metric and
canonical d--connection structures. For our considerations on nonholnomic
Clifford/ spinor structures, the class of metric d--connections plays a
preferred role. That why we emphasize the physical importance of
d--connections $\widehat{\mathbf{D}}$ and $\widehat{\mathbf{D}}^{N}$ instead
of $\mathbf{D}^{B}$ or any other nonmetric d--connections.

Finally, in this section, we note that the d--torsions and d--curvatures on
N--anholo\-no\-mic manifolds can be computed for any type of d--connection
structure, see Theorems \ref{tht} and \ref{thr} and the component formulas (%
\ref{00dtors}) and (\ref{dcurv}).

\subsection{Examples of N--anholonomic spaces:}

For corresponding parametrizations of the N--connection, d--metric and
d--connec\-ti\-on coefficients of a N--anholonomic space, it is possible to
model various classes of (generalized) Lagrange, Finsler and Riemann--Cartan
spaces. We briefly analyze three such nonholonomic geometric structures.

\subsubsection{Lagrange--Finsler geometry}

This class of geometries is usually defined on tangent bundles \cite{00ma} but
it is possible to model such structures on general N--anholonomic manifolds,
in particular in (pseudo) Riemannian and Riemann--Cartan geometry if
nonholonomic frames are introduced into consideration \cite%
{00vhep2,00vs,00vggmaf,00vesalg}. Let us outline the first approach when the
N--anholonomic manifold $\mathbf{V}$ is taken to be just a tangent bundle $%
(TM,\!\pi ,\!M),$ where $M$ is a $n$--dimensional base manifold, $\pi $ is a
surjective projection and $TM$ is the total space. One denotes by $%
\widetilde{TM}=TM\backslash \{0\}$ where $\{0\}$ means the null section of
map $\pi .$

We consider a differentiable fundamental Lagrange function $L(x,y)$ defined
by a map $L:(x,y)\in TM\rightarrow L(x,y)\in \mathbb{R}$ of class $\mathcal{C%
}^{\infty }$ on $\widetilde{TM}$ and continuous on the null section $%
0:M\rightarrow TM$ of $\pi .$ The values $x=\{x^{i}\}$ are local coordinates
on $M$ and $(x,y)=(x^{i},y^{k})$ are local coordinates on $TM.$ For
simplicity, we consider this Lagrangian to be regular, i.e. with
nondegenerated Hessian
\begin{equation}
\ ^{L}g_{ij}(x,y)=\frac{1}{2}\frac{\partial ^{2}L(x,y)}{\partial
y^{i}\partial y^{j}}  \label{00lqf}
\end{equation}%
when $rank\left| g_{ij}\right| =n$ on $\widetilde{TM}$ and the left up ''L''
is an abstract label pointing that the values are defined by the Lagrangian $%
L.$

\begin{definition}
\label{dlg}A Lagrange space is a pair $L^{n}=\left[ M,L(x,y)\right] $ with $%
\ \ ^{L}g_{ij}(x,y)$ being of constant signature over $\widetilde{TM}.$
\end{definition}

The notion of Lagrange space was introduced by J. Kern \cite{00kern} and
elaborated in details in Ref. \cite{00ma} as a natural extension of Finsler
geometry.

\begin{theorem}
\label{tcslg}There are canonical N--connection $\ ^{L}\mathbf{N,}$ almost
complex $^{L}\mathbf{F,}$ d--metric $\ ^{L}\mathbf{g}$ and d--connection $\
^{L}\widehat{\mathbf{D}}$ structures defined by a regular Lagrangian $L(x,y)$
and its Hessian $\ ^{L}g_{ij}(x,y)$ (\ref{00lqf}).
\end{theorem}

\begin{proof} The simplest way to prove this theorem is to take
 do this in local form (using formulas (\ref{00cncl}) and (\ref{00slm}))
 and then to globalize the constructions.
The canonical $\ ^{L}\mathbf{N}$ is defined by certain nonlinear spray
configurations related to the solutions of Euler--Lagrange equations, see
formula (\ref{00cncl}). It is given there the explicit
matrix representation of $^{L}\mathbf{F}$ (\ref{00acs1}) which is a usual
definition of almost complex structure, after $\ ^{L}\mathbf{N}$ and
N--adapted bases have been constructed. The d--metric (\ref{00slm}) is a local
formula for $\ ^{L}\mathbf{g.}$ Finally, the canonical d--connection $\ ^{L}%
\widehat{\mathbf{D}}$ is a usual one but for $\ ^{L}\mathbf{g}$ and $\ ^{L}%
\mathbf{N}$ on $\widetilde{TM}.$
\end{proof}

A similar Theorem can be formulated and proved for the Finsler geometry:

\begin{remark}
\label{rcsfg}A Finsler space defined by a fundamental Finsler function $%
F(x,y),$ being homogeneous of type $F(x,\lambda y)=\lambda F(x,y),$ for
nonzero $\lambda \in \mathbb{R},$ may be considered as a particular case of
Lagrange geometry when $L=F^{2}.$
\end{remark}

From the Theorem \ref{tcslg} \ and Remark \ref{rcsfg}, one follows:

\begin{result}
\label{result1}Any Lagrange mechanics with regular Lagrangian $L(x,y)$ (any
Finsler geometry with fundamental function $F(x,y))$ can be modelled as a
nonhlonomic Riemann--Cartan geometry with canonical structures $\ ^{L}%
\mathbf{N,}$ $\ ^{L}\mathbf{g}$ and $\ ^{L}\widehat{\mathbf{D}}$ ($\ ^{F}%
\mathbf{N,}$ $\ ^{F}\mathbf{g}$ and $\ ^{F}\widehat{\mathbf{D}})$ defined on
a corresponding N--anholonomic manifold $\mathbf{V}.$
\end{result}

It was concluded that any regular Lagrange mechanics/Finsler geometry can be
geometrized/modelled as an almost K\"{a}hler space with canonical
N--connection distribution, see \cite{00ma} and, for N--anholonomic Fedosov
manifolds, \cite{00esv}. Such approaches based on almost complex structures
are related with standard sympletic geometrizations of classical mechanics
and field theory, for a review of results see Ref. \cite{00dl}.

For applications in optics of nonhomogeneous media \cite{00ma} and
gravity (see, for instance, Refs.
\cite{00vhep2,00vs,00vggmaf,00vclalg,00vesalg}), one considers metrics of
type $g_{ij}\sim e^{\lambda (x,y)}\ ^{L}g_{ij}(x,y)\ $\ which can
not be derived from a mechanical Lagrangian but from an effective
''energy'' function. In the so--called generalized Lagrange
geometry, one introduced Sasaki type metrics (\ref{00slm}), see
section \ref{brfg}, with any general coefficients both for the
metric and N--connection.

\subsubsection{N--connections and gravity}

Now we show how N--anholonomic configurations can defined in gravity
theories. In this case, it is convenient to work on a general manifold $%
\mathbf{V},\dim \mathbf{V}=n+m$ enabled with a global N--connection
structure, instead of the tangent bundle $\widetilde{TM}.$

For the N--connection splitting of (pseudo) Riemannian--Cartan spaces of
dimension $(n+m)$ (there were also considered (pseudo) Riemannian
configurations), the Lagrange and Finsler type geometries were modelled by
N--anholonomic structures as exact solutions of gravitational field
equations \cite{00vggmaf,00vs,00vesnc}. Inversely, all approaches to (super)
string gravity theories deal with nontrivial torsion and (super) vielbein
fields which under corresponding parametrizations model N--anholonomic
spaces \cite{00vncsup,00vhs,00vv}. We summarize here some geometric properties of
gravitational models with nontrivial N--anholonomic structure.

\begin{definition}
A N--anholonomic Riemann--Cartan manifold $\ ^{RC}\mathbf{V}$ is defined by
a d--metric $\mathbf{g}$ and a metric d--connection $\mathbf{D}$ structures
adapted to an exact sequence splitting (\ref{exseq}) defined on this
manifold.
\end{definition}

The d--metric structure $\mathbf{g}$ on$\ ^{RC}\mathbf{V}$ is of type (\ref%
{dmetra}) and satisfies the metricity conditions (\ref{mcc}). With
respect to a local coordinate basis, the metric $\mathbf{g}$ is
parametrized by a generic off--diagonal metric ansatz
(\ref{00ansatz}), see section \ref{brfg}. In a particular case, we
can take $\mathbf{D=}\widehat{\mathbf{D}}$ and treat the torsion
$\widehat{\mathbf{T}}$ as a nonholonomic frame effect induced by
nonintegrable N--splitting. For more general applications, we have
to consider additional torsion components, for instance, by the so--called $H$%
--field in string gravity.

Let us denote by $Ric(\mathbf{D})$ and $Sc(\mathbf{D}),$ respectively, the
Ricci tensor and curvature scalar defined by any metric d--connection $%
\mathbf{D}$ and d--metric $\mathbf{g}$ on $\ ^{RC}\mathbf{V,}$ see also the
component formulas (\ref{00dricci}), (\ref{00sdccurv}) \ and (\ref{enstdt})\ in
 Section \ref{brfg}. The Einstein equations are
\begin{equation}
En(\mathbf{D})\doteqdot Ric(\mathbf{D})-\frac{1}{2}\mathbf{g}Sc(\mathbf{D})=%
\mathbf{\Upsilon }  \label{einsta}
\end{equation}%
where the source $\mathbf{\Upsilon }$ reflects any contributions of matter
fields and corrections from, for instance, string/brane theories of gravity.
In a closed physical model, the equation (\ref{einsta}) have to be completed
with equations for the matter fields, torsion contributions and so on (for
instance, in the\ Einstein--Cartan theory one considers algebraic equations
for the torsion and its source)... It should be noted here that because of
nonholonomic structure of $^{RC}\mathbf{V,}$ the tensor $Ric(\mathbf{D})$ is
not symmetric and that $\mathbf{D}\left[ En(\mathbf{D})\right] \neq 0$ which
imposes a more sophisticate form of conservation laws on such spaces with
generic ''local anisotropy'', see discussion in \cite{00vggmaf,00vstav} (this is
similar with the case when the nonholonomic constraints in Lagrange
mechanics modifies the definition of conservation laws). A very important
class of models can be elaborated when $\mathbf{\Upsilon =}diag\left[
\lambda ^{h}(\mathbf{u})g,\lambda ^{v}(\mathbf{u})\mathbf{\ }^{\bullet }g%
\right] ,$ which defines the so--called N--anholonomic Einstein spaces.

\begin{result}
\label{result2}Various classes of vacuum and nonvacuum exact solutions of (%
\ref{einsta}) para\-met\-rized by generic off--diagonal metrics, nonholonomic
vielbeins and Levi--Civita or non--Riemannian connections in Einstein and
extra dimension gravity models define explicit examples of N--anholonomic
Einstein--Cartan (in particular, Einstein) spaces.
\end{result}

Such exact solutions (for instance, with noncommutative, algebroid,
toro\-id\-al, ellipsoid, ... symmetries) have been constructed in Refs. \cite%
{00vhep2,00vs,00esv,00vncl,00vesnc,00vesalg,00vclalg,00vggmaf,00vstav}. We note that a
subclass of N--anholonomic Einstein spaces was related to generic
off--diagonal solutions in general relativity by such nonholonomic
constraints when $Ric(\widehat{\mathbf{D}})=Ric(\mathbf{\bigtriangledown })$
even $\widehat{\mathbf{D}}\neq \bigtriangledown ,$ where $\widehat{\mathbf{D}%
}$ is the canonical d--connection and $\bigtriangledown $ is the
Levi--Civita connection, see formulas (\ref{cdc}) and (\ref{cdca}) in
 section \ref{brfg} and details in Ref. \cite{00vesalg}.

A direction in modern gravity is connected to analogous gravity models when
certain gravitational effects and, for instance, black hole configurations
are modelled by optical and acoustic media, see a recent review or results
in \cite{00blv}. Following our approach on geometric unification of gravity
and Lagrange regular mechanics in terms of N--anholonomic spaces, one holds

\begin{theorem}
\label{tlr}A Lagrange (Finsler) space can be canonically modelled as an
exact solution of the Einstein equations (\ref{einsta}) on a N--anholonomic
Riemann--Cartan space if and only if the canonical N--connection $\ ^{L}%
\mathbf{N} $ ($\ ^{F}\mathbf{N}$)$\mathbf{,}$ d--metric $\ ^{L}\mathbf{g}$ ($%
\ ^{F}\mathbf{g)}$ and d--connection $\ ^{L}\widehat{\mathbf{D}}$ ($\ ^{F}%
\widehat{\mathbf{D}})$ $\ $structures defined by the corresponding
fundamental Lagrange function $L(\mathbf{x,y})$ (Finsler function $F(\mathbf{%
x,y}))$ satisfy the gravitational field equations for certain physically
reasonable sources $\mathbf{\Upsilon .}$
\end{theorem}

\begin{proof}
We sketch the idea: It can be performed in local form by considering the
Einstein tensor (\ref{enstdt}) defined by the $\ ^{L}\mathbf{N}$ ($\ ^{F}%
\mathbf{N}$) in the form (\ref{00cncl}) and $\ ^{L}\mathbf{g}$ ($\ ^{F}\mathbf{%
g)}$ in the form (\ref{00slm}) \ inducing the canonical d--connection $\ ^{L}%
\widehat{\mathbf{D}}$ ($\ ^{F}\widehat{\mathbf{D}}).$ For certain zero or
nonzero $\mathbf{\Upsilon }$, such N--anholonomic configurations may be
defined by exact solutions of the Einstein equations for a d--connection
structure. A number of explicit examples were constructed for N--anholonomic
Einstein spaces
\cite{00vhep2,00vs,00esv,00vncl,00vesnc,00vesalg,00vclalg,00vggmaf,00vstav}.
 \end{proof}

It should be noted that Theorem \ref{tlr} states the explicit conditions when
the Result \ref{result1} holds for N--anholonomic Einstein spaces.

\begin{conclusion}
Generic off--diagonal metric and vielbein structures in gra\-vi\-ty and
regular Lagrange mechanics models can be geometrized in a unified form on
N--anholono\-mic manifolds. In general, such spaces are not spin and this
presents a strong motivation for elaborating the theory of nonholonomic
gerbes and related Clifford/ spinor structures.
\end{conclusion}

Following this Conclusion, it is not surprising that a lot of gravitational
effects (black hole configurations, collapse scenaria, cosmological
anisotropies ....) can be modelled in nonlinear fluid, acoustic or optic
media.

%%%%%%%

\section{For Beginners in Riemann--Finsler Geometry}
\label{brfg}

In this section, we outline some component formulas and equations defining
the local geometry of N--anholonomic spaces, see details in Refs. \cite%
{00vggmaf,00vncl,00vesalg,00ma}. Elementary introductions on Riemann and
Finsler geometry are contained in \cite{00rsb,00sb}.

Locally, a N--connection, see Definition \ref{dnc}, is stated by its
coefficients $N_{i}^{a}(u),$
\begin{equation}
\mathbf{N}=N_{i}^{a}(u)dx^{i}\otimes \partial _{a}  \label{nclf}
\end{equation}%
where the local coordinates (in general, abstract ones both for holonomic
and nonholonomic variables) are split in the form $u=(x,y),$ or $u^{\alpha
}=\left( x^{i},y^{a}\right) ,$ where $i,j,k,\ldots =1,2,\ldots ,n$ and $%
a,b,c,\ldots =n+1,n+2,\ldots ,n+m$ when $\partial _{i}=\partial /\partial
x^{i}$ and $\partial _{a}=\partial /\partial y^{a}.$ The well known class of
linear connections consists on a particular subclass with the coefficients
being linear on $y^{a},$ i.e., $N_{i}^{a}(u)=\Gamma _{bj}^{a}(x)y^{b}.$

An explicit local calculus allows us to write the N--connection curvature (%
\ref{njht}) in the form
\begin{equation*}
\mathbf{\Omega }=\frac{1}{2}\Omega _{ij}^{a}dx^{i}\wedge dx^{j}\otimes
\partial _{a},
\end{equation*}%
with the N--connection curvature coefficients
\begin{equation}
\Omega _{ij}^{a}=\delta _{\lbrack j}N_{i]}^{a}=\delta _{j}N_{i}^{a}-\delta
_{i}N_{j}^{a}=\partial _{j}N_{i}^{a}-\partial
_{i}N_{j}^{a}+N_{i}^{b}\partial _{b}N_{j}^{a}-N_{j}^{b}\partial
_{b}N_{i}^{a}.  \label{00ncurv}
\end{equation}

Any N--connection $\mathbf{N}=N_{i}^{a}(u)$ induces a N--adapted frame
(vielbein) structure
\begin{equation}
\mathbf{e}_{\nu }=\left( e_{i}=\partial _{i}-N_{i}^{a}(u)\partial
_{a},e_{a}=\partial _{a}\right) ,  \label{00dder}
\end{equation}%
and the dual frame (coframe) structure%
\begin{equation}
\mathbf{e}^{\mu }=\left( e^{i}=dx^{i},e^{a}=dy^{a}+N_{i}^{a}(u)dx^{i}\right)
.  \label{00ddif}
\end{equation}%
The vielbeins (\ref{00ddif}) satisfy the nonholonomy (equivalently,
anholonomy) relations
\begin{equation}
\lbrack \mathbf{e}_{\alpha },\mathbf{e}_{\beta }]=\mathbf{e}_{\alpha }%
\mathbf{e}_{\beta }-\mathbf{e}_{\beta }\mathbf{e}_{\alpha }=W_{\alpha \beta
}^{\gamma }\mathbf{e}_{\gamma }  \label{00anhrel}
\end{equation}%
with (antisymmetric) nontrivial anholonomy coefficients $W_{ia}^{b}=\partial
_{a}N_{i}^{b}$ and $W_{ji}^{a}=\Omega _{ij}^{a}.$\footnote{%
One preserves a relation to our previous denotations \cite{00vfs,00vhs} if we
consider that $\mathbf{e}_{\nu }=(e_{i},e_{a})$ and $\mathbf{e}^{\mu
}=(e^{i},e^{a})$ are, respectively, the former $\delta _{\nu }=\delta
/\partial u^{\nu }=(\delta _{i},\partial _{a})$ and $\delta ^{\mu }=\delta
u^{\mu }=(d^{i},\delta ^{a})$ when emphasize that operators (\ref{00dder}) and
(\ref{00ddif}) define, correspondingly, the ``N--elongated'' partial
derivatives and differentials which are convenient for calculations on
N--anholonomic manifolds.} These formulas present a local proof of
Proposition \ref{pddv} when
\begin{equation*}
\mathbf{e}=\{\mathbf{e}_{\nu }\}=(\ e=\{e_{i}\},^{\bullet }e=\{e_{a}\})
\end{equation*}%
and
\begin{equation*}
\widetilde{\mathbf{e}}=\{\mathbf{e}^{\mu }\}=\left( \widetilde{e}%
=\{e^{i}\},\ ^{\bullet }\widetilde{e}=\{e^{a}\}\right) .
\end{equation*}

Let us consider metric structure%
\begin{equation}
\ \breve{g}=\underline{g}_{\alpha \beta }\left( u\right) du^{\alpha }\otimes
du^{\beta }  \label{00metr}
\end{equation}%
defined with respect to a local coordinate basis $du^{\alpha }=\left(
dx^{i},dy^{a}\right) $ by coefficients%
\begin{equation}
\underline{g}_{\alpha \beta }=\left[
\begin{array}{cc}
g_{ij}+N_{i}^{a}N_{j}^{b}h_{ab} & N_{j}^{e}h_{ae} \\
N_{i}^{e}h_{be} & h_{ab}%
\end{array}%
\right] .  \label{00ansatz}
\end{equation}%
Such a metric (\ref{00ansatz})\ is generic off--diagonal, i.e. it
can not be diagonalized by any coordinate transforms if $N_{i}^{a}(u)$
are any general functions. The condition (\ref{algn01}), for $X\rightarrow
e_{i}$ and $\ ^{\bullet }Y\rightarrow \ ^{\bullet }e_{a},$ transform into
\begin{equation*}
\breve{g}(e_{i},\ ^{\bullet }e_{a})=0,\mbox{ equivalently }\underline{g}%
_{ia}-N_{i}^{b}h_{ab}=0
\end{equation*}%
where $\underline{g}_{ia}$ $\doteqdot g(\partial /\partial x^{i},\partial
/\partial y^{a}),$ which allows us to define in a unique form the
coefficients $N_{i}^{b}=h^{ab}\underline{g}_{ia}$ where $h^{ab}$ is inverse
to $h_{ab}.$ We can write the metric $\breve{g}$ with ansatz (\ref{00ansatz})\
in equivalent form, as a d--metric adapted to a N--connection structure, see
Definition \ref{ddm},
\begin{equation}
\mathbf{g}=\mathbf{g}_{\alpha \beta }\left( u\right) \mathbf{e}^{\alpha
}\otimes \mathbf{e}^{\beta }=g_{ij}\left( u\right) e^{i}\otimes
e^{j}+h_{ab}\left( u\right) \ ^{\bullet }e^{a}\otimes \ ^{\bullet }e^{b},
\label{dmetr}
\end{equation}%
where $g_{ij}\doteqdot \mathbf{g}\left( e_{i},e_{j}\right) $ and $%
h_{ab}\doteqdot \mathbf{g}\left( \ ^{\bullet }e_{a},\ ^{\bullet
}e_{b}\right) $ \ and the vielbeins $\mathbf{e}_{\alpha }$ and $\mathbf{e}%
^{\alpha }$ are respectively of type (\ref{00dder}) and (\ref{00ddif}).

We can say that the metric $\ \breve{g}$ (\ref{00metr}) is equivalently
transformed into (\ref{dmetr}) \ by performing a frame (vielbein) transform
\begin{equation*}
\mathbf{e}_{\alpha }=\mathbf{e}_{\alpha }^{\ \underline{\alpha }}\partial _{%
\underline{\alpha }}\mbox{ and }\mathbf{e}_{\ }^{\beta }=\mathbf{e}_{\
\underline{\beta }}^{\beta }du^{\underline{\beta }}.
\end{equation*}%
with coefficients

\begin{eqnarray}
\mathbf{e}_{\alpha }^{\ \underline{\alpha }}(u) &=&\left[
\begin{array}{cc}
e_{i}^{\ \underline{i}}(u) & N_{i}^{b}(u)e_{b}^{\ \underline{a}}(u) \\
0 & e_{a}^{\ \underline{a}}(u)%
\end{array}%
\right] ,  \label{vt1} \\
\mathbf{e}_{\ \underline{\beta }}^{\beta }(u) &=&\left[
\begin{array}{cc}
e_{\ \underline{i}}^{i\ }(u) & -N_{k}^{b}(u)e_{\ \underline{i}}^{k\ }(u) \\
0 & e_{\ \underline{a}}^{a\ }(u)%
\end{array}%
\right] ,  \label{vt2}
\end{eqnarray}%
being linear on $N_{i}^{a}.$ We can consider that a N--anholonomic manifold $%
\mathbf{V}$ provided with metric structure $\breve{g}$ (\ref{00metr})
(equivalently, with d--metric (\ref{dmetr})) is a special type of a manifold
provided with a global splitting into conventional ``horizontal'' and
``vertical'' subspaces (\ref{00whitney}) induced by the ``off--diagonal''
terms $N_{i}^{b}(u)$ and a prescribed type of nonholonomic frame structure (%
\ref{00anhrel}).

A d--connection, see Definition \ref{00ddc}, splits into h-- and v--covariant
derivatives, $\mathbf{D}=D+\ ^{\bullet }D,$ where $D_{k}=\left(
L_{jk}^{i},L_{bk\;}^{a}\right) $ and $\ \ ^{\bullet }D_{c}=\left(
C_{jk}^{i},C_{bc}^{a}\right) $ are correspondingly introduced as h- and
v--parametrizations of (\ref{cond1}),%
\begin{equation*}
L_{jk}^{i}=\left( \mathbf{D}_{k}e_{j}\right) \rfloor e^{i},\quad
L_{bk}^{a}=\left( \mathbf{D}_{k}e_{b}\right) \rfloor
e^{a},~C_{jc}^{i}=\left( \mathbf{D}_{c}e_{j}\right) \rfloor e^{i},\quad
C_{bc}^{a}=\left( \mathbf{D}_{c}e_{b}\right) \rfloor e^{a}.
\end{equation*}%
The components $\mathbf{\Gamma }_{\ \alpha \beta }^{\gamma }=\left(
L_{jk}^{i},L_{bk}^{a},C_{jc}^{i},C_{bc}^{a}\right) $ completely define a
d--connection $\mathbf{D}$ on a N--anholonomic manifold $\mathbf{V}.$

The simplest way to perform a local covariant calculus by applying
d--connecti\-ons is to use N--adapted differential forms like $\mathbf{%
\Gamma }_{\beta }^{\alpha }=\mathbf{\Gamma }_{\beta \gamma }^{\alpha }%
\mathbf{e}^{\gamma }$ with the coefficients defined with respect to (\ref%
{00ddif}) and (\ref{00dder}). One introduces the d--connection 1--form%
\begin{equation*}
\mathbf{\Gamma }_{\ \beta }^{\alpha }=\mathbf{\Gamma }_{\ \beta \gamma
}^{\alpha }\mathbf{e}^{\gamma },
\end{equation*}%
when the N--adapted components of d-connection $\mathbf{D}_{\alpha }=(%
\mathbf{e}_{\alpha }\rfloor \mathbf{D})$ are computed following formulas
\begin{equation}
\mathbf{\Gamma }_{\ \alpha \beta }^{\gamma }\left( u\right) =\left( \mathbf{D%
}_{\alpha }\mathbf{e}_{\beta }\right) \rfloor \mathbf{e}^{\gamma },
\label{cond1}
\end{equation}%
where ''$\rfloor "$ denotes the interior product. This allows us to define
in local form the torsion (\ref{tors1}) $\mathbf{T=\{\mathcal{T}^{\alpha }\},%
}$ where
\begin{equation}
\mathcal{T}^{\alpha }\doteqdot \mathbf{De}^{\alpha }=d\mathbf{e}^{\alpha
}+\Gamma _{\ \beta }^{\alpha }\wedge \mathbf{e}^{\beta }  \label{tors}
\end{equation}%
and curvature (\ref{curv1}) $\mathbf{R}=\{\mathcal{R}_{\ \beta }^{\alpha
}\}, $ where
\begin{equation}
\mathcal{R}_{\ \beta }^{\alpha }\doteqdot \mathbf{D\Gamma }_{\beta }^{\alpha
}=d\mathbf{\Gamma }_{\beta }^{\alpha }-\Gamma _{\ \beta }^{\gamma }\wedge
\mathbf{\Gamma }_{\ \gamma }^{\alpha }.  \label{curv}
\end{equation}

The d--torsions components of a d--connection $\mathbf{D},$ see Theorem \ref%
{tht}, $\ $are computed
\begin{eqnarray}
T_{\ jk}^{i} &=&L_{\ jk}^{i}-L_{\ kj}^{i},\ T_{\ ja}^{i}=-T_{\ aj}^{i}=C_{\
ja}^{i},\ T_{\ ji}^{a}=\Omega _{\ ji}^{a},\   \notag \\
T_{\ bi}^{a} &=&T_{\ ib}^{a}=\frac{\partial N_{i}^{a}}{\partial y^{b}}-L_{\
bi}^{a},\ T_{\ bc}^{a}=C_{\ bc}^{a}-C_{\ cb}^{a}.  \label{00dtors}
\end{eqnarray}%
For instance, $T_{\ jk}^{i}$ and $T_{\ bc}^{a}$ are respectively the
coefficients of the $h(hh)$--torsion $T(X,Y)$ and $v(vv)$--torsion $\mathbf{%
\ }^{\bullet }T(\mathbf{\ }^{\bullet }X,\mathbf{\ }^{\bullet }Y).$

The Levi--Civita linear connection $\bigtriangledown =\{^{\bigtriangledown }%
\mathbf{\Gamma }_{\beta \gamma }^{\alpha }\},$ with vanishing both torsion
and nonmetricity, $\bigtriangledown \breve{g}=0,$ is not adapted to the
global splitting (\ref{00whitney}). There is a preferred, canonical
d--connection structure,$\ \widehat{\mathbf{D}}\mathbf{,}$ $\ $on
N--aholonomic manifold $\mathbf{V}$ constructed only from the metric and
N--con\-nec\-ti\-on coefficients $[g_{ij},h_{ab},N_{i}^{a}]$ and satisfying
the conditions $\widehat{\mathbf{D}}\mathbf{g}=0$ and $\widehat{T}_{\
jk}^{i}=0$ and $\widehat{T}_{\ bc}^{a}=0,$ see Theorem \ref{thcdc}. By
straightforward calculations with respect to the N--adapted bases (\ref{00ddif}%
) and (\ref{00dder}), we can verify that the connection
\begin{equation}
\widehat{\mathbf{\Gamma }}_{\beta \gamma }^{\alpha }=\ ^{\bigtriangledown }%
\mathbf{\Gamma }_{\beta \gamma }^{\alpha }+\ \widehat{\mathbf{P}}_{\beta
\gamma }^{\alpha }  \label{00cdc}
\end{equation}%
with the deformation d--tensor \footnote{$\widehat{\mathbf{P}}_{\beta \gamma
}^{\alpha }$ is a tensor field of type (1,2). As is well known, the sum of a
linear connection and a tensor field of type (1,2) is a new linear
connection.}
\begin{equation}
\widehat{\mathbf{P}}_{\beta \gamma }^{\alpha
}=(P_{jk}^{i}=0,P_{bk}^{a}=e_{b}(N_{k}^{a}),P_{jc}^{i}=-\frac{1}{2}%
g^{ik}\Omega _{\ kj}^{a}h_{ca},P_{bc}^{a}=0)  \label{cdca}
\end{equation}%
satisfies the conditions of the mentioned Theorem. It should be
noted that, in general, the components $\widehat{T}_{\ ja}^{i},\
\widehat{T}_{\ ji}^{a}$ and $\widehat{T}_{\ bi}^{a}$ are not zero.
This is an anholonomic frame (or, equivalently, off--diagonal
metric) effect. With respect to the N--adapted frames, the
coefficients
$\widehat{\mathbf{\Gamma }}_{\ \alpha \beta }^{\gamma }=\left( \widehat{L}%
_{jk}^{i},\widehat{L}_{bk}^{a},\widehat{C}_{jc}^{i},\widehat{C}%
_{bc}^{a}\right) $ are computed:
\begin{eqnarray}
\widehat{L}_{jk}^{i} &=&\frac{1}{2}g^{ir}\left(
e_{k}g_{jr}+e_{j}g_{kr}-e_{r}g_{jk}\right) ,  \label{candcon} \\
\widehat{L}_{bk}^{a} &=&e_{b}(N_{k}^{a})+\frac{1}{2}h^{ac}\left(
e_{k}h_{bc}-h_{dc}\ e_{b}N_{k}^{d}-h_{db}\ e_{c}N_{k}^{d}\right) ,  \notag \\
\widehat{C}_{jc}^{i} &=&\frac{1}{2}g^{ik}e_{c}g_{jk},\ \widehat{C}_{bc}^{a}=%
\frac{1}{2}h^{ad}\left( e_{c}h_{bd}+e_{c}h_{cd}-e_{d}h_{bc}\right) .  \notag
\end{eqnarray}

In some approaches to Finsler geometry \cite{00cbs}, one uses the so--called
Berwald d--connection $\mathbf{D}^{B}$ with the coefficients
\begin{equation}
\ ^{B}\mathbf{\Gamma }_{\ \alpha \beta }^{\gamma }=\left( \ ^{B}L_{jk}^{i}=%
\widehat{L}_{jk}^{i},\ ^{B}L_{bk}^{a}=e_{b}(N_{k}^{a}),\ ^{B}C_{jc}^{i}=0,\
^{B}C_{bc}^{a}=\widehat{C}_{bc}^{a}\right) .  \label{dbc}
\end{equation}%
This d--connection minimally extends the Levi--Civita connection (it is just
the Levi--Civita connection if the integrability conditions are satisfied,
i.e. $\Omega _{\ kj}^{a}=0,$ see (\ref{cdca})). But, in general, for this
d--connection, the metricity conditions are not satisfied, for instance $%
D_{a}g_{ij}\neq 0$ and $D_{i}h_{ab}\neq 0.$

By a straightforward d--form calculus in (\ref{curv}), we can find the
N--adapted components $\mathbf{R}_{\ \beta \gamma \delta }^{\alpha }$ of the
curvature $\mathbf{R=\{\mathcal{R}_{\ \beta }^{\alpha }\}}$ of a
d--connection $\mathbf{D},$ i.e. the d--curvatures from Theorem \ref{thr}:
\begin{eqnarray}
R_{\ hjk}^{i} &=&e_{k}L_{\ hj}^{i}-e_{j}L_{\ hk}^{i}+L_{\ hj}^{m}L_{\
mk}^{i}-L_{\ hk}^{m}L_{\ mj}^{i}-C_{\ ha}^{i}\Omega _{\ kj}^{a},  \notag \\
R_{\ bjk}^{a} &=&e_{k}L_{\ bj}^{a}-e_{j}L_{\ bk}^{a}+L_{\ bj}^{c}L_{\
ck}^{a}-L_{\ bk}^{c}L_{\ cj}^{a}-C_{\ bc}^{a}\Omega _{\ kj}^{c},  \notag \\
R_{\ jka}^{i} &=&e_{a}L_{\ jk}^{i}-D_{k}C_{\ ja}^{i}+C_{\ jb}^{i}T_{\
ka}^{b},  \label{dcurv} \\
R_{\ bka}^{c} &=&e_{a}L_{\ bk}^{c}-D_{k}C_{\ ba}^{c}+C_{\ bd}^{c}T_{\
ka}^{c},  \notag \\
R_{\ jbc}^{i} &=&e_{c}C_{\ jb}^{i}-e_{b}C_{\ jc}^{i}+C_{\ jb}^{h}C_{\
hc}^{i}-C_{\ jc}^{h}C_{\ hb}^{i},  \notag \\
R_{\ bcd}^{a} &=&e_{d}C_{\ bc}^{a}-e_{c}C_{\ bd}^{a}+C_{\ bc}^{e}C_{\
ed}^{a}-C_{\ bd}^{e}C_{\ ec}^{a}.  \notag
\end{eqnarray}

Contracting respectively the components of (\ref{dcurv}), one proves:\
The Ricci tensor $\mathbf{R}_{\alpha \beta }\doteqdot \mathbf{R}_{\ \alpha
\beta \tau }^{\tau }$ is characterized by h- v--components, i.e. d--tensors,%
\begin{equation}
R_{ij}\doteqdot R_{\ ijk}^{k},\ \ R_{ia}\doteqdot -R_{\ ika}^{k},\
R_{ai}\doteqdot R_{\ aib}^{b},\ R_{ab}\doteqdot R_{\ abc}^{c}.
\label{00dricci}
\end{equation}%
It should be noted that this tensor is not symmetric for arbitrary
d--connecti\-ons $\mathbf{D}.$

The scalar curvature of a d--connection is
\begin{equation}
\ ^{s}\mathbf{R}\doteqdot \mathbf{g}^{\alpha \beta }\mathbf{R}_{\alpha \beta
}=g^{ij}R_{ij}+h^{ab}R_{ab},  \label{00sdccurv}
\end{equation}%
defined by a sum the h-- and v--components of (\ref{00dricci}) and d--metric (%
\ref{dmetr}).

The Einstein tensor is defined and computed in standard form
\begin{equation}
\mathbf{G}_{\alpha \beta }=\mathbf{R}_{\alpha \beta }-\frac{1}{2}\mathbf{g}%
_{\alpha \beta }\ ^{s}\mathbf{R}  \label{enstdt}
\end{equation}

For a Lagrange geometry, see Definition \ref{dlg}, by straightforward
component calculations, one can be proved the fundamental results:

\begin{enumerate}
\item The Euler--Lagrange equations%
\begin{equation*}
\frac{d}{d\tau }\left( \frac{\partial L}{\partial y^{i}}\right) -\frac{%
\partial L}{\partial x^{i}}=0
\end{equation*}%
where $y^{i}=\frac{dx^{i}}{d\tau }$ for $x^{i}(\tau )$ depending on
parameter $\tau ,$ are equivalent to the ``nonlinear'' geodesic equations
\begin{equation*}
\frac{d^{2}x^{i}}{d\tau ^{2}}+2G^{i}(x^{k},\frac{dx^{j}}{d\tau })=0
\end{equation*}%
defining paths of the canonical semispray%
\begin{equation*}
S=y^{i}\frac{\partial }{\partial x^{i}}-2G^{i}(x,y)\frac{\partial }{\partial
y^{i}}
\end{equation*}%
where
\begin{equation*}
2G^{i}(x,y)=\frac{1}{2}\ ^{L}g^{ij}\left( \frac{\partial ^{2}L}{\partial
y^{i}\partial x^{k}}y^{k}-\frac{\partial L}{\partial x^{i}}\right)
\end{equation*}%
with $^{L}g^{ij}$ being inverse to (\ref{00lqf}).

\item There exists on $\widetilde{TM}$ a canonical N--connection $\ $%
\begin{equation}
\ ^{L}N_{j}^{i}=\frac{\partial G^{i}(x,y)}{\partial y^{i}}  \label{00cncl}
\end{equation}%
defined by the fundamental Lagrange function $L(x,y),$ which prescribes
nonholonomic frame structures of type (\ref{00dder}) and (\ref{00ddif}), $\ ^{L}%
\mathbf{e}_{\nu }=(e_{i},\ ^{\bullet }e_{k})$ and $\ ^{L}\mathbf{e}^{\mu
}=(e^{i},\ ^{\bullet }e^{k}).$

\item The canonical N--connection (\ref{00cncl}), defining $\ \ ^{\bullet
}e_{i},$ induces naturally an almost complex structure $\mathbf{F}:\chi (%
\widetilde{TM})\rightarrow \chi (\widetilde{TM}),$ where $\chi (\widetilde{TM%
})$ denotes the module of vector fields on $\widetilde{TM},$%
\begin{equation*}
\mathbf{F}(e_{i})=\ \ ^{\bullet }e_{i}\mbox{ and }\mathbf{F}(\ \ ^{\bullet
}e_{i})=-e_{i},
\end{equation*}%
when
\begin{equation}
\mathbf{F}=\ \ ^{\bullet }e_{i}\otimes e^{i}-e_{i}\otimes \ \ ^{\bullet
}e^{i}  \label{00acs1}
\end{equation}%
satisfies the condition $\mathbf{F\rfloor \ F=-I,}$ i. e. $F_{\ \ \beta
}^{\alpha }F_{\ \ \gamma }^{\beta }=-\delta _{\gamma }^{\alpha },$ where $%
\delta _{\gamma }^{\alpha }$ is the Kronecker symbol and ``$\mathbf{\rfloor }
$'' denotes the interior product.

\item On $\widetilde{TM},$ there is a canonical metric structure%
\begin{equation}
\ ^{L}\mathbf{g}=\ ^{L}g_{ij}(x,y)\ e^{i}\otimes e^{j}+\ ^{L}g_{ij}(x,y)\ \
\ ^{\bullet }e^{i}\otimes \ \ ^{\bullet }e^{j}  \label{00slm}
\end{equation}%
constructed as a Sasaki type lift from $M.$

\item There is also a canonical d--connection structure $^{L}\widehat{%
\mathbf{\Gamma }}_{\ \alpha \beta }^{\gamma }$ defined only by the
components of $^{L}N_{j}^{i}$ and $^{L}g_{ij},$ i.e. by the \ coefficients
of metric (\ref{00slm}) which in its turn is induced by a regular Lagrangian.
\ The values $\ ^{L}\widehat{\mathbf{\Gamma }}_{\ \alpha \beta }^{\gamma
}=(\ ^{L}\widehat{L}_{jk}^{i},\ ^{L}\widehat{C}_{bc}^{a})$ are computed just
as similar values from (\ref{candcon}). We note that on $\widetilde{TM}$
there are couples of distinguished sets of h- and v--components.
\end{enumerate}

%%%%%

\section{The Layout of the Book}

This book is organized in three Parts comprising fifteen Chapters.
Every Chapter represents a research paper, begins with an Abstract
and ends with a Bibliography.  We try to follow the original
variants of the selected
 works  but subject the text to some minimal grammar and style
 modifications  if it is necessary.

The Foreword outlines the main results on modelling locally
anisotropic and/or noncommutative structures in modern gravity and
geometric mechanics. Chapter 0 presents an Introduction to the
book: There are stated the main principles and concepts both for
the experts in differential geometry and applications and for the
beginners on Finsler and Lagrange geometry. We  discuss the main
references and results in such directions and present the
corresponding list of references.

\vskip5pt

The Part I consists of three Chapters.

Chapter 1 is devoted to modelling  Finsler--Lagrange and Hamilton--Cartan
 geometries on metric--affine spaces provided with N--connection
 structure. There are defined the Finsler-,  Lagrange-- and
 Hamilton--affine spaces and elaborated complete scheme of their
 classification in terms of N--adapted geometric structures.
 The corresponding Tables are presented in the Appendix section.

 Chapter 2 describes how Finsler--Lagrange metrics and connections
 can be extracted from the metric--affine gravity by introducing
 nonhlonomic distributions and extending the results for
 N--anholonomic manifolds. There are formulated and proved the
 main theorems on constructing exact solutions modelling such
 spacetimes in string gravity and  models with  nontrivial torsion
 and nonmetricity. Some examples of such solutions describing
 configurations with variable cosmological constants and three
 dimensional gravitational solitons propagating self--consistently
 in locally anisotropic spaces are constructed.

In Chapter 3 we construct exact solutions in metric--affine and string
gravity  with noncommutative symmetries defined by nontrivial
N--connection structures. We generalize the methods of generating
such noncommutative solutions in the commutative gauge and Einstein
gravity theories \cite{00vesnc,00vggr,00vcv,00vmur} and show that they can be
performed in  a form generalizing the solutions for the black ellipsoids
\cite{00behg,00bepert,00vesm}. We prove the stability of such locally
anisotropic black holes objects and prove that stability can
be preserved for extensions to solutions in complex gravity.

\vskip5pt

Part II provides an almost complect relief how the so--called "anholonomic
 frame method" of constructing exact solutions in gravity was
 proposed and developed. It reflects a set of nine electronic
 preprints \footnote{the exact references for these preprints
 are given as footnotes to Abstracts at the beginnings of
  Chapters 4--12} and  communications at International Conferences
 concerning developments of a series of works
 \cite{00vlodz,00vapkp,00vpdsw,00vtnut,00vsbdpl,00vs,00vswaw,00vswsdpbh,00behg,00bepert,00vesm,
00dvgrg,00dvgrg,00vesnc,00vmur,00vecalbh,00vcalnut,00vgerb,00vjfg}.

Chapter 4 reflects the results of the first work where, in four dimensional
gravity, an exact generic off--diagonal solution with ellipsoidal symmetry
was constructed. It develops the the results of \cite{00vlodz} and announces
certain preliminary results published latter in \cite{00vhs,00vapkp,00behg,00bepert}.

Chapter 5 is devoted to three dimensional (3D) black holes solutions
with generic local anisotropy. It is well known that the vacuum 3D
(pseudo) Riemannian gravity is trivial because the curvature vanishes
if the Ricci tensor is zero. The firsts nontrivial solutions were
obtained by adding a nonzero cosmological constant, a torsion field
or other contributions, for instance, from string gravity. Our idea
was to generate 3D nonholonomic configurations by deforming the frame
structure for nonholonomic distributions (with anholonomically induced
torsion). The results correct the errors from a previous preprint
(S. Vacaru, gr--qc/ 9811048) caused by testing a Maple program on
generating off--diagonal  exact solutions. In collaboration with
 E.  Gaburov and D. Gont\c ta, one were obtained all formulas in analytic form.

Chapter 6 shows a possible application of 3D nonholonomic exact
solutions in applications of geometric thermodynamics to black hole physics.
It revises a former preprint (S. Vacaru, gr--qc/ 9905053) to the case of
elliptic local anisotropies (a common work together with P. Stavrinos
and D. Gon\c ta). The results were further developed in Refs.
\cite{00vatsp,00vapkp} and partially published in monograph \cite{00vstav}.

 Chapter 7 contains a research of warped configurations with
 generic local anisotropy. Such solutions prove that the running
 hierarchies became anisotropic if generic off--diagonal metric
 terms are included into consideration.

 Chapter 8 introduces some classes of exact black ellipsoid
 solutions in brane gravity constructed by using the N--connection
 formalism. This is a common work together with E. Gaburov.
 Further developments were published in \cite{00vswsdpbh}.

 Chapter 9 elaborates a locally anisotropic models of inflational
 cosmology (a work together with D. Gon\c ta). Such models are
 defined by generic off--diagonal cosmological metrics and can be,
 for instance, with ellipsoid or toroidal symmetry \cite{00vdgrg,00dvgrg}.

 Chapter 10 reviews in detail the "anholonomic frame method" elaborated
 on the base of N--connection formalism and various methods from Finsler
 and Lagrange geometry. There are given the bulk of technical results
 used in Refs. \cite{00vhs,00vapkp,00vpdsw,00vtnut,00vsbdpl,00vs,00vswaw,00vswsdpbh} and
 emphasized the cases of four and five dimensional ellipsoid
 configurations.  The method was developed for the metric--affine
 spaces with N--connection structure (see Chapter 2) and revised for the
 solutions with noncommutative and Lie/Clifford algebroid symmetries
 (see Chapter 3 and Ref. \cite{00vesnc}).

 Chapter 11 develops the "anholonomic frame method" to the cases
 of toroidal configurations. Such solutions are not subjected to
 the restrictions of cosmic censorship criteria because, in
 general, the  nonholonomic structures contain  nontrivial torsion
 coefficients and additional  sources induced by the off--diagonal
 metric terms. Such black  tori solutions exist in five dimensional gravity and for
 nonholonomic configurations they are not restricted by black hole
 uniqueness theorems.

 Chapter 12 extends the results of Chapters 10 and 11 when
 superpositions of ellipsoid and toroidal locally anisotropic
 configurations are constructed in explicit form. There are
 discussed possible applications in modern astrophysics as
 possible topological tests of the Einstein and extra dimension
 gravity.

 \vskip5pt

 In Part III, we mop to several foundations of noncommutative
 Finsler geometry and generalizations. In three Chapters, there are
 elaborated such models following possible realizations of Finsler
 and Lagrange structures as gauge models, Riemann--Cartan or string
 models of gravity provided with N--connection structure.

 Chapter 13 extends the theory of N--connections to the case of
 projective finite modules (i.e. for noncommutative vector
 bundles). With respect to the noncommutative geometry there are
 outlined the necessary well known results from
 \cite{00conesm,00landi,00gbvf,00madore} but with the aim to introduce
 nonholonomic structures. Noncommutative Finsler--gauge theories
 are investigated. There are developed the results elaborated in
 \cite{00vggr,00vcv}.

 Chapter 14 features several fundamental constructions when
 noncommutative Finsler configurations are derived in (super)
 string gravity. We show how locally anisotropic supergravity
 theories are derived in the low energy limit and via anisotropic
 topological compactification. There are analyzed noncommutative
 locally anisotropic field interactions. A model of anisotropic
 gravity is elaborated on noncommutative D--branes. Such
 constructions are provided with explicit examples of exact solutions:
 1) black ellipsoids with noncommutative variables derived from
 string gravity; 2) 2D Finsler structures imbedded noncommutatively
  in string gravity; 3) moving soliton--black
 string configuration; 4) noncommutative anisotropic wormholes and
 strings.

 Chapter 15 deals with the construction of nonholonomic spin geometry
 from the noncommutative point of view. We define noncommutative
 nonholonomic spaces and investigate the Clifford--Lagrange
 (--Finsler) structures.  We prove that any regular fundamental Lagrange
 (Finsler) function induces a corresponding N--anholonomic spinor
  geometry and related nonholonomic Dirac operators. There are
  defined distinguished by N--connection spectral triples and
  proved the main theorems on extracting Finsler--Lagrange
  structures from noncommutative geometry.

  Finally, we note that Chapters 13--15 scan some directions for
  further developments. For instance, the nonholonmic distributions
  can be considered on Hopf structures \cite{00majid}, Lie and
  Clifford algebroids \cite{00vecalbh,00vcalnut} and
  in relation to exact solutions with noncommutative symmetries
  \cite{00vesnc}. We hope that such results will appeal to people
  both interested in noncommutative/ quantum developments of
  Finsler--Lagrange--Hamilton geometries and nonholonomic
  structures in gravity and string theory.

\section{Sources on Finsler Geometry and Applications}

We refer to the most important monographs, original articles
 and survey papers. Some of them sit in the junctions between
 different approaches and new applications. The bibliography is
 not exhaustive and reflects the authors interests and activity.
 The intend is to orient the nonspecialists on Finsler geometry,
  to emphasize some new perspectives and make a bridge
 to modern gravity and string theories and geometric mechanics.
 More specific details and discussions can be found in the references
 presented at the end of Chapters.

The first Finsler metric was considered by B. Riemann in his
famous hability thesis in 1854 \cite{00riemann}. The geometric
approach starts with the P. Finsler thesis work \cite{00fg} in
1918 and the fundamental contributions by L. Berwald \cite{00bw},
a few years latter (see historical remarks and detailed
bibliography in Refs.
\cite{00rund,00mats,00ma1984,00ma,00vmon1,00vstav}).  The first
monograph in the subject was due to E. Cartan \cite{00cart}.

The book  \cite{00rund} by H. Rund was for a long time the most
comprehensive monograph on Finsler geometry.

In the middle of 80ths of the previous century, three new
fundamental monographs stated renewed approaches and developments
of Finsler geometry and applications: 1) The monograph  by R. Miron and M.
Anastasiei \cite{00ma1984} elaborated a common approach to Finsler
and Lagrange spaces following the geometry of nonlinear
connections. Together with a set of further monographs
\cite{00ma,00mhl,00mhf,00mhss,00mhh}, it reflects the results of the
famous Romanian school on Finsler geometry and, in general, higher
order generalizations of the Finsler--Lagrange and
Cartan--Hamilton spaces.  The monograph by G. Asanov \cite{00as1}
developed an approach related to new type of Finsler gauge
symmetries and applications in relativity and field theories (the
further work of his school \cite{00ap,00as2} is related to jet
extensions and generalized nonlinear gauge symmetries). The
monograph by M. Matsumoto  \cite{00mats} reflected the style and
achievements on Finsler geometry in Japan.

Two monographs by A. Bejancu \cite{00bej} and  G. Yu. Bogoslovsky
 \cite{00bog} complete the "80ths wave" on generalizations of
 Finsler geometry related respectively to the geometry of fiber bundles and
 certain bimetric theories of gravity.

 During the last 15 years, the developments on Finsler geometry and applications
 can be conventionally distinguished into 5 main directions and applications
  (with interrelations of various sub--directions; we shall cite the works
  considered to be of  key importance and discuss the items
  to which we contributed with our publications):

\begin{enumerate}
\item  {\sf Generalized Finsler geometries with applications
in geometric mechanics and optimal control theory.} On higher
order generalizations, there were published the monographs
\cite{00mhl,00mhf,00mhss,00mhh} and, related results in optimal
control theory,  \cite{00udr1,00udr2}.

\item {\sf Fisler methods in biology, ecology, diffusion and physics.}
 It was published a series of monographs and collections of selected works like
 \cite{00aim,00amaz,00ant} (see there the main results and detailed references).

\item {\sf Nonmetric Finsler geometry, generalizations, violation
 of local Lorentz symmetry and applications}. The direction originates
 from the L. Berwald and S. S. Chern works on Finsler geometry,
 see details in monographs \cite{00cbs,00sh}. It was a fashion in the
  20-30th years of the previous century to consider possible
  applications of the Riemann--Cartan--Weyl geometry (with nontrivial torsion and
nonmetricity fields) in physics. The Berwald and Chern connections
(with various re--discovering and modifications by Rund, Moor and
others) are typical ones which do not satisfy the compatibility
conditions with the Finsler metric. Sure, they present certain
interest in differential geometry but with more sophisticate applications in physics
\cite{00castroweyl,00brandt,00gallego,00udrn}  (for instance, it is a
quite difficult problem to define spinors on spaces with
nonmetricity and to construct supersymmetric and noncommutative extensions of such
Finsler spaces). Here one should be noted that the E. Cartan
\cite{00cart} approach to Finsler geometry and a number of further
developments \cite{00ma,00vmon1,00vstav} are based on canonical metric
compatible Finsler conections and generalizations.
 In such cases, various real and complex spinor generalizations,
  supersymmetric models and noncommutative extensions are similar
  to those for the Riemann--Cartan geometry but with nonholonomic structures.  The problem
is discussed in details in Chapters 1--3 of this book.

 There are investigated physical models with violation of
 the local Lorentz symmetry \cite{00as1,00as2,00ap,00bog,00bogg}
 being  of  special interest in modern gravity
 \cite{00groj,00kost}. Some authors \cite{00will,00bek} consider
 that such Finsler spacetime and field interaction theories are
 subjected to substantial experimental restrictions but one should
 be noted that their conclusions are with respect to a restricted
 class of theories with nonmetricity and local broken Lorentz
 symmetry without a deep analysis of the N--connection structure.
 Such experimental restrictions do not hold in the metric
 compatible models when the Finsler like structures are defined,
 for instance, as exact solutions in general relativity and string
  gravity theories (see below, in point 5, some additional considerations
  and related references).

\item {\sf Super--Finsler spaces, Finsler--gauge gravity, locally
anisotropic spinors and noncommutative geometry, geometric
kinetics and stochastic processes and conservation laws.}

 A new classification of curved spaces in terms of chains of nearly
autoparallel maps (generalizing the classes of conformal
transforms and geodesic maps) and their invariants is possible
both for the (pseudo) Riemannian and generalized Lagrange spaces
and their supersymmetric generalizations  \cite{00vo,00vmon1}. There
were formulated the conservation laws on Riemann--Cartan--Weyl and
Finsler--Lagrange spaces defined by the basic equations and
invariant conditions for nearly autoparallel maps.  It was also
proven that the field equations of the Finsler--Lagrange (super)
gravity can be formulated as Yang--Mills equations in affine
(super) bundles provided with Cartan type connections and
N--adapted frame structures \cite{00vgonch,00vmon1}.

In monograph \cite{00bej}, there are summarized the A. Bejancu's
results on gauge theories on Finsler spaces and supersymmetric
models on usual manifolds but with supervector fibers and
corresponding nonlinear connection structures. The author followed
the approach to superspaces from \cite{00dewitt} but had not used
any global definition of superbundles and related nonintegrable
super--distributions. It was not possible to do that in rigorous
form without  definition of spinors in Finsler  spaces. The
references \cite{00vncsup,00vmon1,00vcv} contains a comprehensive
formulation of  the geometry of generalized super--Finsler spaces.
The approach  is developed, for instance, by the authors of Ref.
\cite{00essup}.

The idea to use spinor variables in Finsler spaces is due to Y.
Takano (1983) \cite{00t1} (see the  monograph \cite{00vstav} for
details, discussions, and references related to further
contributions by T. Ono and Y. Takano, P. Stavrinos and V. Balan,
who used 2--spinor variables but do not defined and do not proved
the existence of general Clifford structures induced by Finsler
metrics and connections; the
spinor variables constructions, in that monograph, are compared
with the S. Vacaru's approach to Finsler--spinors). The rigorous
definition of locally anisotropic spinors on Finsler and Lagrange
spaces was given in Refs. \cite{00cfs,00vfs} which was a nontrivial
task because on Finsler like spaces there are not even local
rotation symmetries. The differential geometry of spinors for
Finsler, Lagrange and Hamilton spaces and their higher order
generalizations, noncommutative extensions, and their applications
in modern physics is elaborated in Refs.
\cite{00vhs,00vmon1,00vstav,00vv,00vpdsw,00vtnut,00vcv,00vncl,00vclalg,00vcalnut,00vgerb,00vjfg}.
The geometry of generalized Clifford--Finsler spaces, the further
supersymmetric and noncommutative extensions, as well the proof
that Finsler like structures appear in low energy limits of string
theory \cite{00vstrf,00vncsup,00vmon1} demonstrate that there are not
conceptual problems for elaborating Finsler like theories in the
framework of standard models in physics.

We refer also to applications of Finsler geometry in the theory of
stochastic processes and kinetics and thermodynamics in curved
spaces \cite{00vatsp,00vapkp}. Perhaps, the original idea on Finsler
structures on phase spaces came from the A. A. Vlasov monograph
\cite{00vlasov}. The locally anisotropic processes in the language
of Finsler geometry and generalizations were investigated in
parallel by S. Vacaru  \cite{00vstoh1,00vstoh2,00vstoh3,00vstoh4,00vatsp,00vmon1} and by P.
Antonelli, T. Zastavniak and D. Hrimiuc (see details and a number
of applications in Refs. \cite{00amaz,00aim,00ant}).

\item {\sf Generalized Finsler--Lagrange structures in gravity and string
 theory, anholonomic noncommutative and algebroid configurations,
 gravitational gerbes and exact  solutions.}

There is a fundamental result for certain generic off--diagonal
metric ansatz in four and five dimensional gravity: The Einstein
equations for the so--called canonical distinguished connections
are exactly integrable and such very general solutions depends on
classes of functions depending on 2,3 and 4 variables (see details
in Chapters 2, 10 and 11 and Refs. \cite{00vmur,00vesnc,00vesalg}).
Perhaps, this is the most general method of constructing exact
solutions in gravity by using geometric methods and the
N--connection formalism. It was applied in different models of
gravity, in general, with nontrivial torsion and nonmetricity.

Additionally to the formulas and references considered in Sections
0.1 and 0.2, we note that we analyzed the Finsler structures in
explicit form in Refs. \cite{00dvgrg,00vdgrg} and that there are
formulated explicit criteria when the Finsler--Lagrange geometries
can be modelled in metric--affine, Riemann--Cartan, string and
Einstein gravity models by corresponding nonholononmic frame and
N--connection structures (see Chapters 1 and 2 and Refs.
\cite{00vesalg,00vncl}).

 Section 0.3 outlines the main results and
our publications related to modelling locally anisotropic
configurations as exact solutions in gravity and generalization to
noncommutative geometry, Lie/ Clifford algebroids and gerbes.
Finally, we cite the works \cite{00vargas,00panahi} where 'hidden
connections between general relativity and Finsler geometry' are
discussed following alternative methods.

\end{enumerate}

%\nopagebreak

%\samepage

{\small \

 }

%%%%%%%%%%%%%%%%%%%%%%%%%%%%%%%%%%%

\mainmatter

%%%%%%%%%%%%%%%%%%%%%%%

\part{Lagrange Geometry and Finsler--Affine Gravity}

\chapter[Lagrange and Finsler--Affine Gravity]
{Generalized Finsler Geometry in Einstein, String and Metric--Affine Gravity}

{\bf Abstract}
\footnote{\copyright\ S. Vacaru, Generalized Finsler Geometry in Einstein,
 String and  Metric-Affine Gravity, hep--th/ 0310132}

We develop the method of anholonomic frames with associated
nonlinear connection (in brief, N--connection) structure and show
explicitly how geometries with local anisotropy (various type of
Finsler--Lagrange--Cartan--Hamilton spaces) can be modelled on
 the metric--affine spaces. There are formulated the criteria when
such generalized Finsler metrics are effectively defined in the
Einstein, teleparallel, Riemann--Cartan and metric--affine
gravity. We argue that every generic off--diagonal metric (which
can not be diagonalized by coordinate transforms) is related to
specific N--connection configurations. We elaborate the concept
of generalized Finsler--affine geometry for spaces provided with
arbitrary N--connection, metric and linear connection structures
and characterized by gravitational field strengths, i. e. by
nontrivial N--connection curvature, Riemannian curvature, torsion
and nonmetricity. We apply an irreducible decomposition techniques
(in our case with additional N--connection splitting) and study
the dynamics of metric--affine gravity fields generating Finsler
like configurations. The classification of basic eleven classes
of metric--affine spaces with generic local anisotropy is
presented.

\vskip5pt

Pacs:\ 04.50.+h, 02.40.-k,

MSC numbers: 83D05, 83C99, 53B20, 53C60

\section{Introduction}

Brane worlds and related string and gauge theories define the paradigm of
modern physics and have generated enormous interest in higher--dimensional
spacetimes amongst particle and astrophysics theorists (see recent advances
in Refs. \cite{branes,sgr,gauge} and an outline of the gauge idea and
gravity in\ Refs. \cite{mag,ggrav}). The \ unification scheme in the
framework of string/ brane theory indicates that the classical (pseudo)
Riemannian description is not valid on all scales of interactions. It turns
out that low--energy dilaton and axi--dilaton interactions are tractable in
terms of non--Riemannian mathematical structures possessing in particular
anholonomic (super) frame [equivalently, (super) vielbein] fields \cite%
{cartan}, noncommutative geometry \cite{ncg}, quantum group structures \cite%
{majid} all containing, in general, nontrivial torsion and nonmetricity
fields. For instance, in the closest alternatives to general relativity
theory, the teleparallel gravity models \cite{tgm}, the spacetime is of
Witzenbock type with trivial curvature but nontrivial torsion. The frame or
co--frame filed (tetrad, vierbein, in four dimensions, 4D) is the basic
dynamical variable treated as the gauge potential corresponding to the group
of local translations.

Nowadays, it was established a standard point of view that a number of low
energy (super) string and particle physics interactions, at least the
nongravitational ones, are described by (super) gauge potentials interpreted
as linear connections in suitable (super) bundle spaces. The formal identity
between the geometry of fiber bundles \cite{gfb} and gauge theory
is recognized since the works \cite{gfgfb} (see a recent discussing
in connection to a unified description in of interactions in terms of
composite fiber bundles in Ref. \cite{tres}).

The geometry of fiber bundles and the moving frame method originating from
the E.\ Cartan works \cite{cartan} constitute a modern approach to the Finsler
geometry and generalizations (also suggested by E. Cartan \cite{carf} but
finally elaborated in R. Miron and M. Anastasiei works \cite{ma}), see some
earlier and recent developments in Refs. \cite{fg,rund,as,asanovinv,bog}. \
Various type of geometries with local anisotropy (Finsler, Lagrange,
Hamilton, Cartan and their generalizations, according to the terminology
proposed in \cite{ma}), are modelled on (co) vector / tangent bundles and
their higher order generalizations \cite{mat1,mhss} with different
applications in Lagrange and Hamilton mechanics or in generalized Finsler
gravity. Such constructions were defined in low energy limits of (super)
string theory \ and supergravity \cite{vst,vsup} and generalized for spinor
bundles \cite{vsp} and affine-- de Sitter frame bundles \cite{vd} provided
with nonlinear connection (in brief, N--connection) structure of first and
higher order anisotropy.

The gauge and moving frame geometric background is also presented in the
metric--affine gravity (MAG) \cite{mag}. The geometry of this theory is very
general being described by the two--forms of curvature and of torsion and
the one--form of nonmetricity treated respectively as the gravitational
field strengths for the linear connection, coframe and metric. The kinematic
scheme of MAG is well understood at present time as well certain dynamical
aspects of the vacuum configurations when the theory can be reduced to an
effective Einstein--Proca model with nontrivial torsion and nonmetricity %
\cite{oveh,obet2,hm,ghlms}. There were constructed a number of exact
solutions in MAG connecting the theory to modern string gravity and another
extra dimension generalizations \cite{esolmag,cwmag,sbhmag}. Nevertheless,
one very important aspect has not been yet considered. As a gauge theory,
the MAG can be expressed with respect to arbitrary frames and/or coframes.
So, if we introduce frames with associated N--connection structure, the MAG
should incorporate models with generic local anisotropy (Finsler like ones
and their generalizations) which are distinguished by certain prescriptions
for anholonomic frame transforms, N--connection coefficients and metric and
linear connection structures adapted to such anholonomic configurations.
Roughly speaking, the MAG contains the bulk of known generalized Finsler
geometries which can be modelled on metric--affine spaces by defining
splitting on subspaces like on (co) vector/ tangent bundles and considering
certain anholonomically constrained moving frame dynamics and associated
N--connection geometry.

Such metric--affine spaces with local anisotropy are enabled with generic
off--diagonal metrics which can not be diagonalized by any coordinate
transforms. The off--diagonal coefficients can be mapped into the components
of a specific class of anholonomic frames, defining also the coefficients of
the N--connection structure.\ It is possible to redefine equivalently all
geometrical values like tensors, spinors and connections with respect to
 N--adapted anholonomic bases. If the N--connection, metric and linear
connections are chosen for an explicit type of Finsler geometry, such a
geometric structure is modelled on a metric--affine space (we claim that a
Finsler--affine geometry is constructed). The point is to find explicitly by
what type of frames and connections a locally anisotropic structure can be
modelled by exact solutions in the framework of MAG. Such constructions can
be performed in the Einstein--Proca sector of the MAG gravity and they can
be defined even in general relativity theory (see the partners of this paper
with field equations and exact solutions in MAG modelling Finsler like
metrics and generalizations \cite{exsolmag}).

Within the framework of moving frame method \cite{cartan}, we
investigated in a series of works \cite{v1,v2,ncgg,vncggf} the
conditions when various type of metrics with noncommutative
symmetry and/or local anisotropy can be effectively modelled by
anholonomic frames on (pseudo) Riemannian and Riemann--Cartan
spaces \cite{rcg}. We constructed explicit classes of such exact
solutions in general relativity theory and extra dimension gravity
models. They are parametrized by generic off--diagonal metrics
which can not diagonalized by any coordinate transforms but only
by anholonomic frame transforms. The new classes of solutions
describe static black ellipsoid objects, locally anistoropic
configurations with toroidal and/ or ellipsoidal symmetries,
wormholes/ flux tubes and Taub-NUT metrics with polarizied
constants and  warped spinor--soliton--dilaton configurations. For
certain conditions, some classes of such solutions preserve the
four dimensional (4D) local Lorentz symmetry.

Our ongoing effort is to model different classes of geometries following a
general approach to the geometry of (co) vector/tangent bundles and
affine--de Sitter frame bundles \cite{vd} and superbundles \cite{vsup} and
or anisotropic spinor spaces \cite{vsp} provided with N--connection
structures. The basic geometric objects on such spaces are defined by
proper classes of anholonomic frames and associated N--connections and
correspondingly adapted metric and linear connections. There are examples
when certain Finsler like configurations are modelled by some exact
solutions in Einstein or Einstein--Cartan gravity and, inversely (the
outgoing effort), by using the almost Hermitian formulation \cite%
{ma,mhss,vsp} of Lagrange/Hamilton and Finsler/Cartan geometry, we can
consider Einstein and gauge gravity models defined on tangen/cotangent and
vector/covector bundles. Recently, there were also obtained some explicit
results demonstrating that the anholonomic frames geometry has a natural
connection to noncommutative geometry in string/M--theory and noncommutative
gauge models of gravity \cite{ncgg,vncggf} (on existing approaches to
noncommutative geometry and gravity we cite Refs. \cite{ncg}).

We consider torsion fields induced by anholonomic vielbein transforms when
the theory can be extended to a gauge \cite{ggrav}, metric--affine \cite{mag}%
,  a more particular Riemann--Cartan case \cite{rcg}, or to string gravity
with $B$--field \cite{sgr}. We are also interested to define the conditions
when an exact solution possesses hidden noncommutative symmetries, induced
torsion and/or locally anisotropic configurations constructed, for instance,
in the framework of the Einstein theory. This direction of investigation
develops the results obtained in Refs. \cite{v2} and should be distinguished
from our previous works on the geometry of Clifford and spinor structures in
generalized Finsler and Lagrange--Hamilton spacetimes \cite{vsp}. Here we
emphasize that the works \cite{v1,v2,ncgg,vncggf,vsp} were elaborated
following general methods of the geometry of anholonomic frames with
associated N--connections in vector (super) bundles \cite{ma,mhss,vsup}. The
concept of N--connection was proposed in Finsler geometry \cite%
{fg,as,asanovinv,bog,rund,carf}. As a set of coefficients it was firstly
present the E. Cartan's monograph\cite{carf} and then was elaborated in a
more explicit form by A. Kawaguchi \cite{kaw}. It was proven that the
N--connection structures can be defined also on (pseudo) Riemannian spaces
and certain methods work effectively in constructing exact solutions in
Einstein gravity \cite{vsp,v1,v2}.

In order to avoid possible terminology ambiguities, we note that for us the
definition of N--connection is that proposed in global form by W. Barthel in
1963 \cite{barthel} when a N--connection is defined as an exact sequence
related to a corresponding Whitney sum of the vertical and horizontal
subbundles, for instance, \ in a tangent vector bundle. \footnote{%
Instead of a vector bundle we can consider a tangent bundle, or
cotangent/covector ones, or even general manifolds of necessary smooth class
with adapted definitions of global sums of horizonal and vertical subspaces.
The geometry of N--connections is investigated in details in Refs. \cite%
{vilms,ma,vsup,vsp,vd} for various type of spaces.} This concept is
different from that accepted in Ref. \cite{tm} were the term 'nonlinear
connection' is used for tetrads as N--connections which do not transform
inhomogeneously under local frame rotations. That approach invokes nonlinear
realizations of the local spacetime group (see also an early model of gauge
gravity with nonlinear gauge group realizations \cite{ts} and its extensions
to Finsler like \cite{vd} or noncommutative gauge gravity theories \cite%
{ncgg}).

In summary, the aim of the present work is to develop a unified scheme of
anholonomic frames with associated N--connection structure for a large
number of gauge and gravity models (in general, with locally isotropic and
anisotropic interactions and various torsion and nonmetricity contributions)
and effective generalized Finsler--Weyl--Riemann--Cartan geometries derived
from MAG. We elaborate a detailed classification of such spaces with
nontrivial N--connection geometry. The unified scheme and classification
were inspired by a number of exact solutions parametrized by generic
off--diagonal metrics and anholonomic frames in Einstein, Einstein--Cartan
and string gravity. The resulting formalism admits inclusion of locally
anisotropic spinor interactions and extensions to noncommutative geometry
and string/ brane gravity \cite{vst,vsup,v1,v2,ncgg,vncggf}. Thus, the
geometry of metric--affine spaces enabled with an additional N--connection
structure is sufficient not only to model the bulk of physically important
non--Riemannian geometries on (pseudo) Riemannian spaces but also states the
conditions when effective spaces with generic anisotropy can be derived as
exact solutions of gravitational and matter field equations. In the present
work we pay attention  to the geometrical (pre--dynamical) aspects of the
generalized Finsler--affine gravity which constitute a theoretical
background for constructing a number of exact solutions in MAG in the
partner papers \cite{exsolmag}.

The article is organized as follows. We begin, in Sec. 2, with a review of
the main concepts from the metric--affine geometry and the geometry of
anholonomic frames with associated N--connections. We introduce the basic
definitions and formulate and prove the main theorems for the N--connection,
linear connection and metric structures on metric--affine spaces and derive
the formulas for torsion and curvature distinguished by N--connections.
Next, in Sec. 3, we state the main properties of the linear and nonlinear
connections modelling Finsler spaces and their generalizations and consider
how the N--connection structure can be derived from a generic off--diagonal
metric in a metric--affine space. Section 4 is devoted to the definition and
investigation of generalized Finsler--affine spaces. We illustrate how by
corresponding parametrizations of the off--diagonal metrics, anholonomic
frames, N--connections and distinguished connections every type of
generalized Finsler--Lagrange--Cartan--Hamilton geometry can be modelled in
the metric--affine gravity or any its restrictions to the Einstein--Cartan
and general relativity theory. In Sec. 5, we conclude the results and point
out how the synthesis of the Einstein, MAG and generalized Finsler gravity
models can be realized and connected to the modern string gravity. In
Appendix we elaborate a detailed classification of eleven classes of spaces
with generic local anisotropy (i. e. possessing nontrivial N--connection
structure) and various types of curvature, torsion and nonmetricity
distinguished by N--connections.

Our basic notations and conventions combine those from Refs. \cite%
{mag,ma,v1,v2} and contain an interference of traditions from MAG and
generalized Finsler geometry. The spacetime is modelled as a manifold $V^{n+m}
$ of necessary smoothly class of dimension $n+m.$ The Greek indices $\alpha
,\beta ,...$ can split into subclasses like $\alpha =\left( i,a\right) ,$ $%
\beta =\left( j,b\right) ...$ where the Latin indices from the middle of the
alphabet, $i,j,k,...$ run values $1,2,...n$ and the Latin indices from the
beginning of the alphabet, $\ a,b,c,...$ run values $n+1,n+2,$ ..., $n+m.$
We follow the Penrose convention on abstract indices \cite{pen} and use
underlined indices like $\underline{\alpha }=\left( \underline{i},\underline{%
a}\right) ,$ for decompositions with respect to coordinate frames. The
notations for connections $\Gamma _{\ \beta \gamma }^{\alpha },$ metrics $%
g_{\alpha \beta }$ and frames $e_{\alpha }$ and coframes $\vartheta ^{\beta
},$ or other geometrical and physical objects, are the standard ones from
MAG if a nonlinear connection (N--connection) structure is not emphasized on
the spacetime. If a N--connection and corresponding anholonomic frame
structure are prescribed, we use ''boldfaced'' symbols with possible
splitting of the objects and indices like $\mathbf{V}^{n+m},$ $\mathbf{%
\Gamma }_{\ \beta \gamma }^{\alpha }=\left(
L_{jk}^{i},L_{bk}^{a},C_{jc}^{i},C_{bc}^{a}\right) ,$\textbf{\ }$\mathbf{g}%
_{\alpha \beta }=\left( g_{ij},h_{ab}\right) ,$ $\mathbf{e}_{\alpha }=\left(
e_{i},e_{a}\right) ,$ ...being distinguished by N--connection (in brief, we
use the terms d--objects, d--tensor, d--connection in order to say that they
are for a metric--affine space modelling a generalized Finsler, or another
type, anholonomic frame geometry). The symbol ''\ $\doteqdot $'' will be
used is some formulas which state that the relation is introduced ''by
definition'' and the end of proofs will be stated by symbol $\blacksquare .$

\section{Metric--Affine Spaces and Nonlinear Connecti\-ons}

We outline the geometry of anholonomic frames with associated nonlinear
connections (in brief, N--connections) in metric--affine spaces which in
this work are necessary smooth class manifolds, or (co) vector/ tangent
bundles provided  with, in general, independent nonlinear and linear
connections and metrics, and correspondingly derived strengths like
N--connection curvature, Riemannian curvature, torsion and nonmetricity. The
geometric formalism will be applied in the next sections where we shall
prove that every class of (pseudo) Riemannian, Kaluza--Klein,
Einstein--Cartan, metric--affine and generalized Lagrange--Finsler and
Hamilton--Cartan spaces is characterized by corresponding N--connection,
metric and linear connection structures.

\subsection{Linear connections, metrics and anholonomic frames}

\label{standres}We briefly review the standard results on linear connections
and metrics (and related formulas for torsions, curvatures, Ricci and
Einstein tensors and Bianchi identities) defined with respect to arbitrary
anholonomic bases in order to fix a necessary reference which will be
compared with generalized Finsler--affine structures we are going to propose
in the next sections for spaces provided with N--connection. The results are
outlined in a form with conventional splitting into horizontal and vertical
subspaces and sub--indices. We follow the Ref. \cite{stw} but we use Greek
indices and denote a covariant derivative by $D$ preserving the symbol $%
\bigtriangledown $ for the Levi--Civita (metric and torsionless) connection.
Similar formulas can be found, for instance. in Ref. \cite{mtw}.

Let $V^{n+m}$ be a $\left( {n+m}\right) $--dimensional underlying manifold
of necessary smooth class and denote by $TV^{n+m}$ the corresponding tangent
bundle. The local coordinates on $V^{n+m},u=\{u^{\underline{\alpha }}=\left(
x^{\underline{i}},y^{\underline{a}}\right) \}$ conventionally split into two
respective subgroups of ''horizontal'' coordinates (in brief,
h--coordinates), $x=(x^{\underline{i}}),$ and ''vertical'' coordinates
(v--coordinates), $y=\left( y^{\underline{a}}\right) ,$ with respective
indices running the values $\underline{i},\underline{j},...=1,2,...,n$ and $%
\underline{a},\underline{b},...=n+1,n+2,...,n+m.$ The splitting of
coordinates is treated as a formal labelling if any fiber and/or the
N--connection structures are not defined. Such a splitting of abstract
coordinates $u^{\alpha }=(x^{i},y^{a})$ may be considered, for instance, for
a general (pseudo) Riemannian manifold with $x^{i}$ being some 'holonomic''
variables (unconstrained) and $y^{a}$ being ''anholonomic'' variables
(subjected to some constraints), or in order to parametrize locally a vector
bundle $\left( E,\mu ,F,M\right) $ defined by an injective surjection~$\mu
:E\rightarrow M$ from the total space $E$ to the base space $M$ of dimension
$\dim M=n,$ with $F$ being the typical vector space of dimension $\dim F=m.$
For our purposes, we consider that both $M$ and $F$ can be, in general,
provided with metric structures of arbitrary signatures. On vector bundles,
the values $\ x=(x^{i})$ are coordinates on the base and $y=(y^{a})$ are
coordinates in the fiber. If $\dim M=\dim F,$ the vector bundle $E$
transforms into the tangent bundle $TM.$ The same conventional coordinate
notation $u^{\alpha }=(x^{i},y^{a}\rightarrow p_{a})$ can be used for a dual
vector bundle $\left( E,\mu ,F^{\ast },M\right) $ with the typical fiber $%
F^{\ast }$ being a covector space (of 1-forms) dual to $F,$ where $p_{a}$
are local (dual) coordinates. For simplicity, we shall label $y^{a}$ as
general coordinates even for dual spaces if this will not result in
ambiguities. In general, our geometric constructions will be elaborated for
a manifold $V^{n+m}$ (a general metric--affine spaces) with some additional
geometric structures and fibrations to be stated or modelled latter (for
generalized Finsler geometries) on spacetimes under consideration.

At each point $p\in V^{n+m},$ there are defined basis vectors (local frames,
vielbeins) $e_{\alpha }=A_{\alpha }^{\ \underline{\alpha }}(u)\partial _{%
\underline{\alpha }}\in TV^{n+m},$ with $\partial _{\underline{\alpha }%
}=\partial /\partial u^{\underline{\alpha }}$ being tangent vectors to the
local coordinate lines $u^{\underline{\alpha }}=u^{\underline{\alpha }}(\tau
)$ with parameter $\tau .$ In every point $p,$ there is also a dual basis $%
\vartheta ^{\beta }=A_{\ \underline{\beta }}^{\beta }(u)du^{\underline{\beta
}}$ with $du^{\underline{\beta }}$ considered as coordinate one forms. The
duality conditions can be written in abstract form by using the interior
product $\rfloor ,$ $\ e_{\alpha }\rfloor \vartheta ^{\beta }=\delta
_{\alpha }^{\beta },$ or in coordinate form $A_{\alpha }^{\ \underline{%
\alpha }}A_{\ \underline{\alpha }}^{\beta }=\delta _{\alpha }^{\beta },$
where the Einstein rule of summation on index $\underline{\alpha }$ is
considered, $\delta _{\alpha }^{\beta }$ is the Kronecker symbol. The ''not
underlined'' indices $\alpha ,\beta ,...,$ or $i,j,...$ and $a,b,...$ are
treated as abstract labels (as suggested by R. Penrose). We shall underline
the coordinate indices only in the cases when it will be necessary to
distinguish them from the abstract ones.

Any vector and 1--form fields, for instance, $X$ and, respectively, $%
\widetilde{Y}$ on $V^{n+m}$ are decomposed in h-- and v--irreducible
components,
\begin{equation*}
X=X^{\alpha }e_{\alpha }=X^{i}e_{i}+X^{a}e_{a}=X^{\underline{\alpha }%
}\partial _{\underline{\alpha }}=X^{\underline{i}}\partial _{\underline{i}%
}+X^{\underline{a}}\partial _{\underline{a}}
\end{equation*}%
and
\begin{equation*}
\widetilde{Y}=\widetilde{Y}_{\alpha }\vartheta ^{\alpha }=\widetilde{Y}%
_{i}\vartheta ^{i}+\widetilde{Y}_{a}\vartheta ^{a}=\widetilde{Y}_{\underline{%
\alpha }}du^{\underline{\alpha }}=\widetilde{Y}_{\underline{i}}dx^{%
\underline{i}}+\widetilde{Y}_{\underline{a}}dy^{\underline{a}}.
\end{equation*}%
We shall omit labels like ''$\widetilde{}"$ for forms if this will not
result in ambiguities.

\begin{definition}
\label{deflcon}A linear (affine) connection $D$ on $V^{n+m}$ is a linear map
(operator) sending every pair of smooth vector fields $\left( X,Y\right) $
to a vector field $D_{X}Y$ such that
\begin{equation*}
D_{X}\left( sY+Z\right) =sD_{X}Y+D_{X}Z
\end{equation*}%
for any scalar $s=const$ and for any scalar function $f\left( u^{\alpha
}\right) ,$
\begin{equation*}
D_{X}\left( fY\right) =fD_{X}Y+\left( Xf\right) Y\mbox{ and
}D_{X}f=Xf.
\end{equation*}
\end{definition}

$D_{X}Y$ is called the covariant derivative of $Y$ with respect to $X$ (this
is not a tensor). But we can always define a tensor $DY:$ $X\rightarrow
D_{X}Y.$ The value $DY$ is a $\left( 1,1\right) $ tensor field and called
the covariant derivative of $Y.$

With respect to a local basis $e_{\alpha },$ we can define the scalars $%
\Gamma _{\ \beta \gamma }^{\alpha },$ called the components of the linear
connection $D,$ such that
\begin{equation*}
D_{\alpha }e_{\beta }=\Gamma _{\ \beta \alpha }^{\gamma }e_{\gamma }%
\mbox{
and }D_{\alpha }\vartheta ^{\beta }=-\Gamma _{\ \gamma \alpha }^{\beta
}\vartheta ^{\gamma }
\end{equation*}%
were, by definition, $D_{\alpha }\doteqdot D_{e_{\alpha }}$ and because $%
e_{\beta }\vartheta ^{\beta }=const.$

We can decompose
\begin{equation}
D_{X}Y=\left( D_{X}Y\right) ^{\beta }e_{\beta }=\left[ e_{\alpha }(Y^{\beta
})+\Gamma _{\ \gamma \alpha }^{\beta }\vartheta ^{\gamma }\right] e_{\beta
}\doteqdot Y_{\ ;\alpha }^{\beta }X^{\alpha }  \label{covderrul}
\end{equation}%
where $Y_{\ ;\alpha }^{\beta }$ are the components of the tensor $DY.$

It is a trivial proof that any change of basis (vielbein transform), $%
e_{\alpha ^{\prime }}=B_{\alpha ^{\prime }}^{\ \alpha }e_{\alpha },$ with
inverse $B_{\ \alpha }^{\alpha ^{\prime }},$ results in a corresponding
(nontensor) rule of transformation of the components of the linear
connection,%
\begin{equation}
\Gamma _{\ \beta ^{\prime }\gamma ^{\prime }}^{\alpha ^{\prime }}=B_{\
\alpha }^{\alpha ^{\prime }}\left[ B_{\beta ^{\prime }}^{\ \beta }B_{\gamma
^{\prime }}^{\ \gamma }\Gamma _{\ \beta \gamma }^{\alpha }+B_{\gamma
^{\prime }}^{\ \gamma }e_{\gamma }\left( B_{\beta ^{\prime }}^{\ \alpha
}\right) \right] .  \label{lcontr}
\end{equation}

\begin{definition}
A local basis $e_{\beta }$ is anhlonomic (nonholonomic) if there are
satisfied the conditions%
\begin{equation}
e_{\alpha }e_{\beta }-e_{\beta }e_{\alpha }=w_{\alpha \beta }^{\gamma
}e_{\gamma }  \label{1anh}
\end{equation}%
for certain nontrivial anholonomy coefficients $w_{\alpha \beta }^{\gamma
}=w_{\alpha \beta }^{\gamma }(u^{\tau }).$ A such basis is holonomic if $%
w_{\alpha \beta }^{\gamma }\doteqdot 0.$
\end{definition}

For instance, any coordinate basis $\partial _{\alpha }$ is holonomic. Any
holonomic basis can be transformed into a coordinate one by certain
coordinate transforms.

\begin{definition}
\label{deftors}The torsion tensor is a tensor field $\mathcal{T}$ defined by
\begin{equation}
\mathcal{T}\left( X,Y\right) =D_{X}Y-D_{Y}X-[X,Y],  \label{torsa}
\end{equation}%
where $[X,Y]=XY-YX,$ for any smooth vector fields $X$ and $Y.$
\end{definition}

The components $T_{\ \alpha \beta }^{\gamma }$ of a torsion $\mathcal{T}$
with respect to a basis $e_{\alpha }$ are computed by introducing $%
X=e_{\alpha }$ and $Y=e_{\beta }$ in (\ref{torsa}),%
\begin{equation*}
\mathcal{T}\ \left( e_{\alpha },e_{\beta }\right) =D_{\alpha }e_{\beta
}-D_{\beta }e_{\alpha }-[e_{\alpha },e_{\beta }]=T_{\ \alpha \beta }^{\gamma
}e_{\gamma }
\end{equation*}%
where
\begin{equation}
T_{\ \alpha \beta }^{\gamma }=\Gamma _{\ \beta \alpha }^{\gamma }-\Gamma _{\
\alpha \beta }^{\gamma }-w_{\alpha \beta }^{\gamma }.  \label{torsac}
\end{equation}%
We note that with respect to anholonomic frames the coefficients of
anholonomy $w_{\alpha \beta }^{\gamma }$ are contained in the formula for
the torsion coefficients (so any anholonomy induces a specific torsion).
\begin{definition}
\label{defcurv}The Riemann curvature tensor $\mathcal{R}$\ is defined as a
tensor field
\begin{equation}
\mathcal{R}\left( X,Y\right) Z=D_{Y}D_{X}Z-D_{X}D_{Y}Z+D_{[X,Y]}Z.
\label{curva}
\end{equation}
\end{definition}

We can compute the components $R_{\ \beta \gamma \tau }^{\alpha }$of
curvature $\mathcal{R}$, with respect to a basis $e_{\alpha }$ are computed
by introducing $X=e_{\gamma },Y=e_{\tau }$, $Z=e_{\beta }$ in (\ref{curva}).
One obtains
\begin{equation*}
\mathcal{R}\left( e_{\gamma },e_{\tau }\right) e_{\beta }=R_{\ \beta \gamma
\tau }^{\alpha }e_{\alpha }
\end{equation*}%
where
\begin{equation}
R_{\ \beta \gamma \tau }^{\alpha }=e_{\tau }\left( \Gamma _{\ \beta \gamma
}^{\alpha }\right) -e_{\gamma }\left( \Gamma _{\ \beta \tau }^{\alpha
}\right) +\Gamma _{\ \beta \gamma }^{\nu }\Gamma _{\ \nu \tau }^{\alpha
}-\Gamma _{\ \beta \tau }^{\nu }\Gamma _{\ \nu \gamma }^{\alpha }+w_{\gamma
\tau }^{\nu }\Gamma _{\ \beta \nu }^{\alpha }.  \label{curvc}
\end{equation}%
We emphasize that the anholonomy and vielbein coefficients are contained in
the formula for the curvature components (\ref{curva}). With respect to
coordinate frames, $e_{\tau }=\partial _{\tau },$ with $w_{\gamma \tau
}^{\nu }=0,$ we have the usual coordinate formula.
\begin{definition}
The Ricci tensor $\mathcal{R}i$ is a tensor field obtained by contracting
the Riemann tensor,%
\begin{equation}
R_{\ \beta \tau }=R_{\ \beta \tau \alpha }^{\alpha }.  \label{rt}
\end{equation}
\end{definition}

We note that for a general affine (linear) connection the Ricci tensor is
not symmetric $R_{\ \beta \tau }\doteqdot R_{\ \tau \beta }.$
\begin{definition}
\label{defm}A metric tensor is a $\left( 0,2\right) $ symmetric tensor field
\begin{equation*}
\mathit{g}=g_{\alpha \beta }(u^{\gamma })\vartheta ^{\alpha }\otimes
\vartheta ^{\beta }
\end{equation*}%
defining the quadratic (length) linear element,%
\begin{equation*}
ds^{2}=g_{\alpha \beta }(u^{\gamma })\vartheta ^{\alpha }\vartheta ^{\beta
}=g_{\underline{\alpha }\underline{\beta }}(u^{\underline{\gamma }})du^{%
\underline{\alpha }}du^{\underline{\beta }}.
\end{equation*}
\end{definition}

For physical applications, we consider spaces with local Minkowski
signature, when locally, in a point \ $u_{0}^{\underline{\gamma }},$ the
diagonalized metric is $g_{\underline{\alpha }\underline{\beta }}(u_{0}^{%
\underline{\gamma }})=\eta _{\underline{\alpha }\underline{\beta }}=\left(
1,-1,-1,...\right) $ or, for our further convenience, we shall use metrics
with the local diagonal ansatz being defined by any permutation of this
order.
\begin{theorem}
\label{tmetricity}If a manifold $V^{n+m}$ is enabled with a metric structure
$\mathit{g,}$ then there is a unique torsionless connection, the
Levi--Civita connection $D=\bigtriangledown ,$ satisfying the metricity
condition
\begin{equation}
\bigtriangledown \mathit{g}=0.  \label{lcmc}
\end{equation}
\end{theorem}

The proof, as an explicit construction, is given in Ref. \cite{stw}. Here we
present the formulas for the components $\Gamma _{\bigtriangledown \ \beta
\tau }^{\alpha }$ of the connection $\bigtriangledown ,$ computed with
respect to a basis $e_{\tau },$%
\begin{eqnarray}
\Gamma _{\bigtriangledown \ \alpha \beta \gamma } &=&g\left( e_{\alpha
},\bigtriangledown _{\gamma }e_{\beta }\right) =g_{\alpha \tau }\Gamma
_{\bigtriangledown \ \alpha \beta }^{\tau }  \label{lccoef} \\
&=&\frac{1}{2}\left[ e_{\beta }\left( g_{\alpha \gamma }\right) +e_{\gamma
}\left( g_{\beta \alpha }\right) -e_{\alpha }\left( g_{\gamma \beta }\right)
+w_{\ \gamma \beta }^{\tau }g_{\alpha \tau }+w_{\ \alpha \gamma }^{\tau
}g_{\beta \tau }-w_{\ \beta \gamma }^{\tau }g_{\alpha \tau }\right] .  \notag
\end{eqnarray}%
By straightforward calculations , we can check that%
\begin{equation*}
\bigtriangledown _{\alpha }g_{\beta \gamma }=e_{\alpha }\left( g_{\beta
_{\gamma }}\right) -\Gamma _{\bigtriangledown \ \beta \alpha }^{\tau
}g_{\tau _{\gamma }}-\Gamma _{\bigtriangledown \ \gamma \alpha }^{\tau
}g_{\beta \tau }\equiv 0
\end{equation*}%
and, using the formula (\ref{torsac}),
\begin{equation*}
T_{\bigtriangledown \ \alpha \beta }^{\gamma }=\Gamma _{\bigtriangledown \
\beta \alpha }^{\gamma }-\Gamma _{\ \bigtriangledown \alpha \beta }^{\gamma
}-w_{\alpha \beta }^{\gamma }\equiv 0.
\end{equation*}%
We emphasize that the vielbein and anholonomy coefficients are contained in
the formulas for the components of the Levi--Civita connection $\Gamma
_{\bigtriangledown \ \alpha \beta }^{\tau }$ (\ref{lccoef}) given with
respect to an anholonomic basis $e_{\alpha }.$ The torsion of this
connection, by definition, vanishes with respect to all bases, anholonomic
or holonomic ones. With respect to a coordinate base $\partial _{\alpha }$,
the components $\Gamma _{\bigtriangledown \ \alpha \beta \gamma }$ (\ref%
{lccoef}) transforms into the so--called 1-st type Christoffel symbols%
\begin{equation}
\Gamma _{\alpha \beta \gamma }^{\bigtriangledown }=\Gamma _{\alpha \beta
\gamma }^{\{\}}=\{\alpha \beta \gamma \}=\frac{1}{2}\left( \partial _{\beta
}g_{\alpha \gamma }+\partial _{\gamma }g_{\beta \alpha }-\partial _{\alpha
}g_{\gamma \beta }\right) .  \label{christ}
\end{equation}

If a space $V^{n+m}$ posses a metric tensor, we can use $g_{\alpha \beta }$
and the inverse values $g^{\alpha \beta }$ for lowering and upping indices
as well to contract tensor objects.
\begin{definition}
\label{defrset}{\ } \newline
a) The Ricci scalar $R$ is defined
\begin{equation*}
R\doteqdot g^{\alpha \beta }R_{\alpha \beta },
\end{equation*}%
where $R_{\alpha \beta }$ is the Ricci tensor (\ref{rt}).\newline
b) The Einstein tensor $\mathcal{G}$ has the coefficients
\begin{equation*}
G_{\alpha \beta }\doteqdot R_{\alpha \beta }-\frac{1}{2}Rg_{\alpha \beta },
\end{equation*}%
with respect to any anholonomic or anholonomic frame $e_{\alpha }.$\
\end{definition}

We note that $G_{\alpha \beta }$ and $R_{\alpha \beta }$ are symmetric only
for the Levi--Civita connection $\bigtriangledown $ and that $%
\bigtriangledown _{\alpha }G^{\alpha \beta }=0.$

It should be emphasized that for any general affine connection $D$ and
metric $\mathit{g}$ structures the metric compatibility conditions (\ref%
{lcmc}) are not satisfied.
\begin{definition}
\label{dnmf}The nonmetricity field
\begin{equation*}
\ \mathcal{Q}=Q_{\alpha \beta }\ \vartheta ^{\alpha }\otimes \vartheta
^{\beta }
\end{equation*}%
on a space $V^{n+m}$ is defined by a tensor field with the coefficients
\begin{equation}
Q_{\gamma \alpha \beta }\doteqdot -D_{\gamma }g_{\alpha \beta }  \label{nmfa}
\end{equation}%
where the covariant derivative $D$ is defined by a linear connection $\ $%
1--form $\Gamma _{\ \alpha }^{\gamma }=\Gamma _{\ \alpha \beta }^{\gamma
}\vartheta ^{\beta }.$
\end{definition}

In result, we can generalize the concept of (pseudo) Riemann space [defined
only by a locally (pseudo) Euclidean metric inducing the Levi--Civita
connection with vanishing torsion] and Riemann--Cartan space [defined by any
independent metric and linear connection with nontrivial torsion but with
vanishing nonmetricity] (see details in Refs. \cite{mag,rcg}):
\begin{definition}
\label{defmas} A metric--affine space is a manifold of necessary smooth
class provided with independent linear connection and metric structures. In
general, such spaces posses nontrivial curvature, torsion and nonmetricity
(called strength fields).
\end{definition}

We can extend the geometric formalism in order to include into consideration
the Finsler spaces and their generalizations. This is possible by
introducing an additional fundamental geometric object called the
N--connection.

\subsection{Anholonomic frames and associated N--connections}

Let us define the concept of nonlinear connection on a manifold $V^{n+m}.$
\footnote{see Refs. \cite{barthel,ma} for original results and constructions
on vector and tangent bundles. }\
We denote by $\pi ^{T}:TV^{n+m}\rightarrow TV^{n}$ $\ $the differential
of the map $\pi :V^{n+m}\rightarrow V^{n}$ defined as a fiber--preserving
morphism of the tangent bundle $\left( TV^{n+m},\tau _{E},V^{n}\right) $ to $%
V^{n+m}$ and of tangent bundle $\left( TV^{n},\tau ,V^{n}\right) .$ The
kernel of the morphism $\pi ^{T}$ is a vector subbundle of the vector bundle
$\left( TV^{n+m},\tau _{E},V^{n+m}\right) .$ This kernel is denoted $\left(
vV^{n+m},\tau _{V},V^{n+m}\right) $ and called the vertical subbundle over $%
V^{n+m}.$ We denote the inclusion mapping \ by $i:vV^{n+m}\rightarrow
TV^{n+m}$ when the local coordinates of a point $u\in V^{n+m}$ are written $%
u^{\alpha }=\left( x^{i},y^{a}\right) ,$ where the values of indices are $%
i,j,k,...=1,2,...,n$ and $a,b,c,...=n+1,n+2,...,n+m.$

A vector $X_{u}\in TV^{n+m},$ tangent in the point $u\in V^{n+m},$ is
locally represented as $\ (x,y,$ $X,\widetilde{X})$ $=$ $\left(
x^{i},y^{a},X^{i},X^{a}\right) ,$where $\left( X^{i}\right) \in $$\R$$^{n}$
and $\left( X^{a}\right) \in $$\R$$^{m}$ are defined by the equality $%
X_{u}=X^{i}\partial _{i}+X^{a}\partial _{a}$ [$\partial _{\alpha }=\left(
\partial _{i},\partial _{a}\right) $ are usual partial derivatives on
respective coordinates $x^{i}$ and $y^{a}$]. For instance, $\pi ^{T}\left(
x,y,X,\widetilde{X}\right) =\left( x,X\right) $ and the submanifold $%
vV^{n+m} $ contains elements of type $\left( x,y,0,\widetilde{X}\right) $
and the local fibers of the vertical subbundle are isomorphic to $\R$$^{m}.$
Having $\pi ^{T}\left( \partial _{a}\right) =0,$ one comes out that $%
\partial _{a}$ is a local basis of the vertical distribution $u\rightarrow
v_{u}V^{n+m}$ on $V^{n+m},$ which is an integrable distribution.

\begin{definition}
\label{dnlc}A nonlinear connection (N--connection) $\mathbf{N}$ in a space $%
\left( V^{n+m},\pi ,V^{n}\right) $ is defined by the splitting on the left
of the exact sequence
\begin{equation}
0\rightarrow vV^{n+m}\rightarrow TV^{n+m}/vV^{n+m}\rightarrow 0,
\label{1eseq}
\end{equation}%
i. e. a morphism of manifolds $N:TV^{n+m}\rightarrow vV^{n+m}$ such that $%
C\circ i$ is the identity on $vV^{n+m}.$
\end{definition}

The kernel of the morphism $\mathbf{N}$ \ is a subbundle of $\left(
TV^{n+m},\tau _{E},V^{n+m}\right) ,$ it is called the horizontal subspace
(being a subbundle for vector bundle constructions) and denoted by $\left(
hV^{n+m},\tau _{H},V^{n+m}\right) .$ Every tangent bundle $(TV^{n+m},$ $\tau
_{E},$ $V^{n+m})$ provided with a N--connection structure is a Whitney sum
of the vertical and horizontal subspaces (in brief, h- and v-- subspaces),
i. e.
\begin{equation}
TV^{n+m}=hV^{n+m}\oplus vV^{n+m}.  \label{1wihit}
\end{equation}%
It is proven that for every vector bundle $\left( V^{n+m},\pi ,V^{n}\right) $
over a compact manifold $V^{n}$ there exists a nonlinear connection \cite{ma}
(the proof is similar if the bundle structure is modelled on a manifold).%
\footnote{%
We note that the exact sequence (\ref{1eseq}) defines the N--connection in a
global coordinate free form. In a similar form, the N--connection can be
defined for covector bundles or, as particular cases for (co) tangent
bundles. Generalizations for superspaces and noncommutative spaces are
considered respectively in Refs. \cite{vsup} and \cite{ncgg,vncggf}.}

A N--connection $\mathbf{N}$ is defined locally by a set of coefficients $%
N_{i}^{a}(u^{\alpha })$ $=$ $N_{i}^{a}(x^{j},y^{b})$ transforming as
\begin{equation}
N_{i^{\prime }}^{a^{\prime }}\frac{\partial x^{i^{\prime }}}{\partial x^{i}}%
=M_{a}^{a^{\prime }}N_{i}^{a}-\frac{\partial M_{a}^{a^{\prime }}}{\partial
x^{i}}y^{a}  \label{ncontr}
\end{equation}%
under coordinate transforms on the space $\left( V^{n+m},\mu ,M\right) $
when $x^{i^{\prime }}=x^{i^{\prime }}\left( x^{i}\right) $ and $y^{a^{\prime
}}=M_{a}^{a^{\prime }}(x)y^{a}.$ The well known class of linear connections
consists a particular parametization of its coefficients $N_{i}^{a}$ to be
linear on variables $y^{b},$
\begin{equation*}
N_{i}^{a}(x^{j},y^{b})=\Gamma _{bi}^{a}(x^{j})y^{b}.
\end{equation*}

A N--connection structure can be associated to a prescribed ansatz of
vielbein transforms%
\begin{eqnarray}
A_{\alpha }^{\ \underline{\alpha }}(u) &=&\mathbf{e}_{\alpha }^{\ \underline{%
\alpha }}=\left[
\begin{array}{cc}
e_{i}^{\ \underline{i}}(u) & N_{i}^{b}(u)e_{b}^{\ \underline{a}}(u) \\
0 & e_{a}^{\ \underline{a}}(u)%
\end{array}%
\right] ,  \label{1vt1} \\
A_{\ \underline{\beta }}^{\beta }(u) &=&\mathbf{e}_{\ \underline{\beta }%
}^{\beta }=\left[
\begin{array}{cc}
e_{\ \underline{i}}^{i\ }(u) & -N_{k}^{b}(u)e_{\ \underline{i}}^{k\ }(u) \\
0 & e_{\ \underline{a}}^{a\ }(u)%
\end{array}%
\right] ,  \label{1vt2}
\end{eqnarray}%
in particular case $e_{i}^{\ \underline{i}}=\delta _{i}^{\underline{i}}$ and
$e_{a}^{\ \underline{a}}=\delta _{a}^{\underline{a}}$ with $\delta _{i}^{%
\underline{i}}$ and $\delta _{a}^{\underline{a}}$ being the Kronecker
symbols, defining a global splitting of $\mathbf{V}^{n+m}$ into
''horizontal'' and ''vertical'' subspaces with the N--vielbein structure%
\begin{equation*}
\mathbf{e}_{\alpha }=\mathbf{e}_{\alpha }^{\ \underline{\alpha }}\partial _{%
\underline{\alpha }}\mbox{ and }\mathbf{\vartheta }_{\ }^{\beta }=\mathbf{e}%
_{\ \underline{\beta }}^{\beta }du^{\underline{\beta }}.
\end{equation*}%
In this work, we adopt the convention that for the spaces provided with
N--connection structure the geometrical objects can be denoted by
''boldfaced'' symbols if it would be necessary to distinguish such objects
from similar ones for spaces without N--connection. The results from
subsection \ref{standres} can be redefined in order to be compatible with
the N--connection structure and rewritten in terms of ''boldfaced'' values.

A N--connection $\mathbf{N}$ in a space $\mathbf{V}^{n+m}$ is parametrized,
with respect to a local coordinate base,
\begin{equation}
\partial _{\alpha }=(\partial _{i},\partial _{a})\equiv \frac{\partial }{%
\partial u^{\alpha }}=\left( \frac{\partial }{\partial x^{i}},\frac{\partial
}{\partial y^{a}}\right) ,  \label{1pder}
\end{equation}%
and dual base (cobase),
\begin{equation}
d^{\alpha }=(d^{i},d^{a})\equiv du^{\alpha }=(dx^{i},dy^{a}),  \label{1pdif}
\end{equation}%
by its components $N_{i}^{a}(u)=N_{i}^{a}(x,y),$
\begin{equation*}
\mathbf{N}=N_{i}^{a}(u)d^{i}\otimes \partial _{a}.
\end{equation*}%
It is characterized by the N--connection curvature $\mathbf{\Omega }%
=\{\Omega _{ij}^{a}\}$ as a Nijenhuis tensor field $N_{v}\left( X,Y\right) $
associated to $\mathbf{N\ },$
\begin{equation*}
\mathbf{\Omega }=N_{v}=\left[ vX,vY\right] +v\left[ X,Y\right] -v\left[ vX,Y%
\right] -v\left[ X,vY\right] ,
\end{equation*}%
for $X,Y\in \mathcal{X}\left( V^{n+m}\right) $ \cite{vilms} and $\left[ ,%
\right] $ denoting commutators. In local form one has%
\begin{equation*}
\mathbf{\Omega }=\frac{1}{2}\Omega _{ij}^{a}d^{i}\wedge d^{j}\otimes
\partial _{a},
\end{equation*}%
\begin{equation}
\Omega _{ij}^{a}=\delta _{\lbrack j}N_{i]}^{a}=\frac{\partial N_{i}^{a}}{%
\partial x^{j}}-\frac{\partial N_{j}^{a}}{\partial x^{i}}+N_{i}^{b}\frac{%
\partial N_{j}^{a}}{\partial y^{b}}-N_{j}^{b}\frac{\partial N_{i}^{a}}{%
\partial y^{b}}.  \label{1ncurv}
\end{equation}%
The 'N--elongated' operators $\delta _{j}$ from (\ref{1ncurv}) are defined
from a certain vielbein configuration induced by the N--connection, the
N--elongated partial derivatives (in brief, N--derivatives)
\begin{equation}
\mathbf{e}_{\alpha }\doteqdot \delta _{\alpha }=\left( \delta _{i},\partial
_{a}\right) \equiv \frac{\delta }{\delta u^{\alpha }}=\left( \frac{\delta }{%
\delta x^{i}}=\partial _{i}-N_{i}^{a}\left( u\right) \partial _{a},\frac{%
\partial }{\partial y^{a}}\right)  \label{1dder}
\end{equation}%
and the N--elongated differentials (in brief, N--differentials)
\begin{equation}
\mathbf{\vartheta }_{\ }^{\beta }\doteqdot \delta \ ^{\beta }=\left(
d^{i},\delta ^{a}\right) \equiv \delta u^{\alpha }=\left( \delta
x^{i}=dx^{i},\delta y^{a}=dy^{a}+N_{i}^{a}\left( u\right) dx^{i}\right)
\label{1ddif}
\end{equation}%
called also, respectively, the N--frame and N--coframe. \footnote{%
We shall use both type of denotations $\mathbf{e}_{\alpha }\doteqdot \delta
_{\alpha }$ and $\mathbf{\vartheta }_{\ }^{\beta }\doteqdot \delta \
^{\alpha }$ in order to preserve a connection to denotations from Refs. \cite%
{ma,vsup,vsp,v1,v2,vd,vncggf}. The 'boldfaced' symbols $\mathbf{e}_{\alpha }$
and $\mathbf{\vartheta }_{\ }^{\beta }$ are written in order to emphasize
that they define N--adapted vielbeins and the symbols $\delta _{\alpha }$
and $\delta \ ^{\beta }$ will be used for the N--elongated partial
derivatives and, respectively, differentials.
\par
{}}

The N--coframe (\ref{1ddif}) is anholonomic because there are satisfied the
anholonomy relations (\ref{1anh}),
\begin{equation}
\left[ \delta _{\alpha },\delta _{\beta }\right] =\delta _{\alpha }\delta
_{\beta }-\delta _{\beta }\delta _{\alpha }=\mathbf{w}_{\ \alpha \beta
}^{\gamma }\left( u\right) \delta _{\gamma }  \label{1anhr}
\end{equation}%
for which the anholonomy coefficients $\mathbf{w}_{\beta \gamma }^{\alpha
}\left( u\right) $ are computed to have certain nontrivial values
\begin{equation}
\mathbf{w}_{~ji}^{a}=-\mathbf{w}_{~ij}^{a}=\Omega _{ij}^{a},\ \mathbf{w}%
_{~ia}^{b}=-\mathbf{w}_{~ai}^{b}=\partial _{a}N_{i}^{b}.  \label{1anhc}
\end{equation}

We emphasize that the N--connection formalism is a natural one for
investigating physical systems with mixed sets of holonomic--anholonomic
variables. The imposed anholonomic constraints (anisotropies) are
characterized by the coefficients of N--connection which defines a global
splitting of the components of geometrical objects with respect to some
'horizontal' (holonomic) and 'vertical' (anisotropic) directions. In brief,
we shall use respectively the terms h- and/or v--components, h- and/or
v--indices, and h- and/or v--subspaces

A N--connection structure on $\mathbf{V}^{n+m}$ defines the algebra of
tensorial distinguished \ (by N--connection structure) fields $dT\left( T%
\mathbf{V}^{n+m}\right) $ (d--fields, d--tensors, d--objects, if to follow
the terminology from \cite{ma}) on $\mathbf{V}^{n+m}$ introduced as the
tensor algebra $\mathcal{T}=\{\mathcal{T}_{qs}^{pr}\}$ of the distinguished
tangent bundle $\mathcal{V}_{(d)},$ $p_{d}:\ h\mathbf{V}^{n+m}\oplus v%
\mathbf{V}^{n+m}\rightarrow \mathbf{V}^{n+m}.$ An element $\mathbf{t}\in
\mathcal{T}_{qs}^{pr},$ a d--tensor field of type $\left(
\begin{array}{cc}
p & r \\
q & s%
\end{array}%
\right) ,$ can be written in local form as%
\begin{equation*}
\mathbf{t}=t_{j_{1}...j_{q}b_{1}...b_{r}}^{i_{1}...i_{p}a_{1}...a_{r}}\left(
u\right) \delta _{i_{1}}\otimes ...\otimes \delta _{i_{p}}\otimes \partial
_{a_{1}}\otimes ...\otimes \partial _{a_{r}}\otimes d^{j_{1}}\otimes
...\otimes d^{j_{q}}\otimes \delta ^{b_{1}}...\otimes \delta ^{b_{r}}.
\end{equation*}

There are used the denotations $\mathcal{X}\left( \mathcal{V}_{(d)}\right) $
(or $\mathcal{X}(\mathbf{V}^{n+m}{),\wedge }^{p}\left( \mathcal{V}%
_{(d)}\right) $ (or ${\wedge }^{p}\left( \mathbf{V}^{n+m}\right) $ and $%
\mathcal{F}\left( \mathcal{V}_{(d)}\right) $ (or $\mathcal{F}$ $\left(
\mathbf{V}^{n+m}\right) $) for the module of d--vector fields on $\mathcal{V}%
_{(d)}$ (or $\mathbf{V}^{n+m}$ ), the exterior algebra of p--forms on $%
\mathcal{V}_{(d)}$ (or $\mathbf{V}^{n+m})$ and the set of real functions on $%
\mathcal{V}_{(d)}$ (or $\mathbf{V}^{n+m}).$

\subsection{Distinguished linear connection and metric structures}

\label{dlcms} The d--objects on $\mathcal{V}_{(d)}$ are introduced in a
coordinate free form as geometric objects adapted to the N--connection
structure. In coordinate form, we can characterize such objects (linear
connections, metrics or any tensor field) by certain group and coordinate
transforms adapted to the N-connection structure on $\mathbf{V}^{n+m},$ i.
e. to the global space splitting (\ref{1wihit}) into h- and v--subspaces.

\subsubsection{d--connections}

We analyze the general properties of a class of linear connections being
adapted to the N--connection structure (called d--connections).

\begin{definition}
\label{defdcon}A d--connection $\mathbf{D}$ on $\mathcal{V}_{(d)}$ is
defined as a linear connection $D,$ see Definition \ref{deflcon}, on $%
\mathcal{V}_{(d)}$ conserving under a parallelism the global decomposition
of $T\mathbf{V}^{n+m}$ (\ref{1wihit}) into the horizontal subbundle, $h%
\mathbf{V}^{n+m},$ and vertical subbundle, $v\mathbf{V}^{n+m},$ of $\mathcal{%
V}_{(d)}.$
\end{definition}

A N-connection induces decompositions of d--tensor indices into sums of
horizontal and vertical parts, for example, for every d--vector $\mathbf{X}%
\in \mathcal{X}\left( \mathcal{V}_{(d)}\right) $ and 1-form $\widetilde{%
\mathbf{X}}\in \Lambda ^{1}\left( \mathcal{V}_{(d)}\right) $ we have
respectively
\begin{equation*}
X=hX+vX\ \mbox{and \quad
}\widetilde{X}=h\widetilde{X}+v\widetilde{X}.
\end{equation*}%
For simplicity, we shall not use boldface symbols for d--vectors and
d--forms if this will not result in ambiguities. In consequence, we can
associate to every d--covariant derivation $\mathbf{D}_{X}=X\rfloor \mathbf{D%
}$ two new operators of h- and v--covariant derivations, $\mathbf{D}%
_{X}=D_{X}^{[h]}+D_{X}^{[v]},$ defined respectively
\begin{equation*}
D_{X}^{[h]}Y=\mathbf{D}_{hX}Y\quad \mbox{ and \quad }D_{X}^{[v]}Y=\mathbf{D}%
_{vX}Y,
\end{equation*}%
for which the following conditions hold:%
\begin{eqnarray}
\mathbf{D}_{X}Y &=&D_{X}^{[h]}Y+D_{X}^{[v]}Y,  \label{hvder} \\
D_{X}^{[h]}f &=&(hX\mathbf{)}f\mbox{ \quad and\quad
}D_{X}^{[v]}f=(vX)f,\quad  \notag
\end{eqnarray}%
for any $X,Y\in \mathcal{X}\left( E\right) ,f\in \mathcal{F}\left(
V^{n+m}\right) .$

The N--adapted components $\mathbf{\Gamma }_{\beta \gamma }^{\alpha }$ of a
d-connection $\mathbf{D}_{\alpha }=(\delta _{\alpha }\rfloor \mathbf{D})$
are defined by the equations%
\begin{equation*}
\mathbf{D}_{\alpha }\delta _{\beta }=\mathbf{\Gamma }_{\ \alpha \beta
}^{\gamma }\delta _{\gamma },
\end{equation*}%
from which one immediately follows
\begin{equation}
\mathbf{\Gamma }_{\ \alpha \beta }^{\gamma }\left( u\right) =\left( \mathbf{D%
}_{\alpha }\delta _{\beta }\right) \rfloor \delta ^{\gamma }.  \label{1dcon1}
\end{equation}%
The operations of h- and v-covariant derivations, $D_{k}^{[h]}=%
\{L_{jk}^{i},L_{bk\;}^{a}\}$ and $D_{c}^{[v]}=\{C_{jk}^{i},C_{bc}^{a}\}$
(see (\ref{hvder})) are introduced as corresponding h- and
v--parametrizations of (\ref{1dcon1}),%
\begin{eqnarray}
L_{jk}^{i} &=&\left( \mathbf{D}_{k}\delta _{j}\right) \rfloor d^{i},\quad
L_{bk}^{a}=\left( \mathbf{D}_{k}\partial _{b}\right) \rfloor \delta ^{a}
\label{hcov} \\
C_{jc}^{i} &=&\left( \mathbf{D}_{c}\delta _{j}\right) \rfloor d^{i},\quad
C_{bc}^{a}=\left( \mathbf{D}_{c}\partial _{b}\right) \rfloor \delta ^{a}.
\label{vcov}
\end{eqnarray}%
A set of h--components \ (\ref{hcov}) and v--components (\ref{vcov}),
distinguished in the form $\mathbf{\Gamma }_{\ \alpha \beta }^{\gamma }$ $%
=(L_{jk}^{i},$ $L_{bk}^{a},$ $C_{jc}^{i},C_{bc}^{a}),$ completely defines
the local action of a d--connection $\mathbf{D}$ in $\mathbf{V}^{n+m}.$ For
instance, having taken a d--tensor field of type $\left(
\begin{array}{cc}
1 & 1 \\
1 & 1%
\end{array}%
\right) ,$ $\mathbf{t}=t_{jb}^{ia}\delta _{i}\otimes \partial _{a}\otimes
\partial ^{j}\otimes \delta ^{b},$ and a d--vector $\mathbf{X}=X^{i}\delta
_{i}+X^{a}\partial _{a}$ we can write%
\begin{equation*}
\mathbf{D}_{X}\mathbf{t=}D_{X}^{[h]}\mathbf{t+}D_{X}^{[v]}\mathbf{t=}\left(
X^{k}t_{jb|k}^{ia}+X^{c}t_{jb\perp c}^{ia}\right) \delta _{i}\otimes
\partial _{a}\otimes d^{j}\otimes \delta ^{b},
\end{equation*}%
where the h--covariant derivative is
\begin{equation*}
t_{jb|k}^{ia}=\frac{\delta t_{jb}^{ia}}{\delta x^{k}}%
+L_{hk}^{i}t_{jb}^{ha}+L_{ck}^{a}t_{jb}^{ic}-L_{jk}^{h}t_{hb}^{ia}-L_{bk}^{c}t_{jc}^{ia}
\end{equation*}%
and the v--covariant derivative is
\begin{equation*}
t_{jb\perp c}^{ia}=\frac{\partial t_{jb}^{ia}}{\partial y^{c}}%
+C_{hc}^{i}t_{jb}^{ha}+C_{dc}^{a}t_{jb}^{id}-C_{jc}^{h}t_{hb}^{ia}-C_{bc}^{d}t_{jd}^{ia}.
\end{equation*}%
For a scalar function $f\in \mathcal{F}\left( V^{n+m}\right) $ we have
\begin{equation*}
D_{k}^{[h]}=\frac{\delta f}{\delta x^{k}}=\frac{\partial f}{\partial x^{k}}%
-N_{k}^{a}\frac{\partial f}{\partial y^{a}}\mbox{ and }D_{c}^{[v]}f=\frac{%
\partial f}{\partial y^{c}}.
\end{equation*}%
We note that these formulas are written in abstract index form and specify
for d--connections the covariant derivation rule (\ref{covderrul}).

\subsubsection{Metric structures and d--metrics}

We introduce arbitrary metric structures on a space $\mathbf{V}^{n+m}$ and
consider the possibility to adapt them to N--connection structures.

%%%%%%%%%%%%%%%%%%%%%%%%%%%%%%%%%%%%%%%%%

\begin{definition}
A metric structure $\mathbf{g}$ on a space $\mathbf{V}^{n+m}$ is defined as
a symmetric covariant tensor field of type $\left( 0,2\right) ,$ $g_{\alpha
\beta ,}$ being nondegenerate and of constant signature on $\mathbf{V}%
^{n+m}. $
\end{definition}

This Definition is completely similar to Definition \ref{defm} but in our
case it is adapted to the N--connection structure. A N--connection $\mathbf{%
N=}\{N_{\underline{i}}^{\underline{b}}\left( u\right) \}$ and a metric
structure%
\begin{equation}
\mathbf{g}=g_{\underline{\alpha }\underline{\beta }}du^{\underline{\alpha }%
}\otimes du^{\underline{\beta }}  \label{mstr}
\end{equation}%
on $\mathbf{V}^{n+m}$ are mutually compatible if there are satisfied the
conditions
\begin{equation}
\mathbf{g}\left( \delta _{\underline{i}},\partial _{\underline{a}}\right) =0,%
\mbox{ or equivalently,
}g_{\underline{i}\underline{a}}\left( u\right) -N_{\underline{i}}^{%
\underline{b}}\left( u\right) h_{\underline{a}\underline{b}}\left( u\right)
=0,  \label{comp1}
\end{equation}%
where $h_{\underline{a}\underline{b}}\doteqdot \mathbf{g}\left( \partial _{%
\underline{a}},\partial _{\underline{b}}\right) $ and $g_{\underline{i}%
\underline{a}}\doteqdot \mathbf{g}\left( \partial _{\underline{i}},\partial
_{\underline{a}}\right) \,$ resulting in
\begin{equation}
N_{i}^{b}\left( u\right) =h^{ab}\left( u\right) g_{ia}\left( u\right)
\label{nconstr}
\end{equation}%
(the matrix $h^{ab}$ is inverse to $h_{ab};$ for simplicity, we do not
underly \ the indices in the last formula). In consequence, we obtain a
h--v--decomposition of metric (in brief, d--metric)%
\begin{equation}
\mathbf{g}(X,Y)\mathbf{=}h\mathbf{g}(X,Y)+v\mathbf{g}(X,Y),  \label{block1}
\end{equation}%
where the d-tensor $h\mathbf{g}(X,Y)=\mathbf{g}(hX,hY)$ is of type $\left(
\begin{array}{cc}
0 & 0 \\
2 & 0%
\end{array}%
\right) $ and the d-tensor \newline
$\ v\mathbf{g}(X,Y)\mathbf{=h}(vX,vY)$ is of type $\left(
\begin{array}{cc}
0 & 0 \\
0 & 2%
\end{array}%
\right) .$ With respect to a N--coframe (\ref{1ddif}), the d--metric (\ref%
{block1}) is written
\begin{equation}
\mathbf{g}=\mathbf{g}_{\alpha \beta }\left( u\right) \delta ^{\alpha
}\otimes \delta ^{\beta }=g_{ij}\left( u\right) d^{i}\otimes
d^{j}+h_{ab}\left( u\right) \delta ^{a}\otimes \delta ^{b},  \label{1block2}
\end{equation}%
where $g_{ij}\doteqdot \mathbf{g}\left( \delta _{i},\delta _{j}\right) .$
The d--metric (\ref{1block2}) can be equivalently written in
''off--diagonal'' form if the basis of dual vectors consists from the
coordinate differentials (\ref{1pdif}),
\begin{equation}
\underline{g}_{\alpha \beta }=\left[
\begin{array}{cc}
g_{ij}+N_{i}^{a}N_{j}^{b}h_{ab} & N_{j}^{e}h_{ae} \\
N_{i}^{e}h_{be} & h_{ab}%
\end{array}%
\right] .  \label{1ansatz}
\end{equation}%
It is easy to check that one holds the relations%
\begin{equation*}
\mathbf{g}_{\alpha \beta }=\mathbf{e}_{\alpha }^{\ \underline{\alpha }}%
\mathbf{e}_{\beta }^{\ \underline{\beta }}\underline{g}_{\underline{\alpha }%
\underline{\beta }}
\end{equation*}%
or, inversely,
\begin{equation*}
\underline{g}_{\underline{\alpha }\underline{\beta }}=\mathbf{e}_{\
\underline{\alpha }}^{\alpha }\mathbf{e}_{\ \underline{\beta }}^{\beta }%
\mathbf{g}_{\alpha \beta }
\end{equation*}%
as it is stated by respective vielbein transforms (\ref{1vt1}) and (\ref{1vt2}%
).

\begin{remark}
\label{rgod}A metric, for instance, parametrized in the form (\ref{1ansatz})\
is generic off--diagonal if it can not be diagonalized by any coordinate
transforms. If the anholonomy coefficients (\ref{1anhc}) vanish for a such
parametrization, we can define certain coordinate transforms to diagonalize
both the off--diagonal form (\ref{1ansatz}) and the equivalent d--metric (\ref%
{1block2}).
\end{remark}

\begin{definition}
The nonmetricity d--field
\begin{equation*}
\ \mathcal{Q}=\mathbf{Q}_{\alpha \beta }\mathbf{\vartheta }^{\alpha }\otimes
\mathbf{\vartheta }^{\beta }=\mathbf{Q}_{\alpha \beta }\delta \ ^{\alpha
}\otimes \delta ^{\beta }
\end{equation*}%
on a space $\mathbf{V}^{n+m}$ provided with N--connection structure is
defined by a d--tensor field with the coefficients
\begin{equation}
\mathbf{Q}_{\alpha \beta }\doteqdot -\mathbf{Dg}_{\alpha \beta }  \label{1nmf}
\end{equation}%
where the covariant derivative $\mathbf{D}$ is for a d--connection $\mathbf{%
\Gamma }_{\ \alpha }^{\gamma }=\mathbf{\Gamma }_{\ \alpha \beta }^{\gamma }%
\mathbf{\vartheta }^{\beta },$ see (\ref{1dcon1}) with the respective
splitting $\mathbf{\Gamma }_{\alpha \beta }^{\gamma }=\left(
L_{jk}^{i},L_{bk}^{a},C_{jc}^{i},C_{bc}^{a}\right) ,$ as to be adapted to
the N--connection structure.
\end{definition}

This definition is similar to that given for metric--affine spaces (see
definition \ref{dnmf}) and Refs. \cite{mag}, but in our case the
N--connection establishes some 'preferred' N--adapted local frames (\ref%
{1dder}) and (\ref{1ddif}) splitting all geometric objects into irreducible h-
and v--components. A linear connection $D_{X}$ is compatible\textbf{\ } with
a d--metric $\mathbf{g}$ if%
\begin{equation}
D_{X}\mathbf{g}=0,  \label{1mc}
\end{equation}%
$\forall X\mathbf{\in }\mathcal{X}\left( V^{n+m}\right) ,$ i. e. if $%
Q_{\alpha \beta }\equiv 0.$ In a space provided with N--connection
structure, the metricity condition (\ref{1mc}) may split into a set of
compatibility conditions on h- and v-- subspaces. We should consider
separately which of the conditions
\begin{equation}
D^{[h]}(h\mathbf{g)}=0,D^{[v]}(h\mathbf{g)}=0,D^{[h]}(v\mathbf{g)}%
=0,D^{[v]}(v\mathbf{g)}=0  \label{1mca}
\end{equation}%
are satisfied, or not, for a given d--connection $\mathbf{\Gamma }_{\ \alpha
\beta }^{\gamma }.$ For instance, if $D^{[v]}(h\mathbf{g)}=0$ and $D^{[h]}(v%
\mathbf{g)}=0,$ but, in general, $D^{[h]}(h\mathbf{g)}\neq 0$ and $D^{[v]}(v%
\mathbf{g)}\neq 0$ we can consider a nonmetricity d--field (d--nonmetricity)
$\mathbf{Q}_{\alpha \beta }=\mathbf{Q}_{\gamma \alpha \beta }\vartheta
^{\gamma }$ with irreducible h--v--components (with respect to the
N--connection decompositions), $\ \mathbf{Q}_{\gamma \alpha \beta }=\left(
Q_{ijk},Q_{abc}\right) .$

By acting on forms with the covariant derivative $D,$ in a metric--affine
space, we can also define another very important geometric objects (the
'gravitational field potentials', see \cite{mag}):

\begin{equation}
\mbox{ torsion }\ \mathcal{T}^{\alpha }\doteqdot D\vartheta ^{\alpha
}=d\vartheta ^{\alpha }+\Gamma _{\ \beta }^{\gamma }\wedge \vartheta ^{\beta
},\mbox{ see Definition \ref{deftors}}  \label{dta}
\end{equation}%
and
\begin{equation}
\mbox{ curvature }\ \mathcal{R}_{\ \beta }^{\alpha }\doteqdot D\Gamma _{\
\beta }^{\alpha }=d\Gamma _{\ \beta }^{\alpha }-\Gamma _{\ \beta }^{\gamma
}\wedge \Gamma _{\ \gamma }^{\alpha },\mbox{ see Definition \ref{defcurv}}.
\label{dra}
\end{equation}

The Bianchi identities are%
\begin{equation}
DQ_{\alpha \beta }\equiv \mathcal{R}_{\alpha \beta }+\mathcal{R}_{\beta
\alpha },\ D\mathcal{T}^{\alpha }\equiv \mathcal{R}_{\gamma }^{\ \alpha
}\wedge \vartheta ^{\gamma }\mbox{ and }D\mathcal{R}_{\gamma }^{\ \alpha
}\equiv 0,  \label{bi}
\end{equation}%
where we stress the fact that $Q_{\alpha \beta },T^{\alpha }$ and $R_{\beta
\alpha }$ are called also the strength fields of a metric--affine theory.

For spaces provided with N--connections, we write the corresponding formulas
by using ''boldfaced'' symbols and change the usual differential $d$ $\ $%
into N-adapted operator $\delta .$%
\begin{equation}
\ \mathbf{T}^{\alpha }\doteqdot \mathbf{D\vartheta }^{\alpha }=\delta
\mathbf{\vartheta }^{\alpha }+\mathbf{\Gamma }_{\ \beta }^{\gamma }\wedge
\mathbf{\vartheta }^{\beta }  \label{1dt}
\end{equation}%
and
\begin{equation}
\ \mathbf{R}_{\ \beta }^{\alpha }\doteqdot \mathbf{D\Gamma }_{\ \beta
}^{\alpha }=\delta \mathbf{\Gamma }_{\ \beta }^{\alpha }-\mathbf{\Gamma }_{\
\beta }^{\gamma }\wedge \mathbf{\Gamma }_{\ \ \gamma }^{\alpha }  \label{1dc}
\end{equation}%
where the Bianchi identities written in 'boldfaced' symbols split into h-
and v--irreducible decompositions induced by the N--connection. \footnote{%
see similar details in Ref. \cite{ma} for the case of vector/tangent bundles
provided with mutually compatible N--connection, d--connection and d--metric
structure }. We shall examine and compute the general form of torsion
and curvature d--tensors in spaces provided with N--connection structure
in section \ref{torscurv}.

We note that the bulk of works on Finsler geometry and generalizations \cite%
{fg,ma,mhss,rund,as,bog,carf,vsup,vsp,vncggf} consider very general linear
connection and metric fields being adapted to the N--connection structure.
In another turn, the researches on metric--affine gravity \cite{mag,rcg}
concern generalizations to nonmetricity but not N--connections. In this
work, we elaborate a unified moving frame geometric approach to both Finlser
like and metric--affine geometries.

\subsection{ Torsions and curvatures of d--connections}

\label{torscurv}We define and calculate the irreducible components of
torsion and curvature in a space $\mathbf{V}^{n+m}$ provided with additional
N--connection structure (these could be any metric--affine spaces \cite{mag},
 or their particular, like Riemann--Cartan \cite{rcg}, cases with vanishing
nonmetricity and/or torsion, or any (co) vector / tangent bundles like in
Finsler geometry and generalizations).

\subsubsection{d--torsions and N--connections}

We give a definition being equivalent to (\ref{1dt}) but in d--operator form
(the Definition \ref{deftors} was for the spaces not possessing
N--connection structure):

\begin{definition}
The torsion $\mathbf{T}$ of a d--connection $\mathbf{D}=\left(
D^{[h]},D^{[v]}\right) \mathbf{\ }$ in space $\mathbf{V}^{n+m}$ is defined
as an operator (d--tensor field) adapted to the N--connection structure
\begin{equation}
\mathbf{T}\left( X,Y\right) =\mathbf{D}_{X}Y\mathbf{-D}_{Y}X\mathbf{\ -}%
\left[ X,Y\right] \mathbf{.}  \label{1torsion}
\end{equation}
\end{definition}

One holds the following h- and v--decompositions%
\begin{equation}
\mathbf{T}\left( X,Y\right) \mathbf{=T}\left( hX,hY\right) \mathbf{+T}\left(
hX,vY\right) \mathbf{+T}\left( vX,hY\right) \mathbf{+T}\left( vX,vY\right)
\mathbf{.}  \label{hvtorsion}
\end{equation}%
We consider the projections: $h\mathbf{T}\left( X,Y\right) \mathbf{,}v%
\mathbf{T}\left( hX,hY\right) \mathbf{,}h\mathbf{T}\left( hX,hY\right)
\mathbf{,...}$ and say that, for instance, $\mathbf{\ }h\mathbf{T}\left(
hX,hY\right) $ is the h(hh)-torsion of $D$ , $vT\left( hX,hY\right) \mathbf{%
\ }$ is the v(hh)-torsion of $\mathbf{D}$ and so on.

The torsion (\ref{1torsion}) is locally determined by five d--tensor fields,
d--torsions (irreducible N--adapted h--v--decompositions) defined as%
\begin{eqnarray}
T_{jk}^{i} &=&h\mathbf{T}\left( \delta _{k},\delta _{j}\right) \rfloor
d^{i},\quad T_{jk}^{a}=v\mathbf{T}\left( \delta _{k},\delta _{j}\right)
\rfloor \delta ^{a},\quad P_{jb}^{i}=h\mathbf{T}\left( \partial _{b},\delta
_{j}\right) \rfloor d^{i},  \notag \\
\quad P_{jb}^{a} &=&v\mathbf{T}\left( \partial _{b},\delta _{j}\right)
\rfloor \delta ^{a},\quad \ S_{bc}^{a}=v\mathbf{T}\left( \partial
_{c},\partial _{b}\right) \rfloor \delta ^{a}.  \notag
\end{eqnarray}%
Using the formulas (\ref{1dder}), (\ref{1ddif}), and (\ref{1ncurv}), we can
calculate the h--v--components of torsion (\ref{hvtorsion}) for a
d--connection, i. e. we can prove \footnote{%
see also the original proof for vector bundles in \cite{ma}}

\begin{theorem}
\label{tdtors}The torsion $\mathbf{T}_{.\beta \gamma }^{\alpha
}=(T_{.jk}^{i},T_{ja}^{i},T_{.ij}^{a},T_{.bi}^{a},T_{.bc}^{a})$ of a
d--connection\newline
$\mathbf{\Gamma }_{\alpha \beta }^{\gamma }=\left(
L_{jk}^{i},L_{bk}^{a},C_{jc}^{i},C_{bc}^{a}\right) $(\ref{1dcon1}) has
irreducible h- v--components (d--torsions)
\begin{eqnarray}
T_{.jk}^{i} &=&-T_{kj}^{i}=L_{jk}^{i}-L_{kj}^{i},\quad
T_{ja}^{i}=-T_{aj}^{i}=C_{.ja}^{i},\ T_{.ji}^{a}=-T_{.ij}^{a}=\frac{\delta
N_{i}^{a}}{\delta x^{j}}-\frac{\delta N_{j}^{a}}{\delta x^{i}}=\Omega
_{.ji}^{a},  \notag \\
T_{.bi}^{a} &=&-T_{.ib}^{a}=P_{.bi}^{a}=\frac{\partial N_{i}^{a}}{\partial
y^{b}}-L_{.bj}^{a},\
T_{.bc}^{a}=-T_{.cb}^{a}=S_{.bc}^{a}=C_{bc}^{a}-C_{cb}^{a}.\   \label{1dtorsb}
\end{eqnarray}
\end{theorem}

We note that on (pseudo) Riemanian spacetimes the d--torsions can be induced
by the N--connection coefficients and reflect an anholonomic frame
structures. Such objects vanishes when we transfer our considerations with
respect to holonomic bases for a trivial N--connection and zero ''vertical''
dimension.

\subsubsection{d--curvatures and N--connections}

In operator form, the curvature (\ref{1dc}) is stated from the

\begin{definition}
The curvature $\mathbf{R}$ of a d--connection $\mathbf{D}=\left(
D^{[h]},D^{[v]}\right) \mathbf{\ }$ in space $\mathbf{V}^{n+m}$ is defined
as an operator (d--tensor field) adapted to the N--connection structure
\begin{equation}
\mathbf{R}\left( X,Y\right) Z=\mathbf{D}_{X}\mathbf{D}_{Y}Z-\mathbf{D}_{Y}%
\mathbf{D}_{X}Z-\mathbf{D}_{[X,Y]}Z\mathbf{.}  \label{curvaturea}
\end{equation}
\end{definition}

This Definition is similar to the Definition \ref{defcurv} being a
generalization for the spaces provided with N--connection. One holds certain
properties for the h- and v--decompositi\-ons of curvature:%
\begin{equation}
v\mathbf{R}\left( X,Y\right) hZ=0,\ h\mathbf{R}\left( X,Y\right) vZ\mathbf{=}%
0,\ \mathbf{R}\left( X,Y\right) Z=h\mathbf{R}\left( X,Y\right) hZ\mathbf{+}v%
\mathbf{R}\left( X,Y\right) vZ\mathbf{.}  \notag
\end{equation}%
From (\ref{curvaturea}) and the equation $\mathbf{R}\left( X,Y\right)
\mathbf{=-R}\left( Y,X\right) ,$ we get that the curvature of a
d-con\-necti\-on $\mathbf{D}$ in $\mathbf{V}^{n+m}$ is completely determined
by the following six d--tensor fields (d--curvatures):%
\begin{eqnarray}
R_{\ hjk}^{i} &=&d^{i}\rfloor \mathbf{R}\left( \delta _{k},\delta
_{j}\right) \delta _{h},~R_{\ bjk}^{a}=\delta ^{a}\rfloor \mathbf{R}\left(
\delta _{k},\delta _{j}\right) \partial _{b},  \label{curvaturehv} \\
P_{\ jkc}^{i} &=&d^{i}\rfloor \mathbf{R}\left( \partial _{c},\partial
_{k}\right) \delta _{j},~P_{\ bkc}^{a}=\delta ^{a}\rfloor \mathbf{R}\left(
\partial _{c},\partial _{k}\right) \partial _{b},  \notag \\
S_{\ jbc}^{i} &=&d^{i}\rfloor \mathbf{R}\left( \partial _{c},\partial
_{b}\right) \delta _{j},~S_{\ bcd}^{a}=\delta ^{a}\rfloor \mathbf{R}\left(
\partial _{d},\partial _{c}\right) \partial _{b}.  \notag
\end{eqnarray}%
By a direct computation, using (\ref{1dder}), (\ref{1ddif}), (\ref{hcov}), (%
\ref{vcov}) and (\ref{curvaturehv}), we prove

\begin{theorem}
The curvature $\mathbf{R}_{.\beta \gamma \tau }^{\alpha }=(R_{\
hjk}^{i},R_{\ bjk}^{a},P_{\ jka}^{i},P_{\ bka}^{c},S_{\ jbc}^{i},S_{\
bcd}^{a})$ of a d--con\-nec\-ti\-on $\mathbf{\Gamma }_{\alpha \beta
}^{\gamma }=\left( L_{jk}^{i},L_{bk}^{a},C_{jc}^{i},C_{bc}^{a}\right) $(\ref%
{1dcon1}) has the h- v--components (d--curvatures)
\begin{eqnarray}
R_{\ hjk}^{i} &=&\frac{\delta L_{.hj}^{i}}{\delta x^{k}}-\frac{\delta
L_{.hk}^{i}}{\delta x^{j}}%
+L_{.hj}^{m}L_{mk}^{i}-L_{.hk}^{m}L_{mj}^{i}-C_{.ha}^{i}\Omega _{.jk}^{a},
\label{1dcurv} \\
R_{\ bjk}^{a} &=&\frac{\delta L_{.bj}^{a}}{\delta x^{k}}-\frac{\delta
L_{.bk}^{a}}{\delta x^{j}}%
+L_{.bj}^{c}L_{.ck}^{a}-L_{.bk}^{c}L_{.cj}^{a}-C_{.bc}^{a}\ \Omega
_{.jk}^{c},  \notag \\
P_{\ jka}^{i} &=&\frac{\partial L_{.jk}^{i}}{\partial y^{k}}-\left( \frac{%
\partial C_{.ja}^{i}}{\partial x^{k}}%
+L_{.lk}^{i}C_{.ja}^{l}-L_{.jk}^{l}C_{.la}^{i}-L_{.ak}^{c}C_{.jc}^{i}\right)
+C_{.jb}^{i}P_{.ka}^{b},  \notag \\
P_{\ bka}^{c} &=&\frac{\partial L_{.bk}^{c}}{\partial y^{a}}-\left( \frac{%
\partial C_{.ba}^{c}}{\partial x^{k}}+L_{.dk}^{c%
\,}C_{.ba}^{d}-L_{.bk}^{d}C_{.da}^{c}-L_{.ak}^{d}C_{.bd}^{c}\right)
+C_{.bd}^{c}P_{.ka}^{d},  \notag \\
S_{\ jbc}^{i} &=&\frac{\partial C_{.jb}^{i}}{\partial y^{c}}-\frac{\partial
C_{.jc}^{i}}{\partial y^{b}}+C_{.jb}^{h}C_{.hc}^{i}-C_{.jc}^{h}C_{hb}^{i},
\notag \\
S_{\ bcd}^{a} &=&\frac{\partial C_{.bc}^{a}}{\partial y^{d}}-\frac{\partial
C_{.bd}^{a}}{\partial y^{c}}+C_{.bc}^{e}C_{.ed}^{a}-C_{.bd}^{e}C_{.ec}^{a}.
\notag
\end{eqnarray}
\end{theorem}

The components of the Ricci d-tensor
\begin{equation*}
\mathbf{R}_{\alpha \beta }=\mathbf{R}_{\ \alpha \beta \tau }^{\tau }
\end{equation*}%
with respect to a locally adapted frame (\ref{1dder}) has four irreducible h-
v--components, $\mathbf{R}_{\alpha \beta }=\{R_{ij},R_{ia},R_{ai},S_{ab}\},$
where%
\begin{eqnarray}
R_{ij} &=&R_{\ ijk}^{k},\quad R_{ia}=-\ ^{2}P_{ia}=-P_{\ ika}^{k},
\label{1dricci} \\
R_{ai} &=&\ ^{1}P_{ai}=P_{\ aib}^{b},\quad S_{ab}=S_{\ abc}^{c}.  \notag
\end{eqnarray}%
We point out that because, in general, $^{1}P_{ai}\neq ~^{2}P_{ia}$ the
Ricci d--tensor is non symmetric.

Having defined a d--metric of type (\ref{1block2}) in $\mathbf{V}^{n+m},$ we
can introduce the scalar curvature of a d--connection $\mathbf{D,}$
\begin{equation}
{\overleftarrow{\mathbf{R}}}=\mathbf{g}^{\alpha \beta }\mathbf{R}_{\alpha
\beta }=R+S,  \label{1dscal}
\end{equation}%
where $R=g^{ij}R_{ij}$ and $S=h^{ab}S_{ab}$ and define the distinguished
form of the Einstein tensor (the Einstein d--tensor), see Definition \ref%
{defrset},%
\begin{equation}
\mathbf{G}_{\alpha \beta }\doteqdot \mathbf{R}_{\alpha \beta }-\frac{1}{2}%
\mathbf{g}_{\alpha \beta }{\overleftarrow{\mathbf{R}}.}  \label{1deinst}
\end{equation}

The Ricci and Bianchi identities (\ref{bi}) of d--connections are formulated
in h- v- irreducible forms on vector bundle \cite{ma}. The same formulas
hold for arbitrary metric compatible d--connections on $\mathbf{V}^{n+m}$
(for simplicity, we omit such details in this work).

\section{Some Classes of Linear and Nonlinear Connections}

\label{lconnections}The geometry of d--connections in a space $\mathbf{V}%
^{n+m}$ provided with N--connection structure is very reach (works \cite{ma}
and \cite{vsup} contain results on generalized Finsler spaces and
superspaces). If a triple of fundamental geometric objects $(N_{i}^{a}\left(
u\right) ,\mathbf{\Gamma }_{\ \beta \gamma }^{\alpha }\left( u\right) ,$ $%
\mathbf{g}_{\alpha \beta }\left( u\right) )$ is fixed on $\mathbf{V}^{n+m}{,}
$ in general, with respect to N--adapted frames, a multi--connection structure
is defined (with different rules of covariant derivation). In this Section,
we analyze a set of linear connections and associated covariant derivations
being very important for investigating spacetimes provided with anholonomic
frame structure and generic off--diagonal metrics.

\subsection{The Levi--Civita connection and N--connections}

The Levi--Civita connection $\bigtriangledown =\{\mathbf{\Gamma }%
_{\bigtriangledown \beta \gamma }^{\tau }\}$ with coefficients
\begin{equation}
\mathbf{\Gamma }_{\alpha \beta \gamma }^{\bigtriangledown }=g\left( \mathbf{e%
}_{\alpha },\bigtriangledown _{\gamma }\mathbf{e}_{\beta }\right) =\mathbf{g}%
_{\alpha \tau }\mathbf{\Gamma }_{\bigtriangledown \beta \gamma }^{\tau },\,
\label{lccon1}
\end{equation}%
is torsionless,
\begin{equation*}
\mathbf{T}_{\bigtriangledown }^{\alpha }\doteqdot \bigtriangledown \mathbf{%
\vartheta }^{\alpha }=d\mathbf{\vartheta }^{\alpha }+\mathbf{\Gamma }%
_{\bigtriangledown \beta \gamma }^{\tau }\wedge \mathbf{\vartheta }^{\beta
}=0,
\end{equation*}%
and metric compatible, $\bigtriangledown \mathbf{g}=0,$ see see Definition %
\ref{deflcon}. The formula (\ref{lccon1}) states that the operator $%
\bigtriangledown $ can be defined on spaces provided with N--connection
structure (we use 'boldfaced' symbols) but this connection is not adapted to
the N--connection splitting \ (\ref{1wihit}). It is defined as a linear
connection but not as a d--connection, see Definition \ref{defdcon}. The
Levi--Civita connection is usually considered on (pseudo) Riemannian spaces
but it can be also introduced, for instance, in (co) vector/tangent bundles
both with respect to coordinate and anholonomic frames \cite{ma,v1,v2}. One
holds a Theorem similar to the Theorem \ref{tmetricity},

\begin{theorem}
If a space $\mathbf{V}^{n+m}$ is provided with both N--connection $\mathbf{N}
$\ and d--metric $\mathbf{g}$ structures, there is a unique linear symmetric
and torsionless connection $\mathbf{\bigtriangledown },$ being metric
compatible such that $\bigtriangledown _{\gamma }\mathbf{g}_{\alpha \beta
}=0 $ $\ $for $\mathbf{g}_{\alpha \beta }=\left( g_{ij},h_{ab}\right) ,$ see
(\ref{1block2}), with the coefficients
\begin{equation*}
\mathbf{\Gamma }_{\alpha \beta \gamma }^{\bigtriangledown }=\mathbf{g}\left(
\delta _{\alpha },\bigtriangledown _{\gamma }\delta _{\beta }\right) =%
\mathbf{g}_{\alpha \tau }\mathbf{\Gamma }_{\bigtriangledown \beta \gamma
}^{\tau },\,
\end{equation*}%
computed as
\begin{equation}
\mathbf{\Gamma }_{\alpha \beta \gamma }^{\bigtriangledown }=\frac{1}{2}\left[
\delta _{\beta }\mathbf{g}_{\alpha \gamma }+\delta _{\gamma }\mathbf{g}%
_{\beta \alpha }-\delta _{\alpha }\mathbf{g}_{\gamma \beta }+\mathbf{g}%
_{\alpha \tau }\mathbf{w}_{\gamma \beta }^{\tau }+\mathbf{g}_{\beta \tau }%
\mathbf{w}_{\alpha \gamma }^{\tau }-\mathbf{g}_{\gamma \tau }\mathbf{w}%
_{\beta \alpha }^{\tau }\right]  \label{1lcsym}
\end{equation}%
with respect to N--frames $\mathbf{e}_{\beta }\doteqdot \delta _{\beta }$ (%
\ref{1dder}) and N--coframes $\mathbf{\vartheta }_{\ }^{\alpha }\doteqdot
\delta ^{\alpha }$ (\ref{1ddif}).
\end{theorem}

The proof is that from Theorem \ref{tmetricity}, see also Refs. \cite%
{stw,mtw}, with $e_{\beta }\rightarrow \mathbf{e}_{\beta }$ and $\vartheta
^{\beta }\rightarrow \mathbf{\vartheta }^{\beta }$ substituted directly in
formula (\ref{lccoef}).

With respect to coordinate frames $\partial _{\beta }$ (\ref{1pder}) and $%
du^{\alpha }$ (\ref{1pdif}), the metric (\ref{1block2}) transforms
equivalently into (\ref{mstr}) with coefficients (\ref{1ansatz})\ and the
coefficients of (\ref{1lcsym}) transform into the usual Christoffel symbols (%
\ref{christ}). We emphasize that we shall use the coefficients just in the
form (\ref{1lcsym}) in order to compare the properties of different classes
of connections given with respect to N--adapted frames. The coordinate form (%
\ref{christ}) is not ''N--adapted'', being less convenient for geometric
constructions on spaces with anholonomic frames and associated N--connection
structure.

We can introduce the 1-form formalism and express
\begin{equation*}
\mathbf{\Gamma }_{\ \gamma \alpha }^{\bigtriangledown }=\mathbf{\Gamma }%
_{\gamma \alpha \beta }^{\bigtriangledown }\mathbf{\vartheta }^{\beta }
\end{equation*}%
where
\begin{equation}
\mathbf{\Gamma }_{\ \gamma \alpha }^{\bigtriangledown }=\frac{1}{2}\left[
\mathbf{e}_{\gamma }\rfloor \ \delta \mathbf{\vartheta }_{\alpha }-\mathbf{e}%
_{\alpha }\rfloor \ \delta \mathbf{\vartheta }_{\gamma }-\left( \mathbf{e}%
_{\gamma }\rfloor \ \mathbf{e}_{\alpha }\rfloor \ \delta \mathbf{\vartheta }%
_{\beta }\right) \wedge \mathbf{\vartheta }^{\beta }\right] ,
\label{1christa}
\end{equation}%
contains h- v-components, $\mathbf{\Gamma }_{\bigtriangledown \alpha \beta
}^{\gamma }=\left( L_{\bigtriangledown jk}^{i},L_{\bigtriangledown
bk}^{a},C_{\bigtriangledown jc}^{i},C_{\bigtriangledown bc}^{a}\right) ,$
defined similarly to (\ref{hcov}) and (\ref{vcov}) but using the operator $%
\bigtriangledown ,$
\begin{equation*}
L_{\bigtriangledown jk}^{i}=\left( \bigtriangledown _{k}\delta _{j}\right)
\rfloor d^{i},\quad L_{\bigtriangledown bk}^{a}=\left( \bigtriangledown
_{k}\partial _{b}\right) \rfloor \delta ^{a},\ C_{\bigtriangledown
jc}^{i}=\left( \bigtriangledown _{c}\delta _{j}\right) \rfloor d^{i},\quad
C_{\bigtriangledown bc}^{a}=\left( \bigtriangledown _{c}\partial _{b}\right)
\rfloor \delta ^{a}.
\end{equation*}%
In explicit form, the components $L_{\bigtriangledown
jk}^{i},L_{\bigtriangledown bk}^{a},C_{\bigtriangledown jc}^{i}$ and $%
C_{\bigtriangledown bc}^{a}$ are defined by formula (\ref{1christa}) if we
consider N--frame $\mathbf{e}_{\gamma }=\left( \delta _{i}=\partial
_{i}-N_{i}^{a}\partial _{a},\partial _{a}\right) $ and N--coframe $\mathbf{%
\vartheta }^{\beta }=(dx^{i},\delta y^{a}=dy^{a}+ N_{i}^{a}dx^{i})$ and a
d--metric $\mathbf{g}=\left( g_{ij,}h_{ab}\right) .$ In these formulas, we
write $\delta \mathbf{\vartheta }_{\alpha }$ instead of absolute
differentials $d\vartheta _{\alpha }$ from Refs. \cite{mag,rcg} because the
N--connection is considered. The coefficients (\ref{1christa}) transforms
into the usual Levi--Civita (or Christoffel) ones for arbitrary anholonomic
frames $e_{\gamma }$ and $\vartheta ^{\beta }$ and for a metric
\begin{equation*}
g=g_{\alpha \beta }\vartheta ^{\alpha }\otimes \vartheta ^{\beta }
\end{equation*}%
if $\mathbf{e}_{\gamma }\rightarrow e_{\gamma },$ $\mathbf{\vartheta }%
^{\beta }\rightarrow \vartheta ^{\beta }$ and $\delta \mathbf{\vartheta }%
_{\beta }\rightarrow d\vartheta _{\beta }.$

Finally, we note that if the N--connection structure is not trivial, we can
define arbitrary vielbein transforms starting from $\mathbf{e}_{\gamma }$
and $\mathbf{\vartheta }^{\beta },$ i. e. $e_{\alpha }^{[N]}=A_{\alpha }^{\
\alpha ^{\prime }}(u)\mathbf{e}_{\alpha ^{\prime }}$ and $\vartheta
_{\lbrack N]}^{\beta }=A_{\ \beta ^{\prime }}^{\beta }(u)\mathbf{\vartheta }%
^{\beta ^{\prime }}$ (we put the label $[N]$ in order to emphasize that such
object were defined by vielbein transforms starting from certain N--adapted
frames). This way we develop a general anholonomic frame formalism adapted
to the prescribed N--connection structure. If we consider geometric objects
with respect to coordinate frames $\mathbf{e}_{\alpha ^{\prime }}\rightarrow
\partial _{\underline{\alpha }}=\partial /\partial u^{\underline{\alpha }}$
and coframes $\mathbf{\vartheta }^{\beta ^{\prime }}\rightarrow du^{%
\underline{\beta }},$ the N--connection structure is 'hidden' in the
off--diagonal metric coefficients (\ref{1ansatz}) and performed geometric
constructions, in general, are not N--adapted.

\subsection{The canonical d--connection and the Levi--Civita connection}

The Levi--Civita connection $\bigtriangledown $ is constructed only from the
metric coefficients, being torsionless and satisfying the metricity
conditions $\bigtriangledown _{\alpha }g_{\beta \gamma }=0$. Because the
Levi--Civita connection is not adapted to the N--connection structure, we
can not state its coefficients in an irreducible form for the h-- and
v--subspaces. We need a type of d--connection which would be similar to the
Levi--Civita connection but satisfy certain metricity conditions adapted to
the N--connection.

\begin{proposition}
There are metric d--connections $\mathbf{D=}\left( D^{[h]},D^{[v]}\right) $
in a space \ $\mathbf{V}^{n+m},$ see (\ref{hvder}), satisfying the metricity
conditions if and only if
\begin{equation}
D_{k}^{[h]}g_{ij}=0,\ D_{a}^{[v]}g_{ij}=0,\ D_{k}^{[h]}h_{ab}=0,\
D_{a}^{[h]}h_{ab}=0.  \label{1mcas}
\end{equation}
\end{proposition}

The general proof of existence of such metric d--connections on vector
(super) bundles is given in Ref. \cite{ma}. Here we note that the equations (%
\ref{1mcas}) on $\mathbf{V}^{n+m}$ are just the conditions (\ref{1mca}). In
our case the existence may be proved by constructing an explicit example:

\begin{definition}
The canonical d--connection $\widehat{\mathbf{D}}$\ \ $\mathbf{=}\left(
\widehat{D}^{[h]},\widehat{D}^{[v]}\right) ,$ equivalently $\widehat{\mathbf{%
\Gamma }}_{\ \alpha }^{\gamma }=\widehat{\mathbf{\Gamma }}_{\ \alpha \beta
}^{\gamma }\mathbf{\vartheta }^{\beta },$\ is defined by the h--
v--irreducible components $\widehat{\mathbf{\Gamma }}_{\ \alpha \beta
}^{\gamma }=\left( \widehat{L}_{jk}^{i},\widehat{L}_{bk}^{a},\widehat{C}%
_{jc}^{i},\widehat{C}_{bc}^{a}\right) ,$%
\begin{eqnarray}
\widehat{L}_{jk}^{i} &=&\frac{1}{2}g^{ir}\left( \frac{\delta g_{jk}}{\delta
x^{k}}+\frac{\delta g_{kr}}{\delta x^{j}}-\frac{\delta g_{jk}}{\delta x^{r}}%
\right) ,  \label{1candcon} \\
\widehat{L}_{bk}^{a} &=&\frac{\partial N_{k}^{a}}{\partial y^{b}}+\frac{1}{2}%
h^{ac}\left( \frac{\delta h_{bc}}{\delta x^{k}}-\frac{\partial N_{k}^{d}}{%
\partial y^{b}}h_{dc}-\frac{\partial N_{k}^{d}}{\partial y^{c}}h_{db}\right)
,  \notag \\
\widehat{C}_{jc}^{i} &=&\frac{1}{2}g^{ik}\frac{\partial g_{jk}}{\partial
y^{c}},  \notag \\
\widehat{C}_{bc}^{a} &=&\frac{1}{2}h^{ad}\left( \frac{\partial h_{bd}}{%
\partial y^{c}}+\frac{\partial h_{cd}}{\partial y^{b}}-\frac{\partial h_{bc}%
}{\partial y^{d}}\right) .  \notag
\end{eqnarray}%
satisfying the torsionless conditions for the h--subspace and v--subspace,
respectively, $\widehat{T}_{jk}^{i}=\widehat{T}_{bc}^{a}=0.$
\end{definition}

By straightforward calculations with (\ref{1candcon}) we can verify that the
conditions (\ref{1mcas}) are satisfied and that the d--torsions are subjected
to the conditions $\widehat{T}_{jk}^{i}=\widehat{T}_{bc}^{a}=0$ (see section %
\ref{torscurv})). We emphasize that the canonical d--torsion posses
nonvanishing torsion components,%
\begin{equation*}
\widehat{T}_{.ji}^{a}=-\widehat{T}_{.ij}^{a}=\frac{\delta N_{i}^{a}}{\delta
x^{j}}-\frac{\delta N_{j}^{a}}{\delta x^{i}}=\Omega _{.ji}^{a},~\widehat{T}%
_{ja}^{i}=-\widehat{T}_{aj}^{i}=\widehat{C}_{.ja}^{i},~\widehat{T}%
_{.bi}^{a}=-\widehat{T}_{.ib}^{a}=\widehat{P}_{.bi}^{a}=\frac{\partial
N_{i}^{a}}{\partial y^{b}}-\widehat{L}_{.bj}^{a}
\end{equation*}%
induced by $\widehat{L}_{bk}^{a},$ $\widehat{C}_{jc}^{i}$ and N--connection
coefficients $N_{i}^{a}$ and their partial derivatives $\partial
N_{i}^{a}/\partial y^{b}$ (as is to be computed by introducing (\ref{1candcon}%
) in formulas (\ref{1dtorsb})). This is an anholonmic frame effect.

\begin{proposition}
\label{lccdc}The components of the Levi--Civita connection $\mathbf{\Gamma }%
_{\bigtriangledown \beta \gamma }^{\tau }$ and the irreducible components of
the canonical d--connection \ $\widehat{\mathbf{\Gamma }}_{\ \beta \gamma
}^{\tau }$\ are related by formulas%
\begin{equation}
\mathbf{\Gamma }_{\bigtriangledown \beta \gamma }^{\tau }=\left( \widehat{L}%
_{jk}^{i},\widehat{L}_{bk}^{a}-\frac{\partial N_{k}^{a}}{\partial y^{b}},%
\widehat{C}_{jc}^{i}+\frac{1}{2}g^{ik}\Omega _{jk}^{a}h_{ca},\widehat{C}%
_{bc}^{a}\right) ,  \label{lcsyma}
\end{equation}%
where $\Omega _{jk}^{a}$\ \ is the N--connection curvature\ (\ref{1ncurv}).
\end{proposition}

The proof follows from an explicit decomposition of N--adapted frame (\ref%
{1dder}) and N--adapted coframe (\ref{1ddif}) in (\ref{1lcsym}) (equivalently,
in (\ref{1christa})) and re--grouping the components as to distinguish the
h- and v-- irreducible values (\ref{1candcon}) for $\mathbf{g}_{\alpha \beta
}=\left( g_{ij},h_{ab}\right) .$

We conclude from (\ref{lcsyma}) that, in a trivial case, the Levi--Civita
and the canonical d--connection are given by the same h-- v-- components $%
\left( \widehat{L}_{jk}^{i},\widehat{L}_{bk}^{a},\widehat{C}_{jc}^{i},%
\widehat{C}_{bc}^{a}\right) $ if $\Omega _{jk}^{a}=0,\,$\ and $\partial
N_{k}^{a}/\partial y^{b}=0.$ This results in zero anholonomy coefficients (%
\ref{1anhc}) when the anholonomic N--basis is reduced to a holonomic one. It
should be also noted that even in this case some components of the
anholonomically induced by d--connection torsion $\widehat{\mathbf{T}}%
_{\beta \gamma }^{\alpha }$ could be nonzero (see formulas (\ref{1dtorsions})
 for $\widehat{\mathbf{\Gamma }}_{\ \beta \gamma }^{\tau }).$ For
instance, one holds the

\begin{corollary}
\label{ctlc}The d--tensor components
\begin{equation}
\widehat{T}_{.bi}^{a}=-\widehat{T}_{.ib}^{a}=\widehat{P}_{.bi}^{a}=\frac{%
\partial N_{i}^{a}}{\partial y^{b}}-\widehat{L}_{.bj}^{a}  \label{indtors}
\end{equation}%
for a canonical d--connection (\ref{1candcon}) can be nonzero even $\partial
N_{k}^{a}/\partial y^{b}=0$ and $\Omega _{jk}^{a}=0$ and a trivial equality
of the components of the canonical d--connection and of the Levi--Civita
connection, $\mathbf{\Gamma }_{\bigtriangledown \beta \gamma }^{\tau }=$ $%
\widehat{\mathbf{\Gamma }}_{\ \beta \gamma }^{\tau }$ holds  with
respect to coordinate frames.
\end{corollary}

This quite surprising fact follows from the anholonomic character of the
N--connection structure. If a N--connection is defined, there are imposed
specific types of constraints on the frame structure. This is important for
definition of d--connections (being adapted to the N--connection structure)
but not for the Levi--Civita connection which is not a d--connection. Even
such linear connections have the same components with respect to a
N--adapted (co) frame, they are very different geometrical objects because
they are subjected to different rules of transformation with respect to
frame and coordinate transforms. The d--connections' transforms are adapted
to those for the N--connection (\ref{ncontr}) but the Levi--Civita
connection is subjected to general rules of linear connection transforms (%
\ref{lcontr}).\footnote{%
The Corollary \ref{ctlc} is important for constructing various classes of
exact solutions with generic off--diagonal metrics in Einstein gravity, its
higher dimension and/or different gauge, Einstein-\--Cartan and
metric--affine generalizations. Certain type of ansatz were proven to result
in completely integrable gravitational field equations  for the
canonical d--connection (but not for the Levi--Civita one), see details in
Refs. \cite{v1,v2,vd,vncggf}. The induced d--torsion (\ref{indtors}) is
contained in the Ricci d--tensor $R_{ai}=\ ^{1}P_{ai}=P_{a.ib}^{.b},$ see (%
\ref{1dricci}), i. e. in the Einstein d--tensor constructed for the canonical
d--connection. If a class of solutions were obtained for a d--connection, we
can select those subclasses which satisfy the \ condition $\mathbf{\Gamma }%
_{\bigtriangledown \beta \gamma }^{\tau }=$ $\widehat{\mathbf{\Gamma }}_{\
\beta \gamma }^{\tau }$ with respect to a frame of reference. In this case
the nontrivial d--torsion $\widehat{T}_{.bi}^{a}$ (\ref{indtors}) can be
treated as an object constructed from some ''pieces'' of a generic
off--diagonal metric and related to certain components of the N--adapted
anholonomic frames.
\par
{}
\par
{}}

\begin{proposition}
\label{ptorsvlc}A canonical d--connection $\widehat{\mathbf{\Gamma }}_{\
\beta \gamma }^{\tau }$ defined by a N--connection $N_{i}^{a}$ and d--metric
$\mathbf{g}_{\alpha \beta }=\left[ g_{ij},h_{ab}\right] $ has zero
d--torsions \ (\ref{1dtorsions}) if an only if there are satisfied the
conditions $\Omega _{jk}^{a}=0,\widehat{C}_{jc}^{i}=0$ and $\widehat{L}%
_{.bj}^{a}=\partial N_{i}^{a}/\partial y^{b},$ i. e.\\ $\widehat{\mathbf{%
\Gamma }}_{\ \beta \gamma }^{\tau }=\left( \widehat{L}_{jk}^{i},\widehat{L}%
_{bk}^{a}=\partial N_{i}^{a}/\partial y^{b},0,\widehat{C}_{bc}^{a}\right) $
which is equivalent to
\begin{eqnarray}
g^{ik}\frac{\partial g_{jk}}{\partial y^{c}} &=&0,\   \label{cond01} \\
\frac{\delta h_{bc}}{\delta x^{k}}-\frac{\partial N_{k}^{d}}{\partial y^{b}}%
h_{dc}-\frac{\partial N_{k}^{d}}{\partial y^{c}}h_{db} &=&0,  \label{cond02}
\\
\frac{\partial N_{i}^{a}}{\partial x^{j}}-\frac{\partial N_{j}^{a}}{\partial
x^{i}}+N_{i}^{b}\frac{\partial N_{j}^{a}}{\partial y^{b}}-N_{j}^{b}\frac{%
\partial N_{i}^{a}}{\partial y^{b}} &=&0.  \label{1cond03}
\end{eqnarray}%
The Levi--Civita connection defined by the same N--connection and d--metric
structure with respect to N--adapted (co) frames has the components\\ $^{[0]}%
\mathbf{\Gamma }_{\bigtriangledown \beta \gamma }^{\tau }=$ $\widehat{%
\mathbf{\Gamma }}_{\ \beta \gamma }^{\tau }=\left( \widehat{L}_{jk}^{i},0,0,%
\widehat{C}_{bc}^{a}\right) .$
\end{proposition}

\textbf{Proof:} The relations (\ref{cond01})--(\ref{1cond03}) follows from
the condition of vanishing of d--torsion coefficients (\ref{1dtorsions}) when
the coefficients of the canonical d--connection and the Levi--Civita
connection are computed respectively following formulas (\ref{1candcon}) and (%
\ref{lcsyma})

We note a specific separation of variables in the equations (\ref{cond01})--(%
\ref{1cond03}). For instance, the equation (\ref{cond01}) is satisfied by any
$g_{ij}=g_{ij}\left( x^{k}\right) .$ We can search a subclass of
N--connections with $N_{j}^{a}=\delta _{j}N^{a}$, i. e. of 1--forms on the
h--subspace, $\widetilde{N}^{a}=\delta _{j}N^{a}dx^{i}$ which are closed on
this subspace,
\begin{equation*}
\delta \widetilde{N}^{a}=\frac{1}{2}\left( \frac{\partial N_{i}^{a}}{%
\partial x^{j}}-\frac{\partial N_{j}^{a}}{\partial x^{i}}+N_{i}^{b}\frac{%
\partial N_{j}^{a}}{\partial y^{b}}-N_{j}^{b}\frac{\partial N_{i}^{a}}{%
\partial y^{b}}\right) dx^{i}\wedge dx^{j}=0,
\end{equation*}%
satisfying the (\ref{1cond03}). Having defined such $N_{i}^{a}$ and computing
the values $\partial _{c}N_{i}^{a},$ we may try to solve (\ref{cond02})
rewritten as a system of first order partial differential equations
\begin{equation*}
\frac{\partial h_{bc}}{\partial x^{k}}=N_{k}^{e}\frac{\partial h_{bc}}{%
\partial y^{e}}+\partial _{b}N_{k}^{d}\ h_{dc}+\partial _{c}N_{k}^{d}h_{db}
\end{equation*}%
with known coefficients.$\blacksquare $

We can also associate the nontrivial values of $\widehat{\mathbf{T}}_{\beta
\gamma }^{\tau }$ (in particular cases, of $\widehat{T}_{.bi}^{a})$ to be
related to any algebraic equations in the Einstein--Cartan theory or
dynamical equations for torsion like in string or supergravity models. But
in this case we shall prescribe a specific class of anholonomically
constrained dynamics for the N--adapted frames.

Finally, we note that if a (pseudo) Riemannian space is provided with a
generic off--diagonal metric structure (see Remark \ref{rgod}) we can
consider alternatively to the Levi--Civita connection an infinite number of
metric d--connections, details in the section \ref{srgarcg}. Such
d--connections have nontrivial d--torsions $\mathbf{T}_{\beta \gamma }^{\tau
}$ induced by anholonomic frames and constructed from off--diagonal metric
terms and h- and v--components of d--metrics.

\subsection{The set of metric d--connections}

Let us define the set of all possible metric d--connections, satisfying the
conditions (\ref{1mcas}) and being constructed only form $g_{ij},h_{ab}$ and $%
N_{i}^{a}$ and their partial derivatives. Such d--connections satisfy
certain conditions for d--torsions that$\ T_{~jk}^{i}=T_{~bc}^{a}=0$ and
can be generated by two procedures of deformation of the connection
\begin{eqnarray*}
\widehat{\mathbf{\Gamma }}_{\ \alpha \beta }^{\gamma } &\rightarrow &\ ^{[K]}%
\mathbf{\Gamma }_{\ \alpha \beta }^{\gamma }=\mathbf{\Gamma }_{\ \alpha
\beta }^{\gamma }+\ ^{[K]}\mathbf{Z}_{\ \alpha \beta }^{\gamma }%
\mbox{\
(Kawaguchi's metrization \cite{kaw}) }, \\
\mbox{ or } &\rightarrow &^{[M]}\mathbf{\Gamma }_{\ \alpha \beta }^{\gamma }=%
\widehat{\mathbf{\Gamma }}_{\ \alpha \beta }^{\gamma }+\ ^{[M]}\mathbf{Z}_{\
\alpha \beta }^{\gamma }\mbox{\ (Miron's  connections \cite{ma} )}.
\end{eqnarray*}

\begin{theorem}
\ \label{kmp}Every deformation d--tensor (equivalently, distorsion, or
deflection) \
\begin{eqnarray*}
^{\lbrack K]}\mathbf{Z}_{\ \alpha \beta }^{\gamma } &=&\{\ ^{[K]}Z_{\
jk}^{i}=\frac{1}{2}g^{im}D_{j}^{[h]}g_{mk},\ ^{[K]}Z_{\ bk}^{a}=\frac{1}{2}%
h^{ac}D_{k}^{[h]}h_{cb},\  \\
&&\ ^{[K]}Z_{\ ja}^{i}=\frac{1}{2}g^{im}D_{a}^{[v]}g_{mj},\ ^{[K]}Z_{\
bc}^{a}=\frac{1}{2}h^{ad}D_{c}^{[v]}h_{db}\}
\end{eqnarray*}%
\ transforms a d--connection $\mathbf{\Gamma }_{\ \alpha \beta }^{\gamma
}=\left( L_{jk}^{i},L_{bk}^{a},C_{jc}^{i},C_{bc}^{a}\right) $\ (\ref{1dcon1})
into a metric d--connection%
\begin{equation*}
\ ^{[K]}\mathbf{\Gamma }_{\ \alpha \beta }^{\gamma }=\left( L_{jk}^{i}+\
^{[K]}Z_{\ jk}^{i},L_{bk}^{a}+\ ^{[K]}Z_{\ bk}^{a},C_{jc}^{i}+\ ^{[K]}Z_{\
ja}^{i},C_{bc}^{a}+\ ^{[K]}Z_{\ bc}^{a}\right) .
\end{equation*}%
\
\end{theorem}

The proof consists from a straightforward verification which demonstrate
that the conditions (\ref{1mcas}) are satisfied on $\mathbf{V}^{n+m}$ for $\
^{[K]}\mathbf{D=\{^{[K]}\mathbf{\Gamma }_{\alpha \beta }^{\gamma }\}}$ and $%
\mathbf{g}_{\alpha \beta }=\left( g_{ij},h_{ab}\right) .$ We note that the
Kawaguchi's metrization procedure contains additional covariant derivations
of the d--metric coefficients, defined by arbitrary d--connection, not only
N--adapted derivatives of the d--metric and N--connection coefficients as in
the case of the canonical d--connection.

\begin{theorem}
\label{mconnections}For a fixed d--metric structure \ (\ref{1block2}),\ $%
\mathbf{g}_{\alpha \beta }=\left( g_{ij},h_{ab}\right) ,$ on a space $%
\mathbf{V}^{n+m},$ the set of metric d--connections \ \ $^{[M]}\mathbf{%
\Gamma }_{\ \alpha \beta }^{\gamma }=\widehat{\mathbf{\Gamma }}_{\ \alpha
\beta }^{\gamma }+\ ^{[M]}\mathbf{Z}_{\ \alpha \beta }^{\gamma }$\ \ \ is
defined by the deformation d--tensor \
\begin{eqnarray*}
^{\lbrack M]}\mathbf{Z}_{\alpha \beta }^{\gamma } &=&\{\ ^{[M]}Z_{\
jk}^{i}=\ ^{[-]}O_{km}^{li}Y_{lj}^{m},\ ^{[M]}Z_{\ bk}^{a}=\
^{[-]}O_{bd}^{ea}Y_{ej}^{m},\  \\
&&\ ^{[M]}Z_{\ ja}^{i}=\ ^{[+]}O_{jk}^{mi}Y_{mc}^{k},\ ^{[M]}Z_{\ bc}^{a}=\
^{[+]}O_{bd}^{ea}Y_{ec}^{d}\}
\end{eqnarray*}%
where the so--called Obata operators are defined
\begin{equation*}
\ ^{[\pm ]}O_{km}^{li}=\frac{1}{2}\left( \delta _{k}^{l}\delta _{m}^{i}\pm
g_{km}g^{li}\right) \mbox{ and }\ ^{[\pm ]}O_{bd}^{ea}=\frac{1}{2}\left(
\delta _{b}^{e}\delta _{d}^{a}\pm h_{bd}h^{ea}\right)
\end{equation*}%
and \ $Y_{lj}^{m},$\ $Y_{ej}^{m},Y_{mc}^{k},$\ $Y_{ec}^{d}$ are arbitrary
d--tensor fields.
\end{theorem}

The proof consists from a direct verification of the fact that the
conditions (\ref{1mcas}) are satisfied on $\mathbf{V}^{n+m}$ for $\ ^{[M]}%
\mathbf{D=\{^{[M]}\mathbf{\Gamma }_{\ \alpha \beta }^{\gamma }\}.}$ We note
that the relation (\ref{lcsyma}) \ between the Levi--Civita and the
canonical d--connection is a particular case of $^{[M]}\mathbf{Z}_{\alpha
\beta }^{\gamma },$ when $Y_{lj}^{m},$\ $Y_{ej}^{m}$ and $Y_{ec}^{d}$ are
zero, but $Y_{mc}^{k}$ is taken to have $\ ^{[+]}O_{jk}^{mi}Y_{mc}^{k}=\frac{%
1}{2}g^{ik}\Omega _{jk}^{a}h_{ca}.$

There is a very important consequence of the Theorems \ref{kmp} and \ref%
{mconnections}: For a generic off--diagonal metric structure (\ref{1ansatz})
we can derive a N--connection structure $N_{i}^{a}$ with a d--metric $%
\mathbf{g}_{\alpha \beta }=\left( g_{ij},h_{ab}\right) $ (\ref{1block2}). So,
we may consider an infinite number of d--connections $\{\mathbf{D\},}$ all
constructed from the coefficients of the off--diagonal metrics, satisfying
the metricity conditions $\mathbf{D}_{\gamma }\mathbf{g}_{\alpha \beta }=0$
and having partial vanishing torsions, $T_{jk}^{i}=T_{bc}^{a}=0.$ The
covariant calculi associated to the set $\{\mathbf{D\}}$ are adapted to the
N--connection splitting and alternative to the covariant calculus defined by
the Levi--Civita connection $\bigtriangledown ,$ which is not adapted to the
N--connection.

\subsection{Nonmetricity in Finsler Geometry}

Usually, the N--connection, d--connection and d--metric in generalized
Finsler spaces satisfy certain metric compatibility conditions \cite%
{ma,fg,rund,as}. Nevertheless, there were considered some classes of
d--connections (for instance, related to the Berwald d--connection) with
nontrivial components of the nonmetricity d--tensor. Let us consider some
such examples modelled on metric--affine spaces.

\subsubsection{The Berwald d--connection}

\label{berwdcon}A d--connection of Berwald type (see, for instance, Ref. %
\cite{ma} on such configurations in Finsler and Lagrange geometry), $\ ^{[B]}%
\mathbf{\Gamma }_{\ \alpha }^{\gamma }=\ ^{[B]}\widehat{\mathbf{\Gamma }}_{\
\alpha \beta }^{\gamma }\mathbf{\vartheta }^{\beta },$ is defined by h- and
v--irreducible components
\begin{equation}
\ \ ^{[B]}\mathbf{\Gamma }_{\alpha \beta }^{\gamma }=\left( \widehat{L}_{\
jk}^{i},\frac{\partial N_{k}^{a}}{\partial y^{b}},0,\widehat{C}_{\
bc}^{a}\right) ,  \label{berw}
\end{equation}%
with $\widehat{L}_{\ jk}^{i}$ and $\widehat{C}_{\ bc}^{a}$ taken as in (\ref%
{1candcon}), satisfying only partial metricity compatibility conditions for a
d--metric \ (\ref{1block2}),\ $\mathbf{g}_{\alpha \beta }=\left(
g_{ij},h_{ab}\right) $ on space $\mathbf{V}^{n+m}$
\begin{equation*}
\ ^{[B]}D_{k}^{[h]}g_{ij}=0\mbox{ and }\ ^{[B]}D_{c}^{[v]}h_{ab}=0.
\end{equation*}%
This is an example of d--connections which may possess nontrivial
nonmetricity components, $\ ^{[B]}\mathbf{Q}_{\alpha \beta \gamma }=\left(
^{[B]}Q_{cij},\ ^{[B]}Q_{iab}\right) $ with
\begin{equation}
\ ^{[B]}Q_{cij}=\ ^{[B]}D_{c}^{[v]}g_{ij}\mbox{ and }\ ^{[B]}Q_{iab}=\
^{[B]}D_{i}^{[h]}h_{ab}.  \label{berwnm}
\end{equation}%
So, the Berwald d--connection defines a metric--affine space $\mathbf{V}%
^{n+m}$ with N--connection structure.

If $\widehat{L}_{\ jk}^{i}=0$ and $\widehat{C}_{\ bc}^{a}=0,$ we obtain a
Berwald type connection $\ $%
\begin{equation*}
^{\lbrack N]}\mathbf{\Gamma }_{\alpha \beta }^{\gamma }=\left( 0,\frac{%
\partial N_{k}^{a}}{\partial y^{b}},0,0\right)
\end{equation*}%
induced by the N--connection structures. It defines a vertical covariant
derivation $^{[N]}D_{c}^{[v]}$ acting in the v--subspace of $\mathbf{V}%
^{n+m},$ with the coefficients being partial derivatives on v--coordinates $%
y^{a}$ of the N--connection coefficients $N_{i}^{a}$ \cite{vilms}.

We can generalize the Berwald connection (\ref{berw}) to contain any fixed
values of d--torsions $T_{.jk}^{i}$ and $T_{.bc}^{a}$ from the h-
v--decomposition (\ref{1dtorsions}). We can check by a straightforward
calculations that the d--connection%
\begin{equation}
^{\lbrack B\tau ]}\mathbf{\Gamma }_{\ \alpha \beta }^{\gamma }=\left(
\widehat{L}_{\ jk}^{i}+\tau _{\ jk}^{i},\frac{\partial N_{k}^{a}}{\partial
y^{b}},0,\widehat{C}_{\ bc}^{a}+\tau _{\ bc}^{a}\right)  \label{bct}
\end{equation}%
with
\begin{eqnarray}
\tau _{\ jk}^{i} &=&\frac{1}{2}g^{il}\left(
g_{kh}T_{.lj}^{h}+g_{jh}T_{.lk}^{h}-g_{lh}T_{\ jk}^{h}\right)
\label{tauformulas} \\
\tau _{\ bc}^{a} &=&\frac{1}{2}h^{ad}\left( h_{bf}T_{\ dc}^{f}+h_{cf}T_{\
db}^{f}-h_{df}T_{\ bc}^{f}\right)  \notag
\end{eqnarray}%
results in $^{[B\tau ]}\mathbf{T}_{jk}^{i}=T_{.jk}^{i}$ and $^{[B\tau ]}%
\mathbf{T}_{bc}^{a}=T_{.bc}^{a}.$ The d--connection (\ref{bct}) has certain
nonvanishing irreducible nonmetricity components $^{[B\tau ]}\mathbf{Q}%
_{\alpha \beta \gamma }=\left( ^{[B\tau ]}Q_{cij},\ ^{[B\tau
]}Q_{iab}\right) .$

In general, by using the Kawaguchi metrization procedure (see Theorem \ref%
{kmp}) we can also construct metric d--connections with prescribed values of
d--torsions $T_{.jk}^{i}$ and $T_{.bc}^{a},$ or to express, for instance,
the Levi--Civita connection via coefficients of an arbitrary metric
d--connection (see details, for vector bundles, in \cite{ma}).

Similarly to formulas (\ref{1acn}), (\ref{1accn}) and (\ref{1distan}), we can
express a general affine Berwald d--connection $\ ^{[B\tau ]}\mathbf{D,}$ i.
e. $\ ^{[B\tau ]}\mathbf{\Gamma }_{\ \ \alpha }^{\gamma }=\ ^{[B\tau ]}%
\mathbf{\Gamma }_{\ \alpha \beta }^{\gamma }\mathbf{\vartheta }^{\beta },$
via its deformations from the Levi--Civita connection $\mathbf{\Gamma }%
_{\bigtriangledown \ \beta }^{\alpha },$
\begin{equation}
\ ^{[B\tau ]}\mathbf{\Gamma }_{\ \beta }^{\alpha }=\mathbf{\Gamma }%
_{\bigtriangledown \ \beta }^{\alpha }+\ ^{[B\tau ]}\mathbf{Z}_{\ \ \beta
}^{\alpha },  \label{accnb}
\end{equation}%
$\mathbf{\Gamma }_{\bigtriangledown \ \beta }^{\alpha }$ being expressed as (%
\ref{1christa}) (equivalently, defined by (\ref{1lcsym})) and
\begin{eqnarray}
\ ^{[B\tau ]}\mathbf{Z}_{\alpha \beta } &=&\mathbf{e}_{\beta }\rfloor \
^{[B\tau ]}\mathbf{T}_{\alpha }-\mathbf{e}_{\alpha }\rfloor \ ^{[B\tau ]}%
\mathbf{T}_{\beta }+\frac{1}{2}\left( \mathbf{e}_{\alpha }\rfloor \mathbf{e}%
_{\beta }\rfloor \ ^{[B\tau ]}\mathbf{T}_{\gamma }\right) \mathbf{\vartheta }%
^{\gamma }  \label{distanb} \\
&&+\left( \mathbf{e}_{\alpha }\rfloor \ ^{[B\tau ]}\mathbf{Q}_{\beta \gamma
}\right) \mathbf{\vartheta }^{\gamma }-\left( \mathbf{e}_{\beta }\rfloor \
^{[B\tau ]}\mathbf{Q}_{\alpha \gamma }\right) \mathbf{\vartheta }^{\gamma }+%
\frac{1}{2}\ ^{[B\tau ]}\mathbf{Q}_{\alpha \beta }.  \notag
\end{eqnarray}%
defined with prescribed d--torsions $^{[B\tau ]}\mathbf{T}%
_{jk}^{i}=T_{.jk}^{i}$ and $^{[B\tau ]}\mathbf{T}_{bc}^{a}=T_{.bc}^{a}.$
This Berwald d--connection can define a particular subclass of
metric--affine connections being adapted to the N--connection structure and
with prescribed values of d--torsions.

\subsubsection{The canonical/ Berwald metric--affine d--connections}

\label{berwdcona}If the deformations of d--metrics in formulas (\ref{1accn}) $%
\ $and (\ref{accnb}) are considered not with respect to the Levi--Civita
connection $\mathbf{\Gamma }_{\bigtriangledown \ \beta }^{\alpha }$ but with
respect to the canonical d--connection $\widehat{\mathbf{\Gamma }}_{\ \alpha
\beta }^{\gamma }$ with h- v--irreducible coefficients (\ref{1candcon}), we
can construct a set of canonical metric--affine d--connections. Such
metric--affine d--connections $\mathbf{\Gamma }_{\ \ \alpha }^{\gamma }=\
\mathbf{\Gamma }_{\ \alpha \beta }^{\gamma }\mathbf{\vartheta }^{\beta }$
are defined via deformations
\begin{equation}
\ \mathbf{\Gamma }_{\ \beta }^{\alpha }=\widehat{\mathbf{\Gamma }}_{\ \beta
}^{\alpha }+\ \widehat{\mathbf{Z}}_{\ \ \beta }^{\alpha },  \label{mafdc}
\end{equation}%
$\widehat{\mathbf{\Gamma }}_{\ \beta }^{\alpha }$ being the canonical
d--connection (\ref{1dcon1}) and
\begin{eqnarray}
\ \widehat{\mathbf{Z}}_{\alpha \beta } &=&\mathbf{e}_{\beta }\rfloor \
\mathbf{T}_{\alpha }-\mathbf{e}_{\alpha }\rfloor \ \mathbf{T}_{\beta }+\frac{%
1}{2}\left( \mathbf{e}_{\alpha }\rfloor \mathbf{e}_{\beta }\rfloor \ \mathbf{%
T}_{\gamma }\right) \mathbf{\vartheta }^{\gamma }  \label{dmafdc} \\
&&+\left( \mathbf{e}_{\alpha }\rfloor \ ^{[B\tau ]}\mathbf{Q}_{\beta \gamma
}\right) \mathbf{\vartheta }^{\gamma }-\left( \mathbf{e}_{\beta }\rfloor \
\mathbf{Q}_{\alpha \gamma }\right) \mathbf{\vartheta }^{\gamma }+\frac{1}{2}%
\ ^{[B\tau ]}\mathbf{Q}_{\alpha \beta }  \notag
\end{eqnarray}%
where $\mathbf{T}_{\alpha }$ and $\mathbf{Q}_{\alpha \beta }$ are arbitrary
torsion and nonmetricity structures.

A metric--affine d--connection $\mathbf{\Gamma }_{\ \ \alpha }^{\gamma }$
can be also considered as a deformation from the Berwald connection $%
^{[B\tau ]}\mathbf{\Gamma }_{\ \alpha \beta }^{\gamma }$
\begin{equation}
\ \mathbf{\Gamma }_{\ \beta }^{\alpha }=\ ^{[B\tau ]}\mathbf{\Gamma }_{\
\alpha \beta }^{\gamma }+\ ^{[B\tau ]}\ \widehat{\mathbf{Z}}_{\ \ \beta
}^{\alpha },  \label{mafbc}
\end{equation}%
$\ ^{[B\tau ]}\mathbf{\Gamma }_{\ \alpha \beta }^{\gamma }$ being the
Berwald d--connection (\ref{bct}) and
\begin{eqnarray}
\ ^{[B\tau ]}\ \widehat{\mathbf{Z}}_{\ \ \beta }^{\alpha } &=&\mathbf{e}%
_{\beta }\rfloor \ \mathbf{T}_{\alpha }-\mathbf{e}_{\alpha }\rfloor \
\mathbf{T}_{\beta }+\frac{1}{2}\left( \mathbf{e}_{\alpha }\rfloor \mathbf{e}%
_{\beta }\rfloor \ \mathbf{T}_{\gamma }\right) \mathbf{\vartheta }^{\gamma }
\label{dmafbc} \\
&&+\left( \mathbf{e}_{\alpha }\rfloor \ ^{[B\tau ]}\mathbf{Q}_{\beta \gamma
}\right) \mathbf{\vartheta }^{\gamma }-\left( \mathbf{e}_{\beta }\rfloor \
\mathbf{Q}_{\alpha \gamma }\right) \mathbf{\vartheta }^{\gamma }+\frac{1}{2}%
\ ^{[B\tau ]}\mathbf{Q}_{\alpha \beta }  \notag
\end{eqnarray}

The h- and v--splitting of formulas can be computed by introducing N--frames
$\mathbf{e}_{\gamma }=\left( \delta _{i}=\partial _{i}-N_{i}^{a}\partial
_{a},\partial _{a}\right) $ and N--coframes $\mathbf{\vartheta }^{\beta
}=\left( dx^{i},\delta y^{a}=dy^{a}+N_{i}^{a}dx^{i}\right) $ and d--metric $%
\mathbf{g}=\left( g_{ij,}h_{ab}\right) $ into (\ref{1christa}), (\ref{accnb})
and (\ref{distanb}) for the general Berwald d--connections. In a similar
form we can compute splitting by introducing the N--frames and d--metric
into \ (\ref{1dcon1}), (\ref{mafdc}) and (\ref{dmafdc}) \ for the metric
affine canonic d--connections and, respectively, into (\ref{bct}), (\ref%
{mafbc}) and (\ref{dmafbc}) for the metric--affine Berwald d--connections.
For the corresponding classes of d--connections, we can compute the torsion
and curvature tensors by introducing respective connections (\ref{1christa}),
(\ref{1accn}), (\ref{1candcon}), (\ref{berw}), (\ref{bct}), (\ref{accnb}), (%
\ref{mafdc}) and (\ref{mafbc}) into the general formulas for torsion (\ref%
{1dt}) and curvature (\ref{1dc}) on spaces provided with N--connection
structure.

\subsection{N--connections in metric--affine spaces}

In order to elaborate a unified MAG and generalized Finsler spaces scheme,
it is necessary to explain how the N--connection emerge in a metric--affine
space and/or in more particular cases of Riemann--Cartan and (pseudo)
Riemann geometry.

\subsubsection{Riemann geometry as a Riemann--Cartan geometry with
N--connection}

\label{srgarcg}It is well known \ the interpretation of the Riemann--Cartan
geometry as a generalization of the Riemannian geometry by distorsions (of
the Levi--Civita connection) generated by the torsion tensors \cite{rcg}.
Usually, the Riemann--Cartan geometry is described by certain geometric
relations between the torsion tensor, curvature tensor, metric and the
Levi--Civita connection on effective Riemann spaces. We can establish new
relations between the Riemann and Riemann--Cartan geometry if generic
off--diagonal metrics and anholonomic frames of reference are introduced
into consideration. Roughly speaking, a generic off--diagonal metric induces
alternatively to the well known Riemann spaces a certain class of
Riemann--Cartan geometries, with torsions completely defined by
off--diagonal metric terms and related anholonomic frame structures.

\begin{theorem}
\label{teqrgrcg} Any (pseudo) Riemannian spacetime provided with a generic
off--diagonal metric, defining the torsionless and metric Levi--Civita
connection, can be equivalently modelled as a Riemann--Cartan spacetime
provided with a canonical d--connection adapted to N--connection structure.
\end{theorem}

\textbf{Proof:}

Let us consider how the data for a (pseudo) Riemannian generic off--diagonal
metric $g_{\alpha \beta }$ parametrized in the form (\ref{1ansatz}) can
generate a Riemann--Cartan geometry. It is supposed that with respect to any
convenient anholonomic coframes (\ref{1ddif}) the metric is transformed into
a diagonalized form of type (\ref{1block2}), which gives the possibility to
define $N_{i}^{a}$ and $\mathbf{g}_{\alpha \beta }=\left[ g_{ij},h_{ab}%
\right] $ and to compute the aholonomy coefficients $\mathbf{w}_{\ \alpha
\beta }^{\gamma }$ (\ref{1anhc}) and the components of the canonical
d--connection $\widehat{\mathbf{\Gamma }}_{\ \alpha \beta }^{\gamma }=\left(
\widehat{L}_{jk}^{i},\widehat{L}_{bk}^{a},\widehat{C}_{jc}^{i},\widehat{C}%
_{bc}^{a}\right) $ (\ref{1candcon}). This connection has nontrivial
d--torsions\ $\widehat{\mathbf{T}}_{.\beta \gamma }^{\alpha },$ see the
Theorem \ref{tdtors} and Corollary \ref{ctlc}. In general, such d--torsions
are not zero being induced by the values $N_{i}^{a}$ and their partial
derivatives, contained in the former off--diagonal components of the metric (%
\ref{1ansatz}). So, the former Riemannian geometry, with respect to
anholonomic frames with associated N--connection structure, is equivalently
rewritten in terms of a Riemann--Cartan geometry with nontrivial torsion
structure.

We can provide an inverse construction when a diagonal d--metric (\ref%
{1block2}) is given with respect to an anholonomic coframe (\ref{1ddif})
defined from nontrivial values of N--connection coefficients, $N_{i}^{a}.$
The related Riemann--Cartan geometry is defined by the canonical
d--connection $\widehat{\mathbf{\Gamma }}_{\ \alpha \beta }^{\gamma }$
possessing nontrivial d--torsions $\widehat{\mathbf{T}}_{.\beta \gamma
}^{\alpha }.$ The data for this geometry with N--connection and torsion can
be directly transformed [even with respect to the same N--adapted (co)
frames] into the data of related (pseudo) Riemannian geometry by using the
relation (\ref{lcsyma}) between the components of $\widehat{\mathbf{\Gamma }}%
_{\ \alpha \beta }^{\gamma }$ and of the Levi--Civita connection $\mathbf{%
\Gamma }_{\bigtriangledown \beta \gamma }^{\tau }.\blacksquare $

\begin{remark}
{\ } \newline
a) Any generic off--diagonal (pseudo) Riemannian metric $g_{\alpha \beta
}[N_{i}^{a}]\rightarrow $ $\mathbf{g}_{\alpha \beta }=\left[ g_{ij},h_{ab}%
\right] $ induces an infinite number of associated Riemann--Cartan
geometries defined by sets of d--connections $\mathbf{D}=\{\mathbf{\Gamma }%
_{\ \alpha \beta }^{\gamma }\}$ which can be constructed according the
Kawaguchi's and, respectively, Miron's Theorems \ref{kmp} and \ref%
{mconnections}.{\ } \newline
b) For any metric d--connection $\mathbf{D}=\{\mathbf{\Gamma }_{\ \alpha
\beta }^{\gamma }\}$ induced by a generic off--diagonal metric (\ref{1ansatz}%
), we can define alternatively to the standard (induced by the Levi--Civita
connection) the Ricci d--tensor (\ref{1dricci}), $\mathbf{R}_{\alpha \beta },$
and the Einstein d--tensor (\ref{1deinst}), $\mathbf{G}_{\alpha \beta }.$
\end{remark}

We emphasize that all Riemann--Cartan geometries induced by metric
d--connections $\mathbf{D}$ are characterized not only by nontrivial induced
torsions $\mathbf{T}_{.\beta \gamma }^{\alpha }$ but also by corresponding
nonsymmetric Ricci d--tensor, $\mathbf{R}_{\alpha \beta },$ and Einstein
d--tensor, $\mathbf{G}_{\alpha \beta },$ for which $\mathbf{D}_{\gamma }%
\mathbf{G}_{\alpha \beta }\neq 0.$ This is not a surprising fact, because we
transferred the geometrical and physical objects on anholonomic spaces, when
the conservation laws should be redefined as to include the anholonomically
imposed constraints.

Finally, we conclude that for any generic off--diagonal (pseudo)\ Riemannian
metric we have two alternatives:\ 1) to choose the approach defined by the
Levi--Civita connection $\bigtriangledown $, with vanishing torsion and
usually defined conservation laws $\mathbf{\bigtriangledown }_{\gamma }%
\mathbf{G}_{\alpha \beta }^{[\bigtriangledown ]}=0,$ or 2) to diagonalize
the metric effectively, by respective anholonomic transforms, and transfer
the geometric and physical objects into effective Riemann--Cartan geometries
defined by corresponding N--connection and d--connection structures. All
types of such geometric constructions are equivalent. Nevertheless, one
could be defined certain priorities for some physical models like
''simplicity'' of field equations and definition of conservation laws and/or
the possibility to construct exact solutions. We note also that a variant
with induced torsions is more appropriate for including in the scheme
various type of generalized Finsler structures and/or models of (super)
string gravity containing nontrivial torsion fields.

\subsubsection{Metric--affine geometry and N--connections}

A general affine (linear) connection $D=\bigtriangledown +Z=\{\Gamma _{\beta
\alpha }^{\gamma }=\Gamma _{\bigtriangledown \beta \alpha }^{\gamma
}+Z_{\beta \alpha }^{\gamma }\}$
\begin{equation}
\Gamma _{\ \alpha }^{\gamma }=\Gamma _{\alpha \beta }^{\gamma }\vartheta
^{\beta },  \label{1ac}
\end{equation}%
can always be decomposed into the Riemannian $\Gamma _{\bigtriangledown \
\beta }^{\alpha }$ and post--Riemannian $Z_{\ \beta }^{\alpha }$ parts (see
Refs. \cite{mag} and, for irreducible decompositions to the effective
Einstein theory, see Ref. \cite{oveh}),
\begin{equation}
\Gamma _{\ \beta }^{\alpha }=\Gamma _{\bigtriangledown \ \beta }^{\alpha
}+Z_{\ \beta }^{\alpha }  \label{1acc}
\end{equation}%
where the distorsion 1-form $Z_{\ \beta }^{\alpha }$ is expressed in terms
of torsion and nonmetricity,%
\begin{equation}
Z_{\alpha \beta }=e_{\beta }\rfloor T_{\alpha }-e_{\alpha }\rfloor T_{\beta
}+\frac{1}{2}\left( e_{\alpha }\rfloor e_{\beta }\rfloor T_{\gamma }\right)
\vartheta ^{\gamma }+\left( e_{\alpha }\rfloor Q_{\beta \gamma }\right)
\vartheta ^{\gamma }-\left( e_{\beta }\rfloor Q_{\alpha \gamma }\right)
\vartheta ^{\gamma }+\frac{1}{2}Q_{\alpha \beta },  \label{1dista}
\end{equation}
$T_{\alpha }$ is defined as (\ref{dta}) and $Q_{\alpha \beta }\doteqdot
-Dg_{\alpha \beta }.$ \footnote{%
We note that our $\Gamma _{\ \alpha }^{\gamma }$ \ and $Z_{\ \beta }^{\alpha
}$ are respectively the $\Gamma _{\ \alpha }^{\gamma }$ and $N_{\alpha \beta
}$ from Ref. \cite{oveh}; in our works we use the symbol $N$ for
N--connections.}\ For $Q_{\beta \gamma }=0,$ we obtain from (\ref{1dista})
 the distorsion for the Riemannian--Cartan geometry \cite{rcg}.

By substituting arbitrary (co) frames, metrics and linear connections into
N--adapted ones (i. e. performing changes
\begin{equation*}
e_{\alpha }\rightarrow \mathbf{e}_{\alpha },\vartheta ^{\beta }\rightarrow
\mathbf{\vartheta }^{\beta },g_{\alpha \beta }\rightarrow \mathbf{g}_{\alpha
\beta }=\left( g_{ij},h_{ab}\right) ,\Gamma _{\ \alpha }^{\gamma
}\rightarrow \mathbf{\Gamma }_{\ \alpha }^{\gamma }
\end{equation*}%
with $\mathbf{Q}_{\alpha \beta }=\mathbf{Q}_{\gamma \alpha \beta }\mathbf{%
\vartheta }^{\gamma }$ and $\mathbf{T}^{\alpha }$ as in (\ref{1dt})) into
respective formulas (\ref{1ac}), (\ref{1acc}) and (\ref{1dista}), \ we can
define an affine connection $\mathbf{D=\bigtriangledown +Z}=\{\mathbf{\Gamma
}_{\ \beta \alpha }^{\gamma }\}$ with respect to N--adapted (co) frames,
\begin{equation}
\mathbf{\Gamma }_{\ \ \alpha }^{\gamma }=\mathbf{\Gamma }_{\ \alpha \beta
}^{\gamma }\mathbf{\vartheta }^{\beta },  \label{1acn}
\end{equation}%
with
\begin{equation}
\mathbf{\Gamma }_{\ \beta }^{\alpha }=\mathbf{\Gamma }_{\bigtriangledown \
\beta }^{\alpha }+\mathbf{Z}_{\ \ \beta }^{\alpha },  \label{1accn}
\end{equation}%
$\mathbf{\Gamma }_{\bigtriangledown \ \beta }^{\alpha }$ being expressed as (%
\ref{1christa}) (equivalently, defined by (\ref{1lcsym})) and $\mathbf{Z}_{\ \
\beta }^{\alpha }$ expressed as
\begin{equation}
\mathbf{Z}_{\alpha \beta }=\mathbf{e}_{\beta }\rfloor \mathbf{T}_{\alpha }-%
\mathbf{e}_{\alpha }\rfloor \mathbf{T}_{\beta }+\frac{1}{2}\left( \mathbf{e}%
_{\alpha }\rfloor \mathbf{e}_{\beta }\rfloor \mathbf{T}_{\gamma }\right)
\mathbf{\vartheta }^{\gamma }+\left( \mathbf{e}_{\alpha }\rfloor \mathbf{Q}%
_{\beta \gamma }\right) \mathbf{\vartheta }^{\gamma }-\left( \mathbf{e}%
_{\beta }\rfloor \mathbf{Q}_{\alpha \gamma }\right) \mathbf{\vartheta }%
^{\gamma }+\frac{1}{2}\mathbf{Q}_{\alpha \beta }.  \label{1distan}
\end{equation}%
The h-- and v--components of $\mathbf{\Gamma }_{\ \beta }^{\alpha }$ from (%
\ref{1accn}) consists from the components of $\mathbf{\Gamma }%
_{\bigtriangledown \ \beta }^{\alpha }$ (considered for (\ref{1christa})) and
of $\mathbf{Z}_{\alpha \beta }$ with $\mathbf{Z}_{\ \ \gamma \beta }^{\alpha
}=\left( Z_{jk}^{i},Z_{bk}^{a},Z_{jc}^{i},Z_{bc}^{a}\right) .$The values
\begin{equation*}
\mathbf{\Gamma }_{\bigtriangledown \gamma \beta }^{\alpha }+\mathbf{Z}_{\ \
\gamma \beta }^{\alpha }=\left( L_{\bigtriangledown
jk}^{i}+Z_{jk}^{i},L_{\bigtriangledown
bk}^{a}+Z_{bk}^{a},C_{\bigtriangledown
jc}^{i}+Z_{jc}^{i},C_{\bigtriangledown bc}^{a}+Z_{bc}^{a}\right)
\end{equation*}%
are defined correspondingly
\begin{eqnarray*}
L_{\bigtriangledown jk}^{i}+Z_{\ jk}^{i} &=&\left[ (\bigtriangledown
_{k}+Z_{k})\delta _{j}\right] \rfloor d^{i},\quad L_{\bigtriangledown
bk}^{a}+Z_{\ bk}^{a}=\left[ (\bigtriangledown _{k}+Z_{k})\partial _{b}\right]
\rfloor \delta ^{a}, \\
\ C_{\bigtriangledown jc}^{i}+Z_{\ jc}^{i} &=&\left[ \left( \bigtriangledown
_{c}+Z_{c}\right) \delta _{j}\right] \rfloor d^{i},\quad C_{\bigtriangledown
bc}^{a}+Z_{\ bc}^{a}=\left[ \left( \bigtriangledown _{c}+Z_{c}\right)
\partial _{b}\right] \rfloor \delta ^{a}.
\end{eqnarray*}%
and related to (\ref{1distan}) via h- and v--splitting of N--frames $\mathbf{e%
}_{\gamma }=\left( \delta _{i}=\partial _{i}-N_{i}^{a}\partial _{a},\partial
_{a}\right) $ and N--coframes $\mathbf{\vartheta }^{\beta }=\left(
dx^{i},\delta y^{a}=dy^{a}+N_{i}^{a}dx^{i}\right) $ and d--metric $\mathbf{g}%
=\left( g_{ij,}h_{ab}\right) .$

We note that for $\mathbf{Q}_{\alpha \beta }=0,$ the distorsion 1--form $%
\mathbf{Z}_{\alpha \beta }$ defines a Riemann--Cartan geometry adapted to
the N--connection structure.

Let us briefly outline the procedure of definition of N--connections in a
metric--affine space $V^{n+m}$ \ with arbitrary metric and connection
structures $\left( g^{[od]}=\{g_{\alpha \beta }\},\underline{\Gamma }_{\
\beta \alpha }^{\gamma }\right) $ and show how the geometric objects may be
adapted to the N--connection structure.

\begin{proposition}
\label{pmasnc}Every metric--affine space provided with a generic
off--diagonal metric structure admits nontrivial N--connections.
\end{proposition}

\textbf{Proof:} We give an explicit example how to introduce the
N--connection structure. We write the metric with respect to a local
coordinate basis,
\begin{equation*}
g^{[od]}=g_{\underline{\alpha }\underline{\beta }}du^{\underline{\alpha }%
}\otimes du^{\underline{\beta }},
\end{equation*}%
where the matrix $g_{\underline{\alpha }\underline{\beta }}$ contains a
non--degenerated $\left( m\times m\right) $ submatrix $h_{ab},$ for instance
like in ansatz (\ref{1ansatz}). Having fixed the block $h_{ab},$ labelled by
running of indices $a,b,...=n+1,n+2,...,n+m,$ we can define the $\left(
n\times n\right) $ bloc $g_{\underline{i}\underline{j}}$ with indices $%
\underline{i},\underline{j},...=1,2,...n.$ The next step is to find any
nontrivial $N_{i}^{a}$ (the set of coefficients has being defined, we may
omit underlying) and find $N_{\underline{j}}^{\underline{e}}$ from the $%
\left( n\times m\right) $ block relations $g_{\underline{j}\underline{a}}=N_{%
\underline{j}}^{\underline{e}}h_{\underline{a}\underline{e}}.$ This is
always possible if $g_{\underline{\alpha }\underline{\beta }}$ is generic
off--diagonal. The next step is to compute $g_{ij}=g_{\underline{i}%
\underline{j}}-N_{\underline{i}}^{\underline{a}}N_{\underline{j}}^{%
\underline{e}}h_{\underline{a}\underline{e}}$ which gives the possibility to
transform equivalently
\begin{equation*}
g^{[od]}\rightarrow \mathbf{g}=g_{ij}\mathbf{\vartheta }^{i}\otimes \mathbf{%
\vartheta }^{j}+h_{ab}\mathbf{\vartheta }^{a}\otimes \mathbf{\vartheta }^{b}
\end{equation*}%
where
\begin{equation*}
\mathbf{\vartheta }^{i}\doteqdot dx^{i},\ \mathbf{\vartheta }^{a}\doteqdot
\delta y^{a}=dy^{a}+N_{i}^{a}\left( u\right) dx^{i}
\end{equation*}%
are just the N--elongated differentials (\ref{1ddif}) if the local
coordinates associated to the block $h_{ab}$ are denoted by $y^{a}$ and the
rest ones by $x^{i}.$ We impose a global splitting of the metric--affine
spacetime by stating that all geometric objects are subjected to anholonomic
frame transforms with vielbein coefficients of type (\ref{1vt1}) and (\ref%
{1vt2}) defined by $\mathbf{N}=\{N_{i}^{a}\}.$ This way, we define on the
metric--affine space a vector/covector bundle structure if the coordinates $%
y^{a}$ are treated as certain local vector/ covector components.$%
\blacksquare $

We note, that having defined the values $\mathbf{\vartheta }^{\alpha
}=\left( \mathbf{\vartheta }^{i},\mathbf{\vartheta }^{b}\right) $ and their
duals $\mathbf{e}_{\alpha }=\left( \mathbf{e}_{i},\mathbf{e}_{a}\right) ,$
we can compute the linear connection coefficients with respect to N--adapted
(co) frames, $\Gamma _{\ \beta \alpha }^{\gamma }\rightarrow \widetilde{%
\mathbf{\Gamma }}_{\ \beta \alpha }^{\gamma }.$ However, $\widetilde{\mathbf{%
\Gamma }}_{\ \beta \alpha }^{\gamma },$ in general, is not a d--connection,
i. e. it is not \ adapted to the global splitting $T\mathbf{V}^{n+m}=h%
\mathbf{V}^{n+m}\oplus v\mathbf{V}^{n+m}$ defined by N--connection, see
Definition \ref{defdcon}. If the metric and linear connection are not
subjected to any field equations, we are free to consider distorsion tensors
in order to be able to apply the Theorems \ref{kmp} and/or \ref{mconnections}
with the aim to transform $\widetilde{\mathbf{\Gamma }}_{\ \beta \alpha
}^{\gamma }$ into a metric d--connection, or even into a Riemann--Cartan
d--connection. Here, we also note that a metric--affine space, in general,
admits different classes of N--connections with various nontrivial global
splitting $n^{\prime }+m^{\prime }=n+m,$ where $n^{\prime }\neq n.$

We can state from the very beginning that a metric--affine space $\mathbf{V}%
^{n+m}$ is \ provided with d--metric (\ref{1block2}) and d--connection
structure (\ref{1dcon1}) adapted to a class of prescribed vielbein
transforms (\ref{1vt1}) and (\ref{1vt2}) and N--elongated frames (\ref{1dder})
and (\ref{1ddif}). All constructions can be redefined with respect to
coordinate frames (\ref{1pder}) and (\ref{1pdif}) with off--diagonal metric
parametrization (\ref{1ansatz}) and then subjected to another frame and
coordinate transforms hiding the existing N--connection structure and
distinguished character of geometric objects. Such 'distinguished'
metric--affine spaces are characterized by corresponding N--connection
geometries and admit geometric constructions with distinguished objects.
They form a particular subclass of metric--affine spaces admitting
transformations of the general linear connection $\Gamma _{\ \beta \alpha
}^{\gamma }$ into certain classes of d--connections $\mathbf{\Gamma }_{\
\beta \alpha }^{\gamma }.$

\begin{definition}
\label{ddmas}A \ distinguished metric--affine space $\mathbf{V}^{n+m}$ is a
usual metric--affine space additionally enabled with a N--connection
structure $\mathbf{N}=\{N_{i}^{a}\}$ inducing splitting into respective
irreducible horizontal and vertical subspaces of dimensions $n$ and $m.$
This space is provided with independent d--metric (\ref{1block2}) and affine
d--connection (\ref{1dcon1}) structures adapted to the N--connection.
\end{definition}

The metric--affine spacetimes with stated N--connection structure are also
characterized by nontrivial anholonomy relations of type (\ref{1anhr}) with
anholonomy coefficients (\ref{1anhc}). This is a very specific type of
noncommutative symmetry generated by N--adapted (co) frames defining
different anholonomic noncommutative differential calculi (for details with
respect to the Einstein and gauge gravity see Ref. \cite{vnces}).

We construct and analyze explicit examples of metric--affine spacetimes with
associated N--connection (noncommutative) symmetry in Refs. \cite{exsolmag}.
A surprizing fact is that various types of d--metric ansatz (\ref{1block2})
with associated N--elongated frame (\ref{1dder}) and coframe (\ref{1ddif}) (or
equivalently, respective off--diagonal ansatz (\ref{1ansatz})) can be defined
as exact solutions in Einstein gravity of different dimensions and in
metric--affine, or Einstein--Cartan gravity and gauge model realizations.
Such solutions model also generalized Finsler structures.

\section{Generalized Finsler--Affine Spaces}

The aim of this section is to demonstrate that any well known type of
locally anistoropic or locally isotropic spaces can be defined as certain
particular cases of distinguished metric--affine spaces. We use the general
term of ''generalized Finsler--affine spaces'' for all type of geometries
modelled in MAG as generalizations of the Riemann--Cartan--Finsler geometry,
in general, containing nonmetricity fields. A complete classification of
such spaces is given by Tables 1--11 in the Appendix.

\subsection{Spaces with vanishing N--connection curvature}

Three examples of such spaces are given by the well known (pseudo) Riemann,
Riemann--Cartan or Kaluza--Klein manifolds of dimension $\left( n+m\right) $
provided with a gene\-ric off--diagonal metric structure $\underline{g}%
_{\alpha \beta }$ of type (\ref{1ansatz}), of corresponding signature, which
can be reduced equivalently to the block $\left( n\times n\right) \oplus
\left( m\times m\right) $ form (\ref{1block2}) via vielbein transforms (\ref%
{1vt1}). Their N--connection structures may be restricted by the condition $%
\Omega _{ij}^{a}=0,$ see (\ref{1ncurv}).

\subsubsection{Anholonomic (pseudo) Riemannian spaces}

\label{sarg}The (pseudo) Riemannian manifolds, $\mathbf{V}_{R}^{n+m},$
provided with a generic off--diagonal metric and anholonomic frame structure
effectively diagonalizing such a metric is an anholonomic (pseudo)
Riemannian space. The space admits associated N--connection structures with
coefficients induced by generic off--diagonal terms in the metric (\ref%
{1ansatz}). If the N--connection curvature vanishes, the Levi--Civita
connection is closely defined by the same coefficients as the canonical
d--connection (linear connections computed with respect to the N--adapted
(co) frames), see Proposition \ref{lccdc} and related discussions in section %
\ref{lconnections}. Following the Theorem \ref{teqrgrcg}, any (pseudo)
Riemannian space enabled with generic off--diagonal connection structure can
be equivalently modelled as an effective Riemann--Cartan geometry with induced
N--connection and d--torsions.

There were constructed a number of exact 'off--diagonal' solutions of the
Einstein equations \cite{v1,v2,vd}, for instance, in five dimensional
gravity (with various type restrictions to lower dimensions) with nontrivial
N--connection structure with ansatz for metric of type
\begin{eqnarray}
\mathbf{g} &=&\omega \left( x^{i},y^{4}\right)
[g_{1}(dx^{1})^{2}+g_{2}\left( x^{2},x^{3}\right) (dx^{2})^{2}+g_{3}\left(
x^{2},x^{3}\right) (dx^{3})^{2}  \notag \\
&&+h_{4}\left( x^{2},x^{3},y^{4}\right) \left( \delta y^{4}\right)
^{2}+h_{5}\left( x^{2},x^{3},y^{4}\right) \left( \delta y^{5}\right) ^{2}],
\label{pta}
\end{eqnarray}%
for $g_{1}=const,$ where%
\begin{equation*}
\delta y^{a}=dy^{a}+N_{k}^{a}\left( x^{i},y^{4}\right) dy^{a}
\end{equation*}%
with indices $i,j,k...=1,2,3$ and $a=4,5.$ The coefficients $N_{i}^{a}\left(
x^{i},y^{4}\right) $ were searched as a metric ansatz of type (\ref{1ansatz})
transforming equivalently into a certain diagonalized block (\ref{1block2})
would parametrize generic off--diagonal exact solutions. Such effective
N--connections are contained into a corresponding anholonomic moving or
static configuration of tetrads/ pentads (vierbeins/funfbeins) defining a
conventional splitting of coordinates into $n$ holonomic and $m$ anholonomic
ones, where $n+m=4,5.$ The ansatz (\ref{pta}) results in exact solutions of
vacuum and nonvacuum Einstein equations which can be integrated in general
form. Perhaps, all known at present time exact solutions in 3-5 dimensional
gravity can be included as particular cases in (\ref{pta}) and generalized
to anholonomic configurations with running constants and gravitational and
matter polarizations (in general, anisotropic on variable $y^{4})$ of the
metric and frame coefficients.

The vector/ tangent bundle configurations and/or torsion structures can be
effectively modelled on such (pseudo) Riemannian spaces by prescribing a
corresponding class of anholonomic frames. Such configurations are very
different from those, for instance, defined by Killing symmetries and the
induced torsion vanishes after frame transforms to coordinate bases. For a
corresponding parametrizations of $N_{i}^{a}(u)$ and $g_{\alpha \beta },$ we
can model Finsler like structures even in (pseudo) Riemannian spacetimes or
in gauge gravity \cite{vd,ncgg,vncggf}.

The anholonomic Riemannian spaces $\mathbf{V}_{R}^{n+m}$ can be considered as
 a subclass
of distinguished metric--affine spaces $\mathbf{V}^{n+m}$ provided with
N--connection structure, characterized by the condition that nonmetricity
d--filed $\mathbf{Q}_{\alpha \beta \gamma }=0$ and that a certain type of
induced torsions $\mathbf{T}_{\beta \gamma }^{\alpha }$ vanish for the
Levi--Civita connection. We can take a generic off--diagonal metric (\ref%
{1ansatz}), transform it into a d--metric (\ref{1block2}) and compute the h-
and v-components of the canonical d--connection (\ref{1dcon1}) and put them
into the formulas for d--torsions (\ref{1dtorsions}) and d--curvatures (\ref%
{1dcurv}). The vacuum solutions are defined by d--metrics and N--connections
satisfying the condition $\mathbf{R}_{\alpha \beta }=0,$ see the h--,
v--components (\ref{1dricci}).

In order to transform certain geometric constructions defined by the
canonical d--connection into similar ones for the Levi--Civita connection,
we have to constrain the N--connection structure as to have vanishing
N--curvature, $\Omega _{ij}^{a}=0,$ or to see the conditions when the
deformation of Levi--Civita connection to any d--connection result in
non--deformations of the Einstein equation. We obtain a (pseudo) Riemannian
vacuum spacetime with anholonomially induced d--torsion components
$$\widehat{%
T}_{ja}^{i}=-\widehat{T}_{aj}^{i}=\widehat{C}_{.ja}^{i} \mbox{ and }
\widehat{T}%
_{.bi}^{a}=-\widehat{T}_{.ib}^{a}=\partial N_{i}^{a}/\partial y^{b}-\widehat{%
L}_{.bj}^{a}.$$ This torsion can be related algebraically to a
spin source like in the usual Riemann--Cartan gravity if we want
to give an algebraic motivation to the N--connection splitting. We
emphasize that the
N--connection and d--metric coefficients can be chosen in order to model on $%
\mathbf{V}_{R}^{n+m}$ a special subclass of Finsler/ Lagrange structures
(see discussion in section \ref{mfls}).

\subsubsection{Kaluza--Klein spacetimes}

Such higher dimension generalizations of the Einstein gravity are
characterized by a metric ansatz
\begin{equation}
\underline{g}_{\alpha \beta }=\left[
\begin{array}{cc}
g_{ij}(x^{\kappa })+A_{i}^{a}(x^{\kappa })A_{j}^{b}(x^{\kappa
})h_{ab}(x^{\kappa },y^{a}) & A_{j}^{e}(x^{\kappa })h_{ae}(x^{\kappa },y^{a})
\\
A_{i}^{e}(x^{\kappa })h_{be}(x^{\kappa },y^{a}) & h_{ab}(x^{\kappa },y^{a})%
\end{array}%
\right]  \label{anskk}
\end{equation}%
(a particular case of the metric (\ref{1ansatz})) with certain
compactifications on extra dimension coordinates $y^{a}.$ The values $%
A_{i}^{a}(x^{\kappa })$ are considered to define gauge fields after
compactifications (the electromatgnetic potential in the original extension
to five dimensions by Kaluza and Klein, or some non--Abelian gauge fields
for higher dimension generalizations). Perhaps, the ansatz (\ref{anskk}) was
originally introduced in Refs. \cite{salam} (see \cite{ov} as a review of
non--supersymmetry models and \cite{salamsezgin} for supersymmetric
theories).

The coefficients $A_{i}^{a}(x^{\kappa })$ from (\ref{anskk}) are certain
particular parametrizations of the N--connection coefficients $%
N_{i}^{a}(x^{\kappa },y^{a})$ in (\ref{1ansatz}). This suggests a physical
interpretation for the N--connection as a specific nonlinear gauge field
depending both on spacetime and extra dimension coordinates (in general,
noncompactified). In the usual Kaluza--Klein (super) theories, there were
not considered anholonomic transforms to block d--metrics (\ref{1block2})
containing dependencies on variables $y^{a}.$

In some more general approaches, with additional anholonomic structures on
lower dimensional spacetime, there were constructed a set of exact vacuum
five dimensional solutions by reducing ansatz (\ref{anskk}) and their
generalizations of form (\ref{1ansatz}) to d--metric ansatz of type (\ref{pta}%
), see Refs. \cite{v1,v2,ncgg,vncggf,vd,vnces}. Such vacuum and nonvacuum
solutions describe anisotropically polarized Taub--NUT spaces, wormhole/
flux tube confiugurations, moving four dimensional black holes in bulk five
dimensional spacetimes, anisotropically deformed cosmological spacetimes and
various type of locally anisotropic spinor--soliton--dialton interactions in
generalized Kaluza--Klein and string/ brane gravity.

\subsubsection{Teleparallel spaces}

\label{stps}Teleparallel theories are usually defined by two geometrical
constraints \cite{tgm} (here, we introduce them for d--connections and
nonvanishing N--connection structure),%
\begin{equation}
\mathbf{R}_{~\beta }^{\alpha }=\delta \mathbf{\Gamma }_{~\beta }^{\alpha }+%
\mathbf{\Gamma }_{~\gamma }^{\alpha }\wedge \mathbf{\Gamma }_{~\beta
}^{\gamma }=0  \label{dtel1}
\end{equation}%
and
\begin{equation}
\mathbf{Q}_{\alpha \beta }=-\mathbf{Dg}_{\alpha \beta }=-\delta \mathbf{g}%
_{\alpha \beta }+\mathbf{\Gamma }_{~\beta }^{\gamma }\mathbf{g}_{\alpha
\gamma }+\mathbf{\Gamma }_{~\alpha }^{\gamma }\mathbf{g}_{\beta \gamma }=0.
\label{dtel2}
\end{equation}
The conditions (\ref{dtel1}) and (\ref{dtel2}) establish a distant
paralellism in such spaces because the result of a parallel transport of a
vector does not depend on the path (the angles and lengths being also
preserved under parallel transports). It is always possible to find such
anholonomic transforms $e_{\alpha }=A_{\alpha }^{~\underline{\beta }}e_{%
\underline{\beta }}$ and $e_{\underline{\alpha }}=A_{\underline{\alpha }%
}^{~\beta }e_{\beta },$ where $A_{\underline{\alpha }}^{~\beta }$ is inverse
to $A_{\alpha }^{~\underline{\beta }}$ when
\begin{equation*}
\mathbf{\Gamma }_{~\beta }^{\alpha }\rightarrow \mathbf{\Gamma }_{~%
\underline{\beta }}^{\underline{\alpha }}=A_{\underline{\beta }}^{~\beta }%
\mathbf{\Gamma }_{~\beta }^{\alpha }A_{~\alpha }^{\underline{\alpha }%
}+A_{~\gamma }^{\underline{\alpha }}\delta A_{\underline{\beta }}^{~\gamma
}=0
\end{equation*}%
and the transformed local metrics becomes the standard Minkowski,%
\begin{equation*}
g_{\underline{\alpha }\underline{\beta }}=diag\left( -1,+1,....,+1\right)
\end{equation*}%
(it can be fixed any signature). If the (co) frame is considered as the only
dynamical variable, it is called that the space (and choice of gauge) are of
Weitzenbock type. A coframe of type (\ref{1ddif})
\begin{equation*}
\mathbf{\vartheta }_{\ }^{\beta }\doteqdot \left( \delta x^{i}=dx^{i},\delta
y^{a}=dy^{a}+N_{i}^{a}\left( u\right) dx^{i}\right)
\end{equation*}%
is defined by N--connection coefficients. If we impose the condition of
vanishing the N--connection curvature, $\Omega _{ij}^{a}=0,$ see (\ref{1ncurv}%
), the N--connection defines a specific anholonomic dynamics because of
nontrivial anholonomic relations (\ref{1anhr}) with nonzero components (\ref%
{1anhc}).

By embedding teleparallel configurations into metric--affine spaces provided
with N--connection structure we state a distinguished class of (co) frame
fields adapted to this structure and open possibilities to include such
spaces into Finsler--affine ones, see section \ref{stpfa}. For vielbein
fields $\mathbf{e}_{\alpha }^{~\underline{\alpha }}$ and their inverses $%
\mathbf{e}_{~\underline{\alpha }}^{\alpha }$ related to the d--metric (\ref%
{1block2}),%
\begin{equation*}
\mathbf{g}_{\alpha \beta }=\mathbf{e}_{\alpha }^{~\underline{\alpha }}%
\mathbf{e}_{\beta }^{~\underline{\beta }}g_{\underline{\alpha }\underline{%
\beta }}
\end{equation*}%
we define the Weitzenbock d--connection%
\begin{equation}
~^{[W]}\mathbf{\Gamma }_{~\beta \gamma }^{\alpha }=\mathbf{e}_{~\underline{%
\alpha }}^{\alpha }\delta _{\gamma }\mathbf{e}_{\beta }^{~\underline{\alpha }%
},  \label{wcon}
\end{equation}%
where $\delta _{\gamma }$ is the N--elongated partial derivative (\ref{1dder}%
). It transforms in the usual Weitzenbock connection for trivial
N--connections. The torsion of $~^{[W]}\mathbf{\Gamma }_{~\beta \gamma
}^{\alpha }$ is defined
\begin{equation}
~^{[W]}\mathbf{T}_{~\beta \gamma }^{\alpha }=~^{[W]}\mathbf{\Gamma }_{~\beta
\gamma }^{\alpha }-~^{[W]}\mathbf{\Gamma }_{~\gamma \beta }^{\alpha }.
\label{wtors}
\end{equation}%
It posses h-- and v--irreducible components constructed from the components
of a d--metric and N--adapted frames. We can express
\begin{equation*}
~^{[W]}\mathbf{\Gamma }_{~\beta \gamma }^{\alpha }=\mathbf{\Gamma }%
_{\bigtriangledown ~\beta \gamma }^{\alpha }+\mathbf{Z}_{~\beta \gamma
}^{\alpha }
\end{equation*}%
where $\mathbf{\Gamma }_{\bigtriangledown ~\beta \gamma }^{\alpha }$ is the
Levi--Civita connection (\ref{1lcsym}) and the contorsion tensor is
\begin{equation*}
\mathbf{Z}_{\alpha \beta }=\mathbf{e}_{\beta }\rfloor ~^{[W]}\mathbf{T}%
_{\alpha }-\mathbf{e}_{\alpha }\rfloor ~^{[W]}\mathbf{T}_{\beta }+\frac{1}{2}%
\left( \mathbf{e}_{\alpha }\rfloor \mathbf{e}_{\beta }\rfloor ~^{[W]}\mathbf{%
T}_{\gamma }\right) \mathbf{\vartheta }^{\gamma }+\left( \mathbf{e}_{\alpha
}\rfloor \mathbf{Q}_{\beta \gamma }\right) \mathbf{\vartheta }^{\gamma
}-\left( \mathbf{e}_{\beta }\rfloor \mathbf{Q}_{\alpha \gamma }\right)
\mathbf{\vartheta }^{\gamma }+\frac{1}{2}\mathbf{Q}_{\alpha \beta }.
\end{equation*}%
In formulation of teleparallel alternatives to the general relativity it is
considered that $\mathbf{Q}_{\alpha \beta }=0.$

\subsection{Finsler and Finsler--Riemann--Cartan spaces}

\label{ssffc}The first approaches to Finsler spaces \cite{fg,rund} were
developed by generalizing the usual Riemannian metric interval
\begin{equation*}
ds=\sqrt{g_{ij}\left( x\right) dx^{i}dx^{j}}
\end{equation*}%
on a manifold $M$ \ of dimension $n$ into a nonlinear one
\begin{equation}
ds=F\left( x^{i},dx^{j}\right)  \label{finint}
\end{equation}%
defined by the Finsler metric $F$ (fundamental function) on $\widetilde{TM}%
=TM\backslash \{0\}$ (it should be noted an ambiguity in terminology used in
monographs on Finsler geometry and on gravity theories with respect to such
terms as Minkowski space, metric function and so on). It is also considered
a quadratic form on $\R^{2}$ with coefficients
\begin{equation}
\ g_{ij}^{[F]}\rightarrow h_{ab}=\frac{1}{2}\frac{\partial ^{2}F^{2}}{%
\partial y^{i}\partial y^{j}}  \label{finm2}
\end{equation}%
defining a positive definite matrix. The local coordinates are denoted $%
u^{\alpha }=(x^{i},y^{a}\rightarrow y^{i}).$ There are satisfied the
conditions: 1) The Finsler metric on a real manifold $M$ is a function $%
F:TM\rightarrow \R$ which on $\widetilde{TM}=TM\backslash \{0\}$ is of class
$C^{\infty }$ and $F$ is only continuous on the image of the null
cross--sections in the tangent bundle to $M.$ 2) $F\left( x,\chi y\right)
=\chi F\left( x,y\right) $ for every $\R_{+}^{\ast }.$ 3) The restriction of
$F$ to $\widetilde{TM}$ is a positive function. 4) $rank\left[
g_{ij}^{[F]}(x,y)\right] =n.$

There were elaborated a number of models of locally anisotropic spacetime
geometry with broken local Lorentz invariance (see, for instance, those
based on Finsler geometries \cite{as,bog}). In result, in the Ref. \cite%
{will}, it was ambiguously concluded that Finsler gravity models are very
restricted by experimental data. Recently, the subject concerning Lorentz
symmetry violations was revived for instance in brane gravity \cite{groj}
(see a detailed analysis and references on such theoretical and experimental
researches in \cite{kost}). In this case, the Finsler like geometries
broking the local four dimensional Lorentz invariance can be considered as a
possible alternative direction for investigating physical models both with
local anisotropy and violation of local spacetime symmetries. But it should
be noted here that violations of postulates of general relativity is not a
generic property of the so--called ''Finsler gravity''. A subclass of
Finsler geometries and their generalizations could be induced by anholonomic
frames even in general relativity theory and Riemannian--Cartan or gauge
gravity \cite{vd,ncgg,vncggf,vsp}. The idea is that instead of geometric
constructions based on straightforward applications of derivatives of (\ref%
{finm2}), following from a nonlinear interval (\ref{finint}), we should
consider d--metrics (\ref{1block2}) with the coefficients from Finsler
geometry (\ref{finm2}) or their extended variants. In this case, certain
type Finsler configurations can be defined even as exact 'off--diagonal'
solutions in vacuum Einstein gravity or in string gravity.

\subsubsection{Finsler geometry and its almost Kahlerian model}

We outline a modern approach to Finsler geometry \cite{ma} based on the
geometry of nonlinear connections in tangent bundles.

A real (commutative) Finsler space $\mathbf{F}^{n}=\left( M,F\left(
x,y\right) \right) $ can be modelled on a tangent bundle $TM$ enabled with a
Finsler metric $F\left( x^{i},y^{j}\right) $ and a quadratic form $%
g_{ij}^{[F]}$ (\ref{finm2}) satisfying the mentioned conditions and defining
the Christoffel symbols (not those from the usual Riemannian geometry)%
\begin{equation*}
c_{jk}^{\iota }(x,y)=\frac{1}{2}g^{ih}\left( \partial
_{j}g_{hk}^{[F]}+\partial _{k}g_{jh}^{[F]}-\partial _{h}g_{jk}^{[F]}\right) ,
\end{equation*}%
where $\partial _{j}=\partial /\partial x^{j},$ and the Cartan nonlinear
connection
\begin{equation}
\ ^{[F]}\mathbf{N}_{j}^{i}(x,y)=\frac{1}{4}\frac{\partial }{\partial y^{j}}%
\left[ c_{lk}^{\iota }(x,y)y^{l}y^{k}\right] ,  \label{1ncc}
\end{equation}%
where we do not distinguish the v- and h- indices taking on $TM$ \ the same
values.

In Finsler geometry, there were investigated different classes of remarkable
Finsler linear connections introduced by Cartan, Berwald, Matsumoto and
other geometers (see details in Refs. \cite{fg,as,rund}). Here we note that
we can introduce $g_{ij}^{[F]}=g_{ab}$ and $\ ^{[F]}\mathbf{N}_{j}^{i}(x,y)$
in (\ref{1ansatz}) and transfer our considerations to a $\left( n\times
n\right) \oplus \left( n\times n\right) $ blocks of type (\ref{1block2}) for
a metric--affine space $V^{n+n}.$

A usual Finsler space $\mathbf{F}^{n}=\left( M,F\left( x,y\right) \right) $
is completely defined by its fundamental tensor $g_{ij}^{[F]}(x,y)$ and the
Cartan nonlinear connection $\ ^{[F]}\mathbf{N}_{j}^{i}(x,y)$ and any chosen
d--connection structure (\ref{1dcon1}) (see details on different type of
d--connections in section \ref{lconnections}). Additionally, the
N--connection allows us to define an almost complex structure $I$ on $TM$ as
follows%
\begin{equation*}
I\left( \delta _{i}\right) =-\partial /\partial y^{i}\mbox{ and
}I\left( \partial /\partial y^{i}\right) =\delta _{i}
\end{equation*}%
for which $I^{2}=-1.$

The pair $\left( g^{[F]},I\right) $ consisting from a Riemannian metric on a
tangent bundle $TM,$%
\begin{equation}
\mathbf{g}^{[F]}=g_{ij}^{[F]}(x,y)dx^{i}\otimes
dx^{j}+g_{ij}^{[F]}(x,y)\delta y^{i}\otimes \delta y^{j}  \label{1dmetricf}
\end{equation}%
and the almost complex structure $I$ defines an almost Hermitian structure
on $\widetilde{TM}$ associated to a 2--form%
\begin{equation*}
\theta =g_{ij}^{[F]}(x,y)\delta y^{i}\wedge dx^{j}.
\end{equation*}%
This model of Finsler geometry is called almost Hermitian and denoted $%
H^{2n} $ and it is proven \cite{ma} that is almost Kahlerian, i. e. the form
$\theta $ is closed. The almost Kahlerian space $\mathbf{K}^{2n}=\left(
\widetilde{TM},\mathbf{g}^{[F]},I\right) $ is also called the almost
Kahlerian model of the Finsler space $F^{n}.$

On Finsler spaces (and their almost Kahlerian models), one distinguishes the
almost Kahler linear connection of Finsler type, $\mathbf{D}^{[I]}$ on $%
\widetilde{TM}$ with the property that this covariant derivation preserves
by parallelism the vertical distribution and is compatible with the almost
Kahler structure $\left( \mathbf{g}^{[F]},I\right) ,$ i.e.
\begin{equation*}
\mathbf{D}_{X}^{[I]}\mathbf{g}^{[F]}=0\mbox{ and }\mathbf{D}_{X}^{[I]}%
\mathbf{I}=0
\end{equation*}%
for \ every d--vector field on $\widetilde{TM}.$ This d--connection is
defined by the data
\begin{equation}
\ ^{[F]}\widehat{\mathbf{\Gamma }}_{\ \beta \gamma }^{\alpha }=\left( \
^{[F]}\widehat{L}_{jk}^{i},\ ^{[F]}\widehat{L}_{jk}^{i},\ ^{[F]}\widehat{C}%
_{jk}^{i},\ ^{[F]}\widehat{C}_{jk}^{i}\right)  \label{dccfs}
\end{equation}%
with $\ ^{[F]}\widehat{L}_{jk}^{i}$ and $\ ^{[F]}\widehat{C}_{jk}^{i}$
computed by similar formulas in (\ref{1candcon}) by using $g_{ij}^{[F]}$ as
in (\ref{finm2}) and $\ ^{[F]}N_{j}^{i}$ from (\ref{1ncc}).

We emphasize that a Finsler space $\mathbf{F}^{n}$ with a d--metric (\ref%
{1dmetricf}) and Cartan's N--connection structure (\ref{1ncc}), or the
corresponding almost Hermitian (Kahler) model $\mathbf{H}^{2n},$ can be
equivalently modelled on a space of dimension $2n,$ $\mathbf{V}^{n+n},$
provided with an off--diagonal metric (\ref{1ansatz}) and anholonomic frame
structure with associated Cartan's nonlinear connection. Such anholonomic
frame constructions are similar to modelling of the Einstein--Cartan geometry
on (pseudo) Riemannian spaces where the torsion is considered as an
effective tensor field. From this viewpoint a Finsler geometry is a
Riemannian--Cartan geometry defined on a tangent bundle provided with a
respective off--diagonal metric (and a related anholonomic frame structure
with associated N--connection) and with additional prescriptions with
respect to the type of linear connection chosen to be compatible with the
metric and N--connection structures.

\subsubsection{Finsler--Kaluza--Klein spaces}

In Ref. \cite{vncggf} we defined a 'locally anisotropic' toroidal
compactification of the 10 dimensional heterotic string sction \cite{kir}.
We consider here the corresponding anholonomic frame transforms and
off--diagonal metric ansatz. Let $\left( n^{\prime },m^{\prime }\right) $ be
the (holonomic, anholonomic) dimensions of the compactified spacetime (as a
particular case we can state $n^{\prime }+m^{\prime }=4,$ or any integers $%
n^{\prime }+m^{\prime }<10,$ for instance, for brane configurations). There
are used such parametrizations of indices and of vierbeinds: Greek indices $%
\alpha ,\beta ,...\mu ...$ run values for a 10 dimensional spacetime and
split as $\alpha =\left( \alpha ^{\prime },\widehat{\alpha }\right) ,\beta
=\left( \beta ^{\prime },\widehat{\beta }\right) ,...$ when primed indices $%
\alpha ^{\prime },\beta ^{\prime },...\mu ^{\prime }...$ run values for
compactified spacetime and split into h- and v--components like $\alpha
^{\prime }=\left( i^{\prime },a^{\prime }\right) ,$ $\beta ^{\prime }=\left(
j^{\prime },b^{\prime }\right) ,...;$ the frame coefficients are split as
\begin{equation}
e_{\mu }^{~\underline{\mu }}(u)=\left(
\begin{array}{cc}
e_{\alpha ^{\prime }}^{~\underline{\alpha ^{\prime }}}(u^{\beta ^{\prime }})
& A_{\alpha ^{\prime }}^{\widehat{\alpha }}(u^{\beta ^{\prime }})e_{\widehat{%
\alpha }}^{~\underline{\widehat{\alpha }}}(u^{\beta ^{\prime }}) \\
0 & e_{\widehat{\alpha }}^{~\underline{\widehat{\alpha }}}(u^{\beta ^{\prime
}})%
\end{array}%
\right)  \label{vt1a}
\end{equation}%
where $e_{\alpha ^{\prime }}^{~\underline{\alpha ^{\prime }}}(u^{\beta
^{\prime }}),$ in their turn, are taken in the form (\ref{1vt1}),
\begin{equation}
e_{\alpha ^{\prime }}^{~\underline{\alpha ^{\prime }}}(u^{\beta ^{\prime
}})=\left(
\begin{array}{cc}
e_{i^{\prime }}^{~\underline{i^{\prime }}}(x^{j^{\prime }},y^{a^{\prime }})
& N_{i^{\prime }}^{a^{\prime }}(x^{j^{\prime }},y^{a^{\prime }})e_{a^{\prime
}}^{~\underline{a^{\prime }}}(x^{j^{\prime }},y^{a^{\prime }}) \\
0 & e_{a^{\prime }}^{~\underline{a^{\prime }}}(x^{j^{\prime }},y^{a^{\prime
}})%
\end{array}%
\right) .  \label{1frame8}
\end{equation}%
For the metric%
\begin{equation}
\mathbf{g}=\underline{g}_{\alpha \beta }du^{\alpha }\otimes du^{\beta }
\label{auxm01}
\end{equation}%
we have the recurrent ansatz%
\begin{equation}
\underline{g}_{\alpha \beta }=\left[
\begin{array}{cc}
g_{\alpha ^{\prime }\beta ^{\prime }}(u^{\beta ^{\prime }})+A_{\alpha
^{\prime }}^{\widehat{\alpha }}(u^{\beta ^{\prime }})A_{\beta ^{\prime }}^{%
\widehat{\beta }}(u^{\beta ^{\prime }})h_{\widehat{\alpha }\widehat{\beta }%
}(u^{\beta ^{\prime }}) & h_{\widehat{\alpha }\widehat{\beta }}(u^{\beta
^{\prime }})A_{\alpha ^{\prime }}^{\widehat{\alpha }}(u^{\beta ^{\prime }})
\\
h_{\widehat{\alpha }\widehat{\beta }}(u^{\beta ^{\prime }})A_{\beta ^{\prime
}}^{\widehat{\beta }}(u^{\beta ^{\prime }}) & h_{\widehat{\alpha }\widehat{%
\beta }}(u^{\beta ^{\prime }})%
\end{array}%
\right] ,  \label{metr8a}
\end{equation}%
where%
\begin{equation}
g_{\alpha ^{\prime }\beta ^{\prime }}=\left[
\begin{array}{cc}
g_{i^{\prime }j^{\prime }}(u^{\beta ^{\prime }})+N_{i^{\prime }}^{a^{\prime
}}(u^{\beta ^{\prime }})N_{j^{\prime }}^{b^{\prime }}(u^{\beta ^{\prime
}})h_{a^{\prime }b^{\prime }}(u^{\beta ^{\prime }}) & h_{a^{\prime
}b^{\prime }}(u^{\beta ^{\prime }})N_{i^{\prime }}^{a^{\prime }}(u^{\beta
^{\prime }}) \\
h_{a^{\prime }b^{\prime }}(u^{\beta ^{\prime }})N_{j^{\prime }}^{b^{\prime
}}(u^{\beta ^{\prime }}) & h_{a^{\prime }b^{\prime }}(u^{\beta ^{\prime }})%
\end{array}%
\right] .  \label{1metr8}
\end{equation}

After a toroidal compactification on $u^{\widehat{\alpha }}$ with gauge
fields $A_{\alpha ^{\prime }}^{\widehat{\alpha }}(u^{\beta ^{\prime }}),$
generated by the frame transform (\ref{vt1a}), we obtain a metric (\ref%
{auxm01}) like in the usual Kaluza--Klein theory (\ref{anskk}) but
containing the values $g_{\alpha ^{\prime }\beta ^{\prime }}(u^{\beta
^{\prime }}),$\ defined as in (\ref{1metr8}) (in a generic off--diagonal form
similar to (\ref{1ansatz}), labelled by primed indices), which can be induced
as in Finsler geometry. This is possible if $g_{i^{\prime }j^{\prime
}}(u^{\beta ^{\prime }}),h_{a^{\prime }b^{\prime }}(u^{\beta ^{\prime
}})\rightarrow g_{i^{\prime }j^{\prime }}^{[F]}(x^{\prime },y^{\prime })$
(see (\ref{finm2})) and $N_{i^{\prime }}^{a^{\prime }}(u^{\beta ^{\prime
}})\rightarrow N_{j^{\prime }}^{[F]i^{\prime }}(x^{\prime },y^{\prime })$
(see (\ref{1ncc})) inducing a Finsler space with ''primed'' labels for
objects. Such locally anisotropic spacetimes (in this case we emphasized the
Finsler structures) can be generated anisotropic toroidal compactifications
from different models of higher dimension of gravity (string, brane, or
usual Kaluza--Klein theories). \ They define a mixed variant of Finsler and
Kaluza--Klein spaces.

By using the recurrent ansatz (\ref{metr8a}) and (\ref{1metr8}), we can
generate both nontrivial nonmetricity and prescribed torsion structures
adapted to a corresponding N--connection $N_{i^{\prime }}^{a^{\prime }}.$
For instance, (after topological compactification on higher dimension) we
can prescribe in the lower dimensional spacetime certain torsion fields $%
T_{\ i^{\prime }j^{\prime }}^{k^{\prime }}$ and $T_{\ b^{\prime }c^{\prime
}}^{a^{\prime }}$ (they could have a particular relation to the so called $B$%
--fields in string theory, or connected to other models). The next steps are
to compute $\tau _{\ i^{\prime }j^{\prime }}^{k^{\prime }}$ and $\tau _{\
b^{\prime }c^{\prime }}^{a^{\prime }}$ by using formulas (\ref{tauformulas})
and define
\begin{equation}
^{\lbrack B\tau ]}\mathbf{\Gamma }_{\alpha ^{\prime }\beta ^{\prime
}}^{\gamma ^{\prime }}=\left( L_{j^{\prime }k^{\prime }}^{i^{\prime }}=%
\widehat{L}_{\ j^{\prime }k^{\prime }}^{i^{\prime }}+\tau _{\ j^{\prime
}k^{\prime }}^{i^{\prime }},\ L_{.b^{\prime }k^{\prime }}^{a^{\prime }}=%
\frac{\partial N_{k^{\prime }}^{a^{\prime }}}{\partial y^{b^{\prime }}}%
,C_{.j^{\prime }a^{\prime }}^{i^{\prime }}=0,C_{b^{\prime }c^{\prime
}}^{a^{\prime }}=\widehat{C}_{\ b^{\prime }c^{\prime }}^{a^{\prime }}+\tau
_{\ b^{\prime }c^{\prime }}^{a^{\prime }}\right)  \label{berwprim}
\end{equation}%
as in (\ref{bct}) \ (all formulas being with primed indices and $\widehat{L}%
_{\ j^{\prime }k^{\prime }}^{i^{\prime }}$ and $\widehat{C}_{\ b^{\prime
}c^{\prime }}^{a^{\prime }}$ defined as in (\ref{1candcon})). This way we can
generate from Kaluza--Klein/ string theory a Berwald spacetime with
nontrivial N--adapted nonmetricity
\begin{equation*}
^{\lbrack B\tau ]}\mathbf{Q}_{\alpha ^{\prime }\beta ^{\prime }\gamma
^{\prime }}=\ ^{[B\tau ]}\mathbf{Dg}_{\beta ^{\prime }\gamma ^{\prime
}}=\left( ^{[B\tau ]}Q_{c^{\prime }i^{\prime }j^{\prime }},\ ^{[B\tau
]}Q_{i^{\prime }a^{\prime }b^{\prime }}\right)
\end{equation*}%
and torsions \ $^{[B\tau ]}\mathbf{T}_{\ \beta ^{\prime }\gamma ^{\prime
}}^{\alpha ^{\prime }}$ with h- and v-- irreducible components
\begin{eqnarray}
T_{.j^{\prime }k^{\prime }}^{i^{\prime }} &=&-T_{k^{\prime }j^{\prime
}}^{i^{\prime }}=L_{j^{\prime }k^{\prime }}^{i^{\prime }}-L_{k^{\prime
}j^{\prime }}^{i^{\prime }},\quad T_{j^{\prime }a^{\prime }}^{i^{\prime
}}=-T_{a^{\prime }j^{\prime }}^{i^{\prime }}=C_{.j^{\prime }a^{\prime
}}^{i^{\prime }},\ T_{.i^{\prime }j^{\prime }}^{a^{\prime }}=\frac{\delta
N_{i^{\prime }}^{a^{\prime }}}{\delta x^{j^{\prime }}}-\frac{\delta
N_{j^{\prime }}^{a^{\prime }}}{\delta x^{i^{\prime }}},  \notag \\
\quad T_{.b^{\prime }i^{\prime }}^{a^{\prime }} &=&-T_{.i^{\prime }b^{\prime
}}^{a^{\prime }}=\frac{\partial N_{i^{\prime }}^{a^{\prime }}}{\partial
y^{b^{\prime }}}-L_{.b^{\prime }j^{\prime }}^{a^{\prime }},\ T_{.b^{\prime
}c^{\prime }}^{a^{\prime }}=-T_{.c^{\prime }b^{\prime }}^{a^{\prime
}}=C_{b^{\prime }c^{\prime }}^{a^{\prime }}-C_{c^{\prime }b^{\prime
}}^{a^{\prime }}.\   \label{1dtorsions}
\end{eqnarray}%
defined by the h- and v--coefficients of (\ref{berwprim}).

We conclude that if toroidal compactifications are locally anisotropic,
defined by a chain of ansatz containing N--connection, the lower dimensional
spacetime can be not only with torsion structure (like in low energy limit
of string theory) but also with nonmetricity. The anholonomy induced by
N--connection gives the possibility to define a more wide class of linear
connections adapted to the h- and v--splitting.

\subsubsection{Finsler--Riemann--Cartan spaces}

\label{mfls} Such spacetimes are modelled as Riemann--Cartan geometries on a
tangent bundle $TM$ when the metric and anholonomic frame structures
distinguished to be of Finsler type (\ref{1dmetricf}). Both Finsler and
Riemann--Cartan spaces possess nontrivial torsion structures (see section %
\ref{torscurv} for details on definition and computation torsions of locally
anisotropic spaces and Refs. \cite{rcg} for a review of the Einstein--Cartan
gravity). The fundamental geometric objects defining
Finsler--Riemann--Cartan spaces consists in the triple $\left( \mathbf{g}%
^{[F]},\mathbf{\vartheta }_{[F]}^{\alpha },\mathbf{\Gamma }_{[F]\alpha \beta
}^{\gamma }\right) $ where $\mathbf{g}^{[F]}$ is a d--metric (\ref{1dmetricf}%
), $$\mathbf{\vartheta }_{[F]}^{\alpha }=\left( dx^{i},\delta
y^{j}=dy^{j}+N_{[F]k}^{j}\left( x^{l},y^{s}\right) dx^{k}\right) $$ with $%
N_{[F]k}^{j}\left( x^{l},y^{s}\right) $ of type (\ref{1ncc}) and $\mathbf{%
\Gamma }_{[F]\alpha \beta }^{\gamma }$ is an arbitrary d--connection (\ref%
{1dcon1}) on $TM$ (we put the label [F] emphasizing that the N--connection is
a Finsler type one). The torsion $\mathbf{T}_{[F]}^{\alpha }$ and curvature $%
\mathbf{R}_{[F]\beta }^{\alpha }$ d--forms are computed following
respectively the formulas (\ref{1dt}) and \ (\ref{1dc}) but for $\mathbf{%
\vartheta }_{[F]}^{\alpha }$ and $\mathbf{\Gamma }_{[F]\alpha \beta
}^{\gamma }.$

We can consider an inverse modelling of geometries when (roughly speaking)
the Finsler configurations are 'hidden' in Riemann--Cartan spaces. They can
be distinguished for arbitrary Riemann--Cartan manifolds $V^{n+n}$  with
coventional split into ''horizontal'' and ''vertical'' subspaces and
provided with a metric ansatz of type (\ref{1dmetricf}) and with prescribed
procedure of adapting the geometric objects to the Cartan N--connection $%
N_{[F]k}^{j}.$ Of course, the torsion can not be an arbitrary one but
admitting irreducible decompositions with respect to N--frames $\mathbf{e}%
_{\alpha }^{[F]}$ and N--coframes $\mathbf{\vartheta }_{[F]}^{\alpha }$
(see, respectively, the formulas (\ref{1dder}) \ and (\ref{1ddif}) when $%
N_{i}^{a}\rightarrow N_{[F]i}^{j}).\,$\ There were constructed and
investigated different classes of exact solutions of the Einstein equations
with anholonomic variables characterized by anholonomically induced torsions
and modelling Finsler like geometries in (pseudo)\ Riemannian and
Riemann--Cartan spaces (see Refs. \cite{v1,v2,vd}). All constructions from
Finsler--Riemann--Cartan geometry reduce to Finsler--Riemann configurations
(in general, we can see metrics of arbitrary signatures) if $\mathbf{\Gamma }%
_{[F]\alpha \beta }^{\gamma }$ is changed into the Levi--Civita metric
connection defined with respect to anholonomic frames $\mathbf{e}_{\alpha }$
and coframes $\mathbf{\vartheta }^{\alpha }$ when the N--connection
curvature $\Omega _{jk}^{i}$ and the anholonomically induced torsion vanish.

\subsubsection{Teleparallel generalized Finsler geometry}

\label{stpfa}In Refs. \cite{vargtor} the teleparallel Finsler connections,
the Cartan--Einstein unification in the teleparallel approach and related
moving frames with Finsler structures were investigated. In our analysis of
teleparallel geometry we heavily use the results on N--connection geometry
in order to illustrate how the teleparallel and metric affine gravity \cite%
{tgm} can defined as to include generalized Finsler structures. For a
general metric--affine spaces admitting N--connection structure $N_{i}^{a},$
the curvature $\mathbf{R}_{.\beta \gamma \tau }^{\alpha }$ of an arbitrary
d--connection $\mathbf{\Gamma }_{\alpha \beta }^{\gamma }=\left(
L_{jk}^{i},L_{bk}^{a},C_{jc}^{i},C_{bc}^{a}\right) $ splits into h-- and
v--irreversible components, $\mathbf{R}_{.\beta \gamma \tau }^{\alpha
}=(R_{\ hjk}^{i},R_{\ bjk}^{a},P_{\ jka}^{i},P_{\ bka}^{c},S_{\
jbc}^{i},S_{\ bcd}^{a}),$ see (\ref{1dcurv}). In order to include Finsler
like metrics, we state that the N--connection curvature can be nontrivial $%
\Omega _{jk}^{a}\neq 0,$ which is quite different from the condition imposed
in section \ref{stps}. The condition of vanishing of curvature for
teleparallel spaces, see (\ref{dtel1}), is to be stated separately for every
h- v--irreversible component,%
\begin{equation*}
R_{\ hjk}^{i}=0,R_{\ bjk}^{a}=0,P_{\ jka}^{i}=0,P_{\ bka}^{c}=0,S_{\
jbc}^{i}=0,S_{\ bcd}^{a}=0.
\end{equation*}%
We can define certain types of teleparallel Berwald connections (see
sections \ref{berwdcon} and \ref{berwdcona}) with certain nontrivial
components of nonmetricity d--field (\ref{berwnm}) if we modify the metric
compatibility conditions (\ref{dtel2}) into a less strong one when
\begin{equation*}
Q_{kij}=-D_{k}g_{ij}=0\mbox{ and }Q_{abc}=-D_{a}h_{bc}=0
\end{equation*}%
but with nontrivial components
\begin{equation*}
\mathbf{Q}_{\alpha \beta \gamma }=\left( Q_{cij}=-D_{c}g_{ij},\
Q_{iab}=-D_{i}h_{ab}\right) .
\end{equation*}

The class of teleparallel Finsler spaces is distinguished by Finsler
N--connection and d--connection $^{[F]}\mathbf{N}_{j}^{i}(x,y)$ and $^{[F]}%
\widehat{\mathbf{\Gamma }}_{\ \beta \gamma }^{\alpha }=\left( \ ^{[F]}%
\widehat{L}_{jk}^{i},\ ^{[F]}\widehat{L}_{jk}^{i},\ ^{[F]}\widehat{C}%
_{jk}^{i},\ ^{[F]}\widehat{C}_{jk}^{i}\right) ,$ see, respectively, (\ref%
{1ncc}) and (\ref{dccfs}) with vanishing d--curvature components,
\begin{equation*}
~^{[F]}R_{\ hjk}^{i}=0,~^{[F]}P_{\ jka}^{i}=0,~^{[F]}S_{\ jbc}^{i}=0.
\end{equation*}%
We can generate teleparallel Finsler affine structures if it is not imposed
the condition of vanishing of nonmetricity d--field. In this case, there are
considered arbitrary d--connections\ $\mathbf{D}_{\alpha }$ that for the
induced Finsler quadratic form (\ref{1dmetricf}) $\mathbf{g}^{[F]}$
\begin{equation*}
~^{[F]}\mathbf{Q}_{\alpha \beta \gamma }=-\mathbf{D}_{\alpha }\mathbf{g}%
^{[F]}\neq 0
\end{equation*}%
but $\mathbf{R}_{.\beta \gamma \tau }^{\alpha }\left( \mathbf{D}\right) =0.$

The teleparallel--Finsler configurations are contained as particular cases
of Finsler--affine spaces, see section \ref{stpfa}. For vielbein fields $%
\mathbf{e}_{\alpha }^{~\underline{\alpha }}$ and their inverses $\mathbf{e}%
_{~\underline{\alpha }}^{\alpha }$ related to the d--metric (\ref{1dmetricf}),%
\begin{equation*}
\mathbf{g}_{\alpha \beta }^{[F]}=\mathbf{\tilde{e}}_{\alpha }^{~\underline{%
\alpha }}\mathbf{\tilde{e}}_{\beta }^{~\underline{\beta }}g_{\underline{%
\alpha }\underline{\beta }}
\end{equation*}%
we define the Weitzenbock--Finsler d--connection%
\begin{equation}
~^{[WF]}\mathbf{\Gamma }_{~\beta \gamma }^{\alpha }=\mathbf{\tilde{e}}_{~%
\underline{\alpha }}^{\alpha }\delta _{\gamma }\mathbf{\tilde{e}}_{\beta }^{~%
\underline{\alpha }}  \label{wfdcon}
\end{equation}%
where $\delta _{\gamma }$ are the elongated by $^{[F]}\mathbf{N}%
_{j}^{i}(x,y) $ partial derivatives (\ref{1dder}). The torsion of $~^{[WF]}%
\mathbf{\Gamma }_{~\beta \gamma }^{\alpha }$ is defined
\begin{equation}
~^{[WF]}\mathbf{T}_{~\beta \gamma }^{\alpha }=~^{[WF]}\mathbf{\Gamma }%
_{~\beta \gamma }^{\alpha }-~^{[WF]}\mathbf{\Gamma }_{~\gamma \beta
}^{\alpha }  \label{wftors}
\end{equation}%
containing h-- and v--irreducible components being constructed from the
components of a d--metric and N--adapted frames. We can express
\begin{equation*}
~^{[WF]}\mathbf{\Gamma }_{~\beta \gamma }^{\alpha }=\mathbf{\Gamma }%
_{\bigtriangledown ~\beta \gamma }^{\alpha }+\mathbf{\hat{Z}}_{~\beta \gamma
}^{\alpha }+\mathbf{Z}_{~\beta \gamma }^{\alpha }
\end{equation*}%
where $\mathbf{\Gamma }_{\bigtriangledown ~\beta \gamma }^{\alpha }$ is the
Levi--Civita connection (\ref{1lcsym}), $\mathbf{\hat{Z}}_{~\beta \gamma
}^{\alpha }=^{[F]}\widehat{\mathbf{\Gamma }}_{\ \beta \gamma }^{\alpha }-%
\mathbf{\Gamma }_{\bigtriangledown ~\beta \gamma }^{\alpha },$ and the
contorsion tensor is
\begin{equation*}
\mathbf{Z}_{\alpha \beta }=\mathbf{e}_{\beta }\rfloor ~^{[W]}\mathbf{T}%
_{\alpha }-\mathbf{e}_{\alpha }\rfloor ~^{[W]}\mathbf{T}_{\beta }+\frac{1}{2}%
\left( \mathbf{e}_{\alpha }\rfloor \mathbf{e}_{\beta }\rfloor ~^{[W]}\mathbf{%
T}_{\gamma }\right) \mathbf{\vartheta }^{\gamma }+\left( \mathbf{e}_{\alpha
}\rfloor \mathbf{Q}_{\beta \gamma }\right) \mathbf{\vartheta }^{\gamma
}-\left( \mathbf{e}_{\beta }\rfloor \mathbf{Q}_{\alpha \gamma }\right)
\mathbf{\vartheta }^{\gamma }+\frac{1}{2}\mathbf{Q}_{\alpha \beta }.
\end{equation*}%
In the non-Berwald standard approaches to the Finsler--teleparallel gravity
it is considered that $\mathbf{Q}_{\alpha \beta }=0.$

\subsubsection{Cartan geometry}

\label{scs}The theory of Cartan spaces (see, for instance, \cite{rund,kaw1})
\ can be reformulated as a dual to Finsler geometry \cite{mironc} (see
details and references in \cite{mhss}). The Cartan space is constructed on a
cotangent bundle $T^{\ast }M$ similarly to the Finsler space on the tangent
bundle $TM.$

Consider a real smooth manifold $M,$ the cotangent bundle $\left( T^{\ast
}M,\pi ^{\ast },M\right) $ and the manifold $\widetilde{T^{\ast }M}=T^{\ast
}M\backslash \{0\}.$

\begin{definition}
A Cartan space is a pair $C^{n}=\left( M,K(x,p)\right) $ \ such that $%
K:T^{\ast }M\rightarrow \R$ is a scalar function satisfying the following
conditions:

\begin{enumerate}
\item $K$ is a differentiable function on the manifold $\widetilde{T^{\ast }M%
}$ $=T^{\ast }M\backslash \{0\}$ and continuous on the null section of the
projection $\pi ^{\ast }:T^{\ast }M\rightarrow M;$

\item $K$ is a positive function, homogeneous on the fibers of the $T^{\ast
}M,$ i. e. $K(x,\lambda p)=\lambda F(x,p),\lambda \in \R;$

\item The Hessian of $K^{2}$ with elements
\begin{equation}
\check{g}_{[K]}^{ij}(x,p)=\frac{1}{2}\frac{\partial ^{2}K^{2}}{\partial
p_{i}\partial p_{j}}  \label{carm}
\end{equation}%
is positively defined on $\widetilde{T^{\ast }M}.$
\end{enumerate}
\end{definition}

The function $K(x,y)$ and $\check{g}^{ij}(x,p)$ are called \ respectively
the fundamental function and the fundamental (or metric) tensor of the
Cartan space $C^{n}.$ We use symbols like $"\check{g}"$ as to emphasize that
the geometrical objects are defined on a dual space.

One considers ''anisotropic'' (depending on directions, momenta, $p_{i})$
Christoffel symbols. For simplicity, we write the inverse to (\ref{carm}) as
$g_{ij}^{(K)}=\check{g}_{ij}$ and introduce the coefficients
\begin{equation*}
\check{\gamma}_{~jk}^{i}(x,p)=\frac{1}{2}\check{g}^{ir}\left( \frac{\partial
\check{g}_{rk}}{\partial x^{j}}+\frac{\partial \check{g}_{jr}}{\partial x^{k}%
}-\frac{\partial \check{g}_{jk}}{\partial x^{r}}\right) ,
\end{equation*}%
defining the canonical N--connection $\mathbf{\check{N}=\{}\check{N}_{ij}%
\mathbf{\},}$
\begin{equation}
\check{N}_{ij}^{[K]}=\check{\gamma}_{~ij}^{k}p_{k}-\frac{1}{2}\gamma
_{~nl}^{k}p_{k}p^{l}{\breve{\partial}}^{n}\check{g}_{ij}  \label{nccartan}
\end{equation}%
where $~{\breve{\partial}}^{n}=\partial /\partial p_{n}.$ The N--connection $%
\mathbf{\check{N}}=\{\check{N}_{ij}\}$ can be used for definition of an
almost complex structure like in (\ref{1dmetricf}) and introducing on $%
T^{\ast }M$ a d--metric
\begin{equation}
\mathbf{\check{G}}_{[k]}=\check{g}_{ij}(x,p)dx^{i}\otimes dx^{j}+\check{g}%
^{ij}(x,p)\delta p_{i}\otimes \delta p_{j},  \label{dmcar}
\end{equation}%
with $\check{g}^{ij}(x,p)$ taken as (\ref{carm}).

Using the canonical N--connection (\ref{nccartan}) and Finsler metric tensor
(\ref{carm}) (or, equivalently, the d--metric (\ref{dmcar})), we can define
the canonical d--connection $\mathbf{\check{D}=\{\check{\Gamma}}\left(
\check{N}_{[k]}\right) \mathbf{\}}$
\begin{equation*}
\mathbf{\check{\Gamma}}\left( \check{N}_{[k]}\right) =\check{\Gamma}%
_{[k]\beta \gamma }^{\alpha }=\left( \check{H}_{[k]~jk}^{i},\check{H}%
_{[k]~jk}^{i},\check{C}_{[k]~i}^{\quad jk},\check{C}_{[k]~i}^{\quad
jk}\right)
\end{equation*}%
with the coefficients computed
\begin{equation*}
\check{H}_{[k]~jk}^{i}=\frac{1}{2}\check{g}^{ir}\left( \check{\delta}_{j}%
\check{g}_{rk}+\check{\delta}_{k}\check{g}_{jr}-\check{\delta}_{r}\check{g}%
_{jk}\right) ,\ \check{C}_{[k]~i}^{\quad jk}=\check{g}_{is}{\breve{\partial}}%
^{s}\check{g}^{jk}.
\end{equation*}%
The d--connection $\mathbf{\check{\Gamma}}\left( \check{N}_{(k)}\right) $
satisfies the metricity conditions both for the horizontal and vertical
components, i. e. $\mathbf{\check{D}}_{\alpha }\mathbf{\check{g}}_{\beta
\gamma }=0.$

The d--torsions (\ref{1dtorsions}) \ and d--curvatures (\ref{1dcurv}) are
computed like in Finsler geometry but starting from the coefficients in (\ref%
{nccartan}) and (\ref{dmcar}), when the indices $a,b,c...$ run the same
values as indices $i,j,k,...$ and the geometrical objects are modelled as on
the dual tangent bundle. It should be emphasized that in this case all
values $\check{g}_{ij,}\ \check{\Gamma}_{[k]\beta \gamma }^{\alpha }$ and $%
\check{R}_{[k]\beta \gamma \delta }^{.\alpha }$ are defined by a fundamental
function $K\left( x,p\right) .$

In general, we can consider that a Cartan space is provided with a metric $%
\check{g}^{ij}=\partial ^{2}K^{2}/2\partial p_{i}\partial p_{j},$ but the
N--connection and d--connection could be defined in a different manner, even
not be determined by $K.$ If a Cartan space is modelled in a metric--affine
space $V^{n+n},$ with local coordinates $\left( x^{i},y^{k}\right) ,$ we
have to define a procedure of dualization of vertical coordinates, $%
y^{k}\rightarrow p_{k}.$

\subsection{Generalized Lagrange and Hamilton geometries}

The notion of Finsler spaces was extended by J. Kern \cite{ker} and R. Miron %
\cite{mironlg}. It is was elaborated in vector bundle spaces in Refs. \cite%
{ma} and generalized to superspaces \cite{vsup}. We illustrate how such
geometries can be modelled on a space $\mathbf{V}^{n+n}$ provided with
N--connection structure.

\subsubsection{Lagrange geometry and generalizations}

\label{ssslgg}The idea of generalization of the Finsler geometry was to
consider instead of the homogeneous fundamental function $F(x,y)$ in a
Finsler space a more general one, a Lagrangian $L\left( x,y\right) $,
defined as a differentiable mapping $L:(x,y)\in TV^{n+n}\rightarrow
L(x,y)\in \R,$ of class $C^{\infty }$ on manifold $\widetilde{TV}^{n+n}$ and
continuous on the null section $0:V^{n}\rightarrow \widetilde{TV}^{n+n}$ of
the projection $\pi :\widetilde{TV}^{n+n}\rightarrow V^{n}.$ A Lagrangian is
regular if it is differentiable and the Hessian
\begin{equation}
g_{ij}^{[L]}(x,y)=\frac{1}{2}\frac{\partial ^{2}L^{2}}{\partial
y^{i}\partial y^{j}}  \label{9lagm}
\end{equation}%
is of rank $n$ on $V^{n}.$

\begin{definition}
\label{defls}A Lagrange space is a pair $\mathbf{L}^{n}=\left(
V^{n},L(x,y)\right) $ where $V^{n}$ is a smooth real $n$--dimensional
manifold provided with regular Lagrangian \ $L(x,y)$ structure $%
L:TV^{n}\rightarrow \R$ $\ $for which $g_{ij}(x,y)$ from (\ref{9lagm}) has a
constant signature over the manifold $\widetilde{TV}^{n+n}.$
\end{definition}

The fundamental Lagrange function $L(x,y)$ defines a canonical
N--con\-nec\-ti\-on
\begin{equation}
\ ^{[cL]}N_{~j}^{i}=\frac{1}{2}\frac{\partial }{\partial y^{j}}\left[
g^{ik}\left( \frac{\partial ^{2}L^{2}}{\partial y^{k}\partial y^{h}}y^{h}-%
\frac{\partial L}{\partial x^{k}}\right) \right]  \label{cncls}
\end{equation}%
as well a d-metric
\begin{equation}
\mathbf{g}_{[L]}=g_{ij}(x,y)dx^{i}\otimes dx^{j}+g_{ij}(x,y)\delta
y^{i}\otimes \delta y^{j},  \label{dmlag}
\end{equation}%
with $g_{ij}(x,y)$ taken as (\ref{9lagm}). As well we can introduce an almost
K\"{a}hlerian structure and an almost Hermitian model of $\mathbf{L}^{n},$
denoted as $\mathbf{H}^{2n}$ as in the case of Finsler spaces but with a
proper fundamental Lagrange function and metric tensor $g_{ij}.$ The
canonical metric d--connection $\widehat{\mathbf{D}}_{[L]}$ is defined by
the coefficients
\begin{equation}
\ ^{[L]}\widehat{\mathbf{\Gamma }}_{\beta \gamma }^{\alpha }=\left( \ ^{[L]}%
\widehat{L}_{~jk}^{i},\ ^{[L]}\widehat{L}_{~jk}^{i},\ ^{[L]}\widehat{C}%
_{~jk}^{i},\ ^{[L]}\widehat{C}_{~jk}^{i}\right)  \label{lagdcon}
\end{equation}%
computed for $N_{[cL]~j}^{i}$ and by respective formulas (\ref{1candcon})
with $h_{ab}\rightarrow g_{ij}^{[L]}$ and $\widehat{C}_{bc}^{a}\rightarrow $
$\widehat{C}_{ij}^{i}.$ The d--torsions (\ref{1dtorsions}) and d--curvatures (%
\ref{1dcurv}) are determined, in this case, by $\ ^{[L]}\widehat{L}_{~jk}^{i}$
and $\ ^{[L]}\widehat{C}_{~jk}^{i}.$ We also note that instead of $\
^{[cL]}N_{~j}^{i}$ and $\ ^{[L]}\widehat{\mathbf{\Gamma }}_{\ \beta \gamma
}^{\alpha }$ we can consider on a $L^{n}$--space different N--connections $%
N_{~j}^{i},$ d--connections $\mathbf{\Gamma }_{\ \beta \gamma }^{\alpha }$
which are not defined only by $L(x,y)$ and $g_{ij}^{[L]}$ but can be metric,
or non--metric with respect to the Lagrange metric.

The next step of generalization \cite{mironlg} is to consider an arbitrary
metric $g_{ij}\left( x,y\right) $ on $\mathbf{TV}^{n+n}$ (we use boldface
symbols in order to emphasize that the space is enabled with N--connection
structure) instead of (\ref{9lagm}) which is the second derivative of
''anisotropic'' coordinates $y^{i}$ of a Lagrangian.

\begin{definition}
\label{defgls}A generalized Lagrange space is a pair $\mathbf{GL}^{n}=\left(
V^{n},g_{ij}(x,y)\right) $ where $g_{ij}(x,y)$ is a covariant, symmetric and
N--adapted d--tensor field of rank $n$ and of constant signature on $%
\widetilde{TV}^{n+n}.$
\end{definition}

One can consider different classes of N-- and d--connections on $TV^{n+n},$
which are compatible (metric) or non compatible with (\ref{dmlag}) for
arbitrary $g_{ij}(x,y)$ and arbitrary d--metric
\begin{equation}
\mathbf{g}_{[gL]}=g_{ij}(x,y)dx^{i}\otimes dx^{j}+g_{ij}(x,y)\delta
y^{i}\otimes \delta y^{j},  \label{dmgls}
\end{equation}%
We can apply all formulas for d--connections, N--curvatures, d--torsions and
d--curvatures as in sections \ref{dlcms} and \ref{torscurv} but
reconsidering them on $\mathbf{TV}^{n+n},$ by changing \
\begin{equation*}
h_{ab}\rightarrow g_{ij}(x,y),\widehat{C}_{bc}^{a}\rightarrow \widehat{C}%
_{ij}^{i}\mbox{ and } N_{i}^{a}\rightarrow N_{~i}^{k}.
\end{equation*}
Prescribed torsions $T_{\ jk}^{i}$ and $S_{\ jk}^{i}$ can be introduced on $%
\mathbf{GL}^{n}$ by using the d--connection%
\begin{equation}
\ \widehat{\mathbf{\Gamma }}_{\beta \gamma }^{\alpha }=\left( \widehat{L}%
_{[gL]~jk}^{i}+\tau _{\ jk}^{i},\widehat{L}_{[gL]~jk}^{i}+\tau _{\ jk}^{i},%
\widehat{C}_{[gL]~jk}^{i}+\sigma _{\ jk}^{i},\widehat{C}_{[gL]~jk}^{i}+%
\sigma _{\ jk}^{i}\right)  \label{ptdcls}
\end{equation}%
with
\begin{equation}
\tau _{\ jk}^{i}=\frac{1}{2}g^{il}\left(
g_{kh}T_{.lj}^{h}+g_{jh}T_{.lk}^{h}-g_{lh}T_{\ jk}^{h}\right) \mbox{ and }%
\sigma _{\ jk}^{i}=\frac{1}{2}g^{il}\left(
g_{kh}S_{.lj}^{h}+g_{jh}S_{.lk}^{h}-g_{lh}S_{\ jk}^{h}\right)  \notag
\end{equation}%
like we have performed for the Berwald connections (\ref{bct}) with (\ref%
{tauformulas}) and (\ref{berwprim}) but in our case
\begin{equation}
\ ^{[aL]}\widehat{\mathbf{\Gamma }}_{\beta \gamma }^{\alpha }=\left(
\widehat{L}_{[gL]~jk}^{i},\widehat{L}_{[gL]~jk}^{i},\widehat{C}%
_{[gL]~jk}^{i},\widehat{C}_{[gL]~jk}^{i}\right)  \label{dccls}
\end{equation}%
is metric compatible being modelled like on a tangent bundle and with the
coefficients computed as in (\ref{1candcon}) with $h_{ab}\rightarrow
g_{ij}^{[L]}$ and $\widehat{C}_{bc}^{a}\rightarrow $ $\widehat{C}_{ij}^{i},$
by using the d--metric $\mathbf{G}_{[gL]}$ (\ref{dmgls}). The connection (%
\ref{ptdcls}) is a Riemann--Cartan one modelled on effective tangent bundle
provided with N--connection structure.

\subsubsection{Hamilton geometry and generalizations}

\label{shgg}The geometry of Hamilton spaces was defined and investigated by
R. Miron in the papers \cite{mironh} (see details and additional references
in \cite{mhss}). It was developed on the cotangent bundle as a dual geometry
to the geometry of Lagrange spaces. Here we consider their modelling on
couples of spaces $\left( V^{n},\ ^{\ast }V^{n}\right) ,$ or cotangent
bundle $T^{\ast }M,$ where $\ ^{\ast }V^{n}$ is considered as a 'dual'
manifold defined by local coordinates satisfying a duality condition with
respect to coordinates on $V^{n}.$\ We start with the definition of
generalized Hamilton spaces and then consider the particular cases.

\begin{definition}
A generalized Hamilton space is a pair $\mathbf{GH}^{n}=\left( V^{n},\check{g%
}^{ij}(x,p)\right) $ where $V^{n}$ is a real $n$--dimensional manifold and $%
\check{g}^{ij}(x,p)$ is a contravariant, symmetric, nondegenerate of rank $n$
and of constant signature on $\widetilde{T^{\ast }V}^{n+n}.$
\end{definition}

The value $\check{g}^{ij}(x,p)$ is called the fundamental (or metric) tensor
of the space $\mathbf{GH}^{n}.$ One can define such values for every
paracompact manifold $V^{n}.$ In general, a N--connection on $\mathbf{GH}%
^{n} $ is not determined by $\check{g}^{ij}.$ Therefore we can consider an
arbitrary N--connection $\mathbf{\check{N}}=\{\check{N}_{ij}\left(
x,p\right) \}$ and define on $T^{\ast }V^{n+n}$ a d--metric similarly to (%
\ref{1block2}) and/or (\ref{dmlag})
\begin{equation}
{\breve{\mathbf{G}_{[gH]}}}={\breve{g}}_{\alpha \beta }\left( {\breve{u}}%
\right) {\breve{\delta}}^{\alpha }\otimes {\breve{\delta}}^{\beta }={%
\breve{g}}_{ij}\left( {\breve{u}}\right) d^{i}\otimes d^{j}+{\check{g}}%
^{ij}\left( {\breve{u}}\right) {\breve{\delta}}_{i}\otimes {\breve{\delta}}%
_{j},  \label{dmghs}
\end{equation}%
The N--coefficients $\check{N}_{ij}\left( x,p\right) $ and the d--metric
structure (\ref{dmghs}) define an almost K\"{a}hler model of generalized
Hamilton spaces provided with canonical d--connections, d--torsions and
d-curvatures (see respectively the formulas d--torsions (\ref{1dtorsions})
and d--curvatures (\ref{1dcurv}) with the fiber coefficients redefined for
the cotangent bundle $T^{\ast }V$ $^{n+n}$).

A generalized Hamilton space $\mathbf{GH}^{n}$ is called reducible to a
Hamilton one if there exists a Hamilton function $H\left( x,p\right) $ on $%
T^{\ast }V$ $^{n+n}$ such that
\begin{equation}
\check{g}_{[H]}^{ij}(x,p)=\frac{1}{2}\frac{\partial ^{2}H}{\partial
p_{i}\partial p_{j}}.  \label{hsm}
\end{equation}

\begin{definition}
A Hamilton space is a pair $\mathbf{H}^{n}=\left( V^{n},H(x,p)\right) $ \
such that $H:T^{\ast }V^{n}\rightarrow \R$ is a scalar function which
satisfy the following conditions:

\begin{enumerate}
\item $H$ is a differentiable function on the manifold $\widetilde{T^{\ast }V%
}^{n+n}$ $=T^{\ast }V^{n+n}\backslash \{0\}$ and continuous on the null
section of the projection $\pi ^{\ast }:T^{\ast }V^{n+n}\rightarrow V^{n};$

\item The Hessian of $H$ with elements (\ref{hsm}) is positively defined on $%
\widetilde{T^{\ast }V}^{n+n}$ and $\check{g}^{ij}(x,p)$ is nondegenerate
matrix of rank $n$ and of constant signature.
\end{enumerate}
\end{definition}

For Hamilton spaces, the canonical N--connection (defined by $H$ and its
Hessian) is introduced as
\begin{equation}
\ ^{[H]}\check{N}_{ij}=\frac{1}{4}\{\check{g}_{ij},H\}-\frac{1}{2}\left(
\check{g}_{ik}\frac{\partial ^{2}H}{\partial p_{k}\partial x^{j}}+\check{g}%
_{jk}\frac{\partial ^{2}H}{\partial p_{k}\partial x^{i}}\right) ,
\label{ncchs}
\end{equation}%
where the Poisson brackets, for arbitrary functions $f$ and $g$ on $T^{\ast
}V^{n+n},$ act as
\begin{equation*}
\{f,g\}=\frac{\partial f}{\partial p_{i}}\frac{\partial g}{\partial x^{i}}-%
\frac{\partial g}{\partial p_{i}}\frac{\partial p}{\partial x^{i}}.
\end{equation*}%
The canonical metric d--connection $\ ^{[H]}\widehat{\mathbf{D}}$ is defined
by the coefficients
\begin{equation*}
\ ^{[H]}\widehat{\mathbf{\Gamma }}_{\ \beta \gamma }^{\alpha }=\left( \
^{[c]}\check{H}_{~jk}^{i},\ ^{[c]}\check{H}_{~jk}^{i},\ ^{[c]}\check{C}%
_{~jk}^{i},\ ^{[c]}\check{C}_{~jk}^{i}\right)
\end{equation*}%
computed for $\ ^{[H]}\check{N}_{ij}$ and by respective formulas (\ref%
{1candcon}) with $g_{ij}\rightarrow {\breve{g}}_{ij}\left( {\breve{u}}\right)
,$ $h_{ab}\rightarrow {\check{g}}^{ij}$ and $\widehat{L}_{~jk}^{i}%
\rightarrow \ ^{[c]}\widehat{H}_{~jk}^{i},$ $\ \widehat{C}%
_{bc}^{a}\rightarrow $ $\ ^{[c]}\check{C}_{~i}^{\quad jk}$ when
\begin{equation*}
\ ^{[c]}\check{H}_{~jk}^{i}=\frac{1}{2}\check{g}^{is}\left( \check{\delta}%
_{j}\check{g}_{sk}+\check{\delta}_{k}\check{g}_{js}-\check{\delta}_{s}\check{%
g}_{jk}\right) \mbox{ and }\ ^{[c]}\check{C}_{~i}^{\quad jk}=-\frac{1}{2}%
\check{g}_{is}\check{\partial}^{j}\check{g}^{sk}.
\end{equation*}%
In result, we can compute the d--torsions and d--curvatures like on Lagrange
\ or on Cartan spaces. On Hamilton spaces all such objects are defined by
the Hamilton function $H(x,p)$ and indices have to be reconsidered for
co--fibers of the cotangent bundle.

We note that there were elaborated various type of higher order
generalizations (on the higher order tangent and contangent bundles) of the
Finsler--Cartan and Lagrange--Hamilton geometry \cite{mat1} and on higher
order supersymmetric (co) vector bundles in Ref. \cite{vsup}. We can
generalize the d--connection $\ ^{[H]}\widehat{\mathbf{\Gamma }}_{\beta
\gamma }^{\alpha }$ to any d--connection in $\mathbf{H}^{n}$ with prescribed
torsions, like we have done in previous section for Lagrange spaces, see (%
\ref{ptdcls}). This type of Riemann--Cartan geometry is modelled like on a
dual tangent bundle by a Hamilton metric structure (\ref{hsm}),
N--connection $\ ^{[H]}\check{N}_{ij},$ and d--connection coefficients $\
^{[c]}\check{H}_{~jk}^{i}$ and $\ ^{[c]}\check{C}_{~i}^{\quad jk}.$

\subsection{ Nonmetricity and generalized Finsler--affine spaces}

The generalized Lagrange and Finsler geometry may be defined on tangent
bundles by using d--connections and d--metrics satisfying metric
compatibility conditions \cite{ma}. Nonmetricity components can be induced
if Berwald type d--connections are introduced into consideration on
different type of manifolds provided with N--connection structure, see
formulas (\ref{berw}), (\ref{bct}), (\ref{mafbc}) and (\ref{berwprim}).

We define such spaces as generalized Finsler spaces with nonmetricity.

\begin{definition}
\label{defglas}A generalized Lagrange--affine space $\mathbf{GLa}^{n}=\left(
V^{n},g_{ij}(x,y),\ ^{[a]}\mathbf{\Gamma }_{\ \beta }^{\alpha }\right) $ is
defined on manifold $\mathbf{TV}^{n+n},$ provided with an arbitrary
nontrivial N--connection structure $\mathbf{N}=\{N_{j}^{i}\},$\ as a general
Lagrange space $\mathbf{GL}^{n}=\left( V^{n},g_{ij}(x,y)\right) $ (see
Definition \ref{defgls}) enabled with a d--connection structure $\ ^{[a]}%
\mathbf{\Gamma }_{\ \ \alpha }^{\gamma }=\ ^{[a]}\mathbf{\Gamma }_{\ \alpha
\beta }^{\gamma }\mathbf{\vartheta }^{\beta }$ distorted by arbitrary
torsion, $\mathbf{T}_{\beta },$ and nonmetricity, $\mathbf{Q}_{\beta \gamma
},$ d--fields, \
\begin{equation}
\ \ ^{[a]}\mathbf{\Gamma }_{\ \beta }^{\alpha }=\ \ ^{[aL]}\widehat{\mathbf{%
\Gamma }}_{\ \beta }^{\alpha }\ +\ \ ^{[a]}\ \mathbf{Z}_{\ \ \beta }^{\alpha
},  \label{dcglma}
\end{equation}%
where$\ ^{[L]}\widehat{\mathbf{\Gamma }}_{\beta }^{\alpha }$ is the
canonical \ generalized Lagrange d--connection (\ref{dccls}) and
\begin{equation*}
\ \ ^{[a]}\ \mathbf{Z}_{\ \alpha \beta }=\mathbf{e}_{\beta }\rfloor \
\mathbf{T}_{\alpha }-\mathbf{e}_{\alpha }\rfloor \ \mathbf{T}_{\beta }+\frac{%
1}{2}\left( \mathbf{e}_{\alpha }\rfloor \mathbf{e}_{\beta }\rfloor \ \mathbf{%
T}_{\gamma }\right) \mathbf{\vartheta }^{\gamma }+\left( \mathbf{e}_{\alpha
}\rfloor \ \mathbf{Q}_{\beta \gamma }\right) \mathbf{\vartheta }^{\gamma
}-\left( \mathbf{e}_{\beta }\rfloor \ \mathbf{Q}_{\alpha \gamma }\right)
\mathbf{\vartheta }^{\gamma }+\frac{1}{2}\ \mathbf{Q}_{\alpha \beta }.
\end{equation*}
\end{definition}

The d--metric structure on $\mathbf{GLa}^{n}$ is stated by an arbitrary
N--adapted form (\ref{1block2}) but on $\mathbf{TV}^{n+n},$
\begin{equation}
\mathbf{g}_{[a]}=g_{ij}(x,y)dx^{i}\otimes dx^{j}+g_{ij}(x,y)\delta
y^{i}\otimes \delta y^{j}.  \label{dmglas}
\end{equation}

The torsions and curvatures on $\mathbf{GLa}^{n}$ are computed by using
formulas (\ref{1dt}) and (\ref{1dc}) with $\mathbf{\Gamma }_{\ \beta }^{\gamma
}\rightarrow \ ^{[a]}\mathbf{\Gamma }_{\ \beta }^{\alpha },$

\begin{equation}
\ \ ^{[a]}\mathbf{T}^{\alpha }\doteqdot \ ^{[a]}\mathbf{D\vartheta }^{\alpha
}=\delta \mathbf{\vartheta }^{\alpha }+\ ^{[a]}\mathbf{\Gamma }_{\ \beta
}^{\gamma }\wedge \mathbf{\vartheta }^{\beta }  \label{tglma}
\end{equation}%
and
\begin{equation}
\ \ ^{[a]}\mathbf{R}_{\ \beta }^{\alpha }\doteqdot \ ^{[a]}\mathbf{D(\ }%
^{[a]}\mathbf{\Gamma }_{\ \beta }^{\alpha })=\delta (\ ^{[a]}\mathbf{\Gamma }%
_{\ \beta }^{\alpha })-\ ^{[a]}\mathbf{\Gamma }_{\ \beta }^{\gamma }\wedge \
^{[a]}\mathbf{\Gamma }_{\ \ \gamma }^{\alpha }.  \label{cglma}
\end{equation}%
Modelling in $V^{n+n},$ with local coordinates $u^{\alpha }=\left(
x^{i},y^{k}\right) ,$ a tangent bundle structure, we redefine the operators (%
\ref{1ddif}) and (\ref{1dder}) respectively as
\begin{equation}
\mathbf{e}_{\alpha }\doteqdot \delta _{\alpha }=\left( \delta _{i},\tilde{%
\partial}_{k}\right) \equiv \frac{\delta }{\delta u^{\alpha }}=\left( \frac{%
\delta }{\delta x^{i}}=\partial _{i}-N_{i}^{a}\left( u\right) \partial _{a},%
\frac{\partial }{\partial y^{k}}\right)  \label{ddert}
\end{equation}%
and the N--elongated differentials (in brief, N--differentials)
\begin{equation}
\mathbf{\vartheta }_{\ }^{\beta }\doteqdot \delta \ ^{\beta }=\left( d^{i},%
\tilde{\delta}^{k}\right) \equiv \delta u^{\alpha }=\left( \delta
x^{i}=dx^{i},\delta y^{k}=dy^{k}+N_{i}^{k}\left( u\right) dx^{i}\right)
\label{ddift}
\end{equation}%
where Greek indices run the same values, $i,j,...=1,2,...n$ (we shall use
the symbol ''$\sim $'' if one would be necessary to distinguish operators
and coordinates defined on h-- and v-- subspaces).

Let us define the h-- and v--irreducible components of the d--connection $\
^{[a]}\mathbf{\Gamma }_{\ \beta }^{\alpha }$ like in (\ref{hcov}) and (\ref%
{vcov}),%
\begin{equation*}
\ ^{[a]}\widehat{\mathbf{\Gamma }}_{\ \beta \gamma }^{\alpha }=\left( \
^{[L]}\widehat{L}_{~jk}^{i}+\ z_{~jk}^{i},\ ^{[L]}\widehat{L}_{~jk}^{i}\ +\
z_{~jk}^{i},\ ^{[L]}\widehat{C}_{~jk}^{i}+c_{~jk}^{i},\ ^{[L]}\widehat{C}%
_{~jk}^{i}+c_{~jk}^{i}\right)
\end{equation*}%
with the distorsion d--tensor $\ $%
\begin{equation*}
\ ^{[a]}\ \mathbf{Z}_{\ \ \beta }^{\alpha }=\left( z_{~jk}^{i},\
z_{~jk}^{i},c_{~jk}^{i},c_{~jk}^{i}\right)
\end{equation*}%
defined as on a tangent bundle
\begin{eqnarray*}
\ ^{[a]}L_{jk}^{i} &=&\left( ^{[a]}\mathbf{D}_{\delta _{k}}\delta
_{j}\right) \rfloor \delta ^{i}=\left( ^{[L]}\widehat{\mathbf{D}}_{\delta
_{k}}\delta _{j}+\ ^{[a]}\ \mathbf{Z}_{\delta _{k}}\delta _{j}\right)
\rfloor \delta ^{i}=\ ^{[L]}\widehat{L}_{~jk}^{i}+\ z_{~jk}^{i},\quad \\
\ ^{[a]}\tilde{L}_{jk}^{i} &=&\left( ^{[a]}\mathbf{D}_{\delta _{k}}\tilde{%
\partial}_{j}\right) \rfloor \tilde{\partial}^{i}=\left( ^{[L]}\widehat{%
\mathbf{D}}_{k}\tilde{\partial}_{j}+\ ^{[a]}\ \mathbf{Z}_{k}\tilde{\partial}%
_{j}\right) \rfloor \tilde{\partial}^{i}=\ ^{[L]}\widehat{L}_{~jk}^{i}+\
z_{~jk}^{i}, \\
\ \ ^{[a]}C_{jk}^{i} &=&\left( ^{[a]}\mathbf{D}_{\tilde{\partial}_{k}}\delta
_{j}\right) \rfloor \delta ^{i}=\left( ^{[L]}\widehat{\mathbf{D}}_{\tilde{%
\partial}_{k}}\delta _{j}+\ ^{[a]}\ \mathbf{Z}_{\tilde{\partial}_{k}}\delta
_{j}\right) \rfloor \delta ^{i}=\ ^{[L]}\widehat{C}_{~jk}^{i}+c_{~jk}^{i}, \\
\ \ ^{[a]}\tilde{C}_{jk}^{i} &=&\left( ^{[a]}\mathbf{D}_{\tilde{\partial}%
_{k}}\tilde{\partial}_{j}\right) \rfloor \tilde{\partial}^{i}=\left( ^{[L]}%
\widehat{\mathbf{D}}_{\tilde{\partial}_{k}}\tilde{\partial}_{j}+\ ^{[a]}\
\mathbf{Z}_{\tilde{\partial}_{k}}\tilde{\partial}_{j}\right) \rfloor \tilde{%
\partial}^{i}=\ ^{[L]}\widehat{C}_{~jk}^{i}+c_{~jk}^{i},
\end{eqnarray*}%
where for 'lifts' from the h--subspace to the v--subspace we consider that $%
\ ^{[a]}L_{jk}^{i}=\ ^{[a]}\tilde{L}_{jk}^{i}$ and $\ ^{[a]}C_{jk}^{i}=\
^{[a]}\tilde{C}_{jk}^{i}.$ As a consequence, on spaces with modelled tangent
space structure, the d--connections are distinguished as\ $\ \mathbf{\Gamma }%
_{\beta \gamma }^{\alpha }=\left( L_{jk}^{i},C_{jk}^{i}\right) .$

\begin{theorem}
\label{ttgfls}The torsion $\ \ ^{[a]}\mathbf{T}^{\alpha }$ $\ $(\ref{tglma})
of a d--connection $^{[a]}\mathbf{\Gamma }_{\ \beta }^{\alpha }=\left( \
^{[a]}L_{jk}^{i},\ ^{[a]}C_{jc}^{i}\right) $ (\ref{dcglma}) has as
irreducible h- v--components, $\ \ ^{[a]}\mathbf{T}^{\alpha }=\left(
T_{jk}^{i},\tilde{T}_{jk}^{i}\right) ,$ the d--torsions
\begin{eqnarray}
T_{.jk}^{i} &=&-T_{kj}^{i}=\ ^{[L]}\widehat{L}_{~jk}^{i}+\ z_{~jk}^{i}-\
^{[L]}\widehat{L}_{~kj}^{i}-\ z_{~kj}^{i},\quad  \label{dtorsgla} \\
\quad \ \tilde{T}_{jk}^{i} &=&-\ \tilde{T}_{kj}^{i}=\ ^{[L]}\widehat{C}%
_{~jk}^{i}+c_{~jk}^{i}-\ ^{[L]}\widehat{C}_{~kj}^{i}-c_{~kj}^{i}.\   \notag
\end{eqnarray}
\end{theorem}

The proof of this Theorem consists from a standard calculus for
metric--affine spaces of $\ ^{[a]}\mathbf{T}^{\alpha }$ \cite{mag} but with
N--adapted frames. We note that in $\ z_{~jk}^{i}$ and $c_{~kj}^{i}$ it is
possible to include any prescribed values of the d--torsions.

\begin{theorem}
\label{tcgfls}The curvature $\ \ ^{[a]}\mathbf{R}_{\ \beta }^{\alpha }$ $\ $(%
\ref{cglma}) of a d--connection\\ $^{[a]}\mathbf{\Gamma }_{\ \beta }^{\alpha
}=\left( \ ^{[a]}L_{jk}^{i},\ ^{[a]}C_{jc}^{i}\right) $ (\ref{dcglma}) has
 the h- v--components (d--curvatures),\\ $^{[a]}\mathbf{R}_{.\beta \gamma \tau
}^{\alpha }=\{\ ^{[a]}R_{\ hjk}^{i},\ ^{[a]}P_{\ jka}^{i},\ ^{[a]}S_{\
jbc}^{i}\},$
\begin{eqnarray*}
\ ^{[a]}R_{\ hjk}^{i} &=&\frac{\delta \ ^{[a]}L_{.hj}^{i}}{\delta x^{h}}-%
\frac{\delta \ ^{[a]}L_{.hk}^{i}}{\delta x^{j}}+\ ^{[a]}L_{.hj}^{m}\
^{[a]}L_{mk}^{i}-\ ^{[a]}L_{.hk}^{m}\ ^{[a]}L_{mj}^{i}-\
^{[a]}C_{.ho}^{i}\Omega _{.jk}^{o}, \\
\ ^{[a]}P_{\ jks}^{i} &=&\frac{\partial \ ^{[a]}L_{.jk}^{i}}{\partial y^{s}}%
-\left( \frac{\partial \ ^{[a]}C_{.js}^{i}}{\partial x^{k}}+\
^{[a]}L_{.lk}^{i}\ ^{[a]}C_{.js}^{l}-\ ^{[a]}L_{.jk}^{l}\
^{[a]}C_{.ls}^{i}-\ ^{[a]}L_{.sk}^{p}\ ^{[a]}C_{.jp}^{i}\right) \\
&&+\ ^{[a]}C_{.jp}^{i}\ ^{[a]}P_{.ks}^{p}, \\
\ ^{[a]}S_{\ jlm}^{i} &=&\frac{\partial \ ^{[a]}C_{.jl}^{i}}{\partial y^{m}}-%
\frac{\partial \ ^{[a]}C_{.jm}^{i}}{\partial y^{l}}+\ ^{[a]}C_{.jl}^{h}\
^{[a]}C_{.hm}^{i}-\ ^{[a]}C_{.jm}^{h}\ ^{[a]}C_{hl}^{i},
\end{eqnarray*}%
where\ $^{[a]}L_{.hk}^{m}=\ ^{[L]}\widehat{L}_{~jk}^{i}+\ z_{~jk}^{i},\
^{[a]}C_{.jk}^{i}=\ ^{[L]}\widehat{C}_{~jk}^{i}+c_{~jk}^{i},\ \Omega
_{.jk}^{o}=\delta _{j}N_{i}^{o}-\delta _{i}N_{j}^{o}$ and $\
^{[a]}P_{.ks}^{p}=\partial N_{i}^{p}/\partial y^{s}-\ ^{[a]}L_{.ks}^{p}.$
\end{theorem}

The proof consists from a straightforward calculus.

\begin{remark}
\label{rlafs}As a particular case of $\mathbf{GLa}^{n}$, we can define a
Lagrange--affine space $\mathbf{La}^{n}=\left( V^{n},g_{ij}^{[L]}(x,y),\
^{[b]}\mathbf{\Gamma }_{\ \beta }^{\alpha }\right) ,$ provided with a
Lagrange quadratic form $g_{ij}^{[L]}(x,y)$ (\ref{9lagm}) inducing the
canonical N--connection structure $^{[cL]}\mathbf{N}=\{\ ^{[cL]}N_{j}^{i}\}$
(\ref{cncls}) \ as in a Lagrange space $\mathbf{L}^{n}=\left(
V^{n},g_{ij}(x,y)\right) $ (see Definition \ref{defls})) but with a
d--connection structure $\ ^{[b]}\mathbf{\Gamma }_{\ \ \alpha }^{\gamma }=\
^{[b]}\mathbf{\Gamma }_{\ \alpha \beta }^{\gamma }\mathbf{\vartheta }^{\beta
}$ distorted by arbitrary torsion, $\mathbf{T}_{\beta },$ and nonmetricity, $%
\mathbf{Q}_{\beta \gamma },$ d--fields, \
\begin{equation*}
\ \ ^{[b]}\mathbf{\Gamma }_{\ \beta }^{\alpha }=\ ^{[L]}\widehat{\mathbf{%
\Gamma }}_{\beta }^{\alpha }+\ \ ^{[b]}\ \mathbf{Z}_{\ \ \beta }^{\alpha },
\end{equation*}%
where$\ ^{[L]}\widehat{\mathbf{\Gamma }}_{\beta }^{\alpha }$ is the
canonical Lagrange d--connection (\ref{lagdcon}),
\begin{equation*}
\ \ ^{[b]}\ \mathbf{Z}_{\ \ \beta }^{\alpha }=\mathbf{e}_{\beta }\rfloor \
\mathbf{T}_{\alpha }-\mathbf{e}_{\alpha }\rfloor \ \mathbf{T}_{\beta }+\frac{%
1}{2}\left( \mathbf{e}_{\alpha }\rfloor \mathbf{e}_{\beta }\rfloor \ \mathbf{%
T}_{\gamma }\right) \mathbf{\vartheta }^{\gamma }+\left( \mathbf{e}_{\alpha
}\rfloor \ \mathbf{Q}_{\beta \gamma }\right) \mathbf{\vartheta }^{\gamma
}-\left( \mathbf{e}_{\beta }\rfloor \ \mathbf{Q}_{\alpha \gamma }\right)
\mathbf{\vartheta }^{\gamma }+\frac{1}{2}\ \mathbf{Q}_{\alpha \beta },
\end{equation*}%
and the (co) frames $\mathbf{e}_{\beta }$ and $\mathbf{\vartheta }^{\gamma }$
are respectively constructed as in (\ref{1dder}) and (\ref{1ddif}) by using $%
^{[cL]}N_{j}^{i}.$
\end{remark}

\begin{remark}
\label{rfafs}The Finsler--affine spaces $\mathbf{Fa}^{n}=\left(
V^{n},F\left( x,y\right) ,\ ^{[f]}\mathbf{\Gamma }_{\ \beta }^{\alpha
}\right) $ can be introduced by further restrictions of $\mathbf{La}^{n}$ to
a quadratic form $g_{ij}^{[F]}$ (\ref{finm2}) constructed from a Finsler
metric $F\left( x^{i},y^{j}\right) $ inducing the canonical N--connection
structure $\ ^{[F]}\mathbf{N}=\{\ ^{[F]}N_{j}^{i}\}$ (\ref{1ncc})\ as in a
Finsler space $\mathbf{F}^{n}=\left( V^{n},F\left( x,y\right) \right) $ but
with a d--connection structure $\ ^{[f]}\mathbf{\Gamma }_{\ \ \alpha
}^{\gamma }=\ ^{[f]}\mathbf{\Gamma }_{\ \alpha \beta }^{\gamma }\mathbf{%
\vartheta }^{\beta }$ distorted by arbitrary torsion, $\mathbf{T}_{\beta },$
and nonmetricity, $\mathbf{Q}_{\beta \gamma },$ d--fields, \
\begin{equation*}
\ \ ^{[f]}\mathbf{\Gamma }_{\ \beta }^{\alpha }=\ ^{[F]}\widehat{\mathbf{%
\Gamma }}_{\ \beta }^{\alpha }\ +\ \ ^{[f]}\ \mathbf{Z}_{\ \ \beta }^{\alpha
},
\end{equation*}%
where $\ ^{[F]}\widehat{\mathbf{\Gamma }}_{\ \beta }^{\alpha }$ is the
canonical Finsler d--connection (\ref{dccfs}),
\begin{equation*}
\ \ ^{[f]}\ \mathbf{Z}_{\ \ \beta }^{\alpha }=\mathbf{e}_{\beta }\rfloor \
\mathbf{T}_{\alpha }-\mathbf{e}_{\alpha }\rfloor \ \mathbf{T}_{\beta }+\frac{%
1}{2}\left( \mathbf{e}_{\alpha }\rfloor \mathbf{e}_{\beta }\rfloor \ \mathbf{%
T}_{\gamma }\right) \mathbf{\vartheta }^{\gamma }+\left( \mathbf{e}_{\alpha
}\rfloor \ \mathbf{Q}_{\beta \gamma }\right) \mathbf{\vartheta }^{\gamma
}-\left( \mathbf{e}_{\beta }\rfloor \ \mathbf{Q}_{\alpha \gamma }\right)
\mathbf{\vartheta }^{\gamma }+\frac{1}{2}\ \mathbf{Q}_{\alpha \beta },
\end{equation*}%
and the (co) frames $\mathbf{e}_{\beta }$ and $\mathbf{\vartheta }^{\gamma }$
are respectively constructed as in (\ref{1dder}) and (\ref{1ddif}) by using $%
^{[F]}N_{j}^{i}.$
\end{remark}

\begin{remark}
\label{rghas}By similar geometric constructions (see Remarks \ref{rlafs} and %
\ref{rfafs}) on spaces modelling cotangent bundles, we can define generalized
Hamilton--affine spaces $\mathbf{GHa}^{n}=\left( V^{n},\check{g}^{ij}(x,p),\
^{[a]}\mathbf{\check{\Gamma}}_{\ \beta }^{\alpha }\right) $ and theirs
restrictions to Hamilton--affine\\ $\mathbf{Ha}^{n}=(V^{n},\check{g}%
_{[H]}^{ij}(x,p),$ $\ ^{[b]}\mathbf{\check{\Gamma}}_{\ \beta }^{\alpha })$
and Cartan--affine spaces $\mathbf{Ca}^{n}=\left( V^{n},\check{g}%
_{[K]}^{ij}(x,p),\ ^{[c]}\mathbf{\check{\Gamma}}_{\ \beta }^{\alpha }\right)
$ (see sections \ref{shgg} and \ref{scs}) as to contain distorsions induced
by nonmetricity $\mathbf{\check{Q}}_{\alpha \gamma }.$ The geometric objects
have to be adapted to the corresponding N--connection and d--metric/
quadratic form structures (arbitrary $\check{N}_{ij}\left( x,p\right) $ and
d--metric (\ref{dmghs}), $\ ^{[H]}\check{N}_{ij}\left( x,p\right) $ (\ref%
{ncchs}) and quadratic form $\check{g}_{[H]}^{ij}$ (\ref{hsm}) and $\check{N}%
_{ij}^{[K]}$ (\ref{nccartan}) and $\check{g}_{[K]}^{ij}$ (\ref{carm}).
\end{remark}

Finally, in this section, we note that Theorems \ref{ttgfls} and \ref{tcgfls}
can be reformulated in the forms stating procedures of computing d--torsions
and d--curvatures on every type of spaces with nonmetricity and local
anisotropy by adapting the abstract symbol and/or coordinate calculations
with respect to corresponding N--connection, d--metric and canonical
d--connection structures.

\section{Conclusions}

The method of moving anholonomic frames with associated nonlinear connection
(N--connection) structure elaborated in this work on metric--affine spaces
provides a general framework to deal with any possible model of locally
isotropic and/or anisotropic interactions and geometries defined effectively
in the presence of generic off--diagonal metric and linear connection
configurations, in general, subjected to certain anholonomic constraints. As
it has been pointed out, the metric--affine gravity (MAG) contains various
types of generalized Finsler--Lagrange--Hamilton--Cartan geometries which
can be distinguished by a corresponding N--connection structure and metric
and linear connection adapted to the N--connection structure.

As far as the anholonomic frames, nonmetricity and torsion are considered as
fundamental quantities, all mentioned geometries can be included into a
unique scheme which can be developed on arbitrary manifolds, vector and
tangent bundles and their dual bundles (co-bundles) or restricted to
Riemann--Cartan and (pseudo) Riemannian spaces. We observe that a generic
off--diagonal metric (which can not be diagonalized by any coordinate
transform) defining a (pseudo) Riemannian space induces alternatively
various type of Riemann--Cartan and Finsler like configuratons modelled by
respective frame structures. The constructions are generalized if the linear
connection structures are not constrained to metricity conditions. One can
regard this as extensions to metric--affine spaces provided with
N--connection structure modelling also bundle structures and generalized
noncommutative symmetries of metrics and anholonomic frames.

In this paper we have studied the general properties of metric--affine
spaces provided with N--connection structure. We formulated and proved the
main theorems concerning general metric and nonlinear and linear connections
in MAG. There were stated the criteria when the spaces with local isotropy
and/or local anisotropy can be modelled in metric--affine spaces and on
vector/ tangent bundles. We elaborated the concept of generalized
Finsler--affine geometry as a unification of metric--affine (with nontrivial
torsion and nonmetricity) and Finsler like spaces (with nontrivial
N--connection structure and locally anisotropic metrics and connections).

In a general sense, we note that the generalized Finsler--affine geometries
are contained as anhlonomic and noncommutative configurations in extra
dimension gravity models (string and brane models and certain limits to the
Einstein and gauge gravity defined by off--diagonal metrics and anholonomic
constraints). We would like to stress that the N--connection formalism
developed for the metric--affine spaces relates the bulk geometry in string
and/or MAG to gauge theories in vector/tangent bundles and to various type
of non--Riemannian gravity models.

The approach presented here could be advantageous in a triple sense. First,
it provides a uniform treatment of all metric and connection geometries, in
general, with vector/tangent bundle structures which arise in various type
of string and brane gravity models. Second, it defines a complete
classification of the generalized Finsler--affine geometries stated in
Tables 1-11 from the Appendix. Third, it states a new geometric method of
constructing exact solutions with generic off--diagonal metric ansatz,
torsions and nonmetricity, depending on 2--5 variables, in string and
metric--affine gravity, with limits to the Einstein gravity, see Refs. \cite%
{exsolmag}.

%\appendix

\section[Classification of Lagrange--Affine Spaces]
{Appendix:\ Classification of Generalized Finsler--Affine Spaces}

We outline and give a brief characterization of the main classes of
generalized Finsler--affine spaces (see Tables \ref{tablegs}--\ref{tablets}%
). A unified approach to such very different locally isotropic and
anisotropic geometries, defined in the framework of the metric--affine
geometry, can be elaborated only by introducing the concept on N--connection
(see Definition \ref{dnlc}).

The N--connection curvature is computed following the formula $\Omega
_{ij}^{a}=\delta _{\lbrack i}N_{j]}^{a},$ see (\ref{1ncurv}), for any
N--connection $N_{i}^{a}.$ A d--connection $\mathbf{D}=[\mathbf{\Gamma }%
_{\beta \gamma }^{\alpha }]=[L_{\ jk}^{i},L_{\ bk}^{a},C_{\ jc}^{i},C_{\
bc}^{a}]$ (see Definition \ref{defdcon}) defines nontrivial d--torsions $%
\mathbf{T}_{\ \beta \gamma }^{\alpha }=[L_{[\ jk]}^{i},C_{\ ja}^{i},\Omega
_{ij}^{a},T_{\ bj}^{a},C_{\ [bc]}^{a}]$ and d--curvatures $\mathbf{R}_{\
\beta \gamma \tau }^{\alpha }=[R_{\ jkl}^{i},R_{\ bkl}^{a},P_{\
jka}^{i},P_{\ bka}^{c},S_{\ jbc}^{i},S_{\ dbc}^{a}]$ adapted to the
N--connection structure (see, respectively, the formulas (\ref{1dtorsb})\ and
(\ref{1dcurv})). It is considered that a generic off--diagonal metric $%
g_{\alpha \beta }$ (see Remark \ref{rgod}) is associated to a N--connection
structure and reprezented as a d--metric $\mathbf{g}_{\alpha \beta
}=[g_{ij},h_{ab}]$ (see formula (\ref{1block2})). The components of a
N--connection and a d--metric define the canonical d--connection $\mathbf{D}%
=[\widehat{\mathbf{\Gamma }}_{\beta \gamma }^{\alpha }]=[\widehat{L}_{\
jk}^{i},\widehat{L}_{\ bk}^{a},\widehat{C}_{\ jc}^{i},\widehat{C}_{\
bc}^{a}] $ (see (\ref{1candcon})) with the corresponding values of
d--torsions $\widehat{\mathbf{T}}_{\ \beta \gamma }^{\alpha }$ and
d--curvatures $\widehat{\mathbf{R}}_{\ \beta \gamma \tau }^{\alpha }.$ The
nonmetricity d--fields are computed by using formula $\mathbf{Q}_{\alpha
\beta \gamma }=-\mathbf{D}_{\alpha }\mathbf{g}_{\beta \gamma
}=[Q_{ijk},Q_{iab},Q_{ajk},Q_{abc}],$ see (\ref{1nmf}).

\subsection{Generalized Lagrange--affine spaces}

The Table \ref{tablegs} outlines seven classes of geometries modelled in the
framework of metric--affine geometry as spaces with nontrivial N--connection
structure. There are emphasized the configurations:

\begin{enumerate}
\item Metric--affine spaces (in brief, MA) are those stated by Definition %
\ref{defmas} as certain manifolds $V^{n+m}$ of necessary smoothly class\
provided with arbitrary metric, $g_{\alpha \beta },$ and linear connection, $%
\Gamma _{\beta \gamma }^{\alpha },$ structures. For generic off--diagonal
metrics, a MA\ space always admits nontrivial N--connection structures (see
Proposition \ref{pmasnc}). Nevertheless, in general, only the metric field $%
g_{\alpha \beta }$ can be transformed into a d--metric one $\mathbf{g}%
_{\alpha \beta }=[g_{ij},h_{ab}],$ but \ $\Gamma _{\beta \gamma }^{\alpha }$
can be not adapted to the N--connection structure. As a consequence, the
general strength fields $\left( T_{\ \beta \gamma }^{\alpha },R_{\ \beta
\gamma \tau }^{\alpha },Q_{\alpha \beta \gamma }\right) $ can be also not
N--adapted. By using the Kawaguchi's metrization process and Miron's
procedure stated by Theorems \ref{kmp} and \ref{mconnections} we can
consider alternative geometries with d--connections $\mathbf{\Gamma }_{\beta
\gamma }^{\alpha }$ (see Definition \ref{defdcon}) derived from the
components of N--connection and d--metric. Such geometries are adapted to
the N--connection structure. They are characterized by d--torsion $\mathbf{T}%
_{\ \beta \gamma }^{\alpha },$ d--curvature $\mathbf{R}_{\ \beta \gamma \tau
}^{\alpha },$ and nonmetricity d--field $\mathbf{Q}_{\alpha \beta \gamma }.$

\item Distinguished metric--affine spaces (DMA) are defined (see Definition %
\ref{ddmas}) as manifolds $\mathbf{V}^{n+m}$ $\ $provided with N--connection
structure $N_{i}^{a},$ d--metric field (\ref{1block2}) and arbitrary
d--connection $\mathbf{\Gamma }_{\beta \gamma }^{\alpha }.$ In this case,
all strengths $\left( \mathbf{T}_{\ \beta \gamma }^{\alpha },\mathbf{R}_{\
\beta \gamma \tau }^{\alpha },\mathbf{Q}_{\alpha \beta \gamma }\right) $ are
N--adapted.

\item Berwald--affine spaces (BA, see section \ref{berwdcon}) are
metric--affine spaces provided with generic off--diagonal metrics with
associated N--connection structure and with a Ber\-wald d--connection $^{[B]}%
\mathbf{D}=[\ ^{[B]}\mathbf{\Gamma }_{\beta \gamma }^{\alpha }]=[\widehat{L}%
_{\ jk}^{i},\partial _{b}N_{k}^{a},0,\widehat{C}_{\ bc}^{a}]$, see (\ref%
{berw}), for with the d--torsions $\ ^{[B]}\mathbf{T}_{\ \beta \gamma
}^{\alpha }=[\ ^{[B]}L_{\ [jk]}^{i},0,\Omega _{ij}^{a},T_{\ bj}^{a},C_{\
[bc]}^{a}]$ and d--curvatures
\begin{equation*}
^{\lbrack B]}\mathbf{R}_{\ \beta \gamma \tau }^{\alpha }=^{[B]}[R_{\
jkl}^{i},R_{\ bkl}^{a},P_{\ jka}^{i},P_{\ bka}^{c},S_{\ jbc}^{i},S_{\
dbc}^{a}]
\end{equation*}%
\ are computed by introducing the components of $^{[B]}\mathbf{\Gamma }%
_{\beta \gamma }^{\alpha },$ respectively, in formulas (\ref{1dtorsb})\ and (%
\ref{1dcurv}). By definition, this space satisfies the metricity conditions
on the h- and v--subspaces, $Q_{ijk}=0$ and $Q_{abc}=0,$ but, in general,
there are nontrivial nonmetricity d--fields because $Q_{iab}$ and $Q_{ajk}$
are not vanishing (see formulas (\ref{berwnm})).

\item Berwald--affine spaces with prescribed torsion (BAT, see sections \ref%
{berwdcon} and \ref{berwdcona}) are described by a more general class of
d--connection $^{[BT]}\mathbf{\Gamma }_{\beta \gamma }^{\alpha }=[L_{\
jk}^{i},\partial _{b}N_{k}^{a},0,C_{\ bc}^{a}],$ with more general h-- and
v--components, $\ \widehat{L}_{\ jk}^{i}\rightarrow L_{\ jk}^{i}$ and $%
\widehat{C}_{\ bc}^{a}\rightarrow C_{\ bc}^{a},$ inducing prescribed values $%
\tau _{\ jk}^{i}$ and $\tau _{\ bc}^{a}$ in d--torsion\
\begin{equation*}
^{\lbrack BT]}\mathbf{T}_{\ \beta \gamma }^{\alpha }=[L_{\ [jk]}^{i},+\tau
_{\ jk}^{i},0,\Omega _{ij}^{a},T_{\ bj}^{a},C_{\ [bc]}^{a}+\tau _{\ bc}^{a}],
\end{equation*}%
see (\ref{tauformulas})$.$ The components of curvature $^{[BT]}\mathbf{R}_{\
\beta \gamma \tau }^{\alpha }$ have to be computed by introducing $^{[BT]}%
\mathbf{\Gamma }_{\beta \gamma }^{\alpha }$ into (\ref{1dcurv}). There are
nontrivial components of nonmetricity d--fields, $^{[B\tau ]}\mathbf{Q}%
_{\alpha \beta \gamma }=\left( ^{[B\tau ]}Q_{cij},\ ^{[B\tau
]}Q_{iab}\right) .$

\item Generalized Lagrange--affine spaces (GLA, see Definition \ref{defglas}),
 \\  $\mathbf{GLa}^{n}=(V^{n},$\
 $g_{ij}(x,y),$ $\ ^{[a]}\mathbf{\Gamma }_{\ \beta}^{\alpha }),$
 are modelled as distinguished metric--affine spaces of
odd--dimen\-si\-on, $\mathbf{V}^{n+n},$ provided with generic off--diagonal
metrics with associated N--connection inducing a tangent bundle structure.
The d--metric $\mathbf{g}_{[a]}$ (\ref{dmglas}) and the d--connecti\-on $\ \
^{[a]}\mathbf{\Gamma }_{\ \alpha \beta }^{\gamma }$ $=\left( \
^{[a]}L_{jk}^{i},\ ^{[a]}C_{jc}^{i}\right) $ (\ref{dcglma}) are similar to
those for the usual Lagrange spaces (see Definition \ref{defgls}) but with
distorsions $\ ^{[a]}\ \mathbf{Z}_{\ \ \beta }^{\alpha }$ inducing general
nontrivial nonmetricity d--fields $^{[a]}\mathbf{Q}_{\alpha \beta \gamma }.$
The components of d--torsions $^{[a]}\mathbf{T}^{\alpha }=\left( T_{jk}^{i},%
\tilde{T}_{jk}^{i}\right) $and d--curvatures $^{[a]}\mathbf{R}_{.\beta
\gamma \tau }^{\alpha }=\{\ ^{[a]}R_{\ hjk}^{i},\ ^{[a]}P_{\ jka}^{i},\
^{[a]}S_{\ jbc}^{i}\}$ are computed following Theorems \ref{ttgfls} and \ref%
{tcgfls}.

\item Lagrange--affine spaces (LA, see Remark \ \ref{rlafs}), $\mathbf{La}%
^{n}=(V^{n},g_{ij}^{[L]}(x,y),$ $\ ^{[b]}\mathbf{\Gamma }_{\ \beta }^{\alpha
}),$ are provided with a Lagrange quadratic form $g_{ij}^{[L]}(x,y)=\frac{1}{%
2}\frac{\partial ^{2}L^{2}}{\partial y^{i}\partial y^{j}}$ (\ref{9lagm})
inducing the canonical N--connection structure $^{[cL]}\mathbf{N}=\{\
^{[cL]}N_{j}^{i}\}$ (\ref{cncls})\ for a Lagrange space $\mathbf{L}%
^{n}=\left( V^{n},g_{ij}(x,y)\right) $ (see Definition \ref{defls})) but
with a d--connection structure $\ ^{[b]}\mathbf{\Gamma }_{\ \ \alpha
}^{\gamma }=\ ^{[b]}\mathbf{\Gamma }_{\ \alpha \beta }^{\gamma }\mathbf{%
\vartheta }^{\beta }$ distorted by arbitrary torsion, $\mathbf{T}_{\beta },$
and nonmetricity d--fields,$\ \mathbf{Q}_{\beta \gamma \alpha },$ when $%
^{[b]}\mathbf{\Gamma }_{\ \beta }^{\alpha }=\ ^{[L]}\widehat{\mathbf{\Gamma }%
}_{\beta }^{\alpha }+\ \ ^{[b]}\ \mathbf{Z}_{\ \ \beta }^{\alpha }.$ This is
a particular case of GLA spaces with prescribed types of N--connection $%
^{[cL]}N_{j}^{i}$ and d--metric to be like in Lagrange geometry.

\item Finsler--affine spaces (FA, see Remark \ref{rfafs}), $\mathbf{Fa}%
^{n}=\left( V^{n},F\left( x,y\right) ,\ ^{[f]}\mathbf{\Gamma }_{\ \beta
}^{\alpha }\right) ,$ in their turn are introduced by further restrictions
of $\mathbf{La}^{n}$ to a quadratic form $g_{ij}^{[F]}=\frac{1}{2}\frac{%
\partial ^{2}F^{2}}{\partial y^{i}\partial y^{j}}$ (\ref{finm2}) constructed
from a Finsler metric $F\left( x^{i},y^{j}\right) .$ It is induced the
canonical N--connection structure $\ ^{[F]}\mathbf{N}=\{\ ^{[F]}N_{j}^{i}\}$
(\ref{1ncc})\ as in the Finsler space $\mathbf{F}^{n}=\left( V^{n},F\left(
x,y\right) \right) $ but with a d--connection structure $\ ^{[f]}\mathbf{%
\Gamma }_{\ \alpha \beta }^{\gamma }$ distorted by arbitrary torsion, $%
\mathbf{T}_{\beta \gamma }^{\alpha },$ and nonmetricity, $\mathbf{Q}_{\beta
\gamma \tau },$ d--fields, $\ ^{[f]}\mathbf{\Gamma }_{\ \beta }^{\alpha }=\
^{[F]}\widehat{\mathbf{\Gamma }}_{\ \beta }^{\alpha }\ +\ \ ^{[f]}\ \mathbf{Z%
}_{\ \ \beta }^{\alpha },$where $\ ^{[F]}\widehat{\mathbf{\Gamma }}_{\ \beta
\gamma }^{\alpha }$ is the canonical Finsler d--connection (\ref{dccfs}).
\end{enumerate}

\subsection{Generalized Hamilton--affine spaces}

The Table \ref{tableghs} outlines geometries modelled in the framework of
metric--affine geometry as spaces with nontrivial N--connection structure
splitting the space into any conventional a horizontal subspace and vertical
subspace being isomorphic to a dual vector space provided with respective
dual coordinates. We can use respectively the classification from Table \ref%
{tablegs} when the v--subspace is transformed into dual one as we noted in
Remark \ref{rghas} For simplicity, we label such spaces with symbols like $%
\check{N}_{ai}$ instead $N_{i}^{a}$ where ''inverse hat'' points that the
geometric object is defined for a space containing a dual subspaces. The
local h--coordinates are labelled in the usual form, $x^{i},$ with $%
i=1,2,...,n$ but the v--coordinates are certain dual vectors $\check{y}%
^{a}=p_{a}$ with $a=n+1,n+2,...,n+m.$ The local coordinates are denoted $%
\check{u}^{\alpha }=\left( x^{i},\check{y}^{a}\right) =\left(
x^{i},p_{a}\right) .$ The curvature of a N--connection $\check{N}_{ai}$ is
computed as $\check{\Omega}_{iaj}=\delta _{\lbrack i}\check{N}_{j]a}.$ The
h-- v--irreducible components of a general d--connection are parametrized $%
\mathbf{\check{D}}=[\mathbf{\check{\Gamma}}_{\beta \gamma }^{\alpha
}]=[L_{~jk}^{i},L_{a\ \ k}^{\ b},\check{C}_{\ j}^{i\ c},\check{C}_{a}^{\
bc}],$ the d--torsions are $\mathbf{\check{T}}_{\ \beta \gamma }^{\alpha
}=[L_{~[jk]}^{i},L_{a\ \ k}^{\ b},\check{C}_{\ j}^{i\ c},\check{C}_{a}^{\
[bc]}]$ and the d--curvatures $$^{[B]}\mathbf{\check{R}}_{\ \beta \gamma \tau
}^{\alpha }=[R_{\ jkl}^{i},\check{R}_{a\ kl}^{~b},\check{P}_{\ jk}^{i~~a},%
\check{P}_{c~k}^{~b~a},\check{S}_{\ j}^{i~bc},\check{S}_{a\ }^{~dbc}].$$ The
nonmetricity d--fields are stated $\mathbf{\check{Q}}_{\alpha \beta \gamma
}=-\mathbf{\check{D}}_{\alpha }\mathbf{\check{g}}_{\beta \gamma }=[Q_{ijk},%
\check{Q}_{i}^{~ab},\check{Q}_{~jk}^{a},\check{Q}^{abc}].$ There are also
considered additional labels for the Berwald, Cartan and another type
d--connections.

\begin{enumerate}
\item Metric--dual--affine spaces (in brief, MDA) are usual metric--affine
spaces with a prescribed structure of ''dual'' local coordinates.

\item Distinguished metric--dual-affine spaces (DMDA) are provided with
d--metric and d--connection structures adapted to a N--connection $\check{N}%
_{ai}$ defining a global splitting into a usual h--subspace and a
v--dual--subspace being dual to a usual v--subspace.

\item Berwald--dual--affine spaces (BDA) are Berwald--affine spaces with a
dual v--subspa\-ce. Their Berwald d--connection is stated in the form $\ $%
\begin{equation*}
^{\lbrack B]}\mathbf{\check{D}}=[^{[B]}\mathbf{\check{\Gamma}}_{\beta \gamma
}^{\alpha }]=[\widehat{L}_{\ jk}^{i},\partial _{b}\check{N}_{ai},0,\check{C}%
_{a}^{~[bc]}]
\end{equation*}%
with induced d--torsions $\ ^{[B]}\mathbf{\check{T}}_{\ \beta \gamma
}^{\alpha }=[L_{[jk]}^{i},0,\check{\Omega}_{iaj},\check{T}_{a~j}^{~b},%
\check{C}_{a}^{~[bc]}]$ and d--curvatures
\begin{equation*}
^{\lbrack B]}\mathbf{\check{R}}_{\ \beta \gamma \tau }^{\alpha }=[R_{\
jkl}^{i},\check{R}_{a\ kl}^{~b},,\check{P}_{\ jk}^{i~~a},\check{P}%
_{c~k}^{~b~a},\check{S}_{\ j}^{i~bc},\check{S}_{a\ }^{~dbc}]
\end{equation*}%
computed by introducing the components of $^{[B]}\mathbf{\check{\Gamma}}%
_{\beta \gamma }^{\alpha },$ respectively, in formulas (\ref{1dtorsb})\ and (%
\ref{1dcurv}) re--defined for dual v--subspaces. By definition, this
d--connection satisfies the metricity conditions in the h- and v--subspaces,
$Q_{ijk}=0$ and $\check{Q}^{abc}=0$ but with nontrivial components of\\ $^{[B]}%
\mathbf{\check{Q}}_{\alpha \beta \gamma }=-\ ^{[B]}\mathbf{\check{D}}%
_{\alpha }\mathbf{\check{g}}_{\beta \gamma }=[Q_{ijk}=0,\check{Q}%
_{i}^{~ab},\check{Q}_{~jk}^{a},\check{Q}^{abc}=0].$

\item Berwald--dual--affine spaces with prescribed torsion (BDAT) are
described by a more general class of d--connections $\ ^{[BT]}\mathbf{\check{%
\Gamma}}_{\beta \gamma }^{\alpha }=[L_{\ jk}^{i},\partial _{b}\check{N}%
_{ai},0,\check{C}_{a}^{~bc}]$ $,$ inducing prescribed values $\tau _{\
jk}^{i}$ and $\check{\tau}_{a\ }^{~bc}$ for d--torsions\
\begin{equation*}
^{\lbrack BT]}\mathbf{\check{T}}_{\ \beta \gamma }^{\alpha }=[L_{\
[jk]}^{i}+\tau _{\ jk}^{i},0,\check{\Omega}_{iaj}=\delta _{\lbrack i}%
\check{N}_{j]a},T_{a~j}^{~b},\check{C}_{a}^{~[bc]}+\check{\tau}_{a\ }^{~bc}].
\end{equation*}%
The components of d--curvatures
\begin{equation*}
^{\lbrack BT]}\mathbf{\check{R}}_{\ \beta \gamma \tau }^{\alpha }=\ [R_{\
jkl}^{i},\check{R}_{a~\ kl}^{~b},\check{P}_{\ jk}^{i~a},\check{P}_{c\
k}^{~b~a},\check{S}_{\ j}^{i~bc},\check{S}_{a\ }^{~dbc}]
\end{equation*}%
have to be computed by introducing $^{[BT]}\mathbf{\check{\Gamma}}_{\beta
\gamma }^{\alpha }$ into dual form of formulas (\ref{1dcurv}). There are
nontrivial components of nonmetricity d--field, $$^{[B\tau ]}\mathbf{Q}%
_{\alpha \beta \gamma }=-\ ^{[BT]}\mathbf{\check{D}}_{\alpha }\mathbf{\check{%
g}}_{\beta \gamma }=(Q_{ijk}=0,\check{Q}_{i}^{~ab},\check{Q}_{~jk}^{a},%
\check{Q}^{abc}=0).$$

\item Generalized Hamilton--affine spaces (GHA), $\mathbf{GHa}^{n}=\left(
V^{n},\check{g}^{ij}(x,p),\ ^{[a]}\mathbf{\check{\Gamma}}_{\ \beta }^{\alpha
}\right) ,$ are modelled as distinguished metric--affine spaces of
odd--dimension, $\mathbf{V}^{n+n},$ provided with generic off--diagonal
metrics with associated N--connection inducing a cotangent bundle structure.
The d--metric $\mathbf{\check{g}}_{[a]}=[g_{ij},\check{h}^{ab}]$ and the
d--connection $\ \ ^{[a]}\mathbf{\check{\Gamma}}_{\ \alpha \beta }^{\gamma }$
$=(\ ^{[a]}L_{jk}^{i},$ $\ ^{[a]}\check{C}_{i}^{~jc})$ are similar to those
for usual Hamilton spaces (see section \ref{shgg}) but with distorsions $\
^{[a]}\ \mathbf{\check{Z}}_{\ \ \beta }^{\alpha }$ inducing general
nontrivial nonmetricity d--fields $^{[a]}\mathbf{\check{Q}}_{\alpha \beta
\gamma }.$ The components of d--torsion and d--curvature, respectively, $%
^{[a]}\mathbf{\check{T}}_{\ \beta \gamma }^{\alpha }=[L_{\ [jk]}^{i},\check{%
\Omega}_{iaj},\check{C}_{a}^{\ [bc]}]$ and $^{[a]}\mathbf{\check{R}}_{.\beta
\gamma \tau }^{\alpha }=[R_{\ jkl}^{i},\check{P}_{\ jk}^{i~a},\check{S}_{a\
}^{\ dbc}],$ are computed following Theorems \ref{ttgfls} and \ref{tcgfls}
reformulated for cotangent bundle structures.

\item Hamilton--affine spaces (HA, see Remark \ \ref{rghas}), $\mathbf{Ha}%
^{n}=(V^{n},$ $\check{g}_{[H]}^{ij}(x,p),$ $\ ^{[b]}\mathbf{\check{\Gamma}}%
_{\ \beta }^{\alpha }),$ are provided with Hamilton N--connection $^{[H]}%
\check{N}_{ij}\left( x,p\right) $ (\ref{ncchs}) and quadratic form $\check{g}%
_{[H]}^{ij}$ (\ref{hsm}) for a Hamilton space $\mathbf{H}^{n}=\left(
V^{n},H(x,p)\right) $ (see section \ref{shgg})) but with a d--connection
structure $^{[H]}\mathbf{\check{\Gamma}}_{\ \alpha \beta }^{\gamma }=\
^{[H]}[L_{\ jk}^{i},\check{C}_{a}^{\ bc}]$ distorted by arbitrary torsion, $%
\mathbf{\check{T}}_{\ \ \beta \gamma }^{\alpha },$ and nonmetricity
d--fields,$\ \mathbf{\check{Q}}_{\beta \gamma \alpha },$ when $\mathbf{%
\check{\Gamma}}_{\ \beta }^{\alpha }=\ ^{[H]}\widehat{\mathbf{\check{\Gamma}}%
}_{\beta }^{\alpha }+\ \ ^{[H]}\ \mathbf{\check{Z}}_{\ \ \beta }^{\alpha }.$
This is a particular case of GHA spaces with prescribed types of
N--connection $\ ^{[H]}\check{N}_{ij}$ and d--metric$\ \mathbf{\check{g}}%
_{\alpha \beta }^{[H]}=[g_{[H]}^{ij}=\frac{1}{2}\frac{\partial ^{2}H}{%
\partial p_{i}\partial p_{i}}]$ to be like in the Hamilton geometry.

\item Cartan--affine spaces (CA, see Remark \ref{rghas}), $\mathbf{Ca}%
^{n}=\left( V^{n},\check{g}_{[K]}^{ij}(x,p),\ ^{[c]}\mathbf{\check{\Gamma}}%
_{\ \beta }^{\alpha }\right) ,$ are dual to the Finsler spaces $\mathbf{Fa}%
^{n}=\left( V^{n},F\left( x,y\right) ,\ ^{[f]}\mathbf{\Gamma }_{\ \beta
}^{\alpha }\right) .$ The CA spaces are introduced by further restrictions
of $\mathbf{Ha}^{n}$ to a quadratic form $\check{g}_{[C]}^{ij}$ (\ref{carm})
and canonical N--connection $\check{N}_{ij}^{[C]}$ (\ref{nccartan}). They
are like usual Cartan spaces,\ see section \ref{scs}) but contain
distorsions induced by nonmetricity $\mathbf{\check{Q}}_{\alpha \beta \gamma
}.$ The d--metric is parametrized $\mathbf{\check{g}}_{\alpha \beta
}^{[C]}=[g_{[C]}^{ij}=\frac{1}{2}\frac{\partial ^{2}K^{2}}{\partial
p_{i}\partial p_{i}}]$ and the curvature $\ ^{[C]}\check{\Omega}_{iaj}$ of
N--connection $\ ^{[C]}\check{N}_{ia}$ is computed $\ ^{[C]}\check{\Omega}%
_{iaj}=\delta _{\lbrack i}\ ^{[C]}\check{N}_{j]a}.$ The Cartan's
d--connection \ $$^{[C]}\mathbf{\check{\Gamma}}_{\ \alpha \beta }^{\gamma }=\
^{[C]}[L_{\ jk}^{i},L_{\ jk}^{i},\check{C}_{a}^{\ bc},\check{C}_{a}^{\ bc}]$$
possess nontrivial d--torsions $^{[C]}\mathbf{\check{T}}_{\alpha \ }^{\
\beta \gamma }=[L_{\ [jk]}^{i},\check{\Omega}_{iaj},\check{C}_{a}^{\ [bc]}]$
and d--curvatures\\ $^{[C]}\mathbf{\check{R}}_{.\beta \gamma \tau }^{\alpha
}=[R_{\ jkl}^{i},$ $\check{P}_{\ jk}^{i~a},$ $\check{S}_{a\ }^{\ dbc}]$
computed following Theorems \ref{ttgfls} and \ref{tcgfls} reformulated on
cotangent bundles with explicit type of N--connection $\check{N}_{ij}^{[C]}$
d--metric $\mathbf{\check{g}}_{\alpha \beta }^{[C]}$ and d--connection $%
^{[C]}\mathbf{\check{\Gamma}}_{\ \alpha \beta }^{\gamma }.$ The nonmetricity
d--fields are not trivial for such spaces, $\ ^{[C]}\mathbf{\check{Q}}%
_{\alpha \beta \gamma }=-\ ^{[C]}\mathbf{\check{D}}_{\alpha }\mathbf{\check{g%
}}_{\beta \gamma }$ $=$ $[Q_{ijk},$ $\check{Q}_{i}^{~ab},$ $\check{Q}%
_{~jk}^{a},\check{Q}^{abc}].$
\end{enumerate}

\subsection{Teleparallel Lagrange--affine spaces}

We considered the main properties of teleparallel Finsler--affine spaces in
section \ref{stpfa} (see also section \ref{stps} on locally isotropic
teleparallel spaces). Every type of teleparallel spaces is distinguished by
the condition that the curvature tensor vanishes but the torsion plays a
cornerstone role. Modelling generalized Finsler structures on metric--affine
spaces, we do not impose the condition on vanishing nonmetricity (which is
stated for usual teleparallel spaces). For $\mathbf{R}_{\ \beta \gamma \tau
}^{\alpha }=0,$ the classification of spaces from Table \ref{tablegs}
trasforms in that from Table \ref{tabletls}.

\begin{enumerate}
\item Teleparallel metric--affine spaces (in brief, TMA) are usual
metric--affine ones but with vanishing curvature, modelled on manifolds $%
V^{n+m}$ of necessary smoothly class\ provided, for instance, with the
Weitzenbock connection $^{[W]}\Gamma _{\beta \gamma }^{\alpha }$ (\ref{wcon}%
). For generic off--diagonal metrics, a TMA\ space always admits nontrivial
N--connection structures (see Proposition \ref{pmasnc}). We can model
teleparallel geometries with local anisot\-ro\-py by distorting the Levi--Civita
or the canonical d--connection $\mathbf{\Gamma }_{\beta \gamma }^{\alpha }$
(see Definition \ref{defdcon}) both constructed from the components of
N--connection and d--metric. In general, such geometries are characterized
by d--torsion $\mathbf{T}_{\ \beta \gamma }^{\alpha }$ and nonmetricity
d--field $\mathbf{Q}_{\alpha \beta \gamma }$ both constrained to the
condition to result in zero d--curvatures.

\item Distinguished teleparallel metric--affine spaces (DTMA) are manifolds $%
\mathbf{V}^{n+m}$ provided with N--connection structure $N_{i}^{a},$
d--metric field (\ref{1block2}) and d--connection $\mathbf{\Gamma }_{\beta
\gamma }^{\alpha }$ with vanishing d--curvatures defined by
Weitzenbock--affine d--connection $^{[Wa]}\mathbf{\Gamma }_{\beta \gamma
}^{\alpha }=\mathbf{\Gamma }_{\bigtriangledown ~\beta \gamma }^{\alpha }+%
\mathbf{\hat{Z}}_{~\beta \gamma }^{\alpha }+\mathbf{Z}_{~\beta \gamma
}^{\alpha }$ with distorsions by nonmetricity d--fields preserving the
condition of zero values for d--curvatures.

\item Teleparallel Berwald--affine spaces (TBA) are defined by distorsions
of the Weitzenbock connection to any Berwald like structure, $\ ^{[WB]}%
\mathbf{\Gamma }_{\beta \gamma }^{\alpha }=\mathbf{\Gamma }%
_{\bigtriangledown ~\beta \gamma }^{\alpha }+\mathbf{\hat{Z}}_{~\beta \gamma
}^{\alpha }+\mathbf{Z}_{~\beta \gamma }^{\alpha }$ satisfying the condition
that the curvature is zero. All constructions with generic off--diagonal
metrics can be adapted to the N--connection and considered for d--objects.
By definition, such spaces satisfy the metricity conditions in the h- and
v--subspaces, $Q_{ijk}=0$ and $Q_{abc}=0,$ but, in general, there are
nontrivial nonmetricity d--fields because $Q_{iab}$ and $Q_{ajk}$ are not
vanishing (see formulas (\ref{berwnm})).

\item Teleparallel Berwald--affine spaces with prescribed torsion (TBAT) are
defined by a more general class of distorsions resulting in the Weitzenbock
type d--connections, $\ ^{[WB\tau ]}\mathbf{\Gamma }_{\beta \gamma }^{\alpha
}=\mathbf{\Gamma }_{\bigtriangledown ~\beta \gamma }^{\alpha }+\mathbf{\hat{Z%
}}_{~\beta \gamma }^{\alpha }+\mathbf{Z}_{~\beta \gamma }^{\alpha },$ with
more general h-- and v--components, $\ \widehat{L}_{\ jk}^{i}\rightarrow
L_{\ jk}^{i}$ and $\widehat{C}_{\ bc}^{a}\rightarrow C_{\ bc}^{a},$ having
prescribed values $\tau _{\ jk}^{i}$ and $\tau _{\ bc}^{a}$ in d--torsion\ $$%
^{[WB]}\mathbf{T}_{\ \beta \gamma }^{\alpha }=[L_{\ [jk]}^{i},+\tau _{\
jk}^{i},0,\Omega _{ij}^{a},T_{\ bj}^{a},C_{\ [bc]}^{a}+\tau _{\ bc}^{a}]$$
and characterized by the condition $^{[WB\tau ]}\mathbf{R}_{\ \beta \gamma
\tau }^{\alpha }=0$ with notrivial components of nonmetricity $^{[WB\tau ]}%
\mathbf{Q}_{\alpha \beta \gamma }=\left( Q_{cij},\ Q_{iab}\right) .$

\item Teleparallel generalized Lagrange--affine spaces (TGLA) are
distinguished metric--affine spaces of odd--dimension, $\mathbf{V}^{n+n},$
provided with generalized Lagrange d--metric and associated N--connection
inducing a tangent bundle structure with vanishing d--cur\-va\-tu\-re. The\
Weitzenblock--Lagrange d--connection $\ ^{[Wa]}\mathbf{\Gamma }_{\ \alpha
\beta }^{\gamma }$ $=\left( \ ^{[Wa]}L_{jk}^{i},\ ^{[Wa]}C_{jc}^{i}\right)
,~\ $\ where $$^{[WaL]}\mathbf{\Gamma }_{\beta \gamma }^{\alpha }=\mathbf{%
\Gamma }_{\bigtriangledown ~\beta \gamma }^{\alpha }+\mathbf{\hat{Z}}%
_{~\beta \gamma }^{\alpha }+\mathbf{Z}_{~\beta \gamma }^{\alpha }$$ is
defined by a d--metric $\mathbf{g}_{[a]}$ (\ref{dmglas}) $\mathbf{Z}_{\ \
\beta }^{\alpha }$ inducing general nontrivial nonmetrici\-ty d--fields $^{[a]}%
\mathbf{Q}_{\alpha \beta \gamma }$ and $^{[Wa]}\mathbf{R}_{\ \beta \gamma
\tau }^{\alpha }=0.$

\item Teleparallel Lagrange--affine spaces (TLA\ \ref{rlafs}) consist a
subclass of spaces\\ $\mathbf{La}^{n}=\left( V^{n},g_{ij}^{[L]}(x,y),\ ^{[b]}%
\mathbf{\Gamma }_{\ \beta }^{\alpha }\right) $ provided with a Lagrange
quadratic form $g_{ij}^{[L]}(x,y)=\frac{1}{2}\frac{\partial ^{2}L^{2}}{%
\partial y^{i}\partial y^{j}}$ (\ref{9lagm}) inducing the canonical
N--connection structure $^{[cL]}\mathbf{N}=\{\ ^{[cL]}N_{j}^{i}\}$ (\ref%
{cncls})\ for a Lagrange space $\mathbf{L}^{n}=\left(
V^{n},g_{ij}(x,y)\right) $ but with vanishing d--curvature. The
d--connection structure $\ ^{[WL]}\mathbf{\Gamma }_{\ \alpha \beta }^{\gamma
}$ (of Weitzenblock--Lagrange type) is the generated as a distortion by the
Weitzenbock d--torsion, $^{[W]}\mathbf{T}_{\beta },$ and nonmetricity
d--fields,$\ \mathbf{Q}_{\beta \gamma \alpha },$ when $^{[WL]}\mathbf{\Gamma
}_{\ \alpha \beta }^{\gamma }=\mathbf{\Gamma }_{\bigtriangledown ~\beta
\gamma }^{\alpha }+\mathbf{\hat{Z}}_{~\beta \gamma }^{\alpha }+\mathbf{Z}%
_{~\beta \gamma }^{\alpha }.$ This is a generalization of teleparallel
Finsler affine spaces (see section (\ref{stpfa})) when $g_{ij}^{[L]}(x,y)$
is considered instead of $g_{ij}^{[F]}(x,y).$

\item Teleparallel Finsler--affine spaces (TFA) are particular cases of
spaces of type\\ $\mathbf{Fa}^{n}=(V^{n},F\left( x,y\right) ,\ ^{[f]}\mathbf{%
\Gamma }_{\ \beta }^{\alpha }),$ defined by a quadratic form $g_{ij}^{[F]}=%
\frac{1}{2}\frac{\partial ^{2}F^{2}}{\partial y^{i}\partial y^{j}}$ (\ref%
{finm2}) constructed from a Finsler metric $F\left( x^{i},y^{j}\right) .$
They are provided with a canonical N--connection structure $\ ^{[F]}\mathbf{N%
}=\{\ ^{[F]}N_{j}^{i}\}$ (\ref{1ncc})\ as in the Finsler space $\mathbf{F}%
^{n}=\left( V^{n},F\left( x,y\right) \right) $ but with a
Finsler--Weitzenbock d--connection structure $\ ^{[WF]}\mathbf{\Gamma }_{\
\alpha \beta }^{\gamma },$ respective d--torsion, $^{[WF]}\mathbf{T}_{\beta
},$ and nonmetricity, $\mathbf{Q}_{\beta \gamma \tau },$ d--fields, $$\
^{[WF]}\mathbf{\Gamma }_{\ \alpha \beta }^{\gamma }=\mathbf{\Gamma }%
_{\bigtriangledown ~\beta \gamma }^{\alpha }+\mathbf{\hat{Z}}_{~\beta \gamma
}^{\alpha }+\mathbf{Z}_{~\beta \gamma }^{\alpha },$$ where $\ \mathbf{\hat{Z}}%
_{~\beta \gamma }^{\alpha }$ contains distorsions from the canonical Finsler
d--connection (\ref{dccfs}). Such distorsions are constrained to satisfy the
condition of vanishing curvature d--tensors (see section (\ref{stpfa})).
\end{enumerate}

\subsection{Teleparallel Hamilton--affine spaces}

This class of metric--affine spaces is similar to that outlined in previous
subsection, see Table \ref{tabletls} but derived on spaces with dual vector
bundle structure and induced generalized Hamilton--Cartan geometry (section %
\ref{shgg} and Remark \ref{rghas}). We outline the main denotations for such
spaces and note that they are characterized by the condition $\mathbf{%
\check{R}}_{\ \beta \gamma \tau }^{\alpha }=0.$

\begin{enumerate}
\item Teleparallel metric dual affine spaces (in brief, TMDA) define
teleparalles structures on metric--affine spaces provided with generic
off--diagonal metrics and associated N--connections modelling splitting with
effective dual vector bundle structures.

\item Distinguished teleparallel metric dual affine spaces (DTMDA) are
spaces provided with independent d--metric, d--connection structures adapted
to a N--connection in an effective dual vector bundle and resulting in zero
d--curvatures.

\item Teleparallel Berwald dual affine spaces (TBDA) .

\item Teleparallel dual Berwald--affine spaces with prescribed torsion
(TDBAT).

\item Teleparallel dual generalized Hamilton--affine spaces (TDGHA).

\item Teleparallel dual Hamilton--affine spaces (TDHA, see\ section \ref%
{rlafs}).

\item Teleparallel dual Cartan--affine spaces (TDCA).
\end{enumerate}

\subsection{Generalized Finsler--Lagrange spaces}

This class of geometries is modelled on vector/tangent bundles \cite{ma} (see
subsections \ref{ssffc} and \ref{ssslgg}) or on metric--affine spaces
provided with N--connection structure. There are also alternative variants
when metric--affine structures are defined for vector/tangent bundles with
independent generic off--diagonal metrics and linear connection structures.
The standard approaches to generalized Finsler geometries emphasize the
connections satisfying the metricity conditions. Nevertheless, the Berwald
type connections admit certain nonmetricity d--fields. The classification
stated in Table \ref{tablegfls} is similar to that from Table \ref{tablegs}
with that difference that the spaces are defined from the very beginning to
be any vector or tangent bundles. The local coordinates $x^{i}$ are
considered for base subspaces and $y^{a}$ are for fiber type subspaces. We
list the short denotations and main properties of such spaces:

\begin{enumerate}
\item Metric affine vector bundles (in brief, MAVB) are provided with
arbitrary metric $g_{\alpha \beta }~$ and linear connection $\Gamma _{\beta
\gamma }^{\alpha }$ structure. For generic off--diagonal metrics, we can
introduce associated nontrivial N--connection structures. In general, only
the metric field $g_{\alpha \beta }$ can be transformed into a d--metric $%
\mathbf{g}_{\alpha \beta }=[g_{ij},h_{ab}],$ but \ $\Gamma _{\beta \gamma
}^{\alpha }$ may be not adapted to the N--connection structure. As a
consequence, the general strength fields $\left( T_{\ \beta \gamma }^{\alpha
},R_{\ \beta \gamma \tau }^{\alpha },Q_{\alpha \beta \gamma }=0\right) ,$
defined in the total space of the vector bundle are also not N--adapted. We
can consider a metric--affine (MA) structure on the total space if $%
Q_{\alpha \beta \gamma }\neq 0.$

\item Distinguished metric--affine vector bundles (DMAVB) are provided with
N--connec\-ti\-on structure $N_{i}^{a},$ d--metric field and arbitrary
d--connection $\mathbf{\Gamma }_{\beta \gamma }^{\alpha }.$ In this case,
all strengths $(\mathbf{T}_{\ \beta \gamma }^{\alpha },\mathbf{R}_{\ \beta
\gamma \tau }^{\alpha },\mathbf{Q}_{\alpha \beta \gamma }=0) $ are
N--adapted. A distinguished metric--affine (DMA) structure on the total
space is considered if $\mathbf{Q}_{\alpha \beta \gamma }\neq 0.$

\item Berwald metric--affine tangent bundles (BMATB) are provided with
Berwald d--con\-nec\-ti\-on structure $^{[B]}\mathbf{\Gamma }.$ By
definition, this space satisfies the metricity conditions in the h- and
v--subspaces, $Q_{ijk}=0$ and $Q_{abc}=0,$ but, in general, there are
nontrivial nonmetricity d--fields because $Q_{iab}$ and $Q_{ajk}$ do not
vanish (see formulas (\ref{berwnm})).

\item Berwald metric--affine bundles with prescribed torsion (BMATBT) are
described by a more general class of d--connection $^{[BT]}\mathbf{\Gamma }%
_{\beta \gamma }^{\alpha }=[L_{\ jk}^{i},\partial _{b}N_{k}^{a},0,C_{\
bc}^{a}]$ inducing prescribed values $\tau _{\ jk}^{i}$ and $\tau _{\
bc}^{a} $ in d--torsion\
\begin{equation*}
^{\lbrack BT]}\mathbf{T}_{\ \beta \gamma }^{\alpha }=[L_{\ [jk]}^{i},+\tau
_{\ jk}^{i},0,\Omega _{ij}^{a},T_{\ bj}^{a},C_{\ [bc]}^{a}+\tau _{\ bc}^{a}],
\end{equation*}%
see (\ref{tauformulas})$.$ There are nontrivial nonmetricity d--fields, $%
^{[B\tau ]}\mathbf{Q}_{\alpha \beta \gamma }=(Q_{cij},Q_{iab}).$

\item Generalized Lagrange metric--affine bundles (GLMAB) are modelled as $%
\mathbf{GLa}^{n}=(V^{n},$ $g_{ij}(x,y),\ ^{[a]}\mathbf{\Gamma }_{\ \beta
}^{\alpha })$ spaces on tangent bundles provided with generic off--diagonal
metrics with associated N--connection. If the d--connection is a canonical
one, $\widehat{\mathbf{\Gamma }}_{\beta \gamma }^{\alpha },$ the
nonmetricity vanish. But we can consider arbitrary d--connections $\mathbf{%
\Gamma }_{\beta \gamma }^{\alpha }$ with nontrivial nonmetricity d--fields.

\item Lagrange metric--affine bundles (LMAB) are defined on tangent bundles
as spaces $\mathbf{La}^{n}=\left( V^{n},g_{ij}^{[L]}(x,y),\ ^{[b]}\mathbf{%
\Gamma }_{\ \beta }^{\alpha }\right) $ provided with a Lagrange quadratic
form $g_{ij}^{[L]}(x,y)=\frac{1}{2}\frac{\partial ^{2}L^{2}}{\partial
y^{i}\partial y^{j}}$ inducing the canonical N--connection structure $^{[cL]}%
\mathbf{N}=\{\ ^{[cL]}N_{j}^{i}\}$ for a Lagrange space $\mathbf{L}%
^{n}=\left( V^{n},g_{ij}(x,y)\right) $ (see Definition \ref{defls})) but
with a d--connection structure $\ ^{[b]}\mathbf{\Gamma }_{\ \ \alpha
}^{\gamma }=\ ^{[b]}\mathbf{\Gamma }_{\ \alpha \beta }^{\gamma }\mathbf{%
\vartheta }^{\beta }$ distorted by arbitrary torsion, $\mathbf{T}_{\beta },$
and nonmetricity d--fields,$\ \mathbf{Q}_{\beta \gamma \alpha },$ when $%
^{[b]}\mathbf{\Gamma }_{\ \beta }^{\alpha }=\ ^{[L]}\widehat{\mathbf{\Gamma }%
}_{\beta }^{\alpha }+\ \ ^{[b]}\ \mathbf{Z}_{\ \ \beta }^{\alpha }.$ This is
a particular case of GLA spaces with prescribed types of N--connection $%
^{[cL]}N_{j}^{i}$ and d--metric to be like in Lagrange geometry.

\item Finsler metric--affine bundles (FMAB), are modelled on tangent bundles
as spaces $\mathbf{Fa}^{n}=\left( V^{n},F\left( x,y\right) ,\ ^{[f]}\mathbf{%
\Gamma }_{\ \beta }^{\alpha }\right) $ with quadratic form $g_{ij}^{[F]}=%
\frac{1}{2}\frac{\partial ^{2}F^{2}}{\partial y^{i}\partial y^{j}}$ (\ref%
{finm2}) constructed from a Finsler metric $F\left( x^{i},y^{j}\right) .$ It
is induced the canonical N--connection structure $\ ^{[F]}\mathbf{N}=\{\
^{[F]}N_{j}^{i}\}$ as in the Finsler space $\mathbf{F}^{n}=\left(
V^{n},F\left( x,y\right) \right) $ but with a d--connection structure $\
^{[f]}\mathbf{\Gamma }_{\ \alpha \beta }^{\gamma }$ distorted by arbitrary
torsion, $\mathbf{T}_{\beta \gamma }^{\alpha },$ and nonmetricity, $\mathbf{Q%
}_{\beta \gamma \tau },$ d--fields, $\ ^{[f]}\mathbf{\Gamma }_{\ \beta
}^{\alpha }=\ ^{[F]}\widehat{\mathbf{\Gamma }}_{\ \beta }^{\alpha }\ +\ \
^{[f]}\ \mathbf{Z}_{\ \ \beta }^{\alpha },$where $\ ^{[F]}\widehat{\mathbf{%
\Gamma }}_{\ \beta \gamma }^{\alpha }$ is the canonical Finsler
d--connection (\ref{dccfs}).
\end{enumerate}

\subsection{Generalized Hamilton--Cartan spaces}

Such spaces are modelled on vector/tangent dual bundles (see sections
subsections \ref{shgg} and \ref{scs}) as metric--affine spaces provided with
N--connection structure. The classification stated in Table \ref{tableghcs}
is similar to that from Table \ref{tableghs} with that difference that the
geometry is modelled from the very beginning as vector or tangent dual
bundles. The local coordinates $x^{i}$ are considered for base subspaces and
$y^{a}=p_{a}$ are for cofiber type subspaces. So, the spaces from Table \ref%
{tableghcs} are dual to those from Table \ref{tabletfls}, when the
respective Lagrange--Finsler structures are changed into Hamilton--Cartan
structures. We list the short denotations and main properties of such spaces:

\begin{enumerate}
\item The metric--affine dual vector bundles (in brief, MADVB) are defined
by metric--affine independent metric and linear connection structures stated
on dual vector bundles. For generic off--diagonal metrics, there are
nontrivial N--connection structures. The linear connection may be not
adapted to the N--connection structure.

\item Distinguished metric-affine dual vector bundles (DMADVB) are provided
with d--metric and d--connection structures adapted to a N--connection $%
\check{N}_{ai}.$

\item Berwald metric--affine dual bundles (BMADB) are provided with a
Berwald d--con\-nec\-ti\-on
\begin{equation*}
\ ^{[B]}\mathbf{\check{D}}=[\ ^{[B]}\mathbf{\check{\Gamma}}_{\beta \gamma
}^{\alpha }]=[\widehat{L}_{\ jk}^{i},\partial _{b}\check{N}_{ai},0,\check{C}%
_{a}^{~[bc]}].
\end{equation*}%
By definition, on such spaces, there are satisfied the metricity conditions
in the h- and v--subspaces, $Q_{ijk}=0$ and $\check{Q}^{abc}=0$ but with
nontrivial components of $^{[B]}\mathbf{\check{Q}}_{\alpha \beta \gamma }=-\
^{[B]}\mathbf{\check{D}}_{\alpha }\mathbf{\check{g}}_{\beta \gamma
}=[Q_{ijk}=0,\check{Q}_{i}^{~ab},\check{Q}_{~jk}^{a},\check{Q}^{abc}=0].$

\item Berwald metrical--affine dual bundles with prescribed torsion (BMADBT)
are described by a more general class of d--connections $\ ^{[BT]}\mathbf{%
\check{\Gamma}}_{\beta \gamma }^{\alpha }=[L_{\ jk}^{i},\partial _{b}%
\check{N}_{ai},0,\check{C}_{a}^{~bc}]$ inducing prescribed values $\tau _{\
jk}^{i}$ and $\check{\tau}_{a\ }^{~bc}$ for d--torsions\
\begin{equation*}
^{\lbrack BT]}\mathbf{\check{T}}_{\ \beta \gamma }^{\alpha }=[L_{\
[jk]}^{i}+\tau _{\ jk}^{i},0,\check{\Omega}_{iaj}=\delta _{\lbrack i}%
\check{N}_{j]a},T_{a~j}^{~b},\check{C}_{a}^{~[bc]}+\check{\tau}_{a\ }^{~bc}].
\end{equation*}%
There are nontrivial components of nonmetricity d--field, $^{[B\tau ]}%
\mathbf{Q}_{\alpha \beta \gamma }=\ ^{[BT]}\mathbf{\check{D}}_{\alpha }%
\mathbf{\check{g}}_{\beta \gamma }=\left( Q_{ijk}=0,\check{Q}_{i}^{~ab},%
\check{Q}_{~jk}^{a},\check{Q}^{abc}=0\right) .$

\item Generalized metric--affine Hamilton bundles (GMAHB) are modelled on
dual vector bundles as spaces $\mathbf{GHa}^{n}=\left( V^{n},\check{g}%
^{ij}(x,p),\ ^{[a]}\mathbf{\check{\Gamma}}_{\ \beta }^{\alpha }\right) ,$
provided with generic off--diagonal metrics with associated N--connection
inducing a cotangent bundle structure. The d--metric $\mathbf{\check{g}}%
_{[a]}=[g_{ij},\check{h}^{ab}]$ and the d--connection$\ ^{[a]}\mathbf{\check{%
\Gamma}}_{\ \alpha \beta }^{\gamma }$ $=\left( \ ^{[a]}L_{jk}^{i},\ ^{[a]}%
\check{C}_{i}^{~jc}\right) $ are similar to those for usual Hamilton spaces,
with distorsions $\ ^{[a]}\ \mathbf{\check{Z}}_{\ \ \beta }^{\alpha }$
inducing general nontrivial nonmetricity d--fields $^{[a]}\mathbf{\check{Q}}%
_{\alpha \beta \gamma }.$ For canonical configurations, $^{[GH]}\mathbf{%
\check{\Gamma}}_{\ \alpha \beta }^{\gamma },$ we obtain $^{[GH]}\mathbf{%
\check{Q}}_{\alpha \beta \gamma }=0.$

\item Metric--affine Hamilton bundles (MAHB) are defied on dual bundles as
spaces $\mathbf{Ha}^{n}=\left( V^{n},\check{g}_{[H]}^{ij}(x,p),\ ^{[b]}%
\mathbf{\check{\Gamma}}_{\ \beta }^{\alpha }\right) ,$ provided with
Hamilton N--connection $^{[H]}\check{N}_{ij}\left( x,p\right) $ and
qua\-drat\-ic form $\check{g}_{[H]}^{ij}$ for a Hamilton space $\mathbf{H}%
^{n}=\left( V^{n},H(x,p)\right) $ (see section \ref{shgg}) with a
d--connection structure $^{[H]}\mathbf{\check{\Gamma}}_{\ \alpha \beta
}^{\gamma }=\ ^{[H]}[L_{\ jk}^{i},\check{C}_{a}^{\ bc}]$ distorted by
arbitrary torsion, $\mathbf{\check{T}}_{\ \ \beta \gamma }^{\alpha },$ and
nonmetricity d--fields,$\ \mathbf{\check{Q}}_{\beta \gamma \alpha },$ when $%
\mathbf{\check{\Gamma}}_{\ \beta }^{\alpha }=\ ^{[H]}\widehat{\mathbf{\check{%
\Gamma}}}_{\beta }^{\alpha }+\ \ ^{[H]}\ \mathbf{\check{Z}}_{\ \ \beta
}^{\alpha }.$ This is a particular case of GMAHB spaces with prescribed
types of N--connection $\ ^{[H]}\check{N}_{ij}$ and d--metric$\ \mathbf{%
\check{g}}_{\alpha \beta }^{[H]}=[g_{[H]}^{ij}=\frac{1}{2}\frac{\partial
^{2}H}{\partial p_{i}\partial p_{i}}]$ to be like in the Hamilton geometry
but with nontrivial nonmetricity.

\item Metric--affine Cartan bundles (MACB) are modelled on dual tangent
bundles as spaces $\mathbf{Ca}^{n}=\left( V^{n},\check{g}_{[K]}^{ij}(x,p),\
^{[c]}\mathbf{\check{\Gamma}}_{\ \beta }^{\alpha }\right) $ being dual to
the Finsler spaces\textbf{.} They are like usual Cartan spaces,\ see section %
\ref{scs}) but may contain distorsions induced by nonmetricity $\mathbf{%
\check{Q}}_{\alpha \beta \gamma }.$ The d--metric is parametrized $\mathbf{%
\check{g}}_{\alpha \beta }^{[C]}=[g_{[C]}^{ij}=\frac{1}{2}\frac{\partial
^{2}K^{2}}{\partial p_{i}\partial p_{i}}]$ and the curvature $\ ^{[C]}\check{%
\Omega}_{iaj}$ of N--connection $\ ^{[C]}\check{N}_{ia}$ is computed $\
^{[C]}\check{\Omega}_{iaj}=\delta _{\lbrack i}\ ^{[C]}\check{N}_{j]a}.$ The
Cartan's d--connection \ $^{[C]}\mathbf{\check{\Gamma}}_{\ \alpha \beta
}^{\gamma }=\ ^{[C]}[L_{\ jk}^{i},L_{\ jk}^{i},\check{C}_{a}^{\ bc},\check{C}%
_{a}^{\ bc}]$ possess nontrivial d--torsions $^{[C]}\mathbf{\check{T}}%
_{\alpha \ }^{\ \beta \gamma }=[L_{\ [jk]}^{i},\check{\Omega}_{iaj},\check{C}%
_{a}^{\ [bc]}]$ and d--curvatures $^{[C]}\mathbf{\check{R}}_{.\beta \gamma
\tau }^{\alpha }=~^{[C]}[R_{\ jkl}^{i},\check{P}_{\ jk}^{i~a},\check{S}_{a\
}^{\ dbc}]$ computed following Theorems \ref{ttgfls} and \ref{tcgfls}
reformulated on cotangent bundles with explicit type of N--connection $%
\check{N}_{ij}^{[C]}$ d--metric $\mathbf{\check{g}}_{\alpha \beta }^{[C]}$
and d--connection $^{[C]}\mathbf{\check{\Gamma}}_{\ \alpha \beta }^{\gamma
}. $ Distorsions result in d--connection $\mathbf{\check{\Gamma}}_{\ \beta
\gamma }^{\alpha }=\ ^{[C]}\mathbf{\check{\Gamma}}_{\beta \gamma }^{\alpha
}+\ ^{[C]}\ \mathbf{\check{Z}}_{\ \ \beta \gamma }^{\alpha }.$ The
nonmetricity d--fields are not trivial for such spaces.
\end{enumerate}

\subsection{Teleparallel Finsler--Lagrange spaces}

The teleparallel configurations can be modelled on vector and tangent bundles
(the teleparallel Finsler--affine spaces are defined in section \ref{stpfa},
see also section \ref{stps} on locally isotropic teleparallel spaces) were
constructed as subclasses of metric--affine spaces on manifolds of necessary
smoothly class. The classification from Table \ref{tabletfls} is a similar
to that from Table \ref{tabletls} but for direct vector/ tangent bundle
configurations with vanishing nonmetricity. Nevertheless, certain nonzero
nonmetricity d--fields can be present if the Berwald d--connection is
considered or if we consider a metric--affine geometry in bundle spaces.

\begin{enumerate}
\item Teleparallel vector bundles (in brief, TVB) are provided with
independent metric and linear connection structures like in metric--affine
spaces satisfying the condition of vanishing curvature. The N--connection is
associated to generic off--diagonal metrics. The TVB spaces can be provided
with a Weitzenbock connection $^{[W]}\Gamma _{\beta \gamma }^{\alpha }$(\ref%
{wcon}) which can be transformed in a d--connection one with respect to
N--adapted frames. We can model teleparallel geometries with local
anisotropy by distorting the Levi--Civita or the canonical d--connection $%
\mathbf{\Gamma }_{\beta \gamma }^{\alpha }$ (see Definition \ref{defdcon})
both constructed from the components of N--connection and d--metric. In
general, such vector (in particular cases, tangent) bundle geometries are
characterized by d--torsions $\mathbf{T}_{\ \beta \gamma }^{\alpha }$ and
nonmetricity d--fields $\mathbf{Q}_{\alpha \beta \gamma }$ both constrained
to the condition to result in zero d--curvatures.

\item Distinguished teleparallel vector bundles (DTVB, or vect. b.) are
provided with N--connection structure $N_{i}^{a},$\ d--metric field (\ref%
{1block2}) and arbitrary d--connection\ $\mathbf{\Gamma }_{\beta \gamma
}^{\alpha }$ with vanishing d--curvatures. The geometric constructions are
stated by the Weitzen\-bock--affine d--connection $^{[Wa]}\mathbf{\Gamma }%
_{\beta \gamma }^{\alpha }=\mathbf{\Gamma }_{\bigtriangledown ~\beta \gamma
}^{\alpha }+\mathbf{\hat{Z}}_{~\beta \gamma }^{\alpha }+\mathbf{Z}_{~\beta
\gamma }^{\alpha }$ with distorsions by nonmetricity d--fields preserving
the condition of zero values for d--curvatures. The standard constructions
from Finsler geometry and generalizations are with vanishing nonmetricity.

\item Teleparallel Berwald vector bundles (TBVB) are defined by Weitzenbock
connections of Berwald type structure, $\ ^{[WB]}\mathbf{\Gamma }_{\beta
\gamma }^{\alpha }=\mathbf{\Gamma }_{\bigtriangledown ~\beta \gamma
}^{\alpha }+\mathbf{\hat{Z}}_{~\beta \gamma }^{\alpha }+\mathbf{Z}_{~\beta
\gamma }^{\alpha }$ satisfying the condition that the curvature is zero. By
definition, such spaces satisfy the metricity conditions in the h- and
v--subspaces, $Q_{ijk}=0$ and $Q_{abc}=0,$ but, in general, there are
nontrivial nonmetricity d--fields because $Q_{iab}$ and $Q_{ajk}$ do not
vanish (see formulas (\ref{berwnm})).

\item Teleparallel Berwald vector bundles with prescribed torsion (TBVBT)
are defined by a more general class of distorsions resulting in the
Weitzenbock d--connection, $\ ^{[WB\tau ]}\mathbf{\Gamma }_{\beta \gamma
}^{\alpha }=\mathbf{\Gamma }_{\bigtriangledown ~\beta \gamma }^{\alpha }+%
\mathbf{\hat{Z}}_{~\beta \gamma }^{\alpha }+\mathbf{Z}_{~\beta \gamma
}^{\alpha }$ with prescribed values $\tau _{\ jk}^{i}$ and $\tau _{\ bc}^{a}$
in d--torsion,\
\begin{equation*}
^{\lbrack WB]}\mathbf{T}_{\ \beta \gamma }^{\alpha }=[L_{\ [jk]}^{i},+\tau
_{\ jk}^{i},0,\Omega _{ij}^{a},T_{\ bj}^{a},C_{\ [bc]}^{a}+\tau _{\ bc}^{a}],
\end{equation*}%
characterized by the condition $^{[WB\tau ]}\mathbf{R}_{\ \beta \gamma \tau
}^{\alpha }=0$ and nontrivial components of nonmetricity d--field, $%
^{[WB\tau ]}\mathbf{Q}_{\alpha \beta \gamma }=\left( Q_{cij},Q_{iab}\right)
. $

\item Teleparallel generalized Lagrange spaces (TGL) are modelled on tangent
bundles (tang. b.) provided with generalized Lagrange d--metric and
associated N--connecti\-on inducing a tangent bundle structure being enabled
with zero d--curvature. The\ Weitzenblock--Lagrange d--connections $$\ ^{[Wa]}%
\mathbf{\Gamma }_{\ \alpha \beta }^{\gamma }=\left( \ ^{[Wa]}L_{jk}^{i},\
^{[Wa]}C_{jc}^{i}\right) ,~\ ^{[WaL]}\mathbf{\Gamma }_{\beta \gamma
}^{\alpha }=\mathbf{\Gamma }_{\bigtriangledown ~\beta \gamma }^{\alpha }+%
\mathbf{\hat{Z}}_{~\beta \gamma }^{\alpha }+\mathbf{Z}_{~\beta \gamma
}^{\alpha }$$ are defined by a d--metric $\mathbf{g}_{[a]}$ (\ref{dmglas}) $%
\mathbf{Z}_{\ \ \beta }^{\alpha }$ inducing $^{[Wa]}\mathbf{R}_{\ \beta
\gamma \tau }^{\alpha }=0.$ For simplicity, we consider the configurations
when nonmetricity d--fields $^{[Wa]}\mathbf{Q}_{\alpha \beta \gamma }=0$.

\item Teleparallel Lagrange spaces (TL) are modelled on tangent bundles
provided with a Lagrange quadratic form $g_{ij}^{[L]}(x,y)=\frac{1}{2}\frac{%
\partial ^{2}L^{2}}{\partial y^{i}\partial y^{j}}$ (\ref{9lagm}) inducing the
canonical N--connection structure $^{[cL]}\mathbf{N}=\{\ ^{[cL]}N_{j}^{i}\}$
(\ref{cncls})\ for a Lagrange space\\ $\mathbf{L}^{n}=\left(
V^{n},g_{ij}(x,y)\right) $ but with vanishing d--curvature. The
d--connection structure\\ $\ ^{[WL]}\mathbf{\Gamma }_{\ \alpha \beta }^{\gamma
}$ \ (of Weitzenblock--Lagrange type) is the generated as a distortion by
the Weitzenbock d--torsion, $^{[W]}\mathbf{T}_{\beta }$ when $^{[WL]}\mathbf{%
\Gamma }_{\ \alpha \beta }^{\gamma }=\mathbf{\Gamma }_{\bigtriangledown
~\beta \gamma }^{\alpha }+\mathbf{\hat{Z}}_{~\beta \gamma }^{\alpha }+%
\mathbf{Z}_{~\beta \gamma }^{\alpha }.$ For simplicity, we can consider
configurations with zero nonmetricity d--fields, $\ \mathbf{Q}_{\beta \gamma
\alpha }.$

\item Teleparallel Finsler spaces (TF) are modelled on tangent bundles
provided with a quadratic form $g_{ij}^{[F]}=\frac{1}{2}\frac{\partial
^{2}F^{2}}{\partial y^{i}\partial y^{j}}$ (\ref{finm2}) constructed from a
Finsler metric $F\left( x^{i},y^{j}\right) .$ They are also enabled with a
canonical N--connection structure $\ ^{[F]}\mathbf{N}=\{\ ^{[F]}N_{j}^{i}\}$
(\ref{1ncc})\ as in the Finsler space $\mathbf{F}^{n}=\left( V^{n},F\left(
x,y\right) \right) $ but with a Finsler--Weitzenbock d--connection structure$%
\ ^{[WF]}\mathbf{\Gamma }_{\ \alpha \beta }^{\gamma },$ respective
d--torsion, $^{[WF]}\mathbf{T}_{\beta }.$ We can write $$\ ^{[WF]}\mathbf{%
\Gamma }_{\ \alpha \beta }^{\gamma }=\mathbf{\Gamma }_{\bigtriangledown
~\beta \gamma }^{\alpha }+\mathbf{\hat{Z}}_{~\beta \gamma }^{\alpha }+%
\mathbf{Z}_{~\beta \gamma }^{\alpha },$$ where $\ \mathbf{\hat{Z}}_{~\beta
\gamma }^{\alpha }$ contains distorsions from the canonical Finsler
d--connection (\ref{dccfs}). Such distorsions are constrained to satisfy the
condition of vanishing curvature d--tensors (see section (\ref{stpfa})) and,
for simplicity, of vanishing nonmetricity, $\mathbf{Q}_{\beta \gamma \tau
}=0.$
\end{enumerate}

\subsection{Teleparallel Hamilton--Cartan spaces}

This subclass of Hamilton--Cartan spaces is modelled on dual vector/ tangent
bundles being similar to that outlined in Table \ref{tableths} (on
generalized Hamilton--Cartan geometry, see section \ref{shgg} and Remark \ref%
{rghas}) and dual to the subclass outlined in Table \ref{tabletfls}. We
outline the main denotations and properties of such spaces and note that
they are characterized by the condition $\mathbf{\check{R}}_{\ \beta \gamma
\tau }^{\alpha }=0$ and $\mathbf{\check{Q}}_{\ \beta \gamma }^{\alpha }=0$
with that exception that there are nontrivial nonmetricity d--fields for
Berwald configuratons.

\begin{enumerate}
\item Teleparallel dual vector bundles (TDVB, or d. vect. b.) are provided
with generic off--diagonal metrics and associated N--connections. In
general, $\check{Q}_{\ \beta \gamma }^{\alpha }\neq 0.$

\item Distinguished teleparallel dual vector bundles spaces (DTDVB) are
provided with independent d--metric, d--connection structures adapted to a
N--connection in an effective dual vector bundle and resulting in zero
d--curvatures. In general, $\mathbf{\check{Q}}_{\ \beta \gamma }^{\alpha
}\neq 0.$

\item Teleparallel Berwald dual vector bundles (TBDVB) are provided with
Berwald--Weitz\-enbock d--connection structure resulting in vanishing
d--curvature.

\item Teleparallel Berwald dual vector bundles with prescribed d--torsion
(TBDVB) are with d--connections $\ ^{[BT]}\mathbf{\check{\Gamma}}_{\beta
\gamma }^{\alpha }=[L_{\ jk}^{i},\partial _{b}\check{N}_{ai},0,\check{C}%
_{a}^{~bc}]$ inducing prescribed values $\tau _{\ jk}^{i}$ and $\check{\tau}%
_{a\ }^{~bc}$ for d--torsions\ $^{[BT]}\mathbf{\check{T}}_{\ \beta \gamma
}^{\alpha }=[L_{\ [jk]}^{i}+\tau _{\ jk}^{i},0,\check{\Omega}_{iaj}=\delta
_{\lbrack i}\check{N}_{j]a},T_{a~j}^{~b},\check{C}_{a}^{~[bc]}+\check{\tau}%
_{a\ }^{~bc}].$ They are described by certain distorsions to a Weitzenbock
d--connection.

\item Teleparallel generalized Hamilton spaces (TGH) consist a subclass of
generalized Hamilton spaces with vanishing d--curvature structure, defined
on dual tangent bundles (d. tan. b.). They are described by distorsions to a
Weitzenbock d--connection $^{[Wa]}\mathbf{\check{\Gamma}}_{\ \alpha \beta
}^{\gamma }.$ In the simplest case, we consider $^{[Wa]}\mathbf{\check{Q}}%
_{\alpha \beta \gamma }=0.$

\item Teleparallel Hamilton spaces (TH, see section\ \ref{rlafs}), as a
particular subclass of TGH, are provided with d--connection and
N--connection structures corresponding to Hamilton configurations.

\item Teleparallel Cartan spaces (TC) are particular Cartan configurations
with absolut teleparallelism.
\end{enumerate}

\subsection{Distinguished Riemann--Cartan spaces}

A wide class of generalized Finsler geometries can be modelled on
Riemann--Cartan spaces by using generic off--diagonal metrics and associated
N--connection structures. The locally anisotropic metric--affine
configurations from Table \ref{tablegs} transform into a Riemann--Cartan
ones if we impose the condition of metricity. For the Berwald type
connections one could be certain nontrivial nonmetricity d--fields on
intersection of h- and v--subspaces. The local coordinates $x^{i}$ are
considered as certain holonomic ones and $y^{a}$ are anholonomic. We list
the short denotations and main properties of such spaces:

\begin{enumerate}
\item Riemann--Cartan spaces (in brief, RC, see related details in section %
\ref{srgarcg}) are certain manifolds $V^{n+m}$ of necessary smoothly class\
provided with metric structure $g_{\alpha \beta }~$ and linear connection
structure $\Gamma _{\beta \gamma }^{\alpha }$ (constructed as a distorsion
by torsion of the Levi--Civita connection) both satisfying the conditions of
metric compatibility, $Q_{\alpha \beta \gamma }=0.$ For generic
off--diagonal metrics, a RC\ space always admits nontrivial N--connection
structures (see Proposition \ref{pmasnc} reformulated for the case of
vanishing nonmetricity). In general, only the metric field $g_{\alpha \beta
} $ can be transformed into a d--metric one, $\mathbf{g}_{\alpha \beta
}=[g_{ij},h_{ab}],$ but \ $\Gamma _{\beta \gamma }^{\alpha }$ may be not
adapted to the N--connection structure.

\item Distinguished Riemann--Cartan spaces (DRC) are manifolds $\mathbf{V}%
^{n+m}$ provided with N--connection structure $N_{i}^{a},$ d--metric field (%
\ref{1block2}) and d--connection $\mathbf{\Gamma }_{\beta \gamma }^{\alpha }$
(a distorsion of the Levi--Civita connection, or of the canonical
d--connection) satisfying the condition $\mathbf{Q}_{\alpha \beta \gamma
}=0. $ In this case, the strengths $\left( \mathbf{T}_{\ \beta \gamma
}^{\alpha },\mathbf{R}_{\ \beta \gamma \tau }^{\alpha }\right) $ are
N--adapted.

\item Berwald Riemann--Cartan (BRC) are modelled if a N--connection structure
is defined in a Riemann--Cartan space and distorting the connection to a
Berwald d--connection $^{[B]}\mathbf{D}=[\ ^{[B]}\mathbf{\Gamma }_{\beta
\gamma }^{\alpha }]=[\widehat{L}_{\ jk}^{i},\partial _{b}N_{k}^{a},0,%
\widehat{C}_{\ bc}^{a}]$, see (\ref{berw}). By definition, this space
satisfies the metricity conditions in the h- and v--subspaces, $Q_{ijk}=0$
and $Q_{abc}=0,$ but, in general, there are nontrivial nonmetricity
d--fields because $Q_{iab}$ and $Q_{ajk}$ are not vanishing (see formulas (%
\ref{berwnm})). Nonmetricities vanish with respect to holonomic frames.

\item Berwald Riemann--Cartan spaces with prescribed torsion (BRCT) are
defined by a more general class of d--connection $^{[BT]}\mathbf{\Gamma }%
_{\beta \gamma }^{\alpha }=[L_{\ jk}^{i},\partial _{b}N_{k}^{a},0,C_{\
bc}^{a}]$ inducing prescribed values $\tau _{\ jk}^{i}$ and $\tau _{\
bc}^{a} $ in d--torsion\ $^{[BT]}\mathbf{T}_{\ \beta \gamma }^{\alpha
}=[L_{\ [jk]}^{i},+\tau _{\ jk}^{i},0,\Omega _{ij}^{a},T_{\ bj}^{a},C_{\
[bc]}^{a}+\tau _{\ bc}^{a}],$ see (\ref{tauformulas}). The nontrivial
components of nonmetricity d--fields are $^{[B\tau ]}\mathbf{Q}_{\alpha
\beta \gamma }=\left( Q_{cij},Q_{iab}\right) .$ Such components vanish with
respect to holonomic frames.

\item Generalized Lagrange Riemann--Cartan spaces (GLRC) are modelled as
distinguished Riemann--Cartan spaces of odd--dimension, $\mathbf{V}^{n+n},$
provided with generic off--diagonal metrics with associated N--connection
inducing a tangent bundle structure. The d--metric $\mathbf{g}_{[a]}$ (\ref%
{dmglas}) and the d--connection $\ \ ^{[a]}\mathbf{\Gamma }_{\ \alpha \beta
}^{\gamma }$ $=\left( \ ^{[a]}L_{jk}^{i},\ ^{[a]}C_{jc}^{i}\right) $ (\ref%
{dcglma}) are those for the usual Lagrange spaces (see Definition \ref%
{defgls}).

\item Lagrange Riemann--Cartan spaces (LRC, see Remark \ \ref{rlafs}) are
provided with a Lagrange quadratic form $g_{ij}^{[L]}(x,y)=\frac{1}{2}\frac{%
\partial ^{2}L^{2}}{\partial y^{i}\partial y^{j}}$ (\ref{9lagm}) inducing the
canonical N--connection structure $^{[cL]}\mathbf{N}=\{\ ^{[cL]}N_{j}^{i}\}$
(\ref{cncls})\ for a Lagrange space\\ $\mathbf{L}^{n}=\left(
V^{n},g_{ij}(x,y)\right) $ (see Definition \ref{defls})) and, for instance,
with a canonical d--connecti\-on structure $\ ^{[b]}\mathbf{\Gamma }_{\ \
\alpha }^{\gamma }=\ ^{[b]}\mathbf{\Gamma }_{\ \alpha \beta }^{\gamma }%
\mathbf{\vartheta }^{\beta }$ satisfying metricity conditions for the
d--metric defined by $g_{ij}^{[L]}(x,y).$

\item Finsler Riemann--Cartan spaces (FRC, see Remark \ref{rfafs}) are
defined by a quadratic form $g_{ij}^{[F]}=\frac{1}{2}\frac{\partial ^{2}F^{2}%
}{\partial y^{i}\partial y^{j}}$ (\ref{finm2}) constructed from a Finsler
metric $F\left( x^{i},y^{j}\right) .$ It is induced the canonical
N--connection structure $\ ^{[F]}\mathbf{N}=\{\ ^{[F]}N_{j}^{i}\}$ (\ref{1ncc}%
)\ as in the Finsler space $\mathbf{F}^{n}=\left( V^{n},F\left( x,y\right)
\right) $ with $\ ^{[F]}\widehat{\mathbf{\Gamma }}_{\ \beta \gamma }^{\alpha
}$ being the canonical Finsler d--connection (\ref{dccfs}).
\end{enumerate}

\subsection{Distinguished (pseudo) Riemannian spaces}

Sections \ref{srgarcg}\ and \ref{sarg} are devoted to modelling of locally
anisotropic geometric configurations in (pseudo) Riemannian spaces enabled
with generic off--diagonal metrics and associated N--connection structure.
Different classes of generalized Finsler metrics can be embedded in (pseudo)
Riemannian spaces as certain anholonomic frame configurations. Every such
space is characterized by a corresponding off--diagonal metric ansatz and
Levi--Civita connection stated with respect to coordinate frames or,
alternatively (see Theorem \ref{teqrgrcg}), by certain N--connection and
induced d--metric and d--connection structures related to the Levi--Cevita
connection with coefficients defined with respect to N--adapted anholonomic
(co) frames. We characterize every such type of (pseudo) Riemannian spaces
both by Levi--Civita and induced canonical/or Berwald d--connections which
contain also induced (by former off--diagonal metric terms) nontrivial
d--torsion and/or nonmetricity d--fields.

\begin{enumerate}
\item (Pseudo) Riemann spaces (in brief, pR) are certain manifolds $V^{n+m}$
of necessary smoothly class\ provided with generic off--diagonal metric
structure $g_{\alpha \beta }$ of arbitrary signature inducing the unique
torsionless and metric Levi--Civita connection $\Gamma _{\bigtriangledown
\beta \gamma }^{\alpha }.$ We can effectively diagonalize such metrics by
anholonomic frame transforms with associated N--connection structure. We can
also consider alternatively the canonical d--connection $\widehat{\mathbf{%
\Gamma }}_{\beta \gamma }^{\alpha }=[L_{\ jk}^{i},L_{\ bk}^{a},C_{\
jc}^{i},C_{\ bc}^{a}]$ (\ref{1candcon})defined by the coefficients of
d--metric $\mathbf{g}_{\alpha \beta }=[g_{ij},h_{ab}]$ and N--connection $%
N_{i}^{a}.$ We have nontrivial d--torsions $\widehat{\mathbf{T}}_{\beta
\gamma }^{\alpha },$ but $T_{\bigtriangledown \beta \gamma }^{\alpha
}=0,Q_{\alpha \beta \gamma }^{\bigtriangledown }=0$ and $\mathbf{\hat{Q}}%
_{\alpha \beta \gamma }=0.$ The simplest anholonomic configurations are
characterized by associated N--connections with vanishing N--connection
curvature, $\Omega _{ij}^{a}=\delta _{\lbrack i}N_{j]}^{a}=0$. The
d--torsions $\widehat{\mathbf{T}}_{\ \beta \gamma }^{\alpha }=[\widehat{L}%
_{[\ jk]}^{i},\widehat{C}_{\ ja}^{i},\Omega _{ij}^{a},\widehat{T}_{\ bj}^{a},%
\widehat{C}_{\ [bc]}^{a}]$ and d--curvatures $\widehat{\mathbf{R}}_{\ \beta
\gamma \tau }^{\alpha }=[\widehat{R}_{\ jkl}^{i},\widehat{R}_{\ bkl}^{a},%
\widehat{P}_{\ jka}^{i},\widehat{P}_{\ bka}^{c},\widehat{S}_{\ jbc}^{i},%
\widehat{S}_{\ dbc}^{a}]$ \ are computed by introducing the components of $%
\widehat{\mathbf{\Gamma }}_{\beta \gamma }^{\alpha },$ respectively, in
formulas (\ref{1dtorsb})\ and (\ref{1dcurv}).

\item Distinguished (pseudo) Riemannian spaces (DpR) are defined as
manifolds $\mathbf{V}^{n+m}$ $\ $provided with N--connection structure $%
N_{i}^{a},$ d--metric field and d--connection $\mathbf{\Gamma }_{\beta
\gamma }^{\alpha }$ (a distorsion of the Levi--Civita connection, or of the
canonical d--connection) satisfying the condition $\mathbf{Q}_{\alpha \beta
\gamma }=0.$

\item Berwald (pseudo) Riemann spaces (pRB) are modelled if a N--connection
structure is defined by a generic off--diagonal metric. The Levi--Civita
connection is distorted to a Berwald d--connection $^{[B]}\mathbf{D}=[\
^{[B]}\mathbf{\Gamma }_{\beta \gamma }^{\alpha }]=[\widehat{L}_{\
jk}^{i},\partial _{b}N_{k}^{a},0,\widehat{C}_{\ bc}^{a}]$, see (\ref{berw}).
By definition, this space satisfies the metricity conditions in the h- and
v--subspaces, $Q_{ijk}=0$ and $Q_{abc}=0,$ but, in general, there are
nontrivial nonmetricity d--fields because $Q_{iab}$ and $Q_{ajk}$ are not
vanishing (see formulas (\ref{berwnm})). Such nonmetricities vanish with
respect to holonomic frames. The torsion is zero for the Levi--Civita
connection but $^{[B]}\mathbf{T}_{\ \beta \gamma }^{\alpha }=[L_{\
[jk]}^{i},0,\Omega _{ij}^{a},T_{\ bj}^{a},C_{\ [bc]}^{a}]$ is not trivial.

\item Berwald (pseudo) Riemann spaces with prescribed d--torsion (pRBT) are
defined by a more general class of d--connection $^{[BT]}\mathbf{\Gamma }%
_{\beta \gamma }^{\alpha }=[L_{\ jk}^{i},\partial _{b}N_{k}^{a},0,C_{\
bc}^{a}]$ inducing prescribed values $\tau _{\ jk}^{i}$ and $\tau _{\
bc}^{a} $ in d--torsion\ $^{[BT]}\mathbf{T}_{\ \beta \gamma }^{\alpha
}=[L_{\ [jk]}^{i},+\tau _{\ jk}^{i},0,\Omega _{ij}^{a},T_{\ bj}^{a},C_{\
[bc]}^{a}+\tau _{\ bc}^{a}],$ see (\ref{tauformulas}). The nontrivial
components of nonmetricity d--fields are $^{[B\tau ]}\mathbf{Q}_{\alpha
\beta \gamma }=\left( ^{[B\tau ]}Q_{cij},\ ^{[B\tau ]}Q_{iab}\right) .$ Such
components vanish with respect to holonomic frames.

\item Generalized Lagrange (pseudo) Riemannian spaces (pRGL) are modelled as
distinguished Riemann spaces of odd--dimension, $\mathbf{V}^{n+n},$ provided
with generic off--diagonal metrics with associated N--connection inducing a
tangent bundle structure. The d--metric $\mathbf{g}_{[a]}$ (\ref{dmglas})
and the d--connection $\ \ ^{[a]}\mathbf{\Gamma }_{\ \alpha \beta }^{\gamma
} $ $=\left( \ ^{[a]}L_{jk}^{i},\ ^{[a]}C_{jc}^{i}\right) $ (\ref{dcglma})
are those for the usual Lagrange spaces (see Definition \ref{defgls}) but on
a (pseudo) Riemann manifold with prescribed N--connection structure.

\item Lagrange (pseudo) Riemann spaces (pRL) are provided with a Lagrange
quadratic form $g_{ij}^{[L]}(x,y)=\frac{1}{2}\frac{\partial ^{2}L^{2}}{%
\partial y^{i}\partial y^{j}}$ (\ref{9lagm}) inducing the canonical
N--connection structure $^{[cL]}\mathbf{N}=\{\ ^{[cL]}N_{j}^{i}\}$ (\ref%
{cncls})\ for a Lagrange space $\mathbf{L}^{n}=\left(
V^{n},g_{ij}(x,y)\right) $ and, for instance, provided with a canonical
d--connection structure $\ ^{[b]}\mathbf{\Gamma }_{\ \ \alpha }^{\gamma }=\
^{[b]}\mathbf{\Gamma }_{\ \alpha \beta }^{\gamma }\mathbf{\vartheta }^{\beta
}$ satisfying metricity conditions for the d--metric defined by $%
g_{ij}^{[L]}(x,y).$ There is an alternative construction with Levi--Civita
connection.

\item Finsler (pseudo) Riemann (FpR) are defined by a quadratic form $%
g_{ij}^{[F]}=\frac{1}{2}\frac{\partial ^{2}F^{2}}{\partial y^{i}\partial
y^{j}}$ (\ref{finm2}) constructed from a Finsler metric $F\left(
x^{i},y^{j}\right) .$ It is induced the canonical N--connection structure $\
^{[F]}\mathbf{N}=\{\ ^{[F]}N_{j}^{i}\}$ (\ref{1ncc})\ as in the Finsler space
$\mathbf{F}^{n}=\left( V^{n},F\left( x,y\right) \right) $ with $\ ^{[F]}%
\widehat{\mathbf{\Gamma }}_{\ \beta \gamma }^{\alpha }$ being the canonical
Finsler d--connection (\ref{dccfs}).
\end{enumerate}

\subsection{Teleparallel spaces}

Teleparallel spaces were considered in sections \ref{stps} and \ref{stpfa}.
Here we classify what type of locally isotropic and anisotropic structures
can be modelled in by anholonomic transforms of (pseudo) Riemannian spaces to
teleparallel ones. The anholonomic frame structures are with associated
N--connection with the components defined by the off--diagonal metric
coefficients.

\begin{enumerate}
\item Teleparallel spaces (in brief, T) are usual ones with vanishing
curvature, modelled on manifolds $V^{n+m}$ of necessary smoothly class\
provided, for instance, with the Weitzenbock connection $^{[W]}\Gamma
_{\beta \gamma }^{\alpha }$(\ref{wcon}) which can be transformed in a
d--connection one with respect to N--adapted frames. In general, such
geometries are characterized by torsion $^{[W]}T_{\ \beta \gamma }^{\alpha }$
constrained to the condition to result in zero d--curvatures. The simplest
theories are with vanishing nonmetricity.

\item Distinguished teleparallel spaces (DT) are manifolds $\mathbf{V}^{n+m}$
provided with N--con\-nec\-ti\-on structure $N_{i}^{a},$ d--metric field (\ref%
{1block2}) and arbitrary d--connection $\mathbf{\Gamma }_{\beta \gamma
}^{\alpha }$ with vanishing d--curvatures. The geometric constructions are
stated by the Weitzenbock d--connection $^{[Wa]}\mathbf{\Gamma }_{\beta
\gamma }^{\alpha }=\mathbf{\Gamma }_{\bigtriangledown ~\beta \gamma
}^{\alpha }+\mathbf{\hat{Z}}_{~\beta \gamma }^{\alpha }+\mathbf{Z}_{~\beta
\gamma }^{\alpha }$ with distorsions without nonmetricity d--fields
preserving the condition of zero values for d--curvatures.

\item Teleparallel Berwald spaces (TB) are defined by distorsions of the
Weitzenbock connection on a manifold $V^{n+m}$ to any Berwald like
structure, $\ ^{[WB]}\mathbf{\Gamma }_{\beta \gamma }^{\alpha }=\mathbf{%
\Gamma }_{\bigtriangledown ~\beta \gamma }^{\alpha }+\mathbf{\hat{Z}}%
_{~\beta \gamma }^{\alpha }+\mathbf{Z}_{~\beta \gamma }^{\alpha }$
satisfying the condition that the curvature is zero. All constructions with
effective off--diagonal metrics can be adapted to the N--connection and
considered for d--objects. Such spaces satisfy the metricity conditions in
the h- and v--subspaces, $Q_{ijk}=0$ and $Q_{abc}=0,$ but, in general, there
are nontrivial nonmetricity d--fields, $Q_{iab}$ and $Q_{ajk}.$

\item Teleparallel Berwald spaces with prescribed torsion (TBT) are defined
by a more general class of distorsions resulting in the Weitzenbock
d--connection,
\begin{equation*}
\ ^{[WB\tau ]}\mathbf{\Gamma }_{\beta \gamma }^{\alpha }=\mathbf{\Gamma }%
_{\bigtriangledown ~\beta \gamma }^{\alpha }+\mathbf{\hat{Z}}_{~\beta \gamma
}^{\alpha }+\mathbf{Z}_{~\beta \gamma }^{\alpha },
\end{equation*}
having prescribed values $\tau _{\ jk}^{i}$ and $\tau _{\ bc}^{a}$ in
d--torsion\
\begin{equation*}
^{[WB]}\mathbf{T}_{\ \beta \gamma }^{\alpha }=[L_{\ [jk]}^{i},+\tau _{\
jk}^{i},0,\Omega _{ij}^{a},T_{\ bj}^{a},C_{\ [bc]}^{a}+\tau _{\ bc}^{a}]
\end{equation*}
and characterized by the condition $^{[WB\tau ]}\mathbf{R}_{\ \beta \gamma
\tau }^{\alpha }=0$ with certain nontrivial nonmetricity d--fields, $%
^{[WB\tau ]}\mathbf{Q}_{\alpha \beta \gamma }=\left( ^{[WB\tau ]}Q_{cij},\
^{[WB\tau ]}Q_{iab}\right) .$

\item Teleparallel generalized Lagrange spaces (TGL) are modelled as
Riemann--Cartan spaces of odd--dimension, $\mathbf{V}^{n+n},$ provided with
generalized Lagrange d--metric and associated N--connection inducing a
tangent bundle structure with zero d--curvature. The Weitzenblock--Lagrange
d--connection $\ ^{[Wa]}\mathbf{\Gamma }_{\ \alpha \beta }^{\gamma }$ $=( \
^{[Wa]}L_{jk}^{i},$\\ $\ ^{[Wa]}C_{jc}^{i}) , $\ where $^{[Wa]}\mathbf{\Gamma }%
_{\beta \gamma }^{\alpha }=\mathbf{\Gamma }_{\bigtriangledown ~\beta \gamma
}^{\alpha }+\mathbf{\hat{Z}}_{~\beta \gamma }^{\alpha }+\mathbf{Z}_{~\beta
\gamma }^{\alpha },$ are defined by a d--metric $\mathbf{g}_{[a]}$ (\ref%
{dmglas}) with $\mathbf{Z}_{\ \ \beta }^{\alpha }$ inducing zero
nonmetricity d--fields, $^{[a]}\mathbf{Q}_{\alpha \beta \gamma }=0$ and zero
d--curvature, $^{[Wa]}\mathbf{R}_{\ \beta \gamma \tau }^{\alpha }=0.$

\item Teleparallel Lagrange spaces (TL, see section\ \ref{rlafs}) are
Riemann--Cartan spaces $\mathbf{V}^{n+n}$ provided with a Lagrange quadratic
form $g_{ij}^{[L]}(x,y)=\frac{1}{2}\frac{\partial ^{2}L^{2}}{\partial
y^{i}\partial y^{j}}$ (\ref{9lagm}) inducing the canonical N--connection
structure $^{[cL]}\mathbf{N}=\{\ ^{[cL]}N_{j}^{i}\}$ (\ref{cncls})\ for a
Lagrange space $\mathbf{L}^{n}=\left( V^{n},g_{ij}(x,y)\right) $ but with
vanishing d--curvature. The d--connection structure $\ ^{[WL]}\mathbf{\Gamma
}_{\ \alpha \beta }^{\gamma }$ (of Weitzenblock--Lagrange type) is the
generated as a distortion by the Weitzenbock d--torsion, $^{[W]}\mathbf{T}%
_{\beta },$ but zero nonmetricity d--fields,$\ ^{[WL]}\mathbf{Q}_{\beta
\gamma \alpha }=0,$ when $^{[WL]}\mathbf{\Gamma }_{\ \alpha \beta }^{\gamma
}=\mathbf{\Gamma }_{\bigtriangledown ~\beta \gamma }^{\alpha }+\mathbf{\hat{Z%
}}_{~\beta \gamma }^{\alpha }+\mathbf{Z}_{~\beta \gamma }^{\alpha }.$

\item Teleparallel Finsler spaces (TF) are Riemann--Cartan manifolds $%
\mathbf{V}^{n+n}$ defined by a quadratic form $g_{ij}^{[F]}=\frac{1}{2}\frac{%
\partial ^{2}F^{2}}{\partial y^{i}\partial y^{j}}$ (\ref{finm2}) and a
Finsler metric $F\left( x^{i},y^{j}\right) .$ They are provided with a
canonical N--connection structure $\ ^{[F]}\mathbf{N}=\{\ ^{[F]}N_{j}^{i}\}$
(\ref{1ncc})\ as in the Finsler space $\mathbf{F}^{n}=\left( V^{n},F\left(
x,y\right) \right) $ but with a Finsler--Weitzenbock d--connection structure
$\ ^{[WF]}\mathbf{\Gamma }_{\ \alpha \beta }^{\gamma },$ respective
d--torsion, $^{[WF]}\mathbf{T}_{\beta },$ and vanishing nonmetricity, $%
^{[WF]}\mathbf{Q}_{\beta \gamma \tau }=0,$ d--fields, $\ ^{[WF]}\mathbf{%
\Gamma }_{\ \alpha \beta }^{\gamma }=\mathbf{\Gamma }_{\bigtriangledown
~\beta \gamma }^{\alpha }+\mathbf{\hat{Z}}_{~\beta \gamma }^{\alpha }+%
\mathbf{Z}_{~\beta \gamma }^{\alpha },$where $\ \mathbf{\hat{Z}}_{~\beta
\gamma }^{\alpha }$ contains a distorsion from the canonical Finsler
d--connection (\ref{dccfs}).
\end{enumerate}

\newpage

%{\footnotesize
\begin{table}[h]
\begin{center}
%
\end{center}
\caption{Generalized Lagrange--affine spaces}
\label{tablegs}
\end{table}
%}

\newpage

%{\footnotesize
\begin{table}[h]
\begin{center}
%
%
\end{center}
\caption{Generalized Hamilton--affine spaces}
\label{tableghs}
\end{table}
%}

\newpage

%{\footnotesize
\begin{table}[h]
\begin{center}
%
%
\end{center}
\caption{Teleparallel Lagrange--affine spaces}
\label{tabletls}
\end{table}
%}

\newpage

%{\footnotesize
\begin{table}[h]
\begin{center}
%
%
\end{center}
\caption{Teleparallel Hamilton--affine spaces}
\label{tableths}
\end{table}
%}

\newpage

%{\footnotesize
\begin{table}[h]
\begin{center}
%
%
\end{center}
\caption{Generalized Finsler--Lagrange spaces}
\label{tablegfls}
\end{table}
%}

\newpage

%{\footnotesize
\begin{table}[h]
\begin{center}
%
%
\end{center}
\caption{Generalized Hamilton--Cartan spaces}
\label{tableghcs}
\end{table}
%}

\newpage

%{\footnotesize
\begin{table}[h]
\begin{center}
%
%
\end{center}
\caption{Teleparallel Finsler--Lagrange spaces}
\label{tabletfls}
\end{table}
%}

\newpage

%{\footnotesize
\begin{table}[h]
\begin{center}
%
%
\end{center}
\caption{Teleparallel Hamilton--Cartan spaces}
\label{tablehcs}
\end{table}
%}

\newpage

%{\footnotesize
\begin{table}[h]
\begin{center}
%
%
\end{center}
\caption{Distinguished Riemann--Cartan spaces}
\label{tabledrcs}
\end{table}
%}
\

\newpage

%{\footnotesize
\begin{table}[h]
\begin{center}
%
%
\end{center}
\caption{Distinguished (pseudo) Riemannian spaces}
\label{tableprs}
\end{table}
%}

\newpage

%{\footnotesize
\begin{table}[h]
\begin{center}
%
\\ \hline\hline
\end{tabular}%
\end{center}
\caption{Teleparallel spaces}
\label{tablets}
\end{table}
%}

%%%%%%%%%%%%%%%%%%%%%%%%%%%%%%%%%%%%%%%%%%%%%%%%%%%%%%%%%%%%%

\chapter[Off--Diagonal Solutions ...]{A Method of Constructing Off--Diagonal
Solutions in Metric--Affine and String Gravity}

{\bf Abstract}
\footnote{ \copyright\
S. Vacaru, E. Gaburov and D. Gontsa, A Method of Constructing Off-Diagonal
 Solutions in Metric-Affine and String Gravity, hep-th/0310133}

The anholonomic frame method is generalized for non--Riemannian
gravity models defined by string corrections to the general
relativity and metric--affine gravity (MAG) theories. Such
spacetime configurations are modelled as metric--affine spaces
provided with generic off--diagonal metrics (which can not be
diagonalized by coordinate transforms) and anholonomic frames with
associated nonlinear connection (N--connection) structure. We
investigate the field equations of MAG and string gravity with
mixed holonomic and anholonomic variables. There are proved the
main theorems on irreducible reduction to effective
Einstein--Proca equations with respect to anholonomic frames
adapted to N--connections. String corrections induced by the
antisymmetric $H$--fields are considered. There are also proved
the theorems and criteria stating a new method of constructing
exact solutions with generic off--diagonal metric ansatz depending
on 3-5 variables and describing various type of locally
anisotropic gravitational configurations with torsion,
nonmetricity and/or generalized Finsler--affine effective
geometry. We analyze solutions, generated in string gravity, when
generalized Finsler--affine metrics, torsion and nonmetricity
interact with three dimensional solitons.

\vskip5pt

Pacs:\ 04.50.+h, 04.20.JB, 02.40.-k,

MSC numbers: 83D05, 83C15, 83E15, 53B40, 53C07, 53C60

\section{Introduction}

Nowadays, there exists an interest to non--Riemannian descriptions of
gravity interactions derived in the low energy string theory \cite{02sgr}
and/or certain noncommutative \cite{02ncg} and quantum group generalizations %
\cite{02majid} of gravity and field theory. Such effective models can be
expressed in terms of geometries with torsion and nonmetricity in the
framework of metric--affine gravity (MAG) \cite{02mag} and a subclass of such
theories can be expressed as an effective Einstein--Proca gravity derived
via irreducible decompositions \cite{02oveh}.

In a recent work \cite{02vp1} we developed a unified scheme to the geometry of
anholonomic frames with associated nonlinear connection (N--connection)
structure for a large number of gauge and gravity models with locally
isotropic and anisotropic interactions and nontrivial torsion and
nonmetricity contributions and effective generalized
Finsler--Weyl--Riemann--Cartan geometries derived from MAG. The synthesis of
metric--affine and Finsler like theories was inspired by a number of exact
solutions parametrized by generic off--diagonal metrics and anholonomic
frames in Einstein, Einstein--Cartan, gauge and string gravity \cite{02v1,02v1a}%
. The resulting formalism admits inclusion of locally anisotropic spinor
interactions and extensions to noncommutative geometry and string/brane
gravity \cite{02v2,02vnces}. We concluded that the geometry of metric--affine
spaces enabled with an additional N--connection structure is sufficient not
only to model the bulk of physically important non--Riemannian geometries on
(pseudo) Riemannian spaces but also states the conditions when such
effective spaces with generic anisotropy can be defined as certain
generalized Finsler--affine geometric configurations constructed as exact
solutions of field equations. It was elaborated a detailed classification of
such spacetimes provided with N--connection structure.

If in the Ref. \cite{02vp1} we paid attention to the geometrical
(pre--dynamical) aspects of the generalized Finsler--affine configurations
derived in MAG, the aim of this paper (the second partner) is to formulate a
variatonal formalism of deriving field equations on metric--affine spaces
provided with N--connection structure and to state the main theorems for
constructing exact off--diagonal solutions in such generalized
non--Riemannian gravity theories. We emphasize that generalized Finsler
metrics can be generated in string gravity connected to anholonomic
metric--affine configurations. In particular, we investigate how the
so--called Obukhov's equivalence theorem \cite{02oveh} should be modified as
to include various type of Finsler--Lagrange--Hamilton--Cartan metrics, see
Refs. \cite{02fin,02ma,02mhss}. The results of this paper consist a theoretical
background for constructing exact solutions in MAG and string gravity in the
third partner paper \cite{02exsolmag} derived as exact solutions of
gravitational and matter field equations parametrized by generic
off--diagonal metrics (which can not be diagonalized by local coordinate
transforms) and anholonomic frames with associated N--connection structure.
Such solutions depending on 3-5 variables (generalizing to MAG the results
from \cite{02v1,02v1a,02v2,02vnces,02vmethod}) differ substantially from those
elaborated in Refs. \cite{02esolmag}; they define certain extensions to
nontrivial torsion and nonmetricity fields of certain generic off--diagonal
metrics in general relativity theory.

The plan of the paper is as follows: In Sec. 2 we outline the
necessary results on Finsler--affine geometry. Next, in Sec. 3, we
formulate the field equations on metric--affine spaces provided
with N--connection structure. We consider Lagrangians and derive
geometrically the field equations of Finsler--affine gravity. We
prove the main theorems for the Einstein--Proca systems
distinguished by N--connection structure and analyze possible
string gravity corrections by $H$--fields from the bosonic string
theory. There are defined the restrictions on N--connection
structures resulting in Einstein--Cartan and Einstein gravity.
Section 4 is devoted to extension of the anholonomic frame method
in MAG and string gravity. We formulate and prove the main
theorems stating the possibility of constructing exact solutions
parametrized by generic off--diagonal metrics, nontrivial torsion
and nonmetricity structures and possible sources of matter fields.
In Sec. 5 we construct three classes of exact solutions. The first
class of solutions is stated for five subclasses of two
dimensional generalized Finsler geometries modelled in MAG with
possible string corrections. The second class of solutions is for
MAG with effective variable and inhomogeneous cosmological
constant. The third class of solutions are for the string
Finsler--affine gravity (i. e. string gravity containing in
certain limits Finsler like metrics) with possible nonlinear three
dimensional solitonic interactions, Proca fields with almost
vanishing masses, nontrivial torsions and nonmetricity. In Sec. 6
we present the final remarks. In Appendices A, B and C we give
respectively the details on the proof of the Theorem 4.1 (stating
the components of the Ricci tensor for generalized Finsler--affine
spaces), analyze the reduction of nonlinear solutions from five to
four dimensions and present a short characterization of five
classes of generalized Finsler--affine spaces.

Our basic notations and conventions are those from Ref. \cite{02vp1}
and contain an interference of approaches elaborated in MAG and
generalized Finsler geometry. The spacetime is considered to be a
manifold $V^{n+m}$ of necessary smoothly class of dimension $n+m.$
The Greek indices $\alpha ,\beta ,...$ split into subclasses like
$\alpha =\left( i,a\right) ,$ $\beta
=\left( j,b\right) ...$ where the Latin indices $i,j,k,...$ run values $%
1,2,...n$ and $\ a,b,c,...$ run values $n+1,n+2,$ ..., $n+m.$ We follow the
Penrose convention on abstract indices \cite{02pen} and use underlined indices
like $\underline{\alpha }=\left( \underline{i},\underline{a}\right) ,$ for
decompositions with respect to coordinate frames. The symbol ''\ $\doteqdot $%
'' will be used is some formulas will be introduced by definition and the
end of proofs will be stated by symbol $\blacksquare .$ The notations for
connections $\Gamma _{\ \beta \gamma }^{\alpha },$ metrics $g_{\alpha \beta
} $ and frames $e_{\alpha }$ and coframes $\vartheta ^{\beta },$ or another
geometrical and physical objects, are the standard ones from MAG if a
nonlinear connection (N--connection) structure is not emphasized on the
spacetime. If a N--connection and corresponding anholonomic frame structure
are prescribed, we use ''boldfaced'' symbols with possible splitting of the
objects and indices like
$$\mathbf{V}^{n+m},\ \mathbf{\Gamma }_{\ \beta
\gamma }^{\alpha }=\left(
L_{jk}^{i},L_{bk}^{a},C_{jc}^{i},C_{bc}^{a}\right) ,\
\mathbf{g}_{\alpha \beta }=\left( g_{ij},h_{ab}\right),\
\mathbf{e}_{\alpha }=\left( e_{i},e_{a}\right) , ... $$
 being distinguished by N--connection (in brief, there are used the terms
d--objects, d--tensor, d--connection).

\section{Metric--Affine and Generalized Finsler Gravity}

In this section we recall some basic facts on metric--affine spaces provided
with nonlinear connection (N--connection) structure and generalized
Finsler--affine geometry \cite{02vp1}.

The spacetime is modelled as a manifold $V^{n+m}$ of dimension $n+m,$ with $%
n\geq 2$ and $m\geq 1,$ admitting (co) vector/ tangent structures. It is
denoted by $\pi ^{T}:TV^{n+m}\rightarrow TV^{n}$ $\ $the differential of the
map $\pi :V^{n+m}\rightarrow V^{n}$ defined as a fiber--preserving morphism
of the tangent bundle $\left( TV^{n+m},\tau _{E},V^{n}\right) $ to $V^{n+m}$
and of tangent bundle $\left( TV^{n},\tau ,V^{n}\right) .$ We consider also
the kernel of the morphism $\pi ^{T}$ as a vector subbundle of the vector
bundle $\left( TV^{n+m},\tau _{E},V^{n+m}\right) .$ The kernel defines the
vertical subbundle over $V^{n+m},$ s denoted as $\left( vV^{n+m},\tau
_{V},V^{n+m}\right) .$ We parametrize the local coordinates of a point $u\in
V^{n+m}$ as $u^{\alpha }=\left( x^{i},y^{a}\right) ,$ where the values of
indices are $i,j,k,...=1,2,...,n$ and $a,b,c,...=n+1,n+2,...,n+m.$ The
inclusion mapping is written as $i:vV^{n+m}\rightarrow TV^{n+m}.$

A nonlinear connection (N--connection) $\mathbf{N}$ in a space $\left(
V^{n+m},\pi ,V^{n}\right) $ is a morphism of manifolds $N:TV^{n+m}%
\rightarrow vV^{n+m}$ defined by the splitting on the left of the exact
sequence
\begin{equation}
0\rightarrow vV^{n+m}\rightarrow TV^{n+m}/vV^{n+m}\rightarrow 0.
\label{2eseq}
\end{equation}

The kernel of the morphism $\mathbf{N}$ \ is a subbundle of $\left(
TV^{n+m},\tau _{E},V^{n+m}\right) ,$ called the horizontal subspace and
denoted by $\left( hV^{n+m},\tau _{H},V^{n+m}\right) .$ Every tangent bundle
$(TV^{n+m},$ $\tau _{E},$ $V^{n+m})$ provided with a N--connection structure
is a Whitney sum of the vertical and horizontal subspaces (in brief, h- and
v-- subspaces), i. e.
\begin{equation}
TV^{n+m}=hV^{n+m}\oplus vV^{n+m}.  \label{2wihit}
\end{equation}%
We note that the exact sequence (\ref{2eseq}) defines the N--connection in a
global coordinate free form\ resulting in invariant splitting (\ref{2wihit})
(see details in Refs. \cite{02barthel,02ma} stated for vector and tangent
bundles and generalizations on covector bundles, superspaces and
noncommutative spaces \cite{02mhss} and \cite{02v2}).

A N--connection structure prescribes a class of vielbein transforms%
\begin{eqnarray}
A_{\alpha }^{\ \underline{\alpha }}(u) &=&\mathbf{e}_{\alpha }^{\ \underline{%
\alpha }}=\left[
\begin{array}{cc}
e_{i}^{\ \underline{i}}(u) & N_{i}^{b}(u)e_{b}^{\ \underline{a}}(u) \\
0 & e_{a}^{\ \underline{a}}(u)%
\end{array}%
\right] ,  \label{2vt1} \\
A_{\ \underline{\beta }}^{\beta }(u) &=&\mathbf{e}_{\ \underline{\beta }%
}^{\beta }=\left[
\begin{array}{cc}
e_{\ \underline{i}}^{i\ }(u) & -N_{k}^{b}(u)e_{\ \underline{i}}^{k\ }(u) \\
0 & e_{\ \underline{a}}^{a\ }(u)%
\end{array}%
\right] ,  \label{2vt2}
\end{eqnarray}%
in particular case $e_{i}^{\ \underline{i}}=\delta _{i}^{\underline{i}}$ and
$e_{a}^{\ \underline{a}}=\delta _{a}^{\underline{a}}$ with $\delta _{i}^{%
\underline{i}}$ and $\delta _{a}^{\underline{a}}$ being the Kronecker
symbols, defining a global splitting of $\mathbf{V}^{n+m}$ into
''horizontal'' and ''vertical'' subspaces with the N--vielbein structure%
\begin{equation*}
\mathbf{e}_{\alpha }=\mathbf{e}_{\alpha }^{\ \underline{\alpha }}\partial _{%
\underline{\alpha }}\mbox{ and }\mathbf{\vartheta }_{\ }^{\beta }=\mathbf{e}%
_{\ \underline{\beta }}^{\beta }du^{\underline{\beta }}.
\end{equation*}%
We adopt the convention that for the spaces provided with N--connection
structure the geometrical objects can be denoted by ''boldfaced'' symbols if
it would be necessary to distinguish such objects from similar ones for
spaces without N--connection.

A N--connection $\mathbf{N}$ in a space $\mathbf{V}^{n+m}$ is parametrized
by its components $N_{i}^{a}(u)=N_{i}^{a}(x,y),$
\begin{equation*}
\mathbf{N}=N_{i}^{a}(u)d^{i}\otimes \partial _{a}
\end{equation*}%
and characterized by the N--connection curvature
\begin{equation*}
\mathbf{\Omega }=\frac{1}{2}\Omega _{ij}^{a}d^{i}\wedge d^{j}\otimes
\partial _{a},
\end{equation*}%
with N--connection curvature coefficients%
\begin{equation}
\Omega _{ij}^{a}=\delta _{\lbrack j}N_{i]}^{a}=\frac{\partial N_{i}^{a}}{%
\partial x^{j}}-\frac{\partial N_{j}^{a}}{\partial x^{i}}+N_{i}^{b}\frac{%
\partial N_{j}^{a}}{\partial y^{b}}-N_{j}^{b}\frac{\partial N_{i}^{a}}{%
\partial y^{b}}.  \label{2ncurv}
\end{equation}%
On spaces provided with N--connection structure, we have to use
'N--elongated' operators like $\delta _{j}$ in (\ref{2ncurv}) instead of
usual partial derivatives. They are defined by the vielbein configuration
induced by the N--connection, the N--elongated partial derivatives (in
brief, N--derivatives)
\begin{equation}
\mathbf{e}_{\alpha }\doteqdot \delta _{\alpha }=\left( \delta _{i},\partial
_{a}\right) \equiv \frac{\delta }{\delta u^{\alpha }}=\left( \frac{\delta }{%
\delta x^{i}}=\partial _{i}-N_{i}^{a}\left( u\right) \partial _{a},\frac{%
\partial }{\partial y^{a}}\right)  \label{2dder}
\end{equation}%
and the N--elongated differentials (in brief, N--differentials)
\begin{equation}
\mathbf{\vartheta }_{\ }^{\beta }\doteqdot \delta \ ^{\beta }=\left(
d^{i},\delta ^{a}\right) \equiv \delta u^{\alpha }=\left( \delta
x^{i}=dx^{i},\delta y^{a}=dy^{a}+N_{i}^{a}\left( u\right) dx^{i}\right)
\label{2ddif}
\end{equation}%
called also, respectively, the N--frame and N--coframe. There are used both
type of denotations $\mathbf{e}_{\alpha }\doteqdot \delta _{\alpha }$ and $%
\mathbf{\vartheta }_{\ }^{\beta }\doteqdot \delta \ ^{\alpha }$ in order to
preserve a connection to denotations from Refs. \cite{02ma,02v1,02v1a,02v2}. The
'boldfaced' symbols $\mathbf{e}_{\alpha }$ and $\mathbf{\vartheta }_{\
}^{\beta }$ will be considered in order to emphasize that they define
N--adapted vielbeins but the symbols $\delta _{\alpha }$ and $\delta \
^{\beta }$ will be used for the N--elongated partial derivatives and,
respectively, differentials.

The N--coframe (\ref{2ddif}) satisfies the anholonomy relations
\begin{equation}
\left[ \delta _{\alpha },\delta _{\beta }\right] =\delta _{\alpha }\delta
_{\beta }-\delta _{\beta }\delta _{\alpha }=\mathbf{w}_{\ \alpha \beta
}^{\gamma }\left( u\right) \delta _{\gamma }  \label{2anhr}
\end{equation}%
with nontrivial anholonomy coefficients $\mathbf{w}_{\beta \gamma }^{\alpha
}\left( u\right) $ computed as
\begin{equation}
\mathbf{w}_{~ji}^{a}=-\mathbf{w}_{~ij}^{a}=\Omega _{ij}^{a},\ \mathbf{w}%
_{~ia}^{b}=-\mathbf{w}_{~ai}^{b}=\partial _{a}N_{i}^{b}.  \label{2anhc}
\end{equation}

The distinguished objects (by a N--connection on a spaces $\mathbf{V}^{n+m})$
are introduced in a coordinate free form as geometric objects adapted to the
splitting (\ref{2wihit}). In brief, they are called d--objects, d--tensor,
d--connections, d--metrics....

There is an important class of linear connections adapted to the
N--connection structure:

A d--connection $\mathbf{D}$ on a space $\mathbf{V}^{n+m}$ is defined as a
linear connection $D$ conserving under a parallelism the global
decomposition (\ref{2wihit}).

The N--adapted components $\mathbf{\Gamma }_{\beta \gamma }^{\alpha }$ of a
d-connection $\mathbf{D}_{\alpha }=(\delta _{\alpha }\rfloor \mathbf{D})$
are defined by the equations%
\begin{equation*}
\mathbf{D}_{\alpha }\delta _{\beta }=\mathbf{\Gamma }_{\ \alpha \beta
}^{\gamma }\delta _{\gamma },
\end{equation*}%
from which one immediately follows
\begin{equation}
\mathbf{\Gamma }_{\ \alpha \beta }^{\gamma }\left( u\right) =\left( \mathbf{D%
}_{\alpha }\delta _{\beta }\right) \rfloor \delta ^{\gamma }.  \label{2dcon1}
\end{equation}%
The operations of h- and v-covariant derivations, $D_{k}^{[h]}=%
\{L_{jk}^{i},L_{bk\;}^{a}\}$ and $D_{c}^{[v]}=\{C_{jk}^{i},C_{bc}^{a}\}$ are
introduced as corresponding h- and v--parametrizations of (\ref{2dcon1}),%
\begin{equation*}
L_{jk}^{i}=\left( \mathbf{D}_{k}\delta _{j}\right) \rfloor d^{i},\quad
L_{bk}^{a}=\left( \mathbf{D}_{k}\partial _{b}\right) \rfloor \delta
^{a},~C_{jc}^{i}=\left( \mathbf{D}_{c}\delta _{j}\right) \rfloor d^{i},\quad
C_{bc}^{a}=\left( \mathbf{D}_{c}\partial _{b}\right) \rfloor \delta ^{a}.
\end{equation*}%
The components $\mathbf{\Gamma }_{\ \alpha \beta }^{\gamma }=\left(
L_{jk}^{i},L_{bk}^{a},C_{jc}^{i},C_{bc}^{a}\right) $ completely define a
d--connection $\mathbf{D}$ in $\mathbf{V}^{n+m}.$

A metric structure $\mathbf{g}$ on a space $\mathbf{V}^{n+m}$ is defined as
a symmetric covariant tensor field of type $\left( 0,2\right) ,$ $g_{\alpha
\beta ,}$ being nondegenerate and of constant signature on $\mathbf{V}%
^{n+m}. $ A N--connection $\mathbf{N=}\{N_{\underline{i}}^{\underline{b}%
}\left( u\right) \}$ and a metric structure $\mathbf{g}=g_{\underline{\alpha
}\underline{\beta }}du^{\underline{\alpha }}\otimes du^{\underline{\beta }}$
on $\mathbf{V}^{n+m}$ are mutually compatible if there are satisfied the
conditions
\begin{equation*}
\mathbf{g}\left( \delta _{\underline{i}},\partial _{\underline{a}}\right) =0,%
\mbox{ or equivalently,
}g_{\underline{i}\underline{a}}\left( u\right) -N_{\underline{i}}^{%
\underline{b}}\left( u\right) h_{\underline{a}\underline{b}}\left( u\right)
=0,
\end{equation*}%
where $h_{\underline{a}\underline{b}}\doteqdot \mathbf{g}\left( \partial _{%
\underline{a}},\partial _{\underline{b}}\right) $ and $g_{\underline{i}%
\underline{a}}\doteqdot \mathbf{g}\left( \partial _{\underline{i}},\partial
_{\underline{a}}\right) \,$ resulting in
\begin{equation*}
N_{i}^{b}\left( u\right) =h^{ab}\left( u\right) g_{ia}\left( u\right)
\end{equation*}%
(the matrix $h^{ab}$ is inverse to $h_{ab};$ for simplicity, we do not
underline the indices in the last formula). In consequence, we define an
invariant h--v--decomposition of metric (in brief, a d--metric)%
\begin{equation*}
\mathbf{g}(X,Y)\mathbf{=}h\mathbf{g}(X,Y)+v\mathbf{g}(X,Y).
\end{equation*}%
With respect to a N--coframe (\ref{2ddif}), the d--metric is written
\begin{equation}
\mathbf{g}=\mathbf{g}_{\alpha \beta }\left( u\right) \delta ^{\alpha
}\otimes \delta ^{\beta }=g_{ij}\left( u\right) d^{i}\otimes
d^{j}+h_{ab}\left( u\right) \delta ^{a}\otimes \delta ^{b},  \label{2block2}
\end{equation}%
where $g_{ij}\doteqdot \mathbf{g}\left( \delta _{i},\delta _{j}\right) .$
The d--metric (\ref{2block2}) can be equivalently written in
''off--diagonal'' with respect to a coordinate basis defined by usual local
differentials $du^{\alpha }=\left( dx^{i},dy^{a}\right) ,$
\begin{equation}
\underline{g}_{\alpha \beta }=\left[
\begin{array}{cc}
g_{ij}+N_{i}^{a}N_{j}^{b}h_{ab} & N_{j}^{e}h_{ae} \\
N_{i}^{e}h_{be} & h_{ab}%
\end{array}%
\right] .  \label{2ansatz}
\end{equation}

A metric, for instance, parametrized in the form (\ref{2ansatz})\ is generic
off--diagonal if it can not be diagonalized by any coordinate transforms.
The anholonomy coefficients (\ref{2anhc}) do not vanish for the off--diagonal
form (\ref{2ansatz}) and the equivalent d--metric (\ref{2block2}).

The nonmetricity d--field
\begin{equation*}
\ \mathcal{Q}=\mathbf{Q}_{\alpha \beta }\mathbf{\vartheta }^{\alpha }\otimes
\mathbf{\vartheta }^{\beta }=\mathbf{Q}_{\alpha \beta }\delta \ ^{\alpha
}\otimes \delta ^{\beta }
\end{equation*}%
on a space $\mathbf{V}^{n+m}$ provided with N--connection structure is
defined by a d--tensor field with the coefficients
\begin{equation}
\mathbf{Q}_{\alpha \beta }\doteqdot -\mathbf{Dg}_{\alpha \beta }  \label{2nmf}
\end{equation}%
where the covariant derivative $\mathbf{D}$ is for a d--connection (\ref%
{2dcon1}) $\mathbf{\Gamma }_{\ \alpha }^{\gamma }=\mathbf{\Gamma }_{\ \alpha
\beta }^{\gamma }\mathbf{\vartheta }^{\beta }$ with $\mathbf{\Gamma }%
_{\alpha \beta }^{\gamma }=\left(
L_{jk}^{i},L_{bk}^{a},C_{jc}^{i},C_{bc}^{a}\right) .$

A linear connection $D_{X}$ is compatible\textbf{\ } with a d--metric $%
\mathbf{g}$ if%
\begin{equation}
D_{X}\mathbf{g}=0,  \label{2mc}
\end{equation}%
i. e. if $Q_{\alpha \beta }\equiv 0.$ In a space provided with N--connection
structure, the metricity condition (\ref{2mc}) may split into a set of
compatibility conditions on h- and v-- subspaces,
\begin{equation}
D^{[h]}(h\mathbf{g)}=0,D^{[v]}(h\mathbf{g)}=0,D^{[h]}(v\mathbf{g)}%
=0,D^{[v]}(v\mathbf{g)}=0.  \label{2mca}
\end{equation}
For instance, if $D^{[v]}(h\mathbf{g)}=0$ and $D^{[h]}(v\mathbf{g)}=0,$ but,
in general, $D^{[h]}(h\mathbf{g)}\neq 0$ and $D^{[v]}(v\mathbf{g)}\neq 0$ we
have a nontrivial nonmetricity d--field $\mathbf{Q}_{\alpha \beta }=\mathbf{Q%
}_{\gamma \alpha \beta }\vartheta ^{\gamma }$ with irreducible
h--v--components $\mathbf{Q}_{\gamma \alpha \beta }=\left(
Q_{ijk},Q_{abc}\right) .$

In a metric--affine space, by acting on forms with a covariant derivative $%
D, $ we can also define another very important geometric objects (the
'gravitational field potentials', the torsion and, respectively, curvature;
see \cite{02mag}):%
\begin{equation}
\ \mathbf{T}^{\alpha }\doteqdot \mathbf{D\vartheta }^{\alpha }=\delta
\mathbf{\vartheta }^{\alpha }+\mathbf{\Gamma }_{\ \beta }^{\gamma }\wedge
\mathbf{\vartheta }^{\beta }  \label{2dt}
\end{equation}%
and
\begin{equation}
\ \mathbf{R}_{\ \beta }^{\alpha }\doteqdot \mathbf{D\Gamma }_{\ \beta
}^{\alpha }=\delta \mathbf{\Gamma }_{\ \beta }^{\alpha }-\mathbf{\Gamma }_{\
\beta }^{\gamma }\wedge \mathbf{\Gamma }_{\ \ \gamma }^{\alpha }  \label{2dc}
\end{equation}%
For spaces provided with N--connection structures, we consider the same
formulas but for ''boldfaced'' symbols and change the usual differential $d$
$\ $into N-adapted operator $\delta .$

A general affine (linear) connection $D=\bigtriangledown +Z=\{\Gamma _{\beta
\alpha }^{\gamma }=\Gamma _{\bigtriangledown \beta \alpha }^{\gamma
}+Z_{\beta \alpha }^{\gamma }\}$
\begin{equation}
\Gamma _{\ \alpha }^{\gamma }=\Gamma _{\alpha \beta }^{\gamma }\vartheta
^{\beta },  \label{2ac}
\end{equation}%
can always be decomposed into the Riemannian $\Gamma _{\bigtriangledown \
\beta }^{\alpha }$ and post--Riemannian $Z_{\ \beta }^{\alpha }$ parts \cite%
{02mag,02oveh},
\begin{equation}
\Gamma _{\ \beta }^{\alpha }=\Gamma _{\bigtriangledown \ \beta }^{\alpha
}+Z_{\ \beta }^{\alpha }.  \label{2acc}
\end{equation}%
The distorsion 1-form $Z_{\ \beta }^{\alpha }$ from (\ref{2acc}) is expressed
in terms of torsion and nonmetricity,%
\begin{equation}
Z_{\alpha \beta }=e_{\beta }\rfloor T_{\alpha }-e_{\alpha }\rfloor T_{\beta
}+\frac{1}{2}\left( e_{\alpha }\rfloor e_{\beta }\rfloor T_{\gamma }\right)
\vartheta ^{\gamma }+\left( e_{\alpha }\rfloor Q_{\beta \gamma }\right)
\vartheta ^{\gamma }-\left( e_{\beta }\rfloor Q_{\alpha \gamma }\right)
\vartheta ^{\gamma }+\frac{1}{2}Q_{\alpha \beta }  \label{2dista}
\end{equation}%
where $T_{\alpha }$ is defined as (\ref{2dt}) and $Q_{\alpha \beta }\doteqdot
-Dg_{\alpha \beta }.$ (We note that $Z_{\ \beta }^{\alpha }$ are $N_{\alpha
\beta }$ from Ref. \cite{02oveh}, but in our works we use the symbol $N$ for
N--connections .) \ For $Q_{\beta \gamma }=0,$ we obtain from (\ref{2dista})
 the distorsion for the Riemannian--Cartan geometry \cite{02rcg}.

By substituting arbitrary (co) frames, metrics and linear connections into
N--adapted ones,
\begin{equation*}
e_{\alpha }\rightarrow \mathbf{e}_{\alpha },\vartheta ^{\beta }\rightarrow
\mathbf{\vartheta }^{\beta },g_{\alpha \beta }\rightarrow \mathbf{g}_{\alpha
\beta }=\left( g_{ij},h_{ab}\right) ,\Gamma _{\ \alpha }^{\gamma
}\rightarrow \mathbf{\Gamma }_{\ \alpha }^{\gamma },
\end{equation*}%
with $\mathbf{Q}_{\alpha \beta }=\mathbf{Q}_{\gamma \alpha \beta }\mathbf{%
\vartheta }^{\gamma }$ and $\mathbf{T}^{\alpha }$ as in (\ref{2dt}), into
respective formulas (\ref{2ac}), (\ref{2acc}) and (\ref{2dista}), \ we can
define an affine connection $\mathbf{D=\bigtriangledown +Z}=[\mathbf{\Gamma }%
_{\ \beta \alpha }^{\gamma }]$ with respect to N--adapted (co) frames,
\begin{equation}
\mathbf{\Gamma }_{\ \ \alpha }^{\gamma }=\mathbf{\Gamma }_{\ \alpha \beta
}^{\gamma }\mathbf{\vartheta }^{\beta },  \label{2acn}
\end{equation}%
with
\begin{equation}
\mathbf{\Gamma }_{\ \beta }^{\alpha }=\mathbf{\Gamma }_{\bigtriangledown \
\beta }^{\alpha }+\mathbf{Z}_{\ \ \beta }^{\alpha },  \label{2accn}
\end{equation}%
where
\begin{equation}
\mathbf{\Gamma }_{\ \gamma \alpha }^{\bigtriangledown }=\frac{1}{2}\left[
\mathbf{e}_{\gamma }\rfloor \ \delta \mathbf{\vartheta }_{\alpha }-\mathbf{e}%
_{\alpha }\rfloor \ \delta \mathbf{\vartheta }_{\gamma }-\left( \mathbf{e}%
_{\gamma }\rfloor \ \mathbf{e}_{\alpha }\rfloor \ \delta \mathbf{\vartheta }%
_{\beta }\right) \wedge \mathbf{\vartheta }^{\beta }\right] ,
\label{2christa}
\end{equation}%
and
\begin{equation}
\mathbf{Z}_{\alpha \beta }=\mathbf{e}_{\beta }\rfloor \mathbf{T}_{\alpha }-%
\mathbf{e}_{\alpha }\rfloor \mathbf{T}_{\beta }+\frac{1}{2}\left( \mathbf{e}%
_{\alpha }\rfloor \mathbf{e}_{\beta }\rfloor \mathbf{T}_{\gamma }\right)
\mathbf{\vartheta }^{\gamma }+\left( \mathbf{e}_{\alpha }\rfloor \mathbf{Q}%
_{\beta \gamma }\right) \mathbf{\vartheta }^{\gamma }-\left( \mathbf{e}%
_{\beta }\rfloor \mathbf{Q}_{\alpha \gamma }\right) \mathbf{\vartheta }%
^{\gamma }+\frac{1}{2}\mathbf{Q}_{\alpha \beta }.  \label{2distan}
\end{equation}%
The h-- and v--components of $\mathbf{\Gamma }_{\ \beta }^{\alpha }$ from (%
\ref{2accn}) consists from the components of $\mathbf{\Gamma }%
_{\bigtriangledown \ \beta }^{\alpha }$ (considered for (\ref{2christa})) and
of $\mathbf{Z}_{\alpha \beta }$ with $\mathbf{Z}_{\ \ \gamma \beta }^{\alpha
}=\left( Z_{jk}^{i},Z_{bk}^{a},Z_{jc}^{i},Z_{bc}^{a}\right) .$ We note that
for $\mathbf{Q}_{\alpha \beta }=0,$ the distorsion 1--form $\mathbf{Z}%
_{\alpha \beta }$ defines a Riemann--Cartan geometry adapted to the
N--connection structure.

A \ distinguished metric--affine space $\mathbf{V}^{n+m}$ is defined as a
usual metric--affine space additionally enabled with a N--connection
structure $\mathbf{N}=\{N_{i}^{a}\}$ inducing splitting into respective
irreducible horizontal and vertical subspaces of dimensions $n$ and $m.$
This space is provided with independent d--metric (\ref{2block2}) and affine
d--connection (\ref{2dcon1}) structures adapted to the N--connection.\

If a space $\mathbf{V}^{n+m}$ is provided with both N--connection
$\mathbf{N} $\ and d--metric $\mathbf{g}$ structures, there is a
unique linear symmetric and torsionless connection
$\mathbf{\bigtriangledown },$ called the Levi--Civita connection,
being metric compatible such that $\bigtriangledown _{\gamma
}\mathbf{g}_{\alpha \beta }=0$ $\ $for $\mathbf{g}_{\alpha \beta
}=\left( g_{ij},h_{ab}\right) ,$ see (\ref{2block2}), with the
coefficients
\begin{equation*}
\mathbf{\Gamma }_{\alpha \beta \gamma }^{\bigtriangledown }=\mathbf{g}\left(
\delta _{\alpha },\bigtriangledown _{\gamma }\delta _{\beta }\right) =%
\mathbf{g}_{\alpha \tau }\mathbf{\Gamma }_{\bigtriangledown \beta \gamma
}^{\tau },\,
\end{equation*}%
computed as
\begin{equation}
\mathbf{\Gamma }_{\alpha \beta \gamma }^{\bigtriangledown }=\frac{1}{2}\left[
\delta _{\beta }\mathbf{g}_{\alpha \gamma }+\delta _{\gamma }\mathbf{g}%
_{\beta \alpha }-\delta _{\alpha }\mathbf{g}_{\gamma \beta }+\mathbf{g}%
_{\alpha \tau }\mathbf{w}_{\gamma \beta }^{\tau }+\mathbf{g}_{\beta \tau }%
\mathbf{w}_{\alpha \gamma }^{\tau }-\mathbf{g}_{\gamma \tau }\mathbf{w}%
_{\beta \alpha }^{\tau }\right]  \label{92lcsym}
\end{equation}%
with respect to N--frames $\mathbf{e}_{\beta }\doteqdot \delta _{\beta }$ (%
\ref{2dder}) and N--coframes $\mathbf{\vartheta }_{\ }^{\alpha }\doteqdot
\delta ^{\alpha }$ (\ref{2ddif}).

We note that the Levi--Civita connection is not adapted to the N--connection
structure.\ Se, we can not state its coefficients in an irreducible form for
the h-- and v--subspaces. There is a type of d--connections which are
similar to the Levi--Civita connection but satisfying certain metricity
conditions adapted to the N--connection. They are introduced as metric
d--connections $\mathbf{D=}\left( D^{[h]},D^{[v]}\right) $ in a space \ $%
\mathbf{V}^{n+m}$ satisfying the metricity conditions if and only if
\begin{equation}
D_{k}^{[h]}g_{ij}=0,\ D_{a}^{[v]}g_{ij}=0,\ D_{k}^{[h]}h_{ab}=0,\
D_{a}^{[h]}h_{ab}=0.  \label{2mcas}
\end{equation}

Let us consider an important example: The canonical d--connection $\widehat{%
\mathbf{D}}$\ \ $\mathbf{=}\left( \widehat{D}^{[h]},\widehat{D}^{[v]}\right)
,$ equivalently $\widehat{\mathbf{\Gamma }}_{\ \alpha }^{\gamma }=\widehat{%
\mathbf{\Gamma }}_{\ \alpha \beta }^{\gamma }\mathbf{\vartheta }^{\beta },$\
is defined by the h-- v--irreducible components $\widehat{\mathbf{\Gamma }}%
_{\ \alpha \beta }^{\gamma }=(\widehat{L}_{jk}^{i},\widehat{L}_{bk}^{a},$ $%
\widehat{C}_{jc}^{i},\widehat{C}_{bc}^{a}),$ where%
\begin{eqnarray}
\widehat{L}_{jk}^{i} &=&\frac{1}{2}g^{ir}\left( \frac{\delta g_{jk}}{\delta
x^{k}}+\frac{\delta g_{kr}}{\delta x^{j}}-\frac{\delta g_{jk}}{\delta x^{r}}%
\right) ,  \label{2candcon} \\
\widehat{L}_{bk}^{a} &=&\frac{\partial N_{k}^{a}}{\partial y^{b}}+\frac{1}{2}%
h^{ac}\left( \frac{\delta h_{bc}}{\delta x^{k}}-\frac{\partial N_{k}^{d}}{%
\partial y^{b}}h_{dc}-\frac{\partial N_{k}^{d}}{\partial y^{c}}h_{db}\right)
,  \notag \\
\widehat{C}_{jc}^{i} &=&\frac{1}{2}g^{ik}\frac{\partial g_{jk}}{\partial
y^{c}},  \notag \\
\widehat{C}_{bc}^{a} &=&\frac{1}{2}h^{ad}\left( \frac{\partial h_{bd}}{%
\partial y^{c}}+\frac{\partial h_{cd}}{\partial y^{b}}-\frac{\partial h_{bc}%
}{\partial y^{d}}\right) .  \notag
\end{eqnarray}%
satisfying the torsionless conditions for the h--subspace and v--subspace,
respectively, $\widehat{T}_{jk}^{i}=\widehat{T}_{bc}^{a}=0.$

The components of the Levi--Civita connection $\mathbf{\Gamma }%
_{\bigtriangledown \beta \gamma }^{\tau }$ and the irreducible components of
the canonical d--connection \ $\widehat{\mathbf{\Gamma }}_{\ \beta \gamma
}^{\tau }$\ are related by formulas%
\begin{equation}
\mathbf{\Gamma }_{\bigtriangledown \beta \gamma }^{\tau }=\left( \widehat{L}%
_{jk}^{i},\widehat{L}_{bk}^{a}-\frac{\partial N_{k}^{a}}{\partial y^{b}},%
\widehat{C}_{jc}^{i}+\frac{1}{2}g^{ik}\Omega _{jk}^{a}h_{ca},\widehat{C}%
_{bc}^{a}\right) ,  \label{1lcsyma}
\end{equation}%
where $\Omega _{jk}^{a}$\ \ is the N--connection curvature\ (\ref{2ncurv}).

We can define and calculate the irreducible components of torsion and
curvature in a space $\mathbf{V}^{n+m}$ provided with additional
N--connection structure (these could be any metric--affine spaces \cite{02mag}%
, or their particular, like Riemann--Cartan \cite{02rcg}, cases with vanishing
nonmetricity and/or torsion, or any (co) vector / tangent bundles like in
Finsler geometry and generalizations).

The torsion $$\mathbf{T}_{.\beta \gamma }^{\alpha
}=(T_{.jk}^{i},T_{ja}^{i},T_{.ij}^{a},T_{.bi}^{a},T_{.bc}^{a})$$ of a
d--connection $\mathbf{\Gamma }_{\alpha \beta }^{\gamma
}=(L_{jk}^{i},L_{bk}^{a},C_{jc}^{i},C_{bc}^{a})$ (\ref{2dcon1})  has
irreducible h- v--components (d--torsions)
\begin{eqnarray}
T_{.jk}^{i} &=&-T_{kj}^{i}=L_{jk}^{i}-L_{kj}^{i},\quad
T_{ja}^{i}=-T_{aj}^{i}=C_{.ja}^{i},\ T_{.ji}^{a}=-T_{.ij}^{a}=\frac{\delta
N_{i}^{a}}{\delta x^{j}}-\frac{\delta N_{j}^{a}}{\delta x^{i}}=\Omega
_{.ji}^{a},  \notag \\
T_{.bi}^{a} &=&-T_{.ib}^{a}=P_{.bi}^{a}=\frac{\partial N_{i}^{a}}{\partial
y^{b}}-L_{.bj}^{a},\
T_{.bc}^{a}=-T_{.cb}^{a}=S_{.bc}^{a}=C_{bc}^{a}-C_{cb}^{a}.\   \label{2dtorsb}
\end{eqnarray}

We note that on (pseudo) Riemanian spacetimes the d--torsions can be induced
by the N--connection coefficients and reflect an anholonomic frame
structure. Such objects vanish when we transfer our considerations with
respect to holonomic bases for a trivial N--connection and zero ''vertical''
dimension.

The curvature $$\mathbf{R}_{.\beta \gamma \tau }^{\alpha }=(R_{\
hjk}^{i},R_{\ bjk}^{a},P_{\ jka}^{i},P_{\ bka}^{c},S_{\ jbc}^{i},S_{\
bcd}^{a})$$ of a d--con\-nec\-ti\-on $\mathbf{\Gamma }_{\alpha \beta
}^{\gamma }=(L_{jk}^{i},L_{bk}^{a},C_{jc}^{i},C_{bc}^{a})$ (\ref{2dcon1}) has
irreducible h- v--components (d--curvatures)
\begin{eqnarray}
R_{\ hjk}^{i} &=&\frac{\delta L_{.hj}^{i}}{\delta x^{k}}-\frac{\delta
L_{.hk}^{i}}{\delta x^{j}}%
+L_{.hj}^{m}L_{mk}^{i}-L_{.hk}^{m}L_{mj}^{i}-C_{.ha}^{i}\Omega _{.jk}^{a},
\label{2dcurv} \\
R_{\ bjk}^{a} &=&\frac{\delta L_{.bj}^{a}}{\delta x^{k}}-\frac{\delta
L_{.bk}^{a}}{\delta x^{j}}%
+L_{.bj}^{c}L_{.ck}^{a}-L_{.bk}^{c}L_{.cj}^{a}-C_{.bc}^{a}\ \Omega
_{.jk}^{c},  \notag \\
P_{\ jka}^{i} &=&\frac{\partial L_{.jk}^{i}}{\partial y^{k}}-\left( \frac{%
\partial C_{.ja}^{i}}{\partial x^{k}}%
+L_{.lk}^{i}C_{.ja}^{l}-L_{.jk}^{l}C_{.la}^{i}-L_{.ak}^{c}C_{.jc}^{i}\right)
+C_{.jb}^{i}P_{.ka}^{b},  \notag \\
P_{\ bka}^{c} &=&\frac{\partial L_{.bk}^{c}}{\partial y^{a}}-\left( \frac{%
\partial C_{.ba}^{c}}{\partial x^{k}}+L_{.dk}^{c%
\,}C_{.ba}^{d}-L_{.bk}^{d}C_{.da}^{c}-L_{.ak}^{d}C_{.bd}^{c}\right)
+C_{.bd}^{c}P_{.ka}^{d},  \notag \\
S_{\ jbc}^{i} &=&\frac{\partial C_{.jb}^{i}}{\partial y^{c}}-\frac{\partial
C_{.jc}^{i}}{\partial y^{b}}+C_{.jb}^{h}C_{.hc}^{i}-C_{.jc}^{h}C_{hb}^{i},
\notag \\
S_{\ bcd}^{a} &=&\frac{\partial C_{.bc}^{a}}{\partial y^{d}}-\frac{\partial
C_{.bd}^{a}}{\partial y^{c}}+C_{.bc}^{e}C_{.ed}^{a}-C_{.bd}^{e}C_{.ec}^{a}.
\notag
\end{eqnarray}

The components of the Ricci tensor
\begin{equation*}
\mathbf{R}_{\alpha \beta }=\mathbf{R}_{\ \alpha \beta \tau }^{\tau }
\end{equation*}%
with respect to a locally adapted frame (\ref{2dder}) has four irreducible h-
v--components $\mathbf{R}_{\alpha \beta }=(R_{ij},R_{ia},R_{ai},S_{ab}),$
where%
\begin{eqnarray}
R_{ij} &=&R_{\ ijk}^{k},\quad R_{ia}=-\ ^{2}P_{ia}=-P_{\ ika}^{k},
\label{2dricci} \\
R_{ai} &=&\ ^{1}P_{ai}=P_{\ aib}^{b},\quad S_{ab}=S_{\ abc}^{c}.  \notag
\end{eqnarray}%
We point out that because, in general, $^{1}P_{ai}\neq ~^{2}P_{ia}$ the
Ricci d--tensor is non symmetric.

Having defined a d--metric of type (\ref{2block2}) in $\mathbf{V}^{n+m},$ we
can introduce the scalar curvature of a d--connection $\mathbf{D,}$
\begin{equation}
{\overleftarrow{\mathbf{R}}}=\mathbf{g}^{\alpha \beta }\mathbf{R}_{\alpha
\beta }=R+S,  \label{2dscal}
\end{equation}%
where $R=g^{ij}R_{ij}$ and $S=h^{ab}S_{ab}$ and define the distinguished
form of the Einstein tensor (the Einstein d--tensor),
\begin{equation}
\mathbf{G}_{\alpha \beta }\doteqdot \mathbf{R}_{\alpha \beta }-\frac{1}{2}%
\mathbf{g}_{\alpha \beta }{\overleftarrow{\mathbf{R}}.}  \label{2deinst}
\end{equation}

The introduced geometrical objects are extremely useful in definition of
field equations of MAG and string gravity with nontrivial N--connection
structure.

\section{ N--Connections and Field Equations}

The field equations of metric--affine gravity (in brief, MAG) \cite{02mag,02oveh}
can be reformulated with respect to frames and coframes consisting from
mixed holonomic and anholonomic components defined by the N--connection
structure. In this case, various type of (pseudo)\ Riemannian,
Riemann--Cartan and generalized Finsler metrics and additional torsion and
nonmetricity sructures with very general local anisotropy can be embedded
into MAG. It is known that in a metric--affine spacetime the curvature,
torsion and nonmetricity have correspondingly eleven, three and four
irreducible pieces. If the N--connection is defined in a metric--affine
spacetime, every irreducible component of curvature splits additionally into
six h- and v-- components (\ref{2dcurv}), every irreducible component of
torsion splits additionally into five h- and v-- components (\ref{2dtorsb})
and every irreducible component of nonmetricity splits additionally into two
h- and v-- components (defined by splitting of metrics into block ansatz (%
\ref{2block2})).

\subsection{Lagrangians and field equations for Finsler--affine theories}

For an arbitrary d--connection $\mathbf{\Gamma }_{\ \beta }^{\alpha }$ in a
metric--affine space $\mathbf{V}^{n+m}$ provided with N--connecti\-on
structure (for simplicity, we can take $n+m=4)$ one holds the respective
decompositi\-ons for d--torsion and nonmetricity d--field,%
\begin{eqnarray}
^{(2)}\mathbf{T}^{\alpha } &\doteqdot &\frac{1}{3}\mathbf{\vartheta }%
^{\alpha }\wedge \mathbf{T,\mbox{ for }T}\doteqdot \mathbf{e}_{\alpha
}\rfloor \mathbf{T}^{\alpha },  \label{torsdec} \\
^{(3)}\mathbf{T}^{\alpha } &\doteqdot &\frac{1}{3}\ \ast \left( \mathbf{%
\vartheta }^{\alpha }\wedge \mathbf{P}\right) \mathbf{,\mbox{ for }P}%
\doteqdot \ast \left( \mathbf{T}^{\alpha }\wedge \mathbf{\vartheta }_{\alpha
}\right) ,  \notag \\
^{(1)}\mathbf{T}^{\alpha } &\doteqdot &\mathbf{T}^{\alpha }-^{(2)}\mathbf{T}%
^{\alpha }-^{(3)}\mathbf{T}^{\alpha }  \notag
\end{eqnarray}%
and
\begin{eqnarray}
^{(2)}\mathbf{Q}_{\alpha \beta } &\doteqdot &\frac{1}{3}\ast \left( \mathbf{%
\vartheta }_{\alpha }\wedge \mathbf{S}_{\beta }+\mathbf{\vartheta }_{\beta
}\wedge \mathbf{S}_{\alpha }\right) \mathbf{,\ }^{(4)}\mathbf{Q}_{\alpha
\beta }\doteqdot \mathbf{g}_{\alpha \beta }\mathbf{Q,}  \notag \\
^{(3)}\mathbf{Q}_{\alpha \beta } &\doteqdot &\frac{2}{9}\left[ \left(
\mathbf{\vartheta }_{\alpha }\mathbf{e}_{\beta }+\mathbf{\vartheta }_{\beta }%
\mathbf{e}_{\alpha }\right) \rfloor \mathbf{\Lambda }-\frac{1}{2}\mathbf{g}%
_{\alpha \beta }\mathbf{\Lambda }\right] ,  \label{nmdc} \\
^{(1)}\mathbf{Q}_{\alpha \beta } &\doteqdot &\mathbf{Q}_{\alpha \beta }-\
^{(2)}\mathbf{Q}_{\alpha \beta }-\ ^{(3)}\mathbf{Q}_{\alpha \beta }-\ ^{(4)}%
\mathbf{Q}_{\alpha \beta },  \notag
\end{eqnarray}%
where
\begin{eqnarray*}
\mathbf{Q} &\mathbf{\doteqdot }&\frac{1}{4}\mathbf{g}^{\alpha \beta }\mathbf{%
Q}_{\alpha \beta },\ \mathbf{\Lambda }\doteqdot \mathbf{\vartheta }^{\alpha }%
\mathbf{e}^{\beta }\rfloor \left( \mathbf{Q}_{\alpha \beta }-\mathbf{Qg}%
_{\alpha \beta }\right) , \\
\ \mathbf{\Theta }_{\alpha } &\doteqdot &\ast \left[ \left( \mathbf{Q}%
_{\alpha \beta }-\mathbf{Qg}_{\alpha \beta }\right) \wedge \mathbf{\vartheta
}^{\beta }\right] , \\
\mathbf{S}_{\alpha } &\mathbf{\doteqdot }&\ \mathbf{\Theta }_{\alpha }-\frac{%
1}{3}\mathbf{e}_{\alpha }\rfloor \left( \mathbf{\vartheta }^{\beta }\wedge
\mathbf{\Theta }_{\beta }\right)
\end{eqnarray*}%
and the Hodge dual ''$\ast $'' is such that $\mathbf{\eta }\doteqdot \ast 1$
is the volume 4--form and
\begin{equation*}
\mathbf{\eta }_{\alpha }\doteqdot \mathbf{e}_{\alpha }\rfloor \mathbf{\eta
=\ast \vartheta }_{\alpha },\ \mathbf{\eta }_{\alpha \beta }\doteqdot
\mathbf{e}_{\alpha }\rfloor \mathbf{\eta }_{\beta }\mathbf{=\ast }\left(
\mathbf{\vartheta }_{\alpha }\wedge \mathbf{\vartheta }_{\beta }\right) ,\
\mathbf{\eta }_{\alpha \beta \gamma }\doteqdot \mathbf{e}_{\gamma }\rfloor
\mathbf{\eta }_{\alpha \beta },\ \mathbf{\eta }_{\alpha \beta \gamma \tau
}\doteqdot \mathbf{e}_{\tau }\rfloor \mathbf{\eta }_{\alpha \beta \gamma }
\end{equation*}%
with $\mathbf{\eta }_{\alpha \beta \gamma \tau }$ being totally
antisymmetric. In higher dimensions, we have to consider $\mathbf{\eta }%
\doteqdot \ast 1$ as the volume $\left( n+m\right) $--form. For N--adapted
h- and v--constructions, we have to consider couples of 'volume' forms $%
\mathbf{\eta }\doteqdot \left( \eta ^{\lbrack g]}=\ast ^{\lbrack g]}1,\eta
^{\lbrack h]}=\ast ^{\lbrack h]}1\right) $ defined correspondingly by $%
\mathbf{g}_{\alpha \beta }=\left( g_{ij},h_{ab}\right) .$

With respect to N--adapted (co) frames $\mathbf{e}_{\beta }=\left( \delta
_{i},\partial _{a}\right) $ (\ref{2dder}) and $\mathbf{\vartheta }^{\alpha
}=\left( d^{i},\delta ^{a}\right) $ (\ref{2ddif}), the irreducible
decompositions (\ref{torsdec}) split into h- and v--components $^{(A)}%
\mathbf{T}^{\alpha }=\left( ^{(A)}\mathbf{T}^{i},\ ^{(A)}\mathbf{T}%
^{a}\right) $ for every $A=1,2,3,4.$ Because, by definition, $\mathbf{Q}%
_{\alpha \beta }\doteqdot \mathbf{Dg}_{\alpha \beta }$ and $\mathbf{g}%
_{\alpha \beta }=\left( g_{ij},h_{ab}\right) $ is a d--metric field, we
conclude that in a similar form can be decomposed the nonmetricity, $\mathbf{%
Q}_{\alpha \beta }=\left( Q_{ij},Q_{ab}\right).$ The symmetrizations in
formulas (\ref{nmdc}) hide splitting for $^{(1)}\mathbf{Q}_{\alpha \beta
},^{(2)}\mathbf{Q}_{\alpha \beta }$ and $^{(3)}\mathbf{Q}_{\alpha \beta }.$
Nevertheless, the h-- and v-- decompositions can be derived separately on
h-- and v--subspaces by distinguishing the interior product $\rfloor =\left(
\rfloor ^{\lbrack h]},\rfloor ^{\lbrack v]}\right) $ as to have $\mathbf{%
\eta }_{\alpha }=\left( \mathbf{\eta }_{i}=\delta _{i}\rfloor \mathbf{\eta }%
,\ \mathbf{\eta }_{a}=\partial _{a}\rfloor \mathbf{\eta }\right) $...and all
formulas after decompositions with respect to N--adapted frames (co
resulting into a separate relations in h-- and v--subspaces, when $^{(A)}%
\mathbf{Q}_{\alpha \beta }=\left( \ ^{(A)}\mathbf{Q}_{ij},\ ^{(A)}\mathbf{Q}%
_{ab}\right) $ for every $A=1,2,3,4.$

A generalized Finsler--affine theory is described by a Lagrangian%
\begin{equation*}
\mathcal{L}=\mathcal{L}_{GFA}+\mathcal{L}_{mat},
\end{equation*}%
where $\mathcal{L}_{mat}$ represents the Lagrangian of matter fields and%
\begin{eqnarray}
\mathcal{L}_{GFA} &=&\frac{1}{2\kappa }[-a_{0[Rh]}\mathbf{R}^{ij}\wedge \eta
_{ij}-a_{0[Rv]}\mathbf{R}^{ab}\wedge \eta _{ab}-a_{0[Ph]}\mathbf{P}%
^{ij}\wedge \eta _{ij}-a_{0[Pv]}\mathbf{P}^{ab}\wedge \eta _{ab}  \notag \\
&&-a_{0[Sh]}\mathbf{S}^{ij}\wedge \eta _{ij}-a_{0[Sv]}\mathbf{S}^{ab}\wedge
\eta _{ab}-2\lambda _{\lbrack h]}\eta _{\lbrack h]}-2\lambda _{\lbrack
v]}\eta _{\lbrack v]}  \label{actgfa} \\
&&+\mathbf{T}^{i}\wedge \ast ^{\lbrack h]}\left(
\sum\limits_{[A]=1}^{3}a_{[hA]}\ ^{[A]}\mathbf{T}_{i}\right) +\mathbf{T}%
^{a}\wedge \ast ^{\lbrack v]}\left( \sum\limits_{[A]=1}^{3}a_{[vA]}\ ^{[A]}%
\mathbf{T}_{a}\right)  \notag \\
&&+2\left( \sum\limits_{[I]=2}^{4}c_{[hI]}\ ^{[I]}\mathbf{Q}_{ij}\right)
\wedge \mathbf{\vartheta }^{i}\wedge \ast ^{\lbrack h]}\ \mathbf{T}%
^{j}+2\left( \sum\limits_{[I]=2}^{4}c_{[vI]}\ ^{[I]}\mathbf{Q}_{ab}\right)
\wedge \mathbf{\vartheta }^{a}\wedge \ast ^{\lbrack v]}\mathbf{T}^{b}  \notag
\\
&&+\mathbf{Q}_{ij}\wedge \left( \sum\limits_{\lbrack I]=1}^{4}b_{[hI]}\
^{[I]}\mathbf{Q}^{ij}\right) +\mathbf{Q}_{ab}\wedge \left(
\sum\limits_{\lbrack I]=1}^{4}b_{[vI]}\ ^{[I]}\mathbf{Q}^{ab}\right)  \notag
\\
&&+b_{[h5]}\left( ^{[3]}\mathbf{Q}_{ij}\wedge \mathbf{\vartheta }^{i}\right)
\wedge \ast ^{\lbrack h]}\left( ^{[4]}\mathbf{Q}^{kj}\wedge \mathbf{%
\vartheta }_{k}\right) +b_{[v5]}\left( ^{[3]}\mathbf{Q}_{ij}\wedge \mathbf{%
\vartheta }^{i}\right) \wedge \ast ^{\lbrack v]}\left( ^{[4]}\mathbf{Q}%
^{kj}\wedge \mathbf{\vartheta }_{k}\right) ]  \notag
\end{eqnarray}%
\begin{eqnarray*}
&&-\frac{1}{2\rho _{\lbrack Rh]}}\mathbf{R}^{ij}\wedge \ast ^{\lbrack
h]}\{\sum\limits_{[I]=1}^{6}w_{[RhI]}\ (^{[I]}\mathbf{R}_{ij}-\ ^{[I]}%
\mathbf{R}_{ji})+w_{[Rh7]}\mathbf{\vartheta }_{i}\wedge \lbrack \mathbf{e}%
_{k}\rfloor ^{\lbrack h]}\ (\ ^{[5]}\mathbf{R}_{\ \ j}^{k}-\ ^{[5]}\mathbf{R}%
_{j\ }^{\ k})] \\
&&+\sum\limits_{[I]=1}^{5}z_{[RhI]}\ (^{[I]}\mathbf{R}_{ij}+\ ^{[I]}\mathbf{R%
}_{ji})+z_{[Rh6]}\mathbf{\vartheta }_{k}\wedge \lbrack \mathbf{e}_{i}\rfloor
^{\lbrack h]}\ (\ ^{[2]}\mathbf{R}_{\ \ j}^{k}-\ ^{[2]}\mathbf{R}_{j\ }^{\
k})] \\
&&+\sum\limits_{[I]=7}^{9}z_{[RhI]}\mathbf{\vartheta }_{i}\wedge \lbrack
\mathbf{e}_{k}\rfloor ^{\lbrack h]}\ (\ ^{[I-4]}\mathbf{R}_{\ \ j}^{k}-\
^{[I-4]}\mathbf{R}_{j\ }^{\ k})]\}
\end{eqnarray*}%
\begin{eqnarray*}
&&-\frac{1}{2\rho _{\lbrack Rv]}}\mathbf{R}^{ab}\wedge \ast ^{\lbrack
v]}\{\sum\limits_{[I]=1}^{6}w_{[RvI]}\ (^{[I]}\mathbf{R}_{ab}-\ ^{[I]}%
\mathbf{R}_{ba})+w_{[Rv7]}\mathbf{\vartheta }_{a}\wedge \lbrack \mathbf{e}%
_{c}\rfloor ^{\lbrack v]}\ (\ ^{[5]}\mathbf{R}_{\ \ b}^{a}-\ ^{[5]}\mathbf{R}%
_{b\ }^{\ a})] \\
&&+\sum\limits_{[I]=1}^{5}z_{[RvI]}\ (^{[I]}\mathbf{R}_{ab}+\ ^{[I]}\mathbf{R%
}_{ba})+z_{[Rv6]}\mathbf{\vartheta }_{c}\wedge \lbrack \mathbf{e}_{a}\rfloor
^{\lbrack v]}\ (\ ^{[2]}\mathbf{R}_{\ \ b}^{c}-\ ^{[2]}\mathbf{R}_{b\ }^{\
c})] \\
&&+\sum\limits_{[I]=7}^{9}z_{[RvI]}\mathbf{\vartheta }_{a}\wedge \lbrack
\mathbf{e}_{c}\rfloor ^{\lbrack v]}\ (\ ^{[I-4]}\mathbf{R}_{\ \ b}^{c}-\
^{[I-4]}\mathbf{R}_{b\ }^{\ c})]\}
\end{eqnarray*}%
\begin{eqnarray*}
&&-\frac{1}{2\rho _{\lbrack Ph]}}\mathbf{P}^{ij}\wedge \ast ^{\lbrack
h]}\{\sum\limits_{[I]=1}^{6}w_{[PhI]}\ (^{[I]}\mathbf{P}_{ij}-\ ^{[I]}%
\mathbf{P}_{ji})+w_{[Ph7]}\mathbf{\vartheta }_{i}\wedge \lbrack \mathbf{e}%
_{k}\rfloor ^{\lbrack h]}\ (\ ^{[5]}\mathbf{P}_{\ \ j}^{k}-\ ^{[5]}\mathbf{P}%
_{j\ }^{\ k})] \\
&&+\sum\limits_{[I]=1}^{5}z_{[PhI]}\ (^{[I]}\mathbf{P}_{ij}+\ ^{[I]}\mathbf{P%
}_{ji})+z_{[Ph6]}\mathbf{\vartheta }_{k}\wedge \lbrack \mathbf{e}_{i}\rfloor
^{\lbrack h]}\ (\ ^{[2]}\mathbf{P}_{\ \ j}^{k}-\ ^{[2]}\mathbf{P}_{j\ }^{\
k})] \\
&&+\sum\limits_{[I]=7}^{9}z_{[PhI]}\mathbf{\vartheta }_{i}\wedge \lbrack
\mathbf{e}_{k}\rfloor ^{\lbrack h]}\ (\ ^{[I-4]}\mathbf{P}_{\ \ j}^{k}-\
^{[I-4]}\mathbf{P}_{j\ }^{\ k})]\}-
\end{eqnarray*}%
\begin{eqnarray*}
&&\frac{1}{2\rho _{\lbrack Pv]}}\mathbf{P}^{ab}\wedge \ast ^{\lbrack
v]}\{\sum\limits_{[I]=1}^{6}w_{[PvI]}\ (^{[I]}\mathbf{P}_{ab}-\ ^{[I]}%
\mathbf{P}_{ba})+w_{[Pv7]}\mathbf{\vartheta }_{a}\wedge \lbrack \mathbf{e}%
_{c}\rfloor ^{\lbrack v]}\ (\ ^{[5]}\mathbf{P}_{\ \ b}^{a}-\ ^{[5]}\mathbf{P}%
_{b\ }^{\ a})] \\
&&+\sum\limits_{[I]=1}^{5}z_{[PvI]}\ (^{[I]}\mathbf{P}_{ab}+\ ^{[I]}\mathbf{P%
}_{ba})+z_{[Pv6]}\mathbf{\vartheta }_{c}\wedge \lbrack \mathbf{e}_{a}\rfloor
^{\lbrack v]}\ (\ ^{[2]}\mathbf{P}_{\ \ b}^{c}-\ ^{[2]}\mathbf{P}_{b\ }^{\
c})] \\
&&+\sum\limits_{[I]=7}^{9}z_{[PvI]}\mathbf{\vartheta }_{a}\wedge \lbrack
\mathbf{e}_{c}\rfloor ^{\lbrack v]}\ (\ ^{[I-4]}\mathbf{P}_{\ \ b}^{c}-\
^{[I-4]}\mathbf{P}_{b\ }^{\ c})]\}
\end{eqnarray*}%
\begin{eqnarray*}
&&-\frac{1}{2\rho _{\lbrack Sh]}}\mathbf{S}^{ij}\wedge \ast ^{\lbrack
h]}\{\sum\limits_{[I]=1}^{6}w_{[ShI]}\ (^{[I]}\mathbf{S}_{ij}-\ ^{[I]}%
\mathbf{S}_{ji})+w_{[Sh7]}\mathbf{\vartheta }_{i}\wedge \lbrack \mathbf{e}%
_{k}\rfloor ^{\lbrack h]}\ (\ ^{[5]}\mathbf{S}_{\ \ j}^{k}-\ ^{[5]}\mathbf{S}%
_{j\ }^{\ k})] \\
&&+\sum\limits_{[I]=1}^{5}z_{[ShI]}\ (^{[I]}\mathbf{S}_{ij}+\ ^{[I]}\mathbf{S%
}_{ji})+z_{[Sh6]}\mathbf{\vartheta }_{k}\wedge \lbrack \mathbf{e}_{i}\rfloor
^{\lbrack h]}\ (\ ^{[2]}\mathbf{S}_{\ \ j}^{k}-\ ^{[2]}\mathbf{S}_{j\ }^{\
k})] \\
&&+\sum\limits_{[I]=7}^{9}z_{[ShI]}\mathbf{\vartheta }_{i}\wedge \lbrack
\mathbf{e}_{k}\rfloor ^{\lbrack h]}\ (\ ^{[I-4]}\mathbf{S}_{\ \ j}^{k}-\
^{[I-4]}\mathbf{S}_{j\ }^{\ k})]\}-
\end{eqnarray*}%
\begin{eqnarray*}
&&\frac{1}{2\rho _{\lbrack Sv]}}\mathbf{S}^{ab}\wedge \ast ^{\lbrack
v]}\{\sum\limits_{[I]=1}^{6}w_{[SvI]}\ (^{[I]}\mathbf{S}_{ab}-\ ^{[I]}%
\mathbf{S}_{ba})+w_{[Sv7]}\mathbf{\vartheta }_{a}\wedge \lbrack \mathbf{e}%
_{c}\rfloor ^{\lbrack v]}\ (\ ^{[5]}\mathbf{S}_{\ \ b}^{a}-\ ^{[5]}\mathbf{S}%
_{b\ }^{\ a})] \\
&&+\sum\limits_{[I]=1}^{5}z_{[SvI]}\ (^{[I]}\mathbf{S}_{ab}+\ ^{[I]}\mathbf{S%
}_{ba})+z_{[Sv6]}\mathbf{\vartheta }_{c}\wedge \lbrack \mathbf{e}_{a}\rfloor
^{\lbrack v]}\ (\ ^{[2]}\mathbf{S}_{\ \ b}^{c}-\ ^{[2]}\mathbf{S}_{b\ }^{\
c})] \\
&&+\sum\limits_{[I]=7}^{9}z_{[SvI]}\mathbf{\vartheta }_{a}\wedge \lbrack
\mathbf{e}_{c}\rfloor ^{\lbrack v]}\ (\ ^{[I-4]}\mathbf{S}_{\ \ b}^{c}-\
^{[I-4]}\mathbf{S}_{b\ }^{\ c})]\}.
\end{eqnarray*}%
Let us explain the denotations used in (\ref{actgfa}): The signature is
adapted in the form $\left( -+++\right) $ $\ $and there are considered two
Hodge duals, $\ast ^{\lbrack h]}$for h--subspace and $\ast ^{\lbrack v]}$for
v--subspace, and respectively two cosmological constants, $\lambda _{\lbrack
h]}$ and $\lambda _{\lbrack v]}.$ The strong gravity coupling constants $%
\rho _{\lbrack Rh]},\rho _{\lbrack Rv]},\rho _{\lbrack Ph]},....,$ the
constants $a_{0[Rh]},a_{0[Rv]},a_{0[Ph]},...,a_{[hA]},a_{[vA]},...$ $%
c_{[hI]},c_{[vI]},...$ are dimensionless and provided with labels $%
[R],[P],[h],[v],$ emphasizing $\ \ $that the constants are related, for
instance, to respective invariants of curvature, torsion, nonmetricity and
their h- and v--decompositions.

The action (\ref{actgfa}) describes all possible models of Einstein,
Einstein--Cartan and all type of Finsler--Lagrange--Cartan--Hamilton
gravities which can be modelled on metric affine spaces provided with
N--connection structure (i. e. with generic off--diagonal metrics) and
derived from quadratic MAG--type Lagrangians.

We can reduce the number of constants in $\mathcal{L}_{GFA}\rightarrow
\mathcal{L}_{GFA}^{\prime }$ if we select the limit resulting in the usual
quadratic MAG--Lagrangian \cite{02mag} for trivial N--connection structure. In
this case, all constants for h-- and v-- decompositions coincide with those
from MAG without N--connection structure, for instance,%
\begin{equation*}
a_{0}=a_{0[Rh]}=a_{0[Rv]}=a_{0[Ph]}=...,\
a_{[A]}=a_{[hA]}=a_{[vA]}=...,...,c_{[I]}=c_{[hI]}=c_{[vI]},...
\end{equation*}%
The Lagrangian (\ref{actgfa}) can be reduced to a more simple one written in
terms of boldfaced symbols (emphasizing a nontrivial N--connection
structure) provided with Greek indices,

\begin{eqnarray}
\mathcal{L}_{GFA}^{\prime } &=&\frac{1}{2\kappa }[-a_{0}\mathbf{R}^{\alpha
\beta }\wedge \eta _{\alpha \beta }-2\lambda \eta +\mathbf{T}^{i}\wedge \ast
\left( \sum\limits_{\lbrack A]=1}^{3}a_{[A]}\ ^{[A]}\mathbf{T}_{i}\right)
\notag \\
&&+2\left( \sum\limits_{[I]=2}^{4}c_{[I]}\ ^{[I]}\mathbf{Q}_{\alpha \beta
}\right) \wedge \mathbf{\vartheta }^{\alpha }\wedge \ast \ \mathbf{T}^{\beta
}+\mathbf{Q}_{\alpha \beta }\wedge \left( \sum\limits_{\lbrack
I]=1}^{4}b_{[I]}\ ^{[I]}\mathbf{Q}^{\alpha \beta }\right)  \label{actfag1} \\
&&+\mathbf{Q}_{\alpha \beta }\wedge \left( \sum\limits_{\lbrack
I]=1}^{4}b_{[I]}\ ^{[I]}\mathbf{Q}^{\alpha \beta }\right) +b_{[5]}\left(
^{[3]}\mathbf{Q}_{\alpha \beta }\wedge \mathbf{\vartheta }^{\alpha }\right)
\wedge \ast \left( ^{\lbrack 4]}\mathbf{Q}^{\gamma \beta }\wedge \mathbf{%
\vartheta }_{\gamma }\right) ]  \notag \\
&&-\frac{1}{2\rho }\mathbf{R}^{\alpha \beta }\wedge \ast \lbrack
\sum\limits_{\lbrack I]=1}^{6}w_{[I]}\ \mathbf{\ }^{[I]}\mathbf{W}_{\alpha
\beta }+w_{[7]}\mathbf{\vartheta }_{\alpha }\wedge \left( \mathbf{e}_{\gamma
}\rfloor \ \mathbf{\ }^{[5]}\mathbf{W}_{\ \beta }^{\gamma }\right) \   \notag
\\
&&+\sum\limits_{[I]=1}^{5}z_{[I]}\ ^{[I]}\mathbf{Y}_{\alpha \beta }+z_{[6]}%
\mathbf{\vartheta }_{\gamma }\wedge \left( \mathbf{e}_{\alpha }\rfloor \
^{[2]}\mathbf{Y}_{\ \beta }^{\gamma }\right) +\sum\limits_{[I]=7}^{9}z_{[I]}%
\mathbf{\vartheta }_{\alpha }\wedge (\mathbf{e}_{\gamma }\rfloor \ \ ^{[I-4]}%
\mathbf{Y}_{\ \beta }^{\gamma })].  \notag
\end{eqnarray}%
where $\mathbf{\ }^{[I]}\mathbf{W}_{\alpha \beta }=$\ $\ ^{[I]}\mathbf{R}%
_{\alpha \beta }-\ ^{[I]}\mathbf{R}_{\beta \alpha }$ and $\ ^{[I]}\mathbf{Y}%
_{\alpha \beta }=$\ $^{[I]}\mathbf{R}_{\alpha \beta }+\ ^{[I]}\mathbf{R}%
_{\beta \alpha }.$ This action is just for the MAG quadratic theory but with
$\mathbf{e}_{\alpha }$ and $\mathbf{\vartheta }^{\beta }$ being adapted to
the N--connection structure as in (\ref{2dder}) and (\ref{2ddif}) with a
corresponding splitting of geometrical objects.

The field equations of a metric--affine space provided with N--connection
structure, $$\mathbf{V}^{n+m}=\left[ N_{i}^{a},\mathbf{g}_{\alpha \beta
}=\left( g_{ij},h_{ab}\right) ,\mathbf{\Gamma }_{\alpha \beta }^{\gamma
}=\left( L_{jk}^{i},L_{bk}^{a},C_{jc}^{i},C_{bc}^{a}\right) \right] ,$$ can
be obtained by the Noether procedure in its turn being N--adapted to (co)
frames $\mathbf{e}_{\alpha }$ and $\mathbf{\vartheta }^{\beta }.$ At the
first step, we parametrize the generalized Finsler--affine Lagrangian and
matter Lagrangian respectively as
\begin{equation*}
\mathcal{L}_{GFA}^{\prime }=\mathcal{L}_{[fa]}\left( N_{i}^{a},\mathbf{g}%
_{\alpha \beta },\mathbf{\vartheta }^{\gamma },\mathbf{Q}_{\alpha \beta },%
\mathbf{T}^{\alpha },\ \mathbf{R}_{\ \beta }^{\alpha }\right)
\end{equation*}%
and
\begin{equation*}
\mathcal{L}_{mat}=\mathcal{L}_{[m]}\left( N_{i}^{a},\mathbf{g}_{\alpha \beta
},\mathbf{\vartheta }^{\gamma },\mathbf{\Psi ,D\Psi }\right) ,
\end{equation*}%
where $\mathbf{T}^{\alpha }$ and$\ \mathbf{R}_{\ \beta }^{\alpha }$ are the
curvature of arbitrary d--connection $\mathbf{D}$ and $\mathbf{\Psi }$
represents the matter fields as a $p$--form. The action $\mathcal{S}$ on $%
\mathbf{V}^{n+m}$ is written
\begin{equation}
\mathcal{S}=\int \delta ^{n+m}u\sqrt{|\mathbf{g}_{\alpha \beta }|}\left[
\mathcal{L}_{[fa]}+\mathcal{L}_{[m]}\right]  \label{agmafgm}
\end{equation}%
which results in the matter and gravitational (generalized Finsler--affine
type) field equations.

\begin{theorem}
\ The Yang--Mills type field equations of the generalized Finsler--affine
gravity with matter derived by a variational procedure adapted to the
N--connection structure are defined by the system
\begin{eqnarray}
\mathbf{D}\left( \frac{\partial \mathcal{L}_{[m]}}{\partial \left( \mathbf{%
D\Psi }\right) }\right) -\left( -1\right) ^{p}\frac{\partial \mathcal{L}%
_{[m]}}{\partial \mathbf{\Psi }} &=&0,  \label{fefag} \\
\mathbf{D}\left( \frac{\partial \mathcal{L}_{[fa]}}{\partial \mathbf{Q}%
_{\alpha \beta }}\right) +2\frac{\partial \mathcal{L}_{[fa]}}{\partial
\mathbf{g}_{\alpha \beta }} &=&-\mathbf{\sigma }^{\alpha \beta },  \notag \\
\mathbf{D}\left( \frac{\partial \mathcal{L}_{[fa]}}{\partial \mathbf{T}%
^{\alpha }}\right) +2\frac{\partial \mathcal{L}_{[fa]}}{\partial \mathbf{%
\vartheta }^{\alpha }} &=&-\mathbf{\Sigma }_{\alpha },  \notag \\
\mathbf{D}\left( \frac{\partial \mathcal{L}_{[fa]}}{\partial \mathbf{R}_{\
\beta }^{\alpha }}\right) +\mathbf{\vartheta }^{\beta }\wedge \frac{\partial
\mathcal{L}_{[fa]}}{\partial \mathbf{T}^{\alpha }} &=&-\mathbf{\Delta }%
_{\alpha }^{\ \beta },  \notag
\end{eqnarray}%
where the material currents are defined
\begin{equation*}
\mathbf{\sigma }^{\alpha \beta }\doteqdot 2\frac{\mathbf{\delta }\mathcal{L}%
_{[m]}}{\mathbf{\delta g}_{\alpha \beta }},\ \mathbf{\Sigma }_{\alpha
}\doteqdot \frac{\mathbf{\delta }\mathcal{L}_{[m]}}{\mathbf{\delta \vartheta
}^{\alpha }},\ \mathbf{\Delta }_{\alpha }^{\ \beta }=\frac{\mathbf{\delta }%
\mathcal{L}_{[m]}}{\mathbf{\delta \Gamma }_{\ \beta }^{\alpha }}
\end{equation*}%
for variations ''boldfaced'' $\mathbf{\delta }\mathcal{L}_{[m]}/\mathbf{%
\delta }$\ computed with respect to N--adapted (co) frames.
\end{theorem}

The proof of this theorem consists from N--adapted variational calculus. The
equations (\ref{fefag}) transforms correspondingly into ''MATTER, ZEROTH,
FIRST, SECOND'' equations of MAG \cite{02mag} for trivial N--connection
structures.

\begin{corollary}
The system (\ref{fefag}) has respectively the h-- and v--irreducible
components
\begin{equation*}
D^{[h]}\left( \frac{\partial \mathcal{L}_{[m]}}{\partial \left( D^{[h]}%
\mathbf{\Psi }\right) }\right) +D^{[v]}\left( \frac{\partial \mathcal{L}%
_{[m]}}{\partial \left( D^{[v]}\mathbf{\Psi }\right) }\right) -\left(
-1\right) ^{p}\frac{\partial \mathcal{L}_{[m]}}{\partial \mathbf{\Psi }}=0,
\end{equation*}%
\begin{eqnarray}
D^{[h]}\left( \frac{\partial \mathcal{L}_{[fa]}}{\partial Q_{ij}}\right)
+D^{[v]}\left( \frac{\partial \mathcal{L}_{[fa]}}{\partial Q_{ij}}\right) +2%
\frac{\partial \mathcal{L}_{[fa]}}{\partial g_{ij}} &=&-\sigma ^{ij},
\label{fefagd} \\
D^{[h]}\left( \frac{\partial \mathcal{L}_{[fa]}}{\partial Q_{ab}}\right)
+D^{[v]}\left( \frac{\partial \mathcal{L}_{[fa]}}{\partial Q_{ab}}\right) +2%
\frac{\partial \mathcal{L}_{[fa]}}{\partial g_{ab}} &=&-\sigma ^{ab},  \notag
\end{eqnarray}%
\begin{eqnarray*}
D^{[h]}\left( \frac{\partial \mathcal{L}_{[fa]}}{\partial T^{i}}\right)
+D^{[v]}\left( \frac{\partial \mathcal{L}_{[fa]}}{\partial T^{i}}\right) +2%
\frac{\partial \mathcal{L}_{[fa]}}{\partial \vartheta ^{i}} &=&-\Sigma _{i},
\\
D^{[h]}\left( \frac{\partial \mathcal{L}_{[fa]}}{\partial T^{a}}\right)
+D^{[v]}\left( \frac{\partial \mathcal{L}_{[fa]}}{\partial T^{a}}\right) +2%
\frac{\partial \mathcal{L}_{[fa]}}{\partial \vartheta ^{a}} &=&-\Sigma _{a},
\end{eqnarray*}%
\begin{eqnarray*}
D^{[h]}\left( \frac{\partial \mathcal{L}_{[fa]}}{\partial R_{\ j}^{i}}%
\right) +D^{[v]}\left( \frac{\partial \mathcal{L}_{[fa]}}{\partial R_{\
j}^{i}}\right) +\vartheta ^{j}\wedge \frac{\partial \mathcal{L}_{[fa]}}{%
\partial T^{i}} &=&-\mathbf{\Delta }_{i}^{\ j}, \\
D^{[h]}\left( \frac{\partial \mathcal{L}_{[fa]}}{\partial R_{\ b}^{a}}%
\right) +D^{[v]}\left( \frac{\partial \mathcal{L}_{[fa]}}{\partial R_{\
b}^{a}}\right) +\vartheta ^{b}\wedge \frac{\partial \mathcal{L}_{[fa]}}{%
\partial T^{a}} &=&-\mathbf{\Delta }_{a}^{\ b},
\end{eqnarray*}%
where \
\begin{eqnarray*}
\mathbf{\sigma }^{\alpha \beta } &=&\left( \sigma ^{ij},\sigma ^{ab}\right) %
\mbox{ for }\ \sigma ^{ij}\doteqdot 2\frac{\mathbf{\delta }\mathcal{L}_{[m]}%
}{\mathbf{\delta }g_{ij}},\ \sigma ^{ab}\doteqdot 2\frac{\mathbf{\delta }%
\mathcal{L}_{[m]}}{\mathbf{\delta }h_{ab}},\  \\
\mathbf{\Sigma }_{\alpha } &=&\left( \Sigma _{i},\Sigma _{a}\right)
\mbox{
for }\ \Sigma _{i}\doteqdot \frac{\mathbf{\delta }\mathcal{L}_{[m]}}{\mathbf{%
\delta }\vartheta ^{i}},\ \Sigma _{a}\doteqdot \frac{\mathbf{\delta }%
\mathcal{L}_{[m]}}{\mathbf{\delta }\vartheta ^{a}}, \\
\mathbf{\Delta }_{\alpha }^{\ \beta } &=&\left( \Delta _{i}^{\ j},\Delta
_{a}^{\ b}\right) \mbox{ for }\ \Delta _{i}^{\ j}=\frac{\mathbf{\delta }%
\mathcal{L}_{[m]}}{\mathbf{\delta }\Gamma _{\ j}^{i\ \ }},\ \Delta _{a}^{\
b}=\frac{\mathbf{\delta }\mathcal{L}_{[m]}}{\mathbf{\delta }\Gamma _{\
b}^{a\ \ }}\ .
\end{eqnarray*}
\end{corollary}

It should be noted that the complete h-- v--decomposition of the system (\ref%
{fefagd}) can be obtained if we represent the d--connection and curvature
forms as%
\begin{equation*}
\Gamma _{\ j}^{i\ \ }=L_{\ jk}^{i\ \ }dx^{j}+C_{\ ja}^{i\ \ }\delta y^{a}%
\mbox{ and }\Gamma _{\ b}^{a\ \ }=L_{\ bk}^{a\ \ }dx^{k}+C_{\ bc}^{a\ \
}\delta y^{c},
\end{equation*}%
see the d--connection components (\ref{2dcon1}) and%
\begin{eqnarray*}
2R_{\ j}^{i} &=&R_{\ jkl}^{i}dx^{k}\wedge dx^{l}+P_{\ jka}^{i}dx^{k}\wedge
\delta y^{a}+S_{\ jba}^{i}\delta y^{b}\wedge \delta y^{a}, \\
2R_{\ f}^{e} &=&R_{\ fkl}^{e}dx^{k}\wedge dx^{l}+P_{\ fka}^{e}dx^{k}\wedge
\delta y^{a}+S_{\ fba}^{e}\delta y^{a}\wedge \delta y^{a},
\end{eqnarray*}%
see the d--curvature components (\ref{2dcurv}).

\begin{remark}
\label{rfconf}For instance, a Finsler configuration can be modelled on a
metric affine space provided with N--connection structure, $\mathbf{V}^{n+m}=%
\left[ N_{i}^{a},\mathbf{g}_{\alpha \beta }=\left( g_{ij},h_{ab}\right) ,\
^{[F]}\widehat{\mathbf{\Gamma }}_{\alpha \beta }^{\gamma }\right] ,$ if $%
n=m, $ the ansatz for N--connection is of Cartan--Finsler type
\begin{equation*}
N_{j}^{a}\rightarrow \ ^{[F]}N_{j}^{i}=\frac{1}{8}\frac{\partial }{\partial
y^{j}}\left[ y^{l}y^{k}g_{[F]}^{ih}\left( \frac{\partial g_{hk}^{[F]}}{%
\partial x^{l}}+\frac{\partial g_{lh}^{[F]}}{\partial x^{k}}-\frac{\partial
g_{lk}^{[F]}}{\partial x^{h}}\right) \right] ,
\end{equation*}
the d--metric $\mathbf{g}_{\alpha \beta }=\mathbf{g}_{\alpha \beta }^{[F]}$
is defined by (\ref{2block2}) with
\begin{equation*}
g_{ij}^{[F]}=g_{ij}=h_{ij}=\frac{1}{2}\partial ^{2}F/\partial y^{i}\partial
y^{j}
\end{equation*}
and $^{[F]}\widehat{\mathbf{\Gamma }}_{\alpha \beta }^{\gamma }$ is the
Finsler canonical d--connection computed as (\ref{2candcon}). The data should
define an exact solution of the system of field equation (\ref{fefagd})
(equivalently of (\ref{fefag})).
\end{remark}

Similar Remarks hold true for all types of generalized Finsler--affine
spaces considered in Tables 1--11 from Ref. \cite{02vp1}. We shall analyze the
possibility of modelling various type of locally anisotropic geometries by
the Einstein--Proca systems and in string gravity in next subsection.

\subsection{Effective Einstein--Proca systems and N--connections}

Any affine connection can always be decomposed into (pseudo) Riemannian, $%
\Gamma _{\bigtriangledown \ \beta }^{\alpha },$ and post--Riemannian, $Z_{\
\ \beta }^{\alpha },$ parts as $\Gamma _{\ \beta }^{\alpha }=\Gamma
_{\bigtriangledown \ \beta }^{\alpha }+Z_{\ \ \beta }^{\alpha },$ see
formulas (\ref{2acc}) and (\ref{2dista}) (or (\ref{2accn}) and (\ref{2distan})
if any N--connection structure is prescribed). This mean that it is possible
to split \ all quantities of a metric--affine theory into (pseudo)\
Riemannian and post--Riemannian pieces, for instance,
\begin{equation}
R_{\ \beta }^{\alpha }=R_{\bigtriangledown \ \beta }^{\alpha
}+\bigtriangledown Z_{\ \ \beta }^{\alpha }+Z_{\ \ \gamma }^{\alpha }\wedge
Z_{\ \ \beta }^{\gamma }.  \label{dist1}
\end{equation}%
Under certain assumptions one holds the Obukhov's equivalence theorem
according to which the field vacuum metric--affine gravity equations are
equivalent to Einstein's equations with an energy--momentum tensor
determined by a Proca field \cite{02oveh,02obet2}. We can generalize the
constructions and reformulate the equivalence theorem for generalized
Finsler--affine spaces and effective spaces provided with N--connection
structure.

\begin{theorem}
\label{teq}The system of effective field equations of MAG on spaces provided
with N--connection structure (\ref{fefag}) (equivalently, (\ref{fefagd})) \
for certain ansatz for torsion and nonmetricity fields (see (\ref{torsdec})
and (\ref{nmdc}))
\begin{eqnarray}
^{(1)}\mathbf{T}^{\alpha } &=&^{(2)}\mathbf{T}^{\alpha }=0,\ \ ^{(1)}\mathbf{%
Q}_{\alpha \beta }=\ ^{(2)}\mathbf{Q}_{\alpha \beta }=0,  \label{triplemag}
\\
\mathbf{Q} &=&k_{0}\mathbf{\phi ,\ \Lambda =}k_{1}\mathbf{\phi ,\ T=}k_{2}%
\mathbf{\phi ,}  \notag
\end{eqnarray}%
where $k_{0},k_{1},k_{2}=const$ and the Proca 1--form is $\mathbf{\phi =}%
\phi _{\alpha }\mathbf{\vartheta }^{\alpha }=\phi _{i}dx^{i}+\phi _{a}\delta
y^{a},$ reduces to the Einstein--Proca system of equations for the canonical
d--connection $\widehat{\mathbf{\Gamma }}_{\ \alpha \beta }^{\gamma }$ (\ref%
{2candcon}) and massive d--field $\mathbf{\phi }_{\alpha },$%
\begin{eqnarray}
\frac{a_{0}}{2}\mathbf{\eta }_{\alpha \beta \gamma }\wedge \widehat{\mathbf{R%
}}^{\beta \gamma } &=&k\ \mathbf{\Sigma }_{\alpha },  \notag \\
\delta \left( \ast \mathbf{H}\right) +\mu ^{2}\mathbf{\phi }\mathbf{=0,} &&
\label{efmageq}
\end{eqnarray}%
where \ $\mathbf{H\doteqdot }\delta \mathbf{\phi ,}$ the mass$\mathbf{\ }\mu
=const$ and the energy--momentum is given by
\begin{equation*}
\mathbf{\Sigma }_{\alpha }=\mathbf{\Sigma }_{\alpha }^{[\phi ]}+\mathbf{%
\Sigma }_{\alpha }^{[\mathbf{m}]},
\end{equation*}%
\begin{equation*}
\mathbf{\Sigma }_{\alpha }^{[\phi ]}\mathbf{\doteqdot }\frac{z_{4}k_{0}^{2}}{%
2\rho }\{\left( \mathbf{e}_{\alpha }\rfloor \ \mathbf{H}\right) \wedge \ast
\mathbf{H-}\left( \mathbf{e}_{\alpha }\rfloor \ast \mathbf{H}\right) \wedge
\mathbf{H+}\mu ^{2}[\left( \mathbf{e}_{\alpha }\rfloor \ \mathbf{\phi }%
\right) \wedge \ast \mathbf{\phi -}\left( \mathbf{e}_{\alpha }\rfloor \ast
\mathbf{\phi }\right) \wedge \mathbf{\phi }]\}
\end{equation*}%
is the energy--momentum current of the Proca d--field and $\mathbf{\Sigma }%
_{\alpha }^{[\mathbf{\mu }]}$ is the energy--momentum current of the
additional matter d--fields satisfying the corresponding Euler--Largange
equations.
\end{theorem}

The proof of the Theorem is just the reformulation with respect to
N--adapted (co) frames (\ref{2dder}) and (\ref{2ddif}) of similar
considerations in Refs. \cite{02oveh,02obet2}. The constants $k_{0},k_{1}....$
are taken in terms of the gravitational coupling constants like in \cite{02hm}
as to have connection to the usual MAG and Einstein theory for trivial
N--connection structures and for the dimension $m\rightarrow 0.$ We use the
triplet ansatz sector (\ref{triplemag}) \ of MAG theories \cite{02oveh,02obet2}.
It is a remarkable fact that the equivalence Theorem \ref{teq} holds also in
presence of arbitrary N--connections i. e. for all type of anholonomic
generalizations of the Einstein, Einstein--Cartan and Finsler--Lagrange and
Cartan--Hamilton geometries by introducing canonical d--connections (we can
also consider Berwald type d--connections).

\begin{corollary}
\ \label{corfag}In abstract index form, the effective field equations for
the generalized Finsler--affine gravity following from (\ref{efmageq})\ are
written%
\begin{eqnarray}
\widehat{\mathbf{R}}_{\alpha \beta }-\frac{1}{2}\mathbf{g}_{\alpha \beta }%
\overleftarrow{\mathbf{\hat{R}}} &=&\tilde{\kappa}\left( \mathbf{\Sigma }%
_{\alpha \beta }^{[\phi ]}+\mathbf{\Sigma }_{\alpha \beta }^{[\mathbf{m}%
]}\right) ,  \label{efeinst} \\
\widehat{\mathbf{D}}_{\nu }\mathbf{H}^{\nu \mu } &=&\mu ^{2}\mathbf{\phi }%
^{\mu },  \notag
\end{eqnarray}%
with $\mathbf{H}_{\nu \mu }\doteqdot \widehat{\mathbf{D}}_{\nu }\mathbf{\phi
}_{\mu }-\widehat{\mathbf{D}}_{\mu }\mathbf{\phi }_{\nu }+w_{\mu \nu
}^{\gamma }\mathbf{\phi }_{\gamma }$ being the field strengths of the
Abelian Proca field $\mathbf{\phi }^{\mu },\tilde{\kappa}=const,$ and
\begin{equation}
\mathbf{\Sigma }_{\alpha \beta }^{[\phi ]}=\mathbf{H}_{\alpha }^{\ \mu }%
\mathbf{H}_{\beta \mu }-\frac{1}{4}\mathbf{g}_{\alpha \beta }\mathbf{H}_{\mu
\nu }^{\ }\mathbf{H}^{\mu \nu }+\mu ^{2}\mathbf{\phi }_{\alpha }\mathbf{\phi
}_{\beta }-\frac{\mu ^{2}}{2}\mathbf{g}_{\alpha \beta }\mathbf{\phi }_{\mu }%
\mathbf{\phi }^{\mu }.  \label{sourcef}
\end{equation}
\end{corollary}

The Ricci d--tensor $\widehat{\mathbf{R}}_{\alpha \beta }$ and scalar $%
\overleftarrow{\mathbf{\hat{R}}}$ from (\ref{efeinst}) can be decomposed in
irreversible h-- and v--invariant components like (\ref{2dricci}) and (\ref%
{2dscal}),%
\begin{eqnarray}
\widehat{R}_{ij}-\frac{1}{2}g_{ij}\left( \widehat{R}+\widehat{S}\right) &=&%
\tilde{\kappa}\left( \mathbf{\Sigma }_{ij}^{[\phi ]}+\mathbf{\Sigma }_{ij}^{[%
\mathbf{m}]}\right) ,  \label{ep1} \\
\widehat{S}_{ab}-\frac{1}{2}h_{ab}\left( \widehat{R}+\widehat{S}\right) &=&%
\tilde{\kappa}\left( \mathbf{\Sigma }_{ab}^{[\phi ]}+\mathbf{\Sigma }_{ab}^{[%
\mathbf{m}]}\right) ,  \label{ep2} \\
^{1}P_{ai} &=&\tilde{\kappa}\left( \mathbf{\Sigma }_{ai}^{[\phi ]}+\mathbf{%
\Sigma }_{ai}^{[\mathbf{m}]}\right) ,  \label{ep3} \\
\ -^{2}P_{ia} &=&\tilde{\kappa}\left( \mathbf{\Sigma }_{ia}^{[\phi ]}+%
\mathbf{\Sigma }_{ia}^{[\mathbf{m}]}\right) .  \label{ep4}
\end{eqnarray}%
The constants are those from \cite{02oveh} being related to the constants from
(\ref{actfag1}),%
\begin{equation*}
\mu ^{2}=\frac{1}{z_{k}\kappa }\left( -4\beta _{4}+\frac{k_{1}}{2k_{0}}\beta
_{5}+\frac{k_{2}}{k_{0}}\gamma _{4}\right) ,
\end{equation*}%
where
\begin{eqnarray*}
k_{0} &=&4\alpha _{2}\beta _{3}-3(\gamma _{3})^{2}\neq 0,\ k_{1}=9\left(
\frac{1}{2}\alpha _{5}\beta _{5}-\gamma _{3}\gamma _{4}\right) ,\
k_{2}=3\left( 4\beta _{3}\gamma _{4}-\frac{3}{2}\beta _{5}\gamma _{3}\right)
, \\
\alpha _{2} &=&a_{2}-2a_{0},\ \beta _{3}=b_{3}+\frac{a_{0}}{8},\ \beta
_{4}=b_{4}-\frac{3a_{0}}{8},\ \gamma _{3}=c_{3}+a_{0},\ \gamma
_{4}=c_{4}+a_{0}.
\end{eqnarray*}%
If
\begin{equation}
\beta _{4}\rightarrow \frac{1}{4k_{0}}\left( \frac{1}{2}\beta
_{5}k_{1}+k_{2}\gamma _{4}\right) ,  \label{procvan}
\end{equation}%
the mass of Proca field $\mu ^{2}\rightarrow 0.$ The system becomes like the
Einstein--Maxwell one with the source (\ref{sourcef}) defined by the
antisymmetric field $\mathbf{H}_{\mu \nu }^{\ }$ in its turn being
determined by a solution of $\widehat{\mathbf{D}}_{\nu }\widehat{\mathbf{D}}%
^{\nu }\mathbf{\phi }_{\alpha }=0$ (a wave like equation in a curved space
provided with N--connection). Even in this case the nonmetricity and torsion
can be nontrivial, for instance, oscillating (see (\ref{triplemag})).

We note that according the Remark \ref{rfconf}, the system (\ref{efeinst})
defines, for instance, a Finsler configuration if the d--metric $\mathbf{g}%
_{\alpha \beta },$ the d--connection $\widehat{\mathbf{D}}_{\nu }$ and the
N--connection are of Finsler type (or contains as imbedding such objects).

\subsection{Einstein--Cartan gravity and N--connections}

The Einstein--Cartan gravity contains gravitational configurations with
nontrivial N--con\-nec\-ti\-on structure. The simplest model with local
anisotropy is to write on a space $\mathbf{V}^{n+m}$ the Einstein equations
for the canonical d--connection $\widehat{\mathbf{\Gamma }}_{\ \alpha \beta
}^{\gamma }$ (\ref{2candcon}) introduced in the Einstein d--tensor (\ref%
{2deinst}),%
\begin{equation*}
\widehat{\mathbf{R}}_{\alpha \beta }-\frac{1}{2}\mathbf{g}_{\alpha \beta }%
\overleftarrow{\mathbf{\hat{R}}}=\kappa \mathbf{\Sigma }_{\alpha \beta }^{[%
\mathbf{m}]},
\end{equation*}%
or in terms of differential forms,
\begin{equation}
\mathbf{\eta }_{\alpha \beta \gamma }\wedge \widehat{\mathbf{R}}^{\beta
\gamma }=\kappa \mathbf{\Sigma }_{\alpha }^{[\mathbf{m}]}  \label{1einst1}
\end{equation}%
which is a particular case of equations (\ref{efmageq}). The model contains
nontrivial d--torsions, $\widehat{\mathbf{T}}_{\ \alpha \beta }^{\gamma },$
computed by introducing the components of (\ref{2candcon}) into formulas (\ref%
{2dtorsb}). We can consider that specific distributions of ''spin
dust/fluid'' of Weyssenhoff and Raabe type, or any generalizations, adapted
to the N--connection structure, can constitute the source of certain
algebraic equations for torsion (see details in Refs. \cite{02rcg}) or even to
consider generalizations for dynamical equations for torsion like in gauge
gravity theories \cite{02ggrav}. A more special case is defined by the
theories when the d--torsions $\widehat{\mathbf{T}}_{\ \alpha \beta
}^{\gamma }$ are induced by specific frame effects of N--connection
structures. Such models contain all possible distorsions to generalized
Finsler--Lagrange--Cartan spacetimes of the Einstein gravity and emphasize
the conditions when such generalizations to locally anisotropic gravity
preserve the local Lorentz invariance or even model Finsler like
configurations in the framework of general relativity.

Let us express the 1--form of the canonical d--connection $\widehat{\mathbf{%
\Gamma }}_{\ \alpha }^{\gamma }$ as the deformation of the Levi--Civita
connection $\mathbf{\Gamma }_{\bigtriangledown \ \alpha }^{\gamma },$%
\begin{equation}
\widehat{\mathbf{\Gamma }}_{\ \alpha }^{\gamma }=\mathbf{\Gamma }%
_{\bigtriangledown \ \alpha }^{\gamma }+\widehat{\mathbf{Z}}_{\ \alpha
}^{\gamma }  \label{dist2}
\end{equation}%
where%
\begin{equation}
\widehat{\mathbf{Z}}_{\alpha \beta }=\mathbf{e}_{\beta }\rfloor \widehat{%
\mathbf{T}}_{\alpha }-\mathbf{e}_{\alpha }\rfloor \widehat{\mathbf{T}}%
_{\beta }+\frac{1}{2}\left( \mathbf{e}_{\alpha }\rfloor \mathbf{e}_{\beta
}\rfloor \widehat{\mathbf{T}}_{\gamma }\right) \mathbf{\vartheta }^{\gamma }
\label{aux53}
\end{equation}%
being a particular case of formulas (\ref{2accn}) and (\ref{2distan}) when
nonmetricity vanishes, $\mathbf{Q}_{\alpha \beta }=0.$ This induces a
distorsion of the curvature tensor like (\ref{dist1}) but for d--objects,
expressing (\ref{1einst1}) in the form
\begin{equation}
\mathbf{\eta }_{\alpha \beta \gamma }\wedge \mathbf{R}_{\bigtriangledown
}^{\beta \gamma }+\mathbf{\eta }_{\alpha \beta \gamma }\wedge \mathbf{Z}%
_{\bigtriangledown \ }^{\beta \gamma }=\kappa \mathbf{\Sigma }_{\alpha }^{[%
\mathbf{m}]}  \label{einst1a}
\end{equation}%
where%
\begin{equation*}
\mathbf{Z}_{\bigtriangledown \ \gamma }^{\beta }=\bigtriangledown \mathbf{Z}%
_{~\ \gamma }^{\beta }+\mathbf{Z}_{~\ \alpha }^{\beta }\wedge \mathbf{Z}_{~\
\gamma }^{\alpha }.
\end{equation*}

\begin{theorem}
The Einstein equations (\ref{1einst1}) for the canonical d--connection $%
\widehat{\mathbf{\Gamma }}_{\ \alpha }^{\gamma }$ constructed for a
d--metric field $\mathbf{g}_{\alpha \beta }=[g_{ij},h_{ab}]$ (\ref{2block2})
and N--connection $N_{i}^{a}$ is equivalent to the gravitational field
equations for the Einstein--Cartan theory with torsion $\widehat{\mathbf{T}}%
_{\ \alpha }^{\gamma }$ defined by the N--connection, see formulas (\ref%
{2dtorsb}).
\end{theorem}

\textbf{Proof:} The proof is trivial and follows from decomposition (\ref%
{dist2}).

\begin{remark}
Every type of generalized Finsler--Lagrange geometries is characterized by a
corresponding N-- and d--connection and d--metric structures, see Tables
1--11 in Ref. \cite{02vp1}. For the canonical d--connection such locally
anisotropic geometries can be modelled on Riemann--Cartan manifolds as
solutions of (\ref{1einst1}) for a prescribed type of d--torsions (\ref%
{2dtorsb}).
\end{remark}

\begin{corollary}
\label{corcond1}A generalized Finsler geometry can be modelled in a (pseudo)
Riemann spacetime by a d--metric $\mathbf{g}_{\alpha \beta }=[g_{ij},h_{ab}]$
(\ref{2block2}), equivalently by generic off--diagonal metric (\ref{2ansatz}),
satisfying the Einstein equations for the Levi--Civita connection,
\begin{equation}
\mathbf{\eta }_{\alpha \beta \gamma }\wedge \mathbf{R}_{\bigtriangledown
}^{\beta \gamma }=\kappa \mathbf{\Sigma }_{\alpha }^{[\mathbf{m}]}
\label{einst2}
\end{equation}%
if and only if
\begin{equation}
\mathbf{\eta }_{\alpha \beta \gamma }\wedge \mathbf{Z}_{\bigtriangledown \
}^{\beta \gamma }=0.  \label{1cond1}
\end{equation}
\end{corollary}

The proof follows from equations (\ref{einst1a}). We emphasize that the
conditions (\ref{1cond1}) are imposed for the deformations of the Ricci
tensors computed from distorsions of the Levi--Civita connection to the
canonical d--connection. In general, a solution $\mathbf{g}_{\alpha \beta
}=[g_{ij},h_{ab}]$ of the Einstein equations (\ref{einst2}) can be
characterized alternatively by d--connections and N--connections as follows
from relation (\ref{1lcsyma}). The alternative geometric description contains
nontrivial torsion fields. The simplest such anholonomic configurations can
be defined by the condition of vanishing of N--connection curvature (\ref%
{2ncurv}), $\Omega _{ij}^{a}=0,\,$\ but even in such cases there are
nontrivial anholonomy coefficients, see (\ref{2anhc}), $\mathbf{w}_{~ia}^{b}=-%
\mathbf{w}_{~ai}^{b}=\partial _{a}N_{i}^{b},$ and nonvanishing d--torsions (%
\ref{2dtorsb}),
\begin{equation*}
\widehat{T}_{ja}^{i}=-\widehat{T}_{aj}^{i}=\widehat{C}_{.ja}^{i}\mbox{ and }%
\widehat{T}_{.bi}^{a}=-\widehat{T}_{.ib}^{a}=\widehat{P}_{.bi}^{a}=\frac{%
\partial N_{i}^{a}}{\partial y^{b}}-\widehat{L}_{.bj}^{a},
\end{equation*}%
being induced by off--diagonal terms in the metric (\ref{2ansatz}).

\subsection{String gravity and N--connections}

The subjects concerning generalized Finsler (super) geometry, spinors and
(super) strings are analyzed in details in Refs. \cite{02v2}. Here, we
consider the simplest examples when Finsler like geometries can be modelled
in string gravity and related to certain metric--affine structures.

For instance, in the sigma model for bosonic string (see, \cite{02sgr}), the
background connection is taken to be not the Levi--Civita one, but a certain
deformation by the strength (torsion) tensor
\begin{equation*}
H_{\mu \nu \rho }\doteqdot \delta _{\mu }B_{\nu \rho }+\delta _{\rho }B_{\mu
\nu }+\delta _{\nu }B_{\rho \mu }
\end{equation*}%
of an antisymmetric field $B_{\nu \rho },$ defined as
\begin{equation*}
\mathcal{D}_{\mu }=\bigtriangledown _{\mu }+\frac{1}{2}H_{\mu \nu }^{\quad
\rho }.
\end{equation*}%
We consider the $H$--field defined by using N--elongated operators (\ref%
{2dder}) in order to compute the coefficients with respect to anholonomic
frames.

The condition of the Weyl invariance to hold in two dimensions in the lowest
nontrivial approximation in string constant $\alpha ^{\prime },$ see \cite%
{02v2}, turn out to be
\begin{eqnarray*}
R_{\mu \nu } &=&-\frac{1}{4}H_{\mu }^{\ \nu \rho }H_{\nu \lambda \rho
}+2\bigtriangledown _{\mu }\bigtriangledown _{\nu }\Phi , \\
\bigtriangledown _{\lambda }H_{\ \mu \nu }^{\lambda } &=&2\left(
\bigtriangledown _{\lambda }\Phi \right) H_{\ \mu \nu }^{\lambda }, \\
\left( \bigtriangledown \Phi \right) ^{2} &=&\bigtriangledown _{\lambda
}\bigtriangledown ^{\lambda }\Phi +\frac{1}{4}R+\frac{1}{48}H_{\mu \nu \rho
}H^{\mu \nu \rho }.
\end{eqnarray*}%
where $\Phi $ is the dilaton field. For trivial dilaton configurations, $%
\Phi =0,$ we may write
\begin{eqnarray*}
R_{\mu \nu } &=&-\frac{1}{4}H_{\mu }^{\ \nu \rho }H_{\nu \lambda \rho }, \\
\bigtriangledown _{\lambda }H_{\ \mu \nu }^{\lambda } &=&0.
\end{eqnarray*}%
In Refs. \cite{02v2} we analyzed string gravity models derived from
superstring effective actions, for instance, from the 4D Neveu-Schwarz
action. In this paper we consider, for simplicity, a model with zero dilaton
field but with nontrivial $H$--field related to the d--torsions induced by
the N--connection and canonical d--connection.

A class of Finsler like metrics can be derived from the bosonic string
theory if $\mathbf{H}_{\nu \lambda \rho }$ and $\mathbf{B}_{\nu \rho } $ are
related to the d--torsions components, for instance, with $\widehat{\mathbf{T%
}}_{\ \alpha \beta }^{\gamma }.$ Really, we can take an ansatz
\begin{equation*}
\mathbf{B}_{\nu \rho }=\left[ B_{ij},B_{ia},B_{ab}\right]
\end{equation*}%
and consider that
\begin{equation}
\mathbf{H}_{\nu \lambda \rho }=\widehat{\mathbf{Z}}_{\ \nu \lambda \rho }+%
\widehat{\mathbf{H}}_{\nu \lambda \rho }  \label{aux51a}
\end{equation}%
where $\widehat{\mathbf{Z}}_{\ \nu \lambda \rho }$ is the distorsion of the
Levi--Civita connection induced by $\widehat{\mathbf{T}}_{\ \alpha \beta
}^{\gamma },$ see (\ref{aux53}). In this case the induced by N--connection
torsion structure is related to the antisymmetric $H$--field and
correspondingly to the $B$--field from string theory. The equations
\begin{equation}
\bigtriangledown ^{\nu }\mathbf{H}_{\nu \lambda \rho }=\bigtriangledown
^{\nu }(\widehat{\mathbf{Z}}_{\ \nu \lambda \rho }+\widehat{\mathbf{H}}_{\nu
\lambda \rho })=0  \label{aux51}
\end{equation}%
impose certain dynamical restrictions to the N--connection coefficients $%
N_{i}^{a}$ and d--metric $\mathbf{g}_{\alpha \beta }=[g_{ij},h_{ab}]$ \
contained in $\widehat{\mathbf{T}}_{\ \alpha \beta }^{\gamma }.$ If on the
background space it is prescribed the canonical d--connection $\widehat{%
\mathbf{D}}$, we can state a model with (\ref{aux51}) redefined as
\begin{equation}
\widehat{\mathbf{D}}^{\nu }\mathbf{H}_{\nu \lambda \rho }=\widehat{\mathbf{D}%
}^{\nu }(\widehat{\mathbf{Z}}_{\ \nu \lambda \rho }+\widehat{\mathbf{H}}%
_{\nu \lambda \rho })=0,  \label{aux51b}
\end{equation}%
where $\widehat{\mathbf{H}}_{\nu \lambda \rho }$ are computed for stated
values of $\widehat{\mathbf{T}}_{\ \alpha \beta }^{\gamma }.$ For trivial
N--connections when $\widehat{\mathbf{Z}}_{\ \nu \lambda \rho }\rightarrow 0$
and $\widehat{\mathbf{D}}^{\nu }\rightarrow \bigtriangledown ^{\nu },$ the $%
\widehat{\mathbf{H}}_{\nu \lambda \rho }$ transforms into usual $H$--fields.

\begin{proposition}
The dynamics of generalized Finsler--affine string gravity is defined by the
system of field equations%
\begin{eqnarray}
\widehat{\mathbf{R}}_{\alpha \beta }-\frac{1}{2}\mathbf{g}_{\alpha \beta }%
\overleftarrow{\mathbf{\hat{R}}} &=&\tilde{\kappa}\left( \mathbf{\Sigma }%
_{\alpha \beta }^{[\phi ]}+\mathbf{\Sigma }_{\alpha \beta }^{[\mathbf{m}]}+%
\mathbf{\Sigma }_{\alpha \beta }^{[\mathbf{T}]}\right) ,  \label{fagfe} \\
\widehat{\mathbf{D}}_{\nu }\mathbf{H}^{\nu \mu } &=&\mu ^{2}\mathbf{\phi }%
^{\mu },  \notag \\
\widehat{\mathbf{D}}^{\nu }(\widehat{\mathbf{Z}}_{\ \nu \lambda \rho }+%
\widehat{\mathbf{H}}_{\nu \lambda \rho }) &=&0  \notag
\end{eqnarray}%
with $\mathbf{H}_{\nu \mu }\doteqdot \widehat{\mathbf{D}}_{\nu }\mathbf{\phi
}_{\mu }-\widehat{\mathbf{D}}_{\mu }\mathbf{\phi }_{\nu }+w_{\mu \nu
}^{\gamma }\mathbf{\phi }_{\gamma }$ being the field strengths of the
Abelian Proca field $\mathbf{\phi }^{\mu },\tilde{\kappa}=const,$
\begin{equation*}
\mathbf{\Sigma }_{\alpha \beta }^{[\phi ]}=\mathbf{H}_{\alpha }^{\ \mu }%
\mathbf{H}_{\beta \mu }-\frac{1}{4}\mathbf{g}_{\alpha \beta }\mathbf{H}_{\mu
\nu }^{\ }\mathbf{H}^{\mu \nu }+\mu ^{2}\mathbf{\phi }_{\alpha }\mathbf{\phi
}_{\beta }-\frac{\mu ^{2}}{2}\mathbf{g}_{\alpha \beta }\mathbf{\phi }_{\mu }%
\mathbf{\phi }^{\mu },
\end{equation*}%
and%
\begin{equation*}
\mathbf{\Sigma }_{\alpha \beta }^{[\mathbf{T}]}=\mathbf{\Sigma }_{\alpha
\beta }^{[\mathbf{T}]}\left( \widehat{\mathbf{T}},\Phi \right)
\end{equation*}%
contains contributions of $\widehat{\mathbf{T}}$ and $\Phi $ fields.
\end{proposition}

\textbf{Proof:} It follows as an extension of the Corollary \ref{corfag} to
sources induced by string corrections. The system (\ref{fagfe}) should be
completed by the field equations for the matter fields present in $\mathbf{%
\Sigma }_{\alpha \beta }^{[\mathbf{m}]}.$

Finally, we note that the equations (\ref{fagfe}) reduce to equations of
type (\ref{einst1a}) (for Riemann--Cartan configurations with zero
nonmetricity),
\begin{equation*}
\mathbf{\eta }_{\alpha \beta \gamma }\wedge \mathbf{R}_{\bigtriangledown
}^{\beta \gamma }+\mathbf{\eta }_{\alpha \beta \gamma }\wedge \mathbf{Z}%
_{\bigtriangledown \ }^{\beta \gamma }=\kappa \mathbf{\Sigma }_{\alpha }^{[%
\mathbf{T}]},
\end{equation*}%
and to equations of type (\ref{einst2}) and (\ref{1cond1}) (for (pseudo)
Riemannian configurations)
\begin{eqnarray}
\mathbf{\eta }_{\alpha \beta \gamma }\wedge \mathbf{R}_{\bigtriangledown
}^{\beta \gamma } &=&\kappa \mathbf{\Sigma }_{\alpha }^{[\mathbf{T}]},
\label{1cond2} \\
\mathbf{\eta }_{\alpha \beta \gamma }\wedge \mathbf{Z}_{\bigtriangledown \
}^{\beta \gamma } &=&0  \notag
\end{eqnarray}%
with sources defined by torsion (related to N--connection) from string
theory.

\section[The Anholonomic Frame Method]{The Anholonomic Frame Method \newline
in MAG and String Gravity}

In a series of papers, see Refs. \cite{02v1,02v1a,02vnces,02vmethod}, the
anholonomic frame method of constructing exact solutions with generic
off--diagonal metrics (depending on 2-4 variables) in general relativity,
gauge gravity and certain extra dimension generalizations was elaborated. In
this section, we develop the method in MAG and string gravity with
applications to different models of \ five dimensional (in brief, 5D)\
generalized Finsler--affine spaces.

We consider a metric--affine\ space provided with N--connection structure\\ $%
\mathbf{N}=[N_{i}^{4}(u^{\alpha }),$ $N_{i}^{5}(u^{\alpha })]$ where the
local coordinates are labelled $u^{\alpha }=(x^{i},y^{4}=v,y^{5}),$ for $%
i=1,2,3.$ We state the general condition when exact solutions of the field
equations of the generalized Finsler--affine string gravity depending on
holonomic variables $x^{i}$ and on one anholonomic (equivalently,
anisotropic) variable $y^{4}=v$ can be constructed in explicit form. Every
coordinate from a set $u^{\alpha }$ can may be time like, 3D space like, or
extra dimensional. For simplicity, the partial derivatives are denoted $%
a^{\times }=\partial a/\partial x^{1},a^{\bullet }=\partial a/\partial
x^{2},a^{\prime }=\partial a/\partial x^{3},a^{\ast }=\partial a/\partial v.$

The 5D metric
\begin{equation}
\mathbf{g}=\mathbf{g}_{\alpha \beta }\left( x^{i},v\right) du^{\alpha
}\otimes du^{\beta }  \label{metric5}
\end{equation}%
has the metric coefficients $\mathbf{g}_{\alpha \beta }$ parametrized with
respect to the coordinate dual basis by an off--diagonal matrix (ansatz) {\
%%\footnotesize
\begin{equation}
\left[
\begin{array}{ccccc}
g_{1}+w_{1}^{\ 2}h_{4}+n_{1}^{\ 2}h_{5} & w_{1}w_{2}h_{4}+n_{1}n_{2}h_{5} &
w_{1}w_{3}h_{4}+n_{1}n_{3}h_{5} & w_{1}h_{4} & n_{1}h_{5} \\
w_{1}w_{2}h_{4}+n_{1}n_{2}h_{5} & g_{2}+w_{2}^{\ 2}h_{4}+n_{2}^{\ 2}h_{5} &
w_{2}w_{3}h_{4}+n_{2}n_{3}h_{5} & w_{2}h_{4} & n_{2}h_{5} \\
w_{1}w_{3}h_{4}+n_{1}n_{3}h_{5} & w_{2}w_{3}h_{4}+n_{2}n_{3}h_{5} &
g_{3}+w_{3}^{\ 2}h_{4}+n_{3}^{\ 2}h_{5} & w_{3}h_{4} & n_{3}h_{5} \\
w_{1}h_{4} & w_{2}h_{4} & w_{3}h_{4} & h_{4} & 0 \\
n_{1}h_{5} & n_{2}h_{5} & n_{3}h_{5} & 0 & h_{5}%
\end{array}%
\right] ,  \label{ansatz5}
\end{equation}%
} with the coefficients being some necessary smoothly class functions of
type
\begin{eqnarray}
g_{1} &=&\pm 1,g_{2,3}=g_{2,3}(x^{2},x^{3}),h_{4,5}=h_{4,5}(x^{i},v),  \notag
\\
w_{i} &=&w_{i}(x^{i},v),n_{i}=n_{i}(x^{i},v),  \notag
\end{eqnarray}%
where the $N$--coefficients from (\ref{2dder}) and (\ref{2ddif}) are
parametrized $N_{i}^{4}=w_{i}$ and $N_{i}^{5}=n_{i}.$

\begin{theorem}
\label{t5dr}The nontrivial components of the 5D Ricci d--tensors (\ref%
{2dricci}), $\widehat{\mathbf{R}}_{\alpha \beta }=(\widehat{R}_{ij},\widehat{R%
}_{ia},$ $\widehat{R}_{ai},\widehat{S}_{ab}),$ for the d--metric (\ref%
{2block2}) and canonical d--connection $\widehat{\mathbf{\Gamma }}_{\ \alpha
\beta }^{\gamma }$(\ref{2candcon}) both defined by the ansatz (\ref{ansatz5}%
), computed with respect to anholonomic frames (\ref{2dder}) and (\ref{2ddif}%
), consist from h- and v--irreducible components:
\begin{eqnarray}
R_{2}^{2}=R_{3}^{3}=-\frac{1}{2g_{2}g_{3}}[g_{3}^{\bullet \bullet }-\frac{%
g_{2}^{\bullet }g_{3}^{\bullet }}{2g_{2}}-\frac{(g_{3}^{\bullet })^{2}}{%
2g_{3}}+g_{2}^{^{\prime \prime }}-\frac{g_{2}^{^{\prime }}g_{3}^{^{\prime }}%
}{2g_{3}}-\frac{(g_{2}^{^{\prime }})^{2}}{2g_{2}}], &&  \label{1ricci1a} \\
S_{4}^{4}=S_{5}^{5}=-\frac{1}{2h_{4}h_{5}}\left[ h_{5}^{\ast \ast
}-h_{5}^{\ast }\left( \ln \sqrt{|h_{4}h_{5}|}\right) ]^{\ast }\right] , &&
\label{1ricci2a} \\
R_{4i}=-w_{i}\frac{\beta }{2h_{5}}-\frac{\alpha _{i}}{2h_{5}}, &&
\label{1ricci3a} \\
R_{5i}=-\frac{h_{5}}{2h_{4}}\left[ n_{i}^{\ast \ast }+\gamma n_{i}^{\ast }%
\right] , &&  \label{1ricci4a}
\end{eqnarray}%
where
\begin{eqnarray}
\alpha _{i} &=&\partial _{i}{h_{5}^{\ast }}-h_{5}^{\ast }\partial _{i}\ln
\sqrt{|h_{4}h_{5}|},\beta =h_{5}^{\ast \ast }-h_{5}^{\ast }[\ln \sqrt{%
|h_{4}h_{5}|}]^{\ast },\gamma =3h_{5}^{\ast }/2h_{5}-h_{4}^{\ast }/h_{4}
\label{1abc} \\
h_{4}^{\ast } &\neq &0,\text{ }h_{5}^{\ast }\neq 0\
\mbox{ cases
with
vanishing }\text{ }h_{4}^{\ast }\mbox{ and/or }\text{ }h_{5}^{\ast }%
\mbox{
should be analyzed additionally}.  \notag
\end{eqnarray}
\end{theorem}

The proof of Theorem \ref{t5dr} is given in Appendix \ref{appa}.

We can generalize the ansatz (\ref{ansatz5}) by introducing a conformal
factor $\omega (x^{i},v)$ and additional deformations of the metric via
coefficients $\zeta _{\hat{\imath}}(x^{i},v)$ (here, the indices with 'hat'
take values like $\hat{{i}}=1,2,3,5),$ i. e. for metrics of type
\begin{equation}
\mathbf{g}^{[\omega ]}=\omega ^{2}(x^{i},v)\hat{\mathbf{g}}_{\alpha \beta
}\left( x^{i},v\right) du^{\alpha }\otimes du^{\beta },  \label{1cmetric}
\end{equation}%
were the coefficients $\hat{\mathbf{g}}_{\alpha \beta }$ are parametrized by
the ansatz {\scriptsize
\begin{equation}
\left[
\begin{array}{ccccc}
g_{1}+(w_{1}^{\ 2}+\zeta _{1}^{\ 2})h_{4}+n_{1}^{\ 2}h_{5} &
(w_{1}w_{2}+\zeta _{1}\zeta _{2})h_{4}+n_{1}n_{2}h_{5} & (w_{1}w_{3}+\zeta
_{1}\zeta _{3})h_{4}+n_{1}n_{3}h_{5} & (w_{1}+\zeta _{1})h_{4} & n_{1}h_{5}
\\
(w_{1}w_{2}+\zeta _{1}\zeta _{2})h_{4}+n_{1}n_{2}h_{5} & g_{2}+(w_{2}^{\
2}+\zeta _{2}^{\ 2})h_{4}+n_{2}^{\ 2}h_{5} & (w_{2}w_{3}+\zeta _{2}\zeta
_{3})h_{4}+n_{2}n_{3}h_{5} & (w_{2}+\zeta _{2})h_{4} & n_{2}h_{5} \\
(w_{1}w_{3}+\zeta _{1}\zeta _{3})h_{4}+n_{1}n_{3}h_{5} & (w_{2}w_{3}+\zeta
_{2}\zeta _{3})h_{4}+n_{2}n_{3}h_{5} & g_{3}+(w_{3}^{\ 2}+\zeta _{3}^{\
2})h_{4}+n_{3}^{\ 2}h_{5} & (w_{3}+\zeta _{3})h_{4} & n_{3}h_{5} \\
(w_{1}+\zeta _{1})h_{4} & (w_{2}+\zeta _{2})h_{4} & (w_{3}+\zeta _{3})h_{4}
& h_{4} & 0 \\
n_{1}h_{5} & n_{2}h_{5} & n_{3}h_{5} & 0 & h_{5}+\zeta _{5}h_{4}%
\end{array}%
\right] .  \label{1ansatzc}
\end{equation}%
} Such 5D metrics have a second order anisotropy \cite{02v2,02mhss} when the $N$%
--coefficients are paramet\-ri\-zed in the first order anisotropy like $%
N_{i}^{4}=w_{i}$ and $N_{i}^{5}=n_{i}$ (with three anholonomic, $x^{i},$ and
two anholonomic, $y^{4}$ and $y^{5},$ coordinates) and in the second order
anisotropy (on the second 'shell', \ with four holonomic, $(x^{i},y^{5}),$
and one anholonomic,$y^{4},$ coordinates) with $N_{\hat{{i}}}^{5}=\zeta _{%
\hat{{i}}},$ in this work we state, for simplicity, $\zeta _{{5}}=0.$ For
trivial values $\omega =1$ and $\zeta _{\hat{\imath}}=0,$ the metric (\ref%
{1cmetric}) transforms into (\ref{metric5}).

The Theorem \ref{t5dr} can be extended as to include the ansatz (\ref%
{1cmetric}):

\begin{theorem}
\label{t5dra}The nontrivial components of the 5D Ricci d--tensors (\ref%
{2dricci}), $\widehat{\mathbf{R}}_{\alpha \beta }=(\widehat{R}_{ij},\widehat{R%
}_{ia},$ $\widehat{R}_{ai},\widehat{S}_{ab}),$ for the metric (\ref{2block2})
and canonical d--connection $\widehat{\mathbf{\Gamma }}_{\ \alpha \beta
}^{\gamma }$ (\ref{2candcon}) defined by the ansatz (\ref{1ansatzc}), computed
with respect to the anholonomic frames (\ref{2dder}) and (\ref{2ddif}), are
given by the same formulas (\ref{1ricci1a})--(\ref{1ricci4a}) if there are
satisfied the conditions
\begin{equation}
\hat{{\delta }}_{i}h_{4}=0\mbox{\ and\  }\hat{{\delta }}_{i}\omega =0
\label{1conf1}
\end{equation}%
for $\hat{{\delta }}_{i}=\partial _{i}-\left( w_{i}+\zeta _{i}\right)
\partial _{4}+n_{i}\partial _{5}$ when the values $\zeta _{\widetilde{i}%
}=\left( \zeta _{{i}},\zeta _{{5}}=0\right) $ are to be defined as any
solutions of (\ref{1conf1}).
\end{theorem}

The proof of Theorem \ref{t5dra} consists from a straightforward calculation
of the components of the Ricci tensor (\ref{2dricci}) like in Appendix \ref%
{appa}. The simplest way to do this is to compute the deformations by the
conformal factor of the coefficients of the canonical connection (\ref%
{2candcon}) and then to use the calculus for Theorem \ref{t5dr}. \ Such
deformations induce corresponding deformations of the Ricci tensor (\ref%
{2dricci}). \ The condition that we have the same values of the Ricci tensor
for the (\ref{2ansatz}) and (\ref{1ansatzc}) results in equations (\ref{1conf1}%
)  which are compatible, for instance, if for instance, if
\begin{equation}
\omega ^{q_{1}/q_{2}}=h_{4}~(q_{1}\mbox{ and }q_{2}\mbox{ are
integers}),  \label{1confq}
\end{equation}%
and $\zeta _{{i}}$ satisfy the equations \
\begin{equation}
\partial _{i}\omega -(w_{i}+\zeta _{{i}})\omega ^{\ast }=0.  \label{1confeq}
\end{equation}%
\ There are also different possibilities to satisfy the condition (\ref%
{1conf1}). For instance, if $\omega =\omega _{1}$ $\omega _{2},$ we can
consider that $h_{4}=\omega _{1}^{q_{1}/q_{2}}$ $\omega _{2}^{q_{3}/q_{4}}$ $%
\ $for some integers $q_{1},q_{2},q_{3}$ and $q_{4}\blacksquare $

There are some important consequences of the Theorems \ref{t5dr} and \ref%
{t5dra}:

\begin{corollary}
\label{ceint}The non--trivial components of the Einstein tensor [see (\ref%
{2deinst}) for the canonical d--connection] $\widehat{\mathbf{G}}_{\ \beta
}^{\alpha }=\widehat{\mathbf{R}}_{\ \beta }^{\alpha }-\frac{1}{2}%
\overleftarrow{\mathbf{\hat{R}}}\delta _{\beta }^{\alpha }$ for the ansatz (%
\ref{ansatz5}) and (\ref{1ansatzc}) given with respect to the N--adapted (co)
frames are
\begin{equation}
G_{1}^{1}=-\left( R_{2}^{2}+S_{4}^{4}\right)
,G_{2}^{2}=G_{3}^{3}=-S_{4}^{4},G_{4}^{4}=G_{5}^{5}=-R_{2}^{2}.
\label{1einstdiag}
\end{equation}
\end{corollary}

The relations (\ref{1einstdiag}) can be derived following the formulas for
the Ricci tensor (\ref{1ricci1a})--(\ref{1ricci4a}). They impose the condition
that the dynamics of such gravitational fields is defined by two independent
components $R_{2}^{2}$ and $S_{4}^{4}$ and result in

\begin{corollary}
\label{cors}The system of effective 5D Einstein--Proca equations on spaces
provided with N--connection structure (\ref{efeinst}) (equivalently, the
system (\ref{ep1})--(\ref{ep4})\_is compatible for the generic off--diagonal
ansatz (\ref{ansatz5}) and (\ref{1ansatzc}) if the energy--momentum tensor $%
\mathbf{\Upsilon }_{\alpha \beta }=\tilde{\kappa}(\mathbf{\Sigma }_{\alpha
\beta }^{[\phi ]}+\mathbf{\Sigma }_{\alpha \beta }^{[\mathbf{m}]})$ of the
Proca and matter fields given with respect to N-- frames is diagonal and
satisfies the conditions
\begin{equation}
\Upsilon _{2}^{2}=\Upsilon _{3}^{3}=\Upsilon _{2}(x^{2},x^{3},v),\ \Upsilon
_{4}^{4}=\Upsilon _{5}^{5}=\Upsilon _{4}(x^{2},x^{3}),\mbox{ and }\Upsilon
_{1}=\Upsilon _{2}+\Upsilon _{4}.  \label{1emcond}
\end{equation}
\end{corollary}

\begin{remark}
\label{remconds}Instead of the energy--momentum tensor $\mathbf{\Upsilon }%
_{\alpha \beta }=\tilde{\kappa}(\mathbf{\Sigma }_{\alpha \beta }^{[\phi ]}+%
\mathbf{\Sigma }_{\alpha \beta }^{[\mathbf{m}]})$ for the Proca and matter
fields we can consider any source, for instance, with string corrections,
when $\mathbf{\Upsilon }_{\alpha \beta }^{[str]}=\tilde{\kappa}\left(
\mathbf{\Sigma }_{\alpha \beta }^{[\phi ]}+\mathbf{\Sigma }_{\alpha \beta
}^{[\mathbf{m}]}+\mathbf{\Sigma }_{\alpha \beta }^{[\mathbf{T}]}\right) $
like in (\ref{fagfe}) satisfying the conditions (\ref{1emcond}).
\end{remark}

If the conditions of the Corollary \ref{cors}, or Remark \ref{remconds}, are
satisfied, the h- and v-- irreducible components of the 5D Einstein--Proca
equations (\ref{ep1}) and (\ref{ep4}), or of the string gravity equations (%
\ref{fagfe}), for the ansatz (\ref{ansatz5}) and (\ref{1ansatzc}) transform
into the system%
\begin{eqnarray}
R_{2}^{2} &=&R_{3}^{3}=-\frac{1}{2g_{2}g_{3}}[g_{3}^{\bullet \bullet }-\frac{%
g_{2}^{\bullet }g_{3}^{\bullet }}{2g_{2}}-\frac{(g_{3}^{\bullet })^{2}}{%
2g_{3}}+g_{2}^{^{\prime \prime }}-\frac{g_{2}^{^{\prime }}g_{3}^{^{\prime }}%
}{2g_{3}}-\frac{(g_{2}^{^{\prime }})^{2}}{2g_{2}}]=-\Upsilon
_{4}(x^{2},x^{3}),  \label{ep1a} \\
S_{4}^{4} &=&S_{5}^{5}=-\frac{1}{2h_{4}h_{5}}\left[ h_{5}^{\ast \ast
}-h_{5}^{\ast }\left( \ln \sqrt{|h_{4}h_{5}|}\right) ^{\ast }]\right]
=-\Upsilon _{2}(x^{2},x^{3},v).  \label{ep2a} \\
R_{4i} &=&-w_{i}\frac{\beta }{2h_{5}}-\frac{\alpha _{i}}{2h_{5}}=0,
\label{ep3a} \\
R_{5i} &=&-\frac{h_{5}}{2h_{4}}\left[ n_{i}^{\ast \ast }+\gamma n_{i}^{\ast }%
\right] =0.  \label{ep4a}
\end{eqnarray}

A very surprising result is that we are able to construct exact solutions of
the 5D Einstein--Proca equations with anholonomic variables and generic
off--diagonal metrics:

\begin{theorem}
\label{texs}The system of second order nonlinear partial differential
equations\\ (\ref{ep1a})--(\ref{ep4a}) and (\ref{1confeq}) can be solved in
general form if there are given certain values of functions $%
g_{2}(x^{2},x^{3})$ (or, inversely, $g_{3}(x^{2},x^{3})),\ h_{4}\left(
x^{i},v\right) $ (or, inversely, $h_{5}\left( x^{i},v\right) ),$\\ $\omega
\left( x^{i},v\right) $ and of sources $\Upsilon _{2}(x^{2},x^{3},v)$ and $%
\Upsilon _{4}(x^{2},x^{3}).$
\end{theorem}

We outline the main steps of constructing exact solutions and proving this
Theorem.

\begin{itemize}
\item The general solution of equation (\ref{ep1a}) can be written in the
form
\begin{equation}
\varpi =g_{[0]}\exp [a_{2}\widetilde{x}^{2}\left( x^{2},x^{3}\right) +a_{3}%
\widetilde{x}^{3}\left( x^{2},x^{3}\right) ],  \label{1solricci1a}
\end{equation}%
were $g_{[0]},a_{2}$ and $a_{3}$ are some constants and the functions $%
\widetilde{x}^{2,3}\left( x^{2},x^{3}\right) $ define any coordinate
transforms $x^{2,3}\rightarrow \widetilde{x}^{2,3}$ for which the 2D line
element becomes conformally flat, i. e.
\begin{equation}
g_{2}(x^{2},x^{3})(dx^{2})^{2}+g_{3}(x^{2},x^{3})(dx^{3})^{2}\rightarrow
\varpi (x^{2},x^{3})\left[ (d\widetilde{x}^{2})^{2}+\epsilon (d\widetilde{x}%
^{3})^{2}\right] ,  \label{1con10}
\end{equation}%
where $\epsilon =\pm 1$ for a corresponding signature. In coordinates $%
\widetilde{x}^{2,3},$ the equation (\ref{ep1a}) transform into%
\begin{equation*}
\varpi \left( \ddot{\varpi}+\varpi ^{\prime \prime }\right) -\dot{\varpi}%
-\varpi ^{\prime }=2\varpi ^{2}\Upsilon _{4}(\tilde{x}^{2},\tilde{x}^{3})
\end{equation*}%
or%
\begin{equation}
\ddot{\psi}+\psi ^{\prime \prime }=2\Upsilon _{4}(\tilde{x}^{2},\tilde{x}%
^{3}),  \label{auxeq01}
\end{equation}%
for $\psi =\ln |\varpi |.$ The integrals of (\ref{auxeq01}) depends on the
source $\Upsilon _{4}.$ As a particular case we can consider that $\Upsilon
_{4}=0.$ There are three alternative possibilities to generate solutions of (%
\ref{ep1a}). For instance, we can prescribe that $g_{2}=g_{3}$ and get the
equation (\ref{auxeq01}) for $\psi =\ln |g_{2}|=\ln |g_{3}|.$ If we suppose
that $g_{2}^{^{\prime }}=0,$ for a given $g_{2}(x^{2}),$ we obtain from (\ref%
{ep1a})%
\begin{equation*}
g_{3}^{\bullet \bullet }-\frac{g_{2}^{\bullet }g_{3}^{\bullet }}{2g_{2}}-%
\frac{(g_{3}^{\bullet })^{2}}{2g_{3}}=2g_{2}g_{3}\Upsilon _{4}(x^{2},x^{3})
\end{equation*}%
which can be integrated explicitly for given values of $\Upsilon _{4}.$
Similarly, we can generate solutions for a prescribed $g_{3}(x^{3})$ in the
equation
\begin{equation*}
g_{2}^{^{\prime \prime }}-\frac{g_{2}^{^{\prime }}g_{3}^{^{\prime }}}{2g_{3}}%
-\frac{(g_{2}^{^{\prime }})^{2}}{2g_{2}}=2g_{2}g_{3}\Upsilon
_{4}(x^{2},x^{3}).
\end{equation*}%
We note that a transform (\ref{1con10}) is always possible for 2D metrics and
the explicit form of solutions depends on chosen system of 2D coordinates
and on the signature $\epsilon =\pm 1.$ In the simplest case with $\Upsilon
_{4}=0$ the equation (\ref{ep1a}) is solved by arbitrary two functions $%
g_{2}(x^{3})$ and $g_{3}(x^{2}).$

\item For $\Upsilon _{2}=0,$ the equation (\ref{ep2a}) relates two functions
$h_{4}\left( x^{i},v\right) $ and $h_{5}\left( x^{i},v\right) $ following
two possibilities:

a) to compute
\begin{eqnarray}
\sqrt{|h_{5}|} &=&h_{5[1]}\left( x^{i}\right) +h_{5[2]}\left( x^{i}\right)
\int \sqrt{|h_{4}\left( x^{i},v\right) |}dv,~h_{4}^{\ast }\left(
x^{i},v\right) \neq 0;  \notag \\
&=&h_{5[1]}\left( x^{i}\right) +h_{5[2]}\left( x^{i}\right) v,\ h_{4}^{\ast
}\left( x^{i},v\right) =0,  \label{1p2}
\end{eqnarray}%
for some functions $h_{5[1,2]}\left( x^{i}\right) $ stated by boundary
conditions;

b) or, inversely, to compute $h_{4}$ for a given $h_{5}\left( x^{i},v\right)
,h_{5}^{\ast }\neq 0,$%
\begin{equation}
\sqrt{|h_{4}|}=h_{[0]}\left( x^{i}\right) (\sqrt{|h_{5}\left( x^{i},v\right)
|})^{\ast },  \label{1p1}
\end{equation}%
with $h_{[0]}\left( x^{i}\right) $ given by boundary conditions. We note
that the sourceless equation (\ref{ep2a}) is satisfied by arbitrary pairs of
coefficients $h_{4}\left( x^{i},v\right) $ and $h_{5[0]}\left( x^{i}\right)
. $ Solutions with $\Upsilon _{2}\neq 0$ can be found by ansatz of type
\begin{equation}
h_{5}[\Upsilon _{2}]=h_{5},h_{4}[\Upsilon _{2}]=\varsigma _{4}\left(
x^{i},v\right) h_{4},  \label{auxf02}
\end{equation}%
where $h_{4}$ and $h_{5}$ are related by formula (\ref{1p2}), or (\ref{1p1}).
Substituting (\ref{auxf02}), we obtain%
\begin{equation}
\varsigma _{4}\left( x^{i},v\right) =\varsigma _{4[0]}\left( x^{i}\right)
-\int \Upsilon _{2}(x^{2},x^{3},v)\frac{h_{4}h_{5}}{4h_{5}^{\ast }}dv,
\label{auxf02a}
\end{equation}%
where $\varsigma _{4[0]}\left( x^{i}\right) $ are arbitrary functions.

\item The exact solutions of (\ref{ep3a}) for $\beta \neq 0$ are defined
from an algebraic equation, $w_{i}\beta +\alpha _{i}=0,$ where the
coefficients $\beta $ and $\alpha _{i}$ are computed as in formulas (\ref%
{1abc}) by using the solutions for (\ref{ep1a}) and (\ref{ep2a}). The general
solution is
\begin{equation}
w_{k}=\partial _{k}\ln [\sqrt{|h_{4}h_{5}|}/|h_{5}^{\ast }|]/\partial
_{v}\ln [\sqrt{|h_{4}h_{5}|}/|h_{5}^{\ast }|],  \label{3w}
\end{equation}%
with $\partial _{v}=\partial /\partial v$ and $h_{5}^{\ast }\neq 0.$ If $%
h_{5}^{\ast }=0,$ or even $h_{5}^{\ast }\neq 0$ but $\beta =0,$ the
coefficients $w_{k}$ could be arbitrary functions on $\left( x^{i},v\right)
. $ \ For the vacuum Einstein equations this is a degenerated case imposing
the the compatibility conditions $\beta =\alpha _{i}=0,$ which are
satisfied, for instance, if the $h_{4}$ and $h_{5}$ are related as in the
formula (\ref{1p1}) but with $h_{[0]}\left( x^{i}\right) =const.$

\item Having defined $h_{4}$ and $h_{5}$ and computed $\gamma $ from (\ref%
{1abc}) we can solve the equation (\ref{ep4a}) by integrating on variable ''$%
v $'' the equation $n_{i}^{\ast \ast }+\gamma n_{i}^{\ast }=0.$ The exact
solution is
\begin{eqnarray}
n_{k} &=&n_{k[1]}\left( x^{i}\right) +n_{k[2]}\left( x^{i}\right) \int
[h_{4}/(\sqrt{|h_{5}|})^{3}]dv,~h_{5}^{\ast }\neq 0;  \notag \\
&=&n_{k[1]}\left( x^{i}\right) +n_{k[2]}\left( x^{i}\right) \int
h_{4}dv,\qquad ~h_{5}^{\ast }=0;  \label{1n} \\
&=&n_{k[1]}\left( x^{i}\right) +n_{k[2]}\left( x^{i}\right) \int [1/(\sqrt{%
|h_{5}|})^{3}]dv,~h_{4}^{\ast }=0,  \notag
\end{eqnarray}%
for some functions $n_{k[1,2]}\left( x^{i}\right) $ stated by boundary
conditions.
\end{itemize}

The exact solution of (\ref{1confeq}) is given by some arbitrary functions $%
\zeta _{i}=\zeta _{i}\left( x^{i},v\right) $ if \ both $\partial _{i}\omega
=0$ and $\omega ^{\ast }=0,$ we chose $\zeta _{i}=0$ for $\omega =const,$
and
\begin{eqnarray}
\zeta _{i} &=&-w_{i}+(\omega ^{\ast })^{-1}\partial _{i}\omega ,\quad \omega
^{\ast }\neq 0,  \label{1confsol} \\
&=&(\omega ^{\ast })^{-1}\partial _{i}\omega ,\quad \omega ^{\ast }\neq 0,%
\mbox{ for vacuum solutions}.  \notag
\end{eqnarray}%
\bigskip

The Theorem \ref{texs} states a general method of constructing exact
solutions in MAG, of the Einstein--Proca equations and various string
gravity generalizations with generic off--diagonal metrics. Such solutions
are with associated N--connection structure. This method can be also applied
in order to generate, for instance, certain Finsler or Lagrange
configurations as v-irreducible components. The 5D ansatz can not be used to
generate standard Finsler or Lagrange geometries because the dimension of
such spaces can not be an odd number. Nevertheless, the anholonomic frame
method can be applied in order to generate 4D exact solutions containing
Finsler--Lagrange configurations, see Appendix \ref{ss4d}.

\ Summarizing the results for the nondegenerated cases when $h_{4}^{\ast
}\neq 0$ and $h_{5}^{\ast }\neq 0$ and (for simplicity, for a trivial
conformal factor $\omega ),$ we derive an explicit result for 5D exact
solutions with local coordinates $u^{\alpha }=\left( x^{i},y^{a}\right) $
when $x^{i}=\left( x^{1},x^{\widehat{i}}\right) ,x^{\widehat{i}}=\left(
x^{2},x^{3}\right) ,y^{a}=\left( y^{4}=v,y^{a}\right) $ and arbitrary
signatures $\epsilon _{\alpha }=\left( \epsilon _{1},\epsilon _{2},\epsilon
_{3},\epsilon _{4},\epsilon _{5}\right) $ (where $\epsilon _{\alpha }=\pm
1): $

\begin{corollary}
\label{corgsol1}Any off--diagonal metric
\begin{eqnarray}
\delta s^{2} &=&\epsilon _{1}(dx^{1})^{2}+\epsilon _{\widehat{k}}g_{\widehat{%
k}}\left( x^{\widehat{i}}\right) (dx^{\widehat{k}})^{2}+  \notag \\
&&\epsilon _{4}h_{0}^{2}(x^{i})\left[ f^{\ast }\left( x^{i},v\right) \right]
^{2}|\varsigma _{\Upsilon }\left( x^{i},v\right) |\left( \delta v\right)
^{2}+\epsilon _{5}f^{2}\left( x^{i},v\right) \left( \delta y^{5}\right) ^{2},
\notag \\
\delta v &=&dv+w_{k}\left( x^{i},v\right) dx^{k},\ \delta
y^{5}=dy^{5}+n_{k}\left( x^{i},v\right) dx^{k},  \label{gensol1}
\end{eqnarray}%
with coefficients of necessary smooth class, where\ \ $g_{\widehat{k}}\left(
x^{\widehat{i}}\right) $ is a solution of the 2D equation (\ref{ep1a}) for a
given source $\Upsilon _{4}\left( x^{\widehat{i}}\right) ,$%
\begin{equation*}
\varsigma _{\Upsilon }\left( x^{i},v\right) =\varsigma _{4}\left(
x^{i},v\right) =\varsigma _{4[0]}\left( x^{i}\right) -\frac{\epsilon _{4}}{16%
}h_{0}^{2}(x^{i})\int \Upsilon _{2}(x^{\widehat{k}},v)[f^{2}\left(
x^{i},v\right) ]^{2}dv,
\end{equation*}%
and the N--connection coefficients $N_{i}^{4}=w_{i}(x^{k},v)$ and $%
N_{i}^{5}=n_{i}(x^{k},v)$ are
\begin{equation}
w_{i}=-\frac{\partial _{i}\varsigma _{\Upsilon }\left( x^{k},v\right) }{%
\varsigma _{\Upsilon }^{\ast }\left( x^{k},v\right) }  \label{gensol1w}
\end{equation}%
and
\begin{equation}
n_{k}=n_{k[1]}\left( x^{i}\right) +n_{k[2]}\left( x^{i}\right) \int \frac{%
\left[ f^{\ast }\left( x^{i},v\right) \right] ^{2}}{\left[ f\left(
x^{i},v\right) \right] ^{2}}\varsigma _{\Upsilon }\left( x^{i},v\right) dv,
\label{gensol1n}
\end{equation}%
define an exact solution of the system of Einstein equations with holonomic
and anholonomic variables (\ref{ep1a})--(\ref{ep4a}) for arbitrary
nontrivial functions $f\left( x^{i},v\right) $ (with $f^{\ast }\neq 0),$ $%
h_{0}^{2}(x^{i})$, $\varsigma _{4[0]}\left( x^{i}\right) ,n_{k[1]}\left(
x^{i}\right) $ and $\ n_{k[2]}\left( x^{i}\right) ,$ and sources $\Upsilon
_{2}(x^{\widehat{k}},v),\Upsilon _{4}\left( x^{\widehat{i}}\right) $ and any
integration constants and signatures $\epsilon _{\alpha }=\pm 1$ to be
defined by certain boundary conditions and physical considerations.
\end{corollary}

Any metric (\ref{gensol1}) with $h_{4}^{\ast }\neq 0$ and $h_{5}^{\ast }\neq
0$ has the property to be generated by a function of four variables $f\left(
x^{i},v\right) $ with emphasized dependence on the anisotropic coordinate $%
v, $ because $f^{\ast }\doteqdot \partial _{v}f\neq 0$ and by arbitrary
sources $\Upsilon _{2}(x^{\widehat{k}},v),\Upsilon _{4}\left( x^{\widehat{i}%
}\right) .$ The rest of arbitrary functions not depending on $v$
have been obtained in result of integration of partial
differential equations. This fix a specific class of metrics
generated by using the relation (\ref{1p1}) and the first formula
in (\ref{1n}). We can generate also a different class
of solutions with $h_{4}^{\ast }=0$ by considering the second formula in (%
\ref{1p2}) and respective formulas in (\ref{1n}). The ''degenerated'' cases
with $h_{4}^{\ast }=0$ but $h_{5}^{\ast }\neq 0$ and inversely, $h_{4}^{\ast
}\neq 0$ but $h_{5}^{\ast }=0$ are more special and request a proper
explicit construction of solutions. Nevertheless, such type of solutions are
also generic off--diagonal and they could be of substantial interest.

The sourceless case with vanishing $\Upsilon _{2}$ and $\Upsilon _{4}$ is
defined following

\begin{remark}
Any off--diagonal metric (\ref{gensol1}) with $\varsigma _{\Upsilon }=1,$ $%
h_{0}^{2}(x^{i})=$ $h_{0}^{2}=const,$ $w_{i}=0$ and $n_{k}$ computed as in (%
\ref{gensol1n}) but for $\varsigma _{\Upsilon }=1,$ defines a vacuum
solution of 5D Einstein equations for the canonical d--connection (\ref%
{2candcon}) computed for the ansatz (\ref{gensol1}).
\end{remark}

By imposing additional constraints on arbitrary functions from $%
N_{i}^{5}=n_{i}$ and $N_{i}^{5}=w_{i},$ we can select off--diagonal
gravitational configurations with distorsions of the Levi--Civita connection
resulting in canonical d--connections with the same solutions of the vacuum
Einstein equations. For instance, we can model Finsler like geometries in
general relativity, see Corollary \ref{corcond1}. Under similar conditions
the ansatz (\ref{ansatz5}) was used for constructing exact off--diagonal
solutions in the 5D Einstein gravity, see Refs. \cite{02v1,02v1a,02v2}.

Let us consider the procedure of selecting solutions with off--diagonal
metrics from an ansatz (\ref{gensol1}) with trivial N--connection curvature
(such metrics consists a simplest subclass which can be restricted to
(pseudo) Riemannian ones). The corresponding nontrivial coefficients the
N--connection curvature \ (\ref{2ncurv}) are computed
\begin{equation*}
\Omega _{ij}^{4}=\partial _{i}w_{j}-\partial _{j}w_{i}+w_{i}w_{j}^{\ast
}-w_{j}w_{i}^{\ast }\mbox{ and }\Omega _{ij}^{5}=\partial _{i}n_{j}-\partial
_{j}n_{i}+w_{i}n_{j}^{\ast }-w_{j}n_{i}^{\ast }.
\end{equation*}%
So, there are imposed six constraints, $\Omega _{ij}^{4}=\Omega _{ij}^{5}=0,$
for $i,j...=1,2,4$ on six functions $w_{i}$ and $n_{i}$ computed
respectively as (\ref{gensol1n}) and (\ref{gensol1n}) which can be satisfied
by a corresponding subclass of functions $f\left( x^{i},v\right) $ (with $%
f^{\ast }\neq 0),$ $h_{0}^{2}(x^{i}),$ $\varsigma _{4[0]}\left( x^{i}\right)
,n_{k[1]}\left( x^{i}\right) ,$ $\ n_{k[2]}\left( x^{i}\right) $ and $%
\Upsilon _{2}(x^{\widehat{k}},v),\Upsilon _{4}\left( x^{\widehat{i}}\right) $
(in general, we have to solve certain first order partial derivative
equations with may be reduced to algebraic relations by corresponding
parametrizations). For instance, in the vacuum case when $w_{j}=0,$ we
obtain $\Omega _{ij}^{5}=\partial _{i}n_{j}-\partial _{j}n_{i}.$ The
simplest example when condition $\Omega _{\widehat{i}\widehat{j}%
}^{5}=\partial _{\widehat{i}}n_{\widehat{j}}-\partial _{\widehat{j}}n_{%
\widehat{i}}=0,$ with $\widehat{i},\widehat{j}=2,3$ (reducing the metric (%
\ref{gensol1}) to a 4D one trivially embedded into 5D) is satisfied is to
take $n_{3[1]}=n_{3[2]}=0$ in (\ref{gensol1n}) and consider that $f=f\left(
x^{2},v\right) $ with $n_{2[1]}=n_{2[1]}\left( x^{2}\right) $ and $%
n_{2[2]}=n_{2[2]}\left( x^{2}\right) ,$ i. e. by eliminating the dependence
of the coefficients on $x^{3}.$ This also results in a generic off--diagonal
solution, because the anholonomy coefficients (\ref{2anhc}) are not trivial,
for instance, $w_{24}^{5}=n_{2}^{\ast }$ and $w_{14}^{5}=n_{1}^{\ast }.$

Another interesting remark is that even we have reduced the canonical
d--connection to the Levi--Civita one [with respect to N--adapted (co)
frames; this imposes the metric to be (pseudo) Riemannian] by selecting the
arbitrary functions as to have $\Omega _{ij}^{a}=0,$ one could be
nonvanishing d--torsion components like $T_{41}^{5}=P_{41}^{5}$ and $%
T_{41}^{5}=P_{41}^{5}$ in (\ref{2dtorsb}). Such objects, as well the
anholonomy coefficients $w_{24}^{5}$ and $w_{14}^{5}$ (which can be also
considered as torsion like objects) are constructed by taking certain
''scarps'' from the coefficients of off--diagonal metrics and anholonomic
frames. They are induced by the frame anholonomy (like ''torsions'' in
rotating anholonomic systems of reference for the Newton gravity and
mechanics with constraints) and vanish if we transfer the constructions with
respect to any holonomic basis.

The above presented results are for generic 5D off--diagonal metrics,
anholonomic transforms and nonlinear field equations. Reductions to a lower
dimensional theory are not trivial in such cases. We emphasize some specific
points of this procedure in the Appendix \ref{ss4d} (see details in \cite%
{02vmethod}).

\section{Exact Solutions}

\label{1exsol} There were found a set of exact solutions in MAG \cite%
{02esolmag,02obet2,02oveh} describing various configuration of Einstein--Maxwell
of dilaton gravity emerging from low energy string theory, soliton and
multipole solutions and generalized Plebanski--Demianski solutions,
colliding waves and static black hole metrics. In this section we are going
to look for some classes of 4D and 5D solutions of the Einstein--Proca
equations in MAG related to string gravity modelling generalized
Finsler--affine geometries and extending to such spacetimes some our
previous results \cite{02v1,02v1a,02v2}.

\subsection{Finsler--Lagrange metrics in string and metric--affine gra\-vi\-ty}

\label{solmlrc}

As we discussed in section 2, the generalized Finsler--Lagrange spaces can
be modelled in metric--affine spacetimes provided with N--connection
structure. In this subsection, we show how such two dimensional Finsler like
spaces with d--metrics depending on one anisotropic coordinate $y^{4}=v$
(denoted as $\mathbf{F}^{2}=\left[ V^{2},F\left( x^{2},x^{3},y\right) \right]
,$ $\mathbf{L}^{2}=\left[ V^{2},L\left( x^{2},x^{3},y\right) \right] $ and $%
\mathbf{GL}^{2}=\left[ V^{2},g_{ij}\left( x^{2},x^{3},y\right) \right] $
according to Ref. \cite{02vp1}) can be modelled by corresponding diad
transforms on spacetimes with 5D (or 4D) d--metrics being exact solutions of
the field equations for the generalized Finsler--affine string gravity (\ref%
{fagfe}) (as a particular case we can consider the Einstein--Proca system (%
\ref{ep1a})--(\ref{ep4a}) and (\ref{1confeq})). For every particular case of
locally anisotropic spacetime, for instance, outlined in Appendix C, see
Table \ref{1tablegs}, the quadratic form $\tilde{g}_{ij},$ d--metric $\mathbf{%
\tilde{g}}_{\alpha \beta }=\left[ \tilde{g}_{ij},\tilde{g}_{ij}\right] $ and
N--connection $\tilde{N}_{j}^{a}$ one holds

\begin{theorem}
Any 2D locally anisotropic structure given by \ $\mathbf{\tilde{g}}_{\alpha
\beta }$ and $\tilde{N}_{j}^{a}$ can be modelled on the space of exact
solutions of the 5D (or 4D) the generalized Finsler--affine string gravity
system defined by the ansatz (\ref{1ansatzc}) (or (\ref{1ansatzc4})).
\end{theorem}

We give the proof via an explicit construction. Let us consider
\begin{equation*}
\mathbf{g}_{\alpha \beta }=[g_{ij},h_{ab}]=\left[ \omega g_{2}\left(
x^{2},x^{3}\right) ,\omega g_{3}\left( x^{2},x^{3}\right) ,\omega
h_{4}\left( x^{2},x^{3},v\right) ,\omega h_{5}\left( x^{2},x^{3},v\right) %
\right]
\end{equation*}%
for $\omega =\omega \left( x^{2},x^{3},v\right) $ and
\begin{equation*}
N_{i}^{a}=\left[ N_{i}^{4}=w_{i}\left( x^{2},x^{3},v\right)
,N_{i}^{5}=n_{i}\left( x^{2},x^{3},v\right) \right] ,
\end{equation*}%
where indices are running the values $a=4,5$ and $i=2,3$ define an exact 4D
solution of the equations (\ref{fagfe}) (or, in the particular case, of the
system (\ref{ep1a})--(\ref{ep4a}), for simplicity, we put $\omega \left(
x^{2},x^{3},v\right) =1).$ We can relate the data $\left( \mathbf{g}_{\alpha
\beta },N_{i}^{a}\right) $ to any data $\left( \mathbf{\tilde{g}}_{\alpha
\beta },\tilde{N}_{j}^{a}\right) $ via nondegenerate diadic transforms $%
e_{i}^{i^{\prime }}=e_{i}^{i^{\prime }}\left( x^{2},x^{3},v\right) ,\
l_{a}^{i^{\prime }}=l_{a}^{i^{\prime }}\left( x^{2},x^{3},v\right) $ and $%
q_{a}^{i^{\prime }}=q_{a}^{i^{\prime }}\left( x^{2},x^{3},v\right) $ (and
theirs inverse matrices)%
\begin{equation}
g_{ij}=e_{i}^{i^{\prime }}e_{j}^{j^{\prime }}\tilde{g}_{i^{\prime }j^{\prime
}},\ h_{ab}=l_{a}^{i^{\prime }}l_{b}^{i^{\prime }}\tilde{g}_{i^{\prime
}j^{\prime }},N_{i^{\prime }}^{a}=q_{a^{\prime }}^{a}\tilde{N}_{j^{\prime
}}^{a^{\prime }}\ .  \label{aux03}
\end{equation}%
Such transforms may be associated to certain tetradic transforms of the
N--elongated (co) frames ((\ref{2ddif})) (\ref{2dder}). If for the given data $%
\left( \mathbf{g}_{\alpha \beta },N_{i}^{a}\right) $ and $\left( \mathbf{%
\tilde{g}}_{\alpha \beta },\tilde{N}_{j}^{a}\right) $ in (\ref{aux03}), we
can solve the corresponding systems of quadratic algebraic equations \ and
define nondegenerate matrices $\left( e_{i}^{i^{\prime }}\right) ,\left(
l_{a}^{i^{\prime }}\right) $ and $\left( q_{i^{\prime }}^{a}\right) ,$ we
argue that the 2D locally anisotropic spacetime $\left( \mathbf{\tilde{g}}%
_{\alpha \beta },\tilde{N}_{j}^{a}\right) $ (really, it is a 4D
spacetime with generic off--diagonal metric and associated
N--connection structure) can be modelled on by a class of exact
solutions of effective Einstein--Proca equations for
MAG.$\blacksquare $

The d--metric with respect to transformed N--adapted diads is \ written in
the form
\begin{equation}
\mathbf{g=}\tilde{g}_{i^{\prime }j^{\prime }}\mathbf{e}^{i^{\prime }}\otimes
\mathbf{e}^{j^{\prime }}+\tilde{g}_{i^{\prime }j^{\prime }}\mathbf{\tilde{e}}%
^{i^{\prime }}\otimes \mathbf{\tilde{e}}^{j^{\prime }}  \label{blok2b}
\end{equation}%
where
\begin{equation*}
\mathbf{e}^{i^{\prime }}=e_{i}^{i^{\prime }}dx^{i},\ \mathbf{\tilde{e}}%
^{i^{\prime }}=l_{a}^{i^{\prime }}\mathbf{\tilde{e}}^{a},\ \mathbf{\tilde{e}}%
^{a}=dy^{a}+\tilde{N}_{j^{\prime }}^{a}\mathbf{\tilde{e}}_{[N]}^{j^{\prime
}},\ \mathbf{\tilde{e}}_{[N]}^{j^{\prime }}=q_{i}^{j^{\prime }}dx^{i}.
\end{equation*}%
The d--metric (\ref{blok2b}) has the coefficients corresponding to
generalized Finsler--Lagrange spaces and emphasizes that any quadratic form $%
\tilde{g}_{i^{\prime }j^{\prime }}$ from Table \ref{1tablegs} can be related
via an exact solution $\left( g_{ij},h_{ab},N_{i^{\prime }}^{a}\right) .$

We note that we can define particular cases of imbedding with $%
h_{ab}=l_{a}^{i^{\prime }}l_{b}^{i^{\prime }}\tilde{g}_{i^{\prime }j^{\prime
}}$ and $N_{j^{\prime }}^{a}=q_{i^{\prime }}^{a}\tilde{N}_{j^{\prime
}}^{i^{\prime }}$ for a prescribed value of $g_{ij}=\tilde{g}_{i^{\prime
}j^{\prime }}$ and try to model only the quadratic form $\tilde{h}%
_{i^{\prime }j^{\prime }}$ in MAG. Similar considerations were presented for
particular cases of modelling Finsler structures and generalizations in
Einstein and Einstein--Cartan spaces \cite{02v1,02v2}, see the conditions (\ref%
{1cond2}).

\subsection{Solutions in MAG with effective variable cosmological constant}

A class of 4D solutions in MAG with local anisotropy can be derived from (%
\ref{efeinst}) $\ $\ for $\mathbf{\Sigma }_{\alpha \beta }^{[\mathbf{m}]}=0$
and almost vanishing mass $\mu \rightarrow 0$ of the Proca field in the
source $\mathbf{\Sigma }_{\alpha \beta }^{[\mathbf{\phi }]}.$ This holds in
absence of matter fields and when the constant in the action for the
Finsler--affine gravity are subjected to the condition (\ref{procvan}). We
consider that $\phi _{\mu }=\left( \phi _{\widehat{i}}\left( x^{\widehat{k}%
}\right) ,\phi _{a}=0\right) ,$ where $\widehat{i},\widehat{k},...=2,3$ and $%
a,b,...=4,5,$ with respect to a N--adapted coframe (\ref{2ddif}) and choose a
metric ansatz of type (\ref{1ansatz4}) with $g_{2}=1$ and $g_{3}=-1$ which
select a flat h--subspace imbedded into a general anholonomic 4D background
with nontrivial $h_{ab}$ and N--connection structure $N_{i}^{a}.$ The
h--covariant derivatives are $\widehat{D}^{[h]}$ $\phi _{\widehat{i}}=\left(
\partial _{2}\phi _{\widehat{i}},\partial _{3}\phi _{\widehat{i}}\right) $
because the coefficients $\widehat{L}_{\ jk}^{i}$ and $\widehat{C}_{\
ja}^{i} $ are zero in (\ref{2candcon}) and any contraction with $\phi _{a}=0$
results in zero values. In this case the Proca equations, $\widehat{\mathbf{D%
}}_{\nu }\mathbf{H}^{\nu \mu }=\mu ^{2}\mathbf{\phi }^{\mu },$ transform in
a Maxwell like equation,%
\begin{equation}
\partial _{2}(\partial _{2}\phi _{\widehat{i}})-\partial _{3}(\partial
_{3}\phi _{\widehat{i}})=0,  \label{waveeq}
\end{equation}%
for the potential $\phi _{\widehat{i}},$ with the dynamics in the
h--subspace distinguished by a N--connecti\-on structure to be defined latter.
We note that $\phi _{i}$ is not an electromagnetic field, but a component of
the metric--affine gravity related to nonmetricity and torsion. The relation
$\mathbf{Q}=k_{0}\mathbf{\phi ,\ \Lambda =}k_{1}\mathbf{\phi ,\ T=}k_{2}%
\mathbf{\phi }$ from (\ref{triplemag}) transforms into $Q_{\widehat{i}%
}=k_{0}\phi _{\widehat{i}},\Lambda _{\widehat{i}}=k_{1}\phi _{\widehat{i}%
},T_{i}=k_{2}\phi _{\widehat{i}},$ and vanishing $Q_{a},\Lambda _{a}$ and $%
T_{a},$ defined, for instance, by a wave solution of (\ref{waveeq}),
\begin{equation}
\phi _{\widehat{i}}=\phi _{\lbrack 0]\widehat{i}}\cos \left( \varrho
_{i}x^{i}+\varphi _{\lbrack 0]}\right)  \label{wavesoltf}
\end{equation}%
for any constants $\phi _{\lbrack 0]2,3},$ $\varphi _{\lbrack 0]}$ and $%
\left( \varrho _{2}\right) ^{2}-\left( \varrho _{3}\right) ^{2}=0.$ In this
simplified model we have related plane waves of nonmetricity and torsion
propagating on an anholonomic background provided with N--connection. Such
nonmetricity and torsion do not vanish even $\mu \rightarrow 0$ and the
Proca field is approximated by a massless vector field defined in the
h--subspace.

The energy--momentum tensor $\mathbf{\Sigma }_{\alpha \beta }^{[\mathbf{\phi
}]}$ for the massless field (\ref{wavesoltf}) is defined by a nontrivial
value%
\begin{equation*}
H_{23}=\partial _{2}\phi _{3}-\partial _{3}\phi _{2}=\varepsilon
_{23}\lambda _{\lbrack h]}\sin \left( \varrho _{i}x^{i}+\varphi _{\lbrack
0]}\right)
\end{equation*}%
with antisymmetric $\varepsilon _{23},\varepsilon _{23}=1,$ and constant $%
\lambda _{\lbrack h]}$ taken for a normalization $\varepsilon _{23}\lambda
_{\lbrack h]}=$ $\varrho _{2}\phi _{\lbrack 0]3}-\varrho _{3}\phi _{\lbrack
0]2}.$ This tensor is diagonal with respect to N--adapted (co) frames, $%
\mathbf{\Sigma }_{\alpha }^{[\mathbf{\phi }]\beta }=\{\Upsilon _{2},\Upsilon
_{2},0,0\}$ with
\begin{equation}
\Upsilon _{2}\left( x^{2},x^{3}\right) =-\lambda _{\lbrack h]}^{2}\sin
^{2}\left( \varrho _{i}x^{i}+\varphi _{\lbrack 0]}\right) .  \label{source01}
\end{equation}%
So, we have the case from (\ref{1einstdiag4}) and (\ref{1emcond4}) with $%
\Upsilon _{2}\left( x^{2},x^{2},v\right) \rightarrow \Upsilon _{2}\left(
x^{2},x^{2}\right) $ and $\Upsilon _{4},$ i. e.
\begin{equation}
G_{2}^{2}=G_{3}^{3}=-S_{4}^{4}=\Upsilon _{2}\left( x^{2},x^{2}\right)
\mbox{
and }G_{4}^{4}=G_{4}^{4}=-R_{2}^{2}=0.  \label{aux62}
\end{equation}%
There are satisfied the compatibility conditions from Corollary \ref{cors}.
For the above stated ansatz for the d--metric and $\phi $--field, the system
(\ref{efeinst}) $\ $reduces to a particular case of (\ref{ep1a})--(\ref{ep4a}%
), when the first equation is trivially satisfied by $g_{2}=1$ and $g_{3}=-1$
but the second one is
\begin{equation}
S_{4}^{4}=S_{5}^{5}=-\frac{1}{2h_{4}h_{5}}\left[ h_{5}^{\ast \ast
}-h_{5}^{\ast }\left( \ln \sqrt{|h_{4}h_{5}|}\right) ^{\ast }]\right]
=\lambda _{\lbrack h]}^{2}\sin ^{2}\left( \varrho _{i}x^{i}+\varphi
_{\lbrack 0]}\right) .  \label{aux04sol}
\end{equation}%
The right part of this equation is like a ''cosmological constant'', being
nontrivial in the h--subspace and polarized by a nonmetricity and torsion
wave (we can state $x^{2}=t$ and choose the signature $\left( -+---\right)
). $

The exact solution of (\ref{aux04sol}) exists according the Theorem \ref%
{texs} (see formulas (\ref{1p2})--(\ref{auxf02a})). Taking any $%
h_{4}=h_{4}[\lambda _{\lbrack h]}=0]$ and $h_{5}=h_{5}[\lambda _{\lbrack
h]}=0]$ solving the equation with $\lambda _{\lbrack h]}=0,$ for instance,
like in (\ref{1p1}), we can express the general solution with nontrivial
source like
\begin{equation*}
h_{5}[\lambda _{\lbrack h]}]=h_{5},\ h_{4}[\lambda _{\lbrack h]}]=\varsigma
_{\lbrack \lambda ]}\left( x^{i},v\right) h_{4},
\end{equation*}%
where (for an explicit source (\ref{source01}) in (\ref{auxf02a}))%
\begin{equation*}
\varsigma _{\lbrack \lambda ]}\left( t,x^{3},v\right) =\varsigma
_{4[0]}\left( t,x^{3}\right) -\frac{\lambda _{\lbrack h]}^{2}}{4}\sin
^{2}\left( \varrho _{2}t+\varrho _{3}x^{3}+\varphi _{\lbrack 0]}\right) \int
\frac{h_{4}h_{5}}{h_{5}^{\ast }}dv,
\end{equation*}%
where $\varsigma _{4[0]}\left( t,x^{3}\right) =1$ if we want to have $%
\varsigma _{\lbrack \lambda ]}$ for $\lambda _{\lbrack h]}^{2}\rightarrow 0$%
. A particular class of 4D off--diagonal exact solutions with $h_{4,5}^{\ast
}\neq 0$ (see the Corollary \ref{corgsol1} with $x^{2}=t$ stated to be the
time like coordinate and $x^{1}$ considered as the extra 5th dimensional one
to be eliminated for reductions 5D$\rightarrow $4D) is parametrized by the
generic off--diagonal metric
\begin{eqnarray}
\delta s^{2} &=&(dt)^{2}-\left( dx^{3}\right) -h_{0}^{2}(t,x^{3})\left[
f^{\ast }\left( t,x^{3},v\right) \right] ^{2}|\varsigma _{\lbrack \lambda
]}\left( t,x^{3},v\right) |\left( \delta v\right) ^{2}-f^{2}\left(
t,x^{3},v\right) \left( \delta y^{5}\right) ^{2}  \notag \\
\delta v &=&dv+w_{\widehat{k}}\left( t,x^{3},v\right) dx^{\widehat{k}},\
\delta y^{5}=dy^{5}+n_{\widehat{k}}\left( t,x^{3},v\right) dx^{\widehat{k}},
\label{dm5}
\end{eqnarray}%
with coefficients of necessary smooth class, where\ \ $g_{\widehat{k}}\left(
x^{\widehat{i}}\right) $ is a solution of the 2D equation (\ref{ep1a}) for a
given source $\Upsilon _{4}\left( x^{\widehat{i}}\right) ,$%
\begin{equation*}
\varsigma _{\lbrack \lambda ]}\left( t,x^{3},v\right) =1+\frac{\lambda
_{\lbrack h]}^{2}}{16}h_{0}^{2}(t,x^{3})\sin ^{2}\left( \varrho
_{2}t+\varrho _{3}x^{3}+\varphi _{\lbrack 0]}\right) f^{2}\left(
t,x^{3},v\right) ,
\end{equation*}%
and the N--connection coefficients $N_{\widehat{i}}^{4}=w_{\widehat{i}%
}(t,x^{3},v)$ and $N_{\widehat{i}}^{5}=n_{\widehat{i}}(t,x^{3},v)$ are
\begin{equation*}
w_{2,3}=-\frac{\partial _{2,3}\varsigma _{\lbrack \lambda ]}\left(
t,x^{3},v\right) }{\varsigma _{\lbrack \lambda ]}^{\ast }\left(
t,x^{3},v\right) }
\end{equation*}%
and
\begin{equation*}
n_{2,3}\left( t,x^{3},v\right) =n_{2,3[1]}\left( t,x^{3}\right)
+n_{2,3[2]}\left( t,x^{3}\right) \int \frac{\left[ f^{\ast }\left(
t,x^{3},v\right) \right] ^{2}}{\left[ f\left( t,x^{3},v\right) \right] ^{2}}%
\varsigma _{\lbrack \lambda ]}\left( t,x^{3},v\right) dv,
\end{equation*}%
define an exact 4D solution of the system of Einstein--Proca equations (\ref%
{ep1})--(\ref{ep4}) for vanishing mass $\mu \rightarrow 0,$ with holonomic
and anholonomic variables and 1-form field
\begin{equation*}
\phi _{\mu }=\left[ \phi _{\widehat{i}}=\phi _{\lbrack 0]\widehat{i}}\cos
\left( \varrho _{2}t+\varrho _{3}x^{3}+\varphi _{\lbrack 0]}\right) ,\phi
_{4}=0,\phi _{0}=0\right]
\end{equation*}%
for arbitrary nontrivial functions $f\left( t,x^{3},v\right) $ (with $%
f^{\ast }\neq 0),$ $h_{0}^{2}(t,x^{3})$, $n_{k[1,2]}\left( t,x^{3}\right) $
and sources $\Upsilon _{2}\left( t,x^{3}\right) =-\lambda _{\lbrack
h]}^{2}\sin ^{2}\left( \varrho _{2}t+\varrho _{3}x^{3}+\varphi _{\lbrack
0]}\right) $ and $\Upsilon _{4}=0$ and any integration constants to be
defined by certain boundary conditions and additional physical arguments.
For instance, we can consider ellipsoidal symmetries for the set of space
coordinates $\left( x^{3},y^{4}=v,y^{5}\right) $ considered on possibility
to be ellipsoidal ones, or even with topologically nontrivial configurations
like torus, with toroidal coordinates. Such exact solutions emphasize
anisotropic dependencies on coordinate $v$ and do not depend on $y^{5}.$

\subsection{3D solitons in string Finsler--affine gravity}

The d--metric (\ref{dm5}) can be extended as to define a class of exact
solutions of generalized Finsler affine string gravity (\ref{fagfe}), for
certain particular cases describing 3D solitonic configurations.

We start with the the well known ansatz in string theory (see, for instance, %
\cite{02kir}) for the $H$--field (\ref{aux51a}) when
\begin{equation}
\mathbf{H}_{\nu \lambda \rho }=\widehat{\mathbf{Z}}_{\ \nu \lambda \rho }+%
\widehat{\mathbf{H}}_{\nu \lambda \rho }=\lambda _{\lbrack H]}\sqrt{|\mathbf{%
g}_{\alpha \beta }|}\varepsilon _{\nu \lambda \rho }  \label{ans61}
\end{equation}%
where $\varepsilon _{\nu \lambda \rho }$ is completely antisymmetric and $%
\lambda _{\lbrack H]}=const,$ which satisfies the field equations for $%
\mathbf{H}_{\nu \lambda \rho },$ see (\ref{aux51b}). \ The ansatz (\ref%
{ans61}) is chosen for a locally anisotropic background with $\widehat{%
\mathbf{Z}}_{\ \nu \lambda \rho }$ defined by the d--torsions for the
canonical d--connection. So, the values $\widehat{\mathbf{H}}_{\nu \lambda
\rho }$ are constrained to solve the equations (\ref{ans61}) for a fixed
value of the cosmological constant $\lambda _{\lbrack H]}$ effectively
modelling some corrections from string gravity. In this case, the source (\ref%
{source01}) is modified to
\begin{equation*}
\mathbf{\Sigma }_{\alpha }^{[\mathbf{\phi }]\beta }+\mathbf{\Sigma }_{\alpha
}^{[\mathbf{H}]\beta }=\{\Upsilon _{2}+\frac{\lambda _{\lbrack H]}^{2}}{4}%
,\Upsilon _{2}+\frac{\lambda _{\lbrack H]}^{2}}{4},\frac{\lambda _{\lbrack
H]}^{2}}{4},\frac{\lambda _{\lbrack H]}^{2}}{4}\}
\end{equation*}%
and the equations (\ref{aux62}) became more general,
\begin{equation}
G_{2}^{2}=G_{3}^{3}=-S_{4}^{4}=\Upsilon _{2}\left( x^{2},x^{2}\right) +\frac{%
\lambda _{\lbrack H]}^{2}}{4}\mbox{
and }G_{4}^{4}=G_{4}^{4}=-R_{2}^{2}=\frac{\lambda _{\lbrack H]}^{2}}{4},
\label{aux63}
\end{equation}%
or, in component form%
\begin{eqnarray}
R_{2}^{2} &=&R_{3}^{3}=-\frac{1}{2g_{2}g_{3}}[g_{3}^{\bullet \bullet }-\frac{%
g_{2}^{\bullet }g_{3}^{\bullet }}{2g_{2}}-\frac{(g_{3}^{\bullet })^{2}}{%
2g_{3}}+g_{2}^{^{\prime \prime }}-\frac{g_{2}^{^{\prime }}g_{3}^{^{\prime }}%
}{2g_{3}}-\frac{(g_{2}^{^{\prime }})^{2}}{2g_{2}}]=-\frac{\lambda _{\lbrack
H]}^{2}}{4},  \label{eq64a} \\
S_{4}^{4} &=&S_{5}^{5}=-\frac{1}{2h_{4}h_{5}}\left[ h_{5}^{\ast \ast
}-h_{5}^{\ast }\left( \ln \sqrt{|h_{4}h_{5}|}\right) ^{\ast }]\right] =-%
\frac{\lambda _{\lbrack H]}^{2}}{4}+\lambda _{\lbrack h]}^{2}\sin ^{2}\left(
\varrho _{i}x^{i}+\varphi _{\lbrack 0]}\right) .  \label{eq64b}
\end{eqnarray}%
The solution of (\ref{eq64a})\ can be found as in the case for (\ref{auxeq01}%
), when $\psi =\ln |g_{2}|=\ln |g_{3}|$ is a solution of
\begin{equation}
\ddot{\psi}+\psi ^{\prime \prime }=-\frac{\lambda _{\lbrack H]}^{2}}{2},
\label{aux73}
\end{equation}%
where, for simplicity we choose the h--variables $x^{2}=\tilde{x}^{2}$ and $%
x^{3}=\tilde{x}^{3}.$

The solution of (\ref{eq64b}) can be constructed similarly to the equation (%
\ref{aux04sol}) but for a modified source (see Theorem \ref{texs} and
formulas (\ref{1p2})--(\ref{auxf02a})). Taking any $h_{4}=h_{4}[\lambda
_{\lbrack h]}=0,\lambda _{\lbrack H]}=0]$ and $h_{5}=h_{5}[\lambda _{\lbrack
h]}=0,\lambda _{\lbrack H]}=0]$ solving the equation with $\lambda _{\lbrack
h]}=0$ and $\lambda _{\lbrack H]}=0$ like in (\ref{1p1}), we can express the
general solution with nontrivial source like
\begin{equation*}
h_{5}[\lambda _{\lbrack h]},\lambda _{\lbrack H]}]=h_{5},\ h_{4}[\lambda
_{\lbrack h]},\lambda _{\lbrack H]}]=\varsigma _{\lbrack \lambda ,H]}\left(
x^{i},v\right) h_{4},
\end{equation*}%
where (for an explicit source from (\ref{eq64b}) in (\ref{auxf02a}))%
\begin{equation*}
\varsigma _{\lbrack \lambda ,H]}\left( t,x^{3},v\right) =\varsigma
_{4[0]}\left( t,x^{3}\right) -\frac{1}{4}\left[ \lambda _{\lbrack
h]}^{2}\sin ^{2}\left( \varrho _{2}t+\varrho _{3}x^{3}+\varphi _{\lbrack
0]}\right) -\frac{\lambda _{\lbrack H]}^{2}}{4}\right] \int \frac{h_{4}h_{5}%
}{h_{5}^{\ast }}dv,
\end{equation*}%
where $\varsigma _{4[0]}\left( t,x^{3}\right) =1$ if we want to have $%
\varsigma _{\lbrack \lambda ]}$ for $\lambda _{\lbrack h]}^{2},\lambda
_{\lbrack H]}^{2}\rightarrow 0$.

We define a class of 4D off--diagonal exact solutions of the system (\ref%
{fagfe}) with $h_{4,5}^{\ast }\neq 0$ (see the Corollary \ref{corgsol1} with
$x^{2}=t$ stated to be the time like coordinate and $x^{1}$ considered as
the extra 5th dimensional one to be eliminated for reductions 5D$\rightarrow
$4D) is parametrized by the generic off--diagonal metric
\begin{eqnarray}
\delta s^{2} &=&e^{\psi (t,x^{3})}(dt)^{2}-e^{\psi (t,x^{3})}\left(
dx^{3}\right) -f^{2}\left( t,x^{3},v\right) \left( \delta y^{5}\right) ^{2}
\notag \\
&&-h_{0}^{2}(t,x^{3})\left[ f^{\ast }\left( t,x^{3},v\right) \right]
^{2}|\varsigma _{\lbrack \lambda ,H]}\left( t,x^{3},v\right) |\left( \delta
v\right) ^{2},  \label{eq65} \\
\delta v &=&dv+w_{\widehat{k}}\left( t,x^{3},v\right) dx^{\widehat{k}},\
\delta y^{5}=dy^{5}+n_{\widehat{k}}\left( t,x^{3},v\right) dx^{\widehat{k}},
\notag
\end{eqnarray}%
with coefficients of necessary smooth class, where\ \ $g_{\widehat{k}}\left(
x^{\widehat{i}}\right) $ is a solution of the 2D equation (\ref{ep1a}) for a
given source $\Upsilon _{4}\left( x^{\widehat{i}}\right) ,$%
\begin{equation*}
\varsigma _{\lbrack \lambda ,H]}\left( t,x^{3},v\right) =1+\frac{%
h_{0}^{2}(t,x^{3})}{16}\left[ \lambda _{\lbrack h]}^{2}\sin ^{2}\left(
\varrho _{2}t+\varrho _{3}x^{3}+\varphi _{\lbrack 0]}\right) -\frac{\lambda
_{\lbrack H]}^{2}}{4}\right] f^{2}\left( t,x^{3},v\right) ,
\end{equation*}%
and the N--connection coefficients $N_{\widehat{i}}^{4}=w_{\widehat{i}%
}(t,x^{3},v)$ and $N_{\widehat{i}}^{5}=n_{\widehat{i}}(t,x^{3},v)$ are
\begin{equation*}
w_{2,3}=-\frac{\partial _{2,3}\varsigma _{\lbrack \lambda ,H]}\left(
t,x^{3},v\right) }{\varsigma _{\lbrack \lambda ,H]}^{\ast }\left(
t,x^{3},v\right) }
\end{equation*}%
and
\begin{equation*}
n_{2,3}\left( t,x^{3},v\right) =n_{2,3[1]}\left( t,x^{3}\right)
+n_{2,3[2]}\left( t,x^{3}\right) \int \frac{\left[ f^{\ast }\left(
t,x^{3},v\right) \right] ^{2}}{\left[ f\left( t,x^{3},v\right) \right] ^{2}}%
\varsigma _{\lbrack \lambda ,H]}\left( t,x^{3},v\right) dv,
\end{equation*}%
define an exact 4D solution of the system of generalized Finsler--affine
gravity equations (\ref{fagfe}) for vanishing Proca mass $\mu \rightarrow 0,$
with holonomic and anholonomic variables, 1-form field
\begin{equation}
\phi _{\mu }=\left[ \phi _{\widehat{i}}=\phi _{\lbrack 0]\widehat{i}%
}(t,x^{3})\cos \left( \varrho _{2}t+\varrho _{3}x^{3}+\varphi _{\lbrack
0]}\right) ,\phi _{4}=0,\phi _{0}=0\right]  \label{aux71}
\end{equation}%
and nontrivial effective $H$--field $\mathbf{H}_{\nu \lambda \rho }=\lambda
_{\lbrack H]}\sqrt{|\mathbf{g}_{\alpha \beta }|}\varepsilon _{\nu \lambda
\rho }$ for arbitrary nontrivial functions $f\left( t,x^{3},v\right) $ (with
$f^{\ast }\neq 0),$ $h_{0}^{2}(t,x^{3})$, $n_{k[1,2]}\left( t,x^{3}\right) $
and sources
\begin{equation*}
\Upsilon _{2}\left( t,x^{3}\right) =\lambda _{\lbrack H]}^{2}/4-\lambda
_{\lbrack h]}^{2}(t,x^{3})\sin ^{2}\left( \varrho _{2}t+\varrho
_{3}x^{3}+\varphi _{\lbrack 0]}\right) \mbox{ and }\Upsilon _{4}=\lambda
_{\lbrack H]}^{2}/4
\end{equation*}%
and any integration constants to be defined by certain boundary conditions
and additional physical arguments. The function $\phi _{\lbrack 0]\widehat{i}%
}(t,x^{3})$ in (\ref{aux71}) is taken to solve the equation
\begin{equation}
\partial _{2}[e^{-\psi \left( t,x^{3}\right) }\partial _{2}\phi
_{k}]-\partial _{3}[e^{-\psi \left( t,x^{3}\right) }\partial _{3}\phi
_{k}]=L_{ki}^{j}\partial ^{i}\phi _{j}-L_{ij}^{i}\partial ^{j}\phi _{k}
\label{eq74}
\end{equation}%
where $L_{ki}^{j}$ are computed for the d--metric (\ref{eq65}) following the
formulas (\ref{2candcon}). For $\psi =0,$ we obtain just the plane wave
equation (\ref{waveeq}) when $\phi _{\lbrack 0]\widehat{i}}$ and $\lambda
_{\lbrack h]}^{2}(t,x^{3})$ reduce to constant values. We do not fix here
any value of $\psi \left( t,x^{3}\right) $ solving (\ref{aux73}) in order to
define explicitly a particular solution of (\ref{eq74}). We note that for
any value of $\psi \left( t,x^{3}\right) $ we can solve the inhomogeneous
wave equation (\ref{eq74}) by using solutions of the homogeneous case.

For simplicity, we do not present here the explicit value of $\sqrt{|\mathbf{%
g}_{\alpha \beta }|}$ computed for the d--metric (\ref{eq65}) as well the
values for distorsions $\widehat{\mathbf{Z}}_{\ \nu \lambda \rho },$ defined
by d--torsions of the canonical d--connection, see formulas (\ref{aux53})
and (\ref{2dtorsb}) (the formulas are very cumbersome and do not reflect
additional physical properties). Having defined $\widehat{\mathbf{Z}}_{\ \nu
\lambda \rho },$ we can compute
\begin{equation*}
\widehat{\mathbf{H}}_{\nu \lambda \rho }=\lambda _{\lbrack H]}\sqrt{|\mathbf{%
g}_{\alpha \beta }|}\varepsilon _{\nu \lambda \rho }-\widehat{\mathbf{Z}}_{\
\nu \lambda \rho }.
\end{equation*}%
We note that the torsion $\widehat{\mathbf{T}}_{\ \lambda \rho }^{\nu }$
contained in $\widehat{\mathbf{Z}}_{\ \nu \lambda \rho },$ related to string
corrections by the $H$--field, is different from the torsion $\mathbf{T=}%
k_{2}\mathbf{\phi }$ and nontrivial nonmetricity $\mathbf{Q}=k_{0}\mathbf{%
\phi ,\ \Lambda =}k_{1}\mathbf{\phi ,\ }$from the metric--affine part of the
theory, see (\ref{triplemag}).

We can choose the function $f\left( t,x^{3},v\right) $ from (\ref{eq65}), or
(\ref{dm5}), as it would be a solution of the Kadomtsev--Petviashvili (KdP)
equation \cite{02kdp}, i. e. to satisfy
\begin{equation*}
f^{\bullet \bullet }+\epsilon \left( f^{\prime}+6ff^{\ast }+f^{\ast \ast
\ast }\right) ^{\ast }=0,\ \epsilon =\pm 1,
\end{equation*}%
or, for another locally anisotropic background, to satisfy the $(2+1)$%
--dimensional sine--Gordon (SG) equation,
\begin{equation*}
-f^{\bullet \bullet }+f%
%TCIMACRO{\U{b4}}%
%BeginExpansion
{\acute{}}%
%EndExpansion
%TCIMACRO{\U{b4}}%
%BeginExpansion
{\acute{}}%
%EndExpansion
+f^{\ast \ast }=\sin f,
\end{equation*}%
see Refs. \cite{02soliton} on gravitational solitons and theory of solitons.
In this case, we define a nonlinear model of gravitational plane wave and 3D
solitons in the framework of the MAG with string corrections by $H$--field.
Such solutions generalized those considered in Refs. \cite{02v1} for 4D and 5D
gravity.

We can also consider that $F/L=f^{2}\left( t,x^{3},v\right) $ is just the
generation function for a 2D model of Finsler/Lagrange geometry (being of
any solitonic or another type nature). In this case, the geometric
background is characterized by this type locally anisotropic configurations
(for Finsler metrics we shall impose corresponding homogeneity conditions on
coordinates).

\section{ Final Remarks}

In this paper we have investigated the dynamical aspects of metric--affine
gravity (MAG) with certain additional string corrections defined by the
antisymmetric $H$--field when the metric structure is generic off--diagonal
and the spacetime is provided with an anholonomic frame structure with
associated nonlinear connection (N--connection). We analyzed the
corresponding class of Lagrangians and derived the field equations of MAG
and string gravity with mixed holonomic and anholonomic variables. The main
motivation for this work is to determine the place and significance of such
models of gravity which contain as exact solutions certain classes of
metrics and connections modelling Finsler like geometries even in the limits
to the general relativity theory.

The work supports the results of Refs. \cite{02v1,02v1a} where various classes
of exact solutions in Einstein, Einstein--Cartan, gauge and string gravity
modelling Finsler--Lagrange configurations were constructed. We provide an
irreducible decomposition techniques (in our case with additional
N--connection splitting) and study the dynamics of MAG fields generating the
locally anisotropic geometries and interactions classified in Ref. \cite{02vp1}%
. There are proved the main theorems on irreducible reduction to effective
Einstein--Proca equations with string corrections and formulated a new
method of constructing exact solutions.

As explicit examples of the new type of locally anisotropic configurations
in MAG and string gravity, we have elaborated three new classes of exact
solutions depending on 3-4 variables possessing nontrivial torsion and
nonmetricity fields, describing plane wave and three dimensional soliton
interactions and induced generalized Finsler--affine effective
configurations.

Finally, it seems worthwhile to note that such Finsler like configurations
do not violates the postulates of the general relativity theory in the
corresponding limits to the four dimensional Einstein theory because such
metrics transform into exact solutions of this theory. The anisotropies are
modelled by certain anholonomic frame constraints on a (pseudo) Riemannian
spacetime. In this case the restrictions imposed on physical applications of
the Finsler geometry, derived from experimental data on possible limits for
brocken local Lorentz invariance (see, for instance, Ref. \cite{02will}), do
not hold.

%\appendix

\section{Appendix A:\ Proof of Theorem \ref{t5dr}}

\bigskip \label{appa}We give some details on straightforward calculations
outlined in Ref. \cite{02vmethod} for (pseudo) Riemannian and Riemann--Cartan
spaces. In brief, the proof of Theorem \ref{t5dr} is to be performed in this
way:\ Introducing $N_{i}^{4}=w_{i}$ and $N_{i}^{5}=n_{i}$ in (\ref{2dder})
and (\ref{2ddif}) and re--writing (\ref{ansatz5}) into a diagonal (in our
case) block form (\ref{2block2}), we compute the h- and v--irreducible
components of the canonical d--connection (\ref{2candcon}). The next step is
to compute d--curvatures (\ref{2dcurv}) and by contracting of indices to
define the components of the Ricci d--tensor (\ref{2dricci}) which results in
(\ref{1ricci1a})--(\ref{1ricci4a}). We emphasize that such computations can
not be performed directly by applying any Tensor, Maple of Mathematica
macros because, in our case, we consider canonical d--connections instead of
the\ Levi--Civita connection \cite{02cartanmacros}. We give the details of
such calculus related to N--adapted anholonomic frames.

The five dimensional (5D) local coordinates are $x^{i}$ and $y^{a}=\left(
v,y\right) ,$ i. e. $y^{4}=v,$ $y^{5}=y,$ were indices $i,j,k...=1,2,3$ and $%
a,b,c,...=4,5.$ Our reductions to 4D will be considered by excluding
dependencies on the variable $x^{1}$ and for trivial embedding of 4D
off--diagonal ansatz into 5D ones. The signatures of metrics could be
arbitrary ones. In general, the spacetime could be with torsion, but we
shall always be interested to define the limits to (pseudo) Riemannian
spaces.

The d--metric (\ref{2block2}) for an ansatz (\ref{ansatz5}) with $%
g_{1}=const, $ is written
\begin{eqnarray}
\delta s^{2}&=&g_{1}{(dx^{1})}^{2}+g_{2}\left((x^{2},x^{3}\right) {(dx^{2})}%
^{2}+g_{3}\left( x^{k}\right){(dx^{3})}^{2}     \notag \\
&& +h_{4}\left( x^{k},v\right) {%
(\delta v)}^{2}+h_{5}\left( x^{k},v\right){(\delta y)}^{2},  \notag \\
\delta v &=&dv+w_{i}\left( x^{k},v\right) dx^{i},\ \delta y=dy+n_{i}\left(
x^{k},v\right) dx^{i}  \label{ans4d}
\end{eqnarray}%
when the generic off--diagonal metric (\ref{metric5}) is associated to a
N--connection structure $N_{i}^{a}$ with $N_{i}^{4}=w_{i}\left(
x^{k},v\right) $ and $N_{i}^{5}=n_{i}\left( x^{k},v\right) .$ We note that
the metric (\ref{ans4d}) does not depend on variable $y^{5}=y,$ but
emphasize the dependence on ''anisotropic'' variable $y^{4}=v.$

If we regroup (\ref{ans4d}) with respect to true differentials $du^{\alpha
}=\left( dx^{i},dy^{a}\right) $ we obtain just the ansatz (\ref{ansatz5}).
It is a cumbersome task to perform tensor calculations (for instance, of
curvature and Ricci tensors) with such generic off--diagonal ansatz but the
formulas simplify substantially with respect to N--adapted frames of type(%
\ref{2dder}) and (\ref{2ddif}) and for effectively diagonalized metrics like (%
\ref{ans4d}).

So, the \ metric (\ref{metric5}) transform in a diagonal one with respect to
the pentads (frames, funfbeins)%
\begin{equation}
e^{i}=dx^{i},e^{4}=\delta v=dv+w_{i}\left( x^{k},v\right)
dx^{i},e^{5}=\delta y=dy+n_{i}\left( x^{k},v\right) dx^{i}  \label{ddif4a}
\end{equation}%
or
\begin{equation*}
\delta u^{\alpha }=\left( dx^{i},\delta y^{a}=dy^{a}+N_{i}^{a}dx^{i}\right)
\end{equation*}%
being dual to the N--elongated partial derivative operators,
\begin{eqnarray}
e_{1} &=&\delta _{1}=\frac{\partial }{\partial x^{1}}-N_{1}^{a}\frac{%
\partial }{\partial y^{a}}=\frac{\partial }{\partial x^{1}}-w_{1}\frac{%
\partial }{\partial v}-n_{1}\frac{\partial }{\partial y},  \label{dder4a} \\
e_{2} &=&\delta _{2}=\frac{\partial }{\partial x^{2}}-N_{2}^{a}\frac{%
\partial }{\partial y^{a}}=\frac{\partial }{\partial x^{2}}-w_{2}\frac{%
\partial }{\partial v}-n_{2}\frac{\partial }{\partial y},  \notag \\
e_{3} &=&\delta _{3}=\frac{\partial }{\partial x^{3}}-N_{3}^{a}\frac{%
\partial }{\partial y^{a}}=\frac{\partial }{\partial x^{3}}-w_{3}\frac{%
\partial }{\partial v}-n_{3}\frac{\partial }{\partial y},  \notag \\
e_{4} &=&\frac{\partial }{\partial y^{4}}=\frac{\partial }{\partial v},\
e_{5}=\frac{\partial }{\partial y^{5}}=\frac{\partial }{\partial y}  \notag
\end{eqnarray}%
when $\delta _{\alpha }=\frac{\delta }{\partial u^{\alpha }}=\left( \frac{%
\delta }{\partial x^{i}}=\frac{\partial }{\partial x^{i}}-N_{i}^{a}\frac{%
\partial }{\partial y^{a}},\frac{\partial }{\partial y^{b}}\right) .$

The N--elongated partial derivatives of a function $f\left( u^{\alpha
}\right) =f\left( x^{i},y^{a}\right) =f\left( x,r,v,y\right) $ are computed
in the form when the N--elongated derivatives are%
\begin{equation*}
\delta _{2}f=\frac{\delta f}{\partial u^{2}}=\frac{\delta f}{\partial x^{2}}=%
\frac{\delta f}{\partial x}=\frac{\partial f}{\partial x}-N_{2}^{a}\frac{%
\partial f}{\partial y^{a}}=\frac{\partial f}{\partial x}-w_{2}\frac{%
\partial f}{\partial v}-n_{2}\frac{\partial f}{\partial y}=f^{\bullet
}-w_{2}\ f^{\prime }-n_{2}\ f^{\ast }
\end{equation*}%
where
\begin{equation*}
f^{\bullet }=\frac{\partial f}{\partial x^{2}}=\frac{\partial f}{\partial x}%
,\ f^{\prime }=\frac{\partial f}{\partial x^{3}}=\frac{\partial f}{\partial r%
},\ f^{\ast }=\frac{\partial f}{\partial y^{4}}=\frac{\partial f}{\partial v}%
.
\end{equation*}%
The N--elongated differential is
\begin{equation*}
\delta f=\frac{\delta f}{\partial u^{\alpha }}\delta u^{\alpha }.
\end{equation*}%
The N--elongated differential calculus should be applied if we work with
respect to N--adapted frames.

\subsection{Calculation of N--connection curvature}

We compute the coefficients (\ref{2ncurv}) for the d--metric (\ref{ans4d})
(equivalently, the ansatz (\ref{ansatz5})) defining the curvature of
N--connection $N_{i}^{a},$ by substituting $N_{i}^{4}=w_{i}\left(
x^{k},v\right) $ and $N_{i}^{5}=n_{i}\left( x^{k},v\right) ,$ where $i=2,3$
and $a=4,5.$ The result for nontrivial values is
\begin{eqnarray}
\Omega _{23}^{4} &=&-\Omega _{23}^{4}=w_{2}^{\prime }-w_{3}^{\cdot
}-w_{3}w_{2}^{\ast }-w_{2}w_{3}^{\ast },  \label{omega} \\
\Omega _{23}^{5} &=&-\Omega _{23}^{5}=n_{2}^{\prime }-n_{3}^{\cdot
}-w_{3}n_{2}^{\ast }-w_{2}n_{3}^{\ast }.  \notag
\end{eqnarray}%
The canonical d--connection $\widehat{\mathbf{\Gamma }}_{\ \alpha \beta
}^{\gamma }=\left( \widehat{L}_{jk}^{i},\widehat{L}_{bk}^{a},\widehat{C}%
_{jc}^{i},\widehat{C}_{bc}^{a}\right) $ (\ref{2candcon}) defines the
covariant derivative $\widehat{\mathbf{D}},$ satisfying the metricity
conditions $\widehat{\mathbf{D}}_{\alpha }\mathbf{g}_{\gamma \delta }=0$ for
$\mathbf{g}_{\gamma \delta }$ being the metric (\ref{ans4d}) with the
coefficients written with respect to N--adapted frames. $\widehat{\mathbf{%
\Gamma }}_{\ \alpha \beta }^{\gamma }$ has nontrivial d--torsions.

We compute the Einstein tensors for the canonical d--connection $\widehat{%
\mathbf{\Gamma }}_{\ \alpha \beta }^{\gamma }$ defined by the ansatz (\ref%
{ans4d}) with respect to N--adapted frames (\ref{ddif4a}) and (\ref{dder4a}%
). This results in exactly integrable vacuum Einstein equations and certain
type of sources. Such solutions could be with nontrivial torsion for
different classes of linear connections from Riemann--Cartan and generalized
Finsler geometries. So, the anholonomic frame method offers certain
possibilities to be extended to in string gravity where the torsion could be
not zero. But we can always select the limit to Levi--Civita connections, i.
e. to (pseudo) Riemannian spaces by considering additional constraints, see
Corollary \ref{corcond1} and/or conditions (\ref{1cond2}).

\subsection{Calculation of the canonical d--connection}

We compute the coefficients (\ref{2candcon}) for the d--metric (\ref{ans4d})
(equivalently, the ansatz (\ref{ansatz5})) when $g_{jk}=\{g_{j}\}$ and $%
h_{bc}=\{h_{b}\}$ are diagonal and $g_{ik}$ depend only on $x^{2}$ and $%
x^{3} $ but not on $y^{a}.$

We have
\begin{eqnarray}
\delta _{k}g_{ij} &=&\partial _{k}g_{ij}-w_{k}g_{ij}^{\ast }=\partial
_{k}g_{ij},\ \delta _{k}h_{b}=\partial _{k}h_{b}-w_{k}h_{b}^{\ast }
\label{1aux01} \\
\delta _{k}w_{i} &=&\partial _{k}w_{i}-w_{k}w_{i}^{\ast },\ \delta
_{k}n_{i}=\partial _{k}n_{i}-w_{k}n_{i}^{\ast }  \notag
\end{eqnarray}%
resulting in formulas
\begin{equation*}
\widehat{L}_{jk}^{i}=\frac{1}{2}g^{ir}\left( \frac{\delta g_{jk}}{\delta
x^{k}}+\frac{\delta g_{kr}}{\delta x^{j}}-\frac{\delta g_{jk}}{\delta x^{r}}%
\right) =\frac{1}{2}g^{ir}\left( \frac{\partial g_{jk}}{\delta x^{k}}+\frac{%
\partial g_{kr}}{\delta x^{j}}-\frac{\partial g_{jk}}{\delta x^{r}}\right)
\end{equation*}%
The nontrivial values of $\widehat{L}_{jk}^{i}$ are%
\begin{eqnarray}
\widehat{L}_{22}^{2} &=&\frac{g_{2}^{\bullet }}{2g_{2}}=\alpha _{2}^{\bullet
},\ \widehat{L}_{23}^{2}=\frac{g_{2}^{\prime }}{2g_{2}}=\alpha
_{2}^{^{\prime }},\ \widehat{L}_{33}^{2}=-\frac{g_{3}^{\bullet }}{2g_{2}}
\label{1aux02} \\
\widehat{L}_{22}^{3} &=&-\frac{g_{2}^{\prime }}{2g_{3}},\ \widehat{L}%
_{23}^{3}=\frac{g_{3}^{\bullet }}{2g_{3}}=\alpha _{3}^{\bullet },\ \
\widehat{L}_{33}^{3}=\frac{g_{3}^{\prime }}{2g_{3}}=\alpha _{3}^{^{\prime }}.
\notag
\end{eqnarray}%
In a similar form we compute the components%
\begin{equation*}
\widehat{L}_{bk}^{a}=\partial _{b}N_{k}^{a}+\frac{1}{2}h^{ac}\left( \partial
_{k}h_{bc}-N_{k}^{d}\frac{\partial h_{bc}}{\partial y^{d}}-h_{dc}\partial
_{b}N_{k}^{d}-h_{db}\partial _{c}N_{k}^{d}\right)
\end{equation*}%
having nontrivial values
\begin{eqnarray}
\widehat{L}_{42}^{4} &=&\frac{1}{2h_{4}}\left( h_{4}^{\bullet
}-w_{2}h_{4}^{\ast }\right) =\delta _{2}\ln \sqrt{|h_{4}|}\doteqdot \delta
_{2}\beta _{4},  \label{aux02a} \\
\widehat{L}_{43}^{4} &=&\frac{1}{2h_{4}}\left( h_{4}^{^{\prime
}}-w_{3}h_{4}^{\ast }\right) =\delta _{3}\ln \sqrt{|h_{4}|}\doteqdot \delta
_{3}\beta _{4}  \notag
\end{eqnarray}%
\begin{equation}
\widehat{L}_{5k}^{4}=-\frac{h_{5}}{2h_{4}}n_{k}^{\ast },\ \widehat{L}%
_{bk}^{5}=\partial _{b}n_{k}+\frac{1}{2h_{5}}\left( \partial
_{k}h_{b5}-w_{k}h_{b5}^{\ast }-h_{5}\partial _{b}n_{k}\right) ,
\label{aux02b}
\end{equation}

\begin{eqnarray}
\widehat{L}_{4k}^{5} &=&n_{k}^{\ast }+\frac{1}{2h_{5}}\left(
-h_{5}n_{k}^{\ast }\right) =\frac{1}{2}n_{k}^{\ast },  \label{aux02c} \\
\widehat{L}_{5k}^{5} &=&\frac{1}{2h_{5}}\left( \partial
_{k}h_{5}-w_{k}h_{5}^{\ast }\right) =\delta _{k}\ln \sqrt{|h_{4}|}=\delta
_{k}\beta _{4}.  \notag
\end{eqnarray}

We note that
\begin{equation}
\widehat{C}_{jc}^{i}=\frac{1}{2}g^{ik}\frac{\partial g_{jk}}{\partial y^{c}}%
\doteqdot 0  \label{aux02cc}
\end{equation}%
because $g_{jk}=g_{jk}\left( x^{i}\right) $ for the considered ansatz.

The values
\begin{equation*}
\widehat{C}_{bc}^{a}=\frac{1}{2}h^{ad}\left( \frac{\partial h_{bd}}{\partial
y^{c}}+\frac{\partial h_{cd}}{\partial y^{b}}-\frac{\partial h_{bc}}{%
\partial y^{d}}\right)
\end{equation*}%
for $h_{bd}=[h_{4},h_{5}]$ from the ansatz (\ref{ansatz5}) have nontrivial
components
\begin{equation}
\widehat{C}_{44}^{4}=\frac{h_{4}^{\ast }}{2h_{4}}\doteqdot \beta _{4}^{\ast
},\widehat{C}_{55}^{4}=-\frac{h_{5}^{\ast }}{2h_{4}},\widehat{C}_{45}^{5}=%
\frac{h_{5}^{\ast }}{2h_{5}}\doteqdot \beta _{5}^{\ast }.  \label{aux02d}
\end{equation}

The set of formulas (\ref{1aux02})--(\ref{aux02d}) define the nontrivial
coefficients of the canonical d--connection $\widehat{\mathbf{\Gamma }}_{\
\alpha \beta }^{\gamma }=\left( \widehat{L}_{jk}^{i},\widehat{L}_{bk}^{a},%
\widehat{C}_{jc}^{i},\widehat{C}_{bc}^{a}\right) $ (\ref{2candcon}) for the
5D ansatz (\ref{ans4d}).

\subsection{Calculation of torsion coefficients}

We should put the nontrivial values (\ref{1aux02})-- (\ref{aux02d}) into the
formulas for d--torsion \ (\ref{2dtorsb}).

One holds $T_{.jk}^{i}=0$ and $T_{.bc}^{a}=0,$ because of symmetry of
coefficients $L_{jk}^{i}$ and $C_{bc}^{a}.$

We have computed the nontrivial values of $\Omega _{.ji}^{a},$ see \ (\ref%
{omega}) resulting in%
\begin{eqnarray}
T_{23}^{4} &=&\Omega _{23}^{4}=-\Omega _{23}^{4}=w_{2}^{\prime
}-w_{3}^{\bullet }-w_{3}w_{2}^{\ast }-w_{2}w_{3}^{\ast },  \label{aux11} \\
T_{23}^{5} &=&\Omega _{23}^{5}=-\Omega _{23}^{5}=n_{2}^{\prime
}-n_{3}^{\bullet }-w_{3}n_{2}^{\ast }-w_{2}n_{3}^{\ast }.  \notag
\end{eqnarray}

One follows
\begin{equation*}
T_{ja}^{i}=-T_{aj}^{i}=C_{.ja}^{i}=\widehat{C}_{jc}^{i}=\frac{1}{2}g^{ik}%
\frac{\partial g_{jk}}{\partial y^{c}}\doteqdot 0,
\end{equation*}
see (\ref{aux02cc}).

For the components
\begin{equation*}
T_{.bi}^{a}=-T_{.ib}^{a}=P_{.bi}^{a}=\frac{\partial N_{i}^{a}}{\partial y^{b}%
}-L_{.bj}^{a},
\end{equation*}
i. e. for
\begin{equation*}
\widehat{P}_{.bi}^{4}=\frac{\partial N_{i}^{4}}{\partial y^{b}}-\widehat{L}%
_{.bj}^{4}=\partial _{b}w_{i}-\widehat{L}_{.bj}^{4}\mbox{ and }\widehat{P}%
_{.bi}^{5}=\frac{\partial N_{i}^{5}}{\partial y^{b}}-\widehat{L}%
_{.bj}^{5}=\partial _{b}n_{i}-\widehat{L}_{.bj}^{5},
\end{equation*}
we have the nontrivial values%
\begin{eqnarray}
\widehat{P}_{.4i}^{4} &=&w_{i}^{\ast }-\frac{1}{2h_{4}}\left( \partial
_{i}h_{4}-w_{i}h_{4}^{\ast }\right) =w_{i}^{\ast }-\delta _{i}\beta _{4},\
\widehat{P}_{.5i}^{4}=\frac{h_{5}}{2h_{4}}n_{i}^{\ast },  \notag \\
\ \widehat{P}_{.4i}^{5} &=&\frac{1}{2}n_{i}^{\ast },\ \widehat{P}_{.5i}^{5}=-%
\frac{1}{2h_{5}}\left( \partial _{i}h_{5}-w_{i}h_{5}^{\ast }\right) =-\delta
_{i}\beta _{5}.  \label{aux12}
\end{eqnarray}

The formulas\ (\ref{aux11}) and (\ref{aux12}) state the \ nontrivial
coefficients of the canonical d--connection for the chosen ansatz (\ref%
{ans4d}).

\subsection{Calculation of the Ricci tensor}

Let us compute the value $R_{ij}=R_{\ ijk}^{k}$ as in (\ref{2dricci}) for
\begin{equation*}
R_{\ hjk}^{i}=\frac{\delta L_{.hj}^{i}}{\delta x^{k}}-\frac{\delta
L_{.hk}^{i}}{\delta x^{j}}%
+L_{.hj}^{m}L_{mk}^{i}-L_{.hk}^{m}L_{mj}^{i}-C_{.ha}^{i}\Omega _{.jk}^{a},
\end{equation*}%
from (\ref{2dcurv}). It should be noted that $C_{.ha}^{i}=0$ for the ansatz
under consideration, see (\ref{aux02cc}). We compute

\begin{equation*}
\frac{\delta L_{.hj}^{i}}{\delta x^{k}}=\partial
_{k}L_{.hj}^{i}+N_{k}^{a}\partial _{a}L_{.hj}^{i}=\partial
_{k}L_{.hj}^{i}+w_{k}\left( L_{.hj}^{i}\right) ^{\ast }=\partial
_{k}L_{.hj}^{i}
\end{equation*}%
because $L_{.hj}^{i}$ do not depend on variable $y^{4}=v.$

Derivating (\ref{1aux02}), we obtain
\begin{eqnarray*}
\partial _{2}L_{\ 22}^{2} &=&\frac{g_{2}^{\bullet \bullet }}{2g_{2}}-\frac{%
\left( g_{2}^{\bullet }\right) ^{2}}{2\left( g_{2}\right) ^{2}},\ \partial
_{2}L_{\ 23}^{2}=\frac{g_{2}^{\bullet ^{\prime }}}{2g_{2}}-\frac{%
g_{2}^{\bullet }g_{2}^{^{\prime }}}{2\left( g_{2}\right) ^{2}},\ \partial
_{2}L_{\ 33}^{2}=-\frac{g_{3}^{\bullet \bullet }}{2g_{2}}+\frac{%
g_{2}^{\bullet }g_{3}^{\bullet }}{2\left( g_{2}\right) ^{2}}, \\
\ \partial _{2}L_{\ 22}^{3} &=&-\frac{g_{2}^{\bullet ^{\prime }}}{2g_{3}}+%
\frac{g_{2}^{\bullet }g_{3}^{^{\prime }}}{2\left( g_{3}\right) ^{2}}%
,\partial _{2}L_{\ 23}^{3}=\frac{g_{3}^{\bullet \bullet }}{2g_{3}}-\frac{%
\left( g_{3}^{\bullet }\right) ^{2}}{2\left( g_{3}\right) ^{2}},\ \partial
_{2}L_{\ 33}^{3}=\frac{g_{3}^{\bullet ^{\prime }}}{2g_{3}}-\frac{%
g_{3}^{\bullet }g_{3}^{^{\prime }}}{2\left( g_{3}\right) ^{2}}, \\
\partial _{3}L_{\ 22}^{2} &=&\frac{g_{2}^{\bullet ^{\prime }}}{2g_{2}}-\frac{%
g_{2}^{\bullet }g_{2}^{^{\prime }}}{2\left( g_{2}\right) ^{2}},\partial
_{3}L_{\ 23}^{2}=\frac{g_{2}^{ll}}{2g_{2}}-\frac{\left( g_{2}^{l}\right) ^{2}%
}{2\left( g_{2}\right) ^{2}},\partial _{3}L_{\ 33}^{2}=-\frac{g_{3}^{\bullet
^{\prime }}}{2g_{2}}+\frac{g_{3}^{\bullet }g_{2}^{^{\prime }}}{2\left(
g_{2}\right) ^{2}}, \\
\ \partial _{3}L_{\ 22}^{3} &=&-\frac{g_{2}^{^{\prime \prime }}}{2g_{3}}+%
\frac{g_{2}^{\bullet }g_{2}^{^{\prime }}}{2\left( g_{3}\right) ^{2}}%
,\partial _{3}L_{\ 23}^{3}=\frac{g_{3}^{\bullet ^{\prime }}}{2g_{3}}-\frac{%
g_{3}^{\bullet }g_{3}^{^{\prime }}}{2\left( g_{3}\right) ^{2}},\partial
_{3}L_{\ 33}^{3}=\frac{g_{3}^{ll}}{2g_{3}}-\frac{\left( g_{3}^{l}\right) ^{2}%
}{2\left( g_{3}\right) ^{2}}.
\end{eqnarray*}

For these values and (\ref{1aux02}), there are only 2 nontrivial components,

\begin{eqnarray*}
R_{\ 323}^{2} &=&\frac{g_{3}^{\bullet \bullet }}{2g_{2}}-\frac{%
g_{2}^{\bullet }g_{3}^{\bullet }}{4\left( g_{2}\right) ^{2}}-\frac{\left(
g_{3}^{\bullet }\right) ^{2}}{4g_{2}g_{3}}+\frac{g_{2}^{ll}}{2g_{2}}-\frac{%
g_{2}^{l}g_{3}^{l}}{4g_{2}g_{3}}-\frac{\left( g_{2}^{l}\right) ^{2}}{4\left(
g_{2}\right) ^{2}} \\
R_{\ 223}^{3} &=&-\frac{g_{3}^{\bullet \bullet }}{2g_{3}}+\frac{%
g_{2}^{\bullet }g_{3}^{\bullet }}{4g_{2}g_{3}}+\frac{\left( g_{3}^{\bullet
}\right) ^{2}}{4(g_{3})^{2}}-\frac{g_{2}^{ll}}{2g_{3}}+\frac{%
g_{2}^{l}g_{3}^{l}}{4(g_{3})^{2}}+\frac{\left( g_{2}^{l}\right) ^{2}}{%
4g_{2}g_{3}}
\end{eqnarray*}%
with%
\begin{equation*}
R_{22}=-R_{\ 223}^{3}\mbox{ and }R_{33}=R_{\ 323}^{2},
\end{equation*}%
or%
\begin{equation*}
R_{2}^{2}=R_{3}^{3}=-\frac{1}{2g_{2}g_{3}}\left[ g_{3}^{\bullet \bullet }-%
\frac{g_{2}^{\bullet }g_{3}^{\bullet }}{2g_{2}}-\frac{\left( g_{3}^{\bullet
}\right) ^{2}}{2g_{3}}+g_{2}^{\prime \prime }-\frac{g_{2}^{l}g_{3}^{l}}{%
2g_{3}}-\frac{\left( g_{2}^{l}\right) ^{2}}{2g_{2}}\right]
\end{equation*}%
which is (\ref{1ricci1a}).

Now, we consider
\begin{eqnarray*}
P_{\ bka}^{c} &=&\frac{\partial L_{.bk}^{c}}{\partial y^{a}}-\left( \frac{%
\partial C_{.ba}^{c}}{\partial x^{k}}+L_{.dk}^{c%
\,}C_{.ba}^{d}-L_{.bk}^{d}C_{.da}^{c}-L_{.ak}^{d}C_{.bd}^{c}\right)
+C_{.bd}^{c}P_{.ka}^{d} \\
&=&\frac{\partial L_{.bk}^{c}}{\partial y^{a}}%
-C_{.ba|k}^{c}+C_{.bd}^{c}P_{.ka}^{d}
\end{eqnarray*}%
from (\ref{2dcurv}). Contracting indices, we have%
\begin{equation*}
R_{bk}=P_{\ bka}^{a}=\frac{\partial L_{.bk}^{a}}{\partial y^{a}}%
-C_{.ba|k}^{a}+C_{.bd}^{a}P_{.ka}^{d}
\end{equation*}%
Let us denote $C_{b}=C_{.ba}^{c}$ and write%
\begin{equation*}
C_{.b|k}=\delta _{k}C_{b}-L_{\ bk}^{d\,}C_{d}=\partial
_{k}C_{b}-N_{k}^{e}\partial _{e}C_{b}-L_{\ bk}^{d\,}C_{d}=\partial
_{k}C_{b}-w_{k}C_{b}^{\ast }-L_{\ bk}^{d\,}C_{d}.
\end{equation*}%
We express%
\begin{equation*}
R_{bk}=\ _{[1]}R_{bk}+\ _{[2]}R_{bk}+\ _{[3]}R_{bk}
\end{equation*}%
where%
\begin{eqnarray*}
\ _{[1]}R_{bk} &=&\left( L_{bk}^{4}\right) ^{\ast }, \\
\ _{[2]}R_{bk} &=&-\partial _{k}C_{b}+w_{k}C_{b}^{\ast }+L_{\ bk}^{d\,}C_{d},
\\
\ _{[3]}R_{bk} &=&C_{.bd}^{a}P_{.ka}^{d}=C_{.b4}^{4}P_{.k4}^{4}
+C_{.b5}^{4}P_{.k4}^{5}+C_{.b4}^{5}P_{.k5}^{4}+C_{.b5}^{5}P_{.k5}^{5}
\end{eqnarray*}%
and
\begin{eqnarray}
C_{4} &=&C_{44}^{4}+C_{45}^{5}=\frac{h_{4}^{\ast }}{2h_{4}}+\frac{%
h_{5}^{\ast }}{2h_{5}}=\beta _{4}^{\ast }+\beta _{5}^{\ast },  \label{aux6a}
\\
C_{5} &=&C_{54}^{4}+C_{55}^{5}=0  \notag
\end{eqnarray}%
see(\ref{aux02d}) .

We compute%
\begin{equation*}
R_{4k}=\ _{[1]}R_{4k}+\ _{[2]}R_{4k}+\ _{[3]}R_{4k}
\end{equation*}%
with
\begin{eqnarray*}
\ _{[1]}R_{4k} &=&\left( L_{4k}^{4}\right) ^{\ast }=\left( \delta _{k}\beta
_{4}\right) ^{\ast } \\
\ _{[2]}R_{4k} &=&-\partial _{k}C_{4}+w_{k}C_{4}^{\ast }+L_{\
4k}^{4\,}C_{4},L_{\ 4k}^{4\,}=\delta _{k}\beta _{4}\mbox{ see
(\ref{aux02a}} \\
&=&-\partial _{k}\left( \beta _{4}^{\ast }+\beta _{5}^{\ast }\right)
+w_{k}\left( \beta _{4}^{\ast }+\beta _{5}^{\ast }\right) ^{\ast }+L_{\
4k}^{4\,}\left( \beta _{4}^{\ast }+\beta _{5}^{\ast }\right) \\
\ _{[3]}R_{4k}
&=&C_{.44}^{4}P_{.k4}^{4}+C_{.45}^{4}P_{.k4}^{5}+C_{.44}^{5}P_{.k5}^{4}+
C_{.45}^{5}P_{.k5}^{5}
\\
&=&\beta _{4}^{\ast }\left( w_{k}^{\ast }-\delta _{k}\beta _{4}\right)
-\beta _{5}^{\ast }\delta _{k}\beta _{5}
\end{eqnarray*}%
Summarizing, we get%
\begin{equation*}
R_{4k}=w_{k}\left[ \beta _{5}^{\ast \ast }+\left( \beta _{5}^{\ast }\right)
^{2}-\beta _{4}^{\ast }\beta _{5}^{\ast }\right] +\beta _{5}^{\ast }\partial
_{k}\left( \beta _{4}+\beta _{5}\right) -\partial _{k}\beta _{5}^{\ast }
\end{equation*}%
or, for
\begin{equation*}
\beta _{4}^{\ast }=\frac{h_{4}^{\ast }}{2h_{4}},\partial _{k}\beta _{4}=%
\frac{\partial _{k}h_{4}}{2h_{4}},\beta _{5}^{\ast }=\frac{h_{5}^{\ast }}{%
2h_{5}},\beta _{5}^{\ast \ast }=\frac{h_{5}^{\ast \ast }h_{5}-\left(
h_{5}^{\ast }\right) ^{2}}{2\left( h_{5}\right) ^{5}},
\end{equation*}%
we can write%
\begin{equation*}
2h_{5}R_{4k}=w_{k}\left[ h_{5}^{\ast \ast }-\frac{\left( h_{5}^{\ast
}\right) ^{2}}{2h_{5}}-\frac{h_{4}^{\ast }h_{5}^{\ast }}{2h_{4}}\right] +%
\frac{h_{5}^{\ast }}{2}\left( \frac{\partial _{k}h_{4}}{h_{4}}+\frac{%
\partial _{k}h_{5}}{h_{5}}\right) -\partial _{k}h_{5}^{\ast }
\end{equation*}%
which is equivalent to (\ref{1ricci3a})

In a similar way, we compute%
\begin{equation*}
R_{5k}=\ _{[1]}R_{5k}+\ _{[2]}R_{5k}+\ _{[3]}R_{5k}
\end{equation*}%
with
\begin{eqnarray*}
\ _{[1]}R_{5k} &=&\left( L_{5k}^{4}\right) ^{\ast }, \\
\ _{[2]}R_{5k} &=&-\partial _{k}C_{5}+w_{k}C_{5}^{\ast }+L_{\ 5k}^{4\,}C_{4},
\\
\ _{[3]}R_{5k}
&=&C_{.54}^{4}P_{.k4}^{4}+C_{.55}^{4}P_{.k4}^{5}+C_{.54}^{5}P_{.k5}^{4}+
C_{.55}^{5}P_{.k5}^{5}.
\end{eqnarray*}%
We have
\begin{eqnarray*}
R_{5k} &=&\left( L_{5k}^{4}\right) ^{\ast }+L_{\
5k}^{4\,}C_{4}+C_{.55}^{4}P_{.k4}^{5}+C_{.54}^{5}P_{.k5}^{4} \\
&=&\left( -\frac{h_{5}}{h_{4}}n_{k}^{\ast }\right) ^{\ast }-\frac{h_{5}}{%
h_{4}}n_{k}^{\ast }\left( \frac{h_{4}^{\ast }}{2h_{4}}+\frac{h_{5}^{\ast }}{%
2h_{5}}\right) +\frac{h_{5}^{\ast }}{2h_{5}}\frac{h_{5}}{2h_{4}}n_{k}^{\ast
}-\frac{h_{5}^{\ast }}{2h_{4}}\frac{1}{2}n_{k}^{\ast }
\end{eqnarray*}%
which can be written%
\begin{equation*}
2h_{4}R_{5k}=h_{5}n_{k}^{\ast \ast }+\left( \frac{h_{5}}{h_{4}}h_{4}^{\ast }-%
\frac{3}{2}h_{5}^{\ast }\right) n_{k}^{\ast }
\end{equation*}%
i. e. (\ref{1ricci4a})

For the values
\begin{equation*}
P_{\ jka}^{i}=\frac{\partial L_{.jk}^{i}}{\partial y^{k}}-\left( \frac{%
\partial C_{.ja}^{i}}{\partial x^{k}}%
+L_{.lk}^{i}C_{.ja}^{l}-L_{.jk}^{l}C_{.la}^{i}-L_{.ak}^{c}C_{.jc}^{i}\right)
+C_{.jb}^{i}P_{.ka}^{b}
\end{equation*}%
from (\ref{2dcurv}), we obtain zeros because $C_{.jb}^{i}=0$ and $L_{.jk}^{i}$
do not depend on $y^{k}.$ So,
\begin{equation*}
R_{ja}=P_{\ jia}^{i}=0.
\end{equation*}

Taking
\begin{equation*}
S_{\ bcd}^{a}=\frac{\partial C_{.bc}^{a}}{\partial y^{d}}-\frac{\partial
C_{.bd}^{a}}{\partial y^{c}}+C_{.bc}^{e}C_{.ed}^{a}-C_{.bd}^{e}C_{.ec}^{a}.
\end{equation*}%
from (\ref{2dcurv}) and contracting the indices in order to obtain the Ricci
coefficients,%
\begin{equation*}
R_{bc}=\frac{\partial C_{.bc}^{d}}{\partial y^{d}}-\frac{\partial C_{.bd}^{d}%
}{\partial y^{c}}+C_{.bc}^{e}C_{.ed}^{d}-C_{.bd}^{e}C_{.ec}^{d}
\end{equation*}%
with $C_{.bd}^{d}=C_{b}$ already computed, see (\ref{aux6a}), we obtain
\begin{equation*}
R_{bc}=\left( C_{.bc}^{4}\right) ^{\ast }-\partial
_{c}C_{b}+C_{.bc}^{4}C_{4}-C_{.b4}^{4}C_{.4c}^{4}
-C_{.b5}^{4}C_{.4c}^{5}-C_{.b4}^{5}C_{.5c}^{4}-C_{.b5}^{5}C_{.5c}^{5}.
\end{equation*}%
There are nontrivial values,
\begin{eqnarray*}
R_{44} &=&\left( C_{.44}^{4}\right) ^{\ast }-C_{4}^{\ast
}+C_{44}^{4}(C_{4}-C_{44}^{4})-\left( C_{.45}^{5}\right) ^{2} \\
&=&\beta _{4}^{\ast \ast }-\left( \beta _{4}^{\ast }+\beta _{5}^{\ast
}\right) ^{\ast }+\beta _{4}^{\ast }\left( \beta _{4}^{\ast }+\beta
_{5}^{\ast }-\beta _{4}^{\ast }\right) -\left( \beta _{5}^{\ast }\right)
^{\ast } \\
R_{55} &=&\left( C_{.55}^{4}\right) ^{\ast }-C_{.55}^{4}\left(
-C_{4}+2C_{.45}^{5}\right) \\
&=&-\left( \frac{h_{5}^{\ast }}{2h_{4}}\right) ^{\ast }+\frac{h_{5}^{\ast }}{%
2h_{4}}\left( 2\beta _{5}^{\ast }+\beta _{4}^{\ast }-\beta _{5}^{\ast
}\right)
\end{eqnarray*}%
Introducing
\begin{equation*}
\beta _{4}^{\ast }=\frac{h_{4}^{\ast }}{2h_{4}},\beta _{5}^{\ast }=\frac{%
h_{5}^{\ast }}{2h_{5}}
\end{equation*}%
we get%
\begin{equation*}
R_{4}^{4}=R_{5}^{5}=\frac{1}{2h_{4}h_{5}}\left[ -h_{5}^{\ast \ast }+\frac{%
\left( h_{5}^{\ast }\right) ^{2}}{2h_{5}}+\frac{h_{4}^{\ast }h_{5}^{\ast }}{%
2h_{4}}\right]
\end{equation*}%
which is just (\ref{1ricci2a}).

Theorem \ref{t5dr} is proven.

\section{Appendix B:\ Reductions from 5D to 4D}

\label{ss4d}To construct a $5D\rightarrow 4D$ reduction for the ansatz (\ref%
{ansatz5}) and (\ref{1ansatzc}) is to eliminate from formulas the variable $%
x^{1}$ and to consider a 4D space (parametrized by local coordinates $\left(
x^{2},x^{3},v,y^{5}\right) )$ being trivially embedded into 5D space
(parametrized by local coordinates $\left( x^{1},x^{2},x^{3},v,y^{5}\right) $
with $g_{11}=\pm 1,g_{1\widehat{\alpha }}=0,\widehat{\alpha }=2,3,4,5)$ with
possible \ 4D conformal and anholonomic transforms depending only on
variables $\left( x^{2},x^{3},v\right) .$ We suppose that the 4D metric $g_{%
\widehat{\alpha }\widehat{\beta }}$ could be of arbitrary signature. In
order to emphasize that some coordinates are stated just for a such 4D space
we put ''hats'' on the Greek indices, $\widehat{\alpha },\widehat{\beta }%
,... $ \ and on the Latin indices from the middle of alphabet, $\widehat{i},%
\widehat{j},...=2,3,$ where $u^{\widehat{\alpha }}=\left( x^{\widehat{i}%
},y^{a}\right) =\left( x^{2},x^{3},y^{4},y^{5}\right) .$

In result, the Theorems \ref{t5dr} and \ref{t5dra}, Corollaries \ref{ceint}
and \ref{cors} and Theorem \ref{texs} can be reformulated for 4D gravity
with mixed holonomic--anholonomic variables. We outline here the most
important properties of a such reduction.

\begin{itemize}
\item The metric (\ref{metric5}) with ansatz (\ref{ansatz5}) and metric (\ref%
{1cmetric}) with (\ref{1ansatzc}) are respectively transformed on 4D spaces to
the values:

The first type 4D off--diagonal metric is taken
\begin{equation}
\mathbf{g}=\mathbf{g}_{\widehat{\alpha }\widehat{\beta }}\left( x^{\widehat{i%
}},v\right) du^{\widehat{\alpha }}\otimes du^{\widehat{\beta }}
\label{1metric4}
\end{equation}%
with the metric coefficients\textbf{\ }$g_{\widehat{\alpha }\widehat{\beta }%
} $ parametrized

{%%\footnotesize
\begin{equation}
\left[
\begin{array}{cccc}
g_{2}+w_{2}^{\ 2}h_{4}+n_{2}^{\ 2}h_{5} & w_{2}w_{3}h_{4}+n_{2}n_{3}h_{5} &
w_{2}h_{4} & n_{2}h_{5} \\
w_{2}w_{3}h_{4}+n_{2}n_{3}h_{5} & g_{3}+w_{3}^{\ 2}h_{4}+n_{3}^{\ 2}h_{5} &
w_{3}h_{4} & n_{3}h_{5} \\
w_{2}h_{4} & w_{3}h_{4} & h_{4} & 0 \\
n_{2}h_{5} & n_{3}h_{5} & 0 & h_{5}%
\end{array}%
\right] ,  \label{1ansatz4}
\end{equation}%
} where the coefficients are some necessary smoothly class functions of
type:
\begin{eqnarray}
g_{2,3} &=&g_{2,3}(x^{2},x^{3}),h_{4,5}=h_{4,5}(x^{\widehat{k}},v),  \notag
\\
w_{\widehat{i}} &=&w_{\widehat{i}}(x^{\widehat{k}},v),n_{\widehat{i}}=n_{%
\widehat{i}}(x^{\widehat{k}},v);~\widehat{i},\widehat{k}=2,3.  \notag
\end{eqnarray}

The anholonomically and conformally transformed 4D off--diagonal metric is
\begin{equation}
\mathbf{g}=\omega ^{2}(x^{\widehat{i}},v)\hat{\mathbf{g}}_{\widehat{\alpha }%
\widehat{\beta }}\left( x^{\widehat{i}},v\right) du^{\widehat{\alpha }%
}\otimes du^{\widehat{\beta }},  \label{2cmetric4}
\end{equation}%
were the coefficients $\hat{\mathbf{g}}_{\widehat{\alpha }\widehat{\beta }}$
are parametrized by the ansatz
{\small
\begin{equation}
\left[
\begin{array}{cccc}
g_{2}+(w_{2}^{\ 2}+\zeta _{2}^{\ 2})h_{4}+n_{2}^{\ 2}h_{5} &
(w_{2}w_{3}+\zeta _{2}\zeta _{3})h_{4}+n_{2}n_{3}h_{5} & (w_{2}+\zeta
_{2})h_{4} & n_{2}h_{5} \\
(w_{2}w_{3}++\zeta _{2}\zeta _{3})h_{4}+n_{2}n_{3}h_{5} & g_{3}+(w_{3}^{\
2}+\zeta _{3}^{\ 2})h_{4}+n_{3}^{\ 2}h_{5} & (w_{3}+\zeta _{3})h_{4} &
n_{3}h_{5} \\
(w_{2}+\zeta _{2})h_{4} & (w_{3}+\zeta _{3})h_{4} & h_{4} & 0 \\
n_{2}h_{5} & n_{3}h_{5} & 0 & h_{5}+\zeta _{5}h_{4}%
\end{array}%
\right]  \label{1ansatzc4}
\end{equation}%
}
where $\zeta _{\widehat{i}}=\zeta _{\widehat{i}}\left( x^{\widehat{k}%
},v\right) $ and we shall restrict our considerations for $\zeta _{5}=0.$

\item We obtain a quadratic line element
\begin{equation}
\delta s^{2}=g_{2}(dx^{2})^{2}+g_{3}(dx^{3})^{2}+h_{4}(\delta
v)^{2}+h_{5}(\delta y^{5})^{2},  \label{1dmetric4}
\end{equation}%
written with respect to the anholonomic co--frame $\left( dx^{\widehat{i}%
},\delta v,\delta y^{5}\right) ,$ where
\begin{equation}
\delta v=dv+w_{\widehat{i}}dx^{\widehat{i}}\mbox{ and }\delta
y^{5}=dy^{5}+n_{\widehat{i}}dx^{\widehat{i}}  \label{1ddif4}
\end{equation}%
is the dual of $\left( \delta _{\widehat{i}},\partial _{4},\partial
_{5}\right) ,$ where
\begin{equation}
\delta _{\widehat{i}}=\partial _{\widehat{i}}+w_{\widehat{i}}\partial
_{4}+n_{\widehat{i}}\partial _{5}.  \label{1dder4}
\end{equation}

\item If the conditions of the 4D variant of the Theorem \ref{t5dr} are
satisfied, we have the same equations (\ref{ep1a}) --(\ref{ep4a}) were we
substitute $h_{4}=h_{4}\left( x^{\widehat{k}},v\right) $ and $%
h_{5}=h_{5}\left( x^{\widehat{k}},v\right) .$ As a consequence we have $%
\alpha _{i}\left( x^{k},v\right) \rightarrow \alpha _{\widehat{i}}\left( x^{%
\widehat{k}},v\right) ,\beta =\beta \left( x^{\widehat{k}},v\right) $ and $%
\gamma =\gamma \left( x^{\widehat{k}},v\right) $ resulting in $w_{\widehat{i}%
}=w_{\widehat{i}}\left( x^{\widehat{k}},v\right) $ and $n_{\widehat{i}}=n_{%
\widehat{i}}\left( x^{\widehat{k}},v\right) .$

\item The 4D line element with conformal factor (\ref{1dmetric4}) subjected
to an anhlonomic map with $\zeta _{5}=0$ transforms into
\begin{equation}
\delta s^{2}=\omega ^{2}(x^{\widehat{i}%
},v)[g_{2}(dx^{2})^{2}+g_{3}(dx^{3})^{2}+h_{4}(\hat{{\delta }}%
v)^{2}+h_{5}(\delta y^{5})^{2}],  \label{1cdmetric4}
\end{equation}%
given with respect to the anholonomic co--frame $\left( dx^{\widehat{i}},%
\hat{{\delta }}v,\delta y^{5}\right) ,$ where
\begin{equation}
\delta v=dv+(w_{\widehat{i}}+\zeta _{\widehat{i}})dx^{\widehat{i}}%
\mbox{ and
}\delta y^{5}=dy^{5}+n_{\widehat{i}}dx^{\widehat{i}}  \label{1ddif24}
\end{equation}%
is dual to the frame $\left( \hat{{\delta }}_{\widehat{i}},\partial _{4},%
\hat{{\partial }}_{5}\right) $ with
\begin{equation}
\hat{{\delta }}_{\widehat{i}}=\partial _{\widehat{i}}-(w_{\widehat{i}}+\zeta
_{\widehat{i}})\partial _{4}+n_{\widehat{i}}\partial _{5},\hat{{\partial }}%
_{5}=\partial _{5}.  \label{1dder24}
\end{equation}

\item The formulas (\ref{1conf1}) and (\ref{1confeq}) from Theorem \ref{t5dra}
must be modified into a 4D form
\begin{equation}
\hat{{\delta }}_{\widehat{i}}h_{4}=0\mbox{\ and\  }\hat{{\delta }}_{\widehat{%
i}}\omega =0  \label{1conf14}
\end{equation}%
and the values $\zeta _{\widetilde{{i}}}=\left( \zeta \widehat{_{{i}}},\zeta
_{{5}}=0\right) $ are found as to be a unique solution of (\ref{1conf1}); for
instance, if
\begin{equation*}
\omega ^{q_{1}/q_{2}}=h_{4}~(q_{1}\mbox{ and }q_{2}\mbox{ are
integers}),
\end{equation*}%
$\zeta _{\widehat{{i}}}$ satisfy the equations \
\begin{equation}
\partial _{\widehat{i}}\omega -(w_{\widehat{i}}+\zeta _{\widehat{{i}}%
})\omega ^{\ast }=0.  \label{1confeq4}
\end{equation}

\item One holds the same formulas (\ref{1p2})-(\ref{1n}) from the Theorem \ref%
{texs} on the general form of exact solutions with that difference that
their 4D analogs are to be obtained by reductions of holonomic indices, $%
\widehat{i}\rightarrow i,$ and holonomic coordinates, $x^{i}\rightarrow x^{%
\widehat{i}},$ i. e. in the 4D solutions there is not contained the variable
$x^{1}.$

\item The formulae (\ref{1einstdiag}) for the nontrivial coefficients of the
Einstein tensor in 4D stated by the Corollary \ref{ceint} are \ written
\begin{equation}
G_{2}^{2}=G_{3}^{3}=-S_{4}^{4},G_{4}^{4}=G_{5}^{5}=-R_{2}^{2}.
\label{1einstdiag4}
\end{equation}

\item For symmetries of the Einstein tensor (\ref{1einstdiag4}), \ we can
introduce a matter field source with a diagonal energy momentum tensor,
like\ it is stated in the Corollary \ref{cors} by the conditions (\ref%
{1emcond}), which in 4D are transformed into
\begin{equation}
\Upsilon _{2}^{2}=\Upsilon _{3}^{3}=\Upsilon _{2}(x^{2},x^{3},v),\ \Upsilon
_{4}^{4}=\Upsilon _{5}^{5}=\Upsilon _{4}(x^{2},x^{3}).  \label{1emcond4}
\end{equation}
\end{itemize}

The 4D dimensional off--diagonal ansatz may model certain generalized
Lagrange configurations and Lagrange--affine solutions. They can also
include certain 3D Finsler or Lagrange metrics but with 2D frame transforms
of the corresponding quadratic forms and N--connections.

\section{Appendix C:\ Generalized Lagrange--Affine Spa\-ces}

We outline and give a brief characterization of five classes of generalized
Finsler--affine spaces (contained in the Table 1 from Ref. \cite{02vp1}; see
also in that work the details on classification of such geometries). We note
that the N--connection curvature is computed following the formula $\Omega
_{ij}^{a}=\delta _{\lbrack i}N_{j]}^{a},$ see (\ref{2ncurv}), for any
N--connection $N_{i}^{a}.$ A d--connection $\mathbf{D}=[\mathbf{\Gamma }%
_{\beta \gamma }^{\alpha }]=[L_{\ jk}^{i},L_{\ bk}^{a},C_{\ jc}^{i},C_{\
bc}^{a}]$ defines nontrivial d--torsions $\mathbf{T}_{\ \beta \gamma
}^{\alpha }=[L_{[\ jk]}^{i},C_{\ ja}^{i},\Omega _{ij}^{a},T_{\ bj}^{a},C_{\
[bc]}^{a}]$ and d--curvatures $\mathbf{R}_{\ \beta \gamma \tau }^{\alpha
}=[R_{\ jkl}^{i},R_{\ bkl}^{a},P_{\ jka}^{i},P_{\ bka}^{c},S_{\
jbc}^{i},S_{\ dbc}^{a}]$ adapted to the N--connection structure (see,
respectively, the formulas (\ref{2dtorsb})\ and (\ref{2dcurv})). Any generic
off--diagonal metric $g_{\alpha \beta }$ is associated to a N--connection
structure and represented as a d--metric $\mathbf{g}_{\alpha \beta
}=[g_{ij},h_{ab}]$ (see formula (\ref{2block2})). The components of a
N--connection and a d--metric define the canonical d--connection $\mathbf{D}%
=[\widehat{\mathbf{\Gamma }}_{\beta \gamma }^{\alpha }]=$\\ $[\widehat{L}_{\
jk}^{i},\widehat{L}_{\ bk}^{a},\widehat{C}_{\ jc}^{i},\widehat{C}_{\
bc}^{a}] $ (see (\ref{2candcon})) with the corresponding values of
d--torsions $\widehat{\mathbf{T}}_{\ \beta \gamma }^{\alpha }$ and
d--curvatures $\widehat{\mathbf{R}}_{\ \beta \gamma \tau }^{\alpha }.$ The
nonmetricity d--fields are computed by using formula\\ $\mathbf{Q}_{\alpha
\beta \gamma }=-\mathbf{D}_{\alpha }\mathbf{g}_{\beta \gamma
}=[Q_{ijk},Q_{iab},Q_{ajk},Q_{abc}],$ see (\ref{2nmf}).

The Table \ref{1tablegs} outlines five classes of geometries modelled in the
framework of metric--affine geometry as spaces with nontrivial N--connection
structure (for simplicity, we omitted the Berwald configurations, see Ref. %
\cite{02vp1}).

\begin{enumerate}
\item Metric--affine spaces (in brief, MA) are those stated as certain
manifolds $V^{n+m}$ of necessary smoothly class\ provided with arbitrary
metric, $g_{\alpha \beta },$ and linear connection, $\Gamma _{\beta \gamma
}^{\alpha },$ structures. For generic off--diagonal metrics, a MA\ space
always admits nontrivial N--connection structures. Nevertheless, in general,
only the metric field $g_{\alpha \beta }$ can be transformed into a
d--metric one $\mathbf{g}_{\alpha \beta }=[g_{ij},h_{ab}],$ but \ $\Gamma
_{\beta \gamma }^{\alpha }$ can be not adapted to the N--connection
structure. As a consequence, the general strength fields $\left( T_{\ \beta
\gamma }^{\alpha },R_{\ \beta \gamma \tau }^{\alpha },Q_{\alpha \beta \gamma
}\right) $ can be also not N--adapted.

\item Distinguished metric--affine spaces (DMA) are defined as manifolds $%
\mathbf{V}^{n+m}$ $\ $provided with N--connection structure $N_{i}^{a},$
d--metric field (\ref{2block2}) and arbitrary d--connection $\mathbf{\Gamma }%
_{\beta \gamma }^{\alpha }.$ In this case, all strengths $\left( \mathbf{T}%
_{\ \beta \gamma }^{\alpha },\mathbf{R}_{\ \beta \gamma \tau }^{\alpha },%
\mathbf{Q}_{\alpha \beta \gamma }\right) $ are N--adapted.

\item Generalized Lagrange--affine spaces (GLA), $\mathbf{GLa}%
^{n}=(V^{n},g_{ij}(x,y),$ $\ ^{[a]}\mathbf{\Gamma }_{\ \beta }^{\alpha })$ $%
, $ are mo\-delled as distinguished metric--affine spaces of odd--dimension, $%
\mathbf{V}^{n+n},$ provided with generic off--diagonal metrics with
associated N--connection inducing a tangent bundle structure. The d--metric $%
\mathbf{g}_{[a]}$ and the d--connection $\ \ ^{[a]}\mathbf{\Gamma }_{\
\alpha \beta }^{\gamma }$ $=\left( \ ^{[a]}L_{jk}^{i},\
^{[a]}C_{jc}^{i}\right) $ are similar to those for the usual Lagrange spaces
but with distorsions $\ ^{[a]}\ \mathbf{Z}_{\ \ \beta }^{\alpha }$ inducing
general nontrivial nonmetricity d--fields $^{[a]}\mathbf{Q}_{\alpha \beta
\gamma }.$

\item Lagrange--affine spaces (LA), $\mathbf{La}%
^{n}=(V^{n},g_{ij}^{[L]}(x,y),$ $\ ^{[b]}\mathbf{\Gamma }_{\ \beta }^{\alpha
}),$ are provided with a Lagrange quadratic form $g_{ij}^{[L]}(x,y)=\frac{1}{%
2}\frac{\partial ^{2}L^{2}}{\partial y^{i}\partial y^{j}}$ inducing the
canonical N--connection structure $^{[cL]}\mathbf{N}=\{\ ^{[cL]}N_{j}^{i}\}$
for a Lagrange space $\mathbf{L}^{n}=\left( V^{n},g_{ij}(x,y)\right) $ but
with a d--connection structure $\ ^{[b]}\mathbf{\Gamma }_{\ \ \alpha
}^{\gamma }=\ ^{[b]}\mathbf{\Gamma }_{\ \alpha \beta }^{\gamma }\mathbf{%
\vartheta }^{\beta }$ distorted by arbitrary torsion, $\mathbf{T}_{\beta },$
and nonmetricity d--fields,$\ \mathbf{Q}_{\beta \gamma \alpha },$ when $%
^{[b]}\mathbf{\Gamma }_{\ \beta }^{\alpha }=\ ^{[L]}\widehat{\mathbf{\Gamma }%
}_{\beta }^{\alpha }+\ \ ^{[b]}\ \mathbf{Z}_{\ \ \beta }^{\alpha }.$ This is
a particular case of GLA spaces with prescribed types of N--connection $%
^{[cL]}N_{j}^{i}$ and d--metric to be like in Lagrange geometry.

\item Finsler--affine spaces (FA), $\mathbf{Fa}^{n}=\left( V^{n},F\left(
x,y\right) ,\ ^{[f]}\mathbf{\Gamma }_{\ \beta }^{\alpha }\right) ,$ in their
turn are introduced by further restrictions of $\mathbf{La}^{n}$ to a
quadratic form $g_{ij}^{[F]}=\frac{1}{2}\frac{\partial ^{2}F^{2}}{\partial
y^{i}\partial y^{j}}$ constructed from a Finsler metric $F\left(
x^{i},y^{j}\right) .$ It is induced the canonical N--connection structure $\
^{[F]}\mathbf{N}=\{\ ^{[F]}N_{j}^{i}\}$ \ as in the Finsler space $\mathbf{F}%
^{n}=\left( V^{n},F\left( x,y\right) \right) $ but with a d--connection
structure $\ ^{[f]}\mathbf{\Gamma }_{\ \alpha \beta }^{\gamma }$ distorted
by arbitrary torsion, $\mathbf{T}_{\beta \gamma }^{\alpha },$ and
nonmetricity, $\mathbf{Q}_{\beta \gamma \tau },$ d--fields, $\ ^{[f]}\mathbf{%
\Gamma }_{\ \beta }^{\alpha }=\ ^{[F]}\widehat{\mathbf{\Gamma }}_{\ \beta
}^{\alpha }\ +\ \ ^{[f]}\ \mathbf{Z}_{\ \ \beta }^{\alpha },$where $\ ^{[F]}%
\widehat{\mathbf{\Gamma }}_{\ \beta \gamma }^{\alpha }$ is the canonical
Finsler d--connection.
\end{enumerate}

%{\footnotesize
\begin{table}[h]
\begin{center}
\begin{tabular}{|l|l|l|l|}
\hline\hline
Space &
\begin{tabular}{l}
\begin{tabular}{l}
N--connection/ \\
N--curvature%
\end{tabular}
\\
\begin{tabular}{l}
metric/ \\
d--metric%
\end{tabular}%
\end{tabular}
&
\begin{tabular}{l}
(d--)connection/ \\
(d-)torsion%
\end{tabular}
&
\begin{tabular}{l}
(d-)curvature/ \\
(d--)nonmetricity%
\end{tabular}
\\ \hline\hline
1. MA &
\begin{tabular}{l}
$N_{i}^{a},\Omega _{ij}^{a}$ \\
\begin{tabular}{l}
off.d.m. $g_{\alpha \beta },$ \\
$\mathbf{g}_{\alpha \beta }=[g_{ij},h_{ab}]$%
\end{tabular}%
\end{tabular}
&
\begin{tabular}{l}
$\Gamma _{\beta \gamma }^{\alpha }$ \\
$T_{\ \beta \gamma }^{\alpha }$%
\end{tabular}
&
\begin{tabular}{l}
$R_{\ \beta \gamma \tau }^{\alpha }$ \\
$Q_{\alpha \beta \gamma }$%
\end{tabular}
\\ \hline
2. DMA &
\begin{tabular}{l}
$N_{i}^{a},\Omega _{ij}^{a}$ \\
$\mathbf{g}_{\alpha \beta }=[g_{ij},h_{ab}]$%
\end{tabular}
&
\begin{tabular}{l}
$\mathbf{\Gamma }_{\beta \gamma }^{\alpha }$ \\
$\mathbf{T}_{\ \beta \gamma }^{\alpha }$%
\end{tabular}
&
\begin{tabular}{l}
$\mathbf{R}_{\ \beta \gamma \tau }^{\alpha }$ \\
$\mathbf{Q}_{\alpha \beta \gamma }$%
\end{tabular}
\\ \hline
3. GLA &
\begin{tabular}{l}
\begin{tabular}{l}
$\dim i=\dim a$ \\
$N_{i}^{a},\Omega _{ij}^{a}$%
\end{tabular}
\\
\begin{tabular}{l}
off.d.m. $g_{\alpha \beta },$ \\
$\mathbf{g}_{[a]}=[g_{ij},h_{kl}]$%
\end{tabular}%
\end{tabular}
&
\begin{tabular}{l}
$^{\lbrack a]}\mathbf{\Gamma }_{\ \alpha \beta }^{\gamma }$ \\
$^{\lbrack a]}\mathbf{T}_{\ \beta \gamma }^{\alpha }$%
\end{tabular}
&
\begin{tabular}{l}
$^{\lbrack a]}\mathbf{R}_{\ \beta \gamma \tau }^{\alpha }$ \\
$^{\lbrack a]}\mathbf{Q}_{\alpha \beta \gamma }$%
\end{tabular}
\\ \hline
4. LA &
\begin{tabular}{l}
\begin{tabular}{l}
$\dim i=\dim a$ \\
$^{\lbrack cL]}N_{j}^{i},\ ^{[cL]}\Omega _{ij}^{a}$%
\end{tabular}
\\
d--metr.$\mathbf{g}_{\alpha \beta }^{[L]}$%
\end{tabular}
&
\begin{tabular}{l}
$^{\lbrack b]}\mathbf{\Gamma }_{\ \alpha \beta }^{\gamma }$ \\
$^{\lbrack b]}\mathbf{T}_{\ \beta \gamma }^{\alpha }$%
\end{tabular}
&
\begin{tabular}{l}
$^{\lbrack b]}\mathbf{R}_{\ \beta \gamma \tau }^{\alpha }$ \\
$^{\lbrack b]}\mathbf{Q}_{\alpha \beta \gamma }=-~^{[b]}\mathbf{D}_{\alpha }%
\mathbf{g}_{\beta \gamma }^{[L]}$%
\end{tabular}
\\ \hline
5. FA &
\begin{tabular}{l}
\begin{tabular}{l}
$\dim i=\dim a$ \\
$^{\lbrack F]}N_{j}^{i};\ ^{[F]}\Omega _{ij}^{k}$%
\end{tabular}
\\
d--metr.$\mathbf{g}_{\alpha \beta }^{[F]}$%
\end{tabular}
&
\begin{tabular}{l}
$^{\lbrack f]}\mathbf{\Gamma }_{\ \alpha \beta }^{\gamma }$ \\
$^{\lbrack f]}\mathbf{T}_{\ \beta \gamma }^{\alpha }$%
\end{tabular}
&
\begin{tabular}{l}
$^{\lbrack f]}\mathbf{R}_{\ \beta \gamma \tau }^{\alpha }$ \\
$^{\lbrack f]}\mathbf{Q}_{\alpha \beta \gamma }=-~^{[f]}\mathbf{D}_{\alpha }%
\mathbf{g}_{\beta \gamma }^{[F]}$%
\end{tabular}
\\ \hline\hline
\end{tabular}%
\end{center}
\caption{Generalized Finsler/Lagrange--affine spaces}
\label{1tablegs}
\end{table}
%}

%%%%%%%%%%%%%%%%%%%%%%%%%%%%%%%%%%%%%%%%%%%%%%%%%%%%%%%%%%%%%%%%%%%%%%%%%%%%%

\chapter[Noncommutative Black Ellipsoid Solutions]
{Noncommutative Symmetries and Stability of Black
Ellipsoids in Metric--Affine and String Gravity }

{\bf Abstract}
\footnote{\copyright\ S. Vacaru and E. Gaburov,
Noncommutative Symmetries and Stability of Black
Ellipsoids in Metric--Affine and String Gravity, hep-th/0310134}

We construct new classes of exact solutions in metric--affine gravity (MAG)
with string corrections by the antisymmetric $H$--field. The solutions are
parametrized by generic off--diagonal metrics possessing noncommutative
symmetry associated to anholonomy frame relations and related nonlinear
connection (N--connection) structure. We analyze the horizon and geodesic
properties of a class of off--diagonal metrics with deformed spherical
symmetries. The maximal analytic extension of ellipsoid type metrics are
constructed and the Penrose diagrams are analyzed with respect to adapted
frames. We prove that for small deformations (small eccentricities) there
are such metrics that the geodesic behaviour is similar to the Schwarzcshild
one. We conclude that some static and stationary ellipsoid configurations
may describe black ellipsoid objects. The new class of spacetimes do not
possess Killing symmetries even in the limits to the general relativity and,
in consequence, they are not prohibited by black hole uniqueness theorems.
Such static ellipsoid (rotoid) configurations are compatible with the cosmic
cenzorship criteria. We study the perturbations of two classes of static
black ellipsoid solutions of four dimensional gravitational field equations.
The analysis is performed in the approximation of small eccentricity
deformations of the Schwarzschild solution. We conclude that such
anisotropic black hole objects may be stable with respect to the
perturbations parametrized by the Schrodinger equations in the framework of
the one--dimensional inverse scattering theory. We emphasize that the
anholonomic frame method of generating exact solutions is a general one for
off--diagonal metrics (and linear and nonlinear connections) depending on
2-3 variables, in various types of gravity theories.

\section{Introduction}

In the past much effort has been made to construct and investigate exact
solutions of  gravitational field equations with spherical/cylindrical
symmetries and/or with time dependence, paramertrized by metrics
diagonalizable by certain coordinate transforms. Recently, the off--diagonal
metrics were considered in a new manner by diagonalizing them with respect
to anholonomic frames with associated nonlinear connection structure
\cite{03v,03v1,03vsbd,03vspd1}. There were constructed new classes of exact
 solutions of
Einstein's equations in three (3D), four (4D) and five (5D) dimensions. Such
solutions posses a generic geometric local anisotropy (\textit{e.g.} static
black hole and/or cosmological solutions with ellipsoidal or toroidal
symmetry, various soliton--dilaton 2D and 3D configurations in 4D gravity,
and wormholes and flux tubes with anisotropic polarizations and/or running
constants with different extensions to backgrounds of rotation ellipsoids,
elliptic cylinders, bipolar and toroidal symmetry and anisotropy).

A number of ansatz with off--diagonal metrics were investigated in higher
dimensional gravity (see, for instance, the Salam, Strathee, Percacci and
Randjbar--Daemi works \cite{03sal}) which showed that off--diagonal components
in higher dimensional metrics are equi\-valent to including $U(1),SU(2)$ and
$SU(3)$ gauge fields. There are various generalizations of the Kaluza--Klein
gravity when the compactifications of off--diagonal metrics are considered
with the aim to reduce the vacuum 5D gravity to effective Einstein gravity
and Abelian or non--Abelian gauge theories. There were also constructed 4D
exact solutions of Einstein equations with matter fields and cosmological
constants like black torus and black strings induced from some 3D black hole
configurations by considering 4D off--diagonal metrics whose curvature
scalar splits equivalently into a curvature term for a diagonal metric
together with a cosmological constant term and/or a Lagrangian for gauge
(electromagnetic) field \cite{03lemos}.

We can model certain effective (diagonal metric) gravitational and matter
fields interactions for some particular off--diagonal metric ansatz and
redefinitions of Lagrangians. However, in general, the vacuum gravitational
dynamics can not be associated to any matter field contributions. This holds
true even if we consider non--Riemanian generalizations from string and/or
metric--affine gravity (MAG) \cite{03mag}. In this work (being the third
partner of the papers \cite{03vmag1,03vmag2}), we prove that such solutions are
not with usual Killing symmetries but admit certain anholonomic
noncommutative symmetries and preserve such properties if the constructions
are extended to MAG and string gravity (see also \cite{03vncs} for extensions
to complex and/or noncommutative gravity).

There are constructed the maximal analytic extension of a class of static
metrics with deformed spherical symmetry (containing as particular cases
ellipsoid configurations). We analyze the Penrose diagrams and compare the
results with those for the Reissner--Nordstrom solution. Then we state the
conditions when the geodesic congruence with 'ellipsoid' type symmetry can
be reduced to the Schwarzschild configuration. We argue that in this case we
may generate some static black ellipsoid solutions which, for corresponding
parametrizations of off--diagonal metric coefficients, far away from the
horizon, satisfy the asymptotic conditions of the Minkowski spacetime.

For the new classes of ''off--diagonal'' spacetimes possessing
noncommutative symmetries, we extend the methods elaborated to investigate
the perturbations and stability of black hole metrics. The theory of
perturbations of the Schwarzschild spacetime black holes was initiated in
Ref. \cite{03rw}, developed in a series of works, e. g. Refs \cite{03vis,03fried},
and related \cite{03dei} to the theory of inverse scattering and its
ramifications (see, for instance, Refs. \cite{03fad}). The results on the
theory of perturbations and stability of the Schwarzschild,
Reissner--Nordstrom and Kerr solutions are summarized in a monograph \cite%
{03chan}. As alternative treatments of the stability of black holes we cite
Ref. \cite{03mon}.

Our first aim is to investigate such off--diagonal gravitational
configurations in MAG and string gravity (defined by anholonomic frames with
associated nonlinear connection structure) which describe black hole
solutions with deformed horizons, for instance, with a static ellipsoid
hypersurface. The second aim is to study perturbations of black ellipsoids
and to prove that there are such static ellipsoid like configurations which
are stable with respect to perturbations of a fixed type of anisotropy (i.
e. for certain imposed anholonomic constraints). The main idea of a such
proof is to consider small (ellipsoidal, or another type) deformations of
the Schwarzschild metric and than to apply the already developed methods of
the theory of perturbations of classical black hole solutions, with a
re--definition of the formalism for adapted anholonomic frames.

We note that the solutions defining black ellipsoids are very different from
those defining ellipsoidal shapes in general relativity (see Refs. \cite{03es}%
) associated to some perfect--fluid bodies, rotating configurations or to
some families of confocal ellipsoids in Reimannian spaces. Our black
ellipsoid metrics are parametrized by generic off--diagonal ansatz with
anholonomically deformed Killing symmetry and not subjected to uniqueness
theorems. Such ansatz are more general than the class of vacuum solutions
which can not be written in diagonal form \cite{03cans} (see details in Refs. %
\cite{03vth,03vncs}).

The paper is organized as follows: In Sec. 2 we outline the necessary
results on off--diagonal metrics and anhlonomic frames with associated
nonlinear connection structure. We write the system of Einstein--Proca
equations from MAG with string corrections of the antisymmetric $H$--tensor
from bosonic string theory. We introduce a general off--diagonal metric
ansatz and derive the corresponding system of Einstein equations with
anholonomic variables. In Sec. 3 we argue that noncommutative anholonomic
geometries can be associated to real off--diagonal metrics and show two
simple realizations within the algebra for complex matrices. Section 4 is
devoted to the geometry and physics of four dimensional static black
ellipsoids. We illustrate how such solutions can be constructed by using
anholonomic deformations of the Schwarzshild metric, define analytic
extensions of black ellipsoid metrics and analyze the geodesic behaviour of
the static ellipsoid backgrounds. We conclude that black ellipsoid metrics
posses specific noncommutative symmetries. We outline a perturbation theory
of anisotropic black holes and prove the stability of black ellipsoid
objects in Sec. 5. Then, in Sec. 6 we discuss how the method of anholonomic
frame transforms can be related solutions for ellipsoidal shapes and generic
off--diagonal solutions constructed by F. Canfora and H. -J.\ Schmidt. \ We
outline the work and present conclusions in Sec. 7.

There are used the basic notations and conventions stated in Refs. \cite%
{03vmag1,03vmag2}.

\section{Anholonomic Frames and Off--Diagonal Metrics}

We consider a 4D manifold $V^{3+1}$ (for MAG and string gravity with
possible torsion and nonmetricity structures \cite{03mag,03vmag1,03vmag2}) enabled
with local coordinates $u^{\alpha }=\left( x^{i},y^{a}\right) $ where the
indices of type $i,j,k,...$ run values $1$ and $2$ and the indices $%
a,b,c,... $ take values $3$ and $4;$ $\ y^{3}=v=\varphi $ and $y^{4}=t$ are
considered respectively as the ''anisotropic'' and time like coordinates
(subjected to some constraints). It is supposed that such spacetimes can
also admit nontrivial torsion structures induced by certain frame transforms.

The quadratic line element
\begin{equation}
ds^{2}={g}_{\alpha \beta }\left( x^{i},v\right) du^{\alpha }du^{\beta },
\label{3cmetric4}
\end{equation}%
is parametrized by a metric ansatz {%
\begin{equation}
{g}_{\alpha \beta }=\left[
\begin{array}{cccc}
g_{1}+w_{1}^{\ 2}h_{3}+n_{1}^{\ 2}h_{4} & w_{1}w_{2}h_{3}+n_{1}n_{2}h_{4} &
w_{1}h_{3} & n_{1}h_{4} \\
w_{1}w_{2}h_{3}+n_{1}n_{2}h_{4} & g_{2}+w_{2}^{\ 2}h_{3}+n_{2}^{\ 2}h_{4} &
w_{2}h_{3} & n_{2}h_{4} \\
w_{1}h_{3} & w_{2}h_{3} & h_{3} & 0 \\
n_{1}h_{4} & n_{2}h_{4} & 0 & h_{4}%
\end{array}%
\right] ,  \label{12ansatzc4}
\end{equation}%
} with $g_{i}=g_{i}\left( x^{i}\right) ,h_{a}=h_{ai}\left( x^{k},v\right) $
and $n_{i}=n_{i}\left( x^{k},v\right) $ being some functions of necessary
smoothly class or even singular in some points and finite regions. The
coefficients $g_{i}$ depend only on ''holonomic'' variables $x^{i}$ but the
rest of coefficients may also depend on one ''anisotropic'' (anholonomic)
variable $y^{3}=v;$ the ansatz does not depend on the time variable $%
y^{4}=t; $ we shall search for static solutions.

The spacetime may be provided with a general anholonomic frame structure of
tetrads, or vierbiends,
\begin{equation}
e_{\alpha }=A_{\alpha }^{\beta }\left( u^{\gamma }\right) \partial /\partial
u^{\beta },  \label{transftet}
\end{equation}%
satisfying some anholonomy \ relations
\begin{equation}
e_{\alpha }e_{\beta }-e_{\beta }e_{\alpha }=w_{\alpha \beta }^{\gamma
}\left( u^{\varepsilon }\right) e_{\gamma },  \label{1anhol}
\end{equation}%
where $w_{\alpha \beta }^{\gamma }\left( u^{\varepsilon }\right) $ are
called the coefficients of anholonomy. A 'holonomic' frame, for instance, a
coordinate frame, $e_{\alpha }=\partial _{\alpha }=\partial /\partial
u^{\alpha },$ is defined as a set of tetrads satisfying the holonomy
conditions
\begin{equation*}
\partial _{\alpha }\partial _{\beta }-\partial _{\beta }\partial _{\alpha
}=0.
\end{equation*}

We can 'effectively' diagonalize the line element (\ref{3cmetric4}),
\begin{equation}
\delta s^{2}=g_{1}(dx^{1})^{2}+g_{2}(dx^{2})^{2}+h_{3}(\delta
v)^{2}+h_{4}(\delta y^{4})^{2},  \label{2dmetric4}
\end{equation}%
with respect to the anholonomic co--frame
\begin{equation}
\delta ^{\alpha }=(d^{i}=dx^{i},\delta ^{a}=dy^{a}+N_{i}^{a}dx^{i})=\left(
d^{i},\delta v=dv+w_{i}dx^{i},\delta y^{4}=dy^{4}+n_{i}dx^{i}\right)
\label{2ddif4}
\end{equation}%
which is dual to the anholonomic frame
\begin{equation}
\delta _{\alpha }=(\delta _{i}=\partial _{i}-N_{i}^{a}\partial _{a},\partial
_{b})=\left( \delta _{i}=\partial _{i}-w_{i}\partial _{3}-n_{i}\partial
_{4},\partial _{3},\partial _{4}\right) ,  \label{2dder4}
\end{equation}%
where $\partial _{i}=\partial /\partial x^{i}$ and $\partial _{b}=\partial
/\partial y^{b}$ are usual partial derivatives. The tetrads $\delta _{\alpha
}$ and $\delta ^{\alpha }$ are anholonomic because, in general, they satisfy
the anholonomy relations (\ref{1anhol}) with some non--trivial coefficients,
\begin{equation}
w_{ij}^{a}=\delta _{i}N_{j}^{a}-\delta
_{j}N_{i}^{a},~w_{ia}^{b}=-~w_{ai}^{b}=\partial _{a}N_{i}^{b}.  \label{2anh}
\end{equation}%
The anholonomy is induced by the coefficients $N_{i}^{3}=w_{i}$ and $%
N_{i}^{4}=n_{i}$ which ''elongate'' partial derivatives and differentials if
we are working with respect to anholonomic frames. This results in a more
sophisticate differential and integral calculus (as in the tetradic and/or
spinor gravity), but simplifies substantially the tensor computations,
because we are dealing with diagonalized metrics. In order to construct
exact 'off--diagonal' solutions with 4D metrics depending on 3 variables $%
\left( x^{k},v\right) $ it is more convenient to work with respect to
anholonomic frames (\ref{2dder4}) and (\ref{2ddif4}) for diagonalized metrics (%
\ref{2dmetric4}) than to consider directly the \ ansatz (\ref{3cmetric4}) \cite%
{03v,03v1,03vsbd,03vspd1}.

There is a 'preferred' linear connection constructed only from the
components\\ $\left( g_{i},h_{a},N_{k}^{b}\right) $, called the canonical
distinguished linear connection, which is similar to the metric connection
introduced by the Christoffel symbols in the case of holonomic bases, i. e.
being constructed only from the metric components and satisfying the
metricity conditions. It is parametrized by the coefficients,\ $\Gamma _{\
\beta \gamma }^{\alpha }=\left( L_{\ jk}^{i},L_{\ bk}^{a},C_{\ jc}^{i},C_{\
bc}^{a}\right) $ stated with respect to the anholonomic frames (\ref{2dder4})
and (\ref{2ddif4}) as
\begin{eqnarray}
L_{\ jk}^{i} &=&\frac{1}{2}g^{in}\left( \delta _{k}g_{nj}+\delta
_{j}g_{nk}-\delta _{n}g_{jk}\right) ,  \label{2dcon} \\
L_{\ bk}^{a} &=&\partial _{b}N_{k}^{a}+\frac{1}{2}h^{ac}\left( \delta
_{k}h_{bc}-h_{dc}\partial _{b}N_{k}^{d}-h_{db}\partial _{c}N_{k}^{d}\right) ,
\notag \\
C_{\ jc}^{i} &=&\frac{1}{2}g^{ik}\partial _{c}g_{jk},\ C_{\ bc}^{a}=\frac{1}{%
2}h^{ad}\left( \partial _{c}h_{db}+\partial _{b}h_{dc}-\partial
_{d}h_{bc}\right) ,  \notag
\end{eqnarray}%
computed for the ansatz (\ref{12ansatzc4}). This induces a linear covariant
derivative locally adapted to the nonlinear connection structure
(N--connection, see details, for instance, in Refs. \cite{03ma,03v,03vth}). By
straightforward calculations, we can verify that for $D_{\alpha }$ defined
by $\Gamma _{\ \beta \gamma }^{\alpha }$ with the components (\ref{2dcon})
the condition $D_{\alpha }g_{\beta \gamma }=0$ is satisfied.

We note that on (pseudo) Riemannian spaces the N--connection is an object
completely defined by anholonomic frames, when the coefficients of tetradic
transform (\ref{transftet}), $A_{\alpha }^{\beta }\left( u^{\gamma }\right) ,
$ are parametrized explicitly via certain values $\left( N_{i}^{a},\delta
_{i}^{j},\delta _{b}^{a}\right) ,$ where $\delta _{i}^{j}$ $\ $and $\delta
_{b}^{a}$ are the Kronecker symbols. By straightforward calculations we can
compute (see, for instance Ref. \cite{03mtw}) that the coefficients of the
Levi--Civita metric connection
\begin{equation*}
\Gamma _{\alpha \beta \gamma }^{\bigtriangledown }=g\left( \delta _{\alpha
},\bigtriangledown _{\gamma }\delta _{\beta }\right) =g_{\alpha \tau }\Gamma
_{\beta \gamma }^{\bigtriangledown \tau },\,
\end{equation*}%
associated to a covariant derivative operator $\bigtriangledown ,$
satisfying the metricity condition\\ $\bigtriangledown _{\gamma }g_{\alpha
\beta }=0$ for $g_{\alpha \beta }=\left( g_{ij},h_{ab}\right) ,$
\begin{equation}
\Gamma _{\alpha \beta \gamma }^{\bigtriangledown }=\frac{1}{2}\left[ \delta
_{\beta }g_{\alpha \gamma }+\delta _{\gamma }g_{\beta \alpha }-\delta
_{\alpha }g_{\gamma \beta }+g_{\alpha \tau }w_{\gamma \beta }^{\tau
}+g_{\beta \tau }w_{\alpha \gamma }^{\tau }-g_{\gamma \tau }w_{\beta \alpha
}^{\tau }\right] ,  \label{12lcsym}
\end{equation}%
are given with respect to the anholonomic basis (\ref{2ddif4}) by the
coefficients
\begin{equation}
\Gamma _{\beta \gamma }^{\bigtriangledown \tau }=\left( L_{\ jk}^{i},L_{\
bk}^{a}-\frac{\partial N_{k}^{a}}{\partial y^{b}},C_{\ jc}^{i}+\frac{1}{2}%
g^{ik}\Omega _{jk}^{a}h_{ca},C_{\ bc}^{a}\right) ,  \label{1lccon}
\end{equation}%
where $\Omega _{jk}^{a}=\delta _{k}N_{j}^{a}-\delta _{j}N_{k}^{a}.$ The
anholonomic frame structure may induce on (pseudo) Riemannian spacetimes
nontrivial torsion structures. For instance, the canonical connection (\ref%
{2dcon}), in general, has nonvanishing torsion components
\begin{equation}
T_{ja}^{i}=-T_{aj}^{i}=C_{ja}^{i},T_{jk}^{a}=-T_{kj}^{a}=\Omega
_{kj}^{a},T_{bk}^{a}=-T_{kb}^{a}=\partial _{b}N_{k}^{a}-L_{bk}^{a}.
\label{torsion}
\end{equation}%
This is a ''pure'' anholonomic frame effect. We can conclude that the
Einstein theory transforms into an effective Einstein--Cartan theory with
anholonomically induced torsion if the general relativity is formulated with
respect to general anholonomic frame bases. In this paper we shall also
consider distorsions of the Levi--Civita connection induced by nonmetricity.

A very specific property of off--diagonal metrics is that they can define
different classes of linear connections which satisfy the metricity
conditions for a given metric, or inversely, there is a certain class of
metrics which satisfy the metricity conditions for a given linear
connection. \ This result was originally obtained by A. Kawaguchi \cite{03kaw}
(Details can be found in Ref. \cite{03ma}, see Theorems 5.4 and 5.5 in Chapter
III, formulated for vector bundles; here we note that similar proofs hold
also on manifolds enabled with anholonomic frames associated to a
N--connection structure.)

The Levi--Civita connection does not play an exclusive role on
non--Riemannian spaces. For instance, the torsion on spaces provided with
N--connection is induced by anholonomy relation and both linear connections (%
\ref{2dcon}) and (\ref{1lccon}) are compatible with the same metric and
transform into the usual Levi--Civita coefficients for vanishing
N--connection and ''pure'' holonomic coordinates (see related details in
Refs. \cite{03vmag1,03vmag2}). This means that to an off--diagonal metric we can
associated different covariant differential calculi, all being compatible
with the same metric structure (like in noncommutative geometry, which is
not a surprising \ fact because the anolonomic frames satisfy by definition
some noncommutative relations (\ref{1anhol})). In such cases we have to
select a particular type of connection following some physical or
geometrical arguments, or to impose some conditions when there is a single
compatible linear connection constructed only from the metric and
N--coefficients.

The dynamics of generalized Finsler--affine string gravity is defined by the
system of field equations (see Proposition 3.1 in Ref. \cite{03vmag2})%
\begin{eqnarray}
\widehat{\mathbf{R}}_{\alpha \beta }-\frac{1}{2}\mathbf{g}_{\alpha \beta }%
\overleftarrow{\mathbf{\hat{R}}} &=&\tilde{\kappa}\left( \mathbf{\Sigma }%
_{\alpha \beta }^{[\phi ]}+\mathbf{\Sigma }_{\alpha \beta }^{[\mathbf{m}]}+%
\mathbf{\Sigma }_{\alpha \beta }^{[\mathbf{T}]}\right) ,  \label{1fagfe} \\
\widehat{\mathbf{D}}_{\nu }\mathbf{H}^{\nu \mu } &=&\mu ^{2}\mathbf{\phi }%
^{\mu },  \notag \\
\widehat{\mathbf{D}}^{\nu }\mathbf{H}_{\nu \lambda \rho } &=&0  \notag
\end{eqnarray}%
for
\begin{equation*}
\mathbf{H}_{\nu \lambda \rho }=\widehat{\mathbf{Z}}_{\ \nu \lambda \rho }+%
\widehat{\mathbf{H}}_{\nu \lambda \rho }
\end{equation*}%
being the antisymmetric torsion field
\begin{equation*}
\mathbf{H}_{\nu \lambda \rho }=\delta _{\nu }\mathbf{B}_{\lambda \rho
}+\delta _{\rho }\mathbf{B}_{\nu \lambda }+\delta _{\lambda }\mathbf{B}_{\nu
\rho }
\end{equation*}%
of the antisymmetric $\mathbf{B}_{\lambda \rho }$ in bosonic string theory
(for simplicity, we restrict our considerations to the sigma model with $H$%
--field corrections and zero dilatonic field). The covariant derivative $%
\widehat{\mathbf{D}}_{\nu }$ is defined by the coefficients (\ref{2dcon}) (we
use in our references the ''boldfaced'' indices when it is necessary to
emphasize that the spacetime is provided with N--connection structure. The
distorsion $\widehat{\mathbf{Z}}_{\ \nu \lambda \rho }$ of the Levi--Civita
connection, when
\begin{equation*}
\Gamma _{\beta \gamma }^{\tau }=\Gamma _{\bigtriangledown \beta \gamma
}^{\tau }+\widehat{\mathbf{Z}}_{\beta \gamma }^{\tau },
\end{equation*}%
from (\ref{1fagfe}) is defined by the torsion $\widehat{\mathbf{T}}$ with the
components computed for $\widehat{\mathbf{D}}$ by applying the formulas (\ref%
{torsion}),
\begin{equation*}
\widehat{\mathbf{Z}}_{\alpha \beta }=\delta _{\beta }\rfloor \widehat{%
\mathbf{T}}_{\alpha }-\delta _{\alpha }\rfloor \widehat{\mathbf{T}}_{\beta }+%
\frac{1}{2}\left( \delta _{\alpha }\rfloor \delta _{\beta }\rfloor \widehat{%
\mathbf{T}}_{\gamma }\right) \delta ^{\gamma },
\end{equation*}%
see Refs. \cite{03vmag1,03vmag2} on definition of the interior product ''$%
\rfloor $'' and differential forms like $\widehat{\mathbf{T}}_{\beta }$ on
spaces provided with N--connection structure. The tensor $\mathbf{H}_{\nu
\mu }\doteqdot \widehat{\mathbf{D}}_{\nu }\mathbf{\phi }_{\mu }-\widehat{%
\mathbf{D}}_{\mu }\mathbf{\phi }_{\nu }+w_{\mu \nu }^{\gamma }\mathbf{\phi }%
_{\gamma }$ is the field strengths of the Abelian--Proca field $\mathbf{\phi }%
^{\alpha },$ with $\mu ,\tilde{\kappa}=const,$
\begin{equation*}
\mathbf{\Sigma }_{\alpha \beta }^{[\phi ]}=\mathbf{H}_{\alpha }^{\ \mu }%
\mathbf{H}_{\beta \mu }-\frac{1}{4}\mathbf{g}_{\alpha \beta }\mathbf{H}_{\mu
\nu }^{\ }\mathbf{H}^{\mu \nu }+\mu ^{2}\mathbf{\phi }_{\alpha }\mathbf{\phi
}_{\beta }-\frac{\mu ^{2}}{2}\mathbf{g}_{\alpha \beta }\mathbf{\phi }_{\mu }%
\mathbf{\phi }^{\mu },
\end{equation*}%
where the source%
\begin{equation*}
\mathbf{\Sigma }_{\alpha \beta }^{[\mathbf{T}]}=\mathbf{\Sigma }_{\alpha
\beta }^{[\mathbf{T}]}\left( \widehat{\mathbf{T}},\mathbf{H}_{\nu \lambda
\rho }\right)
\end{equation*}%
contains contributions of the torsion fields $\widehat{\mathbf{T}}$ and $%
\mathbf{H}_{\nu \lambda \rho }.$ The field $\mathbf{\phi }_{\alpha }$ is defined by
certain
irreducible components of torsion and nonmetricity in MAG, see  \cite{03mag}
and Theorem 3.2 in  \cite{03vmag2}.

Our aim is to elaborate a method of constructing exact solutions of
equations (\ref{1fagfe}) for vanishing matter fields, $\mathbf{\Sigma }%
_{\alpha \beta }^{[\mathbf{m}]}=0.$ The ansatz for the field $\mathbf{\phi }%
_{\mu }$ is taken in the form
\begin{equation*}
\mathbf{\phi }_{\mu }=\left[ \mathbf{\phi }_{i}\left( x^{k}\right) ,\mathbf{%
\phi }_{a}=0\right]
\end{equation*}%
for $i,j,k...=1,2$ and $a,b,...=3,4.$ The Proca equations $\widehat{\mathbf{D%
}}_{\nu }\mathbf{H}^{\nu \mu }=\mu ^{2}\mathbf{\phi }^{\mu }$ for $\mu
\rightarrow 0$ (for simplicity) transform into
\begin{equation}
\partial _{1}\left[ \left( g_{1}\right) ^{-1}\partial ^{1}\mathbf{\phi }_{k}%
\right] +\partial _{2}\left[ \left( g_{2}\right) ^{-1}\partial ^{2}\mathbf{%
\phi }_{k}\right] =L_{ki}^{j}\partial ^{i}\mathbf{\phi }_{j}-L_{ij}^{i}%
\partial ^{j}\mathbf{\phi }_{k}.  \label{ans21}
\end{equation}%
Two examples of solutions of this equation are considered in Ref. \cite%
{03vmag2}. In this paper, we do not state any particular configurations and
consider that it is possible always to define certain $\mathbf{\phi }%
_{i}\left( x^{k}\right) $ satisfying the wave equation (\ref{ans21}). The
energy--momentum tensor $\mathbf{\Sigma }_{\alpha \beta }^{[\phi ]}$ is
computed for one nontrivial value
\begin{equation*}
H_{12}=\partial _{1}\mathbf{\phi }_{2}-\partial _{2}\mathbf{\phi }_{1}.
\end{equation*}
In result, we can represent the source of the fields $\mathbf{\phi }_{k}$ as
\begin{equation*}
\mathbf{\Sigma }_{\alpha \beta }^{[\phi ]}=\left[ \Psi _{2}\left(
H_{12},x^{k}\right) ,\Psi _{2}\left( H_{12},x^{k}\right) ,0,0\right] .
\end{equation*}

The ansatz for the $H$--field is taken in the form
\begin{equation}
\mathbf{H}_{\nu \lambda \rho }=\widehat{\mathbf{Z}}_{\ \nu \lambda \rho }+%
\widehat{\mathbf{H}}_{\nu \lambda \rho }=\lambda _{\lbrack H]}\sqrt{|\mathbf{%
g}_{\alpha \beta }|}\varepsilon _{\nu \lambda \rho }  \label{2cond03}
\end{equation}%
where $\varepsilon _{\nu \lambda \rho }$ is completely antisymmetric and $%
\lambda _{\lbrack H]}=const.$ This ansatz satisfies the field equations $%
\widehat{\mathbf{D}}^{\nu }\mathbf{H}_{\nu \lambda \rho }=0$ because the
metric $\mathbf{g}_{\alpha \beta }$ is compatible with $\widehat{\mathbf{D}}.
$ The values $\widehat{\mathbf{H}}_{\nu \lambda \rho }$ have to be defined
in a form to satisfy the condition (\ref{2cond03})\ for any $\widehat{\mathbf{%
Z}}_{\ \nu \lambda \rho }$ derived from $\mathbf{g}_{\alpha \beta }$ and, as
a consequence, from (\ref{2dcon}) and (\ref{torsion}), for instance, to
compute them as
\begin{equation*}
\widehat{\mathbf{H}}_{\nu \lambda \rho }=\lambda _{\lbrack H]}\sqrt{|\mathbf{%
g}_{\alpha \beta }|}\varepsilon _{\nu \lambda \rho }-\widehat{\mathbf{Z}}_{\
\nu \lambda \rho }
\end{equation*}%
for defined values of $\widehat{\mathbf{Z}}_{\ \nu \lambda \rho },\lambda
_{\lbrack H]}$ and $\mathbf{g}_{\alpha \beta }.$ In result, we obtain the
effective energy--momentum tensor in the form%
\begin{equation}
\mathbf{\Sigma }_{\alpha }^{[\phi ]\beta }+\mathbf{\Sigma }_{\alpha
}^{[H]\beta }=\left[ \Upsilon _{2}\left( x^{k}\right) +\frac{\lambda
_{\lbrack H]}^{2}}{4},\Upsilon _{2}\left( x^{k}\right) +\frac{\lambda
_{\lbrack H]}^{2}}{4},\frac{\lambda _{\lbrack H]}^{2}}{4},\frac{\lambda
_{\lbrack H]}^{2}}{4}\right] .  \label{sours01}
\end{equation}

For the source (\ref{sours01}), the system of field equations (\ref{1fagfe})
defined for the metric (\ref{2dmetric4}) and connection (\ref{2dcon}), with
respect to anholonomic frames (\ref{2ddif4}) and (\ref{2dder4}), transform
into a system of partial differential equations with anholonomic variables %
\cite{03v,03v1,03vth}, see also details in the section 5.3 in Ref. \cite{03vmag2},
\begin{eqnarray}
R_{1}^{1}=R_{2}^{2}=-\frac{1}{2g_{1}g_{2}}[g_{2}^{\bullet \bullet }-\frac{%
g_{1}^{\bullet }g_{2}^{\bullet }}{2g_{1}}-\frac{(g_{2}^{\bullet })^{2}}{%
2g_{2}}+g_{1}^{^{\prime \prime }}-\frac{g_{1}^{^{\prime }}g_{2}^{^{\prime }}%
}{2g_{2}}-\frac{(g_{1}^{^{\prime }})^{2}}{2g_{1}}] &=&-\frac{\lambda
_{\lbrack H]}^{2}}{4},  \label{2ricci1a} \\
R_{3}^{3}=R_{4}^{4}=-\frac{\beta }{2h_{3}h_{4}}=-\frac{1}{2h_{3}h_{4}}\left[
h_{4}^{\ast \ast }-h_{4}^{\ast }\left( \ln \sqrt{|h_{3}h_{4}|}\right) ^{\ast
}\right]  &=&-\frac{\lambda _{\lbrack H]}^{2}}{4}-\Upsilon _{2}\left(
x^{k}\right) ,  \label{2ricci2a} \\
R_{3i}=-w_{i}\frac{\beta }{2h_{4}}-\frac{\alpha _{i}}{2h_{4}} &=&0,
\label{2ricci3a} \\
R_{4i}=-\frac{h_{4}}{2h_{3}}\left[ n_{i}^{\ast \ast }+\gamma n_{i}^{\ast }%
\right]  &=&0,  \label{2ricci4a}
\end{eqnarray}%
where
\begin{equation}
\alpha _{i}=\partial _{i}h_{4}^{\ast }-h_{4}^{\ast }\partial _{i}\ln \sqrt{%
|h_{3}h_{4}|},~\gamma =3h_{4}^{\ast }/2h_{4}-h_{3}^{\ast }/h_{3},
\label{2abc}
\end{equation}%
and the partial derivatives are denoted $g_{1}^{\bullet }=\partial
g_{1}/\partial x^{1},g_{1}^{^{\prime }}=\partial g_{1}/\partial x^{2}$ and $%
h_{3}^{\ast }=\partial h_{3}/\partial v.$ We \ can additionally impose the
condition $\delta _{i}N_{j}^{a}=\delta _{j}N_{i}^{a}$ in order to have $%
\Omega _{jk}^{a}=0$ which may be satisfied, for instance, if
\begin{equation*}
w_{1}=w_{1}\left( x^{1},v\right) ,n_{1}=n_{1}\left( x^{1},v\right)
,w_{2}=n_{2}=0,
\end{equation*}%
or, inversely, if
\begin{equation*}
w_{1}=n_{1}=0,w_{2}=w_{2}\left( x^{2},v\right) ,n_{2}=n_{2}\left(
x^{2},v\right) .
\end{equation*}%
In this paper we shall select a class of static solutions parametrized by
the conditions
\begin{equation}
w_{1}=w_{2}=n_{2}=0.  \label{2cond1}
\end{equation}

\ The system of equations (\ref{2ricci1a})--(\ref{2ricci4a}) can be integrated
in general form \cite{03vmag2,03vth}. Physical solutions are selected following
some additional boundary conditions, imposed types of symmetries,
nonlinearities and singular behavior and compatibility in the locally
anisotropic limits with some well known exact solutions.

Finally, we note that there is a difference between our approach and the
so--called ''tetradic'' gravity (see basic details and references in \cite%
{03mtw}) when the metric coefficients $g_{\alpha \beta }\left( u^{\gamma
}\right) $ are substituted by tetradic fields $e_{\alpha }^{\underline{%
\alpha }}\left( u^{\gamma }\right) ,$ mutually related by formula $g_{\alpha
\beta }=e_{\alpha }^{\underline{\alpha }}e_{\beta }^{\underline{\beta }}\eta
_{\underline{\alpha }\underline{\beta }}$ with $\eta _{\underline{\alpha }%
\underline{\beta }}$ chosen, for instance, to be the Minkowski metric. In
our case we partially preserve some metric dynamics given by diagonal
effective metric coefficients $\left( g_{i},h_{a}\right) $ but also adapt
the calculus to tetrads respectively defied by $\left( N_{i}^{a},\delta
_{i}^{j},\delta _{b}^{a}\right) ,$ see (\ref{2ddif4}) and (\ref{2dder4}). This
substantially simplifies the method of constructing exact solutions and also
reflects new type symmetries of such classes of metrics.

\section{Anholonomic Noncommutative Symmetries}

\label{ncgg}

The nontrivial anholonomy coefficients, see (\ref{1anhol}) and (\ref{2anh})
induced by off--diagonal metric (\ref{3cmetric4}) (and associated
N--connection) coefficients emphasize a kind of Lie algebra noncommutativity
relation. \ In this section, we analyze a simple realizations of
noncommutative geometry of anholonomic frames within the algebra of complex $%
k\times k$ matrices, $M_{k}(\C,u^{\alpha })$ depending on coordinates $%
u^{\alpha }$ on \ spacetime $V^{n+m}$ connected to complex Lie algebras $%
SL\left( k,\C\right) $ (see Ref. \cite{03vncs} for similar constructions with
the group $SU_{k}).$

We consider matrix valued functions of necessary smoothly class derived from
the anholonomic frame relations (\ref{1anhol}) (being similar to the Lie
algebra relations) with the coefficients (\ref{2anh}) induced by
off--diagonal metric terms in (\ref{12ansatzc4}) and by N--connection
coefficients $N_{i}^{a}.$ We use algebras of complex matrices in order to
have the possibility for some extensions to complex solutions and to relate
the constructions to noncommutative/complex gravity). For commutative
gravity models, the anholonomy coefficients $w_{~\alpha \beta }^{\gamma }$
are real functions but there are considered also complex metrics and tetrads
related to noncommutative gravity \cite{03ncg}.

Let us consider the basic relations for the simplest model of noncommutative
geometry realized with the algebra of complex $\left( k\times k\right) $
noncommutative matrices \cite{03dub}, $M_{k}(\C).$ Any element $M\in M_{k}(\C)$
can be represented as a linear combination of the unit $\left( k\times
k\right) $ matrix $I$ and $\left( k^{2}-1\right) $ hermitian traceless
matrices $q_{\underline{\alpha }}$ with the underlined index $\underline{%
\alpha }$ running values $1,2,...,k^{2}-1,$ i. e.
\begin{equation*}
M=\alpha \ I+\sum \beta ^{\underline{\alpha }}q_{\underline{\alpha }}
\end{equation*}%
for some constants $\alpha $ and $\beta ^{\underline{\alpha }}.$ It is
possible to chose the basis matrices $q_{\underline{\alpha }}$ satisfying
the relations%
\begin{equation}
q_{\underline{\alpha }}q_{\underline{\beta }}=\frac{1}{k}\rho _{\underline{%
\alpha }\underline{\beta }}I+Q_{\underline{\alpha }\underline{\beta }}^{%
\underline{\gamma }}q_{\underline{\gamma }}-\frac{i}{2}f_{~\underline{\alpha
}\underline{\beta }}^{\underline{\gamma }}q_{\underline{\gamma }},
\label{gr1}
\end{equation}%
where $i^{2}=-1$ and the real coefficients satisfy the properties
\begin{equation*}
Q_{\underline{\alpha }\underline{\beta }}^{\underline{\gamma }}=Q_{%
\underline{\beta }\underline{\alpha }}^{\underline{\gamma }},\ Q_{\underline{%
\gamma }\underline{\beta }}^{\underline{\gamma }}=0,\ f_{~\underline{\alpha }%
\underline{\beta }}^{\underline{\gamma }}=-f_{~\underline{\beta }\underline{%
\alpha }}^{\underline{\gamma }},f_{~\underline{\gamma }\underline{\alpha }}^{%
\underline{\gamma }}=0
\end{equation*}%
with $f_{~\underline{\alpha }\underline{\beta }}^{\underline{\gamma }}$
being the structure constants of the Lie group $SL\left( k,\C\right) $ and
the Killing--Cartan metric tensor $\rho _{\underline{\alpha }\underline{%
\beta }}=f_{~\underline{\alpha }\underline{\gamma }}^{\underline{\tau }}f_{~%
\underline{\tau }\underline{\beta }}^{\underline{\gamma }}.$ This algebra
admits a formalism of interior derivatives $\widehat{\partial }_{\underline{%
\gamma }}$ defied by relations
\begin{equation}
\widehat{\partial }_{\underline{\gamma }}q_{\underline{\beta }}=ad\left( iq_{%
\underline{\gamma }}\right) q_{\underline{\beta }}=i[q_{\underline{\gamma }%
},q_{\underline{\beta }}]=f_{~\underline{\gamma }\underline{\beta }}^{%
\underline{\alpha }}q_{\underline{\alpha }}  \label{der1a}
\end{equation}
and%
\begin{equation}
\widehat{\partial }_{\underline{\alpha }}\widehat{\partial }_{\underline{%
\beta }}-\widehat{\partial }_{\underline{\beta }}\widehat{\partial }_{%
\underline{\alpha }}=f_{~\underline{\alpha }\underline{\beta }}^{\underline{%
\gamma }}\widehat{\partial }_{\underline{\gamma }}  \label{jacob}
\end{equation}%
(the last relation follows the Jacoby identity and is quite similar to (\ref%
{1anhol}) but with constant values $f_{~\underline{\alpha }\underline{\beta }%
}^{\underline{\gamma }}).$

Our idea is to associate a noncommutative geometry starting from the
anholonomy relations of frames (\ref{1anhol}) by adding to the structure
constants $f_{~\underline{\alpha }\underline{\beta }}^{\underline{\gamma }}$
the anholonomy coefficients $w_{\ \alpha \gamma }^{[N]\tau }$ (\ref{2anh}) (we
shall put the label [N] if would be necessary to emphasize that the
anholonomic coefficients are induced by a nonlinear connection. Such deformed
structure constants consist from N--connection coefficients $N_{i}^{a}$ and
their first partial derivatives, i. e. they are induced by some
off--diagonal terms in the metric (\ref{12ansatzc4}) being a solution of the
gravitational field equations.

We emphasize that there is a rough analogy between formulas (\ref{jacob})
and (\ref{1anhol}) because the anholonomy coefficients do not satisfy, in
general, the condition $w_{\ \tau \alpha }^{[N]\tau }=0.$ There is also
another substantial difference because the anholonomy relations are defined
for a manifold of dimension $n+m,$ with Greek indices $\alpha ,\beta ,...$
running values from $1$ to $\ n+m$ but the matrix noncommutativity relations
are stated for traceless matrices labelled by underlined indices $\underline{%
\alpha },\underline{\beta },$ running values from $1$ to $k^{2}-1.$ It is
not possible to satisfy the condition $k^{2}-1=n+m$ by using integer numbers
for arbitrary $n+m.$ We may extend the dimension of spacetime from $n+m$ to
any $n^{\prime }\geq n$ and $m^{\prime }\geq m$ when the condition $%
k^{2}-1=n^{\prime }+m^{\prime }$ can be satisfied by a trivial embedding of
the metric (\ref{12ansatzc4}) into higher dimension, for instance, by adding
the necessary number of unities on the diagonal and writing
\begin{equation*}
\widehat{g}_{\underline{\alpha }\underline{\beta }}=\left[
\begin{array}{ccccc}
1 & ... & 0 & 0 & 0 \\
... & ... & ... & ... & ... \\
0 & ... & 1 & 0 & 0 \\
0 & ... & 0 & g_{ij}+N_{i}^{a}N_{j}^{b}h_{ab} & N_{j}^{e}h_{ae} \\
0 & ... & 0 & N_{i}^{e}h_{be} & h_{ab}%
\end{array}%
\right]
\end{equation*}%
and $e_{\underline{\alpha }}^{[N]}=\delta _{\underline{\alpha }}=\left(
1,1,...,e_{\alpha }^{[N]}\right) .$ For simplicity, we preserve the same
type of underlined Greek indices, $\underline{\alpha },\underline{\beta }%
...=1,2,...,k^{2}-1=n^{\prime }+m^{\prime }.$

The anholonomy coefficients $w_{~\alpha \beta }^{[N]\gamma }$ can be
extended with some trivial zero components and for consistency we rewrite
them without labelled indices, $w_{~\alpha \beta }^{[N]\gamma }\rightarrow
W_{~\underline{\alpha }\underline{\beta }}^{\underline{\gamma }}.$ The set
of anholonomy coefficients $w_{~\alpha \beta }^{[N]\gamma }$(\ref{1anhol})
may result in degenerated matrices, for instance for certain classes of
exact solutions of the Einstein equations. So, it  would not be a well
defined construction if we shall substitute the structure Lie algebra
constants directly by $w_{~\alpha \beta }^{[N]\gamma }.$ We can consider a
simple extension $w_{~\alpha \beta }^{[N]\gamma }\rightarrow W_{~\underline{%
\alpha }\underline{\beta }}^{\underline{\gamma }}$ when the coefficients $%
w_{~\underline{\alpha }\underline{\beta }}^{\underline{\gamma }}(u^{%
\underline{\tau }})$ for any fixed value $u^{\underline{\tau }}=u_{[0]}^{%
\underline{\tau }}$ would be some deformations of the\ structure constants
of the Lie algebra $SL\left( k,\C\right) ,$ like
\begin{equation}
W_{~\underline{\alpha }\underline{\beta }}^{\underline{\gamma }}=f_{~%
\underline{\alpha }\underline{\beta }}^{\underline{\gamma }}+w_{~\underline{%
\alpha }\underline{\beta }}^{\underline{\gamma }},  \label{anhb}
\end{equation}%
being nondegenerate. \

Instead of the matrix algebra $M_{k}(\C),$ constructed from constant complex
elements, we have also to introduce dependencies on coordinates $u^{%
\underline{\alpha }}=\left( 0,...,u^{\alpha }\right) ,$ for instance, like a
trivial matrix bundle on $V^{n^{\prime }+m^{\prime }},$ and denote this
space $M_{k}(\C,u^{\underline{\alpha }}).$ Any element $B\left( u^{%
\underline{\alpha }}\right) \in M_{k}(\C,u^{\underline{\alpha }})$ with a
noncommutative structure induced by $W_{~\underline{\alpha }\underline{\beta
}}^{\underline{\gamma }}$ is represented as a linear combination of the unit
$(n^{\prime }+m^{\prime })\times (n^{\prime }+m^{\prime })$ matrix $I$ and
the $[(n^{\prime }+m^{\prime })^{2}-1]$ hermitian traceless matrices $q_{%
\underline{\alpha }}\left( u^{\underline{\tau }}\right) $ with the
underlined index $\underline{\alpha }$ running values $1,2,...,(n^{\prime
}+m^{\prime })^{2}-1,$%
\begin{equation*}
B\left( u^{\underline{\tau }}\right) =\alpha \left( u^{\underline{\tau }%
}\right) \ I+\sum \beta ^{\underline{\alpha }}\left( u^{\underline{\tau }%
}\right) q_{\underline{\alpha }}\left( u^{\underline{\tau }}\right)
\end{equation*}%
under condition that the following relation holds:%
\begin{equation*}
q_{\underline{\alpha }}\left( u^{\underline{\tau }}\right) q_{\underline{%
\beta }}\left( u^{\underline{\gamma }}\right) =\frac{1}{n^{\prime
}+m^{\prime }}\rho _{\underline{\alpha }\underline{\beta }}\left( u^{%
\underline{\nu }}\right) +Q_{\underline{\alpha }\underline{\beta }}^{%
\underline{\gamma }}q_{\underline{\gamma }}\left( u^{\underline{\mu }%
}\right) -\frac{i}{2}W_{~\underline{\alpha }\underline{\beta }}^{\underline{%
\gamma }}q_{\underline{\gamma }}\left( u^{\underline{\mu }}\right)
\end{equation*}%
with the same values of $Q_{\underline{\alpha }\underline{\beta }}^{%
\underline{\gamma }}$ from the Lie algebra for $SL\left( k,\C\right) $ but
with the Killing--Cartan like metric tensor defined by anholonomy
coefficients, i. e. $\rho _{\underline{\alpha }\underline{\beta }}\left( u^{%
\underline{\nu }}\right) = W_{~\underline{\alpha }\underline{\gamma }}^{%
\underline{\tau }}\left( u^{\underline{\alpha }}\right)$\\ $ W_{~\underline{\tau }%
\underline{\beta }}^{\underline{\gamma }}\left( u^{\underline{\alpha }%
}\right) .$ For complex spacetimes, we shall consider that the coefficients $%
N_{\underline{i}}^{\underline{a}}$ and $W_{~\underline{\alpha }\underline{%
\beta }}^{\underline{\gamma }}$ may be some complex valued functions of
necessary smooth (in general, with complex variables) class. In result, the
Killing--Cartan like metric tensor $\rho _{\underline{\alpha }\underline{%
\beta }}$ can be also complex.

We rewrite (\ref{1anhol}) as
\begin{equation}
e_{\underline{\alpha }}e_{\underline{\beta }}-e_{\underline{\beta }}e_{%
\underline{\alpha }}=W_{~\underline{\alpha }\underline{\beta }}^{\underline{%
\gamma }}e_{\underline{\gamma }}  \label{anh1}
\end{equation}%
being equivalent to (\ref{jacob}) and defining a noncommutative anholonomic
structure (for simplicity, we use the same symbols $e_{\underline{\alpha }}$
as for some 'N--elongated' partial derivatives, but with underlined
indices). The analogs of derivation operators (\ref{der1a}) are stated by
using $W_{~\underline{\alpha }\underline{\beta }}^{\underline{\gamma }},$%
\begin{equation}
e_{\underline{\alpha }}q_{\underline{\beta }}\left( u^{\underline{\gamma }%
}\right) =ad\left[ iq_{\underline{\alpha }}\left( u^{\underline{\gamma }%
}\right) \right] q_{\underline{\beta }}\left( u^{\underline{\gamma }}\right)
=i\left[ q_{\underline{\alpha }}\left( u^{\underline{\gamma }}\right) q_{%
\underline{\beta }}\left( u^{\underline{\gamma }}\right) \right] =W_{~%
\underline{\alpha }\underline{\beta }}^{\underline{\gamma }}q_{\underline{%
\gamma }}  \label{nder1}
\end{equation}

The operators (\ref{nder1}) define a linear space of anholonomic derivations
satisfying the conditions (\ref{anh1}), denoted $AderM_{k}(\C,u^{\underline{%
\alpha }}),$ elongated by N--connection and distinguished into irreducible
h-- and v--components, respectively, into $e_{\underline{i}}$ and $e_{%
\underline{b}},$ like $e_{\underline{\alpha }}=\left( e_{\underline{i}%
}=\partial _{\underline{i}}-N_{\underline{i}}^{\underline{a}}e_{\underline{a}%
},e_{\underline{b}}=\partial _{\underline{b}}\right) .$ The space $AderM_{k}(%
\C,u^{\underline{\alpha }})$ is not a left module over \ the algebra $M_{k}(%
\C,u^{\underline{\alpha }})$ which means that there is a a substantial
difference with respect to the usual commutative differential geometry where
a vector field multiplied on the left by a function produces a new vector
field.

The duals to operators (\ref{nder1}), $e^{\underline{\mu }},$ found from $e^{%
\underline{\mu }}\left( e_{\underline{\alpha }}\right) =\delta _{_{%
\underline{\alpha }}}^{\underline{\mu }}I,$ define a canonical basis of
1--forms $e^{\underline{\mu }}$ connected to the N--connection structure. By
using these forms, we can span a left module over $M_{k}(\C,u^{\underline{%
\alpha }})$ following $q_{\underline{\alpha }}e^{\underline{\mu }}\left( e_{%
\underline{\beta }}\right) =q_{\underline{\alpha }}\delta _{_{\underline{%
\beta }}}^{\underline{\mu }}I=q_{\underline{\alpha }}\delta _{_{\underline{%
\beta }}}^{\underline{\mu }}.$ \ For an arbitrary vector field
\begin{equation*}
Y=Y^{\alpha }e_{\alpha }\rightarrow Y^{\underline{\alpha }}e_{\underline{%
\alpha }}=Y^{\underline{i}}e_{\underline{i}}+Y^{\underline{a}}e_{\underline{a%
}},
\end{equation*}%
it is possible to define an exterior differential (in our case being
N--elongated), starting with the action on a function $\varphi $
(equivalent, a 0--form),
\begin{equation*}
\delta \ \varphi \left( Y\right) =Y\varphi =Y^{\underline{i}}\delta _{%
\underline{i}}\varphi +Y^{\underline{a}}\partial _{\underline{a}}\varphi
\end{equation*}%
when%
\begin{equation*}
\left( \delta \ I\right) \left( e_{\underline{\alpha }}\right) =e_{%
\underline{\alpha }}I=ad\left( ie_{\underline{\alpha }}\right) I=i\left[ e_{%
\underline{\alpha }},I\right] =0,\mbox{ i. e. }\delta I=0,
\end{equation*}%
and
\begin{equation}
\delta q_{\underline{\mu }}(e_{\underline{\alpha }})=e_{\underline{\alpha }%
}(e_{\underline{\mu }})=i[e_{\underline{\mu }},e_{\underline{\alpha }}]=W_{~%
\underline{\alpha }\underline{\mu }}^{\underline{\gamma }}e_{\underline{%
\gamma }}.  \label{2aux1}
\end{equation}%
Considering the nondegenerated case, we can invert (\ref{2aux1}) as to obtain
a similar expression with respect to $e^{\underline{\mu }},$%
\begin{equation}
\delta (e_{\underline{\alpha }})=W_{~\underline{\alpha }\underline{\mu }}^{%
\underline{\gamma }}e_{\underline{\gamma }}e^{\underline{\mu }},
\label{1aux2}
\end{equation}%
from which a very important property follows by using the Jacobi identity, $%
\delta ^{2}=0,$ resulting in a possibility to define a usual Grassman
algebra of $p$--forms with the wedge product $\wedge $ stated as%
\begin{equation*}
e^{\underline{\mu }}\wedge e^{\underline{\nu }}=\frac{1}{2}\left( e^{%
\underline{\mu }}\otimes e^{\underline{\nu }}-e^{\underline{\nu }}\otimes e^{%
\underline{\mu }}\right) .
\end{equation*}%
We can write (\ref{1aux2}) as
\begin{equation*}
\delta (e^{\underline{\alpha }})=-\frac{1}{2}W_{~\underline{\beta }%
\underline{\mu }}^{\underline{\alpha }}e^{\underline{\beta }}e^{\underline{%
\mu }}
\end{equation*}%
and introduce the canonical 1--form $e=q_{\underline{\alpha }}e^{\underline{%
\alpha }}$ being coordinate--independent and adapted to the N--connection
structure and satisfying the condition $\delta e+e\wedge e=0.$

In a standard manner, we can introduce the volume element induced by the
canonical Cartan--Killing metric and the corresponding star operator $\star $%
\ (Hodge duality).\ We define the volume element $\sigma $ by using the
complete antisymmetric tensor $\epsilon _{\underline{\alpha }_{1}\underline{%
\alpha }_{2}...\underline{\alpha }_{k^{2}-1}}$as
\begin{equation*}
\sigma =\frac{1}{\left[ (n^{\prime }+m^{\prime })^{2}-1\right] !}\epsilon _{%
\underline{\alpha }_{1}\underline{\alpha }_{2}...\underline{\alpha }%
_{n^{\prime }+m^{\prime }}}e^{\underline{\alpha }_{1}}\wedge e^{\underline{%
\alpha }_{2}}\wedge ...\wedge e^{\underline{\alpha }_{n^{\prime }+m^{\prime
}}}
\end{equation*}%
to which any $\left( k^{2}-1\right) $--form is proportional $\left(
k^{2}-1=n^{\prime }+m^{\prime }\right) .$ The integral of such a form is
defined as the trace of the matrix coefficient in the from $\sigma $ and the
scalar product introduced for any couple of $p$--forms $\varpi $ and $\psi $%
\begin{equation*}
\left( \varpi ,\psi \right) =\int \left( \varpi \wedge \star \psi \right) .
\end{equation*}

Let us analyze how a noncommutative differential form calculus (induced by
an anholonomic structure) can be developed and related to the Hamiltonian
classical and quantum mechanics and Poisson bracket formalism:

For a $p$--form $\varpi ^{\lbrack p]},$ the anti--derivation $i_{Y}$ with
respect to a vector field\\ $Y\in AderM_{k}(\C,u^{\underline{\alpha }})$ can
be defined as in the usual formalism,%
\begin{equation*}
i_{Y}\varpi ^{\lbrack p]}\left( X_{1},X_{2},...,X_{p-1}\right) =\varpi
^{\lbrack p]}\left( Y,X_{1},X_{2},...,X_{p-1}\right)
\end{equation*}%
where $X_{1},X_{2},...,X_{p-1}\in AderM_{k}(\C,u^{\underline{\alpha }}).$ By
a straightforward calculus we can check that for a 2--form $\Xi =\delta $ $e$
one holds
\begin{equation*}
\delta \Xi =\delta ^{2}e=0\mbox{ and }L_{Y}\Xi =0
\end{equation*}%
where the Lie derivative of forms is defined as $L_{Y}\varpi ^{\lbrack
p]}=\left( i_{Y}\ \delta +\delta \ i_{Y}\right) \varpi ^{\lbrack p]}.$

The Hamiltonian vector field $H_{[\varphi ]}$ of an element of algebra $%
\varphi \in M_{k}(\C,u^{\underline{\alpha }})$ is introduced following the
equality $\Xi \left( H_{[\varphi ]},Y\right) =Y\varphi $ which holds for any
vector field. Then, we can define the Poisson bracket of two functions (in a
quantum variant, observables) $\varphi $ and $\chi ,$ $\{\varphi ,\chi
\}=\Xi \left( H_{[\varphi ]},H_{[\chi ]}\right) $ when
\begin{equation*}
\{e_{\underline{\alpha }},e_{\underline{\beta }}\}=\Xi \left( e_{\underline{%
\alpha }},e_{\underline{\beta }}\right) =i[e_{\underline{\alpha }},e_{%
\underline{\beta }}].
\end{equation*}%
This way, a simple version of noncommutative classical and quantum mechanics
(up to a factor like the Plank constant, $\hbar $) is proposed, being
derived by anholonomic relations for a certain class of exact
'off--diagonal' solutions in commutative gravity.

In order to define the algebra of forms $\Omega ^{\ast }\left[ M_{k}(\C,u^{%
\underline{\alpha }})\right] $ over $M_{k}(\C,u^{\underline{\alpha }})$ we
put $\Omega ^{0}=$ $M_{k}$ and write
\begin{equation*}
\delta \varphi \left( e_{\underline{\alpha }}\right) =e_{\underline{\alpha }%
}(\varphi )
\end{equation*}%
for every matrix function $\varphi \in M_{k}(\C,u^{\underline{\alpha }}).$
As a particular case, we have
\begin{equation*}
\delta q^{\underline{\alpha }}\left( e_{\underline{\beta }}\right) =-W_{~%
\underline{\beta }\underline{\gamma }}^{\underline{\alpha }}q^{\underline{%
\gamma }}
\end{equation*}%
where indices are raised and lowered with the anholonomically deformed
metric $\rho _{\underline{\alpha }\underline{\beta }}(u^{\underline{\lambda }%
}).$ This way, we can define the set of 1--forms $\Omega ^{1}\left[ M_{k}(\C%
,u^{\underline{\alpha }})\right] $ to be the set of all elements of the form
$\varphi \delta \beta $ with $\varphi $ and $\beta $ belonging to $M_{k}(\C%
,u^{\underline{\alpha }}).$ The set of all differential forms define a
differential algebra $\Omega ^{\ast }\left[ M_{k}(\C,u^{\underline{\alpha }})%
\right] $ with the couple $\left( \Omega ^{\ast }\left[ M_{k}(\C,u^{%
\underline{\alpha }})\right] ,\delta \right) $ said to be a differential
calculus in $M_{k}(\C,u^{\underline{\alpha }})$ induced by the anholonomy of
certain exact solutions (with off--diagonal metrics and associated
N--connections) in a gravity theory.

We can also find a set of generators $e^{\underline{\alpha }}$ of $\Omega
^{1}\left[ M_{k}(\C,u^{\underline{\alpha }})\right] ,$ as a left/ right
--module completely characterized by the duality equations $e^{\underline{%
\mu }}\left( e_{\underline{\alpha }}\right) =\delta _{_{\underline{\alpha }%
}}^{\underline{\mu }}I$ and related to $\delta q^{\underline{\alpha }},$%
\begin{equation*}
\delta q^{\underline{\alpha }}=W_{~\underline{\beta }\underline{\gamma }}^{%
\underline{\alpha }}q^{\underline{\beta }}q^{\underline{\gamma }}\mbox{ and }%
e^{\underline{\mu }}=q_{\underline{\gamma }}q^{\underline{\mu }}\delta q^{%
\underline{\gamma }}.
\end{equation*}%
Similarly to the formalism presented in details in Ref. \cite{03madore}, we
can elaborate a differential calculus with derivations by introducing a
linear torsionless connection%
\begin{equation*}
\mathcal{D}e^{\underline{\mu }}=-\omega _{\ \underline{\gamma }}^{\underline{%
\mu }}\otimes e^{\underline{\gamma }}
\end{equation*}%
with the coefficients $\omega _{\ \underline{\gamma }}^{\underline{\mu }}=-%
\frac{1}{2}W_{~\underline{\gamma }\underline{\beta }}^{\underline{\mu }}e^{%
\underline{\gamma }},$ resulting in the curvature 2--form%
\begin{equation*}
\mathcal{R}_{\ \underline{\gamma }}^{\underline{\mu }}=\frac{1}{8}W_{~%
\underline{\gamma }\underline{\beta }}^{\underline{\mu }}W_{~\underline{%
\alpha }\underline{\tau }}^{\underline{\beta }}e^{\underline{\alpha }}e^{%
\underline{\tau }}.
\end{equation*}
This is a surprising fact that 'commutative' curved spacetimes provided with
off--diagonal metrics and associated anhlonomic frames and N--connections
may be characterized by a noncommutative 'shadow' space with a Lie algebra
like structure induced by the frame anholonomy. We argue that such metrics
possess anholonomic noncommutative symmetries and conclude that for the
'holonomic' solutions of the Einstein equations, with vanishing $w_{~%
\underline{\alpha }\underline{\beta }}^{\underline{\gamma }},$ any
associated noncommutative geometry or $SL\left( k,\C\right) $ decouples from
the off--diagonal (anholonomic) gravitational background and transforms into
a trivial one defined by the corresponding structure constants of the chosen
Lie algebra. The anholonomic noncommutativity and the related differential
geometry are induced by the anholonomy coefficients. All such structures
reflect a specific type of symmetries of generic off--diagonal metrics and
associated frame/ N--connection structures.

Considering exact solutions of the gravitational field equations, we can
assert that we constructed a class of vacuum or nonvacuum metrics possessing
a specific noncommutative symmetry instead, for instance, of any usual
Killing symmetry. In general, we can introduce a new classification of
spacetimes following anholonomic noncommutative algebraic properties of
metrics and vielbein structures (see Ref. \cite{03vnc,03vncs}). In this paper,
we analyze the simplest examples of such spacetimes.

\newpage

\section{4D Static Black Ellipsoids in MAG and String Gravity}

We outline the black ellipsoid solutions \cite{03velp1,03velp2} and discuss
their associated anholonomic noncommutative symmetries \cite{03vncs}. We note
that such solutions can be extended for the (anti) de Sitter spaces, in
gauge gravity and string gravity with effective cosmological constant \cite%
{03vncfg}. In this paper, the solutions are considered for 'real'
metric--affine spaces and extended to nontrivial cosmological constant.  We
emphasize the possibility to construct solutions with locally ''anisotropic''
cosmological constants (such configurations may be also induced, for
instance, from string/ brane gravity).

\subsection{Anholonomic deformations of the Schwarzschild metric}

We consider a particular case of effectively diagonalized (\ref{2dmetric4})
(and corresponding off--diagonal metric ansatz (\ref{3cmetric4})) when
\begin{eqnarray}
\delta s^{2} &=&[-\left( 1-\frac{2m}{r}+\frac{\varepsilon }{r^{2}}\right)
^{-1}dr^{2}-r^{2}q(r)d\theta ^{2}  \label{sch} \\
&&-\eta _{3}\left( r,\varphi \right) r^{2}\sin ^{2}\theta d\varphi ^{2}+\eta
_{4}\left( r,\varphi \right) \left( 1-\frac{2m}{r}+\frac{\varepsilon }{r^{2}}%
\right) \delta t^{2}]  \notag
\end{eqnarray}%
where the ''polarization'' functions $\eta _{3,4}$ are decomposed on a small
parameter $\varepsilon ,0<\varepsilon \ll 1,$
\begin{eqnarray}
\eta _{3}\left( r,\varphi \right) &=&\eta _{3[0]}\left( r,\varphi \right)
+\varepsilon \lambda _{3}\left( r,\varphi \right) +\varepsilon ^{2}\gamma
_{3}\left( r,\varphi \right) +...,  \label{decom1} \\
\eta _{4}\left( r,\varphi \right) &=&1+\varepsilon \lambda _{4}\left(
r,\varphi \right) +\varepsilon ^{2}\gamma _{4}\left( r,\varphi \right) +...,
\notag
\end{eqnarray}%
and
\begin{equation*}
\delta t=dt+n_{1}\left( r,\varphi \right) dr
\end{equation*}%
for $n_{1}\sim \varepsilon ...+\varepsilon ^{2}$ terms. The functions $%
q(r),\eta _{3,4}\left( r,\varphi \right) $ and $n_{1}\left( r,\varphi
\right) $ will be found as the metric will define a solution of the
gravitational field equations generated by small deformations of the
spherical static symmetry on a small positive parameter $\varepsilon $ (in
the limits $\varepsilon \rightarrow 0$ and $q,\eta _{3,4}\rightarrow 1$ we
have just the Schwarzschild solution for a point particle of mass $m).$ The
metric (\ref{sch}) is a particular case of a class of exact solutions
constructed in \cite{03v,03v1,03vth}. Its complexification by complex valued
N--coefficients is investigated in Ref. \cite{03vncs}.

We can state a particular symmetry of the metric (\ref{sch}) by imposing a
corresponding condition of vanishing of the metric coefficient before the
term $\delta t^{2}.$ For instance, the constraints that
\begin{eqnarray}
\eta _{4}\left( r,\varphi \right) \left( 1-\frac{2m}{r}+\frac{\varepsilon }{%
r^{2}}\right)  &=&1-\frac{2m}{r}+\varepsilon \frac{\Phi _{4}}{r^{2}}%
+\varepsilon ^{2}\Theta _{4}=0,  \label{hor1} \\
\Phi _{4} &=&\lambda _{4}\left( r^{2}-2mr\right) +1  \notag \\
\Theta _{4} &=&\gamma _{4}\left( 1-\frac{2m}{r}\right) +\lambda _{4},  \notag
\end{eqnarray}%
define a rotation ellipsoid configuration if
\begin{equation*}
\lambda _{4}=\left( 1-\frac{2m}{r}\right) ^{-1}(\cos \varphi -\frac{1}{r^{2}}%
)
\end{equation*}%
and
\begin{equation*}
\gamma _{4}=-\lambda _{4}\left( 1-\frac{2m}{r}\right) ^{-1}.
\end{equation*}%
In the first order on $\varepsilon $ one obtains \ a zero value for the
coefficient before $\delta t^{2}$ if
\begin{equation}
r_{+}=\frac{2m}{1+\varepsilon \cos \varphi }=2m[1-\varepsilon \cos \varphi ],
\label{ebh}
\end{equation}%
which is the equation for a 3D ellipsoid like hypersurface with a small
eccentricity $\varepsilon .$ In general, we can consider arbitrary pairs of
functions $\lambda _{4}(r,\theta ,\varphi )$ and $\gamma _{4}(r,\theta
,\varphi )$ (for $\varphi $--anisotropies, \ $\lambda _{4}(r,\varphi )$ and $%
\gamma _{4}(r,\varphi ))$ which may be singular for some values of $r,$ or
on some hypersurvaces $r=r\left( \theta ,\varphi \right) $ ($r=r(\varphi )).$

The simplest way to define the condition of vanishing of the metric
coefficient before the value $\delta t^{2}$ is to choose $\gamma _{4}$ and $%
\lambda _{4}$ as to have $\Theta =0.$ In this case we find from \ (\ref{hor1}%
)%
\begin{equation}
r_{\pm }=m\pm m\sqrt{1-\varepsilon \frac{\Phi }{m^{2}}}=m\left[ 1\pm \left(
1-\varepsilon \frac{\Phi _{4}}{2m^{2}}\right) \right]  \label{hor1a}
\end{equation}%
where $\Phi _{4}\left( r,\varphi \right) $ is taken for $r=2m.$

For a new radial coordinate
\begin{equation}
\xi =\int dr\sqrt{|1-\frac{2m}{r}+\frac{\varepsilon }{r^{2}}|}  \label{int2}
\end{equation}%
and
\begin{equation}
h_{3}=-\eta _{3}(\xi ,\varphi )r^{2}(\xi )\sin ^{2}\theta ,\ h_{4}=1-\frac{2m%
}{r}+\varepsilon \frac{\Phi _{4}}{r^{2}},  \label{sch1q}
\end{equation}%
when $r=r\left( \xi \right) $ is inverse function after integration in (\ref%
{int2}), we write the metric (\ref{sch}) as
\begin{eqnarray}
ds^{2} &=&-d\xi ^{2}-r^{2}\left( \xi \right) q\left( \xi \right) d\theta
^{2}+h_{3}\left( \xi ,\theta ,\varphi \right) \delta \varphi
^{2}+h_{4}\left( \xi ,\theta ,\varphi \right) \delta t^{2},  \label{sch1} \\
\delta t &=&dt+n_{1}\left( \xi ,\varphi \right) d\xi ,  \notag
\end{eqnarray}%
where the coefficient $n_{1}$ is taken to solve the equation (\ref{2ricci4a})
and to satisfy the (\ref{2cond1}). The next step is to state the conditions
when the coefficients of metric (\ref{sch}) define solutions of the Einstein
equations. We put  $g_{1}=-1,g_{2}=-r^{2}\left( \xi \right) q\left( \xi
\right) $ and arbitrary $h_{3}(\xi ,\theta ,\varphi )$ and $h_{4}\left( \xi
,\theta \right) $ in order to find solutions the equations (\ref{2ricci1a})--(%
\ref{2ricci3a}). If $h_{4}$ depends on anisotropic variable $\varphi ,$ the
equation (\ref{2ricci2a}) may be solved if
\begin{equation}
\sqrt{|\eta _{3}|}=\eta _{0}\left( \sqrt{|\eta _{4}|}\right) ^{\ast }
\label{conda}
\end{equation}%
for $\eta _{0}=const.$ Considering decompositions of type (\ref{decom1}) we
put $\eta _{0}=\eta /\varepsilon $ where the constant $\eta $ is taken as to
have $\sqrt{|\eta _{3}|}=1$ in the limits
\begin{equation}
\frac{\left( \sqrt{|\eta _{4}|}\right) ^{\ast }\rightarrow 0}{\varepsilon
\rightarrow 0}\rightarrow \frac{1}{\eta }=const.  \label{condb}
\end{equation}%
These conditions are satisfied if the functions $\eta _{3[0]},$ $\lambda
_{3,4}$ and $\gamma _{3,4}$ are related via relations
\begin{equation*}
\sqrt{|\eta _{3[0]}|}=\frac{\eta }{2}\lambda _{4}^{\ast },\lambda _{3}=\eta
\sqrt{|\eta _{3[0]}|}\gamma _{4}^{\ast }
\end{equation*}%
for arbitrary $\gamma _{3}\left( r,\varphi \right) .$ In this paper we
select only such solutions which satisfy the conditions (\ref{conda}) and (%
\ref{condb}).

For linear infinitesimal extensions on $\varepsilon $ of the Schwarzschild
metric, we write the solution of (\ref{2ricci4a}) as
\begin{equation*}
n_{1}=\varepsilon \widehat{n}_{1}\left( \xi ,\varphi \right)
\end{equation*}%
where
\begin{eqnarray}
\widehat{n}_{1}\left( \xi ,\varphi \right) &=&n_{1[1]}\left( \xi \right)
+n_{1[2]}\left( \xi \right) \int d\varphi \ \eta _{3}\left( \xi ,\varphi
\right) /\left( \sqrt{|\eta _{4}\left( \xi ,\varphi \right) |}\right)
^{3},\eta _{4}^{\ast }\neq 0;  \label{1auxf4} \\
&=&n_{1[1]}\left( \xi \right) +n_{1[2]}\left( \xi \right) \int d\varphi \
\eta _{3}\left( \xi ,\varphi \right) ,\eta _{4}^{\ast }=0;  \notag \\
&=&n_{1[1]}\left( \xi \right) +n_{1[2]}\left( \xi \right) \int d\varphi
/\left( \sqrt{|\eta _{4}\left( \xi ,\varphi \right) |}\right) ^{3},\eta
_{3}^{\ast }=0;  \notag
\end{eqnarray}%
with the functions $n_{k[1,2]}\left( \xi \right) $ to be stated by boundary
conditions.

The data
\begin{eqnarray}
g_{1} &=&-1,g_{2}=-r^{2}(\xi )q(\xi ),  \label{data} \\
h_{3} &=&-\eta _{3}(\xi ,\varphi )r^{2}(\xi )\sin ^{2}\theta ,~h_{4}=1-\frac{%
2m}{r}+\varepsilon \frac{\Phi _{4}}{r^{2}},  \notag \\
w_{1,2} &=&0,n_{1}=\varepsilon \widehat{n}_{1}\left( \xi ,\varphi \right)
,n_{2}=0,  \notag
\end{eqnarray}%
for the metric (\ref{sch}) define a class of solutions of the Einstein
equations for the canonical distinguished connection (\ref{2dcon}), with
non--trivial polarization function $\eta _{3}$ and extended on parameter $%
\varepsilon $ up to the second order (the polarization functions being taken
as to make zero the second order coefficients). Such solutions are generated
by small deformations (in particular cases of rotation ellipsoid symmetry)
of the Schwarschild metric.

We can relate our solutions with some small deformations of the
Schwar\-zschild metric, as well we can satisfy the asymptotically flat
condition, if we chose such functions $n_{k[1,2]}\left( x^{i}\right) $ as $%
n_{k}\rightarrow 0$ for $\varepsilon \rightarrow 0$ and $\eta
_{3}\rightarrow 1.$ These functions have to be selected as to vanish far
away from the horizon, for instance, like $\sim 1/r^{1+\tau },\tau >0,$ for
long distances $r\rightarrow \infty .$

\subsection{Black ellipsoids and anistropic cosmological constants}

\label{belsg}

We can generalize the gravitational field equations to the gravity with
variable cosmological constants $\lambda _{\lbrack h]}\left( u^{\alpha
}\right) $ and $\lambda _{\lbrack v]}\left( u^{\alpha }\right) $ which can
be induced, for instance, from extra dimensions in string/brane gravity,
when the non-trivial components of the Einstein equations are
\begin{equation}
R_{ij}=\lambda _{\lbrack h]}\left( x^{1}\right) g_{ij}\mbox{ and }%
R_{ab}=\lambda _{\lbrack v]}(x^{k},v)g_{ab}  \label{1eq17}
\end{equation}%
where Ricci tensor $R_{\mu \nu }$ with anholonomic variables has two
nontrivial components $R_{ij}$ and $R_{ab},$ and the indices take values $%
i,k=1,2$ and $a,b=3,4$ for $x^{i}=\xi $ and $y^{3}=v=\varphi $ (see
notations from the previous subsection). The equations (\ref{1eq17}) contain
the equations (\ref{2ricci1a}) and (\ref{2ricci2a}) as particular cases when $%
\lambda _{\lbrack h]}\left( x^{1}\right) =\frac{\lambda _{\lbrack H]}^{2}}{4}
$ and $\lambda _{\lbrack v]}(x^{k},v)=\frac{\lambda _{\lbrack H]}^{2}}{4}%
+\Upsilon _{2}\left( x^{k}\right) .$

For an ansatz of type (\ref{2dmetric4})
\begin{eqnarray}
\delta s^{2} &=&g_{1}(dx^{1})^{2}+g_{2}(dx^{2})^{2}+h_{3}\left(
x^{i},y^{3}\right) (\delta y^{3})^{2}+h_{4}\left( x^{i},y^{3}\right) (\delta
y^{4})^{2},  \label{1ansatz18} \\
\delta y^{3} &=&dy^{3}+w_{i}\left( x^{k},y^{3}\right) dx^{i^{\prime }},\quad
\delta y^{4}=dy^{4}+n_{i}\left( x^{k},y^{3}\right) dx^{i},  \notag
\end{eqnarray}%
the Einstein equations (\ref{1eq17}) are written (see \cite{03vth} for details
on computation)
\begin{eqnarray}
R_{1}^{1}=R_{2}^{2}=-\frac{1}{2g_{1}g_{2}}[g_{2}^{\bullet \bullet }-\frac{%
g_{1}^{\bullet }g_{2}^{\bullet }}{2g_{1}}-\frac{(g_{2}^{\bullet })^{2}}{%
2g_{2}}+g_{1}^{^{\prime \prime }}-\frac{g_{1}^{^{\prime }}g_{2}^{^{\prime }}%
}{2g_{2}}-\frac{(g_{1}^{^{\prime }})^{2}}{2g_{1}}] &=&\lambda _{\lbrack
h]}\left( x^{k}\right) ,  \label{ricci1s} \\
R_{3}^{3}=R_{4}^{4}=-\frac{\beta }{2h_{3}h_{4}} &=&\lambda _{\lbrack
v]}(x^{k},v),  \label{ricci2s} \\
R_{3i}=-w_{i}\frac{\beta }{2h_{4}}-\frac{\alpha _{i}}{2h_{4}} &=&0,
\label{ricci3s} \\
R_{4i}=-\frac{h_{4}}{2h_{3}}\left[ n_{i}^{\ast \ast }+\gamma n_{i}^{\ast }%
\right] &=&0.  \label{ricci4s}
\end{eqnarray}%
The coefficients of equations (\ref{ricci1s}) - (\ref{ricci4s}) are given by
\begin{equation}
\alpha _{i}=\partial _{i}{h_{4}^{\ast }}-h_{4}^{\ast }\partial _{i}\ln \sqrt{%
|h_{3}h_{4}|},\qquad \beta =h_{4}^{\ast \ast }-h_{4}^{\ast }[\ln \sqrt{%
|h_{3}h_{4}|}]^{\ast },\qquad \gamma =\frac{3h_{4}^{\ast }}{2h_{4}}-\frac{%
h_{3}^{\ast }}{h_{3}}.  \label{abcs}
\end{equation}%
The various partial derivatives are denoted as $a^{\bullet }=\partial
a/\partial x^{1},a^{^{\prime }}=\partial a/\partial x^{2},a^{\ast }=\partial
a/\partial y^{3}.$ This system of equations can be solved by choosing one of
the ansatz functions (\textit{e.g.} $g_{1}\left( x^{i}\right) $ or $%
g_{2}\left( x^{i}\right) )$ and one of the ansatz functions (\textit{e.g.} $%
h_{3}\left( x^{i},y^{3}\right) $ or $h_{4}\left( x^{i},y^{3}\right) )$ to
take some arbitrary, but physically interesting form. Then, the other ansatz
functions can be analytically determined up to an integration in terms of
this choice. In this way we can generate a lot of different solutions, but
we impose the condition that the initial, arbitrary choice of the ansatz
functions is ``physically interesting'' which means that one wants to make
this original choice so that the generated final solution yield a well
behaved metric.

In this subsection, we show that the data (\ref{data}) can be extended as to
generate exact black ellipsoid solutions with nontrivial polarized
cosmological constant which can be imbedded in string theory. A complex
generalization of the solution (\ref{data}) is analyzed in Ref. \cite{03vncs}
and the case locally isotropic cosmological constant was considered in Ref. %
\cite{03vncfg}.

At the first \ step, we find a class of solutions with $g_{1}=-1$ and $\quad
g_{2}=g_{2}\left( \xi \right) $ solving the equation (\ref{ricci1s}), which
under such parametrizations transforms to
\begin{equation}
g_{2}^{\bullet \bullet }-\frac{(g_{2}^{\bullet })^{2}}{2g_{2}}=2g_{2}\lambda
_{\lbrack h]}\left( \xi \right) .  \label{eqaux1}
\end{equation}%
With respect to the variable $Z=(g_{2})^{2}$ this equation is written as
\begin{equation*}
Z^{\bullet \bullet }+2\lambda _{\lbrack h]}\left( \xi \right) Z=0
\end{equation*}%
which can be integrated in explicit form if $\lambda _{\lbrack h]}\left( \xi
\right) =\lambda _{\lbrack h]0}=const,$
\begin{equation*}
Z=Z_{[0]}\sin \left( \sqrt{2\lambda _{\lbrack h]0}}\xi +\xi _{\lbrack
0]}\right) ,
\end{equation*}%
for some constants $Z_{[0]}$ and $\xi _{\lbrack 0]}$ which means that
\begin{equation}
g_{2}=-Z_{[0]}^{2}\sin ^{2}\left( \sqrt{2\lambda _{\lbrack h]0}}\xi +\xi
_{\lbrack 0]}\right)  \label{aux2p}
\end{equation}%
parametrize in 'real' string gravity a class of solution of (\ref{ricci1s})
for the signature\\ $\left( -,-,-,+\right) .$ For $\lambda _{\lbrack
h]}\rightarrow 0$ we can approximate $g_{2}=r^{2}\left( \xi \right) q\left(
\xi \right) =-\xi ^{2}$ and $Z_{[0]}^{2}=1$ which has compatibility with the
data (\ref{data}). The solution (\ref{aux2p}) with cosmological constant (of
string or non--string origin) induces oscillations in the ''horozontal''
part of the metric written with respect to N--adapted frames. If we put $%
\lambda _{\lbrack h]}\left( \xi \right) $ in (\ref{eqaux1}), we can search
the solution as $g_{2}=u^{2}$ where $u\left( \xi \right) $ solves the linear
equation%
\begin{equation*}
u^{\bullet \bullet }+\frac{\lambda _{\lbrack h]}\left( \xi \right) }{4}u=0.
\end{equation*}%
The method of integration of such equations is given in Ref. \cite{03kamke}.
The explicit forms of solutions depends on function $\lambda _{\lbrack
h]}\left( \xi \right) .$ In this case we have to write \
\begin{equation}
g_{2}=r^{2}\left( \xi \right) q^{[u]}\left( \xi \right) =u^{2}\left( \xi
\right) .  \label{aux3u}
\end{equation}%
For a suitable smooth behavior of $\lambda _{\lbrack h]}\left( \xi \right)
, $ we can generate such $u\left( \xi \right) $ and $r\left( \xi \right) $
when the $r=r\left( \xi \right) $ is the inverse function after integration
in (\ref{int2}).

The next step is to solve the equation (\ref{ricci2s}),%
\begin{equation*}
h_{4}^{\ast \ast }-h_{4}^{\ast }[\ln \sqrt{|h_{3}h_{4}|}]^{\ast }=-2\lambda
_{\lbrack v]}(x^{k},v)h_{3}h_{4}.
\end{equation*}%
For $\lambda =0$ a class of solution is given by any $\widehat{h}_{3}$ and $%
\widehat{h}_{4}$ related as
\begin{equation*}
\widehat{h}_{3}=\eta _{0}\left[ \left( \sqrt{|\hat{h}_{4}|}\right) ^{\ast }%
\right] ^{2}
\end{equation*}%
for a constant $\eta _{0}$ chosen to be negative in order to generate the
signature $\left( -,-,-,+\right) .$ For non--trivial $\lambda ,$ we may
search the solution as
\begin{equation}
h_{3}=\widehat{h}_{3}\left( \xi ,\varphi \right) ~f_{3}\left( \xi ,\varphi
\right) \mbox{ and }h_{4}=\widehat{h}_{4}\left( \xi ,\varphi \right) ,
\label{1sol15}
\end{equation}%
which solves (\ref{ricci2s}) if $f_{3}=1$ for $\lambda _{\lbrack v]}=0$ and
\begin{equation*}
f_{3}=\frac{1}{4}\left[ \int \frac{\lambda _{\lbrack v]}\hat{h}_{3}\hat{h}%
_{4}}{\hat{h}_{4}^{\ast }}d\varphi \right] ^{-1}\mbox{ for }\lambda
_{\lbrack v]}\neq 0.
\end{equation*}

Now it is easy to write down the solutions of equations (\ref{ricci3s})
(being a linear equation for $w_{i})$ and (\ref{ricci4s}) (after two
integrations of $n_{i}$ on $\varphi ),$%
\begin{equation}
w_{i}=\varepsilon \widehat{w}_{i}=-\alpha _{i}/\beta ,  \label{aux3s}
\end{equation}%
were $\alpha _{i}$ and $\beta $ are computed by putting (\ref{1sol15}) $\ $%
into corresponding values from (\ref{abcs}) (we chose the initial conditions
as $w_{i}\rightarrow 0$ for $\varepsilon \rightarrow 0)$ and
\begin{equation*}
n_{1}=\varepsilon \widehat{n}_{1}\left( \xi ,\varphi \right)
\end{equation*}%
where the coefficients
\begin{eqnarray}
\widehat{n}_{1}\left( \xi ,\varphi \right) &=&n_{1[1]}\left( \xi \right)
+n_{1[2]}\left( \xi \right) \int d\varphi \ ~f_{3}\left( \xi ,\varphi
\right) \eta _{3}\left( \xi ,\varphi \right) /\left( \sqrt{|\eta _{4}\left(
\xi ,\varphi \right) |}\right) ^{3},\eta _{4}^{\ast }\neq 0;  \label{aux4s}
\\
&=&n_{1[1]}\left( \xi \right) +n_{1[2]}\left( \xi \right) \int d\varphi \
~f_{3}\left( \xi ,\varphi \right) \eta _{3}\left( \xi ,\varphi \right) ,\eta
_{4}^{\ast }=0;  \notag \\
&=&n_{1[1]}\left( \xi \right) +n_{1[2]}\left( \xi \right) \int d\varphi
/\left( \sqrt{|\eta _{4}\left( \xi ,\varphi \right) |}\right) ^{3},\eta
_{3}^{\ast }=0;  \notag
\end{eqnarray}%
with the functions $n_{k[1,2]}\left( \xi \right) $ to be stated by boundary
conditions.

We conclude that the set of data $g_{1}=-1,$ with non--trivial $g_{2}\left(
\xi \right) ,h_{3},h_{4},w_{i^{\prime }}$ and $n_{1}$ stated respectively by
(\ref{aux2p}), (\ref{1sol15}), (\ref{aux3s}), (\ref{aux4s}) we can define a
black ellipsoid solution with explicit dependence on polarized cosmological
''constants'' $\lambda _{\lbrack h]}\left( x^{1}\right) $ and $\lambda
_{\lbrack v]}(x^{k},v),$ i. e. a metric (\ref{1ansatz18}).

Finally, we analyze the structure of noncommutative symmetries associated to
the (anti) de Sitter black ellipsoid solutions. The metric (\ref{1ansatz18})
with real and/or complex coefficients defining the corresponding solutions
and its analytic extensions also do not posses Killing symmetries being
deformed by anholonomic transforms. For this solution, we can associate
certain noncommutative symmetries following the same procedure as for the
Einstein real/ complex gravity but with additional nontrivial coefficients
of anholonomy and even with nonvanishing coefficients of the nonlinear
connection curvature, $\Omega _{12}^{3}=\delta _{1}N_{2}^{3}-\delta
_{2}N_{1}^{3}.$ Taking the data (\ref{aux3s}) and (\ref{aux4s}) and formulas
(\ref{1anhol}), we compute the corresponding nontrivial anholonomy
coefficients
\begin{eqnarray}
w_{\ 31}^{[N]4} &=&-w_{\ 13}^{[N]4}=\partial n_{1}\left( \xi ,\varphi
\right) /\partial \varphi =n_{1}^{\ast }\left( \xi ,\varphi \right) ,
\label{auxxx2} \\
w_{\ 12}^{[N]4} &=&-w_{\ 21}^{[N]4}=\delta _{1}(\alpha _{2}/\beta )-\delta
_{2}(\alpha _{1}/\beta )  \notag
\end{eqnarray}%
for $\delta _{1}=\partial /\partial \xi -w_{1}\partial /\partial \varphi $
and $\delta _{2}=\partial /\partial \theta -w_{2}\partial /\partial \varphi
, $ with $n_{1}$ defined by (\ref{aux4s}) and $\alpha _{1,2}$ and $\beta $
computed by using the formula (\ref{abcs}) for the solutions (\ref{1sol15}).
We have a 4D exact solution with nontrivial cosmological constant. So, for $%
n+m=4,$ the condition $k^{2}-1=n+m$ can not satisfied by any integer
numbers. We may trivially extend the dimensions like $n^{\prime }=6$ and $%
m^{\prime }=m=2$ and for $k=3$ to consider the \ Lie group $SL\left( 3,\C%
\right) $ noncommutativity with corresponding values of $Q_{\underline{%
\alpha }\underline{\beta }}^{\underline{\gamma }}$\ and structure constants $%
f_{~\underline{\alpha }\underline{\beta }}^{\underline{\gamma }},$ see (\ref%
{gr1}). An extension $w_{~\alpha \beta }^{[N]\gamma }\rightarrow W_{~%
\underline{\alpha }\underline{\beta }}^{\underline{\gamma }}$ may be
performed by stating the N--deformed ''structure'' constants (\ref{anhb}), $%
W_{~\underline{\alpha }\underline{\beta }}^{\underline{\gamma }}=f_{~%
\underline{\alpha }\underline{\beta }}^{\underline{\gamma }}+w_{~\underline{%
\alpha }\underline{\beta }}^{[N]\underline{\gamma }},$ with nontrivial
values of $w_{~\underline{\alpha }\underline{\beta }}^{[N]\underline{\gamma }%
}$ given by (\ref{auxxx2}). We note that the solutions with nontrivial
cosmological constants are with induced torsion with the coefficients
computed by using formulas (\ref{torsion}) and the data (\ref{aux3u}), (\ref%
{1sol15}), (\ref{aux3s}) and (\ref{aux4s}).

\subsection{ Analytic extensions of black ellipsoid metrics}

For the vacuum black ellipsoid metrics the method of analytic extension was
considered in Ref. \cite{03velp1,03velp2}. The coefficients of the metric (\ref%
{sch}) (equivalently (\ref{sch1})) written with respect to the \ anholonomic
frame (\ref{2ddif4}) has a number of similarities with the Schwrzschild and
Reissner--N\"{o}rdstrom solutions. The cosmological ''polarized'' constants
induce some additional factors like $q^{[u]}\left( \xi \right) $ and $%
f_{3}\left( \xi ,\varphi \right) $ (see, respectively, formulas (\ref{aux3u}%
) and (\ref{1sol15})) and modify the N--connection coefficients as in (\ref%
{aux3s}) and (\ref{aux4s}). For a corresponding class of smooth
polarizations, the functions $q^{[u]}$ and $f_{3}$ do not change the
singularity structure of the metric coefficients. If we identify $%
\varepsilon $ with $e^{2},$ we get a static metric with effective
''electric'' charge induced by a small, quadratic on $\varepsilon ,$
off--diagonal metric extension. The coefficients of this metric are similar
to those from the Reissner--N\"{o}rdstrom solution but additionally to the
mentioned frame anholonomy there are additional polarizations by the
functions $q^{[u]},h_{3[0]},f_{3},\eta _{3,4},w_{i}$ and $n_{1}.$ Another
very important property is that the deformed metric was stated to define a
vacuum, or with polarized cosmological constant, solution of the Einstein
equations which differs substantially from the usual Reissner--N\"{o}rdstrom
metric being an exact static solution of the Einstein--Maxwell equations.
For the limits $\varepsilon \rightarrow 0$ and $q,f_{3},h_{3[0]}\rightarrow
1 $ the metric (\ref{sch}) transforms into the usual Schwarzschild metric. A
solution with ellipsoid symmetry can be selected by a corresponding
condition of vanishing of the coefficient before the term $\delta t$ which
defines an ellipsoidal hypersurface like for the Kerr metric, but in our
case the metric is non--rotating. In general, the space may be with frame
induced torsion if we do not impose constraints on $w_{i}$ and $n_{1}$ as to
obtain vanishing nonlinear connection curvature and torsions.

The analytic extension of black ellipsoid solutions with cosmological
constant can be performed similarly both for anholonomic frames with induced
or trivial torsions. We note that the solutions in string theory may contain
a frame induced torsion with the components (\ref{1torsion}) (in general, we
can consider complex coefficients, see Ref. \cite{03vncs}) computed for
nontrivial $N_{i^{\prime }}^{3}=-\alpha _{i^{\prime }}/\beta $ (see (\ref%
{aux3s})) and $N_{1}^{4}=\varepsilon \widehat{n}_{1}\left( \xi ,\varphi
\right) $ (see (\ref{aux4s})). This is an explicit example illustrating that
the anholonomic frame method can be applied also for generating exact
solutions in models of gravity with nontrivial torsion. For such solutions,
we can perform corresponding analytic extensions and define Penrose diagram
formalisms if the constructions are considered with respect to N--elongated
vierbeins.

The metric (\ref{1ansatz18}) has a singular behavior for $r=r_{\pm },$ see (%
\ref{hor1a}). The aim of this subsection is to prove that this way we have
constructed a solution of the Einstein equations with polarized cosmological
constant. This solution possess an ''anisotropic'' horizon being a small
deformation on parameter $\varepsilon $ of the Schwarzschild's solution
horizon. We may analyze the anisotropic horizon's properties for some fixed
''direction'' given in a smooth vicinity of any values $\varphi =\varphi
_{0} $ and $r_{+}=r_{+}\left( \varphi _{0}\right) .$ $\ $The final
conclusions will be some general ones for arbitrary $\varphi $ when the
explicit values of coefficients will have a parametric dependence on angular
coordinate $\varphi .$ The metrics (\ref{sch}), or (\ref{sch1}), and (\ref%
{1ansatz18}) are regular in the regions I ($\infty >r>r_{+}^{\Phi }),$ II ($%
r_{+}^{\Phi }>r>r_{-}^{\Phi })$ and III $\;(r_{-}^{\Phi }>r>0).$ As in the
Schwarzschild, Reissner--N\"{o}rdstrom and Kerr cases these singularities
can be removed by introducing suitable coordinates and extending the
manifold to obtain a maximal analytic extension \cite{03gb,03carter}. We have
similar regions as in the Reissner--N\"{o}rdstrom space--time, but with just
only one possibility $\varepsilon <1$ instead of three relations for static
electro--vacuum cases ($e^{2}<m^{2},e^{2}=m^{2},e^{2}>m^{2};$ where $e$ and $%
m$ are correspondingly the electric charge and mass of the point particle in
the Reissner--N\"{o}rdstrom metric). So, we may consider the usual Penrose's
diagrams as for a particular case of the Reissner--N\"{o}rdstrom space--time
but keeping in mind that such diagrams and horizons have an additional
polarizations and parametrization on an angular coordinate.

We can proceed in steps analogous to those in the Schwarzschild case (see
details, for instance, in Ref. \cite{03haw})) in order to construct the
maximally extended manifold. The first step is to introduce a new coordinate
\begin{equation*}
r^{\Vert }=\int dr\left( 1-\frac{2m}{r}+\frac{\varepsilon }{r^{2}}\right)
^{-1}
\end{equation*}%
for $r>r_{+}^{1}$ and find explicitly the coordinate
\begin{equation}
r^{\Vert }=r+\frac{(r_{+}^{1})^{2}}{r_{+}^{1}-r_{-}^{1}}\ln (r-r_{+}^{1})-%
\frac{(r_{-}^{1})^{2}}{r_{+}^{1}-r_{-}^{1}}\ln (r-r_{-}^{1}),  \label{1r1}
\end{equation}%
where $r_{\pm }^{1}=r_{\pm }^{\Phi }$ with $\Phi =1.$ If $r$ is expressed as
a function on $\xi ,$ than $r^{\Vert }$ can be also expressed as a function
on $\xi $ depending additionally on some parameters.

Defining the advanced and retarded coordinates, $v=t+r^{\Vert }$ and $%
w=t-r^{\Vert },$ with corresponding elongated differentials
\begin{equation*}
\delta v=\delta t+dr^{\Vert }\mbox{ and }\delta w=\delta t-dr^{\Vert }
\end{equation*}%
the metric (\ref{sch1}) takes the form%
\begin{equation}
\delta s^{2}=-r^{2}(\xi )q^{[u]}(\xi )d\theta ^{2}-\eta _{3}(\xi ,\varphi
_{0})f_{3}(\xi ,\varphi _{0})r^{2}(\xi )\sin ^{2}\theta \delta \varphi
^{2}+(1-\frac{2m}{r(\xi )}+\varepsilon \frac{\Phi _{4}(r,\varphi _{0})}{%
r^{2}(\xi )})\delta v\delta w,  \label{met5a}
\end{equation}%
where (in general, in non--explicit form) $r(\xi )$ is a function of type $%
r(\xi )=r\left( r^{\Vert }\right) =$ $r\left( v,w\right) .$ Introducing new
coordinates $(v^{\prime \prime },w^{\prime \prime })$ by%
\begin{equation*}
v^{\prime \prime }=\arctan \left[ \exp \left( \frac{r_{+}^{1}-r_{-}^{1}}{%
4(r_{+}^{1})^{2}}v\right) \right] ,w^{\prime \prime }=\arctan \left[ -\exp
\left( \frac{-r_{+}^{1}+r_{-}^{1}}{4(r_{+}^{1})^{2}}w\right) \right]
\end{equation*}%
Defining $r$ by
\begin{equation*}
\tan v^{\prime \prime }\tan w^{\prime \prime }=-\exp \left[ \frac{%
r_{+}^{1}-r_{-}^{1}}{2(r_{+}^{1})^{2}}r\right] \sqrt{\frac{r-r_{+}^{1}}{%
(r-r_{-}^{1})^{\chi }}},\chi =\left( \frac{r_{+}^{1}}{r_{-}^{1}}\right) ^{2}
\end{equation*}
and multiplying (\ref{met5a}) on the conformal factor we obtain
\begin{eqnarray}
\delta s^{2} &=&-r^{2}q^{[u]}(r)d\theta ^{2}-\eta _{3}(r,\varphi
_{0})f_{3}(r,\varphi _{0})r^{2}\sin ^{2}\theta \delta \varphi ^{2}
\label{el2b} \\
&&+64\frac{(r_{+}^{1})^{4}}{(r_{+}^{1}-r_{-}^{1})^{2}}(1-\frac{2m}{r(\xi )}%
+\varepsilon \frac{\Phi _{4}(r,\varphi _{0})}{r^{2}(\xi )})\delta v^{\prime
\prime }\delta w^{\prime \prime },  \notag
\end{eqnarray}%
As particular cases, we may chose $\eta _{3}\left( r,\varphi \right) $ as
the condition of vanishing of the metric coefficient before $\delta
v^{\prime \prime }\delta w^{\prime \prime }$ will describe a horizon
parametrized by a resolution ellipsoid hypersurface. We emphasize that
quadratic elements (\ref{met5a}) and (\ref{el2b}) have respective
coefficients as the metrics investigated in\ Refs. \cite{03velp1,03velp2} but
the polarized cosmological constants introduce not only additional
polarizing factors $q^{[u]}(r)$ and $f_{3}(r,\varphi _{0})$ but also
elongate the anholonomic frames in a different manner.

The maximal \ extension of the Schwarzschild metric deformed by a small
parameter $\varepsilon $ (for ellipsoid configurations treated as the
eccentricity), i. e. \ the extension of the metric (\ref{1ansatz18}), is
defined by taking (\ref{el2b}) as the metric on the maximal manifold on
which this metric is of smoothly class $C^{2}.$ The Penrose diagram of this
static but locally anisotropic space--time, for any fixed angular value $%
\varphi _{0}$ is similar to the Reissner--Nordstrom solution, for the case $%
e^{2}\rightarrow \varepsilon $ and $e^{2}<m^{2}$(see, for instance, Ref. %
\cite{03haw})). There are an infinite number of asymptotically flat regions of
type I, connected by intermediate regions II and III, where there is still
an irremovable singularity at $r=0$ for every region III. We may travel from
a region I to another ones by passing through the 'wormholes' made by
anisotropic deformations (ellipsoid off--diagonality of metrics, or
anholonomy) like in the Reissner--Nordstrom universe because $\sqrt{%
\varepsilon }$ may model an effective electric charge. One can not turn back
in a such travel. Of course, this interpretation holds true only for a
corresponding smoothly class of polarization functions. For instance, if the
cosmological constant is periodically polarized from a string model, see the
formula (\ref{eqaux1}), one could be additional resonances, aperiodicity and
singularities.

It should be noted that the metric (\ref{el2b}) can be analytic every were
except at $r=r_{-}^{1}.$ We may eliminate this coordinate degeneration by
introducing another new coordinates%
\begin{equation*}
v^{\prime \prime \prime }=\arctan \left[ \exp \left( \frac{%
r_{+}^{1}-r_{-}^{1}}{2n_{0}(r_{+}^{1})^{2}}v\right) \right] ,w^{\prime
\prime \prime }=\arctan \left[ -\exp \left( \frac{-r_{+}^{1}+r_{-}^{1}}{%
2n_{0}(r_{+}^{1})^{2}}w\right) \right] ,
\end{equation*}%
where the integer $n_{0}\geq (r_{+}^{1})^{2}/(r_{-}^{1})^{2}.$ In \ these
coordinates, the metric is analytic every were except at $r=r_{+}^{1}$ where
it is degenerate.$\,$\ This way the space--time manifold can be covered by
an analytic atlas by using coordinate carts defined by $(v^{\prime \prime
},w^{\prime \prime },\theta ,\varphi )$ and $(v^{\prime \prime \prime
},w^{\prime \prime \prime },\theta ,\varphi ).$ Finally, we note that the
analytic extensions of the deformed metrics were performed with respect to
anholonomc frames which distinguish such constructions from those dealing
only with holonomic coordinates, like for the usual Reissner--N\"{o}rdstrom
and Kerr metrics. We stated the conditions when on 'radial' like coordinates
we preserve the main properties of the well know black hole solutions but in
our case the metrics are generic off--diagonal and with vacuum gravitational
polarizations.

\subsection{Geodesics on static polarized ellipsoid backgrounds}

We analyze the geodesic congruence of the metric (\ref{1ansatz18}) with the
data (\ref{data}) modified by polarized cosmological constant, for
simplicity, being linear on $\varepsilon ,$by introducing the effective
Lagrangian (for instance, like in Ref. \cite{03chan})%
\begin{eqnarray}
2L &=&g_{\alpha \beta }\frac{\delta u^{\alpha }}{ds}\frac{\delta u^{\beta }}{%
ds}=-\left( 1-\frac{2m}{r}+\frac{\varepsilon }{r^{2}}\right) ^{-1}\left(
\frac{dr}{ds}\right) ^{2}-r^{2}q^{[u]}(r)\left( \frac{d\theta }{ds}\right)
^{2}  \label{lagrb} \\
&&-\eta _{3}(r,\varphi )f_{3}(r,\varphi )r^{2}\sin ^{2}\theta \left( \frac{%
d\varphi }{ds}\right) ^{2}+\left( 1-\frac{2m}{r}+\frac{\varepsilon \Phi _{4}%
}{r^{2}}\right) \left( \frac{dt}{ds}+\varepsilon \widehat{n}_{1}\frac{dr}{ds}%
\right) ^{2},  \notag
\end{eqnarray}%
for $r=r(\xi ).$

The corresponding Euler--Lagrange equations,
\begin{equation*}
\frac{d}{ds}\frac{\partial L}{\partial \frac{\delta u^{\alpha }}{ds}}-\frac{%
\partial L}{\partial u^{\alpha }}=0
\end{equation*}%
are%
\begin{eqnarray}
&&\frac{d}{ds}\left[ -r^{2}q^{[u]}(r)\frac{d\theta }{ds}\right] =-\eta
_{3}f_{3}r^{2}\sin \theta \cos \theta \left( \frac{d\varphi }{ds}\right)
^{2},  \label{lag2b} \\
&&\frac{d}{ds}\left[ -\eta _{3}f_{3}r^{2}\frac{d\varphi }{ds}\right] =-(\eta
_{3}f_{3})^{\ast }\frac{r^{2}}{2}\sin ^{2}\theta \left( \frac{d\varphi }{ds}%
\right) ^{2}     \notag \\
&&
+\frac{\varepsilon }{2}\left( 1-\frac{2m}{r}\right) \left[ \frac{%
\Phi _{4}^{\ast }}{r^{2}}\left( \frac{dt}{ds}\right) ^{2}+\widehat{n}%
_{1}^{\ast }\frac{dt}{ds}\frac{d\xi }{ds}\right]  \notag
\end{eqnarray}%
and%
\begin{equation}
\frac{d}{ds}\left[ (1-\frac{2m}{r}+\frac{\varepsilon \Phi _{4}}{r^{2}}%
)\left( \frac{dt}{ds}+\varepsilon \widehat{n}_{1}\frac{d\xi }{ds}\right) %
\right] =0,  \label{lag3}
\end{equation}%
where, for instance, $\Phi _{4}^{\ast }=\partial $ $\Phi _{4}/\partial
\varphi $ we have omitted the variations for $d\xi /ds$ which may be found
from (\ref{lagrb}). The system of equations (\ref{lagrb})--(\ref{lag3})
transform into the usual system of geodesic equations for the Schwarzschild
space--time if $\varepsilon \rightarrow 0$ and $q^{[u]},\eta
_{3},f_{3}\rightarrow 1$ which can be solved exactly \cite{03chan}. For
nontrivial values of the parameter $\varepsilon $ and polarizations $\eta
_{3},f_{3}$ even to obtain some decompositions of solutions on $\varepsilon $
for arbitrary $\eta _{3}$ and $n_{1[1,2]},$ see (\ref{1auxf4}), is a
cumbersome task. In spite of the fact that with respect to anholonomic
frames the metrics (\ref{sch1}) and/or (\ref{1ansatz18}) has their
coefficients \ being very similar to the Reissner--Nordstom solution. The
geodesic behavior, in our anisotropic cases, is more sophisticate because
of anholonomy, polarization of constants and coefficients and ''elongation''
of partial derivatives. For instance, the equations (\ref{lag2b}) \ state
that there is not any angular on $\varphi ,$ conservation law if $(\eta
_{3}f_{3})^{\ast }\neq 0,$ even for $\varepsilon \rightarrow 0$ (which holds
both for the Schwarzschild and Reissner--Nordstom metrics). One follows from
the equation (\ref{lag3}) the existence of an energy like integral of
motion, $E=E_{0}+$ $\varepsilon E_{1},$ with%
\begin{eqnarray*}
E_{0} &=&\left( 1-\frac{2m}{r}\right) \frac{dt}{ds} \\
E_{1} &=&\frac{\Phi _{4}}{r^{2}}\frac{dt}{ds}+\left( 1-\frac{2m}{r}\right)
\widehat{n}_{1}\frac{d\xi }{ds}.
\end{eqnarray*}

The introduced anisotropic deformations of congruences of Schwarzschild's
space--time geodesics maintain the known behavior in the vicinity of the
horizon hypersurface defined by the condition of vanishing of the
coefficient $\left( 1-2m/r+\varepsilon \Phi _{4}/r^{2}\right) $ in (\ref%
{el2b}). The simplest way to prove this is to consider radial null geodesics
in the ''equatorial plane'', which satisfy the condition (\ref{lagrb}) with $%
\theta =\pi /2,d\theta /ds=0,d^{2}\theta /ds^{2}=0$ and $d\varphi /ds=0,$
from which follows that%
\begin{equation*}
\frac{dr}{dt}=\pm \left( 1-\frac{2m}{r}+\frac{\varepsilon _{0}}{r^{2}}%
\right) \left[ 1+\varepsilon \widehat{n}_{1}d\varphi \right] .
\end{equation*}%
The integral of this equation, for every fixed value $\varphi =\varphi _{0}$
is
\begin{equation*}
t=\pm r^{\Vert }+\varepsilon \int \left[ \frac{\Phi _{4}(r,\varphi _{0})-1}{%
2\left( r^{2}-2mr\right) ^{2}}-\widehat{n}_{1}(r,\varphi _{0})\right] dr
\end{equation*}%
where the coordinate $r^{\Vert }$ is defined in equation (\ref{1r1}). In this
formula the term proportional to $\varepsilon $ can have non--singular
behavior for a corresponding class of polarizations $\lambda _{4},$ see the
formulas (\ref{hor1}). Even the explicit form of the integral depends on the
type of polarizations $\eta _{3}(r,\varphi _{0}),$ $f_{3}(r,\varphi _{0})$
and values $n_{1[1,2]}(r),$ which results in some small deviations of the
null--geodesics, we may conclude that for an in--going null--ray the
coordinate time $t$ increases from $-\infty $ to $+\infty $ as $r$ decreases
from $+\infty $ to $r_{+}^{1},$ decreases from $+\infty $ to $-\infty $ as $%
r $ further decreases from $r_{+}^{1}$ to $r_{-}^{1},$ and increases again
from $-\infty $ to a finite limit as $r$ decreases from $r_{-}^{1}$ to zero.
We have a similar behavior as for the Reissner--Nordstrom solution but with
some additional anisotropic contributions being proportional to $\varepsilon
.$ Here we also note that as $dt/ds$ tends to $+\infty $ for $r\rightarrow
r_{+}^{1}+0$ and to $-\infty $ as $r_{-}+0,$ any radiation received from
infinity appear to be infinitely red--shifted at the crossing of the event
horizon and infinitely blue--shifted at the crossing of the Cauchy horizon.

The mentioned properties of null--geodesics allow us to conclude that the
metric (\ref{sch}) (equivalently, (\ref{sch1})) with the data (\ref{data})
and their maximal analytic extension (\ref{el2b}) really define a black hole
static solution which is obtained by anisotropic small deformations on $%
\varepsilon $ and renormalization by $\eta _{3}f_{3}$ of the Schwarzchild
solution (for a corresponding type of deformations the horizon of such black
holes is defined by ellipsoid hypersurfaces). We call such objects as black
ellipsoids, or black rotoids. They exists in the framework of general
relativity as certain solutions of the Einstein equations defined by static
generic off--diagonal metrics and associated anholonomic frames or can be
induced by polarized cosmological constants. This property disinguishes them
from similar configurations of Reissner--Norstrom type (which are static
electrovacuum solutions of the Einstein--Maxwell equations) and of Kerr type
rotating configurations, with ellipsoid horizon, also defined by
off--diagonal vacuum metrics (here we emphasized that the spherical
coordinate system is associated to a holonomic frame which is a trivial case
of anholonomic bases). By introducing the polarized cosmological constants,
the anholonomic character of N--adapted frames allow to construct solutions
being very different from the black hole solutions in (anti) de Sitter
spacetimes. We selected here a class of solutions where cosmological factors
correspond to some additional polarizations but do not change the
singularity structure of black ellipsoid solutions.

The metric (\ref{sch}) and its analytic extensions do not posses Killing
symmetries being deformed by anholonomic transforms. Nevertheless, we can
associate to such solutions certain noncommutative symmetries \cite{03vncs}.
Taking the data (\ref{data}) and formulas (\ref{2anh}), we compute the
corresponding nontrivial anholonomy coefficients
\begin{equation}
w_{\ 42}^{[N]5}=-w_{\ 24}^{[N]5}=\partial n_{2}\left( \xi ,\varphi \right)
/\partial \varphi =n_{2}^{\ast }\left( \xi ,\varphi \right)  \label{auxxx1}
\end{equation}%
with $n_{2}$ defined by (\ref{data}). Our solutions are for 4D
configuration. So for $n+m=4,$ the condition $k^{2}-1=n+m$ can not satisfied
in integer numbers. We may trivially extend the dimensions like $n^{\prime
}=6$ and $m^{\prime }=m=2 $ and for $k=3$ to consider the \ Lie group $%
SL\left( 3,\C\right) $ noncommutativity with corresponding values of $Q_{%
\underline{\alpha }\underline{\beta }}^{\underline{\gamma }}$\ and structure
constants $f_{~\underline{\alpha }\underline{\beta }}^{\underline{\gamma }},$
see (\ref{gr1}). An extension $w_{~\alpha \beta }^{[N]\gamma }\rightarrow
W_{~\underline{\alpha }\underline{\beta }}^{\underline{\gamma }}$ may be
performed by stating the N--deformed ''structure'' constants (\ref{anhb}), $%
W_{~\underline{\alpha }\underline{\beta }}^{\underline{\gamma }}=f_{~%
\underline{\alpha }\underline{\beta }}^{\underline{\gamma }}+w_{~\underline{%
\alpha }\underline{\beta }}^{[N]\underline{\gamma }},$ with only two
nontrivial values of $w_{~\underline{\alpha }\underline{\beta }}^{[N]%
\underline{\gamma }}$ given by (\ref{auxxx1}). In a similar manner we can
compute the anholonomy coefficients for the black ellipsoid metric with
cosmological constant contributions (\ref{1ansatz18}).

\section{Perturbations of Anisotropic Black Holes}

The stability of black ellipsoids was proven in Ref. \cite{03vels}. A similar
proof may hold true for a class of metrics with anholonomic noncommutative
symmetry and possible complexification of some off--diagonal metric and
tetradics coefficients \cite{03vncs}. In this section we reconsider the
perturbation formalism and stability proofs for rotoid metrics defined by
polarized cosmological constants.

\subsection{Metrics describing anisotropic perturbations}

We consider a four dimensional pseudo--Riemannian quadratic linear element
\begin{eqnarray}
ds^{2} = &\Omega (r,\varphi )&\left[ -\left( 1-\frac{2m}{r}+\frac{\varepsilon
}{r^{2}}\right) ^{-1}dr^{2}-r^{2}q^{[v]}(r)d\theta ^{2}-\eta
_{3}^{[v]}(r,\theta ,\varphi )r^{2}\sin ^{2}\theta \delta \varphi ^{2}\right]
\notag \\
&+ &\left[ 1-\frac{2m}{r}+\frac{\varepsilon }{r^{2}}\eta (r,\varphi )\right]
\delta t^{2},  \label{metric1p} \\
\eta _{3}^{[v]}(r,\theta ,\varphi ) &=&\eta _{3}(r,\theta ,\varphi
)f_{3}(r,\theta ,\varphi )  \notag
\end{eqnarray}%
with
\begin{equation*}
\delta \varphi =d\varphi +\varepsilon w_{1}(r,\varphi )dr,\mbox{
and }\delta t=dt+\varepsilon n_{1}(r,\varphi )dr,
\end{equation*}%
where the local coordinates are denoted $u=\{u^{\alpha }=\left( r,\theta
,\varphi ,t\right) \}$ (the Greek indices $\alpha ,\beta ,...$ will run the
values 1,2,3,4), $\varepsilon $ is a small parameter satisfying the
conditions $0\leq \varepsilon \ll 1$ (for instance, an eccentricity for some
ellipsoid deformations of the spherical symmetry) and the functions $\Omega
(r,\varphi ),q(r),$ $\eta _{3}(r,\theta ,\varphi )$ and $\eta (\theta
,\varphi )$ are of necessary smooth class. The metric (\ref{metric1p}) is
static, off--diagonal and transforms into the usual Schwarzschild solution
if $\varepsilon \rightarrow 0$ and $\Omega ,q^{[v]},\eta
_{3}^{[v]}\rightarrow 1.$ For vanishing cosmological constants, it describes
at least two classes of static black hole solutions generated as small
anhlonomic deformations of the Schwarzschild solution \cite%
{03v,03v1,03velp1,03velp2,03vncfg,03vels} but models also nontrivial
 vacuum polarized
cosmological constants.

We can apply the perturbation theory for the metric (\ref{metric1p}) (not
paying a special attention to some particular parametrization of
coefficients for one or another class of anisotropic black hole solutions)
and analyze its stability by using the results of Ref. \cite{03chan} for a
fixed anisotropic direction, i. e. by imposing certain anholonomic frame
constraints for an angle $\varphi =\varphi _{0}$ but considering possible
perturbations depending on three variables $(u^{1}=x^{1}=r,u^{2}=x^{2}=%
\theta ,$ $u^{4}=t).$ We suppose that if we prove that there is a stability
on perturbations for a value $\varphi _{0},$ we can analyze in a similar
manner another values of $\varphi .$ A more general perturbative theory with
variable anisotropy on coordinate $\varphi ,$ i. e. with dynamical
anholonomic constraints, connects the approach with a two dimensional
inverse problem which makes the analysis more sophisticate. There have been
not elaborated such analytic methods in the theory of black holes.

It should be noted that in a study of perturbations of any spherically
symmetric system \ and, for instance, of small ellipsoid deformations,
without any loss of generality, we can restrict our considerations to
axisymmetric modes of perturbations. Non--axisymmetric modes of
perturbations with an $e^{in\varphi }$ dependence on the azimutal angle $%
\varphi $ $\ (n$ being an integer number) can be deduced from modes of
axisymmetric perturbations with $n=0$ by suitable rotations since there are
not preferred axes in a spherically symmetric background. The ellipsoid like
deformations may be included into the formalism as some low frequency and
constrained perturbations.

We develop the black hole perturbation and stability theory as to include
into consideration off--diagonal metrics with the coefficients polarized by
cosmological constants. This is the main difference comparing to the paper %
\cite{03vels}. For simplicity, in this section, we restrict our study only to
fixed values of the coordinate $\varphi $ assuming that anholonomic
deformations are proportional to a small parameter $\varepsilon ;$ we shall
investigate the stability of solutions only by applying the one dimensional
inverse methods.

\newpage

We state a quadratic metric element
\begin{eqnarray}
ds^{2} &=&-e^{2\mu _{1}}(du^{1})^{2}-e^{2\mu _{2}}(du^{2})^{2}-e^{2\mu
_{3}}(\delta u^{3})^{2}+e^{2\mu _{4}}(\delta u^{4})^{2},  \notag \\
\delta u^{3} &=&d\varphi -q_{1}dx^{1}-q_{2}dx^{2}-\omega dt,  \label{metric2}
\\
\delta u^{4} &=&dt+n_{1}dr  \notag
\end{eqnarray}%
where
\begin{eqnarray}
\mu _{\alpha }(x^{k},t) &=&\mu _{\alpha }^{(\varepsilon )}(x^{k},\varphi
_{0})+\delta \mu _{\alpha }^{(\varsigma )}(x^{k},t),  \label{coef1} \\
q_{i}(x^{k},t) &=&q_{i}^{(\varepsilon )}(r,\varphi _{0})+\delta
q_{i}^{(\varsigma )}(x^{k},t),\ \omega (x^{k},t)=0+\delta \omega
^{(\varsigma )}(x^{k},t)  \notag
\end{eqnarray}%
with
\begin{eqnarray}
e^{2\mu _{1}^{(\varepsilon )}} &=&\Omega (r,\varphi _{0})(1-\frac{2m}{r}+%
\frac{\varepsilon }{r^{2}})^{-1},\ e^{2\mu _{2}^{(\varepsilon )}}=\Omega
(r,\varphi _{0})q^{[v]}(r)r^{2},  \label{coef2} \\
e^{2\mu _{3}^{(\varepsilon )}} &=&\Omega (r,\varphi _{0})r^{2}\sin
^{2}\theta \eta _{3}^{[v]}(r,\varphi _{0}),\ e^{2\mu _{4}^{(\varepsilon
)}}=1-\frac{2m}{r}+\frac{\varepsilon }{r^{2}}\eta (r,\varphi _{0}),  \notag
\end{eqnarray}%
and some non--trivial values for $q_{i}^{(\varepsilon )}$ and $\varepsilon
n_{i},$%
\begin{eqnarray*}
q_{i}^{(\varepsilon )} &=&\varepsilon w_{i}(r,\varphi _{0}), \\
n_{1} &=&\varepsilon \left( n_{1[1]}(r)+n_{1[2]}(r)\int_{0}^{\varphi
_{0}}\eta _{3}(r,\varphi )d\varphi \right) .
\end{eqnarray*}%
We have to distinguish two types of small deformations from the spherical
symmetry. The first type of deformations, labelled with the index $%
(\varepsilon )$ are generated by some $\varepsilon $--terms which define a
fixed ellipsoid like configuration and the second type ones, labelled with
the index $(\varsigma ),$ are some small linear fluctuations of the metric
coefficients

The general formulas for the Ricci and Einstein tensors for metric elements
of class (\ref{metric2}) with $w_{i},n_{1}=0$ are given in \cite{03chan}. We
compute similar values with respect to anholnomic frames, when, for a
conventional splitting $u^{\alpha }=(x^{i},y^{a}),$ the coordinates $x^{i}$
and $y^{a}$ are treated respectively as holonomic and anholonomic ones. In
this case the partial derivatives $\partial /\partial x^{i}$ must be changed
into certain 'elongated' ones
\begin{eqnarray*}
\frac{\partial }{\partial x^{1}} &\rightarrow &\frac{\delta }{\partial x^{1}}%
=\frac{\partial }{\partial x^{1}}-w_{1}\frac{\partial }{\partial \varphi }%
-n_{1}\frac{\partial }{\partial t}, \\
\frac{\partial }{\partial x^{2}} &\rightarrow &\frac{\delta }{\partial x^{2}}%
=\frac{\partial }{\partial x^{2}}-w_{2}\frac{\partial }{\partial \varphi },
\end{eqnarray*}%
see details in Refs \cite{03vth,03vsbd,03velp1,03velp2}. In the ansatz (\ref{metric2}%
), the anholonomic contributions of $w_{i}$ are included in the coefficients
$q_{i}(x^{k},t).$ For convenience, we give present bellow the necessary
formulas for $R_{\alpha \beta }$ (the Ricci tensor) and $G_{\alpha \beta }$
(the Einstein tensor) computed for the ansatz (\ref{metric2}) with three
holonomic coordinates $\left( r,\theta ,\varphi \right) $ and on anholonomic
coordinate $t$ (in our case, being time like), with the partial derivative
operators
\begin{equation*}
\partial _{1}\rightarrow \delta _{1}=\frac{\partial }{\partial r}w_{1}\frac{%
\partial }{\partial \varphi }-n_{1}\frac{\partial }{\partial t},\delta _{2}=%
\frac{\partial }{\partial \theta }-w_{2}\frac{\partial }{\partial \varphi }%
,\partial _{3}=\frac{\partial }{\partial \varphi },
\end{equation*}%
and for a fixed value $\varphi _{0}.$

A general perturbation of an anisotropic black--hole described by a
quadratic line element (\ref{metric2}) \ results in some small quantities of
the first order $\omega $ and $q_{i},$ inducing a dragging of frames and
imparting rotations, and in some functions $\mu _{\alpha }$ with small
increments $\delta \mu _{\alpha },$ which do not impart rotations. Some
coefficients contained in such values are proportional to $\varepsilon ,$
another ones are considered only as small quantities. The perturbations of
metric are of two generic types: axial and polar one. We shall investigate
them separately in the next two subsection after we shall have computed the
coefficients of the Ricci tensor.

We compute the coefficients of the the Ricci tensor as
\begin{equation*}
R_{\beta \gamma \alpha }^{\alpha }=R_{\beta \gamma }
\end{equation*}%
and of the Einstein tensor as%
\begin{equation*}
G_{\beta \gamma }=R_{\beta \gamma }-\frac{1}{2}g_{\beta \gamma }R
\end{equation*}%
for $R=g^{\beta \gamma }R_{\beta \gamma }.$ Straightforward computations for
the quadratic line element (\ref{metric2}) give%
\begin{eqnarray}
R_{11} &=&-e^{-2\mu _{1}}[\delta _{11}^{2}(\mu _{3}+\mu _{4}+\mu
_{2})+\delta _{1}\mu _{3}\delta _{1}(\mu _{3}-\mu _{1})+\delta _{1}\mu
_{2}\delta _{1}(\mu _{2}-\mu _{1})+  \label{riccip} \\
&&\delta _{1}\mu _{4}\delta _{1}(\mu _{4}-\mu _{1})]-e^{-2\mu _{2}}[\delta
_{22}^{2}\mu _{1}+\delta _{2}\mu _{1}\delta _{2}(\mu _{3}+\mu _{4}+\mu
_{1}-\mu _{2})]+  \notag \\
&&e^{-2\mu _{4}}[\partial _{44}^{2}\mu _{1}+\partial _{4}\mu _{1}\partial
_{4}(\mu _{3}-\mu _{4}+\mu _{1}+\mu _{2})]-\frac{1}{2}e^{2(\mu _{3}-\mu
_{1})}[e^{-2\mu _{2}}Q_{12}^{2}+e^{-2\mu _{4}}Q_{14}^{2}],  \notag
\end{eqnarray}%
\begin{eqnarray*}
R_{12} &=&-e^{-\mu _{1}-\mu _{2}}[\delta _{2}\delta _{1}(\mu _{3}+\mu
_{2})-\delta _{2}\mu _{1}\delta _{1}(\mu _{3}+\mu _{1})-\delta _{1}\mu
_{2}\partial _{4}(\mu _{3}+\mu _{1}) \\
&&+\delta _{1}\mu _{3}\delta _{2}\mu _{3}+\delta _{1}\mu _{4}\delta _{2}\mu
_{4}]+\frac{1}{2}e^{2\mu _{3}-2\mu _{4}-\mu _{1}-\mu _{2}}Q_{14}Q_{24},
\end{eqnarray*}%
\begin{equation*}
R_{31}=-\frac{1}{2}e^{2\mu _{3}-\mu _{4}-\mu _{2}}[\delta _{2}(e^{3\mu
_{3}+\mu _{4}-\mu _{1}-\mu _{2}}Q_{21})+\partial _{4}(e^{3\mu _{3}-\mu
_{4}+\mu _{2}-\mu _{1}}Q_{41})],
\end{equation*}%
\begin{eqnarray*}
R_{33} &=&-e^{-2\mu _{1}}[\delta _{11}^{2}\mu _{3}+\delta _{1}\mu _{3}\delta
_{1}(\mu _{3}+\mu _{4}+\mu _{2}-\mu _{1})]- \\
&&e^{-2\mu _{2}}[\delta _{22}^{2}\mu _{3}+\partial _{2}\mu _{3}\partial
_{2}(\mu _{3}+\mu _{4}-\mu _{2}+\mu _{1})]+\frac{1}{2}e^{2(\mu _{3}-\mu
_{1}-\mu _{2})}Q_{12}^{2}+ \\
&&e^{-2\mu _{4}}[\partial _{44}^{2}\mu _{3}+\partial _{4}\mu _{3}\partial
_{4}(\mu _{3}-\mu _{4}+\mu _{2}+\mu _{1})]-\frac{1}{2}e^{2(\mu _{3}-\mu
_{4})}[e^{-2\mu _{2}}Q_{24}^{2}+e^{-2\mu _{1}}Q_{14}^{2}],
\end{eqnarray*}%
\begin{eqnarray*}
R_{41} &=&-e^{-\mu _{1}-\mu _{4}}[\partial _{4}\delta _{1}(\mu _{3}+\mu
_{2})+\delta _{1}\mu _{3}\partial _{4}(\mu _{3}-\mu _{1})+\delta _{1}\mu
_{2}\partial _{4}(\mu _{2}-\mu _{1}) \\
&&-\delta _{1}\mu _{4}\partial _{4}(\mu _{3}+\mu _{2})]+\frac{1}{2}e^{2\mu
_{3}-\mu _{4}-\mu _{1}-2\mu _{2}}Q_{12}Q_{34},
\end{eqnarray*}%
\begin{equation*}
R_{43}=-\frac{1}{2}e^{2\mu _{3}-\mu _{1}-\mu _{2}}[\delta _{1}(e^{3\mu
_{3}-\mu _{4}-\mu _{1}+\mu _{2}}Q_{14})+\delta _{2}(e^{3\mu _{3}-\mu
_{4}+\mu _{1}-\mu _{2}}Q_{24})],
\end{equation*}%
\begin{eqnarray*}
R_{44} &=&-e^{-2\mu _{4}}[\partial _{44}^{2}(\mu _{1}+\mu _{2}+\mu
_{3})+\partial _{4}\mu _{3}\partial _{4}(\mu _{3}-\mu _{4})+\partial _{4}\mu
_{1}\partial _{4}(\mu _{1}-\mu _{4})+ \\
&&\partial _{4}\mu _{2}\partial _{4}(\mu _{2}-\mu _{4})]+e^{-2\mu
_{1}}[\delta _{11}^{2}\mu _{4}+\delta _{1}\mu _{4}\delta _{1}(\mu _{3}+\mu
_{4}-\mu _{1}+\mu _{2})]+ \\
&&e^{-2\mu _{2}}[\delta _{22}^{2}\mu _{4}+\delta _{2}\mu _{4}\delta _{2}(\mu
_{3}+\mu _{4}-\mu _{1}+\mu _{2})]-\frac{1}{2}e^{2(\mu _{3}-\mu
_{4})}[e^{-2\mu _{1}}Q_{14}^{2}+e^{-2\mu _{2}}Q_{24}^{2}],
\end{eqnarray*}%
where the rest of coefficients are defined by similar formulas with a
corresponding changing of indices and partial derivative operators, $%
R_{22}, $ $R_{42}$ and $R_{32}$ is like $R_{11},R_{41}$ and $R_{31}$ with
with changing the index $1\rightarrow 2.$ The values $Q_{ij}$ and $Q_{i4}$
are defined respectively%
\begin{equation*}
Q_{ij}=\delta _{j}q_{i}-\delta _{i}q_{j}\mbox{ and
}Q_{i4}=\partial _{4}q_{i}-\delta _{i}\omega .
\end{equation*}

The nontrivial coefficients of the Einstein tensor are
\begin{eqnarray}
G_{11} &=&e^{-2\mu _{2}}[\delta _{22}^{2}(\mu _{3}+\mu _{4})+\delta _{2}(\mu
_{3}+\mu _{4})\delta _{2}(\mu _{4}-\mu _{2})+\delta _{2}\mu _{3}\delta
_{2}\mu _{3}]-  \notag \\
&&e^{-2\mu _{4}}[\partial _{44}^{2}(\mu _{3}+\mu _{2})+\partial _{4}(\mu
_{3}+\mu _{2})\partial _{4}(\mu _{2}-\mu _{4})+\partial _{4}\mu _{3}\partial
_{4}\mu _{3}]+  \notag \\
&&e^{-2\mu _{1}}[\delta _{1}\mu _{4}+\delta _{1}(\mu _{3}+\mu _{2})+\delta
_{1}\mu _{3}\delta _{1}\mu _{2}]-  \label{einstp} \\
&&\frac{1}{4}e^{2\mu _{3}}[e^{-2(\mu _{1}+\mu _{2})}Q_{12}^{2}-e^{-2(\mu
_{1}+\mu _{4})}Q_{14}^{2}+e^{-2(\mu _{2}+\mu _{3})}Q_{24}^{2}],  \notag
\end{eqnarray}%
\begin{eqnarray*}
G_{33} &=&e^{-2\mu _{1}}[\delta _{11}^{2}(\mu _{4}+\mu _{2})+\delta _{1}\mu
_{4}\delta _{1}(\mu _{4}-\mu _{1}+\mu _{2})+\delta _{1}\mu _{2}\delta
_{1}(\mu _{2}-\mu _{1})]+ \\
&&e^{-2\mu _{2}}[\delta _{22}^{2}(\mu _{4}+\mu _{1})+\delta _{2}(\mu
_{4}-\mu _{2}+\mu _{1})+\delta _{2}\mu _{1}\partial _{2}(\mu _{1}-\mu _{2})]-
\\
&&e^{-2\mu _{4}}[\partial _{44}^{2}(\mu _{1}+\mu _{2})+\partial _{4}\mu
_{1}\partial _{4}(\mu _{1}-\mu _{4})+\partial _{4}\mu _{2}\partial _{4}(\mu
_{2}-\mu _{4})+\partial _{4}\mu _{1}\partial _{4}\mu _{2}]+ \\
&&\frac{3}{4}e^{2\mu _{3}}[e^{-2(\mu _{1}+\mu _{2})}Q_{12}^{2}-e^{-2(\mu
_{1}+\mu _{4})}Q_{14}^{2}-e^{-2(\mu _{2}+\mu _{3})}Q_{24}^{2}],
\end{eqnarray*}%
\begin{eqnarray*}
G_{44} &=&e^{-2\mu _{1}}[\delta _{11}^{2}(\mu _{3}+\mu _{2})+\delta _{1}\mu
_{3}\delta _{1}(\mu _{3}-\mu _{1}+\mu _{2})+\delta _{1}\mu _{2}\delta
_{1}(\mu _{2}-\mu _{1})]- \\
&&e^{-2\mu _{2}}[\delta _{22}^{2}(\mu _{3}+\mu _{1})+\delta _{2}(\mu
_{3}-\mu _{2}+\mu _{1})+\delta _{2}\mu _{1}\partial _{2}(\mu _{1}-\mu _{2})]-%
\frac{1}{4}e^{2(\mu _{3}-\mu _{1}-\mu _{2})}Q_{12}^{2} \\
&&+e^{-2\mu _{4}}[\partial _{4}\mu _{3}\partial _{4}(\mu _{1}+\mu
_{2})+\partial _{4}\mu _{1}\partial _{4}\mu _{2}]-\frac{1}{4}e^{2(\mu
_{3}-\mu _{4})}[e^{-2\mu _{1}}Q_{14}^{2}-e^{-2\mu _{2}}Q_{24}^{2}].
\end{eqnarray*}%
The component $G_{22}$ is to be found from $G_{11}$ by changing the index $%
1\rightarrow 2.$ We note that the formulas (\ref{einstp}) transform into
similar ones from Ref. \cite{03vels} if $\delta _{2}\rightarrow \partial _{2}.$

\subsection{Axial metric perturbations}

Axial perturbations are characterized by non--vanishing $\omega $ and $q_{i}$
which satisfy the equations
\begin{equation*}
R_{3i}=0,
\end{equation*}%
see the explicit formulas for such coefficients of the Ricci tensor in (\ref%
{riccip}). The resulting equations governing axial perturbations, $\delta
R_{31}=0,$ $\delta R_{32}=0,$ are respectively%
\begin{eqnarray}
\delta _{2}\left( e^{3\mu _{3}^{(\varepsilon )}+\mu _{4}^{(\varepsilon
)}-\mu _{1}^{(\varepsilon )}-\mu _{2}^{(\varepsilon )}}Q_{12}\right)
&=&-e^{3\mu _{3}^{(\varepsilon )}-\mu _{4}^{(\varepsilon )}-\mu
_{1}^{(\varepsilon )}+\mu _{2}^{(\varepsilon )}}\partial _{4}Q_{14},
\label{eq1} \\
\delta _{1}\left( e^{3\mu _{3}^{(\varepsilon )}+\mu _{4}^{(\varepsilon
)}-\mu _{1}^{(\varepsilon )}-\mu _{2}^{(\varepsilon )}}Q_{12}\right)
&=&e^{3\mu _{3}^{(\varepsilon )}-\mu _{4}^{(\varepsilon )}+\mu
_{1}^{(\varepsilon )}-\mu _{2}^{(\varepsilon )}}\partial _{4}Q_{24},  \notag
\end{eqnarray}%
where
\begin{equation}
Q_{ij}=\delta _{i}q_{j}-\delta _{j}q_{i},Q_{i4}=\partial _{4}q_{i}-\delta
_{i}\omega  \label{eq1a}
\end{equation}%
and for $\mu _{i}$ there are considered unperturbed values $\mu
_{i}^{(\varepsilon )}.$ Introducing the values of coefficients (\ref{coef1})
and (\ref{coef2}) \ and assuming that the perturbations have a time
dependence of type $\exp (i\sigma t)$ for a real constant $\sigma ,$ \ we
rewrite the equations (\ref{eq1})
\begin{eqnarray}
\frac{1+\varepsilon \left( \Delta ^{-1}+3r^{2}\phi /2\right) }{r^{4}\sin
^{3}\theta (\eta _{3}^{[v]})^{3/2}}\delta _{2}Q^{(\eta )} &=&-i\sigma \delta
_{r}\omega -\sigma ^{2}q_{1},  \label{eq2a} \\
\frac{\Delta }{r^{4}\sin ^{3}\theta (\eta _{3}^{[v]})^{3/2}}\delta
_{1}\left\{ Q^{(\eta )}\left[ 1+\frac{\varepsilon }{2}\left( \frac{\eta -1}{%
\Delta }-r^{2}\phi \right) \right] \right\} &=&i\sigma \partial _{\theta
}\omega +\sigma ^{2}q_{2}  \label{eq2b}
\end{eqnarray}%
for
\begin{equation*}
Q^{(\eta )}(r,\theta ,\varphi _{0},t)=\Delta Q_{12}\sin ^{3}\theta =\Delta
\sin ^{3}\theta (\partial _{2}q_{1}-\delta _{1}q_{2}),\Delta =r^{2}-2mr,
\end{equation*}%
where $\phi =0$ for solutions with $\Omega =1$ and $\phi (r,\varphi )=\eta
_{3}^{[v]}\left( r,\theta ,\varphi \right) \sin ^{2}\theta ,$ i. e. $\eta
_{3}\left( r,\theta ,\varphi \right) \sim \sin ^{-2}\theta $ for solutions
with $\Omega =1+\varepsilon ....$

We can exclude the function $\omega $ and define an equation for $Q^{(\eta
)} $ if we take the sum of the (\ref{eq2a}) subjected by the action of
operator $\partial _{2}$ and of the (\ref{eq2b}) subjected by the action of
operator $\delta _{1}.$ Using the relations (\ref{eq1a}), we write%
\begin{eqnarray*}
r^{4}\delta _{1}\left\{ \frac{\Delta }{r^{4}(\eta _{3}^{[v]})^{3/2}}\left[
\delta _{1}\left[ Q^{(\eta )}+\frac{\varepsilon }{2}\left( \frac{\eta -1}{%
\Delta }-r^{2}\right) \phi \right] \right] \right\} + && \\
\sin ^{3}\theta \partial _{2}\left[ \frac{1+\varepsilon (\Delta
^{-1}+3r^{2}\phi /2)}{\sin ^{3}\theta (\eta _{3}^{[v]})^{3/2}}\delta
_{2}Q^{(\eta )}\right] +\frac{\sigma ^{2}r^{4}}{\Delta \eta _{3}^{3/2}}%
Q^{(\eta )} &=&0.
\end{eqnarray*}%
The solution of this equation is searched in the form $Q^{(\eta
)}=Q+\varepsilon Q^{(1)}$ which results in
\begin{equation}
r^{4}\partial _{1}\left( \frac{\Delta }{r^{4}(\eta _{3}^{[v]})^{3/2}}%
\partial _{1}Q\right) +\sin ^{3}\theta \partial _{2}\left( \frac{1}{\sin
^{3}\theta (\eta _{3}^{[v]})^{3/2}}\partial _{2}Q\right) +\frac{\sigma
^{2}r^{4}}{\Delta (\eta _{3}^{[v]})^{3/2}}Q=\varepsilon A\left( r,\theta
,\varphi _{0}\right) ,  \label{eq3}
\end{equation}%
where%
\begin{eqnarray*}
A\left( r,\theta ,\varphi _{0}\right) &=&r^{4}\partial _{1}\left( \frac{%
\Delta }{r^{4}(\eta _{3}^{[v]})^{3/2}}n_{1}\right) \frac{\partial Q}{%
\partial t}-r^{4}\partial _{1}\left( \frac{\Delta }{r^{4}(\eta
_{3}^{[v]})^{3/2}}\partial _{1}Q^{(1)}\right) \\
&&-\sin ^{3}\theta \delta _{2}\left[ \frac{1+\varepsilon (\Delta
^{-1}+3r^{2}\phi /2)}{\sin ^{3}\theta (\eta _{3}^{[v]})^{3/2}}\delta
_{2}Q^{(1)}-\frac{\sigma ^{2}r^{4}}{\Delta (\eta _{3}^{[v]})^{3/2}}Q^{(1)}%
\right] ,
\end{eqnarray*}%
with a time dependence like $\exp [i\sigma t]$

It is possible to construct different classes of solutions of the equation (%
\ref{eq3}). At the first step we find the solution for $Q$ when $\varepsilon
=0.$ Then, for a known value of $Q\left( r,\theta ,\varphi _{0}\right) $
from
\begin{equation*}
Q^{(\eta )}=Q+\varepsilon Q^{(1)},
\end{equation*}%
we can define $Q^{(1)}$ from the equations (\ref{eq2a}) and (\ref{eq2b}) by
considering the values proportional to $\varepsilon $ which can be written
\begin{eqnarray}
\partial _{1}Q^{(1)} &=&B_{1}\left( r,\theta ,\varphi _{0}\right) ,
\label{eq4} \\
\partial _{2}Q^{(1)} &=&B_{2}\left( r,\theta ,\varphi _{0}\right) .  \notag
\end{eqnarray}%
The integrability condition of the system (\ref{eq4}), $\partial
_{1}B_{2}=\partial _{2}B_{1}$ imposes a relation between the polarization
functions $\eta _{3},\eta ,w_{1}$and $n_{1}$ (for a corresponding class of
solutions). In order to prove that there are stable anisotropic
configurations of anisotropic black hole solutions, we may consider a set of
polarization functions when $A\left( r,\theta ,\varphi _{0}\right) =0$ and
the solution with $Q^{(1)}=0$ is admitted. This holds, for example, if
\begin{equation*}
\Delta n_{1}=n_{0}r^{4}(\eta _{3}^{[v]})^{3/2},\ n_{0}=const.
\end{equation*}%
In this case the axial perturbations are described by the equation
\begin{equation}
(\eta _{3}^{[v]})^{3/2}r^{4}\partial _{1}\left( \frac{\Delta }{r^{4}(\eta
_{3}^{[v]})^{3/2}}\partial _{1}Q\right) +\sin ^{3}\theta \delta _{2}\left(
\frac{1}{\sin ^{3}\theta }\delta _{2}Q\right) +\frac{\sigma ^{2}r^{4}}{%
\Delta }Q=0  \label{eq5}
\end{equation}%
which is obtained from (\ref{eq3}) for $\eta _{3}^{[v]}=\eta
_{3}^{[v]}\left( r,\varphi _{0}\right) ,$ or for $\phi (r,\varphi _{0})=\eta
_{3}^{[v]}\left( r,\theta ,\varphi _{0}\right) \sin ^{2}\theta .$

In the limit $\eta _{3}^{[v]}\rightarrow 1$ the solution of equation (\ref%
{eq5}) is investigated in details in Ref. \cite{03chan}. \ Here, we prove that
in a similar manner we can define exact solutions for non--trivial values of
$\eta _{3}^{[v]}.$ \ The variables $r$ and $\theta $ can be separated if we
substitute
\begin{equation*}
Q(r,\theta ,\varphi _{0})=Q_{0}(r,\varphi _{0})C_{l+2}^{-3/2}(\theta ),
\end{equation*}%
where $C_{n}^{\nu }$ are the Gegenbauer functions generated by the equation%
\begin{equation*}
\left[ \frac{d}{d\theta }\sin ^{2\nu }\theta \frac{d}{d\theta }+n\left(
n+2\nu \right) \sin ^{2\nu }\theta \right] C_{n}^{\nu }(\theta )=0.
\end{equation*}%
The function $C_{l+2}^{-3/2}(\theta )$ is related to the second derivative
of the Legendre function $P_{l}(\theta )$ by formulas%
\begin{equation*}
C_{l+2}^{-3/2}(\theta )=\sin ^{3}\theta \frac{d}{d\theta }\left[ \frac{1}{%
\sin \theta }\frac{dP_{l}(\theta )}{d\theta }\right] .
\end{equation*}%
The separated part of (\ref{eq5}) depending on radial variable with a fixed
value $\varphi _{0}$ transforms into the equation%
\begin{equation}
(\eta _{3}^{[v]})^{3/2}\Delta \frac{d}{dr}\left( \frac{\Delta }{r^{4}(\eta
_{3}^{[v]})^{3/2}}\frac{dQ_{0}}{dr}\right) +\left( \sigma ^{2}-\frac{\mu
^{2}\Delta }{r^{4}}\right) Q_{0}=0,  \label{eq6}
\end{equation}%
where $\mu ^{2}=(l-1)(l+2)$ for $l=2,3,...$ A further simplification is
possible for $\eta _{3}^{[v]}=\eta _{3}^{[v]}(r,\varphi _{0})$ if we
introduce in the equation (\ref{eq6}) a new radial coordinate
\begin{equation*}
r_{\#}=\int (\eta _{3}^{[v]})^{3/2}(r,\varphi _{0})r^{2}dr
\end{equation*}%
and a new unknown function $Z^{(\eta )}=r^{-1}Q_{0}(r).$ The equation for $%
Z^{(\eta )}$ is an Schrodinger like one--dimensional wave equation%
\begin{equation}
\left( \frac{d^{2}}{dr_{\#}^{2}}+\frac{\sigma ^{2}}{(\eta _{3}^{[v]})^{3/2}}%
\right) Z^{(\eta )}=V^{(\eta )}Z^{(\eta )}  \label{eq7}
\end{equation}%
with the potential
\begin{equation}
V^{(\eta )}=\frac{\Delta }{r^{5}(\eta _{3}^{[v]})^{3/2}}\left[ \mu ^{2}-r^{4}%
\frac{d}{dr}\left( \frac{\Delta }{r^{4}(\eta _{3}^{[v]})^{3/2}}\right) %
\right]  \label{eq7a}
\end{equation}%
and polarized parameter
\begin{equation*}
\widetilde{\sigma }^{2}=\sigma ^{2}/(\eta _{3}^{[v]})^{3/2}.
\end{equation*}%
This equation transforms into the so--called Regge--Wheeler equation if $%
\eta _{3}^{[v]}=1.$ For instance, for the Schwarzschild black hole such
solutions are investigated and tabulated for different values of $l=2,3$ and
$4$ in Ref. \cite{03chan}.

We note that for static anisotropic black holes with nontrivial anisotropic
conformal factor, $\Omega =1+\varepsilon ...,$ even $\eta _{3}$ may depend
on angular variable $\theta $ because of condition that $\phi (r,\varphi
_{0})=\eta _{3}^{[v]}\left( r,\theta ,\varphi _{0}\right) \sin ^{2}\theta $
the equation (\ref{eq5}) transforms directly in (\ref{eq7}) with $\mu =0$
without any separation of variables $r$ and $\theta .$ It is not necessary
in this case to consider the Gegenbauer functions because $Q_{0}$ does not
depend on $\theta $ which corresponds to a solution with $l=1.$

We may transform (\ref{eq7}) into the usual form,
\begin{equation*}
\left( \frac{d^{2}}{dr_{\star }^{2}}+\sigma ^{2}\right) Z^{(\eta )}=%
\widetilde{V}^{(\eta )}Z^{(\eta )}
\end{equation*}%
if we introduce the variable
\begin{equation*}
r_{\star }=\int dr_{\#}(\eta _{3}^{[v]})^{-3/2}\left( r_{\#},\varphi
_{0}\right)
\end{equation*}%
for $\widetilde{V}^{(\eta )}=(\eta _{3}^{[v]})^{3/2}V^{(\eta )}.$ So, the
polarization function $\eta _{3}^{[v]},$ describing static anholonomic
deformations of the Schwarzshild black hole, ''renormalizes'' the potential
in the one--dimensional Schrodinger wave--equation governing axial
perturbations of such objects.

We conclude that small static ''ellipsoid'' like deformations and
polarizations of constants of spherical black holes (the anisotropic
configurations being described by generic off--diagonal metric ansatz) do
not change the type of equations for axial petrubations: one modifies the
potential barrier,%
\begin{equation*}
V^{(-)}=\frac{\Delta }{r^{5}}\left[ \left( \mu ^{2}+2\right) r-6m\right]
\longrightarrow \widetilde{V}^{(\eta )}
\end{equation*}%
and re--defines the radial variables%
\begin{equation*}
r_{\ast }=r+2m\ln \left( r/2m-1\right) \longrightarrow r_{\star }(\varphi
_{0})
\end{equation*}%
with a parametric dependence on anisotropic angular coordinate which is
caused by the existence of a deformed static horizon.

\subsection{Polar metric perturbations}

The polar perturbations are described by non--trivial increments of the
diagonal metric coefficients, $\delta \mu _{\alpha }=\delta \mu _{\alpha
}^{(\varepsilon )}+\delta \mu _{\alpha }^{(\varsigma )},$ for
\begin{equation*}
\mu _{\alpha }^{(\varepsilon )}=\nu _{\alpha }+\delta \mu _{\alpha
}^{(\varepsilon )}
\end{equation*}%
where $\delta \mu _{\alpha }^{(\varsigma )}(x^{k},t)$ parametrize time
depending fluctuations which are stated to be the same both for spherical
and/or spheroid configurations and $\delta \mu _{\alpha }^{(\varepsilon )}$
is a static deformation from the spherical symmetry. Following notations (%
\ref{coef1}) and (\ref{coef2}) we write%
\begin{equation*}
e^{v_{1}}=r/\sqrt{|\Delta |},e^{v_{2}}=r\sqrt{|q^{[v]}(r)|}%
,e^{v_{3}}=rh_{3}\sin \theta ,e^{v_{4}}=\Delta /r^{2}
\end{equation*}%
and
\begin{equation*}
\delta \mu _{1}^{(\varepsilon )}=-\frac{\varepsilon }{2}\left( \Delta
^{-1}+r^{2}\phi \right) ,\delta \mu _{2}^{(\varepsilon )}=\delta \mu
_{3}^{(\varepsilon )}=-\frac{\varepsilon }{2}r^{2}\phi ,\delta \mu
_{4}^{(\varepsilon )}=\frac{\varepsilon \eta }{2\Delta }
\end{equation*}%
where $\phi =0$ for the solutions with $\Omega =1.$

Examining the expressions for $R_{4i},R_{12,}R_{33}$ and $G_{11}$ (see
respectively (\ref{riccip}) and (\ref{einstp})\ ) we conclude that the
values $Q_{ij}$ appear quadratically which can be ignored in a linear
perturbation theory. Thus the equations for the axial and the polar
perturbations decouple. Considering only linearized expressions, both for
static $\varepsilon $--terms and fluctuations depending on time about the
Schwarzschild values we obtain the equations%
\begin{eqnarray}
\delta _{1}\left( \delta \mu _{2}+\delta \mu _{3}\right) +\left(
r^{-1}-\delta _{1}\mu _{4}\right) \left( \delta \mu _{2}+\delta \mu
_{3}\right) -2r^{-1}\delta \mu _{1} &=&0\quad \left( \delta R_{41}=0\right) ,
\notag \\
\delta _{2}\left( \delta \mu _{1}+\delta \mu _{3}\right) +\left( \delta \mu
_{2}-\delta \mu _{3}\right) \cot \theta &=&0\quad \left( \delta
R_{42}=0\right) ,  \notag \\
\delta _{2}\delta _{1}\left( \delta \mu _{3}+\delta \mu _{4}\right) -\delta
_{1}\left( \delta \mu _{2}-\delta \mu _{3}\right) \cot \theta - &&  \notag \\
\left( r^{-1}-\delta _{1}\mu _{4}\right) \delta _{2}(\delta \mu _{4})-\left(
r^{-1}+\delta _{1}\mu _{4}\right) \delta _{2}(\delta \mu _{1}) &=&0\quad
\left( \delta R_{42}=0\right) ,  \notag \\
e^{2\mu _{4}}\{2\left( r^{-1}+\delta _{1}\mu _{4}\right) \delta _{1}(\delta
\mu _{3})+r^{-1}\delta _{1}\left( \delta \mu _{3}+\delta \mu _{4}-\delta \mu
_{1}+\delta \mu _{2}\right) + &&  \label{peq1} \\
\delta _{1}\left[ \delta _{1}(\delta \mu _{3})\right] -2r^{-1}\delta \mu
_{1}\left( r^{-1}+2\delta _{1}\mu _{4}\right) \}-2e^{-2\mu _{4}}\partial
_{4}[\partial _{4}(\delta \mu _{3})]+ &&  \notag \\
r^{-2}\{\delta _{2}[\delta _{2}(\delta \mu _{3})]+\delta _{2}\left( 2\delta
\mu _{3}+\delta \mu _{4}+\delta \mu _{1}-\delta \mu _{2}\right) \cot \theta
+2\delta \mu _{2}\} &=&0\quad \left( \delta R_{33}=0\right) ,  \notag \\
e^{-2\mu _{1}}[r^{-1}\delta _{1}(\delta \mu _{4})+\left( r^{-1}+\delta
_{1}\mu _{4}\right) \delta _{1}\left( \delta \mu _{2}+\delta \mu _{3}\right)
- &&  \notag \\
2r^{-1}\delta \mu _{1}\left( r^{-1}+2\delta _{1}\mu _{4}\right) ]-e^{-2\mu
_{4}}\partial _{4}[\partial _{4}(\delta \mu _{3}+\delta \mu _{2})] &&  \notag
\\
+r^{-2}\{\delta _{2}[\delta _{2}(\delta \mu _{3})]+\delta _{2}\left( 2\delta
\mu _{3}+\delta \mu _{4}-\delta \mu _{2}\right) \cot \theta +2\delta \mu
_{2}\} &=&0\quad \left( \delta G_{11}=0\right) .  \notag
\end{eqnarray}

The values of type $\delta \mu _{\alpha }=\delta \mu _{\alpha
}^{(\varepsilon )}+\delta \mu _{\alpha }^{(\varsigma )}$ from (\ref{peq1})
contain two components: the first ones are static, proportional to $%
\varepsilon ,$ and the second ones may depend on time coordinate $t.$ We
shall assume that the perturbations $\delta \mu _{\alpha }^{(\varsigma )}$
have a time--dependence $\exp [\sigma t]$ $\ $so that the partial time
derivative $"\partial _{4}"$ is replaced by the factor $i\sigma .$ In order
to treat both type of increments in a similar fashion we may consider that
the values labelled with $(\varepsilon )$ also oscillate in time like $\exp
[\sigma ^{(\varepsilon )}t]$ but with a very small (almost zero) frequency $%
\sigma ^{(\varepsilon )}\rightarrow 0.$ There are also actions of
''elongated'' partial derivative operators like
\begin{equation*}
\delta _{1}\left( \delta \mu _{\alpha }\right) =\partial _{1}\left( \delta
\mu _{\alpha }\right) -\varepsilon n_{1}\partial _{4}\left( \delta \mu
_{\alpha }\right) .
\end{equation*}%
To avoid a calculus with complex values we associate the terms proportional $%
\varepsilon n_{1}\partial _{4}$ to amplitudes of type $\varepsilon
in_{1}\partial _{4}$ and write this operator as
\begin{equation*}
\delta _{1}\left( \delta \mu _{\alpha }\right) =\partial _{1}\left( \delta
\mu _{\alpha }\right) +\varepsilon n_{1}\sigma \left( \delta \mu _{\alpha
}\right) .
\end{equation*}%
For the ''non-perturbed'' Schwarzschild values, which are static, the
operator $\delta _{1}$ reduces to $\partial _{1},$ i.e. $\delta
_{1}v_{\alpha }=\partial _{1}v_{\alpha }.$ Hereafter we shall consider that
the solution of the system (\ref{peq1}) consists \ from a superposition of
two linear solutions, $\delta \mu _{\alpha }=\delta \mu _{\alpha
}^{(\varepsilon )}+\delta \mu _{\alpha }^{(\varsigma )}$; the first class of
solutions for increments will be provided with index $(\varepsilon ),$
corresponding to the frequency $\sigma ^{(\varepsilon )}$ and the second
class will be for the increments with index $(\varsigma )$ and correspond to
the frequency $\sigma ^{(\varsigma )}.$ We shall write this as $\delta \mu
_{\alpha }^{(A)}$ and $\sigma _{(A)}$ for the labels $A=\varepsilon $ or $%
\varsigma $ and suppress the factors $\exp [\sigma ^{(A)}t]$ in our
subsequent considerations. The system of equations (\ref{peq1}) will be
considered for both type of increments.

We can separate the variables by substitutions (see the method in Refs. \cite%
{03fried,03chan})%
\begin{eqnarray}
\delta \mu _{1}^{(A)} &=&L^{(A)}(r)P_{l}(\cos \theta ),\quad \delta \mu
_{2}^{(A)}=\left[ T^{(A)}(r)P_{l}(\cos \theta )+V^{(A)}(r)\partial
^{2}P_{l}/\partial \theta ^{2}\right] ,  \label{auxc1} \\
\delta \mu _{3}^{(A)} &=&\left[ T^{(A)}(r)P_{l}(\cos \theta )+V^{(A)}(r)\cot
\theta \partial P_{l}/\partial \theta \right] ,\quad \delta \mu
_{4}^{(A)}=N^{(A)}(r)P_{l}(\cos \theta )  \notag
\end{eqnarray}%
and reduce the system of equations (\ref{peq1}) \ to%
\begin{eqnarray}
\delta _{1}\left( N^{(A)}-L^{(A)}\right) &=&\left( r^{-1}-\partial _{1}\nu
_{4}\right) N^{(A)}+\left( r^{-1}+\partial _{1}\nu _{4}\right) L^{(A)},
\notag \\
\delta _{1}L^{(A)}+\left( 2r^{-1}-\partial _{1}\nu _{4}\right) N^{(A)} &=&-
\left[ \delta _{1}X^{(A)}+\left( r^{-1}-\partial _{1}\nu _{4}\right) X^{(A)}%
\right] ,  \label{peq2a}
\end{eqnarray}%
and
\begin{eqnarray}
2r^{-1}\delta _{1}\left( N^{(A)}\right)
-l(l+1)r^{-2}e^{-2v_{4}}N^{(A)}-2r^{-1}(r^{-1}+2\partial _{1}\nu
_{4})L^{(A)}-2(r^{-1}+ &&  \label{peq2b} \\
\partial _{1}\nu _{4})\delta _{1}\left[ N^{(A)}+(l-1)(l+2)V^{(A)}/2\right]
-(l-1)(l+2)r^{-2}e^{-2v_{4}}\left( V^{(A)}-L^{(A)}\right) - &&  \notag \\
2\sigma _{(A)}^{2}e^{-4v_{4}}\left[ L^{(A)}+(l-1)(l+2)V^{(A)}/2\right] &=&0,
\notag
\end{eqnarray}%
where we have introduced new functions
\begin{equation*}
X^{(A)}=\frac{1}{2}(l-1)(l+2)V^{(A)}
\end{equation*}%
and considered the relation
\begin{equation*}
T^{(A)}-V^{(A)}+L^{(A)}=0\quad (\delta R_{42}=0).
\end{equation*}%
We can introduce the functions
\begin{eqnarray}
\widetilde{L}^{(A)} &=&L^{(A)}+\varepsilon \sigma _{(A)}\int
n_{1}L^{(A)}dr,\quad \widetilde{N}^{(A)}=N^{(A)}+\varepsilon \sigma
_{(A)}\int n_{1}N^{(A)}dr,  \label{tilds} \\
\widetilde{T}^{(A)} &=&N^{(A)}+\varepsilon \sigma _{(A)}\int
n_{1}N^{(A)}dr,\quad \widetilde{V}^{(A)}=V^{(A)}+\varepsilon \sigma
_{(A)}\int n_{1}V^{(A)}dr,  \notag
\end{eqnarray}%
for which
\begin{equation*}
\partial _{1}\widetilde{L}^{(A)}=\delta _{1}\left( L^{(A)}\right) ,\partial
_{1}\widetilde{N}^{(A)}=\delta _{1}\left( N^{(A)}\right) ,\partial _{1}%
\widetilde{T}^{(A)}=\delta _{1}\left( T^{(A)}\right) ,\partial _{1}%
\widetilde{V}^{(A)}=\delta _{1}\left( V^{(A)}\right) ,
\end{equation*}%
and, this way it is possible to substitute in (\ref{peq2a}) and (\ref{peq2b}%
) the elongated partial derivative $\delta _{1}$ by the usual one acting on
''tilded'' radial increments.

By straightforward calculations (see details in Ref. \cite{03chan}) one can
check that the functions
\begin{equation*}
Z_{(A)}^{(+)}=r^{2}\frac{6mX^{(A)}/r(l-1)(l+2)-L^{(A)}}{r(l-1)(l+2)/2+3m}
\end{equation*}%
satisfy one--dimensional wave equations similar to (\ref{eq7}) for $Z^{(\eta
)}$ with $\eta _{3}=1,$ when $r_{\star }=r_{\ast },$%
\begin{eqnarray}
\left( \frac{d^{2}}{dr_{\ast }^{2}}+\sigma _{(A)}^{2}\right) \widetilde{Z}%
_{(A)}^{(+)} &=&V^{(+)}Z_{(A)}^{(+)},  \label{peq3} \\
\widetilde{Z}_{(A)}^{(+)} &=&Z_{(A)}^{(+)}+\varepsilon \sigma _{(A)}\int
n_{1}Z_{(A)}^{(+)}dr,  \notag
\end{eqnarray}%
where%
\begin{eqnarray}
V^{(+)} &=&\frac{2\Delta }{r^{5}[r(l-1)(l+2)/2+3m]^{2}}\times \{9m^{2}\left[
\frac{r}{2}(l-1)(l+2)+m\right]  \label{pot3} \\
&&+\frac{1}{4}(l-1)^{2}(l+2)^{2}r^{3}\left[ 1+\frac{1}{2}(l-1)(l+2)+\frac{3m%
}{r}\right] \}.  \notag
\end{eqnarray}%
For $\varepsilon \rightarrow 0,$ the equation (\ref{peq3}) transforms in the
usual Zerilli equation \cite{03zerilli,03chan}.

To complete the solution we give the formulas for the ''tilded'' $L$--, $X$%
-- and $N$--factors,%
\begin{eqnarray}
\widetilde{L}^{(A)} &=&\frac{3m}{r^{2}}\widetilde{\Phi }^{(A)}-\frac{%
(l-1)(l+2)}{2r}\widetilde{Z}_{(A)}^{(+)},  \label{peg3a} \\
\widetilde{X}^{(A)} &=&\frac{(l-1)(l+2)}{2r}(\widetilde{\Phi }^{(A)}+%
\widetilde{Z}_{(A)}^{(+)}),  \notag \\
\widetilde{N}^{(A)} &=&\left( m-\frac{m^{2}+r^{4}\sigma _{(A)}^{2}}{r-2m}%
\right) \frac{\widetilde{\Phi }^{(A)}}{r^{2}}-\frac{(l-1)(l+2)r}{%
2(l-1)(l+2)+12m}\frac{\partial \widetilde{Z}_{(A)}^{(+)}}{\partial r_{\#}}
\notag \\
&&-\frac{(l-1)(l+2)}{\left[ r(l-1)(l+2)+6m\right] ^{2}}\times  \notag \\
&&\left\{ \frac{12m^{2}}{r}+3m(l-1)(l+2)+\frac{r}{2}(l-1)(l+2)\left[
(l-1)(l+2)+2\right] \right\} ,  \notag
\end{eqnarray}%
where
\begin{equation*}
\widetilde{\Phi }^{(A)}=(l-1)(l+2)e^{\nu _{4}}\int \frac{e^{-\nu _{4}}%
\widetilde{Z}_{(A)}^{(+)}}{(l-1)(l+2)r+6m}dr.
\end{equation*}%
Following \ the relations (\ref{tilds}) we can compute the corresponding
''untilded'' values an put them in (\ref{auxc1}) in order to find the
increments of fluctuations driven by the system of equations (\ref{peq1}).
For simplicity, we omit the rather cumbersome final expressions.

The formulas (\ref{peg3a}) together with a solution of the wave equation (%
\ref{peq3}) complete the procedure of definition of formal solutions for
polar perturbations. In Ref. \cite{03chan} there are tabulated the data for
the potential (\ref{pot3}) for different values of $l$ and $\left(
l-1\right) (l+2)/2.$ In the anisotropic case the explicit form of solutions
is deformed by terms proportional to $\varepsilon n_{1}\sigma .$ The static
ellipsoidal like deformations can be modelled by the formulas obtained in the
limit $\sigma _{(\varepsilon )}\rightarrow 0.$

\subsection{The stability of polarized black ellipsoids}

The problem of stability of anholonomically deformed Schwarzschild metrics
to external perturbation is very important to be solved in order to
understand if such static black ellipsoid like objects may exist in general
relativity and its cosmological constant generalizations. We address the
question: Let be given any initial values for a static locally anisotropic
configuration confined to a finite interval of $r_{\star },$ for axial
perturbations, and $r_{\ast },$ for polar perturbations, will one remain
bounded such perturbations at all times of evolution? The answer to this
question is to obtained similarly to Refs. \cite{03chan} and \cite{03vels} with
different type of definitions of functions g $Z^{(\eta )}$ and $%
Z_{(A)}^{(+)} $ for different type of black holes.

We have proved that even for anisotropic configurations every type of
perturbations are governed by one dimensional wave equations of the form%
\begin{equation}
\frac{d^{2}Z}{d\rho }+\sigma ^{2}Z=VZ  \label{eq8}
\end{equation}%
where $\rho $ is a radial type coordinate, $Z$ is a corresponding $Z^{(\eta
)}$ or $Z_{(A)}^{(+)}$ with respective smooth real, independent of $\sigma
>0 $ potentials $\widetilde{V}^{(\eta )}$ or $V^{(-)}$ with bounded
integrals. For such equations a solution $Z(\rho ,\sigma ,\varphi _{0})$
satisfying the boundary conditions \ $Z\rightarrow e^{i\sigma \rho
}+R(\sigma )e^{-i\sigma \rho }\quad (\rho \rightarrow +\infty )$ and $%
Z\rightarrow T(\sigma )e^{i\sigma \rho }\qquad (\rho \rightarrow -\infty )$
(the first expression corresponds to an incident wave of unit amplitude from
$+\infty $ giving rise to a reflected wave of amplitude $R(\sigma )$ at $%
+\infty $ and the second expression is for a transmitted wave of \ amplitude
$T(\sigma )$ at $-\infty ),$ provides a basic complete set of wave functions
which allows to obtain a stable evolution. For any initial perturbation that
is smooth and confined to finite interval of $\rho ,$ we can write the
integral%
\begin{equation*}
\psi (\rho ,0)=\left( 2\pi \right) ^{-1/2}\int_{-\infty }^{+\infty }\widehat{%
\psi }(\sigma ,0)Z(\rho ,\sigma )d\sigma
\end{equation*}%
and define the evolution of perturbations,%
\begin{equation*}
\psi (\rho ,t)=\left( 2\pi \right) ^{-1/2}\int_{-\infty }^{+\infty }\widehat{%
\psi }(\sigma ,0)e^{i\sigma t}Z(\rho ,\sigma )d\sigma .
\end{equation*}%
The Schrodinger theory states the conditions%
\begin{equation*}
\int_{-\infty }^{+\infty }|\psi (\rho ,0)|^{2}d\rho =\int_{-\infty
}^{+\infty }|\widehat{\psi }(\sigma ,0)|^{2}d\sigma =\int_{-\infty
}^{+\infty }|\psi (\rho ,0)|^{2}d\rho ,
\end{equation*}%
from which the boundedness of $\psi (\rho ,t)$ follows for all $t>0.$

In our consideration we have replaced the time partial derivative $\partial
/\partial t$ by $i\sigma ,$ which was represented by the approximation of
perturbations to be periodic like $e^{i\sigma t}.$ This is connected with a
time--depending variant of (\ref{eq8}), like%
\begin{equation*}
\frac{\partial ^{2}Z}{\partial t^{2}}=\frac{\partial ^{2}Z}{\partial \rho
^{2}}-VZ.
\end{equation*}%
Multiplying this equation on $\partial \overline{Z}/\partial t,$ where $%
\overline{Z}$ denotes the complex conjugation, and integrating on parts, we
obtain
\begin{equation*}
\int_{-\infty }^{+\infty }\left( \frac{\partial \overline{Z}}{\partial t}%
\frac{\partial ^{2}Z}{\partial t^{2}}+\frac{\partial Z}{\partial \rho }\frac{%
\partial ^{2}\overline{Z}}{\partial t\partial \rho }+VZ\frac{\partial
\overline{Z}}{\partial t}\right) d\rho =0
\end{equation*}%
providing the conditions of convergence of necessary integrals. This
equation added to its complex conjugate results in a constant energy
integral,
\begin{equation*}
\int_{-\infty }^{+\infty }\left( \left| \frac{\partial Z}{\partial t}\right|
^{2}+\left| \frac{\partial Z}{\partial \rho }\right| ^{2}+V\left| Z\right|
^{2}\right) d\rho =const,
\end{equation*}%
which bounds the expression $|\partial Z/\partial t|^{2}$ and excludes an
exponential growth of any bound\-ed solution of the equation (\ref{eq8}). We
note that this property holds for every type of ''ellipsoidal'' like
deformation of the potential, $V\rightarrow V+\varepsilon V^{(1)},$ with
possible dependencies on polarization functions as we considered in (\ref%
{eq7a}) and/or (\ref{pot3}).

The general properties of the one--dimensional Schrodinger equations related
to perturbations of holonomic and anholonomic solutions of the Einstein
equations allow us to conclude that there are locally anisotropic static
configurations which are stable under linear deformations.

In a similar manner we may analyze perturbations (axial or polar) governed
by a two--dimensional Schrodinger waive equation like
\begin{equation*}
\frac{\partial ^{2}Z}{\partial t^{2}}=\frac{\partial ^{2}Z}{\partial \rho
^{2}}+A(\rho ,\varphi ,t)\frac{\partial ^{2}Z}{\partial \varphi ^{2}}-V(\rho
,\varphi ,t)Z
\end{equation*}%
for some functions of necessary smooth class. The stability in this case is
proven if exists an (energy) integral
\begin{equation*}
\int_{0}^{\pi }\int_{-\infty }^{+\infty }\left( \left| \frac{\partial Z}{%
\partial t}\right| ^{2}+\left| \frac{\partial Z}{\partial \rho }\right|
^{2}+\left| A\frac{\partial Z}{\partial \rho }\right| ^{2}+V\left| Z\right|
^{2}\right) d\rho d\varphi =const
\end{equation*}%
which bounds $|\partial Z/\partial t|^{2}$ $\ $for two--dimensional
perturbations. For simplicity, we omitted such calculus in this work.

We emphasize that this way we can also prove the stability of perturbations
along ''aniso\-tro\-pic'' directions of arbitrary anholonomic deformations
of the Schwarzschild solution which have non--spherical horizons and can be
covered by a set of finite regions approximated as small, ellipsoid like,
deformations of some spherical hypersurfaces. We may analyze the geodesic
congruence on every deformed sub-region of necessary smoothly class and
proof the stability as we have done for the resolution ellipsoid horizons.
In general, we may consider horizons of with non--trivial topology, like
vacuum black tori, or higher genus anisotropic configurations. This is not
prohibited by the principles of topological censorship \cite{03ptc} if we are
dealing with off--diagonal metrics and associated anholonomic frames \cite{03v}%
. The vacuum anholonomy in such cases may be treated as an effective matter
which change the conditions of topological theorems.

\section[Two Additional Examples of Off--Diagonal Solutions]
{Two Additional Examples of Off--Diagonal Exact Solutions}

There are some classes of exact solutions which can be modelled by anholonomic
frame transforms and generic off--diagonal metric ansatz and related to
configurations constructed by using another methods \cite{03es,03cans}. We
analyze in this section two classes of such 4D spacetimes.

\subsection{Anholonomic ellipsoidal shapes}

The present status of ellipsoidal shapes in general relativity associated to
some perfect--fluid bodies, rotating configurations or to some families of
confocal ellipsoids in Riemannian spaces is examined in details in Ref. \cite%
{03es}. We shall illustrate in this subsection how such configurations may be
modelled by generic off--diagonal metrics and/or as spacetimes with
anisotropic cosmological constant. The off--diagonal coefficients will be
subjected to certain anholonomy conditions resulting (roughly speaking) in
effects similar to those of perfect--fluid bodies.

We consider a metric ansatz with conformal factor like in (\ref{metric1p})%
\begin{eqnarray}
\delta s^{2} &=&\Omega (\theta ,\nu )\left[ g_{1}d\theta ^{2}+g_{2}(\theta
)d\varphi ^{2}+h_{3}(\theta ,\nu )\delta \nu ^{2}+h_{4}(\theta ,\nu )\delta
t^{2}\right] ,  \notag \\
\delta \nu &=&d\nu +w_{1}(\theta ,\nu )d\theta +w_{2}(\theta ,\nu )d\varphi ,
\label{ansatzes} \\
\delta t &=&dt+n_{1}(\theta ,\nu )d\theta +n_{2}(\theta ,\nu )d\varphi ,
\notag
\end{eqnarray}%
where the coordinated $\left( x^{1}=\theta ,x^{2}=\varphi \right) $ are
holonomic and the coordinate $y^{3}=\nu $ and the timelike coordinate $%
y^{4}=t$ are 'anisotropic' ones. For a particular parametrization when
\begin{eqnarray}
\Omega &=&\Omega _{\lbrack 0]}(\nu )=v\left( \rho \right) \rho ^{2},\
g_{1}=1,  \label{dataes} \\
g_{2} &=&g_{2[0]}=\sin ^{2}\theta ,h_{3}=h_{3[0]}=1,h_{4}=h_{4[0]}=-1,
\notag \\
w_{1} &=&0,\ w_{2}=w_{2[0]}(\theta )=\sin ^{2}\theta ,\ n_{1}=0,\
n_{2}=n_{2[0]}(\theta )=2R_{0}\cos \theta  \notag
\end{eqnarray}%
and the coordinate $\nu $ is defined related to $\rho $ as
\begin{equation*}
d\nu =\int \left| \frac{f\left( \rho \right) }{v\left( \rho \right) }\right|
^{1/2}\frac{d\rho }{\rho },
\end{equation*}%
we obtain the metric element for a special case spacetimes with co--moving
ellipsoidal symmetry defined by an axially symmetric, rigidly rotating
perfect--fluid configuration with confocal inside ellipsoidal symmetry (see
formula (4.21) and related discussion in Ref. \cite{03es}, where the status of
constant $R_{0}$ and functions $v\left( \rho \right) $ and $f\left( \rho
\right) $ are explicitly defined).

By introducing nontrivial ''polarization'' functions $q^{[v]}\left( \theta
\right) $ and $\eta _{3,4}(\theta ,\nu )$ for which
\begin{equation*}
g_{2}=g_{2[0]}q^{[v]}\left( \theta \right) ,\ h_{3,4}=\eta _{3,4}(\theta
,\nu )h_{3,4[0]}
\end{equation*}%
we can state the conditions when the ansatz (\ref{ansatzes}) defines a) an
off--diagonal ellipsoidal shape or b) an ellipsoidal configuration induced
by anisotropically polarized cosmological constant.

Let us consider the case a). The Theorem 2 from Ref. \cite{03vth} and the
formula (72) in Ref. \cite{03vmag2} (see also  the Appendix in \cite{03vncs})
states that any metric of type (\ref{ansatzes}) is vacuum if $\Omega
^{p_{1}/p_{2}}=h_{3}$ for some integers $p_{1}$ and $p_{2},$ the factor $%
\Omega =\Omega _{\lbrack 1]}(\theta ,\nu )\Omega _{\lbrack 0]}(\nu )$
satisfies the condition
\begin{equation*}
\partial _{i}\Omega -\left( w_{i}+\zeta _{i}\right) \partial _{\nu }\Omega =0
\end{equation*}%
for any additional deformation functions $\zeta _{i}(\theta ,\nu )$ and the
coefficients
\begin{equation}
g_{1}=1,g_{2}=g_{2[0]}q^{[v]}\left( \theta \right) ,\ h_{3,4}=\eta
_{3,4}(\theta ,\nu )h_{3,4[0]},w_{i}(\theta ,\nu ),n_{i}(\theta ,\nu )
\label{data5}
\end{equation}%
satisfy the equations (\ref{2ricci1a})--(\ref{2ricci4a}). The procedure of
constructing such exact solutions is very similar to the considered in
subsection 4.1 for black ellipsoids. For anholonomic ellipsoidal shapes
(they are characterized by nontrivial anholonomy coefficients (\ref{2anh})
and respectively induced noncommutative symmetries) we have to put as
''boundary'' condition in integrals of type (\ref{1auxf4}) just to have $%
n_{1}=0,\ n_{2}=n_{2[0]}(\theta )=2R_{0}\cos \theta $ from data (\ref{dataes}%
) in the limit when dependence on ''anisotropic'' variable $\nu $ vanishes.
The functions $w_{i}(\theta ,\nu )$ and $n_{i}(\theta ,\nu )$ must be
subjected to additional constraints if we wont to construct ellipsoidal
shape configurations with zero anholonomically induced torsion (\ref{1torsion}%
) and N--connection curvature, $\Omega _{jk}^{a}=\delta _{k}N_{j}^{a}-\delta
_{j}N_{k}^{a}=0.$

b) The simplest way to construct an ellipsoidal shape configuration induced
by anisotropic cosmological constant is to find data (\ref{data5}) solving
the equations (\ref{1eq17}) following the procedure defined in subsection
4.2. \ We note that we can solve the equation (\ref{eqaux1}) for $%
g_{2}=g_{2}\left( \theta \right) =\sin ^{2}\theta $ with $q^{[v]}\left(
\theta \right) =1$ if $\lambda _{\lbrack h]0}=1/2,$ see solution (\ref{aux2p}%
) with $\xi \rightarrow \theta .$ For simplicity, we can consider that $%
\lambda _{\lbrack v]}=0.$ Such type configurations contain, in general,
anholonomically induced torsion.

We conclude, that by using the anholonomic frame method we can generate
ellipsoidal shapes (in general, with nontrivial polarized cosmological
constants and induced torsions). Such solutions are similar to corresponding
rotation configurations in general relativity with rigidly rotating
perfect-fluid sources. The rough analogy consists in the fact that by
certain frame constraints induced by off--diagonal metric terms we can model
gravitational--matter like metrics. In previous section we proved the
stability of black ellipsoids for small excentricities. Similar
investigations for ellipsoidal shapes is a task for future (because the
shapes could be with arbitrary excentricity). In Ref. \cite{03es}, there were
discussed points of matchings of locally rotationally symmetric spacetimes
to Taub--NUT metrics. We emphasize that this topic was also specifically
elaborated by using anholonomic frame transforms in Refs. \cite{03vspd1}.

\subsection{Generalization of Canfora--Schmidt solutions}

In general, the solutions generated by anholonomic transforms cannot be
reduced to a diagonal transform only by coordinate transforms (this is
stated in our previous works \cite%
{03v,03v1,03vsbd,03vspd1,03vncs,03vth,03velp1,03velp2,03vncfg,03vels,03vncs},
 see also Refs. \cite%
{03vd} for modelling Finsler like geometries in (pseudo) Riemannian
spacetimes). \ We discus here how 4D off--diagonal ansatz (\ref{12ansatzc4})
generalize the solutions obtained in Ref. \cite{03cans} by a corresponding
parametrization of coordinates as $x^{1}=x,x^{2}=t,y^{3}=\nu =y$ and $%
y^{4}=p.$ If we consider for (\ref{12ansatzc4}) (equivalently, for (\ref%
{2dmetric4}) ) the non-trivial data%
\begin{eqnarray}
g_{1} &=&g_{1[0]}=1,\ g_{2}=g_{2[0]}(x^{1})=-B\left( x\right) P(x)^{2}-C(x),
\notag \\
h_{3} &=&h_{3[0]}(x^{1})=A\left( x\right) >0,\ h_{4}=h_{4[0]}(x^{1})=B\left(
x\right) ,  \notag \\
w_{i} &=&0,n_{1}=0,n_{2}=n_{2[0]}(x^{1})=P(x)/B\left( x\right)  \label{data6}
\end{eqnarray}%
we obtain just the ansatz (12) from Ref. \cite{03cans} (in this subsection we
use a different label for coordinates) which, for instance, for $%
B+C=2,B-C=\ln |x|,P=-1/(B-C)$ with $e^{-1}<\sqrt{|x|}<e$ for a constant $e,$
defines an exact 4D solution of the Einstein equation (see metric (27) from %
\cite{03cans}). By introducing 'polarization' functions $\eta _{k}=\eta
_{k}(x^{i})$ [when $i,k,...=1,2]$ and $\eta _{a}=\eta _{a}(x^{i},\nu )$
[when $a,b,...=3,4$] we can generalize the data (\ref{data6}) as to have
\begin{equation*}
g_{k}(x^{i})=\eta _{k}(x^{i})g_{k[0]},\ h_{a}(x^{i},\nu )=\eta
_{a}(x^{i},\nu )h_{a[0]}
\end{equation*}%
and certain nontrivial values $w_{i}=w_{i}(x^{i},\nu )$ and $%
n_{i}=n_{i}(x^{i},\nu )$ solving the Einstein equations with anholonomic
variables (\ref{2ricci1a})--(\ref{2ricci4a}). We can easy find new classes of
exact solutions, for instance, for $\eta _{1}=1$ and $\eta _{2}=\eta
_{2}(x^{1}).$ In this case $g_{1}=1$ and the function $g_{2}(x^{1}) $ is any
solution of the equation
\begin{equation}
g_{2}^{\bullet \bullet }-\frac{(g_{2}^{\bullet })^{2}}{2g_{2}}=0
\label{partic1}
\end{equation}%
(see equation (\ref{eqaux1}) for $\lambda _{\lbrack v]}=0),$ $g_{2}^{\bullet
}=\partial g_{2}/\partial x^{1}$ which is solved as a particular case if $%
g_{2}=(x^{1})^{2}.$ This impose certain conditions on $\eta _{2}(x^{1})$ if
we wont to take $g_{2[0]}(x^{1})$ just as in (\ref{data6}). For more general
solutions with arbitrary $\eta _{k}(x^{i}),$ we have to take solutions of
equation (\ref{2ricci1a}) and not of a particular case like (\ref{partic1}).

We can generate solutions of (\ref{2ricci2a}) for any $\eta _{a}(x^{i},\nu )$
satisfying the condition (\ref{conda}), $\sqrt{|\eta _{3}|}=\eta _{0}\left(
\sqrt{|\eta _{4}|}\right) ^{\ast },\eta _{0}=const.$ For instance, we can
take arbitrary $\eta _{4}$ and using elementary derivations with $\eta
_{4}^{\ast }=\partial \eta _{4}/\partial \nu $ and a nonzero constant $\eta
_{0},$ to define $\sqrt{|\eta _{3}|}.$ For the vacuum solutions, we can put $%
w_{i}=0$ because $\beta =\alpha _{i}=0$ (see \ formulas (\ref{2ricci2a}) and (%
\ref{2abc})). In this case the solutions of (\ref{2ricci3a}) are trivial.
Having defined $\eta _{a}(x^{i},\nu )$ we can integrate directly the
equation (\ref{2ricci4a}) and find $n_{i}(x^{i},\nu )$ like in formula (\ref%
{1auxf4}) with fixed value $\varepsilon =1$ and considering dependence on all
holonomic variables,
\begin{eqnarray}
n_{i}\left( x^{k},\nu \right) &=&n_{i[1]}\left( x^{k}\right) +n_{i[2]}\left(
x^{k}\right) \int d\nu \ ~\eta _{3}\left( x^{k},\nu \right) /\left( \sqrt{%
|\eta _{4}\left( x^{k},\nu \right) |}\right) ^{3},\eta _{4}^{\ast }\neq 0;
\notag \\
&=&n_{i[1]}\left( x^{k}\right) +n_{i[2]}\left( x^{k}\right) \int d\nu \ \eta
_{3}\left( x^{k},\nu \right) ,\eta _{4}^{\ast }=0;  \label{formau} \\
&=&n_{i[1]}\left( x^{k}\right) +n_{i[2]}\left( x^{k}\right) \int d\nu
/\left( \sqrt{|\eta _{4}\left( x^{k},\nu \right) |}\right) ^{3},\eta
_{3}^{\ast }=0.  \notag
\end{eqnarray}%
These values will generalize the data (\ref{data6}) if we identify $%
n_{1[1]}\left( x^{k}\right) =0$ and $n_{1[1]}\left( x^{k}\right)
=n_{2[0]}(x^{1})=P(x)/B\left( x\right) .$ The solutions with vanishing
induced torsions and zero nonlinear connection curvatures are to be selected
by choosing $n_{i}\left( x^{k},\nu \right) $ and $\eta _{3}\left( x^{k},\nu
\right) $ (or $\eta _{4}\left( x^{k},\nu \right) )$ as to reduce the
canonical connection (\ref{2dcon}) to the Levi--Civita connection (as we
discussed in the end of Section 2).

The solution defined by the data (\ref{data6}) is compared in Ref. \cite%
{03cans} with the Kasner diagonal solution which define the simplest models of
anisotropic cosmology. The metrics obtained by F. Canfora and H.-J. Schmidt
(CS) is generic off--diagonal and can not written in diagonal form by
coordinate transforms. We illustrated that the CS metrics can be effectively
diagonalized with respect to N--adapted anholonomic frames (like a more
general ansatz (\ref{12ansatzc4}) can be reduced to (\ref{2dmetric4})) and that
by anholonomic frame transforms of the CS metric we can generate new classes
of generic off--diagonal  solutions. Such spacetimes may describe certain
models of anisotropic and/or inhomogeneous cosmologies (see, for instance,
Refs. \cite{03vd} were we considered a model of Friedman--Robertson--Walker
metric with ellipsoidal symmetry). The anholonomic generalizations of CS\
metrics are with nontrivial noncommutative symmetry because the anholonomy
coefficients (\ref{2anh}) (see also (\ref{anhb})) are not zero being defined
by nontrivial values (\ref{formau}).

\newpage

\section{Outlook and Conclusions}

The work is devoted to investigation of a new class of exact solutions in
metric--affine and string gravity describing static back rotoid (ellipsoid)
\ and shape configurations possessing hidden noncommutative symmetries.
There are generated also certain generic off--diagonal cosmological metrics.

We consider small, with nonlinear gravitational polarization, static
deformations of the Schwarschild black hole solution (in particular cases,
to some resolution ellipsoid like configurations) preserving the horizon and
geodesic behavior but slightly deforming the spherical constructions. It
was proved that there are such parameters of the exact solutions of the
Einstein equations defined by off--diagonal metrics with ellipsoid symmetry
constructed in Refs. \cite{03v,03v1,03vth,03velp1,03velp2,03vels} as the vacuum
solutions positively define static ellipsoid black hole configurations.

We illustrate that the new class of static ellipsoidal black hole solutions
posses some similarities with the Reissner--Nordstrom metric if the metric's
coefficients are defined with respect to correspondingly adapted anholonomic
frames. \ The parameter of ellipsoidal deformation results in an effective
electromagnetic charge induced by off--diagonal vacuum gravitational
interactions. We note that effective electromagnetic charges and
Reissner--Nordstrom metrics induced by interactions in the bulk of extra
dimension gravity were considered in brane gravity \cite{03maartens}. In our
works we proved that such Reissner--Nordstrom like ellipsoid black hole
configurations may be constructed even in the framework of vacuum Einstein
gravity. It should be emphasized that the static ellipsoid black holes
posses spherical topology and satisfy the principle of topological
censorship \cite{03haw1}. Such solutions are also compatible with the black
hole uniqueness theorems \cite{03ut}. In the asymptotical limits at least for
a very small eccentricity such black ellipsoid metrics transform into the
usual Schwarzschild one. We have proved that the stability of static
ellipsoid black holes can be proved similarly by considering small
perturbations of the spherical black holes \cite{03velp1,03velp2} even the
solutions are extended to certain classes of spacetimes with anisotropically
polarized cosmological constants. (On the stability of the Schwarzschild
solution see details in Ref. \cite{03chan}.)

The off--diagonal metric coefficients induce a specific spacetime distorsion
comparing to the solutions with metrics diagonalizable by coordinate
transforms. So, it is necessary to compare the off--diagonal ellipsoidal
metrics with those describing the distorted diagonal black hole solutions
(see the vacuum case in Refs. \cite{03ms} and an extension to the case of
non--vanishing electric fields \cite{03fk}). For the ellipsoidal cases, the
distorsion of spacetime can be of vacuum origin caused by some anisotropies
(anholonomic constraints) related to off--diagonal terms. In the case of
''pure diagonal'' distorsions such effects follow from the fact that the
vacuum Einstein equations are not satisfied in some regions because of
presence of matter.

The off--diagonal gravity may model some gravity--matter like interactions
like in Kaluza--Klein theory (for some particular configurations and
topological compactifications) but, in general, the off--diagonal vacuum
gravitational dynamics can not be associated to any effective matter
dynamics. So, we may consider that the anholonomic ellipsoidal deformations
of the Schwarzschild metric are some kind of anisotropic off--diagonal
distorsions modelled by certain vacuum gravitational fields with the
distorsion parameteres (equivalently, vacuum gravitational polarizations)
depending both on radial and angular coordinates.

There is a common property that, in general, both classes of off--diagonal
anisotropic and ''pure'' diagonal distorsions (like in Refs. \cite{03ms})
result in solutions which are not asymptotically flat. However, it is
possible to find asymptotically flat extensions even for ellipsoidal
configurations by introducing the corresponding off--diagonal terms (the
asymptotic conditions for the diagonal distorsions are discussed in Ref. %
\cite{03fk}; to satisfy such conditions one has to include some additional
matter fields in the exterior portion of spacetime).

We analyzed the conditions when the anholonomic frame method can model
ellipsoid shape configurations. It was demonstrated that the off--diagonal
metric terms and respectively associated nonlinear connection coefficients
may model ellipsoidal shapes being similar to those derived from solutions
with rotating perfect fluids (roughly speaking, a corresponding frame
anholnomy/ anisotropy may result in modelling of specific matter interactions
but with polarizations of constants, metric coefficients and related frames).

In order to point to some possible observable effects, we note that for the
ellipsoidal metrics with the Schwarzschild asymptotic, the ellipsoidal
character could result in some observational effects in the vicinity of the
horizon (for instance, scattering of particles on a static ellipsoid; we can
compute anisotropic matter accretion effects on an ellipsoidal black hole
put in the center of a galactic being of ellipsoidal or another
configuration). A point of further investigations could be the anisotropic
ellipsoidal collapse when both the matter and spacetime are of ellipsoidal
generic off--diagonal symmetry and/or shape configurations (former
theoretical and computational investigations were performed only for rotoids
with anisotropic matter and particular classes of perturbations of the
Schwarzshild solutions \cite{03st}). For very small eccentricities, we may not
have any observable effects like perihelion shift or light bending if we
restrict our investigations only to the Schwarzshild--Newton asymptotic.

We present some discussion on mechanics and thermodynamics of ellipsoidal
black holes. For the static black ellipsoids with flat asymptotic, we can
compute the area of the ellipsoidal horizon, associate an entropy and
develop a corresponding black ellipsoid thermodynamics. This can be done
even for stable black torus configurations. But this is a very rough
approximation because, in general, we are dealing with off--diagonal metrics
depending anisotropically on two/three coordinates. Such solutions are with
anholonomically deformed Killing horizons and should be described by a
thermodynamics (in general, both non-equilibrium and irreversible) of black
ellipsoids self--consistently embedded into an off--diagonal anisotropic
gravitational vacuum. This is a ground for numerous new conceptual issues to
be developed and related to anisotropic black holes and the anisotropic
kinetics and thermodynamics \cite{03v1} as well to a framework of isolated
anisotropic horizons \cite{03asht} which is a matter of our further
investigations. As an example of a such new concept, we point to a
noncommutative dynamics which can be associated to black ellipsoids.

We emphasize that it is a remarkable fact that, in spite of appearance
complexity, the perturbations of static off--diagonal vacuum gravitational
configurations are governed by similar types of equations as for diagonal
holonomic solutions. Perhaps in a similar manner (as a future development of
this work) by using locally adapted ''N--elongated'' partial derivatives we
can prove stability of very different classes of exact solutions with
ellipsoid, toroidal, dilaton and spinor--soliton symmetries constructed in
Refs. \cite{03v,03v1,03vth,03velp1,03velp2,03vels}. The origin of this mystery is
located in the fact that by anholnomic transforms we effectively
diagonalized the off--diagonal metrics by ''elongating'' some partial
derivatives. This way the type of equations governing the perturbations is
preserved but, for small deformations, the systems of linear equations for
fluctuations became ''slightly'' nondiagonal and with certain tetradic
modifications of partial derivatives and differentials.

It is known that in details the question of relating the particular
integrals of such systems associated to systems of linear differential
equations is investigated in Ref. \cite{03chan}. For anholonomic
configurations, one holds the same relations between the potentials $%
\widetilde{V}^{(\eta )}$ and $V^{(-)}$ and wave functions $Z^{(\eta )}$ and $%
Z_{(A)}^{(+)}$ with that difference that the physical values and formulas
where polarized by some anisotropy functions $\eta _{3}(r,\theta ,\varphi
),\Omega (r,\varphi ),$ $q(r),\eta (r,\varphi ),$ $w_{1}(r,\varphi )$ and $%
n_{1}(r,\varphi )$ and deformed on a small parameter $\varepsilon .$ It is
not clear that a similar procedure could be applied in general for proofs of
stability of ellipsoidal shapes but it would be true for small deformations
from a supposed to be stable primordial configuration.

\newpage

We conclude that there are static black ellipsoid vacuum configurations as
well induced by nontrivially polarized cosmological constants which are
stable with respect to one dimensional perturbations, axial and/or polar
ones, governed by solutions of the corresponding one--dimensional
Schrodinger equations. The problem of stability of such objects with respect
to two, or three, dimensional perturbations, and the possibility of modelling
such perturbations in the framework of a two--, or three--, dimensional
inverse scattering problem is a topic of our further investigations. The
most important problem to be solved is to find a geometrical interpretation
for the anholonomic Schrodinger mechanics of stability to the anholonomic
frame method and to see if we can extend the approach at least to the two
dimensional scattering equations.

%%%%%%%%%%%%%%%%%%%%%%%%%%%%%%

\subsection*{Acknowledgements}

~~ The work is partially supported by a NATO/Portugal fellowship at CENTRA,
Instituto Superior Tecnico, Lisbon. The author is grateful for support to
the organizes of the International Congress of Mathematical Physics, ICMP
2003, and of the Satelite\ Meeting OPORTO 2003. He would like to thank J.
Zsigrai for valuable discussions.

%%%%%%%%%%%%%%%%%%%%%%%%%%%%%%%%%%%%%%%%%%%%%%%%%%%%%%%%%%%%%%%%%%%%%%%%%%%%%

\part{Generic Off--Diagonal Exact Solutions}

\chapter[Locally Anisotropic Black Holes ]
{Locally Anisotropic Black Holes in Einstein Gravity }

{\bf Abstract}
\footnote{\copyright\
 S. Vacaru, Locally Anisotropic Black Holes in Einstein Gravity,
 gr--qc/ 0001020}

By applying the method of moving frames  modelling one and two dimensional
  local anisotropies we construct  new solutions of Einstein equations
 on  pseudo--Riemannian spacetimes. The first class of solutions
 describes non--trivial deformations of static spherically symmetric
 black holes to locally anisotropic ones which have elliptic (in three
 dimensions) and  ellipsoidal, toroidal and  elliptic and another forms
 of cylinder  symmetries (in four dimensions). The  second class consists
 from  black holes with oscillating  elliptic horizons.

\section{Introduction}

In recent years, there has been great interest in investigation of
gravitational models with anisotropies and applications in modern cosmology
and astrophysics. There are possible locally anisotropic inflational and
black hole like solutions of Einstein equations in the framework of
so--called generalized Finsler--Kaluza--Klein models \cite{04v1} and in
low--energy locally anisotropic limits of (super) string theories \cite{04v2}.

In this paper we shall restrict ourselves to a more limited problem of
definition of black hole solutions with local anisotropy in the framework of
the Einstein theory (in three and four dimensions). Our purpose is to
construct solutions of gravitational field equations by imposing symmetries
differing in appearance from the static spherical one (which uniquely
results in the Schwarzschild solution) and search for solutions with
configurations of event horizons like rotation ellipsoids, torus and
ellipsoidal and cylinders. We shall proof that there are possible elliptic
oscillations in time of horizons.

In order to simplify the procedure of solution and investigate more deeply
the physical implications of general relativistic models with local
anisotropy we shall transfer our analysis with respect to anholonomic frames
which are equivalently characterized by nonlinear connection (N--connection)
structures \cite{04barthel,04cartan,04ma,04v1,04v2}. This geometric approach
 is very
useful for construction of metrics with prescribed symmetries of horizons
and definition of conditions when such type black hole like solutions could
be selected from an integral variety of the Einstein field equations with a
corresponding energy--momentum tensor. We argue that, in general, the
symmetries of solutions are not completely determined by the field equations
and coordinate conditions but there are also required some physical
motivations for choosing of corresponding classes of systems of reference
(prescribed type of local anisotropy and symmetries of horizons) with
respect to which the 'picture' of interactions, symmetries and conservation
laws is drawn in the simplest form.

The paper is organized as follows: In section 2 we introduce metrics and
anholonomic frames with local anisotropies admitting equivalent
N--connection structures. We write down the Einstein equations with respect
to such locally anisotropic frames. In section 3 we analyze the general
properties of the system of gravitational field equations for an ansatz for
metrics with local anisotropy. In section 4 we generalize the three
dimensional static black hole solution to the case with elliptic horizon and
proof that there are possible elliptic oscillations in time of locally
anisotropic black holes. The section 5 is devoted to four dimensional
locally anisotropic static solutions with rotation ellipsoidal, toroidal and
cylindrical like horizons and consider elliptic oscillations in time. In the
last section we make some final remarks.

\section{Anholonomic frames and N--con\-nec\-ti\-ons}

\setcounter{equation}{0}

In this section we outline the necessary results on spacetime differential
geometry \cite{04haw} and anholonomic frames induced by N--connection
structures \cite{04ma,04v1,04v2}. We examine an ansatz for locally anisotropic
(pseudo) Riemannian metrics with respect to coordinate bases and illustrate
a substantial geometric simplification and reduction of the number of
coefficients of geometric objects and field equations after linear
transforms to anholonomic bases defined by coefficients of a corresponding
N--connection. The Einstein equations are rewritten in an invariant form
with respect to such locally anisotropic bases.

Consider a class of pseudo--Riemannian metrics
$$
g=g_{\underline{\alpha }\underline{\beta }}\left( u^\varepsilon \right) ~du^{%
\underline{\alpha }}\otimes du^{\underline{\beta }}
$$
in a $n+m$ dimensional spacetime $V^{(n+m)},\
  (n=2$ and $m=1,2$), with components
\begin{equation}
\label{1metric}g_{\underline{\alpha }\underline{\beta }}=\left[
\begin{array}{cc}
g_{ij}+N_i^aN_j^bh_{ab} & N_j^eh_{ae} \\
N_i^eh_{be} & h_{ab}
\end{array}
\right] ,
\end{equation}
where $g_{ij}=g_{ij}\left( u^\alpha \right) $ and $h_{ab}=h_{ab}\left(
u^\alpha \right) $ are respectively some symmetric $n\times n$ and $m\times
m $ dimensional matrices, $N_j^e=N_j^e\left( u^\beta \right) $ is a $n\times
m $ matrix, and the $n+m$ dimensional local coordinates are
 provide with general Greek indices and denoted $u^\beta
=(x^i,y^a).$  The Latin indices $i,j,k,...$ in (\ref{1metric}) run values
$1,2$ and $a,b,c,..$ run values $3,4$ and we note that both type of isotropic,
 $x^i,$ and the so--called anisotropic, $y^a,$ coordinates could be space or
time like ones. We  underline indices in order to emphasize that
components are given with respect to a coordinate (holonomic) basis
\begin{equation}
\label{hb}e_{\underline{\alpha }}=\partial _{\underline{\alpha }}=\partial
/\partial u^{\underline{\alpha }}
\end{equation}
and/or its dual
\begin{equation}
\label{dhb}e^{\underline{\alpha }}=du^{\underline{\alpha }}.
\end{equation}

The class of metrics (\ref{1metric}) transform into a $(n\times n)\oplus
(m\times m)$ block form
\begin{equation}
\label{1dmetric}g=g_{ij}\left( u^\varepsilon \right) ~dx^i\otimes
dx^j+h_{ab}\left( u^\varepsilon \right) \left( \delta y^a\right) ^2\otimes
\left( \delta y^a\right) ^2
\end{equation}
if one chooses a frame of basis vectors
\begin{equation}
\label{3dder}\delta _\alpha =\delta /\partial u^\alpha =\left( \delta
/\partial x^i=\partial _i-N_i^a\left( u^\varepsilon \right) \partial
_a,\partial _b\right) ,
\end{equation}
where $\partial _i=\partial /x^i$ and $\partial _a=\partial /\partial y^a,$
with the dual basis being
\begin{equation}
\label{3ddif}\delta ^\alpha =\delta u^\alpha =\left( dx^i,\delta
y^a=dy^a+N_i^a\left( u^\varepsilon \right) dx^i\right) .
\end{equation}

The set of coefficients $N=\{N_i^a\left( u^\varepsilon \right) \}$ from (\ref
{3dder}) and (\ref{3ddif}) could be associated to components of a nonlinear
connection (in brief, N--connection) structure defining a local
decomposition of spacetime into $n$ isotropic directions $x^i$ and one or
two anisotropic directions $y^a.$ The global definition of N--connection is
due to W. Barthel \cite{04barthel} (the rigorous mathematical definition of
N--connection is possible on the language of exact sequences of vector, or
tangent, subbundles) and this concept is largely applied in Finsler geometry
and its generalizations \cite{04cartan,04ma}. It was concluded \cite{04v1,04v2}
 that
N--connection structures are induced under non--trivial dynamical
compactifications of higher dimensions in (super) string and (super) gravity
theories and even in general relativity if we are dealing with anholonomic
frames.

A N--connection is characterized by its curvature, N--curvature,
\begin{equation}
\label{3ncurv}\Omega _{ij}^a=\partial _iN_j^a-\partial _jN_i^a+N_i^b\partial
_bN_j^a-N_j^b\partial _bN_i^a.
\end{equation}
As a particular case we obtain a linear connection field $\Gamma
_{ib}^a\left( x^i\right) $ if $N_i^a(x^i,y^a) = \Gamma _{ib}^a\left( x^i,
y^a \right) $ \cite{04ma,04v1}.

For nonvanishing values of $\Omega _{ij}^a$ the basis (\ref{3dder}) is
anholonomic and satisfies the conditions
$$
\delta _\alpha \delta _\beta -\delta _\beta \delta _\alpha =w_{~\alpha \beta
}^\gamma \delta _\gamma ,
$$
where the anholonomy coefficients $w_{~\alpha \beta }^\gamma $ are defined
by the components of N--connection,
\begin{eqnarray}
w_{~ij}^k & = & 0,w_{~aj}^k=0,w_{~ia}^k=0,w_{~ab}^k=0,w_{~ab}^c=0,
\nonumber\\
w_{~ij}^a & = &
-\Omega _{ij}^a,w_{~aj}^b=-\partial _aN_i^b,w_{~ia}^b=\partial _aN_i^b.
\nonumber
\end{eqnarray}

We emphasize that the elongated by N--connection operators (\ref{3dder}) and (%
\ref{3ddif}) must be used, respectively, instead of local operators of
partial derivation (\ref{hb}) and differentials (\ref{dhb}) if some
differential calculations are performed with respect to any anholonomic
bases locally adapted to a fixed N--connection structure (in brief, we shall
call such local frames as la--bases or la--frames, where, in brief,
 la-- is from locally anisotropic).

The torsion, $T\left( \delta _\gamma ,\delta _\beta \right) =T_{~\beta
\gamma }^\alpha \delta _\alpha ,$ and curvature, $R\left( \delta _\tau
,\delta _\gamma \right) \delta _\beta =R_{\beta ~\gamma \tau }^{~\alpha
}\delta _\alpha ,$ tensors of a linear connection $\Gamma _{~\beta \gamma
}^\alpha $ are introduced in a usual manner and, respectively, have the
components
\begin{equation}
\label{2torsion}T_{~\beta \gamma }^\alpha =\Gamma _{~\beta \gamma }^\alpha
-\Gamma _{~\gamma \beta }^\alpha +w_{~\beta \gamma }^\alpha
\end{equation}
and
\begin{equation}
\label{1curvature}R_{\beta ~\gamma \tau }^{~\alpha }=\delta _\tau \Gamma
_{~\beta \gamma }^\alpha -\delta _\gamma \Gamma _{~\beta \delta }^\alpha
+\Gamma _{~\beta \gamma }^\varphi \Gamma _{~\varphi \tau }^\alpha -\Gamma
_{~\beta \tau }^\varphi \Gamma _{~\varphi \gamma }^\alpha +\Gamma _{~\beta
\varphi }^\alpha w_{~\gamma \tau }^\varphi .
\end{equation}

The Ricci tensor is defined
\begin{equation}
\label{ricci}R_{\beta \gamma }=R_{\beta ~\gamma \alpha }^{~\alpha }
\end{equation}
and the scalar curvature is
\begin{equation}
\label{scalarcurvature}R=g^{\beta \gamma }R_{\beta \gamma } .
\end{equation}

The Einstein equations with respect to a la--basis (\ref{3ddif}) are written
\begin{equation}
\label{einsteq1}R_{\beta \gamma }-\frac R2g_{\beta \gamma }=k\Upsilon
_{\beta \gamma },
\end{equation}
where the energy--momentum d--tensor $\Upsilon _{\beta \gamma }$ includes
the cosmological constant terms and possible contributions of torsion (\ref
{2torsion}) and matter and $k$ is the coupling constant. For a symmetric
linear connection the torsion field can be considered as induced by the
anholonomy coefficients. For dynamical torsions there are necessary
additional field equations, see, for instance, the case of locally
anisotropic gauge like theories \cite{04vg}.

The geometrical objects with respect to a la--bases are distinguished by the
corresponding N--connection structure and called (in brief) d--tensors,
d--metrics (\ref{1dmetric}), linear d--connections and so on \cite{04ma,04v1,04v2}.

A linear d--connection $D$ on a spacetime $V,$
$$
D_{\delta _\gamma }\delta _\beta =\Gamma _{~\beta \gamma }^\alpha \left(
x^k,y^a\right) \delta _\alpha ,
$$
is parametrized by non--trivial horizontal (isotropic) -- vertical
(anisotropic), in brief, h--v--components,
\begin{equation}
\label{3dcon}\Gamma _{~\beta \gamma }^\alpha =\left(
L_{~jk}^i,L_{~bk}^a,C_{~jc}^i,C_{~bc}^a\right) .
\end{equation}
Some d--connection and d--metric structures are compatible if there are
satisfied the conditions
$$
D_\alpha g_{\beta \gamma }=0.
$$
For instance, the canonical compatible d--connection
$$
^c\Gamma _{~\beta \gamma }^\alpha =\left(
^cL_{~jk}^i,^cL_{~bk}^a,^cC_{~jc}^i,^cC_{~bc}^a\right)
$$
is defined by the coefficients of d--metric (\ref{1dmetric}), $g_{ij}\left(
x^i,y^a\right) $ and $h_{ab}\left( x^i,y^a\right) ,$ and of N--connection, $%
N_i^a=N_i^a\left( x^i,y^b\right) ,$
\begin{eqnarray}
^cL_{~jk}^i & = & \frac 12g^{in}\left( \delta _kg_{nj}+\delta _jg_{nk}-\delta
_ng_{jk}\right) , \label{cdcon} \\
^cL_{~bk}^a & = & \partial _bN_k^a+\frac 12h^{ac}\left( \delta
_kh_{bc}-h_{dc}\partial _bN_i^d-h_{db}\partial _cN_i^d\right) ,
\nonumber \\
^cC_{~jc}^i & = & \frac 12g^{ik}\partial _cg_{jk}, \nonumber \\
^cC_{~bc}^a & = & \frac 12h^{ad}\left( \partial _ch_{db}+\partial
_bh_{dc}-\partial _dh_{bc}\right) .  \nonumber
\end{eqnarray}
The coefficients of the canonical d--connection generalize with respect to
la--bases the well known Christoffel symbols.

For a d--connection (\ref{3dcon}) we can compute the non--trivial components
of d--torsion (\ref{2torsion})
\begin{eqnarray}
T_{.jk}^i & = & T_{jk}^i=L_{jk}^i-L_{kj}^i,\quad
T_{ja}^i=C_{.ja}^i,T_{aj}^i=-C_{ja}^i, \nonumber \\
T_{.ja}^i & = & 0,\quad T_{.bc}^a=S_{.bc}^a=C_{bc}^a-C_{cb}^a,
\label{1dtors} \\
T_{.ij}^a & = &
-\Omega _{ij}^a,\quad T_{.bi}^a= \partial _b  N_i^a
-L_{.bj}^a,\quad T_{.ib}^a=-T_{.bi}^a. \nonumber
\end{eqnarray}

In a similar manner, putting non--vanishing coefficients (\ref{3dcon}) into
the formula for curvature (\ref{1curvature}), we can compute the coefficients
of d--curvature
$$
R\left( \delta _\tau ,\delta _\gamma \right) \delta _\beta = R_{\beta
~\gamma\tau }^{~\alpha }\delta _\alpha ,%
$$
split into h--, v--invariant components,
\begin{eqnarray}
R_{h.jk}^{.i} & = & \delta _kL_{.hj}^i-\delta_jL_{.hk}^i
 +  L_{.hj}^mL_{mk}^i-L_{.hk}^mL_{mj}^i-C_{.ha}^i\Omega _{.jk}^a,
\nonumber \\
R_{b.jk}^{.a} & = & \delta _kL_{.bj}^a-\delta_jL_{.bk}^a
  +  L_{.bj}^cL_{.ck}^a-L_{.bk}^cL_{.cj}^a-C_{.bc}^a\Omega _{.jk}^c,
\nonumber \\
P_{j.ka}^{.i} & = & \partial _kL_{.jk}^i +C_{.jb}^iT_{.ka}^b
 -  ( \partial _kC_{.ja}^i+L_{.lk}^iC_{.ja}^l -
L_{.jk}^lC_{.la}^i-L_{.ak}^cC_{.jc}^i ), \nonumber \\
P_{b.ka}^{.c} & = & \partial _aL_{.bk}^c +C_{.bd}^cT_{.ka}^d
 - ( \partial _kC_{.ba}^c+L_{.dk}^{c\,}C_{.ba}^d
- L_{.bk}^dC_{.da}^c-L_{.ak}^dC_{.bd}^c ) \nonumber \\
S_{j.bc}^{.i} & = & \partial _cC_{.jb}^i-\partial _bC_{.jc}^i
 +  C_{.jb}^hC_{.hc}^i-C_{.jc}^hC_{hb}^i, \nonumber \\
S_{b.cd}^{.a} & = &\partial _dC_{.bc}^a-\partial
_cC_{.bd}^a+C_{.bc}^eC_{.ed}^a-C_{.bd}^eC_{.ec}^a. \nonumber
\end{eqnarray}

The components of the Ricci tensor (\ref{ricci}) with respect to locally
adapted frames (\ref{3dder}) and (\ref{3ddif}) (in this case, d--tensor) are
as follows:
\begin{eqnarray}
R_{ij} & = & R_{i.jk}^{.k},\quad
 R_{ia}=-^2P_{ia}=-P_{i.ka}^{.k},\label{3dricci} \\
R_{ai} &= & ^1P_{ai}=P_{a.ib}^{.b},\quad R_{ab}=S_{a.bc}^{.c}. \nonumber
\end{eqnarray}

We point out that because, in general, $^1P_{ai}\neq ~^2P_{ia}$ the Ricci
d--tensor is non symmetric. This is a consequence of anholonomy of la--bases.

Having defined a d-metric of type (\ref{1dmetric}) on spacetime $V$ we can
compute the scalar curvature (\ref{scalarcurvature}) of a d-connection $D,$%
\begin{equation}
\label{1dscalar}{\overleftarrow{R}}=G^{\alpha \beta }R_{\alpha \beta }=%
\widehat{R}+S,
\end{equation}
where $\widehat{R}=g^{ij}R_{ij}$ and $S=h^{ab}S_{ab}.$

Now, by introducing the values of (\ref{3dricci}) and (\ref{1dscalar}) into
equations (\ref{einsteq1}), the Einstein equations with respect to a
la--basis seen to be%
\begin{eqnarray}
R_{ij}-\frac 12\left( \widehat{R}+S\right) g_{ij} & = &
k\Upsilon _{ij}, \label{1einsteq2} \\
S_{ab}-\frac 12\left( \widehat{R}+S\right) h_{ab} & = & k\Upsilon _{ab},
 \nonumber \\
^1P_{ai} & = & k\Upsilon _{ai}, \nonumber \\
^2P_{ia} & = & -k\Upsilon _{ia}, \nonumber
\end{eqnarray}
where $\Upsilon _{ij},\Upsilon _{ab},\Upsilon _{ai}$ and $\Upsilon _{ia}$
are the components of the energy--momentum d--tensor field $\Upsilon _{\beta
\gamma }$ (which includes possible cosmological constants, contributions of
anholonomy d--torsions (\ref{1dtors}) and matter) and $k$ is the coupling
constant. For simplicity, we omitted the upper left index $c$ pointing that
for the Einstein theory the Ricci d--tensor and curvature scalar should be
computed by applying the coefficients of canonical d--connection (\ref{cdcon}%
).

\section{An ansatz for la--metrics}

\setcounter{equation}{0}

Let us consider a four dimensional (in brief, 4D) spacetime $V^{(2+2)}$
 (with two isotropic plus two anisotropic local coordinates)
 provided with a metric  (\ref{1metric})
 (of  signature\\ (-,+,+,+), or  (+,+,+,-), (+,+,-,+))
 parametrized by a symmetric matrix of type
\begin{equation}
\label{ansatz2}\left[
\begin{array}{cccc}
g_1+q_1{}^2h_3+n_1{}^2h_4 & 0 & q_1h_3 & n_1h_4 \\
0 & g_2+q_2{}^2h_3+n_2{}^2h_4 & q_2h_3 & n_2h_4 \\
q_1h_3 & q_2h_3 & h_3 & 0 \\
n_1h_4 & n_2h_4 & 0 & h_4
\end{array}
\right]
\end{equation}
with components being some functions
$$
g_i=g_i(x^j),q_i=q_i(x^j,z),n_i=n_i(x^j,z),h_a=h_a(x^j,z)%
$$
of necessary smoothly class. With respect to a la--basis (\ref{3ddif}) this
ansatz results in diagonal $2\times 2$ h-- and v--metrics for a d--metric (%
\ref{1dmetric}) (for simplicity, we shall consider only diagonal 2D
nondegenerated metrics  because for such dimensions every symmetric
matrix can be diagonalized).

An equivalent diagonal d--metric (\ref{1dmetric}) is obtained for the
associated N--connection with coefficients being functions on three
coordinates $(x^i,z),$%
\begin{eqnarray}
N_1^3&=&q_1(x^i,z),\ N_2^3=q_2(x^i,z), \label{ncoef} \\
N_1^4&=&n_1(x^i,z),\ N_2^4=n_2(x^i,z). \nonumber
\end{eqnarray}
For simplicity, we shall use brief denotations of partial derivatives, like $%
\dot a$$=\partial a/\partial x^1,a^{\prime }=\partial a/\partial x^2,$ $%
a^{*}=\partial a/\partial z$ $\dot a^{\prime }$$=\partial ^2a/\partial
x^1\partial x^2,$ $a^{**}=\partial ^2a/\partial z\partial z.$

\newpage

The non--trivial components of the Ricci d--tensor (\ref{3dricci}) ( for the
ansatz (\ref{ansatz2})) when $R_1^1 = R_2^2$ and $S_3^3 = S_4^4,$ are
computed
\begin{eqnarray} \label{ricci1}
R_1^1&=&\frac 1{2g_1g_2}
[-(g_1^{^{\prime \prime }}+{\ddot g}_2)+\frac 1{2g_2}\left( {\dot g}%
_2^2+g_1^{\prime }g_2^{\prime }\right) +
\frac 1{2g_1}\left( g_1^{\prime \ 2}+%
\dot g_1\dot g_2\right) ], \\
\label{1ricci2}
S_3^3&=&\frac 1{h_3h_4}[-h_4^{**}+\frac 1{2h_4}(h_4^{*})^2+%
\frac 1{2h_3}h_3^{*}h_4^{*}], \\
 &{} & \nonumber \\
 \label{ricci3}
P_{3i}&=&\frac{q_i}2[\left( \frac{h_3^{*}}{h_3}\right) ^2-
\frac{h_3^{**}}{h_3}+ \frac{h_4^{*}}{2h_4^{\ 2}}-
\frac{h_3^{*}h_4^{*}}{2h_3h_4}]  \\
 &{}&
+\frac 1{2h_4}[\frac{\dot h_4}{2h_4}h_4^{*}-
\dot h_4^{*}+\frac{\dot h_3}{2h_3}h_4^{*}], \nonumber  \\
 &{} & \nonumber \\
 P_{4i}&=&-\frac{h_4}{2h_3}n_i^{**}. \label{ricci4}
\end{eqnarray}

The curvature scalar $\overleftarrow{R}$ (\ref{1dscalar}) is defined by two
non-trivial components $\widehat{R}=2R_1^1$ and $S=2S_3^3.$

The system of Einstein equations (\ref{1einsteq2}) transforms into
\begin{eqnarray}
R_1^1&=&-\kappa \Upsilon _3^3=-\kappa \Upsilon _4^4,
\label{einsteq3a} \\
S_3^3&=&-\kappa \Upsilon _1^1=-\kappa \Upsilon _2^2, \label{einsteq3b}\\
P_{3i}&=& \kappa \Upsilon _{3i}, \label{einsteq3c} \\
P_{4i}&=& \kappa \Upsilon _{4i}, \label{einsteq3d}
\end{eqnarray}
where the values of $R_1^1,S_3^3,P_{ai},$ are taken respectively from (\ref
{ricci1}), (\ref{1ricci2}), (\ref{ricci3}), (\ref{ricci4}).

We note that we can define the N--coefficients (\ref{ncoef}), $q_i(x^k,z)$
and $n_i(x^k,z),$ by solving the equations (\ref{einsteq3c}) and (\ref
{einsteq3d}) if the functions $h_i(x^k,z)$ are known as solutions of the
equations (\ref{einsteq3b}).

Let us analyze the basic properties of equations (\ref{einsteq3b})--(\ref
{einsteq3d}) (the h--equa\-ti\-ons will be considered for 3D and 4D
in the next sections). The v--component of the Einstein equations (%
\ref{einsteq3a})
\begin{equation}
\label{heq}\frac{\partial ^2h_4}{\partial z^2} - \frac 1{2h_4}\left( \frac{%
\partial h_4}{\partial z}\right) ^2 -\frac 1{2h_3}\left( \frac{\partial h_3}{%
\partial z}\right) \left( \frac{\partial h_4}{\partial z}\right) - \frac
\kappa 2\Upsilon _1h_3h_4=0 \nonumber
\end{equation}
(here we write down the partial derivatives on $z$ in explicit form) follows
from (\ref{1ricci2}) and (\ref{einsteq3b}) and relates some first and second
order partial on $z$ derivatives of diagonal components $h_a(x^i,z)$ of a
v--metric with a source $\kappa \Upsilon _1(x^i,z)=\kappa \Upsilon
_1^1=\kappa \Upsilon _2^2$ in the h--subspace. We can consider as unknown
the function $h_3(x^i,z)$ (or, inversely, $h_4(x^i,z))$ for some compatible
values of $h_4(x^i,z)$ (or $h_3(x^i,z))$ and source $\Upsilon _1(x^i,z).$

By introducing a new variable $\beta =h_4^{*}/h_4$ the equation (\ref{heq})
transforms into
\begin{equation}
\label{heq1}\beta ^{*}+\frac 12\beta ^2-\frac{\beta h_3^{*}}{2h_3}-2\kappa
\Upsilon _1h_3=0
\end{equation}
which relates two functions $\beta \left( x^i,z\right) $ and $h_3\left(
x^i,z\right) .$ There are two possibilities: 1) to define $\beta $ (i. e. $%
h_4)$ when $\kappa \Upsilon _1$ and $h_3$ are prescribed and, inversely 2)
to find $h_3$ for given $\kappa \Upsilon _1$ and $h_4$ (i. e. $\beta );$ in
both cases one considers only ''*'' derivatives on $z$--variable
(coordinates $x^i$ are treated as parameters).

\begin{enumerate}
\item  In the first case the explicit solutions of (\ref{heq1}) have to be
constructed by using the integral varieties of the general Riccati equation
\cite{04kamke} which by a corresponding redefinition of variables, $%
z\rightarrow z\left( \varsigma \right) $ and $\beta \left( z\right)
\rightarrow \eta \left( \varsigma \right) $ (for simplicity, we omit here
the dependencies on $x^i)$ could be written in the canonical form
$$
\frac{\partial \eta }{\partial \varsigma }+\eta ^2+\Psi \left( \varsigma
\right) =0
$$
where $\Psi $ vanishes for vacuum gravitational fields. In vacuum cases the
Riccati equation reduces to a Bernoulli equation which (we can use the
former variables) for $s(z)=\beta ^{-1}$ transforms into a linear
differential (on $z)$ equation,
\begin{equation}
\label{heq1a}s^{*}+\frac{h_3^{*}}{2h_3}s-\frac 12=0.
\end{equation}

\item  In the second (inverse) case when $h_3$ is to be found for some
prescribed $\kappa \Upsilon _1$ and $\beta $ the equation (\ref{heq1}) is to
be treated as a Bernoulli type equation,
\begin{equation}
\label{heq2}h_3^{*}=-\frac{4\kappa \Upsilon _1}\beta (h_3)^2+\left( \frac{%
2\beta ^{*}}\beta +\beta \right) h_3
\end{equation}
which can be solved by standard methods. In the vacuum case the squared on $%
h_3$ term vanishes and we obtain a linear differential (on $z)$ equation.
\end{enumerate}

A particular interest presents those solutions of the equation (\ref{heq1})
which via 2D conformal transforms with a factor $\omega =\omega (x^i,z)$ are
equivalent to a diagonal h--metric on $x$--variables, i.e. one holds the
parametrization
\begin{equation}
\label{conf4d}h_3=\omega (x^i,z)\ a_3\left( x^i\right) \mbox{ and }%
h_4=\omega (x^i,z)\ a_4\left( x^i\right) ,
\end{equation}
where $a_3\left( x^i\right) $ and $a_4\left( x^i\right) $ are some arbitrary
functions (for instance, we can impose the condition that they describe some
2D soliton like or black hole solutions). In this case $\beta =\omega
^{*}/\omega $ and for $\gamma =\omega ^{-1}$ the equation (\ref{heq1})
transforms into
\begin{equation}
\label{2confeq}\gamma \ \gamma ^{**}=-2\kappa \Upsilon _1a_3\left( x^i\right)
\end{equation}
with the integral variety determined by
\begin{equation}
\label{confeqsol}z=\int \frac{d\gamma }{\sqrt{\left| -4k\Upsilon
_1a_3(x^i)\ln |\gamma |+C_1(x^i)\right| }}+C_2(x^i),
\end{equation}
where it is considered that the source $\Upsilon _1$ does not depend on $z.$

Finally, we conclude that the v--metrics are defined by the integral
varieties of corresponding Riccati and/or Bernoulli equations with respect
to $z$--variables with the h--coordinates $x^i$ treated as parameters.

\section{3D black la--holes}

\setcounter{equation}{0}

Let us analyze some basic properties of 3D spacetimes $V^{(2+1)}$
 (we emphasize that in approach $(2+1)$ points to a splitting
 into two isotropic and  one anisotropic directions and not to usual
  2D space plus one time like coordinates;  in general anisotropies could
 be associate to both space and/or time like coordinates)
 provided with d--metrics of type
\begin{equation}
\label{dmetr3}\delta s^2=g_1\left( x^k\right) \left( dx^1\right)
^2+g_2\left( x^k\right) \left( dx^2\right) ^2+h_3(x^i,z)\left( \delta
z\right) ^2,
\end{equation}
where $x^k$ are 2D coordinates, $y^3=z$ is the anisotropic coordinate and
$$
\delta z=dz+N_i^3(x^k,z)dx^i.
$$
The N--connection coefficients are
\begin{equation}
\label{ncoef3}N_1^3=q_1(x^i,z),\ N_2^3=q_2(x^i,z).
\end{equation}

The non--trivial components of the Ricci d--tensor (\ref{3dricci}), for the
ansatz (\ref{ansatz2}) with $h_4 =1$ and $n_i =0$, $R_1^1=R_2^2$ and $%
P_{3i}, $ are
\begin{eqnarray} \label{ricci1_3}
R_1^1&=& \frac 1{2g_1g_2} \ [-(g_1^{^{\prime \prime }}
+{\ddot g}_2)  +\frac 1{2g_2}\left( {\dot g}_2^2
+g_1^{\prime }g_2^{\prime }\right) +
\frac 1{2g_1}\left( g_1^{\prime \ 2}+%
\dot g_1\dot g_2\right) ], \\
\label{ricci3_3}
P_{3i} &=& \frac{q_i}2[\left( \frac{h_3^{*}}{h_3}\right) ^2-
\frac{h_3^{**}}{h_3} ]
\end{eqnarray}
(for 3D the component $S_3^3\equiv 0,$ see (\ref{1ricci2})).

The curvature scalar $\overleftarrow{R}$ (\ref{1dscalar}) is $\overleftarrow{R%
}=\widehat{R}=2R_1^1.$

The system of Einstein equations (\ref{1einsteq2}) transforms into
\begin{eqnarray}
R_1^1&=&-\kappa \Upsilon _3^3,
\label{einsteq3a3} \\
P_{3i}&=& \kappa \Upsilon _{3i}, \label{einsteq3c3}
\end{eqnarray}
which is compatible for energy--momentum d--tensors with $%
\Upsilon_1^1=\Upsilon _2^2=0;$ the values of $R_1^1$ and $P_{3i}$ are taken
respectively from (\ref{ricci1_3}) and (\ref{ricci3_3}).

By using the equation (\ref{einsteq3c3}) we can define the N--coefficients (%
\ref{ncoef3}), $q_i(x^k,z),$ if the function $h_3(x^k,z)$ and the components
$\Upsilon _{3i}$ of the energy--momentum d--tensor are given. We note that
the equations (\ref{ricci3_3}) are solved for arbitrary functions $%
h_3=h_3(x^k)$ and $q_i=q_i(x^k,z)$ if $\Upsilon _{3i}=0$ and in this case
the component of d--metric $h_3(x^k)$ is not contained in the system of 3D
field equations.

\subsection{Static elliptic horizons}

Let us consider a class of 3D d-metrics which local anisotropy which are
similar to Banados--Teitelboim--Zanelli (BTZ) black holes \cite{04btz}.

The d--metric is parametrized
\begin{equation}
\label{dim3}\delta s^2=g_1\left( \chi ^1,\chi ^2\right) (d\chi ^1)^2+\left(
d\chi ^2\right) ^2-h_3\left( \chi ^1,\chi ^2,t\right) \ \left( \delta
t\right) ^2,
\end{equation}
where $\chi ^1=r/r_h$ for $r_h=const,$ $\chi ^2=\theta /r_a$ if $r_a=\sqrt{%
|\kappa \Upsilon _3^3|}\neq 0$ and $\chi ^2=\theta $ if $\Upsilon _3^3=0,$ $%
y^3=z=t,$ where $t$ is the time like coordinate. The Einstein equations (\ref
{einsteq3a3}) and (\ref{einsteq3c3}) transforms respectively into
\begin{equation}
\label{hbh1a3}\frac{\partial ^2g_1}{\partial (\chi ^2)^2}-\frac
1{2g_1}\left( \frac{\partial g_1}{\partial \chi ^2}\right) ^2-2\kappa
\Upsilon _3^3g_1=0
\end{equation}
and
\begin{equation}
\label{hbh1c3}\left[ \frac 1{h_3}\frac{\partial ^2h_3}{\partial z^2}-\left(
\frac 1{h_3}\frac{\partial h_3}{\partial z}\right) ^2\right] q_i=-\kappa
\Upsilon _{3i}.
\end{equation}
By introducing new variables
\begin{equation}
\label{p-var3}p=g_1^{\prime }/g_1\mbox{ and }s=h_3^{*}/h_3
\end{equation}
where the 'prime' in this subsection denotes the partial derivative $%
\partial /\chi ^2,$ the equations (\ref{hbh1a3}) and (\ref{hbh1c3})
transform into
\begin{equation}
\label{hbh2a3}p^{\prime }+\frac{p^2}2+2\epsilon =0
\end{equation}
and
\begin{equation}
\label{hbh2c3}s^{*}q_i=\kappa \Upsilon _{3i},
\end{equation}
where the vacuum case should be parametrized for $\epsilon =0$ with $\chi
^i=x^i$ and $\epsilon =1(-1)$ for the signature $1(-1)$ of the anisotropic
coordinate.

A class of solutions of 3D Einstein equations for arbitrary $q_i=q_i(\chi
^k,t)$ and $\Upsilon _{3i}=0$ is obtained if $s=s(\chi ^i).$ After
integration of the second equation from (\ref{p-var3}), we find
\begin{equation}
\label{hbh2c3s}h_3(\chi ^k,t)=h_{3(0)}(\chi ^k)\exp \left[ s_{(0)}\left(
\chi ^k\right) t\right]
\end{equation}
as a general solution of the system (\ref{hbh2c3}) with vanishing right
part. Static solutions are stipulated by $q_i=q_i(\chi ^k)$ and $%
s_{(0)}(\chi ^k)=0.$

The integral curve of (\ref{hbh2a3}), intersecting a point $\left( \chi
_{(0)}^2,p_{(0)}\right) ,$ considered as a differential equation on $\chi ^2$
is defined by the functions \cite{04kamke}%
\begin{eqnarray}
p &=&
\frac{p_{(0)}}{1+\frac{p_{(0)}}2
\left( \chi ^2-\chi _{(0)}^2\right) },\qquad   \epsilon =0; \label{eq3a} \\
p & = &
 \frac{p_{(0)}-2\tanh \left( \chi ^2- \chi _{(0)}^2\right) }{1+\frac{p_{(0)}}2
  \tanh \left( \chi ^2-\chi _{(0)}^2\right) },\qquad  %
  \epsilon >0; \label{eq3b}  \\
    p & = & \frac{p_{(0)}-2\tan \left( \chi ^2-\chi _{(0)}^2\right) }
     {1+\frac{p_{(0)}}2\tan \left( \chi ^2     -\chi _{(0)}^2\right)
     },\qquad
 \epsilon <0.   \label{eq3c} %
\end{eqnarray}

Because the function $p$ depends also parametrically on variable $\chi ^1$
we must consider functions $\chi _{(0)}^2=\chi _{(0)}^2\left( \chi ^1\right)
$ and $p_{(0)}=p_{(0)}\left( \chi ^1\right) .$

For simplicity, here we elucidate the case $\epsilon <0.$ The general
formula for the nontrivial component of h--metric is to be obtained after
integration on $\chi ^1$ of (\ref{eq3c}) (see formula (\ref{p-var3}))%
$$
g_1\left( \chi ^1,\chi ^2\right) =g_{1(0)}\left( \chi ^1\right) \left\{ \sin
[\chi ^2-\chi _{(0)}^2\left( \chi ^1\right) ]+\arctan \frac 2{p_{(0)}\left(
\chi ^1\right) }\right\} ^2,
$$
for $p_{(0)}\left( \chi ^1\right) \neq 0,$ and
\begin{equation}
\label{btzlh3}g_1\left( \chi ^1,\chi ^2 \right) =g_{1(0)}\left( \chi
^1\right) \ \cos ^2[\chi ^2-\chi _{(0)}^2\left( \chi ^1\right) ]
\end{equation}
for $p_{(0)}\left( \chi ^1\right) =0,$ where $g_{1(0)}\left( \chi
^1\right),\chi _{(0)}^2\left( \chi ^1\right) $ and $p_{(0)}\left( \chi
^1\right) $ are some functions of necessary smoothness class on variable $%
\chi ^1=x^1/\sqrt{\kappa \varepsilon },$ when $\varepsilon $ is the energy
density. If we consider $\Upsilon _{3i}=0$ and a nontrivial diagonal
components of energy--momentum d--tensor, $\Upsilon _\beta ^\alpha
=diag[0,0,-\varepsilon],$ the N--connection coefficients $q_i(\chi ^i, t)$
could be arbitrary functions.

For simplicity, in our further considerations we shall apply the solution (%
\ref{btzlh3}).

The d--metric (\ref{dim3}) with the coefficients (\ref{btzlh3}) and (\ref
{hbh2c3s}) gives a general description of a class of solutions with generic
local anisotropy of the Einstein equations (\ref{1einsteq2}).

Let us construct static black la--hole solutions for $s_{(0)}\left( \chi
^k\right) =0$ in (\ref{hbh2c3s}).

In order to construct an explicit la--solution we have to chose some
coefficients $h_{3(0)}(\chi ^k),g_{1(0)}\left( \chi ^1\right) $ and $\chi
_0\left( \chi ^1\right) $ from some physical considerations. For instance,
the Schwarzschild solution is selected from a general 4D metric with some
general coefficients of static, spherical symmetry by relating the radial
component of metric with the Newton gravitational potential. In this
section, we construct a locally anisotropic BTZ like solution by supposing
that it is conformally equivalent to the BTZ solution if one neglects
anisotropies on angle $\theta ),$
$$
g_{1(0)}\left( \chi ^1\right) =\left[ r\left( -M_0+\frac{r^2}{l^2}\right)
\right] ^{-2},
$$
where $M_0=const>0$ and $-1/l^2$ is a constant (which is to be considered
the cosmological from the locally isotropic limit. The time--time
coefficient of d--metric is chosen
\begin{equation}
\label{btzlva3}h_3\left( \chi ^1,\chi ^2\right) =r^{-2}\lambda _3\left( \chi
^1,\chi ^2\right) \cos ^2[\chi ^2-\chi _{(0)}^2\left( \chi ^1\right) ].
\end{equation}

If we chose in (\ref{btzlva3})
$$
\lambda _3={(-M_0+\frac{r^2}{l^2})}^2,
$$
when the constant
$$
r_h=\sqrt{M_0}l
$$
defines the radius of a circular horizon, the la--solution is conformally
equivalent, with the factor $r^{-2}\cos ^2[\chi ^2-\chi _{(0)}^2\left( \chi
^1\right) ], $ to the BTZ solution embedded into a anholonomic background
given by arbitrary functions $q_i(\chi ^i,t)$ which are defined by some
initial conditions of gravitational la--background polarization.

A more general class of la--solutions could be generated if we put, for
instance,
$$
\lambda _3\left( \chi ^1,\chi ^2\right) ={(-M}\left( \theta \right) {+\frac{%
r^2}{l^2})}^2,
$$
with
$$
{M}\left( \theta \right) =\frac{M_0}{(1+e\cos \theta )^2},
$$
where $e<1.$ This solution has a horizon, $\lambda _3=0,$ parametrized by an
ellipse
$$
r=\frac{r_h}{1+e\cos \theta }
$$
with parameter $r_h$ and eccentricity $e.$

We note that our solution with elliptic horizon was constructed for a
diagonal energy--momentum d-tensor with nontrivial energy density but
without cosmological constant. On the other hand the BTZ solution was
constructed for a generic 3D cosmological constant. There is not a
contradiction here because the la--solutions can be considered for a
d--tensor $\Upsilon _\beta ^\alpha =diag[p_1-1/l^2,p_2-1/l^2,-\varepsilon
-1/l^2]$ with $p_{1,2}=1/l^2$ and $\varepsilon _{(eff)}=\varepsilon +1/l^2$
(for $\varepsilon =const$ the last expression defines the effective constant
$r_a).$ The locally isotropic limit to the BTZ black hole could be realized
after multiplication on $r^2$ and by approximations $e\simeq 0,$ $\cos
[\theta -\theta _0\left( \chi ^1\right) ]\simeq 1$ and $q_i(x^k,t)\simeq 0.$

\subsection{Oscillating elliptic horizons}

The simplest way to construct 3D solutions of the Einstein equations with
oscillating in time horizon is to consider matter states with constant
nonvanishing values of $\Upsilon _{31}=const.$ In this case the coefficient $%
h_3$ could depend on $t$--variable. For instance, we can chose such initial
values when
\begin{equation}
\label{btzlva3osc1}h_3(\chi ^1,\theta ,t)=r^{-2}\left( -M\left( t\right) +%
\frac{r^2}{l^2}\right) \cos ^2[\theta -\theta _0\left( \chi ^1\right) ]
\end{equation}
with
$$
M=M_0\exp \left( -\widetilde{p}t\right) \sin \widetilde{\omega }t,
$$
or, for an another type of anisotropy,
\begin{equation}
\label{btzlva3osc2}h_3(\chi ^1,\theta ,t)=r^{-2}\left( -M_0+\frac{r^2}{l^2}%
\right) \cos ^2\theta \ \sin ^2[\theta -\theta _0\left( \chi ^1,t\right) ]
\end{equation}
with
$$
\cos \theta _0\left( \chi ^1,t\right) =e^{-1}\left( \frac{r_a}r\cos \omega
_1t-1\right) ,
$$
when the horizon is given parametrically, %
$$
r=\frac{r_a}{1+e\cos \theta }\cos \omega _1t,
$$
where the new constants (comparing with those from the previous subsection)
are fixed by some initial and boundary conditions as to be $\widetilde{p}>0,$
and $\widetilde{\omega }$ and $\omega _1$ are treated as some real numbers.

For a prescribed value of $h_3(\chi ^1,\theta ,t)$ with non--zero source $%
\Upsilon _{31},$ in the equation (\ref{einsteq3c3}), we obtain
\begin{equation}
\label{ncon3osc}q_1(\chi ^1,\theta ,t)=\kappa \Upsilon _{31}\left( \frac{%
\partial ^2}{\partial t^2}\ln |h_3(\chi ^1,\theta ,t)|\right) ^{-1}.
\end{equation}

A solution (\ref{dmetr3}) of the Einstein equations (\ref{einsteq3a3}) and (%
\ref{einsteq3c3}) with $g_2(\chi ^i)=1$ and $g_1(\chi ^1,\theta )$ and $%
h_3(\chi ^1,\theta ,t)$ given respectively by formulas (\ref{btzlh3}) and (%
\ref{btzlva3osc1}) describe a 3D evaporating black la--hole solution with
circular oscillating in time horizon. An another type of solution, with
elliptic oscillating in time horizon, could be obtained if we choose (\ref
{btzlva3osc2}). The non--trivial coefficient of the N--connection must be
computed following the formula (\ref{ncon3osc}).

\section{4D la--solutions}

\setcounter{equation}{0}

\subsection{Basic properties}

The purpose of this section is the construction of d--metrics which are
conformally equivalent to some la--deformations of black hole, torus and
cylinder like solutions in general relativity. We shall analyze 4D d-metrics
of type
\begin{equation}
\label{dmetr4}\delta s^2 = g_1\left( x^k\right) \left( dx^1\right) ^2+
\left(dx^2\right) ^2 + h_3(x^i,z)\left( \delta z\right) ^2 +
h_4(x^i,z)\left( \delta y^4 \right) ^2.
\end{equation}

The Einstein equations (\ref{einsteq3a}) with the Ricci h--tensor (\ref
{ricci1}) and diagonal energy momentum d--tensor transforms into
\begin{equation}
\label{hbh1}\frac{\partial ^2g_1}{\partial (x^2)^2}-\frac 1{2g_1}\left(
\frac{\partial g_1}{\partial x^2}\right) ^2-2\kappa \Upsilon _3^3g_1=0.
\end{equation}
By introducing a dimensionless coordinate, $\chi ^2=x^2/\sqrt{|\kappa
\Upsilon _3^3|},$ and the variable $p=g_1^{\prime }/g_1,$ where by 'prime'
in this section is considered the partial derivative $\partial /\chi ^2,$
the equation (\ref{hbh1}) transforms into
\begin{equation}
\label{hbh2}p^{\prime }+\frac{p^2}2+2\epsilon =0,
\end{equation}
where the vacuum case should be parametrized for $\epsilon =0$ with $\chi
^i=x^i$ and $\epsilon =1(-1).$ The equations (\ref{hbh1}) and (\ref{hbh2})
are, correspondingly, equivalent to the equations (\ref{hbh1a3}) and (\ref
{hbh2a3}) with that difference that in this section we are dealing with 4D
coefficients and values. The solutions for the h--metric are parametrized
like (\ref{eq3a}), (\ref{eq3b}), and (\ref{eq3c}) and the coefficient $%
g_1(\chi ^i)$ is given by a similar to (\ref{btzlh3}) formula (for
simplicity, here we elucidate the case $\epsilon <0)$ which for $%
p_{(0)}\left( \chi ^1\right) =0$ transforms into
\begin{equation}
\label{btzlh4}g_1\left( \chi ^1,\chi ^2\right) =g_{1(0)}\left( \chi
^1\right) \ \cos ^2[\chi ^2-\chi _{(0)}^2\left( \chi ^1\right) ],
\end{equation}
where $g_1\left( \chi ^1\right) ,\chi _{(0)}^2\left( \chi ^1\right) $ and $%
p_{(0)}\left( \chi ^1\right) $ are some functions of necessary smoothness
class on variable $\chi ^1=x^1/\sqrt{\kappa \varepsilon },$ $\varepsilon $
is the energy density. The coefficients $g_1\left( \chi ^1,\chi ^2\right) $ (%
\ref{btzlh4}) and $g_2\left( \chi ^1,\chi ^2\right) =1$ define a h--metric.
The next step is the construction of h--components of d--metrics, $%
h_a=h_a(\chi ^i,z),$ for different classes of symmetries of anisotropies.

The system of equations (\ref{einsteq3b}) with the vertical Ricci d--tensor
component (\ref{1ricci2}) is satisfied by arbitrary functions
\begin{equation}
\label{hdm2var}h_3=a_3(\chi ^i)\mbox{ and }h_4=a_4(\chi ^i).
\end{equation}
For v--metrics depending on three coordinates $(\chi ^i,z)$ the
v--components of the Einstein equations transform into (\ref{heq}) which
reduces to (\ref{heq1}) for prescribed values of $h_3(\chi ^i,z),\,$ and,
inversely, to (\ref{heq2}) if $h_4(\chi ^i,z)$ is prescribed. For h--metrics
being conformally equivalent to (\ref{hdm2var}) (see transforms (\ref{conf4d}%
)) we are dealing to equations of type (\ref{2confeq}) with integral
varieties (\ref{confeqsol}).

\subsection{Rotation Hypersurfaces Horizons}

We proof that there are static black hole and cylindrical like solutions of
the Einstein equations with horizons being some 3D rotation hypersurfaces.
The space components of corresponding d--metrics are conformally equivalent
to some locally anisotropic deformations of the spherical symmetric
Schwarzschild and cylindrical Weyl solutions. We note that for some classes
of solutions the local anisotropy is contained in non--perturbative
anholonomic structures.

\subsubsection{Rotation ellipsoid configuration}

There two types of rotation ellipsoids, elongated and flattened ones. We
examine both cases of such horizon configurations.

\vskip0.2cm

\paragraph{\qquad Elongated rotation ellipsoid coordinates:}

${~}$\\ ${\qquad}$
 An elongated rotation ellipsoid hypersurface is given by the formula
 \cite{04korn}
\begin{equation}
\label{relhor}\frac{\widetilde{x}^2+\widetilde{y}^2}{\sigma ^2-1}+\frac{%
\widetilde{z}^2}{\sigma ^2}=\widetilde{\rho }^2,
\end{equation}
where $\sigma \geq 1$ and $\widetilde{\rho }$ is similar to the radial
coordinate in the spherical symmetric case.

The space 3D coordinate system is defined%
$$
\widetilde{x}=\widetilde{\rho}\sinh u\sin v\cos \varphi ,\ \widetilde{y}=%
\widetilde{\rho}\sinh u\sin v\sin \varphi ,\ \widetilde{z}=\widetilde{\rho}%
\cosh u\cos v,
$$
where $\sigma =\cosh u,(0\leq u<\infty ,\ 0\leq v\leq \pi ,\ 0\leq \varphi
<2\pi ). $\ The hypersurface metric is
\begin{eqnarray}
g_{uu} &=& g_{vv}=\widetilde{\rho}^2\left( \sinh ^2u+\sin ^2v\right) ,
 \label{hsuf1} \\
g_{\varphi \varphi } &=&\widetilde{\rho}^2\sinh ^2u\sin ^2v.
 \nonumber
\end{eqnarray}

Let us introduce a d--metric
\begin{equation}
\label{rel1}\delta s^2 = g_1(u,v)du^2+dv^2 + h_3\left( u,v,\varphi \right)
\left( \delta t\right) ^2+h_4\left( u,v,\varphi \right) \left( \delta
\varphi \right) ^2,
\end{equation}
where $\delta t$ and $\delta \varphi $ are N--elongated differentials.

As a particular solution (\ref{btzlh4}) for the h--metric we choose the
coefficient
\begin{equation}
\label{relh1h}g_1(u,v)=\cos ^2v.
\end{equation}
The $h_3(u,v,\varphi )=h_3(u,v,\widetilde{\rho }
\left( u,v,\varphi \right) )$
is considered as
\begin{equation}
\label{relh1}h_3(u,v,\widetilde{\rho })=\frac 1{\sinh ^2u+\sin ^2v}\frac{%
\left[ 1-\frac{r_g}{4\widetilde{\rho }}\right] ^2}{\left[ 1+\frac{r_g}{4%
\widetilde{\rho }}\right] ^6}.
\end{equation}
In order to define the $h_4$ coefficient solving the Einstein equations, for
simplicity with a diagonal energy--momentum d--tensor for vanishing pressure
we must solve the equation (\ref{heq1}) which transforms into a linear
equation (\ref{heq1a}) if $\Upsilon _1=0.$ In our case $s\left( u,v,\varphi
\right) =\beta ^{-1}\left( u,v,\varphi \right) ,$ where $\beta =\left(
\partial h_4/\partial \varphi \right) /h_4,$ must be a solution of
$$
\frac{\partial s}{\partial \varphi }+\frac{\partial \ln \sqrt{\left|
h_3\right| }}{\partial \varphi }\ s=\frac 12.
$$
After two integrations (see \cite{04kamke}) the general solution for $%
h_4(u,v,\varphi ),$ is
\begin{equation}
\label{relh1a}h_4(u,v,\varphi )=a_4\left( u,v\right) \exp \left[
-\int\limits_0^\varphi F(u,v,z)\ dz\right] ,
\end{equation}
where%
$$
F(u,v,z)=1/\{\sqrt{|h_3(u,v,z)|}[s_{1(0)}\left( u,v\right) +\frac
12\int\limits_{z_0\left( u,v\right) }^z\sqrt{|h_3(u,v,z)|}dz]\},
$$
$s_{1(0)}\left( u,v\right) $ and $z_0\left( u,v\right) $ are some functions
of necessary smooth class. We note that if we put $h_4=a_4(u,v)$ the
equations (\ref{einsteq3b}) are satisfied for every $h_3=h_3(u,v,\varphi ).$

Every d--metric (\ref{rel1}) with coefficients of type (\ref{relh1h}), (\ref
{relh1}) and (\ref{relh1a}) solves the Einstein equations (\ref{einsteq3a}%
)--(\ref{einsteq3d}) with the diagonal momentum d--tensor
$$
\Upsilon _\beta ^\alpha =diag\left[ 0,0,-\varepsilon =-m_0,0\right] ,
$$
when $r_g=2\kappa m_0;$ we set the light constant $c=1.$ If we choose
$$
a_4\left( u,v\right) =\frac{\sinh ^2u\ \sin ^2v}{\sinh ^2u+\sin ^2v}
$$
our solution is conformally equivalent (if not considering the time--time
component) to the hypersurface metric (\ref{hsuf1}). The condition of
vanishing of the coefficient (\ref{relh1}) parametrizes the rotation
ellipsoid for the horizon%
$$
\frac{\widetilde{x}^2+\widetilde{y}^2}{\sigma ^2-1}+\frac{\widetilde{z}^2}{%
\sigma ^2}=\left( \frac{r_g}4\right) ^2,
$$
where the radial coordinate is redefined via relation\ $\widetilde{r}=%
\widetilde{\rho }\left( 1+\frac{r_g}{4\widetilde{\rho }}\right) ^2. $ After
multiplication on the conformal factor
$$
\left( \sinh ^2u+\sin ^2v\right) \left[ 1+\frac{r_g}{4\widetilde{\rho }}%
\right] ^4,
$$
approximating $g_1(u,v)=\cos ^2v\approx 1,$ in the limit of locally
isotropic spherical symmetry,%
$$
\widetilde{x}^2+\widetilde{y}^2+\widetilde{z}^2=r_g^2,
$$
the d--metric (\ref{rel1}) reduces to
$$
ds^2=\left[ 1+\frac{r_g}{4\widetilde{\rho }}\right] ^4\left( d\widetilde{x}%
^2+d\widetilde{y}^2+d\widetilde{z}^2\right) -\frac{\left[ 1-\frac{r_g}{4%
\widetilde{\rho }}\right] ^2}{\left[ 1+\frac{r_g}{4\widetilde{\rho }}\right]
^2}dt^2
$$
which is just the Schwarzschild solution with the redefined radial coordinate
when the space component becomes conformally Euclidean.

So, the d--metric (\ref{rel1}), the coefficients of N--connection being
solutions of (\ref{einsteq3c}) and (\ref{einsteq3d}), describe a static 4D
solution of the Einstein equations when instead of a spherical symmetric
horizon one considers a locally anisotropic deformation to the hypersurface
of rotation elongated ellipsoid.

\vskip0.2cm

\paragraph{\qquad Flattened rotation ellipsoid coordinates}

${~}$
\\ ${\qquad}$ In a similar fashion we can construct a static 4D black hole
solution with the horizon parametrized by a flattened rotation ellipsoid
\cite{04korn},
$$
\frac{\widetilde{x}^2+\widetilde{y}^2}{1+\sigma ^2}+\frac{\widetilde{z}^2}{%
\sigma ^2}=\widetilde{\rho }^2,
$$
where $\sigma \geq 0$ and $\sigma =\sinh u.$

The space 3D special coordinate system is defined%
$$
\widetilde{x}=\widetilde{\rho}\cosh u\sin v\cos \varphi ,\ \widetilde{y}=%
\widetilde{\rho}\cosh u\sin v\sin \varphi ,\ \widetilde{z}=\widetilde{\rho}%
\sinh u\cos v,
$$
where $0\leq u<\infty ,\ 0\leq v\leq \pi ,\ 0\leq \varphi <2\pi .$

The hypersurface metric is
\begin{eqnarray}
g_{uu} &=& g_{vv}=\widetilde{\rho}^2\left( \sinh ^2u+\cos ^2v\right) ,
 \nonumber \\
g_{\varphi \varphi } &=&\widetilde{\rho}^2\sinh ^2u\cos ^2v.
 \nonumber
\end{eqnarray}
In the rest the black hole solution is described by the same formulas as in
the previous subsection but with respect to new canonical coordinates for
flattened rotation ellipsoid.

\subsubsection{Cylindrical, Bipolar and Toroidal Configurations}

We consider a d--metric of type (\ref{dmetr4}). As a coefficient for
h--metric we choose $g_1(\chi ^1,\chi ^2)=\left( \cos \chi ^2\right) ^{2}$
which solves the Einstein equations (\ref{einsteq3a}). The energy momentum
d--tensor is chosen to be diagonal, $\Upsilon _\beta ^\alpha
=diag[0,0,-\varepsilon ,0]$ with $\varepsilon \simeq m_0=\int m_{(lin)}dl,$
where $\varepsilon _{(lin)}$ is the linear 'mass' density. The coefficient $%
h_3\left( \chi ^i,z\right) $ will be chosen in a form similar to (\ref{relh1}%
),%
$$
h_3\simeq \left[ 1-\frac{r_g}{4\widetilde{\rho }}\right] ^2/\left[ 1+\frac{%
r_g}{4\widetilde{\rho }}\right] ^6
$$
for a cylindrical elliptic horizon. We parametrize the second v--component
as $h_4=a_4(\chi ^1,\chi ^2)$ when the equations (\ref{einsteq3b}) are
satisfied for every $h_3=h_3(\chi ^1,\chi ^2,z).$

\vskip0.2cm

\paragraph{\qquad Cylindrical coordinates:}

${~}$
\\ $\qquad$
Let us construct a solution of the Einstein equation with the horizon
having the symmetry of ellipsoidal cylinder given by hypersurface formula
\cite{04korn}
$$
\frac{\widetilde{x}^2}{\sigma ^2}+\frac{\widetilde{y}^2}{\sigma ^2-1}=\rho
_{*}^2,\ \widetilde{z}=\widetilde{z},
$$
where $\sigma \geq 1.$ The 3D radial coordinate $\widetilde{r}$ is to be
computed from $\widetilde{\rho }^2=\rho _{*}^2+\widetilde{z}^2.$

The 3D space coordinate system is defined%
$$
\widetilde{x}=\rho _{*}\cosh u\cos v,\ \widetilde{y}=\rho _{*}\sinh u\sin
v\sin ,\ \widetilde{z}=\widetilde{z},
$$
where $\sigma =\cosh u,\ (0\leq u<\infty ,\ 0\leq v\leq \pi ).$

The hypersurface metric is
\begin{equation}
\label{melcy}g_{uu}=g_{vv}=\rho _{*}^2\left( \sinh ^2u+\sin ^2v\right)
,g_{zz}=1.
\end{equation}

A solution of the Einstein equations with singularity on an ellipse is given
by
\begin{eqnarray}
h_3 &=&
\frac 1{\rho _{*}^2\left( \sinh ^2u+\sin ^2v\right) }\times \frac{\left[
1-\frac{r_g}{4\widetilde{\rho }}\right] ^2}
{\left[ 1+\frac{r_g}{4\widetilde{\rho }}\right] ^6},  \nonumber \\
h_4 &=& a_4=\frac 1{\rho _{*}^2\left( \sinh ^2u+\sin ^2v\right) },
\nonumber
\end{eqnarray}
where $\widetilde{r}=\widetilde{\rho }\left( 1+\frac{r_g}{4\widetilde{\rho }}%
\right) ^2.$ The condition of vanishing of the time--time coefficient $h_3$
parametrizes the hypersurface equation of the horizon%
$$
\frac{\widetilde{x}^2}{\sigma ^2}+\frac{\widetilde{y}^2}{\sigma ^2-1}=\left(
\frac{\rho _{*(g)}}4\right) ^2,\ \widetilde{z}=\widetilde{z},
$$
where $\rho _{*(g)}=2\kappa m_{(lin)}.$

By multiplying the d--metric on the conformal factor
$$
\rho _{*}^2\left( \sinh ^2u+\sin ^2v\right)
\left[ 1+\frac{r_g}{4\widetilde{\rho }}\right] ^4,
$$
where $r_g=\int \rho _{*(g)}dl$ (the integration is taken along the
ellipse), for $\rho _{*}\rightarrow 1,$ in the local isotropic limit, $\sin
v\approx 0, $ the space component transforms into (\ref{melcy}).

\vskip0.2cm

\paragraph{\qquad Bipolar coordinates:}

${~}$\\ ${\qquad}$
 Let us construct 4D solutions of the Einstein equation with the horizon
having the symmetry of the bipolar hypersurface given by the formula \cite
{04korn}%
$$
\left( \sqrt{\widetilde{x}^2+\widetilde{y}^2}-\frac{\widetilde{\rho }}{\tan
\sigma }\ \right) ^2+\widetilde{z}^2=\frac{\widetilde{\rho }^2}{\sin
^2\sigma },
$$
which describes a hypersurface obtained under the rotation of the circles
$$
\left( \widetilde{y}-\frac{\widetilde{\rho }}{\tan \sigma }\right) ^2+%
\widetilde{z}^2=\frac{\widetilde{\rho }^2}{\sin ^2\sigma }
$$
around the axes $Oz$; because $|c\tan \sigma |<|\sin \sigma |^{-1},$ the
circles intersect the axes $Oz.$ The 3D space coordinate system is defined%
\begin{eqnarray}
\widetilde{x} &=&
\frac{\widetilde{\rho}\sin \sigma \cos \varphi }{\cosh \tau -\cos\sigma },
 \qquad
\widetilde{y} =
\frac{\widetilde{\rho}\sin \sigma \sin \varphi }{\cosh\tau -\cos \sigma },
 \nonumber \\
\widetilde{z} & = &\frac{\widetilde{r}\sinh \tau }{\cosh \tau
-\cos \sigma }\
\left( -\infty <\tau <\infty ,0\leq \sigma <\pi ,0\leq \varphi <2\pi \right).
 \nonumber
\end{eqnarray}
The hypersurface metric is
\begin{equation}
\label{mbipcy}g_{\tau \tau }=g_{\sigma \sigma }=\frac{\widetilde{\rho }^2}{%
\left( \cosh \tau -\cos \sigma \right) ^2},g_{\varphi \varphi }=\frac{%
\widetilde{\rho }^2\sin ^2\sigma }
{\left( \cosh \tau -\cos \sigma \right) ^2}.
\end{equation}

A solution of the Einstein equations with singularity on a circle is given
by
$$
h_3=\left[ 1-\frac{r_g}{4\widetilde{\rho }}\right] ^2/\left[ 1+\frac{r_g}{4%
\widetilde{\rho }}\right] ^6\mbox{ and }h_4=a_4=\sin ^2\sigma ,
$$
where $\widetilde{r}=\widetilde{\rho }\left( 1+\frac{r_g}{4\widetilde{\rho }}%
\right) ^2.$ The condition of vanishing of the time--time coefficient $h_3$
parametrizes the hypersurface equation of the horizon%
$$
\left( \sqrt{\widetilde{x}^2+\widetilde{y}^2}-\frac{r_g}2\ c\tan \sigma
\right) ^2+\widetilde{z}^2=\frac{r_g^2}{4\sin ^2\sigma },
$$
where $r_g=\int \rho _{*(g)}dl$ (the integration is taken along the circle),
$\rho _{*(g)}=2\kappa m_{(lin)}.$

By multiplying the d--metric on the conformal factor
\begin{equation}
\label{confbip}\frac 1{\left( \cosh \tau -\cos \sigma \right) ^2}\left[ 1+%
\frac{r_g}{4\widetilde{\rho }}\right] ^4,
\end{equation}
for $\rho _{*}\rightarrow 1,$ in the local isotropic limit, $\sin v\approx
0, $ the space component transforms into (\ref{mbipcy}).

\vskip0.2cm

\paragraph{\qquad Toroidal coordinates:}

${~}$\\ ${\qquad}$
Let us consider solutions of the Einstein equations with toroidal
symmetry of horizons. The hypersurface formula of a torus is \cite{04korn}%
$$
\left( \sqrt{\widetilde{x}^2+\widetilde{y}^2}-\widetilde{\rho }\ c\tanh
\sigma \right) ^2+\widetilde{z}^2=
\frac{\widetilde{\rho }^2}{\sinh ^2\sigma }.
$$
The 3D space coordinate system is defined%
\begin{eqnarray}
\widetilde{x} &=&
\frac{\widetilde{\rho}\sinh \tau \cos \varphi }{\cosh \tau -\cos\sigma },
 \qquad
\widetilde{y} = \frac{\widetilde{\rho}\sin \sigma \sin \varphi }{\cosh
\tau -\cos \sigma }, \nonumber \\
\widetilde{z} &=& \frac{\widetilde{\rho}\sinh \sigma }{\cosh
\tau -\cos \sigma }\
\left( -\pi <\sigma <\pi ,0\leq \tau <\infty ,0\leq \varphi <2\pi \right) .
 \nonumber
\end{eqnarray}
The hypersurface metric is
\begin{equation}
\label{mtor}g_{\sigma \sigma }=g_{\tau \tau }=\frac{\widetilde{\rho }^2}{%
\left( \cosh \tau -\cos \sigma \right) ^2},
g_{\varphi \varphi }=\frac{\widetilde{\rho }^2\sin ^2\sigma }
{\left( \cosh \tau -\cos \sigma \right) ^2}.
\end{equation}

This, another type of solution of the Einstein equations with singularity on
a circle, is given by
$$
h_3=\left[ 1-\frac{r_g}{4\widetilde{\rho }}\right] ^2/\left[ 1+\frac{r_g}{4%
\widetilde{\rho }}\right] ^6\mbox{ and }h_4=a_4=\sinh ^2\sigma ,
$$
where $\widetilde{r}=\widetilde{\rho }\left( 1+\frac{r_g}{4\widetilde{\rho }}%
\right) ^2.$ The condition of vanishing of the time--time coefficient $h_3$
parametrizes the hypersurface equation of the horizon%
$$
\left( \sqrt{\widetilde{x}^2+\widetilde{y}^2}-\frac{r_g}{2\tanh \sigma }%
c\right) ^2+\widetilde{z}^2=\frac{r_g^2}{4\sinh ^2\sigma },
$$
where $r_g=\int \rho _{*(g)}dl$ (the integration is taken along the circle),
$\rho _{*(g)}=2\kappa m_{(lin)}.$

By multiplying the d--metric on the conformal factor (\ref{confbip}), for $%
\rho _{*}\rightarrow 1,$ in the local isotropic limit, $\sin v\approx 0, $
the space component transforms into (\ref{mtor}).

\subsection{A Schwarzschild like la--solution}

The d--metric of type (\ref{rel1}) is taken
\begin{equation}
\label{schla}\delta s^2=g_1(\chi ^1,\theta )d(\chi ^1)^2+d\theta
^2+h_3\left( \chi ^1,\theta ,\varphi \right) \left( \delta t\right)
^2+h_4\left( \chi ^1,\theta ,\varphi \right) \left( \delta \varphi \right)
^2,
\end{equation}
where on the horizontal subspace $\chi ^1=\rho /r_a$ is the dimensionless
radial coordinate (the constant $r_a$ will be defined below), $\chi
^2=\theta $ and in the vertical subspace $y^3=z=t$ and $y^4=\varphi .$ The
energy--momentum d--tensor is taken to be diagonal $\Upsilon _\beta ^\alpha
=diag[0,0,-\varepsilon ,0].$ The coefficient $g_1$ is chosen to be a
solution of type (\ref{btzlh4})%
$$
g_1\left( \chi ^1,\theta \right) =\cos ^2\theta .
$$
For
$$
h_4=\sin ^2\theta \mbox{ and }h_3\left( \rho \right) =-\frac{\left[
1-r_a/4\rho \right] ^2}{\left[ 1+r_a/4\rho \right] ^6},
$$
where $r=\rho \left( 1+\frac{r_g}{4\rho }\right) ^2,r^2=x^2+y^2+z^2,$ $%
r_a\dot =r_g$ is the Schwarzschild gravitational radius, the d--metric (\ref
{schla}) describes a la--solution of the Einstein equations which is
conformally equivalent, with the factor $\rho ^2\left( 1+r_g/4\rho \right)
^2,$ to the Schwarzschild solution (written in coordinates $\left( \rho
,\theta ,\varphi ,t\right) ),$ embedded into a la--background given by
non--trivial values of $q_i(\rho ,\theta ,t)$ and $n_i(\rho ,\theta ,t).$ In
the anisotropic case we can extend the solution for anisotropic (on angle $%
\theta )$ gravitational polarizations of point particles masses, $m=m\left(
\theta \right) ,$ for instance in elliptic form, when
$$
r_a\left( \theta \right) =\frac{r_g}{\left( 1+e\cos \theta \right) }
$$
induces an ellipsoidal dependence on $\theta $ of  the radial coordinate,%
$$
\rho =\frac{r_g}{4\left( 1+e\cos \theta \right) }.
$$
We can also consider arbitrary solutions with $r_a=r_a\left( \theta ,t\right)
$ of oscillation type, $r_a\simeq \sin \left( \omega _1t\right) ,$ or
modelling the mass evaporation, $r_a\simeq \exp [-st],s=const>0.$

So, fixing a physical solution for $h_3(\rho ,\theta ,t),$ for instance,
$$
h_3(\rho ,\theta ,t)=-\frac{\left[ 1-r_a\exp [-st]/4\rho \left( 1+e\cos
\theta \right) \right] ^2}{\left[ 1+r_a\exp [-st]/4\rho \left( 1+e\cos
\theta \right) \right] ^6},
$$
where $e=const<1,$ and computing the values of $q_i(\rho ,\theta ,t)$ and $%
n_i(\rho ,\theta ,t)$ from (\ref{einsteq3c}) and (\ref{einsteq3d}),
corresponding to given $h_3$ and $h_4,$ we obtain a la--generalization of
the Schwarzschild metric.

We note that fixing this type of anisotropy,  in the locally isotropic limit
we obtain not just the Schwarzschild metric but a conformally transformed
one, multiplied on the factor $1/\rho ^2\left( 1+r_g/4\rho \right) ^4.$

\section{Final remarks}

We have presented new classes of three and four dimensional black hole
solutions with local anisotropy which are given both with respect to a
coordinate basis or to an anholonomic frame defined by a N--connection
structure. We proved that for a corresponding ansatz such type of solutions
can be imbedded into the usual (three or four dimensional) Einstein gravity.
It was demonstrated that in general relativity there are admitted static,
but anisotropic (with nonspheric symmetry), and elliptic oscillating in time
black hole like configurations with horizons of events being elliptic (in
three dimensions) and rotation ellipsoidal, elliptic cylinder, toroidal and
another type of closed hypersurfaces or cylinders.

From the results obtained, it appears that the components of metrics with
generic local anisotropy are somehow undetermined from field equations if
the type of symmetry and a correspondence with locally isotropic limits are
not imposed. This is the consequence of the fact that in general relativity
only a part of components of the metric field (six from ten in four
dimensions and three from six in three dimensions) can be treated as
dynamical variables. This is caused by the Bianchi identities which hold on
(pseudo) Riemannian spaces. The rest of components of metric should be
defined from some symmetry prescriptions on the type of locally anisotropic
solutions and corresponding anholonomic frames and, if existing,
compatibility with the locally isotropic limits when some physically
motivated coordinate and/or boundary conditions are enough to state and
solve the Cauchy problem.

Some of the problems discussed so far might be solved by considering
theories containing non--trivial torsion fields like metric--affine and
gauge gravity and for so--called generalized Finsler--Kaluza--Klein models.
More general solutions connected with locally anisotropic low energy limits
in string/M--theory and supergravity could be also generated by applying the
method of computation with respect to anholonomic (super) frames adapted to
a N--connection structure. This topic is currently under study.

\subsection*{Acknowledgments}

The author would like to thank P J Steinhardt, N G Turok and V A Rubakov for
support of his participation at NATO ASI ''Structure Formation in the
Universe'', July 26 -- August 6, 1999 (held at Isaac Newton Institute for
Mathematical Sciences, Cambridge University, UK), where this work was
communicated. He is also very grateful to H Dehnen for hospitality and
useful discussions during his visit at Konstantz University, Germany.

  %%%%%%%%%%%%%%%%%%%%%%%%%%%%%%%%%%%%%%%%%%%%%%%%%%%%%%%%%%%%%%%%%%%%%%%%%%%%%

\chapter[Anholonomic Triads and Black Hole Solutions]
{Anholonomic Triads and New Classes of  (2+1)-Dimensional Black Hole Solutions }

{\bf Abstract}
\footnote{ \copyright\ S. Vacaru, P. Stavrinos and  E. Gaburov,
Anholonomic Triads and New Classes of  (2+1)-Dimensional Black Hole
Solutions, gr--qc/0106068}

We apply the method of moving anholonomic frames in order to construct new
classes of solutions of the Einstein equations on (2+1)--dimensional
pseudo--Riemannian spaces. The anholonomy associated to a class of
off--diagonal metrics results in alternative classes of black hole solutions
which are constructed following distinguished (by nonlinear connection
structure) linear connections and metrics. There are investigated black
holes with deformed horizons and renormalized locally an\-iso\-trop\-ic
constants. We speculate on properties of such anisotropic black holes with
characteristics defined by anholonomic frames and anisotropic interactions
of matter and gravity. The thermodynamics of locally anisotropic black holes
is discussed in connection with a possible statistical mechanics background
based on locally anisotropic variants of Chern--Simons theories.

\section{Introduction}

In recent years there has occurred a substantial interest to the
(2+1)--dimensional gravity and black holes and possible connections of such
objects with string/M--theory. Since the first works of Deser, Jackiv and 't
Hooft \cite{05djh} and Witten \cite{05w} on three dimensional gravity and the
seminal solution for (2+1)--black holes constructed by Ba\~nados,
Teitelboim, and Zanelli (BTZ) \cite{05btz} the gravitational models in three
dimensions have become a very powerful tool for exploring the foundations of
classical and quantum gravity, black hole physics, as well the geometrical
properties of the spaces on which the low--dimensional physics takes place %
\cite{05cm}.

On the other hand, the low--dimensional geometries could be
considered as an arena for elaboration of new methods of solution
of gravitational field equations. One of peculiar features of
general relativity in 2+1 dimensions is that the bulk of physical
solutions of Einstein equations are constructed for a negative
cosmological constant and on a space of constant curvature. There
are not such limitations if anholonomic frames modelling locally
anisotropic (la) interactions of gravity and matter are
considered.

In our recent works \cite{05v3} we emphasized the importance of
definition of frames of reference in general relativity in
connection with new methods of construction of solutions of the
Einstein equations. The former priority given to holonomic frames
holds good for the 'simplest' spherical symmetries and is less
suitable for construction of solutions with 'deformed' symmetries,
for instance, of static black holes with elliptic (or ellipsoidal
and/or torus) configurations of horizons. Such type of 'deformed',
locally anisotropc, solutions of the Einstein equations are easily
to be derived from the ansatz of metrics diagonalized with respect
to some classes of anholonomic frames induced by locally
anisotropic 'elongations' of partial derivatives. After the task
has been solved in anholonomic variables it can be removed with
respect to usual coordinate bases when the metric becomes
off--diagonal and the (for instance, elliptic) symmetry is hidden
in some nonlinear dependencies of the metric components.

The specific goal of the present work is to formulate the
(2+1)--dimensional gravity theory with respect to anholonomic
frames with associated nonlinear connection (N--connection)
structure and to construct and investigate some new classes of
solutions of Einstein equations on locally anisotropic spacetimes
(modelled as usual pseudo--Riemannian spaces provided with an
anholonomic frame structure). A material of interest are the
properties of the locally anisotropic elastic media and rotating
null fluid and anisotropic collapse described by gravitational
field equations with locally anisotropic matter. We investigate
black hole solutions that arise from coupling in a
self--consistent manner the three dimensional (3D)
pseudo--Riemannian geometry and its anholonomic deformations to
the physics of locally anisotropic fluids formulated with respect
to anholonomic frames of reference. For certain special cases the
locally anisotropic matter gives the BTZ black holes with/or not
rotation and electrical charge and variants of their anisotropic
generalizations. For other cases, the resulting solutions are
generic black holes with ''locally anisotropic hair''.

It should be emphasized, that general anholonomic frame transforms
with associated N--connection structure result in deformation of
both metric and linear connection structures. One can be generated
spaces with nontrivial nonmetricity and torsion fields. There are
subclasses of deformations when the condition of metric and
connection compatibility is preserved and the torsion fields are
effectively induced by the anholonomic frame structure. On such
spaces we can work with the torsionless Levi--Civita connection or
(equivalently, but in a more general geometric form) with certain
linear connections with effective torsion. With respect to
anholonomic frames (this can be naturally adapted to the nonlinear
connection structure), the Ricci tensor can be nonsymmetric and
the conservation lows are to be formulated in more sophisticate
form. This is similar to the nonholonomic mechanics with various
types of constraints on dynamics (in modern literature, one uses
two equivalent terms, nonholonomic and/or anholonomic). Such
geometric constructions are largely used in generalized
Lagrange--Hamilton and Finsler--Cartan geometry \cite{05ma}, but for
nonholonomic manifolds (i.e. manifolds provided with nonintegrable
distributions, in the simplest case defining a nonlinear
connection) such generalized geometries can be modelled by
nonholnomic frames and their deformations on (pseudo) Riemannian
spaces, see details in Refs. \cite{05v5}. This work is devoted to a
study of such 3D nonholonomic frame deformations and their
possible physical implications in lower--dimensional gravity.

We note that the anisotropic gravitational field has very unusual
properties. For instance, the vacuum solutions of Einstein
anisotropic gravitational field equations could describe
anisotropic black holes with elliptic symmetry. Some subclasses of
such locally anisotropic spaces are teleparallel (with non--zero
induced torsion but with vanishing curvature tensor) another are
characterized by nontrivial, induced from general relativity on
anholonomic frame bundle, N--connection and Riemannian curvature
and anholonomy induced torsion. In a more general approach the
N--connection and torsion are induced also from the condition that
the  metric and nonlinear connection must solve the Einstein
equations.

The paper is organized as follows: In the next section we briefly review the
locally anisotropic gravity in (2+1)--dimensions. Conformal transforms with
anisotropic factors and corresponding classes of solutions of Einstein
equations with dynamical equations for N--connection coefficients are
examined in Sec. 3. In Sec. 4 we derive the energy--momentum tensors for
locally an\-iso\-tro\-pic elastic media and rotating null fluids. Sec. 5 is
devoted to the local anisotropy of (2+1)--dimensional solutions of Einstein
equations with anisotropic matter. The nonlinear self--polarization of
anisotropic vacuum gravitational fields and matter induced polarizations and
related topics on anisotropic black hole solutions are considered in Sec. 6.
We derive some basic formulas for thermodynamics of anisotropic black holes
in Sec. 7. The next Sec. 8 provides a statistical mechanics background for
locally anisotropic thermodynamics starting from the locally anisotropic
variants of Chern--Simons and Wess--Zumino--Witten models of locally
anisotropic gravity. Finally, in Sec. 9 we conclude and discuss the obtained
results.

\section{Anholonomic Frames and 3D Gravity}

In this Section we wish to briefly review and reformulate the Cartan's
method of moving frames \cite{05cartan1} for investigation of gravitational
and matter field interactions with mixed subsets of holonomic
(unconstrained) and anholonomic (constrained, equivalently, locally
anisotropic, in brief, la) variables \cite{05v3}. Usual tetradic (frame, or
vielbein) approaches to general relativity, see, for instance, \cite{05mtw,05haw}%
, consider 'non--mixed' cases when all basic vectors are anholonomic or
transformed into coordinate (holonomic) ones. We note that a more general
geometric background for locally anisotropic interactions and locally
anisotropic spacetimes, with applications in physics, was elaborated by
Miron and Anastasiei \cite{05ma} in their generalized Finsler and Lagrange
geometry; further developments for locally anisotropic spinor bundles and
locally anisotropic superspaces are contained in Refs \cite{05v1,05v2}. Here we
restrict our constructions only to three dimensional (3D) pseudo--Riemannian
spacetimes provided with a global splitting characterized by two holonomic
and one anholonomic coordinates.

\subsection{Anholonomic frames and nonlinear connections}

We model the low dimensional spacetimes as a smooth (i. e. class $C^\infty )$
3D (pseudo) Riemannian manifolds $V^{(3)}$ being Hausdorff, paracompact and
connected and enabled with the fundamental structures of symmetric metric $%
g_{\alpha \beta },$ with signature $(-,+,+)$ and of linear, in general
nonsymmetric (if we consider anholonomic frames), metric connection $\Gamma
_{~\beta \gamma }^\alpha $ defining the covariant derivation $D_\alpha .$
The indices of geometrical objects on $V^{(3)}$ are stated with respect to a
frame vector field (triad, or dreibien) $e_\alpha $ and its dual $e^\alpha .$
A holonomic frame structure on 3D spacetime could be given by a local
coordinate base
\begin{equation}  \label{2pder}
\partial _\alpha =\partial /\partial u^\alpha ,
\end{equation}
consisting from usual partial derivatives on local coordinates $u=\{u^\alpha
\}$ and the dual basis
\begin{equation}  \label{2pdif}
d^\alpha =du^\alpha ,
\end{equation}
consisting from usual coordinate differentials $du^\alpha .$

An arbitrary holonomic frame $e_\alpha $ could be related to a coordinate
one by a local linear transform $e_\alpha =A_\alpha ^{~\beta }(u) \partial
_\beta ,$ for which the matrix $A_\alpha ^{~\beta }$ is nondegenerate and
there are satisfied the holonomy conditions
\[
e_\alpha e_\beta -e_\beta e_\alpha =0.
\]

Let us consider a 3D metric parametrized into $\left( 2+1\right) $
components
\begin{equation}  \label{3ansatz}
g_{\alpha \beta }=\left[
\begin{array}{cc}
g_{ij}+N_i^{\bullet }N_j^{\bullet }h_{\bullet \bullet } & N_j^{\bullet
}h_{\bullet \bullet } \\
N_i^{\bullet }h_{\bullet \bullet } & h_{\bullet \bullet }%
\end{array}
\right]
\end{equation}
given with respect to a local coordinate basis (\ref{2pdif}), $du^\alpha
=\left( dx^i,dy\right) ,$ where the Greek indices run values $1,2,3,$ the
Latin indices $i,j,k,...$ from the middle of the alphabet run values for $%
n=1,2,...$ and the Latin indices from the beginning of the alphabet, $%
a,b,c,...,$ run values for $m=3,4,....$ if we wont to consider imbedding of
3D spaces into higher dimension ones. The coordinates $x^i$ are treated as
isotropic ones and the coordinate $y^{\bullet }=y$ is considered anholonomic
(anisotropic). For 3D we denote that $a,b,c,...=\bullet ,$ $y^{\bullet
}\rightarrow y,$ $h_{ab}\rightarrow h_{\bullet \bullet }=h$ and $%
N_i^a\rightarrow N_i^{\bullet }=w_i.$ The coefficients $g_{ij}=g_{ij}\left(
u\right) ,h_{\bullet \bullet }=h\left( u\right) $ and $N_i^{\bullet }=N_i(u)$
are supposed to solve the 3D Einstein gravitational field equations. The
metric (\ref{3ansatz}) can be rewritten in a block $(2\times 2)\oplus 1$ form
\begin{equation}  \label{dm}
g_{\alpha \beta }=\left(
\begin{array}{cc}
g_{ij}(u) & 0 \\
0 & h(u)%
\end{array}
\right)
\end{equation}
with respect to the anholonomic basis (frame, anisotropic basis)
\begin{equation}  \label{4dder}
\delta _\alpha = (\delta _i,\partial _{\bullet }) = \frac \delta {\partial
u^\alpha } = \left( \delta _i=\frac \delta {\partial x^i}=\frac \partial
{\partial x^i}-N_i^{\bullet }\left( u\right) \frac \partial {\partial
y},\partial _{\bullet }=\frac \partial {\partial y}\right)
\end{equation}
and its dual anholonomic frame
\begin{equation}  \label{4ddif}
\delta ^\beta = \left( d^i,\delta ^{\bullet }\right) =\delta u^\beta =
\left( d^i=dx^i,\delta ^{\bullet }=\delta y=dy+N_k^{\bullet }\left( u\right)
dx^k\right) .  \nonumber
\end{equation}
where the coefficients $N_j^{\bullet }\left( u\right) $ from (\ref{4dder})
and (\ref{4ddif}) could be treated as the components of an associated
nonlinear connection (N--connection) structure \cite{05barth,05ma,05v1,05v2} which
was considered in Finsler and generalized Lagrange geometries and applied in
general relativity and Kaluza--Klein gravity for construction of new classes
of solutions of Einstein equations by using the method of moving anholonomic
frames \cite{05v3}. On 3D (pseudo)--Riemannian spaces the coefficients $%
N_j^{\bullet}$ define a triad of basis vectors (dreibein) with respect to
which the geometrical objects (tensors, connections and spinors) are
decomposed into holonomic (with indices $i,j,...$) and anholonomic (provided
with $\bullet$--index) components.

A local frame (local basis) structure $\delta _\alpha $ on $V^{(3)}\to
V^{(2+1)}$ (by $(2+1)$ we denote the N--connection splitting into $2$
holonomic and $1$ anholonomic variables in explicit form; this decomposition
differs from the usual two space and one time--like parametrizations) is
characterized by its anholonomy coefficients $w_{~\beta \gamma }^\alpha $
defined from relations
\begin{equation}  \label{anholon}
\delta _\alpha \delta _\beta -\delta _\beta \delta _\alpha =w_{~\alpha \beta
}^\gamma \delta _\gamma .
\end{equation}
The rigorous mathematical definition of N--connection is based on the
formalism of horizontal and vertical subbundles and on exact sequences in
vector bundles \cite{05barth,05ma}. In this work we introduce a N--connection as
a distribution which for every point $u=(x,y)\in V^{(2+1)}$ defines a local
decomposition of the tangent space
\[
T_uV^{(2+1)}=H_uV^{(2)}\oplus V_uV^{(1)}.
\]
into horizontal subspace, $H_uV^{(2)},$ and vertical (an\-iso\-tro\-py)
subspace, $V_uV^{(1)},$ which is given by a set of coefficients $%
N_j^{\bullet }\left( u^\alpha \right) .$ A N--connection is characterized by
its curvature
\begin{equation}  \label{4ncurv}
\Omega _{ij}^{\bullet }=\frac{\partial N_i^{\bullet }}{\partial x^j}-\frac{%
\partial N_j^{\bullet }}{\partial x^i}+N_i^{\bullet }\frac{\partial
N_j^{\bullet }}{\partial y}-N_j^{\bullet }\frac{\partial N_i^{\bullet }}{%
\partial y}.
\end{equation}
The class of usual linear connections can be considered as a particular case
of N--connecti\-ons when
\[
N_j^{\bullet }(x,y)=\Gamma _{\bullet j}^{\bullet }(x)y^{\bullet }.
\]
The elongation (by N--connection) of partial derivatives and differentials
in the adapted to the N--connection operators (\ref{4dder}) and (\ref{4ddif})
reflects the fact that on the (pseudo) Riemannian spacetime $V^{(2+1)}$ it
is modelled a generic local anisotropy characterized by anholonomy relations (%
\ref{anholon}) when the anholonomy coefficients are computed as follows
\begin{eqnarray}
w_{~ij}^k & = & 0,w_{~\bullet j}^k=0,w_{~i\bullet}^k=0, w_{~\bullet
\bullet}^k=0,w_{~\bullet \bullet}^\bullet=0,  \label{2anhol} \\
w_{~ij}^\bullet & = & -\Omega _{ij}^\bullet, w_{~\bullet j}^\bullet =
-\partial _\bullet N_i^\bullet, w_{~i\bullet }^\bullet = \partial _\bullet
N_i^\bullet.  \nonumber
\end{eqnarray}
The frames (\ref{4dder}) and (\ref{4ddif}) are locally adapted to the
N--connection structure, define a local anisotropy and, in brief, are called
anholonomic bases. A N--connection structure distinguishes (d) the
geometrical objects into horizontal and vertical components, i. e. transform
them into d--objects which are briefly called d--tensors, d--metrics and
d--connections. Their components are defined with respect to an anholonomic
basis of type (\ref{4dder}), its dual (\ref{4ddif}), or their tensor products
(d--linear or d--affine transforms of such frames could also be considered).
For instance, a covariant and contravariant d--tensor $Q,$ is expressed
\begin{equation}
Q = Q_{~\beta }^\alpha \delta _\alpha \otimes \delta ^\beta = Q_{~j}^i\delta
_i\otimes d^j+ Q_{~\bullet }^i\delta _i\otimes \delta ^{\bullet} +
Q_{~j}^{\bullet }\partial _{\bullet }\otimes d^j+ Q_{~\bullet }^{\bullet}
\partial _{\bullet }\otimes \delta ^{\bullet }.  \nonumber
\end{equation}
Similar decompositions on holonomic--anholonomic, conventionally on
horizontal (h) and vertical (v) components, hold for connection, torsion and
curvature components adapted to the N--connection structure.

\subsection{Compatible N- and d--connections and metrics}

A linear d--connection $D$ on a locally anisotropic spacetime $V^{(2+1)},$
 \\ $D_{\delta _\gamma }\delta _\beta =\Gamma _{~\beta \gamma }^\alpha \left(
x,y\right) \delta _\alpha , $ is given by its h--v--components,
\begin{equation}
\Gamma _{~\beta \gamma }^\alpha =\left( L_{~jk}^i,L_{~\bullet k}^{\bullet
},C_{~j\bullet }^i,C_{~\bullet \bullet }^{\bullet }\right)  \nonumber
\end{equation}
where
\begin{equation}
D_{\delta _k}\delta _j = L_{~jk}^i\delta _i, D_{\delta _k}\partial _\bullet
= L_{\bullet k}^\bullet \partial _\bullet, D_{\partial _\bullet }\delta _j =
C_{~j\bullet }^i\delta _i, D_{\delta_\bullet }\partial _\bullet = C_{\bullet
\bullet }^\bullet \partial _\bullet.
\end{equation}
A metric on $V^{(2+1)}$ with its coefficients parametrized as (\ref{3ansatz})
can be written in distinguished form (\ref{dm}), as a metric d--tensor (in
brief, d--metric), with respect to an anholonomic base (\ref{4ddif}), i. e.
\begin{equation}  \label{2dmetric}
\delta s^2 = g_{\alpha \beta }\left( u\right) \delta ^\alpha \otimes \delta
^\beta = g_{ij}(x,y)dx^idx^j+h(x,y)(\delta y)^2.
\end{equation}
Some N--connection, d--connection and d--metric structures are compatible if
there are satisfied the conditions
\[
D_\alpha g_{\beta \gamma }=0.
\]
For instance, a canonical compatible d--connection
\[
^c\Gamma _{~\beta \gamma }^\alpha =\left( ^cL_{~jk}^i,^cL_{~\bullet
k}^{\bullet },^cC_{~j\bullet }^i,^cC_{~\bullet \bullet }^{\bullet }\right)
\]
is defined by the coefficients of d--metric (\ref{2dmetric}), $g_{ij}\left(
x,y\right) $ and $h\left( x,y\right) ,$ and by the coefficients of
N--connection,
\begin{eqnarray}
^cL_{~jk}^i & = & \frac 12g^{in}\left( \delta _kg_{nj}+ \delta
_jg_{nk}-\delta _ng_{jk}\right) ,  \nonumber \\
^cL_{~\bullet k}^\bullet & = & \partial _\bullet N_k^\bullet+\frac 12h^{-1}
\left( \delta _kh- 2 h \partial _\bullet N_i^\bullet \right) ,  \nonumber \\
^cC_{~j\bullet }^i & = & \frac 12g^{ik}\partial _\bullet g_{jk},  \nonumber
\\
^cC_{~\bullet \bullet}^\bullet & = & \frac 12h^{-1} \left( \partial _\bullet
h \right).  \label{4dcon}
\end{eqnarray}
The coefficients of the canonical d--connection generalize for locally
anisotropic spacetimes the well known Christoffel symbols. By a local linear
non--degenerate transform to a coordinate frame we obtain the coefficients
of the usual (pseudo) Riemannian metric connection. For a canonical
d--connection (\ref{4dcon}), hereafter we shall omit the left--up index
''c'', the components of canonical torsion,
\begin{eqnarray}
&T\left( \delta _\gamma ,\delta _\beta \right) &= T_{~\beta \gamma }^\alpha
\delta _\alpha ,  \nonumber \\
&T_{~\beta \gamma }^\alpha &= \Gamma _{~\beta \gamma }^\alpha - \Gamma
_{~\gamma \beta }^\alpha +w_{~\beta \gamma }^\alpha  \nonumber
\end{eqnarray}
are expressed via d--torsions
\begin{eqnarray}
T_{.jk}^i & = & T_{jk}^i=L_{jk}^i-L_{kj}^i,\quad
T_{j\bullet}^i=C_{.j\bullet}^i,T_{\bullet j}^i=-C_{j \bullet}^i,  \nonumber
\\
T_{.bc}^a &= &S_{.bc}^a=C_{bc}^a-C_{cb}^a\to S_{.\bullet \bullet}^\bullet
\equiv 0,  \label{2dtorsions} \\
T_{.ij}^\bullet & = & -\Omega _{ij}^\bullet, \quad T_{.\bullet i}^\bullet =
\partial _\bullet N_i^\bullet -L_{.\bullet i}^\bullet ,\quad T_{.i\bullet
}^\bullet = -T_{.\bullet i}^\bullet  \nonumber
\end{eqnarray}
which reflects the anholonomy of the corresponding locally anisotropic frame
of reference on $V^{(2+1)};$ they are induced effectively. With respect to
holonomic frames the d--torsions vanishes. Putting the non--vanishing
coefficients (\ref{4dcon}) into the formula for curvature
\begin{eqnarray}
R\left( \delta _\tau ,\delta _\gamma \right) \delta _\beta &= & R_{\beta
~\gamma\tau }^{~\alpha }\delta _\alpha ,  \nonumber \\
R_{\beta ~\gamma \tau }^{~\alpha } & = & \delta _\tau \Gamma _{~\beta \gamma
}^\alpha -\delta _\gamma \Gamma _{~\beta \delta }^\alpha + \Gamma _{~\beta
\gamma }^\varphi \Gamma _{~\varphi \tau }^\alpha -\Gamma _{~\beta \tau
}^\varphi \Gamma _{~\varphi \gamma }^\alpha + \Gamma _{~\beta \varphi
}^\alpha w_{~\gamma \tau }^\varphi  \nonumber
\end{eqnarray}
we compute the components of canonical d--curvatures
\begin{eqnarray}
R_{h.jk}^{.i} & = & \delta _kL_{.hj}^i-\delta_jL_{.hk}^i +
L_{.hj}^mL_{mk}^i- L_{.hk}^mL_{mj}^i-C_{.h\bullet }^i\Omega _{.jk}^\bullet,
\nonumber \\
R_{\bullet .jk}^{.\bullet} & = & \delta _kL_{.\bullet j}^\bullet
-\delta_jL_{.\bullet k}^\bullet -C_{.\bullet \bullet}^\bullet \Omega
_{.jk}^\bullet,  \label{1dcurvatures} \\
P_{j.k\bullet}^{.i} & = & \delta _k L_{.jk}^i + C_{.j\bullet
}^iT_{.k\bullet}^\bullet - ( \delta
_kC_{.j\bullet}^i+L_{.lk}^iC_{.j\bullet}^l - L_{.jk}^lC_{.l\bullet}^i -
L_{.\bullet k}^\bullet C_{.j\bullet}^i ),  \nonumber \\
P_{\bullet.k\bullet}^{.\bullet} & = & \partial _\bullet L_{.\bullet
k}^\bullet + C_{.\bullet \bullet}^\bullet T_{.k\bullet}^\bullet - ( \delta
_kC_{.\bullet \bullet }^\bullet - L_{.\bullet k}^\bullet C_{.\bullet \bullet
}^\bullet ),  \nonumber \\
S_{j.bc}^{.i} & = & \partial _cC_{.jb}^i-\partial _bC_{.jc}^i +
C_{.jb}^hC_{.hc}^i-C_{.jc}^hC_{hb}^i \to S_{j.\bullet \bullet}^{.i} \equiv 0
,  \nonumber \\
S_{b.cd}^{.a} & = &\partial _dC_{.bc}^a-\partial _cC_{.bd}^a +
C_{.bc}^eC_{.ed}^a-C_{.bd}^eC_{.ec}^a \to S_{\bullet.\bullet
\bullet}^{.\bullet} \equiv 0.  \nonumber
\end{eqnarray}
The h--v--decompositions for the torsion, (\ref{2dtorsions}), and curvature, (%
\ref{1dcurvatures}), are invariant under local coordinate transforms adapted
to a prescribed N--connection structure.

\subsection{Anholonomic constraints and Einstein equations}

The Ricci d--tensor $R_{\beta \gamma }=R_{\beta ~\gamma \alpha }^{~\alpha }$
has the components
\begin{eqnarray}
R_{ij} & = & R_{i.jk}^{.k},\quad R_{i\bullet
}=-^2P_{ia}=-P_{i.k\bullet}^{.k},  \label{4dricci} \\
R_{\bullet i} &= & ^1P_{\bullet i}= P_{\bullet .i\bullet}^{.\bullet},\quad
R_{ab}=S_{a.bc}^{.c}\to S_{\bullet \bullet} \equiv 0  \nonumber
\end{eqnarray}
and, in general, this d--tensor is non symmetric. We can compute the scalar
curvature $\overleftarrow{R}=g^{\beta \gamma }R_{\beta \gamma }$ of a
d-connection $D,$%
\begin{equation}  \label{2dscalar}
{\overleftarrow{R}}=\widehat{R}+S,
\end{equation}
where $\widehat{R}=g^{ij}R_{ij}$ and $S=h^{ab}S_{ab}\equiv 0$ for one
dimensional anisotropies. By introducing the values (\ref{4dricci}) and (\ref%
{2dscalar}) into the usual Einstein equations
\begin{equation}  \label{2einst1}
G_{\alpha \beta }+\Lambda g_{\alpha \beta }=k\Upsilon _{\beta \gamma },
\end{equation}
where
\begin{equation}  \label{einstdt}
G_{\alpha \beta }=R_{\beta \gamma }-\frac 12g_{\beta \gamma }R
\end{equation}
is the Einstein tensor, written with respect to an anholonomic frame of
reference, we obtain the system of field equations for locally anisotropic
gravity with N--connection structure \cite{05ma}:
\begin{eqnarray}
R_{ij}-\frac 12\left( \widehat{R} - 2\Lambda \right) g_{ij} & = & k\Upsilon
_{ij},  \label{einsteq2a} \\
-\frac 12\left( \widehat{R}-2\Lambda\right) h_{\bullet \bullet} & = &
k\Upsilon _{\bullet \bullet},  \label{einsteq2b} \\
^1P_{\bullet i} & = & k\Upsilon _{\bullet i},  \label{einsteq2c} \\
^2P_{i\bullet} & = & -k\Upsilon _{i\bullet},  \label{einsteq2d}
\end{eqnarray}
where $\Upsilon _{ij},\Upsilon _{\bullet \bullet },\Upsilon _{\bullet i}$
and $\Upsilon _{i\bullet }$ are the components of the energy--momentum
d--tensor field $\Upsilon _{\beta \gamma }$ which includes the cosmological
constant terms and possible contributions of d--torsions and matter, and $k$
is the coupling constant.

The bulk of nontrivial locally isotropic solutions in 3D gravity were
constructed by considering a cosmological constant $\Lambda =-1/l^2,$ with
and equivalent vacuum energy--momentum $\Upsilon _{\beta \gamma }^{(\Lambda
)}=-\Lambda g_{\beta \gamma }.$

\subsection{Some ansatz for d--metrics}

\subsubsection{Diagonal d-metrics}

Let us introduce on 3D locally anisotropic spacetime $V^{(2+1)}$ the local
coordinates\newline
$(x^1,x^2,y),$ where $y$ is considered as the anisotropy coordinate, and
parametrize the d--metric (\ref{2dmetric}) in the form
\begin{equation}  \label{dm2}
\delta s^2=a\left( x^i\right) \left( dx^1\right) ^2+b\left( x^i\right)
(dx^2)^2+h\left( x^i,y\right) (\delta y)^2,
\end{equation}
where
\[
\delta y=dy + w_1(x^i,y)dx^1+w_2(x^i,y)dx^2,
\]
i. e. $N_i^\bullet =w_i(x^i,y).$

With respect to the coordinate base (\ref{2pder}) the d--metric (\ref{2dmetric}%
) transforms into the ansatz
\begin{equation}  \label{m2}
g_{\alpha \beta }=\left[
\begin{array}{ccc}
a+w_1^{\ 2}h & w_1w_2h & w_1h \\
w_1w_2h & b+w_2^{\ 2}h & w_2h \\
w_1h & w_2h & h%
\end{array}
\right] .
\end{equation}

The nontrivial components of the Ricci d--tensor (\ref{4dricci}) are computed
\begin{eqnarray}  \label{1ricci1d}
2abR_1^1 &=& 2abR_2^2 -\ddot b+\frac 1{2b}\dot b^2+\frac 1{2a}\dot a\dot
b+\frac 1{2b}a^{\prime }b^{\prime }-a^{\prime \prime }+\frac 1{2a}(a^{\prime
})^2  \nonumber
\end{eqnarray}
where the partial derivatives are denoted, for instance, $\dot h=\partial
h/\partial x^1,h^{\prime }=\partial h/\partial x^2$ and $h^{*}=\partial
h/\partial y.$ The scalar curvature is $R=2R_1^1.$

The Einstein d--tensor has a nontrivial component
\[
G_3^3=-hR_1^1.
\]

In the vacuum case with $\Lambda =0,$ the Einstein equ\-a\-tions (\ref%
{einsteq2a})--(\ref{einsteq2d}) are satisfied by arbitrary functions $%
a\left(x^i\right) ,b\left( x^i\right) $ solving the equation
\begin{equation}  \label{vaceq}
-\ddot b+\frac 1{2b}\dot b^2+\frac 1{2a}\dot a\dot b+\frac 1{2b}a^{\prime
}b^{\prime }-a^{\prime \prime }+\frac 1{2a}(a^{\prime })^2=0
\end{equation}
and arbitrary function $h\left( x^i,y\right).$ Such functions should be
defined following some boundary conditions in a manner as to have
compatibility with the locally isotropic limit.

\subsubsection{Off--diagonal d--metrics}

For our further investigations it is convenient to consider d--metrics of
type
\begin{equation}  \label{dm2a}
\delta s^2=g\left( x^i\right) \left( dx^1\right) ^2+2dx^1dx^2+h\left(
x^i,y\right) (\delta y)^2.
\end{equation}
The nontrivial components of the Ricci d--tensor are
\begin{equation}  \label{ricci2d}
R_{11} =\frac 12g\frac{\partial ^2g}{\partial (x^2)^2},\qquad R_{12}
=R_{21}=\frac 12\frac{\partial ^2g}{\partial (x^2)^2},
\end{equation}
when the scalar curvature is $R=2R_{12}$ and the nontrivial component of the
Einstein d--tensor is
\[
G_{33}=-\frac h2\frac{\partial ^2g}{\partial (x^2)^2}.
\]

We note that for the both d--metric ansatz (\ref{dm2}) and (\ref{dm2a}) and
corresponding coefficients of Ricci d--tensor, (\ref{1ricci1d}) and (\ref%
{ricci2d}), the h--components of the Einstein d--tensor vanishes
for arbitrary values of metric coefficients, i. e. $G_{ij}=0.$ In
absence of matter such ansatz admit arbitrary nontrivial
anholonomy (N--connection and N--curvature) coefficients
(\ref{2anhol}) because the values $w_{i}$ are not contained in the
3D vacuum Einstein equations. The h--component of the
d--metric, $h(x^{k},y),$ and the coefficients of d--connection, $%
w_{i}(x^{k},y),$ are to be defined by some boundary conditions (for
instance, by a compatibility with the locally isotropic limit) and
compatibility conditions between nontrivial values of the cosmological
constant and energy--momentum d--tensor.

\section{Conformal Transforms with Anisotropic Factors}

One of pecular proprieties of the d--metric ansatz (\ref{dm2}) and (\ref%
{dm2a}) is that there is only one non--trivial component of the
Einstein d--tensor, $G_{33}.$ Because the values $P_{3i}$ and
$P_{i3}$ for the equations (\ref{einsteq2b}) and (\ref{einsteq2c})
vanish identically the coefficients of N--connection, $w_i,$ are
not contained in the Einstein equations and could take arbitrary
values. For static anisotropic configurations the solutions
constructed in Sections IV and V can be considered as 3D black
hole like objects embedded in a locally anisotropic background
with prescribed anholonomic frame (N--connection) structure.

In this Section we shall proof that there are d--metrics for which the
Einstein equations reduce to some dynamical equations for the N--connection
coefficients.

\subsection{Conformal transforms of d--metrics}

A conformal transform of a d--metric
\begin{equation}  \label{conft}
\left( g_{ij},h_{ab}\right) \longrightarrow \left( \widetilde{g}_{ij}=\Omega
^2\left( x^i,y\right) g_{ij},\widetilde{h}_{ab}= \Omega ^2\left(x^i,y\right)
h_{ab}\right)
\end{equation}
with fixed N--connection structure, $\widetilde{N}_i^a=N_i^a,$ deforms the
coefficients of canonical d--connection,
\[
\widetilde{\Gamma }_{\ \beta \gamma }^\alpha =\Gamma _{\ \beta \gamma
}^\alpha +\widehat{\Gamma }_{\ \beta \gamma }^\alpha ,
\]
where the coefficients of deformation d--tensor $\widehat{\Gamma }_{\ \beta
\gamma }^\alpha =\{\widehat{L}_{jk}^i,\widehat{L}_{bk}^a,\widehat{C}_{jc}^i,%
\widehat{C}_{bc}^a\}$ are computed by introducing the values (\ref{conft})
into (\ref{4dcon}),
\begin{eqnarray}  \label{1defcon}
\widehat{L}_{jk}^i &=& \delta _j^i\psi _k+\delta _k^i\psi
_j-g_{jk}g^{in}\psi _n, \qquad \widehat{L}_{bk}^a = \delta _b^a\psi _k, \\
\widehat{C}_{jc}^i &=& \delta _j^i\psi _c,\ \widehat{C}_{bc}^a = \delta
_b^a\psi _c+\delta _c^a\psi _b-h_{bc}h^{ae}\psi _e  \nonumber
\end{eqnarray}
with $\delta _j^i$ and $\delta _b^a$ being corresponding Kronecker symbols
in h-- and v--subspaces and
\[
\psi _i=\delta _i\ln \Omega \mbox{ and }\psi _a=\partial _a\ln \Omega .
\]
In this subsection we present the general formulas for a $n$--dimensional $h$%
--subspace, with indices $i,j,k...=1,2,...n,$ and $m$--dimensional $v$%
--subspace, with indices $a,b,c,...=1,2,...m.$

The d--connection deformations (\ref{1defcon}) induce conformal deformations
of the Ricci d--tensor (\ref{4dricci}),
\begin{eqnarray}
\widetilde{R}_{hj} &=& R_{hj}+\widehat{R}_{[1]hj}+\widehat{R}_{[2]hj},\
\widetilde{R}_{ja}= R_{ja}+\widehat{R}_{ja},  \nonumber \\
\widetilde{R}_{bk} &= & R_{bk}+ \widehat{R}_{bk},\ \widetilde{S}_{bc} =
S_{bc}+\widehat{S}_{bc},  \nonumber
\end{eqnarray}
where the deformation Ricci d--tensors are
\begin{eqnarray}
\widehat{R}_{[1]hj} &=& \partial _i\widehat{L}_{hj}^i-\partial _j\widehat{L}%
_h+\widehat{L}_{\ hj}^mL_m+L_{\ hj}^m\widehat{L}_m+ \widehat{L}_{\ hj}^m%
\widehat{L}_m -\widehat{L}_{\ hi}^mL_{mj}^i-L_{\ hi}^m\widehat{L}_{mj}^i-
\widehat{L}_{\ hi}^m\widehat{L}_{mj}^i,  \nonumber \\
\widehat{R}_{[2]hj} &=& N_i^a\partial _a\widehat{L}_{hj}^i-N_j^a\partial _a%
\widehat{L}_h+\widehat{C}_{ha}^iR_{\ ji}^a;  \label{driccid} \\
\widehat{R}_{ja} &=& -\partial _a\widehat{L}_j+\delta _i\widehat{C}%
_{ja}^i+L_{ki}^i\widehat{C}_{ja}^k-L_{ji}^k\widehat{C}_{ka}^i-L_{ai}^b%
\widehat{C}_{jb}^i -\widehat{C}_{\ jb}^iP_{\ ia}^b-C_{\ jb}^i\widehat{P}_{\
ia}^b- \widehat{C}_{\ jb}^i\widehat{P}_{\ ia}^b,  \nonumber \\
\widehat{R}_{bk} &=& \partial _a\widehat{L}_{\ bk}^a-\delta _k\widehat{C}_b+
L_{\ bk}^a\widehat{C}_a +\widehat{C}_{\ bd}^aP_{\ ka}^d + C_{\ bd}^a\widehat{%
P}_{\ ka}^d+ \widehat{C}_{\ bd}^a\widehat{P}_{\ ka}^d,  \nonumber \\
\widehat{S}_{bc} &=& \partial _a\widehat{C}_{\ bc}^a-\partial _c\widehat{C}%
_b+ \widehat{C}_{\ bc}^eC_e+C_{\ bc}^e\widehat{C}_e+ \widehat{C}_{\ bc}^e%
\widehat{C}_e - \widehat{C}_{\ ba}^eC_{\ ec}^a - C_{\ ba}^e\widehat{C}_{\
ec}^a- \widehat{C}_{\ ba}^e\widehat{C}_{\ ec}^a,  \nonumber
\end{eqnarray}
when $\widehat{L}_h=\widehat{L}_{hi}^i$ and $\widehat{C}_b=\widehat{C}_{\
be}^e.$

\subsection{An ansatz with adapted conformal factor and N--connec\-ti\-on}

We consider a 3D metric
\begin{equation}  \label{3ansatzc}
g_{\alpha \beta }=\left[
\begin{array}{ccc}
\Omega ^2(a-w_1^{\ 2}h) & -w_1w_2h\Omega ^2 & -w_1h\Omega ^2 \\
-w_1w_2h\Omega ^2 & \Omega ^2(b-w_2^{\ 2}h) & -w_2h\Omega ^2 \\
-w_1h\Omega ^2 & -w_2h\Omega ^2 & -h\Omega ^2%
\end{array}
\right]
\end{equation}
where $a=a(x^i),b=b\left( x^i\right) ,w_i=w_i(x^k,y),\Omega =\Omega \left(
x^k,y\right) \geq 0$ and $h=h\left( x^k,y\right) $ when the conditions
\[
\psi _i=\delta _i\ln \Omega =\frac \partial {\partial x^i}\ln \Omega -w_i\ln
\Omega =0
\]
are satisfied. With respect to anholonomic bases (\ref{4ddif}) the (\ref%
{3ansatzc}) transforms into the d--metric
\begin{equation}  \label{dansatzc}
\delta s^2 = \Omega ^2(x^k,y)[a(x^k)(dx^1)^2+b(x^k)(dx^1)^2 +h\left(
x^k,y\right) (\delta y)^2].
\end{equation}

By straightforward calculus, by applying consequently the formulas (\ref%
{1dcurvatures})--(\ref{einsteq2d}) we find that there is a non--trivial
coefficient of the Ricci d--tensor (\ref{4dricci}), of the deformation
d--tensor (\ref{driccid}),
\[
\widehat{R}_{j3}=\psi _3\cdot \delta _j\ln \sqrt{|h|},
\]
which results in non--trivial components of the Einstein d-tensor (\ref%
{einstdt}),
\[
G_3^3=-hR_1^1 \mbox{ and } P_{\bullet i}=-\psi _3\cdot \delta _j\ln \sqrt{|h|%
},
\]
where $R_1^1$ is given by the formula (\ref{1ricci1d}).

We can select a class of solutions of 3D Einstein equations with $P_{\bullet
j}=0$ but with the horizontal components of metric depending on anisotropic
coordinate $y,$ via conformal factor $\Omega (x^{k},y),$ and dynamical
components of the N--connection, $w_{i},$ if we choose
\[
h(x^{k},y)=\pm \Omega ^{2}(x^{k},y)
\]
and state
\begin{equation}
w_{i}(x^{k},y)=\partial _{i}\ln |\ln \Omega |.  \label{conformnc}
\end{equation}
Finally, we not that for the ansatz (\ref{3ansatzc}) (equivalently (\ref%
{dansatzc})) the coefficients of N--connection have to be found as dynamical
values by solving the Einstein equations.

\section{Matter Energy Momentum D--Tensors}

\subsection{Variational definition of energy-momentum d--tensors}

For locally isotropic spacetimes the symmetric energy momentum tensor is to
be computed by varying on the metric (see, for instance, Refs. \cite{05haw,05mtw}%
) the matter action
\[
S=\frac 1c\int {\cal L}\sqrt{|g|}dV,
\]
where ${\cal L}$ is the Lagrangian of matter fields, $c$ is the light
velocity and $dV$ is the infinitesimal volume, with respect to the inverse
metric $g^{\alpha \beta }$. By definition one states that the value
\begin{equation}  \label{emtens}
\frac 12\sqrt{|g|}T_{\alpha \beta }=\frac{\partial (\sqrt{|g|}{\cal L})}{%
\partial g^{\alpha \beta }}-\frac \partial {\partial u^\tau }\frac{\partial (%
\sqrt{|g|}{\cal L})}{\partial g^{\alpha \beta }/\partial u^\tau }
\end{equation}
is the symmetric energy--momentum tensor of matter fields. With respect to
anholonomic frames (\ref{4dder}) and (\ref{4ddif}) there are imposed
constraints of type
\[
g_{ib}-N_i^{\bullet }h=0
\]
in order to obtain the block representation for d--metric (\ref{dm}). Such
constraints, as well the substitution of partial derivatives into
N--elongated, could result in nonsymmetric energy--momentum d--tensors $%
\Upsilon _{\alpha \beta }$ which is compatible with the fact that on a
locally anisotropic spacetime the Ricci d--tensor could be nonsymmetric.

The gravitational--matter field interactions on locally anisotropic
spa\-ce\-ti\-mes are described by dynamical models with imposed constraints
(a generalization of anholonomic analytic mechanics for gravitational field
theory). The physics of systems with mixed holonomic and anholonomic
variables states additional tasks connected with the definition of
conservation laws, interpretation of non--symmetric energy--momentum tensors
$\Upsilon _{\alpha \beta}$ on locally anisotropic spacetimes and relation of
such values with, for instance, the non--symmetric Ricci d--tensor. In this
work we adopt the convention that for locally anisotropic gravitational
matter field interactions the non--symmetric Ricci d--tensor induces a
non--symmetric Einstein d--tensor which has as a source a corresponding
non--symmetric matter energy--momentum tensor. The values $\Upsilon _{\alpha
\beta}$ should be computed by a variational calculus on locally anisotropic
spacetime as well by imposing some constraints following the symmetry of
anisotropic interactions and boundary conditions.

In the next subsection we shall investigate in explicit form some cases of
definition of energy momentum tensor for locally anisotropic matter on
locally anisotropic spacetime.

\subsection{Energy--Momentum D--Tensors for Anisotropic Media}

Following DeWitt approach \cite{05dw} and recent results on dynamical collapse
and hair of black holes of Husain and Brown \cite{05hus}, we set up a
formalism for deriving energy--momentum d--tensors for locally anisotropic
matter.

Our basic idea for introducing a local anisotropy of matter is to rewrite
the energy--momentum tensors with respect to locally adapted frames and to
change the usual partial derivations and differentials into corresponding
operators (\ref{4dder}) and (\ref{4ddif}), ''elongated'' by N--connection. The
energy conditions (weak, dominant, or strong) in a locally anisotropic
background have to be analyzed with respect to a locally anisotropic basis.

We start with DeWitt's action written in locally anisotropic spacetime,
\[
S\left[ g_{\alpha \beta },z^{\underline{i}}\right] =-\int\limits_V\delta ^3u%
\sqrt{-g}\rho \left( z^{\underline{i}},q_{\underline{j}\underline{k}}\right)
,
\]
as a functional on region $V,$ of the locally anisotropic metric $g_{\alpha
\beta }$ and the Lagrangian coordinates $z^{\underline{i}}=z^{\underline{i}%
}\left( u^\alpha \right) $ (we use underlined indices $\underline{i},%
\underline{j},...=1,2$ in order to point out that the 2--dimensional matter
space could be different from the locally anisotropic spacetime). The
functions $z^{\underline{i}}=z^{\underline{i}}\left( u^\alpha \right) $ are
two scalar locally anisotropic fields whose locally anisotropic gradients
(with partial derivations substituted by operators (\ref{2pder})) are
orthogonal to the matter world lines and label which particle passes through
the point $u^\alpha .$ The action $S\left[ g_{\alpha \beta },z^{\underline{i}%
}\right] $ is the proper volume integral of the proper energy density $\rho $
in the rest anholonomic frame of matter. The locally anisotropic density $%
\rho \left( z^{\underline{i}},q_{\underline{j}\underline{k}}\right) $
depends explicitly on $z^{\underline{i}}$ and on matter space d--metric $q^{%
\underline{i}\underline{j}}=\left( \delta _\alpha z^{\underline{i}}\right)
g^{\alpha \beta }\left( \delta _\beta z^{\underline{j}}\right) ,$ which is
interpreted as the inverse d--metric in the rest anholonomic frame of the
matter.

Using the d--metric $q^{\underline{i}\underline{j}}$ and locally anisotropic
fluid velocity $V^\alpha ,$ defined as the future pointing unit d--vector
orthogonal to d--gradients $\delta _\alpha z^{\underline{i}},$ the locally
anisotropic spacetime d--metric (\ref{2dmetric}) of signature (--,+,+) may be
written in the form
\[
g_{\alpha \beta }=-V_\alpha V_\beta +q_{\underline{j}\underline{k}}\delta
_\alpha z^{\underline{j}}\delta _\beta z^{\underline{k}}
\]
which allow us to define the energy--momentum d--tensor for elastic locally
anisotropic medium as
\begin{equation}  \label{emem}
\Upsilon _{\beta \gamma } \equiv -\frac 2{\sqrt{-g}}\frac{\delta S}{\delta
g^{\beta \gamma }} \rho V_\beta V_\gamma + t_{\underline{j}\underline{k}%
}\delta _\beta z^{\underline{j}} \delta _\gamma z^{\underline{k}} ,
\nonumber
\end{equation}
where the locally anisotropic matter stress d--tensor $t_{\underline{j}%
\underline{k}}$ is expressed as
\begin{eqnarray}
t_{\underline{j}\underline{k}}= 2\frac{\delta \rho } {\partial q^{\underline{%
j}\underline{k}}}- \rho q_{\underline{j}\underline{k}}= \frac 2{\sqrt{q}}
\frac{\delta \left( \sqrt{q}\rho \right) } {\partial q^{\underline{j}%
\underline{k}}}.  \label{dem1}
\end{eqnarray}

Here one should be noted that on locally anisotropic spaces
\[
D_\alpha \Upsilon ^{\alpha \beta }=D_\alpha \left( R^{\alpha \beta }-\frac
12g^{\alpha \beta }R\right) =J^\beta \neq 0
\]
and this expression must be treated as a generalized type of conservation
law with a geometric source $J^\beta $ for the divergence of locally
anisotropic matter d--tensor \cite{05ma}.

The stress--energy--momentum d--tensor for locally an\-iso\-trop\-ic elastic
medium is defined by applying N--elongated operators $\delta _\alpha $ of
partial derivatives (\ref{2pder}),
\begin{equation}
T_{\alpha \beta } = -\frac 2{\sqrt{-g}}\frac{\delta S}{\delta g^{\alpha
\beta }} = -\rho g_{\alpha \beta }+ 2\frac{\partial \rho }{\partial q^{%
\underline{i} \underline{j}}}\delta _\alpha z^{\underline{j}} \delta _\beta
z^{\underline{k}} = -V_\alpha V_\beta +\tau _{^{\underline{i} \underline{j}%
}}\delta _\alpha z^{\underline{i}} \delta _\beta z^{\underline{j}},
\nonumber
\end{equation}
where we introduce the matter stress d--tensor
\[
\tau _{^{\underline{i}\underline{j}}}=2\frac{\partial \rho }{\partial q^{%
\underline{i}\underline{j}}}-\rho q_{\underline{i}\underline{j}}=\frac 2{%
\sqrt{q}}\frac{\partial \left( \sqrt{q}\rho \right) }{\partial q^{\underline{%
i}\underline{j}}}.
\]
The obtained formulas generalize for spaces with nontrivial N--connection
structures the results on isotropic and anisotropic media on locally
isotropic spacetimes.

\subsection{Isotropic and anisotropic media}

The {\it isotropic elastic}, but in general locally anisotropic medium is
introduced as one having equal all principal pressures with stress d--tensor
being for a perfect fluid and the density $\rho =\rho \left( n\right) ,$
where the proper density (the number of particles per unit proper volume in
the material rest anholonomic frame) is $n=\underline{n}\left( z^{\underline{%
i}}\right) /\sqrt{q};$ the value $\underline{n}\left( z^{\underline{i}%
}\right) $ is the number of particles per unit coordinate cell $\delta ^3z.$
With respect to a locally anisotropic frame, using the identity
\[
\frac{\partial \rho \left( n\right) }{\partial q^{\underline{j}\underline{k}}%
}=\frac n2\frac{\partial \rho }{\partial n}q_{\underline{j}\underline{k}}
\]
in (\ref{dem1}), the energy--momentum d--tensor (\ref{emem}) %(3.1)
for a isotropic elastic locally anisotropic medium becomes
\[
\Upsilon _{\beta \gamma }=\rho V_\beta V_\gamma +\left( n\frac{\partial \rho
}{\partial n}-\rho \right) \left( g_{\beta \gamma }+V_\beta V_\gamma \right)
.
\]
This medium looks like isotropic with respect to anholonomic frames but, in
general, it is locally anisotropic.

The {\it anisotropic elastic} and locally anisotropic medium has not equal
principal pressures. In this case we have to introduce (1+1) decompositions
of locally anisotropic matter d--tensor $q_{\underline{j}\underline{k}}$%
\[
q_{\underline{j}\underline{k}}=\left(
\begin{array}{cc}
\alpha ^2+\beta ^2 & \beta \\
\beta & \sigma%
\end{array}
\right) ,
\]
and consider densities $\rho \left( n_{\underline{1}},n_{\underline{2}%
}\right) ,$ where $n_{\underline{1}}$ and $n_{\underline{2}}$ are
respectively the particle numbers per unit length in the directions given by
bi--vectors $v_{\underline{j}}^{\underline{1}}$ and $v_{\underline{j}}^{%
\underline{2}}.$ Substituting
\[
\frac{\partial \rho \left( n_{\underline{1}},n_{\underline{2}}\right) }{%
\partial h^{\underline{j}\underline{k}}}=\frac{n_{\underline{1}}}2\frac{%
\partial \rho }{\partial n_{\underline{1}}}v_{\underline{j}}^{\underline{1}%
}v_{\underline{k}}^{\underline{1}}+\frac{n_{\underline{2}}}2\frac{\partial
\rho }{\partial n_{\underline{2}}}v_{\underline{j}}^{\underline{2}}v_{%
\underline{k}}^{\underline{2}}
\]
into (\ref{dem1}), %(3.2),
which gives
\[
t_{\underline{j}\underline{k}}=\left( n_{\underline{1}}\frac{\partial \rho }{%
\partial n_{\underline{1}}}-\rho \right) v_{\underline{j}}^{\underline{1}}v_{%
\underline{k}}^{\underline{1}}+\left( n_{\underline{2}}\frac{\partial \rho }{%
\partial n_{\underline{2}}}-\rho \right) v_{\underline{j}}^{\underline{2}}v_{%
\underline{k}}^{\underline{2}},
\]
we obtain from (\ref{emem}) %(3.1)
the energy--momentum d--tensor for the anisotropic locally aniso\-tro\-pic
matter
\begin{equation}
\Upsilon _{\beta \gamma } = \rho V_\beta V_\gamma + \left( n_{\underline{1}}%
\frac{\partial \rho }{\partial n_{\underline{1}}}- \rho \right) v_{%
\underline{j}}^{\underline{1}} v_{\underline{k}}^{\underline{1}}+\left( n_{%
\underline{2}}\frac{\partial \rho }{\partial n_{\underline{2}}}- \rho
\right) v_{\underline{j}}^{\underline{2}} v_{\underline{k}}^{\underline{2}}.
\nonumber
\end{equation}
So, the pressure $P_1=\left( n_{\underline{1}}\frac{\partial \rho }{\partial
n_{\underline{1}}}-\rho \right) $ in the direction $v_{\underline{j}}^{%
\underline{1}}$ differs from the pressure\newline
$P_2=\left( n_{\underline{2}}\frac{\partial \rho }{\partial n_{\underline{2}}%
}-\rho \right) $ in the direction $v_{\underline{j}}^{\underline{2}}.$ For
instance, if for the (2+1)--dimensional locally anisotropic spacetime we
impose the conditions $\Upsilon _1^1=\Upsilon _2^2\neq \Upsilon _3^3,$ when
\[
\rho =\rho \left( n_{\underline{1}}\right) ,z^{\underline{1}}\left( u^\alpha
\right) =r,z^{\underline{2}}\left( u^\alpha \right) =\theta ,
\]
$r$ and $\theta $ are correspondingly radial and angle coordinates on
locally anisotropic spacetime, we have
\begin{equation}
\Upsilon _1^1=\Upsilon _2^2=\rho ,\Upsilon _3^3=\left( n_{\underline{1}}%
\frac{\partial \rho }{\partial n_{\underline{1}}}-\rho \right) .
\end{equation}

We shall also consider the variant when the coordinated $\theta $ is
anisotropic $(t$ and $r$ being isotropic). In this case we shall impose the
conditions $\Upsilon _1^1\neq \Upsilon _2^2=\Upsilon _3^3$ for
\[
\rho =\rho \left( n_{\underline{1}}\right) ,z^{\underline{1}}\left( u^\alpha
\right) =t,z^{\underline{2}}\left( u^\alpha \right) =r
\]
and
\begin{equation}  \label{dtema}
\Upsilon _1^1=\left( n_{\underline{1}}\frac{\partial \rho }{\partial n_{%
\underline{1}}}-\rho \right) ,\Upsilon _2^2=\Upsilon _3^3=\rho ,.
\end{equation}

The anisotropic elastic locally anisotropic medium described here satisfies
respectively weak, dominant, or strong energy conditions only if the
corresponding restrictions are placed on the equation of state considered
with respect to an anholonomic frame (see Ref. \cite{05hus} for similar
details in locally isotropic cases). For example, the weak energy condition
is characterized by the inequalities $\rho \geq 0$ and $\partial \rho
/\partial n_{\underline{1}}\geq 0.$

\subsection{Spherical symmetry with respect to holonomic and anholonomic
frames}

In radial coordinates $\left( t,r,\theta \right) $ (with $-\infty \leq
t<\infty ,$ $0\leq r<\infty ,$ $0\leq \theta \leq 2\pi )$ for a spherically
symmetric 3D metric (\ref{m2}) % (2.9),%
\begin{equation}  \label{m3}
ds^2=-f\left( r\right) dt^2+\frac 1{f\left( r\right) }dr^2+r^2d\theta ^2,
\end{equation}
with the energy--momentum tensor (\ref{emtens}) written
\begin{equation}
T_{\alpha \beta } = \rho \left( r\right) \left( v_\alpha w_\beta + v_\beta
w_\alpha \right) + P\left( r\right) \left( g_{\alpha \beta} + v_\alpha
w_\beta +v_\beta w_\alpha \right) ,  \nonumber
\end{equation}
where the null d--vectors $v_\alpha $ and $w_\beta $ are defined by
\begin{eqnarray}
V_\alpha & = & \left( \sqrt{f},-\frac 1{\sqrt{f}},0\right) =\frac 1{\sqrt{2}%
}\left( v_\alpha +w_\alpha \right) ,  \nonumber \\
q_\alpha & = & \left( 0,\frac 1{\sqrt{f}},0\right) =\frac 1{\sqrt{2}}\left(
v_\alpha -w_\alpha \right) .  \nonumber
\end{eqnarray}

In order to investigate the dynamical spherically symmetric \cite{05cm}
collapse solutions it is more convenient to use the coordinates $\left(
v,r,\theta \right) ,$ where the advanced time coordinate $v$ is defined by $%
dv=dt+\left( 1/f\right) dr.$ The metric (\ref{m3}) may be written
\begin{equation}  \label{m4}
ds^2=-e^{2\psi \left( v,r\right) }F\left( v,r\right) dv^2+2e^{\psi \left(
v,r\right) }dvdr+r^2\theta ^2,
\end{equation}
where the mass function $m\left( v,r\right) $ is defined by $F\left(
v,r\right) =1-2m\left( v,r\right) /r.$ Usually, one considers the case $\psi
\left( v,r\right) =0$ for the type II \cite{05haw} energy--momentum d--tensor
\begin{equation}
T_{\alpha \beta } = \frac 1{2\pi r^2}\frac{\delta m}{\partial v}v_\alpha
v_\beta + \rho \left( v,r\right) \left( v_\alpha w_\beta +v_\beta w_\alpha
\right) + P\left( v,r\right) \left( g_{\alpha \beta }+ v_\alpha w_\beta +
v_\beta w_\alpha \right)  \nonumber
\end{equation}
with the eigen d--vectors $v_\alpha =\left( 1,0,0\right) $ and
$w_\alpha =\left( F/2,-1,0\right) $ and the non--vanishing
components
\begin{eqnarray}
T_{vv} & = & \rho \left( v,r\right) \left( 1-\frac{2m\left( v,r\right) }%
r\right) +\frac 1{2\pi r^2}\frac{\delta m\left( v,r\right) } {\partial v},
\label{dem4b} \\
T_{vr} & = & -\rho \left( v,r\right) ,\quad T_{\theta \theta }=P\left(
v,r\right) g_{\theta \theta }.  \nonumber
\end{eqnarray}

To describe a locally isotropic collapsing pulse of radiation one may use
the metric
\begin{equation}  \label{m4a}
d s^2 = \left[ \Lambda r^2+m\left( v\right) \right] dv^2 + 2dvdr - j\left(
v\right) dvd\theta +r^2 d \theta ^2,
\end{equation}
with the Einstein field equations (\ref{2einst1}) reduced to
\[
\frac{\partial m\left( v\right) }{dv}=2\pi \rho \left( v\right) ,\frac{%
\partial j\left( v\right) }{dv}=2\pi \omega \left( v\right)
\]
having non--vanishing com\-po\-nents of the energy--mo\-mentum d--tensor
(for a rotating null locally anisotropic fluid),
\begin{equation}  \label{dem4a}
T_{vv}=\frac{\rho \left( v\right) }r+\frac{j\left( v\right) \omega \left(
v\right) }{2r^3},\quad T_{v\theta }=-\frac{\omega \left( v\right) }r,
\end{equation}
where $\rho \left( v\right) $ and $\omega \left( v\right) $ are arbitrary
functions.

In a similar manner we can define energy--momentum d--tensors for various
systems of locally anisotropic distributed matter fields; all values have to
be re--defined with respect to anholonomic bases of type (\ref{4dder}) and (%
\ref{4ddif}). For instance, let us consider the angle $\theta $ as the
anisotropic variable. In this case we have to 'elongate' the differentials,
\[
d\theta \rightarrow \delta \theta =d\theta +w_i\left( v,r,\theta \right)
dx^i,
\]
for the metric (\ref{m4}) (or (\ref{m4a})), by transforming it into a
d--metric, substitute all partial derivatives into N--elongated ones,
\[
\partial _i\rightarrow \delta _i=\partial _i-w_i\left( v,r,\theta \right)
\frac \partial {\partial \theta },
\]
and 'N--extend' the operators defining the Riemanni\-an, Ricci, Einstein and
energy--mo\-ment\-um tensors $T_{\alpha \beta },$ transforming them into
respective d--tensors. We compute the components of the energy--momentum
d--tensor for elastic media as the coefficients of usual energy--momentum
tensor redefined with respect to locally anisotropic frames,
\begin{eqnarray}
\Upsilon _{11} &=& T_{11}+\left( w_1\right) ^2T_{33}, \Upsilon _{33}=T_{33}
\label{demelm} \\
\Upsilon _{22} &=& T_{22}+ \left( w_2\right) ^2 T_{33}, \Upsilon _{12} =
\Upsilon _{21}=T_{21}+w_2w_1T_{33},  \nonumber \\
\Upsilon _{i3}&=& T_{i3}+w_iT_{33},\Upsilon _{3i}=T_{3i}+w_iT_{33},
\nonumber
\end{eqnarray}
where the $T_{\alpha \beta }$ are given by the coefficients (\ref{dem4b})
(or (\ref{dem4a})). If the isotropic energy--momentum tensor does not
contain partial derivatives, the corresponding d--tensor is also symmetric
which is less correlated with the possible antisymmetry of the Ricci tensor
(for such configurations we shall search solutions with vanishing
antisymmetric components).

\section{3D Solutions Induced by Anisotropic Matter}

We investigate a new class of solutions of (2+1)--dimensional Einstein
equations coupled with anisotropic matter \cite{05cm,05hus,05btz,05chan,05ross}
 which
describe locally anisotropic collapsing configurations.

Let us consider the locally isotropic metric
\begin{equation}  \label{m5}
\widehat{g}_{\alpha \beta }=\left[
\begin{array}{ccc}
g\left( v,r\right) & 1/2 & 0 \\
1/2 & 0 & 0 \\
0 & 0 & r^2%
\end{array}
\right]
\end{equation}
which solves the locally isotopic variant of Einstein equations (\ref{2einst1}%
) if
\begin{equation}  \label{m5a}
g\left( v,r\right) =-[1-2g(v)-2h(v)r^{1-k}-\Lambda r^2],
\end{equation}
where the functions $g(v)$ and $h(v)$ define the mass function
\[
m(r,v)=g(v)r+h(v)r^{2-k}+\frac \Lambda 2r^2
\]
satisfying the dominant energy conditions
\[
P\geq 0,\rho \geq P,T_{ab}w^aw^b>0
\]
if
\[
\frac{dm}{dv}=\frac{dg}{dv}r+\frac{dh}{dv}r^{2-k}>0.
\]
Such solutions of the Vaidya type with locally isotropic null
fluids have been considered in Ref. \cite{05hus}.

\subsection{Solutions with generic local anisotropy in spherical coordinates}

By introducing a new time--like variable
\[
t=v+\int \frac{dr}{g\left( v,r\right) }
\]
the metric(\ref{m5}) can be transformed in diagonal form
\begin{equation}  \label{m6}
ds^2=-g\left( t,r\right) dt^2+\frac 1{g\left( t,r\right) }dr^2+r^2d\theta ^2
\end{equation}
which describe the locally isotropic collapse of null fluid matter.

A variant of locally anisotropic inhomogeneous collapse could be
modelled, for instance, by the N--elongation of the variable
$\theta $ in (\ref{m6}) and considering solutions of vacuum
Einstein equations for the d--metric (a particular case of
(\ref{dm3}))
\begin{equation}
ds^{2}=-g\left( t,r\right) dt^{2}+\frac{1}{g\left( t,r\right) }%
dr^{2}+r^{2}\delta \theta ^{2},  \label{dm3}
\end{equation}
where
\[
\delta \theta =d\theta +w_{1}\left( t,r,\theta \right) dt+w_{2}\left(
t,r,\theta \right) dr.
\]
The coefficients $g\left( t,r\right) ,1/g\left( t,r\right) $ and $r^{2}$ of
the d--metric were chosen with the aim that in the locally isotropic limit,
when $w_{i}\rightarrow 0,$ we shall obtain the 3D metric (\ref{m6}). We note
that the gravitational degrees of freedom are contained in nonvanishing
values of the Ricci d--tensor (\ref{4dricci}),
\begin{equation}
R_{1}^{1}=R_{2}^{2}=\frac{1}{2g^{3}}[(\frac{\partial g}{\partial r}%
)^{2}-g^{3}\frac{\partial ^{2}g}{\partial t^{2}}-g\frac{\partial ^{2}g}{%
\partial r^{2}}],  \label{2ricci2}
\end{equation}
of the N--curvature (\ref{4ncurv}),
\[
\Omega _{12}^{3}=-\Omega _{21}^{3}=\frac{\partial w_{1}}{\partial r}-\frac{%
\partial w_{2}}{\partial t}-w_{2}\frac{\partial w_{1}}{\partial \theta }%
+w_{1}\frac{\partial w_{2}}{\partial \theta },
\]
and d--torsion (\ref{2dtorsions})
\[
P_{13}^{3}=\frac{1}{2}\left( 1+r^{4}\right) \frac{\partial w_{1}}{\partial
\theta },\ P_{23}^{3}=\frac{1}{2}\left( 1+r^{4}\right) \frac{\partial w_{2}}{%
\partial \theta }-r^{3}.
\]

We can construct a solution of 3D Einstein equations with cosmological
constant $\Lambda $ (\ref{2einst1}) and energy momentum d--tensor $\Upsilon
_{\alpha \beta },$ when $\Upsilon _{ij}=T_{ij}+w_iw_jT_{33},\Upsilon
_{3j}=T_{3j}+w_jT_{33}$ and $\Upsilon _{33}=T_{33}$ when $T_{\alpha \beta }$
is given by a d--tensor of type (\ref{dtema}), $T_{\alpha \beta }=\{n_{%
\underline{1}}\frac{\partial \rho }{\partial n_{\underline{1}}},P,0\}$ with
anisotropic matter pressure $P.$ A self--consistent solution is given by
\begin{equation}  \label{auxw1}
\Lambda = \kappa n_{\underline{1}}\frac{\partial \rho } {\partial n_{%
\underline{1}}}=\kappa P, \mbox{ and } h=\frac{\kappa \rho }{R_1^1+\Lambda }
\end{equation}
where $R_1^1$ is computed by the formula (\ref{2ricci2}) for arbitrary values
$g\left( t,r\right) .$ For instance, we can take the $g(\nu ,r)$ from (\ref%
{m5a}) with $\nu \rightarrow t=\nu +\int g^{-1}(\nu ,r)dr.$

For $h=r^2,$ the relation (\ref{auxw1}) results in an equation for $g(t,r),$
\[
(\frac{\partial g}{\partial r})^2-g^3\frac{\partial ^2g}{\partial t^2}-g%
\frac{\partial ^2g}{\partial r^2}=2g^3\left( \frac{\kappa \rho }{r^2}%
-\Lambda \right) .
\]
The static configurations are described by the equation
\begin{equation}  \label{auxw2}
gg^{\prime \prime }-(g^{\prime })^2+\varpi (r)g^3=0,
\end{equation}
where
\[
\varpi (r)=2\left( \frac{\kappa \rho \left( r\right) }{r^2}-\Lambda \right)
\]
and the prime denote the partial derivative $\partial /\partial r.$ There
are four classes (see (\cite{05kamke})) of solutions of the equation (\ref%
{auxw2}), which depends on constants of the relation
\[
(\ln |g|)^{\prime }=\pm \sqrt{2|\varpi (r)|(C_1\mp g)},
\]
where the minus (plus) sign under square root is taken for $\varpi (r)>0$ $%
(\varpi (r)<0)$ and the constant $C_1$ can be negative, $C_1=-c^2,$ or
positive, $C_1=c^2.$ In explicit form the solutions are
\begin{equation}  \label{faze}
g(r)=\left\{
\begin{array}{rcl}
c^{-2}\cosh ^{-2}\left[ \frac c2\sqrt{2|\varpi (r)|}\left( r-C_2\right) %
\right] & , &  \\
\mbox{ for } \varpi (r)>0,C_1=c^2 & ; &  \\
c^{-2}\sinh ^{-2}\left[ \frac c2\sqrt{2|\varpi (r)|}\left( r-C_2\right) %
\right] & , &  \\
\mbox{ for } \varpi (r)<0,C_1=c^2 & ; &  \\
c^{-2}\sin ^{-2}\left[ \frac c2\sqrt{2|\varpi (r)|}\left( r-C_2\right) %
\right] \neq 0 & , &  \\
\mbox{ for } \varpi (r)<0,C_1=-c^2 & ; &  \\
-2{\varpi (r)}^{-1}(r-C_2)^{-2}{\qquad}{\qquad} & , &  \\
\mbox{ for } \varpi (r)<0,C_1=0, &  &
\end{array}
\right.
\end{equation}
where $C_2=const.$ The values of constants are to be found from boundary
conditions. In dependence of prescribed type of matter density distribution
and of values of cosmological constant one could fix one of the four classes
of obtained solutions with generic local anisotropy of 3D Einstein equations.

The constructed in this section static solutions of 3D Einstein equations
are locally anisotropic alternatives (with proper phases of anisotropic
polarizations of gravitational field) to the well know BTZ solution. Such
configurations are possible if anholonomic frames with associated
N--connection structures are introduced into consideration.

\subsection{An anisotropic solution in ($\protect\nu ,r,\protect\theta )$%
--coordinates}

For modelling a spherical collapse with generic local anisotropy
we use the
d--metric (\ref{dm2a}) by stating the coordinates $x^1=v,x^2=r$ and $%
y=\theta .$ The equations (\ref{2einst1}) are solved if
\[
\kappa \rho (v,r)=\Lambda \mbox{ and } \kappa P(v,r)=-\Lambda -\frac 12\frac{%
\partial ^2g}{\partial r^2}
\]
for
\[
g=\frac \kappa \Lambda \left[ \rho (v,r)\left( 1-\frac{2m(v,r)}r\right)
+\frac 1{2\pi r^2}\frac{\delta m(v,r)}{\delta v}\right] .
\]
Such metrics depend on classes of functions. They can be extended
 ellipsoidal configurations and additional polarizations.

\subsection{A solution for rotating two locally anisotropic fluids}

The anisotropic configuration from the previous subsection admits a
generalization to a two fluid elastic media, one of the fluids being of
locally anisotropic rotating configuration. For this model we consider an
anisotropic extension of the metric (\ref{m4a}) and of the sum of
energy--momentum tensors (\ref{dem4b}) and (\ref{dem4a}). The coordinates
are parametrized $x^1=v,x^2=r,y=\theta $ and the d--metric is given by the
ansatz
\[
g_{ij}=\left(
\begin{array}{cc}
g(v,r) & 1 \\
1 & 0%
\end{array}
\right) \mbox{ and }h=h(v,r,\theta ).
\]
The nontrivial components of the Einstein d--tensor is
\[
G_{33}=-\frac 12h\frac{\partial ^2g}{\partial r^2}.
\]
We consider a non--rotating fluid component with nontrivial energy--momentum
components
\begin{equation}  \label{dem5b}
{}^{(1)}T_{vv} = {}^{(1)} \rho \left( v,r\right) \left( 1-\frac{2
{}^{(1)}m\left( v,r\right) }r\right) + \frac 1{2\pi r^2}\frac{\delta
{}^{(1)} m\left( v,r\right) }{\partial v}, \ {}^{(1)}T_{vr} = - {}^{(1)}\rho
\left( v,r\right).  \nonumber
\end{equation}
and a rotating null locally anisotropic fluid with energy--momentum
components
\begin{equation}  \label{dem5a}
^{(2)}T_{vv} = \frac{^{(2)}\rho \left( v\right) }r + \frac{%
^{(2)}j\left(v\right) \ ^{(2)}\omega \left( v\right) }{2r^3}, \
^{(2)}T_{v\theta } = -\frac{^{(2)}\omega \left( v\right) }r.  \nonumber
\end{equation}
The nontrivial components of energy momentum d--tensor $\Upsilon _{\alpha
\beta} = {}^{(1)} \Upsilon _{\alpha \beta} + {}^{(2)} \Upsilon _{\alpha
\beta}$ (associated in the locally anisotropic limit to (\ref{dem4a}) and/or
(\ref{dem4b})) are computed by using the formulas (\ref{dem5a}), (\ref{dem5b}%
) and (\ref{demelm}).

The Einstein equations are solved by the set of func\-ti\-ons
\[
g(v,r), {}^{(1)}\rho \left( v,r\right), {}^{(1)}m\left( v,r\right),
{}^{(2)}\rho \left( v\right), {}^{(2)}j\left( v\right), {}^{(2)}\omega
\left( v\right)
\]
satisfying the conditions
\[
g(v,r) = \frac \kappa \Lambda \left[ ^{(1)}T_{vv}+\ ^{(2)}T_{vv}\right] , %
\mbox{ and } \Lambda = \frac 12\frac{\partial ^2g}{\partial r^2}=\kappa \
^{(1)}T_{vr},
\]
where $h\left( v,r,\theta \right) $ is an arbitrary function which
results in nontrivial solutions for the N--connection coefficients
$w_i\left( v,r,\theta \right) $ if $\Lambda \neq 0.$ In the
locally isotropic limit, for $^{(1)}\rho ,^{(1)}m=0,$ we could
take $g(v,r)=g_1(v)+\Lambda r^2,w_1=-j(v)/(2r^2)$ and $w_2=0$
which results in a solution of the Vaidya type with locally
isotropic null fluids \cite{05chan}.

The main conclusion of this subsection is that we can model the 3D collapse
of inhomogeneous null fluid by using vacuum locally anisotropic
configurations polarized by an anholonomic frame in a manner as to reproduce
in the locally isotropic limit the usual BTZ geometry.

We end this section with the remark that the locally isotropic collapse of
dust without pressure was analyzed in details in Ref. \cite{05ross}.

\section[Gravitational Polarizations and Black Holes]
{Gravitational Anisotropic Polarizations \newline and Black Holes}

If we introduce in consideration anholonomic frames, locally anisotropic
black hole configurations are possible even for vacuum locally anisotropic
spacetimes without matter. Such solutions could have horizons with deformed
circular symmetries (for instance, elliptic one) and a number of unusual
properties comparing with locally isotropic black hole solutions. In this
Section we shall analyze two classes of such solutions. Then we shall
consider the possibility to introduces matter sources and analyze such
configurations of matter energy density distribution when the gravitational
locally anisotropic polarization results into constant renormalization of
constants of BTZ solution.

\subsection{Non--rotating black holes with ellipsoidal horizon}

We consider a metric (\ref{3ansatzc}) for local coordinates $%
(x^{1}=r,x^{2}=\theta ,y=t),$ where $t$ is the time--like coordinate and the
coefficients are parametrized
\begin{equation}
a(x^{i})=a\left( r\right) ,b(x^{i})=b(r,\theta )  \label{elips1}
\end{equation}
and
\begin{equation}
h(x^{i},y)=h\left( r,\theta \right) .  \label{elips1a}
\end{equation}
The functions $a(r)$ and $\ b\left( r,\theta \right) $ and the coefficients
of nonlinear connection $w_{i}(r,\theta ,t)$ will be found as to satisfy the
vacuum Einstein equations (\ref{vaceq}) with arbitrary function $h(x^{i},y)$
(\ref{elips1a}) stated in the form in order to have compatibility with the
BTZ solution in the locally isotropic limit.

We consider a particular case of d-metrics (\ref{dansatzc}) with
coefficients like (\ref{elips1}) and (\ref{elips1a}) when
\begin{equation}
h(r,\theta )=4\Lambda ^{3}(\theta )\left( 1-\frac{r_{+}^{2}(\theta )}{r^{2}}%
\right) ^{3}  \label{hel3}
\end{equation}
where, for instance,
\begin{equation}
r_{+}^{2}(\theta )=\frac{p^{2}}{\left[ 1+\varepsilon \cos \theta \right] ^{2}%
}  \label{elipsrad}
\end{equation}
is taken as to construct a 3D solution of vacuum Einstein equations with
generic local anisotropy having the horizon given by the parametric equation
\[
r^{2}=r_{+}^{2}(\theta )
\]
describing a ellipse with parameter $p$ and eccentricity $\varepsilon .$ We
have\ to identify
\[
p^{2}=r_{+[0]}^{2}=-M_{0}/\Lambda _{0},
\]
where\ $r_{+[0]},M_{0}$ and $\Lambda _{0}$ are respectively the
horizon radius, mass parameter and cosmological constant of the
non--rotating BTZ solution \cite{05btz} if we wont to have a
connection with locally isotropic limit with $\varepsilon
\rightarrow 0.$ We can consider that the elliptic horizon
(\ref{elipsrad}) is modelled by the anisotropic mass
\[
M\left( \theta \right) =M_{0}/\left[ 1+\varepsilon \cos \theta \right] ^{2}.
\]

For the coefficients (\ref{elips1}) the equations (\ref{vaceq}) simplifies
into
\begin{equation}  \label{vaceqel1}
-\ddot b+\frac 1{2b}\dot b^2+\frac 1{2a}\dot a\dot b=0,
\end{equation}
where (in this subsection) $\dot b=\partial b/\partial r.$ The general
solution of (\ref{vaceqel1}), for a given function $a(r)$ is defined by two
arbitrary functions $b_{[0]}(\theta )$ and $b_{[1]}(\theta )$ (see \cite%
{05kamke}),
\[
b(r,\theta )=\left[ b_{[0]}(\theta )+b_{[1]}(\theta )\int \sqrt{|a(r)|}dr%
\right] ^2.
\]

If we identify
\[
b_{[0]}(\theta )=2\frac{\Lambda (\theta )}{\sqrt{|\Lambda _{0}|}}%
r_{+}^{2}(\theta )\mbox{ and }b_{[1]}(\theta )=-2\frac{\Lambda (\theta )}{%
\Lambda _{0}},
\]
we construct a d--metric locally anisotropic solution of vacuum Einstein
equations
\begin{eqnarray}
&&\delta s^{2}=\Omega ^{2}\left( r,\theta \right)  \label{elipbh}  \\
&& \left[ 4r^{2}|\Lambda
_{0}|dr^{2}+\frac{4}{|\Lambda _{0}|}\Lambda ^{2}(\theta )\left[
r_{+}^{2}(\theta )-r^{2}\right] ^{2}d\theta ^{2}-\frac{4}{|\Lambda _{0}|r^{2}%
}\Lambda ^{3}(\theta )\left[ r_{+}^{2}(\theta )-r^{2}\right] ^{3}\delta t^{2}%
\right] , \nonumber
\end{eqnarray}
where
\[
\delta t=dt+w_{1}(r,\theta )dr+w_{2}(r,\theta )d\theta
\]
is to be associated to a N--connection structure
\[
w_{r}=\partial _{r}\ln |\ln \Omega |\mbox{ and }w_{\theta }=\partial
_{\theta }\ln |\ln \Omega |
\]
with $\Omega ^{2}=\pm h(r,\theta ),$ where $h(r,\theta )$ is taken from (\ref%
{hel3}). In the simplest case we can consider a constant effective
cosmological constant $\Lambda (\theta )\simeq \Lambda _{0}.$

\newpage

The matrix
\[
g_{\alpha \beta }=\Omega ^{2}\left[
\begin{array}{ccc}
a-w_{1}^{\ 2}h & -w_{1}w_{2}h & -w_{1}h \\
-w_{1}w_{2}h & b-w_{2}^{\ 2}h & -w_{2}h \\
-w_{1}h & -w_{2}h & -h%
\end{array}
\right] .
\]
parametrizes a class of solutions of 3D vacuum Einstein equations with
generic local anisotro\-py and nontrivial N--connection curvature (\ref%
{4ncurv}), which describes black holes with variable mass parameter $M\left(
\theta \right) $ and elliptic horizon. As a matter of principle, by fixing
necessary functions $b_{[0]}(\theta )$ and $b_{[1]}(\theta )$ we can
construct solutions with effective (polarized by the vacuum anisotropic
gravitational field) variable cosmological constant $\Lambda (\theta ).$ We
emphasize that this type of anisotropic black hole solutions have been
constructed by solving the vacuum Einstein equations without cosmological
constant. Such type of constants or varying on $\theta $ parameters were
introduced as some values characterizing anisotropic polarizations of vacuum
gravitational field and this approach can be developed if we are considering
anholonomic frames on (pseudo) Riemannian spaces. For the examined
anisotropic model the cosmological constant is induced effectively in
locally isotropic limit via specific gravitational field vacuum
polarizations.

\subsection{Rotating black holes with running in time constants}

A new class of solutions of vacuum Einstein equations is generated by a
d--metric (\ref{dm2}) written for local coordinates $(x^1=r,x^2=t,y=\theta
), $ where as the anisotropic coordinate is considered the angle variable $%
\theta $ and the coefficients are parametrized
\begin{equation}  \label{elips4}
a(x^i)=a\left( r\right) ,b(x^i)=b(r,t)
\end{equation}
and
\begin{equation}  \label{elips4a}
h(x^i,y)=h\left( r,t\right) .
\end{equation}

Let us consider a 3D metric
\begin{equation}  \label{metrbtzc}
ds^2 = 4\frac{\psi ^2}{r^2}dr^2-\frac{N_{[s]}^4\psi ^4}{r^4}dt^2 +\frac{%
N_{(s)}^2\psi ^6}{r^4}\left[ d\theta +N_{[\theta ]}dt\right] ^2  \nonumber
\end{equation}
which is conformally equivalent (if multiplied to the conformal factor $%
4N_{(s)}^2\psi ^4/r^4)$ to the rotating BTZ solution with

\newpage

\begin{eqnarray}  \label{auxform1}
N_{[s]}^2(r) &=& -\Lambda _0\frac{r^2}{\psi ^2}\left( r^2-r_{+[0]}^2\right)
,N_{[\theta ]}(r)=-\frac{J_0}{2\psi },  \nonumber \\
\psi ^2(r) &=& r^2-\frac 12\left( \frac{M_0}{\Lambda _0}+r_{+[0]}^2\right),
\nonumber \\
r_{+[0]}^2 &=& -\frac{M_0}{\Lambda _0}\sqrt{1+\Lambda _0\left( \frac{J_0}{M_0%
}\right) ^2},  \nonumber
\end{eqnarray}
where $J_0$ is the rotation moment and $\Lambda _0$ and $M_0$ are
respectively the cosmological and mass BTZ constants.

A d--metric (\ref{dm2}) defines a locally anisotropic extension of (\ref%
{metrbtzc}) if the solution of (\ref{vaceqel1}), in variables $%
(x^1=r,x^2=t), $ with coefficients (\ref{elips4}) and (\ref{elips4a}), is
written
\begin{equation}
b(r,t) = - \left[ b_{[0]}(t)+b_{[1]}(t)\int \sqrt{\left| a(r)\right| }dr%
\right]^2 = -\Lambda ^2(t)\left[ r_{+}^2(t)-r^2\right] ^2,  \nonumber
\end{equation}
for
\[
a(r)=4\Lambda _0r^2,b_{[0]}(t)=\Lambda (t)r_{+}^2(t),b_{[1]}(t)=2\Lambda (t)/%
\sqrt{|\Lambda _0|}
\]
with $\Lambda (t)\sim \Lambda _0$ and $r_{+}(t)\sim r_{+[0]}$ being some
running in time values.

The functions $a(r)$ and $\ b\left( r,t\right) $ and the coefficients of
nonlinear connection $w_i(r,t,\theta )$ must solve the vacuum Einstein
equations (\ref{vaceq}) with arbitrary function $h(x^i,y)$ (\ref{elips1a})
stated in the form in order to have a relation with the BTZ solution for
rotating black holes in the locally isotropic limit. This is possible if we
choose
\begin{equation}
w_1(r,t) = -\frac{J\left( t\right) }{2\psi (r,t)}, \qquad h(r,t) = \frac{%
4N_{[s]}^2(r,t)\psi ^6(r,t)}{r^4},  \nonumber
\end{equation}
for an arbitrary function $w_2(r,t,\theta )$ with $N_{[s]}(r,t)$ and $\psi
(r,t)$ computed by the same formulas (\ref{auxform1}) with the constant
substituted into running values,
\[
\Lambda _0\rightarrow \Lambda (t),M_0\rightarrow M\left( t\right)
,J_0\rightarrow J(t).
\]

We can model a dissipation of 3D black holes, by anisotropic gravitational
vacuum polarizations if for instance,
\[
r_{+}^2(t)\simeq r_{+[0]}^2\exp [-\lambda t]
\]
for $M(t)=M_0\exp [-\lambda t]$ with $M_0$ and $\lambda $ being some
constants defined from some ''experimental'' data or a quantum model for 3D
gravity. The gravitational vacuum admits also polarizations with exponential
and/or oscillations in time for $\Lambda (t)$ and/or of $M(t).$

\subsection{Anisotropic Renormalization of Constants}

The BTZ black hole \cite{05btz} in ``Schwarzschild'' coordinates is described
by the metric
\begin{equation}  \label{a1}
ds^2=-(N^{\perp })^2dt^2+f^{-2}dr^2+r^2\left( d\phi +N^\phi dt\right) ^2
\end{equation}
with lapse and shift functions and radial metric
\begin{eqnarray}  \label{a2}
N^{\perp } &=& f=\left( -M+{\frac{r^2}{\ell ^2}}+{\frac{J^2}{4r^2}}\right)
^{1/2}, \\
N^\phi &=& -{\frac J{2r^2}}\qquad (|J|\le M\ell ).  \nonumber
\end{eqnarray}
which satisfies the ordinary vacuum field equations of (2+1)-dimensional
general relativity (\ref{2einst1}) with a cosmological constant $\Lambda
=-1/\ell ^2$.

If we are considering anholonomic frames, the matter fields ''deform'' such
solutions not only by presence of a energy--momentum tensor in the right
part of the Einstein equations but also via anisotropic polarizations of the
frame fields. In this Section we shall construct a subclass of d--metrics (%
\ref{dm3}) selecting by some particular distributions of matter energy
density $\rho (r)$ and pressure $P(r)$ solutions of type (\ref{a1}) but with
renormalized constants in (\ref{a2}),
\begin{equation}  \label{renconst}
M\rightarrow \overline{M}=\alpha ^{(M)}M,J\rightarrow \overline{J}=\alpha
^{(J)}J,\Lambda \rightarrow \overline{\Lambda }=\alpha ^{(\Lambda )}\Lambda ,
\end{equation}
where the receptivities $\alpha ^{(M)},\alpha ^{(J)}$ and $\alpha ^{(\Lambda
)}$ are considered, for simplicity, to be constant (and defined
''experimentally'' or computed from a more general model of quantum 3D
gravity) and tending to a trivial unity value in the locally isotropic
limit. The d--metric generalizing (\ref{a1}) is stated in the from

\begin{equation}  \label{a3}
\delta s^2=-F\left( r\right) ^{-1}dt^2+F\left( r\right) dr^2+r^2\delta
\theta ^2
\end{equation}
where
\[
F(r)=\left( -\overline{M}-\overline{\Lambda }r^2+{\frac{J^2}{4r^2}}\right),\
\delta \theta =d\theta +w_1dt\mbox{ and }w_1=-{\frac{\overline{J}}{2r^2}.}
\]
The d--metric (\ref{a3}) is a static variant of d--metric (\ref{dm3}) when
the solution (\ref{faze}) is constructed for a particular function
\[
\overline{\varpi }(r)=2\left( \frac{\kappa \overline{\rho }\left( r\right) }{%
r^2}-\overline{\Lambda }\right)
\]
is defined by corresponding matter distribution $\overline{\rho }\left(
r\right) $ when the function $F(r)$ is the solution of equations (\ref{auxw2}%
) with coefficient $\overline{\varpi }(r)$ before $F^3,$ i. e.
\[
FF^{\prime \prime }-(F^{\prime })^2+\varpi (r)F^3=0.
\]

The d--metric (\ref{a3}) is singular when $r\!=\overline{\!r}_{\pm }$, where
\begin{equation}  \label{rada}
\overline{r}_{\pm }^2=-{\frac{\overline{M}}{2\overline{\Lambda }}}\left\{
1\pm \left[ 1+\overline{\Lambda }\left( {\frac{\overline{J}}{\overline{M}}}%
\right) ^2\right] ^{1/2}\right\} ,
\end{equation}
i.e.,
\[
\overline{M}=-\overline{\Lambda }(\overline{r}_{+}^2+\overline{r}%
_{-}^2),\quad \overline{J}={\frac{2\overline{r}_{+}\overline{r}_{-}}{%
\overline{\ell }}}\ ,\overline{\Lambda }=-1/\overline{\ell }^2.
\]

In locally isotropic gravity the surface gravity was computed \cite{05Wald2}
\[
\sigma ^2=-{\frac 12}D^\alpha \chi ^\beta D_\alpha \chi _\beta ={\frac{%
r_{+}^2-r_{-}^2}{\ell ^2r_{+}}},
\]
where the vector $\chi =\partial _v-N^\theta (r_{+})\partial _\theta $ is
orthogonal to the Killing horizon defined by the surface equation $%
r\!=\!r_{+}.$ For locally anisotropic renormalized (overlined) values we
have
\[
\overline{\chi }=\delta _\nu =\partial _\nu -w_1(\overline{r}_{+})\partial
_\theta
\]
and
\[
\overline{\sigma }^2=-{\frac 12}D^\alpha \overline{\chi }^\beta D_\alpha
\overline{\chi }_\beta =\overline{\Lambda }{\frac{\overline{r}_{-}^2-%
\overline{r}_{+}^2}{\overline{r}_{+}}}.
\]

The renormalized values allow us to define a corresponding thermodynamics of
locally anisotropic black holes.

\subsection{Ellipsoidal black holes with running in time constants}

The anisotropic black hole solution of 3D vacuum Einstein equations (\ref%
{elipbh}) with elliptic horizon can be generalized for a case with
varying in time cosmological constant $\Lambda _0(t).$ For this
class of solutions we choose the local coordinates
$(x^1=r,x^2=\theta ,y=t)$ and a d--metric of type
(\ref{dansatzc}),
\begin{equation}  \label{dansatzc1}
\delta s^2 = \Omega _{(el)}^2(r,\theta ,t)[a(r)(dr)^2+b(r,\theta )(d\theta
)^2 +h(r,\theta ,t)(\delta t)^2],
\end{equation}
where
\[
h(r,\theta ,t)=-\Omega _{(el)}^2(r,\theta ,t)=-\frac{4\Lambda ^3\left(
\theta \right) }{|\Lambda _0(t)|r^2}\left[ r_{+}^2(\theta ,t)-r^2\right] ^3,
\]
for
\begin{equation}
r_{+}^2(\theta ,t) = \frac{p(t)}{(1+\varepsilon \cos \theta )^2},
\mbox{ and
} p(t) = r_{+(0)}^2(\theta ,t)=-M_0/\Lambda _0(t)  \nonumber
\end{equation}
and it is considered that $\Lambda _0(t)\simeq \Lambda _0$ for static
configurations.

The d--metric (\ref{dansatzc1}) is a solution of 3D vacuum Einstein
equations if the 'elongated' differential
\[
\delta t=dt+w_r(r,\theta ,t)dr+w_\theta (r,\theta ,t)d\theta
\]
has the N--connection coefficients are computed following the condition (\ref%
{conformnc}),
\[
w_r=\partial _r\ln |\ln \Omega _{(el)}|\mbox{ and }w_\theta =\partial
_\theta \ln |\ln \Omega _{(el)}|.
\]

The functions $a(r)$ and $b(r,\theta )$ from (\ref{dansatzc1}) are arbitrary
ones of type (\ref{elips1}) satisfying the equations (\ref{vaceqel1}) which
in the static limit could be fixed to transform into static locally
anisotropic elliptic configurations. The time dependence of $\Lambda _0(t)$
has to be computed, for instance, from a higher dimension theory or from
experimental data.

\section{On the Thermodynamics of Anisotropic Black Ho\-les}

A general approach to the anisotropic black holes should be based on a kind
of nonequilibrium thermodynamics of such objects imbedded into locally
anisotropic gravitational (locally anisotropic ether) continuous, which is a
matter of further investigations (see the first works on the theory of
locally anisotropic kinetic processes and thermodynamics in curved spaces %
\cite{05v4}).

In this Section, we explore the simplest type of locally anisotropic black
holes with anisotropically renormalized constants being in thermodynamic
equilibrium with the locally anisotropic spacetime ''bath'' for suitable
choices of N--connection coefficients. We do not yet understand the detailed
thermodynamic behavior of locally anisotropic black holes but believe one
could define their thermodynamics in the neighborhoods of some equilibrium
states when the horizons are locally anisotropically deformed and constant
with respect to an anholonomic frame.

In particular, for a class of BTZ like locally anisotropic spacetimes with
horizons radii (\ref{rada}) we can still use the first law of thermodynamics
to determine an entropy with respect to some fixed anholonomic bases (\ref%
{4ddif}) and (\ref{4dder}) (here we note that there are developed some
approaches even to the thermodynamics of usual BTZ black holes and that
uncertainty is to be transferred in our considerations, see discussions and
references in \cite{05cm}).

In the approximation that the locally anisotropic spacetime receptivities $%
\alpha ^{(m)},\alpha ^{(J)}$ and $\alpha ^{(\Lambda )}$ do not depend on
coordinates we have similar formulas as in locally isotropic gravity for the
locally anisotropic black hole temperature at the boundary of a cavity of
radius $r_H,$%
\begin{equation}  \label{temp}
\overline{T}=-\frac{\overline{\sigma }}{2\pi \left( \overline{M}+\overline{%
\Lambda }r_H^2\right) ^{1/2}},
\end{equation}
and entropy
\begin{equation}  \label{entropy}
\overline{S}=4\pi \overline{r}_{+}
\end{equation}
in Plank units.

For a elliptically deformed locally anisotropic black hole with the outer
horizon $r_{+}\left( \theta \right) $ given by the formula (\ref{elipsrad})
the Bekenstein--Hawking entropy,
\[
S^{(a)}=\frac{L_{+}}{4G_{(gr)}^{(a)}},
\]
were
\[
L_{+}=4\int\limits_0^{\pi /2}r_{+}\left( \theta \right) d\theta
\]
is the length of ellipse's perimeter and $G_{(gr)}^{(a)}$ is the three
dimensional gravitational coupling constant in locally anisotropic media,
has the value
\[
S^{(a)}=\frac{2p}{G_{(gr)}^{(a)}\sqrt{1-\varepsilon ^2}}arctg\sqrt{\frac{%
1-\varepsilon }{1+\varepsilon }}.
\]
If the eccentricity vanishes, $\varepsilon =0,$ we obtain the locally
isotropic formula with $p$ being the radius of the horizon circumference,
but the constant $G_{(gr)}^{(a)}$ could be locally anisotropically
renormalized.

In dependence of dispersive or amplification character of locally
anisotropic ether with $\alpha ^{(m)},\alpha ^{(J)}$ and $\alpha ^{(\Lambda
)}$ being less or greater than unity we can obtain temperatures of locally
anisotropic black holes less or greater than that for the locally isotropic
limit. For example, we get anisotropic temperatures $T^{(a)}(\theta )$ if
locally anisotropic black holes with horizons of type (\ref{elipsrad}) are
considered.

If we adapt the Euclidean path integral formalism of Gibbons and Hawking \cite%
{05gh} to locally anisotropic spacetimes, by performing calculations with
respect to an anholonomic frame, we develop a general approach to the
locally anisotropic black hole irreversible thermodynamics. For locally
anisotropic backgrounds with constant receptivities we obtain similar to %
\cite{05btz94,05cl,05cm} but anisotropically renormalized formulas.

Let us consider the Euclidean variant of the d--metric (\ref{a3})
\begin{equation}  \label{a3e}
\delta s_E^2=\left( F_E\right) d\tau ^2+\left( F_E\right)
^{-1}dr^2+r^2\delta \theta ^2
\end{equation}
where $t=i\tau $ and the Euclidean lapse function is taken with locally
anisotropically renormalized constants, as in (\ref{renconst}) (for
simplicity, there is analyzed a non--rotating locally anisotropic black
hole), $F=\left( -\overline{M}-\overline{\Lambda }r^2\right),$ which leads
to the root $\overline{r}_{+}=\left[ -\overline{M}/\overline{\Lambda }\right]
^{1/2}.$ By applying the coordinate transforms
\begin{eqnarray}  \label{coordt}
x & = & \left( 1-\left( \frac{{\overline r}_{+}}r\right) ^2\right) ^{1/2} \
\cos \left( -{\overline \Lambda} {\overline r}_{+}\tau \right) \exp \left(
\sqrt{|{\overline \Lambda} |}{\overline r}_{+} \theta \right) ,  \nonumber \\
y & = & \left( 1-\left( \frac{{\overline r}_{+}}r\right) ^2\right) ^{1/2} \
\sin \left(-{\overline \Lambda} {\overline r}_{+} \tau \right) \exp \left(
\sqrt{|{\overline \Lambda }|} {\overline r}_{+} \theta \right) ,  \nonumber
\\
z & = & \left( \left( \frac{{\overline r}_{+}}r\right) ^2- 1\right)
^{1/2}\exp \left(\sqrt{|{\overline \Lambda}|} {\overline r}_{+}\theta
\right) ,  \nonumber
\end{eqnarray}
the d--metric (\ref{a3e}) is rewritten in a standard upper half--space $%
\left( z>0\right) $ representation of locally anisotropic hyperbolic
3--space,
\[
\delta s_E^2 = -\frac 1{\overline \Lambda }(z^2 dz^2+dy^2+\delta z^2).
\]

The coordinate transform (\ref{coordt}) is non--singular at the $z$--axis $r=%
\overline{r}_{+}$ if we require the periodicity
\[
\left( \theta ,\tau \right) \sim \left( \theta ,\tau +\overline{\beta }%
_0\right)
\]
where
\begin{equation}  \label{invtemp}
\overline{\beta }_0=\frac 1{\overline{T}_0}=-\frac{2\pi }{\overline{\Lambda }%
\ \overline{r}_{+}}
\end{equation}
is the inverse locally anisotropically renormalized temperature, see (\ref%
{temp}).

To get the locally anisotropically renormalized entropy from the Euclidean
locally anisotropic path integral we must define a locally anisotropic
extension of the grand canonical partition function
\begin{equation}  \label{partfunct}
\overline{Z}=\int \left[ dg\right] e^{\overline{I}_E[g]},
\end{equation}
where $\overline{I}_E$ is the Euclidean locally anisotropic action. We
consider as for usual locally isotropic spaces the classical approximation $%
\overline{Z}\sim \exp \{\overline{I}_E[\overline{g}]\},$ where as the
extremal d--metric $\overline{g}$ is taken (\ref{a3e}). In (\ref{partfunct})
there are included boundary terms at $\overline{r}_{+}$ and $\infty $ (see
the basic conclusions and detailed discussions for the locally isotropic
case \cite{05btz94,05cl,05cm} which are also true with respect to anholonomic
bases).

For an inverse locally anisotropic temperature $\overline{\beta }_0$ the
action from (\ref{partfunct}) is
\[
\overline{I}_E[\overline{g}]=4\pi \overline{r}_{+}-\overline{\beta }_0M
\]
which corresponds to the locally anisotropic entropy (\ref{entropy}) being a
locally anisotropic renormalization of the standard Bekenstein entropy.

\section{Chern--Simons Theories and Locally Anisotropic Gra\-vi\-ty}

In order to compute the first quantum corrections to the locally anisotropic
path integral (\ref{partfunct}), inverse locally anisotropic temperature (%
\ref{invtemp}) and locally anisotropic entropy (\ref{entropy}) we take the
advantage of the Chern--Simons formalism generalized for (2+1)--dimensional
locally anisotropic spacetimes.

By using the locally anisotropically renormalized cosmological constant $%
\overline{\Lambda }$ and adapting the Achucarro and Townsend \cite{05at}
construction to anholonomic frames we can define two SO(2,1) gauge locally
anisotropic fields
\[
A^{\underline{a}}=\omega ^{\underline{a}}+ \frac 1{\sqrt{\left| {\overline
\Lambda }\right| }} e^{\underline{a}} \mbox{ and } \widetilde{A}^{\underline{%
a}}=\omega ^{\underline{a}}-\frac 1{\sqrt{\left| {\overline \Lambda} \right|
}}e^{\underline{a}}
\]
where the index\underline{ }$\underline{a}$ enumerates an anholonomic triad $%
e^{\underline{a}}=e_\mu ^{\underline{a}}\delta x^\mu $ and $\omega ^{%
\underline{a}}=\frac 12\epsilon ^{\underline{a}\underline{b}\underline{c}%
}\omega _{\mu \underline{b}\underline{c}}\delta x^\mu $ is a spin
d--connection (d--spinor calculus is developed in \cite{05v1}). The
first--order action for locally anisotropic gravity is written
\begin{equation}  \label{1act1}
\overline{I}_{grav}=\overline{I}_{CS}[A]-\overline{I}_{CS}[\widetilde{A}]
\end{equation}
with the Chern--Simons action for a (2+1)--dimensional vector bundle $%
\widetilde{E}$ provided with N--connection structure,
\begin{equation}  \label{actcs}
\overline{I}_{CS}[A]=\frac{\overline{k}}{4\pi }\int\nolimits_{\widetilde{E}%
}Tr\left( A\land \delta A+\frac 23A\land A\land A\right)
\end{equation}
where the coupling constant $\overline{k}=\sqrt{\left| \overline{\Lambda }%
\right| }/(4\sqrt{2}G_{(gr)}) $ and $G_{(gr)}$ is the gravitational
constant. The one d--form from (\ref{actcs}) $A=A_\mu ^{\underline{a}}T_{%
\underline{a}}\delta x^\mu $ is a gauge d--field for a Lie algebra with
generators $\left\{ T_{\underline{a}}\right\} .$ Following \cite{05cr} we
choose
\[
\left( T_{\underline{a}}\right) _{\underline{b}}^{\quad \underline{c}%
}=-\epsilon _{\underline{a}\underline{b}\underline{d}}\eta ^{\underline{d}%
\underline{c}},~\eta _{\underline{a}\underline{b}}=diag\left( -1,1,1\right)
,~\epsilon _{\underline{0}\underline{1}\underline{2}}=1
\]
and considering $Tr$ as the ordinary matrix trace we write
\begin{eqnarray}
[T_{\underline{a}},T_{\underline{b}}]= f_{\underline{a}\underline{b}}^{\quad
\underline{c}}T_{\underline{c}}= \epsilon _{\underline{a}\underline{b}
\underline{d}}\eta ^{\underline{d}\underline{c}}T_{\underline{c}}, ~TrT_{%
\underline{a}}T_{\underline{b}}= 2\eta _{\underline{a}\underline{b}},
\nonumber \\
~g_{\mu \nu }= 2\eta _{\underline{a}\underline{b}}e_\mu ^{\underline{a}%
}e_\nu ^{\underline{b}},~\eta ^{\underline{a}\underline{d}} \eta ^{%
\underline{b}\underline{e}} f_{\underline{a}\underline{b}}^{\quad \underline{%
c}} f_{\underline{d}\underline{e}}^{\quad \underline{s}}= -2\eta ^{%
\underline{c}\underline{s}}.  \nonumber
\end{eqnarray}

If the manifold $\widetilde{E}$ is closed the action (\ref{1act1}) is
invariant under locally anisotropic gauge transforms
\[
\widetilde{A}\rightarrow A=q^{-1}\widetilde{A}q+q^{-1}\delta q.
\]
This invariance is broken if $\widetilde{E}$ has a boundary $\partial
\widetilde{E}.$ In this case we must add to (\ref{actcs}) a boundary term,
written in $\left( v,\theta \right) $--coordinates as
\begin{equation}  \label{act2}
\overline{I}_{CS}^{\prime }=-\frac{\overline{k}}{4\pi }\int\nolimits_{%
\partial \widetilde{E}}TrA_\theta A_v,
\end{equation}
which results in a term proportional to the standard chiral
Wess--Zumino--Witten (WZW) action \cite{05m,05e}:
\[
\left( \overline{I}_{CS}+\overline{I}_{CS}^{\prime }\right) [A]=\left(
\overline{I}_{CS}+\overline{I}_{CS}^{\prime }\right) [\overline{A}]-%
\overline{k}\ \overline{I}_{WZW}^{+}[q,\overline{A}]
\]
where
\begin{equation}  \label{1act3}
\overline{I}_{WZW}^{+}[q,\overline{A}]=\frac 1{4\pi }\int\nolimits_{\partial
\widetilde{E}}Tr\left( q^{-1}\delta _\theta q\right) \left( q^{-1}\delta
_vq\right)
\end{equation}
\[
+\frac 1{2\pi }\int\nolimits_{\partial \widetilde{E}}Tr\left( q^{-1}\delta
_vq\right) \left( q^{-1}\overline{A}_\theta q\right) +\frac 1{12\pi
}\int\nolimits_{\widetilde{E}}Tr\left( q^{-1}\delta q\right) ^3.
\]

With respect to a locally anisotropic base the gauge locally anisotropic
field satisfies standard boundary conditions
\[
A_\theta ^{+}=A_v^{+}=\widetilde{A}_\theta ^{+}= \widetilde{A}_v^{+}=0.
\]

By applying the action (\ref{1act1}) with boundary terms (\ref{act2}) and (%
\ref{1act3}) we can formulate a statistical mechanics approach to the
(2+1)--dimensional locally anisotropic black holes with locally
anisotropically renormalized constants when the locally anisotropic entropy
of the black hole can be computed as the logarithm of microscopic states at
the anisotropically deformed horizon. In this case the Carlip's results \cite%
{05cr,05gm} could be generalized for locally anisotropic black holes. We present
here the formulas for one--loop corrected locally anisotropic temperature (%
\ref{temp}) and locally anisotropic entropy (\ref{entropy})
\[
\overline{\beta }_0=-\frac \pi {4\overline{\Lambda }\hbar G_{(gr)}\
\overline{r}_{+}}\left( 1+\frac{8\hbar G_{(gr)}}{\sqrt{|\overline{\Lambda }|}%
}\right) \mbox{ and } \overline{S}^{(a)}=\frac{\pi \overline{r}_{+}}{2\hbar
G_{(gr)}}\left( 1+\frac{8\hbar G_{(gr)}}{\sqrt{|\overline{\Lambda }|}}%
\right) .
\]
We do not yet have a general accepted approach even to the thermodynamics
and its statistical mechanics foundation of locally isotropic black holes
and this problem is not solved for locally anisotropic black holes for which
one should be associated a model of nonequilibrium thermodynamics.
Nevertheless, the formulas presented in this section allows us a calculation
of basic locally anisotropic thermodynamical values for equilibrium locally
anisotropic configurations by using locally anisotropically renormalized
constants.

\section{Conclusions and Discussion}

In this paper we have aimed to justify the use of moving frame method in
construction of metrics with generic local anisotropy, in general relativity
and its modifications for higher and lower dimension models \cite{05v3,05v4}.

We argued that the 3D gravity reformulated with respect to anholonomic
frames (with two holonomic and one anholonomic coordinates) admits new
classes of solutions of Einstein equations, in general, with nonvanishing
cosmological constants. Such black hole like and another type ones, with
deformed horizons, variation of constants and locally anisotropic
gravitational polarizations in the vacuum case induced by anholonomic moving
triads with associated nonlinear connection structure, or (in the presence
of 3D matter) by self--consistent distributions of matter energy density and
pressure and dreibein (3D moving frame) fields.

The solutions considered in the present paper have the following
properties:\ 1) they are exact solutions of 3D Einstein equations;\ 2) the
integration constants are to be found from boundary conditions and
compatibility with locally isotropic limits;\ 3) having been rewritten in
'pure' holonomic variables the 3D metrics are off--diagonal;\ 4) it is
induced a nontrivial torsion structure which vanishes in holonomic
coordinates;\ for vacuum solutions the 3D gravity is transformed into a
teleparallel theory;\ 5) such solutions are characterized by nontrivial
nonlinear connection curvature.

The arguments in this paper extend the results in the literature on the
black hole thermodynamics by elucidating the fundamental questions of
formulation of this theory for anholonomic gravitational systems with local
frame anisotropy. We computed the entropy and temperature of black holes
with elliptic horizons and/or with anisotropic variation and
renormalizations of constants.

We also showed that how the 3D gravity models with anholonomic constraints
can be transformed into effective Chern--Simons theories and following this
priority we computed the locally anisotropic quantum corrections for the
entropy and temperature of black holes.

Our results indicate that there exists a kind of universality of inducing
locally anisot\-rop\-ic interactions in physical theories formulated in
mixed holonomic--anholonomic variables:\ the spacetime geometry and
gravitational field are effectively polarized by imposed constraints which
could result in effective renormalization and running of interaction
constants.

Finally, we conclude that problem of definition of adequate systems of
reference for a prescribed type of symmetries of interactions could be of
nondynamical nature if we fix at the very beginning the class of admissible
frames and symmetries of solutions, but could be transformed into a
dynamical task if we deform symmetries (for instance, a circular horizon
into a elliptic one) and try to find self--consistently a corresponding
anholomic frame for which the metric is diagonal but with generic
anisotropic structure).

\subsection*{Acknowledgements}

%\acknowledgments
The S. V. work was supported by a Legislative Award of the
California University at Fresno and a sabbatical grant supported
by the Spanish Ministry of Education and Science.

%%%%%%%%%%%%%%%%%%%%%%%%%%%%%%%%%%%%%%%%%%%%%%%%%%%%%%%%%%%%%%%%%%%%%%%%%%%%%

\chapter[Anholonomic Thermodynamic Geometry]
{Anholonomic Frames and Thermodynamic Geometry  of 3D  Black Holes }

{\bf Abstract}
\footnote{ \copyright\  S. Vacaru, P. Stavrinos and  D. Gon\c ta,
Anholonomic Frames and Thermodynamic Geometry  of 3D  Black Holes, gr-qc/0106069}

We study new classes three dimensional  black hole solutions of Einstein
 equations written in two holonomic and one anholonomic variables with respect to
  anholonomic frames.  Thermodynamic properties of such $(2+1)$--black holes with
 generic local  aniso\-tro\-py (for instance, with elliptic horizons) are studied
 by applying geometric methods.
 The corresponding thermodynamic systems are  three dimensional with entropy $S$ being a
  hypersurface function on  mass $M,$ anisotropy angle $\theta$  and eccentricity of elliptic
 deformations $\varepsilon.$ Two--dimensional curved thermodynamic
 geometries for locally anistropic deformed black holes are constructed
 after integration on anisotropic parameter $\theta$. Two approaches,
 the first one  based on two--dimensional hypersurface parametric
 geometry and the second one developed in a Ruppeiner--Mrugala--Janyszek
 fashion, are analyzed.  The thermodynamic  curvatures are computed and
 the critical points of curvature vanishing are defined.

\section{Introduction}

This is the second paper in a series in which we examine black holes for
spacetimes with generic local anisotropy. Such spacetimes are usual
pseudo--Riemannian spaces for which an anholonomic frame structure  By using
moving anholonomic frames one can construct solutions of Einstein equations
with deformed spherical  symmetries (for instance, black holes with elliptic
horizons  (in three dimensions, 3D), black tori and another type
configurations) which are locally anisotropic \cite{06v3,06v6}.

In the first paper \cite{06v4} (hereafter referred to as Paper I)
we analyzed the low--dimensi\-on\-al locally anisotropic gravity (we
shall use terms like local\-ly an\-isotropic gravity, locally
an\-isotropic spacetime, locally anisotropic geometry, locally
an\-isotropic black holes and so on) and constructed new classes
of locally anisotropic $(2+1)$--dimensional black hole solutions.
We emphasize that in this work the splitting $(2+1)$  points not
to a space--time decomposition, but to a spacetime distribution in
two isotropic and one anisotropic coordinate.

In particular, it was shown following \cite{06v2} how black holes can recast
in a new fashion in generalized Kaluza--Klein spaces and emphasized that
such type solutions can be considered in the framework of usual Einstein
gravity on anholonomic manifolds. We discussed the physical properties of $%
(2+1)$--dimensional black holes with locally anisotropic matter, induced by
a rotating null fluid and by an inhomogeneous and non--static collapsing
null fluid, and examined the vacuum polarization of locally anisotropic
spacetime by non--rotating black holes with ellipsoidal horizon and by
rotating locally anisotropic black holes with time oscillating and
ellipsoidal horizons. It was concluded that a general approach to the
locally anisotropic black holes should be based on a kind of nonequilibrium
thermodynamics of such objects imbedded into locally anisotropic spacetime
background. Nevertheless, we proved that for the simplest type of locally
anisotropic black holes theirs thermodynamics could be defined in the
neighborhoods of some equilibrium states when the horizons are deformed but
constant with respect to a frame base locally adapted to a nonlinear
connection structure which model a locally anisotropic configuration.

In this paper we will specialize to the geometric thermodynamics of, for
simplicity non--rotating, locally anisotropic black holes with elliptical
horizons. We follow the notations and results from the Paper I which are
reestablished in a manner compatible in the locally isotropic thermodynamic
\cite{06cai} and spacetime \cite{06btz} limits with the
Banados--Teitelboim--Zanelli (BTZ) black hole. This new approach (to black
hole physics) is possible for locally anisotropic spacetimes and is based on
classical results \cite{06gp,06klim,06kre,06sien}.

Since the seminal works of Bekenstein \cite{06bek}, Bardeen, Carter and
Hawking \cite{06bch} and Hawking \cite{06haw1}, black holes were shown to have
properties very similar to those of ordinary thermodynamics. One was treated
the surface gravity on the event horizon as the temperature of the black
hole and proved that a quarter of the event horizon area corresponds to the
entropy of black holes. At present time it is widely believed that a black
hole is a thermodynamic system (in spite of the fact that one have been
developed a number of realizations of thermodynamics involving radiation)
and the problem of statistical interpretation of the black hole entropy is
one of the most fascinating subjects of modern investigations in
gravitational and string theories.

In parallel to the 'thermodynamilazation' of black hole physics
one have developed a new approach to the classical thermodynamics
based of Riemannian geometry and its generalizations (a review on
this subject is contained in Ref. \cite{06rup}). Here is to be
emphasized that geometrical methods have always played an
important role in thermodynamics (see, for instance, a work by
Blaschke \cite{06bla} from 1923). Buchdahl used in 1966 a Euclidean
metric in thermodynamics \cite{06buch} and then Weinhold considered
a sort of Riemannian metric \cite{06wein}. It is considered that
the Weinhold's metric has not physical interpretation in the
context of purely equilibrium thermodynamics \cite{06rup0,06rup} and
Ruppeiner introduced a new metric (related via the temperature
$T$ as the conformal factor with the Weinhold's metric).

The thermodynamical geometry was generalized in various directions, for
instance, by Janyszek and Mrugala \cite{06jm1,06jm2,06m3} even to discussions of
applications of Finsler geometry in thermodynamic fluctuation theory and for
nonequilibrium thermodynamics \cite{06sien}.

Our goal will be to provide a characterization of thermodynamics of $(2+1)$%
--dimen\-si\-on\-al locally anisotropic black holes with elliptical (constant in
time) horizon obtained in \cite{06v3,06v4}. From one point of view we shall
consider the thermodynamic space of such objects (locally anisotropic black
holes in local equilibrium with locally anisotropic spacetime ether) to
depend on parameter of anisotropy, the angle $\theta ,$ and on deformation
parameter, the eccentricity $\varepsilon .$ From another point, after we
shall integrate the formulas on $\theta ,$ the thermodynamic geometry will
be considered in a usual two--dimensional Ruppeiner--Mrugala--Janyszek
fashion. The main result of this work are the computation of thermodynamic
curvatures and the proof that constant in time elliptic locally anisotropic
black holes have critical points of vanishing of curvatures (under both
approaches to two--dimensional thermodynamic geometry) for some values of
eccentricity, i. e. for under corresponding deformations of locally
anisotropic spacetimes.

The paper is organized as follows. In Sec. 2, we briefly review
the geometry pseudo--Riemannian spaces provided with anholonimic
frame and
associated nonlinear connection structure and present the $(2+1)$%
--dimensional constant in time elliptic black hole solution. In
Sec. 3, we state the thermodynamics of nearly equilibrium
stationary locally anisotropic black holes and establish the
basic thermodynamic law and relations. In Sec. 4 we develop two
approaches to the thermodynamic geometry of locally anisotropic
black holes, compute thermodynamic curvatures and the equations
for critical points of vanishing of curvatures for some values of
eccentricity. In Sec. 5, we draw a discussion and conclusions.

\newpage

\section{Locally Anisotropic Spacetimes and Black Holes}

In this section we outline for further applications the basic results on $%
(2+1)$--dimensional locally anisotropic spacetimes and locally anisotropic
black hole solutions \cite{06v3,06v4}.

\subsection{Anholonomic frames and nonlinear connections in $(2+1)$%
--dimensi\-on\-al spacetimes}

A (2+1)--dimensional locally anisotropic spacetime is defined as a 3D
pseudo--Rie\-man\-ni\-an space provided with a structure of anholonomic
frame with
two holonomic coordinates $x^i,i=1,2$ and one anholonomic coordinate $y,$
for which $u=(x,y)=\{u^\alpha =(x^i,y)\},$ the Greek indices run values $%
\alpha =1,2,3,$ when $u^3=y.$ We shall use also underlined indices, for
instance $\underline{\alpha },\underline{i},$ in order to emphasize that
some tensors are given with respect to a local coordinate base $\partial _{%
\underline{\alpha }}=\partial /\partial u^{\underline{\alpha }}.$

An anholonomic frame structure of triads (dreibein) is given by a set of
three independent basis fields
\[
e_\alpha (u)=e_\alpha ^{\underline{\alpha }}(u)\partial _{\underline{\alpha }%
}
\]
which satisfy the relations
\[
e_\alpha e_\beta -e_\beta e_\alpha =w_{\ \alpha \beta }^\gamma e_\gamma ,
\]
where $w_{\ \alpha \beta }^\gamma =w_{\ \alpha \beta }^\gamma
(u)$ are called anholonomy coefficients.

We investigate anholonomic structures with mixed holonomic and
anholonomic tri\-ads when
\[
e_\alpha ^{\underline{\alpha }}(u)=\{e_j^{\underline{i}}=\delta _j^i,e_j^{%
\underline{3}}=N_j^3(u)=w_i(u),e_3^{\underline{3}}=1\}.
\]
In this case we have to apply 'elongated' by N--coefficients operators
instead of usual local coordinate basis $\partial _\alpha =\partial
/\partial u^\alpha $ and $d^\alpha =du^\alpha ,$ (for simplicity we shall
omit underling of indices if this does not result in ambiguities):
\begin{eqnarray}
\delta _\alpha & = & (\delta _i,\partial _{(y)})=\frac \delta {\partial
u^\alpha }   \label{2.1} \\
& = & \left( \frac \delta {\partial x^i}=\frac \partial {\partial x^i}-
w_i\left( x^j,y\right) \frac \partial {\partial y}, \partial _{(y)} = \frac %
\partial {\partial y}\right)  \nonumber
\end{eqnarray}
and their duals
\begin{eqnarray}
\delta ^\beta & = & \left( d^i,\delta ^{(y)}\right) = \delta u^\beta
\label{2.2} \\
& = & \left( d^i = dx^i, \delta ^{(y)} = \delta y=dy+w_k\left( x^j,y\right)
dx^k\right) .  \nonumber
\end{eqnarray}

The coefficients $N=\{N_i^3\left( x,y\right) =w_i\left(
x^j,y\right) \},$ are associated to a nonlinear connection (in
brief, N--connection, see \cite {06barthel}) structure which on
pseudo-Riemannian spaces defines a locally anisotropic, or
equivalently, mixed holo\-no\-mic--anholonomic structure. The
geometry of N--connection was investigated for vector bundles and
generalized Finsler geometry \cite{06ma} and for superspaces and
locally anisotropic (super)gravity and string theory \cite{06v2}
with applications in general relativity, extra dimension gravity
and formulation of locally anisotropic kinetics and
thermodynamics on curved spacetimes \cite{06v3,06v4,06v6}. In this paper
(following the Paper I) we restrict our considerations to the
simplest case with one anholonomic (anisotropic) coordinate when
the N--connection is associated to a subclass of anholonomic
triads (\ref{2.1}), and/or (\ref{2.2}), defining some locally
anisotropic frames (in brief, anholonomic basis, anholonomic
frames).

With respect to a fixed structure of locally anisotropic bases and their
tensor products we can construct distinguished, by N--connection, tensor
algebras and various geometric objects (in brief, one writes d--tensors,
d--metrics, d--connections and so on).

A symmetrical locally anisotropic metric, or d--metric, could be written
with respect to an anhlonomic basis (\ref{2.2}) %(2.2)
as
\begin{eqnarray}
\delta s^2 &= & g_{\alpha \beta }\left( u^\tau \right) \delta u^\alpha
\delta u^\beta   \label{2.3} \\
& = & g_{ij}\left( x^k,y\right) dx^idx^j+h\left( x^k,y\right) \left( \delta
y\right) ^2.  \nonumber
\end{eqnarray}
We note that the anisotropic coordinate $y$ could be both type time--like $%
(y=t,$ or space--like coordinate, for instance, $y=r,$ radial coordinate, or
$y=\theta ,$ angular coordinate).

\subsection{Non--rotating black holes with ellipsoidal horizon}

Let us consider a 3D locally anisotropic spacetime provided with local space
coordinates $x^{1}=r,\ x^{2}=\theta $ when as the anisotropic direction is
chosen the time like coordinate, $y=t.$ We proved (see the Paper I) that a
d--metric of type (\ref{2.3}), % (2.3),
\begin{equation}
\delta s^{2}=\Omega ^{2}\left( r,\theta \right) \left[ a(r)dr^{2}+b\left(
r,\theta \right) d\theta ^{2}+h(r,\theta )\delta t^{2}\right] ,  \label{2.4}
\end{equation}
where
\begin{eqnarray*}
\delta t &=&dt+w_{1}\left( r,\theta \right) dr+w_{2}(r,\theta )d\theta , \\
w_{1} &=&\partial _{r}\ln \left| \ln \Omega \right| ,w_{2}=\partial _{\theta
}\ln \left| \ln \Omega \right| ,
\end{eqnarray*}
for $\Omega ^{2}=\pm h(r,\theta ),$ satisfies the system of vacuum locally
anisotropic gravitational equations with cosmological constant $\Lambda
_{\lbrack 0]},$%
\[
R_{\alpha \beta }-\frac{1}{2}g_{\alpha \beta }R-\Lambda _{\lbrack
0]}g_{\alpha \beta }=0
\]
if
\[
a\left( r\right) =4r^{2}|\Lambda _{0}|,b(r,\theta )=\frac{4}{|\Lambda _{0}|}%
\Lambda ^{2}(\theta )\left[ r_{+}^{2}(\theta )-r^{2}\right] ^{2}
\]
and
\begin{equation}
h\left( r,\theta \right) =-\frac{4}{|\Lambda _{0}|r^{2}}\Lambda ^{3}(\theta )%
\left[ r_{+}^{2}(\theta )-r^{2}\right] ^{3}.  \label{2.5}
\end{equation}
The functions $a(r),b\left( r,\theta \right) $ and $h(x^{i},y)$
and the coefficients of nonlinear connection\\ $w_{i}(r,\theta
,t)$ (for this class of solutions being arbitrary prescribed
functions) were defined as to have compatibility with the locally
isotropic limit.

We construct a black hole like solution with elliptical horizon $%
r^2=r_{+}^2(\theta ),$ on which the function (\ref{2.5}) vanishes  if we
chose
\begin{equation}  \label{2.7}
r_{+}^2(\theta )=\frac{p^2}{\left[ 1+\varepsilon \cos \theta \right] ^2}.
\end{equation}
where $p$ is the ellipse parameter and $\dot \varepsilon $ is the
eccentricity. We have\ to identify
\[
p^2=r_{+[0]}^2=-M_0/\Lambda _0,
\]
where\ $r_{+[0]},M_0$ and $\Lambda _0$ are respectively the
horizon radius, mass parameter and the cosmological constant of
the non--rotating BTZ solution \cite{06btz} if we wont to have a
connection with locally isotropic limit with $\varepsilon
\rightarrow 0.$ In the simplest case we can consider that the
elliptic horizon (\ref{2.7}) is modelled by an anisotropic mass
\begin{equation}  \label{3.1}
M\left( \theta ,\varepsilon \right) =\frac{M_0}{2\pi \left( 1+\varepsilon
\cos \left( \theta -\theta _0\right) \right) ^2}=\frac{r_{+}^2}{2\pi }
\end{equation}
and constant effective cosmological constant, $\Lambda (\theta )\simeq
\Lambda _0.$ The coefficient $2\pi \,$ was introduced in order to have the
limit
\begin{equation}  \label{3.2}
\lim _{\varepsilon \rightarrow 0}2\int\limits_0^\pi M\left( \theta
,\varepsilon \right) d\theta =M_0.
\end{equation}
Throughout this paper, the units $c=\hbar =k_B=1$ will be used, but we shall
consider that for an locally anisotropically renormalized gravitational
constant $8G_{(gr)}^{(a)}\neq 1,$ see \cite{06v4}.

\section{On the Thermodynamics of Elliptical Black Ho\-les}

In this paper we will be interested in thermodynamics of locally anisotropic
black holes defined by a d--metric (\ref{2.4}). %(2.4).

The Hawking temperature $T\left( \theta ,\varepsilon \right) $ of a locally
an\-iso\-trop\-ic black hole is anisotropic and is computed by using the
anisotropic mass %(3.1):%
(\ref{3.1}):
\begin{equation}  \label{3.3}
T\left( \theta ,\varepsilon \right) =\frac{M\left( \theta ,\varepsilon
\right) }{2\pi r_{+}\left( \theta ,\varepsilon \right) }=\frac{r_{+}\left(
\theta ,\varepsilon \right) }{4\pi ^2}>0.
\end{equation}

The two parametric analog of the Bekenstein--Hawking entropy is to be
defined as
\begin{equation}  \label{3.4}
S\left( \theta ,\varepsilon \right) =4\pi r_{+}=\sqrt{32\pi ^3}\sqrt{M\left(
\theta ,\varepsilon \right) }
\end{equation}

The introduced thermodynamic quantities obey the first law of thermodynamics
(under the supposition that the system is in local equilibrium under the
variation of parameters $\left( \theta ,\varepsilon \right) $)
\begin{equation}  \label{3.5}
\Delta M\left( \theta ,\varepsilon \right) =T\left( \theta ,\varepsilon
\right) \Delta S,
\end{equation}
where the variation of entropy is
\[
\Delta S=4\pi \Delta r_{+}=4\pi \frac{1 }{\sqrt{M\left( \theta ,\varepsilon
\right) }}\left( \frac{\partial M}{\partial \theta }\Delta \theta +\frac{%
\partial M}{\partial \varepsilon }\Delta \varepsilon \right) .
\]
According to the formula $C=\left( \partial m/\partial T\right) $ we can
compute the heat capacity
\[
C=2\pi r_{+}\left( \theta ,\varepsilon \right) =2\pi \sqrt{M\left( \theta
,\varepsilon \right) }.
\]
Because of $C>0$ always holds the temperature is increasing with the mass.

The formulas (\ref{3.1})--(\ref{3.5}) %(3.1)--(3.5)
can be integrated on angular variable $\theta $ in order to obtain some
thermodynamic relations for black holes with elliptic horizon depending only
on deformation parameter, the eccentricity $\varepsilon .$

For a elliptically deformed black hole with the outer horizon $r_{+}$ given
by formula (\ref{3.4}) %(3.4)
the depending on eccentricity\cite{06v4} Bekenstein--Hawking entropy is
computed as
\[
S^{(a)}\left( \varepsilon \right) =\frac{L_{+}}{4G_{(gr)}^{(a)}},
\]
were
\[
L_{+}\left( \varepsilon \right) =4\int\limits_0^{\pi /2}r_{+}\left( \theta
,\varepsilon \right) d\theta
\]
is the length of ellipse's perimeter and $G_{(gr)}^{(a)}$ is the three
dimensional gravitational coupling constant in locally anisotropic media
(the index $\left( a\right) $ points to locally anisotropic
renormalizations), and has the value
\begin{equation}  \label{3.6}
S^{(a)}\left( \varepsilon \right) =\frac{2p}{G_{(gr)}^{(a)}\sqrt{%
1-\varepsilon ^2}}arctg\sqrt{\frac{1-\varepsilon }{1+\varepsilon }}.
\end{equation}
If the eccentricity vanishes, $\varepsilon =0,$ we obtain the locally
isotropic formula with $p$ being the radius of the horizon circumference,
but the constant $G_{(gr)}^{(a)}$ could be locally anisotropic renormalized.

The total mass of a locally anisotropic black hole of eccentricity $%
\varepsilon $ is found by integrating (\ref{3.1}) %(3.1)
on angle $\theta :$%
\begin{equation}  \label{3.7}
M\left( \varepsilon \right) =\frac{M_0}{\left( 1-\varepsilon ^2\right) ^{3/2}%
}
\end{equation}
which satisfies the condition (\ref{3.2}). %(3.2).

The integrated on angular variable $\theta $ temperature $T\left(
\varepsilon \right) $ is to by defined by using $T\left( \theta ,\varepsilon
\right) $ from (\ref{3.3}), %(3.3),
\begin{equation}  \label{3.8}
T\left( \varepsilon \right) =4\int\limits_0^{\pi /2}T\left( \theta
,\varepsilon \right) d\theta =\frac{2\sqrt{M_0}}{\pi ^2\sqrt{1-\varepsilon ^2%
}}arctg\sqrt{\frac{1-\varepsilon }{1+\varepsilon }.}
\end{equation}

Formulas (\ref{3.6})--(\ref{3.8}) %(3.6)-(3.8)
describes the thermodynamics of $\varepsilon $--deformed black holes.

Finally, in this section, we note that a black hole with elliptic horizon is
to be considered as a thermodynamic subsystem placed into the anisotropic
ether bath of spacetime. To the locally anisotropic ether one associates a
continuous locally anisotropic medium assumed to be in local equilibrium.
The locally anisotropic black hole subsystem is considered as a subsystem
described by thermodynamic variables which are continuous field on variables
$\left( \theta ,\varepsilon \right) ,$ or in the simplest case when one have
integrated on $\theta ,$ on $\varepsilon .$ It will be our first task to
establish some parametric thermodynamic relations between the mass $m\left(
\theta ,\varepsilon \right) $ (equivalently, the internal locally
anisotropic black hole energy), temperature $T\left( \theta ,\varepsilon
\right) $ and entropy $S\left( \theta ,\varepsilon \right) .$

\section[Thermodynamic Metrics and Curvatures]
{Thermodynamic Metrics and Curvatures of An\-iso\-tro\-pic Black Holes}

We emphasize in this paper two approaches to the thermodynamic
geometry of nearly equilibrium locally anisotropic black holes
based on their thermodynamics. The first one is to consider the
thermodynamic space as depending locally on two parameters
$\theta$ and $\varepsilon$ and to compute the corresponding metric
and curvature following standard formulas from curved
bi--dimensional hypersurface Riemannian geometry. The second
possibility is to take as basic the Ruppeiner metric in the
thermodynamic space with coordinates $(M, \varepsilon ),$ in a
manner proposed in Ref. \cite{06cai} with that difference that as
the extensive coordinate is taken the black hole eccentricity
$\varepsilon$ (instead of the usual angular momentum $J$ for
isotropic $(2+1)$--black holes). Of course, in this case we shall
background our thermodynamic geometric constructions starting from
the
relations %(3.6)--(3.8).
(\ref{3.6})--(\ref{3.8}).

\subsection{The thermodynamic parametric geometry}

Let us consider the thermodynamic parametric geometry of the elliptic
(2+1)--dimen\-si\-onal black hole based on its thermodynamics
 given by formulas (\ref{3.1})--(\ref{3.5}). %(3.1)--(3.5).

Rewriting equations (\ref{3.5}, %(3.5),
we have
\[
\Delta S=\beta \left( \theta ,\varepsilon \right) \Delta M\left( \theta
,\varepsilon \right) ,
\]
where $\beta \left( \theta ,\varepsilon \right) =1/T\left( \theta
,\varepsilon \right) $ is the inverse to temperature (\ref{3.3}). %(3.3).
This case is quite different from that from \cite{06cai,06ferrara} where there
are considered, respectively, BTZ and dilaton black holes (by introducing
Ruppeiner and Weinhold thermodynamic metrics). Our thermodynamic space is
defined by a hypersurface given by parametric dependencies of mass and
entropy. Having chosen as basic the relative entropy function,
\[
\varsigma =\frac{S\left( \theta ,\varepsilon \right) }{4\pi \sqrt{M_0}}=%
\frac 1{1+\varepsilon \cos \theta },
\]
in the vicinity of a point $P=(0,0),$ when, for simplicity, $\theta _0=0,$
our hypersurface is given locally by conditions
\[
\varsigma =\varsigma \left( \theta ,\varepsilon \right) \mbox{ and }
grad_{|P}\varsigma =0.
\]
For the components of bi--dimensional metric on the hypersurface
we have
\begin{eqnarray}
g_{11} & = & 1+\left( \frac{\partial \varsigma } {\partial \theta }\right)
^2, \ g_{12} = \left( \frac{\partial \varsigma }{\partial \theta }\right)
\left(\frac{\partial \varsigma }{\partial \varepsilon }\right) ,  \nonumber
\\
g_{22} & = & 1+ \left( \frac{\partial \varsigma }{\partial \varsigma }%
\right) ^2,  \nonumber
\end{eqnarray}
The nonvanishing component of curvature tensor in the vicinity of the point $%
P=(0,0)$ is
\[
R_{1212}=\frac{\partial ^2\varsigma }{\partial \theta ^2}\frac{\partial
^2\zeta }{\partial \varepsilon ^2}-\left( \frac{\partial ^2\varsigma }{%
\partial \varepsilon \partial \theta }\right) ^2
\]
and the curvature scalar is
\begin{equation}  \label{4.1}
R=2R_{1212}.
\end{equation}

By straightforward calculations we can find the condition of vanishing of
the curvature (\ref{4.1}) %(4.1)
when
\begin{equation}  \label{4.2}
\varepsilon _{\pm }=\frac{-1\pm (2-\cos ^2\theta )}{\cos \theta \left(
3-\cos ^2\theta \right) }.
\end{equation}
So, the parametric space is separated in subregions with elliptic
eccentricities $0<\varepsilon _{\pm }<0$ and $\theta $ satisfying conditions
(\ref{4.2}). %(4.2).

Ruppeiner suggested that the curvature of thermodynamic space is a measure
of the smallest volume where classical thermodynamic theory based on the
assumption of a uniform environment could conceivably work and that near the
critical point it is expected this volume to be proportional to the scalar
curvature \cite{06rup}. There were also proposed geometric equations relating
the thermodynamic curvature via inverse relations to free energy. Our
definition of thermodynamic metric and curvature in parametric spaces
differs from that of Ruppeiner or Weinhold and it is obvious that relations
of type (\ref{4.2}) %(4.2)
(stating the conditions of vanishing of curvature) could be related with
some conditions for stability of thermodynamic space under variations of
eccentricity $\varepsilon $ and anisotropy angle $\theta .$ This
interpretation is very similar to that proposed by Janyszek and Mrugala \cite
{06jm1} and supports the viewpoint that the first law of thermodynamics makes
a statement about the first derivatives of the entropy, the second law is
for the second derivatives and the curvature is a statement about the third
derivatives. This treatment holds good also for the parametric thermodynamic
spaces for locally anisotropic black holes.

\subsection{Thermodynamic Metrics and Eccentricity of Black Hole}

A variant of thermodynamic geometry of locally an\-iso\-trop\-ic black holes
could be grounded on integrated on anisotropy angle $\theta $ formulas (\ref
{3.6})--(\ref{3.8}). %(3.6)-(3.8).
The Ruppeiner metric of elliptic black holes in coordinates $\left(
M,\varepsilon \right) $ is
\begin{equation}  \label{4.3}
ds_R^2=-\left( \frac{\partial ^2S}{\partial M^2}\right) _\varepsilon
dM^2-\left( \frac{\partial ^2S}{\partial \varepsilon ^2}\right)
_Md\varepsilon ^2.
\end{equation}
For our further analysis we shall use dimensionless values $\mu =M\left(
\varepsilon \right) /M_0$ and $\zeta =S^{(a)}G_{gr}^{(a)}/2p$ and consider
instead of (\ref{4.3}) %(4.3)
the thermodynamic diagonal metrics $g_{ij}\left( a^1,a^2\right)$ $ =g_{ij}(\mu
,\varepsilon )$ with components
\begin{equation}  \label{4.4}
g_{11}=-\frac{\partial ^2\zeta }{\partial \mu ^2}=-\zeta _{,11}\mbox{\ and \
}g_{22}=-\frac{\partial ^2\zeta }{\partial \varepsilon ^2}=-\zeta _{,22},
\end{equation}
where by comas we have denoted partial derivatives.

The expressions (\ref{3.6}) %(3.6)
and (\ref{3.6}) %(3.7)
are correspondingly rewritten as
\[
\zeta =\frac 1{\sqrt{1-\varepsilon ^2}}arctg\sqrt{\frac{1-\varepsilon }{%
1+\varepsilon }}
\]
and
\[
\mu =\left( 1-\varepsilon ^2\right) ^{-3/2}.
\]

By straightforward calculations we obtain
\begin{eqnarray}
\zeta _{,11} & = & -\frac 19\left( 1-\varepsilon ^2\right) ^{5/2}arctg\sqrt{%
\frac{1-\varepsilon }{1+\varepsilon }}  \nonumber \\
{\ } & + & \frac 1{9\varepsilon }\left( 1-\varepsilon ^2\right) ^3+\frac 1{%
18\varepsilon ^4}\left( 1-\varepsilon ^2\right) ^4  \nonumber
\end{eqnarray}
and
\[
\zeta _{,22}=\frac{1+2\varepsilon ^2}{\left( 1-\varepsilon ^2\right) ^{5/2}}%
arctg\sqrt{\frac{1-\varepsilon }{1+\varepsilon }}-\frac{3\varepsilon }{%
\left( 1-\varepsilon ^2\right) ^2}.
\]

The thermodynamic curvature of metrics of type (\ref{4.4}) %(4.4)
can be written in terms of second and third derivatives \cite{06jm1} by using
third and second order determinants:
\begin{eqnarray}
R & = & \frac 12\left|
\begin{array}{ccc}
-\zeta _{,11} & 0 & -\zeta _{,22} \\
-\zeta _{,111} & -\zeta _{,112} & 0 \\
-\zeta _{,112} & 0 & -\zeta _{,222}
\end{array}
\right| \times \left|
\begin{array}{cc}
-\zeta _{,11} & 0 \\
0 & -\zeta _{,22}
\end{array}
\right| ^{-2}  \nonumber \\
{} & = & -\frac 12\left( \frac 1{\zeta _{,11}}\right) _{,2}\times \left(
\frac{\zeta _{,11}}{\zeta _{,22}}\right) _{,2}.   \label{4.5}
\end{eqnarray}
The conditions of vanishing of thermodynamic curvature (\ref{4.5}) %(4.5)
are as follows
\begin{equation}  \label{4.6}
\zeta _{,112}\left( \varepsilon _1\right) =0 \mbox{ or } \left( \frac{\zeta
_{,11}}{\zeta _{,22}}\right) _{,2}\left( \varepsilon _2\right) =0
\end{equation}
for some values of eccentricity, $\varepsilon =\varepsilon _1$ or $%
\varepsilon =\varepsilon _2,$ satisfying conditions $0<\varepsilon _1<1$ and
$0<\varepsilon _2<1.$ For small deformations of black holes, i.e. for small
values of eccentricity, we can approximate $\varepsilon _1\approx 1/\sqrt{5.5%
}$ and $\varepsilon _2\approx 1/(18\lambda ),$ where $\lambda $ is a
constant for which $\zeta _{,11}=\lambda \zeta _{,22}$ and the condition $%
0<\varepsilon _2<1$ is satisfied. We omit general formulas for curvature (%
\ref{4.5}) %(4.5)
and conditions (\ref{4.6}), % (4.6),
when the critical points $\varepsilon _1$ and/or $\varepsilon _2$ must be
defined from nonlinear equations containing $arctg\sqrt{\frac{1-\varepsilon
}{1+\varepsilon }}$ and powers of $\left( 1-\varepsilon ^2\right) $ and $%
\varepsilon .$

\section{Discussion and Conclusions}

In closing, we would like to discuss the meaning of geometric thermodynamics
following from locally an\-iso\-trop\-ic black holes.

(1) {\it Nonequilibrium thermodynamics of locally\newline
anisotropic black holes in  locally anisotropic spacetimes}. In this paper
and in the Paper I \cite{06v4}  we concluded that the thermodynamics in
locally anisotropic spacetimes has a generic nonequilibrium character and
could be developed in a geometric fashion following the approach proposed by
S. Sieniutycz, P. Salamon and R. S. Berry \cite{06sien,06sal}. This is a new
branch of black hole thermodynamics which should be based on locally
anisotropic nonequilibrium thermodynamics and kinetics \cite{06v6}.

(2) {\it Locally Anisotropic Black holes thermodynamics in vicinity of
equilibrium points}. The usual thermodynamical approach in the
Bekenstein--Hawking manner is valid for anisotropic black holes for a
subclass of such physical systems when the hypothesis of local equilibrium
is physically motivated and corresponding renormalizations, by locally
anisotropic spacetime parameters, of thermodynamical values are defined.

(3) {\it The geometric thermodynamics of locally an\-iso\-trop\-ic black
holes with constant in time elliptic horizon} was formulated following two
approaches: for a parametric thermodynamic space depending on anisotropy
angle $\theta$ and eccentricity $\varepsilon$ and in a standard
Ruppeiner--Mrugala--Janyszek fashion, after integration on anisotropy $\theta
$ but maintaining locally anisotropic spacetime deformations on $\varepsilon
.$

(4) {\it The thermodynamic curvatures of locally\newline
anisotropic black holes} were shown to have critical values of eccentricity
when the scalar curvature vanishes. Such type of thermodynamical systems are
rather unusual and a corresponding statistical model is not that for
ordinary systems composed by classical or quantum like gases.

(5) {\it Thermodynamic systems with constraints} require a new
geometric structure in addition to the thermodynamical metrics
which is that of nonlinear connection. We note this object must
be introduced both in spacetime geometry and in thermodynamic
geometry if generic anisotropies and constrained field and/or
thermodynamic behavior are analyzed.

\subsection*{Acknowledgements}
The S. V. work was supported both by a California State University
Legislative Award and a sabbatical visit supported by the Spanish
Ministry of Education and Science.

%%%%%%%%%%%%%%%%%%%%%%%%%%%%%%%%%%%%%%%%%%%%%%%%%%%%%%%%%%%%%%%%%%%%%%%%%%%%%

\chapter[Anisotropic and Running Hierarchies]
{Off--Diagonal 5D Metrics and Mass Hierarchies with Anisotropies
 and Running Constants  }

{\bf Abstract}
\footnote{\copyright\
 S. Vacaru, Off--Diagonal 5D Metrics and Mass Hierarchies with
 Anisotropies and Running Constants, hep-ph/0106268}

The gravitational equations of the three dimensional (3D) brane world are
investigated for both off--diagonal and warped 5D metrics which can be
diagonalized with respect to some anholonomic frames when the gravitational
and matter fields dynamics are described by mixed sets of holonomic and
anholonomic variables. We construct two new classes of exact solutions of
Kaluza--Klein gravity which generalize the Randall--Sundrum metrics to
configurations with running on the 5th coordinate gravitational constant and
anisotropic dependencies of effective 4D constants on time and/or space
variables. We conclude that by introducing gauge fields as off--diagonal
components of 5D metrics, or by considering  anholonomic frames modelling
some anisotropies in extra dimension  spacetime, we induce anisotropic
tensions (gravitational  polarizations) and running of constants on the
branes. This way  we can generate the TeV scale as a hierarchically
suppressed anisotropic mass scale and the Newtonian and general
relativistic gravity are reproduced with adequate precisions but  with
corrections which depend anisotropically on some  coordinates.

\vskip30pt

Recent approaches to String/M--theory and particle physics are based on the
idea that our universe is realized as a three brane, modelling a four
dimensional, 4D, pseudo--Riemannian spacetime, embedded in the 5D anti--de
Sitter ($AdS_5$) bulk spacetime. In such models the extra dimension need not
be small (they could be even infinite) if a nontrivial warped geometric
configuration, being essential for solving the mass hierarchy problem and
localization of gravity, can ''bound'' the matter fields on a 3D subspace on
which we live at low energies, the gravity propagating, in general, in a
higher dimension spacetime (see Refs.: \cite{07stringb} for string gravity
papers; \cite{07arkani} for extra dimension particle fields and gravity
phenomenology with effective Plank scale; \cite{07rs} for the simplest and
comprehensive models proposed by Randall and Sundrum; here we also point the
early works \cite{07akama} in this line and cite \cite{07shir} as some further
developments with supresymmetry, black hole solutions and cosmological
scenarios).

In higher di\-men\-sional gravi\-ty much at\-tention has been paid to
off--diago\-nal metrics beginning the Salam, Strathee and Perracci work \cite
{07sal} which showed that including off--diagonal components in higher
dimensional metrics is equi\-valent to including $U(1),SU(2)$ and $SU(3)$
gauge fields. Recently, the off--diagonal metrics were considered in a new
fashion by applying the method of anholonomic frames with associated
nonlinear connections \cite{07v} which allowed us to construct new classes of
solutions of Einstein's equations in three (3D), four (4D) and five (5D)
dimensions, with generic local anisotropy ({\it e.g.} static black hole and
cosmological solutions with ellipsoidal or torus symmetry, soliton--dilaton
2D and 3D configurations in 4D gravity, and wormhole and flux tubes with
anisotropic polarizations and/or running on the 5th coordinate constants
with different extensions to backgrounds of rotation ellipsoids, elliptic
cylinders, bipolar and torus symmetry and anisotropy.

The point of this paper is to argue that if the 5D gravitational
interactions are parametrized by off--diagonal metrics with a warped factor,
which could be related with an anholonomic higher dimensional gravitational
dynamics and/or with the fact that the gauge fields are included into a
Salam--Strathee--Peracci manner, the fundamental Plank scale $M_{4+d}$ in $%
4+d$ dimensions can be not only considerably smaller than the the effective
Plank scale, as in the usual locally isotropic Randall--Sundrum (in brief,
RS) scenarios, but the effective Plank constant is also anisotropically
polarized which could have profound consequences for elaboration of
gravitational experiments and for models of the very early universes.

We will give two examples with one additional dimension ($d=1$) when an
extra dimension gravitational anisotropic polarization on a space coordinate
is emphasized or, in the second case, a running of constants in time is
modelled. We will show that effective gravitational Plank scale is
determined by the higher-dimensional curvature and anholonomy of pentad
(funfbein, of frame basis) fields rather than the size of the extra
dimension. Such curvatures and anholonomies are not in conflict with the
local four-dimensional Poincare invariance.

We will present a higher dimensional scenario which provides a new RS like
approach generating anisotropic mass hierarchies. We consider that the 5D
metric is both not factorizable and off--diagonal when the four-dimensional
metric is multiplied by a ``warp'' factor which is a rapidly changing
function of an additional dimension and depend anisotropically on a space
direction and runs in the 5-th coordinate.

Let us consider a 5D pseudo--Riemannian spacetime provided with local
coordinates $u^\alpha =(x^i,y^a)=(x^1=x,x^2=f,x^3=y,y^4=s,y^5=p), $ where $%
\left( s,p\right) =\left( z,t\right) $ (Case I) or, inversely, $\left(
s,p\right) =\left( t,z\right) $ (Case II) -- or more compactly $u=(x,y)$ --
where the Greek indices are conventionally split into two subsets $x^i$ and $%
y^a$ labelled respectively by Latin indices of type $i,j,k,...=1,2,3$ and $%
a,b,...=4,5.$ The local coordinate bases, $\partial _\alpha =(\partial
_i,\partial _a),$ and their duals, $d^\alpha =\left( d^i,d^a\right) ,$ are
defined respectively as

\newpage

\begin{equation}
\partial _\alpha \equiv \frac \partial {du^\alpha }=(\partial _i=\frac %
\partial {dx^i},\partial _a=\frac \partial {dy^a}) \mbox{ and } d^\alpha
\equiv du^\alpha =(d^i=dx^i,d^a=dy^a).  \label{3pdif}
\end{equation}

For the 5D (pseudo) Riemannian interval $dl^{2}=G_{\alpha \beta }du^{\alpha
}du^{\beta }$ we choose the metric coefficients $G_{\alpha \beta }$ (with
respect to the coordinate frame (\ref{3pdif})) to be parametrized by a
off--diagonal matrix (ansatz)
\begin{equation}
\left[
\begin{array}{ccccc}
g+w_{1}^{\ 2}h_{4}+n_{1}^{\ 2}h_{5} & w_{1}w_{2}h_{4}+n_{1}n_{2}h_{5} &
w_{1}w_{3}h_{4}+n_{1}n_{3}h_{5} & w_{1}h_{4} & n_{1}h_{5} \\
w_{1}w_{2}h_{4}+n_{1}n_{2}h_{5} & 1+w_{2}^{\ 2}h_{4}+n_{2}^{\ 2}h_{5} &
w_{2}w_{3}h_{4}+n_{2}n_{3}h_{5} & w_{2}h_{4} & n_{2}h_{5} \\
w_{1}w_{3}h_{4}+n_{1}n_{3}h_{5} & w_{3}w_{2}h_{4}+n_{2}n_{3}h_{5} &
g+w_{3}^{\ 2}h_{4}+n_{3}^{\ 2}h_{5} & w_{3}h_{4} & n_{3}h_{5} \\
w_{1}h_{4} & w_{2}h_{4} & w_{3}h_{4} & h_{4} & 0 \\
n_{1}h_{5} & n_{2}h_{5} & n_{3}h_{5} & 0 & h_{5}
\end{array}
\right]   \label{1ansatz0}
\end{equation}
where the coefficients are some necessary smoothly class functions of type:
\begin{eqnarray*}
g &=&g(f,y)=a(f)\ b(y),h_{4}=h_{4}(f,y,s)=\eta _{4}(f,y)g(f,y)q_{4}(s), \\
h_{5} &=&h_{5}(f,y,s)=g(f,y)q_{5}(s),w_{i}=w_{i}(f,y,s),n_{i}=n_{i}(f,y,s).
\end{eqnarray*}

The metric (\ref{1ansatz0}) can be equivalently rewritten in the form
\begin{equation}
\delta l^{2}=g_{ij}\left( f,y\right) dx^{i}dx^{i}+h_{ab}\left( f,y,s\right)
\delta y^{a}\delta y^{b},  \label{3dmetric}
\end{equation}
with diagonal coefficients
\begin{equation}
g_{ij}=\left[
\begin{array}{lll}
g & 0 & 0 \\
0 & 1 & 0 \\
0 & 0 & g
\end{array}
\right] \mbox{ and }h_{ab}=\left[
\begin{array}{ll}
h_{4} & 0 \\
0 & h_{5}
\end{array}
\right]   \label{ansatzd}
\end{equation}
if instead the coordinate bases (\ref{3pdif}) one introduce the anholonomic
frames (anisotropic bases)
\begin{equation}
{\delta }_{\alpha }\equiv \frac{\delta }{du^{\alpha }}=(\delta _{i}=\partial
_{i}-N_{i}^{b}(u)\ \partial _{b},\partial _{a}=\frac{\partial }{dy^{a}}),%
\delta ^{\alpha }\equiv \delta u^{\alpha }=(\delta
^{i}=dx^{i},\delta ^{a}=dy^{a}+N_{k}^{a}(u)\ dx^{k})  \label{5ddif}
\end{equation}
where the $N$--coefficients are parametrized $N_{i}^{4}=w_{i}$ and $%
N_{i}^{5}=n_{i}.$

In this paper we consider a slice of $AdS_5$ provided with an anholnomic
frame structure (\ref{5ddif}) satisfying the relations $\delta _\alpha \delta
_\beta -\delta _\beta \delta _\alpha =W_{\alpha \beta }^\gamma \delta
_\gamma ,$ with nontrivial anholonomy coefficients
\begin{eqnarray}
W_{ij}^k &=&0,W_{aj}^k=0,W_{ia}^k=0,W_{ab}^k=0,W_{ab}^c=0,     \nonumber \\
 W_{ij}^a &=& \delta
_iN_j^a-\delta _jN_i^a,W_{bj}^a=-\partial _bN_j^a,W_{ia}^b=\partial _aN_j^b.
 \nonumber
 \end{eqnarray}

We assume there exists a solution of 5D Einstein equations with 3D brane
configuration that effectively respects the local 4D Poincare invariance
with respect to anholonomic frames (\ref{5ddif}) and that the metric ansatz (%
\ref{1ansatz0}) (equivalently, (\ref{ansatzd})) transforms into the usual RS
solutions
\begin{equation}
ds^2=e^{-2k|f|}\eta _{\underline{\mu }\underline{\nu }}dx^{\underline{\mu }%
}dx^{\underline{\nu }}+df^2  \label{solrs}
\end{equation}
for the data:\ $a(f) = e^{-2k|f|},\ k=const,\ b(y)=1, \eta _4(f,y) = 1,
q_4(s)=q_5(s)=1, w_i = 0,n_i=0, $ where $\eta _{\underline{\mu }\underline{%
\nu }}$ and $x^{\underline{\mu }}$ are correspondingly the diagonal metric
and Cartezian coordinates in 4D Minkowski spacetime and the
extra-dimensional coordinate $f$ is to be identified $f=r_c\phi ,$ ($%
r_c=const$ is the compactification radius, $0\leq f\leq \pi r_c$) like in
the first work \cite{07rs} (or 'f'' is just the coordinate 'y' in the second
work \cite{07rs}).

The set-up for our model is a single 3D brane with positive tension,
subjected to some anholonomic constraints, embedded in a 5D bulk spacetime
provided with a off--diagonal metric (\ref{1ansatz0}). In order to carefully
quantize the system, and treat the non--normalizable modes which will appear
in the Kaluza-Klein reduction, it is useful to work with respect to
anholonomic frames were the metric is diagonalized by corresponding
anholonomic transforms and is necessary to work in a finite volume by
introducing another brane at a distance $\pi r_c$ from the brane of
interest, and taking the branes to be the boundaries of a finite 5th
dimension. We can remove the second brane from the physical set-up by taking
the second brane to infinity.

The action for our anholonomic funfbein (pentadic) system is
\begin{eqnarray}
S &=&S_{gravity}+S_{brane}+S_{brane^{\prime }}  \label{anhbr} \\
S_{gravity} &=&\int \delta ^{4}x\int \delta f\sqrt{-G}\{-\Lambda
(f)+2M^{3}R\}, \nonumber \\
S_{brane}&=&\int \delta ^{4}x\sqrt{-g_{brane}}\{V_{brane}+%
{\cal L}_{brane}\},  \nonumber
\end{eqnarray}
where $R$ is the 5D Ricci scalar made out of the 5D metric, $G_{\alpha \beta
}$, and $\Lambda $ and $V_{brane}$ are cosmological terms in the bulk  and
boundary respectively. We write down $\delta ^{4}x$ and $\delta f,$ instead
of usual differentials $d^{4}x$ and $df,$ in order to emphasize that the
variational calculus should be performed by using N--elongated partial
derivatives and differentials (\ref{5ddif}). The coupling to the branes and
their fields and the related orbifold boundary conditions for vanishing
N--coefficients are described in Refs. \cite{07rs} and \cite{07me1}.

The Einstein equations,
$$R_{\beta }^{\alpha }-\frac{1}{2}\delta _{\beta
}^{\alpha }R=\Upsilon _{\beta }^{\alpha },$$
 for a diagonal
energy--moment\-um tensor $\Upsilon _{\alpha }^{\beta }=\left[
\Upsilon _{1},\Upsilon _{2},\Upsilon _{3},\Upsilon _{4},\Upsilon
_{5}\right] $ and following from the action (\ref{anhbr}) and for
the ansatz (\ref{1ansatz0}) (equivalently, (\ref{ansatzd})) with
$g=a(f)b(y)$ transform into
\begin{eqnarray}
\frac{1}{a}\left[ a_{1}^{\prime \prime }-\frac{(a^{\prime })^{2}}{2a}\right]
+\frac{\beta }{h_{4}h_{5}} &=&2\Upsilon _{1},\ \frac{(a^{\prime })^{2}}{2a}+%
\frac{P(y)}{a}+\frac{\beta }{h_{4}h_{5}}=2\Upsilon _{2}(f),  \label{einst} \\
\frac{1}{a}\left[ a^{\prime \prime }-\frac{(a^{\prime })^{2}}{2a}\right] +%
\frac{P(y)}{a} &=&2\Upsilon _{4},\ w_{i}\beta +\alpha _{i}=0,\ n_{i}^{\ast
\ast }+\gamma n_{i}^{\ast }=0,  \nonumber
\end{eqnarray}
where
\begin{eqnarray}
\alpha _{1} &=&\ {h_{5}^{\ast }}^{^{\bullet }}-\frac{{h_{5}^{\ast }}}{2}%
\left( \frac{h_{4}^{\bullet }}{h_{4}}+\frac{h_{5}^{\bullet }}{h_{5}}\right),
\ \alpha _{2}={h_{5}^{\ast }}^{^{\prime }}-\frac{{h_{5}^{\ast }}}{2}\left(
\frac{h_{4}^{^{\prime }}}{h_{4}}+\frac{h_{5}^{^{\prime }}}{h_{5}}\right),
\nonumber \\
\alpha _{3}&=&{h_{5}^{\ast }}^{\#}-\frac{{h_{5}^{\ast }}}{2}\left( \frac{%
h_{4}^{\#}}{h_{4}}+\frac{h_{5}^{\#}}{h_{5}}\right) ,  \nonumber \\
\beta  &=&h_{5}^{\ast \ast }-\frac{(h_{5}^{\ast })^{2}}{2h_{5}}-\frac{%
h_{5}^{\ast }h_{4}^{\ast }}{2h_{4}},P=\frac{1}{b^{2}}\left[ b^{\#\#}-\frac{%
(b^{\#})^{2}}{b}\right] ,\ \gamma =\frac{3}{2}\frac{h_{5}}{h_{5}}^{\ast }-%
\frac{h_{4}}{h_{4}}^{\ast },  \label{gamma}
\end{eqnarray}
the partial derivatives are denoted: $h^{\bullet }=\partial h/\partial
x^{1},h^{\prime }=\partial h/\partial x^{2},h^{\#}=\partial h/\partial x^{3},
$ $h^{\ast }=\partial h/\partial s.$

Our aim is to construct a metric
\begin{equation}
\delta s^{2}=g\left( f,y\right) [dx^{2}+dy^{2}+\eta _{4}\left( f,y\right)
\delta s^{2}+q_{5}(s)\delta p^{2}]+df^{2},  \label{1sol1b}
\end{equation}
with the anholonomic frame components defined by 'elongation' of
differentials, $\delta s=ds+w_{2}df+w_{3}dy,$ $\delta
p=dp+n_{1}dx+n_{2}df+n_{3}dy,$ and the ''warp'' factor being written in a
form similar to the RS solution
\begin{equation}
g\left( f,y\right) =a(f)b(y)=\exp [-2k_{f}|f|-2k_{y}|y|],  \label{warpan}
\end{equation}
which defines anisotropic RS like solutions of 5D Einstein equations with
variation on the 5th coordinate cosmological constant in the bulk and
possible variations of induced on the brane cosmological constants.

By straightforward calculations we can verify that a class of exact
solutions of the system of equations (\ref{einst}) for $P(y)=0\,$ (see (\ref
{gamma})) :
\[
h_{4}=g(f,y),h_{5}=g(f,y)\rho ^{2}(f,y,s),
\]
were
\begin{eqnarray*}
\rho (f,y,s) &=&|\cos \tau _{+}\left( f,y\right) |,\ \tau _{+}=\sqrt{\left(
\Upsilon _{4}-\Upsilon _{2}\right) g(f,y)},\Upsilon _{4}>\Upsilon _{2}; \\
&=&\exp [-\tau _{-}\left( f,y\right) s],\tau _{-}=\sqrt{\left( \Upsilon
_{2}-\Upsilon _{4}\right) g(f,y)},\Upsilon _{4}<\Upsilon _{2}; \\
&=&|c_{1}(f,y)+sc_{2}(f,y)|^{2},\Upsilon _{4}=\Upsilon _{2},
\end{eqnarray*}
\ and
\begin{eqnarray*}
w_{i} &=&-\partial _{i}(\ln |\rho ^{\ast }|)/(\ln |\rho ^{\ast }|)^{\ast },
\\
n_{i} &=&n_{i[0]}(f,y)+n_{i[1]}(f,y)\int \exp [-3\rho ]ds,
\end{eqnarray*}
with functions  $c_{1,2}(f,y)$ and $n_{i[0,1]}(f,y)$  to be stated by some
boundary conditions.  We emphasize that the constants $k_{f}$ and $k_{y}$
have to be defined from some experimental data.

The solution (\ref{1sol1b}) transforms into the usual RS solution (\ref{solrs}%
) if $k_{y}=0,n_{i[0,1]}(f,y)=0,\Lambda =\Lambda _{0}=const$ and $\Upsilon
_{2}\rightarrow \Upsilon _{2[0]}=-{\frac{\Lambda _{0}}{4M^{3}};}\ \Upsilon
_{1},\Upsilon _{3},\Upsilon _{4},\Upsilon _{5}\rightarrow \Upsilon _{\lbrack
0]}=\frac{V_{brane}}{4M^{3}}\delta (f)+\frac{V_{brane^{\prime }}}{4M^{3}}%
\delta (f-\pi r_{c}),$ which holds only when the boundary and bulk
cosmological terms are related by formulas $V_{brane}=-V_{brane^{\prime
}}=24M^{3}k_{f},~~\Lambda _{0}=-24M^{3}k_{f}^{2};\ $ we use values with the
index $[0]$ in order to emphasize that they belong to the usual (holonomic)
RS solutions. In the anholonomic case with ''variation of constants'' we
shall not impose such relations.

We note that using the metric (\ref{1sol1b}) with anisotropic warp factor (%
\ref{warpan}) it is easy to identify the massless gravitational fluctuations
about our classical solutions like in the usual RS cases but performing (in
this work) all computations with respect to anholonomic frames. All
off--diagonal fluctuations of the anholonomic diagonal metric are massive
and excluded from the low-energy effective theory.

We see that the physical mass scales are set by an anisotropic
symmetry--breaking scale, $v(y)\equiv e^{-k_{y}|y|}e^{-k_{f}r_{c}\pi }v_{0}.$%
\ This result the conclusion: any mass parameter $m_{0}$ on the visible
3-brane in the fundamental higher-dimensional theory with Salam--Strathee
--Peracci gauge interactions and/or effective anholonomic frames will
correspond to an anisotropic dependence on coordinate $y$ of the physical
mass $m(y)\equiv e^{-k_{y}|y|}e^{-kr_{c}\pi }m_{0}$ when measured with the
metric $\overline{g}_{\mu \nu }$ that appears in the effective Einstein
action, since all operators get re-scaled according to their four-dimensional
conformal weight. If $e^{kr_{c}\pi }$ is of order $10^{15}$, this mechanism
can produces TeV physical mass scales from fundamental mass parameters not
far from the Planck scale, $10^{19}$ GeV. Because this geometric factor is
an exponential, we clearly do not require very large hierarchies among the
fundamental parameters, $v_{0},k,M,$ and $\mu _{c}\equiv 1/r_{c}$; in fact,
we only require $kr_{c}\approx 50$. These conclusions were made in Refs.
\cite{07rs} with respect to diagonal (isotropic) metrics. But the physical
consequences could radically change if the off--diagonal metrics with
effective anholonomic frames and gauge fields are considered. In this case
we have additional dependencies on variable $y$ which make the fundamental
spacetime geometry to be locally anisotropic, polarized via dependencies
both on coordinate $y$  receptivity $k_{y}.$ We emphasize that our $y$
coordinate is not that from \cite{07rs}.

The phenomenological implications of these anisotropic scenarios for future
collider searches could be very distinctive: the geometry of experiments
will play a very important role. In such anisotropic models we also have a
roughly weak scale splitting with a relatively small number of excitations
which can be kinematically accessible at accelerators.

We also reconsider in an anisotropic fashion the derivation of the 4D
effective Planck scale $M_{Pl}$ given in Ref. \cite{07rs}. The 4D graviton
zero mode follows from the solution, Eq. (\ref{1sol1b}), by replacing the
Minkowski metric by a effective 4D metric $\overline{g}_{\mu \nu }$ which it
is described by an effective action following from substitution into Eq. (%
\ref{anhbr}),
\begin{equation}
S_{eff}\supset \int \delta ^{4}x\int_{0}^{\pi
r_{c}}df~2M^{3}r_{c}e^{-2k_{f}|f|}e^{-2k_{y}|y|}\sqrt{\overline{g}}~%
\overline{R},  \label{effaction}
\end{equation}
where $\overline{R}$ denotes the four-dimensional Ricci scalar made out of $%
\overline{g}_{\mu \nu }(x)$, in contrast to the five-dimensional Ricci
scalar, $R$, made out of $G_{MN}(x,f)$. We use the symbol $\delta ^{4}x$ in (%
\ref{anhbr}) in order to emphasize that our integration is adapted to the
anholonomic structu5e stated by the differentials (\ref{5ddif}). We also can
explicitly perform the $f$ integral in (\ref{effaction}) to obtain a purely
4D action and to derive
\begin{equation}
M_{Pl}^{2}=2M^{3}\int_{0}^{\pi r_{c}}dfe^{-2k_{f}|f|}=\frac{M^{3}}{k}%
e^{-2k_{y}|y|}[1-e^{-2k_{f}r_{c}\pi }].  \label{effplanck}
\end{equation}
We see that there is a well-defined value for $M_{Pl}$, even in the $%
r_{c}\rightarrow \infty $ limit, but which may have an anisotropic
dependence on one of the 4D coordinates, in the stated parametrizations
denoted by $y$. Nevertheless, we can get a sensible effective anisotropic 4D
theory, with the usual Newtonian force law, even in the infinite radius
limit, in contrast to the product--space expectation that $%
M_{Pl}^{2}=M^{3}r_{c}\pi $.

In consequence of (\ref{effplanck}), the gravitational potential behaves
anisotropically as
\[
V(r)=G_{N}{\frac{m_{1}m_{2}}{r}}\left( 1+\frac{e^{-2k_{y}|y|}}{r^{2}k_{f}^{2}%
}\right)
\]
i.e. our models produce effective 4D theories of gravity with local
anisotropy. The leading term due to the bound state mode is the usual
Newtonian potential; the Kaluza Klein anholonomic modes generate an
extremely anisotropically suppressed correction term, for $k_{f}$ taking the
expected value of order the fundamental Planck scale and $r$ of the size
tested with gravity.

Let us conclude the paper: It is known that we can consistently exist with
an infinite 5th dimension, without violating known tests of gravity \cite{07rs}%
. The scenarios consist of two or a single 3--brane, (a piece of) $AdS_5$ in
the bulk, and an appropriately tuned tension on the brane. But if we
consider off--diagonal 5D metrics like in Ref. \cite{07sal}, which was used
for including of $U(1),SU(2)$ and $SU(3)$ gauge fields, or, in a different
but similar fashion, for construction of generic anisotropic, partially
anholonomic, solutions (like static black holes with ellipsoidal horizons,
static black tori and anisotropic wormholes) in Einstein and extra dimension
gravity, \cite{07v} the RS theories become substantially locally anisotropic.
One obtains variations of constants on the 5th coordinate and possible
anisotropic oscillations in time (in the first our model), or on space
coordinate (in the second our model). Here it should be emphasized that the
anisotropic oscillations (in time or in a space coordinate) are defined by
the constant component of the cosmological constant (which in our model can
generally run on the 5th coordinate). This sure is related to the the
cosmological constant problem which in this work is taken as a given one,
with an approximation of linear dependence on the 5th coordinate, and not
solved. In the other hand a new, anisotropic, solution to the hierarchy
problem is supposed to be subjected to experimental verification.

Finally, we note that many interesting questions remain to be investigated.
Having constructed another, anisotropic, valid alternative to conventional
4D gravity, it is important to analyze the astrophysical and cosmological
implications. These anisotropic scenarios might even provide a new
perspective for solving unsolved issues in string/M-theory, quantum gravity
and cosmology. \vskip 4pt

The author thanks D. Singleton, E. Gaburov and D. Gon\c ta for
collaboration and discussing of results. The author is grateful to P.
Stavrinos for hospitality and support. The work is partially supported by
''The 2000--2001 California State University Legislative Award''.

%%%%%%%%%%%%%%%%%%%%%%%%%%%%%%%%%%%%%%%%%%%%%%%%%%%%%%%%%%%%%%%%%%%%%%%%%%%%%

\chapter[Anisotropic Black Holes ]
{Anisotropic Black Holes in Einstein and Brane Gravity  }

{\bf Abstract}
\footnote{\copyright\ S. Vacaru and E. Gaburov, Anisotropic Black
Holes in Einstein and Brane Gravity, hep-th/0108065}

We consider exact solutions of Einstein equations defining static black
holes paramet\-riz\-ed by off--diagonal metrics which by anholonomic mappings
can be equivalently transformed into some diagonal metrics with
coefficients being very similar to those from the Schwarzschild and/or
Reissner-N\"ordstrom solutions with anisotropic renormalizations of
constants. We emphasize that such classes of solutions, for instance,  with
ellipsoidal symmetry of horizons, can be constructed even in  general
relativity theory if off--diagonal metrics and anholonomic frames are
introduced into considerations. Such solutions do not violate the Israel's
uniqueness theorems on static black hole configurations \cite{08israel}
because at long radial distances one holds the usual Schwarzschild limit. We
show that anisotropic deformations of  the Reissner-N\"ordstrom metric can
be an exact solution on the brane, re-interpreted as a black hole with an
effective electromagnetic like  charge anisotropically induced and polarized
by higher dimension  gravitational interactions.

\vskip20pt

The idea of extra--dimension is gone through a renewal in connection to
string/M--theory \cite{08hw} which in low energy limits results in models of
brane gravity and/or high energy physics. It was proven that the matter
fields could be localized on a 3--brane in $1+3+n$ dimensions, while gravity
can propagate in the $n$ extra dimensions which can be large (see, e. g.,
\cite{08add}) and even not compact, as in the 5-dimensional (in brief, 5D)
warped space models of Randall and Sundrum \cite{08rs} (in brief RS, see also
early versions \cite{08a}).

The bulk of solutions of 5D Einstein equations and their reductions to 4D
were constructed by using static diagonal metrics and extensions to
solutions with rotations given with respect to holonomic coordinate frames
of references. On the other hand much attention has been paid to
off--diagonal metrics in higher dimensional gravity beginning the Salam,
Strathee and Petracci work \cite{08salam} which proved that including
off--diagonal components in higher dimensional metrics is equivalent to
including of $U(1),SU(2)$ and $SU(3)$ gauge fields. Recently, it was shown
in Ref. \cite{08v1} that if we consider off--diagonal metrics which can be
equivalently diagonalized to some corresponding anholonomic frames, the RS
theories become substantially locally anisotropic with variations of
constants on extra dimension coordinate or with anisotropic angular
polarizations of effective 4D constants, induced by higher dimension
gravitational interactions.

If matter on a such anisotropic 3D branes collapses under gravity without
rotating to form a black hole, then the metric on the brane-world should be
close to some anisotropic deformations of the Schwarzschild metric at
astrophysical scales in order to preserve the observationally tested
predictions of general relativity. We emphasize that it is possible to
construct anisotropic deformations of spherical symmetric black hole
solutions to some static configurations with ellipsoidal or toroidal
horizons even in the framework of 4D and in 5D Einstein theory if
off--diagonal metrics and associated anholonomic frames and nonlinear
connections are introduced into consideration \cite{08v2}.

Collapse to locally isotropic black holes in the Randall-Sundrum brane-world
scenario was studied by Chamblin et al. \cite{08chr} (see also \cite{08ehm,08gkr}
and a review on the subject \cite{08maartens}). The item of definition of
black hole solutions have to be reconsidered if we are dealing with
off--diagonal metrics, anholonomic frames both in general relativity and on
anisotropic branes.

In this Letter, we give four classes of exact black hole solutions which
describes ellipsoidal static deformations with anisotropic polarizations and
running of constants of the Schwarzschild and Reissner-N\"ordstrom
solutions. We analyze the conditions when such type anisotropic solutions
defined on 3D branes have their analogous in general relativity.

The 5D pseudo--Riemannian spacetime is provided with local coordinates $%
u^\alpha =(x^i,y^a)=(x^1=f,x^2,x^3,y^4=s,y^5=p),$ where $f$ is the extra
dimension coordinate, $(x^2,x^3)$ are some space coordinates and $(s=\varphi
,p=t)$ (or inversely, $(s=t,p=\varphi )$) are correspondingly some angular
and time like coordinates (or inversely). We suppose that indices run
corresponding values: $i,j,k,...=1,2,3$ and $a,b,c,...=4,5.$ The local
coordinate bases $\partial _\alpha =(\partial _i,\partial _a),$ and their
duals, $d^\alpha =\left( d^i,d^a\right) ,$ are defined respectively as

\newpage

\begin{equation}
\partial _\alpha \equiv \frac \partial {du^\alpha }=(\partial _i=\frac %
\partial {dx^i},\partial _a=\frac \partial {dy^a}) \mbox{ and } d^\alpha
\equiv du^\alpha =(d^i=dx^i,d^a=dy^a).  \label{4pdif}
\end{equation}

For the 5D line element $dl^{2}=G_{\alpha \beta }du^{\alpha }du^{\beta }$ we
choose the metric coefficients $G_{\alpha \beta }$ (with respect to the
coordinate frame (\ref{4pdif})) to be parametrized by a off--diagonal matrix
(ansatz) {\small
\begin{equation}
\left[
\begin{array}{ccccc}
1+w_{1}^{\ 2}h_{4}+n_{1}^{\ 2}h_{5} & w_{1}w_{2}h_{4}+n_{1}n_{2}h_{5} &
w_{1}w_{3}h_{4}+n_{1}n_{3}h_{5} & w_{1}h_{4} & n_{1}h_{5} \\
w_{1}w_{2}h_{4}+n_{1}n_{2}h_{5} & g_{2}+w_{2}^{\ 2}h_{4}+n_{2}^{\ 2}h_{5} &
w_{2}w_{3}h_{4}+n_{2}n_{3}h_{5} & w_{2}h_{4} & n_{2}h_{5} \\
w_{1}w_{3}h_{4}+n_{1}n_{3}h_{5} & w_{3}w_{2}h_{4}+n_{2}n_{3}h_{5} &
g_{3}+w_{3}^{\ 2}h_{4}+n_{3}^{\ 2}h_{5} & w_{3}h_{4} & n_{3}h_{5} \\
w_{1}h_{4} & w_{2}h_{4} & w_{3}h_{4} & h_{4} & 0 \\
n_{1}h_{5} & n_{2}h_{5} & n_{3}h_{5} & 0 & h_{5}
\end{array}
\right]   \label{2ansatz0}
\end{equation}
} where the coefficients are some necessary smoothly class functions of
type:
\[
g_{2,3}=g_{2,3}(x^{2},x^{3}),h_{4,5}=h_{4,5}(x^{1},x^{2},x^{3},s),
w_{i}=w_{i}(x^{1},x^{2},x^{3},s),n_{i}=n_{i}(x^{1},x^{2},x^{3},s).
\]

The line element (\ref{2ansatz0}) can be equivalently rewritten in the form
\begin{equation}
\delta l^2=g_{ij}\left( x^2,x^3\right) dx^idx^i+h_{ab}\left(
x^1,x^2,x^3,s\right) \delta y^a\delta y^b,  \label{4dmetric}
\end{equation}
with diagonal coefficients $g_{ij}= diag[1,g_2,g_3]$ and $h_{ab}= diag
[h_4,h_5]$ if instead the coordinate bases  (\ref{4pdif}) one introduce the
anholonomic frames (anisotropic bases)
\begin{equation}
{\delta }_\alpha \equiv \frac \delta {du^\alpha }=(\delta _i=\partial
_i-N_i^b(u)\ \partial _b,\partial _a=\frac \partial {dy^a}),\
\delta ^\alpha \equiv \delta u^\alpha = (\delta ^i=dx^i,\delta
^a=dy^a+N_k^a(u)\ dx^k)  \label{6ddif}
\end{equation}
where the $N$--coefficients are parametrized $N_i^4=w_i$ and $N_i^5=n_i$
(on anholonomic frame method see details in \cite{08v1}).

The nontrivial components of the 5D vacuum Einstein equations, $R_{\alpha
}^{\beta }=0,$ for the ansatz (\ref{4dmetric}) given with respect to
anholonomic frames (\ref{6ddif}) are
\begin{eqnarray}
R_{2}^{2}=R_{3}^{3}=-\frac{1}{2g_{2}g_{3}}[g_{3}^{\bullet \bullet }-\frac{%
g_{2}^{\bullet }g_{3}^{\bullet }}{2g_{2}}-\frac{(g_{3}^{\bullet })^{2}}{%
2g_{3}}+g_{2}^{^{\prime \prime }}-\frac{g_{2}^{^{\prime }}g_{3}^{^{\prime }}%
}{2g_{3}}-\frac{(g_{2}^{^{\prime }})^{2}}{2g_{2}}] &=&0,  \label{3ricci1a} \\
R_{4}^{4}=R_{5}^{5}=-\frac{\beta }{2h_{4}h_{5}} &=&0,  \label{1ricci1b} \\
R_{4i}=-w_{i}\frac{\beta }{2h_{5}}-\frac{\alpha _{i}}{2h_{5}} &=&0,
\label{ricci1c} \\
R_{5i}=-\frac{h_{5}}{2h_{4}}\left[ n_{i}^{\ast \ast }+\gamma n_{i}^{\ast }%
\right]  &=&0,  \label{2ricci1d}
\end{eqnarray}
where
\[
\alpha _{i}=\partial _{i}{h_{5}^{\ast }}-h_{5}^{\ast }\partial _{i}\ln \sqrt{%
|h_{4}h_{5}|},\beta =h_{5}^{\ast \ast }-h_{5}^{\ast }[\ln \sqrt{|h_{4}h_{5}|}%
]^{\ast },\gamma =(3h_{5}/2h_{4})-h_{4}^{\ast }/h_{4},
\]
the partial derivatives are denoted like $a\symbol{94}=\partial a/\partial
x^{1},h^{\bullet }=\partial h/\partial x^{2},f^{\prime }=\partial f/\partial
x^{2}$ and $f^{\ast }=\partial f/\partial s.$

The system of second order nonlinear partial equations (\ref{3ricci1a})--(\ref
{2ricci1d}) can be solved in general form:

The equation (\ref{3ricci1a}) relates two functions $g_2(x^2,x^3)$ and $%
g_3(x^2,x^3).$ It is solved, for instance, by arbitrary two functions $%
g_2(x^2)$ and $g_3(x^3),$ or by $g_2=g_3=g_{[0]}\exp [a_2x^2+a_3x^3],$ were $%
g_{[0]},a_2$ and $a_3$ are some constants. For a given parametrization of $%
g_2=b_2(x^2)c_2(x^3)$ we can find a decomposition in series for $%
g_3=b_3(x^2)c_3(x^3)$ (in the inverse case a multiple parametrization is
given for $g_3$ and we try to find $g_2);$ for simplicity we omit such
cumbersome formulas. We emphasize that we can always redefine the variables $%
\left( x^2,x^3\right) ,$ or (equivalently) we can perform a 2D conformal
transform to the flat 2D line element
\[
g_2(x^2,x^3)(dx^2)^2+g_3(x^2,x^3)(dx^3)^2\rightarrow (dx^2)^2+(dx^3)^2,
\]
for which the solution of (\ref{3ricci1a}) becomes trivial.

The next step is to find solutions of the equation (\ref{1ricci1b}) which
relates two functions $h_4\left( x^i,s\right) $ and $h_5\left( x^i,s\right) $%
. This equation is satisfied by arbitrary pairs of coefficients $h_4\left(
x^i,s\right) $ and $h_{5[0]}\left( x^i\right) .$ If dependencies of $h_5$ on
anisotropic variable $s$ are considered, there are two possibilities:

a) to compute
\begin{eqnarray*}
\sqrt{|h_5|} &=&h_{5[1]}\left( x^i\right) +h_{5[2]}\left( x^i\right) \int
\sqrt{|h_4\left( x^i,s\right) |}ds,~h_4^{*}\left( x^i,s\right) \neq 0; \\
&=&h_{5[1]}\left( x^i\right) +h_{5[2]}\left( x^i\right) s,h_4^{*}\left(
x^i,s\right) =0,
\end{eqnarray*}
for some functions $h_{5[1,2]}\left( x^i\right) $ stated by boundary
conditions;

b) or, inversely, to compute $h_4$ for a given $h_5\left( x^i,s\right)
,h_5^{*}\neq 0,$%
\begin{equation}
\sqrt{|h_4|}=h_{[0]}\left( x^i\right) (\sqrt{|h_5\left( x^i,s\right) |})^{*},
\label{2p1}
\end{equation}
with $h_{[0]}\left( x^i\right) $ given by boundary conditions.

Having the values of functions $h_4$ and $h_5,$ we can define the
coefficients $w_i\left( x^i,s\right) $ and $n_i\left( x^i,s\right) :$

The exact solutions of (\ref{ricci1c}) is found by solving linear algebraic
equation on $w_k,$
\begin{equation}
w_k=\partial _k\ln [\sqrt{|h_4h_5|}/|h_5^{*}|]/\partial _s\ln [\sqrt{|h_4h_5|%
}/|h_5^{*}|],  \label{5w}
\end{equation}
for $\partial _s=\partial /\partial s$ and $h_5^{*}\neq 0.$ If $h_5^{*}=0$
the coefficients $w_k$ could be arbitrary functions on $\left( x^i,s\right)
. $

Integrating two times on variable $s$ we find the exact solution of (\ref
{2ricci1d}),
\begin{eqnarray}
n_k &=&n_{k[1]}\left( x^i\right) +n_{k[2]}\left( x^i\right) \int [h_4/(\sqrt{%
|h_5|})^3]ds,~h_5^{*}\neq 0;  \nonumber \\
&=&n_{k[1]}\left( x^i\right) +n_{k[2]}\left( x^i\right) \int
h_4ds,~h_5^{*}=0;  \label{4n} \\
&=&n_{k[1]}\left( x^i\right) +n_{k[2]}\left( x^i\right) \int [1/(\sqrt{|h_5|}%
)^3]ds,~h_4^{*}\neq 0,  \nonumber
\end{eqnarray}
for some functions $n_{k[1,2]}\left( x^i\right) $ stated by boundary
conditions.

We shall construct some classes of exact solutions of 5D and 4D vacuum
Einstein equations describing anholonomic deformations of black hole
solutions of the Reissner-N\"{o}rdstrom and Schwarzschild metrics. We
consider two systems of 3D space coordinates:

a) The isotropic spherical coordinates $(\rho ,\theta ,\varphi ),$
\thinspace where the isotropic radial coordinate $\rho $ is related with the
usual radial coordinate $r$ via relation $r=\rho \left( 1+r_g/4\rho \right)
^2$ for $r_g=2G_{[4]}m_0/c^2$ being the 4D gravitational radius of point
particle of mass $m_0,$ $G_{[4]}=1/M_{P[4]}^2$ is the 4D Newton constant
expressed via Plank mass $M_{P[4]}$ which following modern string/brane
theories can considered as a value induced from extra dimensions,  we shall
put the light speed constant $c=1$ (this system of coordinates is
considered, for instance, for the so--called isotropic representation of the
Schwarzschild solution \cite{08ll}).

b) The rotation ellipsoid coordinates (in our case isotropic, in brief
re--coordinates) \cite{08korn} $(u,v,\varphi )$ with $0\leq u<\infty ,0\leq
v\leq \pi ,0\leq \varphi \leq 2\pi ,$ where $\sigma =\cosh u=4\rho /r_g\geq
1 $ are related with  the isotropic 3D Cartezian coordinates $(\tilde{x} =
\sinh u\sin v\cos \varphi ,\tilde{y}=\sinh u\sin v\sin \varphi , \tilde{z}
=\cosh u\cos v)$ and define an elongated rotation ellipsoid hypersurface $%
\left( \tilde{x}^2+\tilde{y}^2\right) /(\sigma ^2-1)+\tilde{z}^2/\sigma ^2=1.
$

By straightforward calculations we can verify that we can generate from the
ansatz (\ref{2ansatz0}) four classes of exact solutions of the system (\ref
{3ricci1a})--(\ref{2ricci1d}):

\begin{enumerate}
\item  The anisotropic Reissner-N\"{o}rdstrom black hole solutions with
polarizations on extra dimension and 3D space coordinates  are parametrized
by the data
\begin{equation}
g_2 =\left( \frac{1-\frac{r_g}{4\rho }}{1+\frac{r_g}{4\rho }}\right) \frac %
1{\left[ \rho ^2+a\rho /(1+\frac{r_g}{4\rho })^2+b/(1+\frac{r_g}{4\rho })^4%
\right] },g_3=1;  \label{sol1a}
\end{equation}
\begin{equation}
h_5 = -\frac 1{\rho ^2\left( 1+\frac{r_g}{4\rho }\right) ^4}[1+\frac{%
a\sigma _m\left( f,\rho ,\theta ,\varphi \right) }{\rho \left( 1+\frac{r_g}{%
4\rho }\right) ^2}+\frac{b\sigma _q\left( f,\rho ,\theta ,\varphi \right) }{%
\rho ^2\left( 1+\frac{r_g}{4\rho }\right) ^4}],      \nonumber
\end{equation}
\begin{equation}
 h_4 = \sin ^2\theta \left[
\left( \sqrt{\left| h_5(f,\rho ,\theta ,\varphi )\right| }\right) \right]
^2\quad \mbox{(see (\ref{2p1}))};  \nonumber
\end{equation}
where $a,b$ are constants and $\sigma _m\left( f,\rho ,\theta ,\varphi
\right) $ and $\sigma _q\left( f,\rho ,\theta ,\varphi \right) $ are called
respectively mass and charge polarizations and the coordinates are $\left(
x^i,y^a\right) =\left( f,\rho ,\theta ,t,\varphi \right) .$

\item  The anisotropic Reissner-N\"{o}rdstrom black hole solutions with
extra dimension and time running of constants are parametrized by the data
\begin{eqnarray}
g_2 &=&\left( \frac{1-\frac{r_g}{4\rho }}{1+\frac{r_g}{4\rho }}\right) \frac %
1{\left[ \rho ^2+a\rho /(1+\frac{r_g}{4\rho })^2+b/(1+\frac{r_g}{4\rho })^4%
\right] },g_3=1;  \label{2sol1b} \\
h_4 &=&-\frac 1{\rho ^2\left( 1+\frac{r_g}{4\rho }\right) ^4}[1+\frac{%
a\sigma _m\left( f,\rho ,\theta ,t\right) }{\rho \left( 1+\frac{r_g}{4\rho }%
\right) ^2}+\frac{b\sigma _q\left( f,\rho ,\theta ,t\right) }{\rho ^2\left(
1+\frac{r_g}{4\rho }\right) ^4}],h_5=\sin ^2\theta ,  \nonumber
\end{eqnarray}
where $a,b$ are constants and $\sigma _m\left( f,\rho ,\theta ,\varphi
\right) $ and $\sigma _q\left( f,\rho ,\theta ,\varphi \right) $ are called
respectively mass and charge polarizations and the coordinates are $\left(
x^i,y^a\right) =\left( f,\rho ,\theta ,\varphi ,t\right) .$

\item  The ellipsoidal Schwarzschild like black hole solutions with
polarizations on extra dimension and 3D space coordinates  are parametrized
by the data $g_2=g_3 =1$ and
\begin{eqnarray}
h_5 &=& -\frac{r_g^2}{16}\frac{\cosh ^2u}{\left( 1+\cosh u\right) ^4}\left(
\frac{\cosh u_m(f,u,v,\varphi )-\cosh u}{\cosh u_m(f,u,v,\varphi )+\cosh u}%
\right) ^2,\  \nonumber \\
 h_4 &=&\frac{\sinh ^2u\sin ^2v}{\sinh ^2u+\sin ^2v}\left[ \left(
\sqrt{\left| h_5(f,u,v,\varphi )\right| }\right) \right] ^2,  \label{sol2a}
\end{eqnarray}
where $\sigma _m=\cosh u_m$ and the coordinates are $\left( x^i,y^a\right)
=\left( f,u,v,\varphi ,t\right) .$

\item  The ellipsoidal Schwarzschild like black hole solutions with extra
dimension and time running of constants are parametrized by the data $%
g_2=g_3=1$ and
\begin{equation}
h_4 = -\frac{r_g^2}{16}\frac{\cosh ^2u}{\left( 1+\cosh u\right) ^4}\left(
\frac{\cosh u_m(f,\rho ,\theta ,t)-\cosh u}{\cosh u_m(f,\rho ,\theta
,t)+\cosh u}\right) ^2, h_5 = \frac{\sinh ^2u\sin ^2v}{\sinh ^2u+\sin ^2v},
\label{sol2b}
\end{equation}
where $\sigma _m=\cosh u_m$ and the coordinates are $\left( x^i,y^a\right)
=\left( f,u,v,t,\varphi \right) .$
\end{enumerate}

The N--coefficients $w_i$ and $n_i$ for the solutions (\ref{sol1a})--(\ref
{sol2b}) are computed respectively following formulas (\ref{5w}) and (\ref{1n}%
) (we omit the final expressions in this paper).

The mathematical form of the solutions (\ref{sol1a}) and (\ref{2sol1b}), with
constants\\ $a=-2m/M_p^2$ and $b=Q,$ is very similar to that of the
Reissner-N\"{o}rdstrom solution from RS gravity \cite{08maartens}, but
multiplied on a conformal factor $\left( 1+\frac{r_g}{4\rho }\right)
^{-4}\rho ^{-2},$ with renormalized factors $\sigma _m$ and $\sigma _q$ and
{\em without electric charge} being present. The induced 4D gravitational
''receptivities'' $\sigma _m$ and $\sigma _q$ in (\ref{sol1a}) emphasize
dependencies on coordinates $\left( f,\rho ,\theta ,\varphi \right) ,$ where
$s=\varphi $ is the anisotropic coordinate. In a similar fashion one induces
running on time and the 5th coordinate, and anisotropic polarizations on $%
\rho $ and $\theta ,$ of constants for the solution (\ref{2sol1b}).

Instead the Reissner-N\"{o}rdstrom-type correction to the Schwarzschild
potential the mentioned polarizations can be thought as defined by some
nonlinear higher dimension gravitational interactions and anholonomic frame
constraints for anisotropic Reissner-N\"{o}rdstrom black hole configurations
with a {\em `tidal charge'} $Q$ arising from the projection onto the brane
of free gravitational field effects in the bulk. These effects are
transmitted via the bulk Weyl tensor, off--diagonal components of the metric
and by anholonomic frames. The Schwarzschild potential $\Phi =-M/(M_{{\rm p}%
}^2r)$, where $M_{{\rm p}}$ is the effective Planck mass on the brane, is
modified to
\begin{equation}
\Phi =-{\frac{M\sigma _m}{M_{{\rm p}}^2r}}+{\frac{Q\sigma _q}{2r^2}}\,,
\label{npot}
\end{equation}
where the `tidal charge' parameter $Q$ may be positive or negative. The
possibility to modify anisotropically the Newton law via effective
anisotropic masses $M\sigma _m,$ or by anisotropic effective 4D Plank
constants, renormalized like $\sigma _m/M_{{\rm p}}^2,$ was recently
emphasized in Ref. \cite{08v1}. In this paper we state that there are possible
additional renormalizations of the ''effective'' electric charge, $Q\sigma
_q.$ For diagonal metrics we put $\sigma _m=\sigma _q=1$  and by
multiplication on corresponding conformal factors and  with respect to
holonomic frames we recover the locally isotropic results from Refs. \cite
{08maartens}. We must also impose the condition that the 5D spacetime is
modelled as a $AdS_5$ slice provided with an anholonomic frame structure.

The renormalized tidal charge $Q\sigma _q$ affects the geodesics and the
gravitational potential, so that indirect limits may be placed on it by
observations. Nevertheless, current observational limits on $|Q\sigma _q|$
are rather weak, since the correction term in Eq.~(\ref{npot}) decreases off
rapidly with increasing $r$, and astrophysical measurements (lensing and
perihelion precession) probe mostly (weak-field) solar scales.

Now we analyze the properties of solutions (\ref{sol2a}) and (\ref{sol2b}).
They describe Schwarz\-schild like solutions with the horizon forming a
rotation ellipsoid horizon. For the general relativity such solutions were
constructed in Refs. \cite{08v2}. Here, it should be emphasized that static
anisotropic deformations of the Schwarzschild metric are described by
off--diagonal metrics and corresponding conformal transforms. At large
radial distances from the horizon the anisotropic configurations transform
into the usual one with spherical symmetry. That why the solutions with
anisotropic rotation ellipsoidal horizons do not contradict the well known
Israel and Carter theorems \cite{08israel} which were proved in the assumption
of spherical symmetry at asymptotic. Anisotropic 4D black hole solutions
follow from the data (\ref{sol2a}) and (\ref{sol2b}) if you state some
polarizations depending only on 3D space coordinates ($u,v,\varphi ),$ or on
some of them. In this paper we show that in 5D there are warped to 4D static
ellipsoidal like solutions with constants renormalized anisotropically on
some 3D space coordinates and on extra dimension coordinate (in the class of
solutions (\ref{sol2a})) and running of constants on time and the 5th
coordinate, with possible additional polarizations on some 3D coordinates
(in the class of solutions (\ref{sol2b})).

A geometric approach to the Randall-Sundrum scenario has been developed by
Shiromizu et al. \cite{08sms} (see also \cite{08bdul}), and proves to be a
useful starting point for formulating the problem and seeing clear lines of
approach. In this work we considered a variant of anholonomic RS geometry.
The vacuum solutions (\ref{sol1a})--(\ref{sol2b}) localized on the brane
must satisfy the 5D equation in the Shiromizu et al. representation if in 4D
some sources are considered as to be induced from extra dimension gravity.

The method of anholonomic frames covers the results on linear
extensionss of the Schwarzschild horizon into the bulk
\cite{08gian}. The solutions presented in this paper are
nonlinearly induced, are based on very general method of
construction exact solutions in extra dimension gravity and
generalize also the Reissner-N\"{o}rdstrom solution from RS
gravity. The obtained solutions are locally anisotopic but,
nevertheless, they posses local 4D Lorentz symmetry, which is
explicitly emphasized with respect to anholonomic frames. There
are possible constructions with broken Lorentz symmetry as in
\cite{08csaki} (if we impose not a locally isotropic limit of our
solutions, but an anisotropic static one). We omit such
considerations here.

In conclusion we formulate a prescription for mapping 4D general relativity
solutions with diagonal metrics to 4D and 5D solutions of brane world: {\em %
a general relativity vacuum solution gives rise to a vacuum brane--world
solution in 5D gravity given with similar coefficients of metrics  but
defined with respect to some anholonomic frames and with  anisotropic
renormalization of fundamental constants; such type of  solutions are
parametrized by off--diagonal metrics if of type  (\ref{2ansatz0}) if they
are re--defined with respect to  coordinate frames }.

\vskip3pt {\bf Acknowledgements:} The authors thank D.\ Gon\c ta for
discussion and collaboration. S. V. is grateful to P. Stavrinos
and D. Singleton for support and hospitality. The work is
supported both by ''The 2000--2001 California State University
Legislative Award'' and a NATO/Portugal fellowship grant at the
Instituto Superior Tecnico,  Lisbon.

%%%%%%%%%%%%%%%%%%%%%%%%%%%%%%%%%%%%%%%%%%%%%%%%%%%%%%%%%%%%%%%%%%%%%%%%%%%%%

\chapter[Anisotropic Brane Inflation  ]
{Off--Diagonal Metrics and Anisotropic Brane Inflation  }

{\bf Abstract}
\footnote{\copyright\  S. Vacaru and D. Gon\c ta,
 Off--Diagonal Metrics and Anisotropic  Brane Inflation, hep--th/ 0109114}

We study anisotropic reheating (entropy production) on 3D brane from a
decaying bulk scalar field in the brane--world picture of the Universe by
considering off--diagonal metrics and anholonomic frames. We show that a
significant amount of, in general, anisotropic dark radiation is produced in
this process unless only the modes which satisfy a specific relation are
excited. We conclude that subsequent entropy production within the brane is
required in general before primordial nucleosynthesis and that the presence
of off--diagonal components (like in the Salam, Strathee and Petracci works
\cite{09salam}) of the bulk metric modifies the processes of entropy
production which could substantially change the brane--world picture of the
Universe.

\vskip20pt

The brane world picture of the Universe \cite{09rs} resulted in a number of
works on brane world cosmology \cite{09wc1,09wc2} and inflationary solutions and
scenaria \cite{09inf1,09inf2,09inf3,09inf4,09inf5}. Such solutions have been
constructed by using diagonal cosmological metrics with respect to holonomic
coordinate frames.

In Kaluza--Klein gravity there were also used off--diagonal five dimensional
(in brief, 5D) metrics beginning Salam and Strathee and Perrachi works \cite{09salam}
which suggested to treat the off--diagonal components as some
coefficients including $U(1), SU(2)$ and $SU(3)$ gauge fields. Recently, the
off--diagonal metrics were considered in a new fashion both in Einstein and
brane gravity \cite{09v1,09v2}, by applying the method of anholonomic frames
with associated nonlinear connection, which resulted in a new method of
construction of exact solutions of Einstein equations describing, for
instance, static black hole and cosmological solutions with ellipsoidal or
torus symmetry, soliton--dilaton and wormhole--flux tube configurations with
anisotropic polarizations and/or running of constants.

The aim of this paper is to investigate reheating after
anisotropic inflation in the brane world with generic local
anisotropy induced by off--diagonal metrics in the bulk
\cite{09v2}. In this scenario, our locally anisotropic Universe is
described on a 4D boundary (3D anisotropic brane) of
$Z_2$--symmetric 5D space--time with a gravitational vacuum
polarization constant and its computed renormalized effective
value. In the locally isotropic limit the constant of
gravitational vacuum polarization results in a negative
cosmological constant $\Lambda_5\equiv -6k^2$, where $k$ is a
positive constant. Our approach is in the spirit of
Horava--Witten theory \cite{09horava,09lukas} and recovers the
Einstein gravity around the brane with positive tension
\cite{09rs,09tanaka,09sms}, the considerations being extended with
respect to anholonomic frames.

Theories of gravity and/or high energy physics must satisfy a
number of cosmological tests including cosmological inflation
\cite{09inf} which for brane models could directed by anisotropic
renormalizations of parameters \cite{09v2}. We shall develop a
model of anisotropic inflation scenarios satisfying the next three
requirements:\ 1) it is characterized by a sufficiently long
quasi--exponential expansion driven by vacuum--like energy
density of the potential energy of a scalar field;\ 2) the
termination of accelerated anisotropic expansion is associated
with an entropy production or reheating to satisfy the conditions
for the initial state of the classical hot Big Bang cosmology,
slightly anisotropically deformed, before the primordial
nucleosynthesis \cite{09ns} and 3)\ generation of primordial
fluctuations with desired amplitude and spectrum \cite{09fluc}.

We assume the 5D vacuum Einstein equations written with respect to
anholonomic frames which for diagonal metrics with respect to holonomic
frames contains a negative cosmological constant $\Lambda _5$ and a 3D brane
at the 5th coordinate $w=0$ about which the space--time is $Z_2$ symmetric
and consider a quadratic line interval
\begin{equation}
\delta s^2=\Omega ^2(t,w,y)[dx^2+g_2\left( t,w\right) dt^2+g_3\left(
t,w\right) dw^2+h_4(t,w,y)\delta y^2+h_5\left( t,w\right) dz^2],
\label{dmetr1}
\end{equation}
where the 'elongated' differential
 $\delta y=dy+\zeta _2(t,w,y)dt+\zeta _3(t,w,y)dw,$
 together with $dx,dt$ and $dw$ define an anholonomic co--frame basis $%
(dx,dt,dw,\delta y,$ $\delta z=dz)$ which is dual to the anholonomic frame
basis \cite{09v1,09v2} $
(\delta _1=\frac \partial {\partial x},\delta _2=\frac \partial
{\partial t}-\zeta _2\frac \partial {\partial y},\delta _3=\frac \partial
{\partial t}-\zeta _3\frac \partial {\partial y},\partial _4=\frac \partial
{\partial y},\partial _5=\frac \partial {\partial z});$\
we denote the 4D space--time coordinates as $(x,t,w,y,z)$ with $t$ being the
time like variable. The metric ansatz for the interval (\ref{dmetr1}) is
off--diagonal with respect to the usual coordinate basis $\left(
dx,dt,dw,dy,dz\right).$

As a particular case we can parametrize from (\ref{dmetr1}) the metric near
an locally isotropic brane like a flat Robertson--Walker metric with the
scale factor $a(t)$ \cite{09inf5} if we state the values
\[
\Omega ^2=(aQ)^2,g_2=-(aQ)^{-2}N^2,g_3=(aQ)^{-2},h_4=1,h_5=1,\zeta _{2,3}=0,
\]
for
\begin{eqnarray}
N^2(t,w) &=&Q^{-2}(t,w) [ \cosh (2kw)+{\frac 12}k^{-2}\left( {H^2+\dot{H}%
}\right) \left( {\cosh (2kw)-1}\right)    \nonumber \\
 && -{\frac{{1+{\frac 12}k^{-2}\left( {%
2H^2+\dot{H}}\right) }}{{\sqrt{1+k^{-2}H^2+Ca^{-4}}}}}\sinh (2k|w|)]
\nonumber \\
Q^2(t,w) &=&\cosh (2kw)+{\frac 12}k^{-2}H^2\left( {\cosh (2kw)-1}\right) -%
\sqrt{1+k^{-2}H^2+Ca^{-4}},  \label{NQ}
\end{eqnarray}
when the bulk is in a vacuum state with a negative cosmological constant $%
\Lambda _5$, $C$ is an integration constant \cite{09mukoyama}. One takes $%
N=Q=1 $ on the brane $w=0$. The function $H\left( t\right) $ and constants $%
k $ and $C$ from (\ref{NQ}) are related with the evolution equation on the
brane in this case is given by
\begin{equation}
H^2 = \left( \frac{\dot{a}}a\right) ^2={\frac{{\kappa _5^4\sigma }}{{18}}}%
\rho _{{\rm tot}}+{\frac{{\Lambda _4}}3}+{\frac{{\kappa _5^4}}{{36}}}\rho _{%
{\rm tot}}^2-{\frac{{k^2C}}{{a^4}}},\
\Lambda _4 \equiv {\frac 12}\left( {\Lambda _5+{\frac{{\kappa _5^2}}6}%
\sigma ^2}\right) ,  \label{feq}
\end{equation}
where $\kappa _5^2$ is the 5D gravitational constant related with the 5D
reduced Planck scale, $M_5$, by $\kappa _5^2=M_5^{-3}$; $\sigma $ is the
brane tension, the total energy density on the brane is denoted by $\rho _{%
{\rm tot}}$, and the last term of (\ref{feq}) represents the dark radiation
with $C$ being an integration constant \cite{09sms,09bki,09mukoyama}. We recover
the standard Friedmann equation with a vanishing cosmological constant at
low energy scales if $\sigma =6k/\kappa _5^2$ and $\kappa _4^2=\kappa
_5^4\sigma /6=\kappa _5^2k$, where $\kappa _4^2$ is the 4D gravitational
constant related with the 4D reduced Planck scale, $M_4$, as $\kappa
_4^2=M_4^{-2}$. We find that $M_4^2=M_5^3/k$. If we take $k=M_4$, all the
fundamental scales in the theory take the same value, i. e. $k=M_4=M_5.$ The
the scale above is stated by constant $k$ which the nonstandard term
quadratic in $\rho _{{\rm tot}}$ is effective in (\ref{feq}). One suppose
\cite{09inf5} that $k$ is much larger than the scale of inflation so that such
quadratic corrections are negligible.

For simplicity, in locally isotropic cases one assumes that the bulk metric
is governed by $\Lambda _5$ and neglect terms suppressed by $k^{-1}$ and $%
Ca^{-4}$ and writes
\begin{equation}
ds_5^2=-e^{-2k|w|}dt^2+e^{-2k|w|}a^2(t)\left( {dx^2+dy^2+dz^2}\right) +dw^2.
\label{met}
\end{equation}

The conclusion of Refs \cite{09v2} is that the presence of off--diagonal
components in the bulk 5D metric results in locally anisotropic
renormalizations of fundamental constants and modification of the Newton low
on the brane. The purpose of this paper is to analyze the basic properties
of models of anisotropic inflation on 3D brane with induced from the bulk
local anisotropy of metrics of type (\ref{dmetr1}) which define cosmological
solutions of 5D vacuum Einstein equations depending on variables $\left(
w,t,y\right) ,$ being anisotropic on coordinate $y$ (see details on
construction of various classes of solutions by applying the method of
moving anholonomic frames in Ref \cite{09v1,09v2}).

For simplicity, we shall develop a model of inflation on 3D brane with
induced from the bulk local anisotropy by considering the ansatz
\begin{equation}
\delta s_5^2 =e^{-2(k|w|+k_y|y|)}a^2(t)[dx^2-dt^2]+e^{-2k_y|y|}dw^2+
e^{-2(k|w|+k_y|y|)}a^2(t)({\delta y}^2+{z^2)}  \label{meta}
\end{equation}
which is a particular case of the metric (\ref{dmetr1}) with\
$\Omega ^2(w,t,y)=e^{-2(k|w|+k_y|y|)}a^2(t),{g}_2=-1,
g_3=e^{2k|w|},h_4=1,h_5=1$
and ${\zeta }_2=k/k_y,{\zeta }_3=(da/dt)/k_ya$ taken as the ansatz (\ref
{meta}) would be an exact solution of 5D vacuum Einstein equations. The
constants $k$ and $k_y$ have to be established experimentally. We emphasize
that the metric (\ref{meta}) is induced alternatively on the brane from the
5D anholonomic gravitational vacuum with off--diagonal metrics. With respect
to anholonomic frames it has some diagonal coefficients being similar to
those from (\ref{met}) but these metrics are very different in nature and
describes two types of branes: the first one is with generic off--diagonal
metrics and induced local anisotropy, the second one is locally isotropic
defined by a brane configuration and the bulk cosmological constant. For
anisotropic models, the respective constants can be treated as some
'receptivities' of the bulk gravitational vacuum polarization.

The next step is to investigate a scenarios of anisotropic inflation driven
by a bulk scalar field $\phi $ with a 4D potential $V[\phi ]$ \cite{09bbinf,09hs}%
. We shall study the evolution of $\phi $ after anisotropic brane
inflation expecting that reheating is to proceed in the same way
as in 4D theory with anholonomic modification (a similar idea is
proposed in Ref. \cite{09bas} but for locally isotropic branes). We
suppose that the scalar field is homogenized in 3D space as a
result of inflation, it depends only on $t$ and $w$ and
anisotropically on $y$ and consider a situation when $\phi $
rapidly oscillates around $\phi =0$ by expressing $V[\phi
]=m^2\phi ^2/2$. The field $\phi (t,w,y)$ is non--homogeneous
because of induced space--time anisotropy. Under such assumptions
he Klein--Gordon equation in the background of metric
(\ref{meta}) is written
\begin{eqnarray}
&& \Box _5\phi (f,t,y)- V^{\prime }[\phi (f,t,y)] =\frac 1{\sqrt{|g|}}
[\delta _t\left( \sqrt{|g|}g^{22}\delta _t\phi \right) +\delta _w\left( \sqrt{%
|g|}g^{33}\delta _f\phi \right)      \nonumber \\
&& +\partial _y\left( \sqrt{|g|}h^{44}\partial
_y\phi \right)] - V^{\prime }[\phi ]=0,  \label{kg}
\end{eqnarray}
where $\delta _w =\frac \partial {\partial w}-
\zeta _2\frac \partial {\partial y},\delta _t=\frac \partial {\partial t}-\zeta _3\frac \partial {\partial y},
 $, $\Box _5$ is the d'Alambert operator and $|g|$ is the determinant of
the matrix of coefficients of metric given with respect to the anholonomic
frame (in Ref. \cite{09inf5} the operator $\Box _5$ is alternatively
constructed by using the metric (\ref{met})).

The energy release of $\phi $ is modelled by introducing
phenomenologically a
dissipation terms defined by some constants $\Gamma _D^w,\Gamma _D^y$ and $%
\Gamma _B$ representing the energy release to the brane and to the entire
space,
\begin{equation}
\Box _5\phi (w,t,y)-V^{\prime }[\phi (w,t,y)]=\frac{\Gamma _D^w}{2k}\delta
(w)\frac 1N\delta _t\phi +\frac{\Gamma _D^y}{2k_y}\delta (y)\frac 1N\delta
_t\phi +\Gamma _B\frac 1N\delta _t\phi .  \label{kgv}
\end{equation}

Following (\ref{kg}) and (\ref{kgv}) together with the $Z_2$ symmetries on
coordinates $w$ and $y$, we have
\begin{equation}
\delta _w\phi ^{+}=-\delta _w\phi ^{-}=\frac{\Gamma _D^w}{4k}\delta _t\phi
(0,y,t),\partial _y\phi ^{+}=-\partial _y\phi ^{-}=\frac{\Gamma _D^y}{4k_y}%
\delta _t\phi (w,0,t),  \label{phiw}
\end{equation}
where superscripts $+$ and $-$ imply values at $w,y\longrightarrow +0$ and $%
-0$, respectively. In this model we have two types of warping
factors, on coordinates $w$ and $y.$ The constant $k_y$
characterize the gravitational anisotropic polarization in the
direction $y.$

Comparing the formulas (\ref{kgv})$\,$and (\ref{phiw}) with similar ones
from Ref. \cite{09inf5} we conclude the the induced from the bulk brane
anisotropy could result in additional dissipation terms like that
proportional to $\Gamma _D^y$. This modifies the divergence of divergence $%
T_{~~A;C}^{(\phi )C}$ of the energy--momentum tensor $T_{MN}^{(\phi )}$ of
the scalar field $\phi $: Taking
\[
T_{MN}^{(\phi )}={{\delta }_M}\phi {{\delta }_N}\phi -g_{MN}\left( {{\frac 12%
}g^{PQ}{\delta }_P\phi {\delta }_Q\phi +V[\phi ]}\right) ,
\]
with five dimensional indices, $M,N,...=1,2,...,5$ and anholonomic partial
derivative operators ${{\delta }_P}$ being dual to $\delta ^P$ we compute
\begin{eqnarray}
T_{~~A;C}^{(\phi )C} &=& \left\{ \Box _5\phi (w,t,y)-V^{\prime }[\phi
(w,t,y)]\right\} \phi _{,A}  \nonumber \\
&=&\left[ \frac{\Gamma _D^w}{2k}\delta (w)\frac
1N\delta _t\phi +\frac{\Gamma _D^y}{2k_y}\delta (y)\frac 1N\delta _t\phi
+\Gamma _B\frac 1N\delta _t\phi \right] \delta _A\phi .  \label{divergence}
\end{eqnarray}
We can integrate the $A=0$ component of (\ref{divergence}) from
$w=-\epsilon $ to $w=+\epsilon $ near the brane, than we
integrate from $y=-\epsilon _1$ to $y=+\epsilon _1,$ in the zero
order in $\epsilon $ and $\epsilon _1,$ we find from (\ref{phiw})
that
\begin{equation}
\frac{\delta \rho _\phi (0,0,t)}{\partial t}=-(3H+\Gamma _B)(\delta _t\phi
)^2(0,0,t)-J_\phi (0,0,t),  \label{rhophieq}
\end{equation}
with
$\rho _\phi \equiv \frac 12(\delta _t\phi )^2+V[\phi ],~~J_\phi \equiv -%
\frac{\delta _t\phi }{\sqrt{|g|}}\delta _f\left( \sqrt{|g|}\delta _f\phi
\right) -\frac{\delta _t\phi }{\sqrt{|g|}}\delta _y\left( \sqrt{|g|}\delta
_y\phi \right),$
which states that the energy dissipated by the $\Gamma _D^f$ and $\Gamma _D^y
$ terms on anisotropic brane is entirely compensated by the energy flows
(locally isotropic and anisotropic) onto the brane. In this paper we shall
model anisotropic inflation by considering that $\phi $ looks like
homogeneous with respect to anholonomic frames; the local anisotropy and
induced non--homogeneous effects are modelled by additional terms like $%
\Gamma _D^y$ and elongated partial operators with a further integration on
variable $y.$

Now we analyze  how both the isotropic and anisotropic energy released from $%
\phi $ affects evolution of our brane Universe by analyzing gravitational
field equations \cite{09sms,09hs} written with respect to anholonomic frames. We
consider that the total energy--momentum tensor has a similar structure as
in holonomic coordinates but with the some anholonomic variables, including
the contribution of bulk cosmological constant,
\[
T_{MN}=-\kappa _5^{-2}\Lambda _5g_{MN}+T_{MN}^{(\phi )}+S_{MN}\delta (w),
\]
where $S_{MN}$ is the stress tensor on the brane and the capital
Latin indices $M,N,...$ run values $1,2,...5$ (we follow the
denotations from \cite{09inf5} with that difference that the
coordinates are reordered and stated with respect to anholonomic
frames). One introduces a  further decomposition as $S_{\mu \nu
}=-\sigma q_{\mu \nu }+\tau _{\mu \nu },$ where $\tau _{\mu \nu
}$ represents the energy--momentum tensor of the radiation fields
produced by the decay of $\phi $ and it is of the form $\tau _\nu
^\mu ={\rm diag}(2p_r,-\rho _r,p_r,0)$ with $p_r=\rho _r/3$ which
defines an anisotropic distribution of matter because of
anholonomy of the frame of reference.

We can remove the considerations on an anisotropic brane (hypersurface) by
using a unit vector $n_M=(0,0,1,0,0)$ normal to the brane for which the
extrinsic curvature of a $w=const$ hypersurface is given by $%
K_{MN}=q_M^Pq_N^Qn_{Q;P}$ with $q_{MN}=g_{MN}-n_Mn_N$. Applying the Codazzi
equation and the 5D Einstein equations with anholonomic variables
\cite{09v1,09v2}, we find
\begin{equation}
D_\nu K_\mu ^\nu -D_\mu K=\kappa _5^2T_{MN}n^Nq_\mu ^M=\kappa _5^2T_{\mu
w}=\kappa _5^2(\delta _t\phi )(\delta _w\phi )\delta _\mu ^2,  \label{k1}
\end{equation}
where $\delta _\mu ^1$ is the Kronecker symbol, small Greek indices
parametrize coodinates on the brane, $D_\nu $ is the 4D covariant derivative
with respect to the metric $q_{\mu \nu }$. The above equation reads
\begin{equation}
D_\nu K_0^{\nu +}-D_0K^{+}=\kappa _5^2\left[ {\frac{\Gamma _D}{{4k}}(}\delta
_t\phi )^2(0,t,0)+{\frac{\Gamma _D}{{4}k_y}(}\partial _y\phi
)^2(0,t,0)\right] ,  \label{k1a}
\end{equation}
near the brane  $w\longrightarrow +0$ and neglecting
non--homogeneous behavior, by putting $y=0.$ We have
\begin{equation}
D_\nu K_\mu ^{\nu +}-D_\mu K^{+}=-{\frac{{\kappa _5^2}}2}D_\nu S_\mu ^\nu =-{%
\frac{{\kappa _5^2}}2}D_\nu \tau _\mu ^\nu .  \label{eqa1}
\end{equation}
which follows from the junction condition and  $Z_2$--symmetry with\\ $K_{\mu
\nu }^{+}=-{\frac{{\kappa _5^2}}2}\left( {S_{\mu \nu }-{\frac 13}q_{\mu \nu
}S}\right) .$ Using (\ref{k1a}) and (\ref{eqa1}), we get
\[
D_\nu \tau _\mu ^\nu =-{\frac{\Gamma _D}{2k}(}\delta _t\phi )^2\delta _\mu
^2-{\frac{\Gamma _D^y}{2k_y}(}\partial _y\phi )^2\delta _\mu ^3,
\]
i. e.,
\[
\delta _t{\rho _r}=-3H(\rho _r+p_r)+{\frac{\Gamma _D}{2k}(}\delta _t\phi )^2+%
{\frac{\Gamma _D^y}{2k_y}(}\partial _y\phi )^2=-4H\rho _r+{\frac{\Gamma _D}{%
2k}(}\delta _t\phi )^2+{\frac{\Gamma _D^y}{2k_y}(}\partial _y\phi )^2,
\]
on the anisotropic brane. This equation describe the reheating in an
anisotropic  perturbation theory (for  inflation in 4D theory see \cite{09prt}%
).

The 4D Einstein equations with the Einstein tensor  $G_{~\mu }^{(4)\nu }$
were proven  \cite{09hs} to have the form
\[
G_{~\mu }^{(4)\nu }=\kappa _4^2\left( {T_{~\mu }^{(s)\nu }+\tau _\mu ^\nu }%
\right) +\kappa _5^4\pi _\mu ^\nu -E_\mu ^\nu ,
\]
with $
T_{~\mu }^{(s)\nu }\equiv {\frac 1{{6k}}}\left[ {4q^{\nu \zeta }(\delta }%
_\mu {\phi ){(\delta }_\zeta {\phi )}+\left( {{\frac 32{(\delta }_\zeta {%
\phi )}}^2-{\frac 52}q^{\xi \zeta }{(\delta }_\xi {\phi )(\delta }_\zeta {%
\phi )}-\frac 32m^2\phi ^2}\right) q_\mu ^\nu }\right], $
 where  $\pi _\mu ^\nu $ contains
  terms quadratic in $\tau _\alpha ^\beta $ which are higher order in
 $\rho_r/(kM_4)^2$ and are consistently neglected in our analysis.
$E_\mu ^\nu \equiv C_{\mu w}^{w\nu }$ is a component of the 5D Weyl
 tensor $C_{PQ}^{MN}$ treated as a source of  dark radiation \cite{09mukoyama}.

With respect to anholonomic frames the 4D Bianchi identities are written in
the usual manner,
\begin{equation}
D_\nu G_{~\mu }^{(4)\nu }=\kappa _4^2\left( {D_\nu T_{~\mu }^{(s)\nu }+D_\nu
\tau _\mu ^\nu }\right) -D_\nu E_\mu ^\nu =0,  \label{bianchi}
\end{equation}
with that difference that $D_\nu G_{~\mu }^{(4)\nu }=0$ only for holonomic
frames but in the anholonomic cases, for general constraints one could be
 $D_\nu G_{~\mu }^{(4)\nu }\neq 0$ \cite{09v1,09v2}. In this paper we shall
consider such constraints for which the equalities (\ref{bianchi}) hold
which  yield
\begin{eqnarray*}
D_\nu E_2^\nu &=& -{\frac{{\kappa _4^2}}{{4k}}}{\frac \delta {{\partial t}}}%
\left[ {{(}\delta _t\phi )^2-{(}\delta _w\phi )^2-{(\partial }}_y{\phi
)^2+m^2\phi ^2}\right]   \\
 && -\frac{2\kappa _4^2H}k{{(}\delta _t\phi )^2}-{\frac{{%
\kappa _4^2}}{2k}}\Gamma _D{{(}\delta _t\phi )^2}-{\frac{{\kappa _4^2}}{2k}}%
_y\Gamma _D^y{(\partial }_y{\phi )^2}.
\end{eqnarray*}
Putting on the anisotropic brane  ${{(}\delta _w\phi )^2=}\Gamma _D^2{{(}%
\delta _t\phi )^2}/(16k^2)$ and\\ ${(\partial }_y{\phi )^2=(}\Gamma _D^y)^2{{(}%
\delta _t\phi )^2}/(16k_y^2)$ similarly to Ref. \cite{09inf5}, by substituting
usual partial derivatives into 'elongated' ones and introducing $\varphi
(t)\equiv \phi (0,t.y)/\sqrt{2k},$ $b=a(t)e^{-k_y|y|}$ and $\varepsilon
\equiv E_2^2/\kappa _4^2$ we prove the evolution equations in the brane
universe $w=0$

\newpage

\begin{eqnarray*}
H^2 &=&\left( {{\frac{\delta _tb}b}}\right) ^2={\frac{{\kappa _4^2}}3}\left(
\rho _\varphi +\rho _r+\varepsilon \right) ,~~~\rho _\varphi \equiv {{\frac
12(}\delta _t\varphi )^2+\frac 12m^2\varphi ^2=\frac{\rho _\phi }{2k}},~~~ \\
{\delta _t}\rho _\varphi  &=&-(3H+\Gamma _B){{(}\delta _t\varphi )^2}%
-J_\varphi ,~~~J_\varphi \equiv \frac{J_\phi }{2k}, \\
{\delta _t\rho _r} &=&-4H\rho _r+\Gamma _D{{(}\delta _t\varphi )^2}+\Gamma
_D^y{{(}\delta _t\varphi )^2}, \\
{\delta _t\varepsilon } &=&-4H\varepsilon -(H+\Gamma _D+\Gamma _D^y-\Gamma
_B){{(}\delta _t\varphi )^2}+J_\varphi .
\end{eqnarray*}

We find the solution of (\ref{kg}) in the background  (\ref{meta}) in the
way suggested by \cite{09gw,09inf5} by introducing an additional factor
depending on anisotropic variable $y,$
\[
\phi (t,w)=\sum\limits_{n+n^{\prime }}{c_{n+n^{\prime }}T_{n+n^{\prime
}}(t)Y_n(w)Y_{n^{\prime }}(y)}+H.C.
\]
with
\begin{eqnarray*}
T_{n+n^{\prime }}(t) &\cong &a^{-{\frac 32}}(t)e^{-i(m_n+m_{n^{\prime }})t},
\\
Y_n(w) &=&e^{2k|w|}\left[ {J_\nu ^{}\left( {{{\frac{{m_n}}k}}e^{k|w|}}%
\right) +b_nN_\nu ^{}\left( {{{\frac{{m_n}}k}}e^{k|w|}}\right)
}\right],\\
Y_{n^{\prime }}(y) & = & e^{2k_y|y|}\left[ {J_\nu ^{}\left( {{{\frac{%
m_{n^{\prime }}}k}}e^{k_y|y|}}\right) +b_{n^{\prime }}N_\nu ^{}\left( {{{%
\frac{{m_n}}k}}e^{k_y|y|}}\right) }\right] ,~
\end{eqnarray*}
for $~\nu =2\sqrt{1+{\frac{{m^2}}{{4k^2}}}}\cong 2+{\frac{{m^2}}{{4k^2}}},$
and considering that the field oscillates rapidly in cosmic expansion time
scale. The values  $m_n$ and $m_{n^{\prime }}$ are some  constants which may
take continuous values in the case of a single brane and $b_n$ and  ${%
b_{n^{\prime }}}$ are some constants determined by the boundary conditions, $%
\delta _w\phi =0$ at $w=0$ and $\partial _y\phi =0$ at $y=0.$ We write
\begin{eqnarray*}
b_n &=&\left[ 2J_\nu \left( \frac{m_n}k\right) +\frac{m_n}kJ_\nu ^{\prime
}\left( \frac{m_n}k\right) \right] \left[ 2N_\nu \left( \frac{m_n}k\right) +%
\frac{m_n}kN_\nu ^{\prime }\left( \frac{m_n}k\right) \right] ^{-1}, \\
b_{n^{\prime }} &=&\left[ 2J_\nu \left( \frac{m_{n^{\prime }}}{k_y}\right) +%
\frac{m_{n^{\prime }}}{k_y}J_\nu ^{\prime }\left( \frac{m_{n^{\prime }}}{k_y}%
\right) \right] \left[ 2N_\nu \left( \frac{m_{n^{\prime }}}{k_y}\right) +%
\frac{m_{n^{\prime }}}{k_y}N_\nu ^{\prime }\left( \frac{m_{n^{\prime }}}{k_y}%
\right) \right] ^{-1}.
\end{eqnarray*}
The effect of dissipation on the boundary conditions is given by

\newpage

\begin{eqnarray*}
b_n &\cong &\left[ \left( 2+\frac{im_n\Gamma _D}{2k^2}\right) J_\nu \left(
\frac{m_n}k\right) +\frac{m_n}kJ_\nu ^{\prime }\left( \frac{m_n}k\right)
\right] \times      \\
&& \left[ \left( 2+\frac{im_n\Gamma _D}{2k^2}\right) N_\nu \left( \frac{%
m_n}k\right) +\frac{m_n}kN_\nu ^{\prime }\left( \frac{m_n}k\right) \right]
^{-1}, \\
b_{n^{\prime }} &\cong &\left[ \left( 2+\frac{im_{n^{\prime }}\Gamma _D^y}{%
2k_y^2}\right) J_\nu \left( \frac{m_{n^{\prime }}}{k_y}\right) +\frac{%
m_{n^{\prime }}}{k_y}J_\nu ^{\prime }\left( \frac{m_{n^{\prime }}}{k_y}%
\right) \right] \times      \\
&&
 \left[ \left( 2+\frac{im_{n^{\prime }}\Gamma _D^y}{2k_y^2}%
\right) N_\nu \left( \frac{m_{n^{\prime }}}{k_y}\right) +\frac{m_{n^{\prime
}}}{k_y}N_\nu ^{\prime }\left( \frac{m_{n^{\prime }}}{k_y}\right) \right]
^{-1},
\end{eqnarray*}
where use has been made of $\dot{\partial _tT_{n+n^{\prime }}}(t)\cong
-im_{n+n^{\prime }}T_{n+n^{\prime }}(t)$.

For simplicity, let analyze the case when a single oscillation mode exists,
neglect explicit dependence of $\varphi $ on variable $y$ (the effect of
anisotropy being modelled by terms like $m_{n^{\prime }},$ $k_y$ and  $%
\Gamma _D^y$and compare our results with those for isotropic inflation \cite
{09inf5}. In this case we approximate $\delta \varphi \simeq \dot{\varphi},$
where dot is used for the partial derivative $\partial _t.$ We find
\begin{equation}
J_\varphi =(m_n^2+m_{n^{\prime }}^2-m^2)\varphi \dot{\varphi}.
\label{jvarphi}
\end{equation}

The evolution of the dark radiation is approximated  in the regime when $%
\varphi (t)$ oscillates rapidly, parametrized as
\[
\varphi (t)=\varphi _i\left( \frac{a(t)}{a(t_i)}\right) ^{-3/2}e^{-i\lambda
_{n+n^{\prime }}(t-t_i)},\ \lambda _{n+n^{\prime }}\equiv m_n+m_{n^{\prime
}}-\frac i2\Gamma _B,
\]
with $m_n+m_{n^{\prime }}\gg H$ and $\Gamma _B$ being positive constants
which  assumes that only a single oscillation mode exists.

Then the evolution equation of $\varepsilon (t)\equiv \kappa _4^{-2}E_2^2$
is given by
\[
\frac{\partial \varepsilon }{\partial t}=-4H\varepsilon -(\Gamma _D+\Gamma
_D^y+H-\Gamma _B)\dot{\varphi}^2-(m^2-m_n^2-m_{n^{\prime }}^2)\varphi \dot{%
\varphi}.
\]
The next approximation is to consider $\varphi $ as oscillating rapidly in
the expansion time scale by averaging the right-hand-side of evolution
equations over an oscillation period. Using $\overline{\dot{\varphi}^2}%
(t)=(m_n^2+m_{n^{\prime }}^2)\overline{\varphi ^2}(t)$ and $\overline{%
\varphi \dot{\varphi}}(t)=-(3H+\Gamma _B)\overline{\varphi
^2}(t)/2$, we obtain the following set of evolution equations in
the anisotropic brane universe $w=0,$ for small non--homogeneities
on $y,$ where the bar denotes average over the oscillation period.
\begin{eqnarray}
H^2 &=&\left( {{\frac{{\dot{a}}}a}}\right) ^2\cong {\frac{{\kappa _4^2}}3}%
\left( \rho _\varphi +\rho _r+\varepsilon \right) ,  \nonumber \\
\frac{\partial \rho _\varphi }{\partial t} &=&-\frac 12(3H+\Gamma
_B)(m^2+m_n^2+m_{n^{\prime }}^2)\overline{\varphi ^2},  \nonumber \\
{\frac{{\partial \rho _r}}{{\partial t}}} &=&-4H\rho _r+\left( \Gamma
_Dm_n^2+\Gamma _D^ym_{n^{\prime }}^2\right) \overline{\varphi ^2},
\label{reqb} \\
{\frac{{\partial \varepsilon }}{{\partial t}}} &=&-4H\varepsilon -(\Gamma
_D+H-\Gamma _B)m_n^2\overline{\varphi ^2}-(\Gamma _D^y+H-\Gamma
_B)m_{n^{\prime }}^2\overline{\varphi ^2}      \nonumber \\
&&+\frac 12(3H+\Gamma
_B)(m^2-m_n^2-m_{n^{\prime }}^2)\overline{\varphi ^2},  \label{eeqb}
\end{eqnarray}
with $
\overline{\varphi ^2}(t)\equiv \overline{\varphi _i^2}\left( \frac{a(t)}{%
a(t_i)}\right) ^{-3}e^{-\Gamma _B(t-t_i)}.$

The system (\ref{reqb}) and (\ref{eeqb}) was analyzed in Ref. \cite{09inf5}
for the case when $m_{n^{\prime }}$ and $\Gamma _D^y$ vanishes:

It was concluded that if  $m_n\geq m,$  we do not recover standard cosmology
on the brane after inflation. The same holds true in the presence of
anisotropic terms.

In the locally isotropic case it was proven that if $m_n\ll n$  the last
term of (\ref{eeqb}) is dominant and we find more dark radiation than
ordinary radiation unless $\Gamma _B$ is extremely small with $\Gamma
_B/\Gamma _D<m_n^2/m^2\ll 1$. The presence of anisotropic values $%
m_{n^{\prime }}$ and $\Gamma _D^y$ can violate this condition.

The cases $m_n,m_{n^{\prime }}\lesssim m$ are  the most delicate
cases because the final amount of dark radiation can be either
positive or negative depending on the details of the model
parameters and the type of anisotropy. The amount of extra
radiation--like matter have to be hardly constrained \cite{09ns} if
we wont a successful primordial nucleosynthesis.  In order to have
sufficiently small $\varepsilon $ compared with $\rho _r$ after
reheating without resorting to subsequent both isotropic and
anisotropic entropy production within the brane, the magnitude of
creation
terms of $\varepsilon $ should be vanishingly small at the reheating epoch $%
H\simeq \Gamma _B$. This hold if  only there are presented isotropic and/or
anisotropic modes which  satisfies the inequality
\begin{equation}
\left| 2\Gamma _B(m^2-m_n^2-m_{n^{\prime }}^2)-\Gamma _Dm_n^2-\Gamma
_D^ym_{n^{\prime }}^2\right| \ll \Gamma _Dm_n^2+\Gamma _D^ym_{n^{\prime }}^2.
\label{cond}
\end{equation}

We conclude that the relation (\ref{cond}) should be satisfied for the
graceful exit of anisotropic brane inflation driven by a bulk scalar field $%
\phi $. The presence of off--diagonal components of the metric in the bulk
which induces brane anisotropies could modify the process of
nucleosynthesis.

In summary we have analyzed a model of anisotropic inflation generated
 by a 5D off--diagonal metric in the bulk. We studied entropy production
on the anisotropic 3D brane from a decaying bulk scalar field $\phi $
by considering anholonomic frames and introducing dissipation terms
 to its  equation of motion phenomenologically. We illustrated that
 the dark radiation is significantly produced at the same time unless the
 inequality (\ref{cond}) is satisfied. Comparing our results with
 a similar model in locally isotropic background we found that
 off--diagonal metric components and anisotropy results in additional
 dissipation terms and coefficients which could substantially
modify the scenario of inflation but could not to fall the
qualitative
 isotropic possibilities for well defined cases with specific
 form of isotropic and anisotropic dissipation.
Although we have analyzed only a  case of anisotropic metric with a
specific form of the dissipation, we expect our conclusion is generic and
applicable to other forms of anisotropies and dissipation,
 because it is essentially an outcome of the anholonomic frame method
 and Bianchi identities (\ref{bianchi}). We therefore conclude that
 in the brane--world picture of the Universe it is very important what
 type of metrics and frames we consider, respectively, diagonal
 or off--diagonal and  holonomic or anholonomic (i. e. locally isotropic
 or anisotropic). In all cases there are conditions to be imposed
 on anisotropic parameters and polarizations when
 the dominant part of the entropy we observe experimentally originates
within the brane rather than in the locally anisotropic  bulk.
 Such extra dimensional vacuum gravitational anisotropic
 polarizations of cosmological inflation parameters may be observed
 experimentally.

\vskip 0.5cm %\acknowledgements{
The authors are grateful to D. Singleton and E. Gaburov for useful
communications. The work of S. V.  is partially supported by the
  "The 2000--2001 California State University Legislative Award".
 %}

%%%%%%%%%%%%%%%%%%%%%%%%%%%%%%%%%%%%%%%%%%%%%%%%%%%%%%%%%%%%%%%%%%%%%%%%%%%%%

\chapter[A New Method of Constructing Exact Solutions   ]
{A New Method of Constructing Black Hole Solutions in Einstein and 5D Gravity   }

{\bf Abstract}
\footnote{\copyright\  S. Vacaru, A New Method of Constructing Black Hole
 Solutions in Einstein  and 5D Dimension Gravity, hep-th/0110250}

It is formulated a new 'anholonomic frame' method of constructing
 exact solutions of Einstein equations with off--diagonal
metrics in 4D and 5D gravity. The previous approaches and results
\cite{10v,10v2,10vf,10v1} are summarized and generalized  as three
theorems which state the conditions when two types of ansatz
result in integrable gravitational field equations.  There are
constructed and analyzed different classes of anisotropic and/or
warped vacuum 5D and 4D metrics describing ellipsoidal black
holes with static anisotropic horizons and possible anisotropic
gravitational polarizations and/or running  constants. We
conclude that warped metrics can be defined in 5D vacuum gravity
without postulating any brane configurations with specific energy
momentum tensors. Finally, the 5D and 4D anisotropic Einstein
spaces with cosmological constant are investigated.

\section{Introduction}

During the last three years large extra dimensions and brane worlds attract
a lot of attention as possible new paradigms for gravity, particle physics
and string/M--theory. As basic references there are considered Refs.
\cite{10stringb}, for string gravity papers, the Refs. \cite{10arkani}, for extra
dimension particle fields, and gravity phenomenology with effective Plank
scale and \cite{10rs}, for the simplest and comprehensive models proposed by
Randall and Sundrum (in brief, RS; one could also find in the same line some
early works \cite{10akama} as well to cite, for instance, \cite{10shir} for
further developments with supersymmetry, black hole solutions and
cosmological scenaria).

The new ideas are based on the assumption that our Universe is realized as a
three dimensional (in brief, 3D) brane, modelling a 4D pseudo--Riemannian
spacetime, embedded in the 5D anti--de Sitter ($AdS_5$) bulk spacetime. It
was proved in the RS papers \cite{10rs} that in such models the extra
dimensions could be not compactified (being even infinite) if a nontrivial
warped geometric configuration is defined. Some warped factors are essential
for solving the mass hierarchy problem and localization of gravity which
at low energies can ''bound'' the matter fields on a 3D subspace. In
general, the gravity may propagate in extra dimensions.

In connection to modern string and brane gravity it is very
important to develop new methods of constructing  exact solutions
of gravitational field equations in the bulk of extra dimension
spacetime and to develop new applications in particle physics,
astrophysics and cosmology. This paper is devoted to elaboration
of a such method and investigation of new classes of anisotropic
black hole solutions.

In higher dimensional gravity much attention has been paid to the
off--diagonal metrics beginning the Salam, Strathee and Peracci works
 \cite{10sal} which showed that including off--diagonal components in higher
dimensional metrics is equivalent to including $U(1),SU(2)$ and $SU(3)$
gauge fields. They considered a parametrization of metrics of type
\begin{equation}
g_{\alpha \beta }=\left[
\begin{array}{ll}
g_{ij}+N_i^aN_j^bh_{ab} & N_j^eh_{ae} \\
N_i^eh_{be} & h_{ab}
\end{array}
\right]  \label{salam}
\end{equation}
where the Greek indices run values $1,2,...,n+m,$ the Latin indices $%
i,j,k,...$ from the middle of the alphabet run values $1,2,...,n$ (usually,
in Kaluza--Klein theories one put $n=4)$ and the Latin indices from the
beginning of the alphabet, $a,b,c,...,$ run values $n+1,n+2,...,n+m$ taken
for extra dimensions. The local coordinates on higher dimensional spacetime
are denoted $u^\alpha =\left( x^i,y^a\right) $ which defines respectively
the local coordinate frame (basis), co--frame (co--basis, or dual basis)
\begin{eqnarray}
\partial _\alpha &=&\frac \partial {\partial u^\alpha }=\left( \partial _i=%
\frac \partial {\partial x^i},\partial _a=\frac \partial {\partial y^a}%
\right) ,  \label{3pder} \\
d^\alpha &=&du^\alpha =\left( d^i=dx^i,d^a=dy^a\right) .  \label{5pdif}
\end{eqnarray}
The coefficients $g_{ij}=g_{ij}\left( u^\alpha \right) ,h_{ab}=h_{ab}\left(
u^\alpha \right) $ and $N_i^a=N_i^a\left( u^\alpha \right) $ should be
defined by a solution of the Einstein equations (in some models of
Kaluza--Klein gravity \cite{10ow} one considers the Einstein--Yang--Mills
fields) for extra dimension gravity.

The metric (\ref{salam}) can be rewritten in a block $(n\times n)\oplus
(m\times m)$ form
\begin{equation}
g_{\alpha \beta }=\left(
\begin{array}{ll}
g_{ij} & 0 \\
0 & h_{ab}
\end{array}
\right)  \label{diagm}
\end{equation}
with respect to some anholonomic frames (N--elongated basis), co--frame
(N--elongated co--basis),
\begin{eqnarray}
\delta _\alpha &=&\frac \delta {\partial u^\alpha }=\left( \delta
_i=\partial _i-N_i^b\partial _b,\delta _a=\partial _a\right) ,  \label{5dder}
\\
\delta ^\alpha &=&\delta u^\alpha =\left( \delta ^i=d^i=dx^i,\delta
^a=dy^a+N_i^adx^i\right) ,  \label{17ddif}
\end{eqnarray}
which satisfy the anholonomy relations
\[
\delta _\alpha \delta _\beta -\delta _\beta \delta _\alpha =w_{\alpha \beta
}^\gamma \delta _\gamma
\]
with the anholonomy coefficients computed as
\begin{equation}
w_{ij}^k=0,w_{aj}^k=0,w_{ab}^k=0,w_{ab}^c=0,w_{ij}^a=\delta _iN_j^a-\delta
_jN_i^a,w_{ja}^b=-w_{aj}^b=\partial _aN_j^b.  \label{anholonomy}
\end{equation}

In Refs. \cite{10sal} the coefficients $N_i^a$ (hereafter, N--coefficients)
were treated as some $U(1),$ $SU(2)$ or $SU(3)$ gauge fields (depending on
the extra dimension $m).$ There are another classes of gravity models which
are constructed on vector (or tangent) bundles generalizing the Finsler
geometry \cite{10ma}. In such approaches the set of functions $N_i^a$ were
stated to define a structure of nonlinear connection and the variables $y^a$
were taken to parametrize fibers in some bundles. In the theory of locally
anisotropic (super) strings and supergravity, and gauge generalizations of
the so--called Finsler--Kaluza--Klein gravity the coefficients $N_i^a$ were
suggested to be found from some alternative string models in low energy
limits or from gauge and spinor variants of gravitational field equations
with anholonomic frames and generic local anisotropy \cite{10vf}.

The Salam, Strathee and Peracci \cite{10sal} idea on a gauge field like status
of the coefficients of off--diagonal metrics in extra dimension gravity
was developed in a new fashion by applying the method of anholonomic frames
with associated nonlinear connections just on the (pseudo) Riemannian spaces
\cite{10v,10v2}. The approach allowed to construct new classes of solutions of
Einstein's equations in three (3D), four (4D) and five (5D) dimensions with
generic local anisotropy ({\it e.g.} static black hole and cosmological
solutions with ellipsoidal or toroidal symmetry, various soliton--dilaton 2D
and 3D configurations in 4D gravity, and wormhole and flux tubes with
anisotropic polarizations and/or running on the 5th coordinate constants
with different extensions to backgrounds of rotation ellipsoids, elliptic
cylinders, bipolar and toroidal symmetry and anisotropy).

Recently, it was shown in Refs. \cite{10v1} that if we consider off--diagonal
metrics which can be equivalently diagonalized with respect to corresponding
anholonomic frames, the RS theories become substantially locally anisotropic
with variations of constants on extra dimension coordinate or with
anisotropic angular polarizations of effective 4D constants, induced by
higher dimension and/or anholonomic gravitational interactions.

The basic idea on the application of the anholonomic frame method
for constructing exact solutions of the Einstein equations is to
define such N--coefficients when a given type of off--diagonal
metric is diagonalized with respect to some anholonomic frames
(\ref{5dder}) and the Einstein equations, re--written in mixed
holonomic and anholonomic variables, transform into a system of
partial differential equations with separation of variables which
admit exact solutions. This approach differs from the usual
tetradic method where the differential forms and frame bases are
all 'pure' holonomic or 'pure'' anholonomic. In our case the
N--coefficients and associated N--elongated partial derivatives
(\ref{5dder}) are chosen as to be some undefined values which at
the final step are fixed as to separate variables and satisfy the
Einstein equations.

The first aim of this paper is to formulate three theorems (and to suggest
the way of their proof) for two off--diagonal metric ansatz which admit
anholonomic transforms resulting in a substantial simplification of the
system of Einstein equations in 5D and 4D gravity. The second aim is to
consider four applications of the anholonomic frame method in order to
construct new classes of exact solutions describing ellipsoidal black holes
with anisotropies and running of constants. We emphasize that is possible to
define classes of warped on the extra dimension coordinate metrics which are
 exact solutions of 5D vacuum gravity. We analyze basic
physical properties of such solutions. We also investigate 5D spacetimes
with anisotropy and cosmological constants.

We use the term 'locally anisotropic' spacetime (or 'anisotropic' spacetime)
for a 5D (4D) pseudo-Riemannian spacetime provided with an anholonomic frame
structure with mixed holonomic and anholonomic variables. The anisotropy of
gravitational interactions is modelled by off--diagonal metrics, or,
equivalently, by theirs diagonalized analogs given with respect to
anholonomic frames.

The paper is organized as follow:\ In Sec. II we formulate three
theorems for two types of off--diagonal metric ansatz, construct
the corresponding exact solutions of 5D vacuum Einstein equations
and illustrate the possibility of extension by introducing matter
fields (the necessary geometric background and some proofs are
presented in the Appendix). We also consider the conditions when
the method generates 4D metrics. In Sec. III we construct two
classes of 5D anisotropic black hole solutions with rotation
ellipsoid horizon and consider  subclasses and reparemetization
of such solutions in order to generate new ones. Sec. IV is
devoted to 4D ellipsoidal black hole solutions. In Sec. V we
extend the method for anisotropic 5D and 4D spacetimes with
cosmological constant, formulate two theorems on basic properties
of the system of field equations and theirs solutions, and give
an example of 5D anisotropic black solution with cosmological
constant. Finally, in Sec. VI, we conclude and discuss the
obtained results.

\section{Off--Diagonal Metrics in Extra Dimension Gra\-vity}

The bulk of solutions of 5D Einstein equations and their
reductions to 4D (like the Schwarzshild solution and brane
generalizations \cite{10bh}, metrics with cylindrical and toroidal
symmetry \cite{10lemos}, the Friedman--Robertson--Walker metric
and brane generalizations \cite{10wc}) were constructed by using
diagonal metrics and extensions to solutions with rotation, all
given with respect to holonomic coordinate frames of references.
This Section is devoted to a geometrical and nonlinear partial
derivation equations formalism which deals with more general,
generic off--diagonal metrics with respect to coordinate frames,
and anholonomic frames. It summarizes and generalizes various
particular cases and ansatz used for construction of exact
solutions of the Einstein gravitational field equations in 3D, 4D
and 5D gravity \cite{10v,10v2,10v1}.

\subsection{The first ansatz for vacuum Einstein equations}

Let us consider a 5D pseudo--Riemannian spacetime provided with local
coordinates $u^\alpha =(x^i,y^4=v,y^5),$ for $i=1,2,3.$ Our aim is to prove
that a metric ansatz of type (\ref{salam}) can be diagonalized by some
anholonomic transforms with the N--coefficients $N_a^i=N_a^i\left(
x^i,v\right) $ depending on variables $\left( x^i,v\right) $ and to define
the corresponding system of vacuum Einstein equations in the bulk. The exact
solutions of the Einstein equations to be constructed will depend on the
so--called holonomic variables $x^i$ and on one anholonomic (equivalently,
anisotropic) variable $y^4=v.$ In our further considerations every
coordinate from a set $u^\alpha $ can be stated to be time like, 3D space
like or extra dimensional.

For simplicity, the partial derivatives will be denoted like $a^{\times
}=\partial a/\partial x^{1},a^{\bullet }=\partial a/\partial
x^{2},$ $a^{^{\prime }}=\partial a/\partial x^{3},a^{\ast }=\partial a/\partial
v.$

We begin our approach by considering a 5D quadratic line element
\begin{equation}
ds^{2}=g_{\alpha \beta }\left( x^{i},v\right) du^{\alpha }du^{\beta }
\label{2metric}
\end{equation}
with the metric coefficients $g_{\alpha \beta }$ parametrized (with respect
to the coordinate frame (\ref{5pdif})) by an off--diagonal matrix (ansatz)

{%%\footnotesize
\begin{equation}
\left[
\begin{array}{ccccc}
g_{1}+w_{1}^{\ 2}h_{4}+n_{1}^{\ 2}h_{5} & w_{1}w_{2}h_{4}+n_{1}n_{2}h_{5} &
w_{1}w_{3}h_{4}+n_{1}n_{3}h_{5} & w_{1}h_{4} & n_{1}h_{5} \\
w_{1}w_{2}h_{4}+n_{1}n_{2}h_{5} & g_{2}+w_{2}^{\ 2}h_{4}+n_{2}^{\ 2}h_{5} &
w_{2}w_{3}h_{4}+n_{2}n_{3}h_{5} & w_{2}h_{4} & n_{2}h_{5} \\
w_{1}w_{3}h_{4}+n_{1}n_{3}h_{5} & w_{2}w_{3}h_{4}+n_{2}n_{3}h_{5} &
g_{3}+w_{3}^{\ 2}h_{4}+n_{3}^{\ 2}h_{5} & w_{3}h_{4} & n_{3}h_{5} \\
w_{1}h_{4} & w_{2}h_{4} & w_{3}h_{4} & h_{4} & 0 \\
n_{1}h_{5} & n_{2}h_{5} & n_{3}h_{5} & 0 & h_{5}
\end{array}
\right] ,  \label{4ansatz}
\end{equation}
} where the coefficients are some necessary smoothly class functions of
type:
\begin{eqnarray}
g_{1} &=&\pm 1,g_{2,3}=g_{2,3}(x^{2},x^{3}),h_{4,5}=h_{4,5}(x^{i},v),
\nonumber \\
w_{i} &=&w_{i}(x^{i},v),n_{i}=n_{i}(x^{i},v).  \nonumber
\end{eqnarray}

\begin{lemma}
The quadratic line element (\ref{2metric}) with metric coefficients (\ref
{4ansatz}) can be diagonalized,
\begin{equation}
\delta
s^{2}=[g_{1}(dx^{1})^{2}+g_{2}(dx^{2})^{2}+g_{3}(dx^{3})^{2}+h_{4}(\delta
v)^{2}+h_{5}(\delta y^{5})^{2}],  \label{5dmetric}
\end{equation}
with respect to the anholonomic co--frame $\left( dx^{i},\delta v,\delta
y^{5}\right) ,$ where
\begin{equation}
\delta v=dv+w_{i}dx^{i}\mbox{ and }\delta y^{5}=dy^{5}+n_{i}dx^{i}
\label{1ddif1}
\end{equation}
which is dual to the frame $\left( \delta _{i},\partial _{4},\partial
_{5}\right) ,$ where
\begin{equation}
\delta _{i}=\partial _{i}+w_{i}\partial _{4}+n_{i}\partial _{5}.
\label{1dder1}
\end{equation}
\end{lemma}

In the Lemma 1 the $N$--coefficients from (\ref{5dder}) and (\ref{17ddif}) are
parametrized like $N_{i}^{4}=w_{i}$ and $N_{i}^{5}=n_{i}.$

The proof of the Lemma 1 is a trivial computation if we substitute the
values of (\ref{1ddif1}) into the quadratic line element (\ref{5dmetric}).
Re-writing the metric coefficients with respect to the coordinate basis (\ref
{5pdif}) we obtain just the quadratic line element (\ref{2metric}) with the
ansatz (\ref{4ansatz}).

In the Appendix A we outline the basic formulas from the geometry of
anholonomic frames with mixed holonomic and anholonomic variables and
associated nonlinear connections on (pseudo) Riemannian spaces.

Now we can formulate the

\begin{theorem}
The nontrivial components of the 5D vacuum Einstein equations, $R_{\alpha
}^{\beta }=0,$ (see (\ref{1einsteq3}) in the Appendix) for the metric (\ref
{5dmetric}) given with respect to anholonomic frames (\ref{1ddif1}) and (\ref
{1dder1}) are written in a form with separation of variables:
\begin{eqnarray}
R_{2}^{2}=R_{3}^{3}=-\frac{1}{2g_{2}g_{3}}[g_{3}^{\bullet \bullet }-\frac{%
g_{2}^{\bullet }g_{3}^{\bullet }}{2g_{2}}-\frac{(g_{3}^{\bullet })^{2}}{%
2g_{3}}+g_{2}^{^{\prime \prime }}-\frac{g_{2}^{^{\prime }}g_{3}^{^{\prime }}%
}{2g_{3}}-\frac{(g_{2}^{^{\prime }})^{2}}{2g_{2}}] &=&0,  \label{4ricci1a} \\
S_{4}^{4}=S_{5}^{5}=-\frac{\beta }{2h_{4}h_{5}} &=&0,  \label{3ricci2a} \\
R_{4i}=-w_{i}\frac{\beta }{2h_{5}}-\frac{\alpha _{i}}{2h_{5}} &=&0,
\label{3ricci3a} \\
R_{5i}=-\frac{h_{5}}{2h_{4}}\left[ n_{i}^{\ast \ast }+\gamma n_{i}^{\ast }%
\right] &=&0,  \label{3ricci4a}
\end{eqnarray}
where
\begin{equation}
\alpha _{i}=\partial _{i}{h_{5}^{\ast }}-h_{5}^{\ast }\partial _{i}\ln \sqrt{%
|h_{4}h_{5}|},\beta =h_{5}^{\ast \ast }-h_{5}^{\ast }[\ln \sqrt{|h_{4}h_{5}|}%
]^{\ast },\gamma =3h_{5}^{\ast }/2h_{5}-h_{4}^{\ast }/h_{4}.  \label{3abc}
\end{equation}
\end{theorem}

Here the separation of variables means: 1) we can define a function $%
g_{2}(x^{2},x^{3})$ for a given $g_{3}(x^{2},x^{3}),$ or inversely, to
define a function $g_{2}(x^{2},x^{3})$ for a given $g_{3}(x^{2},x^{3}),$
from equation (\ref{4ricci1a}); 2) we can define a function $%
h_{4}(x^{1},x^{2},x^{3},v)$ for a given $h_{5}(x^{1},x^{2},x^{3},v),$ or
inversely, to define a function $h_{5}(x^{1},x^{2},x^{3},v)$ for a given $%
h_{4}(x^{1},x^{2},x^{3},v),$ from equation (\ref{3ricci2a}); 3-4) having the
values of $h_{4}$ and $h_{5},$ we can compute the coefficients (\ref{3abc})
which allow to solve the algebraic equations (\ref{3ricci3a}) and to
integrate two times on $v$ the equations (\ref{3ricci4a}) which allow to find
respectively the coefficients $w_{i}(x^{k},v)$ and $n_{i}(x^{k},v).$

The proof of Theorem 1 is a straightforward tensorial and differential
calculus for the components of Ricci tensor (\ref{4dricci}) as it is outlined
in the Appendix A. We omit such cumbersome calculations in this paper.

\subsection{The second ansatz for vacuum Einstein equations}

We can consider a generalization of the constructions from the previous
subsection by introducing a conformal factor $\Omega (x^{i},v)$ and
additional deformations of the metric via coefficients $\zeta _{\hat{\imath}%
}(x^{i},v)$ (indices with 'hat' take values like $\hat{{i}}=1,2,3,5).$ The
new metric is written like
\begin{equation}
ds^{2}=\Omega ^{2}(x^{i},v)\hat{{g}}_{\alpha \beta }\left( x^{i},v\right)
du^{\alpha }du^{\beta },  \label{4cmetric}
\end{equation}
were the coefficients $\hat{{g}}_{\alpha \beta }$ are parametrized by the
ansatz {\scriptsize
\begin{equation}
\left[
\begin{array}{ccccc}
g_{1}+(w_{1}^{\ 2}+\zeta _{1}^{\ 2})h_{4}+n_{1}^{\ 2}h_{5} &
(w_{1}w_{2}+\zeta _{1}\zeta _{2})h_{4}+n_{1}n_{2}h_{5} & (w_{1}w_{3}+\zeta
_{1}\zeta _{3})h_{4}+n_{1}n_{3}h_{5} & (w_{1}+\zeta _{1})h_{4} & n_{1}h_{5}
\\
(w_{1}w_{2}+\zeta _{1}\zeta _{2})h_{4}+n_{1}n_{2}h_{5} & g_{2}+(w_{2}^{\
2}+\zeta _{2}^{\ 2})h_{4}+n_{2}^{\ 2}h_{5} & (w_{2}w_{3}+\zeta _{2}\zeta
_{3})h_{4}+n_{2}n_{3}h_{5} & (w_{2}+\zeta _{2})h_{4} & n_{2}h_{5} \\
(w_{1}w_{3}+\zeta _{1}\zeta _{3})h_{4}+n_{1}n_{3}h_{5} & (w_{2}w_{3}+\zeta
_{2}\zeta _{3})h_{4}+n_{2}n_{3}h_{5} & g_{3}+(w_{3}^{\ 2}+\zeta _{3}^{\
2})h_{4}+n_{3}^{\ 2}h_{5} & (w_{3}+\zeta _{3})h_{4} & n_{3}h_{5} \\
(w_{1}+\zeta _{1})h_{4} & (w_{2}+\zeta _{2})h_{4} & (w_{3}+\zeta _{3})h_{4}
& h_{4} & 0 \\
n_{1}h_{5} & n_{2}h_{5} & n_{3}h_{5} & 0 & h_{5}+\zeta _{5}h_{4}
\end{array}
\right]   \label{4ansatzc}
\end{equation}
}

Such 5D pseudo--Riemannian metrics are considered to have second order
anisotropy \cite{10vf,10ma}. For trivial values $\Omega =1$ and $\zeta _{\hat{%
\imath}}=0,$ the squared line interval (\ref{4cmetric}) transforms into (\ref
{2metric}).

\begin{lemma}
The quadratic line element (\ref{4cmetric}) with metric coefficients (\ref
{4ansatzc}) can be diagonalized,
\begin{equation}
\delta s^{2}=\Omega
^{2}(x^{i},v)[g_{1}(dx^{1})^{2}+g_{2}(dx^{2})^{2}+g_{3}(dx^{3})^{2}+h_{4}(%
\hat{{\delta }}v)^{2}+h_{5}(\delta y^{5})^{2}],  \label{2cdmetric}
\end{equation}
with respect to the anholonomic co--frame $\left( dx^{i},\hat{{\delta }}%
v,\delta y^{5}\right) ,$ where
\begin{equation}
\delta v=dv+(w_{i}+\zeta _{i})dx^{i}+\zeta _{5}\delta y^{5}\mbox{ and }%
\delta y^{5}=dy^{5}+n_{i}dx^{i}  \label{11ddif2}
\end{equation}
which is dual to the frame $\left( \hat{{\delta }}_{i},\partial _{4},\hat{{%
\partial }}_{5}\right) ,$ where
\begin{equation}
\hat{{\delta }}_{i}=\partial _{i}-(w_{i}+\zeta _{i})\partial
_{4}+n_{i}\partial _{5},\hat{{\partial }}_{5}=\partial _{5}-\zeta
_{5}\partial _{4}.  \label{11dder2}
\end{equation}
\end{lemma}

In the Lemma 2 the $N$--coefficients from (\ref{3pder}) and (\ref{5dder}) are
parametrized in the first order anisotropy (with three anholonomic, $x^{i},$
and two anholonomic, $y^{4}$ and $y^{5},$ coordinates) like $N_{i}^{4}=w_{i}$
and $N_{i}^{5}=n_{i}$ and in the second order anisotropy (on the second
'shell', \ with four anholonomic, $(x^{i},y^{5}),$ and one anholonomic,$%
y^{4},$ coordinates) with $N_{\hat{{i}}}^{5}=\zeta _{\hat{{i}}},$ in this
work we state, for simplicity, $\zeta _{\hat{{i}}}=0.$

The Theorem 1 can be extended as to include the generalization to the second
ansatz:

\begin{theorem}
The nontrivial components of the 5D vacuum Einstein equations, $R_{\alpha
}^{\beta }=0,$ (see (\ref{1einsteq3}) in the Appendix) for the metric (\ref
{2cdmetric}) given with respect to anholonomic frames (\ref{11ddif2}) and (\ref
{11dder2}) are written in \ the same form as in the system (\ref{4ricci1a})--(%
\ref{3ricci4a}) with the additional conditions that
\begin{equation}
\hat{{\delta }}_{i}h_{4}=0\mbox{\ and\  }\hat{{\delta }}_{i}\Omega =0
\label{2conf1}
\end{equation}
and the values $\zeta _{\hat{{i}}}=\left( \zeta _{{i}},\zeta _{{5}}=0\right)
$ are found as to be a unique solution of (\ref{2conf1}); for instance, if
\begin{equation}
\Omega ^{q_{1}/q_{2}}=h_{4}~(q_{1}\mbox{ and }q_{2}\mbox{ are integers}),
\label{2confq}
\end{equation}
$\zeta _{{i}}$ satisfy the equations \
\begin{equation}
\partial _{i}\Omega -(w_{i}+\zeta _{{i}})\Omega ^{\ast }=0.  \label{3confeq}
\end{equation}
\end{theorem}

The proof of Theorem 2 consists from a straightforward calculation of the
components of the Ricci tensor (\ref{4dricci}) as it is outlined in the
Appendix. The simplest way is to use the calculus for Theorem 1 and then to
compute deformations of the canonical d--connection (\ref{4dcon}). \ Such
deformations induce corresponding deformations of the Ricci tensor (\ref
{4dricci}). \ The condition that we have the same values of the Ricci tensor
for the (\ref{4ansatz}) and (\ref{4ansatzc}) results in equations (\ref{2conf1}%
) and (\ref{3confeq}) which are compatible, for instance, if $\Omega
^{q_{1}/q_{2}}=h_{4}.$\ There are also another possibilities to satisfy the
condition (\ref{2conf1}), for instance, if $\Omega =\Omega _{1}$ $\Omega
_{2}, $ we can consider that $h_{4}=\Omega _{1}^{q_{1}/q_{2}}$ $\Omega
_{2}^{q_{3}/q_{4}}$ $\ $for some integers $q_{1},q_{2},q_{3}$ and $q_{4}.$

\subsection{General solutions}

The surprising result is that we are able to construct exact solutions of
the 5D vacuum Einstein equations for both types of the ansatz (\ref{4ansatz})
and (\ref{4ansatzc}):

\begin{theorem}
The system of second order nonlinear partial differential equations (\ref
{4ricci1a})--(\ref{3ricci4a}) and (\ref{3confeq}) can be solved in general form
if there are given some values of functions $g_{2}(x^{2},x^{3})$ (or $%
g_{3}(x^{2},x^{3})),h_{4}\left( x^{i},v\right) $ (or $h_{5}\left(
x^{i},v\right) )$ and $\Omega \left( x^{i},v\right) :$

\begin{itemize}
\item  The general solution of equation (\ref{4ricci1a}) can be written in
the form
\begin{equation}
\varpi =g_{[0]}\exp [a_{2}\widetilde{x}^{2}\left( x^{2},x^{3}\right) +a_{3}%
\widetilde{x}^{3}\left( x^{2},x^{3}\right) ],  \label{2solricci1a}
\end{equation}
were $g_{[0]},a_{2}$ and $a_{3}$ are some constants and the functions $%
\widetilde{x}^{2,3}\left( x^{2},x^{3}\right) $ define coordinate transforms $%
x^{2,3}\rightarrow \widetilde{x}^{2,3}$ for which the 2D line element
becomes conformally flat, i. e.
\begin{equation}
g_{2}(x^{2},x^{3})(dx^{2})^{2}+g_{3}(x^{2},x^{3})(dx^{3})^{2}\rightarrow
\varpi \left[ (d\widetilde{x}^{2})^{2}+\epsilon (d\widetilde{x}^{3})^{2}%
\right] .  \label{2con10}
\end{equation}

\item  The equation (\ref{3ricci2a}) relates two functions $h_{4}\left(
x^{i},v\right) $ and $h_{5}\left( x^{i},v\right) $. There are two
possibilities:

a) to compute
\begin{eqnarray}
\sqrt{|h_{5}|} &=&h_{5[1]}\left( x^{i}\right) +h_{5[2]}\left( x^{i}\right)
\int \sqrt{|h_{4}\left( x^{i},v\right) |}dv,~h_{4}^{\ast }\left(
x^{i},v\right) \neq 0;  \nonumber \\
&=&h_{5[1]}\left( x^{i}\right) +h_{5[2]}\left( x^{i}\right) v,h_{4}^{\ast
}\left( x^{i},v\right) =0,  \label{2p2}
\end{eqnarray}
for some functions $h_{5[1,2]}\left( x^{i}\right) $ stated by boundary
conditions;

b) or, inversely, to compute $h_{4}$ for a given $h_{5}\left( x^{i},v\right)
,h_{5}^{\ast }\neq 0,$%
\begin{equation}
\sqrt{|h_{4}|}=h_{[0]}\left( x^{i}\right) (\sqrt{|h_{5}\left( x^{i},v\right)
|})^{\ast },  \label{3p1}
\end{equation}
with $h_{[0]}\left( x^{i}\right) $ given by boundary conditions.

\item  The exact solutions of (\ref{3ricci3a}) for $\beta \neq 0$ is
\begin{equation}
w_{k}=\partial _{k}\ln [\sqrt{|h_{4}h_{5}|}/|h_{5}^{\ast }|]/\partial
_{v}\ln [\sqrt{|h_{4}h_{5}|}/|h_{5}^{\ast }|],  \label{4w}
\end{equation}
with $\partial _{v}=\partial /\partial v$ and $h_{5}^{\ast }\neq 0.$ If $%
h_{5}^{\ast }=0,$ or even $h_{5}^{\ast }\neq 0$ but $\beta =0,$ the
coefficients $w_{k}$ could be arbitrary functions on $\left( x^{i},v\right)
. $ \ For vacuum Einstein equations this is a degenerated case which imposes
the the compatibility conditions $\beta =\alpha _{i}=0,$ which are
satisfied, for instance, if the $h_{4}$ and $h_{5}$ are related as in the
formula (\ref{3p1}) but with $h_{[0]}\left( x^{i}\right) =const.$

\item  The exact solution of (\ref{3ricci4a}) is
\begin{eqnarray}
n_{k} &=&n_{k[1]}\left( x^{i}\right) +n_{k[2]}\left( x^{i}\right) \int
[h_{4}/(\sqrt{|h_{5}|})^{3}]dv,~h_{5}^{\ast }\neq 0;  \nonumber \\
&=&n_{k[1]}\left( x^{i}\right) +n_{k[2]}\left( x^{i}\right) \int
h_{4}dv,\qquad ~h_{5}^{\ast }=0;  \label{2n} \\
&=&n_{k[1]}\left( x^{i}\right) +n_{k[2]}\left( x^{i}\right) \int [1/(\sqrt{%
|h_{5}|})^{3}]dv,~h_{4}^{\ast }=0,  \nonumber
\end{eqnarray}
for some functions $n_{k[1,2]}\left( x^{i}\right) $ stated by boundary
conditions.

\item  The exact solution of (\ref{3confeq}) is given by some arbitrary
functions $\zeta _{i}=\zeta _{i}\left( x^{i},v\right) $ if \ both $\partial
_{i}\Omega =0$ and $\Omega ^{\ast }=0,$ we chose $\zeta _{i}=0$ for $\Omega
=const,$ and
\begin{eqnarray}
\zeta _{i} &=&-w_{i}+(\Omega ^{\ast })^{-1}\partial _{i}\Omega ,\quad \Omega
^{\ast }\neq 0,  \label{2confsol} \\
&=&(\Omega ^{\ast })^{-1}\partial _{i}\Omega ,\quad \Omega ^{\ast }\neq 0,%
\mbox{ for vacuum solutions}.  \nonumber
\end{eqnarray}
\end{itemize}
\end{theorem}

\bigskip We note that a transform (\ref{2con10}) is always possible for 2D
metrics and the explicit form of solutions depends on chosen system of 2D
coordinates and on the signature $\epsilon =\pm 1.$ In the simplest case the
equation (\ref{4ricci1a}) is solved by arbitrary two functions $g_{2}(x^{3})$
and $g_{3}(x^{2}).$ The equation (\ref{3ricci2a}) is satisfied by arbitrary
pairs of coefficients $h_{4}\left( x^{i},v\right) $ and $h_{5[0]}\left(
x^{i}\right) .$

The proof of Theorem 3 is given in the Appendix B.

\subsection{Consequences of Theorems 1--3}

We consider three important consequences of the Lemmas and Theorems
formulated in this Section:

\begin{corollary}
The non--trivial diagonal components of the Einstein tensor, $G_{\beta
}^{\alpha }=R_{\beta }^{\alpha }-\frac{1}{2}R\quad \delta _{\beta }^{\alpha
},$ for the metric (\ref{5dmetric}), given with respect to anholonomic
N--bases, are
\begin{equation}
G_{1}^{1}=-\left( R_{2}^{2}+S_{4}^{4}\right)
,G_{2}^{2}=G_{3}^{3}=-S_{4}^{4},G_{4}^{4}=G_{5}^{5}=-R_{2}^{2}.
\label{2einstdiag}
\end{equation}
So, the dynamics of the system is defined by two values $R_{2}^{2}$ and $%
S_{4}^{4}.$ The rest of non--diagonal components of the Ricci (Einstein
tensor) are compensated by fixing corresponding values of N--coefficients.
\end{corollary}

The formulas (\ref{2einstdiag}) are obtained following the relations for the
Ricci tensor (\ref{4ricci1a})--(\ref{3ricci4a}).

\begin{corollary}
We can extend the system of 5D vacuum Einstein equations (\ref{4ricci1a})--(%
\ref{3ricci4a}) by introducing matter fields for which the energy--momentum
tensor $\Upsilon _{\alpha \beta }$ given with respect to anholonomic frames
satisfy the conditions
\begin{equation}
\Upsilon _{1}^{1}=\Upsilon _{2}^{2}+\Upsilon _{4}^{4},\Upsilon
_{2}^{2}=\Upsilon _{3}^{3},\Upsilon _{4}^{4}=\Upsilon _{5}^{5}.
\label{2emcond}
\end{equation}
\end{corollary}

We note that, in general, the tensor $\Upsilon _{\alpha \beta }$ for the
non--vacuum Einstein equations,
\[
R_{\alpha \beta }-\frac{1}{2}g_{\alpha \beta }R=\kappa \Upsilon _{\alpha
\beta },
\]
is not symmetric because with respect to anholonomic frames there are
imposed constraints which makes non symmetric the Ricci and Einstein tensors
(the symmetry conditions hold only with respect to holonomic, coordinate
frames; for details see the Appendix and the formulas (\ref{1einsteq2})).

For simplicity, in our further investigations we shall consider only
diagonal matter sources, given with respect to anholonomic frames,
satisfying the conditions
\begin{equation}
\kappa \Upsilon _{2}^{2}=\kappa \Upsilon _{3}^{3}=\Upsilon _{2},\kappa
\Upsilon _{4}^{4}=\kappa \Upsilon _{5}^{5}=\Upsilon _{4},\mbox{ and }%
\Upsilon _{1}=\Upsilon _{2}+\Upsilon _{4},  \label{diagemt}
\end{equation}
where $\kappa $ is the gravitational coupling constant. In this case the
equations (\ref{4ricci1a}) and (\ref{3ricci2a}) are respectively generalized
to
\begin{equation}
R_{2}^{2}=R_{3}^{3}=-\frac{1}{2g_{2}g_{3}}[g_{3}^{\bullet \bullet }-\frac{%
g_{2}^{\bullet }g_{3}^{\bullet }}{2g_{2}}-\frac{(g_{3}^{\bullet })^{2}}{%
2g_{3}}+g_{2}^{^{\prime \prime }}-\frac{g_{2}^{^{\prime }}g_{3}^{^{\prime }}%
}{2g_{3}}-\frac{(g_{2}^{^{\prime }})^{2}}{2g_{2}}]=-\Upsilon _{4}
\label{2ricci1b}
\end{equation}
and
\begin{equation}
S_{4}^{4}=S_{5}^{5}=-\frac{\beta }{2h_{4}h_{5}}=-\Upsilon _{2}.
\label{ricci2b}
\end{equation}

\begin{corollary}
The class of metrics (\ref{4cmetric}) satisfying vacuum Einstein equations\\ (%
\ref{4ricci1a})--(\ref{3ricci4a}) and (\ref{3confeq}) contains as particular
cases some solutions when the\\ Schwar\-z\-schild potential $\Phi =-M/(M_{{\rm p}%
}^{2}r)$, where $M_{{\rm p}}$ is the effective Planck mass on the brane, is
modified to
\[
\Phi =-{\frac{M\sigma _{m}}{M_{{\rm p}}^{2}r}}+{\frac{Q\sigma _{q}}{2r^{2}}}%
\,,
\]
where the `tidal charge' parameter $Q$ may be positive or negative.
\end{corollary}

As proofs of this corollary we can consider the Refs \cite{10v1} where the
possibility to modify anisotropically the Newton law via effective
anisotropic masses $M\sigma _{m},$ or by anisotropic effective 4D Plank
constants, renormalized like $\sigma _{m}/M_{{\rm p}}^{2},$ and with
''effective'' electric charge, $Q\sigma _{q}$ was recently emphasized (see
also the end of Section III in this paper). For diagonal metrics, in the
locally isotropic limit, we put the effective polarizations $\sigma
_{m}=\sigma _{q}=1.$

\subsection{Reduction from 5D to 4D gravity}

The above presented results are for generic off--diagonal metrics of
gravitational fields, anholonomic transforms and nonlinear field equations.
Reductions to a lower dimensional theory are not trivial in such cases. We
give a detailed analysis of this procedure.

The simplest way to construct a $5D\rightarrow 4D$ reduction for the ansatz (%
\ref{4ansatz}) and (\ref{4ansatzc}) is to eliminate from formulas the variable
$x^{1}$ and to consider a 4D space (parametrized by local coordinates $%
\left( x^{2},x^{3},v,y^{5}\right) )$ being trivially embedded into 5D space
(parametrized by local coordinates $\left( x^{1},x^{2},x^{3},v,y^{5}\right) $
with $g_{11}=\pm 1,g_{1\underline{\alpha }}=0,\underline{\alpha }=2,3,4,5)$
with further possible conformal and anholonomic transforms depending only on
variables $\left( x^{2},x^{3},v\right) .$ We admit that the 4D metric $g_{%
\underline{\alpha }\underline{\beta }}$ could be of arbitrary signature. In
order to emphasize that some coordinates are stated just for a such 4D space
we underline the Greek indices, $\underline{\alpha },\underline{\beta },...$
\ and the Latin indices from the middle of alphabet, $\underline{i},%
\underline{j},...=2,3,$ where $u^{\underline{\alpha }}=\left( x^{\underline{i%
}},y^{a}\right) =\left( x^{2},x^{3},y^{4},y^{5}\right) .$

In result, the analogs of Lemmas 1and 2, Theorems 1-3 and Corollaries 1-3
can be reformulated for 4D gravity with mixed holonomic--anholonomic
variables. We outline here the most important properties of a such reduction.

\begin{itemize}
\item  The line element (\ref{2metric}) with ansatz (\ref{4ansatz}) and the
line element (\ref{2metric}) with (\ref{4ansatzc}) are respectively
transformed on 4D space to the values:

The first type 4D quadratic line element is taken
\begin{equation}
ds^{2}=g_{\underline{\alpha }\underline{\beta }}\left( x^{\underline{i}%
},v\right) du^{\underline{\alpha }}du^{\underline{\beta }}  \label{2metric4}
\end{equation}
with the metric coefficients $g_{\alpha \beta }$ parametrized (with respect
to the coordinate frame (\ref{5pdif}) in 4D) by an off--diagonal matrix
(ansatz)

{%%\footnotesize
\begin{equation}
\left[
\begin{array}{cccc}
g_{2}+w_{2}^{\ 2}h_{4}+n_{2}^{\ 2}h_{5} & w_{2}w_{3}h_{4}+n_{2}n_{3}h_{5} &
w_{2}h_{4} & n_{2}h_{5} \\
w_{2}w_{3}h_{4}+n_{2}n_{3}h_{5} & g_{3}+w_{3}^{\ 2}h_{4}+n_{3}^{\ 2}h_{5} &
w_{3}h_{4} & n_{3}h_{5} \\
w_{2}h_{4} & w_{3}h_{4} & h_{4} & 0 \\
n_{2}h_{5} & n_{3}h_{5} & 0 & h_{5}
\end{array}
\right] ,  \label{2ansatz4}
\end{equation}
} where the coefficients are some necessary smoothly class functions of
type:
\begin{eqnarray}
g_{2,3} &=&g_{2,3}(x^{2},x^{3}),h_{4,5}=h_{4,5}(x^{\underline{k}},v),
\nonumber \\
w_{\underline{i}} &=&w_{\underline{i}}(x^{\underline{k}},v),n_{\underline{i}%
}=n_{\underline{i}}(x^{\underline{k}},v);~\underline{i},\underline{k}=2,3.
\nonumber
\end{eqnarray}

The anholonomically and conformally transformed 4D line element is
\begin{equation}
ds^{2}=\Omega ^{2}(x^{\underline{i}},v)\hat{{g}}_{\underline{\alpha }%
\underline{\beta }}\left( x^{\underline{i}},v\right) du^{\underline{\alpha }%
}du^{\underline{\beta }},  \label{4cmetric4}
\end{equation}
were the coefficients $\hat{{g}}_{\underline{\alpha }\underline{\beta }}$
are parametrized by the ansatz {\scriptsize
\begin{equation}
\left[
\begin{array}{cccc}
g_{2}+(w_{2}^{\ 2}+\zeta _{2}^{\ 2})h_{4}+n_{2}^{\ 2}h_{5} &
(w_{2}w_{3}++\zeta _{2}\zeta _{3})h_{4}+n_{2}n_{3}h_{5} & (w_{2}+\zeta
_{2})h_{4} & n_{2}h_{5} \\
(w_{2}w_{3}++\zeta _{2}\zeta _{3})h_{4}+n_{2}n_{3}h_{5} & g_{3}+(w_{3}^{\
2}+\zeta _{3}^{\ 2})h_{4}+n_{3}^{\ 2}h_{5} & (w_{3}+\zeta _{3})h_{4} &
n_{3}h_{5} \\
(w_{2}+\zeta _{2})h_{4} & (w_{3}+\zeta _{3})h_{4} & h_{4} & 0 \\
n_{2}h_{5} & n_{3}h_{5} & 0 & h_{5}+\zeta _{5}h_{4}
\end{array}
\right] .  \label{22ansatzc4}
\end{equation}
}where $\zeta _{\underline{i}}=\zeta _{\underline{i}}\left( x^{\underline{k}%
},v\right) $ and we shall restrict our considerations for $\zeta _{5}=0.$

\item  In the 4D analog of Lemma 1 we have
\begin{equation}
\delta s^{2}=[g_{2}(dx^{2})^{2}+g_{3}(dx^{3})^{2}+h_{4}(\delta
v)^{2}+h_{5}(\delta y^{5})^{2}],  \label{3dmetric4}
\end{equation}
with respect to the anholonomic co--frame $\left( dx^{\underline{i}},\delta
v,\delta y^{5}\right) ,$ where
\begin{equation}
\delta v=dv+w_{\underline{i}}dx^{\underline{i}}\mbox{ and }\delta
y^{5}=dy^{5}+n_{\underline{i}}dx^{\underline{i}}  \label{3ddif4}
\end{equation}
which is dual to the frame $\left( \delta _{\underline{i}},\partial
_{4},\partial _{5}\right) ,$ where
\begin{equation}
\delta _{\underline{i}}=\partial _{\underline{i}}+w_{\underline{i}}\partial
_{4}+n_{\underline{i}}\partial _{5}.  \label{3dder4}
\end{equation}

\item  In the conditions of the 4D variant of Theorem 1 we have the same
equations (\ref{4ricci1a})--(\ref{3ricci4a}) were we must put $%
h_{4}=h_{4}\left( x^{\underline{k}},v\right) $ and $h_{5}=h_{5}\left( x^{%
\underline{k}},v\right) .$ As a consequence we have that $\alpha _{i}\left(
x^{k},v\right) \rightarrow \alpha _{\underline{i}}\left( x^{\underline{k}%
},v\right) ,\beta =\beta \left( x^{\underline{k}},v\right) $ and $\gamma
=\gamma \left( x^{\underline{k}},v\right) $ which result that $w_{\underline{%
i}}=w_{\underline{i}}\left( x^{\underline{k}},v\right) $ and $n_{\underline{i%
}}=n_{\underline{i}}\left( x^{\underline{k}},v\right) .$

\item  The respective formulas from Lemma 2, for $\zeta _{5}=0,$ transform
into
\begin{equation}
\delta s^{2}=\Omega ^{2}(x^{\underline{i}%
},v)[g_{2}(dx^{2})^{2}+g_{3}(dx^{3})^{2}+h_{4}(\hat{{\delta }}%
v)^{2}+h_{5}(\delta y^{5})^{2}],  \label{cdmetric4}
\end{equation}
with respect to the anholonomic co--frame $\left( dx^{\underline{i}},\hat{{%
\delta }}v,\delta y^{5}\right) ,$ where
\begin{equation}
\delta v=dv+(w_{\underline{i}}+\zeta _{\underline{i}})dx^{\underline{i}}%
\mbox{ and }\delta y^{5}=dy^{5}+n_{\underline{i}}dx^{\underline{i}}
\label{2ddif24}
\end{equation}
which is dual to the frame $\left( \hat{{\delta }}_{\underline{i}},\partial
_{4},\hat{{\partial }}_{5}\right) ,$ where
\begin{equation}
\hat{{\delta }}_{\underline{i}}=\partial _{\underline{i}}-(w_{\underline{i}%
}+\zeta _{\underline{i}})\partial _{4}+n_{\underline{i}}\partial _{5},\hat{{%
\partial }}_{5}=\partial _{5}.  \label{2dder24}
\end{equation}

\item  The formulas (\ref{2conf1}) and (\ref{3confeq}) from Theorem 2 must be
modified into a 4D form
\begin{equation}
\hat{{\delta }}_{\underline{i}}h_{4}=0\mbox{\ and\  }\hat{{\delta }}_{%
\underline{i}}\Omega =0  \label{2conf14}
\end{equation}
and the values $\zeta _{\hat{{i}}}=\left( \zeta \underline{_{{i}}},\zeta _{{5%
}}=0\right) $ are found as to be a unique solution of (\ref{2conf1}); for
instance, if
\[
\Omega ^{q_{1}/q_{2}}=h_{4}~(q_{1}\mbox{ and }q_{2}\mbox{ are integers}),
\]
$\zeta _{\underline{{i}}}$ satisfy the equations \
\begin{equation}
\partial _{\underline{i}}\Omega -(w_{\underline{i}}+\zeta \underline{_{{i}}}%
)\Omega ^{\ast }=0.  \label{2confeq4}
\end{equation}

\item  One holds the same formulas (\ref{2p2})-(\ref{2n}) from the Theorem 3
on the general form of exact solutions with that difference that their 4D
analogs are to be obtained by reductions of holonomic indices, $\underline{i}%
\rightarrow i,$ and holonomic coordinates, $x^{i}\rightarrow x^{\underline{i}%
},$ i. e. in the 4D solutions there is not contained the variable $x^{1}.$

\item  The formulae (\ref{2einstdiag}) for the nontrivial coefficients of the
Einstein tensor in 4D stated by the Corollary 1 are \ written
\begin{equation}
G_{2}^{2}=G_{3}^{3}=-S_{4}^{4},G_{4}^{4}=G_{5}^{5}=-R_{2}^{2}.
\label{2einstdiag4}
\end{equation}

\item  For symmetries of the Einstein tensor (\ref{2einstdiag4}) \ we can
introduce a matter field source with a diagonal energy momentum tensor,
like\ it is stated in the Corollary 2 by the conditions (\ref{2emcond}),
which in 4D are transformed into
\begin{equation}
\Upsilon _{2}^{2}=\Upsilon _{3}^{3},\Upsilon _{4}^{4}=\Upsilon _{5}^{5}.
\label{2emcond4}
\end{equation}

\item  In 4D Einstein gravity we are not having violations of the Newton law
as it was state in Corollary 3 for 5D. Nevertheless, off--diagonal and
anholonomic frames can induce an anholonomic \ particle and field dynamics,
for instance, with deformations of horizons of black holes, which can be
modelled by an effective anisotropic renormalization of constants if some
conditions are satisfied \cite{10v,10v2}.
\end{itemize}

There were constructed and analyzed various classes of exact
solutions of the Einstein equations (both in the vacuum, reducing
to the system (\ref {4ricci1a}), (\ref{3ricci2a}), (\ref{3ricci3a})
and (\ref{3ricci4a}) and non--vacuum, reducing to (\ref{2ricci1b}),
(\ref{ricci2b}), (\ref{3ricci3a}) and (\ref{3ricci4a}), cases) in
3D, 4D and 5D gravity \cite{10v,10v1}. The aim of the next Sections
III -- V is to prove that such solutions contain warped factors
which in the vacuum case are induced by a second order anisotropy.
We shall analyze some classes of such exact solutions with
running constants and/or their anisotropic polarizations induced
from extra dimension gravitational interactions.

\section{5D Ellipsoidal Black Holes}

Our goal is to apply the anholonomic frame method as to construct such exact
solutions of vacuum 5D Einstein equations as they will be static ones but,
for instance, with ellipsoidal horizon for a diagonal metric given with
respect to some well defined anholonomic frames. If such metrics are
redefined with respect to usual coordinate frames, they are described by
some particular cases of off--diagonal ansatz of type (\ref{4ansatz}), or (%
\ref{4ansatzc}) which results in a very sophysticate form of the Einstein
equations. That why it was not possible to construct such solutions in the
past, before elaboration of the anholonomic frame method with associated
nonlinear connection structure which allows to find exact solutions of the
Einstein equations for very general off--diagonal metric ansatz.

By using anholonomic transforms the Schwarzschild and Reissner-N\"{o}rdstrom
solutions were generalized in anisotropic forms with deformed horizons,
anisoropic polarizations and running constants both in the Einstein and
extra dimension gravity (see Refs. \cite{10v,10v1}). It was shown that there are
possible anisotropic solutions which preserve the local Lorentz symmetry.
and that at large radial distances from the horizon the anisotropic
configurations transform into the usual one with spherical symmetry. So, the
solutions with anisotropic rotation ellipsoidal horizons do not contradict
the well known Israel and Carter theorems \cite{10israel} which were proved in
the assumption of spherical symmetry at asymptotic. The vacuum metrics
presented here differ from anisotropic black hole solutions investigated in
Refs. \cite{10v,10v1}.

\subsection{The Schwarzschild solution in ellipsoidal coordinates}

Let us consider the system of {\it \ isotropic spherical coordinates} $(\rho
,\theta ,\varphi ),$ \thinspace where the isotropic radial coordinate $\rho $
is related with the usual radial coordinate $r$ via the relation $r=\rho
\left( 1+r_{g}/4\rho \right) ^{2}$ for $r_{g}=2G_{[4]}m_{0}/c^{2}$ being the
4D gravitational radius of a point particle of mass $m_{0},$ $%
G_{[4]}=1/M_{P[4]}^{2}$ is the 4D Newton constant expressed via Plank mass $%
M_{P[4]}$ (following modern string/brane theories, $M_{P[4]}$ can be
considered as a value induced from extra dimensions). We put the light speed
constant $c=1.$ This system of coordinates is considered for the so--called
isotropic representation of the Schwarzschild solution \cite{10ll}
\begin{equation}
dS^{2}=\left( \frac{\widehat{\rho }-1}{\widehat{\rho }+1}\right)
^{2}dt^{2}-\rho _{g}^{2}\left( \frac{\widehat{\rho }+1}{\widehat{\rho }}%
\right) ^{4}\left( d\widehat{\rho }^{2}+\widehat{\rho }^{2}d\theta ^{2}+%
\widehat{\rho }^{2}\sin ^{2}\theta d\varphi ^{2}\right) ,  \label{1schw}
\end{equation}
where, for our further considerations, we re--scaled the isotropic radial
coordinate as $\widehat{\rho }=\rho /\rho _{g},$ with $\rho _{g}=r_{g}/4.$
The metric (\ref{1schw}) is a vacuum static solution of 4D Einstein equations
with spherical symmetry describing the gravitational field of a point
particle of mass $m_{0}.$ It has a singularity for $r=0$ and a spherical
horizon for $r=r_{g},$ or, in re--scaled isotropic coordinates, for $%
\widehat{\rho }=1.$ We emphasize that this solution is parametrized by a
diagonal metric given with respect to holonomic coordinate frames.

We also introduce the {\it \ rotation ellipsoid coordinates} (in our case
considered as alternatives to the isotropic radial coordinates) \cite{10korn} $%
(u,\lambda ,\varphi )$ with $0\leq u<\infty ,0\leq \lambda \leq \pi ,0\leq
\varphi \leq 2\pi ,$ where $\sigma =\cosh u\geq 1$ are related with the
isotropic 3D Cartezian coordinates
\begin{equation}
(\tilde{x}=\widetilde{\rho }\sinh u\sin \lambda \cos \varphi ,\tilde{y}=%
\widetilde{\rho }\sinh u\sin \lambda \sin \varphi ,\tilde{z}=\widetilde{\rho
}\cosh u\cos \lambda )  \label{1rec}
\end{equation}
and define an elongated rotation ellipsoid hypersurface
\begin{equation}
\left( \tilde{x}^{2}+\tilde{y}^{2}\right) /(\sigma ^{2}-1)+\tilde{z}%
^{2}/\sigma ^{2}=\widetilde{\rho }^{2}. \label{1reh}
\end{equation}
with $\sigma =\cosh u.$ The 3D metric on a such hypersurface is
\[
dS_{(3D)}^{2}=g_{u u}du^{2}+g_{\lambda \lambda }d\lambda
^{2}+g_{\varphi \varphi }d\varphi ^{2},
\]
where
\[
g_{u u}=g_{\lambda \lambda }=\widetilde{\rho }^{2}\left( \sinh
^{2}u+\sin ^{2}\lambda \right) ,g_{\varphi \varphi
}=\widetilde{\rho }^{2}\sinh ^{2}u\sin ^{2}\lambda .
\]

We can relate the rotation ellipsoid coordinates $\left( u,\lambda ,\varphi
\right) $ from (\ref{1rec}) with the isotrop\-ic radial coordinates $\left(
\widehat{\rho },\theta ,\varphi \right) $, scaled by the constant $\rho
_{g}, $ from (\ref{1schw}) as
\[
\widetilde{\rho }=1,\sigma =\cosh u=\widehat{\rho }
\]
and deform the Schwarzschild metric by introducing ellipsoidal
coordinates and a new  horizon defined by the condition that
vanishing of the metric coefficient before $dt^2$ describe an
elongated rotation ellipsoid hypersurface (\ref{1reh}),
\begin{eqnarray}
dS_{(S)}^{2} &=&\left( \frac{\cosh u-1}{\cosh u+1}\right) ^{2}dt^{2}-\rho
_{g}^{2}\left( \frac{\cosh u+1}{\cosh u}\right) ^{4}(\sinh ^{2}u+\sin
^{2}\lambda )  \label{1schel} \\
&&\times \lbrack du^{2}+d\lambda ^{2}+\frac{\sinh ^{2}u~\sin ^{2}\lambda }{%
\sinh ^{2}u+\sin ^{2}\lambda }d\varphi ^{2}].  \nonumber
\end{eqnarray}
The ellipsoidally deformed metric (\ref{1schel}) does not satisfy
the vacuum Einstein equations, but at long distances from the
horizon it transforms into the usual Schwarzchild solution
(\ref{1schw}).

For our further considerations we introduce two Classes (A and B)
of 4D auxiliary pseudo--Riemannian metrics, also given in
ellipsoid coordinates, being some conformal transforms of
(\ref{1schel}), like
\[
dS_{(S)}^{2}=\Omega _{A,B}\left( u,\lambda \right) dS_{(A,B)}^{2}
\]
but which are not supposed to be solutions of the Einstein equations:

\begin{itemize}
\item  Metric of Class A:
\begin{equation}
dS_{(A)}^{2}=-du^{2}-d\lambda ^{2}+a(u,\lambda )d\varphi ^{2}+b(u,\lambda
)dt^{2}],  \label{1auxm1}
\end{equation}
where
\[
a(u,\lambda )=-\frac{\sinh ^{2}u~\sin ^{2}\lambda }{\sinh ^{2}u+\sin
^{2}\lambda }\mbox{ and }b(u,\lambda )=-\frac{(\cosh u-1)^{2}\cosh ^{4}u}{%
\rho _{g}^{2}(\cosh u+1)^{6}(\sinh ^{2}u+\sin ^{2}\lambda )},
\]
which results in the metric (\ref{1schel}) by multiplication on the conformal
factor
\begin{equation}
\Omega _{A}\left( u,\lambda \right) =\rho _{g}^{2}\frac{(\cosh u+1)^{4}}{%
\cosh ^{4}u}(\sinh ^{2}u+\sin ^{2}\lambda ).  \label{1auxm1c}
\end{equation}

\item  Metric of Class B:
\begin{equation}
dS^{2}=g(u,\lambda )\left( du^{2}+d\lambda ^{2}\right) -d\varphi
^{2}+f(u,\lambda )dt^{2},  \label{1auxm2}
\end{equation}
where
\[
g(u,\lambda )=-\frac{\sinh ^{2}u+\sin ^{2}\lambda }{\sinh ^{2}u~\sin
^{2}\lambda }\mbox{ and }f(u,\lambda )=\frac{(\cosh u-1)^{2}\cosh ^{4}u}{%
\rho _{g}^{2}(\cosh u+1)^{6}\sinh ^{2}u\sin ^{2}\lambda },
\]
which results in the metric (\ref{1schel}) by multiplication on the conformal
factor
\[
\Omega _{B}\left( u,\lambda \right) =\rho _{g}^{2}\frac{(\cosh u+1)^{4}}{%
\cosh ^{4}u}\sinh ^{2}u\sin ^{2}\lambda .
\]
\end{itemize}

Now it is possible to generate exact solutions of the Einstein equations
with rotation ellipsoid horizons and anisotropic polarizations and running
of constants by performing corresponding anholonomic transforms as the
solutions will have an horizon parametrized by a hypersurface like rotation
ellipsoid and gravitational (extra dimensional or nonlinear 4D)
renormalization of the constant $\rho _{g}$ of the Schwarzschild solution, $%
\rho _{g}\rightarrow \overline{\rho }_{g}=\omega \rho _{g},$ where the
dependence of the function $\omega $ on some holonomic or anholonomic
coordinates depend on the type of anisotropy. For some solutions we can
treat $\omega $ as a factor modelling running of the gravitational constant,
induced, induced from extra dimension, in another cases we may consider $%
\omega $ as a nonlinear gravitational polarization which model some
anisotropic distributions of masses and matter fields and/or anholonomic
vacuum gravitational interactions.

\subsection{Ellipsoidal 5D metrics of Class A}

In this subsection we consider four classes of 5D vacuum solutions which are
related to the metric of Class A (\ref{1auxm1}) and to the Schwarzschild
metric in ellipsoidal coordinates (\ref{1schel}).

Let us parametrize the 5D coordinates as $\left( x^{1}=\chi
,x^{2}=u,x^{3}=\lambda ,y^{4}=v,y^{5}=p\right) ,$ where the solutions with
the so--called $\varphi $--anisotropy will be constructed for $\left(
v=\varphi ,p=t\right) $ and the solutions with $t$--anisotropy will be
stated for $\left( v=t,p=\varphi \right) $ (in brief, we shall write $\ $\
respective $\varphi $--solutions and $t$--solutions).

\subsubsection{Class A solutions with ansatz (\ref{4ansatz}):}

We take an off--diagonal metric ansatz of type (\ref{4ansatz}) (equivalently,
(\ref{2metric})) by representing
\[
g_{1}=\pm 1,g_{2}=-1,g_{3}=-1,h_{4}=\eta _{4}(x^{i},v)h_{4(0)}(x^{i})%
\mbox{
and }h_{5}=\eta _{5}(x^{i},v)h_{5(0)}(x^{i}),
\]
where $\eta _{4,5}(x^{i},v)$ are corresponding ''gravitational
renormalizations'' of the metric coefficients $h_{4,5(0)}(x^{i}).$ For $%
\varphi $--solutions we state $h_{4(0)}=a(u,\lambda )$ and $%
h_{5(0)}=b(u,\lambda )$ (inversely, for \ $t$--solutions, $%
h_{4(0)}=b(u,\lambda )$ and $h_{5(0)}=a(u,\lambda )).$ \

Next we consider a renormalized gravitational 'constant' $\overline{\rho }%
_{g}=\omega \rho _{g},$ were for $\varphi $--solutions\ the receptivity $%
\omega =\omega \left( x^{i},v\right) $ is included in the gravitational
polarization $\eta _{5}$ as $\eta _{5}=\left[ \omega \left( x^{i},\varphi
\right) \right] ^{-2},$ or for $t$--solutions is included in $\eta _{4},$
when $\eta _{4}=\left[ \omega \left( x^{i},t\right) \right] ^{-2}.$ We can
construct an exact solution of the 5D vacuum Einstein equations if, for
explicit dependencies on anisotropic coordinate, the metric coefficients $%
h_{4}$ and $h_{5}$ are related by formula (\ref{3p1}) with $h_{[0]}\left(
x^{i}\right) =h_{(0)}=const$ (see the Theorem 3, with statements on formulas
(\ref{3p1}) and (\ref{5w})), which in its turn imposes a corresponding
relation between $\eta _{4}$ and $\eta _{5},$%
\[
\eta _{4}h_{4(0)}(x^{i})=h_{(0)}^{2}h_{5(0)}(x^{i})\left[ \left( \sqrt{|\eta
_{5}|}\right) ^{\ast }\right] ^{2}.
\]
In result, we express the polarizations $\eta _{4}$ and $\eta _{5}$ via the
value of receptivity $\omega ,$
\begin{equation}
\eta _{4}\left( \chi ,u,\lambda ,\varphi \right) =h_{(0)}^{2}\frac{%
b(u,\lambda )}{a(u,\lambda )}\left\{ \left[ \omega ^{-1}\left( \chi
,u,\lambda ,\varphi \right) \right] ^{\ast }\right\} ^{2},\eta _{5}\left(
\chi ,u,\lambda ,\varphi \right) =\omega ^{-2}\left( \chi ,u,\lambda
,\varphi \right) ,  \label{1etap}
\end{equation}
for $\varphi $--solutions , and
\begin{equation}
\eta _{4}\left( \chi ,u,\lambda ,t\right) =\omega ^{-2}\left( \chi
,u,\lambda ,t\right) ,\eta _{5}\left( \chi ,u,\lambda ,t\right) =h_{(0)}^{-2}%
\frac{b(u,\lambda )}{a(u,\lambda )}\left[ \int dt\omega ^{-1}\left( \chi
,u,\lambda ,t\right) \right] ^{2},  \label{1etat}
\end{equation}
for $t$--solutions, where $a(u,\lambda )$ and $b(u,\lambda )$ are those from
(\ref{1auxm1}).

For vacuum configurations, following the discussions of formula (\ref{5w}) in
Theorem 3, we put $w_{i}=0.$ The next step is to find the values of $n_{i}$
by introducing $h_{4}=\eta _{4}h_{4(0)}$ and $h_{5}=\eta _{5}h_{5(0)}$ into
the formula \ (\ref{2n}), which, for convenience, is expressed via general
coefficients $\eta _{4}$ and $\eta _{5},$ with the functions $n_{k[2]}\left(
x^{i}\right) $ redefined as to contain the values $h_{(0)}^{2},$ $%
a(u,\lambda )$ and $b(u,\lambda )$
\begin{eqnarray}
n_{k} &=&n_{k[1]}\left( x^{i}\right) +n_{k[2]}\left( x^{i}\right) \int [\eta
_{4}/(\sqrt{|\eta _{5}|})^{3}]dv,~\eta _{5}^{\ast }\neq 0;  \label{1nel} \\
&=&n_{k[1]}\left( x^{i}\right) +n_{k[2]}\left( x^{i}\right) \int \eta
_{4}dv,\qquad ~\eta _{5}^{\ast }=0;  \nonumber \\
&=&n_{k[1]}\left( x^{i}\right) +n_{k[2]}\left( x^{i}\right) \int [1/(\sqrt{%
|\eta _{5}|})^{3}]dv,~\eta _{4}^{\ast }=0.  \nonumber
\end{eqnarray}

\bigskip By introducing the formulas (\ref{1etap}) for $\varphi $--solutions
(or (\ref{1etat}) for $t$--solutions) and fixing some boundary condition, in
order to state the values of coefficients $n_{k[1,2]}\left( x^{i}\right) $
we can express the ansatz components $n_{k}\left( x^{i},\varphi \right) $ as
integrals of some functions of $\omega \left( x^{i},\varphi \right) $ and $%
\partial _{\varphi }\omega \left( x^{i},\varphi \right) $ (or, we can
express the ansatz components $n_{k}\left( x^{i},t\right) $ as integrals of
some functions of $\omega \left( x^{i},t\right) $ and $\partial _{t}\omega
\left( x^{i},t\right) ).$ We do not present an explicit form of such
formulas because they depend on the type of receptivity $\omega =\omega
\left( x^{i},v\right) ,$ which must be defined experimentally, or from some
quantum models of gravity in the quasi classical limit. We preserved a
general dependence on coordinates $x^{i}$ which reflect the fact that there
is a freedom in fixing holonomic coordinates (for instance, on ellipsoidal
hypersurface and its extensions to 4D and 5D spacetimes). \ For simplicity,
we write that $n_{i}$ are some functionals of $\{x^{i},\omega \left(
x^{i},v\right) ,\omega ^{\ast }\left( x^{i},v\right) \}$
\[
n_{i}\{x,\omega ,\omega ^{\ast }\}=n_{i}\{x^{i},\omega \left( x^{i},v\right)
,\omega ^{\ast }\left( x^{i},v\right) \}.
\]

In conclusion, we constructed two exact solutions of the 5D vacuum Einstein
equations, defined by the ansatz (\ref{4ansatz}) with coordinates and
coefficients stated by the data:
\begin{eqnarray}
\mbox{$\varphi$--solutions} &:&(x^{1}=\chi ,x^{2}=u,x^{3}=\lambda
,y^{4}=v=\varphi ,y^{5}=p=t),g_{1}=\pm 1,  \nonumber \\
g_{2} &=&-1,g_{3}=-1,h_{4(0)}=a(u,\lambda ),h_{5(0)}=b(u,\lambda ),%
\mbox{see
(\ref{1auxm1})};  \nonumber \\
h_{4} &=&\eta _{4}(x^{i},\varphi )h_{4(0)}(x^{i}),h_{5}=\eta
_{5}(x^{i},\varphi )h_{5(0)}(x^{i}),  \nonumber \\
\eta _{4} &=&h_{(0)}^{2}\frac{b(u,\lambda )}{a(u,\lambda )}\left\{ \left[
\omega ^{-1}\left( \chi ,u,\lambda ,\varphi \right) \right] ^{\ast }\right\}
^{2},\eta _{5}=\omega ^{-2}\left( \chi ,u,\lambda ,\varphi \right) ,
\nonumber \\
w_{i} &=&0,n_{i}\{x,\omega ,\omega ^{\ast }\}=n_{i}\{x^{i},\omega \left(
x^{i},\varphi \right) ,\omega ^{\ast }\left( x^{i},\varphi \right) \}.
\label{1sol5p1}
\end{eqnarray}
and
\begin{eqnarray}
\mbox{$t$--solutions} &:&(x^{1}=\chi ,x^{2}=u,x^{3}=\lambda
,y^{4}=v=t,y^{5}=p=\varphi ),g_{1}=\pm 1,  \nonumber \\
g_{2} &=&-1,g_{3}=-1,h_{4(0)}=b(u,\lambda ),h_{5(0)}=a(u,\lambda ),%
\mbox{see
(\ref{1auxm1})};  \nonumber \\
h_{4} &=&\eta _{4}(x^{i},t)h_{4(0)}(x^{i}),h_{5}=\eta
_{5}(x^{i},t)h_{5(0)}(x^{i}),  \nonumber \\
\eta _{4} &=&\omega ^{-2}\left( \chi ,u,\lambda ,t\right) ,\eta
_{5}=h_{(0)}^{-2}\frac{b(u,\lambda )}{a(u,\lambda )}\left[ \int dt~\omega
^{-1}\left( \chi ,u,\lambda ,t\right) \right] ^{2},  \nonumber \\
w_{i} &=&0,n_{i}\{x,\omega ,\omega ^{\ast }\}=n_{i}\{x^{i},\omega \left(
x^{i},t\right) ,\omega ^{\ast }\left( x^{i},t\right) \}.  \label{1sol5t1}
\end{eqnarray}

Both types of solutions have a horizon parametrized by a rotation ellipsoid
hypersurface (as the condition of vanishing of \ the ''time'' metric
coefficient states, i. e. when the function $b(u,\lambda )=0)$. $\ $These
solutions are generically anholonomic (anisotropic) because in the locally
isotropic limit, when $\eta _{4},\eta _{5},$ $\omega \rightarrow 1$ and $%
n_{i}\rightarrow 0,$ they reduce to the coefficients of the
metric (\ref{1auxm1}). The last one is not an exact solution of
4D vacuum Einstein equations, but it is a conformal transform of
the 4D Schwarzschild solution with a further trivial extension to
5D. With respect to the anholonomic frames adapted to the
coefficients $n_{i}$ (see (\ref{1ddif1})), the obtained solutions
have diagonal metric coefficients being very similar to the
Schwarzschild metric (\ref{1schel}) written in ellipsoidal
coordinates. We can treat such solutions as black hole ones with
a point particle mass put in one of the focuses of rotation
ellipsoid hypersurface (for flattened ellipsoids the mass should
be placed on the circle described by ellipse's focuses under
rotation; we omit such details in this work which were presented
for 4D gravity in Ref. \cite{10v}).

 The initial data for anholonomic frames and the chosen
configuration of gravitational interactions in the bulk lead to
deformed ''ellipsoidal'' horizons even for static configurations.
The solutions admit anisotropic polarizations on ellipsoidal and
angular coordinates $\left( u,\lambda \right) $ and running of
constants on time $t$ and/or on extra dimension coordinate $\chi
$. Such renormalizations of constants are defined by the nonlinear
configuration of the 5D vacuum gravitational field and depend on
introduced receptivity function $\omega \left( x^{i},v\right) $
which is to be considered an intrinsic characteristics of the 5D
vacuum gravitational 'ether', emphasizing the possibility \ of
nonlinear self--polarization of gravitational fields.

Finally, we note that the data (\ref{1sol5p1}) and (\ref{1sol5t1}) parametrize
two very different classes of solutions. The first one is for static 5D
vacuum black hole configurations with explicit dependence on anholonomic
coordinate $\varphi $ and possible renormalizations on the rest of 3D space
coordinates $u$ and $\lambda $ and on the 5th coordinate $\chi .$ The second
class of solutions are similar to the static solutions but with an
emphasized anholonomic time running of constants and with possible
anisotropic dependencies on coordinates $(u,\lambda ,\chi ).$

\subsubsection{Class A solutions with ansatz (\ref{4ansatzc}):}

We construct here 5D vacuum $\varphi $-- and $t$--solutions parametrized by
an ansatz with conformal factor $\Omega (x^{i},v)$ (see (\ref{4ansatzc}) and (%
\ref{2cdmetric})). Let us consider conformal factors parametrized as $\Omega
=\Omega _{\lbrack 0]}(x^{i})\Omega _{\lbrack 1]}(x^{i},v).$ We can generate
from the data (\ref{1sol5p1}) (or (\ref{1sol5t1})) an exact solution of vacuum
Einstein equations if there are satisfied the conditions (\ref{2confq}) and (%
\ref{2confsol}), i. e.
\[
\Omega _{\lbrack 0]}^{q_{1}/q_{2}}\Omega _{\lbrack 1]}^{q_{1}/q_{2}}=\eta
_{4}h_{4(0)},
\]
for some integers $q_{1}$ and $q_{2},$ and there are defined the second
anisotropy coefficients
\[
\zeta _{i}=\left( \partial _{i}\ln |\Omega _{\lbrack 0]}\right) |)~\left(
\ln |\Omega _{\lbrack 1]}|\right) ^{\ast }+\left( \Omega _{\lbrack 1]}^{\ast
}\right) ^{-1}\partial _{i}\Omega _{\lbrack 1]}.
\]
So, taking a $\varphi $-- or $t$--solution with corresponding values of $%
h_{4}=\eta _{4}h_{4(0)},$\ for some $q_{1}$ and $q_{2},$ we obtain
new exact solutions, called in brief, $\varphi _{c}$-- or
$t_{c}$--solutions (with the index ''c'' pointing to an ansatz
with conformal factor), of the vacuum 5D Einstein equations given
in explicit form by the data:

\newpage

\begin{eqnarray}
\mbox{$\varphi_c$--solutions} &:&(x^{1}=\chi ,x^{2}=u,x^{3}=\lambda
,y^{4}=v=\varphi ,y^{5}=p=t),g_{1}=\pm 1,  \nonumber \\
g_{2} &=&-1,g_{3}=-1,h_{4(0)}=a(u,\lambda ),h_{5(0)}=b(u,\lambda ),%
\mbox{see
(\ref{1auxm1})};  \nonumber \\
h_{4} &=&\eta _{4}(x^{i},\varphi )h_{4(0)}(x^{i}),h_{5}=\eta
_{5}(x^{i},\varphi )h_{5(0)}(x^{i}),  \nonumber \\
\eta _{4} &=&h_{(0)}^{2}\frac{b(u,\lambda )}{a(u,\lambda )}\left\{ \left[
\omega ^{-1}\left( \chi ,u,\lambda ,\varphi \right) \right] ^{\ast }\right\}
^{2},\eta _{5}=\omega ^{-2}\left( \chi ,u,\lambda ,\varphi \right) ,
 \label{1sol5pc} \\
w_{i} &=&0,n_{i}\{x,\omega ,\omega ^{\ast }\}=n_{i}\{x^{i},\omega \left(
x^{i},\varphi \right) ,\omega ^{\ast }\left( x^{i},\varphi \right) \},\Omega
=\Omega _{\lbrack 0]}(x^{i})\Omega _{\lbrack 1]}(x^{i},\varphi )
 \nonumber  \\
\zeta _{i} &=&\left( \partial _{i}\ln |\Omega _{\lbrack 0]}\right) |)~\left(
\ln |\Omega _{\lbrack 1]}|\right) ^{\ast }+\left( \Omega _{\lbrack 1]}^{\ast
}\right) ^{-1}\partial _{i}\Omega _{\lbrack 1]},  \nonumber \\
\eta _{4}a &=&\Omega _{\lbrack
0]}^{q_{1}/q_{2}}(x^{i})\Omega _{\lbrack 1]}^{q_{1}/q_{2}}(x^{i},\varphi ).
\nonumber
\end{eqnarray}
and
\begin{eqnarray}
\mbox{$t_c$--solutions} &:&(x^{1}=\chi ,x^{2}=u,x^{3}=\lambda
,y^{4}=v=t,y^{5}=p=\varphi ),g_{1}=\pm 1,  \nonumber \\
g_{2} &=&-1,g_{3}=-1,h_{4(0)}=b(u,\lambda ),h_{5(0)}=a(u,\lambda ),%
\mbox{see
(\ref{1auxm1})};  \nonumber \\
h_{4} &=&\eta _{4}(x^{i},t)h_{4(0)}(x^{i}),h_{5}=\eta
_{5}(x^{i},t)h_{5(0)}(x^{i}),  \nonumber \\
\eta _{4} &=&\omega ^{-2}\left( \chi ,u,\lambda ,t\right) ,\eta
_{5}=h_{(0)}^{-2}\frac{b(u,\lambda )}{a(u,\lambda )}\left[ \int dt~\omega
^{-1}\left( \chi ,u,\lambda ,t\right) \right] ^{2},  \label{1sol5tc}  \\
w_{i} &=&0,n_{i}\{x,\omega ,\omega ^{\ast }\}=n_{i}\{x^{i},\omega \left(
x^{i},t\right) ,\omega ^{\ast }\left( x^{i},t\right) \},\Omega =\Omega
_{\lbrack 0]}(x^{i})\Omega _{\lbrack 1]}(x^{i},t) \nonumber  \\
\zeta _{i} &=&\left( \partial _{i}\ln |\Omega _{\lbrack 0]}\right) |)~\left(
\ln |\Omega _{\lbrack 1]}|\right) ^{\ast }+\left( \Omega _{\lbrack 1]}^{\ast
}\right) ^{-1}\partial _{i}\Omega _{\lbrack 1]},\eta _{4}a=\Omega _{\lbrack
0]}^{q_{1}/q_{2}}(x^{i})\Omega _{\lbrack 1]}^{q_{1}/q_{2}}(x^{i},t).
\nonumber
\end{eqnarray}

These solutions have two very interesting properties: 1) they admit a warped
factor on the 5th coordinate, like $\Omega _{\lbrack 1]}^{q_{1}/q_{2}}\sim
\exp [-k|\chi |],$ which in our case is constructed for an anisotropic 5D
vacuum gravitational configuration and not following a brane configuration
like in Refs. \cite{10rs}; 2) we can impose such conditions on the receptivity
$\omega \left( x^{i},v\right) $ as to obtain in the locally isotropic limit
just the Schwarzschild metric (\ref{1schel}) trivially embedded into the 5D
spacetime.

Let us analyze the second property in details. We have to chose the
conformal factor as to be satisfied three conditions:
\begin{equation}
\Omega _{\lbrack 0]}^{q_{1}/q_{2}}=\Omega _{A},\Omega _{\lbrack
1]}^{q_{1}/q_{2}}\eta _{4}=1,\Omega _{\lbrack 1]}^{q_{1}/q_{2}}\eta _{5}=1,
\label{1cond1a}
\end{equation}
were $\Omega _{A}$ is that from (\ref{2confq}). The last two conditions are
possible if
\begin{equation}
\eta _{4}^{-q_{1}/q_{2}}\eta _{5}=1,  \label{2cond2}
\end{equation}
which selects a specific form of receptivity $\omega \left( x^{i},v\right) .$
\ Putting into (\ref{2cond2}) the values $\eta _{4}$ and $\eta _{5}$
respectively from (\ref{1sol5pc}), or (\ref{1sol5tc}), we obtain some
differential, or integral, relations of the unknown $\omega \left(
x^{i},v\right) ,$ which results that
\begin{eqnarray}
\omega \left( x^{i},\varphi \right) &=&\left( 1-q_{1}/q_{2}\right)
^{-1-q_{1}/q_{2}}\left[ h_{(0)}^{-1}\sqrt{|a/b|}\varphi +\omega _{\lbrack
0]}\left( x^{i}\right) \right] ,\mbox{ for }\varphi _{c}\mbox{--solutions};
\nonumber   \\
\omega \left( x^{i},t\right) &=&\left[ \left( q_{1}/q_{2}-1\right) h_{(0)}%
\sqrt{|a/b|}t+\omega _{\lbrack 1]}\left( x^{i}\right) \right]
^{1-q_{1}/q_{2}},\mbox{ for }t_{c}\mbox{--solutions},   \label{4cond1}
\end{eqnarray}
for some arbitrary functions $\omega _{\lbrack 0]}\left( x^{i}\right) $ and $%
\omega _{\lbrack 1]}\left( x^{i}\right) .$ \ So, receptivities of particular
form like (\ref{4cond1}) allow us to obtain in the locally isotropic limit
just the Schwarzschild metric.

We conclude this subsection by \ the remark: the \ vacuum 5D metrics solving
the Einstein equations describe a nonlinear gravitational dynamics which
under some particular boundary conditions and parametrizations of metric's
coefficients can model anisotropic solutions transforming, in a
corresponding locally isotropic limit, in some well known exact solutions
like Schwarzschild, Reissner-N\"{o}rdstrom, Taub NUT, various type of
wormhole, solitonic and disk solutions (see details in Refs. \cite{10v,10v2,10v1}%
). Here we emphasize that, in general, an anisotropic solution (parametrized
by an off--diagonal ansatz) could not have a locally isotropic limit to a
diagonal metric with respect to some holonomic coordinate frames. By some
boundary conditions and suggested type of horizons, singularities,
symmetries and topological configuration such solutions model new classes of
black hole/tori, wormholes and another type of solutions which defines a
generic anholonomic gravitational field dynamics and has not locally
isotropic limits.

\subsection{Ellipsiodal 5D metrics of Class B}

In this subsection we construct and analyze another two classes of 5D vacuum
solutions which are related to the metric of Class B (\ref{1auxm2}) and which
can be reduced to the Schwarzschild metric in ellipsoidal coordinates (\ref
{1schel}) by corresponding parametrizations of receptivity $\omega \left(
x^{i},v\right) $. We emphasize that because the function $g(u,\lambda )$
from (\ref{1auxm2}) is not a solution of equation (\ref{4ricci1a}) we
introduce an auxiliary factor $\varpi $ $(u,\lambda )$ for which $\varpi g$
became a such solution, then we consider conformal factors parametrized as $%
\Omega =\varpi ^{-1}$ $\Omega _{\lbrack 2]}\left( x^{i},v\right) $ and find
solutions parametrized by the ansatz (\ref{4ansatzc}) and anholonomic metric
interval (\ref{2cdmetric}).

Because the method of definition of such solutions is similar to that from
previous subsection, in our further considerations we shall omit intermediary
computations and present directly the data which select the respective
configurations for $\varphi _{c}$--solutions and $t_{c}$--solutions.

The Class B of 5D solutions with conformal factor are parametrized by the
data:
\begin{eqnarray}
\mbox{$\varphi_c$--solutions} &:&(x^{1}=\chi ,x^{2}=u,x^{3}=\lambda
,y^{4}=v=\varphi ,y^{5}=p=t),g_{1}=\pm 1,  \nonumber \\
g_{2} &=&g_{3}=\varpi (u,\lambda )g(u,\lambda ),h_{4(0)}=-\varpi (u,\lambda
), \nonumber\\
h_{5(0)}&=&\varpi (u,\lambda )f(u,\lambda ),\mbox{see
(\ref{1auxm2})};  \nonumber \\
\varpi &=&g^{-1}\varpi _{0}\exp [a_{2}u+a_{3}\lambda ],~\varpi
_{0},a_{2},a_{3}=const;~\mbox{see
(\ref{2solricci1a})}  \nonumber \\
h_{4} &=&\eta _{4}(x^{i},\varphi )h_{4(0)}(x^{i}),h_{5}=\eta
_{5}(x^{i},\varphi )h_{5(0)}(x^{i}),  \nonumber \\
\eta _{4} &=&-h_{(0)}^{2}f(u,\lambda )\left\{ \left[ \omega ^{-1}\left( \chi
,u,\lambda ,\varphi \right) \right] ^{\ast }\right\} ^{2},\eta _{5}=\omega
^{-2}\left( \chi ,u,\lambda ,\varphi \right) ,  \label{1sol5p} \\
w_{i} &=&0,n_{i}\{x,\omega ,\omega ^{\ast }\}=n_{i}\{x^{i},\omega \left(
x^{i},\varphi \right) ,\omega ^{\ast }\left( x^{i},\varphi \right) \},\Omega
=\varpi ^{-1}(u,\lambda )\Omega _{\lbrack 2]}(x^{i},\varphi )  \nonumber \\
\zeta _{i} &=&\partial _{i}\ln |\varpi |)~\left( \ln |\Omega _{\lbrack
2]}|\right) ^{\ast }+\left( \Omega _{\lbrack 2]}^{\ast }\right)
^{-1}\partial _{i}\Omega _{\lbrack 2]},\eta _{4}=-\varpi
^{-q_{1}/q_{2}}(x^{i})\Omega _{\lbrack 2]}^{q_{1}/q_{2}}(x^{i},\varphi ).
\nonumber
\end{eqnarray}
and
\begin{eqnarray}
\mbox{$t_c$--solutions} &:&(x^{1}=\chi ,x^{2}=u,x^{3}=\lambda
,y^{4}=v=t,y^{5}=p=\varphi ),g_{1}=\pm 1,  \nonumber \\
g_{2} &=&g_{3}=\varpi (u,\lambda )g(u,\lambda ),h_{4(0)}=\varpi (u,\lambda
)f(u,\lambda ), \nonumber \\
h_{5(0)}&=& -\varpi (u,\lambda ),\mbox{see
(\ref{1auxm2})};  \nonumber \\
\varpi &=&g^{-1}\varpi _{0}\exp [a_{2}u+a_{3}\lambda ],~\varpi
_{0},a_{2},a_{3}=const,~\mbox{see
(\ref{2solricci1a})}  \nonumber \\
h_{4} &=&\eta _{4}(x^{i},t)h_{4(0)}(x^{i}),h_{5}=\eta
_{5}(x^{i},t)h_{5(0)}(x^{i}),  \nonumber \\
\eta _{4} &=&\omega ^{-2}\left( \chi ,u,\lambda ,t\right) ,\eta
_{5}=-h_{(0)}^{-2}f(u,\lambda )\left[ \int dt~\omega ^{-1}\left( \chi
,u,\lambda ,t\right) \right] ^{2},  \label{1sol5t} \\
w_{i} &=&0,n_{i}\{x,\omega ,\omega ^{\ast }\}=n_{i}\{x^{i},\omega \left(
x^{i},t\right) ,\omega ^{\ast }\left( x^{i},t\right) \},\Omega =\varpi
^{-1}(u,\lambda )\Omega _{\lbrack 2]}(x^{i},t)  \nonumber \\
\zeta _{i} &=&\partial _{i}(\ln |\varpi |)~\left( \ln |\Omega _{\lbrack
2]}|\right) ^{\ast }+\left( \Omega _{\lbrack 2]}^{\ast }\right)
^{-1}\partial _{i}\Omega _{\lbrack 2]},\eta _{4}=-\varpi
^{-q_{1}/q_{2}}(x^{i})\Omega _{\lbrack 2]}^{q_{1}/q_{2}}(x^{i},t).  \nonumber
\end{eqnarray}
where the coefficients $n_{i}$ can be found explicitly by introducing the
corresponding values $\eta _{4}$ and $\eta _{5}$ in formula (\ref{1nel}).

By a procedure similar to the solutions of Class A (see previous subsection)
we can find the conditions when the solutions (\ref{1sol5p}) and (\ref{1sol5t}%
) will have in the locally anisotropic limit the Schwarzschild solutions,
which impose corresponding parametrizations and dependencies on $\Omega
_{\lbrack 2]}(x^{i},v)$ and $\omega \left( x^{i},v\right) $ like (\ref
{1cond1a}) and (\ref{4cond1}). We omit these formulas because, in general, for
anholonomic configurations and nonlinear solutions there are not hard
arguments to prefer any holonomic limits of such off--diagonal metrics.

Finally, in this Section, we remark that for the considered
classes of ellipsoidal black hole solutions the so--called
$tt$--components of metric contain modifications of the
Schwarzschild potential
\[
\Phi =-\frac{M}{M_{P[4]}^{2}r}\mbox{ into }\Phi =-\frac{M\omega \left(
x^{i},v\right) }{M_{P[4]}^{2}r},
\]
where $M_{P[4]}$ is the usual 4D Plank constant, and this is given with
respect to the corresponding anholonomic frame of reference. The receptivity $%
\omega \left( x^{i},v\right) $ could model corrections warped on extra
dimension coordinate, $\chi ,$ which for our solutions are induced by
anholonomic vacuum gravitational interactions in the bulk and not from a
brane configuration in $AdS_{5}$ spacetime. In the vacuum case $k$ is a
constant characterizing the receptivity for bulk vacuum gravitational
polarizations.

\section{4D Ellipsoidal Black Holes}

For the ansatz (\ref{2ansatz4}), without conformal factor, some
classes of ellipsoidal solutions of 4D Einstein equations were
constructed in Ref. \cite{10v} with further generalizations and
applications to brane physics \cite{10v1} . The goal of this
Section is to consider some alternative variants, both with and
without conformal factors and for different coordinate
parametrizations and types of anisotropies. The bulk of 5D
solutions from the previous Section are reduced into
corresponding 4D ones if one eliminates the 5th coordinate $\chi $
from\ the formulas and  the off--diagonal ansatz (\ref{2ansatz4})
and (\ref {22ansatzc4}) are considered.

\subsection{Ellipsiodal 5D metrics of Class A}

Let us parametrize the 4D coordinates as $(x^{\underline{i}},y^{a})=\left(
x^{2}=u,x^{3}=\lambda ,y^{4}=v,y^{5}=p\right) ;$ for the $\varphi $%
--solutions we shall take $\left( v=\varphi ,p=t\right) $ and for
the solutions $t$--solutions we shall consider $\left(
v=t,p=\varphi \right) $. Following the prescription from
subsection IIE we can write down the data for solutions without
proofs and computations.

\subsubsection{Class A solutions with ansat (\ref{2ansatz4}):}

The off--diagonal metric ansatz of type (\ref{2ansatz4}) (equivalently, (\ref
{2metric})) \ with the data
\begin{eqnarray}
\mbox{$\varphi$--solutions} &:&(x^{2}=u,x^{3}=\lambda ,y^{4}=v=\varphi
,y^{5}=p=t)  \nonumber \\
g_{2} &=&-1,g_{3}=-1,h_{4(0)}=a(u,\lambda ),h_{5(0)}=b(u,\lambda ),%
\mbox{see
(\ref{1auxm1})};  \nonumber \\
h_{4} &=&\eta _{4}(u,\lambda ,\varphi )h_{4(0)}(u,\lambda ),h_{5}=\eta
_{5}(u,\lambda ,\varphi )h_{5(0)}(u,\lambda ),  \nonumber \\
\eta _{4} &=&h_{(0)}^{2}\frac{b(u,\lambda )}{a(u,\lambda )}\left\{ \left[
\omega ^{-1}\left( u,\lambda ,\varphi \right) \right] ^{\ast }\right\}
^{2},\eta _{5}=\omega ^{-2}\left( u,\lambda ,\varphi \right) ,  \nonumber \\
w_{\underline{i}} &=&0,n_{\underline{}i}\{x,\omega ,\omega ^{\ast
}\}=n_{i}\{u,\lambda ,\omega \left( u,\lambda ,\varphi \right) ,\omega
^{\ast }\left( u,\lambda ,\varphi \right) \}.  \label{1sol4p1}
\end{eqnarray}
and

\newpage

\begin{eqnarray}
\mbox{$t$--solutions} &:&(x^{2}=u,x^{3}=\lambda ,y^{4}=v=t,y^{5}=p=\varphi )
\nonumber \\
g_{2} &=&-1,g_{3}=-1,h_{4(0)}=b(u,\lambda ),h_{5(0)}=a(u,\lambda ),%
\mbox{see
(\ref{1auxm1})};  \nonumber \\
h_{4} &=&\eta _{4}(u,\lambda ,t)h_{4(0)}(u,\lambda ),h_{5}=\eta
_{5}(u,\lambda ,t)h_{5(0)}(u,\lambda ),  \nonumber \\
\eta _{4} &=&\omega ^{-2}\left( u,\lambda ,t\right) ,\eta _{5}=h_{(0)}^{-2}%
\frac{b(u,\lambda )}{a(u,\lambda )}\left[ \int dt~\omega ^{-1}\left(
u,\lambda ,t\right) \right] ^{2},  \nonumber \\
w_{\underline{i}} &=&0,n_{\underline{i}}\{x,\omega ,\omega ^{\ast }\}=n_{%
\underline{i}}\{u,\lambda ,\omega \left( u,\lambda ,t\right) ,\omega ^{\ast
}\left( u,\lambda ,t\right) \}.  \label{1sol4t1}
\end{eqnarray}
where the $n_{\underline{i}}$ are computed
\begin{eqnarray}
n_{\underline{k}} &=&n_{\underline{k}[1]}\left( u,\lambda \right)
+n_{\underline{k}[2]}\left( u,\lambda \right) \int [\eta
_{4}/(\sqrt{|\eta _{5}|})^{3}]dv,~\eta _{5}^{\ast }\neq 0;
\label{1nem4} \\
&=&n_{\underline{k}[1]}\left( u,\lambda \right)
+n_{\underline{k}[2]}\left( u,\lambda \right) \int
\eta _{4}dv,\qquad ~\eta _{5}^{\ast }=0;  \nonumber \\
&=&n_{\underline{k}[1]}\left( u,\lambda \right)
+n_{\underline{k}[2]}\left( u,\lambda \right) \int
[1/(\sqrt{|\eta _{5}|})^{3}]dv,~\eta _{4}^{\ast }=0.  \nonumber
\end{eqnarray}
 These solutions have the same ellipsoidal
symmetries and properties stated for their 5D analogs
(\ref{1sol5p1}) and for (\ref{1sol5t1}) with that difference that
there are not any warped factors and extra dimension
dependencies. We emphasize that the solutions defined by the
formulas (\ref{1sol4p1}) and (\ref{1sol4t1}) do not result in a
locally isotropic limit into an exact solution having diagonal
coefficients with respect to some holonomic coordinate frames.
The data introduced in this subsection are for generic 4D vacuum
solutions of the Einstein equations parametrized by off--diagonal
metrics. The renormalization of constants and metric coefficients
have a 4D nonlinear vacuum gravitational origin and reflects a
corresponding anholonomic dynamics.

\subsubsection{Class A solutions with ansatz (\ref{22ansatzc4}):}

The 4D vacuum $\varphi $-- and $t$--solutions parametrized by an ansatz with
conformal factor $\Omega (u,\lambda ,v)$ (see (\ref{22ansatzc4}) and (\ref
{cdmetric4})). Let us consider conformal factors parametrized as $\Omega
=\Omega _{\lbrack 0]}(u,\lambda )\Omega _{\lbrack 1]}(u,\lambda ,v).$ The
data are

\newpage

\begin{eqnarray}
\mbox{$\varphi_c$--solutions} &:&(x^{2}=u,x^{3}=\lambda ,y^{4}=v=\varphi
,y^{5}=p=t)  \nonumber \\
g_{2} &=&-1,g_{3}=-1,h_{4(0)}=a(u,\lambda ),h_{5(0)}=b(u,\lambda ),%
\mbox{see
(\ref{1auxm1})};  \nonumber \\
h_{4} &=&\eta _{4}(u,\lambda ,\varphi )h_{4(0)}(u,\lambda ),h_{5}=\eta
_{5}(u,\lambda ,\varphi )h_{5(0)}(u,\lambda ),  \nonumber \\
\eta _{4} &=&h_{(0)}^{2}\frac{b(u,\lambda )}{a(u,\lambda )}\left\{ \left[
\omega ^{-1}\left( u,\lambda ,\varphi \right) \right] ^{\ast }\right\}
^{2},\eta _{5}=\omega ^{-2}\left( u,\lambda ,\varphi \right) ,
\label{1sol4pc} \\
w_{i} &=&0,n_{i}\{x,\omega ,\omega ^{\ast }\}=n_{i}\{u,\lambda ,\omega
\left( u,\lambda ,\varphi \right) ,\omega ^{\ast }\left( u,\lambda ,\varphi
\right) \},  \nonumber \\
\Omega &=&\Omega _{\lbrack 0]}(u,\lambda )\Omega _{\lbrack
1]}(u,\lambda ,\varphi ),\
\zeta _{i} =\left( \partial _{i}\ln |\Omega _{\lbrack 0]}\right) |)~\left(
\ln |\Omega _{\lbrack 1]}|\right) ^{\ast }+\left( \Omega _{\lbrack 1]}^{\ast
}\right) ^{-1}\partial _{i}\Omega _{\lbrack 1]},   \nonumber \\
\eta _{4}a &=& \Omega _{\lbrack
0]}^{q_{1}/q_{2}}(u,\lambda )\Omega _{\lbrack 1]}^{q_{1}/q_{2}}(u,\lambda
,\varphi ).  \nonumber
\end{eqnarray}
and
\begin{eqnarray}
\mbox{$t_c$--solutions} &:&(x^{2}=u,x^{3}=\lambda ,y^{4}=v=t,y^{5}=p=\varphi
)  \nonumber \\
g_{2} &=&-1,g_{3}=-1,h_{4(0)}=b(u,\lambda ),h_{5(0)}=a(u,\lambda ),%
\mbox{see
(\ref{1auxm1})};  \nonumber \\
h_{4} &=&\eta _{4}(u,\lambda ,t)h_{4(0)}(u,\lambda ),h_{5}=\eta
_{5}(u,\lambda ,t)h_{5(0)}(u,\lambda ),  \nonumber \\
\eta _{4} &=&\omega ^{-2}\left( u,\lambda ,t\right) ,\eta _{5}=h_{(0)}^{-2}%
\frac{b(u,\lambda )}{a(u,\lambda )}\left[ \int dt~\omega ^{-1}\left(
u,\lambda ,t\right) \right] ^{2},  \label{1sol4tc} \\
w_{i} &=&0,n_{i}\{x,\omega ,\omega ^{\ast }\}=n_{i}\{u,\lambda ,\omega
\left( u,\lambda ,t\right) ,\omega ^{\ast }\left( u,\lambda ,t\right)
\},  \nonumber \\
\Omega &=&\Omega _{\lbrack 0]}(u,\lambda )\Omega _{\lbrack 1]}(u,\lambda ,t),
\zeta _{i} =\left( \partial _{i}\ln |\Omega _{\lbrack 0]}\right) |)~\left(
\ln |\Omega _{\lbrack 1]}|\right) ^{\ast }+\left( \Omega _{\lbrack 1]}^{\ast
}\right) ^{-1}\partial _{i}\Omega _{\lbrack 1]},    \nonumber \\
\eta _{4}a&=&\Omega _{\lbrack
0]}^{q_{1}/q_{2}}(u,\lambda )\Omega _{\lbrack 1]}^{q_{1}/q_{2}}(u,\lambda
,t),  \nonumber
\end{eqnarray}
where the coefficients the $n_{\underline{i}}$ are given by the same
formulas (\ref{1nem4}).

Contrary to the solutions (\ref{1sol4p1}) and for (\ref{1sol4t1}) theirs
conformal anholonomic transforms, respectively, (\ref{1sol4pc}) and (\ref
{1sol4tc}), can be subjected to such parametrizations of the conformal factor
and conditions on the receptivity $\omega \left( u,\lambda ,v\right) $ as to
obtain in the locally isotropic limit just the Schwarzschild metric (\ref
{1schel}). These conditions are stated for $\Omega _{\lbrack
0]}^{q_{1}/q_{2}}=\Omega _{A},$ $\Omega _{\lbrack 1]}^{q_{1}/q_{2}}\eta
_{4}=1,$ $\Omega _{\lbrack 1]}^{q_{1}/q_{2}}\eta _{5}=1,$were $\Omega _{A}$
is that from (\ref{2confq}), which is possible if $\eta
_{4}^{-q_{1}/q_{2}}\eta _{5}=1,$which selects a specific form of the
receptivity $\omega .$ \ Putting the values $\eta _{4}$ and $\eta _{5},$
respectively, from (\ref{1sol4pc}), or (\ref{1sol4tc}), we obtain some
differential, or integral, relations of the unknown $\omega \left(
x^{i},v\right) ,$ which results that
\begin{eqnarray*}
&&\omega \left( u,\lambda ,\varphi \right) =\left( 1-q_{1}/q_{2}\right)
^{-1-q_{1}/q_{2}}\left[ h_{(0)}^{-1}\sqrt{|a/b|}\varphi +\omega _{\lbrack
0]}\left( u,\lambda \right) \right] ,\mbox{for }\varphi _{c}%
\mbox{--solutions}; \\
&&\omega \left( u,\lambda ,t\right) =\left[ \left( q_{1}/q_{2}-1\right)
h_{(0)}\sqrt{|a/b|}t+\omega _{\lbrack 1]}\left( u,\lambda \right) \right]
^{1-q_{1}/q_{2}},\mbox{for }t_{c}\mbox{--solutions},
\end{eqnarray*}
for some arbitrary functions $\omega _{\lbrack 0]}\left( u,\lambda \right) $
and $\omega _{\lbrack 1]}\left( u,\lambda \right) .$ The  formulas
for $\omega \left( u,\lambda ,\varphi \right) $ and $\omega \left( u,\lambda
,t\right) $ are 4D reductions of the formulas (\ref{1cond1a}) and (\ref{4cond1}%
).

\subsection{Ellipsiodal 4D metrics of Class B}

We construct another two classes of 4D vacuum solutions which are related to
the metric of Class B (\ref{1auxm2}) and which can be reduced to the
Schwarzschild metric in ellipsoidal coordinates (\ref{1schel}) by
corresponding parametrizations of receptivity $\omega \left( u,\lambda
,v\right) $. The solutions contain a 2D conformal factor $\varpi $ $%
(u,\lambda )$ for which $\varpi g$ becomes a solution of (\ref{4ricci1a}) and
a 4D conformal factor parametrized as $\Omega =\varpi ^{-1}$ $\Omega
_{\lbrack 2]}\left( u,\lambda ,v\right) $ in \ order to set the
constructions into the ansatz (\ref{22ansatzc4}) and anholonomic metric
interval (\ref{cdmetric4}).

The data selecting the 4D configurations for $\varphi _{c}$--solutions and $%
t_{c}$--solutions:

\begin{eqnarray}
\mbox{$\varphi_c$--solutions} &:&(x^{2}=u,x^{3}=\lambda ,y^{4}=v=\varphi
,y^{5}=p=t)  \nonumber \\
g_{2} &=&g_{3}=\varpi (u,\lambda )g(u,\lambda ),    \nonumber \\
h_{4(0)}&=& -\varpi (u,\lambda
),h_{5(0)}=\varpi (u,\lambda )f(u,\lambda ),\mbox{see
(\ref{1auxm2})};  \nonumber \\
\varpi &=&g^{-1}\varpi _{0}\exp [a_{2}u+a_{3}\lambda ],~\varpi
_{0},a_{2},a_{3}=const;~\mbox{see
(\ref{2solricci1a})}  \nonumber \\
h_{4} &=&\eta _{4}(u,\lambda ,\varphi )h_{4(0)}(u,\lambda ),h_{5}=\eta
_{5}(u,\lambda ,\varphi )h_{5(0)}(u,\lambda ),  \nonumber \\
\eta _{4} &=&-h_{(0)}^{2}f(u,\lambda )\left\{ \left[ \omega ^{-1}\left(
u,\lambda ,\varphi \right) \right] ^{\ast }\right\} ^{2},\eta _{5}=\omega
^{-2}\left( u,\lambda ,\varphi \right) ,  \label{1sol4p} \\
w_{i} &=&0,n_{i}\{x,\omega ,\omega ^{\ast }\}=n_{i}\{u,\lambda ,\omega
\left( u,\lambda ,\varphi \right) ,\omega ^{\ast }\left( u,\lambda ,\varphi
\right) \}, \nonumber \\
\Omega &=& \varpi ^{-1}(u,\lambda )\Omega _{\lbrack 2]}(u,\lambda
,\varphi ),\
\zeta _{\underline{i}} = \partial _{\underline{i}}\ln |\varpi |)~\left( \ln
|\Omega _{\lbrack 2]}|\right) ^{\ast }+\left( \Omega _{\lbrack 2]}^{\ast
}\right) ^{-1}\partial _{\underline{i}}\Omega _{\lbrack 2]}, \nonumber \\
\eta _{4}&=& -\varpi ^{-q_{1}/q_{2}}(u,\lambda )\Omega _{\lbrack
2]}^{q_{1}/q_{2}}(u,\lambda ,\varphi ).  \nonumber
\end{eqnarray}
and
\begin{eqnarray}
\mbox{$t_c$--solutions} &:&(x^{2}=u,x^{3}=\lambda ,y^{4}=v=t,y^{5}=p=\varphi
)  \nonumber \\
g_{2} &=&g_{3}=\varpi (u,\lambda )g(u,\lambda ),   \nonumber \\
h_{4(0)}&=&\varpi (u,\lambda
)f(u,\lambda ),h_{5(0)}=-\varpi (u,\lambda ),\mbox{see
(\ref{1auxm2})};  \nonumber \\
\varpi &=&g^{-1}\varpi _{0}\exp [a_{2}u+a_{3}\lambda ],~\varpi
_{0},a_{2},a_{3}=const,~\mbox{see
(\ref{2solricci1a})}  \nonumber \\
h_{4} &=&\eta _{4}(u,\lambda ,t)h_{4(0)}(x^{i}),h_{5}=\eta _{5}(u,\lambda
,t)h_{5(0)}(x^{i}),  \nonumber
 \end{eqnarray}
 \begin{eqnarray}
\eta _{4} &=&\omega ^{-2}\left( u,\lambda ,t\right) ,\eta
_{5}=-h_{(0)}^{-2}f(u,\lambda )\left[ \int dt~\omega ^{-1}\left( u,\lambda
,t\right) \right] ^{2},  \label{1sol4t} \\
w_{i} &=&0,n_{i}\{x,\omega ,\omega ^{\ast }\}=n_{i}\{u,\lambda ,\omega
\left( u,\lambda ,t\right) ,\omega ^{\ast }\left( u,\lambda ,t\right)
\},   \nonumber
\end{eqnarray}
\begin{eqnarray}
\Omega &=& \varpi ^{-1}(u,\lambda )\Omega _{\lbrack 2]}(u,\lambda ,t),
\zeta _{i} =\partial _{i}(\ln |\varpi |)~\left( \ln |\Omega _{\lbrack
2]}|\right) ^{\ast }+\left( \Omega _{\lbrack 2]}^{\ast }\right)
^{-1}\partial _{i}\Omega _{\lbrack 2]}, \nonumber \\
\eta _{4}&=&-\varpi
^{-q_{1}/q_{2}}(u,\lambda )\Omega _{\lbrack 2]}^{q_{1}/q_{2}}(u,\lambda ,t).
\nonumber
\end{eqnarray}
where the coefficients $n_{i}$ can be found explicitly by introducing the
corresponding values $\eta _{4}$ and $\eta _{5}$ in formula (\ref{1nel}).

For the 4D Class B solutions one can be imposed some conditions (see
previous subsection) when the solutions (\ref{1sol4p}) and (\ref{1sol4t}) have
in the locally anisotropic limit the Schwarzschild solution, which imposes
some specific parametrizations and dependencies on $\Omega _{\lbrack
2]}(u,\lambda ,v)$ and $\omega \left( u,\lambda ,v\right) $ like (\ref
{1cond1a}) and (\ref{4cond1}). We omit these considerations because for
aholonomic configurations and nonlinear solutions there are not arguments to
prefer any holonomic limits of such off--diagonal metrics.

We conclude this Section by noting that for the considered classes of
ellipsoidal black hole 4D solutions the so--called $t$--component of metric
contains modifications of the Schwarzschild potential
\[
\Phi =-\frac{M}{M_{P[4]}^{2}r}\mbox{ into }\Phi =-\frac{M\omega \left(
u,\lambda ,v\right) }{M_{P[4]}^{2}r},
\]
where $M_{P[4]}$ is the usual 4D Plank constant; the metric coefficients are
given with respect to the corresponding anholonomic frame of reference. In 4D
anholonomic gravity the receptivity $\omega \left( u,\lambda ,v\right) $ is
considered to renormalize the mass constant. Such gravitational
self-polarizations are induced by anholonomic vacuum gravitational
interactions. They should be defined experimentally or computed following a
model of quantum gravity.

\section{The Cosmological Constant and Anisotropy}

In this Section we analyze the general properties of anholonomic
Einstein equations in 5D and 4D gravity with cosmological
constant and construct a 5D exact solution with cosmological
constant.

\subsection{4D and 5D Anholnomic Einstein spaces}

There is a difference between locally anisotropic 4D and 5D
gravity. The first theory admits an ''isotropic'' 4D cosmological
constant $\Lambda _{\lbrack 4]}=\Lambda $ even for anisotropic
gravitational configurations. The second, 5D, theory admits
extensions of vacuum anistoropic solutions to those with a
cosmological constant only for anisotropic 5D sources
parametrized like $\Lambda _{\lbrack 5]\alpha \beta }=(2\Lambda
g_{11},\Lambda g_{\underline{\alpha }\underline{\beta }})$ (see
the Corollary 4 below). We emphasize that the conclusions from
this subsection refer to the two classes of ansatz (\ref{4ansatz})
and (\ref{4ansatzc}).

The simplest way to consider a source into the 4D Einstein
equations, both with or not anistoropy, is to consider a
gravitational constant $\Lambda $ and to write the field equations
\begin{equation}
G_{\underline{\beta }}^{\underline{\alpha }}=\Lambda _{\lbrack 4]}\delta _{%
\underline{\beta }}^{\underline{\alpha }}  \label{einst4cc}
\end{equation}
which means that we introduced a ''vacuum'' energy--momentum tensor $\kappa
\Upsilon _{\underline{\beta }}^{\underline{\alpha }}=\Lambda _{\lbrack
4]}\delta _{\underline{\beta }}^{\underline{\alpha }}$ which is diagonal
with respect to anholonomic frames and the conditions (\ref{2emcond4})
transforms into $\Upsilon _{2}^{2}=\Upsilon _{3}^{3}=\Upsilon
_{4}^{4}=\Upsilon _{5}^{5}=\kappa ^{-1}\Lambda .$ According to A. Z. Petrov
\cite{10petrov} the spaces described by solutions of the Einstein equations
\[
R_{\underline{\alpha }\underline{\beta }}=\Lambda g_{\underline{\alpha }%
\underline{\beta }},\Lambda =const
\]
are called the Einstein spaces. With respect to anisotropic frames we shall
use the term anholonomic (equivalently, anisotropic) Einstein spaces.

In order to extend the equations (\ref{einst4cc}) to 5D gravity we have to
take into consideration the compatibility conditions for the
energy--momentum tensors (\ref{2emcond}).

\begin{corollary}
We are able to satisfy the conditions of the Corollary 2 if \ we consider a
5D diagonal source $\Upsilon _{\beta }^{\alpha }=\{2\Lambda ,$ $\Upsilon _{%
\underline{\beta }}^{\underline{\alpha }}=\Lambda \delta _{\underline{\beta }%
}^{\underline{\alpha }}\},$ for an anisotropic 5D cosmological constant source $%
(2\Lambda g_{11},\Lambda g_{\underline{\alpha }\underline{\beta
}}).$ The 5D Einstein equations with anisotropic cosmological
"constants", for ansatz (\ref {4ansatz}) are written in the form
\begin{equation}
R_{2}^{2}=S_{4}^{4}=-\Lambda .  \label{eecvcosm}
\end{equation}
These equations without coordinate $x^{1}$ and $g_{11}$ hold for the (\ref
{2ansatz4}). We can extend the constructions for the ansatz with conformal
factors, (\ref{4ansatzc}) and (\ref{22ansatzc4}) by considering additional
coefficients $\zeta _{i}$ satisfying the equations (\ref{3confeq}) and (\ref
{2confeq4}) for non vanishing values of $w_{i}.$
\end{corollary}

The proof follows from Corollaries 1 and 2 formulated respectively to 4D and
5D gravity (see formulas (\ref{2einstdiag4}) and (\ref{2emcond4}) and,
correspondingly, (\ref{2einstdiag}) and (\ref{2emcond})).

\begin{theorem}
The nontrivial components of the 5D Einstein equations with\\ anisotro\-pic
cosmological constant, $R_{11}=2\Lambda g_{11}$ and $R_{\underline{\alpha }%
\underline{\beta }}=\Lambda g_{\underline{\alpha }\underline{\beta }},$ for
the ansatz (\ref{4ansatzc}) and anholonomic metric (\ref{2cdmetric}) given
with respect to anholonomic frames (\ref{11ddif2}) and (\ref{11dder2}) are
written in a form with separation of variables:
\begin{eqnarray}
g_{3}^{\bullet \bullet }-\frac{g_{2}^{\bullet }g_{3}^{\bullet }}{2g_{2}}-%
\frac{(g_{3}^{\bullet })^{2}}{2g_{3}}+g_{2}^{^{\prime \prime }}-\frac{%
g_{2}^{^{\prime }}g_{3}^{^{\prime }}}{2g_{3}}-\frac{(g_{2}^{^{\prime }})^{2}%
}{2g_{2}} &=&2\Lambda g_{2}g_{3},  \label{ricci1const} \\
h_{5}^{\ast \ast }-h_{5}^{\ast }[\ln \sqrt{|h_{4}h_{5}|}]^{\ast }
&=&2\Lambda h_{4}h_{5},  \label{ricci2const} \\
w_{i}\beta +\alpha _{i} &=&0,  \label{ricci3const} \\
n_{i}^{\ast \ast }+\gamma n_{i}^{\ast } &=&0,  \label{ricci4const} \\
\partial _{i}\Omega -(w_{i}+\zeta _{{i}})\Omega ^{\ast } &=&0.
\label{confeql}
\end{eqnarray}
where
\begin{equation}
\alpha _{i}=\partial _{i}{h_{5}^{\ast }}-h_{5}^{\ast }\partial _{i}\ln \sqrt{%
|h_{4}h_{5}|},\beta =2\Lambda h_{4}h_{5},\gamma =3h_{5}^{\ast
}/2h_{5}-h_{4}^{\ast }/h_{4}.  \label{abcl4}
\end{equation}
\end{theorem}

The Theorem 4 is a generalization of the Theorem 2 for energy--momentum
tensors induced by the an anisotropic 5D constant. The proof follows from (%
\ref{4ricci1a})--(\ref{3ricci4a}) and (\ref{3confeq}), revised as to
satisfy the formulas\ (\ref{2ricci1b}) and (\ref{ricci2b}) with
that substantial difference that $\beta \neq 0$ and in this case,
in general, $w_{i}\neq 0.$ We conclude that in the presence of a
nonvanishing cosmological constant the
equations (\ref{4ricci1a}) and (\ref{3ricci2a}) transform respectively into (%
\ref{ricci1const}) and \ (\ref{ricci2const}) \ which have a more
general nonlinearity because of  the $2\Lambda g_{2}g_{3}$ and
$2\Lambda
h_{4}h_{5}$ terms. For instance, the solutions with $g_{2}=const$ and $%
g_{3}=const$ (and $h_{4}=const$ and $h_{5}=const)$ are not admitted. This
makes more sophisticate the procedure of definition of $g_{2}$ for a given $%
g_{3}$ (or inversely, of definition of $g_{3}$ for a given $g_{2})$ from (%
\ref{ricci1const}) [similarly of construction $h_{4}$ for a given
$h_{5}$ from (\ref{ricci2const}) and inversely], nevertheless,
the separation of variables is not affected by introduction of
cosmological constant and there is a number of possibilities to
generate new exact solutions.

The general properties of solutions of the system (\ref{ricci1const})--(\ref
{confeql}) are stated by the

\begin{theorem}
The system of second order nonlinear partial differential equations (\ref
{ricci1const})-(\ref{ricci4const}) and (\ref{confeql}) can be solved in
general form if there are given some values of functions $g_{2}(x^{2},x^{3})$
(or $g_{3}(x^{2},x^{3})),h_{4}\left( x^{i},v\right) $ (or $h_{5}\left(
x^{i},v\right) )$ and $\Omega \left( x^{i},v\right) :$

\begin{itemize}
\item  The general solution of equation (\ref{ricci1const}) is to be found
from the equation
\begin{equation}
\varpi \varpi ^{\bullet \bullet }-(\varpi ^{\bullet })^{2}+\varpi \varpi
^{^{\prime \prime }}-(\varpi ^{^{\prime }})^{2}=2\Lambda \varpi ^{3}.
\label{1auxr1}
\end{equation}
for a coordinate transform coordinate transforms $x^{2,3}\rightarrow
\widetilde{x}^{2,3}\left( u,\lambda \right) $ for which
\[
g_{2}(u,\lambda )(du)^{2}+g_{3}(u,\lambda )(d\lambda )^{2}\rightarrow \varpi
\left[ (d\widetilde{x}^{2})^{2}+\epsilon (d\widetilde{x}^{3})^{2}\right]
,\epsilon =\pm 1
\]
and $\varpi ^{\bullet }=\partial \varpi /\partial \widetilde{x}^{2}$ and $%
\varpi ^{^{\prime }}=\partial \varpi /\partial \widetilde{x}^{3}.$

\item  The equation (\ref{ricci2const}) relates two functions $h_{4}\left(
x^{i},v\right) $ and $h_{5}\left( x^{i},v\right) $ with $h_{5}^{\ast }\neq
0. $ If the function $h_{5}$ is given we can find $h_{4}$ as a solution of
\begin{equation}
h_{4}^{\ast }+\frac{2\Lambda }{\tau }(h_{4})^{2}+2\left( \frac{\tau ^{\ast }%
}{\tau }-\tau \right) h_{4}=0,  \label{1auxr2c}
\end{equation}

where $\tau =h_{5}^{\ast }/2h_{5}.$

\item  The exact solutions of (\ref{ricci3const}) for $\beta \neq 0$ is
\begin{eqnarray}
w_{k} &=&-\alpha _{k}/\beta ,  \label{1aw} \\
&=&\partial _{k}\ln [\sqrt{|h_{4}h_{5}|}/|h_{5}^{\ast }|]/\partial _{v}\ln [%
\sqrt{|h_{4}h_{5}|}/|h_{5}^{\ast }|],  \nonumber
\end{eqnarray}
for $\partial _{v}=\partial /\partial v$ and $h_{5}^{\ast }\neq 0.$

\item  The exact solution of (\ref{ricci4const}) is
\begin{eqnarray}
n_{k} &=&n_{k[1]}\left( x^{i}\right) +n_{k[2]}\left( x^{i}\right) \int
[h_{4}/(\sqrt{|h_{5}|})^{3}]dv,  \label{1nlambda} \\
&=&n_{k[1]}\left( x^{i}\right) +n_{k[2]}\left( x^{i}\right) \int [1/(\sqrt{%
|h_{5}|})^{3}]dv,~h_{4}^{\ast }=0,  \nonumber
\end{eqnarray}
for some functions $n_{k[1,2]}\left( x^{i}\right) $ stated by boundary
conditions.

\item  The exact solution of (\ref{3confeq}) is given by
\begin{equation}
\zeta _{i}=-w_{i}+(\Omega ^{\ast })^{-1}\partial _{i}\Omega ,\quad \Omega
^{\ast }\neq 0,  \label{1aconf4}
\end{equation}
\end{itemize}
\end{theorem}

We note that by a corresponding re--parametrizations of the conformal factor
$\Omega \left( x^{i},v\right) $ we can reduce (\ref{1auxr1}) to
\begin{equation}
\varpi \varpi ^{\bullet \bullet }-(\varpi ^{\bullet })^{2}=2\Lambda \varpi
^{3}  \label{1redaux}
\end{equation}
which has an exact solution $\varpi =\varpi \left( \widetilde{x}%
^{2}\right) $ to be found from
\[
(\varpi ^{\bullet })^{2}=\varpi ^{3}\left( C\varpi ^{-1}+4\Lambda \right)
,C=const,
\]
(or, inversely, to reduce to
\[
\varpi \varpi ^{^{\prime \prime }}-(\varpi ^{^{\prime }})^{2}=2\Lambda
\varpi ^{3}
\]
with exact solution $\varpi =\varpi \left( \widetilde{x}^{3}\right) $ found
from
\[
(\varpi ^{\prime })^{2}=\varpi ^{3}\left( C\varpi ^{-1}+4\Lambda \right)
,C=const).
\]
The inverse problem of definition of $h_{5}$ for a given $h_{4}$ can be
solved in explicit form when $h_{4}^{\ast }=0,$ $h_{4}=h_{4(0)}(x^{i}).$ In
this case we have to solve
\begin{equation}
h_{5}^{\ast \ast }+\frac{(h_{5}^{\ast })^{2}}{2h_{5}}-2\Lambda
h_{4(0)}h_{5}=0,  \label{1auxr2ccp}
\end{equation}
which admits exact solutions by reduction to a Bernulli equation.

The proof of Theorem 5 is outlined in Appendix C.

The conditions of the Theorem 4 and 5 can be reduced to 4D anholonomic
spacetimes with ''isotropic'' cosmological constant $\Lambda .$ To do this
we have to eliminate dependencies on the coordinate $x^{1}$ and to consider
the 4D ansatz without $g_{11}$ term as it was stated in the subsection II E.

\subsection{5D anisotropic black holes with cosmological constant}

We give an example of generalization of anisotropic black hole
solutions of Class A , constructed in the Section III, as they
will contain the
cosmological constant $\Lambda ;$ we extend the solutions given by the data (%
\ref{1sol5pc}).

Our new 5D $\varphi $-- solution is parametrized by an ansatz with conformal
factor $\Omega (x^{i},v)$ (see (\ref{4ansatzc}) and (\ref{2cdmetric})) as $%
\Omega =\varpi ^{-1}(u)\Omega _{\lbrack 0]}(x^{i})\varpi
^{-1}(u)\Omega _{\lbrack 1]}(x^{i},v).$ The factor $\varpi (u)$
is chosen  to be a
solution of (\ref{1redaux}). This conformal data must satisfy the conditions (%
\ref{2confq}) and (\ref{2confsol}), i. e.
\[
\varpi ^{-q_{1}/q_{2}}\Omega _{\lbrack 0]}^{q_{1}/q_{2}}\Omega _{\lbrack
1]}^{q_{1}/q_{2}}=\eta _{4}\varpi h_{4(0)}
\]
for some integers $q_{1}$ and $q_{2},$ where $\eta _{4}$ is found as $%
h_{4}=\eta _{4}\varpi h_{4(0)}$ is a solution of equation (\ref{1auxr2c}).
 The factor
 $\Omega _{\lbrack 0]}(x^{i})$ could be chosen as to obtain in the locally
isotropic limit and $\Lambda \rightarrow 0$ the Schwarzschild
metric in ellipsoidal coordinates (\ref{1schel}). Putting
$h_{5}=\eta _{5}\varpi h_{5(0)},$ $\eta _{5}h_{5(0)}$ in the
ansatz for (\ref{1sol5pc}), for which we compute the value $\tau
=h_{5}^{\ast }/2h_{5},$ we obtain from (\ref
{1auxr2c}) an equation for $\eta _{4},$%
\[
\eta _{4}^{\ast }+\frac{2\Lambda }{\tau }\varpi h_{4(0)}(\eta
_{4})^{2}+2\left( \frac{\tau ^{\ast }}{\tau }-\tau \right) \eta _{4}=0
\]
which is a Bernulli equation \cite{10kamke} and admit an exact solution, in
general, in non explicit form, $\eta _{4}=\eta _{4}^{[bern]}(x^{i},v,\Lambda
,\varpi ,\omega ,a,b),$ were we emphasize the functional dependencies on
functions $\varpi ,\omega ,a,b$ and cosmological constant $\Lambda .$ Having
defined $\eta _{4[bern]},$ $\eta _{5}$ and $\varpi ,$ we can compute the $%
\alpha _{i}$--$,\beta -,$ and $\gamma $--coefficients, expressed as $$\alpha
_{i}=\alpha _{i}^{[bern]}(x^{i},v,\Lambda ,\varpi ,\omega ,a,b),\beta =\beta
^{\lbrack bern]}(x^{i},v,\Lambda ,\varpi ,\omega ,a,b)$$ and $\gamma =\gamma
^{\lbrack bern]}(x^{i},v,\Lambda ,\varpi ,\omega ,a,b),$ following the
formulas (\ref{abcl4}).

The next step is to find
\[
w_{i}=w_{i}^{[bern]}(x^{i},v,\Lambda ,\varpi ,\omega ,a,b)\mbox{ and }%
n_{i}=n_{i}^{[bern]}(x^{i},v,\Lambda ,\varpi ,\omega ,a,b)
\]
as for the general solutions (\ref{1aw}) and (\ref{1nlambda}).

At the final step we are able to compute the the second anisotropy
coefficients
\[
\zeta _{i}=-w_{i}^{[bern]}+\left( \partial _{i}\ln |\varpi ^{-1}\Omega
_{\lbrack 0]}\right) |)~\left( \ln |\Omega _{\lbrack 1]}|\right) ^{\ast
}+\left( \Omega _{\lbrack 1]}^{\ast }\right) ^{-1}\partial _{i}\Omega
_{\lbrack 1]},
\]
which depends on an arbitrary function $\Omega _{\lbrack 0]}(u,\lambda ).$
If we state $\Omega _{\lbrack 0]}(u,\lambda )=\Omega _{A},$ as for $\Omega
_{A}$ from (\ref{1auxm2}), see similar details with respect to formulas (\ref
{1cond1a}), (\ref{2cond2}) and (\ref{4cond1}).

The data for the exact solutions with cosmological constant for $v=\varphi $
can be stated in the form
\begin{eqnarray}
\mbox{$\varphi_c$--solutions} &:&(x^{1}=\chi ,x^{2}=u,x^{3}=\lambda
,y^{4}=v=\varphi ,y^{5}=p=t),g_{1}=\pm 1,  \nonumber \\
g_{2} &=&\varpi (u),g_{3}=\varpi (u),  \nonumber \\
h_{4(0)}&=&a(u,\lambda
),h_{5(0)}=b(u,\lambda ),\mbox{see (\ref{1auxm1}) and  (\ref{1redaux})};
\nonumber \\
h_{4} &=&\eta _{4}(x^{i},\varphi )\varpi (u)h_{4(0)}(x^{i}),h_{5}=\eta
_{5}(x^{i},\varphi )\varpi (u)h_{5(0)}(x^{i}),  \nonumber \\
\eta _{4} &=&\eta _{4}^{[bern]}(x^{i},v,\Lambda ,\varpi ,\omega ,a,b),\eta
_{5}=\omega ^{-2}\left( \chi ,u,\lambda ,\varphi \right) ,  \label{1slambdap1}
\\
w_{i} &=&w_{i}^{[bern]}(x^{i},v,\Lambda ,\varpi ,\omega
,a,b),n_{i}\{x,\omega ,\omega ^{\ast }\}=n_{i}^{[bern]}(x^{i},v,\Lambda
,\varpi ,\omega ,a,b),  \nonumber \\
\Omega &=&\varpi ^{-1}(u)\Omega _{\lbrack 0]}(x^{i})\Omega _{\lbrack
1]}(x^{i},\varphi ),\eta _{4}a=\Omega _{\lbrack
0]}^{q_{1}/q_{2}}(x^{i})\Omega _{\lbrack 1]}^{q_{1}/q_{2}}(x^{i},\varphi ).
\nonumber \\
\zeta _{i} &=&-w_{i}^{[bern]}+\left( \partial _{i}\ln |\varpi ^{-1}\Omega
_{\lbrack 0]}\right) |)~\left( \ln |\Omega _{\lbrack 1]}|\right) ^{\ast
}+\left( \Omega _{\lbrack 1]}^{\ast }\right) ^{-1}\partial _{i}\Omega
_{\lbrack 1]}.  \nonumber
\end{eqnarray}

We note that a solution with $v=t$ can be constructed as to
generalize (\ref {1sol5tc}) in order to contain $\Lambda .$ We can
not present such data in explicit form because in this case we
have to define $\eta _{5}$ by integrating an equation like
(\ref{ricci2const}) for $h_{5},$ for a given $h_{4},$ with
$h_{4}^*\neq 0$ which can not be integrated in explicit form.

The solution (\ref{1slambdap1}) has has the same the two very
interesting properties as the solution (\ref{1sol5pc}):\
 1) it admits a warped factor on the 5th coordinate, like $%
\Omega _{\lbrack 1]}^{q_{1}/q_{2}}\sim \exp [-k|\chi |],$ which
in this case is constructed for an anisotropic 5D vacuum
gravitational configuration with anisotropic cosmological
constant and does not follow from a brane configuration like in
Refs. \cite{10rs}; 2) we can impose such conditions on the
receptivity $\omega \left( x^{i},\varphi \right) $ as to obtain
in the locally isotropic limit just the Schwarzschild metric
(\ref{1schel}) trivially embedded into the 5D spacetime (the
procedure is the same as in the subsection IIIB).

Finally, we note that in a similar manner like in the Sections III and IV we
can construct another classes of anisotropic black holes solutions in 5D and
4D spacetimes with cosmological constants, being of Class A or Class B, with
anisotropic $\varphi $--coordinate, or anisotropic $t$--coordinate. We omit
the explicit data which are some nonlinear anholonomic generalizations of
those solutions.

\section{Conclusions}

We formulated a new method of constructing exact solutions of
Einstein equations with off--diagonal metrics in 4D and 5D
gravity. We introduced ahnolonomic transforms which diagonalize
metrics and simplify  the system of  gravitational field
equations.   The method works also for gravitational
configurations with cosmological constants and for non--trivial
matter sources. We constructed different classes of new exact
solutions of the Einstein equations is 5D and 4D gravity which
describe a generic anholonomic (anisotropic) dynamics modelled by
off--diagonal metrics and anholonomic frames with mixed holonomic
and anholonomic variables. They extend the class of exact
solutions with linear extensions to the bulk 5D gravity
\cite{10gian}.

We emphasized such exact solutions which can be associated to
some black hole like configurations in 5D and 4D gravity. We
consider that the constructed off--diagonal metrics define
anisotropic black holes because they have a static horizon
parametrized by a rotation ellipsoid hypersurface, they are
singular in focuses of ellipsoid (or on the circle of focuses,
for flattened ellipsoids) and they reduce in the locally
anisotropic limit, with holonomic coordinates, to the
Schwarzschild solution in ellipsoidal coordinates, or to some
conformal transforms of the  Schwarzschild metric.

The new classes of solutions admit variations of constants (in
time and extra dimension coordinate) and anholonomic
gravitational polarizations of masses which are induced by
nonlinear gravitational interactions in the bulk of 5D gravity
and by a constrained (anholonomic) dynamics of the fields in the
4D gravity. There are possible solutions with warped factors
which are defined by some\ vacuum 5D gravitational interactions
in the bulk and not by a specific brane configuration with
energy--momentum tensor source. \ We emphasized anisotropies
which in the effective 4D spacetime preserve the local Lorentz
invariance but the method allows constructions with violation of
local Lorentz symmetry like in Refs.  \cite{10csaki}. In order to
generate such solutions we should admit that the metric
coefficients depends, for instance, anisotropically on extra
dimension coordinate. \

It should be noted that the anholonomic frame method deals with
generic off--diagonal metrics and nonlinear systems of equations
and allows to construct substantially nonlinear solutions. In
general, such solutions could not have a locally isotropic limit
with a holonomic analog. We can understand the physical
properties of such solutions by analyzing both the metric
coefficients stated with respect to an adapted anholonomic frame
of reference and by a study of the coefficients defining such
frames.

There is a subclass of static anisotropic black holes solutions,
with static ellipsoidal horizons, which do not violate the well
known Israel and Carter theorems \cite{10israel} on spherical
symmetry of solutions in asymptotically flat spacetimes. Those
theorems were proved in the radial symmetry asymptotic limit and
for holonomic coordinates. There is not a much difference between
\ 3D static spherical and ellipsoidal horizons at long distances.
In other turn, the statements of the mentioned theorems do not
refers to generic off--diagonal gravitational metrics,
anholonomic frames and anholonomic deformations of symmetries.

Finally, we note that the anholonomic frame method may have a
number of applications in modern brane and string/M--theory
gravity because it defines a general formalism of constructing
exact solutions with off--diagonal metrics. It results in such
prescriptions on anholonomic ''mappings'' of some known locally
isotropic solutions from a gravity/string theory that new types of
anisotropic solutions are generated:

{\em A vacuum, or non-vacuum, solution, and metrics conformally
equivalent to a such solution, parametrized by a diagonal matrix
given with respect to a holonomic (coordinate) base, contained in
a trivial form of ansatz (\ref {4ansatz}), or (\ref{4ansatzc}), can
be generalized to an anisotropic solution with similar but
anisotropically renormalized physical constants and diagonal
metric coefficients given with respect to adapted anholonomic
frames; the new anholonomic metric defines an exact solution of a
simplified form of \ the Einstein equations
(\ref{4ricci1a})--(\ref{3ricci4a}) and (\ref {3confeq}); such
solutions are parametrized by off--diagonal metrics if they are
re--defined with respect to coordinate frames }.

\subsection*{Acknowledgements}

The author thanks D. Singleton, E. Gaburov, D. Gon\c ta and Nadejda
Vicol for collaboration and discussing of results. The work is
supported both by a 2000--2001 California State University
Legislative Award and a NATO/Portugal fellowship grant at the
Technical University of Lisbon.

%\appendix

\section{A: Anholonomic Frames and Nonlinear Connections}

For convenience, we outline here the basic formulas for connections,
curvatures and, induced by anholonomic frames, torsions on (pseudo)
Riemannian spacetimes provided with N--coefficient bases (\ref{5dder}) and (%
\ref{17ddif}) \cite{10v,10v1}. The N--coefficients define an associated nonlinear
connection (in brief, N--connection) structure. On (pseudo)--Riemannian
spacetimes the  N--connection structure can be treated as a ''pure''
anholonomic frame effect which is induced if we are dealing with mixed sets
of holonomic--anholonomic basis vectors. When we are transferring our
considerations only to coordinate frames (\ref{3pder}) and (\ref{5pdif}) the
N--connection coefficients are removed into both off--diagonal and diagonal
components of the metric like in (\ref{4ansatz}). In some cases the
N--connection (anholonomic) structure is to be stated in a non--dynamical
form by definition of some initial (boundary) conditions for the frame
structure, following some prescribed symmetries of the gravitational--matter
field interactions, or , in another cases, a subset of N--coefficients have
to be treated as some dynamical variables defined as to satisfy the Einstein
equations.

\subsection{D--connections, d--torsions and d--curvatures}

If a pseudo--Riemannian spacetime is enabled with a N--connection structure,
the components of geometrical objects (for instance, linear connections and
tensors) are distinguished into horizontal components (in brief
h--components, labelled by indices like $i,j,k,...)$ and vertical components
(in brief v--components, labelled by indices like $a,b,c,..).$ One call such
objects, distinguished (d) by the N--connection structure, as d--tensors,
d--connections, d--spinors and so on \cite{10ma,10vf,10v}.

\subsubsection{D--metrics and d-connections:}

A metric of type (\ref{5dmetric}), in general, with arbitrary coefficients $%
g_{ij}\left( x^k,y^a\right) $ and\\ $h_{ab}\left( x^k,y^a\right) $ defined
with respect to a N--elongated basis (\ref{17ddif}) is called a d--metric.

A linear connection $D_{\delta _{\gamma }}\delta _{\beta }=\Gamma _{\ \beta
\gamma }^{\alpha }\left( x,y\right) \delta _{\alpha },$ associated to an
operator of covariant derivation $D,$ is compatible with a metric $g_{\alpha
\beta }$ and N--connection structure on a 5D pseudo--Riemannian spacetime if
$D_{\alpha }g_{\beta \gamma }=0.$ The linear d--connection is parametrized
by irreducible h--v--components,\ $\Gamma _{\ \beta \gamma }^{\alpha
}=\left( L_{\ jk}^{i},L_{\ bk}^{a},C_{\ jc}^{i},C_{\ bc}^{a}\right) ,$ where
\begin{eqnarray}
L_{\ jk}^{i} &=&\frac{1}{2}g^{in}\left( \delta _{k}g_{nj}+\delta
_{j}g_{nk}-\delta _{n}g_{jk}\right) ,  \label{5dcon} \\
L_{\ bk}^{a} &=&\partial _{b}N_{k}^{a}+\frac{1}{2}h^{ac}\left( \delta
_{k}h_{bc}-h_{dc}\partial _{b}N_{k}^{d}-h_{db}\partial _{c}N_{k}^{d}\right) ,
\nonumber \\
C_{\ jc}^{i} &=&\frac{1}{2}g^{ik}\partial _{c}g_{jk},\ C_{\ bc}^{a}=\frac{1}{%
2}h^{ad}\left( \partial _{c}h_{db}+\partial _{b}h_{dc}-\partial
_{d}h_{bc}\right) .  \nonumber
\end{eqnarray}
This defines a canonical linear connection (distinguished by a
N--connection, in brief, the canonical d--connection) which is similar to
the metric connection introduced by Christoffel symbols in the case of
holonomic bases.

\subsubsection{D--torsions and d--curvatures:}

The anholonomic coefficients $W_{\ \alpha \beta }^{\gamma }$ and
N--elongated derivatives give nontrivial coefficients for the torsion
tensor, $T(\delta _{\gamma },\delta _{\beta })=T_{\ \beta \gamma }^{\alpha
}\delta _{\alpha },$ where
\begin{equation}
T_{\ \beta \gamma }^{\alpha }=\Gamma _{\ \beta \gamma }^{\alpha }-\Gamma _{\
\gamma \beta }^{\alpha }+w_{\ \beta \gamma }^{\alpha },  \label{3torsion}
\end{equation}
and for the curvature tensor, $R(\delta _{\tau },\delta _{\gamma })\delta
_{\beta }=R_{\beta \ \gamma \tau }^{\ \alpha }\delta _{\alpha },$ where
\begin{eqnarray}
R_{\beta \ \gamma \tau }^{\ \alpha } &=&\delta _{\tau }\Gamma _{\ \beta
\gamma }^{\alpha }-\delta _{\gamma }\Gamma _{\ \beta \tau }^{\alpha }
\nonumber \\
&&+\Gamma _{\ \beta \gamma }^{\varphi }\Gamma _{\ \varphi \tau }^{\alpha
}-\Gamma _{\ \beta \tau }^{\varphi }\Gamma _{\ \varphi \gamma }^{\alpha
}+\Gamma _{\ \beta \varphi }^{\alpha }w_{\ \gamma \tau }^{\varphi }.
\label{2curvature}
\end{eqnarray}
We emphasize that the torsion tensor on (pseudo) Riemannian spacetimes is
induced by anholonomic frames, whereas its components vanish with respect to
holonomic frames. All tensors are distinguished (d) by the N--connection
structure into irreducible h--v--components, and are called d--tensors. For
instance, the torsion, d--tensor has the following irreducible,
nonvanishing, h--v--components,\ $T_{\ \beta \gamma }^{\alpha }=\{T_{\
jk}^{i},C_{\ ja}^{i},S_{\ bc}^{a},T_{\ ij}^{a},T_{\ bi}^{a}\},$ where
\begin{eqnarray}
T_{.jk}^{i} &=&T_{jk}^{i}=L_{jk}^{i}-L_{kj}^{i},\quad
T_{ja}^{i}=C_{.ja}^{i},\quad T_{aj}^{i}=-C_{ja}^{i},  \nonumber \\
T_{.ja}^{i} &=&0,\quad T_{.bc}^{a}=S_{.bc}^{a}=C_{bc}^{a}-C_{cb}^{a},
\label{2dtors} \\
T_{.ij}^{a} &=&-\Omega _{ij}^{a},\quad T_{.bi}^{a}=\partial
_{b}N_{i}^{a}-L_{.bi}^{a},\quad T_{.ib}^{a}=-T_{.bi}^{a}  \nonumber
\end{eqnarray}
(the d--torsion is computed by substituting the h--v--components of the
canonical d--connection (\ref{5dcon}) and anholonomic coefficients (\ref
{anholonomy}) into the formula for the torsion coefficients (\ref{3torsion}%
)), where
\[
\Omega _{ij}^{a}=\delta _{j}N_{i}^{a}-\delta _{i}N_{j}^{a}
\]
is called the N--connection curvature (N--curvature).

The curvature d-tensor has the following irreducible, non-vanishing,
h--v--components\ $R_{\beta \ \gamma \tau }^{\ \alpha
}=%
\{R_{h.jk}^{.i},R_{b.jk}^{.a},P_{j.ka}^{.i},P_{b.ka}^{.c},S_{j.bc}^{.i},S_{b.cd}^{.a}\},
$\ where
\begin{eqnarray}
R_{h.jk}^{.i} &=&\delta _{k}L_{.hj}^{i}-\delta
_{j}L_{.hk}^{i}+L_{.hj}^{m}L_{mk}^{i}-L_{.hk}^{m}L_{mj}^{i}-C_{.ha}^{i}%
\Omega _{.jk}^{a},  \label{2dcurvatures} \\
R_{b.jk}^{.a} &=&\delta _{k}L_{.bj}^{a}-\delta
_{j}L_{.bk}^{a}+L_{.bj}^{c}L_{.ck}^{a}-L_{.bk}^{c}L_{.cj}^{a}-C_{.bc}^{a}%
\Omega _{.jk}^{c},  \nonumber \\
P_{j.ka}^{.i} &=&\partial _{a}L_{.jk}^{i}+C_{.jb}^{i}T_{.ka}^{b}-(\delta
_{k}C_{.ja}^{i}+L_{.lk}^{i}C_{.ja}^{l}-L_{.jk}^{l}C_{.la}^{i}-L_{.ak}^{c}C_{.jc}^{i}),
\nonumber \\
P_{b.ka}^{.c} &=&\partial _{a}L_{.bk}^{c}+C_{.bd}^{c}T_{.ka}^{d}-(\delta
_{k}C_{.ba}^{c}+L_{.dk}^{c\
}C_{.ba}^{d}-L_{.bk}^{d}C_{.da}^{c}-L_{.ak}^{d}C_{.bd}^{c}),  \nonumber \\
S_{j.bc}^{.i} &=&\partial _{c}C_{.jb}^{i}-\partial
_{b}C_{.jc}^{i}+C_{.jb}^{h}C_{.hc}^{i}-C_{.jc}^{h}C_{hb}^{i},  \nonumber \\
S_{b.cd}^{.a} &=&\partial _{d}C_{.bc}^{a}-\partial
_{c}C_{.bd}^{a}+C_{.bc}^{e}C_{.ed}^{a}-C_{.bd}^{e}C_{.ec}^{a}  \nonumber
\end{eqnarray}
(the d--curvature components are computed in a similar fashion by using the
formula for curvature coefficients (\ref{2curvature})).

\subsection{Einstein equations with holonomic--anholonomic va\-ri\-ab\-les}

In this subsection we write and analyze the Einstein equations on
5D (pseudo) Riemannian spacetimes provided with anholonomic frame
structures and associated N--connections.

\subsubsection{Einstein equations with matter sources}

The Ricci tensor $R_{\beta \gamma }=R_{\beta ~\gamma \alpha }^{~\alpha }$
has the d--components
\begin{eqnarray}
R_{ij} &=&R_{i.jk}^{.k},\quad R_{ia}=-^2P_{ia}=-P_{i.ka}^{.k},
\label{5dricci} \\
R_{ai} &=&^1P_{ai}=P_{a.ib}^{.b},\quad R_{ab}=S_{a.bc}^{.c}.  \nonumber
\end{eqnarray}
In general, since $^1P_{ai}\neq ~^2P_{ia}$, the Ricci d-tensor is
non-symmetric (this could be with respect to anholonomic frames of
reference). The scalar curvature of the metric d--connection, $%
\overleftarrow{R}=g^{\beta \gamma }R_{\beta \gamma },$ is computed
\begin{equation}
{\overleftarrow{R}}=G^{\alpha \beta }R_{\alpha \beta }=\widehat{R}+S,
\label{3dscalar}
\end{equation}
where $\widehat{R}=g^{ij}R_{ij}$ and $S=h^{ab}S_{ab}.$

By substituting (\ref{5dricci}) and (\ref{3dscalar}) into the 5D Einstein
equations
\begin{equation}
R_{\alpha \beta }-\frac 12g_{\alpha \beta }R=\kappa \Upsilon _{\alpha \beta
},  \label{15einstein}
\end{equation}
where $\kappa $ and $\Upsilon _{\alpha \beta }$ are respectively the
coupling constant and the energy--momentum tensor we obtain the
h-v-decomposition by N--connection of the Einstein equations
\begin{eqnarray}
R_{ij}-\frac 12\left( \widehat{R}+S\right) g_{ij} &=&\kappa \Upsilon _{ij},
\label{2einsteq2} \\
S_{ab}-\frac 12\left( \widehat{R}+S\right) h_{ab} &=&\kappa \Upsilon _{ab},
\nonumber \\
^1P_{ai}=\kappa \Upsilon _{ai},\ ^2P_{ia} &=&\kappa \Upsilon _{ia}.
\nonumber
\end{eqnarray}
The definition of matter sources with respect to anholonomic frames is
considered in Refs. \cite{10vf,10v}.

\subsubsection{5D vacuum Einstein equations}

The vacuum 5D, locally anisotropic gravitational field equations, in
invariant h-- v--components, are written
\begin{eqnarray}
R_{ij} &=&0,S_{ab}=0,  \label{1einsteq3} \\
^1P_{ai} &=&0,\ ^2P_{ia}=0.  \nonumber
\end{eqnarray}

The main `trick' of the anholonomic frames method for integrating the
Einstein equations in general relativity and various (super) string and
higher / lower dimension gravitational theories is to find the coefficients $%
N_{j}^{a}$ such that the block matrices $g_{ij}$ and $h_{ab}$ are
diagonalized \cite{10vf,10v,10v1}. This greatly simplifies computations. With
respect to such anholonomic frames the partial derivatives are N--elongated
(locally anisotropic).

\section{B: Proof of the Theorem 3}

We prove step by step the items of the Theorem 3.

The first statement with respect to the solution of
(\ref{4ricci1a}) is a connected with the well known result from 2D
(pseudo)\ Riemannian gravity that every 2D metric can be
redefined by using coordinate transforms into a conformally flat
one.

The equation (\ref{3ricci2a}) can be treated as a second order differential
equation on variable $v,$ with parameters $x^i,$ for the unknown function $%
h_5(x^i,v)$ if the value of $h_4(x^i,v)$ is given (or inversely
as a first order differential equation on variable $v,$ with
parameters $x^i,$ for the unknown function $h_4(x^i,v)$ if the
value of $h_5(x^i,v)$ is given). The formulas (\ref{3p1}) and
(\ref{2p2}) are consequences of integration on $v$ of the equation
(\ref{3ricci2a}) being considered also the degenerated cases when
$h_5^{*}=0$ or $h_4^{*}=0.$

Having defined the values $h_{4}$ and $h_{5},$ we can compute the values the
coefficients $\alpha _{i},\beta $ and $\gamma $ (\ref{3abc}) and find the
coefficients $w_{i}$ and $n_{i}$ The first set (\ref{5w}) for $w_{i}$ is a
solution of three independent first order algebraic equations (\ref{3ricci3a}%
) with known coefficients $\alpha _{i}$ and $\beta $. The second set of
solutions (\ref{2n}) for $n_{i}$ is found after two integrations on the
anisotropic variable $v$ of the independent equations (\ref{3ricci4a}) with
known $\gamma $ (the variables $x^{i}$ being considered as parameters). In
the formulas (\ref{2n}) we distinguish also the degenerated cases when $%
h_{5}^{\ast }=0$ or $h_{4}^{\ast }=0.$

Finally, we note that the formula (\ref{2confsol}) is a simple
algebraic consequence from (\ref{3confeq}).

The Theorem 3 has been proven.

\section{C: Proof of Theorem 5}

We emphasize the first two items:

\begin{itemize}
\item  The equation (\ref{ricci1const}) imposes a constraint on coefficients
of a diagonal 2D metric parametrized by coordinates $x^{2}=u$ and $%
x^{3}=\lambda .$ By coordinate transforms $x^{2,3}\rightarrow \widetilde{x}%
^{2,3}\left( u,\lambda \right) ,$ see for instance, \cite{10petrov} we can
reduce 2D every metric
\[
ds_{[2]}^{2}=g_{2}(u,\lambda )du^{2}+g_{3}(u,\lambda )d\lambda ^{2}
\]
to a conformally flat one
\[
ds_{[2]}^{2}=\varpi (\widetilde{x}^{2},\widetilde{x}^{3})\left[ d(\widetilde{%
x}^{2})^{2}+\epsilon d(\widetilde{x}^{3})^{2}\right] ,\epsilon =\pm 1.
\]
with conformal factor $\varpi
(\widetilde{x}^{2},\widetilde{x}^{3}),$ for which
(\ref{ricci1const}) transforms into (\ref{1auxr1}) with new 'dot'
and
'prime' derivatives $\varpi ^{\bullet }=\partial \varpi /\partial \widetilde{%
x}^{2}$ and $\varpi ^{^{\prime }}=\partial \varpi /\partial \widetilde{x}%
^{3}.$ It is not possible to find an explicit form of the general solution
of (\ref{1auxr1}). If we approximate, for instance, that $\varpi =\varpi
\left( \widetilde{x}^{2}\right) ,$ the equation
\[
\varpi \varpi ^{\bullet \bullet }-(\varpi ^{\bullet })^{2}=2\Lambda \varpi
^{3}
\]
has an exact solution (see 6.127 in \cite{10kamke}) which can be found from a
Bernulli equation
\[
(\varpi ^{\bullet })^{2}=\varpi ^{3}\left( C\varpi ^{-1}+4\Lambda \right)
,C=const,
\]
which allow us to find $\widetilde{x}^{2}(\varpi ),$ or, in non explicit
form $\varpi =\varpi \left( \widetilde{x}^{2}\right) .$ We can chose a such
solution as a background one and by using conformal factors $\Omega (%
\widetilde{x}^{2},\widetilde{x}^{3}),$ transforming $\varpi (\widetilde{x}%
^{2},\widetilde{x}^{3})$ into $\varpi \left( \widetilde{x}^{2}\right) $ we
can generate solutions of the 5D Einstein equations with anisotropic
cosmological constant by inducing second order anisotropy $\zeta _{i}.$ The
case when $\varpi =\varpi \left( \widetilde{x}^{3}\right) $ is to be
obtained in a similar manner by changing the 'dot' derivative into 'prime'
derivative.

\item  The equation (\ref{ricci2const}) does not admit $h_{5}^{\ast }=0$
because in this case we must have $h_{5}=0.$ For a given value of
$h_{5},$
introducing a new variable $\tau =h_{5}^{\ast }/2h_{5}$ we can transform (%
\ref{ricci2const}) into a first order nonlinear equation for $h_{4}$ (\ref
{auxr2p}),which can be transformed \cite{10kamke} to a Ricatti, then to a
Bernulli equation which admits exact solutions. We note that the holonomic
coordinates are considered as parameters. The inverse problem, to find $%
h_{5} $ for a given $h_{4}$ is more complex because is connected
with solution of a second order nonlinear differential equation
\begin{equation}
h_{5}^{\ast \ast }+\frac{(h_{5}^{\ast })^{2}}{2h_{5}}-\frac{h_{4}^{\ast }}{%
2h_{4}}h_{5}^{\ast }-2\Lambda h_{4}h_{5}=0,  \label{auxr2p}
\end{equation}
which can not integrated in general form. Nevertheless, a very
general class of solutions can be found explicitly if
$h_{4}^{\ast }=0,$ i. e. if $h_{4}$
depend only on holonomic coordinates. In this case the equation (\ref{auxr2p}%
) can be reduced to a Bernulli equation \cite{10kamke} which admits exact
solutions.

\item  The formulas (\ref{1aw}), (\ref{1nlambda}) and (\ref{1aconf4}) solving
respectively (\ref{ricci3const}), (\ref{ricci4const}) and (\ref{confeql})
are proven similarly as for the Theorem 3 with that difference that in the
presence of the cosmological term $h_{5}^{\ast }\neq 0,\beta \neq 0$ and, in
general, $w_{i}\neq 0.$
\end{itemize}

The Theorem 5 has been proven.

%%%%%%%%%%%%%%%%%%%%%%%%%%%%%%%%%%%%%%%%%%%%%%%%%%%%%%%%%%%%%%%%%%%%%%%%%%%%%

\chapter[Anisotropic Black Tori Solutions]
{Black Tori Solutions in Einstein and 5D Gravity }

{\bf Abstract}
\footnote{\copyright\  S. Vacaru, Black Tori Solutions in Einstein and 5D Gravity,
 hep-th/0110284}

The 'anholonomic frame' method \cite{11v,11v2,11vth} is applied for
constructing new classes of exact solutions of vacuum Einstein
equations with off--diagonal metrics in 4D and 5D gravity. We
examine several black tori solutions generated by anholonomic
transforms with non--trivial topology of the Schwarzschild metric,
which have a static toroidal horizon. We define ansatz and
parametrizations which contain warping factors, running constants
(in time and extra dimension coordinates) and effective nonlinear
gravitational polarizations. Such anisotropic vacuum toroidal
metrics, the first example was given in \cite{11v}, differ
substantially from the well known toroidal black holes
\cite{11lemos} which were constructed as non--vacuum solutions of
the Einstein--Maxwell gravity with cosmological constant.
Finally, we analyze two anisotropic 5D and 4D black tori solutions
with cosmological constant.

\section{Introduction}

Black hole - torus systems \cite{11putten} and toroidal black holes
\cite{11lemos,11v} became objects of astrophysical interest since it
was shown that they are inevitable outcome of complete
gravitational collapse of a massive star, cluster of stars, or
can be present in the center of galactic systems.

Black hole and black tori solutions appear naturally as exact
solutions in general relativity and extra dimension gravity
theories. Such solutions can be constructed in both
asymptotically flat spacetiems and in spacetimes with
cosmological constant,  posses a specific supersymmetry and could
be with toroidal, cylindrical or planar topology \cite{11lemos}.

String theory suggests that we may live in a fundamentally higher
dimensional spacetime \cite{11stringb}. The recent approaches  are
based on the assumption that our Universe is realized as a three
dimensional (in brief, 3D) brane, modelling a 4D
pseudo--Riemannian spacetime, embedded in the 5D anti--de Sitter
($AdS_{5}$) bulk spacetime. It was proposed  in the Rundall and
Sundrum (RS) papers \cite{11rs} that  such models could be with
relatively large extra dimension as a way to solve the hierarchy
problem in high energy physics.

In the present paper we explore possible black tori solutions in
5D and 4D gravity. We obtain a new class of exact solutions to
the 5D vacuum Einstein equations in the bulk, which have toroidal
horizons and are related via anholonomic transforms with toroidal
deformations of the Schwarzschild solutions. The solutions could
be with warped factors, running constants and anisotropic
gravitational polarizations. We than consider 4D black tori
solutions and generalize both 5D and 4D constructions for
spacetimes with cosmological constant.

We also discuss implications of existence of such anisotropic
 black tori solutions with non-trivial topology to the extra
 dimension gravity and general relativity theory. We prove that
 warped metrics can be obtained from vacuum 5D gravity and
 not only from a brane configurations with specific
 energy--momentum tensor.

We apply the  Salam, Strathee and Peracci \cite{11sal} idea on a
gauge field like status of the coefficients of off--diagonal
metrics in extra dimension gravity and develop it in a new
fashion by applying the method of anholonomic frames with
associated nonlinear connections  on 5D and 4D (pseudo) Riemannian
spaces \cite{11v,11v2,11vth}.

We use the term 'locally anisotropic' spacetime (or 'anisotropic' spacetime)
for a 5D (4D) pseudo-Riemannian spacetime provided with an anholonomic frame
structure with mixed holonomic and anholonomic variables. The anisotropy of
gravitational interactions is modelled by off--diagonal metrics, or,
equivalently, by theirs diagonalized analogs given with respect to
anholonomic frames.

The paper is organized as follow:\ In Sec. II we consider two
off--diagonal metric ansatz, construct the corresponding exact
solutions of 5D vacuum Einstein equations and illustrate the
possibility of extension by introducing matter fields and
 the cosmological constant term.  In Sec. III we construct two classes of 5D
anisotropic black tori solutions and consider  subclasses and
reparemetizations of such solutions in order to generate new ones.
Sec. IV is devoted to 4D black tori solutions. In Sec. V we extend
the approach for anisotropic 5D and 4D spacetimes with
cosmological constant  and give two examples of 5D and 4D
anisotropic black tori solution. Finally, in Sec. VI, we conclude
 and discuss the obtained results.

\newpage

\section{Off--Diagonal Metric Ansatz}

We introduce the basic denotations and two ansatz for off--diagonal 5D
metrics (see details in Refs. \cite{11v,11v2,11vth}) to be applied in definition
of anisotropic black tori solutions.

Let us consider a 5D pseudo--Riemannian spacetime provided with local
coordinates $u^{\alpha }=(x^{i},y^{4}=v,y^{5}),$ for indices like $%
i,j,k,..=1,2,3$ and $a,b,...=4,5.$ The $x^{i}$--coordinates are called
holonomic and $y^{a}$--coordinates are called anholonomic (anisotrop\-ic);
they are given respectively with respect to some holonomic and anholonomic
subframes (see the formulae (\ref{1dder1}) and (\ref{3anh})). Every coordinate
$x^{i}$ or $y^{a}$ could be a time \ like, 3D space, or the 5th (extra
dimensional) coordinate; we shall fix on necessity different
parametrizations.

We investigate two classes of 5D metrics:

The first type of metrics are given by a line element
\begin{equation}
ds^{2}=g_{\alpha \beta }\left( x^{i},v\right) du^{\alpha }du^{\beta }
\label{3metric}
\end{equation}
with the metric coefficients $g_{\alpha \beta }$ parametrized with respect
to the coordinate co--frame $du^{\alpha },$ being dual to $\partial _{\alpha
}=\partial /\partial u^{\alpha },$ $\ $by an off--diagonal matrix (ansatz)

{%%\footnotesize
\begin{equation}
\left[
\begin{array}{ccccc}
g_{1}+w_{1}^{\ 2}h_{4}+n_{1}^{\ 2}h_{5} & w_{1}w_{2}h_{4}+n_{1}n_{2}h_{5} &
w_{1}w_{3}h_{4}+n_{1}n_{3}h_{5} & w_{1}h_{4} & n_{1}h_{5} \\
w_{1}w_{2}h_{4}+n_{1}n_{2}h_{5} & g_{2}+w_{2}^{\ 2}h_{4}+n_{2}^{\ 2}h_{5} &
w_{2}w_{3}h_{4}+n_{2}n_{3}h_{5} & w_{2}h_{4} & n_{2}h_{5} \\
w_{1}w_{3}h_{4}+n_{1}n_{3}h_{5} & w_{2}w_{3}h_{4}+n_{2}n_{3}h_{5} &
g_{3}+w_{3}^{\ 2}h_{4}+n_{3}^{\ 2}h_{5} & w_{3}h_{4} & n_{3}h_{5} \\
w_{1}h_{4} & w_{2}h_{4} & w_{3}h_{4} & h_{4} & 0 \\
n_{1}h_{5} & n_{2}h_{5} & n_{3}h_{5} & 0 & h_{5}
\end{array}
\right] ,  \label{5ansatz}
\end{equation}
} where the coefficients are some necessary smoothly class functions of
type:
\begin{eqnarray}
g_{1} &=&\pm 1,g_{2,3}=g_{2,3}(x^{2},x^{3}),h_{4,5}=h_{4,5}(x^{i},v),
\nonumber \\
w_{i} &=&w_{i}(x^{i},v),n_{i}=n_{i}(x^{i},v).  \nonumber
\end{eqnarray}

The second type of metrics are given by a line element (with a
conformal factor $\Omega (x^{i},v)$ and additional deformations
of the metric via coefficients $\zeta _{\hat{\imath}}(x^{i},v),$
indices with 'hat' take values like $\hat{{i}}=1,2,3,5))$ written
as
\begin{equation}
ds^{2}=\Omega ^{2}(x^{i},v)\hat{{g}}_{\alpha \beta }\left( x^{i},v\right)
du^{\alpha }du^{\beta },  \label{5cmetric}
\end{equation}
were the coefficients $\hat{{g}}_{\alpha \beta }$ are parametrized by the
ansatz {\scriptsize
\begin{equation}
\left[
\begin{array}{ccccc}
g_{1}+(w_{1}^{\ 2}+\zeta _{1}^{\ 2})h_{4}+n_{1}^{\ 2}h_{5} &
(w_{1}w_{2}+\zeta _{1}\zeta _{2})h_{4}+n_{1}n_{2}h_{5} & (w_{1}w_{3}+\zeta
_{1}\zeta _{3})h_{4}+n_{1}n_{3}h_{5} & (w_{1}+\zeta _{1})h_{4} & n_{1}h_{5}
\\
(w_{1}w_{2}+\zeta _{1}\zeta _{2})h_{4}+n_{1}n_{2}h_{5} & g_{2}+(w_{2}^{\
2}+\zeta _{2}^{\ 2})h_{4}+n_{2}^{\ 2}h_{5} & (w_{2}w_{3}+\zeta _{2}\zeta
_{3})h_{4}+n_{2}n_{3}h_{5} & (w_{2}+\zeta _{2})h_{4} & n_{2}h_{5} \\
(w_{1}w_{3}+\zeta _{1}\zeta _{3})h_{4}+n_{1}n_{3}h_{5} & (w_{2}w_{3}+\zeta
_{2}\zeta _{3})h_{4}+n_{2}n_{3}h_{5} & g_{3}+(w_{3}^{\ 2}+\zeta _{3}^{\
2})h_{4}+n_{3}^{\ 2}h_{5} & (w_{3}+\zeta _{3})h_{4} & n_{3}h_{5} \\
(w_{1}+\zeta _{1})h_{4} & (w_{2}+\zeta _{2})h_{4} & (w_{3}+\zeta _{3})h_{4}
& h_{4} & 0 \\
n_{1}h_{5} & n_{2}h_{5} & n_{3}h_{5} & 0 & h_{5}+\zeta _{5}h_{4}
\end{array}
\right]  \label{5ansatzc}
\end{equation}
}

For trivial values $\Omega =1$ and $\zeta _{\hat{\imath}}=0,$ the  line
interval (\ref{5cmetric}) transforms into (\ref{3metric}).

The quadratic line element (\ref{3metric}) with metric coefficients (\ref
{5ansatz}) can be diagonalized,
\begin{equation}
\delta
s^{2}=[g_{1}(dx^{1})^{2}+g_{2}(dx^{2})^{2}+g_{3}(dx^{3})^{2}+h_{4}(\delta
v)^{2}+h_{5}(\delta y^{5})^{2}],  \label{6dmetric}
\end{equation}
with respect to the anholonomic co--frame $\left( dx^{i},\delta v,\delta
y^{5}\right) ,$ where
\begin{equation}
\delta v=dv+w_{i}dx^{i}\mbox{ and }\delta y^{5}=dy^{5}+n_{i}dx^{i}
\label{2ddif1}
\end{equation}
which is dual to the frame $\left( \delta _{i},\partial _{4},\partial
_{5}\right) ,$ where
\begin{equation}
\delta _{i}=\partial _{i}+w_{i}\partial _{4}+n_{i}\partial _{5}.
\label{2dder1}
\end{equation}
The bases (\ref{2ddif1}) and (\ref{2dder1}) are considered to satisfy some
anholonomic relations of type
\begin{equation}
\delta _{i}\delta _{j}-\delta _{j}\delta _{i}=W_{ij}^{k}\delta _{k}
\label{3anh}
\end{equation}
for some non--trivial values of anholonomy coefficients $W_{ij}^{k}.$ We
obtain a holonomic (coordinate) base if the coefficients $W_{ij}^{k}$ vanish.

The quadratic line element (\ref{5cmetric}) with metric coefficients (\ref
{5ansatzc}) can be also diagonalized,
\begin{equation}
\delta s^{2}=\Omega
^{2}(x^{i},v)[g_{1}(dx^{1})^{2}+g_{2}(dx^{2})^{2}+g_{3}(dx^{3})^{2}+h_{4}(%
\hat{{\delta }}v)^{2}+h_{5}(\delta y^{5})^{2}],  \label{3cdmetric}
\end{equation}
but with respect to another anholonomic co--frame $\left( dx^{i},\hat{{%
\delta }}v,\delta y^{5}\right) ,$ with
\begin{equation}
\delta v=dv+(w_{i}+\zeta _{i})dx^{i}+\zeta _{5}\delta y^{5}\mbox{ and }%
\delta y^{5}=dy^{5}+n_{i}dx^{i}  \label{21ddif2}
\end{equation}
which is dual to the frame $\left( \hat{{\delta }}_{i},\partial _{4},\hat{{%
\partial }}_{5}\right) ,$ where
\begin{equation}
\hat{{\delta }}_{i}=\partial _{i}-(w_{i}+\zeta _{i})\partial
_{4}+n_{i}\partial _{5},\hat{{\partial }}_{5}=\partial _{5}-\zeta
_{5}\partial _{4}.  \label{21dder2}
\end{equation}

The nontrivial components of the 5D Ricci tensor, $R_{~\alpha }^{\beta },$
for the metric (\ref{6dmetric}) given with respect to anholonomic frames (\ref
{2ddif1}) and (\ref{2dder1}) are
\begin{eqnarray}
R_{2}^{2} &=&R_{3}^{3}=-\frac{1}{2g_{2}g_{3}}[g_{3}^{\bullet \bullet }-\frac{%
g_{2}^{\bullet }g_{3}^{\bullet }}{2g_{2}}-\frac{(g_{3}^{\bullet })^{2}}{%
2g_{3}}+g_{2}^{^{\prime \prime }}-\frac{g_{2}^{^{\prime }}g_{3}^{^{\prime }}%
}{2g_{3}}-\frac{(g_{2}^{^{\prime }})^{2}}{2g_{2}}],  \label{5ricci1a} \\
R_{4}^{4} &=&R_{5}^{5}=-\frac{\beta }{2h_{4}h_{5}},  \label{4ricci2a} \\
R_{4i} &=&-w_{i}\frac{\beta }{2h_{5}}-\frac{\alpha _{i}}{2h_{5}},
\label{4ricci3a} \\
R_{5i} &=&-\frac{h_{5}}{2h_{4}}\left[ n_{i}^{\ast \ast }+\gamma n_{i}^{\ast }%
\right]  \label{14ricci4a}
\end{eqnarray}
where
\begin{equation}
\alpha _{i}=\partial _{i}{h_{5}^{\ast }}-h_{5}^{\ast }\partial _{i}\ln \sqrt{%
|h_{4}h_{5}|},\beta =h_{5}^{\ast \ast }-h_{5}^{\ast }[\ln \sqrt{|h_{4}h_{5}|}%
]^{\ast },\gamma =3h_{5}^{\ast }/2h_{5}-h_{4}^{\ast }/h_{4}.  \label{4abc}
\end{equation}

For simplicity, the partial derivatives are denoted like $a^{\times
}=\partial a/\partial x^{1},a^{\bullet }=\partial a/\partial
x^{2},a^{^{\prime }}=\partial a/\partial x^{3},a^{\ast }=\partial a/\partial
v.$\bigskip

We obtain the same values of the Ricci tensor for the second ansatz (\ref
{3cdmetric}) if there are satisfied the conditions
\begin{equation}
\hat{{\delta }}_{i}h_{4}=0\mbox{\ and\  }\hat{{\delta }}_{i}\Omega =0
\label{3conf1}
\end{equation}
and the values $\zeta _{\hat{{i}}}=\left( \zeta _{{i}},\zeta _{{5}}=0\right)
$ are found as to be a unique solution of (\ref{3conf1}); for instance, if
\begin{equation}
\Omega ^{q_{1}/q_{2}}=h_{4}~(q_{1}\mbox{ and }q_{2}\mbox{ are integers}),
\label{3confq}
\end{equation}
the coefficients $\zeta _{{i}}$ must solve the equations \
\begin{equation}
\partial _{i}\Omega -(w_{i}+\zeta _{{i}})\Omega ^{\ast }=0.  \label{4confeq}
\end{equation}

\bigskip The system of 5D vacuum Einstein equations, $R_{~\alpha }^{\beta
}=0,$ reduces to a system of nonlinear equations with separation of
variables,
\[
R_{2}^{2}=0,~R_{4}^{4}=0,~R_{4i}=0,R_{5i}=0,
\]
which together with (\ref{4confeq}) can be solved in general form \cite{11vth}:
For any given values of $g_{2}$ (or $g_{3})$, $h_{4}$ (or $h_{5})$ and $%
\Omega ,$ and stated boundary conditions we can define \ consequently the
set of metric coefficients $g_{3}$(or $g_{2})$, $h_{4}$ (or $%
h_{4}),w_{i},n_{i}$ and $\zeta _{{i}}.$

The introduced ansatz can be used also for constructing solutions of 5D and
4D Einstein equations with nontrivial energy-momentum tensor
\[
R_{\alpha \beta }-\frac{1}{2}g_{\alpha \beta }R=\kappa \Upsilon _{\alpha
\beta }.
\]

The non--trivial diagonal components of the Einstein tensor, $G_{\beta
}^{\alpha }=R_{\beta }^{\alpha }-\frac{1}{2}R\delta _{\beta }^{\alpha },$
for the metric (\ref{6dmetric}), given with respect to anholonomic frames,
are
\begin{equation}
G_{1}^{1}=-\left( R_{2}^{2}+S_{4}^{4}\right)
,G_{2}^{2}=G_{3}^{3}=-S_{4}^{4},G_{4}^{4}=G_{5}^{5}=-R_{2}^{2}.
\label{3einstdiag}
\end{equation}
So, we can extend the system of 5D vacuum Einstein equations  by introducing
matter fields for which the energy--momentum tensor $\Upsilon _{\alpha \beta
}$ given with respect to anholonomic frames satisfy the conditions
\begin{equation}
\Upsilon _{1}^{1}=\Upsilon _{2}^{2}+\Upsilon _{4}^{4},\Upsilon
_{2}^{2}=\Upsilon _{3}^{3},\Upsilon _{4}^{4}=\Upsilon _{5}^{5}.
\label{3emcond}
\end{equation}

We note that, in general, the tensor $\Upsilon _{\alpha \beta }$
may be not symmetric because with respect to anholonomic frames
there are imposed constraints which makes non symmetric the Ricci
and Einstein tensors \ \cite{11v,11v2,11vth}.

In the simplest case we can consider a ''vacuum'' source induced by a
non--vanishing 4D cosmological constant, $\Lambda .$ In order to satisfy the
conditions (\ref{3emcond}) the source induced by $\Lambda $ should be in the
form $\kappa \Upsilon _{\alpha \beta }=(2\Lambda g_{11},\Lambda g_{%
\underline{\alpha }\underline{\beta }}),$ where underlined indices $%
\underline{\alpha },\underline{\beta },...$ run 4D values $2,3,4,5.$ We note
that in 4D anholonomic gravity the source $\kappa \Upsilon _{\underline{\alpha }%
\underline{\beta }}=\Lambda g_{\underline{\alpha
}\underline{\beta }}$ satisfies the equalities $\Upsilon
_{2}^{2}=\Upsilon _{3}^{3}=\Upsilon _{4}^{4}=\Upsilon _{5}^{5}.$

By straightforward computations we obtain that the nontrivial components of
the 5D Einstein equations with anisotropic cosmological constant, $%
R_{11}=2\Lambda g_{11}$ and $R_{\underline{\alpha }\underline{\beta }%
}=\Lambda g_{\underline{\alpha }\underline{\beta }},$ for the
ansatz (\ref {5ansatzc}) and anholonomic metric (\ref{3cdmetric})
given with respect to anholonomic frames (\ref{21ddif2}) and
(\ref{21dder2}), are written in a form with separated  variables:
\begin{eqnarray}
g_{3}^{\bullet \bullet }-\frac{g_{2}^{\bullet }g_{3}^{\bullet }}{2g_{2}}-%
\frac{(g_{3}^{\bullet })^{2}}{2g_{3}}+g_{2}^{^{\prime \prime }}-\frac{%
g_{2}^{^{\prime }}g_{3}^{^{\prime }}}{2g_{3}}-\frac{(g_{2}^{^{\prime }})^{2}%
}{2g_{2}} &=&2\Lambda g_{2}g_{3},  \label{ein1} \\
h_{5}^{\ast \ast }-h_{5}^{\ast }[\ln \sqrt{|h_{4}h_{5}|}]^{\ast }
&=&2\Lambda h_{4}h_{5},  \label{ein2} \\
w_{i}\beta +\alpha _{i} &=&0,  \label{ein3} \\
n_{i}^{\ast \ast }+\gamma n_{i}^{\ast } &=&0,  \label{ein4} \\
\partial _{i}\Omega -(w_{i}+\zeta _{{i}})\Omega ^{\ast } &=&0.  \label{einc}
\end{eqnarray}
where
\begin{equation}
\alpha _{i}=\partial _{i}{h_{5}^{\ast }}-h_{5}^{\ast }\partial _{i}\ln \sqrt{%
|h_{4}h_{5}|},\beta =2\Lambda h_{4}h_{5},\gamma =3h_{5}^{\ast
}/2h_{5}-h_{4}^{\ast }/h_{4}.  \label{abcc}
\end{equation}
In the vacuum case (with $\Lambda =0)$ these equations are compatible if $%
\beta =\alpha _{i}=0$ which results that $w_{i}\left(
x^{i},v\right) $ could be arbitrary functions; this reflects a
freedom in definition of the
holonomic coordinates. For simplicity, for vacuum solutions we shall put $%
w_{i}=0.$ Finally, we remark that we can ''select'' 4D Einstein solutions
from an ansatz (\ref{5ansatz}) or (\ref{5ansatzc}) by considering that the
metric coefficients do not depend on variable $x^{1},$ which mean that in
the system of equations (\ref{ein1})--(\ref{einc}) we have to deal with 4D
values $w_{\underline{i}}\left( x^{\underline{k}},v\right) ,n_{\underline{i}%
}\left( x^{\underline{k}},v\right) ,\zeta _{{i}}\left( x^{\underline{k}%
},v\right) ,$ and $h_{4}\left( x^{\underline{k}},v\right) ,h_{5}\left( x^{%
\underline{k}},v\right) ,\Omega \left( x^{\underline{k}},v\right) .$

\section{5D Black Tori}

Our goal is to apply the anholonomic frame method as to construct such exact
solutions of vacuum (and with cosmological constant) 5D Einstein equations
as they have a static toroidal horizon for a metric ansatz (\ref{5ansatz}) or
(\ref{5ansatzc}) which can be diagonalized with respect to some well defined
anholonomic frames. Such solutions are defined as some anholonomic
transforms of the Schwarzschild solution to a toroidal configuration with
non--trivial topology. \ In general form, they could be defined with warped
factors, running constants (in time and extra dimension coordinate) and
nonlinear polarizations.

\subsection{Toroidal deformations of the Schwarzschild metric}

Let us consider the system of {\it \ isotropic spherical coordinates} $(\rho
,\theta ,\varphi ),$ \thinspace where the isotropic radial coordinate $\rho $
is related with the usual radial coordinate $r$ via the relation $r=\rho
\left( 1+r_{g}/4\rho \right) ^{2}$ for $r_{g}=2G_{[4]}m_{0}/c^{2}$ being the
4D gravitational radius of a point particle of mass $m_{0},$ $%
G_{[4]}=1/M_{P[4]}^{2}$ is the 4D Newton constant expressed via Plank mass $%
M_{P[4]}$ (following modern string/brane theories, $M_{P[4]}$ can be
considered as a value induced from extra dimensions). We put the light speed
constant $c=1.$ This system of coordinates is considered for the so--called
isotropic representation of the Schwarzschild solution \cite{11ll}
\begin{equation}
ds^{2}=\left( \frac{\widehat{\rho }-1}{\widehat{\rho }+1}\right)
^{2}dt^{2}-\rho _{g}^{2}\left( \frac{\widehat{\rho }+1}{\widehat{\rho }}%
\right) ^{4}\left( d\widehat{\rho }^{2}+\widehat{\rho }^{2}d\theta ^{2}+%
\widehat{\rho }^{2}\sin ^{2}\theta d\varphi ^{2}\right) ,  \label{2schw}
\end{equation}
where, for our further considerations, we re--scaled the isotropic radial
coordinate as $\widehat{\rho }=\rho /\rho _{g},$ with $\rho _{g}=r_{g}/4.$
The metric (\ref{2schw}) is a vacuum static solution of 4D Einstein equations
with spherical symmetry describing the gravitational field of a point
particle of mass $m_{0}.$ It has a singularity for $r=0$ and a spherical
horizon for $r=r_{g},$ or, in re--scaled isotropic coordinates, for $%
\widehat{\rho }=1.$ We emphasize that this solution is parametrized by a
diagonal metric given with respect to holonomic coordinate frames.

We also introduce the {\it \ toroidal coordinates} (in our case considered
as alternatives to the isotropic radial coordinates) \cite{11korn} $(\sigma
,\tau ,\varphi ),$ running values $-\pi \leq \sigma <\pi ,0\leq \tau \leq
\infty ,0\leq \varphi <2\pi ,$ which are related with the isotropic 3D
Cartezian coordinates via transforms
\begin{equation}
\tilde{x}=\frac{\widetilde{\rho }\sinh \tau }{\cosh \tau -\cos \sigma }\cos
\varphi ,\tilde{y}=\frac{\widetilde{\rho }\sinh \tau }{\cosh \tau -\cos
\sigma }\sin \varphi ,\tilde{z}=\frac{\widetilde{\rho }\sinh \sigma }{\cosh
\tau -\cos \sigma }  \label{12rec}
\end{equation}
and define a toroidal hypersurface
\[
\left( \sqrt{\tilde{x}^{2}+\tilde{y}^{2}}-\widetilde{\rho }\frac{\cosh \tau
}{\sinh \tau }\right) ^{2}+\tilde{z}^{2}=\frac{\widetilde{\rho }^{2}}{\sinh
^{2}\tau }.
\]
The 3D metric on a such toroidal hypersurface is
\[
ds_{(3D)}^{2}=g_{\sigma \sigma }d\sigma ^{2}+g_{\tau \tau }d\tau
^{2}+g_{\varphi \varphi }d\varphi ^{2},
\]
where
\[
g_{\sigma \sigma }=g_{\tau \tau }=\frac{\widetilde{\rho }^{2}}{\left( \cosh
\tau -\cos \sigma \right) ^{2}},g_{\varphi \varphi }=\frac{\widetilde{\rho }%
^{2}\sinh ^{2}\tau }{\left( \cosh \tau -\cos \sigma \right) ^{2}}.
\]

We can relate the toroidal coordinates $\left( \sigma ,\tau ,\varphi \right)
$ from (\ref{1rec}) with the isotropic radial coordinates $\left( \widehat{%
\rho },\theta ,\varphi \right) $, scaled by the constant $\rho _{g},$ from (%
\ref{2schw}) as
\[
\widetilde{\rho }=1,\sinh ^{-1}\tau =\widehat{\rho }
\]
and transform the Schwarzschild solution into a new metric with
toroidal coordinates \ by changing the 3D radial line element
into the toroidal one and stating the $tt$--coefficient of the
metric to have a toroidal horizon. The resulting metric is
\begin{equation}
ds_{(S)}^{2}=\left( \frac{\sinh \tau -1}{\sinh \tau +1}\right)
^{2}dt^{2}-\rho _{g}^{2}\frac{\left( \sinh \tau +1\right) ^{4}}{\left( \cosh
\tau -\cos \sigma \right) ^{2}}\left( d\sigma ^{2}+d\tau ^{2}+\sinh ^{2}\tau
d\varphi ^{2}\right) ],  \label{1schtor}
\end{equation}
Such deformed Schwarzschild like toroidal metric is not an exact
solution of the vacuum Einstein equations, but at long radial
distances it transform into usual Schwarzschild solution with the
3D line element parametrized by toroidal coordinates.

For our further considerations we introduce two Classes (A and B)
of 4D auxiliary pseudo--Riemannian metrics, also given in toroidal
coordinates, being some conformal transforms of (\ref{1schtor}),
like
\[
ds_{(S)}^{2}=\Omega _{A,B}\left( \sigma ,\tau \right) ds_{(A,B)}^{2}
\]
but which are not supposed to be solutions of the Einstein equations:

\begin{itemize}
\item  Metric of Class A:
\begin{equation}
ds_{(A)}^{2}=-d\sigma ^{2}-d\tau ^{2}+a(\tau )d\varphi ^{2}+b(\sigma ,\tau
)dt^{2}],  \label{2auxm1}
\end{equation}
where
\[
a(\tau )=-\sinh ^{2}\tau \mbox{ and }b(\sigma ,\tau )=-\frac{\left( \sinh
\tau -1\right) ^{2}\left( \cosh \tau -\cos \sigma \right) ^{2}}{\rho
_{g}^{2}\left( \sinh \tau +1\right) ^{6}},
\]
which results in the metric (\ref{1schtor}) by multiplication on the
conformal factor
\begin{equation}
\Omega _{A}\left( \sigma ,\tau \right) =\rho _{g}^{2}\frac{\left( \sinh \tau
+1\right) ^{4}}{\left( \cosh \tau -\cos \sigma \right) ^{2}}.  \label{2auxm1c}
\end{equation}

\item  Metric of Class B:
\begin{equation}
ds_{(B)}^{2}=g(\tau )\left( d\sigma ^{2}+d\tau ^{2}\right) -d\varphi
^{2}+f(\sigma ,\tau )dt^{2},  \label{2auxm2}
\end{equation}
where
\[
g(\tau )=-\sinh ^{-2}\tau \mbox{ and }f(\sigma ,\tau )=\rho _{g}^{2}\left(
\frac{\sinh ^{2}\tau -1}{\cosh \tau -\cos \sigma }\right) ^{2},
\]
which results in the metric (\ref{1schtor}) by multiplication on the
conformal factor
\[
\Omega _{B}\left( \sigma ,\tau \right) =\rho _{g}^{-2}\frac{\left( \cosh
\tau -\cos \sigma \right) ^{2}}{\left( \sinh \tau +1\right) ^{2}}.
\]
\end{itemize}

We shall use the metrics (\ref{1schtor}), (\ref{2auxm1}) and
(\ref{2auxm2}) in order to  generate exact solutions of the
Einstein equations with toroidal horizons and anisotropic
polarizations and running of constants by performing
corresponding anholonomic transforms as the solutions will have a
horizon parametrized by a torus hypersurface  and gravitational
(extra dimensional, or nonlinear 4D)
renormalizations  of the constant $\rho _{g}$ of the Schwarzschild solution, $%
\rho _{g}\rightarrow \overline{\rho }_{g}=\omega \rho _{g},$
where the dependence of the function $\omega $ on some holonomic
or anholonomic coordinates will depend on the type of anisotropy.
For some solutions we shall treat $\omega $ as a factor modelling
running of the gravitational constant,
induced, induced from extra dimension, in another cases we will consider $%
\omega $ as a nonlinear gravitational polarization which models
some anisotropic distributions of masses and matter fields and/or
anholonomic vacuum gravitational interactions.

\subsection{Toroidal 5D metrics of Class A}

In this subsection we consider four classes of 5D vacuum
solutions which are related to the metric of Class A
(\ref{2auxm1}) and to the toroidally deformed  Schwarzschild metric
 (\ref{1schtor}).

Let us parametrize the 5D coordinates as $\left( x^{1}=\chi ,x^{2}=\sigma
,x^{3}=\tau ,y^{4}=v,y^{5}=p\right) ,$ where the solutions with the
so--called $\varphi $--anisotropy will be constructed for $\left( v=\varphi
,p=t\right) $ and the solutions with $t$--anisotropy will be stated for $%
\left( v=t,p=\varphi \right) $ (in brief, we write \ respectively, $\varphi $%
--solutions and $t$--solutions).

\subsubsection{Class A of vacuum solutions with ansatz (\ref{5ansatz}):}

We take an off--diagonal metric ansatz of type (\ref{5ansatz}) (equivalently,
(\ref{3metric})) by representing
\[
g_{1}=\pm 1,g_{2}=-1,g_{3}=-1,h_{4}=\eta _{4}(\sigma ,\tau
,v)h_{4(0)}(\sigma ,\tau )\mbox{
and }h_{5}=\eta _{5}(\sigma ,\tau ,v)h_{5(0)}(\sigma ,\tau ),
\]
where $\eta _{4,5}(\sigma ,\tau ,v)$ are corresponding ''gravitational
renormalizations'' of the metric coefficients $h_{4,5(0)}(\sigma ,\tau ).$
For $\varphi $--solutions we state $h_{4(0)}=a(\tau )$ and $%
h_{5(0)}=b(\sigma ,\tau )$ (inversely, for \ $t$--solutions, $%
h_{4(0)}=b(\sigma ,\tau )$ and $h_{5(0)}=a(\sigma ,\tau )).$ \

Next we consider a renormalized gravitational 'constant' $\overline{\rho }%
_{g}=\omega \rho _{g},$ were for $\varphi $--solutions\ the receptivity $%
\omega =\omega \left( \sigma ,\tau ,v\right) $ is included in the
gravitational polarization $\eta _{5}$ as $\eta _{5}=\left[ \omega \left(
\sigma ,\tau ,\varphi \right) \right] ^{-2},$ or for $t$--solutions is
included in $\eta _{4},$ when $\eta _{4}=\left[ \omega \left( \sigma ,\tau
,t\right) \right] ^{-2}.$ We can construct an exact solution of the 5D
vacuum Einstein equations if, for explicit dependencies on anisotropic
coordinate, the metric coefficients $h_{4}$ and $h_{5}$ are related by
 the equation (\ref{ein2}), which in its turn imposes a corresponding relation
between $\eta _{4}$ and $\eta _{5},$%
\[
\eta _{4}h_{4(0)}=h_{(0)}^{2}h_{5(0)}\left[ \left( \sqrt{|\eta _{5}|}\right)
^{\ast }\right] ^{2},~h_{(0)}^{2}=const.
\]
In result, we express the polarizations $\eta _{4}$ and $\eta _{5}$ via the
value of receptivity $\omega ,$
\begin{equation}
\eta _{4}\left( \chi ,\sigma ,\tau ,\varphi \right) =h_{(0)}^{2}\frac{%
b(\sigma ,\tau )}{a(\tau )}\left\{ \left[ \omega ^{-1}\left( \chi ,\sigma
,\tau ,\varphi \right) \right] ^{\ast }\right\} ^{2},\eta _{5}\left( \chi
,\sigma ,\tau ,\varphi \right) =\omega ^{-2}\left( \chi ,\sigma ,\tau
,\varphi \right) ,  \label{2etap}
\end{equation}
for $\varphi $--solutions , and
\begin{equation}
\eta _{4}\left( \chi ,\sigma ,\tau ,t\right) =\omega ^{-2}\left( \chi
,\sigma ,\tau ,t\right) ,\eta _{5}\left( \chi ,\sigma ,\tau ,t\right)
=h_{(0)}^{-2}\frac{b(\sigma ,\tau )}{a(\tau )}\left[ \int dt\omega
^{-1}\left( \chi ,\sigma ,\tau ,t\right) \right] ^{2},  \label{2etat}
\end{equation}
for $t$--solutions, where $a(\tau )$ and $b(\sigma ,\tau )$ are those from (%
\ref{2auxm1}).

For vacuum configurations, following (\ref{ein3}), we put $w_{i}=0.$ The
next step is to find the values of $n_{i}$ by introducing $h_{4}=\eta
_{4}h_{4(0)}$ and $h_{5}=\eta _{5}h_{5(0)}$ into the formula \ (\ref{ein4}),
which, for convenience, is expressed via general coefficients $\eta _{4}$
and $\eta _{5}.$ After two integrations on variable $v,$ we obtain the exact
solution
\begin{eqnarray}
n_{k}(\sigma ,\tau ,v) &=&n_{k[1]}\left( \sigma ,\tau \right)
+n_{k[2]}\left( \sigma ,\tau \right) \int [\eta _{4}/(\sqrt{|\eta _{5}|}%
)^{3}]dv,~\eta _{5}^{\ast }\neq 0;  \label{2nel} \\
&=&n_{k[1]}\left( \sigma ,\tau \right) +n_{k[2]}\left( \sigma ,\tau \right)
\int \eta _{4}dv,\qquad ~\eta _{5}^{\ast }=0;  \nonumber \\
&=&n_{k[1]}\left( \sigma ,\tau \right) +n_{k[2]}\left( \sigma ,\tau \right)
\int [1/(\sqrt{|\eta _{5}|})^{3}]dv,~\eta _{4}^{\ast }=0,  \nonumber
\end{eqnarray}
with the functions $n_{k[2]}\left( \sigma ,\tau \right) $ are defined as to
contain the values $h_{(0)}^{2},$ $a(\tau )$ and $b(\sigma ,\tau ).$

\bigskip By introducing the formulas (\ref{2etap}) for $\varphi $--solutions
(or (\ref{2etat}) for $t$--solutions) and fixing some boundary condition, in
order to state the values of coefficients $n_{k[1,2]}\left( \sigma ,\tau
\right) $ we can express the ansatz components $n_{k}\left( \sigma ,\tau
,\varphi \right) $ as integrals of some functions of $\omega \left( \sigma
,\tau ,\varphi \right) $ and $\partial _{\varphi }\omega \left( \sigma ,\tau
,\varphi \right) $ (or, we can express the ansatz components $n_{k}\left(
\sigma ,\tau ,t\right) $ as integrals of some functions of $\omega \left(
\sigma ,\tau ,t\right) $ and $\partial _{t}\omega \left( \sigma ,\tau
,t\right) ).$ We do not present an explicit form of such formulas because
they depend on the type of receptivity $\omega =\omega \left( \sigma ,\tau
,v\right) ,$ which must be defined experimentally, or from some quantum
models of gravity in the quasi classical limit. We preserved a general
dependence on coordinates $\left( \sigma ,\tau \right) $ which reflect the
fact that there is a freedom in fixing holonomic coordinates (for instance,
on a toroidal hypersurface and its extensions to 4D and 5D spacetimes). \
For simplicity, we write that $n_{i}$ are some functionals of $\{\sigma
,\tau ,\omega \left( \sigma ,\tau ,v\right) ,\omega ^{\ast }\left( \sigma
,\tau ,v\right) \}$
\[
n_{i}\{\sigma ,\tau ,\omega ,\omega ^{\ast }\}=n_{i}\{\sigma ,\tau ,\omega
\left( \sigma ,\tau ,v\right) ,\omega ^{\ast }\left( \sigma ,\tau ,v\right)
\}.
\]

In conclusion, we constructed two exact solutions of the 5D vacuum Einstein
equations, defined by the ansatz (\ref{5ansatz}) with coordinates and
coefficients stated by the data:
\begin{eqnarray}
\mbox{$\varphi$--solutions} &:&(x^{1}=\chi ,x^{2}=\sigma ,x^{3}=\tau
,y^{4}=v=\varphi ,y^{5}=p=t),g_{1}=\pm 1,  \nonumber \\
g_{2} &=&-1,g_{3}=-1,h_{4(0)}=a(\tau ),h_{5(0)}=b(\sigma ,\tau ),%
\mbox{see
(\ref{2auxm1})};  \nonumber \\
h_{4} &=&\eta _{4}(\sigma ,\tau ,\varphi )h_{4(0)},h_{5}=\eta _{5}(\sigma
,\tau ,\varphi )h_{5(0)},  \nonumber \\
\eta _{4} &=&h_{(0)}^{2}\frac{b(\sigma ,\tau )}{a(\tau )}\left\{ \left[
\omega ^{-1}\left( \chi ,\sigma ,\tau ,\varphi \right) \right] ^{\ast
}\right\} ^{2},\eta _{5}=\omega ^{-2}\left( \chi ,\sigma ,\tau ,\varphi
\right) ,  \nonumber \\
w_{i} &=&0,n_{i}\{\sigma ,\tau ,\omega ,\omega ^{\ast }\}=n_{i}\{\sigma
,\tau ,\omega \left( \sigma ,\tau ,\varphi \right) ,\omega ^{\ast }\left(
\sigma ,\tau ,\varphi \right) \}.  \label{2sol5p1}
\end{eqnarray}
and
\begin{eqnarray}
\mbox{$t$--solutions} &:&(x^{1}=\chi ,x^{2}=\sigma ,x^{3}=\tau
,y^{4}=v=t,y^{5}=p=\varphi ),g_{1}=\pm 1,  \nonumber \\
g_{2} &=&-1,g_{3}=-1,h_{4(0)}=b(\sigma ,\tau ),h_{5(0)}=a(\tau ),%
\mbox{see
(\ref{2auxm1})};  \nonumber \\
h_{4} &=&\eta _{4}(\sigma ,\tau ,t)h_{4(0)},h_{5}=\eta _{5}(\sigma ,\tau
,t)h_{5(0)},  \nonumber \\
\eta _{4} &=&\omega ^{-2}\left( \chi ,\sigma ,\tau ,t\right) ,\eta
_{5}=h_{(0)}^{-2}\frac{b(\sigma ,\tau )}{a(\tau )}\left[ \int dt~\omega
^{-1}\left( \chi ,\sigma ,\tau ,t\right) \right] ^{2},  \nonumber \\
w_{i} &=&0,n_{i}\{\sigma ,\tau ,\omega ,\omega ^{\ast }\}=n_{i}\{\sigma
,\tau ,\omega \left( \sigma ,\tau ,t\right) ,\omega ^{\ast }\left( \sigma
,\tau ,t\right) \}.  \label{2sol5t1}
\end{eqnarray}

Both types of solutions have a horizon parametrized by torus hypersurface
(as the condition of vanishing of \ the ''time'' metric coefficient states,
i. e. when the function $b(\sigma ,\tau )=0)$. $\ $These solutions are
generically anholonomic (anisotropic) because in the locally isotropic
limit, when $\eta _{4},\eta _{5},$ $\omega \rightarrow 1$ and $%
n_{i}\rightarrow 0,$ they reduce to the coefficients of the metric (\ref
{2auxm1}). The last one is not an exact solution of 4D vacuum Einstein
equations, but it is a conformal transform of the 4D Schwarzschild metric
deformed to a torus horizon, with a further trivial extension to 5D. With
respect to the anholonomic frames adapted to the coefficients $n_{i}$ (see (%
\ref{2ddif1})), the obtained solutions have diagonal metric coefficients
being very similar to the metric (\ref{1schtor}) written in toroidal
coordinates. We can treat such solutions as black tori with the mass
distributed linearly on the circle which can not transformed in a point, in
the center of torus.

The solutions are constructed as to have singularities on the mentioned
circle. \ The initial data for anholonomic frames and the chosen
configuration of gravitational interactions in the bulk lead to deformed
toroidal horizons even for static configurations. The solutions admit
anisotropic polarizations on toroidal coordinates $\left( \sigma ,\tau
\right) $ and running of constants on time $t$ and/or on extra dimension
coordinate $\chi $. Such renormalizations of constants are defined by the
nonlinear configuration of the 5D vacuum gravitational field and depend on
introduced receptivity function $\omega \left( \sigma ,\tau ,v\right) $
which is to be considered an intrinsic characteristics of the 5D vacuum
gravitational 'ether', emphasizing the possibility \ of nonlinear
self--polarization of gravitational fields.

Finally, we point that the data (\ref{2sol5p1}) and (\ref{2sol5t1})
parametrize two very different classes of solutions. The first one is for
static 5D vacuum black tori configurations with explicit dependence on
anholonomic coordinate $\varphi $ and possible renormalizations on the rest
of 3D space coordinates $\sigma $ and $\tau $ and on the 5th coordinate $%
\chi .$ The second class of solutions is similar to the static ones but with
an emphasized anholonomic running on time of constants and with possible
anisotropic dependencies on coordinates $(\sigma ,\tau ,\chi ).$

\subsubsection{Class A of vacuum solutions with ansatz (\ref{5ansatzc}):}

We construct here 5D vacuum $\varphi $-- and $t$--solutions
parametrized by an ansatz with conformal factor $\Omega (\sigma
,\tau ,v)$ (see (\ref {5ansatzc}) and (\ref{3cdmetric})). Let us
consider conformal factors parametrized as $\Omega =\Omega
_{\lbrack 0]}(\sigma ,\tau )\Omega _{\lbrack 1]}(\sigma ,\tau
,v).$ We can generate from the data (\ref{2sol5p1}) (or (\ref
{2sol5t1})) an exact solution of vacuum Einstein equations if
there are satisfied the conditions (\ref{3conf1}), (\ref{3confq})
and (\ref{4confeq}), i. e.
\[
\Omega _{\lbrack 0]}^{q_{1}/q_{2}}\Omega _{\lbrack 1]}^{q_{1}/q_{2}}=\eta
_{4}h_{4(0)},
\]
for some integers $q_{1}$ and $q_{2},$ and there are defined the second
anisotropy coefficients
\[
\zeta _{i}=\left( \partial _{i}\ln |\Omega _{\lbrack 0]}\right) |)~\left(
\ln |\Omega _{\lbrack 1]}|\right) ^{\ast }+\left( \Omega _{\lbrack 1]}^{\ast
}\right) ^{-1}\partial _{i}\Omega _{\lbrack 1]}.
\]
So, taking a $\varphi $-- or $t$--solution with corresponding values of $%
h_{4}=\eta _{4}h_{4(0)},$\ for some $q_{1}$ and $q_{2},$ we obtain new exact
solutions, called in brief, $\varphi _{c}$-- or $t_{c}$--solutions (with the
index ''c'' pointing to an ansatz with conformal factor), of the vacuum 5D
Einstein equations given in explicit form by the data:
\begin{eqnarray}
\mbox{$\varphi_c$--solutions} &:&(x^{1}=\chi ,x^{2}=\sigma ,x^{3}=\tau
,y^{4}=v=\varphi ,y^{5}=p=t),g_{1}=\pm 1,  \nonumber \\
g_{2} &=&-1,g_{3}=-1,h_{4(0)}=a(\tau ),h_{5(0)}=b(\sigma ,\tau ),%
\mbox{see
(\ref{2auxm1})};  \nonumber \\
h_{4} &=&\eta _{4}(\sigma ,\tau ,\varphi )h_{4(0)},h_{5}=\eta _{5}(\sigma
,\tau ,\varphi )h_{5(0)},  \nonumber \\
\eta _{4} &=&h_{(0)}^{2}\frac{b(\sigma ,\tau )}{a(\tau )}\left\{ \left[
\omega ^{-1}\left( \chi ,\sigma ,\tau ,\varphi \right) \right] ^{\ast
}\right\} ^{2},\eta _{5}=\omega ^{-2}\left( \chi ,\sigma ,\tau ,\varphi
\right) , \label{2sol5pc} \\
w_{i} &=&0,n_{i}\{\sigma ,\tau ,\omega ,\omega ^{\ast }\}=n_{i}\{\sigma
,\tau ,\omega \left( \sigma ,\tau ,\varphi \right) ,\omega ^{\ast }\left(
\sigma ,\tau ,\varphi \right) \},    \nonumber  \\
\zeta _{i} &=&\left( \partial _{i}\ln |\Omega _{\lbrack 0]}\right) |)~\left(
\ln |\Omega _{\lbrack 1]}|\right) ^{\ast }+\left( \Omega _{\lbrack 1]}^{\ast
}\right) ^{-1}\partial _{i}\Omega _{\lbrack 1]},  \nonumber  \\
\eta _{4}a &=&\Omega _{\lbrack 0]}^{q_{1}/q_{2}}(\sigma ,\tau )
\Omega _{\lbrack 1]}^{q_{1}/q_{2}}(\sigma ,\tau ,\varphi ),\
\Omega =\Omega _{\lbrack 0]}(\sigma ,\tau )
\Omega _{\lbrack 1]}(\sigma ,\tau ,\varphi )   \nonumber
\end{eqnarray}
and
\begin{eqnarray}
\mbox{$t_c$--solutions} &:&(x^{1}=\chi ,x^{2}=\sigma ,x^{3}=\tau
,y^{4}=v=t,y^{5}=p=\varphi ),g_{1}=\pm 1,  \nonumber \\
g_{2} &=&-1,g_{3}=-1,h_{4(0)}=b(\sigma ,\tau ),h_{5(0)}=a(\tau ),%
\mbox{see
(\ref{2auxm1})};  \nonumber \\
h_{4} &=&\eta _{4}(\sigma ,\tau ,t)h_{4(0)},h_{5}=\eta _{5}(\sigma ,\tau
,t)h_{5(0)},  \nonumber \\
\eta _{4} &=&\omega ^{-2}\left( \chi ,\sigma ,\tau ,t\right) ,\eta
_{5}=h_{(0)}^{-2}\frac{b(\sigma ,\tau )}{a(\tau )}\left[ \int dt~\omega
^{-1}\left( \chi ,\sigma ,\tau ,t\right) \right] ^{2},  \label{2sol5tc} \\
w_{i} &=&0,n_{i}\{\sigma ,\tau ,\omega ,\omega ^{\ast }\}=n_{i}\{\sigma
,\tau ,\omega \left( \sigma ,\tau ,t\right) ,\omega ^{\ast }\left( \sigma
,\tau ,t\right) \},
 \nonumber  \\
\zeta _{i} &=&\left( \partial _{i}\ln |\Omega _{\lbrack 0]}\right) |)~\left(
\ln |\Omega _{\lbrack 1]}|\right) ^{\ast }+\left( \Omega _{\lbrack 1]}^{\ast
}\right) ^{-1}\partial _{i}\Omega _{\lbrack 1]},  \nonumber  \\
\eta _{4}a &=& \Omega _{\lbrack 0]}^{q_{1}/q_{2}}(\sigma ,\tau )
\Omega _{\lbrack 1]}^{q_{1}/q_{2}}(\sigma
,\tau ,t),\ \Omega =\Omega _{\lbrack 0]}(\sigma ,\tau )
\Omega _{\lbrack 1]}(\sigma ,\tau ,t)    \nonumber
\end{eqnarray}

These solutions have two very interesting properties: 1) they admit a warped
factor on the 5th coordinate, like $\Omega _{\lbrack 1]}^{q_{1}/q_{2}}\sim
\exp [-k|\chi |],$ which in our case is constructed for an anisotropic 5D
vacuum gravitational configuration and not following a brane configuration
like in Refs. \cite{11rs}; 2) we can impose such conditions on the receptivity
$\omega \left( \sigma ,\tau ,v\right) $ as to obtain in the locally
isotropic limit just the toroidally deformed Schwarzschild metric (\ref
{1schtor}) trivially embedded into the 5D spacetime.

We analyze the second property in details. We have to chose the conformal
factor as to be satisfied three conditions:
\begin{equation}
\Omega _{\lbrack 0]}^{q_{1}/q_{2}}=\Omega _{A},\Omega _{\lbrack
1]}^{q_{1}/q_{2}}\eta _{4}=1,\Omega _{\lbrack 1]}^{q_{1}/q_{2}}\eta _{5}=1,
\label{2cond1a}
\end{equation}
were $\Omega _{A}$ is that from (\ref{2auxm1c}). The last two conditions are
possible if
\begin{equation}
\eta _{4}^{-q_{1}/q_{2}}\eta _{5}=1,  \label{3cond2}
\end{equation}
which selects a specific form of receptivity $\omega \left( x^{i},v\right) .$
\ Putting into (\ref{3cond2}) the values $\eta _{4}$ and $\eta _{5}$
respectively from (\ref{2sol5pc}), or (\ref{2sol5tc}), we obtain some
differential, or integral, relations of the unknown $\omega \left( \sigma
,\tau ,v\right) ,$ which results that
\begin{eqnarray}
\omega \left( \sigma ,\tau ,\varphi \right) &=&\left( 1-q_{1}/q_{2}\right)
^{-1-q_{1}/q_{2}}\left[ h_{(0)}^{-1}\sqrt{|a/b|}\varphi +\omega _{\lbrack
0]}\left( \sigma ,\tau \right) \right] ,\mbox{ for }\varphi _{c}%
\mbox{--solutions}; \nonumber  \\
\omega \left( \sigma ,\tau ,t\right) &=&\left[ \left( q_{1}/q_{2}-1\right)
h_{(0)}\sqrt{|a/b|}t+\omega _{\lbrack 1]}\left( \sigma ,\tau \right) \right]
^{1-q_{1}/q_{2}},\mbox{ for }t_{c}\mbox{--solutions},   \label{5cond1}
\end{eqnarray}
for some arbitrary functions $\omega _{\lbrack 0]}\left( \sigma ,\tau
\right) $ and $\omega _{\lbrack 1]}\left( \sigma ,\tau \right) .$ \ So, a
receptivity of particular form like (\ref{5cond1}) allow us to obtain in
the locally isotropic limit just the toroidally deformed Schwarzschild
metric.

We conclude this subsection:\ the  vacuum 5D metrics solving the
Einstein equations describe a nonlinear gravitational dynamics
which under some particular boundary conditions and
parametrizations of metric's coefficients can model anisotropic,
topologically not--trivial, solutions transforming, in a
corresponding locally isotropic limit, in some toroidal or
ellipsoidal deformations of the well known exact solutions like
Schwarzschild, Reissner-N\"{o}rdstrom, Taub NUT, various type of
wormhole, solitonic and disk solutions (see details in Refs.
\cite{11v,11v2,11vth}). We emphasize that, in general, an anisotropic
solution (parametrized by an off--diagonal ansatz) could not have
a locally isotropic limit to a diagonal metric with respect to
some holonomic coordinate frames. This was proved in explicit form
by choosing a configuration with toroidal symmetry.

\subsection{Toroidal 5D metrics of Class B}

In this subsection we construct and analyze another two classes of 5D vacuum
solutions which are related to the metric of Class B (\ref{2auxm2}) and which
can be reduced to the toroidally deformed Schwarzschild metric (\ref{1schtor})
by corresponding parametrizations of receptivity $\omega \left( \sigma ,\tau
,v\right) $. We emphasize that because the function $g(\sigma ,\tau )$ from (%
\ref{2auxm2}) is not a solution of equation(\ref{ein1}) we introduce an
auxiliary factor $\varpi $ $(\sigma ,\tau )$ for which $\varpi g$ became a
such solution, then we consider conformal factors parametrized as $\Omega
=\varpi ^{-1}(\sigma ,\tau )$ $\Omega _{\lbrack 2]}\left( \sigma ,\tau
,v\right) $ and find solutions parametrized by the ansatz (\ref{5ansatzc})
and anholonomic metric interval (\ref{3cdmetric}).

Because the method of definition of such solutions is similar to that from
previous subsection, in our further considerations we shall omit
computations and present directly the data which select the respective
configurations for $\varphi _{c}$--solutions and $t_{c}$--solutions.

The Class B of 5D solutions with conformal factor are parametrized by the
data:

\begin{eqnarray}
\mbox{$\varphi_c$--solutions} &:&(x^{1}=\chi ,x^{2}=\sigma ,x^{3}=\tau
,y^{4}=v=\varphi ,y^{5}=p=t),g_{1}=\pm 1,  \nonumber \\
g_{2} &=&g_{3}=\varpi (\sigma ,\tau )g(\sigma ,\tau ), \nonumber \\
h_{4(0)}&=& -\varpi
(\sigma ,\tau ),h_{5(0)}=\varpi (\sigma ,\tau )f(\sigma ,\tau ),%
\mbox{see  (\ref{2auxm2})};  \nonumber \\
\varpi &=&g^{-1}(\sigma ,\tau )\varpi _{0}\exp [a_{2}\sigma +a_{3}\tau
],~\varpi _{0},a_{2},a_{3}=const;  \nonumber \\
h_{4} &=&\eta _{4}(\sigma ,\tau ,\varphi )h_{4(0)},h_{5}=\eta _{5}(\sigma
,\tau ,\varphi )h_{5(0)},  \nonumber \\
\eta _{4} &=&-h_{(0)}^{2}f(\sigma ,\tau )\left\{ \left[ \omega ^{-1}\left(
\chi ,\sigma ,\tau ,\varphi \right) \right] ^{\ast }\right\} ^{2},\eta
_{5}=\omega ^{-2}\left( \chi ,\sigma ,\tau ,\varphi \right) ,  \label{2sol5p}
\\
w_{i} &=&0,n_{i}\{\sigma ,\tau ,\omega ,\omega ^{\ast }\}=n_{i}\{\sigma
,\tau ,\omega \left( \sigma ,\tau ,\varphi \right) ,\omega ^{\ast }\left(
\sigma ,\tau ,\varphi \right) \},   \nonumber \\
\zeta _{i} &=&\partial _{i}\ln |\varpi |)~\left( \ln |\Omega _{\lbrack
2]}|\right) ^{\ast }+\left( \Omega _{\lbrack 2]}^{\ast }\right)
^{-1}\partial _{i}\Omega _{\lbrack 2]}, \nonumber \\
\eta _{4}&=& -\varpi ^{-q_{1}/q_{2}}(\sigma ,\tau )
\Omega _{\lbrack 2]}^{q_{1}/q_{2}}(\sigma ,\tau ,\varphi ),
\Omega =\varpi ^{-1}(\sigma ,\tau )\Omega _{\lbrack 2]}(\sigma ,\tau ,\varphi )
 \nonumber
\end{eqnarray}
and
\begin{eqnarray}
\mbox{$t_c$--solutions} &:&(x^{1}=\chi ,x^{2}=\sigma ,x^{3}=\tau
,y^{4}=v=t,y^{5}=p=\varphi ),g_{1}=\pm 1,  \nonumber \\
g_{2} &=&g_{3}=\varpi (\sigma ,\tau )g(\sigma ,\tau ), \nonumber \\
h_{4(0)}&=& \varpi
(\sigma ,\tau )f(\sigma ,\tau ),h_{5(0)}=-\varpi (\sigma ,\tau ),%
\mbox{see (\ref{2auxm2})};  \nonumber \\
\varpi &=&g^{-1}(\sigma ,\tau )\varpi _{0}\exp [a_{2}\sigma +a_{3}\tau
],~\varpi _{0},a_{2},a_{3}=const,  \nonumber \\
h_{4} &=&\eta _{4}(\sigma ,\tau ,t)h_{4(0)},h_{5}=\eta _{5}(\sigma ,\tau
,t)h_{5(0)},  \nonumber \\
\eta _{4} &=&\omega ^{-2}\left( \chi ,\sigma ,\tau ,t\right) ,\eta
_{5}=-h_{(0)}^{-2}f(\sigma ,\tau )\left[ \int dt~\omega ^{-1}\left( \chi
,\sigma ,\tau ,t\right) \right] ^{2},  \label{2sol5t} \\
w_{i} &=&0,n_{i}\{\sigma ,\tau ,\omega ,\omega ^{\ast }\}=n_{i}\{\sigma
,\tau ,\omega \left( \sigma ,\tau ,t\right) ,\omega ^{\ast }\left( \sigma
,\tau ,t\right) \},    \nonumber \\
\zeta _{i} &=&\partial _{i}(\ln |\varpi |)~\left( \ln |\Omega _{\lbrack
2]}|\right) ^{\ast }+\left( \Omega _{\lbrack 2]}^{\ast }\right)
^{-1}\partial _{i}\Omega _{\lbrack 2]},     \nonumber \\
\eta _{4}&=& -\varpi
^{-q_{1}/q_{2}}(\sigma ,\tau )\Omega _{\lbrack 2]}^{q_{1}/q_{2}}(\sigma
,\tau ,t),\ \Omega =\varpi ^{-1}(\sigma ,\tau )
\Omega _{\lbrack 2]}(\sigma ,\tau ,t),   \nonumber
\end{eqnarray}
where the coefficients $n_{i}$ can be found explicitly by introducing the
corresponding values $\eta _{4}$ and $\eta _{5}$ in formula (\ref{2nel}).

By a procedure similar to the solutions of Class A (see previous subsection)
we can find the conditions when the solutions (\ref{2sol5p}) and (\ref{2sol5t}%
) will have in the locally anisotropic limit the toroidally deformed
Schwarzschild solutions, which impose corresponding parametrizations and
dependencies on $\Omega _{\lbrack 2]}(\sigma ,\tau ,v)$ and $\omega \left(
\sigma ,\tau ,v\right) $ like (\ref{2cond1a}) and (\ref{5cond1}). We omit
these formulas because, in general, for anholonomic configurations and
nonlinear solutions there are not hard arguments to prefer any holonomic
limits of such off--diagonal metrics.

Finally, in this Section, we remark that for the considered classes of black
tori solutions the so--called $t$--components of metric contain
modifications of the Schwarzschild potential
\[
\Phi =-\frac{M}{M_{P[4]}^{2}r}\mbox{ into }\Phi =-\frac{M\omega \left(
\sigma ,\tau ,v\right) }{M_{P[4]}^{2}r},
\]
where $M_{P[4]}$ is the usual 4D Plank constant, and this is given with
respect to the corresponding anholonomic frame of reference. The receptivity $%
\omega \left( \sigma ,\tau ,v\right) $ could model corrections warped on
extra dimension coordinate, $\chi ,$ which for our solutions are induced by
anholonomic vacuum gravitational interactions in the bulk and not from a
brane configuration in $AdS_{5}$ spacetime. In the vacuum case $k$ is a
constant which characterizes the receptivity for bulk vacuum gravitational
polarizations.

\section{4D Black Tori}

For the ansatz (\ref{5ansatz}), with trivial conformal factor, a
black torus solution of 4D vacuum Einstein equations was
constructed in Ref. \cite{11v}. \ The goal of this Section is to
consider some alternative variants,  with trivial or nontrivial
conformal factors and for different coordinate parametrizations
and types of anisotropies. The bulk of 5D solutions from the
previous Section are reduced into corresponding 4D ones if we
eliminate the 5th coordinate $\chi $ from\ the the off--diagonal
ansatz (\ref {5ansatz}) and (\ref{5ansatzc}) and corresponding
formulas and
 solutions.

\subsection{Toroidal 4D vacuum metrics of Class A}

Let us parametrize the 4D coordinates as $(x^{\underline{i}},y^{a})=\left(
x^{2}=\sigma ,x^{3}=\tau ,y^{4}=v,y^{5}=p\right) ;$ for the $\varphi $%
--solutions we shall take $\left( v=\varphi ,p=t\right) $ and for the
solutions $t$--solutions will consider $\left( v=t,p=\varphi \right) $. For
simplicity, we write down the data for solutions without proofs and
computations.

\subsubsection{Class A of vacuum solutions with ansatz (\ref{5ansatz}):}

The off--diagonal metric ansatz of type (\ref{5ansatz}) (equivalently, (\ref
{6dmetric})) \ with the data
\newpage
\begin{eqnarray}
\mbox{$\varphi$--solutions} &:&(x^{2}=\sigma ,x^{3}=\tau ,y^{4}=v=\varphi
,y^{5}=p=t)  \nonumber \\
g_{2} &=&-1,g_{3}=-1,h_{4(0)}=a(\tau ),h_{5(0)}=b(\sigma ,\tau ),%
\mbox{see
(\ref{2auxm1})};  \nonumber \\
h_{4} &=&\eta _{4}(\sigma ,\tau ,\varphi )h_{4(0)},h_{5}=\eta _{5}(\sigma
,\tau ,\varphi )h_{5(0)},  \nonumber \\
\eta _{4} &=&h_{(0)}^{2}\frac{b(\sigma ,\tau )}{a(\tau )}\left\{ \left[
\omega ^{-1}\left( \sigma ,\tau ,\varphi \right) \right] ^{\ast }\right\}
^{2},\eta _{5}=\omega ^{-2}\left( \sigma ,\tau ,\varphi \right) ,  \nonumber
\\
w_{\underline{i}} &=&0,n_{\underline{i}}\{\sigma ,\tau ,\omega ,\omega
^{\ast }\}=n_{\underline{i}}\{\sigma ,\tau ,\omega \left( \sigma ,\tau
,\varphi \right) ,\omega ^{\ast }\left( \sigma ,\tau ,\varphi \right) \}.
\label{2sol4p1}
\end{eqnarray}
and
\begin{eqnarray}
\mbox{$t$--solutions} &:&(x^{2}=\sigma ,x^{3}=\tau
,y^{4}=v=t,y^{5}=p=\varphi )  \nonumber \\
g_{2} &=&-1,g_{3}=-1,h_{4(0)}=b(\sigma ,\tau ),h_{5(0)}=a(\tau ),%
\mbox{see
(\ref{2auxm1})};  \nonumber \\
h_{4} &=&\eta _{4}(\sigma ,\tau ,t)h_{4(0)},h_{5}=\eta _{5}(\sigma ,\tau
,t)h_{5(0)},  \nonumber \\
\eta _{4} &=&\omega ^{-2}\left( \sigma ,\tau ,t\right) ,\eta
_{5}=h_{(0)}^{-2}\frac{b(\sigma ,\tau )}{a(\tau )}\left[ \int dt~\omega
^{-1}\left( \sigma ,\tau ,t\right) \right] ^{2},  \nonumber \\
w_{\underline{i}} &=&0,n_{\underline{i}}\{\sigma ,\tau ,\omega ,\omega
^{\ast }\}=n_{\underline{i}}\{\sigma ,\tau ,\omega \left( \sigma ,\tau
,t\right) ,\omega ^{\ast }\left( \sigma ,\tau ,t\right) \}.  \label{2sol4t1}
\end{eqnarray}
where the $n_{\underline{i}}$ are computed
\begin{eqnarray}
n_{k}\left( \sigma ,\tau ,v\right) &=&n_{k[1]}\left( \sigma ,\tau \right)
+n_{k[2]}\left( \sigma ,\tau \right) \int [\eta _{4}/(\sqrt{|\eta _{5}|}%
)^{3}]dv,~\eta _{5}^{\ast }\neq 0;  \label{2nem4} \\
&=&n_{k[1]}\left( \sigma ,\tau \right) +n_{k[2]}\left( \sigma ,\tau \right)
\int \eta _{4}dv,\qquad ~\eta _{5}^{\ast }=0;  \nonumber \\
&=&n_{k[1]}\left( \sigma ,\tau \right) +n_{k[2]}\left( \sigma ,\tau \right)
\int [1/(\sqrt{|\eta _{5}|})^{3}]dv,~\eta _{4}^{\ast }=0.  \nonumber
\end{eqnarray}
when the integration variable is taken $v=\varphi ,$ for (\ref{2sol4p1}), or $%
v=t,$ for (\ref{2sol4t1}). These solutions have the same toroidal symmetries
and properties stated for their 5D analogs (\ref{2sol5p1}) and for (\ref
{2sol5t1}) with that difference that there are not any warped factors and
extra dimension dependencies. Such solutions defined by the formulas (\ref
{2sol4p1}) and (\ref{2sol4t1}) do not result in a locally isotropic limit into
an exact solution having diagonal coefficients with respect to some
holonomic coordinate frames. The data introduced in this subsection are for
generic 4D vacuum solutions of the Einstein equations parametrized by
off--diagonal metrics. The renormalization of constants and metric
coefficients have a 4D nonlinear vacuum gravitational nature and reflects a
corresponding anholonomic dynamics.

\subsubsection{Class A of vacuum solutions with ansatz (\ref{5ansatzc}):}

The 4D vacuum $\varphi $-- and $t$--solutions parametrized by an ansatz with
conformal factor $\Omega (\sigma ,\tau ,v)$ (see (\ref{5ansatzc}) and (\ref
{3cdmetric})). Let us consider conformal factors parametrized as $\Omega
=\Omega _{\lbrack 0]}(\sigma ,\tau )\Omega _{\lbrack 1]}(\sigma ,\tau ,v).$
The data are
\begin{eqnarray}
\mbox{$\varphi_c$--solutions} &:&(x^{2}=\sigma ,x^{3}=\tau ,y^{4}=v=\varphi
,y^{5}=p=t)  \nonumber \\
g_{2} &=&-1,g_{3}=-1,h_{4(0)}=a(\tau ),h_{5(0)}=b(\sigma ,\tau ),%
\mbox{see
(\ref{2auxm1})};  \nonumber \\
h_{4} &=&\eta _{4}(\sigma ,\tau ,\varphi )h_{4(0)},h_{5}=\eta _{5}(\sigma
,\tau ,\varphi )h_{5(0)}, \Omega =\Omega _{\lbrack 0]}(\sigma ,\tau )
\Omega _{\lbrack 1]}(\sigma ,\tau ,\varphi ),
 \nonumber \\
\eta _{4} &=&h_{(0)}^{2}\frac{b(\sigma ,\tau )}{a(\tau )}\left\{ \left[
\omega ^{-1}\left( \sigma ,\tau ,\varphi \right) \right] ^{\ast }\right\}
^{2},\eta _{5}=\omega ^{-2}\left( \sigma ,\tau ,\varphi \right) ,
\label{2sol4pc} \\
w_{i} &=&0,n_{i}\{\sigma ,\tau ,\omega ,\omega ^{\ast
}\}=n_{i}\{\sigma ,\tau ,\lambda ,\omega \left( \sigma ,\tau
,\varphi \right) ,\omega ^{\ast }\left( \sigma ,\tau ,\varphi
\right) \},  \nonumber \\
\zeta _{i} &=&\left( \partial _{i}\ln |\Omega _{\lbrack 0]}\right) |)~\left(
\ln |\Omega _{\lbrack 1]}|\right) ^{\ast }+\left( \Omega _{\lbrack 1]}^{\ast
}\right) ^{-1}\partial _{i}\Omega _{\lbrack 1]},\nonumber \\
\eta _{4}a &=& \Omega _{\lbrack 0]}^{q_{1}/q_{2}}(\sigma ,\tau )
\Omega _{\lbrack 1]}^{q_{1}/q_{2}}(\sigma
,\tau ,\varphi ).  \nonumber
\end{eqnarray}
and
\begin{eqnarray}
\mbox{$t_c$--solutions} &:&(x^{2}=\sigma ,x^{3}=\tau
,y^{4}=v=t,y^{5}=p=\varphi )  \nonumber \\
g_{2} &=&-1,g_{3}=-1,h_{4(0)}=b(\sigma ,\tau ),h_{5(0)}=a(\tau ),%
\mbox{see  (\ref{2auxm1})};  \nonumber \\
h_{4} &=&\eta _{4}(\sigma ,\tau ,t)h_{4(0)},h_{5}=\eta
_{5}(\sigma ,\tau ,t)h_{5(0)}, \Omega =\Omega _{\lbrack
0]}(\sigma ,\tau )\Omega
_{\lbrack 1]}(\sigma ,\tau ,t), \nonumber \\
\eta _{4} &=&\omega ^{-2}\left( \sigma ,\tau ,t\right) ,\eta
_{5}=h_{(0)}^{-2}\frac{b(\sigma ,\tau )}{a(\tau )}\left[ \int dt~\omega
^{-1}\left( \sigma ,\tau ,t\right) \right] ^{2},  \label{2sol4tc} \\
w_{i} &=&0,n_{i}\{\sigma ,\tau ,\omega ,\omega ^{\ast
}\}=n_{i}\{\sigma ,\tau ,\omega \left( \sigma ,\tau ,t\right)
,\omega ^{\ast }\left( \sigma ,\tau ,t\right) \},
 \nonumber \\
\zeta _{i} &=&\left( \partial _{i}\ln |\Omega _{\lbrack 0]}\right) |)~\left(
\ln |\Omega _{\lbrack 1]}|\right) ^{\ast }+\left( \Omega _{\lbrack 1]}^{\ast
}\right) ^{-1}\partial _{i}\Omega _{\lbrack 1]},   \nonumber \\
\eta _{4}a &= & \Omega _{\lbrack
0]}^{q_{1}/q_{2}}(\sigma ,\tau )\Omega _{\lbrack 1]}^{q_{1}/q_{2}}(\sigma
,\tau ,t),  \nonumber
\end{eqnarray}
where the coefficients the $n_{\underline{i}}$ are given by the same
formulas (\ref{2nem4}).

Contrary to the solutions (\ref{2sol4p1}) and for (\ref{2sol4t1}) theirs
conformal anholonomic transforms, respectively, (\ref{2sol4pc}) and (\ref
{2sol4tc}), can be subjected to such parametrizations of the conformal factor
and conditions on the receptivity $\omega \left( \sigma ,\tau ,v\right) $ as
to obtain in the locally isotropic limit just the toroidally deformed
Schwarzschild metric (\ref{1schtor}). These conditions are stated for $\Omega
_{\lbrack 0]}^{q_{1}/q_{2}}=\Omega _{A},$ $\Omega _{\lbrack
1]}^{q_{1}/q_{2}}\eta _{4}=1,$ $\Omega _{\lbrack 1]}^{q_{1}/q_{2}}\eta
_{5}=1,$were $\Omega _{A}$ is that from (\ref{2auxm1c}), which is possible if
$\eta _{4}^{-q_{1}/q_{2}}\eta _{5}=1,$which selects a specific form of the
receptivity $\omega .$ \ Putting the values $\eta _{4}$ and $\eta _{5},$
respectively, from (\ref{2sol4pc}), or (\ref{2sol4tc}), we obtain some
differential, or integral, relations of the unknown $\omega \left( \sigma
,\tau ,v\right) ,$ which results that
\begin{eqnarray*}
\omega \left( \sigma ,\tau ,\varphi \right) &=&\left( 1-q_{1}/q_{2}\right)
^{-1-q_{1}/q_{2}}\left[ h_{(0)}^{-1}\sqrt{|a/b|}\varphi +\omega _{\lbrack
0]}\left( \sigma ,\tau \right) \right] ,\mbox{ for }\varphi _{c}%
\mbox{--solutions}; \\
\omega \left( \sigma ,\tau ,t\right) &=&\left[ \left( q_{1}/q_{2}-1\right)
h_{(0)}\sqrt{|a/b|}t+\omega _{\lbrack 1]}\left( \sigma ,\tau \right) \right]
^{1-q_{1}/q_{2}},\mbox{ for }t_{c}\mbox{--solutions},
\end{eqnarray*}
for some arbitrary functions $\omega _{\lbrack 0]}\left( \sigma ,\tau
\right) $ and $\omega _{\lbrack 1]}\left( \sigma ,\tau \right) .$ The
 formulas for $\omega \left( \sigma ,\tau ,\varphi \right) $ and $%
\omega \left( \sigma ,\tau ,t\right) $ are 4D reductions of the formulas (%
\ref{2cond1a}) and (\ref{5cond1}).

\subsection{Toroidal 4D vacuum metrics of Class B}

We construct another two classes of 4D vacuum solutions which are related to
the metric of Class B (\ref{2auxm2}) and can be reduced to the toroidally
deformed Schwarzshild metric (\ref{1schtor}) by corresponding
parametrizations of receptivity $\omega \left( \sigma ,\tau ,v\right) $. The
solutions contain a 2D conformal factor $\varpi (\sigma ,\tau )$ for which $%
\varpi g$ becomes a solution of (\ref{ein1}) and a 4D conformal factor
parametrized as $\Omega =\varpi ^{-1}$ $\Omega _{\lbrack 2]}\left( \sigma
,\tau ,v\right) $ in \ order to set the constructions into the ansatz (\ref
{5ansatzc}) and anholonomic metric interval (\ref{3cdmetric}).

The data selecting the 4D configurations for $\varphi _{c}$--solutions and $%
t_{c}$--solutions:

\begin{eqnarray}
\mbox{$\varphi_c$--solutions} &:&(x^{2}=\sigma ,x^{3}=\tau ,y^{4}=v=\varphi
,y^{5}=p=t)  \nonumber \\
g_{2} &=&g_{3}=\varpi (\sigma ,\tau )g(\sigma ,\tau ),   \nonumber \\
h_{4(0)}&=& -\varpi
(\sigma ,\tau ),h_{5(0)}=\varpi (\sigma ,\tau )f(\sigma ,\tau ),%
\mbox{see (\ref{2auxm2})};  \nonumber \\
\varpi &=&g^{-1}\varpi _{0}\exp [a_{2}\sigma +a_{3}\tau ],~\varpi
_{0},a_{2},a_{3}=const;  \nonumber \\
h_{4} &=&\eta _{4}(\sigma ,\tau ,\varphi )h_{4(0)},h_{5}=\eta
_{5}(\sigma ,\tau ,\varphi )h_{5(0)}, \Omega =\varpi ^{-1}(\sigma
,\tau )\Omega
_{\lbrack 2]}(\sigma ,\tau ,\varphi )  \nonumber \\
\eta _{4} &=&-h_{(0)}^{2}f(\sigma ,\tau )\left\{ \left[ \omega ^{-1}\left(
\sigma ,\tau ,\varphi \right) \right] ^{\ast }\right\} ^{2},\eta _{5}=\omega
^{-2}\left( \sigma ,\tau ,\varphi \right) ,  \label{2sol4p} \\
w_{i} &=&0,n_{i}\{\sigma ,\tau ,\omega ,\omega ^{\ast }\}=n_{i}\{\sigma
,\tau ,\omega \left( \sigma ,\tau ,\varphi \right) ,\omega ^{\ast }\left(
\sigma ,\tau ,\varphi \right) \}, \nonumber \\
\zeta _{\underline{i}} &=&\partial _{\underline{i}}\ln |\varpi |)~\left( \ln
|\Omega _{\lbrack 2]}|\right) ^{\ast }+\left( \Omega _{\lbrack 2]}^{\ast
}\right) ^{-1}\partial _{\underline{i}}\Omega _{\lbrack 2]},  \nonumber \\
\eta _{4}&=&-\varpi ^{-q_{1}/q_{2}}(\sigma ,\tau )\Omega _{\lbrack
2]}^{q_{1}/q_{2}}(\sigma ,\tau ,\varphi )  \nonumber
\end{eqnarray}
and
\newpage
\begin{eqnarray}
\mbox{$t_c$--solutions} &:&(x^{2}=\sigma ,x^{3}=\tau
,y^{4}=v=t,y^{5}=p=\varphi )  \nonumber \\
g_{2} &=&g_{3}=\varpi (\sigma ,\tau )g(\sigma ,\tau ),  \nonumber \\
h_{4(0)}&=& \varpi (\sigma ,\tau )f(\sigma ,\tau ),
h_{5(0)}=-\varpi (\sigma ,\tau ), \mbox{see (\ref{2auxm2})};  \nonumber \\
\varpi &=&g^{-1}\varpi _{0}\exp [a_{2}\sigma +a_{3}\tau ],~\varpi
_{0},a_{2},a_{3}=const, \nonumber \\
h_{4} &=&\eta _{4}(\sigma ,\tau ,t)h_{4(0)},h_{5}=\eta
_{5}(\sigma ,\tau ,t)h_{5(0)}, \Omega =\varpi ^{-1}(\sigma ,\tau
)\Omega _{\lbrack
2]}(\sigma ,\tau ,t)  \nonumber \\
\eta _{4} &=&\omega ^{-2}\left( \sigma ,\tau ,t\right) ,\eta
_{5}=-h_{(0)}^{-2}f(\sigma ,\tau )\left[ \int dt~\omega ^{-1}\left( \sigma
,\tau ,t\right) \right] ^{2},  \label{2sol4t} \\
w_{i} &=&0,n_{i}\{\sigma ,\tau ,\omega ,\omega ^{\ast }\}=n_{i}\{\sigma
,\tau ,\omega \left( \sigma ,\tau ,t\right) ,\omega ^{\ast }\left( \sigma
,\tau ,t\right) \}, \nonumber \\
\zeta _{i} &=&\partial _{i}(\ln |\varpi |)~\left( \ln |\Omega _{\lbrack
2]}|\right) ^{\ast }+\left( \Omega _{\lbrack 2]}^{\ast }\right)
^{-1}\partial _{i}\Omega _{\lbrack 2]},   \nonumber \\
\eta _{4}&=& -\varpi ^{-q_{1}/q_{2}}(\sigma ,\tau )
\Omega _{\lbrack 2]}^{q_{1}/q_{2}}(\sigma
,\tau ,t).  \nonumber
\end{eqnarray}
where the coefficients $n_{i}$ can be found explicitly by introducing the
corresponding values $\eta _{4}$ and $\eta _{5}$ in formula (\ref{2nel}).

For the 4D Class B solutions, some conditions can be imposed (see previous
subsection) when the solutions (\ref{2sol4p}) and (\ref{2sol4t}) have in the
locally anisotropic limit the toroidally deformed Schwarzschild solution,
which imposes some specific parametrizations and dependencies on $\Omega
_{\lbrack 2]}(\sigma ,\tau ,v)$ and $\omega \left( \sigma ,\tau ,v\right) $
like (\ref{2cond1a}) and (\ref{5cond1}). We omit these considerations because
for anholonomic configurations and nonlinear solutions there are not
arguments to prefer any holonomic limits of such off--diagonal metrics.

We conclude this Section by noting that for the constructed classes of 4D
black tori solutions the so--called $t$--component of metric contains
modifications of the Schwarzschild potential
\[
\Phi =-\frac{M}{M_{P[4]}^{2}r}\mbox{ into }\Phi =-\frac{M\omega \left(
\sigma ,\tau ,v\right) }{M_{P[4]}^{2}r},
\]
where $M_{P[4]}$ is the usual 4D Plank constant; the metric coefficients are
given with respect to the corresponding anholonomic frame of reference. In 4D
anholonomic gravity the receptivity $\omega \left( \sigma ,\tau ,v\right) $
is considered to renormalize the mass constant. Such gravitational
self-polarizations are induced by anholonomic vacuum gravitational
interactions. They should be defined experimentally or computed following a
model of quantum gravity.

\section{The Cosmological Constant and Anisotropy}

In this Section we analyze the general properties of anholonomic Einstein
equations in 5D and 4D gravity with cosmological constant and consider two
examples of 5D and 4D exact solutions.

A non--vanishing $\Lambda $ term in the system of Einstein's equations
results in substantial differences because t $\beta \neq 0$ and, in this
case, one could be $w_{i}\neq 0;$ The equations (\ref{ein1}) and (\ref{ein2}%
) are of more general nonlinearity because of presence of the $2\Lambda
g_{2}g_{3}$ and $2\Lambda h_{4}h_{5}$ terms. In this case, the solutions
with $g_{2}=const$ and $g_{3}=const$ (and $h_{4}=const$ and $h_{5}=const)$
are not admitted. This makes more sophisticate the procedure of definition
of $g_{2}$ for a stated $g_{3}$ (or inversely, of definition of $g_{3}$ for
a stated $g_{2})$ from (\ref{ein1}) [similarly of constructing $h_{4}$ for a
given $h_{5}$ from (\ref{ein2}) and inversely], nevertheless, the separation
of variables is not affected by introduction of cosmological constant and
there is a number of possibilities to generate exact solutions.

The general properties of solutions of the system (\ref{ein1})--(\ref{einc}%
), with cosmological constant $\Lambda ,$ are stated in the form:

\begin{itemize}
\item  The general solution of equation (\ref{ein1}) is to be found from the
equation
\begin{equation}
\varpi \varpi ^{\bullet \bullet }-(\varpi ^{\bullet })^{2}+\varpi \varpi
^{^{\prime \prime }}-(\varpi ^{^{\prime }})^{2}=2\Lambda \varpi ^{3}.
\label{2auxr1}
\end{equation}
for a coordinate transform coordinate transforms $x^{2,3}\rightarrow
\widetilde{x}^{2,3}\left( u,\lambda \right) $ for which
\[
g_{2}(\sigma ,\tau )(d\sigma )^{2}+g_{3}(\sigma ,\tau )(d\tau
)^{2}\rightarrow \varpi \left[ (d\widetilde{x}^{2})^{2}+\epsilon (d%
\widetilde{x}^{3})^{2}\right] ,\epsilon =\pm 1
\]
and $\varpi ^{\bullet }=\partial \varpi /\partial \widetilde{x}^{2}$ and $%
\varpi ^{^{\prime }}=\partial \varpi /\partial \widetilde{x}^{3}.$

\item  The equation (\ref{ein2}) relates two functions $h_{4}\left( \sigma
,\tau ,v\right) $ and $h_{5}\left( \sigma ,\tau ,v\right) $ with $%
h_{5}^{\ast }\neq 0.$ If the function $h_{5}$ is given we can find $h_{4}$
as a solution of
\begin{equation}
h_{4}^{\ast }+\frac{2\Lambda }{\pi }(h_{4})^{2}+2\left( \frac{\pi ^{\ast }}{%
\pi }-\pi \right) h_{4}=0,  \label{2auxr2c}
\end{equation}

where $\pi =h_{5}^{\ast }/2h_{5}.$

\item  The exact solutions of (\ref{ein3}) for $\beta \neq 0$ is
\begin{eqnarray}
w_{k} &=&-\alpha _{k}/\beta ,  \label{2aw} \\
&=&\partial _{k}\ln [\sqrt{|h_{4}h_{5}|}/|h_{5}^{\ast }|]/\partial _{v}\ln [%
\sqrt{|h_{4}h_{5}|}/|h_{5}^{\ast }|],  \nonumber
\end{eqnarray}
for $\partial _{v}=\partial /\partial v$ and $h_{5}^{\ast }\neq 0.$

\item  The exact solution of (\ref{ein4}) is
\begin{eqnarray}
n_{k} &=&n_{k[1]}\left( \sigma ,\tau \right) +n_{k[2]}\left( \sigma ,\tau
\right) \int [h_{4}/(\sqrt{|h_{5}|})^{3}]dv,  \label{2nlambda} \\
&=&n_{k[1]}\left( \sigma ,\tau \right) +n_{k[2]}\left( \sigma ,\tau \right)
\int [1/(\sqrt{|h_{5}|})^{3}]dv,~h_{4}^{\ast }=0,  \nonumber
\end{eqnarray}
for some functions $n_{k[1,2]}\left( \sigma ,\tau \right) $ stated by
boundary conditions.

\item  The exact solution of (\ref{einc}) is given by
\begin{equation}
\zeta _{i}=-w_{i}+(\Omega ^{\ast })^{-1}\partial _{i}\Omega ,\quad \Omega
^{\ast }\neq 0,  \label{2aconf4}
\end{equation}
\end{itemize}

We note that by a corresponding re--parametrizations of the conformal
factor\\
$\Omega \left( \sigma ,\tau ,v\right) $ we can reduce (\ref{2auxr1}) to
\begin{equation}
\varpi \varpi ^{\bullet \bullet }-(\varpi ^{\bullet })^{2}=2\Lambda \varpi
^{3}  \label{2redaux}
\end{equation}
which gives and exact solution $\varpi =\varpi \left( \widetilde{x}%
^{2}\right) $ found from
\[
(\varpi ^{\bullet })^{2}=\varpi ^{3}\left( C\varpi ^{-1}+4\Lambda \right)
,C=const,
\]
(or, inversely, to reduce to
\[
\varpi \varpi ^{^{\prime \prime }}-(\varpi ^{^{\prime }})^{2}=2\Lambda
\varpi ^{3}
\]
with exact solution $\varpi =\varpi \left( \widetilde{x}^{3}\right) $ found
from
\[
(\varpi ^{\prime })^{2}=\varpi ^{3}\left( C\varpi ^{-1}+4\Lambda \right)
,C=const).
\]
The inverse problem of definition of $h_{5}$ for a given $h_{4}$ can be
solved in explicit form when $h_{4}^{\ast }=0,$ $h_{4}=h_{4(0)}(\sigma ,\tau
).$ In this case we have to solve
\begin{equation}
h_{5}^{\ast \ast }+\frac{(h_{5}^{\ast })^{2}}{2h_{5}}-2\Lambda
h_{4(0)}h_{5}=0,  \label{2auxr2ccp}
\end{equation}
which admits exact solutions by reduction to a Bernulli equation.

The outlined properties of solutions with cosmological constant hold also
for 4D anholonomic spacetimes with ''isotropic'' cosmological constant $%
\Lambda .$ To transfer general solutions from 5D to 4D we have to
eliminate dependencies on the coordinate $x^{1}$ and to consider
the 4D ansatz without $g_{11}$ term.

\subsection{A 5D anisotropic black torus solution with cosmological constant}

We give an example of generalization of ansiotropic black hole solutions of
Class A , constructed in the Section III as they will contain the
cosmological constant $\Lambda ;$ we extend the solutions given by the data (%
\ref{2sol5pc}).

Our new 5D $\varphi $-- solution is parametrized by an ansatz with conformal
factor $\Omega (x^{i},v)$ (see (\ref{5ansatzc}) and (\ref{3cdmetric})) as $%
\Omega =\varpi ^{-1}(\sigma )\Omega _{\lbrack 0]}(\sigma ,\tau
)\Omega _{\lbrack 1]}(\sigma ,\tau ,v).$ The factor $\varpi
(\sigma ,\tau )$ is chosen as to be a solution of (\ref{2redaux}).
This conformal data must satisfy the condition (\ref{3confq}), i.
e.
\[
\varpi ^{-q_{1}/q_{2}}\Omega _{\lbrack 0]}^{q_{1}/q_{2}}\Omega _{\lbrack
1]}^{q_{1}/q_{2}}=\eta _{4}\varpi h_{4(0)}
\]
for some integers $q_{1}$ and $q_{2},$ where $\eta _{4}$ is found as $%
h_{4}=\eta _{4}\varpi h_{4(0)}$ satisfies the equation (\ref{2auxr2c}) and $%
\Omega _{\lbrack 0]}(\sigma ,\tau )$ could be chosen as to obtain
in the locally isotropic limit and $\Lambda \rightarrow 0$ the
toroidally deformed Schwarzschild metric (\ref{1schtor}). Choosing
$h_{5}=\eta _{5}\varpi h_{5(0)},$ $\eta _{5}h_{5(0)}$ is for the
ansatz for (\ref{2sol5pc}), for which we compute the value $\pi
=h_{5}^{\ast }/2h_{5},$ we obtain from (\ref
{2auxr2c}) an equation for $\eta _{4},$%
\[
\eta _{4}^{\ast }+\frac{2\Lambda }{\pi }\varpi h_{4(0)}(\eta
_{4})^{2}+2\left( \frac{\pi ^{\ast }}{\pi }-\pi \right) \eta _{4}=0
\]
which is a Bernulli equation \cite{11kamke} and admit an exact solution, in
general, in non explicit form, $\eta _{4}=\eta _{4}^{[bern]}(\sigma ,\tau
,v,\Lambda ,\varpi ,\omega ,a,b),$ were we emphasize the functional
dependencies on functions $\varpi ,\omega ,a,b$ and cosmological constant $%
\Lambda .$ Having defined $\eta _{4[bern]},$ $\eta _{5}$ and
$\varpi ,$ we can compute the $\alpha _{i}$--$,\beta -,$ and
$\gamma $--coefficients, expressed as $$\alpha _{i}=\alpha
_{i}^{[bern]}(\sigma ,\tau ,v,\Lambda ,\varpi ,\omega ,a,b),\beta
=\beta ^{\lbrack bern]}(\sigma ,\tau ,v,\Lambda ,\varpi ,\omega
,a,b)$$ and $\gamma =\gamma ^{\lbrack bern]}(\sigma ,\tau
,v,\Lambda ,\varpi ,\omega ,a,b),$ following the formulas
(\ref{4abc}).

The next step is to find
\[
w_{i}=w_{i}^{[bern]}(\sigma ,\tau ,v,\Lambda ,\varpi ,\omega ,a,b)%
\mbox{ and
}n_{i}=n_{i}^{[bern]}(\sigma ,\tau ,v,\Lambda ,\varpi ,\omega ,a,b)
\]
as for the general solutions (\ref{2aw}) and (\ref{2nlambda}).

At the final step we are able to compute the the second anisotropy
coefficients
\[
\zeta _{i}=-w_{i}^{[bern]}+(\partial _{i}\ln |\varpi ^{-1}\Omega _{\lbrack
0]}|)~\left( \ln |\Omega _{\lbrack 1]}|\right) ^{\ast }+\left( \Omega
_{\lbrack 1]}^{\ast }\right) ^{-1}\partial _{i}\Omega _{\lbrack 1]},
\]
which depends on an arbitrary function $\Omega _{\lbrack 0]}(\sigma ,\tau ).$
If we state $\Omega _{\lbrack 0]}(\sigma ,\tau )=\Omega _{A},$ as for $%
\Omega _{A}$ from (\ref{2auxm2}), see similar details with respect to
formulas (\ref{2cond1a}), (\ref{3cond2}) and (\ref{5cond1}).

The data for the exact solutions with cosmological constant for $v=\varphi $
can be stated in the form
\begin{eqnarray}
\mbox{$\varphi_c$--solutions} &:&(x^{1}=\chi ,x^{2}=\sigma ,x^{3}=\tau
,y^{4}=v=\varphi ,y^{5}=p=t),g_{1}=\pm 1,  \nonumber \\
g_{2} &=&\varpi (\sigma ),g_{3}=\varpi (\sigma ),  \nonumber \\
h_{4(0)} &=& a(\tau ),h_{5(0)}=b(\sigma ,\tau ),
\mbox{see (\ref{2auxm1}) and  (\ref{2redaux})}; \nonumber \\
h_{4} &=&\eta _{4}(\sigma ,\tau ,\varphi )\varpi (\sigma
)h_{4(0)},h_{5}=\eta _{5}(\sigma ,\tau ,\varphi )\varpi (\sigma )h_{5(0)},
\nonumber \\
\eta _{4} &=&\eta _{4}^{[bern]}(\sigma ,\tau ,v,\Lambda ,\varpi ,\omega
,a,b),\eta _{5}=\omega ^{-2}\left( \chi ,\sigma ,\tau ,\varphi \right) ,
\label{2slambdap1} \\
w_{i} &=&w_{i}^{[bern]}(\sigma ,\tau ,v,\Lambda ,\varpi ,\omega ,a,b),
\nonumber \\
n_{i}&=&n_{i}^{[bern]}(\sigma
,\tau ,v,\Lambda ,\varpi ,\omega ,a,b),  \nonumber \\
\Omega &=&\varpi ^{-1}(\sigma )\Omega _{\lbrack 0]}(\sigma ,\tau )\Omega
_{\lbrack 1]}(\sigma ,\tau ,\varphi ),\eta _{4}a=\Omega _{\lbrack
0]}^{q_{1}/q_{2}}(\sigma ,\tau )\Omega _{\lbrack 1]}^{q_{1}/q_{2}}(\sigma
,\tau ,\varphi ).  \nonumber \\
\zeta _{i} &=&-w_{i}^{[bern]}+\left( \partial _{i}\ln |\varpi ^{-1}\Omega
_{\lbrack 0]}\right) |)~\left( \ln |\Omega _{\lbrack 1]}|\right) ^{\ast
}+\left( \Omega _{\lbrack 1]}^{\ast }\right) ^{-1}\partial _{i}\Omega
_{\lbrack 1]}.  \nonumber
\end{eqnarray}

We note that a solution with $v=t$ can be constructed as to generalize (\ref
{2sol5tc}) to the presence of $\Lambda .$ We can not present such data in
explicit form because in this case we have to define $\eta _{5}$ by solving
a solution like (\ref{ein2}) for $h_{5},$ for a given $h_{4},$ which can not
be integrated in explicit form.

The solution (\ref{2slambdap1}) preserves the two  interesting properties
of (\ref{2sol5pc}): 1) it admits a warped factor on the 5th coordinate, like $%
\Omega _{\lbrack 1]}^{q_{1}/q_{2}}\sim \exp [-k|\chi |],$ which
in this case is constructed for an anisotropic 5D vacuum
gravitational configuration with anisotropic cosmological
constant but not following a brane configuration like in Refs.
\cite{11rs}; 2) we can impose such conditions on the receptivity
$\omega \left( \sigma ,\tau ,\varphi \right) $ as to obtain in
the locally isotropic limit just the toroidally deformed
Schwarzschild metric (\ref{1schtor}) trivially embedded into the
5D spacetime.

\subsection{A 4D anisotropic black torus solution with cosmological constant}

The simplest way to construct a such solution is to take the data (\ref
{2slambdap1}), for $v=\varphi $, to eliminate the variable $\chi $ and to
reduce the 5D indices to 4D ones. We obtain the 4D data:
\newpage
\begin{eqnarray}
\mbox{$\varphi_c$--solutions} &:&(x^{2}=\sigma ,x^{3}=\tau ,y^{4}=v=\varphi
,y^{5}=p=t),  \nonumber \\
g_{2} &=&\varpi (\sigma ),g_{3}=\varpi (\sigma ),   \nonumber \\
h_{4(0)} &=& a(\tau ),h_{5(0)}=b(\sigma ,\tau ),
 \mbox{see (\ref{2auxm1}) and  (\ref{2redaux})};  \nonumber \\
h_{4} &=&\eta _{4}(\sigma ,\tau ,\varphi )\varpi (\sigma
)h_{4(0)},h_{5}=\eta _{5}(\sigma ,\tau ,\varphi )\varpi (\sigma )h_{5(0)},
\nonumber \\
\eta _{4} &=&\eta _{4}^{[bern]}(\sigma ,\tau ,v,\Lambda ,\varpi ,\omega
,a,b),\eta _{5}=\omega ^{-2}\left( \sigma ,\tau ,\varphi \right) ,
\label{sot4tpanis} \\
w_{i} &=&w_{i}^{[bern]}(\sigma ,\tau ,v,\Lambda ,\varpi ,\omega
,a,b),  \nonumber \\
n_{i}&=&n_{i}^{[bern]}(\sigma
,\tau ,v,\Lambda ,\varpi ,\omega ,a,b),  \nonumber \\
\Omega &=&\varpi ^{-1}(\sigma )\Omega _{\lbrack 0]}(\sigma ,\tau )\Omega
_{\lbrack 1]}(\sigma ,\tau ,\varphi ),\eta _{4}a=\Omega _{\lbrack
0]}^{q_{1}/q_{2}}(\sigma ,\tau )\Omega _{\lbrack 1]}^{q_{1}/q_{2}}(\sigma
,\tau ,\varphi ),  \nonumber \\
\zeta _{i} &=&-w_{i}^{[bern]}+\left( \partial _{i}\ln |\varpi ^{-1}\Omega
_{\lbrack 0]}\right) |)~\left( \ln |\Omega _{\lbrack 1]}|\right) ^{\ast
}+\left( \Omega _{\lbrack 1]}^{\ast }\right) ^{-1}\partial _{i}\Omega
_{\lbrack 1]}.  \nonumber
\end{eqnarray}
The solution (\ref{sot4tpanis}) describes a static black torus
solution in 4D gravity with cosmological constant, $\Lambda.$ The
parameters of solutions depends on the $\Lambda$ as well are
renormalized by nonlinear anholonomic gravitational interactions.
We can consider that the mass associated to such toroidal
configuration can be anisotropically distributed in the interior
of the torus and gravitationally polarized.

Finally, we note that in a similar manner like in the Sections III and IV we
can construct another classes of anisotropic black holes solutions in 5D and
4D spacetimes with cosmological constants, being of Class A or Class B, with
anisotropic $\varphi $--coordinate, or anisotropic $t$--coordinate. We omit
the explicit data which are some nonlinear anholonomic generalizations of
those solutions.

\section{Conclusions and Discussion}

We have shown that static black tori solutions can be constructed
both in vacuum Einstein and five dimensional (5D) gravity. The
solutions are parametrized by  off--diagonal metric ansatz which
can diagonalized with respect to corresponding anholonomic frames
with mixtures of holonomic and anholonomic variables. Such
metrics contain a toroidal horizon being some deformations with
non-trivial topology of the Schwarzschild black hole solution.

The solutions were constructed by using the anholonomic frame
method \cite{11v,11v2,11vth} which results in a very substantial
simplification of the Einstein equations which admit general
integrals for solutions.

The constructed black tori metrics depend on classes of two
dimensional and three dimensional functions which reflect the
freedom in definition of toroidal coordinates as well the
possibility to state by boundary conditions various
configurations with running constants, anisotropic gravitational
polarizations and (in presence of extra dimensions) with warping
geometries. The new toroidal solutions can be extended for
 spacetimes with cosmological constant.

In view of existence of such solutions, the old problem of the
status of frames in gravity theories rises once again, now in the
context of "effective" diagonalization of off--diagonal metrics
by using anholonomic transforms.  The bulk of solutions with
spherical, cylindrical and plane symmetries were constructed in
gravitational theories of diverse dimensions by using diagonal
metrics (sometimes with off--diagonal terms) given with respect
to "pure" coordinate frames. Such solutions can be equivalently
re--defined with respect to arbitrary frames of reference and
usually the  problem of fixing some reference bases in order to
state the boundary conditions is an important physical problem but
not a dynamical one. This problem becomes more sophisticate when
we deal with generic off--diagonal metrics and anholonomic frames.
In this case some 'dynamical, components of metrics can be
transformed into "non--dynamical" components of frame bases,
which, following  a more rigorous
 mathematical approach, reflects a constrained nonlinear
 dynamics  for gravitational and matter fields with  both holonomic
 (unconstrained)   and anholonomic (constrained)  variables. In result there are
  more possibilities in definition of classes of exact solutions
  with non--trivial topology, anisotropies and nonlinear
  interactions.

The solutions obtained in this paper contain as particular cases
(for corresponding parametrizations of considered ansatz) the
'black ring' metrics with event horizon of topology $S^1 \times
S^2$  analyzed in Refs. \cite{11emp}. In our case we emphasized the
presence of off--diagonal terms which results in warping,
anisotropy and running of constants. Here it should be noted that
 the generic nonlinear character of the Einstein equations written
  with respect to  anholonomic frames connected with diagonalization
  of off--diagonal   metrics allow us
 to construct different classes of exact 5D and 4D solutions with the same
 or different topology; such  solutions can define very different
  vacuum gravitational and gravitational--matter field
  configurations.

The method and results presented in this paper  provide a
prescription on anholonomic  transforming of some known locally
isotropic solutions from a gravity/string theory into
corresponding classes of anisoropic solutions of the same, or of
an extended theory:

{\em A vacuum, or non-vacuum, solution, and metrics conformally
equivalent to a known solution, parametrized by a diagonal matrix
given with respect to a holonomic (coordinate) base, contained as
a trivial form of ansatz (\ref {5ansatz}), or (\ref{5ansatzc}), can
 be transformed into a metric with non-trivial topological horizons and
 then generalized to be an anisotropic solution with similar but
anisotropically renormalized physical constants and diagonal
metric coefficients, given with respect to adapted anholonomic
frames; the new anholonomic metric defines an exact solution of a
simplified form of \ the Einstein equations
(\ref{ein1})--(\ref{einc}) and (\ref {4confeq}); such types of
solutions are parametrized by off--diagonal metrics if they are
re--defined with respect to usual coordinate frames}.

We emphasize that the anholonomic frame method and constructed
black tori solutions conclude in a general formalism of generating
exact solutions with off--diagonal metrics in  gravity theories
and  may have a number of applications in modern astrophysics and
string/M--theory gravity.

\subsection*{Acknowledgements}

The author thanks J.P.S. Lemos and D. Singleton for support and
collaboration. The work is supported both by a 2000--2001
California State University Legislative Award and a NATO/Portugal
fellowship grant at the Technical University of Lisbon.

%%%%%%%%%%%%%%%%%%%%%%%%%%%%%%%%%%%%%%%%%%%%%%%%%%%%%%%%%%%%%%%%%%%%%%%%%%%%%

\chapter[4D Ellipsoidal -- Toroidal Systems]
{Ellipsoidal Black Hole -- Black Tori Systems in 4D Gravity}

{\bf Abstract}
\footnote{\copyright\  S. Vacaru, Ellipsoidal Black Hole -- Black Tori Systems
        in 4D Gravity,  hep-th/0111166}

We construct new classes of exact solutions of the 4D vacuum
Einstein equations which describe  ellipsoidal black holes, black
tori and combined black hole -- black tori configurations. The
solutions can be static or with anisotropic polarizations and
running constants. They are defined by off--diagonal metric ansatz
which may be diagonalized with respect to  anholonomic moving
frames.  We examine physical properties of such anholonomic
gravitational configurations and discuss why the anholonomy may
remove the restriction that horizons must be with spherical
 topology.

\section{Introduction}

Torus configurations of matter around black hole -- neutron star objects are
intensively investigated in modern astrophysics \cite{12putten}. One considers
that such tori may radiate gravitational radiation powered by the spin
energy of the black hole in the presence of non--axisymmetries; long
gamma--ray bursts from rapidly spinning black hole--torus systems may
represent hypernovae or black hole--neutron star coalescence. Thus the topic
of constructing of exact vacuum and non--vacuum solutions with non--trivial
topology in the framework of general relativity and extra dimension
gravitational theories becomes of special importance and interest.

In the early 1990s, new solutions with non--spherical black hole horizons
(black tori) were found \cite{12lemos} for different states of matter and for
locally anti-de Sitter space times; for a recent review, see \cite{12rev}.
Static ellipsoidal black hole, black tori, anisotropic wormhole and Taub NUT
metrics and solitonic solutions of the vacuum and non--vacuum Einstein
equations were constructed in Refs. \cite{12v,12v2}. Non--trivial topology
configurations (for instance, black rings) are intensively studied in extra
dimension gravity \cite{12emp,12vth,12vtor}.

For four dimensional gravity (4D), it is considered that a number of
classical theorems \cite{12israel} impose that a stationary, asymptotically
flat, vacuum black hole solution is completely characterized by its mass and
spin and event horizons of non--spherical topology are forbidden \cite{12haw};
see \cite{12cen} for further discussion of this issue.

Nevertheless, there were constructed various classes of exact solutions in
4D and 5D gravity with non--trivial topology, anisotropies, solitonic
configurations, running constants and warped factors, under certain
conditions defining static configurations in 4D vacuum gravity. Such metrics
were parametrized by off--diagonal ansazt (for coordinate frames) which can
be effectively diagonalized with respect to certain anholonomic frames with
mixtures of holonomic and anholonomic variables. The system of vacuum
Einstein equations for such ansatz becomes exactly integrable and describe a
new "anholonomic nonlinear dynamics" of vacuum gravitational fields, which
posses generic local anisotropy. The new classes of solutions may have
locally isotropic limits, or can be associated to metric coefficients of
some well known, for instance, black hole, cylindrical, or wormhole soutions.

There is one important question if such anholonomic (anisotropic) solutions
can exist only in extra dimension gravity, with some specific effective
reductions to lower dimensions, or the anholonomic transforms generate a new
class of solutions even in general relativity theory which might be not
restricted by the conditions of Israel--Carter--Robinson uniqueness and
Hawking cosmic cenzorship theorems \cite{12israel,12haw}?

In the present paper, we explore possible 4D ellipsoidal black hole -- black
torus systems which are defined by generic off--diagonal matrices and
describe anholonomic vacuum gravitational configurations. We present a new
class of exact solutions of 4D vacuum Einstein equations which can be
associated to some exact solutions with ellipsoidal/toroidal horizons and
signularities, and theirs superpositions, being of static configuration, or,
in general, with nonlinear gravitational polarization and running constants.
We also discuss implications of these anisotropic solutions to gravity
theories and ponder possible ways to solve the problem with topologically
non--trivial and deformed horizons.

The organization of this paper is as follows:\ In Sec. II, we consider
ellipsoidal and torus deformations and anistoropic conformal transforms of
the Schwarzschild metric. We introduce an off--diagonal ansatz which can be
diagonalized by anholonomic transforms and compute the non--trivial
components of the vacuum Einstein equations in Sec. III. In Sec. IV, we
construct and analyze three types of exact static solutions with
ellisoidal--torus horizons. Sec. V is devoted to generalization of such
solutions for configurations with running constants and anisotropic
polarizations. The conclusion and discussion are presented in Sec. VI.

\section{Ellipsoidal/Torus Deformations of Metrics}

In this Section we analyze\ anholonomic transforms with ellipsoidal/torus
deformations of the Schwarzschild solution to some off--diagonal metrics. We
define the conditions when the \ new 'deformed' metrics are exact solutions
of vacuum Einstein equations.

The Schwarzschild solution may be written in {\it \ isotropic spherical
coordinates} $(\rho ,\theta ,\varphi )$ \thinspace \cite{12ll}
\begin{eqnarray}
dS^{2} &=&-\rho _{g}^{2}\left( \frac{\widehat{\rho }+1}{\widehat{\rho }}%
\right) ^{4}\left( d\widehat{\rho }^{2}+\widehat{\rho }^{2}d\theta ^{2}+%
\widehat{\rho }^{2}\sin ^{2}\theta d\varphi ^{2}\right)  \label{3schw} \\
&&+\left( \frac{\widehat{\rho }-1}{\widehat{\rho }+1}\right) ^{2}dt^{2},
\nonumber
\end{eqnarray}
where the isotropic radial coordinate $\rho $ is related with the usual
radial coordinate $r$ via the relation $r=\rho \left( 1+r_{g}/4\rho \right)
^{2}$ for $r_{g}=2G_{[4]}m_{0}/c^{2}$ being the 4D gravitational radius of a
point particle of mass $m_{0},$ $G_{[4]}=1/M_{P[4]}^{2}$ is the 4D Newton
constant expressed via Plank mass $M_{P[4]}.$ In our further considerations,
we put the light speed constant $c=1$ and re--scale the isotropic radial
coordinate as $\widehat{\rho }=\rho /\rho _{g},$ with $\rho _{g}=r_{g}/4.$
The metric (\ref{3schw}) is a vacuum static solution of 4D Einstein equations
with spherical symmetry describing the gravitational field of a point
particle of mass $m_{0}.$ It has a singularity for $r=0$ and a spherical
horizon for $r=r_{g},$ or, in re--scaled isotropic coordinates, for $%
\widehat{\rho }=1.$ We emphasize that this solution is parametrized by a
diagonal metric given with respect to holonomic coordinate frames.

We may introduce a new 'exponential' radial coordinate $\varsigma =\ln
\widehat{\rho }$ and write the Schwarzschild metric as
\begin{eqnarray}
ds^{2} &=&-\rho _{g}^{2}b\left( \varsigma \right) \left( d\varsigma
^{2}+d\theta ^{2}+\sin ^{2}\theta d\varphi ^{2}\right) +a\left( \varsigma
\right) dt^{2},  \label{schw1} \\
a\left( \varsigma \right) &=&\left( \frac{\exp \varsigma -1}{\exp \varsigma
+1}\right) ^{2},b\left( \varsigma \right) =\frac{\left( \exp \varsigma
+1\right) ^{4}}{(\exp \varsigma )^{2}}.  \label{scw1c}
\end{eqnarray}
The condition of vanishing of coefficient $a\left( \varsigma \right) ,\exp
\varsigma =1,$ defines the horizon 3D spherical hypersurface
\[
\varsigma =\varsigma \left[ \widehat{\rho }\left( \sqrt{x^{2}+y^{2}+z^{2}}%
\right) \right] ,
\]
where $x,y$ and $z$ are usual Cartezian coordinates.

The 3D spherical line element
\[
ds_{(3)}^{2}=d\varsigma ^{2}+d\theta ^{2}+\sin ^{2}\theta d\varphi ^{2},
\]
may be written in arbitrary ellipsoidal, or toroidal, coordinates which
transforms the spherical horizon equation into very sophisticate relations
(with respect to new coordinates).

Our idea is to deform (renormalize) the coefficients (\ref{scw1c}), $a\left(
\varsigma \right) \rightarrow A\left( \varsigma ,\theta \right) $ and $%
b\left( \varsigma \right) \rightarrow B\left( \varsigma ,\theta \right) ,$
as they would define a rotation ellipsoid and/or a toroidal horizon and
symmetry (for simplicity, we shall consider the elongated ellipsoid
configuration; the flattened ellipsoids may be analyzed in a similar
manner). \ But such a diagonal metric with respect to ellipsoidal, or
toroidal, local coordinate frame does not solve the vacuum Einstein
equations. In order to generate a new vacuum solution we have to
''elongate'' the differentials $d\varphi $ and $dt,$ i. e. to introduce some
''anholonomic transforms'' (see details in \cite{12vth}), like
\begin{eqnarray*}
d\varphi &\rightarrow &\delta \varphi +w_{\varsigma }\left( \varsigma
,\theta ,v\right) d\varsigma +w_{\theta }\left( \varsigma ,\theta ,v\right)
d\theta , \\
dt &\rightarrow &\delta t+n_{\varsigma }\left( \varsigma ,\theta ,v\right)
d\varsigma +n_{\theta }\left( \varsigma ,\theta ,v\right) d\theta ,
\end{eqnarray*}
for $v=\varphi $ (static configuration), or $v=t$ (running in time
configuration) and find the conditions when $w$- and $n$--coefficients and
the renormalized metric coefficients define off--diagonal metrics solving
the Einstein equations and possessing some ellipsoidal and/or toroidal
horizons and symmetries.

We shall define the 3D space ellipsoid -- toroidal configuration in this
manner: in the center of Cartezian coordinates we put an rotation ellipsoid
elongated along axis $z$ (its intersection by the $xy$--coordinate plane
describes a circle of radius $\rho _{g}^{[e]}=\sqrt{x^{2}+y^{2}}\sim \rho
_{g});$ the ellipsoid is surrounded by a torus with the same $z$ axis of
symmetry, when $-z_{0}\leq z\leq z_{0},$ and the intersections of the torus
with the $xy$--coordinate plane describe two circles of radia $\rho
_{g}^{[t]}-z_{0}=\sqrt{x^{2}+y^{2}}$ and $\rho _{g}^{[t]}+z_{0}=\sqrt{%
x^{2}+y^{2}};$ the parameters $\rho _{g}^{[e]},\rho _{g}^{[t]}$ and $z_{0}$
are chosen as to define not intersecting toroidal and ellipsoidal horizons,
i. e. the conditions
\begin{equation}
\rho _{g}^{[t]}-z_{0}>\rho _{g}^{[e]}>0  \label{scale}
\end{equation}
are imposed.

\subsection{Ellipsoidal Configurations}

We shall consider the {\it \ rotation ellipsoid coordinates} \cite{12korn} $%
(u,\lambda ,\varphi )$ with $0\leq u<\infty ,0\leq \lambda \leq \pi ,0\leq
\varphi \leq 2\pi ,$ where $\sigma =\cosh u\geq 1,$ are related with the
isotropic 3D Cartezian coordinates $\left( x,y,z\right) $ as
\begin{eqnarray}
(x &=&\widetilde{\rho }\sinh u\sin \lambda \cos \varphi ,  \label{22rec} \\
y &=&\widetilde{\rho }\sinh u\sin \lambda \sin \varphi ,z=\widetilde{\rho }%
\cosh u\cos \lambda )  \nonumber
\end{eqnarray}
and define an elongated rotation ellipsoid hypersurface
\begin{equation}
\left( x^{2}+y^{2}\right) /(\sigma ^{2}-1)+\tilde{z}^{2}/\sigma ^{2}=%
\widetilde{\rho }^{2}.  \label{2reh}
\end{equation}
with $\sigma =\cosh u.$ The 3D metric on a such hypersurface is
\[
dS_{(3D)}^{2}=g_{uu}du^{2}+g_{\lambda \lambda }d\lambda ^{2}+g_{\varphi
\varphi }d\varphi ^{2},
\]
where
\begin{eqnarray*}
g_{uu} &=&g_{\lambda \lambda }=\widetilde{\rho }^{2}\left( \sinh ^{2}u+\sin
^{2}\lambda \right) , \\
g_{\varphi \varphi } &=&\widetilde{\rho }^{2}\sinh ^{2}u\sin ^{2}\lambda .
\end{eqnarray*}

We can relate the rotation ellipsoid coordinates\newline
$\left( u,\lambda ,\varphi \right) $ from (\ref{22rec}) with the isotropic
radial coordinates $\left( \widehat{\rho },\theta ,\varphi \right) $, scaled
by the constant $\rho _{g}, $ from (\ref{3schw}), $\ $equivalently with
coordinates $\left( \varsigma ,\theta ,\varphi \right) $ from (\ref{schw1}),
as
\[
\widetilde{\rho }=1,\cosh u=\widehat{\rho }=\exp \varsigma
\]
and deform the Schwarzschild metric by introducing ellipsoidal coordinates
and a new horizon defined by the condition that vanishing of the metric
coefficient before $dt^{2}$ describe an elongated rotation ellipsoid
hypersurface (\ref{2reh}),
\begin{eqnarray}
ds_{E}^{2} &=&-\rho _{g}^{2}\left( \frac{\cosh u+1}{\cosh u}\right)
^{4}(\sinh ^{2}u+\sin ^{2}\lambda )  \label{2schel} \\
&&\times \lbrack du^{2}+d\lambda ^{2}+\frac{\sinh ^{2}u~\sin ^{2}\lambda }{%
\sinh ^{2}u+\sin ^{2}\lambda }d\varphi ^{2}]  \nonumber \\
& &+\left( \frac{\cosh u-1}{\cosh u+1}\right) ^{2}dt^{2}.  \nonumber
\end{eqnarray}
The ellipsoidaly deformed metric (\ref{2schel}) does not satisfy the vacuum
Einstein equations, but at long distances from the horizon it transforms
into the usual Schwarzschild solution (\ref{3schw}).

We introduce two Classes (A and B) of 4D auxiliary pseudo--Riemannian
metrics, also given in ellipsoid coordinates, being some conformal
transforms of (\ref{2schel}), like
\[
ds_{E}^{2}=\Omega _{A(B)E}\left( u,\lambda \right) ds_{A(B)E}^{2}
\]
which also are not supposed to be solutions of the Einstein equations:

Metric of Class A:
\begin{equation}
ds_{(AE)}^{2}=-du^{2}-d\lambda ^{2}+a_{E}(u,\lambda )d\varphi
^{2}+b_{E}(u,\lambda )dt^{2}],  \label{3auxm1}
\end{equation}
where
\begin{eqnarray}
a_{E}(u,\lambda ) &=&-\frac{\sinh ^{2}u~\sin ^{2}\lambda }{\sinh ^{2}u+\sin
^{2}\lambda },  \label{auxm1s} \\
b_{E}(u,\lambda ) &=&\frac{(\cosh u-1)^{2}\cosh ^{4}u}{\rho _{g}^{2}(\cosh
u+1)^{6}(\sinh ^{2}u+\sin ^{2}\lambda )},  \nonumber
\end{eqnarray}
which results in the metric (\ref{2schel}) by multiplication on the conformal
factor
\begin{equation}
\Omega _{AE}\left( u,\lambda \right) =\rho _{g}^{2}\frac{(\cosh u+1)^{4}}{%
\cosh ^{4}u}(\sinh ^{2}u+\sin ^{2}\lambda ).  \label{3auxm1c}
\end{equation}

Metric of Class B:
\begin{equation}
ds_{BE}^{2}=g_{E}(u,\lambda )\left( du^{2}+d\lambda ^{2}\right) -d\varphi
^{2}+f_{E}(u,\lambda )dt^{2},  \label{3auxm2}
\end{equation}
where
\begin{eqnarray}
g_{E}(u,\lambda ) &=&-\frac{\sinh ^{2}u+\sin ^{2}\lambda }{\sinh ^{2}u~\sin
^{2}\lambda },  \nonumber \\
f_{E}(u,\lambda ) &=&\frac{(\cosh u-1)^{2}\cosh ^{4}u}{\rho _{g}^{2}(\cosh
u+1)^{6}\sinh ^{2}u\sin ^{2}\lambda },  \label{auxm2f}
\end{eqnarray}
which results in the metric (\ref{2schel}) by multiplication on the conformal
factor
\[
\Omega _{BE}\left( u,\lambda \right) =\rho _{g}^{2}\frac{(\cosh u+1)^{4}}{%
\cosh ^{4}u}\sinh ^{2}u\sin ^{2}\lambda .
\]

In Ref. \cite{12vth} we proved that there are anholonomic transforms of the
metrics (\ref{2schel}), (\ref{3auxm1}) and (\ref{3auxm2}) which results in
exact ellipsoidal black hole solutions of the vacuum Einstein equations.

\subsection{Toroidal Configurations}

Fixing a scale parameter $\rho _{g}^{[t]}$ which satisfies the conditions (%
\ref{scale}) we define the {\it \ toroidal coordinates} $(\sigma ,\tau
,\varphi )$ (we emphasize that in in this paper we use different letters for
ellipsoidal and toroidal coordinates introduced in Ref. \cite{12korn}). These
coordinates run the values $-\pi \leq \sigma <\pi ,0\leq \tau \leq \infty
,0\leq \varphi <2\pi .$ They are related with the isotropic 3D Cartezian
coordinates via transforms
\begin{eqnarray}
\tilde{x} &=&\frac{\widetilde{\rho }\sinh \tau }{\cosh \tau -\cos \sigma }%
\cos \varphi ,  \label{rect} \\
\tilde{y} &=&\frac{\widetilde{\rho }\sinh \tau }{\cosh \tau -\cos \sigma }%
\sin \varphi ,\tilde{z}=\frac{\widetilde{\rho }\sinh \sigma }{\cosh \tau
-\cos \sigma }  \nonumber
\end{eqnarray}
and define a toroidal hypersurface
\[
\left( \sqrt{\tilde{x}^{2}+\tilde{y}^{2}}-\widetilde{\rho }\frac{\cosh \tau
}{\sinh \tau }\right) ^{2}+\tilde{z}^{2}=\frac{\widetilde{\rho }^{2}}{\sinh
^{2}\tau }.
\]
The 3D metric on a such toroidal hypersurface is
\[
ds_{(3D)}^{2}=g_{\sigma \sigma }d\sigma ^{2}+g_{\tau \tau }d\tau
^{2}+g_{\varphi \varphi }d\varphi ^{2},
\]
where
\begin{eqnarray*}
g_{\sigma \sigma } &=&g_{\tau \tau }=\frac{\widetilde{\rho }^{2}}{\left(
\cosh \tau -\cos \sigma \right) ^{2}}, \\
g_{\varphi \varphi } &=&\frac{\widetilde{\rho }^{2}\sinh ^{2}\tau }{\left(
\cosh \tau -\cos \sigma \right) ^{2}}.
\end{eqnarray*}

We can relate the toroidal coordinates $\left( \sigma ,\tau ,\varphi \right)
$ from (\ref{rect}) with the isotropic radial coordinates $\left( \widehat{%
\rho }^{[t]},\theta ,\varphi \right) $, scaled by the constant $\rho
_{g}^{[t]},$ as
\[
\widetilde{\rho }=1,\sinh ^{-1}\tau =\widehat{\rho }^{[t]}
\]
and transform the Schwarzschild solution into a new metric with toroidal
coordinates \ by changing the 3D radial line element into the toroidal one
and stating the $tt$--coefficient of the metric to have a toroidal horizon.
The resulting metric is
\begin{eqnarray}
ds_{T}^{2} &=&-\left( \rho _{g}^{[t]}\right) ^{2}\frac{\left( \sinh \tau
+1\right) ^{4}}{\left( \cosh \tau -\cos \sigma \right) ^{2}} \times
\label{2schtor} \\
&& \left( d\sigma ^{2}+d\tau ^{2}+\sinh ^{2}\tau d\varphi ^{2}\right) +
\left( \frac{\sinh \tau -1}{\sinh \tau +1}\right) ^{2}dt^{2},  \nonumber
\end{eqnarray}
Such a deformed Schwarzschild like toroidal metric is not an exact solution
of the vacuum Einstein equations, but at long radial distances it transforms
into the usual Schwarzschild solution with effective horizon $\rho _{g}^{[t]}$
with the 3D line element parametrized by toroidal coordinates.

We introduce two Classes (A and B) of 4D auxiliary pseudo--Riemannian
metrics, also given in toroidal coordinates, being some conformal transforms
of (\ref{2schtor}), like
\[
ds_{T}^{2}=\Omega _{A(B)T}\left( \sigma ,\tau \right) ds_{A(B)T}^{2}
\]
but which are not supposed to be solutions of the Einstein equations:

Metric of Class A:
\begin{equation}
ds_{AT}^{2}=-d\sigma ^{2}-d\tau ^{2}+a_{T}(\tau )d\varphi ^{2}+b_{T}(\sigma
,\tau )dt^{2},  \label{auxm1t}
\end{equation}
where
\begin{eqnarray}
a_{T}(\tau ) &=&-\sinh ^{2}\tau ,  \nonumber \\
b_{T}(\sigma ,\tau ) &=&\frac{\left( \sinh \tau -1\right) ^{2}\left( \cosh
\tau -\cos \sigma \right) ^{2}}{\rho _{g}^{[t]2}\left( \sinh \tau +1\right)
^{6}},  \label{auxm1tb}
\end{eqnarray}
which results in the metric (\ref{2schtor}) by multiplication on the
conformal factor
\begin{equation}
\Omega _{AT}\left( \sigma ,\tau \right) =\rho _{g}^{[t]2}\frac{\left( \sinh
\tau +1\right) ^{4}}{\left( \cosh \tau -\cos \sigma \right) ^{2}}.
\label{auxm1tc}
\end{equation}

Metric of Class B:
\begin{equation}
ds_{BT}^{2}=g_{T}(\tau )\left( d\sigma ^{2}+d\tau ^{2}\right) -d\varphi
^{2}+f_{T}(\sigma ,\tau )dt^{2},  \label{auxm2t}
\end{equation}
where
\begin{eqnarray*}
g_{T}(\tau ) &=&-\sinh ^{-2}\tau , \\
f_{T}(\sigma ,\tau ) &=&\rho _{g}^{[t]2}\left( \frac{\sinh ^{2}\tau -1}{%
\cosh \tau -\cos \sigma }\right) ^{2},
\end{eqnarray*}
which results in the metric (\ref{2schtor}) by multiplication on the
conformal factor
\begin{equation}
\Omega _{BT}\left( \sigma ,\tau \right) =\left( \rho _{g}^{[t]}\right) ^{-2}%
\frac{\left( \cosh \tau -\cos \sigma \right) ^{2}}{\left( \sinh \tau
+1\right) ^{2}}.  \label{auxm2tbc}
\end{equation}

In Ref. \cite{12vtor} we used the metrics (\ref{2schtor}), (\ref{auxm1t}) and (%
\ref{auxm2t}) in order to generate exact solutions of the Einstein equations
with toroidal horizons and anisotropic polarizations and running constants
by performing corresponding anholonomic transforms.

\section{The Metric Ansatz and Vacuum Einstein Equations}

Let us denote the local system of coordinates as $u^{\alpha }=\left(
x^{i},y^{a}\right) ,$ where $\ x^{1}=u$ and $x^{2}=\lambda $ for ellipsoidal
coordinates ($x^{1}=\sigma $ and $x^{2}=\tau $ for toroidal coordinates) and
$y^{3}=v=\varphi $ and $y^{4}=t$ for the so--called $\varphi $--anisotropic
configurations ($y^{4}=v=t$ and $y^{5}=\varphi $ for the so--called $t$%
--anisotropic configurations). Our spacetime is modelled as a 4D
pseudo--Riemannian space of signature $\left( -,-,-,+\right) $ (or $\left(
-,-,+,-\right) ),$ which in general may be enabled with an anholonomic frame
structure (tetrads, or vierbiend) $e_{\alpha }=A_{\alpha }^{\beta }\left(
u^{\gamma }\right) \partial /\partial u^{\beta }$ subjected to some
anholonomy \ relations
\begin{equation}
e_{\alpha }e_{\beta }-e_{\beta }e_{\alpha }=W_{\alpha \beta }^{\gamma
}\left( u^{\varepsilon }\right) e_{\gamma },  \label{3anhol}
\end{equation}
where $W_{\alpha \beta }^{\gamma }\left( u^{\varepsilon }\right) $ are
called the coefficients of anholonomy.

The anholonomically and conformally transformed 4D line element is
\begin{equation}
ds^{2}=\Omega ^{2}(x^{i},v)\hat{{g}}_{\alpha \beta }\left( x^{i},v\right)
du^{\alpha }du^{\beta },  \label{5cmetric4}
\end{equation}
were the coefficients $\hat{{g}}_{\alpha \beta }$ are parametrized by the
ansatz {\scriptsize
\begin{equation}
\left[
\begin{array}{cccc}
g_{1}+\zeta _{1}^{\ 2}h_{3}+n_{3}^{\ 2}h_{4} & \zeta _{1}\zeta
_{2}h_{3}+n_{1}n_{2}h_{4} & \zeta _{1}h_{3} & n_{1}h_{4} \\
\zeta _{1}\zeta _{2}h_{3}+n_{1}n_{2}h_{4} & g_{2}+\zeta _{2}^{\
2}h_{3}+n_{3}^{\ 2}h_{4} & +\zeta _{2}h_{3} & n_{2}h_{4} \\
\zeta _{1}h_{3} & \zeta _{2}h_{3} & h_{3} & 0 \\
n_{1}h_{4} & n_{2}h_{4} & 0 & h_{4}
\end{array}
\right],  \label{3ansatzc4}
\end{equation}
} with $g_{i}=g_{i}\left( x^{i}\right) ,h_{a}=h_{ai}\left( x^{k},v\right)
,n_{i}=n_{i}\left( x^{k},v\right) ,$ $\zeta _{i}=\zeta _{i}\left(
x^{k},v\right) $ $\ $\ and $\Omega =\Omega \left( x^{k},v\right) $ being
some functions of necessary smoothly class or even singular in some points
and finite regions. So, the $g_{i}$--components of our ansatz depend only on
''holonomic'' variables $x^{i}$ and the rest of coefficients may also depend
on ''anisotropic'' (anholonomic) variable $y^{3}=v;$ our ansatz does not
depend on the second anisotropic variable $y^{4}.$

We may diagonalize the line element
\begin{equation}
\delta s^{2}=\Omega ^{2}[g_{1}(dx^{1})^{2}+g_{2}(dx^{2})^{2}+h_{3}(\delta
v)^{2}+h_{4}(\delta y^{4})^{2}],  \label{4dmetric4}
\end{equation}
with respect to the anholonomic co--frame \newline
$\delta ^{\alpha }=\left( dx^{i},\delta v,\delta y^{4}\right) ,$ where
\begin{equation}
\delta v=dv+\zeta _{i}dx^{i}\mbox{ and }\delta y^{4}=dy^{4}+n_{i}dx^{i},
\label{4ddif4}
\end{equation}
which is dual to the frame $\delta _{\alpha }=\left( \delta _{i},\partial
_{4},\partial _{5}\right) ,$ where
\begin{equation}
\delta _{i}=\partial _{i}+\zeta _{i}\partial _{3}+n_{i}\partial _{4}.
\label{4dder4}
\end{equation}
The tetrads $\delta _{\alpha }$ and $\delta ^{\alpha }$ are anholonomic
because, in general, they satisfy some non--trivial anholonomy relations (%
\ref{3anhol}). The anholonomy is induced by the coefficients $\zeta _{i}$ and
$n_{i}$ which ''elongate'' partial derivatives and differentials if we are
working with respect to anholonomic frames. This result in a more
sophisticate differential and integral calculus (a usual situation in
'tetradic' and 'spinor' gravity), but simplifies substantially tensor
computations, because we are dealing with diagonalized metrics.

The vacuum Einstein equations for the (\ref{3ansatzc4}) (equivalently, for (%
\ref{4dmetric4})), $R_{\alpha }^{\beta }=0,$ computed with respect to
anholonomic frames (\ref{4ddif4}) and (\ref{4dder4}), transforms into a system
of partial differential equations \cite{12v,12vth,12vtor}:
\begin{eqnarray}
R_{1}^{1}=R_{2}^{2}=-\frac{1}{2g_{1}g_{2}}[g_{2}^{\bullet \bullet }-\frac{%
g_{1}^{\bullet }g_{2}^{\bullet }}{2g_{1}}-\frac{(g_{2}^{\bullet })^{2}}{%
2g_{2}} &&  \nonumber \\
+g_{1}^{^{\prime \prime }}-\frac{g_{1}^{^{\prime }}g_{2}^{^{\prime }}}{2g_{2}%
}-\frac{(g_{1}^{^{\prime }})^{2}}{2g_{1}}] &=&0,  \label{6ricci1a} \\
R_{3}^{3}=R_{4}^{4}=\frac{-1}{2h_{3}h_{4}}\left[ h_{4}^{\ast \ast
}-h_{4}^{\ast }\left( \ln \sqrt{|h_{3}h_{4}|}\right) ^{\ast }\right] &=&0,
\label{5ricci2a} \\
R_{4i}=-\frac{h_{4}}{2h_{3}}\left[ n_{i}^{\ast \ast }+\gamma n_{i}^{\ast }%
\right] &=&0,  \label{24ricci4a}
\end{eqnarray}
where
\begin{equation}
\gamma =3h_{4}^{\ast }/2h_{4}-h_{3}^{\ast }/h_{3},  \label{5abc}
\end{equation}
and the partial derivatives are written in brief like $g_{1}^{\bullet }=\partial
g_{1}/\partial x^{1},g_{1}^{^{\prime }}=\partial g_{1}/\partial x^{2}$ and $%
h_{3}^{\ast }=\partial h_{3}/\partial v.$ The coefficients $\zeta _{{i}}$
are found as to consider non--trivial conformal factors $\Omega :$ we
compensate by $\zeta _{{i}}$ possible conformal deformations of the Ricci
tensors, computed with respect to anholonomic frames. The conformal
invariance of such anholonomic transforms holds if
\begin{equation}
\Omega ^{q_{1}/q_{2}}=h_{3}~(q_{1}\mbox{ and }q_{2}\mbox{ are
integers}),  \label{4confq}
\end{equation}
and there are satisfied the equations
\begin{equation}
\partial _{i}\Omega -\zeta _{{i}}\Omega ^{\ast }=0.  \label{5confeq}
\end{equation}

The system of equations (\ref{6ricci1a})--(\ref{24ricci4a}) and (\ref{5confeq})
can be integrated in general form \cite{12vth}. Physical solutions are defined
from some additional boundary conditions, imposed types of symmetries,
nonlinearities and singular behavior and compatibility in locally
anisotropic limits with some well known exact solutions.

In this paper we give some examples of ellipsoidal and toroidal solutions
and investigate some classes of metrics for combined ellipsoidal black hole
-- black tori configurations.

\section{Static Black Hole -- Black Torus Metrics}

We analyzed in detail the method of anholonomic frames and constructed 4D
and 5D ellipsoidal black hole and black tori solutions in Refs. \cite
{12v,12vth,12vtor}. In this Section we give same new examples of metrics
describing \ one static 4D black hole \ or one static 4D black torus
configurations. Then we extend the constructions for metrics describing
combined variants of black hole -- black torus solutions. We shall analyze
solutions with trivial and non--trivial conformal factors.

In this section the 4D local coordinates are written as $\left(
x^{1},x^{2},y^{3}=v=\varphi ,y^{4}=t\right) ,$ where we take $%
x^{i}=(u,\lambda )$ for ellipsoidal configurations and $x^{i}=(\sigma ,\tau
) $ for toroidal configurations. Here we note that, we can introduce a
''general'' 2D space ellipsoidal coordinate system, $u=u(\sigma ,\tau )$ and
$\lambda =\tau ,$\ for both ellipsoidal and toroidal configurations if, for
instance, we identify the ellipsoidal coordinate $\lambda $ with the
toroidal $\tau ,$ and relate $u$ with $\sigma $ and $\tau $ as
\[
\sinh u=\frac{1}{\cosh \tau -\cos \sigma }.
\]
In the vicinity of $\tau =0$ we can approximate $\cosh \tau \approx 1$ and to
write $u=u\left( \sigma \right) $and $\lambda =\tau .$ For $\tau \gg 1$ we
have
\[
\sinh u\approx \frac{1}{\cosh \tau }\left( 1+\frac{1}{\cos \sigma }\right) .
\]
In general, we consider that the ''holonomic'' coordinates are some
functions $x^{i}=x^{i}\left( \sigma ,\tau \right) =\widetilde{x}^{i}\left(
u,\lambda \right) $ for which the 2D line element can be written in
conformal metric form,
\[
ds_{[2]}^{2}=-\mu ^{2}\left( x^{i}\right) \left[ \left( dx^{1}\right)
^{2}+\left( dx^{2}\right) ^{2}\right] .
\]
For simplicity, we consider 4D coordinate parametrizations when the angular
coordinate $\varphi $ and the time like coordinate $t$ are not affected by
any transforms of $x$--coordinates.

\subsection{Static anisotropic black hole/torus solutions}

\subsubsection{An example of ellipsoidal\ black hole configuration}

The simplest way to generate a static but anisotropic ellipsoidal black hole
solution with an anholonomically diagonalized metric (\ref{4dmetric4}) is \
to take a metric of type (\ref{3auxm1}), to ''elongate'' its differentials,
\begin{eqnarray*}
d\varphi &\rightarrow &\delta \varphi =d\varphi +\zeta _{i}\left(
x^{k},\varphi \right) dx^{i}, \\
dt &\rightarrow &\delta t=dt+n_{i}\left( x^{k},\varphi \right) dx^{i},
\end{eqnarray*}
than to multiply on a conformal factor
\[
\Omega ^{2}\left( x^{k},\varphi \right) =\omega ^{2}\left( x^{k},\varphi
\right) \Omega _{AE}^{2}(x^{k}),
\]
the factor $\omega ^{2}\left( x^{k},\varphi \right) $ is obtained by
re--scaling the constant $\rho _{g}$ from (\ref{3auxm1c}),
\begin{equation}
\rho _{g}\rightarrow \overline{\rho }_{g}=\omega \left( x^{k},\varphi
\right) \rho _{g},  \label{oresc}
\end{equation}
in the simplest case we can consider only ''angular'' on $\varphi $
anisotropies. Then we 'renormalize' (by introducing $x^{i}$ coordinates) the
$g_{1},g_{2}$ and $h_{3}$ coefficients,
\begin{eqnarray}
g_{1,2} &=&-1\rightarrow -\mu ^{2}\left( x^{i}\right) ,  \label{parsol1} \\
h_{3} &=&h_{3[0]}=a_{E}(u,\lambda )\rightarrow h_{3}=-\Omega ^{-2}\left(
x^{k},\varphi \right) ,  \label{parsol1a}
\end{eqnarray}
we fix a relation of type (\ref{4confq}), and take $h_{4}=$ $%
h_{4[0]}=b_{E}(x^{i}).$ The anholonomically transformed metric is
pa\-ra\-metrized in the form
\begin{eqnarray}
\delta s^{2} &=&\Omega ^{2}\{-\mu ^{2}\left( x^{i}\right) \left[
(dx^{1})^{2}+(dx^{2})^{2}\right]  \label{sol1} \\
&&-\Omega ^{-2}\left( x^{k},\varphi \right) (\delta
v)^{2}+b_{E}(x^{i})(\delta y^{4})^{2}\},  \nonumber
\end{eqnarray}
where $\mu ,\zeta _{i}$ and $n_{i}$ are to be defined respectively from the
equations (\ref{6ricci1a}), (\ref{5confeq}) and (\ref{24ricci4a}). We note that
the equation (\ref{5ricci2a}) is already solved because in our case $%
h_{4}^{\ast }=0.$

The equation (\ref{6ricci1a}), with partial derivations on coordinates $x^{i}
$ and parametrizations (\ref{parsol1}) has the general solution
\begin{equation}
\mu ^{2}=\mu _{\lbrack 0]}^{2}\exp \left[ c_{[1]}x^{1}\left( u,\lambda
\right) +c_{[2]}x^{2}\left( u,\lambda \right) \right] ,  \label{f1}
\end{equation}
where $\mu _{\lbrack 0]},c_{[1]}$ and $c_{[2]}$ are some constants which
should be defined from boundary conditions and by fixing a corresponding 2D
system of coordinates; we pointed that we may redefine the factor (\ref{f1})
in 'pure' ellipsoidal coordinates $\left( u,\lambda \right) .$

The general solution of (\ref{5confeq}) for renormalization (\ref{oresc}) and
parametrization\\ (\ref{parsol1a}) is
\begin{eqnarray}
\zeta _{i}\left( x^{k},\varphi \right) &=&\left( \omega ^{\ast }\right)
^{-1}\partial _{i}\omega +\partial _{i}\ln |\Omega _{AE}|/\left( \ln |\omega
|\right) ^{\ast },  \label{e1} \\
&=&\partial _{i}\ln |\Omega _{AE}|/\left( \ln |\omega |\right) ^{\ast }%
\mbox{ for }\omega =\omega \left( \varphi \right) .  \nonumber
\end{eqnarray}
For a given $h_{3}$ with $h_{4}^{\ast }=0,$ we can compute the coefficient $%
\gamma $ from (\ref{5abc}). After two integrations on $\varphi $ in (\ref
{24ricci4a}) we find
\begin{equation}
n_{i}\left( x^{k},\varphi \right) =n_{i[0]}\left( x^{k}\right)
+n_{i[1]}\left( x^{k}\right) \int \omega ^{-2}d\varphi .  \label{n1}
\end{equation}
The set of functions (\ref{f1}), (\ref{e1}) and (\ref{n1}) for any given $%
\Omega _{AE}\left( x^{i}\right) $ and $\omega \left( x^{k},\varphi \right) $
defines an exact static solution of the vacuum Einstein equations
parametrized by an off--diagonal metric of type (\ref{sol1}). This solution
have an ellipsoidal horizon defined by the condition of vanishing of the
coefficient $h_{4[0]}=b_{E}(x^{i}),$ see the coefficients for the auxiliary
metric (\ref{3auxm1}) and an anisotropic effective constant (\ref{oresc}).
This is a general solution depending on arbitrary functions $\omega \left(
x^{k},\varphi \right) $ and $n_{i[0,1]}\left( x^{k}\right) $ and constants $%
\mu _{\lbrack 0]},c_{[1]}$ and $c_{[2]}$ which have to be stated
from some additional physical arguments.

For instance, if we wont to impose the condition that our solution, far away
from the ellipsoidal horizon, transform into the Schwarzschild solution with
an effective anisotropic ''mass'', or a renormalized gravitational Newton
constant, we may put $\mu _{\lbrack 0]}=1$ and fix the $x^{i}$--coordinates
and constants $c_{[1,2]}$ as to obtain the linear interval
\[
ds_{[2]}^{2}=-\left[ du^{2}+d\lambda ^{2}\right] .
\]
The coefficients $n_{i[0,1]}\left( x^{k}\right) $ and $\omega \left(
x^{k},\varphi \right) $ may be taken as at long distances from the horizon
one holds the limits $n_{i[0,1]}\left( x^{k}\right) \rightarrow 0$ and $%
\zeta _{i}\left( x^{k},\varphi \right) \rightarrow 0$ for $\omega \left(
x^{k},\varphi \right) \rightarrow 0.$ In this case, at asymptotic, our
solution will transform into a Schwarzschild like solution with
''renormalized'' parameter $\overline{\rho }_{g}\rightarrow const.$

Nevertheless, we consider that it is not obligatory to select only such type
of ellipsoidal solutions (with imposed asymptotic spherical symmetry)
parametrized by metrics of class (\ref{sol1}). The system of vacuum
gravitational equations (\ref{6ricci1a})--(\ref{5confeq}) for the ansatz (\ref
{sol1}) defines a nonlinear static configuration (an alternative vacuum
Einstein configuration with ellipsoidal horizon) which, in general, is not
equivalent to the Schwarzschild vacuum. This points to some specific
properties of the gravitational vacuum which follow from the nonlinear
character of the Einstein equations. In quantum field theory the nonlinear
effects may result in unitary non--equivalent vacua; in classical
gravitational theories we could obtain a similar behavior if we are dealing
with off--diagonal metrics and anholonomic frames.

\begin{figure*}
\includegraphics[scale=0.8]{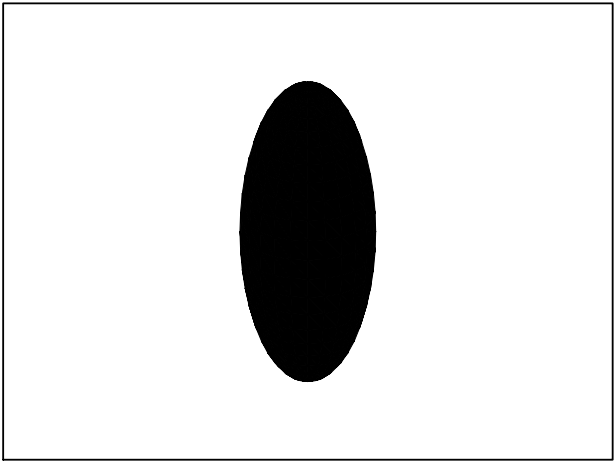}
\caption{\label{ellipsoid}Ellipsoidal Configuration}
\end{figure*}

The constructed new static vacuum solution (\ref{sol1}) for a 4D ellipsoidal
black hole is stated by the coefficients
\begin{eqnarray}
g_{1,2} &=&-1,\mu =1,\overline{\rho }_{g}=\omega \left( x^{k},\varphi
\right) \rho _{g},\Omega ^{2}=\omega ^{2}\Omega _{AE}^{2},  \nonumber \\
h_{3} &=&-\Omega ^{-2}\left( x^{k},\varphi \right) ,h_{4}=b_{E}(x^{i}), (%
\mbox{ see (\ref{3auxm1}),(\ref{3auxm1c})}),  \nonumber \\
\zeta _{i} &=&\left( \omega ^{\ast }\right) ^{-1}\partial _{i}\omega
+\partial _{i}\ln |\Omega _{AE}|/\left( \ln |\omega |\right) ^{\ast },
\nonumber \\
n_{i} &=&n_{i[0]}\left( x^{k}\right) +n_{i[1]}\left( x^{k}\right) \int
\omega ^{-2}d\varphi .  \label{data1}
\end{eqnarray}
These data define an ellipsoidal configuration, see Fig.
\ref{ellipsoid}.

%%%%%%%%%%%%%%%%%%%%%%%%%%%%%%%%%%%%
% Figure 1: Ellipsoidal Configuration
%%%%%%%%%%%%%%%%%%%%%%%%%%%%%%%%%%%

Finally, we remark that we have generated a vacuum ellipsoidal gravitational
configuration starting from the metric (\ref{3auxm1}), i. e. we constructed
an ellipsoidal $\varphi $--solution of Class A (see details on
classification in \cite{12vth}). In a similar manner we can define anholonomic
deformations of the metric (\ref{3auxm2}) and renormalization of conformal
factor $\Omega _{BE}\left( u,\lambda \right) $ in order to construct an
ellipsoidal $\varphi $--solution of Class B. We omit such considerations in
this paper but present, in the next subsection, an example of toroidal $%
\varphi $--solution of Class B.

\subsubsection{An example of toroidal\ black hole configuration}

We start with the metric (\ref{auxm2t}), ''elongate'' its differentials $%
d\varphi \rightarrow \delta \varphi $ and $dt\rightarrow \delta t$ and than
multiply on a conformal factor
\[
\Omega ^{2}\left( x^{k},\varphi \right) =\varpi ^{2}\left( x^{k},\varphi
\right) \Omega _{BT}^{2}(x^{k})g_{T}\left( \tau \right) ,
\]
see (\ref{auxm2tbc}) which is connected with the renormalization of constant
$\rho _{g}^{[t]},$
\begin{equation}
\rho _{g}^{[t]}\rightarrow \overline{\rho }_{g}^{[t]}=\varpi \left(
x^{k},\varphi \right) \rho _{g}^{[t]}.  \label{osect}
\end{equation}
For toroidal configurations it is naturally to use 2D toroidal holonomic
coordinates $x^{i}=(\sigma ,\tau ).$

\begin{figure*}
\includegraphics[scale=0.8]{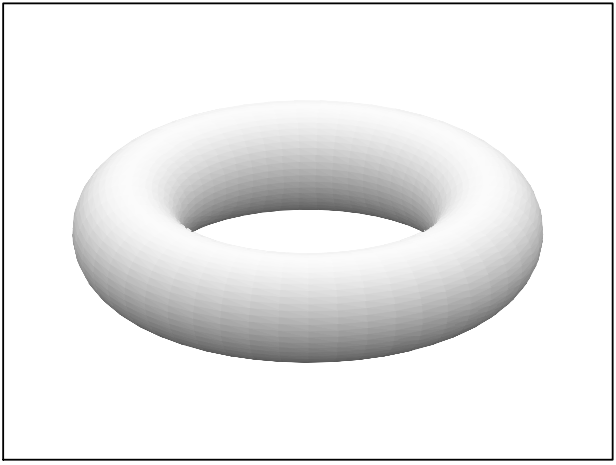}
\caption{\label{1torus}Toroidal Configuration}
\end{figure*}

The anholonomically transformed metric is pa\-ra\-met\-rized in the form
\begin{eqnarray}
\delta s^{2} &=&\Omega ^{2}\{-\left[ d\sigma ^{2}+d\tau ^{2}\right] -\eta
_{3}\left( \sigma ,\tau ,\varphi \right) g_{T}^{-1}\left( \tau \right)
\delta \varphi ^{2}  \nonumber \\
&&+f_{T}(\sigma ,\tau )g_{T}^{-1}\left( \tau \right) \delta t^{2}\}.
\label{sol2}
\end{eqnarray}
We state the coefficients
\[
h_{3}=-\eta _{3}\left( \sigma ,\tau ,\varphi \right) g_{T}^{-1}\left( \tau
\right) \mbox{ and }h_{4}=f_{T}(\sigma ,\tau )g_{T}^{-1}\left( \tau \right)
,
\]
where the polarization
\[
\eta _{3}\left( \sigma ,\tau ,\varphi \right) =\varpi ^{-2}\left( \sigma
,\tau ,\varphi \right) \Omega _{BT}^{-2}(\sigma ,\tau )
\]
is found from the condition (\ref{4confq}) as $h_{3}=-\Omega ^{-2}.$ The
equation (\ref{5ricci2a}) is solved by arbitrary couples $h_{3}\left( \sigma
,\tau ,\varphi \right) $ $\ $and $h_{4}(\sigma ,\tau )$ when $h_{4}^{\ast
}=0.$ The procedure of definition of $\zeta _{i}\left( \sigma ,\tau ,\varphi
\right) $ and $n_{i}\left( \sigma ,\tau ,\varphi \right) $ is similar to
that from the previous subsection. We present the final results as the data
\begin{eqnarray}
g_{1,2} &=&-1,\overline{\rho }_{g}=\varpi \left( \sigma ,\tau ,\varphi
\right) \rho _{g},\Omega ^{2}=\varpi ^{2}\Omega _{BT}^{2}g_{T}\left( \tau
\right) ,  \nonumber \\
h_{3} &=&-\eta _{3}\left( \sigma ,\tau ,\varphi \right) g_{T}^{-1}\left(
\tau \right) ,h_{4}=f_{T}(\sigma ,\tau )g_{T}^{-1}\left( \tau \right) ,
\nonumber \\
\eta _{3} &=&\varpi ^{-2}\left( \sigma ,\tau ,\varphi \right) \Omega
_{BT}^{-2}(\sigma ,\tau ), (\mbox{ see (\ref{auxm2t}),
(\ref{auxm2tbc})}),  \nonumber \\
\zeta _{i} &=&\left( \varpi ^{\ast }\right) ^{-1}\partial _{i}\varpi
+\partial _{i}\ln |\Omega _{BT}\sqrt{g_{T}}|/\left( \ln |\varpi |\right)
^{\ast },  \label{data2} \\
n_{i} &=&n_{i[0]}\left( \sigma ,\tau \right) +n_{i[1]}\left( \sigma ,\tau
\right) \int \varpi ^{-2}d\varphi  \nonumber
\end{eqnarray}
for the ansatz (\ref{sol2}) which defines an exact static
solution of the vacuum Einstein equations with toroidal symmetry,
of Class B, with anisotropic dependence on coordinate $\varphi ,$
see the torus configuration from Fig.  \ref{1torus}. The
off--diagonal solution is non--trivial for anisotropic linear
distributions of mass on the circle contained in the torus ring,
or alternatively, if there is a renormalized gravitational
constant with anisotropic dependence on angle $\varphi .$ This
class of solutions have a toroidal horizon defined by the
condition of vanishing of the coefficient $h_{4}$ which holds if
$f_{T}(\sigma ,\tau )=0.$
The functions $\varpi \left( \sigma ,\tau ,\varphi \right) $ and $%
n_{i[0,1]}\left( \sigma ,\tau \right) $ may be stated in a form that at long
distance from the toroidal horizon the (\ref{sol2}) with data (\ref{data2}%
) will have asymptotic like the Schwarzschild metric. We can also consider
alternative toroidal vacuum configurations. We note that instead of
relations like $h_{3}=-\Omega ^{-2}$ we can use every type $h_{3}\sim \Omega
^{p/q},$ like is stated by (\ref{4confq}); it depends on what type of
nonlinear configuration and asymptotic limits we wont to obtain.

%%%%%%%%%%%%%%%%%%%%%%%%%%%%%%%%%%%%
% Figure 2: Torus Configuration
%%%%%%%%%%%%%%%%%%%%%%%%%%%%%%%%%%%

We remark also that in a similar manner we can generate toroidal
configurations of Class A, starting from the auxiliary metric (\ref{auxm1t}).
In the next subsection we elucidate this possibility by interfering it
with a Class B ellipsoidal configuration.

\subsection{Static Ellipsoidal Black Hole -- Black Torus solutions}

There are different possibilities to combine static ellipsoidal black hole
and black torus solutions as they will give configurations with two
horizons. In this subsection we analyze two such variants. We consider a 2D
system of holonomic coordinates $x^{i},$ which may be used both on the
'ellipsoidal' and 'toroidal' sectors via transforms like $u=u(x^{i}),\lambda
=\tau \left( x^{i}\right) $ and $\sigma =\sigma \left( x^{i}\right) .$

\subsubsection{Ellipsoidal--torus black configurations of Class BA}

We construct a 4D vacuum metric with posses two type of horizons,
ellipsoidal and toroidal one, having both type characteristics like a metric
of Class B for ellipsoidal configurations and a metric of Class A for
toroidal configurations (we conventionally call this ellipsoidal torus
metric to be of Class BA). \ The ansatz is taken
\begin{eqnarray}
\delta s^{2} &=&\Omega ^{2}\{-\mu ^{2}\left( x^{i}\right) \left[
(dx^{1})^{2}+(dx^{2})^{2}\right]  \label{sol3} \\
&&-\eta _{3}\left( x^{k},\varphi \right) a_{T}\left( x^{i}\right) \delta
\varphi ^{2}+\frac{b_{T}(x^{i})f_{E}(x^{i})}{g_{E}(x^{i})}\delta t^{2}\},
\nonumber
\end{eqnarray}
with
\begin{eqnarray*}
\Omega ^{2} &=&\omega ^{2}\left( x^{k},\varphi \right) \varpi ^{2}\left(
x^{k},\varphi \right) \Omega _{AT}^{2}\left( x^{i}\right) \Omega
_{BE}^{2}\left( x^{i}\right) , \\
\eta _{3} &=&-a_{T}^{-1}\left( x^{i}\right) \Omega ^{-2},h_{3}=-\eta
_{3}\left( x^{k},\varphi \right) a_{T}\left( x^{i}\right) , \\
h_{4} &=&b_{T}(x^{i})f_{E}(x^{i})/g_{E}(x^{i}), \\
\mu ^{2} &=&\mu _{\lbrack 0]}^{2}\exp \left[ c_{[1]}x^{1}+c_{[2]}x^{2}\right]
.
\end{eqnarray*}
So, in general we may having both type of anisotropic renormalizations of
constants $\rho _{g}$ and $\rho _{g}^{[t]}$ as in (\ref{oresc}) and (\ref
{osect}). The prolongations of differentials $\delta \varphi $ and $\delta t$
are defined by the\ coefficients

\begin{eqnarray*}
\zeta _{i}\left( x^{k},\varphi \right) &=&\left( \Omega ^{\ast }\right)
^{-1}\partial _{i}\Omega , \\
n_{i}\left( x^{k},\varphi \right) &=&n_{i[0]}\left( x^{k}\right)
+n_{i[1]}\left( x^{k}\right) \int \omega ^{-2}\varpi ^{-2}d\varphi .
\end{eqnarray*}
The constants $\mu _{\lbrack 0]}^{2},c_{[1,2]},$ functions $\omega
^{2}\left( x^{k},\varphi \right) ,\varpi ^{2}\left( x^{k},\varphi \right) $
and $n_{i[0,1]}\left( x^{k}\right) $ \ and relation $h_{3}\sim \Omega ^{p/q}$%
\ may be selected as to obtain at asymptotic a Schwarzschild like behavior.
The metric (\ref{sol3}) has two horizons, a toroidal one, defined by the
condition $b_{T}(x^{i})=0,$ and an ellipsoidal one, defined by the condition
$f_{E}(x^{i})=0$ (see respectively these functions in (\ref{auxm1tb}) \ and (%
\ref{auxm2f})).

\begin{figure*}
\includegraphics[scale=0.8]{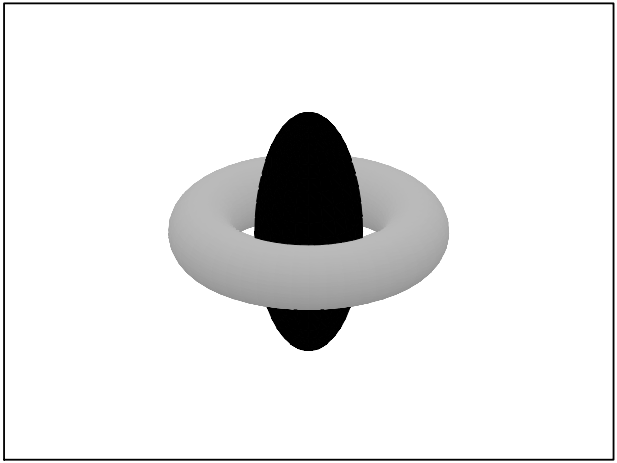}
\caption{\label{eltorus}Ellipsoidal--Torus Configuration}
\end{figure*}

The ellipsoidal--torus configuration is illustrated in  Fig.
\ref{eltorus}.

%%%%%%%%%%%%%%%%%%%%%%%%%%%%%%%%%%%%
% Figure 3: Ellipsoidal--Torus Configuration
%%%%%%%%%%%%%%%%%%%%%%%%%%%%%%%%%%%

We can consider different combinations of ellipsoidal black hole an black
torus metrics in order to construct solutions of Class AA, AB and BB (we
omit such similar constructions).

\subsubsection{A second example of ellipsoidal black hole -- black torus
system}

In the simplest case we can construct a solution with an ellipsoidal and
toroidal horizon which have a trivial conformal factor $\Omega $ and
vanishing coefficients $\zeta _{i}=0$ (see (\ref{5confeq})). Establishing a
global 3D toroidal space coordinate system, we consider the ansatz
\begin{eqnarray}
\delta s^{2} &=&\{-\left[ d\sigma ^{2}+d\tau ^{2}\right] -\eta _{3}\left(
\sigma ,\tau ,\varphi \right) h_{3[0]}\left( \sigma ,\tau \right) \delta
\varphi ^{2}  \label{sol4} \\
&&+\eta _{4}\left( \sigma ,\tau ,\varphi \right) h_{4[0]}\left( \sigma ,\tau
\right) \delta t^{2}\},  \nonumber
\end{eqnarray}
where (in order to construct a Class AA solution) we put
\begin{eqnarray*}
h_{3[0]} &=&a_{E}\left( \sigma ,\tau \right) a_{T}\left( \sigma ,\tau
\right) ,h_{4[0]}=b_{E}\left( \sigma ,\tau \right) b_{T}\left( \sigma ,\tau
\right) , \\
\eta _{4} &=&\omega ^{-2}\left( \sigma ,\tau ,\varphi \right) \varpi
^{-2}\left( \sigma ,\tau ,\varphi \right) ,
\end{eqnarray*}
considering anisotropic renormalizations of constants as in (\ref{oresc})
and (\ref{osect}). The polarization $\eta _{3}$ is to be found from the
relation
\begin{equation}
h_{3}=h_{[0]}^{2}[(\sqrt{|h_{4}|})^{\ast }]^{2},h_{[0]}^{2}=const,
\label{q1}
\end{equation}
which defines a solution of equation (\ref{5ricci2a}) for $h_{4}^{\ast }\neq
0 $ , when $h_{3}=-\eta _{3}h_{3[0]}$ and $h_{4}=\eta _{4}h_{4[0]}.$
Substituting the last values in (\ref{q1}) we get
\[
|\eta _{3}|=h_{[0]}^{2}\frac{b_{E}b_{T}}{a_{E}a_{T}}\left( \frac{\omega
^{\ast }+\varpi ^{\ast }}{\omega \varpi }\right) ^{2}.
\]
Then, computing the coefficient $\gamma ,$ see (\ref{5abc}), after two
integrations on $\varphi $ we find
\begin{eqnarray*}
&& n_{i}\left( \sigma ,\tau ,\varphi \right) =n_{i[0]}\left(
\sigma ,\tau
\right) +n_{i[1]}\left( \sigma ,\tau \right) \int [\eta _{3}/\left( \sqrt{%
|\eta _{3}|}\right) ^{3}]d\varphi \\
&&=n_{i[0]}\left( \sigma ,\tau \right) +\tilde{n}_{i[1]}\left(
\sigma ,\tau \right) \int \omega \varpi \left( \omega ^{\ast
}+\varpi ^{\ast }\right) ^{2}d\varphi ,
\end{eqnarray*}
where we re-defined the function $n_{i[1]}\left( \sigma ,\tau
\right) $ into a new $\tilde{n}_{i[1]}\left( \sigma ,\tau \right)
$ by including all factors and constants like $h_{[0]}^{2},$
$b_{E},b_{T},a_{E}$ and $a_{T}.$

The constructed solution (\ref{sol4}) does not has as locally isotropic
limit the Schwarzs\-child metric. It has also a toroidal and ellipsoidal
horizons defined by the conditions of vanishing of $b_{E}$ and $b_{T},$ but
this solution is different from the metric (\ref{sol3}): it has a trivial
conformal factor and vanishing coefficients $\zeta _{i}$ which means that in
this case we are having a splitting of dynamics into three holonomic and one
anholonomic coordinate. We can select such functions $n_{i[0,1]}\left(
\sigma ,\tau \right) ,$ $\omega \left( \sigma ,\tau ,\varphi \right) $ and $%
\varpi \left( \sigma ,\tau ,\varphi \right) ,$ when at asymptotic one
obtains the Minkowski metric.

\section{Anisotropic Polarizations and Running Const\-ants}

In this Section we consider non--static vacuum anho\-lo\-nomic
ellipsoidal and/or toroidal configurations de\-pen\-ding
explicitly on time variable $t$ and on holonomic coordinates
$x^{i},$ but not on angular coordinate $\varphi .$ Such solutions
are generated by dynamical an\-ho\-lo\-nomic deformations and
conformal transforms of the Schwarzschild metric. For simplicity,
we analyze only Class A and AA solutions.

The coordinates are parametrized: $x^{i}$ are holonomic ones, in
particular, $x^{i}=\left( u,\lambda \right) ,$ for ellipsoidal
configurations, and $x^{i}=\left( \sigma ,\tau \right) ,$ for
toroidal configurations; $y^{3}=v=t$ and $y^{4}=\varphi .$ The
metric ansatz is stated in the form
\begin{eqnarray}
\delta s^{2} &=&\Omega ^{2}\left( x^{i},t\right) [-(dx^{1})^{2}-(dx^{2})^{2}
\nonumber \\
&&+h_{3}\left( x^{i},t\right) \delta t^{2}+h_{4}\left( x^{i},t\right) \delta
\varphi ^{2}],  \label{ansatz3}
\end{eqnarray}
where the differentials are elongated
\begin{eqnarray*}
d\varphi &\rightarrow &\delta \varphi =d\varphi +\zeta _{i}\left(
x^{k},t\right) dx^{i}, \\
dt &\rightarrow &\delta t=dt+n_{i}\left( x^{k},t\right) dx^{i}.
\end{eqnarray*}
The ansatz (\ref{ansatz3}) is related with some ellipsoidal and/ or toroidal
anholonomic deformations of the Schwarzschild metric (see respectively, (\ref
{2schel}), (\ref{3auxm1}), (\ref{3auxm2}) and (\ref{2schtor}), (\ref{auxm1t}), (%
\ref{auxm2t})) via time running renormalizations of ellipsoidal and toroidal
constants (instead of the static ones, (\ref{oresc}) and (\ref{osect})),
\begin{equation}
\rho _{g}\rightarrow \widehat{\rho }_{g}=\omega \left( x^{k},t\right) \rho
_{g},  \label{oresca}
\end{equation}
and
\begin{equation}
\rho _{g}^{[t]}\rightarrow \widehat{\rho }_{g}^{[t]}=\varpi \left(
x^{k},t\right) \rho _{g}^{[t]}.  \label{osecta}
\end{equation}
As particular cases we shall consider trivial values $\Omega ^{2}=1.$ The
horizons of such classes of solutions are defined by the condition of
vanishing of the coefficient $h_{3}\left( x^{i},t\right) .$

\subsection{Ellipsoidal/toroidal solutions with running constants}

\subsubsection{Trivial conformal factors, $\Omega ^{2}=1$}

The simplest way to generate a $t$--depending ellipsoidal (or
toroidal) configuration is to take the metric (\ref{3auxm1}) (or
(\ref{auxm1t})) and to renormalize the constant as (\ref{oresca})
(or (\ref{osecta})). In result we obtain a metric (\ref{ansatz3})
with $\Omega ^{2}=1,$ $h_{3}=\eta _{3}\left( x^{i},t\right)
h_{3[0]}\left( x^{i}\right) $ and $h_{4}=h_{4[0]}\left(
x^{i}\right) ,$ where
\begin{eqnarray*}
\eta _{3} &=&\omega ^{-2}\left( u,\lambda ,t\right) ,h_{3[0]}=b_{E}\left(
u,\lambda \right) ,h_{4[0]}=a_{E}\left( u,\lambda \right) , \\
(\eta _{3} &=&\varpi ^{-2}\left( \sigma ,\tau ,t\right)
,h_{3[0]}=b_{T}\left( \sigma ,\tau \right) ,h_{4[0]}=a_{T}\left( \tau
\right) ).
\end{eqnarray*}
The equation (\ref{5ricci2a}) is satisfied by these data because $h_{4}^{\ast
}=0$ and the condition (\ref{5confeq}) holds for $\zeta _{i}=0.$ The
coefficient $\gamma $ from (\ref{5abc}) is defined only by polarization $\eta
_{3},$ which allow us to write the integral of (\ref{24ricci4a}) as
\[
n_{i}=n_{i[0]}\left( x^{i}\right) +n_{i[1]}\left( x^{i}\right) \int \eta
_{3}\left( x^{i},t\right) dt.
\]
The corresponding ellipsoidal (or toroidal) configuration may be transformed
into asymptotically Minkowschi metric if the functions $\omega ^{-2}\left(
u,\lambda ,t\right) $ (or $\varpi ^{-2}\left( \sigma ,\tau ,t\right) )$ and $%
n_{i[0,1]}\left( x^{i}\right) $ are such way determined by
boundary conditions that $\eta _{3}\rightarrow const$ and
$n_{i[0,1]}\left( x^{i}\right) \rightarrow 0,$ far away from the
horizons, which are defined by the conditions $b_{E}\left(
u,\lambda \right) =0$ (or $b_{T}\left( \sigma ,\tau \right) =0).$

Such vacuum gravitational configurations may be considered as to posses
running of gravitational constants in a local spacetime region. For
instance, in Ref \cite{12v} we suggested the idea that a vacuum gravitational
soliton may renormalize effectively the constants, but at asymptotic we
have static configurations.

\subsubsection{Non--trivial conformal factors}

The previous configuration can not be related directly with the
Schwarzschild metric (we used \ its conformal transforms). A more direct
relation is possible if we consider non--trivial conformal factors. For
ellipsoidal (or toroidal) configurations we renromalize (as in (\ref{oresca}%
), or (\ref{osecta})) the conformal factor (\ref{3auxm1c}) (or (\ref{auxm1tc}%
)),
\begin{eqnarray*}
\Omega ^{2}\left( x^{k},t\right) &=&\omega ^{2}\left( x^{k},t\right) \Omega
_{AE}^{2}(x^{k})b_{E}^{-1}\left( x^{k}\right) , \\
(\Omega ^{2}\left( x^{k},t\right) &=&\varpi ^{2}\left( x^{k},t\right) \Omega
_{AT}^{2}(x^{k})b_{T}^{-1}\left( x^{k}\right) ).
\end{eqnarray*}
In order to satisfy the condition (\ref{4confq}) we choose $h_{3}=\Omega
^{-2} $ but $h_{4}=h_{4[0]}$ as in previous subsection: this solves the
equation (\ref{5ricci2a}). The non--trivial values of $\zeta _{i}$ and $n_{i}$
are defined from (\ref{5confeq}) and (\ref{24ricci4a}),

\begin{eqnarray*}
\zeta _{i}\left( x^{k},t\right) &=&\left( \Omega ^{\ast }\right)
^{-1}\partial _{i}\Omega , \\
n_{i}\left( x^{k},t\right) &=&n_{i[0]}\left( x^{k}\right) +n_{i[1]}\left(
x^{k}\right) \int h_{3}\left( x^{i},t\right) dt.
\end{eqnarray*}
We note that the conformal factor $\Omega ^{2}$ is singular on horizon,
which is defined by the condition of vanishing of the coefficient $h_{3},$
i. e. of $b_{E}$ (or $b_{T}$). By a corresponding parametrization of
functions $\omega ^{2}\left( x^{k},t\right) $ (or $\varpi ^{2}\left(
x^{k},t\right) )$ and $n_{i[0,1]}\left( x^{k}\right) $ we may generate
asymptotically flat solutions, very similar to the Schwarzschild solution,
which have anholonomic running constants in a local region of spacetime.

\subsection{Black Ellipsoid -- Torus Metrics with Running Constants}

Now we consider nonlinear superpositions of the previous metrics as to
construct solutions with running constants and two horizons (one ellipsoidal
and another toroidal).

\subsubsection{ Trivial conformal factor, $\Omega ^{2}=1$}

The simplest way to generate such metrics\ with two horizons is to
establish, for instance, a common toroidal system of coordinate,
to take the ellipsoidal and toroidal metrics constructed in
subsection V.A.1 and to multiply correspondingly their
coefficients. The corresponding data, defining a new solution for
the ansazt (\ref{ansatz3}), \ are
\begin{eqnarray}
g_{1,2} &=&-1,\widehat{\rho }_{g}=\omega \left( x^{k},t\right) \rho _{g},%
\widehat{\rho }
_{g}^{[t]}=\varpi \left( x^{k},t\right) \rho
_{g}^{[t]},\Omega =1,  \nonumber \\
h_{3} &=&\eta _{3}\left( x^{i},t\right) h_{3[0]}\left( x^{i}\right) ,\eta
_{3}=\omega ^{-2}\left( x^{k},t\right) \varpi ^{-2}\left( x^{k},t\right) ,
\nonumber \\
h_{3[0]} &=&b_{E}\left( x^{k}\right) b_{T}\left( x^{k}\right)
,h_{4}=h_{4[0]}=a_{E}\left( x^{i}\right) a_{T}(x^{i}),(%
  \nonumber \\
\zeta _{i} &=&0,n_{i}=n_{i[0]}\left( x^{k}\right) +n_{i[1]}\left(
x^{k}\right) \int \omega ^{-2}\varpi ^{-2}dt,  \label{data7}
\end{eqnarray}
where the  functions $a_{E},a_{T}$ and $b_{E}, b_{T}$ are given
by formulas  (\ref{auxm1s}) and (\ref{auxm1tb}).
 Analyzing the data (\ref{data7}) we conclude
that we have two horizons, when $b_{E}\left( x^{k}\right) =0$ and
$b_{T}\left( x^{k}\right) =0,$ parametrized respectively as
ellipsoidal and torus hypersurfaces. The boundary conditions on
running constants and on off--diagonal components of the metric
may be imposed as the solution would result in an asymptotic flat
metric. In a finite region of spacetime we may consider various
dependencies in time.

\subsubsection{Non--trivial conformal factor}

In a similar manner, we can multiply the conformal factors and coefficients
of the metrics from subsection V.A.2 in order to generate a solution
parametrized by the (\ref{ansatz3}) with nontrivial conformal factor $\Omega
$ and non-vanishing coefficients $\zeta _{i}.$ The data are
\begin{eqnarray}
g_{1,2} &=&-1,\widehat{\rho }_{g}=\omega \left( x^{k},t\right) \rho _{g},%
\widehat{\rho }_{g}^{[t]}=\varpi \left( x^{k},t\right) \rho _{g}^{[t]},
\label{data8} \\
\Omega ^{2} &=&\omega ^{2}\left( x^{k},t\right) \varpi ^{2}\left(
x^{k},t\right) \Omega _{AE}^{2}(x^{k})\times   \nonumber \\
&&\Omega _{AT}^{2}(x^{k})b_{E}^{-1}\left( x^{k}\right) b_{T}^{-1}\left(
x^{k}\right) ,(\mbox{ see (\ref{auxm1s}),(\ref{auxm1tb})}),  \nonumber \\
h_{3} &=&\Omega ^{-2},h_{3[0]}=b_{E}\left( x^{k}\right) b_{T}\left(
x^{k}\right) ,  \nonumber \\
h_{4} &=&h_{4[0]}=a_{E}\left( x^{i}\right) a_{T}(x^{i}),\zeta _{i}\left(
x^{k},t\right) =\left( \Omega ^{\ast }\right) ^{-1}\partial _{i}\Omega ,
\nonumber \\
n_{i} &=&n_{i[0]}\left( x^{k}\right) +n_{i[1]}\left( x^{k}\right) \int
\omega ^{-2}\varpi ^{-2}dt.  \nonumber
\end{eqnarray}
The data (\ref{data8}) define a new type of solution than that
given by (\ref {data7}). It this case there is a singular on
horizons conformal factor. The behavior nearly horizons is very
complicated. By corresponding parametrizations of functions
$\omega \left( x^{k},t\right) ,$\ $\varpi \left( x^{k},t\right) $
and $n_{i[0,1]}\left( x^{k}\right) ,$ which approximate $\omega
,\varpi \rightarrow const$ and $\zeta _{i},n_{i}\rightarrow 0$ we
may obtain a stationary flat asymptotic.

Finally, we note that instead of Class AA solutions with anisotropic and
running constants we may generate solutions with two horizons (ellipsoidal
and toroidal) by considering nonlinear superpositions, anholonomic
deformations, conformal transforms and combinations of solutions of Classes
A, B. The method of construction is similar to that considered in this
Section.

\section{Conclusions and Discussion}

We constructed new classes of exact solutions of vacuum Einstein
equations by considering anholonomic deformations and conformal
transforms of the Schwarzschild black hole metric. The solutions
posses ellipsoidal and/ or toroidal horizons and symmetries and
could be with anisotropic renormalizations and running constants.
Some of such solutions define static configurations and have
Schwarzschild like (in general, multiplied to a conformal factor)
asymptotically flat limits. The new metrics are parametrized by
off--diagonal metrics which can be diagonalized with respect to
certain anholonomic frames. The coefficients of diagonalized
metrics are similar to the Schwarzschild metric coefficients but
describe deformed horizons and contain additional dependencies on
one 'anholonomic' coordinate.

We consider that such vacuum gravitational configurations with non--trivial
topology and deformed horizons define a new class of ellipsoidal black hole
and black torus objects and/or their combinations.

Toroidal and ellipsoidal black hole solutions were constructed for different
models of extra dimension gravity and in the four dimensional (4D) gravity
with cosmological constant and specific configurations of matter
\cite{12lemos,12rev,12emp}. There were defined also vacuum configurations for such
objects \cite{12v,12v2,12vth,12vtor}. However, we must solve the very important
problems of physical interpretation of solutions with anholonomy and to
state their compatibility with the black hole uniqueness theorems \cite
{12israel} and the principle of topological censorship \cite{12haw,12cen}.

It is well known that the Schwarzschild metric is no longer the unique
asymptotically flat static solution if the 4D gravity is derived as an
effective theory from extra dimension like in recent Randall and Sundrum
theories (see basic results and references in \cite{12rs}). The Newton law may
be modified at sub-millimeter scales and there are possible configurations
with violation of local Lorentz symmetry \cite{12csaki}. Guided by modern
conjectures with extra dimension gravity and string/M--theory, we have to
answer the question:\ it is possible to give a physical meaning to the
solutions constructed in this paper only from a viewpoint of a generalized
effective 4D Einstein theory, or they also can be embedded into the
framework of general relativity theory?

It should be noted that the Schwarzschild solution was constructed as the
unique static solution with spherical symmetry which was connected to the
Newton spherical gravitational potential $\sim 1/r$ and defined as to result
in the Minkowski flat spacetime, at long distances. This potential describes
the static gravitational field of a point particle with ''isotropic'' mass $%
m_{0}.$ The spherical symmetry is imposed at the very beginning and it is
not a surprising fact that the spherical topology and spherical symmetry of
horizons are obtained for well defined states of matter with specific energy
conditions and in the vacuum limits. Here we note that the spherical
coordinates and systems of reference are holonomic ones and the considered
ansatz for the Schwarzschild metric is diagonal in the more ''natural''
spherical coordinate frame.

We can approach in a different manner the question of constructing 4D
static vacuum metrics. We might introduce into consideration off--diagonal
ansatz, prescribe instead of the spherical symmetry a deformed one
(ellipsoidal, toroidal, or their superposition) and try to check if a such
configurations may be defined by a metric as to satisfy the 4D vacuum Einstein
equations. Such metrics were difficult to be found because of cumbersome
calculus if dealing with off--diagonal ansatz. But the problem was
substantially simplified by an equivalent transferring of calculations with
respect to anholonomic frames \cite{12v,12vth,12vtor}. Alternative exact static
solutions, with ellipsoidal and toroidal horizons (with possible extensions
for nonlinear polarizations and running constants), were constructed and
related to some anholonomic and conformal transforms of the Schwarzschild
metric.

It is not difficult to suit such solutions with the asymptotic limit to the
locally isotropic Minkowschi spacetime: ''an egg and/or a ring look like
spheres far away from their non--trivial horizons''. The unsolved question
is that what type of modified Newton potentials should be considered in this
case as they would be compatible with non--spherical symmetries of
solutions? The answer may be that at short distances the masses and
constants are renormalized by specific nonlinear vacuum gravitational
interactions which can induce anisotropic effective masses, ellipsoidal or
toroidal polarizations and running constants. For instance, the Laplace
equation for the Newton potential can be solved in ellipsoidal coordinates
\cite{12ll}: this solution could be a background for constructing ellipsoidal
Schwarzschild like metrics. Such nonlinear effects should be treated, in some
approaches, as certain quasi--classical approximations for some 4D quantum
gravity models, or related to another type of theories of extra dimension
classical or quantum gravity.

Independently of the type of little, or more, internal structure of black
holes with non--spherical horizons we search for physical justification, it
is a fact that exact vacuum solutions with prescribed non--spherical
symmetry of horizons can be constructed even in the framework of general
relativity theory. Such solutions are parametrized by off--diagonal metrics,
described equivalently, in a more simplified form, with respect to
associated anholonomic frames; they define some anholonomic vacuum
gravitational configurations of corresponding symmetry and topology.
Considering certain characteristic initial value problems we can select
solutions which at asymptotic result in the Minkowschi flat spacetime, or
into an anti--de Sitter (AdS) spacetime, and have a causal behavior of
geodesics with the equations solved with respect to anholonomic frames.

It is known that the topological censorship principle was reconsidered for
AdS black holes \cite{12cen}. But such principles and uniqueness black hole
theorems have not yet been proven for spacetimes defined by generic
off--diagonal metrics with prescribed non--spherical symmetries and horizons
and with associated anholonomic frames with mixtures of holonomic and
anholonomic variables. It is clear that we do not violate the conditions of
such theorems for those solutions which are locally anisotropic and with
nontrivial topology in a finite region of spacetime and have locally
isotropic flat and trivial topology limits. We can select for physical
considerations only the solutions which satisfy the conditions of the
mentioned restrictive theorems and principles but with respect to well
defined anholonomic frames with holonomic limits. As to more sophisticate
nonlinear vacuum gravitational configurations with global non--trivial
topology we conclude that there are required a more deep analysis and new
physical interpretations.

The off--diagonal metrics and associated anholonomic frames extend the class
of vacuum gravitational configurations as to be described by a nonlinear,
anholonomic and anisotropic dynamics which, in general, may not have any
well known locally isotropic and holonomic limits. The formulation and proof
of some uniqueness theorems and principles of topological censorship as well
analysis of physical consequences of such anholonomic vacuum solutions is
very difficult. We expect that it is possible to reconsider the statements
of the Israel--Carter--Robinson and Hawking theorems with respect to
anholonomic frames and spacetimes with non--spherical topology and
anholonomically deformed spherical symmetries. These subjects are currently
under our investigation.

\subsection*{Acknowledgements}

The work is partially supported by a
2000--2001 California State University Legislative Award and by a
NATO/Portugal fellowship grant at the Technical University of
Lisbon. The author is grateful to J. P. S. Lemos,  D. Singleton
and E. Gaburov for support and collaboration.

%%%%%%%%%%%%%%%%%%%%%%%%%%%%%%%%%%%%%%%%%%%%%%%%%%%%%%%%%%%%%%%%%%%%%%%%%%%%%

\part{ Noncommutative Riemann--Lagrange Geometry }

\chapter[Noncommutative Finsler  Gravity]
{Noncommutative Finsler Geometry, Gauge Fields and Gravity}

{\bf Abstract}
\footnote{\copyright\
  S. Vacaru, Noncommutative Finsler Geometry, Gauge Fields and Gravity,
math-ph/0205023}

The work extends the A. Connes' noncommutative geometry to spaces
with generic local anisotropy. We apply the E. Cartan's
anholonomic frame approach to geometry models and physical
theories and develop the nonlinear connection formalism for
projective module spaces.  Examples of noncommutative generation
of anholonomic Riemann, Finsler and Lagrange spaces are analyzed.
We also present a research on noncommutative Finsler--gauge
theories, generalized Finsler gravity and anholonomic (pseudo)
Riemann geometry which appear naturally if anholonomic frames
(vierbeins) are defined in the context of string/M--theory and
extra dimension Riemann gravity.
\vskip5pt.

\section{Introduction}

In the last twenty years, there has been an increasing interest in
noncommutative and/or quantum geometry with applications both in
mathematical and particle physics. It is now generally considered
that at very high energies, the spacetime can not be described by
a usual manifold structure. Because of quantum fluctuations, it
is difficult to define localized points and the quantum spacetime
structure is supposed to posses generic noncommutative, nonlocal
and locally anisotropic properties. Such ideas originate from the
suggestion that the spacetime coordinates do not commute at a
quantum level \cite{13snyder}, they are present in the modern
string theory \cite{13deligne,13polchinski} and background the
noncommutative physics and geometry \cite{13connes} and quantum
geometry \cite{13qg}.

Many approaches can be taken to introducing noncommutative
geometry and
developing noncommutative physical theories (Refs. \cite%
{13connes,13douglas,13dubois,13gracia,13konechny,13landi,13madore,13varilly}
emphasize some basic monographs and reviews). \ This paper has
three aims: \ First of all we would like to give an exposition of
some basic facts on anholonomic frames and associated nonlinear
connection structures both on commutative and noncommutative
spaces (respectively modelled in vector bundles and in projective
modules of finite type). \ Our second goal is to state the
conditions when different variants of Finsler, Lagrange and
generalized Lagrange geometries, in commutative and
noncommutative forms, can be defined by corresponding frame,
metric and connection structures. The third aim is to construct
and analyze properties of gauge and gravitational noncommutative
theories with generic local anisotropy and to prove that such
models can be elaborated in the framework of noncommutative
approaches to Riemannian gravity theories.

This paper does not concern the topic of Finsler like commutative
and noncommutative structures in string/M-theories (see the Ref.
\cite{13vfncs}, which can be considered as a string partner of this
work).

We are inspired by the geometrical ideas from a series of
monographs and works by E. Cartan \cite{13cartan} where a unified
moving frame approach to the Riemannian and Finsler geometry,
Einstein gravity and Pffaf systems, bundle spaces and spinors, as
well the preliminary ideas on nonlinear connections and various
generalizations of gravity theories were developed. By
considering anholonomic frames on (pseudo) Riemannian manifolds
and in tangent and vector bundles, we can model very sophisticate
geometries with local anisotropy. We shall apply the concepts and
methods developed by the
Romanian school on Finsler geometry and generalizations \cite%
{13miron,13ma,13bejancu,13vmon1} from which we leaned that the Finsler
and Cartan like geometries may be modelled on vector (tangent) and
covector (cotangent) bundles if the constructions are adapted to
the corresponding nonlinear connection structure via anholonomic
frames. In this case the geometric ''picture'' and physical
models have a number of common points with those from the usual
Einstein--Cartan theory and/or extra dimension (pseudo)
Riemannian geometry. As general references on Finsler geometry and
applications we cite the monographs \cite%
{13finsler,13miron,13ma,13bejancu,13vmon1,13vmon2}) and point the fact that
the bulk of works on Finsler geometry and generalizations
emphasize differences with the usual Riemannian geometry rather
than try to approach them from a unified viewpoint (as we propose
in this paper).

By applying the formalism of nonlinear connections (in brief,
N--connection) and adapted anholonomic frames in vector bundles
and superbundles we extended the geometry of Clifford structures
and spinors for generalized Finsler
spaces and their higher order extensions in vector--covector bundles \cite%
{13vspinors,13vmon2}, constructed and analyzed different models of
gauge theories and gauge gravity with generic anisotropy
\cite{13vgauge}, defined an anisotropic stochastic calculus in
bundle and superbundle spaces provided with nonlinear connection
structure \cite{13vstoch,13vmon1}, with a number of applications in
the theory of anisotropic kinetic and thermodynamic processes
\cite{13vankin}, developed supersymmetric theories with local
anisotropy \cite{13vsuper,13vmon1,13vstring} and proved that Finsler
like (super)
geometries are contained alternatively in modern string theory \cite%
{13vstring,13vmon1}. One should be emphasized here that in our
approach we have not proposed any ''exotic'' locally anisotropic
string theories modifications but demonstrated that anisotropic
structures, Finsler like or another ones, may appear
alternatively to the Riemannian geometry, or even can be modelled
in the framework of a such geometry, in the low energy limit of
the string theory, because we are dealing with frame, vierbein,
constructions.

The most surprising fact was that the Finsler like structures
arise in the usual (pseudo) Riemannian geometry of lower and
higher dimensions and even in the Einstein gravity. References
\cite{13vexsol} contain investigations of a number of exact
solutions in modern gravity theories (Einstein, Kaluza--Klein and
string/brane gravity) which describe locally anisotropic
wormholes, Taub NUT spaces, black ellipsoid/torus solutions,
solitonic and another type configurations. It was proposed a new
consequent method of constructing exact solutions of the Einstein
equations for off--diagonal metrics, in spaces of dimension
$d>2,$ depending on three and more isotropic and anisotropic
variables which are effectively diagonalized by anholonomic frame
transforms. The vacuum and matter field equations are reduced to
very simplified systems of partial differential equations which
can be integrated in general form \cite{13vmethod}.

A subsequent research in\ Riemann--Finsler and noncommutative
geometry and phy\-sics requires the investigation of the fact if
the A. Connes functional analytic approach to noncommutative
geometry and gravity may be such way generalized as to include
the Finsler, and of another type anisotropy, spaces. The first
attempt was made in Refs. \cite{13vnonc} where some models of
noncommutative gauge gravity (in the commutative limit being
equivalent to the Einstein gravity, or to different
generalizations to de Sitter, affine, or Poincare gauge gravity
with, or not, nonlinear realization of the gauge groups) were
analyzed. \ Further developments in formulation of noncommutative
geometries with anholonomic and anisotropic structures and their
applications in modern particle physics lead to a rigorous study
of the geometry of anholonomic noncommutative frames with
associated N--connection structure, to which are devoted our
present researches.

The paper has the following structure: in section 2 we present
the necessary definitions and results on the functional approach
to commutative and noncommutative geometry. Section 3 is devoted
to the geometry of vector bundles and theirs noncommutative
generalizations as finite projective modules. We define the
nonlinear connection in commutative and noncommutative spaces,
introduce locally anisotropic Clifford/spinor structures and
consider the gravity and gauge theories from the viewpoint of
anholonomic frames with associated nonlinear connection
structures. In section 4 we prove that various type of gravity
theories with generic anisotropy, constructed on anholonomic
Riemannian spaces and their Kaluza--Klein and Finsler like
generalizations can be derived from the A. Connes' functional
approach to noncommutative geometry by applying the canonical
triple formalism but extended to vector bundles provided with
nonlinear connection structure. In section 5, we elaborate and
investigate noncommutative gauge like gravity models (which in
different limits contain the standard Einstein's general
relativity and various its anisotropic and gauge
generalizations). The approach holds true also for (pseudo)
Riemannian metrics, but is based on noncommutative extensions of
the frame and connection formalism. This variant is preferred
instead of the usual metric models which seem to be more
difficult to be tackled in the framework of noncommutative
geometry if we are dealing with pseudo--Euclidean signatures and
with complex and/or nonsymmetic metrics. Finally, we present a
discussion and conclusion of the results in section 6.

\section{Commutative and Noncommutative Spaces}

The A. Connes' functional analytic approach \cite{13connes} to the
noncommutative topology and geometry is based on the theory of
noncommutative $C^{\ast }$--algebras. Any commutative $C^{\ast
}$--algebra can be realized as the $C^{\ast }$--algebra of
complex valued functions over locally compact Hausdorff space. A
noncommutative $C^{\ast }$--algebra can be thought of as the
algebra of continuous functions on some 'noncommutative
space' (see details in Refs. \ \cite%
{13connes,13dubois,13gracia,13landi,13madore,13varilly}). \

Commutative gauge and gravity theories stem from the notions of
connections (linear and nonlinear), metrics and frames of
references on manifolds and vector bundle spaces. The possibility
of extending such theories to some noncommutative models is based
on the Serre--Swan theorem \cite{13swan} stating that there is a
complete equivalence between the category of (smooth) vector
bundles over a smooth compact space (with bundle maps) and the
category of projective modules of finite type over commutative
algebras and module morphisms. \ Following that theorem, the
space $\Gamma \left( E\right) $ of smooth sections of a vector
bundle $E$ over a compact space is a projective module of finite
type over the algebra $C\left( M\right) $ of
smooth functions over $M$ and any finite projective $C\left( M\right) $%
--module can be realized as the module of sections of some vector
bundle
over $M.$ This construction may be extended if a noncommutative algebra $%
\mathcal{A}$ is taken as the starting ingredient: the
noncommutative
analogue of vector bundles are projective modules of finite type over $%
\mathcal{A}$. This way one developed a theory of linear
connections which culminates in the definition of Yang--Mills
type actions or, by some much more general settings, one
reproduced Lagrangians for the Standard model with its Higgs
sector or different type of gravity and Kaluza--Klein models
(see, for instance, Refs \cite{13cl,13ch1,13landi1,13madore}).

This section is devoted to the theory of nonlinear connections in
projective \ modules of finite type over a noncommutative algebra
$\mathcal{A}$. We shall introduce the basic definitions and
present the main results connected with anhlolonomic frames and
metric structures in such noncommutative spaces.

\subsection{Algebras of functions and (non) commutative spaces}

The general idea of noncommutative geometry is to shift from
spaces to the algebras of functions defined on them. In this
subsection, we give some general facts about algebras of
continuous functions on topological spaces, analyze the concept
of modules as bundles and define the nonlinear connections. We
present mainly the objects we shall need later on while
referring to \cite%
{13connes,13douglas,13dubois,13gracia,13konechny,13landi,13madore,13varilly} for
details. $\ $

We start with $\ $some necessary definitions on $C^{\ast
}$--algebras and compact operators

In this work any algebra $\mathcal{A}$ is an algebra over the
field of complex numbers $\C$, i. e. $\mathcal{A}$ is a vector
space over $\C$ when
the objects like $\alpha a\pm \beta b,$ with $a,b\in \mathcal{A}$ and $%
\alpha ,\beta \in \C,$ make sense. Also, there is defined (in
general) a noncommutative product $\mathcal{A}$ $\times
\mathcal{A}$ $\rightarrow
\mathcal{A}$ when for every elements $\left( a,b\right) $ and $a,b,$ $%
\mathcal{A}$ $\times \mathcal{A}$ $\ni \left( a,b\right)
\rightarrow ab\in
\mathcal{A}$ the conditions of distributivity,%
\begin{equation*}
a(b+c)=ab+ac,~(a+b)c=ac+bc,
\end{equation*}
for any $a,b,c\in \mathcal{A},$ in general, $ab\neq ba.$ It is
assumed that there is a unity $I\in \mathcal{A}.$

The algebra $\mathcal{A}$ is considered to be a so--called ''$\ast $%
--algebra'', for which an (antilinear) involution $\ast :\mathcal{A}$ $%
\rightarrow \mathcal{A}$ is defined by the properties%
\begin{equation*}
a^{\ast \ast }=a,~\left( ab\right) ^{\ast }=b^{\ast }a^{\ast
},~\left( \alpha a+\beta b\right) ^{\ast }=\overline{\alpha
}a^{\ast }+\overline{\beta }b^{\ast },
\end{equation*}%
where the bar operation denotes the usual complex conjugation.

One also considers $\mathcal{A}$ to be a normed algebra with a norm \ $%
\left| \left| \cdot \right| \right| $ $:\mathcal{A}$ $\rightarrow
\R,$ where
$\R$ the real number field, satisfying the properties%
\begin{eqnarray*}
\left| \left| \alpha a\right| \right| &=&|\alpha |\left| \left|
a\right| \right| ;~\left| \left| a\right| \right| \geq 0,~\left|
\left| a\right|
\right| =0\Leftrightarrow a=0; \\
\left| \left| a+b\right| \right| &\leq &\left| \left| a\right|
\right| +\left| \left| b\right| \right| ;~\left| \left| ab\right|
\right| \leq \left| \left| a\right| \right| \left| \left|
b\right| \right| .
\end{eqnarray*}%
This allows to define the 'norm' or 'uniform' topology when an $\varepsilon $%
--neighborhood of any $a\in \mathcal{A}$ is given by%
\begin{equation*}
U\left( a,\varepsilon \right) =\left\{ b\in \mathcal{A},\left|
\left| a-b\right| \right| <\varepsilon \right\} ,\varepsilon >0.
\end{equation*}

A Banach algebra is a normed algebra which is complete in the
uniform topology and a Banach $\ast $--algebra is a normed $\ast
$--algebra which is complete and such that $\left| \left| a^{\ast
}\right| \right| =\left|
\left| a\right| \right| $ for every $a\in \mathcal{A}.$ We can define now a $%
C^{\ast }$--algebra $\mathcal{A}$ as a Banach $\ast $--algebra
with the norm satisfying the additional identity $\left| \left|
a^{\ast }a\right| \right| =\left| \left| a\right| \right| ^{2}$
for every $a\in \mathcal{A}.$

We shall use different commutative and noncommutative algebras:

By $\mathcal{C}(M)$ one denotes the algebra of continuous
functions on a compact Hausdorf topological space $M,$ with $\ast
$ treated as the complex conjugation and the norm given by the
supremum norm, $||f||_{\infty }=\sup_{x\in M}|f(x)|.$ If the
space $M$ is only locally compact, one writes
$\mathcal{C}_{0}(M)$ for the algebra of continuous functions
vanishing at infinity (this algebra has no unit).

The $\mathcal{B(H)}$ is used for the noncommutative algebra of
bounded operators on an infinite dimensional Hilbert space
$\mathcal{H}$ with the involution $\ast $ given by the adjoint
and the norm defined as the operator
norm%
\begin{equation*}
||A||=\sup \left\{ ||A\zeta ||;\zeta \in \mathcal{H},~A\in \mathcal{B(H)}%
,~||\zeta ||\leq 1\right\} .
\end{equation*}

One considers the noncommutative algebra $M_{n}\left( \C\right) $ of $%
n\times n$ matrices $T$ with complex entries, when $T^{\ast }$ is
considered
as the Hermitian conjugate of $T.$ We may define a norm as%
\begin{equation*}
||T||=\{\mbox{the positive square root of the largest eigenvalue
of }\ T^{\ast }T\}
\end{equation*}%
or as
\begin{equation*}
||T||^{\prime }=\sup [T_{ij}],~T=\{T_{ij}\}.
\end{equation*}%
The last definition does not define a $C^{\ast }$--norm, but both
norms are
equivalent as Banach norm because they define the same topology on $%
M_{n}\left( \C\right) .$

A left (right) ideal $\mathcal{T}$ is a subalgebra $\mathcal{A}\in \mathcal{T%
}$ $\ $\ if $a\in \mathcal{A}$ and $b\in \mathcal{T}$ $\ $imply
that $ab\in \mathcal{T}$ $\ (ba\in \mathcal{T}).$ A two sided
ideal is a subalgebra
(subspace) which is both a left and right ideal. An ideal $\mathcal{T}$ $\ $%
\ is called maximal if there is not other ideal of the same type
which contain it. For a Banach $\ast $--algebra $\mathcal{A}$ and
two--sided $\ast
$--ideal $\mathcal{T}$ (which is closed in the norm topology) we can make $%
\mathcal{A}/\mathcal{T}$ $\ $\ a Banach $\ast $--algebra. This
allows to
define the quotient $\mathcal{A}/\mathcal{T}$ $\ $to be a $C^{\ast }$%
--algebra if $\mathcal{A}$ \ is a $C^{\ast }$--algebra. A $C^{\ast }$%
--algebra is called simple if it has no nontrivial two--sided
ideals. A two--sided ideal is called essential in a $C^{\ast
}$--algebra if any other non--zero ideal in this algebra has a
non--zero intersection with it.

One \ defines the resolvent set $r(a)$ of an element $a\in
\mathcal{A}$ as a the subset of complex numbers given by
$r(a)=\{\lambda \in \C|a-\lambda I$ is invertible\}. The
resolvent of $a$ at any $\lambda \in $ $r(a)$ is given by the
inverse $\left( a-\lambda I\right) ^{-1}.$ The spectrum $\sigma
\left( a\right) $ of an element $a$ is introduced as the
complement of $r(a)$ $\ $in $\C.$ For $C^{\ast }$--algebras the
spectrum of any element \ is a nonempty compact subset of $\C.$
The spectral radius $\rho \left( a\right) $ of $a\in \mathcal{A}$
$\ $\ is defined $\rho \left( a\right) =\sup
\{|\lambda |,\lambda \in r(a)\};$ for $\mathcal{A}$ being a $C^{\ast }$%
--algebra, one obtains $\rho \left( a\right) =||a||$ for every
$a\in \mathcal{A}.$ This distinguishes the $C^{\ast }$--algebras
as those for which the norm may be uniquely determined by the
algebraic structure. One considers self--adjoint elements for
which $a=a^{\ast },$ such elements have real spectra and satisfy
the conditions $\sigma (a)\subseteq \left[ -||a||,||a||\right] $
and $\sigma (a^{2})\subseteq \left[ 0|,||a||\right] .$ An element
$a$ is positive, $\ $i. e. $a>0,$ if its spectrum belongs to the
positive half--line. This is possible if and only if $a=bb^{\ast
}$ for some $b\in \mathcal{A}.$

One may consider $\ast $--morphisms between two $C^{\ast }$--algebras $%
\mathcal{A}$ and $\mathcal{B}$ as some $\C$--linear maps $\pi :\mathcal{A}$ $%
\rightarrow $ $\mathcal{B}$ which are subjected to the additional conditions%
\begin{equation*}
\pi (ab)=\pi (a)\pi (b),~\pi (a^{\ast })=\pi (a)^{\ast }
\end{equation*}%
which imply that $\pi $ are positive and continuous and that $\pi (\mathcal{A%
})$ is a $C^{\ast }$--subalgebra of $\mathcal{B}$ (see, for instance, \cite%
{13landi}). We note that a $\ast $--morphism which is bijective as
a map defines a $\ast $--isomorphism for which the inverse map
$\pi ^{-1}$ is automatically a $\ast $--morphism.

In order to construct models of noncommutative geometry one uses
representations of a $C^{\ast }$--algebra $\mathcal{A}$ as pairs
$\left( \mathcal{H},\pi \right) $ where $\mathcal{H}$ is a
Hilbert space and $\pi $
is a $\ast $--morphism $\pi :$ $\mathcal{A\rightarrow B}\left( \mathcal{H}%
\right) $ with $\mathcal{B}\left( \mathcal{H}\right) $ being the $C^{\ast }$%
--algebra of bounded operators on $\mathcal{H}.$ There are
different
particular cases of representations: A representation $\left( \mathcal{H}%
,\pi \right) $ is faithful if $\ker \pi =\{0\},$ i. e. $\pi $ is a $\ast $%
--isomorphism between $\mathcal{A}$ and $\pi (\mathcal{A})$ which
holds if and only if $||\pi (a)||=||a||$ for any $a\in
\mathcal{A}$ or $\pi (a)>0$ for all $a>0.$ A representation is
irreducible if the only closed subspaces of $\mathcal{H}$ which
are invariant under the action of $\pi (\mathcal{A})$ are the
trivial subspaces $\{0\}$ and $\mathcal{H}.$ It can be proven that
if the set of the elements in $\mathcal{B}\left(
\mathcal{H}\right) $ commute with each element in $\pi
(\mathcal{A}),$ i. e. the set consists of multiples of the
identity operator, the representation is irreducible. Here we
note that two representations $\left( \mathcal{H}_{1},\pi
_{1}\right) $ and $\left( \mathcal{H}_{2},\pi _{2}\right) $ are
said to be (unitary) equivalent if there exists a unitary
operator $U:\mathcal{H}_{1}\rightarrow
\mathcal{H}_{2}$ such that $\pi _{1}(a)=U^{\ast }\pi _{2}(a)U$ for every $%
a\in \mathcal{A}.$

A subspace (subalgebra) $\mathcal{T}$ of the $C^{\ast }$--algebra $\mathcal{A%
}$ is a primitive ideal if $\mathcal{T}=\ker \pi $ for some
irreducible representation $\left( \mathcal{H},\pi \right) $ of
$\mathcal{A}.$ In this case $\mathcal{T}$ $\ $is automatically a
closed two--sided ideal. One say that $\mathcal{A}$ is a
primitive $C^{\ast }$--algebra if $\mathcal{A}$ has a faithful
irreducible representation on some Hilbert space for which the
set $\{0\}$ is a primitive ideal. One denotes by $\Pr
im\mathcal{A}$ the set of all primitive ideals of a $C^{\ast
}$--algebra $\mathcal{A}.$

Now we recall some basic definitions and properties of compact
operators on Hilbert spaces \cite{13reed}:

Let us first consider the class of operators which may be thought
as some infinite dimensional matrices acting on an infinite
dimensional Hilbert
space $\mathcal{H}.$ More exactly, an operator on the Hilbert space $%
\mathcal{H}$ is said to be of finite rank if the orthogonal
component of its null space is finite dimensional. An operator
$T$ on $\mathcal{H}$ which can be approximated in norm by finite
rank operators is called compact. It can be characterized by the
property that for every $\varepsilon >0$ there is a finite
dimensional subspace $E\subset \mathcal{H}:||T_{|E^{\perp
}}||<\varepsilon ,$ where the orthogonal subspace $E^{\perp }$ is
of finite
codimension in $\mathcal{H}.$ $\ $This way we may define the set $\mathcal{%
K(H)}$ of \ all compact operators on the Hilbert spaces which is
the largest
two--sided ideal in the $C^{\ast }$--algebra $\mathcal{B}\left( \mathcal{H}%
\right) $ of all bounded operators. This set is also a $C^{\ast
}$--algebra with no unit, since the operator $I$ on an infinite
dimensional Hilbert
space is not compact, it is the only norm closed and two--sided when $%
\mathcal{H}$ is separable. We note that the defining representation of $%
\mathcal{K(H)}$ by itself is irreducible and it is the only
irreducible representation up to equivalence.

For an arbitrary $C^{\ast }$--algebra $\mathcal{A}$ acting
irreducibly on a
Hilbert space $\mathcal{H}$ and having non--zero intersection with $\mathcal{%
K(H)}$ one holds $\mathcal{K(H)}$ $\subseteq \mathcal{A}.$ In the
particular
case of finite dimensional Hilbert spaces, for instance, for $\mathcal{H}=\C%
^{n},$ we may write $\mathcal{B}\left( \C^{n}\right) =\mathcal{K}(\C%
^{n})=M_{n}\left( \C\right) ,$ which is the algebra of $n\times
n$ matrices with complex entries. Such algebra has only one
irreducible representation (the defining one).

\subsection{Commutative spaces}

Let us denote by $\mathcal{C}$ a fixed commutative $C^{\ast
}$--algebra with unit and by $\widehat{\mathcal{C}}$ the
corresponding structure space defined \ as the space of
equivalence classes of irreducible representations of
$\mathcal{C}$ ( $\widehat{\mathcal{C}}$ does not contains the
trivial representation $\mathcal{C}$ $\rightarrow \{0\}).$ One
can define a
non--trivial $\ast $--linear multiplicative functional $\phi :\mathcal{C}%
\rightarrow \C$ with the property that $\phi \left( ab\right)
=\phi \left(
a\right) \phi \left( b\right) $ for any $a$ and $b$ from $\mathcal{C}$ and $%
\phi (I)=1$ for every $\phi \in $ $\widehat{\mathcal{C}}.$ Every
such
multiplicative functional defines a character of $\mathcal{C},$ i. e. $%
\widehat{\mathcal{C}}$ is also the space of all characters of
$\mathcal{C}.$

The Gel'fand topology \ is the one with point wise convergence on $\mathcal{C%
}.$ A sequence $\left\{ \phi _{\varpi }\right\} _{\varpi \in \Xi
}$ of elements of $\widehat{\mathcal{C}}$, where \ $\Xi $ is any
directed set, \
converges to $\phi (c)\in \widehat{\mathcal{C}}$ if and only if for any $%
c\in \mathcal{C},$ the sequence $\left\{ \phi _{\varpi
}(c)\right\} _{\varpi
\in \Xi }$ converges to $\phi (c)$ in the topology of $\C.$ If the algebra $%
\mathcal{C}$ has a unite, $\widehat{\mathcal{C}}$ is a compact
Hausdorff space (a topological space is Hausdorff if for any two
points of the space there are two open disjoint neighborhoods
each containing one of the point,
see Ref. \cite{13kel}). The space $\widehat{\mathcal{C}}$ is only compact if $%
\mathcal{C}$ is without unit. This way the space
$\widehat{\mathcal{C}}$ (called the Gel'fand space) is made a
topological space. We may also consider $\widehat{\mathcal{C}}$
$\ $as a space of maximal ideals, two sided, of $\mathcal{C}$ $\
$instead of the space of irreducible representations. If there is
no unit, the ideals to be considered should be regular (modular),
see details in Ref. \cite{13dix}. Considering $\phi \in \C,$
we can decompose $\mathcal{C}=Ker\left( \phi \right) \oplus \C,$ where $%
Ker\left( \phi \right) $ is an ideal of codimension one and so is
a maximal
ideal of $\mathcal{C}.$ Considered in terms of maximal ideals, the space $%
\widehat{\mathcal{C}}$ is given the Jacobson topology,
equivalently, hull kernel topology (see next subsection for
general definitions for both commutative and noncommutative
spaces), producing a space which is homeomorphic to the one
constructed by means of the Gel'fand topology.

Let us consider an example when the algebra $\mathcal{C}$
generated by $s$
commuting self--adjoint elements $x_{1},...x_{s}.$ The structure space $%
\widehat{\mathcal{C}}$ can be identified with a compact subset of
$\R^{s}$ by the map $\phi (c)\in \widehat{\mathcal{C}}\rightarrow
\left[ \phi (x_{1}),...,\phi (x_{s})\right] \in \R^{s}.$ This map
has a joint spectrum of $x_{1},...x_{s}$ as the set of all
$s$--tuples of eigenvalues corresponding to common eigenvectors.

In general, we get an interpretation of elements $\mathcal{C}$ as $\C$%
--valued continuous functions on $\widehat{\mathcal{C}}.$ The
Gel'fand--Naimark theorem (see, for instance, \cite{13dix}) states
that all
continuous functions on $\widehat{\mathcal{C}}$ are of the form $\widehat{c}%
(\phi )=\phi \left( c\right) ,$ which defines the so--called
Gel'fand
transform for every $\phi (c)\in \widehat{\mathcal{C}}$ and the map $%
\widehat{c}:$ $\widehat{\mathcal{C}}\rightarrow \C$ being
continuous for
each $c.$ A transform $c\rightarrow \widehat{c}$ is isometric for every $%
c\in \mathcal{C}$ if $||\widehat{c}||_{\infty }=$ $||c||,$ with $%
||...||_{\infty }$ defined at the supremum norm on
$\mathcal{C}\left( \widehat{\mathcal{C}}\right) .$

The Gel'fand transform can be extended for an arbitrary locally
compact
topological space $M$ for which there exists a natural $C^{\ast }$--algebra $%
\mathcal{C}$ $(M).$ On can be identified both set wise and
topologically the Gel'fand space $\widehat{\mathcal{C}}(M)$ \ and
the space $M$ itself through
the evaluation map%
\begin{equation*}
\phi _{x}:\mathcal{C}(M)\rightarrow \C,~\phi _{x}(f)=f(x)
\end{equation*}%
for each $x\in M,$ where $\phi _{x}\in \widehat{\mathcal{C}}(M)$
gives a complex homomorphism. Denoting by $\mathcal{I}_{x}=\ker
\phi _{x},$ which is the maximal ideal of $\mathcal{C}$ $(M)$
consisting of all functions vanishing at $x,$ one proves
\cite{13dix} that the map $\phi _{x}$ is a homomorphism of $M$ onto
$\widehat{\mathcal{C}}(M),$ and, equivalently, every maximal
ideal of $\mathcal{C}$ $(M)$ is of the form $\mathcal{I}_{x}$ for
some $x\in M.$

We conclude this subsection: There is a one--to--one
correspondence between the $\ast $--isomorphism classes of
commutative $C^{\ast }$--algebras and the homomorphism classes of
locally compact Hausdorff spaces (such commutative $C^{\ast
}$--algebras with unit correspond to compact Hausdorff spaces).
This correspondence defines a complete duality between the
category of (locally) compact Hausdorff spaces and (proper, when
a map $\ f$ relating two locally compact Hausdorff spaces
$f:X\rightarrow Y$ has the property that $f^{-1}\left( K\right) $
is a compact subset of $X$ when $K$ is a compact subset of $Y,$
and ) continuous maps and the category of commutative (non
necessarily) unital $C^{\ast }$--algebras and $\ast
$--homomorphisms.
In result, any commutative $C^{\ast }$--algebra can be realized as the $%
C^{\ast }$--algebra of complex valued functions over a (locally)
compact Hausdorff space. It should be mentioned that the space
$M$ is a metrizable topological space, i. e. its topology comes
from a metric, if and only if the $C^{\ast }$--algebra is norm
separable (it admits a dense in norm countable subset). This
space is connected topologically if the corresponding algebra has
no projectors which are self--adjoint, $p^{\ast }=p $ and satisfy
the idempotentity condition $p^{2}=p.$

We emphasize that the constructions considered for commutative
algebras cannot be directly generalized for noncommutative
$C^{\ast }$--algebras.

\subsection{Noncommutative spaces}

For a given noncommutative $C^{\ast }$--algebra, there is more
than one
candidate for the analogue of the topological space $M.$ Following Ref. \cite%
{13landi} (see there the proofs of results and Appendices), we
consider two possibilities:

\begin{itemize}
\item To use the space $\widehat{\mathcal{A}}$ , \ called the structure
space of $\ $the noncommutative $C^{\ast }$--algebra
$\mathcal{A},$ which is
the space of all unitary equivalence classes of irreducible $\ast $%
--representations.

\item To use the space $\Pr im\mathcal{A},$ called the primitive spectrum of
$\mathcal{A},$ which is the space of kernels of irreducible $\ast $%
--representations (any element of $\Pr im\mathcal{A}$ is
automatically a two--sided $\ast $--ideal of $\mathcal{A)}$.
\end{itemize}

The spaces $\widehat{\mathcal{A}}$ and $\Pr im\mathcal{A}$ agree
for a commutative $C^{\ast }$--algebra, for instance,
$\widehat{\mathcal{A}}$ may be very complicate while $\Pr
im\mathcal{A}$ consisting of a single point.

Let us examine a simple example of generalization to
noncommutative $C^{\ast }$--algebra given by the $2\times 2$
complex matrix algebra
\begin{equation*}
M_{2}(\C)=\{\left[
\begin{array}{cc}
a_{11} & a_{12} \\
a_{21} & a_{22}%
\end{array}%
\right] ,~a_{ij}\in \C\}.
\end{equation*}%
The commutative subalgebra of diagonal matrices
$\mathcal{C}=\{diag[\lambda _{1},\lambda _{2}],~\lambda _{1,2}\in
\C\}$ has a structure space consisting of two points given by the
characters $\ \phi _{1,2}(\left[
\begin{array}{cc}
\lambda _{1} & 0 \\
0 & \lambda _{2}%
\end{array}%
\right] )=\lambda _{1,2}.$ These two characters extend as pure
states to the
full algebra $M_{2}(\C)$ by the maps $\widetilde{\phi }_{1,2}:M_{2}(\C%
)\rightarrow \C,$%
\begin{equation*}
\widetilde{\phi }_{1}\left( \left[
\begin{array}{cc}
a_{11} & a_{12} \\
a_{21} & a_{22}%
\end{array}%
\right] \right) =a_{11},~\widetilde{\phi }_{2}\left( \left[
\begin{array}{cc}
a_{11} & a_{12} \\
a_{21} & a_{22}%
\end{array}%
\right] \right) =a_{22}.
\end{equation*}%
Further details are given in Appendix B to Ref. \cite{13landi}.

It is possible to define natural topologies on $\widehat{\mathcal{A}}$ and $%
\Pr im\mathcal{A},$ for instance, by means of a closure
operation. \ For a subset $Q\subset \Pr im\mathcal{A},$ the
closure $\overline{Q}$ is by definition the subset of all
elements in $\Pr im\mathcal{A}$ containing the intersection $\cap
Q$ of the elements of $Q,~\overline{Q}\doteqdot \left\{
\mathcal{I}\in \Pr im\mathcal{A}:\cap Q\subseteq
\mathcal{I}\right\} .$ It is possible to check that such subsets
satisfy the Kuratowski topology axioms and this way defined
topology on $\Pr im\mathcal{A}$ is called the Jacobson topology
or hull--kernel topology, for which $\cap Q$ is the kernel of $Q$
and $~\overline{Q}$ is the hull of $\cap Q$ (see \cite{13landi,13dix}
on the properties of this type topological spaces).

\section{Nonlinear Connections in \newline Noncommutative Spa\-ces}

In this subsection we define the nonlinear connections in module
spaces, i. e. in noncommutative spaces. The concept on nonlinear
connection came from Finsler geometry (as a set of coefficients
it is present in the works of E. Cartan \cite{13cartan}, then the
concept was elaborated in a more explicit fashion by A. Kawaguchi
\cite{13kaw}). The global formulation in commutative spaces is due
to W. Barthel \ \cite{13barthel} and it was developed in details
for vector, covector and higher order bundles
\cite{13ma,13miron,13bejancu},
spinor bundles \cite{13vspinors,13vmon2}, superspaces and superstrings \cite%
{13vsuper,13vmon1,13vstring} and in the theory of exact off--diagonal
solutions of the Einstein equations \cite{13vexsol,13vmethod}. The
concept of nonlinear connection can be extended in a similar
manner from commutative to noncommutative spaces if a
differential calculus is fixed on a noncommutative vector (or
covector) bundle.

\subsection{Modules as bundles}

A vector bundle $E\rightarrow M$ over a manifold $M$ is completely
characterized by the space $\mathcal{E}=\Gamma \left( E,M\right)
$ over its smooth sections defined as a (right) module over the
algebra of $C^{\infty }\left( M\right) $ of smooth functions over
$M.$ It is known the Serre--Swan theorem \cite{13swan} which states
that locally trivial, finite--dimensional complex vector bundles
over a compact Hausdorff space $M$ correspond
canonically to finite projective modules over the algebra \ $\mathcal{A}%
=C^{\infty }\left( M\right).$ Inversely, for $\mathcal{E}$ being
a finite projective modules over $C^{\infty }\left( M\right) ,$
the fiber $E_{m}$ of
the associated bundle $E$ over the point $x\in M$ is the space $E_{x}=%
\mathcal{E}/\mathcal{EI}_{x}$ where the ideal is given by%
\begin{equation*}
\mathcal{I}_{x}=\ker \{\xi _{x}:C^{\infty }\left( M\right) \rightarrow \C%
;\xi _{x}(x)=f(x)\}=\{f\in C^{\infty }\left( M\right) |f(x)=0\}\in \mathcal{C%
}\left( M\right) .
\end{equation*}%
If the algebra $\mathcal{A}$ is taken to play the role of smooth
functions
on a noncomutative, instead of the commutative algebra smooth functions $%
C^{\infty }\left( M\right) $, the analogue of a vector bundle is
provided by a projective module of finite type (equivalently,
finite projective module) over $\mathcal{A}.$ On considers the
proper construction of projective modules of finite type
generalizing the Hermitian bundles as well the notion of Hilbert
module when $\mathcal{A}$ is a $C^{\ast }$--algebra in the
Appendix C of Ref. \cite{13landi}.

A vector space $\mathcal{E}$ over the complex number field $\C$
can be defined also as a right module of an algebra $\mathcal{A}$
over $\C$ \ which carries a right representation of
$\mathcal{A},$ when for every map of elements $\mathcal{E}$
$\times \mathcal{A}\ni \left( \eta ,a\right) \rightarrow \eta
a\in \mathcal{E}$ one hold the properties
\begin{equation*}
\lambda (ab)=(\lambda a)b,~\lambda (a+b)=\lambda a+\lambda
b,~(\lambda +\mu )a=\lambda a+\mu a
\end{equation*}%
fro every $\lambda ,\mu \in \mathcal{E}$ and $a,b\in \mathcal{A}.$

Having two $\mathcal{A}$--modules $\mathcal{E}$ and
$\mathcal{F},$ a
morphism of $\mathcal{E}$ into $\mathcal{F}$ is \ any linear map $\rho :%
\mathcal{E}$ $\rightarrow $ $\mathcal{F}$ $\ $which is also $\mathcal{A}$%
--linear, i. e. $\rho (\eta a)=\rho (\eta )a$ for every $\eta \in
\mathcal{E} $ and $a\in \mathcal{A}.$

We can define in a similar (dual) manner the left modules and
theirs morphisms which are distinct from the right ones for
noncommutative algebras
$\mathcal{A}.$ A bimodule over an algebra $\mathcal{A}$ is a vector space $%
\mathcal{E}$ which carries both a left and right module
structures. We may define the opposite algebra $\mathcal{A}^{o}$
with elements $a^{o}$ being in bijective correspondence with the
elements \ $a\in \mathcal{A}$ while the multiplication is given
by $\mathcal{\,}a^{o}b^{o}=\left( ba\right) ^{o}.$A right
(respectively, left) $\mathcal{A}$--module $\mathcal{E}$ is
connected
to a left (respectively right) $\mathcal{A}^{o}$--module via relations $%
a^{o}\eta =\eta a^{o}$ (respectively, $a\eta =\eta a).$

One introduces the enveloping algebra $\mathcal{A}^{\varepsilon }=\mathcal{A}%
\otimes _{\C}\mathcal{A}^{o};$ any $\mathcal{A}$--bimodule
$\mathcal{E}$ can be regarded as a right [left]
$\mathcal{A}^{\varepsilon }$--module by setting $\eta \left(
a\otimes b^{o}\right) =b\eta a$ $\quad \left[ \left( a\otimes
b^{o}\right) \eta =a\eta b\right] .$

For a (for instance, right) module $\mathcal{E}$ , we may
introduce a family of elements $\left( e_{t}\right) _{t\in T}$
parametrized by any (finite or infinite) directed set $T$ for
which any element $\eta \in \mathcal{E}$ is expressed as a
combination (in general, in \ more than one manner) $\eta
=\sum\nolimits_{t\in T}e_{t}a_{t}$ with $a_{t}\in \mathcal{A}$
and only a finite number of non vanishing terms in the sum. A
family $\left( e_{t}\right) _{t\in T}$ is free if it consists
from linearly independent elements and defines a basis if any
element $\eta \in \mathcal{E}$ can be written as a unique
combination (sum). One says a module to be free if it admits a
basis. The module $\mathcal{E}$ is said to be of finite type if \
it is finitely generated, i. e. it admits a generating family of
finite cardinality.

Let us consider the module $\mathcal{A}^{K}\doteqdot \C^{K}\otimes _{\C}%
\mathcal{A}.$ The elements of this module can be thought as
$K$--dimensional vectors with entries in $\mathcal{A}$ and
written uniquely as a linear combination $\eta
=\sum\nolimits_{t=1}^{K}e_{t}a_{t}$ were the basis $e_{t}$
identified with the canonical basis of $\C^{K}.$ This is a free
and finite type module. In general, we can have bases of different
cardinality. However, if a module $\mathcal{E}$ \ is of finite
type there is always an integer $K$ and a module surjection $\rho
:\mathcal{A}^{K}\rightarrow \mathcal{E}$ with a base being a
image of a free basis, $\epsilon _{j}=\rho (e_{j});j=1,2,...,K.$

In general, it is not possible to solve the constraints among the
basis
elements as to get a free basis. \ The simplest example is to take a sphere $%
S^{2}$ and the Lie algebra of smooth vector fields on it, $\mathcal{G}=%
\mathcal{G}(S^{2})$ which is a module of finite type over
$C^{\infty }\left( S^{2}\right) ,$ with the basis defined by
$X_{i}=\sum_{j,k=1}^{3}\varepsilon _{ijk}x_{k}\partial /\partial
x^{k};i,j,k=1,2,3,$ and coordinates $x_{i}$ such that
$\sum_{j=1}^{3}x_{j}^{2}=1.$ The introduced basis is not free
because $\sum_{j=1}^{3}x_{j}X_{j}=0;$ there are not global vector field on $%
S^{2}$ which could form a basis of $\mathcal{G}(S^{2}).$ This
means that the tangent bundle $TS^{2}$ is not trivial.

We say that a right $\mathcal{A}$--module $\mathcal{E}$ is
projective if for
every surjective module morphism $\rho :\mathcal{M}$ $\rightarrow $ $%
\mathcal{N}$ splits, i. e. there exists a module morphism \ $s:\mathcal{E}$ $%
\rightarrow $ $\mathcal{M}$ such that $\rho \circ
s=id_{\mathcal{E}}.$ There are different definitions of
porjective modules (see Ref. \cite{13landi} on properties of such
modules). Here we note the property that if a $\mathcal{A}
$--module $\mathcal{E}$ is projective, there exists a free module $\mathcal{F%
}$ and a module $\mathcal{E}^{\prime }$ (being a priory
projective) such that $\mathcal{F}=\mathcal{E}\oplus
\mathcal{E}^{\prime }.$

For the right $\mathcal{A}$--module $\mathcal{E}$ being
projective and of finite type with surjection $\rho
:\mathcal{A}^{K}\rightarrow \mathcal{E}$ and following the
projective property we can find a lift $\widetilde{\lambda
}:\mathcal{E}$ $\rightarrow $ $\mathcal{A}^{K}$ such that $\rho
\circ \widetilde{\lambda }=id_{\mathcal{E}}.$ There is a proof of
the property that the module $\mathcal{E}$ is projective of
finite type over $\mathcal{A}$
if and only if there exists an idempotent $p\in End_{\mathcal{A}}\mathcal{A}%
^{K}=M_{K}(\mathcal{A}),$ $p^{2}=p,$ the $M_{K}(\mathcal{A})$
denoting the
algebra of $K\times K$ matrices with entry in $\mathcal{A},$ such that $%
\mathcal{E}=p\mathcal{A}^{K}.$ We may associate the elements of
$\mathcal{E}$ to $K$--dimensional column vectors whose elements
are in $\mathcal{A},$ the collection \ of which are invariant
under the map $p,$ $\mathcal{E}$ $=\{\xi =(\xi _{1},...,\xi
_{K});\xi _{j}\in \mathcal{A},~p\xi =\xi \}.$ For simplicity, we
shall use the term finite projective to mean projective of finite
type.

The noncommutative variant of the theory of vector bundles may be
constructed by using the Serre and Swan theorem \cite{13swan,13landi}
which states that for a compact finite dimensional manifold $M,$
a $C^{\infty }\left( M\right) $--module $\mathcal{E}$ is
isomorphic to a module $\Gamma \left( E,M\right) $ of smooth
sections of a bundle $E\rightarrow M,$ if and only if it is
finite projective. If $E$ is a complex vector bundle over a
compact manifold $M$ of dimension $n,$ there exists a finite cover $%
\{U_{i},i=1,...,n\}$ of $M$ such that $E_{|U_{i}}$ is trivial.
Thus, the integer $K$ which determines the rank of the free
bundle from which to project onto sections of the bundle is
determined by the equality $N=mn$ where $m$ is the rank of the
bundle (i. e. of the fiber) and $n$ is the dimension of $M.$

\subsection{The commutative nonlinear connection geometry}

Let us remember the definition and the main results on nonlinear
connections in commutative vector bundles as in Ref. \cite{13ma}.

\subsubsection{Vector bundles, Riemannian spaces and nonlinear connections}

We consider a vector bundle $\xi =\left( E,\mu ,M\right) $ whose
fibre is $\R
$$^{m}$ and $\mu ^{T}:TE\rightarrow TM$ denotes the differential of the map $%
\mu :E\rightarrow M.$ The map $\mu ^{T}$ is a fibre--preserving
morphism of the tangent bundle $\left( TE,\tau _{E},E\right) $ to
$E$ and of tangent bundle $\left( TM,\tau ,M\right) $ to $M.$ The
kernel of the morphism $\mu ^{T}$ is a vector subbundle of the
vector bundle $\left( TE,\tau _{E},E\right).$ This kernel is
denoted $\left( VE,\tau _{V},E\right) $ and called the vertical
subbundle over $E.$ We denote by $i:VE\rightarrow TE$
the inclusion mapping and the local coordinates of a point $u\in E$ by $%
u^{\alpha }=\left( x^{i},y^{a}\right) ,$ where indices
$i,j,k,...=1,2,...,n$ and $a,b,c,...=1,2,...,m.$

A vector $X_{u}\in TE,$ tangent in the point $u\in E,$ is locally
represented $\left( x,y,X,\widetilde{X}\right) =\left(
x^{i},y^{a},X^{i},X^{a}\right) ,$ where $\left( X^{i}\right) \in
$$\R$$^{n}$
and $\left( X^{a}\right) \in $$\R$$^{m}$ are defined by the equality $%
X_{u}=X^{i}\partial _{i}+X^{a}\partial _{a}$ [$\partial _{\alpha
}=\left(
\partial _{i},\partial _{a}\right) $ are usual partial derivatives on
respective coordinates $x^{i}$ and $y^{a}$]. For instance, $\mu
^{T}\left( x,y,X,\widetilde{X}\right) =\left( x,X\right) $ and
the submanifold $VE$ contains elements of type $\left(
x,y,0,\widetilde{X}\right) $ and the local fibers of the vertical
subbundle are isomorphic to $\R$$^{m}.$ Having $\mu ^{T}\left(
\partial _{a}\right) =0,$ one comes out that $\partial _{a}$ is a
local basis of the vertical distribution $u\rightarrow V_{u}E$ on
$E,$ which is an integrable distribution.

A nonlinear connection (in brief, N--connection) in the vector
bundle $\xi =\left( E,\mu ,M\right) $ is the splitting on the
left of the exact sequence
\begin{equation*}
0\rightarrow VE\rightarrow TE/VE\rightarrow 0,
\end{equation*}%
i. e. a morphism of vector bundles $N:TE\rightarrow VE$ such that
$C\circ i$ is the identity on $VE.$

The kernel of the morphism $N$ is a vector subbundle of $\left(
TE,\tau
_{E},E\right) ,$ it is called the horizontal subbundle and denoted by $%
\left( HE,\tau _{H},E\right) .$ Every vector bundle $\left(
TE,\tau _{E},E\right) $ provided with a N--connection structure
is Whitney sum of the vertical and horizontal subbundles, i. e.
\begin{equation}
TE=HE\oplus VE.  \label{3wihit}
\end{equation}
It is proven that for every vector bundle $\xi =\left( E,\mu
,M\right) $ over a compact manifold $M$ there exists a nonlinear
connection \cite{13ma}.

Locally a N--connection $N$ is parametrized by a set of
coefficients\newline $N_{i}^{a}(u^{\alpha
})=N_{i}^{a}(x^{j},y^{b})$ which transforms as
\begin{equation*}
N_{i^{\prime }}^{a^{\prime }}\frac{\partial x^{i^{\prime }}}{\partial x^{i}}%
=M_{a}^{a^{\prime }}N_{i}^{a}-\frac{\partial M_{a}^{a^{\prime
}}}{\partial x^{i}}y^{a}
\end{equation*}%
under coordinate transforms on the vector bundle $\xi =\left(
E,\mu
,M\right) ,$%
\begin{equation*}
x^{i^{\prime }}=x^{i^{\prime }}\left( x^{i}\right) \mbox{ and
}y^{a^{\prime }}=M_{a}^{a^{\prime }}(x)y^{a}.
\end{equation*}

If a N--connection structure is defined on $\xi ,$ the operators
of local partial derivatives $\partial _{\alpha }=\left( \partial
_{i},\partial _{a}\right) $ and differentials $d^{\alpha
}=du^{\alpha }=\left( d^{i}=dx^{i},d^{a}=dy^{a}\right) $ should
be elongated as to adapt the local basis (and dual basis)
structure to the Whitney decomposition of the vector
bundle into vertical and horizontal subbundles, (\ref{3wihit}):%
\begin{eqnarray}
\partial _{\alpha } &=&\left( \partial _{i},\partial _{a}\right) \rightarrow
\delta _{\alpha }=\left( \delta _{i}=\partial
_{i}-N_{i}^{b}\partial
_{b},\partial _{a}\right) ,  \label{6dder} \\
d^{\alpha } &=&\left( d^{i},d^{a}\right) \rightarrow \delta
^{\alpha }=\left( d^{i},\delta ^{a}=d^{a}+N_{i}^{b}d^{i}\right)
.  \label{7ddif}
\end{eqnarray}%
The transforms can be considered as some particular case of frame
(vielbein) transforms of type
\begin{equation*}
\partial _{\alpha }\rightarrow \delta _{\alpha }=e_{\alpha }^{\beta
}\partial _{\beta }\mbox{ and }d^{\alpha }\rightarrow \delta
^{\alpha }=(e^{-1})_{\beta }^{\alpha }\delta ^{\beta },
\end{equation*}%
$e_{\alpha }^{\beta }(e^{-1})_{\beta }^{\gamma }=\delta _{\alpha
}^{\gamma }, $ when the ''tetradic'' coefficients $\delta
_{\alpha }^{\beta }$ are induced by using the Kronecker symbols
$\delta _{a}^{b},\delta _{j}^{i}$ and $N_{i}^{b}.$

The bases $\delta _{\alpha }$ and $\delta ^{\alpha }$ satisfy in
general some anholonomy conditions, for instance,
\begin{equation}
\delta _{\alpha }\delta _{\beta }-\delta _{\beta }\delta _{\alpha
}=W_{\alpha \beta }^{\gamma }\delta _{\gamma },  \label{4anhol}
\end{equation}%
where $W_{\alpha \beta }^{\gamma }$ are called the anholonomy
coefficients.

Tensor fields on a vector bundle $\xi =\left( E,\mu ,M\right) $
provided with N--connection structure $N,$ we shall write $\xi
_{N},$ may be decomposed with in N--adapted form with respect to
the bases $\delta _{\alpha }$ and $\delta ^{\alpha },$ and their
tensor products. For instance, for a tensor of rang (1,1)
$T=\{T_{\alpha }^{~\beta }=\left(
T_{i}^{~j},T_{i}^{~a},T_{b}^{~j},T_{a}^{~b}\right) \}$ we have
\begin{equation}
T=T_{\alpha }^{~\beta }\delta ^{\alpha }\otimes \delta _{\beta
}=T_{i}^{~j}d^{i}\otimes \delta _{i}+T_{i}^{~a}d^{i}\otimes
\partial _{a}+T_{b}^{~j}\delta ^{b}\otimes \delta
_{j}+T_{a}^{~b}\delta ^{a}\otimes
\partial _{b}.  \label{1dten}
\end{equation}

Every N--connection with coefficients $N_{i}^{b}$ $\
$automatically generates a linear connection on $\xi $ as $\Gamma
_{\alpha \beta }^{(N)\gamma }=\{N_{bi}^{a}=\partial
N_{i}^{a}(x,y)/\partial y^{b}\}$ which defines a covariant
derivative $D_{\alpha }^{(N)}A^{\beta }=\delta _{\alpha }A^{\beta
}+\Gamma _{\alpha \gamma }^{(N)\beta }A^{\gamma }.$

Another important characteristic of a N--connection is its
curvature $\Omega =\{\Omega _{ij}^{a}\}$ with the coefficients
\begin{equation*}
\Omega _{ij}^{a}=\delta _{j}N_{i}^{a}-\delta
_{i}N_{j}^{a}=\partial _{j}N_{i}^{a}-\partial
_{i}N_{j}^{a}+N_{i}^{b}N_{bj}^{a}-N_{j}^{b}N_{bi}^{a}.
\end{equation*}

In general, on a vector bundle we consider arbitrary linear
connection and, for instance, metric structure adapted to the
N--connection decomposition into vertical and horizontal
subbundles (one says that such objects are distinguished by the
N--connection, in brief, d--objects, like the d-tensor
(\ref{1dten}), d--connection, d--metric:

\begin{itemize}
\item the coefficients of linear d--connections $\Gamma =\{\Gamma _{\alpha
\gamma }^{\beta }=\left(
L_{jk}^{i},L_{bk}^{a},C_{jc}^{i},C_{ac}^{b}\right) \}$ are
defined for an arbitrary covariant derivative $D$ on $\xi $ being
adapted to the $N$--connection structure as $D_{\delta _{\alpha
}}(\delta _{\beta })=\Gamma _{\beta \alpha }^{\gamma }\delta
_{\gamma }$ with the coefficients being invariant under
horizontal and vertical decomposition
\begin{equation*}
\quad D_{\delta _{i}}(\delta _{j})=L_{ji}^{k}\delta
_{k},~D_{\delta _{i}}(\partial _{a})=L_{ai}^{b}\partial
_{b},~D_{\partial _{c}}(\delta _{j})=C_{jc}^{k}\delta
_{k},~~D_{\partial _{c}}(\partial _{a})=C_{ac}^{b}\partial _{b}.
\end{equation*}

\item the d--metric structure $G=g_{\alpha \beta }\delta ^{a}\otimes \delta
^{b}$ which has the invariant decomposition as $g_{\alpha \beta
}=\left(
g_{ij},g_{ab}\right) $ following from%
\begin{equation}
G=g_{ij}(x,y)d^{i}\otimes d^{j}+g_{ab}(x,y)\delta ^{a}\otimes
\delta ^{b}. \label{7dmetric}
\end{equation}
\end{itemize}

We may impose the condition that a d--metric and a d--connection
are compatible, i. e. there are satisfied the conditions
\begin{equation}
D_{\gamma }g_{\alpha \beta }=0.  \label{1metrcond}
\end{equation}

With respect to the anholonomic frames (\ref{6dder}) and
(\ref{7ddif}), there is a linear connection, called the canonical
distinguished linear connection, which is similar to the metric
connection introduced by the Christoffel symbols in the case of
holonomic bases, i. e. being constructed
only from the metric components and satisfying the metricity conditions (\ref%
{1metrcond}). It is parametrized by the coefficients,\ $\Gamma _{\
\beta \gamma }^{\alpha }=\left( L_{\ jk}^{i},L_{\ bk}^{a},C_{\
jc}^{i},C_{\ bc}^{a}\right) $ with the coefficients
\begin{eqnarray}
L_{\ jk}^{i} &=&\frac{1}{2}g^{in}\left( \delta _{k}g_{nj}+\delta
_{j}g_{nk}-\delta _{n}g_{jk}\right) ,  \label{6dcon} \\
L_{\ bk}^{a} &=&\partial _{b}N_{k}^{a}+\frac{1}{2}h^{ac}\left(
\delta _{k}h_{bc}-h_{dc}\partial _{b}N_{k}^{d}-h_{db}\partial
_{c}N_{k}^{d}\right) ,
\notag \\
C_{\ jc}^{i} &=&\frac{1}{2}g^{ik}\partial _{c}g_{jk},\ C_{\ bc}^{a}=\frac{1}{%
2}h^{ad}\left( \partial _{c}h_{db}+\partial _{b}h_{dc}-\partial
_{d}h_{bc}\right) .  \notag
\end{eqnarray}%
We note that on Riemannian spaces the N--connection is an object
completely
defined by anholonomic frames, when the coefficients of frame transforms, $%
e_{\alpha }^{\beta }\left( u^{\gamma }\right) ,$ are parametrized
explicitly via certain values $\left( N_{i}^{a},\delta
_{i}^{j},\delta _{b}^{a}\right) , $ where $\delta _{i}^{j}$ $\
$and $\delta _{b}^{a}$ are the Kronecker symbols. By
straightforward calculations we can compute that the coefficients
of the Levi--Civita metric connection
\begin{equation*}
\Gamma _{\alpha \beta \gamma }^{\bigtriangledown }=g\left( \delta
_{\alpha },\bigtriangledown _{\gamma }\delta _{\beta }\right)
=g_{\alpha \tau }\Gamma _{\beta \gamma }^{\bigtriangledown \tau
},\,
\end{equation*}%
associated to a covariant derivative operator $\bigtriangledown ,$
satisfying the metricity condition\\ $\bigtriangledown _{\gamma
}g_{\alpha \beta }=0$ for $g_{\alpha \beta }=\left(
g_{ij},h_{ab}\right) ,$
\begin{equation}
\Gamma _{\alpha \beta \gamma }^{\bigtriangledown
}=\frac{1}{2}\left[ \delta _{\beta }g_{\alpha \gamma }+\delta
_{\gamma }g_{\beta \alpha }-\delta _{\alpha }g_{\gamma \beta
}+g_{\alpha \tau }W_{\gamma \beta }^{\tau }+g_{\beta \tau
}W_{\alpha \gamma }^{\tau }-g_{\gamma \tau }W_{\beta \alpha
}^{\tau }\right] ,  \label{2lcsym}
\end{equation}%
are given with respect to the anholonomic basis (\ref{7ddif}) by
the coefficients
\begin{equation}
\Gamma _{\beta \gamma }^{\bigtriangledown \tau }=\left( L_{\
jk}^{i},L_{\ bk}^{a},C_{\ jc}^{i}+\frac{1}{2}g^{ik}\Omega
_{jk}^{a}h_{ca},C_{\ bc}^{a}\right) .  \label{2lccon}
\end{equation}%
A specific property of off--diagonal metrics is that they can
define different classes of linear connections which satisfy the
metricity conditions for a given metric, or inversely, there is a
certain class of metrics which satisfy the metricity conditions
for a given linear connection. \ This result was originally
obtained by A. Kawaguchi \cite{13kaw} (Details can be found in Ref.
\cite{13ma}, see Theorems 5.4 and 5.5 in Chapter III, formulated
for vector bundles; here we note that similar proofs hold also on
manifolds enabled with anholonomic frames associated to a
N--connection structure).

With respect to anholonomic frames, we can not distinguish the
Levi--Civita connection as the unique both metric and torsionless
one. For instance, both linear connections (\ref{6dcon}) and
(\ref{2lccon}) contain anholonomically induced torsion
coefficients, are compatible with the same metric and transform
into the usual Levi--Civita coefficients for vanishing
N--connection and ''pure'' holonomic coordinates. This means that
to an off--diagonal metric in general relativity one may be
associated different covariant differential calculi, all being
compatible with the same metric structure (like in the
non--commutative geometry, which is not a surprising \ fact
because the anolonomic frames satisfy by definition some
non--commutative relations (\ref{4anhol})). In such cases we have
to select a particular type of connection following some physical
or geometrical arguments, or to impose some conditions when there
is a single compatible linear connection constructed only from
the metric and N--coefficients. We
note that if $\Omega _{jk}^{a}=0$ the connections (\ref{6dcon}) and (\ref%
{2lccon}) coincide, i. e. $\Gamma _{\ \beta \gamma }^{\alpha
}=\Gamma _{\beta \gamma }^{\bigtriangledown \alpha }.$

\subsubsection{D--torsions and d--curvatures:}

The anholonomic coefficients $W_{\ \alpha \beta }^{\gamma }$ and
N--elongated derivatives give nontrivial coefficients for the
torsion tensor, $T(\delta _{\gamma },\delta _{\beta })=T_{\ \beta
\gamma }^{\alpha }\delta _{\alpha },$ where
\begin{equation}
T_{\ \beta \gamma }^{\alpha }=\Gamma _{\ \beta \gamma }^{\alpha
}-\Gamma _{\ \gamma \beta }^{\alpha }+w_{\ \beta \gamma }^{\alpha
},  \label{4torsion}
\end{equation}%
and for the curvature tensor, $R(\delta _{\tau },\delta _{\gamma
})\delta _{\beta }=R_{\beta \ \gamma \tau }^{\ \alpha }\delta
_{\alpha },$ where
\begin{eqnarray}
R_{\beta \ \gamma \tau }^{\ \alpha } &=&\delta _{\tau }\Gamma _{\
\beta \gamma }^{\alpha }-\delta _{\gamma }\Gamma _{\ \beta \tau
}^{\alpha }  \notag
\\
&&+\Gamma _{\ \beta \gamma }^{\varphi }\Gamma _{\ \varphi \tau
}^{\alpha }-\Gamma _{\ \beta \tau }^{\varphi }\Gamma _{\ \varphi
\gamma }^{\alpha }+\Gamma _{\ \beta \varphi }^{\alpha }w_{\
\gamma \tau }^{\varphi }. \label{3curvature}
\end{eqnarray}%
We emphasize that the torsion tensor on (pseudo) Riemannian
spacetimes is induced by anholonomic frames, whereas its
components vanish with respect to holonomic frames. All tensors
are distinguished (d) by the N--connection structure into
irreducible (horizont\-al--vertical) h--v--components, and are
called d--tensors. For instance, the torsion, d--tensor has the
following irreducible, nonvanishing, h--v--components,\\ $T_{\
\beta \gamma }^{\alpha }=\{T_{\ jk}^{i},C_{\ ja}^{i},S_{\
bc}^{a},T_{\ ij}^{a},T_{\ bi}^{a}\},$ where
\begin{eqnarray}
T_{.jk}^{i} &=&T_{jk}^{i}=L_{jk}^{i}-L_{kj}^{i},\quad
T_{ja}^{i}=C_{.ja}^{i},\quad T_{aj}^{i}=-C_{ja}^{i},  \notag \\
T_{.ja}^{i} &=&0,\quad
T_{.bc}^{a}=S_{.bc}^{a}=C_{bc}^{a}-C_{cb}^{a},
\label{3dtors} \\
T_{.ij}^{a} &=&-\Omega _{ij}^{a},\quad T_{.bi}^{a}=\partial
_{b}N_{i}^{a}-L_{.bi}^{a},\quad T_{.ib}^{a}=-T_{.bi}^{a}  \notag
\end{eqnarray}%
(the d--torsion is computed by substituting the
h--v--compo\-nents of the
canonical d--connection (\ref{6dcon}) and anholonomy coefficients(\ref{4anhol}%
) into the formula for the torsion coefficients (\ref{4torsion})).

The curvature d-tensor has the following irreducible,
non-vanishing, h--v--compon\-ents\ $R_{\beta \ \gamma \tau }^{\
\alpha
}=%
\{R_{h.jk}^{.i},R_{b.jk}^{.a},P_{j.ka}^{.i},P_{b.ka}^{.c},S_{j.bc}^{.i},S_{b.cd}^{.a}\},
$\ where
\begin{eqnarray}
R_{h.jk}^{.i} &=&\delta _{k}L_{.hj}^{i}-\delta
_{j}L_{.hk}^{i}+L_{.hj}^{m}L_{mk}^{i}-L_{.hk}^{m}L_{mj}^{i}-C_{.ha}^{i}%
\Omega _{.jk}^{a},  \label{3dcurvatures} \\
R_{b.jk}^{.a} &=&\delta _{k}L_{.bj}^{a}-\delta
_{j}L_{.bk}^{a}+L_{.bj}^{c}L_{.ck}^{a}-L_{.bk}^{c}L_{.cj}^{a}-C_{.bc}^{a}%
\Omega _{.jk}^{c},  \notag \\
P_{j.ka}^{.i} &=&\partial
_{a}L_{.jk}^{i}+C_{.jb}^{i}T_{.ka}^{b}-(\delta
_{k}C_{.ja}^{i}+L_{.lk}^{i}C_{.ja}^{l}-L_{.jk}^{l}C_{.la}^{i}-L_{.ak}^{c}C_{.jc}^{i}),
\notag \\
P_{b.ka}^{.c} &=&\partial
_{a}L_{.bk}^{c}+C_{.bd}^{c}T_{.ka}^{d}-(\delta
_{k}C_{.ba}^{c}+L_{.dk}^{c\
}C_{.ba}^{d}-L_{.bk}^{d}C_{.da}^{c}-L_{.ak}^{d}C_{.bd}^{c}),  \notag \\
S_{j.bc}^{.i} &=&\partial _{c}C_{.jb}^{i}-\partial
_{b}C_{.jc}^{i}+C_{.jb}^{h}C_{.hc}^{i}-C_{.jc}^{h}C_{hb}^{i},  \notag \\
S_{b.cd}^{.a} &=&\partial _{d}C_{.bc}^{a}-\partial
_{c}C_{.bd}^{a}+C_{.bc}^{e}C_{.ed}^{a}-C_{.bd}^{e}C_{.ec}^{a}
\notag
\end{eqnarray}
(the d--curvature components are computed in a similar fashion by
using the formula for curvature coefficients (\ref{3curvature})).

\subsubsection{Einstein equations in d--variables}

In this subsection we write and analyze the Einstein equations on
spaces provided with anholonomic frame structures and associated
N--connections.

The Ricci tensor $R_{\beta \gamma }=R_{\beta ~\gamma \alpha
}^{~\alpha }$ has the d--components
\begin{eqnarray}
R_{ij} &=&R_{i.jk}^{.k},\quad R_{ia}=-^2P_{ia}=-P_{i.ka}^{.k},
\label{6dricci} \\
R_{ai} &=&^1P_{ai}=P_{a.ib}^{.b},\quad R_{ab}=S_{a.bc}^{.c}.
\notag
\end{eqnarray}
In general, since $^1P_{ai}\neq ~^2P_{ia}$, the Ricci d-tensor is
non-symmetric (this could be with respect to anholonomic frames of
reference). The scalar curvature of the metric d--connection, $%
\overleftarrow{R}=g^{\beta \gamma }R_{\beta \gamma },$ is computed
\begin{equation}
{\overleftarrow{R}}=G^{\alpha \beta }R_{\alpha \beta
}=\widehat{R}+S, \label{4dscalar}
\end{equation}
where $\widehat{R}=g^{ij}R_{ij}$ and $S=h^{ab}S_{ab}.$

By substituting (\ref{6dricci}) and (\ref{4dscalar}) into the
Einstein equations
\begin{equation}
R_{\alpha \beta }-\frac{1}{2}g_{\alpha \beta }R=\kappa \Upsilon
_{\alpha \beta },  \label{25einstein}
\end{equation}%
where $\kappa $ and $\Upsilon _{\alpha \beta }$ are respectively
the coupling constant and the energy--momentum tensor we obtain
the h-v-decomposition by N--connection of the Einstein equations
\begin{eqnarray}
R_{ij}-\frac{1}{2}\left( \widehat{R}+S\right) g_{ij} &=&\kappa
\Upsilon
_{ij},  \label{3einsteq2} \\
S_{ab}-\frac{1}{2}\left( \widehat{R}+S\right) h_{ab} &=&\kappa
\Upsilon
_{ab},  \notag \\
^{1}P_{ai}=\kappa \Upsilon _{ai},\ ^{2}P_{ia} &=&\kappa \Upsilon
_{ia}. \notag
\end{eqnarray}%
The definition of matter sources with respect to anholonomic
frames is considered in Refs. \cite{13vspinors,13vmon1,13ma}.

The vacuum 5D, locally anisotropic gravitational field equations,
in invariant h-- v--components, are written
\begin{eqnarray}
R_{ij} &=&0,S_{ab}=0,  \label{2einsteq3} \\
^{1}P_{ai} &=&0,\ ^{2}P_{ia}=0.  \notag
\end{eqnarray}

We emphasize that vector bundles and even the (pseudo) Riemannian
space-times admit non--trivial torsion components, if
off--diagonal metrics and anholomomic frames are introduced into
consideration. This is a ''pure'' anholonomic frame effect: the
torsion vanishes for the Levi--Civita connection stated with
respect to a coordinate frame, but even this metric connection
contains some torsion coefficients if it is defined with respect
to anholonomic frames (this follows from the $W$--terms in
(\ref{12lcsym})). For (pseudo) Riemannian spaces we conclude that
the Einstein theory transforms into an effective Einstein--Cartan
theory with anholonomically induced torsion if the general
relativity is formulated with respect to general frame bases
(both holonomic and anholonomic).

The N--connection geometry can be similarly formulated for a tangent bundle $%
TM$ of a manifold $M$ (which is used in Finsler and Lagrange geometry \cite%
{13ma}), on cotangent bundle $T^{\ast }M$ and higher order bundles
(higher order Lagrange and Hamilton geometry \cite{13miron}) as
well in the geometry of locally anisotropic superspaces
\cite{13vsuper}, superstrings \cite{13vstr2}, anisotropic spinor
\cite{13vspinors} and gauge \ \cite{13vgauge} theories or even on
(pseudo) Riemannian spaces provided with anholonomic frame
structures \cite{13vmon2}.

\subsection{Nonlinear connections in projective modules}

{\quad }The nonlinear connection (N--connection) for
noncommutative spaces can be defined similarly to commutative
spaces by considering instead of usual vector bundles theirs
noncommutative analogs defined as finite projective modules over
noncommutative algebras. The explicit constructions depend on the
type of differential calculus we use for definition of tangent
structures and theirs maps.

In general, there can be several differential calculi over a given algebra $%
\mathcal{A}$ (for a more detailed discussion within the context of
noncommutative geometry see Refs. \cite{13connes,13madore,13dubois}; a
recent approach is connected with Lie superalgebra structures on
the space of multiderivations \cite{13giun}). \ For simplicity, in
this work we fix a differential calculus on $\mathcal{A},$ which
means that we choose a
(graded) algebra $\Omega ^{\ast }(\mathcal{A})=\cup _{p}\Omega ^{p}(\mathcal{%
A})$ which gives a differential structure to $\mathcal{A}.$ The elements of $%
\Omega ^{p}(\mathcal{A})$ are called $p$--forms. There is a
linear map $d$ which takes $p$--forms into $(p+1)$--forms and
which satisfies a graded
Leibniz rule as well the condition $d^{2}=0.$ By definition $\Omega ^{0}(%
\mathcal{A})=\mathcal{A}.$

The differential $df$ of a real or complex variable on a vector
bundle $\xi _{N}$
\begin{eqnarray*}
df &=&\delta _{i}f~dx^{i}+\partial _{a}f~\delta y^{a}, \\
\delta _{i}f~ &=&\partial _{i}f-N_{i}^{a}\partial _{a}f~,~\delta
y^{a}=dy^{a}+N_{i}^{a}dx^{i}
\end{eqnarray*}%
in the noncommutative case is replaced by a distinguished
commutator
(d--commutator)%
\begin{equation*}
\overline{d}f=\left[ F,f\right] =\left[ F^{[h]},f\right] +\left[ F^{[v]},f%
\right]
\end{equation*}%
where the operator $F^{[h]}$ $\ (F^{[v]})$ is acting on the
horizontal (vertical) projective submodule being defined by some
fixed differential
calculus $\Omega ^{\ast }(\mathcal{A}^{[h]})$ ($\Omega ^{\ast }(\mathcal{A}%
^{[v]}))$ on the so--called horizontal (vertical) $\mathcal{A}^{[h]}$ ($%
\mathcal{A}^{[v]})$ algebras.

Let us consider instead of a vector bundle $\xi $ $\ $an $\mathcal{A}$%
--module $\mathcal{E}$ being projective and of finite type. For a
fixed
differential calculus on $\mathcal{E}$ we define the tangent structures $T%
\mathcal{E}$ \ and $TM.\,$ A nonlinear connection$\,N$ in an $\mathcal{A}$%
--module $\mathcal{E}$ is defined by an exact sequence of finite projective $%
\mathcal{A}$--moduli
\begin{equation*}
0\rightarrow V\mathcal{E}\rightarrow
T\mathcal{E}/V\mathcal{E}\rightarrow 0,
\end{equation*}%
where all subspaces are constructed as in the commutative case
with that difference that the vector bundle objects are
substituted by theirs projective modules equivalents. A
projective module provided with N--connection structures will be
denoted as $\mathcal{E}_{N}.$ All objects on a $\mathcal{E}_{N}$
\ have a distinguished invariant character with respect to the
horizontal and vertical subspaces.

To understand how the N--connection structure may be taken into
account on noncommutative spaces we analyze in the next
subsection an example.

\subsection{Commutative and noncommutative gauge d--fields}

Let us consider a vector bundle $\xi _{N}$ and a another vector bundle $%
\beta =\left( B,\pi ,\xi _{N}\right) $ with $\pi :B\rightarrow
\xi _{N}$ with a typical $k$-dimensional vector fiber. In local
coordinates a linear connection (a gauge field) in $\beta $ is
given by a collection of
differential operators%
\begin{equation*}
\bigtriangledown _{\alpha }=D_{\alpha }+B_{\alpha }(u),
\end{equation*}%
acting on $T\xi _{N}$ where
\begin{equation*}
D_{\alpha }=\delta _{\alpha }\pm \Gamma _{\cdot \alpha }^{\cdot }%
\mbox{ with }D_{i}=\delta _{i}\pm \Gamma _{\cdot i}^{\cdot
}\mbox{ and }D_{a}=\partial _{a}\pm \Gamma _{\cdot a}^{\cdot }
\end{equation*}
is a d--connection in $\xi _{N}$ ($\alpha =1,2,...,n+m),$ with
$\delta
_{\alpha }$ N--elongated as in (\ref{6dder}), $u=(x,y)\in \xi _{N}$ and $%
B_{\alpha }$ are $k\times k$--matrix valued functions. For every
vector field
\begin{equation*}
X=X^{\alpha }(u)\delta _{\alpha }=X^{i}(u)\delta
_{i}+X^{a}(u)\partial _{a}\in T\xi _{N}
\end{equation*}
we can consider the operator
\begin{equation}
X^{\alpha }(u)\bigtriangledown _{\alpha }(f\cdot s)=f\cdot
\bigtriangledown _{X}s+\delta _{X}f\cdot s  \label{1rul1c}
\end{equation}%
for any section $s\in \mathcal{B}$ \ and function $f\in C^{\infty
}(\xi
_{N}),$ where%
\begin{equation*}
\delta _{X}f=X^{\alpha }\delta _{\alpha }~\mbox{ and
}\bigtriangledown _{fX}=f\bigtriangledown _{X}.
\end{equation*}%
In the simplest definition we assume that there is a Lie algebra $\mathcal{GL%
}B$ that acts on associative algebra $B$ by means of infinitesimal
automorphisms (derivations). This means that we have linear operators $%
\delta _{X}:B\rightarrow B$ which linearly depend on $X$ and satisfy%
\begin{equation*}
\delta _{X}(a\cdot b)=(\delta _{X}a)\cdot b+a\cdot (\delta _{X}b)
\end{equation*}%
for any $a,b\in B.$ The mapping $X\rightarrow \delta _{X}$ is a
Lie algebra homomorphism, i. e. $\delta _{\lbrack X,Y]}=[\delta
_{X},\delta _{Y}].$

Now we consider respectively instead of vector bundles $\xi $ and
$\beta $ the finite \ projective $\mathcal{A}$--module
$\mathcal{E}_{N},$ provided
with N--connection structure, and the finite projective $\mathcal{B}$%
--module $\mathcal{E}_{\beta }.$

A d--connection $\bigtriangledown _{X\text{ }}$on
$\mathcal{E}_{\beta }$ is by definition a set of linear
d--operators, adapted to the N--connection
structure, depending linearly on $X$ and satisfying the Leibniz rule%
\begin{equation}
\bigtriangledown _{X}(b\cdot e)=b\cdot \bigtriangledown
_{X}(e)+\delta _{X}b\cdot e  \label{1rul1n}
\end{equation}%
for any $e\in \mathcal{E}_{\beta }$ and $b\in \mathcal{B}.$ The rule (\ref%
{1rul1n}) is a noncommutative generalization of (\ref{1rul1c}). We
emphasize that both operators $\bigtriangledown _{X}$ and $\delta
_{X}$ are distinguished by the N--connection structure and that
the difference of two such linear d--operators, $\bigtriangledown
_{X}-\bigtriangledown _{X}^{\prime }$ commutes with action of $B$
on $\mathcal{E}_{\beta },$ which is an endomorphism of
$\mathcal{E}_{\beta }.$ Hence, if we fix some fiducial
connection $\bigtriangledown _{X}^{\prime }$ (for instance, $%
\bigtriangledown _{X}^{\prime }=D_{X})$ on $\mathcal{E}_{\beta }$
an arbitrary connection has the form
\begin{equation*}
\bigtriangledown _{X}=D_{X}+B_{X},
\end{equation*}%
where $B_{X}\in End_{B}\mathcal{E}_{\beta }$ depend linearly on
$X.$

The curvature of connection $\bigtriangledown _{X}$ is a
two--form $F_{XY}$ which values linear operator in $\mathcal{B}$
and measures a deviation of mapping $X\rightarrow
\bigtriangledown _{X}$ from being a Lie algebra
homomorphism,%
\begin{equation*}
F_{XY}=[\bigtriangledown _{X},\bigtriangledown
_{Y}]-\bigtriangledown _{\lbrack X,Y]}.
\end{equation*}%
The usual curvature d--tensor is defined as
\begin{equation*}
F_{\alpha \beta }=\left[ \bigtriangledown _{\alpha
},\bigtriangledown _{\beta }\right] -\bigtriangledown _{\lbrack
\alpha ,\beta ]}.
\end{equation*}

The simplest connection on a finite projective $\mathcal{B}$--module $%
\mathcal{E}_{\beta }$ is to be specified by a projector $P:\mathcal{B}%
^{k}\otimes \mathcal{B}^{k}$ when the d--operator $\delta _{X}$
acts
naturally on the free module $\mathcal{B}^{k}.$ The operator $%
\bigtriangledown _{X}^{LC}=P\cdot \delta _{X}\cdot P$ $\ $\ is
called the Levi--Civita operator and satisfy the condition
$Tr[\bigtriangledown _{X}^{LC},\phi ]=0$ for any endomorphism
$\phi \in End_{B}\mathcal{E}_{\beta }.$ From this identity, and
from the fact that any two connections differ by an endomorphism
that
\begin{equation*}
Tr[\bigtriangledown _{X},\phi ]=0
\end{equation*}%
for an arbitrary connection $\bigtriangledown _{X}$ and an
arbitrary endomorphism $\phi ,$ that instead of $\bigtriangledown
_{X}^{LC}$ we may consider equivalently the canonical
d--connection, constructed only from d-metric and N--connection
coefficients.

\section{ Distinguished Spectral Triples}

In this section we develop the basic ingredients introduced by A. Connes %
\cite{13connes} to define the analogue of differential calculus for
noncommutative distinguished algebras. The N--connection
structures distinguish a commutative or a noncommutative spaces
into horizontal and vertical subspaces. The geometric objects
possess a distinguished invariant character with respect to a
such splitting. The basic idea in definition of spectral triples
generating locally anisotropic spaces (Rimannian spaces with
anholonomic structure, or, for more general constructions,
Finsler and Lagrange spaces) is to consider pairs of noncommutative algebras
 $\mathcal{A}%
_{[d]}=(\mathcal{A}_{[h]},\mathcal{A}_{[v]}),$ given by
respective pairs of elements $a=\left( a_{[h]},a_{[v]}\right) \in
\mathcal{A}_{[d]},$ called also distinguished algebras (in brief,
d--algebras), together with d-operators
$D_{[d]}=(D_{[h]},D_{[v]})$ on a Hilbert space $\mathcal{H}$ (for
simplicity we shall consider one Hilbert space, but a more general
construction can be provided for Hilbert d-spaces,
$\mathcal{H}_{[d]}=\left(
\mathcal{H}_{[h]},\mathcal{H}_{[v]}\right) .$

The formula of Wodzicki--Adler--Manin--Guillemin residue (see, for instance, %
\cite{13landi}) may be written for vector bundles provided with
N--connection
structure. It is necessary to introduce the N--elongated differentials (\ref%
{6dder}) in definition of the measure: Let $Q$ be a
pseudo--differential
operator of order $-n$ acting on sections of a complex vector bundle $%
E\rightarrow M$ over an $n$--dimensional compact Riemannian
manifold $M.$ The residue $ResQ$ of $Q$ is defined by the formula
\begin{equation*}
{Re}sQ=:\frac{1}{n\left( 2\pi \right) ^{n}}\int\limits_{S^{\ast
}M}tr_{E}\sigma _{-n}(Q)\delta \mu ,
\end{equation*}%
where $\sigma _{-n}(Q)$ is the principal symbol (a matrix--valued
function on $T^{\ast }M$ which is homogeneous of degree $-n$ in
the fiber coordinates), \ the integral is taken over the unit
co--sphere $S^{\ast }M=\{(x,y)\in T^{\ast }M:||y||=1\}\subset
T^{\ast }M,$ the $tr_{E}$ is the matrix trace over ''internal
indices'' and the measures $\delta \mu =dx^{i}\delta y^{a}.$

A spectral d--triple $\left[
\mathcal{A}_{[d]},\mathcal{H},D_{[d]}\right] $ is given by an
involutive d--algebra of d--operators $D^{[d]}$ consisting from
pairs of bounded operators $D_{[h]}$ and $D_{[v]}$ $\ $on the
Hilbert
space $\mathcal{H},$ together with the self--adjoint operation $%
D_{[d]}=D_{[d]}^{\ast }$ for respective $h$- and $v$--components on $%
\mathcal{H}$ being satisfied the properties:

\begin{enumerate}
\item The resolvents $(D_{[h]}-\lambda _{\lbrack h]})^{-1}$ and $%
(D_{[v]}-\lambda _{\lbrack v]})^{-1},$ $\lambda _{\lbrack
h]},\lambda _{\lbrack v]}\in \R,$ are compact operators on
$\mathcal{H};$

\item The commutators $\left[ D_{[h,]},a_{[h]}\right] \doteqdot
D_{[h]}a_{[h]}-a_{[h]}D_{[h]}\in \mathcal{B}(\mathcal{H})$ and
$\left[
D_{[v,]},a_{[v]}\right] \doteqdot D_{[v]}a_{[v]}-a_{[v]}D_{[v]}\in \mathcal{B%
}(\mathcal{H})$ for any $a\in \mathcal{A}_{[d]},$ where by $\mathcal{B}(%
\mathcal{H})$ we denote the algebra of bounded operators on
$\mathcal{H}.$
\end{enumerate}

The $h(v)$--component of a d--triple is said to be even if there is a ${\Z}%
_{2}$--grading for $\mathcal{H},$ i. e. an operator $\Upsilon $ on $\mathcal{%
H}$ such that
\begin{equation*}
\Upsilon =\Upsilon ^{\ast },\Upsilon ^{2}=1,~\Upsilon
D_{[h(v)]}-D_{[h(v)]}\Upsilon =0,~\Upsilon a-a\Upsilon =0
\end{equation*}%
for every $a\in \mathcal{A}_{[d]}.$ If such a grading does not exist, the $%
h(v)$--component of a d--triple is said to be odd.

\subsection{Canonical triples over vector bundles}

The basic examples of spectral triples in connections with
noncommutative field theory and geometry models were constructed
by means of the Dirac operator on a closed $n$--dimensional
Riemannian spin manifold $\left( M,g\right) $ \cite{13connes,13cl}. \
In order to generate by using functional methods some anisotropic
geometries, it is necessary to generalize the approach to vector
and covector bundles provided with compatible N--connection,
d--connection and metric structures. The theory of spinors on
locally anisotropic spaces was developed in Refs.
\cite{13vspinors,13vmon2}. \ This section is devoted to the spectral
d--triples defined by the Dirac operators on closed regions of
$\left( n+m\right) $--dimensional spin--vector manifolds. We note
that if we deal with off--diagonal metrics and/or anholonomic
frames there is an infinite number of d--connections which are
compatible with d--metric and N--connection structures, see
discussion and details in Ref. \cite{13vexsol}. For simplicity, we
restrict
our consideration only to the Euclidean signature of metrics of type (\ref%
{7dmetric}) (on attempts to define triples with Minkowskian
signatures see, for instance, Refs. \cite{13hawkins}).

For a spectral d--triple $\left[
\mathcal{A}_{[d]},\mathcal{H},D_{[d]}\right] $ associated to a
vector bundle $\xi _{N}$ one takes the components:

\begin{enumerate}
\item $\mathcal{A}_{[d]}=\mathcal{F}(\xi _{N})$ is the algebra of complex
valued functions on $\xi _{N}.$

\item $\mathcal{H}=L^{2}(\xi _{N},S)$ is the Hilbert space of square
integrable sections of the irreducible d--spinor bundle (of rank $%
2^{(n+m)/2} $ over $\xi _{N}$ \cite{13vspinors,13vmon2}. The scalar product in $%
L^{2}(\xi _{N},S)$ is the defined by the measure associated to
the d--metric
(\ref{7dmetric}),%
\begin{equation*}
(\psi ,\phi )=\int \delta \mu (g)\overline{\psi }(u)\phi (u)
\end{equation*}%
were the bar indicates to the complex conjugation and the scalar
product in d--spinor space is the natural one in
$\C^{2[n/2]}\oplus \C^{2[m/2]}.$

\item $D$ is a Dirac d--operator associated to one of the d--metric
compatible d--connecti\-on, for instance, with the Levi--Civita
connection,
canonical d--connection or another one, denoted with a general symbol $%
\Gamma =\Gamma _{\mu }\delta u^{\mu }.$
\end{enumerate}

We note that the elements of the algebra $\mathcal{A}_{[d]}$ acts
as
multiplicative operators on $\mathcal{H},$%
\begin{equation*}
(a\psi )(u)=:f(u)\psi (u),
\end{equation*}%
for every $a\in \mathcal{A}_{[d]},\psi \in \mathcal{H}.$

\subsubsection{Distinguished spinor structures \label{spinorsubsection}}

Let us analyze the connection between d--spinor structures and
spectral d--triples over a vector bundle $\xi _{N}.$ One consider
a $(n+m)$--bein (frame) decomposition of the d--metric $g_{\alpha
\beta }$ (\ref{7dmetric})
(and its inverse $g^{\alpha \beta }),$%
\begin{equation*}
g^{\alpha \beta }(u)=e_{\underline{\alpha }}^{\alpha
}(u)e_{\underline{\beta
}}^{\beta }(u)\eta ^{\underline{\alpha }\underline{\beta }},~\eta _{%
\underline{\alpha }\underline{\beta }}=e_{\underline{\alpha
}}^{\alpha }(u)e_{\underline{\beta }}^{\beta }(u)g_{\alpha \beta
},
\end{equation*}%
$\eta _{\underline{\alpha }\underline{\beta }}$ it the diagonal Euclidean $%
(n+m)$--metric, which is adapted to the N--connection structure
because the coefficients $g_{\alpha \beta }$ are defined with
respect to the dual N--distinguish\-ed basis (\ref{7ddif}). We can
define compatible with this
decomposition d--connections $\Gamma _{\underline{\beta }\mu }^{\underline{%
\alpha }}$ (for instance, the Levi--Civita connection, which with
respect to anholonomic frames contains torsions components, or
the canonical d--connection), defined by
\begin{equation*}
D_{\mu }e_{\underline{\beta }}=\Gamma _{\underline{\beta }\mu }^{\underline{%
\alpha }}e_{\underline{\alpha }},
\end{equation*}%
as the solution of the equations%
\begin{equation*}
\delta _{\mu }e_{v}^{\underline{\nu }}-\delta _{\nu }e_{\mu }^{\underline{%
\nu }}=\Gamma _{\underline{\beta }\mu }^{\underline{\nu }}e_{v}^{\underline{%
\beta }}-\Gamma _{\underline{\beta }\nu }^{\underline{\nu }}e_{\mu }^{%
\underline{\beta }}.
\end{equation*}

We define by $C\left( \xi _{N}\right) $ the Clifford bundle over
\ $\xi _{N}$ with the fiber at $u\in \xi _{N}$ being just the
complexified Clifford d--algebra $Cliff_{\C}\left( T_{u}^{\ast
}\xi _{N}\right) ,$ $T_{u}^{\ast }\xi _{N}$ being dual to
$T_{u}\xi _{N},$ and $\Gamma \lbrack \xi _{N},C\left( \xi
_{N}\right) ]$ is the module of corresponding sections. By
defining the maps
\begin{equation*}
\gamma \left( \delta ^{\alpha }\right) =\left( \gamma \left(
d^{i}\right) ,\gamma \left( \delta ^{a}\right) \right) \doteqdot
\gamma ^{\alpha }(u)=\gamma ^{\underline{\alpha
}}e_{\underline{\alpha }}^{\alpha
}(u)=\left( \gamma ^{\underline{\alpha }}e_{\underline{\alpha }%
}^{i}(u),\gamma ^{\underline{\alpha }}e_{\underline{\alpha
}}^{a}(u)\right) ,
\end{equation*}%
extended as an algebra map by $\mathcal{A}_{[d]}$--linearity, we
construct
an algebra morphism%
\begin{equation}
\gamma :\Gamma \left( \xi _{N},C\left( \xi _{N}\right) \right)
\rightarrow \mathcal{B}(\mathcal{H}).  \label{gammamorph}
\end{equation}%
The indices of the ''curved'' $\gamma ^{\alpha }(u)$ and ''flat'' $\gamma ^{%
\underline{\alpha }}$ gamma matrices can be lowered by using
respectively
the d-metric components $g_{\alpha \beta }$ $(u)$ and $\eta _{\underline{%
\alpha }\underline{\beta }},$ i. e. $\gamma _{\beta }(u)$
=$\gamma ^{\alpha
}(u)$ $g_{\alpha \beta }$ $(u)$ and $\ \gamma _{\underline{\beta }}=\gamma ^{%
\underline{\alpha }}\eta _{\underline{\alpha }\underline{\beta
}}.$ We take the gamma matrices to be Hermitian and to obey the
relations,
\begin{eqnarray*}
\gamma _{\alpha }\gamma _{\beta }+\gamma _{\beta }\gamma _{\alpha
} &=&-2g_{\alpha \beta }~\left( \gamma _{i}\gamma _{j}+\gamma
_{j}\gamma _{i}=-2g_{ij},~\gamma _{a}\gamma _{b}+\gamma
_{b}\gamma _{a}=-2g_{ab}\right)
, \\
\gamma _{\underline{\alpha }}\gamma _{\underline{\beta }}+\gamma _{%
\underline{\beta }}\gamma _{\underline{\alpha }} &=&-2\eta _{\underline{%
\alpha }\underline{\beta }}.
\end{eqnarray*}%
Every d--connection $\Gamma _{\underline{\beta }\mu
}^{\underline{\nu }}$ can be shifted as a d--covariant operator
$\bigtriangledown _{\mu }^{[S]}=\left( \bigtriangledown
_{i}^{[S]},\bigtriangledown
_{a}^{[S]}\right) $ on the bundle of d--spinors,%
\begin{equation*}
\bigtriangledown _{\mu }^{[S]}=\delta _{\mu }+\frac{1}{2}\Gamma _{\underline{%
\alpha }\underline{\beta }\mu }\gamma ^{\underline{\alpha }}\gamma ^{%
\underline{\beta }},~\Gamma _{\mu }^{[S]}=\frac{1}{2}\Gamma _{\underline{%
\alpha }\underline{\beta }\mu }\gamma ^{\underline{\alpha }}\gamma ^{%
\underline{\beta }},
\end{equation*}%
which defines the Dirac d--operator
\begin{equation}
_{\lbrack d]}D=:\gamma \circ \bigtriangledown _{\mu
}^{[S]}=\gamma ^{\alpha
}(u)\left( \delta _{\mu }+\Gamma _{\mu }^{[S]}\right) =\gamma ^{\underline{%
\alpha }}e_{\underline{\alpha }}^{\mu }\left( \delta _{\mu
}+\Gamma _{\mu }^{[S]}\right) .  \label{dirac}
\end{equation}%
Such formulas were introduced in Refs. \cite{13vspinors} for
distinguished spinor bundles (of first and higher order). In this
paper we revise them in connection to spectral d--triples and
noncommutative geometry. \ On such spaces one also holds a
variant of Lichnerowicz formula \cite{13berline}\ for
the square of the Dirac d--operator%
\begin{equation}
_{\lbrack d]}D^{2}=\bigtriangledown ^{\lbrack S]}+\frac{1}{4}{\overleftarrow{%
R}},  \label{lich}
\end{equation}%
where the formulas for the scalar curvature ${\overleftarrow{R}}$
is given in (\ref{4dscalar}) and
\begin{equation*}
\bigtriangledown ^{\lbrack S]}=-g^{\mu \nu }\left(
\bigtriangledown _{\mu }^{[S]}\bigtriangledown _{v}^{[S]}-\Gamma
_{\mu \nu }^{\rho }\bigtriangledown _{\rho }^{[S]}\right) .
\end{equation*}

In a similar manner as in Ref. \cite{13landi} but reconsidering all
computations on a vector bundle $\xi _{N}$ we can prove that for
every d--triple $\left[
\mathcal{A}_{[d]},\mathcal{H},D_{[d]}\right] $ one holds the
properties:

\begin{enumerate}
\item The vector bundle $\xi _{N}$ is the structure space of the algebra $%
\overline{\mathcal{A}}_{[d]}$ of continuous functions on $\xi
_{N}$ (the bar here points to the norm closure of
$\mathcal{A}_{[d]}).$

\item The geodesic distance $\rho $ between two points $p_{1},p_{2}\in $ $%
\xi _{N}$ is defined by using the Dirac d--operator,%
\begin{equation*}
\rho \left( p_{1},p_{2}\right) =\sup_{f\in }\left\{
|f(p)-f(q)|:~||[D_{[d]},f]||\leq 1\right\} .
\end{equation*}

\item The Dirac d--operator also defines the Riemannian measure on $\xi
_{N}, $%
\begin{equation*}
\int_{\xi _{N}}f=c\left( n+m\right) tr_{\Gamma }\left(
f|D_{[d]}|^{-(n+m)}\right)
\end{equation*}%
for every $f\in \mathcal{A}_{[d]}$ and $c\left( n+m\right)
=2^{[n+m-(n+m)/2-1]}\pi ^{(n+m)/2}(n+m)\Gamma \left(
\frac{n+m}{2}\right) ,\Gamma $ being the gamma function.
\end{enumerate}

The spectral d--triple \ formalism has the same properties as the
usual one with that difference that we are working on spaces
provided with N--connection structures and the bulk of
constructions and objects are distinguished by this structure.

\subsubsection{Noncommutative differential forms}

To construct a differential algebra of forms out a spectral
d--triple $\left[ \mathcal{A}_{[d]},\mathcal{H},D_{[d]}\right] $
one follows universal graded differential d--algebras \ defined
as couples of universal ones, respectively associated to the
$h$-- and $v$--components of some splitting to subspaces defined
by N--connection structures. Let $\mathcal{A}_{[d]}$ be an
associative d--algebra (for simplicity, with unit) over the field
of complex numbers $\C.$ The universal d--algebra of differential
forms $\Omega \mathcal{A}_{[d]}=\oplus _{p}\Omega
^{p}\mathcal{A}_{[d]}$ is introduced as a graded d--algebra when
$\Omega ^{0}\mathcal{A}_{[d]}=\mathcal{A}_{[d]}$ and the space
$\Omega ^{1}\mathcal{A}_{[d]}$ of one--forms is generated as a
left $\mathcal{A}_{[d]}$--module by symbols of degree $\delta
a,a\in
\mathcal{A}_{[d]}$ satisfying the properties%
\begin{equation*}
\delta (ab)=(\delta a)b+a\delta b\mbox{ and }\delta (\alpha
a+\beta b)=\alpha (\delta a)+\beta \delta b
\end{equation*}%
from which follows $\delta 1=0$ which in turn implies $\delta
\C=0.$ These relations state the Leibniz rule for the map
\begin{equation*}
\delta :\mathcal{A}_{[d]}\rightarrow \Omega ^{1}\mathcal{A}_{[d]}
\end{equation*}%
An element $\varpi \in \Omega ^{1}\mathcal{A}_{[d]}$ is expressed
as a
finite sum of the form%
\begin{equation*}
\varpi
=\sum\limits_{\underline{i}}a_{\underline{i}}b_{\underline{i}}
\end{equation*}%
for $a_{\underline{i}},b_{\underline{i}}\in \mathcal{A}_{[d]}.$ The left $%
\mathcal{A}_{[d]}$--module $\Omega ^{1}\mathcal{A}_{[d]}$ can be
also endowed with a structure of right
$\mathcal{A}_{[d]}$--module if the
elements are imposed to satisfy the conditions%
\begin{equation*}
\left( \sum\limits_{\underline{i}}a_{\underline{i}}\delta b_{\underline{i}%
}\right) c=:\sum\limits_{\underline{i}}a_{\underline{i}}(\delta b_{%
\underline{i}})c=\sum\limits_{\underline{i}}a_{\underline{i}}\delta (b_{%
\underline{i}}c)-\sum\limits_{\underline{i}}a_{\underline{i}}b_{\underline{i}%
}\delta c.
\end{equation*}

Given a spectral d--triple $\left[ \mathcal{A}_{[d]},\mathcal{H},D_{[d]}%
\right] ,$ one constructs and exterior d--algebra of forms by
means of a suitable representation of the universal algebra
$\Omega \mathcal{A}_{[d]}$ in the d--algebra of bounded operators
on $\mathcal{H}$ by considering the
map%
\begin{eqnarray*}
\pi &:&\Omega \mathcal{A}_{[d]}\rightarrow \mathcal{B}(\mathcal{H}), \\
\pi \left( a_{0}\delta a_{1}...\delta a_{p}\right) &=&:a_{0}\left[ D,a_{1}%
\right] ...\left[ D,a_{p}\right]
\end{eqnarray*}%
which is a homomorphism since both $\delta $ and $\left[
D,.\right] $ are
distinguished derivations on $\mathcal{A}_{[d]}.$ More than that, since $%
\left[ D,a\right] ^{\ast }=-\left[ D,a^{\ast }\right] ,$ we have
$\pi \left(
\varpi \right) ^{\ast }=\pi \left( \varpi ^{\ast }\right) $ for any d--form $%
\varpi \in \Omega \mathcal{A}_{[d]}$ and $\pi $ being a $\ast $%
--homomorphism.

Let $J_{0}=:\oplus _{p}J_{0\text{\ }}^{p}$ be the graded
two--sided ideal of $\Omega \mathcal{A}_{[d]}$ given by
$J_{0\text{\ }}^{p}=:\{\pi \left( \varpi \right) =0\}$ when
$J=J_{0}+\delta J_{0}$ is a graded differential two--sided ideal
of $\Omega \mathcal{A}_{[d]}.$ At the next step we can
define the graded differential algebra of Connes' forms over the d--algebra $%
\mathcal{A}_{[d]}$ as%
\begin{equation*}
\Omega _{D}\mathcal{A}_{[d]}=:\Omega \mathcal{A}_{[d]}/J\simeq
\pi \left( \Omega \mathcal{A}_{[d]}\right) /\pi \left( \delta
J_{0}\right) .
\end{equation*}%
It is naturally graded by the degrees of $\Omega
\mathcal{A}_{[d]}$ and $J$
with the space of $p$--forms being given by $\Omega _{D}^{p}\mathcal{A}%
_{[d]}=\Omega ^{p}\mathcal{A}_{[d]}/J^{p}.$ Being $J$ a
differential ideal,
the exterior differential $\delta $ defines a differential on $\Omega _{D}%
\mathcal{A}_{[d]},$%
\begin{equation*}
\delta :\Omega _{D}^{p}\mathcal{A}_{[d]}\rightarrow \Omega _{D}^{p+1}%
\mathcal{A}_{[d]},~\delta \lbrack \varpi ]=:\left[ \delta \varpi
\right]
\end{equation*}%
with $\varpi \in \Omega _{D}^{p}\mathcal{A}_{[d]}$ and $[\varpi
]$ being the corresponding class in $\Omega
_{D}^{p}\mathcal{A}_{[d]}.$

We conclude that the theory of distinguished d--forms generated by
d--algebras, as well of the graded differential d--algebra of
Connes' forms, is constructed in a usual form (see Refs.
\cite{13connes,13landi}) but for two subspaces (the horizontal and
vertical ones) defined by a N--connection structure.

\subsubsection{The exterior d--algebra}

The differential d--form formalism when applied to the canonical d--triple\\ $%
\left[ \mathcal{A}_{[d]},\mathcal{H},D_{[d]}\right] $ over an
ordinary vector bundle $\xi _{N}$ provided with N--connection
structure reproduce the usual exterior d--algebra over this
vector bundle. Consider our d--triple
on a closed $\left( n+m\right) $--dimensional Riemannian $spin^{c}\,$%
manifold as described in subsection \ref{spinorsubsection} when $\mathcal{A}%
_{[d]}=\mathcal{F}(\xi _{N})$ is the algebra of smooth complex
valued functions on $\xi _{N}$ and $\mathcal{H}=L^{2}(\xi
_{N},S)$ is the Hilbert space of square integrable sections of
the irreducible d--spinor bundle (of
rank $2^{(n+m)/2}$ over $\xi _{N}.$ We can identify%
\begin{equation}
\pi \left( \delta f\right) =:[_{[d]}D,f]=\gamma ^{\mu }(u)\delta
_{\mu }f=\gamma \left( \delta _{\xi _{N}}f\right)  \label{pi1}
\end{equation}%
for every $f\in \mathcal{A}_{[d]},$ see the formula for the Dirac
d--operator (\ref{dirac}), \ where $\gamma $ is the d--algebra morphism \ (%
\ref{gammamorph}) and $\delta _{\xi _{N}}$ denotes the usual
exterior
derivative on $\xi _{N}.$ In a more general case, with $f_{[i]}\in \mathcal{A%
}_{[d]},[i]=[1],...,[p],$ we can write
\begin{equation}
\pi \left( f_{[0]}\delta f_{[1]}...\delta f_{[p]}\right)
=:f_{[0]}[_{[d]}D,f_{[1]}]...[_{[d]}D,f_{[p]}]=\gamma \left(
f_{[0]}\delta _{\xi _{N}}f_{[1]}\cdot ...\cdot \delta _{\xi
_{N}}f_{[p]}\right) , \label{pi2}
\end{equation}%
where the d--differentials $\delta _{\xi _{N}}f_{[1]}$ are
regarded as sections of the Clifford d--bundle $C_{1}(\xi _{N}),$
while $f_{[i]}$ can be thought of as sections of $C_{0}(\xi
_{N})$ and the $dot$ $\cdot $ the Clifford product in the fibers
of $C(\xi _{N})=\oplus _{k}C_{k}(\xi _{N}),$ see details in Refs.
\cite{13vspinors,13vmon2}.

A generic differential 1--form on $\xi _{N}$ can be written as $%
\sum_{[i]}f_{0}^{[i]}\delta _{\xi _{N}}f_{1}^{[i]}$ with $%
f_{0}^{[i]},f_{1}^{[i]}$ $\in \mathcal{A}_{[d]}.$ Following the definitions (%
\ref{pi1}) and (\ref{pi2}), we can identify the distinguished
Connes' 1--forms $\Omega _{D}^{p}\mathcal{A}_{[d]}$ with the
usual distinguished differential 1--forms, i. e.
\begin{equation*}
\Omega _{D}^{p}\mathcal{A}_{[d]}\simeq \Lambda ^{p}\left( \xi
_{N}\right) .
\end{equation*}

For each $u\in \xi _{N},$ we can introduce a natural filtration
for the Clifford d--algebra, $C_{u}(\xi _{N})=\cup C_{u}^{(p)},$
where $C_{u}^{(p)}$ is spanned by products of type $\chi
_{\lbrack 1]}\cdot \chi _{\lbrack 2]}\cdot ...\cdot \chi
_{\lbrack p^{\prime }]},p^{\prime }\leq p,~\chi _{\lbrack i]}\in
T_{u}^{\ast }\xi _{N}.$ One defines a natural graded
d--algebra,%
\begin{equation}
grC_{u}=:\sum_{p}gr_{p}C_{u},~gr_{p}C_{u}=C_{u}^{(p)}/C_{u}^{(p-1)},
\label{pi3}
\end{equation}%
for which the natural projection is called the symbol map,%
\begin{equation*}
\sigma _{p}:C_{u}^{(p)}\rightarrow gr_{p}C_{u}.
\end{equation*}%
The natural graded d--algebra is canonical isomorphic to the
complexified exterior d--algebra $\Lambda _{\C}\left( T_{u}^{\ast
}\xi _{N}\right) ,$ the
isomorphism being defined as%
\begin{equation}
\Lambda _{\C}\left( T_{u}^{\ast }\xi _{N}\right) \ni \chi
_{\lbrack 1]}\wedge \chi _{\lbrack 2]}\wedge ...\wedge \chi
_{\lbrack p]}\rightarrow \sigma _{p}\left( \chi _{\lbrack
1]}\cdot \chi _{\lbrack 2]}\cdot ...\cdot \chi _{\lbrack
p]}\right) \in gr_{p}C_{u}.  \label{pi4}
\end{equation}

As a consequence of formulas (\ref{pi3}) and (\ref{pi4}), for a
canonical d--triple\\ $\left[
\mathcal{A}_{[d]},\mathcal{H},D_{[d]}\right] $ over the
vector bundle $\xi _{N},$ one follows the property: a pair of operators $%
Q_{1}$ and $Q_{2}$ on $\mathcal{H}$ is of the form $Q_{1}=\pi
(\varpi )$ and
$Q_{2}=\pi (\delta \varpi )$ for some universal form $\varpi \in \Omega ^{p}%
\mathcal{A}_{[d]},$ if and only if there are sections $\rho _{1}$ of $%
C^{(p)} $ and $\rho _{2}$ of $C^{(p+1)}$ such that%
\begin{equation*}
Q_{1}=\gamma \left( \rho _{1}\right) \mbox{ and }Q_{2}=\gamma
\left( \rho _{2}\right)
\end{equation*}%
for which%
\begin{equation*}
\delta _{\xi _{N}}\sigma _{p}\left( \rho _{1}\right) =\sigma
_{p+1}\left( \rho _{2}\right) .
\end{equation*}

The introduced symbol map defines the canonical isomorphism
\begin{equation}
\sigma _{p}:\Omega _{D}^{p}\mathcal{A}_{[d]}\simeq \Gamma \left( \Lambda _{\C%
}^{p}T^{\ast }\xi _{N}\right)  \label{form1}
\end{equation}%
which commutes with the differential. With this isomorphism the
inner product on $\Omega _{D}^{p}\mathcal{A}_{[d]}$ (the scalar
product of forms) is proportional to the Riemannian inner product
of distinguished $p$--forms
on $\xi _{N},$%
\begin{equation}
<\varpi _{1},\varpi _{2}>_{p}=\left( -1\right) ^{p}\frac{2^{(n+m)/2+1-(n+m)}%
\pi ^{-(n+m)/2}}{(n+m)\Gamma \left( (n+m)/2\right) }\int_{\xi
_{N}}\varpi _{1}\wedge \ast \varpi _{2}  \label{form2}
\end{equation}%
for every $\varpi _{1},\varpi _{2}\in \Omega
_{D}^{p}\mathcal{A}_{[d]}\simeq \Gamma \left( \Lambda
_{\C}^{p}T^{\ast }\xi _{N}\right) .$

The proofs of formulas (\ref{form1}) and (\ref{form2}) are
similar to those given in \cite{13landi} for $\xi _{N}=M.$

\subsection{Noncommutative Geometry and Anholonomic Gravity}

We introduce the concepts of generalized Lagrange and Finsler
geometry and outline the conditions when such structures can be
modelled on a Riemannian space by using anholnomic frames.

\subsubsection{Anisotropic spacetimes}

Different classes of commutative anisotropic spacetimes are
modelled by corresponding parametriztions of some compatible (or
even non--compatible) N--connection, d--connection and d--metric
structures on (pseudo) Riemannian spaces, tangent (or cotangent)
bundles, vector (or covector) bundles and their higher order
generalizations in their usual manifold,
supersymmetric, spinor, gauge like or another type approaches (see Refs. %
\cite{13vexsol,13miron,13ma,13bejancu,13vspinors,13vgauge,13vmon1,13vmon2}). Here
we revise the basic definitions and formulas which will be used
in further noncommutative embedding and generalizations.

\paragraph{Anholonomic structures on Riemannian spaces:}

We can generate an anholonomic (equivalently, anisotropic)
structure on a Rieman space of dimension $(n+m)$ space $\ $(let
us denote this space $\ V^{(n+m)}$ and call it as a anholonomic
Riemannian space) by fixing an
anholonomic frame basis and co-basis with associated N--connection $%
N_{i}^{a}(x,y),$ respectively, as (\ref{6dder}) and (\ref{7ddif})
which splits the local coordinates $u^{\alpha }=(x^{i},y^{a})$
into two classes: $n$ holonomic coorinates, $x^{i},$ and $m$
anholonomic coordinates,$\,$\ $y^{a}.$ The d--metric
(\ref{7dmetric}) on $V^{(n+m)}$,
\begin{equation}
G^{[R]}=g_{ij}(x,y)dx^{i}\otimes dx^{j}+g_{ab}(x,y)\delta
y^{a}\otimes \delta y^{b}  \label{1dmetrr}
\end{equation}%
written with respect to a usual coordinate basis $du^{\alpha
}=\left(
dx^{i},dy^{a}\right) ,$%
\begin{equation*}
ds^{2}=\underline{g}_{\alpha \beta }\left( x,y\right) du^{\alpha
}du^{\beta }
\end{equation*}%
is a generic off--diagonal \ Riemannian metric parametrized as%
\begin{equation}
\underline{g}_{\alpha \beta }=\left[
\begin{array}{cc}
g_{ij}+N_{i}^{a}N_{j}^{b}g_{ab} & h_{ab}N_{i}^{a} \\
h_{ab}N_{j}^{b} & g_{ab}%
\end{array}%
\right] .  \label{1odm}
\end{equation}%
Such type of metrics were largely investigated in the Kaluza--Klein gravity %
\cite{13salam}, but also in the Einstein gravity \cite{13vexsol}. An
off--diagonal metric (\ref{1odm}) can be reduced to a block
$\left( n\times n\right) \oplus \left( m\times m\right) $ form
$\left( g_{ij},g_{ab}\right) , $ and even effectively
diagonalized in result of a superposition of ahnolonomic
N--transforms. It can be defined as an exact solution of the
Einstein equations. With respect to anholonomic frames, in
general, the Levi--Civita connection obtains a torsion component
(\ref{12lcsym}). Every class of off--diagonal metrics can be
anholonomically equivalent to another ones for which it is not
possible to a select the Levi--Civita metric defied as the unique
torsionless and metric compatible linear connection. \ The
conclusion is that if anholonomic frames of reference, which
authomatically induce the torsion via anholonomy coefficients,
are considered on a Riemannian space we have to postulate
explicitly what type of linear connection (adapted both to the
anholonomic frame and metric structure) is chosen in order to
construct a Riemannian geometry and corresponding physical
models. For instance, we may postulate the connection
(\ref{2lccon}) or the d--connection (\ref{6dcon}). Both these
connections are metric compatible and transform into the usual
Christoffel symbols if the N--connection vanishes, i. e. the
local frames became holonomic. But, in general, anholonomic
frames and off--diagonal Riemannian metrics are connected with
anisotropic configurations which allow, in principle, to
model even Finsler like structures in (pseudo) Riemannian spaces \cite%
{13vankin,13vexsol}.

\paragraph{Finsler geometry and its almost Kahlerian model:}

The modern approaches to Finsler geometry are outlined in Refs. \cite%
{13finsler,13ma,13miron,13bejancu,13vmon1,13vmon2}. Here we emphasize that a
Finsler metric can be defined on a tangent bundle $TM$ with local coordinates $%
u^{\alpha }=(x^{i},y^{a}\rightarrow y^{i})$ of dimension $2n,$
with a d--metric (\ref{7dmetric}) for which the Finsler metric, i.
e. the quadratic form
\begin{equation*}
g_{ij}^{[F]}=g_{ab}=\frac{1}{2}\frac{\partial ^{2}F^{2}}{\partial
y^{i}\partial y^{j}}
\end{equation*}%
is positive definite, is defined in this way: \ 1) A Finsler
metric on a real manifold $M$ is a function $F:TM\rightarrow \R$ which on $\widetilde{TM}%
=TM\backslash \{0\}$ is of class $C^{\infty }$ and $F$ is only
continuous on
the image of the null cross--sections in the tangent bundle to $M.$ 2) $%
F\left( x,\lambda y\right) =\lambda F\left( x,\lambda y\right) $ for every $%
\R_{+}^{\ast }.$ 3) The restriction of $F$ to $\widetilde{TM}$ is
a positive function. 4) $rank\left[ g_{ij}^{[F]}(x,y)\right] =n.$

The Finsler metric $F(x,y)$ and the quadratic form $g_{ij}^{[F]}$
can be used to define the Christoffel symbols (not those from the
usual Riemannian
geometry)%
\begin{equation*}
c_{jk}^{\iota }(x,y)=\frac{1}{2}g^{ih}\left( \partial
_{j}g_{hk}+\partial _{k}g_{jh}-\partial _{h}g_{jk}\right)
\end{equation*}%
which allows to define the Cartan nonlinear connection as
\begin{equation}
N_{j}^{i}(x,y)=\frac{1}{4}\frac{\partial }{\partial y^{j}}\left[
c_{lk}^{\iota }(x,y)y^{l}y^{k}\right]  \label{2ncc}
\end{equation}%
where we may not distinguish the v- and h- indices taking on $TM$
the same values.

In Finsler geometry there were investigated different classes of
remarkable Finsler linear connections introduced by Cartan,
Berwald, Matsumoto and other ones (see details in Refs.
\cite{13finsler,13ma,13bejancu}). Here we note
that we can introduce $g_{ij}^{[F]}=g_{ab}$ and $N_{j}^{i}(x,y)$ in (\ref%
{7dmetric}) and construct a d--connection via formulas
(\ref{6dcon}).

A usual Finsler space $F^{n}=\left( M,F\left( x,y\right) \right)
$ is completely defined by its fundamental tensor
$g_{ij}^{[F]}(x,y)$ and Cartan nonlinear connection
$N_{j}^{i}(x,y)$ and its chosen d--connection structure. But the
N--connection allows us to define an almost complex
structure $I$ on $TM$ as follows%
\begin{equation*}
I\left( \delta _{i}\right) =-\partial /\partial y^{i}\mbox{ and
}I\left(
\partial /\partial y^{i}\right) =\delta _{i}
\end{equation*}%
for which $I^{2}=-1.$

The pair $\left( G^{[F]},I\right) $ consisting from a Riemannian metric on $%
TM,$%
\begin{equation}
G^{[F]}=g_{ij}^{[F]}(x,y)dx^{i}\otimes
dx^{j}+g_{ij}^{[F]}(x,y)\delta y^{i}\otimes \delta y^{j}
\label{2dmetricf}
\end{equation}%
and the almost complex structure $I$ defines an almost Hermitian
structure
on $\widetilde{TM}$ associated to a 2--form%
\begin{equation*}
\theta =g_{ij}^{[F]}(x,y)\delta y^{i}\wedge dx^{j}.
\end{equation*}%
This model of Finsler geometry is called almost Hermitian and denoted $%
H^{2n} $ and it is proven \cite{13ma} that is almost Kahlerian, i.
e. the form
$\theta $ is closed. The almost Kahlerian space $K^{2n}=\left( \widetilde{TM}%
,G^{[F]},I\right) $ is also called the almost Kahlerian model of
the Finsler space $F^{n}.$

On Finsler (and their almost Kahlerian models) spaces one
distinguishes the almost Kahler linear connection of Finsler
type, $D^{[I]}$ on $\widetilde{TM} $ with the property that this
covariant derivation preserves by parallelism the vertical
distribution and is compatible with the almost Kahler structure
$\left( G^{[F]},I\right) ,$ i.e.
\begin{equation*}
D_{X}^{[I]}G^{[F]}=0\mbox{ and }D_{X}^{[I]}I=0
\end{equation*}%
for \ every d--vector field on $\widetilde{TM}.$ This
d--connection is defined by the data
\begin{equation*}
\Gamma =\left(
L_{jk}^{i},L_{bk}^{a}=0,C_{ja}^{i}=0,C_{bc}^{a}\rightarrow
C_{jk}^{i}\right)
\end{equation*}%
with $L_{jk}^{i}$ and $C_{jk}^{i}$ computed as in the formulas
(\ref{6dcon}) by using $g_{ij}^{[F]}$ and $N_{j}^{i}$ from
(\ref{2ncc}).

We emphasize that a Finsler space $F^{n}$ with a d--metric
(\ref{2dmetricf}) and Cartan's N--connection structure
(\ref{2ncc}), or the corresponding almost Hermitian (Kahler) model
$H^{2n},$ can be equivalently modelled on a Riemannian space of
dimension $2n$ provided with an off--diagonal Riemannian metric
(\ref{1odm}). From this viewpoint a Finsler geometry is a
corresponding Riemannian geometry with a respective off--diagonal
metric (or, equivalently, with an anholonomic frame structure
with associated N--connection) and a corresponding prescription
for the type of linear connection chosen to be compatible with
the metric and N--connection structures.

\paragraph{Lagrange and generalized Lagrange geometry:}

The Lagrange spaces were introduced in order to generalize the
fundamental
concepts in mechanics \cite{13kern} and investigated in Refs. \cite{13ma} (see %
\cite{13vspinors,13vgauge,13vsuper,13vstr2,13vmon1,13vmon2} for their spinor,
gauge and supersymmetric generalizations).

A Lagrange space $L^{n}=\left( M,L\left( x,y\right) \right) $ is
defined as a pair which consists of a real, smooth
$n$--dimensional manifold $M$ and regular Lagrangian
$L:TM\rightarrow \R.$ Similarly as for Finsler spaces one
introduces the symmetric d--tensor field
\begin{equation}
g_{ij}^{[L]}=g_{ab}=\frac{1}{2}\frac{\partial ^{2}L}{\partial
y^{i}\partial y^{j}}.  \label{1mfl}
\end{equation}%
So, the Lagrangian $L(x,y)$ is like the square of the fundamental
Finsler metric, $F^{2}(x,y),$ but not subjected to any
homogeneity conditions.

In the rest me can introduce similar concepts of almost Hermitian
(Kahlerian) models of Lagrange spaces as for the Finsler spaces,
by using the similar definitions and formulas as in the previous
subsection, but changing $g_{ij}^{[F]}\rightarrow g_{ij}^{[L]}.$

R. Miron introduced the concept of generalized Lagrange space,
GL--space (see details in \cite{13ma}) and a corresponding
N--connection geometry on $TM$ when the fundamental metric
function $g_{ij}=g_{ij}\left( x,y\right) $ is a general one, not
obligatory defined as a second derivative from a Lagrangian as in
(\ref{1mfl}). The corresponding almost Hermitian (Kahlerian)
models of GL--spaces were investigated and applied in order to
elaborate generalizations of gravity and gauge theories
\cite{13ma,13vgauge}.

Finally, a few remarks on definition of gravity models with
generic local anisotropy on anholonomic Riemannian, Finsler or
(generalized) Lagrange spaces and vector bundles. So, by choosing
a d-metric (\ref{7dmetric}) (in particular cases (\ref{1dmetrr}),
or (\ref{2dmetricf}) with $g_{ij}^{[F]},$ or $g_{ij}^{[L]})$ we
may compute the coefficients of, for instance, d--connection
(\ref{6dcon}), d--torsion (\ref{3dtors}) and (\ref{3dcurvatures})
and even to write down the explicit form of Einstein equations (\ref%
{3einsteq2}) which define such geometries. For instance, in a series of works %
\cite{13vankin,13vexsol,13vmon2} we found explicit solutions when
Finsler like and another type anisotropic configurations are
modelled in anisotropic kinetic theory and irreversible
thermodynamics and even in Einstein or
low/extra--dimension gravity as exact solutions of the vacuum (\ref{3einsteq2}%
) and nonvacuum (\ref{2einsteq3}) Einstein equations. From the
viewpoint of the geometry of anholonomic frames is not much
difference between the usual Riemannian geometry and its Finsler
like generalizations. The explicit form and parametrizations of
coefficients of metric, linear connections, torsions, curvatures
and Einstein equations in all types of mentioned geometric models
depends on the type of anholomic frame relations and
compatibility metric conditions between the associated
N--connection structure and linear connections we fixed. Such
structures can be correspondingly picked up from a noncommutative
functional model, for instance from some almost Hermitian
structures over projective modules and/or generalized to some
noncommutative configurations.

\subsection{Noncommutative Finsler like gravity models}

We shall briefly describe two possible approaches to the
construction of gravity models with generic anisotropy following
from noncommutative geometry which while agreeing for the
canonical d--triples associated with vector bundles provided with
N--connection structure. Because in the previous section we
proved that the Finsler geometry and its extensions are
effectively modelled by anholonomic structures on Riemannian
manifolds (bundles) we shall only emphasize the basic ideas how
from the beautiful result by Connes \cite{13connes,13connes1} we may
select an anisotropic gravity (possible alternative approaches to
noncommutative gravity are examined in Refs.
\cite{13ch1,13madore,13hawkins,13landi1,13landi2}; by introducing
anholonomic frames with associated N--connections those models
also can be transformed into certain anisotropic ones; we omit
such considerations in the present work).

\subsubsection{Anisotropic gravity a la Connes--Deximier--Wodzicki}

The first scheme to construct gravity models in noncommutative
geometry (see details in \cite{13connes,13landi}) may be extend for
vector bundles provided with N--connection structure (i. e. to
projective finite distinguished moduli) and in fact to
reconstruct the full anisotorpic (for instance, Finsler) geometry
from corresponding distinguishing of the Diximier trace and the
Wodzicki residue.

Let us consider a smooth compact vector bundle $\xi _{N}$ without
boundary and of dimension $n+m$ and $D_{[t]}$ as a ''symbol'' for
a time being operator and denote $\mathcal{A}_{[d]}=C^{\infty
}\left( \xi _{N}\right) .$ For a unitary representation $\left[
\mathcal{A}_{\pi },D_{\pi }\right] $ $\ $of the couple $\left(
\mathcal{A}_{[d]},D_{[t]}\right) $ as operators on an
Hilbert space $\mathcal{H}_{\pi }$ provided with a real structure operator $%
J_{\pi },$ such that $\left[ \mathcal{A}_{\pi },D_{\pi },\mathcal{H},J_{\pi }%
\right] $ $\ $satisfy all axioms of a real spectral d--triple.
Then, one holds the properties:

\begin{enumerate}
\item There is a unique Riemannian d--metric $g_{\pi }$ on $\xi _{N}$ such
the geodesic distance in the total space of the vector bundle
between every
two points $u_{[1]}$ and $u_{[2]}$ is given by%
\begin{equation*}
d\left( u_{[1]},u_{[2]}\right) =\sup_{a\in
\mathcal{A}_{[d]}}\left\{ \left| a(u_{[1]})-a(u_{[2]})\right|
:\left\| D_{\pi },\pi \left( a\right) \right\|
_{\mathcal{B}(\mathcal{H}_{\pi })}\leq 1\right\} .
\end{equation*}

\item The d--metric $g_{\pi }$ depends only on the unitary equivalence class
of the representations $\pi .$ The fibers of the map $\pi
\rightarrow g_{\pi }$ form unitary equivalence classes of
representations to metrics define a finite collection of affine
spaces $\mathcal{A}_{\sigma }$ parametrized by the spin
structures $\sigma $ on $\xi _{N}.$ These spin structures depends
on the type of d--metrics we are using in $\xi _{N}.$

\item The action functional given by the Diximier trace%
\begin{equation*}
G\left( D_{[t]}\right) =tr_{\varpi }\left( D_{[t]}^{n+m-2}\right)
\end{equation*}%
is a positive quadratic d--form with a unique minimum $\pi
_{\sigma }$ for each $\mathcal{A}_{\sigma }.$ At the minimum, the
values of $G\left( D_{[t]}\right) $ coincides with the Wodzicki
residue of $D_{\sigma }^{n+m-2}$ and is proportional to the
Hilbert--Einstein action for a fixed
d--connection,%
\begin{eqnarray*}
G\left( D_{\sigma }\right) ={Re}s_{W}\left( D_{\sigma
}^{n+m-2}\right)
&=&:\frac{1}{\left( n+m\right) \left( 2\pi \right) ^{n+m}}%
\int\limits_{S^{\ast }\xi _{N}}tr\left[ \sigma _{-(n+m)}\left(
u,u^{\prime
}\right) \delta u\delta u^{\prime }\right] \\
&=&c_{n+m}\int\limits_{\xi _{N}}\overleftarrow{R}\delta u,
\end{eqnarray*}%
where%
\begin{equation*}
c_{n+m}=\frac{n+m-2}{12}\frac{2^{[(n+m)/2]}}{\left( 4\pi \right)
^{(n+m)/2}\Gamma \left( \frac{n+m}{2}+1\right) },
\end{equation*}%
$\sigma _{-(n+m)}\left( u,u^{\prime }\right) $ is the part of
order $-(n+m)$ of the total symbol of $D_{\sigma }^{n+m-2},$
$\overleftarrow{R}$ is the scalar curvature (\ref{4dscalar}) on
$\xi _{N}$ and $tr$ is a normalized Clifford trace.

\item It is defined a representation of $\left( \mathcal{A}%
_{[d]},D_{[t]}\right) $ for every minimum $\pi _{\sigma }$ on the
Hilbert space of square integrable d--spinors
$\mathcal{H}=L^{2}(\xi _{N},S_{\sigma
})$ where $\mathcal{A}_{[d]}$ acts by multiplicative operators and $%
D_{\sigma }$ is the Dirac operator of chosen d--connection. If
there is no real structure $J,$ one has to replace $spin\,\ $by
$spin^{c}$ (for d--spinors investigated in Refs.
\cite{13vspinors,13vmon1,13vmon2}). In this case there is not \ a
uniqueness and the minimum of the functional $G\left(
D\right) $ is reached on a linear subspace of $\mathcal{A}_{\sigma }$ with $%
\sigma $ a fixed $spin^{c}$ structure. This subspace is parametrized by the $%
U\left( 1\right) $ gauge potentials entering in the $spin^{c}$
Dirac operator (the rest properties hold).
\end{enumerate}

The properties 1-4 are proved in a similar form as in \cite%
{13kalau,13hawkins,13landi}, but all computations are distinguished by
the N--connection structure and a fixed type of d--connection (we
omit such details). We can generate an anholonomic Riemannian,
Finsler or Lagrange
gravity depending on the class of d--metrics ((\ref{1dmetrr}), (\ref{2dmetricf}%
), (\ref{1mfl}), or a general one for vector bundles
(\ref{7dmetric})) we choose.

\subsubsection{Spectral anisotropic Gravity}

Consider a canonical d--triple $\left[
\mathcal{A}_{[d]}=C^{\infty }\left( \xi _{N}\right)
,\mathcal{H}=L^{2}\left( \xi _{N}\right) ,_{[d]}D\right] $ \
defined in subsection \ref{spinorsubsection} for a vector bundle
$\xi _{N},$ where $_{[d]}D$ is the Dirac d--operator
(\ref{dirac}) defined for a d--connection on \ $\xi _{N}.$ We are
going to compute the action
\begin{equation}
S_{G}\left( _{[d]}D,\Lambda \right) =tr_{\mathcal{H}}\left[ \chi
\left( \frac{_{\lbrack d]}D^{2}}{\Lambda ^{2}}\right) \right] ,
\label{action}
\end{equation}%
depending on the spectrum of $_{[d]}D,$ were $tr_{\mathcal{H}}$
is the usual trace in the Hilbert space, $\Lambda $ is the cutoff
parameter and $\chi $ will be closed as a suitable cutoff
function which cut off all eigenvalues of $_{[d]}D^{2}$ larger
than $\Lambda ^{2}.$ By using the Lichnerowicz formula, in our
case with operators for a vector bundle, and the heat kernel
expansion (similarly as for the proof summarized in Ref.
\cite{13landi})
\begin{equation*}
S_{G}\left( _{[d]}D,\Lambda \right) =\sum_{k\geq
0}f_{k}a_{k}\left( _{[d]}D^{2}/\Lambda ^{2}\right) ,
\end{equation*}%
were the coefficients $f_{k}$ are computed
\begin{equation*}
f_{0}=\int\limits_{0}^{\infty }\chi \left( z\right)
zdz,~~f_{2}=\int\limits_{0}^{\infty }\chi \left( z\right)
dz,~~~f_{2(k^{\prime }+2)}=\left( -1\right) ^{k^{\prime }}\chi
^{(k^{\prime })}\left( 0\right) ,k^{\prime }\geq 0,
\end{equation*}%
$\chi ^{(k^{\prime })}$ denotes the $k^{\prime }$th derivative on
its
argument, the so--called non--vanishing Seeley--de Witt coefficients $%
a_{k}\left( _{[d]}D^{2}/\Lambda ^{2}\right) $ are defined for
even values of
$k$ as integrals%
\begin{equation*}
a_{k}\left( _{[d]}D^{2}/\Lambda ^{2}\right) =\int\limits_{\xi
_{N}}a_{k}\left( u;_{[d]}D^{2}/\Lambda ^{2}\right) \sqrt{g}\delta
u
\end{equation*}%
with the first three subintegral functions given by%
\begin{eqnarray*}
a_{0}\left( u;_{[d]}D^{2}/\Lambda ^{2}\right) &=&\Lambda
^{4}\left( 4\pi
\right) ^{-(n+m)/2}trI_{2^{[(n+m)/2]}}, \\
a_{2}\left( u;_{[d]}D^{2}/\Lambda ^{2}\right) &=&\Lambda
^{2}\left( 4\pi \right) ^{-(n+m)/2}\left(
-\overleftarrow{R}/6+E\right) trI_{2^{[(n+m)/2]}},
\\
a_{4}\left( u;_{[d]}D^{2}/\Lambda ^{2}\right) &=&\left( 4\pi
\right)
^{-(n+m)/2}\frac{1}{360}(-12D_{\mu }D^{\mu }\overleftarrow{R}+5%
\overleftarrow{R}^{2}-2R_{\mu \nu }R^{\mu \nu } \\
&&-\frac{7}{4}R_{\mu \nu \alpha \beta }R^{\mu \nu \alpha \beta
}-60RE+180E^{2}+60D_{\mu }D^{\mu
}\overleftarrow{E})trI_{2^{[(n+m)/2]}},
\end{eqnarray*}%
and $\overleftarrow{E}=:_{[d]}D^{2}-\bigtriangledown ^{\lbrack S]}={%
\overleftarrow{R}/4},$ see (\ref{lich}). We can use for the
function $\chi $ the characteristic value of the interval
$[0,1],$ namely $\chi \left( z\right) =1$ for $z\leq 1$ and $\chi
\left( z\right) =0$ for $z\geq 1,$
possibly 'smoothed out' at $z=1,$ we get%
\begin{equation*}
f_{0}=1/2,f_{2}=1,f_{2(k^{\prime }+2)}=0,k^{\prime }\geq 0.
\end{equation*}

We compute (a similar calculus is given in \cite{13landi}; we only
distinguish
the curvature scalar, the Ricci and curvature d--tensor) the action (\ref%
{action}),%
\begin{equation*}
S_{G}\left( _{[d]}D,\Lambda \right) =\Lambda
^{4}\frac{2^{(n+m)/2-1}}{\left( 4\pi \right)
^{(n+m)/2}}\int\limits_{\xi _{N}}\sqrt{g}\delta u+\frac{\Lambda
^{2}}{6}\frac{2^{(n+m)/2-1}}{\left( 4\pi \right)
^{(n+m)/2}}\int\limits_{\xi _{N}}\sqrt{g}\overleftarrow{R}\delta
u.
\end{equation*}%
This action is dominated by the first term with a huge
cosmological
constant. But this constant can be eliminated \cite{13landi2} if the function $%
\chi \left( z\right) $ is replaced by $\widetilde{\chi }\left(
z\right) =\chi \left( z\right) -\alpha \chi \left( \beta z\right)
$ with any two numbers $\alpha $ and $\beta $ such that $\alpha
=\beta ^{2}$ and $\beta
\geq 0,\beta \neq 1.$ The final form of the action becomes%
\begin{equation}
S_{G}\left( _{[d]}D,\Lambda \right) =\left( 1-\frac{\alpha }{\beta ^{2}}%
\right) f_{2}\frac{\Lambda ^{2}}{6}\frac{2^{(n+m)/2-1}}{\left(
4\pi \right) ^{(n+m)/2}}\int\limits_{\xi
_{N}}\sqrt{g}\overleftarrow{R}\delta u+O\left( (\Lambda
^{2})^{0}\right) .  \label{2act1}
\end{equation}

From the action (\ref{2act1}) we can generate different models of
anholonomic Riemannian, Finsler or Lagrange gravity depending on
the class of d--metrics ((\ref{1dmetrr}), (\ref{2dmetricf}),
(\ref{1mfl}), or a general one for vector bundles (\ref{7dmetric}))
we parametrize for computations. But this construction has a
problem connected with ''spectral invariance versus
diffeomorphysm invariance on manifolds or vector bundles. Let us denote by $%
spec\left( \xi _{N},_{[d]}D\right) $ the spectrum of the Dirac
d--operator with each eigenvalue repeated according to its
multiplicity. Two vector
bundles $\xi _{N}$ and $\xi _{N}^{\prime }$ are called isospectral if $%
spec\left( \xi _{N},_{[d]}D\right) =spec\left( \xi _{N}^{\prime
},_{[d]}D\right) ,$ which defines an invariant transform of the action (\ref%
{action}). There are manifolds (and in consequence vector
bundles) which are isospectral without being isometric (the
converse is obviously true). This is known as a fact that one
cannot 'hear the shape of a drum \cite{13drum} because the spectral
invariance is stronger that usual diffeomorphysm invariance.

In spirit of spectral gravity, the eigenvalues of the Dirac
operator are diffeomorphic invariant functions of the geometry
and therefore true observable in general relativity. As we have
shown in this section they can be taken as a set of variables for
invariant descriptions to the anholonomic dynamics of the
gravitational field with (or not) local anisotropy in different
approaches of anholonomic Riemannian gravity and Finsler like
generalizations. But in another turn there exist isospectral
vector bundles which fail to be isometric. Thus, the eigenvalues
of the Dirac operator cannot be used to distinguish among such
vector bundles (or manifolds). A rigorous analysis is also
connected with the type of d--metric and d--connection structures
we prescribe for our geometric and physical models.

Finally, we remark that there are different models of gravity with
noncommutative setting (see, for instance, Refs. \cite%
{13ch1,13madore,13hawkins,13landi1,13landi2,13cl,13kalau}). By introducing
nonlinear connections in a respective commutative or
noncommutative variant we can transform such theories to be
anholonomic, i. e. locally anisotropic, in different approaches
with (pseudo) Riemannian geometry and Finsler/Lagrange or
Hamilton extensions.

\newpage

\section{Noncommutative Finsler--Gauge Theories}

The bulk of noncommutative models extending both locally
isotropic and anisotropic gravity theories are confrunted with
the problem of definition of noncommutative variants of
pseudo--Eucliedean and pseudo--Riemannian metrics. The problem is
connected with the fact of generation of noncommutative metric
structures via the Moyal results in complex and noncommutative
metrics. In order to avoid this difficulty we elaborated a model
of noncommutative gauge gravity (containing as particular case the
Einstein general relativity theory) starting from a variant of
gauge gravity being equivalent to the Einstein gravity and
emphasizing in a such approach the tetradic (frame) and
connection structures, but not the metric configuration (see
Refs. \cite{13vnonc}). The metric for such theories is induced from
the frame structure which can be holonomic or anholonomic. The
aim of this section is to generalize our results on
noncommutative gauge gravity as to include also possible
anisotropies in different variants of gauge realization of
anholonomic Einstein and Finsler like generalizations formally
developed in Refs. \cite{13vgauge,13vmon1,13vmon2}.

A still presented drawback of noncommutative geometry and physics
is that there is not yet formulated a generally accepted approach
to interactions of elementary particles coupled to gravity. There
are improved Connes--Lott and Chamsedine--Connes models of
nocommutative geometry \cite{13connes1,13cl} which yielded action
functionals typing together the gravitational and Yang--Mills
interactions and gauge bosons the Higgs sector (see also the approaches \cite%
{13hawkins} and, for an outline of recent results, \cite{13majid}).

In the last years much work has been made in noncommutative
extensions of
physical theories (see reviews and original results in Refs. \cite%
{13douglas,13soch}). It was not possible to formulate gauge theories
on noncommutative spaces \cite{13cds,13sw,13js,13mssw} with Lie algebra
valued infinitesimal transformations and with Lie algebra valued
gauge fields. In order to avoid the problem it was suggested to
use enveloping algebras of the Lie algebras for setting this type
of gauge theories and showed that in spite of the fact that such
enveloping algebras are infinite--dimensional one can restrict
them in a way that it would be a dependence on the Lie algebra
valued parameters, the Lie algebra valued gauge fields and their
spacetime derivatives only.

We follow the method of restricted enveloping algebras \cite{13js}
and construct gauge gra\-vitational theories by stating
corresponding structures with semisimple or nonsemisimple Lie
algebras and their extensions. We consider power series of
generators for the affine and non linear realized de Sitter gauge
groups and compute the coefficient functions of all the higher
powers of the generators of the gauge group which are functions of
the coefficients of the first power. Such constructions are based
on the Seiberg--Witten map \cite{13sw} and on the formalism of
$\ast $--product formulation of the algebra \cite{13w} when for
functional objects, being functions of commuting variables, there
are associated some algebraic noncommutative properties encoded
in the $\ast $--product.

The concept of gauge theory on noncommutative spaces was
introduced in a geometric manner \cite{13mssw} by defining the
covariant coordinates without speaking about derivatives and this
formalism was developed for quantum planes \cite{13wz}. In this
section we shall prove the existence for noncommutative spaces of
gauge models of gravity which agrees with usual gauge gravity
theories being equivalent, or extending, the general relativity
theory (see works \cite{13pd,13ts} for locally isotropic and
anisotropic spaces and corresponding reformulations and
generalizations respectively for anholonomic frames \cite{13vd} and
locally anisotropic (super) spaces
\cite{13vgauge,13vsuper,13vstring,13vmon1}) in the limit of commuting
spaces.

\subsection{Star--products and enveloping algebras in noncommutative spaces}

For a noncommutative space the coordinates ${\hat u}^i,$
$(i=1,...,N)$ satisfy some noncommutative relations
\begin{equation}  \label{ncr}
[{\hat u}^i,{\hat u}^j]=\left\{
\begin{array}{rcl}
& i\theta ^{ij}, & \theta ^{ij}\in \C,\mbox{ canonical structure;
} \\
& if_k^{ij}{\hat u}^k, & f_k^{ij}\in \C,\mbox{ Lie
structure; } \\
& iC_{kl}^{ij}{\hat u}^k{\hat u}^l, & C_{kl}^{ij}\in \C ,%
\mbox{ quantum
plane }%
\end{array}
\right.
\end{equation}
where $\C$ denotes the complex number field.

The noncommutative space is modelled as the associative algebra of
$\C;$\ this algebra is freely generated by the coordinates modulo
ideal $\mathcal{R}
$ generated by the relations (one accepts formal power series)\ $\mathcal{A}%
_{u}=\C[[{\hat u}^1,...,{\hat u}^N]]/\mathcal{R}.$ One restricts attention %
\cite{13jssw} to algebras having the (so--called,
Poincare--Birkhoff--Witt) property that any element of
$\mathcal{A}_{u}$ is defined by its coefficient
function and vice versa,%
\begin{equation*}
\widehat{f}=\sum\limits_{L=0}^{\infty }f_{i_{1},...,i_{L}}:{\hat{u}}%
^{i_{1}}\ldots {\hat{u}}^{i_{L}}:\quad \mbox{ when
}\widehat{f}\sim \left\{ f_{i}\right\} ,
\end{equation*}%
where $:{\hat{u}}^{i_{1}}\ldots {\hat{u}}^{i_{L}}:$ denotes that
the basis
elements satisfy some prescribed order (for instance, the normal order $%
i_{1}\leq i_{2}\leq \ldots \leq i_{L},$ or, another example, are
totally symmetric). The algebraic properties are all encoded in
the so--called diamond $(\diamond )$ product which is defined by
\begin{equation*}
\widehat{f}\widehat{g}=\widehat{h}~\sim ~\left\{ f_{i}\right\}
\diamond \left\{ g_{i}\right\} =\left\{ h_{i}\right\} .
\end{equation*}

In the mentioned approach to every function $f(u)=f(u^{1},\ldots
,u^{N})$ of commuting variables $u^{1},\ldots ,u^{N}$ one
associates an element of algebra $\widehat{f}$ when the commuting
variables are substituted by anticommuting ones,
\begin{equation}
f(u)=\sum f_{i_{1}\ldots i_{L}}u^{1}\cdots u^{N}\rightarrow \widehat{f}%
=\sum\limits_{L=0}^{\infty }f_{i_{1},...,i_{L}}:{\hat{u}}^{i_{1}}\ldots {%
\hat{u}}^{i_{L}}:  \notag
\end{equation}%
when the $\diamond $--product leads to a bilinear $\ast
$--product of
functions (see details in \cite{13mssw})%
\begin{equation*}
\left\{ f_{i}\right\} \diamond \left\{ g_{i}\right\} =\left\{
h_{i}\right\} \sim \left( f\ast g\right) \left( u\right) =h\left(
u\right) .
\end{equation*}

The $*$--product is defined respectively for the cases (\ref{ncr})
\begin{equation*}
f*g=\left\{
\begin{array}{rcl}
\exp [{\frac i2}{\frac \partial {\partial u^i}}{\theta
}^{ij}\frac \partial {\partial {u^{\prime }}^j}]f(u)g(u^{\prime
}){|}_{u^{\prime }\to u}, &  &
\\
\exp [\frac i2u^kg_k(i\frac \partial {\partial u^{\prime
}},i\frac \partial
{\partial u^{\prime \prime }})]f(u^{\prime })g(u^{\prime \prime }){|}%
_{u^{\prime \prime }\to u}^{u^{\prime }\to u}, &  &  \\
q^{{\frac 12}(-u^{\prime }{\frac \partial {\partial u^{\prime
}}}v{\frac
\partial {\partial v}}+u{\frac \partial {\partial u}}v^{\prime }{\frac
\partial {\partial v^{\prime }}})}f(u,v)g(u^{\prime },v^{\prime }){|}%
_{v^{\prime }\to v}^{u^{\prime }\to u}, &  &
\end{array}
\right.
\end{equation*}
where there are considered values of type%
\begin{eqnarray}
&e^{ik_n\widehat{u}^n}& e^{ip_{nl}\widehat{u}^n}
=e^{i\{k_n+p_n+\frac
12g_n\left( k,p\right) \}\widehat{u}^n,}  \label{gdecomp} \\
&g_n\left( k,p\right)& = -k_ip_jf_{\ n}^{ij}+\frac 16k_ip_j\left(
p_k-k_k\right) f_{\ m}^{ij}f_{\ n}^{mk}+...,  \notag \\
&e^Ae^B& = e^{A+B+\frac 12[A,B]+\frac 1{12}\left(
[A,[A,B]]+[B,[B,A]]\right) }+...  \notag
\end{eqnarray}
and for the coordinates on quantum (Manin) planes one holds the relation $%
uv=qvu.$

A non--abelian gauge theory on a noncommutative space is given by
two algebraic structures, the algebra $\mathcal{A}_{u}$ and a
non--abelian Lie
algebra $\mathcal{A}_{I}$ of the gauge group with generators $%
I^{1},...,I^{S} $ and the relations
\begin{equation}
\lbrack I^{\underline{s}},I^{\underline{p}}]=if_{~\underline{t}}^{\underline{%
s}\underline{p}}I^{\underline{t}}.  \label{commutators1}
\end{equation}%
In this case both algebras are treated on the same footing and
one denotes
the generating elements of the big algebra by $\widehat{u}^{i},$%
\begin{equation}
\widehat{z}^{\underline{i}}=\{\widehat{u}^{1},...,\widehat{u}%
^{N},I^{1},...,I^{S}\},\mathcal{A}_{z}=\C[[\widehat{u}^1,...,%
\widehat{u}^{N+S}]]/\mathcal{R}  \notag
\end{equation}%
and the $\ast $--product formalism is to be applied for the whole algebra $%
\mathcal{A}_{z}$ when there are considered functions of the
commuting variables $u^{i}\ (i,j,k,...=1,...,N)$ and
$I^{\underline{s}}\ (s,p,...=1,...,S).$

For instance, in the case of a canonical structure for the space variables $%
u^{i}$ we have
\begin{eqnarray}
(F\ast G)(u) &=&e^{\frac{i}{2}\left( \theta ^{ij}\frac{\partial
}{\partial u^{\prime i}}\frac{\partial }{\partial u^{\prime
\prime j}}+t^{s}g_{s}\left( i\frac{\partial }{\partial t^{\prime
}},i\frac{\partial }{\partial t^{\prime \prime }}\right) \right)
\times F\left( u^{\prime },t^{\prime }\right) G\left( u^{\prime
\prime },t^{\prime \prime }\right) \mid _{t^{\prime }\rightarrow
t,t^{\prime \prime }\rightarrow t}^{u^{\prime }\rightarrow
u,u^{\prime \prime }\rightarrow u}.}  \label{csp1} \\
&&  \notag
\end{eqnarray}%
This formalism was developed in \cite{13jssw} for general Lie
algebras. In this section we consider those cases when in the
commuting limit one obtains the gauge gravity and general
relativity theories or some theirs anisotropic generalizations..

\subsection{Enveloping algebras for gauge gravity connections}

In order to construct gauge gravity theories on noncommutative
space we define the gauge fields as elements the algebra
$\mathcal{A}_{u}$ that form representation of the generator
$I$--algebra for the de Sitter gauge group. For commutative
spaces it is known \cite{13pd,13ts,13vgauge} that an equivalent
re--expression of the Einstein theory as a gauge like theory
implies, for both locally isotropic and anisotropic spacetimes,
the nonsemisimplicity of the gauge group, which leads to a
nonvariational theory in the total space of the bundle of locally
adapted affine frames (to this class one belong the gauge
Poincare theories;\ on metric--affine and gauge gravity models see
original results and reviews in \cite{13ut}). By using auxiliary
biliniear forms, instead of degenerated Killing form for the
affine structural group, on fiber spaces, the gauge models of
gravity can be formulated to be variational. After projection on
the base spacetime, for the so--called Cartan connection form,
the Yang--Mills equations transforms equivalently into the
Einstein equations for general relativity \cite{13pd}. A variational
gauge gravitational theory can be also formulated by using a
minimal
extension of the affine structural group ${\mathcal{A}f}_{3+1}\left( {\R}%
\right) $ to the de Sitter gauge group $S_{10}=SO\left(
4+1\right) $ acting on ${\R}^{4+1}$ space.

\subsubsection{Nonlinear gauge theories of de Sitter group in commutative
spaces}

Let us consider the de Sitter space $\Sigma ^{4}$ as a
hypersurface given by the equations $\eta _{AB}u^{A}u^{B}=-l^{2}$
in the four dimensional flat space enabled with diagonal metric
$\eta _{AB},$\\ $\eta _{AA}=\pm 1$ (in this section
$A,B,C,...=1,2,...,5),$ where $\{u^{A}\}$ are global Cartesian
coordinates in $\R^{5};l>0$ is the curvature of de Sitter space.
The de Sitter group $S_{\left( \eta \right) }=SO_{\left( \eta
\right) }\left( 5\right) $ is defined as the isometry group of
$\Sigma ^{5}$--space with $6$ generators of Lie algebra
${\mathit{s}o}_{\left( \eta \right) }\left( 5\right) $ satisfying
the commutation relations
\begin{eqnarray}
\left[ M_{AB},M_{CD}\right] &=&\eta _{AC}M_{BD}-\eta
_{BC}M_{AD}-\eta
_{AD}M_{BC}+\eta _{BD}M_{AC}.  \label{dsc} \\
&&  \notag
\end{eqnarray}

Decomposing indices $A,B,...$ as $A=\left( \underline{\alpha
},5\right)
,B=\left( \underline{\beta },5\right) ,...,$ the metric $\eta _{AB}$ as $%
\eta _{AB}=\left( \eta _{\underline{\alpha }\underline{\beta
}},\eta
_{55}\right) ,$ and operators $M_{AB}$ as $M_{\underline{\alpha }\underline{%
\beta }}=\mathcal{F}_{\underline{\alpha }\underline{\beta }}$ and $P_{%
\underline{\alpha }}=l^{-1}M_{5\underline{\alpha }},$ we can write (\ref{dsc}%
) as
\begin{eqnarray}
\left[ \mathcal{F}_{\underline{\alpha }\underline{\beta }},\mathcal{F}_{%
\underline{\gamma }\underline{\delta }}\right] &=&\eta _{\underline{\alpha }%
\underline{\gamma }}\mathcal{F}_{\underline{\beta
}\underline{\delta }}-\eta
_{\underline{\beta }\underline{\gamma }}\mathcal{F}_{\underline{\alpha }%
\underline{\delta }}+\eta _{\underline{\beta }\underline{\delta }}\mathcal{F}%
_{\underline{\alpha }\underline{\gamma }}-\eta _{\underline{\alpha }%
\underline{\delta }}\mathcal{F}_{\underline{\beta
}\underline{\gamma }},
\notag \\
\left[ P_{\underline{\alpha }},P_{\underline{\beta }}\right] &=&-l^{-2}%
\mathcal{F}_{\underline{\alpha }\underline{\beta }},\left[ P_{\underline{%
\alpha }},\mathcal{F}_{\underline{\beta }\underline{\gamma }}\right] =\eta _{%
\underline{\alpha }\underline{\beta }}P_{\underline{\gamma }}-\eta _{%
\underline{\alpha }\underline{\gamma }}P_{\underline{\beta }},
\label{dsca}
\\
&&  \notag
\end{eqnarray}%
where we decompose the Lie algebra ${\mathit{s}o}_{\left( \eta
\right) }\left( 5\right) $ into a direct sum,
${\mathit{s}o}_{\left( \eta \right) }\left( 5\right)
={\mathit{s}o}_{\left( \eta \right) }(4)\oplus V_{4},$ where
$V_{4}$ is the vector space stretched on vectors
$P_{\underline{\alpha }}.$ We remark that $\Sigma ^{4}=S_{\left(
\eta \right) }/L_{\left( \eta \right) },$ where $L_{\left( \eta
\right) }=SO_{\left( \eta \right) }\left(
4\right) .$ For $\eta _{AB}=diag\left( 1,-1,-1,-1\right) $ and $%
S_{10}=SO\left( 1,4\right) ,L_{6}=SO\left( 1,3\right) $ is the
group of Lorentz rotations.

In this paper the generators $I^{\underline{a}}$ and structure constants $%
f_{~\underline{t}}^{\underline{s}\underline{p}}$ from
(\ref{commutators1})
are parametr\-iz\-ed just to obtain de Sitter generators and commutations (\ref%
{dsca}).

The action of the group $S_{\left( \eta \right) }$ can be realized by using $%
4\times 4$ matrices with a parametrization distinguishing the subgroup $%
L_{\left( \eta \right) }:$%
\begin{equation}  \label{parametriz}
B=bB_L,
\end{equation}
where%
\begin{equation*}
B_L=\left(
\begin{array}{cc}
L & 0 \\
0 & 1%
\end{array}
\right) ,
\end{equation*}
$L\in L_{\left( \eta \right) }$ is the de Sitter bust matrix
transforming the vector $\left( 0,0,...,\rho \right) \in {\R}^5$
into the arbitrary point $\left( V^1,V^2,...,V^5\right) \in
\Sigma _\rho ^5\subset \mathcal{R}^5$ with curvature $\rho,$
$(V_A V^A=-\rho ^2, V^A=t^A\rho ).$ Matrix $b$ can be expressed as
\begin{equation*}
b=\left(
\begin{array}{cc}
\delta _{\quad \underline{\beta }}^{\underline{\alpha }}+\frac{t^{\underline{%
\alpha }}t_{\underline{\beta }}}{\left( 1+t^5\right) } & t^{\underline{%
\alpha }} \\
t_{\underline{\beta }} & t^5%
\end{array}
\right) .
\end{equation*}

The de Sitter gauge field is associated with a
${\mathit{s}o}_{\left( \eta \right) }\left( 5\right) $--valued
connection 1--form
\begin{equation}  \label{dspot}
\widetilde{\Omega }=\left(
\begin{array}{cc}
\omega _{\quad \underline{\beta }}^{\underline{\alpha }} &
\widetilde{\theta
}^{\underline{\alpha }} \\
\widetilde{\theta }_{\underline{\beta }} & 0%
\end{array}
\right) ,
\end{equation}
where $\omega _{\quad \underline{\beta }}^{\underline{\alpha }}\in
so(4)_{\left( \eta \right) },$ $\widetilde{\theta
}^{\underline{\alpha }}\in
\mathcal{R}^4,\widetilde{\theta }_{\underline{\beta }}\in \eta _{\underline{%
\beta }\underline{\alpha }}\widetilde{\theta }^{\underline{\alpha
}}.$

Because $S_{\left( \eta \right) }$--transforms mix the components
of the
matrix $\omega _{\quad \underline{\beta }}^{\underline{\alpha }}$ and $%
\widetilde{\theta }^{\underline{\alpha }}$ fields in
(\ref{dspot}) (the introduced para\-met\-ri\-za\-ti\-on is
invariant on action on $SO_{\left( \eta \right) }\left( 4\right)
$ group we cannot identify $\omega _{\quad
\underline{\beta }}^{\underline{\alpha }}$ and $\widetilde{\theta }^{%
\underline{\alpha }},$ respectively, with the connection $\Gamma
_{~\beta \gamma }^{\alpha }$ and the fundamental form $\chi
^{\alpha }$ in a metric--affine spacetime). To avoid this
difficulty we consider \cite{13ts} a nonlinear gauge realization of
the de Sitter group $S_{\left( \eta \right) }, $ namely, we
introduce into consideration the nonlinear gauge field
\begin{equation}
\Gamma =b^{-1}{\widetilde{\Omega }}b+b^{-1}db=\left(
\begin{array}{cc}
\Gamma _{~\underline{\beta }}^{\underline{\alpha }} & \theta ^{\underline{%
\alpha }} \\
\theta _{\underline{\beta }} & 0%
\end{array}%
\right) ,  \label{1npot}
\end{equation}%
where
\begin{eqnarray}
\Gamma _{\quad \underline{\beta }}^{\underline{\alpha }}
&=&\omega _{\quad
\underline{\beta }}^{\underline{\alpha }}-\left( t^{\underline{\alpha }}Dt_{%
\underline{\beta }}-t_{\underline{\beta }}Dt^{\underline{\alpha
}}\right)
/\left( 1+t^{5}\right) ,  \notag \\
\theta ^{\underline{\alpha }} &=&t^{5}\widetilde{\theta
}^{\underline{\alpha
}}+Dt^{\underline{\alpha }}-t^{\underline{\alpha }}\left( dt^{5}+\widetilde{%
\theta }_{\underline{\gamma }}t^{\underline{\gamma }}\right)
/\left(
1+t^{5}\right) ,  \notag \\
Dt^{\underline{\alpha }} &=&dt^{\underline{\alpha }}+\omega
_{\quad \underline{\beta }}^{\underline{\alpha
}}t^{\underline{\beta }}.  \notag
\end{eqnarray}

The action of the group $S\left( \eta \right) $ is nonlinear,
yielding transforms
\begin{equation*}
\Gamma ^{\prime }=L^{\prime }\Gamma \left( L^{\prime }\right)
^{-1}+L^{\prime }d\left( L^{\prime }\right) ^{-1},\theta ^{\prime
}=L\theta ,
\end{equation*}%
where the nonlinear matrix--valued function
\begin{equation*}
L^{\prime }=L^{\prime }\left( t^{\alpha },b,B_{T}\right)
\end{equation*}%
is defined from $B_{b}=b^{\prime }B_{L^{\prime }}$ (see the parametrization (%
\ref{parametriz})). The de Sitter algebra with generators
(\ref{dsca}) and
nonlinear gauge transforms of type (\ref{1npot}) is denoted $\mathcal{A}%
_{I}^{(dS)}.$

\subsubsection{\ De Sitter nonlinear gauge gravity and Einstein and Finsler
like gravity}

Let us consider the de Sitter nonlinear gauge gravitational connection (\ref%
{1npot}) rewritten in the form
\begin{equation}
\Gamma =\left(
\begin{array}{cc}
\Gamma _{\quad \underline{\beta }}^{\underline{\alpha }} & l_{0}^{-1}\chi ^{%
\underline{\alpha }} \\
l_{0}^{-1}\chi _{\underline{\beta }} & 0%
\end{array}%
\right)  \label{1a}
\end{equation}%
where
\begin{eqnarray}
\Gamma _{\quad \underline{\beta }}^{\underline{\alpha }}
&=&\Gamma _{\quad
\underline{\beta }\mu }^{\underline{\alpha }}\delta u^{\mu },  \notag \\
\Gamma _{\quad \underline{\beta }\mu }^{\underline{\alpha }}
&=&\chi _{\quad
\alpha }^{\underline{\alpha }}\chi _{\quad \beta }^{\underline{\beta }%
}\Gamma _{\quad \beta \gamma }^{\alpha }+\chi _{\quad \alpha }^{\underline{%
\alpha }}\delta _{\mu }\chi _{\quad \underline{\beta }}^{\alpha },\chi ^{%
\underline{\alpha }}=\chi _{\quad \mu }^{\underline{\alpha
}}\delta u^{\mu }, \notag
\end{eqnarray}%
and
\begin{equation*}
G_{\alpha \beta }=\chi _{\quad \alpha }^{\underline{\alpha }}\chi
_{\quad \beta }^{\underline{\beta }}\eta _{\underline{\alpha
}\underline{\beta }},
\end{equation*}%
$\eta _{\underline{\alpha }\underline{\beta }}=\left(
1,-1,...,-1\right) $ and $l_{0}$ is a dimensional constant. As
$\Gamma _{\quad \beta \gamma }^{\alpha }$ we take the Christoffel
symbols for the Einstein theory, or every type of d--connection
(\ref{6dcon}) for an anisotropic spacetime. Correspondingly,
$G_{\alpha \beta }$ can be the pseudo--Rieamannian metric in
general relativity or any d--metric (\ref{7dmetric}), which can be
particularized for the anholonomic Einstein gravity
(\ref{1dmetrr}) \ or for a Finsler type gravity (\ref{2dmetricf}).

The curvature of (\ref{1a}),
\begin{equation*}
\mathcal{R}^{(\Gamma )}=d\Gamma +\Gamma \bigwedge \Gamma ,
\end{equation*}%
can be written
\begin{equation}
\mathcal{R}^{(\Gamma )}=\left(
\begin{array}{cc}
\mathcal{R}_{\quad \underline{\beta }}^{\underline{\alpha }}+l_{0}^{-1}\pi _{%
\underline{\beta }}^{\underline{\alpha }} &
l_{0}^{-1}T^{\underline{\alpha }}
\\
l_{0}^{-1}T^{\underline{\beta }} & 0%
\end{array}%
\right) ,  \label{2a}
\end{equation}%
where
\begin{equation*}
\pi _{\underline{\beta }}^{\underline{\alpha }}=\chi ^{\underline{\alpha }%
}\bigwedge \chi _{\underline{\beta }},\mathcal{R}_{\quad \underline{\beta }%
}^{\underline{\alpha }}=\frac{1}{2}\mathcal{R}_{\quad
\underline{\beta }\mu \nu }^{\underline{\alpha }}\delta u^{\mu
}\bigwedge \delta u^{\nu },
\end{equation*}%
and
\begin{equation*}
\mathcal{R}_{\quad \underline{\beta }\mu \nu }^{\underline{\alpha }}=\chi _{%
\underline{\beta }}^{\quad \beta }\chi _{\alpha }^{\quad \underline{\alpha }%
}R_{\quad \beta _{\mu \nu }}^{\alpha }.
\end{equation*}%
with the $R_{\quad \beta {\mu \nu }}^{\alpha }$ being the
metric--affine (for Ein\-stein-\-Car\-tan--Weyl spaces), or the
(pseu\-do) Riemannian curvature, or for anisotropic spaces the
d--curvature (\ref{3dcurvatures}). The de Sitter gauge group is
semisimple and we are able to construct a variational gauge
gravitational theory with the Lagrangian
\begin{equation*}
L=L_{\left( G\right) }+L_{\left( m\right) }
\end{equation*}%
where the gauge gravitational Lagrangian is defined
\begin{equation*}
L_{\left( G\right) }=\frac{1}{4\pi }Tr\left( \mathcal{R}^{(\Gamma
)}\bigwedge \ast _{G}\mathcal{R}^{(\Gamma )}\right)
=\mathcal{L}_{\left( G\right) }\left| G\right| ^{1/2}\delta
^{n+m}u,
\end{equation*}%
with
\begin{equation}
\mathcal{L}_{\left( G\right) }=\frac{1}{2l^{2}}T_{\quad \mu \nu }^{%
\underline{\alpha }}T_{\underline{\alpha }}^{\quad \mu \nu }+\frac{1}{%
8\lambda }\mathcal{R}_{\quad \underline{\beta }\mu \nu }^{\underline{\alpha }%
}\mathcal{R}_{\quad \underline{\alpha }}^{\underline{\beta }\quad
\mu \nu }{}-\frac{1}{l^{2}}\left( {\overleftarrow{R}}\left(
\Gamma \right) -2\lambda _{1}\right) ,  \notag
\end{equation}%
$\delta ^{4}u$ being the volume element, $T_{\quad \mu \nu }^{\underline{%
\alpha }}=\chi _{\quad \alpha }^{\underline{\alpha }}T_{\quad \mu
\nu
}^{\alpha }$ (the gravitational constant $l^{2}$ satisfies the relations $%
l^{2}=2l_{0}^{2}\lambda ,\lambda _{1}=-3/l_{0}),\quad Tr$ denotes
the trace on $\underline{\alpha },\underline{\beta }$ indices,
and the matter field Lagrangian is defined
\begin{equation*}
L_{\left( m\right) }=-1\frac{1}{2}Tr\left( \Gamma \bigwedge \ast _{G}%
\mathcal{I}\right) =\mathcal{L}_{\left( m\right) }\left| G\right|
^{1/2}\delta ^{n+m}u,
\end{equation*}%
where
\begin{equation*}
\mathcal{L}_{\left( m\right) }=\frac{1}{2}\Gamma _{\quad \underline{\beta }%
\mu }^{\underline{\alpha }}S_{\quad \alpha }^{\underline{\beta
}\quad \mu }-t_{\quad \underline{\alpha }}^{\mu }l_{\quad \mu
}^{\underline{\alpha }}.
\end{equation*}%
The matter field source $\mathcal{J}$ is obtained as a variational
derivation of $\mathcal{L}_{\left( m\right) }$ on $\Gamma $ and is
parametrized as
\begin{equation*}
\mathcal{J}=\left(
\begin{array}{cc}
S_{\quad \underline{\beta }}^{\underline{\alpha }} & -l_{0}t^{\underline{%
\alpha }} \\
-l_{0}t_{\underline{\beta }} & 0%
\end{array}%
\right)
\end{equation*}%
with $t^{\underline{\alpha }}=t_{\quad \mu }^{\underline{\alpha
}}\delta u^{\mu }$ and $S_{\quad \underline{\beta
}}^{\underline{\alpha }}=S_{\quad \underline{\beta }\mu
}^{\underline{\alpha }}\delta u^{\mu }$ being respectively the
canonical tensors of energy--momentum and spin density.

Varying the action
\begin{equation*}
S=\int \delta ^{4}u\left( \mathcal{L}_{\left( G\right)
}+\mathcal{L}_{\left( m\right) }\right)
\end{equation*}%
on the $\Gamma $--variables (1a), we obtain the
gau\-ge--gra\-vi\-ta\-ti\-on\-al field equations, in general,
with local anisotropy,
\begin{equation}
d\left( \ast \mathcal{R}^{(\Gamma )}\right) +\Gamma \bigwedge
\left( \ast \mathcal{R}^{(\Gamma )}\right) -\left( \ast
\mathcal{R}^{(\Gamma )}\right) \bigwedge \Gamma =-\lambda \left(
\ast \mathcal{J}\right) ,  \label{3a}
\end{equation}%
were the Hodge operator $\ast $ is used.

Specifying the variations on $\Gamma _{\quad \underline{\beta }}^{\underline{%
\alpha }}$ and $\chi $--variables, we rewrite (\ref{3a})
\begin{eqnarray}
\widehat{\mathcal{D}}\left( \ast \mathcal{R}^{(\Gamma )}\right) &&+\frac{%
2\lambda }{l^{2}}(\widehat{\mathcal{D}}\left( \ast \pi \right)
+\chi \bigwedge \left( \ast T^{T}\right) -\left( \ast T\right)
\bigwedge \chi
^{T})=-\lambda \left( \ast S\right) ,  \notag \\
\widehat{\mathcal{D}}\left( \ast T\right) &-&\left( \ast \mathcal{R}%
^{(\Gamma )}\right) \bigwedge \chi -\frac{2\lambda }{l^{2}}\left(
\ast \pi \right) \bigwedge \chi =\frac{l^{2}}{2}\left( \ast
t+\frac{1}{\lambda }\ast \tau \right) ,  \notag
\end{eqnarray}%
where
\begin{eqnarray}
T^{t} &=&\{T_{\underline{\alpha }}=\eta _{\underline{\alpha }\underline{%
\beta }}T^{\underline{\beta }},~T^{\underline{\beta
}}=\frac{1}{2}T_{\quad \mu \nu }^{\underline{\beta }}\delta
u^{\mu }\bigwedge \delta u^{\nu }\},
\notag \\
\chi ^{T} &=&\{\chi _{\underline{\alpha }}=\eta _{\underline{\alpha }%
\underline{\beta }}\chi ^{\underline{\beta }},~\chi ^{\underline{\beta }%
}=\chi _{\quad \mu }^{\widehat{\beta }}\delta u^{\mu }\},\qquad \widehat{%
\mathcal{D}}=d+\widehat{\Gamma },  \notag
\end{eqnarray}%
($\widehat{\Gamma }$ acts as $\Gamma _{\quad \underline{\beta }\mu }^{%
\underline{\alpha }}$ on indices $\underline{\gamma },\underline{\delta }%
,... $ and as $\Gamma _{\quad \beta \mu }^{\alpha }$ on indices
$\gamma ,\delta ,...).$ The value $\tau $ defines the
energy--momentum tensor of the
gauge gravitational field $\widehat{\Gamma }:$%
\begin{equation*}
\tau _{\mu \nu }\left( \widehat{\Gamma }\right)
=\frac{1}{2}Tr\left(
\mathcal{R}_{\mu \alpha }\mathcal{R}_{\quad \nu }^{\alpha }-\frac{1}{4}%
\mathcal{R}_{\alpha \beta }\mathcal{R}^{\alpha \beta }G_{\mu \nu
}\right) .
\end{equation*}

Equations (\ref{3a}) make up the complete system of variational
field equations for nonlinear de Sitter gauge anisotropic gravity.

We note that we can obtain a nonvariational Poincare gauge
gravitational theory if we consider the contraction of the gauge
potential (\ref{1a}) to a potential with values in the Poincare
Lie algebra
\begin{equation}
\Gamma =\left(
\begin{array}{cc}
\Gamma _{\quad \widehat{\beta }}^{\widehat{\alpha }} & l_{0}^{-1}\chi ^{%
\widehat{\alpha }} \\
l_{0}^{-1}\chi _{\widehat{\beta }} & 0%
\end{array}%
\right) \rightarrow \Gamma =\left(
\begin{array}{cc}
\Gamma _{\quad \widehat{\beta }}^{\widehat{\alpha }} & l_{0}^{-1}\chi ^{%
\widehat{\alpha }} \\
0 & 0%
\end{array}%
\right) .  \label{4a}
\end{equation}%
A similar gauge potential was considered in the formalism of
linear and affine frame bundles on curved spacetimes by Popov and
Daikhin \cite{13pd}. They treated (\ref{4a}) as the Cartan
connection form for affine gauge like gravity and by using 'pure'
geometric methods proved that the Yang--Mills equations of their
theory are equivalent, after projection on the base, to the
Einstein equations. The main conclusion for a such approach to
Einstein gravity is that this theory admits an equivalent
formulation as a gauge model but with a nonsemisimple structural
gauge group. In order to have a variational theory on the total
bundle space it is necessary to introduce an auxiliary bilinear
form on the typical fiber, instead of degenerated Killing form;
the coefficients of auxiliary form disappear after projection on
the base. An alternative variant is to consider a gauge
gravitational theory when the gauge group was minimally extended
to the de\ Sitter one with nondegenerated Killing form. The
nonlinear realizations have to be introduced if we consider in a
common fashion both the frame (tetradic) and connection components
 included as the coefficients of the potential (\ref{1a}).
  Finally, we note that the models of de Sitter gauge gravity
were generalized for Finsler and Lagrange theories in Refs.
\cite{13vgauge,13vmon1}.

\subsubsection{Enveloping nonlinear de Sitter algebra valued connection}

Let now us consider a noncommutative space. In this case the
gauge fields are elements of the algebra $\widehat{\psi }\in
\mathcal{A}_{I}^{(dS)}$ that
form the nonlinear representation of the de Sitter algebra ${\mathit{s}o}%
_{\left( \eta \right) }\left( 5\right) $ when the whole algebra is denoted $%
\mathcal{A}_{z}^{(dS)}.$ Under a nonlinear de Sitter
transformation the
elements transform as follows%
\begin{equation*}
\delta \widehat{\psi }=i\widehat{\gamma }\widehat{\psi
},\widehat{\psi }\in \mathcal{A}_{u},\widehat{\gamma }\in
\mathcal{A}_{z}^{(dS)}.
\end{equation*}%
So, the action of the generators (\ref{dsca}) on $\widehat{\psi
}$ is
defined as this element is supposed to form a nonlinear representation of $%
\mathcal{A}_{I}^{(dS)}$ and, in consequence, $\delta
\widehat{\psi }\in \mathcal{A}_{u}$ despite $\widehat{\gamma }\in
\mathcal{A}_{z}^{(dS)}.$ It
should be emphasized that independent of a representation the object $%
\widehat{\gamma }$ takes values in enveloping de Sitter algebra
and not in a Lie algebra as would be for commuting spaces. The
same holds for the connections that we introduce \cite{13mssw}, in
order to define covariant
coordinates,%
\begin{equation*}
\widehat{U}^{\nu }=\widehat{u}^{v}+\widehat{\Gamma }^{\nu },\widehat{\Gamma }%
^{\nu }\in \mathcal{A}_{z}^{(dS)}.
\end{equation*}

The values $\widehat{U}^{\nu }\widehat{\psi }$ transform
covariantly,
\begin{equation*}
\delta \widehat{U}^{\nu }\widehat{\psi }=i\widehat{\gamma }\widehat{U}^{\nu }%
\widehat{\psi },
\end{equation*}%
if and only if the connection $\widehat{\Gamma }^{\nu }$
satisfies the
transformation law of the enveloping nonlinear realized de Sitter algebra,%
\begin{equation*}
\delta \widehat{\Gamma }^{\nu }\widehat{\psi }=-i[\widehat{u}^{v},\widehat{%
\gamma }]+i[\widehat{\gamma },\widehat{\Gamma }^{\nu }],
\end{equation*}%
where $\delta \widehat{\Gamma }^{\nu }\in
\mathcal{A}_{z}^{(dS)}.$ The enveloping algebra--valued
connection has infinitely many component fields. Nevertheless, it
was shown that all the component fields can be induced from
a Lie algebra--valued connection by a Seiberg--Witten map (\cite{13sw,13js} and %
\cite{13bsst} for $SO(n)$ and $Sp(n)).$ In this subsection we show
that similar constructions could be proposed for nonlinear
realizations of de
Sitter algebra when the transformation of the connection is considered%
\begin{equation*}
\delta \widehat{\Gamma }^{\nu }=-i[u^{\nu },^{\ast }~\widehat{\gamma }]+i[%
\widehat{\gamma },^{\ast }~\widehat{\Gamma }^{\nu }].
\end{equation*}%
For simplicity, we treat in more detail the canonical case with
the star
product (\ref{csp1}). The first term in the variation $\delta \widehat{%
\Gamma }^{\nu }$ gives
\begin{equation*}
-i[u^{\nu },^{\ast }~\widehat{\gamma }]=\theta ^{\nu \mu }\frac{\partial }{%
\partial u^{\mu }}\gamma .
\end{equation*}%
Assuming that the variation of $\widehat{\Gamma }^{\nu }=\theta
^{\nu \mu
}Q_{\mu }$ starts with a linear term in $\theta ,$ we have%
\begin{equation*}
\delta \widehat{\Gamma }^{\nu }=\theta ^{\nu \mu }\delta Q_{\mu
},~\delta
Q_{\mu }=\frac{\partial }{\partial u^{\mu }}\gamma +i[\widehat{\gamma }%
,^{\ast }~Q_{\mu }].
\end{equation*}%
We follow the method of calculation from the papers
\cite{13mssw,13jssw} and expand the star product (\ref{csp1}) in
$\theta $ but not in $g_{a}$ and
find to first order in $\theta ,$%
\begin{equation}
\gamma =\gamma _{\underline{a}}^{1}I^{\underline{a}}+\gamma _{\underline{a}%
\underline{b}}^{1}I^{\underline{a}}I^{\underline{b}}+...,Q_{\mu }=q_{\mu ,%
\underline{a}}^{1}I^{\underline{a}}+q_{\mu ,\underline{a}\underline{b}%
}^{2}I^{\underline{a}}I^{\underline{b}}+...  \label{seriesa}
\end{equation}%
where $\gamma _{\underline{a}}^{1}$ and $q_{\mu
,\underline{a}}^{1}$ are of
order zero in $\theta $ and $\gamma _{\underline{a}\underline{b}}^{1}$ and $%
q_{\mu ,\underline{a}\underline{b}}^{2}$ are of second order in
$\theta .$ The expansion in $I^{\underline{b}}$ leads to an
expansion in $g_{a}$ of the $\ast $--product because the higher
order $I^{\underline{b}}$--derivatives
vanish. For de Sitter case as $I^{\underline{b}}$ we take the generators (%
\ref{dsca}), see commutators (\ref{commutators1}), with the
corresponding de
Sitter structure constants $f_{~\underline{d}}^{\underline{b}\underline{c}%
}\simeq f_{~\underline{\beta }}^{\underline{\alpha
}\underline{\beta }}$ (in our further identifications with
spacetime objects like frames and connections we shall use Greek
indices).

The result of calculation of variations of (\ref{seriesa}), by
using $g_{a}$
to the order given in (\ref{gdecomp}), is%
\begin{eqnarray}
\delta q_{\mu ,\underline{a}}^{1} &=&\frac{\partial \gamma _{\underline{a}%
}^{1}}{\partial u^{\mu }}-f_{~\underline{a}}^{\underline{b}\underline{c}%
}\gamma _{\underline{b}}^{1}q_{\mu ,\underline{c}}^{1},  \notag \\
\delta Q_{\tau } &=&\theta ^{\mu \nu }\partial _{\mu }\gamma _{\underline{a}%
}^{1}\partial _{\nu }q_{\tau ,\underline{b}}^{1}I^{\underline{a}}I^{%
\underline{b}}+...,  \notag \\
\delta q_{\mu ,\underline{a}\underline{b}}^{2} &=&\partial _{\mu }\gamma _{%
\underline{a}\underline{b}}^{2}-\theta ^{\nu \tau }\partial _{\nu }\gamma _{%
\underline{a}}^{1}\partial _{\tau }q_{\mu ,\underline{b}}^{1}-2f_{~%
\underline{a}}^{\underline{b}\underline{c}}\{\gamma _{\underline{b}%
}^{1}q_{\mu ,\underline{c}\underline{d}}^{2}+\gamma _{\underline{b}%
\underline{d}}^{2}q_{\mu ,\underline{c}}^{1}\}.  \notag
\end{eqnarray}

Next, we introduce the objects $\varepsilon ,$ taking the values
in de
Sitter Lie algebra and $W_{\mu },$ being enveloping de Sitter algebra valued,%
\begin{equation*}
\varepsilon =\gamma _{\underline{a}}^{1}I^{\underline{a}}\mbox{
and
}W_{\mu }=q_{\mu ,\underline{a}\underline{b}}^{2}I^{\underline{a}}I^{%
\underline{b}},
\end{equation*}%
with the variation $\delta W_{\mu }$ satisfying the equation
\cite{13mssw,13jssw}
\begin{equation*}
\delta W_{\mu }=\partial _{\mu }(\gamma _{\underline{a}\underline{b}}^{2}I^{%
\underline{a}}I^{\underline{b}})-\frac{1}{2}\theta ^{\tau \lambda
}\{\partial _{\tau }\varepsilon ,\partial _{\lambda }q_{\mu
}\}+{}i[\varepsilon ,W_{\mu }]+i[(\gamma _{\underline{a}\underline{b}}^{2}I^{%
\underline{a}}I^{\underline{b}}),q_{\nu }].
\end{equation*}%
This equation has the solution (found in \cite{13mssw,13sw})%
\begin{equation}
\gamma _{\underline{a}\underline{b}}^{2}=\frac{1}{2}\theta ^{\nu
\mu
}(\partial _{\nu }\gamma _{\underline{a}}^{1})q_{\mu ,\underline{b}%
}^{1},~q_{\mu ,\underline{a}\underline{b}}^{2}=-\frac{1}{2}\theta
^{\nu \tau
}q_{\nu ,\underline{a}}^{1}\left( \partial _{\tau }q_{\mu ,\underline{b}%
}^{1}+R_{\tau \mu ,\underline{b}}^{1}\right)  \notag
\end{equation}%
where
\begin{equation*}
R_{\tau \mu ,\underline{b}}^{1}=\partial _{\tau }q_{\mu ,\underline{b}%
}^{1}-\partial _{\mu }q_{\tau ,\underline{b}}^{1}+f_{~\underline{d}}^{%
\underline{e}\underline{c}}q_{\tau ,\underline{e}}^{1}q_{\mu ,\underline{e}%
}^{1}
\end{equation*}%
could be identified with the coefficients $\mathcal{R}_{\quad \underline{%
\beta }\mu \nu }^{\underline{\alpha }}$ of de Sitter nonlinear
gauge gravity
curvature (see formula (\ref{2a})) if in the commutative limit $q_{\mu ,%
\underline{b}}^{1}\simeq \left(
\begin{array}{cc}
\Gamma _{\quad \underline{\beta }}^{\underline{\alpha }} & l_{0}^{-1}\chi ^{%
\underline{\alpha }} \\
l_{0}^{-1}\chi _{\underline{\beta }} & 0%
\end{array}%
\right) $ (see (\ref{1a})).

The below presented procedure can be generalized to all the
higher powers of $\theta $.

\subsection{Noncommutative Gravity Covariant Gauge Dynamics}

\subsubsection{First order corrections to gravitational curvature}

The constructions from the previous section are summarized by the
conclusion
that the de Sitter algebra valued object $\varepsilon =\gamma _{\underline{a}%
}^1\left( u\right) I^{\underline{a}}$ determines all the terms in
the
enveloping algebra%
\begin{equation*}
\gamma =\gamma _{\underline{a}}^1I^{\underline{a}}+\frac 14\theta
^{\nu \mu
}\partial _\nu \gamma _{\underline{a}}^1\ q_{\mu ,\underline{b}}^1\left( I^{%
\underline{a}}I^{\underline{b}}+I^{\underline{b}}I^{\underline{a}}\right)
+...
\end{equation*}
and the gauge transformations are defined by $\gamma _{\underline{a}%
}^1\left( u\right) $ and $q_{\mu ,\underline{b}}^1(u),$ when
\begin{equation*}
\delta _{\gamma ^1}\psi =i\gamma \left( \gamma ^1,q_\mu ^1\right)
*\psi .
\end{equation*}
For de Sitter enveloping algebras one holds the general formula
for
compositions of two transformations%
\begin{equation*}
\delta _\gamma \delta _\varsigma -\delta _\varsigma \delta
_\gamma =\delta _{i(\varsigma *\gamma -\gamma *\varsigma )}
\end{equation*}
which is also true for the restricted transformations defined by $\gamma ^1,$%
\begin{equation*}
\delta _{\gamma ^1}\delta _{\varsigma ^1}-\delta _{\varsigma
^1}\delta _{\gamma ^1}=\delta _{i(\varsigma ^1*\gamma ^1-\gamma
^1*\varsigma ^1)}.
\end{equation*}

Applying the formula (\ref{csp1}) we calculate%
\begin{eqnarray*}
\lbrack \gamma ,^{\ast }\zeta ] &=&i\gamma _{\underline{a}}^{1}\zeta _{%
\underline{b}}^{1}f_{~\underline{c}}^{\underline{a}\underline{b}}I^{%
\underline{c}}+\frac{i}{2}\theta ^{\nu \mu }\{\partial _{v}\left( \gamma _{%
\underline{a}}^{1}\zeta _{\underline{b}}^{1}f_{~\underline{c}}^{\underline{a}%
\underline{b}}\right) q_{\mu ,\underline{c}} \\
&&+{}\left( \gamma _{\underline{a}}^{1}\partial _{v}\zeta _{\underline{b}%
}^{1}-\zeta _{\underline{a}}^{1}\partial _{v}\gamma _{\underline{b}%
}^{1}\right) q_{\mu ,\underline{b}}f_{~\underline{c}}^{\underline{a}%
\underline{b}}+2\partial _{v}\gamma _{\underline{a}}^{1}\partial
_{\mu }\zeta
_{\underline{b}}^{1}\}I^{\underline{d}}I^{\underline{c}}.
\end{eqnarray*}%
Such commutators could be used for definition of tensors
\cite{13mssw}
\begin{equation}
\widehat{S}^{\mu \nu }=[\widehat{U}^{\mu },\widehat{U}^{\nu }]-i\widehat{%
\theta }^{\mu \nu },  \label{tensor1}
\end{equation}%
where $\widehat{\theta }^{\mu \nu }$ is respectively stated for
the canonical, Lie and quantum plane structures. Under the
general enveloping
algebra one holds the transform%
\begin{equation*}
\delta \widehat{S}^{\mu \nu }=i[\widehat{\gamma
},\widehat{S}^{\mu \nu }].
\end{equation*}%
For instance, the canonical case is characterized by%
\begin{eqnarray}
S^{\mu \nu } &=&i\theta ^{\mu \tau }\partial _{\tau }\Gamma ^{\nu
}-i\theta ^{\nu \tau }\partial _{\tau }\Gamma ^{\mu }+\Gamma
^{\mu }\ast \Gamma ^{\nu
}-\Gamma ^{\nu }\ast \Gamma ^{\mu }  \notag \\
&=&\theta ^{\mu \tau }\theta ^{\nu \lambda }\{\partial _{\tau
}Q_{\lambda }-\partial _{\lambda }Q_{\tau }+Q_{\tau }\ast
Q_{\lambda }-Q_{\lambda }\ast Q_{\tau }\}.  \notag
\end{eqnarray}%
By introducing the gravitational gauge strength (curvature)
\begin{equation}
R_{\tau \lambda }=\partial _{\tau }Q_{\lambda }-\partial
_{\lambda }Q_{\tau }+Q_{\tau }\ast Q_{\lambda }-Q_{\lambda }\ast
Q_{\tau },  \label{qcurv}
\end{equation}%
which could be treated as a noncommutative extension of de Sitter
nonlinear gauge gravitational curvature (2a), we calculate
\begin{equation}
R_{\tau \lambda ,\underline{a}}=R_{\tau \lambda
,\underline{a}}^{1}+\theta
^{\mu \nu }\{R_{\tau \mu ,\underline{a}}^{1}R_{\lambda \nu ,\underline{b}%
}^{1}{}-\frac{1}{2}q_{\mu ,\underline{a}}^{1}\left[ (D_{\nu
}R_{\tau \lambda
,\underline{b}}^{1})+\partial _{\nu }R_{\tau \lambda ,\underline{b}}^{1}%
\right] \}I^{\underline{b}},  \notag
\end{equation}%
where the gauge gravitation covariant derivative is introduced,%
\begin{equation*}
(D_{\nu }R_{\tau \lambda ,\underline{b}}^{1})=\partial _{\nu
}R_{\tau
\lambda ,\underline{b}}^{1}+q_{\nu ,\underline{c}}R_{\tau \lambda ,%
\underline{d}}^{1}f_{~\underline{b}}^{\underline{c}\underline{d}}.
\end{equation*}%
Following the gauge transformation laws for $\gamma $ and $q^{1}$
we find
\begin{equation*}
\delta _{\gamma ^{1}}R_{\tau \lambda }^{1}=i\left[ \gamma ,^{\ast
}R_{\tau \lambda }^{1}\right]
\end{equation*}%
with the restricted form of $\gamma .$

Such formulas were proved in references \cite{13sw} for usual gauge
(nongravitational) fields. Here we reconsidered them for
gravitational gauge fields.

\subsubsection{Gauge covariant gravitational dynamics}

Following the nonlinear realization of de Sitter algebra and the $\ast $%
--formalism we can formulate a dynamics of noncommutative spaces.
Derivatives can be introduced in such a way that one does not
obtain new relations for the coordinates. In this case a Leibniz
rule can be defined that
\begin{equation*}
\widehat{\partial }_{\mu }\widehat{u}^{\nu }=\delta _{\mu }^{\nu
}+d_{\mu \sigma }^{\nu \tau }\ \widehat{u}^{\sigma }\
\widehat{\partial }_{\tau }
\end{equation*}%
where the coefficients $d_{\mu \sigma }^{\nu \tau }=\delta
_{\sigma }^{\nu
}\delta _{\mu }^{\tau }$ are chosen to have not new relations when $\widehat{%
\partial }_{\mu }$ acts again to the right hand side. In consequence one
holds the $\ast $--derivative formulas
\begin{equation}
{\partial }_{\tau }\ast f=\frac{\partial }{\partial u^{\tau }}f+f\ast {%
\partial }_{\tau },~[{\partial }_{l},{{}^{\ast }}(f\ast g)]=([{\partial }%
_{l},{{}^{\ast }}f])\ast g+f\ast ([{\partial }_{l},{}^{\ast }g])
\notag
\end{equation}%
and the Stokes theorem%
\begin{equation*}
\int [\partial _{l},f]=\int d^{N}u[\partial _{l},^{\ast }f]=\int d^{N}u\frac{%
\partial }{\partial u^{l}}f=0,
\end{equation*}%
where, for the canonical structure, the integral is defined,%
\begin{equation*}
\int \widehat{f}=\int d^{N}uf\left( u^{1},...,u^{N}\right) .
\end{equation*}

An action can be introduced by using such integrals. For
instance, for a
tensor of type (\ref{tensor1}), when%
\begin{equation*}
\delta \widehat{L}=i\left[ \widehat{\gamma },\widehat{L}\right] ,
\end{equation*}
we can define a gauge invariant action%
\begin{equation*}
W=\int d^Nu\ Tr\widehat{L},~\delta W=0,
\end{equation*}
were the trace has to be taken for the group generators.

For the nonlinear de Sitter gauge gravity a proper action is
\begin{equation*}
L=\frac{1}{4}R_{\tau \lambda }R^{\tau \lambda },
\end{equation*}%
where $R_{\tau \lambda }$ is defined by (\ref{qcurv}) (in the
commutative limit we shall obtain the connection (\ref{1a})). In
this case the dynamic of noncommutative space is entirely
formulated in the framework of quantum field theory of gauge
fields. In general, we are dealing with anisotropic gauge
gravitational interactions. The method works for matter fields as
well to restrictions to the general relativity theory.

\section{Outlook and Conclusions}

In this work we have extended the A. Connes' approach to
noncommutative geometry by introducing into consideration
anholonomic frames and locally anisotropic structures. We defined
nonlinear connections for finite projective module spaces
(noncommutative generalization of vector bundles) and related
this geometry with the E. Cartan's moving frame method.

We have explicitly shown that the functional analytic approach and
noncommutative $C^{\ast }$--algebras may be transformed into
arena of modelling geometries and physical theories with generic
local anisotropy, for instance, the anholonomic Riemannian
gravity and generalized Finsler like geometries. The formalism of
spectral triples elaborated for vector bundles provided with
nonlinear connection structure allows a functional and algebraic
generation of new types of anholonomic/ anisotropic interactions.

A novel future in our work is that by applying anholonomic
transforms associated to some nonlinear connections we may
generate various type of spinor, gauge and gravity models,
subjected to some anholonomic constraints and/or with generic
anisotropic interactions, which can be included in noncommutative
field theory.

We can address a number of questions which were put or solved
partially in this paper and may have further generalizations:

One of the question is how to combine the noncommutative geometry
contained in string theory with locally anisotropic
configurations arising in the low energy limits. It is known that
the nonsymmetric background field results in effective
noncommutative coordinates. In other turn, a (super) frame set
consisting from mixed subsets of holonomic and anholonomic
vectors may result in an anholonomic geometry with associated
nonlinear connection structure. A further work is to investigate
the conditions when from a string theory one appears explicit
variants of commutative--anisotropic, commutative--isotropic,
noncommutative--isotropic and, finally,
nocommutative--an\-iso\-tro\-pic geometries.

A second question is connected with the problem of definition of
noncommutative (pseudo)\ Riemannian metric structures which is
connected with nonsymmetric and/or complex metrics. We have
elaborated variants of noncommutative gauge gravity with
noncommutative representations of the affine and de Sitter
algebras which contains in the commutative limit an Yang--Mills
theory (with nonsemisimple structure group) being equivalent to
the Einstein theory. The gauge connection in such theories is
constructed from the frame and linear connection coefficients.
Metrics, in this case, arise as some effective configurations
which avoid problems with their noncommutative definition. The
approach can be generated as to include anholonomic frames and,
in consequence, to define anisotropic variants of commutative and
noncommutative gauge gravity with the Einstein type or Finsler
generalizations.

Another interesting open question is to establish a relation
between quantum groups and geometries with anisotropic models of
gravity and field theories. Different variants of quantum
generalizations for anholonomic frames with associated nonlinear
connection structures are possible.

Finally, we give some historical remarks. An approach to Finsler
and spinor like spaces of infinite dimensions (in Banach and/or
Hilbert spaces) and to nonsymmetric locally anisotropic metrics
was proposed by some authors belonging to the Romanian school on
Finsler geometry and generalizations (see, Refs.
\cite{13ma,13hrimiuc,13anastasiei,13atanasiu}). It could not be finalized
before elaboration of the\ A. Connes' models of noncommutative
geometry and gravity and before  definition of Clifford and spinor
distinguished structures \cite{13vspinors,13vmon2},  formulation of
supersymmetric variants of Finsler spaces \cite{13vsuper} and
establishing theirs relation to string theory
\cite{13vstring,13vstr2,13vmon1}. This paper concludes a noncommutative
interference and a development of the mentioned results.

\subsection*{Acknowledgements}
The work was supported by a NATO/Portugal fellowship at CENTRA,
Instituto Superior Tecnico, Lisbon and sabbatical from the
Ministry of Education and Science of Spain.

%%%%%%%%%%%%%%%%%%%%%%%%%%%%%%%%%%%%%%%%%%%%%%%%%%%%%%%%%%%%%%%%%%%%%%%%%%%%%

\chapter[(Non) Commutative Finsler Geometry and Strings]
{(Non) Commutative Finsler Geometry from String/ M--Theory }

{\bf Abstract}
\footnote{\copyright\
 S. Vacaru, (Non) Commutative Finsler Geometry from String/M-Theory,
 hep-th/0211068}

We synthesize and extend the previous ideas about appearance of both
noncommutative and Finsler geometry in string theory with nonvanishing
B--field and/or anholonomic (super) frame structures \cite%
{14vstring,14vstr2,14vnonc,14vncf}. There are investigated the limits to the
Einstein gravity and string generalizations containing locally anisotropic
structures modelled by moving frames. The relation of anholonomic frames and
nonlinear connection geometry to M--theory and possible noncommutative
versions of locally anisotropic supergravity and D--brane physics is
discussed. We construct and analyze new classes of exact solutions with
noncommutative local anisotropy describing anholonomically deformed black
holes (black ellipsoids) in string gravity, embedded Finsler--string two
dimensional structures, solitonically moving black holes in extra dimensions
and wormholes with noncommutativity and anisotropy induced from string
theory.

\section{Introduction}

The idea that string/M--theory results in a noncommutative limit of field
theory and spacetime geometry is widely investigated by many authors both
from mathematical and physical perspectives \cite{14strncg,14sw,14connes1} (see,
for instance, the reviews \cite{14dn}). It is now generally accepted that
noncommutative geometry and quantum groups \cite{14nc,14qg,14majid} play a
fundamental role in further developments of high energy particle physics and
gravity theory.

First of all we would like to give an exposition of some basic facts about
the geometry of anholonomic frames (vielbeins) and associated nonlinear
connection (N--connection) structures which emphasize surprisingly some new
results: We will consider N--connecti\-ons in commutative geometry and we will
show that locally anisotropic spacetimes (anholonomic Riemannian, Finsler
like and their generalizations) can be obtained from the string/M--theory.
We shall discuss the related low energy limits to Einstein and gauge
gravity. Our second goal is to extend A. Connes' differential noncommutative
geometry as to include geometries with anholonomic frames and N--connections
and to prove that such 'noncommutative anisotropies' also arise very
naturally in the framework of strings and extra dimension gravity. We will
show that the anholonomic frame method is very useful in investigating of
new symmetries and nonperturbative states and for constructing new exact
solutions in string gravity with anholonomic and/or noncommutative
variables. We remember here that some variables are considered anholonomic
(equivalently, nonholonomic) if they are subjected to some constraints
(equivalently, anholonomy conditions).

Almost all of the physics paper dealing with the notion of (super) frame in
string theory do not use the well developed apparatus of E. Cartan's 'moving
frame' method \cite{14cartan} which gave an unified approach to the Riemannian
and Finsler geometry, to bundle spaces and spinors, to the geometric theory
of systems of partial equations and to Einstein (and the so--called
Einstein--Cartan--Weyl) gravity. It is considered that very ''sophisticate''
geometries like the Finsler and Cartan ones, and theirs generalizations, are
less related to real physical theories. In particular, the bulk of frame
constructions in string and gravity theories are given by coefficients
defined with respect to coordinate frames or in abstract form with respect
to some general vielbein bases. It is completely disregarded the fact that
via anholonomic frames on (pseudo) Riemannian manifolds and on (co) tangent
and (co) vector bundles we can model different geometries and interactions
with local anisotropy even in the framework of generally accepted classical
and quantum theories. For instance, there were constructed a number of exact
solutions in general relativity and its lower/higher dimension extensions
with generic local anisotropy, which under certain conditions define Finsler
like geometries \cite{14vexsol,14vankin,14vbel,14vsingl,14vsingl1,14vsolsp}.
 It was
demonstrated that anholonomic geometric constructions are inevitable in the
theory of anisotropic stochastic, kinetic and thermodynamic processes in
curved spacetimes \cite{14vankin} and proved that Finsler like (super)
geometries are contained alternatively in modern string theory \cite%
{14vstring,14vstr2,14vmon1}.

We emphasize that we have not proposed any ''exotic'' locally
anisotropic modifications of string theory and general relativity
but demonstrated that such anisotropic structures, Finsler like
or another type ones, may appear alternatively to the Riemannian
geometry, or even can be modelled in the framework of a such
geometry, in the low energy limit of the string theory, if we are
dealing with frame (vielbein) constructions. One of our main goals
is to give an accessible exposition of some important notions and
results of N--connection geometry and to show how they can be
applied to concrete problems in string theory, noncommutative
geometry and gravity. We hope to convince a reader--physicist,
who knows that 'the B--field' in string theory may result in
noncommutative geometry, that the anholonomic (super) frames
could define nonlinear connections and Finsler like commutative
and/ or noncommutative geometries in string theory and (super)
gravity and this holds true in certain limits to general
relativity.

We address the present work to physicists who would like to learn about some
new geometrical methods and to apply them to mathematical problems arising
at the forefront of modern theoretical physics. We do not assume that such
readers have very deep knowledge in differential geometry and nonlinear
connection formalism (for convenience, we give an Appendix outlining the
basic results on the geometry of commutative spaces provided with
N--connection structures \cite{14ma,14vmon1,14vmon2}) but consider that they are
familiar with some more geometric approaches to gravity \cite{14haw,14mtw} and
string theories \cite{14deligne}.

Finally, we note that the first attempts to relate Riemann--Finsler spaces
(and spaces with anisotropy of another type) to noncommutative geometry and
physics were made in Refs. \cite{14vnonc} where some models of noncommutative
gauge gravity (in the commutative limit being equivalent to the Einstein
gravity, or to different generalizations to de Sitter, affine, or Poincare
gauge gravity with, or not, nonlinear realization of the gauge groups) were
analyzed. \ Further developments of noncommutative geometries with
anholonomic/ anisotropic structures and their applications in modern
particle physics lead to a rigorous \ study of the geometry of
noncommutative anholonomic frames with associated N--connection structure %
\cite{14vncf} (that work should be considered as the non--string partner of
the present paper).

The paper has the following structure:

In Section 2 we consider stings in general manifolds and bundles provided
with anholonomic frames and associated nonlinear connection structures and
analyze the low energy string anholonomic field equations. \ The conditions
when anholonomic Einstein or Finsler like gravity models can be derived from
string theory are stated.

Section 3 outlines the geometry of locally anisotropic supergravity models
contained in superstring theory. Superstring effective actions and
anisotropic toroidal compactifications are analyzed. The corresponding
anholonomic field equations with distinguishing of anholonomic
Riemannian--Finesler (super) gravities are derived.

In Section 4 we formulate the theory of noncommutative anisotropic scalar
and gauge fields interactions and examine their anholonomic symmetries.

In Section 5 we emphasize how noncommutative anisotropic structures are
embedded in string/M--theory and discuss their connection to anholonomic
geometry.

Section 6 is devoted to locally anisotropic gravity models generated on
noncommutative D--branes.

In Section 7 we construct four classes of exact solutions with
noncommutative and locally anisotropic structures. We analyze solutions
describing locally anisotropic black holes in string theory, define a class
of Finsler--string structures containing two dimensional Finsler metrics,
consider moving solitonic string--black hole configurations and give an
examples of anholonomic noncommutative wormhole solution induced from string
theory.

Finally, in Section 8, some additional comments and questions for further
developments are presented. The Appendix outlines the necessary results from
the geometry of nonlinear connections and generalized Finsler--Riemannian
spaces.

\section[Riemann--Finsler String Gravity]
{String Theory and Commutative \newline Riemann--Fin\-sler Gravity}

The string gravitational effects are computed from corresponding low--energy
effective actions and moving equations of stings in curved spacetimes (on
string theory, see monographs \cite{14deligne}). The basic idea is to consider
propagation of a string not only of a flat 26--dimensional space with
Minkowski metric $\eta _{\mu \nu }$ but also its propagation in a background
more general manifold with metric tensor $g_{\mu \nu }$ from where one
derived string--theoretic corrections to general relativity when the vacuum
Einstein\ equations $R_{\mu \nu }=0$ correspond to vanishing of the
one--loop beta function in corresponding sigma model. More rigorous theories
were formulated by adding an antisymmetric tensor field $B_{\mu \nu },$ the
dilaton field $\Phi $ and possible other background fields, by introducing
supersymmetry, higher loop corrections and another generalizations. It
should be noted here that propagation of (super) strings may be considered
on arbitrary (super) manifolds. For instance, in Refs. \cite%
{14vstring,14vstr2,14vmon1}, the corresponding background (super) spaces were
treated as (super) bundles provided with nonlinear connection
(N--connection) structure and, in result, there were constructed some types
of generalized (super) Finsler corrections to the usual Einstein and to
locally anisotropic (Finsler type, or theirs generalizations) gravity
theories.

The aim of this section is to demonstrate that anisotropic corrections and
extensions may be computed both in Einstein and string gravity [derived for
string propagation in usual (pseudo) Riemannian backgrounds] if the approach
is developed following a more rigorous geometrical formalism with
off--diagonal metrics and anholonomic frames. We note that (super) frames
[vielbeins] were used in general form, for example, in order to introduce
spinors and supersymmetry in sting theory but the anholonomic transforms
with mixed holonomic--anholonomic variables, resulting in diagonalization of
off--diagonal (super) metrics and effective anisotropic structures, were not
investigated in the previous literature on string/M--theory.

\subsection{Strings in general manifolds and bundles}

\subsubsection{Generalized nonlinear sigma models (some basics)}

The first quantized string theory was constructed in flat
Minkowski spacetime of dimension $k\geq 4.$ \ Then the analysis
was extended to more general manifolds with (pseudo) Riemannian
metric $\underline{g}_{\mu \nu },$ antisymmetric $B_{\mu \nu }$
and dilaton field $\Phi $ and possible other
background fields, including  tachyonic matter associated to a field $%
U $ in a tachyon state. The starting point in investigating the string
dynamics in the background of these fields is the generalized nonlinear
sigma model action for the maps $u:\Sigma \rightarrow M$ \ from a two
dimensional surface $\Sigma $ to a spacetime manifold $M$ of dimension $k,$%
\begin{equation}
S=S_{\underline{g},B}+S_{\Phi }+S_{U},  \label{3act1}
\end{equation}%
with%
\begin{eqnarray*}
S_{\underline{g},B}[u,g] &=&\frac{1}{8\pi l^{2}}\int\limits_{\Sigma }d\mu
_{g}\partial _{A}u^{\mu }\partial _{B}u^{\nu }\left[ g_{[2]}^{AB}\underline{g%
}_{\mu \nu }(u)+\varepsilon ^{AB}B_{\mu \nu }(u)\right] , \\
S_{\Phi }[u,g] &=&\frac{1}{2\pi }\int\limits_{\Sigma }d\mu _{g}R_{g}\Phi
(u),~S_{U}[u,g]=\frac{1}{4\pi }\int\limits_{\Sigma }d\mu _{g}U(u),
\end{eqnarray*}%
where $B_{\mu \nu }$ is the pullback of a two--form $B=B_{\mu \nu }du^{\mu
}\wedge du^{\nu }$ under the map $u,\,$written out in local coordinates $%
u^{\mu };~g_{[2]AB}$ is the metric on the two dimensional surface $\Sigma $
(indices $A,B=0,1);~$ $\varepsilon ^{AB}=\overline{\varepsilon }^{AB}/\sqrt{%
\det |g_{AB}|},\overline{\varepsilon }^{01}=-\overline{\varepsilon }^{10}=1;$
the integration measure $d\mu _{g}$ is defined by the coefficients of the
metric $g_{AB},$ $R_{g}$ is the Gauss curvature of $\Sigma .$ The constants
in the action are related as%
\begin{equation*}
k=\frac{1}{4\pi \alpha ^{\prime }}=\frac{1}{8\pi \ell ^{2}},\alpha ^{\prime
}=2\ell ^{2}
\end{equation*}%
where $\alpha ^{\prime }$ is the Regge slope parameter $\alpha ^{\prime }$
and $\ell \sim 10^{-33}cm$ is the Planck length scale. The metric
coefficients $\underline{g}_{\mu \nu }(u)$ are defined by the quadratic
metric element given with respect to the coordinate co--basis $d^{\mu
}=du^{\mu }$ (being dual to the local coordinate basis $\partial _{\mu
}=\partial /\partial _{\mu }),$%
\begin{equation}
ds^{2}=\underline{g}_{\mu \nu }(u)du^{\mu }du^{\nu }.  \label{4metric}
\end{equation}

The parameter $\ell $ is a very small length--scale, compared to
experimental scales $L_{\exp }\sim 10^{-17}$ accessible at present. This
defines the so--called low energy, or $\alpha ^{\prime }$--expansion. A
perturbation theory may be carried out as usual by letting $u=u_{0}+\ell
u_{[1]}$ for some reference configuration $u_{0}$ and considering expansions
of the fields $\underline{g},B$ and $\Phi ,$ for instance,
\begin{equation}
\underline{g}_{\mu \nu }(u)=\underline{g}_{\mu \nu }(u_{0})+\ell \partial
_{\alpha }\underline{g}_{\mu \nu }(u_{0})u_{[1]}^{\alpha }+\frac{1}{2}\ell
^{2}\partial _{\alpha }\partial _{\beta }\underline{g}_{\mu \nu
}(u_{0})u_{[1]}^{\alpha }u_{[1]}^{\beta }+...  \label{ser1}
\end{equation}%
This reveals that the quantum field theory defined by the action (\ref{3act1}%
) is with an infinite number of couplings; the independent couplings of this
theory correspond to the successive derivatives of the fields $\underline{g}%
,B$ and $\Phi $ at the expansion point $u_{0}.$ Following an analysis of the
general structure of the Weyl dependence of Green functions in the quantum
field theory, standard regularizations schemes (see, for instance, Refs. %
\cite{14deligne}) and conditions of vanishing of Weyl anomalies,
computing the $\beta $--functions, one derive the low energy
string effective actions and field equations.

\subsubsection{Anholonomic frame transforms of background metrics}

Extending the general relativity principle to the string theory, we should
consider that the string dynamics in the background of fields $\underline{g}%
,B$ and $\Phi $ and possible another ones, defined in the low energy limit
by certain effective actions and moving equations, does not depend on
changing of systems of coordinates, $u^{\alpha ^{\prime }}\rightarrow
u^{\alpha ^{\prime }}\left( u^{\alpha }\right) ,$ for a fixed local basis
(equivalently, system, frame, or vielbein) of reference, $e_{\alpha }\left(
u\right) ,$ on spacetime $M$ \ (for which, locally, $u=u^{\alpha }e_{\alpha
}=$ $u^{\alpha ^{\prime }}e_{\alpha ^{\prime }},$ $e_{\alpha ^{\prime
}}=\partial u^{\alpha }/\partial u^{\alpha ^{\prime }}e_{\alpha },$ usually
one considers local coordinate bases when $e_{\alpha }=\partial /\partial
u^{\alpha })$ as well the string dynamics should not depend on changing of
frames like $e_{\underline{\alpha }}\rightarrow e_{\underline{\alpha }%
}^{~\alpha }\left( u\right) e_{\alpha },$ parametrized by non--degenerated
matrices $e_{\underline{\alpha }}^{~\alpha }\left( u\right) .$

Let us remember some details connected with the geometry of moving frames in
(pseudo) Riemannian spaces \cite{14cartan} and discuss its applications in
string theory, where the orthonormal frames were introduced with the aim to
eliminate non--trivial dependencies on the metric $\underline{g}_{\mu \nu }$
and on the background field $u_{0}^{\mu }$ which appears in elaboration of
the covariant background expansion method \ for the nonlinear sigma models %
\cite{14deligne,14friedan}. Such orthonormal frames, in the framework of a $%
SO\left( 1,k-1\right) $ like gauge theory are stated by the conditions%
\begin{eqnarray}
\underline{g}_{\mu \nu }\left( u\right) &=&e_{\mu }^{~\underline{\mu }%
}\left( u\right) e_{\nu }^{~\underline{\nu }}\left( u\right) \eta _{%
\underline{\mu }\underline{\nu }},  \label{tetrad} \\
e_{\mu }^{~\underline{\mu }}e_{\ \underline{\mu }}^{\nu } &=&\delta _{\mu
}^{\nu },\quad e_{\mu }^{~\underline{\mu }}e_{\ \underline{\nu }}^{\mu
}=\delta _{\underline{\nu }}^{\underline{\mu }},  \notag
\end{eqnarray}%
where $\eta _{\underline{\mu }\underline{\nu }}=diag\left(
-1,+1,...,+1\right) $ is the flat Minkowski metric \thinspace and $\delta
_{\mu }^{\nu },\delta _{\underline{\nu }}^{\underline{\mu }}$ are
Kronecker's delta symbols. One considers the covariant derivative $D_{\mu }$
with respect to an affine connection $\Gamma $ and a corresponding spin
connection $\omega _{\mu ~\underline{\beta }}^{~\underline{\alpha }}$ for
which the frame $e_{\mu }^{~\underline{\mu }}$ is covariantly constant,%
\begin{equation*}
D_{\mu }e_{\nu }^{~\underline{\alpha }}\equiv \partial _{\mu }e_{\nu }^{~%
\underline{\alpha }}-\Gamma _{~\mu \nu }^{\alpha }e_{\alpha }^{~\underline{%
\alpha }}+\omega _{\mu ~\underline{\beta }}^{~\underline{\alpha }}e_{\alpha
}^{~\underline{\beta }}=0.
\end{equation*}%
One also uses the covariant derivative
\begin{equation}
\mathcal{D}_{\mu }e_{\nu }^{~\underline{\alpha }}=D_{\mu }e_{\nu }^{~%
\underline{\alpha }}+\frac{1}{2}H_{\mu \nu }^{\quad \rho }e_{\rho }^{~%
\underline{\alpha }}  \label{cd}
\end{equation}%
including the torsion tensor $H_{\mu \nu \rho }$ which is the field strength
of the field $B_{\nu \rho },$ given by $H=dB,$ or, in component notation,
\begin{equation}
H_{\mu \nu \rho }\equiv \partial _{\mu }B_{\nu \rho }+\partial _{\nu
}B_{\rho \mu }+\partial _{\nu }B_{\rho \mu }.  \label{str1}
\end{equation}%
All tensors may be written with respect to an orthonormal frame basis, for
instance,
\begin{equation*}
H_{\mu \nu \rho }=e_{\mu }^{~\underline{\mu }}e_{\nu }^{~\underline{\nu }%
}e_{\rho }^{~\underline{\rho }}H_{\underline{\mu }\underline{\nu }\underline{%
\rho }}
\end{equation*}%
and
\begin{equation*}
\mathcal{R}_{\mu \nu \rho \sigma }=e_{\mu }^{~\underline{\mu }}e_{\nu }^{~%
\underline{\nu }}e_{\rho }^{~\underline{\rho }}e_{\sigma }^{~\underline{%
\sigma }}R_{\underline{\mu }\underline{\nu }\underline{\rho }\underline{%
\sigma }},
\end{equation*}%
where the curvature $\mathcal{R}_{\mu \nu \rho \sigma }$ of the connection $%
\mathcal{D}_{\mu },$ defined as
\begin{equation*}
(\mathcal{D}_{\mu }\mathcal{D}_{\nu }-\mathcal{D}_{\nu }\mathcal{D}_{\mu
})\xi ^{\rho }\doteqdot \lbrack \mathcal{D}_{\mu }\mathcal{D}_{\nu }]\xi
^{\rho }=H_{\ \mu \nu }^{\sigma }\mathcal{D}_{\sigma }\xi ^{\rho }+\mathcal{R%
}_{\ \sigma \mu \nu }^{\rho }\xi ^{\sigma },
\end{equation*}%
can be expressed in terms of the Riemannian tensor $R_{\mu \nu \rho \sigma }$
and the torsion tensor $H_{\ \mu \nu }^{\sigma },$%
\begin{equation*}
\mathcal{R}_{\mu \nu \rho \sigma }=R_{\mu \nu \rho \sigma }+\frac{1}{2}%
D_{\rho }H_{\sigma \mu \nu }-\frac{1}{2}D_{\sigma }H_{\rho \mu \nu }+\frac{1%
}{4}H_{\rho \mu \alpha }H_{\sigma \nu }^{\quad \alpha }-\frac{1}{4}H_{\sigma
\mu \alpha }H_{\rho \nu }^{\quad \alpha }.
\end{equation*}

Let us consider a generic off--diagonal metric, a non-degenerated matrix of
dimension $k\times k$ with the coefficients $\underline{g}_{\mu \nu }(u)$ \
defined with respect to a local coordinate frame like in (\ref{4metric}).
This metric can transformed into a block $\left( n\times n\right) \oplus
\left( m\times m\right) $ form, for $k=n+m,\,$\
\begin{equation*}
\underline{g}_{\mu \nu }(u)\rightarrow \{ g_{ij}(u),h_{ab}\left(
u\right) \}
\end{equation*}%
if we perform a frame map with the vielbeins%
\begin{eqnarray}
e_{\mu }^{~\underline{\mu }}(u) &=&\left(
\begin{array}{cc}
e_{i}^{~\underline{i}}(x^{j},y^{a}) & N_{i}^{a}(x^{j},y^{a})e_{a}^{~%
\underline{a}}(x^{j},y^{a}) \\
0 & e_{a}^{~\underline{a}}(x^{j},y^{a})%
\end{array}%
\right)  \label{vielbtr} \\
e_{\ \underline{\nu }}^{\mu }(u) &=&\left(
\begin{array}{cc}
e_{\ \underline{i}}^{i}(x^{j},y^{a}) & -N_{k}^{a}(x^{j},y^{a})e_{\
\underline{i}}^{k}(x^{j},y^{a}) \\
0 & e_{\ \underline{a}}^{a}(x^{j},y^{a})%
\end{array}%
\right)  \notag
\end{eqnarray}%
which conventionally splits the spacetime into two subspaces: the first
subspace is parametrized by coordinates $x^{i}$ $\ $\ provided with indices
of type $i,j,k,...$ running values from $1$ to $n$ and the second subspace
is parametrized by coordinates $y^{a}$ provided with indices of type $%
a,b,c,...$ running values from $1$ to $m.$ This splitting is induced by the
coefficients $N_{i}^{a}(x^{j},y^{a}).$ For simplicity, we shall write the
local coordinates as $u^{\alpha }=\left( x^{i},y^{a}\right) ,$ or $u=\left(
x,y\right) .$

The coordinate bases $\partial _{\alpha }=\left( \partial _{i},\partial
_{a}\right) $ and theirs duals $d^{\alpha }=du^{\alpha }=\left(
d^{i}=dx^{i},d^{a}=dy^{a}\right) $ are transformed under maps (\ref{vielbtr}%
) as $\ $%
\begin{equation*}
\partial _{\alpha }\rightarrow e_{\underline{\alpha }}=e_{\ \underline{%
\alpha }}^{\alpha }(u)\partial _{\alpha },d^{\alpha }\rightarrow e^{%
\underline{\alpha }}=e_{\alpha }^{~\underline{\alpha }}(u)d^{\alpha },
\end{equation*}%
or, in 'N--distinguished' form,%
\begin{eqnarray}
e_{\underline{i}} &=&e_{\ \underline{i}}^{i}\partial _{i}-N_{k}^{a}e_{\
\underline{i}}^{k}\partial _{a},e_{\underline{a}}=e_{\ \underline{a}%
}^{a}\partial _{a},  \label{dder1a} \\
e^{\underline{i}} &=&e_{i}^{~\underline{i}}d^{i},~e^{\underline{a}%
}=N_{i}^{a}e_{a}^{~\underline{a}}d^{i}+e_{a}^{~\underline{a}}d^{a}.
\label{ddif1a}
\end{eqnarray}%
The quadratic line element (\ref{4metric}) may be written equivalently in the
form
\begin{equation}
ds^{2}=g_{\underline{i}\underline{j}}(x,y)e^{\underline{i}}e^{\underline{j}%
}+h_{\underline{a}\underline{b}}(x,y)e^{\underline{a}}e^{\underline{b}}
\label{dmetric1a}
\end{equation}%
with the metric $\underline{g}_{\mu \nu }(u)$ parametrized in the form%
\begin{equation}
\underline{g}_{\alpha \beta }=\left[
\begin{array}{cc}
g_{ij}+N_{i}^{a}N_{j}^{b}h_{ab} & h_{ab}N_{i}^{a} \\
h_{ab}N_{j}^{b} & h_{ab}%
\end{array}%
\right] .  \label{6ansatz}
\end{equation}

If we choose $e_{i}^{~\underline{i}}(x^{j},y^{a})=\delta _{i}^{~\underline{i}%
}$ and $e_{a}^{~\underline{a}}(x^{j},y^{a})=\delta _{a}^{~\underline{a}},$
we may not distinguish the 'underlined' and 'non--underlined' indices. The
operators (\ref{dder1a}) and (\ref{ddif1a}) transform respectively into the
operators of 'N--elongated' partial derivatives and differentials
\begin{eqnarray}
e_{i} &=&\delta _{i}=\partial _{i}-N_{i}^{a}\partial _{a},e_{a}=\partial
_{a},  \label{viel1} \\
e^{i} &=&d^{i},~e^{a}=\delta ^{a}=d^{a}+N_{i}^{a}d^{i}  \notag
\end{eqnarray}%
(which means that the anholonomic frames (\ref{dder1a}) and (\ref{ddif1a})
generated by vielbein transforms (\ref{vielbtr}) \ are, in general,
anholonomic; see the respective formulas (\ref{6dder}), (\ref{7ddif}) and (\ref%
{4anhol}) in the Appendix) and the quadratic line element (\ref{dmetric1a})
trasforms in a d--metric element (see (\ref{7dmetric}) in the Appendix).

The physical treatment of the vielbein transforms (\ref{vielbtr}) and
associated $N$--coefficients depends on the types of constraints
(equivalently, anholonomies) we impose on the string dynamics and/or on the
considered curved background. There were considered different possibilities:

\begin{itemize}
\item Ansatz of type (\ref{6ansatz}) were used in Kaluza--Klein gravity \cite%
{14salam}, as well in order to describe toroidal Kaluza--Klein reductions in
string theory (see, for instance, \cite{14kir})). \ The coefficients $%
N_{i}^{a},$ usually written as $A_{i}^{a},$ are considered as the potentials
of some, in general, non--Abelian gauge fields, which in such theories are
generated by a corresponding compactification. In this case, the coordinates
$x^{i}$ can be used for the four dimensional spacetime and the coordinates $%
y^{a}$ are for extra dimensions.

\item Parametrizations of type (\ref{6ansatz}) were considered in order to
elaborate an unified approach on vector/tangent bundles to
Finsler geometry
and its generalizations \cite%
{14ma,14miron,14bejancu,14vspinors,14vmon1,14vsuper,14vstring,14vstr2}.
 The coefficients $%
N_{i}^{a}$ were supposed to define a nonlinear connection \
(N--connection) structure in corresponding (super) bundles and
the metric coefficients $g_{ij}(u)$ and $g_{ab}\left( u\right) $
were taken for a
corresponding Finsler metric, or its generalizations (see formulas (\ref%
{fmetric}), (\ref{2ncc}), (\ref{2dmetricf}), (\ref{1mfl}) and related
discussions in Appendix). The coordinates $x^{i}$ were defined on base
manifolds and the coordinates $y^{a}$ were used for fibers of bundles.

\item In a series of papers \cite%
{14vexsol,14vankin,14vmethod,14vbel,14vsingl,14vsingl1,14vsolsp} the concept of
N--connection was introduced for (pseudo) Riemannian spaces
provided with off--diagonal metrics and/or anholonomic frames. In
a such approach the coefficients $N_{i}^{a}$ are associated to an
anholonomic frame structure describing a gravitational and matter
fields dynamics with mixed holonomic and anholonomic variables.
The coordinates $x^{i}$ are defined with respect to the subset of
holonomic frame vectors, but $y^{a}$ are given with respect to
the subset of anholonomic, N--ellongated, frame vectors. It was
proven that by using vielbein transforms of type (\ref{vielbtr})
the off--diagonal
metrics could be diagonalized and, for a very large class of ansatz of type (%
\ref{6ansatz}), with the coefficients depending on 2,3 or 4 coordinate
variables, it was shown that the corresponding vacuum and non--vacuum
Einstein equations may be integrated in general form. This allowed an
explicit construction of new classes of exact solutions parametrized by
off--diagonal metrics with some anholonomically deformed symmetries. Two new
and very surprising conclusions were those that the Finsler like (and
another type) anisotropies may be modelled even in the framework of the
general relativity theory and its higher/lower dimension modifications, as
some exact solutions of the Einstein equations, and that the anholonomic
frame method is very efficient for constructing such solutions.
\end{itemize}

There is an important property of the off--diagonal metrics $\underline{g}%
_{\mu \nu }$ (\ref{6ansatz}) which does not depend on the type of
space (a pseudo--Riemannian manifold, or a vector/tangent bundle)
this metric is given. With respect to the coordinate frames it is
defined a unique torsionless and metric compatible linear
connection derived as the usual Christoffel symbols (or the Levi
Civita connection). If anholonomic frames are introduced into
consideration, we can define an infinite number of metric
connections constructed from the coefficients of off--diagonal
metrics and induced by the anholonomy coefficients (see formulas
(\ref{12lcsym}) and (\ref{2lccon}) and the related discussion from
Appendix); this property is also mentioned in the monograph
\cite{14mtw} (pages 216, 223, 261) for anholonomic frames but
without any particularities related to associated N--connection
structures. In this case there is an infinite number of metric
compatible linear connections, constructed from metric and
vielbein coefficients, all of them having non--trivial torsions
and transforming into the usual Christoffel symbols for $N_{i}^{a}%
\rightarrow 0$ and $m\rightarrow 0.$ For off--diagonal metrics
considered, and even diagonalized, with respect to anholonomic
frames and associated N--connections, we can not select a linear
connection being both torsionless and metric. The problem of
establishing of a physical linear connection structure
constructed from metric/frame coefficients is to be solved
together with that of fixing of a system of reference on a curved
spacetime which is not a pure dynamical task but depends on the
type of prescribed constraints, symmetries and boundary
conditions are imposed on interacting fields and/or string
dynamics.

In our further consideration we shall suppose that both a metric $\underline{%
g}_{\mu \nu }$ (equivalently, a set $\{ g_{ij},g_{ab},N_{i}^{a} \}
$) and metric linear connection $\Gamma _{~\beta \gamma }^{\alpha
},$ i.e. satisfying the conditions $D_{\alpha }g_{\alpha \beta
}=0,$ exist in the background spacetime. Such spaces will be
called locally anisotropic (equivalently, anolonomic) because the
anholonomic frames structure imposes locally a kind of anisotropy
with respective constraints on string and effective string
dynamics. For such configurations the torsion, induced as an
anholonomic frame effect, vanishes only with respect coordinate
frames. Here we note that in the string theory there are also
another type of torsion contributions to linear connections like
$H_{\ \mu \nu }^{\sigma },$ see formula (\ref{cd}).

\subsubsection{Anholonomic background field quantization method}

We revise the perturbation theory around general field
configurations for background spaces provided with anholonomic
frame structures (\ref{dder1a}) and (\ref{ddif1a}), $\delta
_{\alpha }=(\delta _{i}=\partial _{i}-N_{i}^{a}\partial
_{a},\partial _{a})$ and $\delta ^{\alpha }=(d^{i},\delta
^{a}=d^{a}+N_{i}^{a}d^{i}),$ with associated N--connections,
$N_{i}^{a},$ and $\{ g_{ij},h_{ab}\} $ (\ref{dmetric1a}) adapted
to
such structures (distinguished metrics, or d--metrics, see formula (\ref%
{7dmetric})). The linear connection in such locally anisotropic backgrounds
is considered to be compatible both to the metric and N--connection
structure (for simplicity, being a d--connection or an anholonomic variant
of Levi Civita connection, both with nonvanishing torsion, see formulas (\ref%
{6dcon}), (\ref{12lcsym}), (\ref{2lccon}), and (\ref{3dtors}), and related
discussions in the Appendix). The general rule for the tensorial calculus on
a space provided with N--connection structure is to split indices $\alpha
,\beta ,...$ into ''horozontal'', $i,j,...,$ and ''vertical'', $a,b,...,$
subsets and to apply for every type of indices the corresponding operators
of N--adapted partial and covariant derivations.

The anisotropic sigma model is to be formulated by anholonomic transforms of
the metric, $\underline{g}_{\mu \nu }\rightarrow \{g_{ij},h_{ab}\},$ partial
derivatives and differentials, $\partial _{\alpha }\rightarrow \delta
_{\alpha }$ and $d^{\alpha }\rightarrow \delta ^{a},$ volume elements, $d\mu
_{g}\rightarrow \delta \mu _{g}$ in the action (\ref{3act1})

\begin{equation}
S=S_{g_{N},B}+S_{\Phi }+S_{U},  \label{act1a}
\end{equation}%
with%
\begin{eqnarray*}
S_{g_{N},B}[u,g] &=&\frac{1}{8\pi l^{2}}\int\limits_{\Sigma }\delta \mu
_{g}\{g^{AB}\left[ \partial _{A}x^{i}\partial _{B}x^{j}g_{ij}(x,y)+\partial
_{A}x^{a}\partial _{B}x^{b}h_{ab}(x,y)\right] \\
&&+\varepsilon ^{AB}\partial _{A}u^{\mu }\partial _{B}u^{\nu }B_{\mu \nu
}(u)\}, \\
S_{\Phi }[u,g] &=&\frac{1}{2\pi }\int\limits_{\Sigma }\delta \mu
_{g}R_{g}\Phi (u),~S_{U}[u,g]=\frac{1}{4\pi }\int\limits_{\Sigma }\delta \mu
_{g}U(u),
\end{eqnarray*}%
where the coefficients $B_{\mu \nu }$ are computed for a two--form $B=B_{\mu
\nu }\delta u^{\mu }\wedge \delta u^{\nu }.$

The perturbation theory has to be developed by changing the usual partial
derivatives into N--elongated ones, for instance, the decomposition (\ref%
{ser1}) is to be written
\begin{equation*}
\underline{g}_{\mu \nu }(u)=\underline{g}_{\mu \nu }(u_{0})+\ell \delta
_{\alpha }\underline{g}_{\mu \nu }(u_{0})u_{[1]}^{\alpha }+\ell ^{2}\delta
_{\alpha }\delta _{\beta }\underline{g}_{\mu \nu }(u_{0})u_{[1]}^{\alpha
}u_{[1]}^{\beta }+\ell ^{2}\delta _{\beta }\delta _{\alpha }\underline{g}%
_{\mu \nu }(u_{0})u_{[1]}^{\alpha }u_{[1]}^{\beta }+...,
\end{equation*}%
where we should take into account the fact that the operators $\delta
_{\alpha }$ do not commute but satisfy certain anholonomy relations (see (%
\ref{4anhol})\ \ in Appendix).

The action (\ref{act1a}) is invariant under the group of diffeomorphisms on $%
\Sigma $ and $M$ (on spacetimes provided with N--connections the
diffeomorphisms may be adapted to such structures) and posses a $U(1)_{B}$
gauge invariance, acting by $B\rightarrow B+\delta \gamma $ for some $\gamma
\in \Omega ^{(1)}\left( M\right) ,$ where $\Omega ^{(1)}$ denotes the space
of 1--forms on $M.$ Wayl's conformal transformations of $\Sigma $ leave $%
S_{g_{N},B}$ invariant but result in anomalies under quantization. $S_{\Phi
} $ and $S_{U}$ fail to be conformal invariant even classically. We discuss
the renormalization of quantum filed theory defined by the action (\ref%
{act1a}) for general fields $g_{ij},h_{ab},N_{\mu }^{a},B_{\mu \nu }$ and $%
\Phi .$ We shall not discus in this work the effects of the tachyon field.

The string corrections to gravity (in both locally isotropic and locally
anisotropic cases) may be computed following some regularizatons schemes
preserving the classical symmetries and determining the general structure of
the Weyl dependence of Green functions specified by the action (\ref{act1a})
in terms of fixed background fields $g_{ij},h_{ab},N_{\mu }^{a},B_{\mu \nu }$
and $\Phi .$ One can consider un--normalized correlation functions of
operators $\phi _{1},...,\phi _{p},$ instead of points $\xi _{1},...,\xi
_{p}\in \Sigma $ \cite{14deligne}.

By definition of the stress tensor $T_{AB},$ under conformal transforms on
the two dimensional hypersurface, $g_{[2]}\rightarrow \exp [2\delta \sigma
]g_{[2]}$ with support away from $\xi _{1},...,\xi _{p},$ we have%
\begin{equation*}
\triangle _{\sigma }<\phi _{1}...\phi _{p}>_{g_{[2]}}=\frac{1}{2\pi }%
\int_{\Sigma }\delta \mu _{g}\triangle \sigma <T_{A}^{~A}\phi _{1}...\phi
_{p}>_{g_{[2]}},
\end{equation*}%
when assuming throughout that correlation functions are covariant under the
diffeomorphisms on $\Sigma ,$ $\bigtriangledown ^{A}T_{AB}=0.$ The value $%
T_{A}^{~A}$ receives contributions from the explicit conformal
non--invariance of $S_{\Phi },$ from conformal (Weyl) anomalies which are
local functions of $u,$ i.e. dependent on $u$ and on finite order
derivatives on $u,$ and polynomial in the derivatives of $u.$ For spaces
provided with N--connection structures we should consider N--elongated
partial derivatives, choose a N--adapted linear connection structure with
some coefficients $\Gamma _{~\mu \nu }^{\alpha }$ (for instance the Levi
Civita connection (\ref{12lcsym}), or d--connection (\ref{6dcon})). The basic
properties of $T_{A}^{~A}$ are the same as for trivial values of $N_{i}^{a}$ %
\cite{14deligne}, which allows us to write directly that
\begin{eqnarray*}
T_{A}^{~A} &=&g^{AB}[\partial _{A}x^{i}\partial _{B}x^{j}\beta
_{ij}^{g,N}(x,y)+\partial _{A}x^{i}\partial _{B}y^{b}\beta
_{ib}^{g,N}(x,y)+\partial _{A}y^{a}\partial _{B}x^{j}\beta _{aj}^{g,N}(x,y)
\\
&&+\partial _{A}y^{a}\partial _{B}y^{b}\beta _{ab}^{g,N}(x,y)+\varepsilon
^{AB}\partial _{A}u^{\alpha }\partial _{B}u^{\beta }\beta _{\alpha \beta
}^{B}(x,y)+\beta ^{\Phi }(x,y)R_{g},
\end{eqnarray*}%
where the functions $\beta _{\alpha \beta }^{g,N}=\{\beta _{ij}^{g,N},\beta
_{ab}^{g,N}\},\beta _{\alpha \beta }^{B}$ and $\beta ^{\Phi }(x,y)$ are
called beta functions. On general grounds, the expansions of $\beta $%
--functions are of type
\begin{equation*}
\beta (x,y)=\sum_{r=0}^{\infty }\ell ^{2r}\beta ^{\lbrack 2r]}(x,y).
\end{equation*}

One considers expanding up to and including terms with two derivatives on
the fields including expansions up to order $r=0$ of $\beta _{\alpha
\beta }^{g,N}$ and $\beta _{\alpha \beta }^{B}$ and orders $s=0,2$ for $%
\beta ^{\Phi }.$ In this approximation, after cumbersome but simple
calculations (similar to those given in \cite{14deligne}, in our case on
locally anisotropic backgrounds)%
\begin{eqnarray*}
\beta _{ij}^{g,N} &=&a_{1[1]}R_{ij}+a_{2[1]}g_{ij}+a_{3[1]}g_{ij}\widehat{R}%
+a_{4[1]}H_{i\rho \sigma }^{[N]}H_{j}^{[N]\rho \sigma
}+a_{5[1]}g_{ij}H_{\rho \sigma \tau }^{[N]}H^{[N]\rho \sigma \tau } \\
&&+a_{6[1]}D_{i}D_{j}\Phi +a_{7[1]}g_{ij}D^{2}\Phi +a_{8[1]}g_{ij}D^{\rho
}\Phi D_{\rho }\Phi ,
\end{eqnarray*}%
\begin{eqnarray}
\beta _{ib}^{g,N} &=&a_{1[2]}R_{ib}+a_{4[2]}H_{i\rho \sigma
}^{[N]}H_{b}^{[N]\rho \sigma }+a_{6[2]}D_{i}D_{b}\Phi ,  \label{beta1} \\
\beta _{aj}^{g,N} &=&a_{1[3]}R_{aj}+a_{4[3]}H_{a\rho \sigma
}^{[N]}H_{j}^{[N]\rho \sigma }+a_{6[3]}D_{a}D_{j}\Phi ,  \notag
\end{eqnarray}%
\begin{eqnarray*}
\beta _{ab}^{g,N}
&=&a_{1[4]}S_{ab}+a_{2[4]}h_{ab}+a_{3[4]}h_{ab}S+a_{4[4]}H_{a\rho \sigma
}^{[N]}H_{b}^{[N]\rho \sigma }+a_{5[4]}h_{ab}H_{\rho \sigma \tau
}^{[N]}H^{[N]\rho \sigma \tau } \\
&&+a_{6[4]}D_{a}D_{b}\Phi +a_{7[4]}h_{ab}D^{2}\Phi +a_{8[4]}h_{ab}D^{\rho
}\Phi D_{\rho }\Phi ,
\end{eqnarray*}%
\begin{eqnarray*}
\beta _{\alpha \beta }^{B} &=&b_{1}D^{\lambda }H_{\lambda \mu \nu
}^{[N]}+b_{2}(D^{\lambda }\Phi )H_{\lambda \mu \nu }^{[N]}, \\
&& \\
\beta ^{\Phi } &=&c_{0}+\ell ^{2}\left[ c_{1[1]}\widehat{R}%
+c_{1[2]}S+c_{2}D^{2}\Phi +c_{3}\left( D^{\lambda }\Phi \right) D_{\lambda
}\Phi +c_{4}H_{\rho \sigma \tau }^{[N]}H^{[N]\rho \sigma \tau }\right] ,
\end{eqnarray*}%
where $R_{\alpha \beta }=\{R_{ij},R_{ib},R_{aj},S_{ab}\}$ and $%
\overleftarrow{R}=\{\widehat{R},S\}$ are given respectively by the formulas (%
\ref{6dricci}) and (\ref{4dscalar}) and the $B$--strength $H_{\lambda \mu \nu
}^{[N]}$ is computed not by using partial derivatives, like in (\ref{str1}),
but with N--adapted partial derivatives,
\begin{equation}
H_{\mu \nu \rho }^{[N]}\equiv \delta _{\mu }B_{\nu \rho }+\delta _{\nu
}B_{\rho \mu }+\delta _{\nu }B_{\rho \mu }.  \label{htors}
\end{equation}%
The formulas for $\beta $--functions (\ref{beta1}) are adapted to the
N--connection structure being expressed via invariant decompositions for the
Ricci d--tensor and curvature scalar; every such invariant object was
provided with proper constants. In order to have physical compatibility with
the case $N\rightarrow 0$ we should take%
\begin{eqnarray*}
a_{z[1]} &=&a_{z[2]}=a_{z[3]}=a_{z[4]}=a_{z},~z=1,2,...,8; \\
c_{1[1]} &=&c_{1[2]}=c_{1,}
\end{eqnarray*}%
where $a_{z}$ and $c_{1}$ are the same as in the usual string theory,
computed from the 1-- and 2--loop $\ell $--dependence of graphs ($a_{2}=0,$ $%
a_{6}=1,$ $a_{7}=a_{8}=0$ and $b_{2}=1/2,$ $c_{3}=2)$ and by using the
background field method (in order to define the values $%
a_{1},a_{3},a_{4},a_{5},b_{1}$ and $c_{1},c_{2},c_{4}).$

\subsection{Low energy string anholonomic field equations}

\label{dcovrule}The effective action, as the generating functional for
1--particle irreducible Feynman diagrams in terms of a functional integral,
can be obtained following the background quantization method adapted, in our
constructions, to sigma models on spacetimes with N--connection structure.
On such spaces, we can also make use of the Riemannian coordinate expansion,
but taking into account that the coordinates are defined with respect to
N--adapted bases and that the covariant derivative $D$ is of type (\ref{6dcon}%
), (\ref{12lcsym}) or (\ref{2lccon}), i. e. is d--covariant, defined by a
d--connection.

For two infinitesimally closed points $u_{0}^{\mu }=u^{\mu }(\tau _{0})$ and
$u^{\mu }(\tau ),$ with $\tau $ being a parameter on a curve connected the
points, we denote $\zeta ^{\alpha }=du^{\alpha }/d\tau _{\mid 0}$ and write $%
u^{\mu }=e^{\ell \zeta }u_{0}^{\mu }.$ We can consider diffeomorphism
invariant d--covariant expansions of d--tensors in powers of $\ell ,$ for
instance,
\begin{eqnarray*}
\Phi (u) &=&\Phi (u_{0})+\ell D_{\alpha }[\Phi (u)\zeta ^{\alpha }]_{\mid
u=u_{0}}+\frac{\ell ^{2}}{2}D_{\alpha }D_{\beta }[\Phi (u)\zeta ^{\alpha
}\zeta ^{\beta }]_{\mid u=u_{0}}+o\left( \ell ^{3}\right) , \\
A_{\alpha \beta }\left( u\right) &=&A_{\alpha \beta }\left( u_{0}\right)
+\ell D_{\alpha }[A_{\alpha \beta }(u)\zeta ^{\alpha }]_{\mid u=u_{0}}+\frac{%
\ell ^{2}}{2}\{D_{\alpha }D_{\beta }[A_{\mu \nu }(u)\zeta ^{\alpha }\zeta
^{\beta } \\
&&-\frac{1}{3}R_{~\alpha \mu \beta }^{[N]\rho }(u)A_{\rho \nu }(u)-\frac{1}{3%
}R_{~\alpha \nu \beta }^{[N]\rho }(u)A_{\rho \mu }(u)]\}_{\mid
u=u_{0}}+o\left( \ell ^{3}\right) ,
\end{eqnarray*}%
where the Riemannian curvature d--tensor $R_{~\alpha \mu \beta }^{[N]\rho
}=%
\{R_{h.jk}^{.i},R_{b.jk}^{.a},P_{j.ka}^{.i},P_{b.ka}^{.c},S_{j.bc}^{.i},S_{b.cd}^{.a}\}
$ has the invariant components given by the formulas
(\ref{3dcurvatures}) from Appendix. \ Putting such expansions in
the action for the nonlinear
sigma model (\ref{act1a}), we obtain the decomposition%
\begin{equation*}
S_{g_{N},B}[u,g]=S_{g_{N},B}[u_{0},g]+\ell \int\limits_{\Sigma }\delta \mu
_{g}\zeta ^{\beta }S_{\beta }[u_{0},g]+\overline{S}[u,\zeta ,g],
\end{equation*}%
where $S_{\beta }$ is given by the variation
\begin{equation*}
S_{\beta }[u_{0},g]=(\det |g|)^{-1/2}\frac{\triangle S[e^{\chi }u_{0},g]}{%
\triangle \chi ^{\beta }}\mid _{\chi =0}
\end{equation*}%
and the last term $\overline{S}$ is an expansion on $\ell ,$%
\begin{equation*}
\overline{S}=\overline{S}_{[0]}+\ell \overline{S}_{[1]}+\ell ^{2}\overline{S}%
_{[2]}+o\left( \ell ^{3}\right) ,
\end{equation*}%
with%
\begin{eqnarray}
\overline{S}_{[0]} &=&\frac{1}{8\pi }\int\limits_{\Sigma }\delta \mu
_{g}\{g^{AB}[g_{ij}(u_{0})\mathcal{D}_{A}^{\ast }\zeta ^{i}\mathcal{D}%
_{B}^{\ast }\zeta ^{j}+h_{ab}(u_{0})\mathcal{D}_{A}^{\ast }\zeta ^{a}%
\mathcal{D}_{B}^{\ast }\zeta ^{b}]  \label{actser} \\
&&+\mathcal{R}_{\mu \nu \rho \sigma }^{[N]}(u_{0})[g^{AB}-\varepsilon
^{AB}]\partial _{A}u_{0}^{\mu }\partial _{B}u_{0}^{\rho }\zeta ^{\nu }\zeta
^{\sigma }\},  \notag \\
\overline{S}_{[1]} &=&\frac{1}{24\pi }\int\limits_{\Sigma }\delta \mu
_{g}H_{\mu \nu \rho }^{[N]}\varepsilon ^{AB}\zeta ^{\mu }\mathcal{D}%
_{A}^{\ast }\zeta ^{\nu }\mathcal{D}_{B}^{\ast }\zeta ^{\rho },  \notag \\
\overline{S}_{[2]} &=&\frac{1}{8\pi }\int\limits_{\Sigma }\delta \mu _{g}\{%
\frac{g^{AB}}{3}R_{\mu \nu \rho \sigma }^{[N]}\zeta ^{\nu }\zeta ^{\rho }%
\mathcal{D}_{A}^{\ast }\zeta ^{\mu }\mathcal{D}_{B}^{\ast }\zeta ^{\sigma }
\notag \\
&&-\frac{\varepsilon ^{AB}}{2}\mathcal{R}_{\mu \nu \rho \sigma }^{[N]}\zeta
^{\nu }\zeta ^{\rho }\mathcal{D}_{A}^{\ast }\zeta ^{\mu }\mathcal{D}%
_{B}^{\ast }\zeta ^{\sigma }+2D_{\alpha }D_{\beta }\Phi (u_{0})\zeta
^{\alpha }\zeta ^{\beta }R_{g}\}.  \notag
\end{eqnarray}%
The operator $\mathcal{D}_{A}^{\ast }\zeta ^{\nu }$ from (\ref{actser}) is
defined according the rule%
\begin{equation*}
\mathcal{D}_{A}^{\ast }\zeta ^{\nu }=D_{A}^{\ast }\zeta ^{\nu }+\frac{1}{2}%
H_{\quad \mu \rho }^{[N]\sigma }g_{AB}\varepsilon ^{BC}\partial _{C}u^{\mu
}\zeta ^{\rho },
\end{equation*}%
with $D_{A}^{\ast }$ being the covariant derivative on $T^{\ast }\Sigma
\otimes TM$ pulled back to $\Sigma $ by the map $u^{\alpha }$ and acting as
\begin{equation*}
D_{A}^{\ast }\partial _{B}u^{\nu }=\bigtriangledown _{A}\partial _{B}u^{\nu
}+\Gamma _{~\mu \nu }^{\alpha }\partial _{B}u^{\nu }\partial _{A}u^{\nu },
\end{equation*}%
with a h-- and v--invariant decomposition $\Gamma _{\ \beta \gamma }^{\alpha
}=\{L_{\ jk}^{i},L_{\ bk}^{a},C_{\ jc}^{i},C_{\ bc}^{a}\},$ see (\ref{6dcon})
from Appendix, and the operator $\ \mathcal{R}_{\mu \nu \rho \sigma }^{[N]}$
is computed as
\begin{equation*}
\mathcal{R}_{\mu \nu \rho \sigma }^{[N]}=R_{\mu \nu \rho \sigma }^{[N]}+%
\frac{1}{2}D_{\rho }H_{\sigma \mu \nu }^{[N]}-\frac{1}{2}D_{\sigma }H_{\rho
\mu \nu }^{[N]}+\frac{1}{4}H_{\rho \mu \alpha }^{[N]}H_{\sigma \nu
}^{[N]\alpha }-\frac{1}{4}H_{\sigma \mu \alpha }^{[N]}H_{\rho \nu
}^{[N]\alpha }.
\end{equation*}

A comparative analysis of the expansion (\ref{actser}) with a similar one
for $N=0$ from the usual nonlinear sigma model (see, for instance, \cite%
{14deligne}) define the 'geometric d--covariant rule': \ we may apply the same
formulas as in the usual covariant expansions but with that difference that
1) the usual spacetime partial derivatives and differentials are substituted
by N--elongated ones; 2) the Christoffell symbols of connection are changed
into certain d--connection ones, of type (\ref{6dcon}), (\ref{12lcsym}) or (\ref%
{2lccon}); 3) the torsion $H_{\sigma \mu \nu }^{[N]}$ is computed via
N--elongated partial derivatives as in (\ref{htors}) and 4) the curvature $%
R_{\mu \nu \rho \sigma }^{[N]}$ is split into horizontal--vertical, in
brief, h--v--invariant, components according the the formulas (\ref%
{3dcurvatures}). The geometric d--covariant rule allows us to transform
directly the formulas for spacetime backgrounds with metrics written with
respect to coordinate frames into the respective formulas with N--elongated
terms and splitting of indices into h-- and v-- subsets.

\subsubsection{Low energy string anisotropic field equations and effective
action}

Following the geometric d--covariant rule we may apply the results of the
holonomic sigma models in order to define the coefficients $%
a_{1},a_{3},a_{4},a_{5},b_{1}$ and $c_{1},c_{2},c_{4}$ of beta functions (%
\ref{beta1}) and to obtain the following equations of (in our case,
anholonomic) string dynamics,%
\begin{eqnarray*}
2\beta _{ij}^{g,N} &=&R_{ij}-\frac{1}{4}H_{i\rho \sigma
}^{[N]}H_{j}^{[N]\rho \sigma }+2D_{i}D_{j}\Phi =0, \\
2\beta _{ib}^{g,N} &=&R_{ib}-\frac{1}{4}H_{i\rho \sigma
}^{[N]}H_{b}^{[N]\rho \sigma }+2D_{i}D_{b}\Phi =0,
\end{eqnarray*}%
\begin{eqnarray*}
2\beta _{aj}^{g,N} &=&R_{aj}-\frac{1}{4}H_{a\rho \sigma
}^{[N]}H_{j}^{[N]\rho \sigma }+2D_{a}D_{j}\Phi =0, \\
2\beta _{ab}^{g,N} &=&S_{ab}-\frac{1}{4}H_{a\rho \sigma
}^{[N]}H_{b}^{[N]\rho \sigma }+2D_{a}D_{b}\Phi =0,
\end{eqnarray*}%
\begin{eqnarray}
2\beta _{\alpha \beta }^{B} &=&-\frac{1}{2}D^{\lambda }H_{\lambda \mu \nu
}^{[N]}+(D^{\lambda }\Phi )H_{\lambda \mu \nu }^{[N]}=0,  \label{streq} \\
&&  \notag \\
2\beta ^{\Phi } &=&\frac{n+m-26}{3}+\ell ^{2}\left[ \frac{1}{12}H_{\rho
\sigma \tau }^{[N]}H^{[N]\rho \sigma \tau }-\widehat{R}-S-4D^{2}\Phi
+4\left( D^{\lambda }\Phi \right) D_{\lambda }\Phi \right] =0,  \notag
\end{eqnarray}%
where $n+m$ denotes the total dimension of a spacetime with $n$ holonomic
and $m$ anholonomic variables. It should be noted that \ $\beta ^{g,N}=\beta
^{B}=0$ imply the condition that $\beta ^{\Phi }=const,$ which is similar to
the holonomic strings. The only way to satisfy $\beta ^{\Phi }=0$ with
integers $n$ and $m$ is to take $n+m=26.$

The equations (\ref{streq}) are similar to the Einstein equations for the
locally anisotropic gravity (see (\ref{3einsteq2}) in Appendix) with the
matter energy--momentum d--tensor defined from the string theory. \ From
this viewpoint the fields $B_{\alpha \beta }$ and $\Phi $ can be viewed as
certain matter fields and the effective field equations (\ref{streq}) can be
derived from action%
\begin{equation}
S\left( g_{ij},h_{ab},N_{i}^{a},B_{\mu \nu },\Phi \right) =\frac{1}{2\kappa
^{2}}\int \delta ^{26}u\sqrt{|\det g_{\alpha \beta }|}e^{-2\Phi }\left[
\widehat{R}+S+4(D\Phi )^{2}-\frac{1}{12}H^{2}\right] ,  \label{2act3}
\end{equation}%
where $\kappa $ is a constant and, for instance, $D\Phi =D_{\alpha }\Phi ,$ $%
H^{2}=H_{\mu }H^{\mu }$ and the critical dimension $n+m=26$ is taken. For $%
N\rightarrow 0$ and $m\rightarrow 0$ the metric $g_{\alpha \beta }$ is
called the string metric. We shall call $g_{\alpha \beta }$ the string
d--metric for nontrivial values of $N.$

Instead of action (\ref{2act3}), a more standard action, for arbitrary
dimensions, \ can be obtained via a conformal transform of d--metrics of
type (\ref{7dmetric}),
\begin{equation*}
g_{\alpha \beta }\rightarrow \widetilde{g}_{\alpha \beta }=e^{-4\Phi /\left(
n+m-2\right) }g_{\alpha \beta }.
\end{equation*}%
The action in d--metric $\widetilde{g}_{\alpha \beta }$ (by analogy with the
locally isotropic backgrounds we call it the Einstein d--metric) is written%
\begin{eqnarray*}
S\left( \widetilde{g}_{ij},\widetilde{h}_{ab},N_{i}^{a},B_{\mu \nu },\Phi
\right) &=&\frac{1}{2\kappa ^{2}}\int \delta ^{26}u\sqrt{|\det \widetilde{g}%
_{\alpha \beta }|}[\widetilde{\widehat{R}}+\widetilde{S} \\
&&+\frac{4}{n+m-2}(D\Phi )^{2}-\frac{1}{12}e^{-8\Phi /\left( n+m-2\right)
}H^{2}].
\end{eqnarray*}%
This action, for $N\rightarrow 0$ and $m\rightarrow 0,$ is known in
supergravity theory as a part of Chapline--Manton action, see Ref. \cite%
{14deligne} and for the so--called locally anisotropic supergravity, \cite%
{14vstr2,14vmon1}. When we deal with superstirngs, the susperstring calculations
to the mentioned orders give the same results as the bosonic string except
the dimension. For anholonomic backrounds we have to take into account the
nontrivial contributions of $N_{i}^{a}$ and splitting into h-- and v--parts.

\subsubsection{Ahnolonomic Einstein and Finsler gravity from string theory}

\label{efs}It is already known that the $B$--field can be used for
generation of different types of noncommutative geometries from string
theories (see original results and reviews in Refs. \cite%
{14strncg,14dn,14connes1,14sw,14vncf}). Under certain conditions such $B$--field
configurations may result in different variants of geometries with local
anistropy like anholonomic Riemannian geometry, Finsler like spaces and
their generalizations. There is also an alternative possibility when locally
anisotropic interactions are modelled by anholonomic frame fields with
arbitrary $B$--field contributions. In this subsection, we investigate both
type of anisotropic models contained certain low energy limits of string
theory.

\paragraph{B--fields and anholonomic Einstein--Finsler structures}

{\ }\newline
The simplest way to generate an anholonomic structure in a low energy limit
of string theory is to consider a background metric $g_{\mu \nu }=\left(
\begin{array}{cc}
g_{ij} & 0 \\
0 & h_{ab}%
\end{array}%
\right) $ with symmetric Christoffel symbols $\{_{\beta \gamma }^{\alpha }\}$
and such $B_{\mu \nu },$ with corresponding $H_{\mu \nu \rho }^{[N]}$ from (%
\ref{htors}), as there are the nonvanishing values $H_{\mu \nu }^{[N]\rho
}=\{H_{ij}^{[N]a},H_{bj}^{[N]a}=-H_{jb}^{[N]a}\}.$ The next step is to
consider a covariant operator $\mathcal{D}_{\mu }=D_{\mu }^{\{\}}+\frac{1}{2}%
H_{\mu \nu }^{[N]\rho }$ (\ref{cd}), where $\frac{1}{2}H_{\mu \nu }^{[N]\rho
}$ is identified with the torsion (\ref{4torsion}). This way the torsion $%
H_{\mu \nu \rho }^{[N]}$ is associated to an aholonomic frame structure with
non--trivial $W_{ij}^{a}=\delta _{i}N_{j}^{a}-\delta
_{j}N_{i}^{a},~W_{ai}^{b}=-W_{ia}^{b}=-\partial _{a}N_{i}^{b}$ (\ref%
{anholncoef}), when $B_{\mu \nu }$ is parametrized in the form $B_{\mu \nu
}=\{B_{ij}=-B_{ji},B_{bj}=-B_{jb}\}$ by identifying
\begin{equation*}
g_{\mu \nu }W_{\gamma \beta }^{\nu }=\delta _{\mu }B_{\gamma \beta },
\end{equation*}%
i. .e.
\begin{equation}
h_{ca}W_{ij}^{a}=\partial _{c}B_{ij}\mbox{ and }h_{ca}W_{bj}^{a}=\partial
_{c}B_{bj}.  \label{2aux01}
\end{equation}%
Introducing the formulas for the anholonomy coefficients (\ref{anholncoef})
into (\ref{2aux01}), we find some formulas relating partial derivatives $%
\partial _{\alpha }N_{j}^{a}$ and the coefficients $N_{j}^{a}$ with partial
derivatives of $\{B_{ij},B_{bj}\},$%
\begin{eqnarray}
h_{ca}\left( \partial _{i}N_{j}^{a}-N_{i}^{b}\partial _{b}N_{j}^{a}-\partial
_{j}N_{i}^{a}+N_{j}^{b}\partial _{b}N_{i}^{a}\right) &=&\partial _{c}B_{ij},
\notag \\
-h_{ca}\partial _{b}N_{j}^{a} &=&\partial _{c}B_{bj}.  \label{2aux02}
\end{eqnarray}%
So, given any data $\left( h_{ca},N_{i}^{a}\right) $ we can define from the
system of first order partial derivative equations (\ref{2aux02}) \ the
coefficients $B_{ij}$ and $B_{bj},$ or, inversely, from the data $\left(
h_{ca},B_{ij},B_{bj}\right) $ we may construct some non--trivial values $%
N_{j}^{a}.$ We note that the metric coefficients $g_{ij}$ and the $B$--field
components $B_{ab}=-B_{ba}$ could be arbitrary ones, in the simplest
approach we may put $B_{ab}=0.$

The formulas (\ref{2aux02}) define the conditions when a
$B$--field may be transformed into a N--connection structure, or
inversely, a N--connection
can be associated to a $B$--field for a prescribed d--metric structure $%
h_{ca},$ (\ref{7dmetric}).

The next step is to decide what type of d--connection we consider on our
background spacetime. If the values $\{g_{ij},h_{ca}\}$ and $W_{\gamma \beta
}^{\nu }$ (defined by $N_{i}^{b}$ as in (\ref{anholncoef}), but also induced
from $\{B_{ij},B_{bj}\}$ following (\ref{2aux02})) are introduced in formulas
(\ref{12lcsym}) we construct a Levi Civita d--connection $\mathcal{D}_{\mu }$
with nontrivial torsion induced by anholonomic frames with associated
nonlinear connection structure. This spacetime is provided with a d--metric (%
\ref{7dmetric}), $g_{\alpha \beta }=\{g_{ij},h_{ca}\},$ which is compatible
with $\mathcal{D}_{\mu },$ i. e. $\mathcal{D}_{\mu }g_{\alpha \beta }=0.$
The coefficients of $\mathcal{D}_{\mu }$ with respect to anholonomic frames (%
\ref{6dder}) and (\ref{7ddif}), $\Gamma _{\beta \gamma }^{\bigtriangledown
\tau },$ can be computed in explicit form by using formulas (\ref{12lcsym}).
It is proven in the Appendix that on spacetimes provided with anholonomic
structures the Levi Civita connection is not a priority one being both
metric and torsion vanishing. We can construct an infinite number of metric
connections, for instance, the canonical d--connection with the coefficients
(\ref{6dcon}), or, equivalently, following formulas (\ref{2lccon}), to
substitute from the coefficients (\ref{12lcsym}) the values $\frac{1}{2}%
g^{ik}\Omega _{jk}^{a}h_{ca},$ where the coefficients of N--connection
curvature are defined by $N_{i}^{a}$ as in (\ref{4ncurv}). In general, all
such type of linear connections are with nontrivial torsion because of
anholonomy coefficients.

We may generate by $B$--fields an anholonomic (pseudo) Riemannian geometry
if (for given values of $g_{\alpha \beta }=\{g_{ij},h_{ca}\}$ and $N_{i}^{a},
$ satisfying the conditions (\ref{2aux02})) the metric is considered in the
form (\ref{1odm}) \ with respect to coordinate frames, or, equivalently, in
the form (\ref{7dmetric}) with respected to N--adapted frame (\ref{7ddif}).
The metric has to satisfy the gravitational field equations (\ref{3einsteq2})
for the Einstein gravity of arbitrary dimensions with holonomic--anholonomic
variables, or the equations (\ref{streq}) if the gravity with anholonomic
constraints is induced in a low energy string dynamics. We emphasize that
the Ricci d--tensor coefficients from $\beta $--functions (\ref{streq})
should be computed by using the formulas (\ref{6dricci}), derived from those
for d--curvatures (\ref{3dcurvatures}) and for d--torsions (\ref{3dtors}) for
a chosen variant of d--connection coefficients, for instance, (\ref{6dcon})
or (\ref{12lcsym}).

\newpage

We note here that a number of particular ansatz of form (\ref{1odm}) were
considered in Kaluza--Klein gravity \cite{14salam} for different type of
compactifications. In Refs. \cite{14vexsol,14vbel,14vsingl,14vsingl1,14vsolsp} there
were constructed and investigated a number exact solutions with
off--diagonal metrics and anholonomic frames with associated N--connection
structures in the Einstein gravity of different dimensions (see also the
Section \ref{1exsol}).

Now, we discuss the possibility to generate a Finsler geometry from string
theory. We note that the standard definition of Finsler quadratic form
$$g_{ij}^{[F]}=(1/2)\partial ^{2}F/\partial y^{i}\partial y^{j}$$
is considered to be positively definite (see (\ref{fmetric}) in
Appendix). There are different possibilities to include Finsler
like structures in string theories. For instance, we can consider
quadratic forms with non--constant signatures and to generate
(pseudo) Finsler geometries [similarly to (pseudo)
Eucliedean/Riemannian metrics], or, as a second approach, to
consider some embedding of Finsler d--metrics (\ref{2dmetricf}) of
signature $\left( ++...+\right) $ into a 26 dimensional
pseudo--Riemannian anholonomic background with signature $\left(
-++...+\right) .$ In the last case, a particular class of Finsler
background d--metrics may be chosen in the form
\begin{equation}
G^{[F]}=-dx^{0}\otimes dx^{0}+dx^{1}\otimes dx^{1}+g_{i^{\prime }j^{\prime
}}^{[F]}(x,y)dx^{i^{\prime }}\otimes dx^{j^{\prime }}+g_{i^{\prime
}j^{\prime }}^{[F]}(x,y)\delta y^{i^{\prime }}\otimes \delta y^{j^{\prime }}
\label{dmfstr}
\end{equation}%
where $i^{\prime },j^{\prime },...$run values $1,2,..,n^{\prime }\leq 12$
for bosonic strings. The coefficients $g_{i^{\prime }j^{\prime }}^{[F]}$ are
of type (\ref{fmetric}) $\ $or may take the value $+\delta _{i^{\prime
}j^{\prime }}$ for some values of $i\neq i^{\prime },j\neq j^{\prime }.$ We
may consider some static Finsler backgrounds if $g_{i^{\prime }j^{\prime
}}^{[F]}$ do not depend on coordinates $\left( x^{0},x^{1}\right) ,$ but, in
general, we are not imposed to restrict ourselves only to such
constructions. The $N$--coefficients from $\delta y^{i^{\prime
}}=dy^{i^{\prime }}+N_{j^{\prime }}^{i^{\prime }}dx^{i^{\prime }}$ must be
of the form (\ref{2ncc}) if we wont to generate in the low energy string
limit a Finsler structure with Cartan nonlinear connection (there are
possible different variants of nonlinear and distinguished nonlinear
connections, see details in Refs. \cite{14finsler,14ma,14bejancu} and Appendix).

Let us consider in details how a Finsler metric can be included in a low
energy string dynamics. We take a Finsler metric $F$ which generate the
metric coefficients $g_{i^{\prime }j^{\prime }}^{[F]}$ and the N--connection
coefficients $N_{j^{\prime }}^{[F]i^{\prime }},$ respectively, via formulas (%
\ref{fmetric}) and (\ref{2ncc}). The Cartan's N--connection structure $%
N_{j^{\prime }}^{[F]i^{\prime }}$ may be induced by a $B$--field if there
are some nontrivial values, let us denote them $\{B_{ij}^{[F]},B_{bj}^{[F]}%
\},$ which satisfy the conditions (\ref{2aux02}). This way the $B$--field is
expressed via a Finsler metric $F\left( x,y\right) $ and induces a d--metric
(\ref{dmfstr}). This Finsler structure follows from a low energy string
dynamics if the Ricci d--tensor $R_{\alpha \beta
}=\{R_{ij},R_{ia},R_{ai},R_{ab}\}$ (\ref{6dricci}) and the torsion
$$H_{\mu
\nu }^{[N]\rho }=\{H_{ij}^{[N]a},H_{bj}^{[N]a}=-H_{jb}^{[N]a}\}$$
related with $N_{j^{\prime }}^{[F]i^{\prime }}$as in
(\ref{2aux01}), all computed for d--metric (\ref{dmfstr}) are
solutions of the motion equations (\ref{streq}) for any value of
the dilaton field $\Phi .$ In the Section \ \ref{1exsol} we shall
consider an explicit example of a string--Finsler metric.

Here it should be noted that instead of a Finsler structure, in a similar
manner, we may select from a string locally anisotropic dynamics a Lagrange
structure if the metric coefficients $g_{i^{\prime }j^{\prime }}$ are
generated by a Lagrange function $L(x,y)$ (\ref{1mfl}). The N--connection may
be an arbitrary one, or of a similar Cartan form. We omit such constructions
in this paper.

\paragraph{Anholonomic Einstein--Finsler structures for arbitrary B--fields}

{\ }\newline
Locally anisotropic metrics may be generated by anholonomic frames with
associated N--connections which are not induced by some $B$--field
configurations.

For an anholonomic (pseudo) Riemannian background we consider an ansatz of
form (\ref{1odm}) which by anholonomic transform can be written as an
equivalent d--metric (\ref{7dmetric}). The coefficients $N_{i}^{a}$ and $%
B_{\mu \nu }$ are related only via the string motion equations (\ref{streq})
which must be satisfied by the Ricci d--tensor (\ref{6dricci}) computed, for
instance, for the canonical d--connection (\ref{6dcon}).

A Finsler like structure, not induced directly by $B$--fields, may be
emphasized if the d-metric is taken in the form (\ref{dmfstr}), but the $\ $%
\ values $\delta y^{i^{\prime }}=dy^{i^{\prime }}+N_{j^{\prime }}^{i^{\prime
}}dx^{i^{\prime }}$ being elongated by some $N_{j^{\prime }}^{i^{\prime }}$
are not obligatory constrained by the conditions (\ref{2aux02}). Of course,
the Finsler metric $F$ and $B_{\mu \nu }$ are not completely independent;
these fields must be chosen as to generate a solution of string--Finsler
equations (\ref{streq}).

In a similar manner we can model as some alternative low energy limits of
the string theory, with corresponding nonlinear sigma models, different
variants of spacetime geometries with anholonomic and N--connection
structures, derived on manifold or vector bundles when the metric, linear
and N--connection structures are proper for a Lagrange, generalized Lagrange
or anholonomic Riemannian geometry \cite%
{14ma,14finsler,14bejancu,14kern,14vexsol,14vankin,14vbel,14vgauge,14vsingl}.

\section{Superstrings and Anisotropic Supergravity}

The bosonic string theory, from which in the low energy limits we may
generate different models of anholonomic Riemannian--Finsler gravity,
suffers from at least four major problems: 1) there are tachyonic states
which violates the physical causality and divergence of transitions
amplitudes; 2) there are not included any fermionic states transforming
under a spinor representation of the spacetime Lorentz group; 3) it is not
clear why Yang--Mills gauge particles arise in both type of closed and open
string theories and to what type of strings should be given priority; 4)
experimentally there are 4 dimensions and not 26 as in the bosonic string
theory: it must be understood why the remaining dimensions are almost
invisible.

The first three problems may be resolved by introducing certain additional
dynamical degrees of freedom on the string worldsheet which results in
fermionic string states in the physical Hibert space and modifies the
critical dimension of spacetime. \ One tries to solve the forth problem by
developing different models of compactification.

There are distinguished five, consistent, tachyon free, spacetime
supersymmetric string theories in flat Minkowski spacetime (see, for
instance, \cite{14deligne,14kir} for basic results and references on types I,
IIA, IIB, Heterotic $Spin(32)/Z_{2}$ and Heterotic $E_{8}\times E_{8}$
string theories). The (super) string and (super) gravity theories in
geralized Finsler like, in general, supersymmetric backgrounds provided with
N--connection structure, and corresponding anisotropic superstring
perturbation theories, were investigated in \ Refs. \cite{14vsuper,14vstr2,14vmon1}%
. The goal of this Section is to illustrate how anholonomic type structures
arise in the low energy limits of the mentioned string theories if the
backgrounds are considered with certain anholonomic frame and off--diagonal
metric structures. We shall consider the conditions when generalized Finsler
like geometries arise in (super) string theories.

We would like to start with the example of the two--dimensional $\mathcal{N}%
=1$ supergravity coupled to the dimension 1 superfields, containing a
bosonic coordinate $X^{\mu }$ and two fermionic coordinates, one
left--moving $\psi ^{\mu }$ and one right moving $\overline{\psi }^{\mu }$
(we use the symbol $\mathcal{N}$ for the supersimmetric dimension which must
be not confused with the symbol $N$ for a N--connection structure). We note
that the two dimensional $\mathcal{N}=1$ supergravity multiplet contains the
metric and a gravitino $\chi _{A}.$In order to develop models in backgrounds
distinguished by a N--connection structure, we have to consider splitting
into h-- and v--components, i. e. to write $X^{\mu }=\left(
X^{i},X^{a}\right) $ and $\psi ^{\mu }=\left( \psi ^{i},\psi ^{a}\right) ,%
\overline{\psi }^{\mu }=\left( \overline{\psi }^{i},\overline{\psi }%
^{a}\right) .$ The spinor differential geometry on anisotropic spacetimes
provided with N--connections (in brief, d--spinor geometry) was developed in
Refs. \cite{14vspinors,14vmon2}. Here we shall present only the basic formulas,
emphasizing the fact that the coefficients of d--spinors have the usual
spinor properties on separated h-- (v-) subspaces.

The simplest distinguished superstring model can be developed from an analog
of the bosonic Polyakov action,%
\newpage
\begin{eqnarray}
S_{P} &=&\frac{1}{4\pi \alpha ^{\prime }}\int\limits_{\Sigma }\delta \mu
_{g}\{g^{AB}\left[ \partial _{A}X^{i}\partial _{B}X^{j}g_{ij}+\partial
_{A}X^{a}\partial _{B}X^{b}h_{ab}\right]  \label{actps} \\
&&+\frac{i}{2}[\psi ^{k}\gamma ^{A}\partial _{A}\psi ^{k}+\psi ^{a}\gamma
^{A}\partial _{A}\psi ^{a}]+\frac{i}{2}\left( \chi _{A}\gamma ^{B}\gamma
^{A}\psi ^{k}\right) \left( \partial _{B}X^{k}-\frac{i}{4}\chi _{B}\psi
^{k}\right)  \notag \\
&&+\frac{i}{2}\left( \chi _{A}\gamma ^{B}\gamma ^{A}\psi ^{a}\right) \left(
\partial _{B}X^{a}-\frac{i}{4}\chi _{B}\psi ^{a}\right) \}  \notag
\end{eqnarray}%
being invariant under transforms (i. e. being $\mathcal{N}=1$ left--moving $%
\left( 1,0\right) $ supersymmetric)%
\begin{eqnarray*}
\bigtriangleup g_{AB} &=&i\epsilon \left( \gamma _{A}\chi _{B}+\gamma
_{B}\chi _{A}\right) ,~\bigtriangleup \chi _{A}=2\bigtriangledown
_{A}\epsilon , \\
\bigtriangleup X^{i} &=&i\epsilon \psi ^{i},~\bigtriangleup \psi ^{k}=\gamma
^{A}\left( \partial _{A}X^{k}-\frac{i}{2}\chi _{A}\psi ^{k}\right) \epsilon
,~\bigtriangleup \overline{\psi }^{i}=0, \\
\bigtriangleup X^{a} &=&i\epsilon \psi ^{a},~\bigtriangleup \psi ^{a}=\gamma
^{A}\left( \partial _{A}X^{a}-\frac{i}{2}\chi _{A}\psi ^{a}\right) \epsilon
,~\bigtriangleup \overline{\psi }^{a}=0,
\end{eqnarray*}%
where the gamma matrices $\gamma _{A}$ and the covariant differential
operator $\bigtriangledown _{A}$ are defined on the two dimensional surface,
$\epsilon $ is a left--moving Majorana--Weyl spinor. There is also a similar
right--moving $\left( 0,1\right) $ supersymmetry involving a right moving
Majorana--Weyl spinor $\overline{\epsilon }$ and the fermions $\overline{%
\psi }^{\mu }$ which means that the model has a $\left( 1,1\right) $
supersymmetry. The superconformal gauge for the action (\ref{actps}) is
defined as%
\begin{equation*}
g_{AB}=e^{\Phi }\delta _{AB},~\chi _{A}=\gamma _{A}\zeta ,
\end{equation*}%
for a constant Majorana spinor $\zeta .$ This action has also certain matter
like supercurents $i\psi ^{\mu }\partial X^{\mu }$ and $i\overline{\psi }%
^{\mu }\overline{\partial }X^{\mu }.$

We remark that the so--called distinguished gamma matrices (d--matrices), $%
\gamma ^{\alpha }=\left( \gamma ^{i},\gamma ^{a}\right) $ and related spinor
calculus are derived from $\gamma $--decompositions of the h-- and v--
components of d--metrics \ $g^{\alpha \beta }=\{g^{ij},h^{ab}\}$ (\ref%
{7dmetric})%
\begin{equation*}
\gamma ^{i}\gamma ^{j}+\gamma ^{j}\gamma ^{i}=-2g^{ij},~\gamma ^{a}\gamma
^{b}+\gamma ^{b}\gamma ^{a}=-2h^{ab},
\end{equation*}%
see details in Refs. \cite{14vspinors,14vmon2}.

In the next subsections we shall distinguish more realistic superstring
actions than (\ref{actps}) following the geometric d--covariant rule
introduced in subsection \ref{dcovrule}, when the curved spacetime geometric
objects like metrics, connections, tensors, spinors, ... as well the partial
and covariant derivatives and differentials are decomposed in invariant h--
and v--components, adapted to the N--connection structure. This will allow
us to extend directly the results for superstring low energy isotropic
actions to backgrounds with local anisotropy.

\subsection{Locally anisotropic supergravity theories}

We indicate that many papers on supergravity theories in various dimensions
are reprinted in a set of two volumes \cite{14salamsezgin}. The bulk of
supergravity models contain locally anisotropic configurations which can be
emphasized by some vielbein transforms (\ref{vielbtr}) and metric anzatz (%
\ref{6ansatz}) with associated N--connection. For corresponding
parametrizations of the d--metric coefficients, $g_{\alpha \beta
}(u)=\{g_{ij},h_{ab}\},$ N--connection, $N_{i}^{a}(x,y),$ and d--connection,
$\Gamma _{\ \beta \gamma }^{\alpha }=\left( L_{\ jk}^{i},L_{\ bk}^{a},C_{\
jc}^{i},C_{\ bc}^{a}\right) ,$ with possible superspace generalizations, we
can generate (pseudo) Riemannian off--diagonal metrics, Finsler or Lagrange
(super) geometries. \ In this subsection, we analyze the anholonomic frame
transforms of some supergravity actions which can be coupled to superstring
theory.

We note that the field components will be organized according to multiplets
of $Spin\left( 1,10\right) .$ We shall use 10 dimensional spacetime indices $%
\alpha ,\beta ...=0,1,2,...,9$ or 11 dimensional ones $\overline{\alpha },%
\overline{\beta }...=0,1,2,...,9,10.$ The coordinate $u^{10\text{ }}$could
be considered as a compactified one, or distinguished in a non--compactified
manner, by the N--connection structure. There is a general argument \cite%
{14nahm} is that 11 is the largest possible dimension in which supersymmetric
multiplets can exist with spin less, or equal to 2, with \ a single local
supersymmetry. We write this as $n+m=11,$ which points to possible
splittings of indices like $\overline{\alpha }=\left( \overline{i},\overline{%
a}\right) $ where $\overline{i}$ and $\overline{a}$ run respectively $n$ and
$m$ values. A consistent superstring theory holds if $n+m=10.$ In this case,
indices are to be decomposed as $\alpha =\left( i,a\right) .$ For
simplicity, we shall consider that a metric tensor in $n+m=11$ dimensions
decomposes as $g_{\overline{\alpha }\overline{\beta }}\left( u^{\mu
},u^{10}\right) \rightarrow g_{\overline{\alpha }\overline{\beta }}\left(
u^{\mu }\right) $ and that in low energy approximation the fields are
locally anisotropically interacting and independent on $u^{10}.$ The
antisymmetric rank 3 tensor is taken to decompose as $A_{\overline{\alpha }%
\overline{\beta }\overline{\gamma }}\left( u^{\mu },u^{10}\right)
\rightarrow A_{\overline{\alpha }\overline{\beta }\overline{\gamma }}(u^{\mu
}).$ A fitting with superstring theory is to be obtained if $\left(
A_{\alpha \beta \gamma }^{[3]},B_{\mu \nu }\right) \rightarrow A_{\overline{%
\alpha }\overline{\beta }\overline{\gamma }}$ and consider for spinors
''dilatino'' fields $\left( \chi _{\mu }^{~\tau },\lambda _{\tau }\right)
\rightarrow \chi _{\overline{\mu }}^{~\tau },$ see, for instance, Refs. \cite%
{14deligne} for details on couplings of supergravity and low energy
superstrings.

\subsubsection{$\mathcal{N}=1,n+m=11$ anisotropic supergravity}

The field content of $\mathcal{N}=1$ and 11 dimensional supergravity is
given by $g_{\overline{\alpha }\overline{\beta }}$ (graviton), $A_{\overline{%
\alpha }\overline{\beta }\overline{\gamma }}$ (U(1) gauge fields) and $\chi
_{\overline{\mu }}^{\alpha }$ (gravitino). The dimensional reduction is
stated by $g_{\alpha 10}=g_{10\alpha }=A_{\alpha }^{[1]}$ and $%
g_{10~10}=e^{-2\Phi },$ where the coefficients are given with respect to an
N--elongated basis. We suppose that an effective action
\begin{equation*}
S(g_{ij},h_{ab},N_{i}^{a},B_{\mu \nu },\Phi )=\frac{1}{2\kappa ^{2}}\int
\delta \mu _{\lbrack g,h]}e^{-2\Phi }\left[ -\widehat{R}-S+4(D\Phi )^{2}-%
\frac{1}{12}H^{2}\right] ,
\end{equation*}%
is to be obtained if the values $A_{\alpha }^{[1]},A_{\alpha \beta \gamma
}^{[3]},\chi _{\mu }^{~\tau },\lambda _{\tau }$ vanish. For $N\rightarrow
0,m\rightarrow 0$ this action results from the so--called NS sector of the
superstring theory, being related to the sigma model action (\ref{2act3}). A
full $\mathcal{N}=1$ and 11 dimensional locally anisotropic supergravity can
be constructed similarly to the locally isotropic case \cite{14cjs} but
considering that $H^{[N]}=\delta B$ and $F^{[N]}=\delta A$ are computed as
differential forms with respect to N--elongated differentials (\ref{7ddif}),
\begin{eqnarray}
S\left( g_{ij},h_{ab},N_{i}^{a},A_{\alpha },\chi \right) &=&-\frac{1}{%
2\kappa ^{2}}\int \delta \mu _{\lbrack g,h]}[\widehat{R}+S-\frac{\kappa ^{2}%
}{12}F^{2}+\kappa ^{2}\overline{\chi }_{\overline{\mu }}\Gamma ^{\overline{%
\mu }\overline{\nu }\overline{\lambda }}D_{\overline{\nu }}\chi _{\overline{%
\lambda }}  \label{act4} \\
&&+\frac{\sqrt{2}\kappa ^{3}}{384}\left( \overline{\chi }_{\overline{\mu }%
}\Gamma ^{\overline{\mu }\overline{\nu }\overline{\rho }\overline{\sigma }%
\overline{\tau }\overline{\lambda }}\overline{\chi }_{\overline{\lambda }%
}+12\Gamma ^{\overline{\rho }\overline{\nu }\overline{\sigma }}\chi ^{%
\overline{\tau }}\right) \left( F+\widehat{F}\right) _{\overline{\nu }%
\overline{\rho }\overline{\sigma }\overline{\tau }}]  \notag \\
&&-\frac{\sqrt{2}\kappa }{81\times 56}\int A\wedge F\wedge F,  \notag
\end{eqnarray}%
where $\Gamma ^{\overline{\mu }\overline{\nu }\overline{\rho }\overline{%
\sigma }\overline{\tau }\overline{\lambda }}=\Gamma ^{\lbrack \overline{\mu }%
}\Gamma ^{\overline{\nu }}...\Gamma ^{\overline{\lambda }]}$ is the standard
notation for gamma matrices for 11 dimensional spacetimes, the field $%
\widehat{F}=F+\chi $--terms and $D_{\overline{\nu }}$ is the covariant
derivative with respect to $\frac{1}{2}\left( \omega +\widehat{\omega }%
\right) $ where
\begin{equation*}
\widehat{\omega }_{\overline{\mu }\overline{\alpha }\overline{\beta }%
}=\omega _{\overline{\mu }\overline{\alpha }\overline{\beta }}+\frac{1}{8}%
\chi ^{\overline{\nu }}\Gamma _{\overline{\nu }\overline{\mu }\overline{%
\alpha }\overline{\beta }\overline{\rho }}\chi ^{\overline{\rho }}
\end{equation*}%
with $\omega _{\overline{\mu }\overline{\alpha }\overline{\beta }}$ being
the spin connection determined by its equation of motion. We put the same
coefficients in the action (\ref{act4}) as in the locally isotropic case as
to have compatibility for such limits. Every object (tensors, connections,
connections) has a N--distinguished invariant character with indices split
into h-- and v--subsets. For simplicity we omit here further decompositions
of fields with splitting of \ indices.

\subsubsection{Type IIA anisotropic supergravity}

The action for a such model can be deduced from (\ref{act4}) if $A_{\alpha
\beta \gamma }=\kappa ^{1/4}A_{\alpha \beta \gamma }^{[3]}$ and $A_{\alpha
\beta 10}=\kappa ^{-1}B_{\alpha \beta }$ with further h-- and v--
decompositions of \ indices. The bosonic part of the type IIA locally
anisotropic supergravity is described by
\begin{eqnarray}
S\left( g_{ij},h_{ab},N_{i}^{a},\Phi ,A^{(1)},A^{(3)}\right) &=&-\frac{1}{%
2\kappa ^{2}}\int \delta \mu _{\lbrack g,h]}\{e^{-2\Phi }[\widehat{R}%
+S-4(D\Phi )^{2}+\frac{1}{12}H^{2}]  \notag \\
&&+\sqrt{\kappa }G_{[A]}+\frac{\sqrt{\kappa }}{12}F^{2}-\frac{\kappa ^{-3/2}%
}{288}\int B\wedge F\wedge F\},  \label{act4b}
\end{eqnarray}%
with $G_{[A]}=\delta A^{(1)},H=\delta B$ and $F=\delta A^{(3)}.$ This action
may be written directly from the locally isotropic analogous following the
d--covariant geometric rule.

\subsubsection{Type IIB, n+m=10, $\mathcal{N}=2$ anisotropic supergravity}

In a similar manner, geometrically, for d--objects, we may compute possible
anholnomic effects from an action describing a model of locally anisotropic
supergravity with a super Yang--Mills action (the bosonic part)%
\begin{equation}
S_{IIB}=-\frac{1}{2\kappa ^{2}}\int \delta \mu _{\lbrack g,h]}e^{-2\Phi }[%
\widehat{R}+S+4(D\Phi )^{2}-\frac{1}{12}\widetilde{H}^{2}-\frac{1}{4}F_{\mu
\nu }^{\widehat{a}}F^{\widehat{a}\mu \nu }],  \label{act5}
\end{equation}%
when the super--Yang--Mills multiplet is stated by the action
\begin{equation*}
S_{YM}=\frac{1}{\kappa }\int \delta \mu _{\lbrack g,h]}e^{-2\Phi }[-\frac{1}{%
4}F_{\mu \nu }^{\widehat{a}}F^{\widehat{a}\mu \nu }-\frac{1}{2}\overline{%
\psi }^{\widehat{a}}\Gamma ^{\mu }D_{\mu }\psi ^{\widehat{a}}].
\end{equation*}%
In these actions
\begin{equation*}
A=A_{\mu }^{\widehat{a}}t^{\widehat{a}}\delta u^{\mu }
\end{equation*}%
is the gauge d--field of $E_{8}\times E_{8}$ or $Spin\left( 32\right) /Z_{2}$
group (with generators $t^{\widehat{a}}$ labelled by the index $\widehat{a}),$
having the strength
\begin{equation*}
F=\delta A+g_{F}A\wedge A=\frac{1}{2}F_{\mu \nu }^{\widehat{a}}t^{\widehat{a}%
}\delta u^{\mu }\wedge \delta u^{\nu },
\end{equation*}%
$g_{F}$ being the coupling constant, and $\psi $ is the gaugino of $%
E_{8}\times E_{8}$ or $Spin\left( 32\right) /Z_{2}$ group (details on
constructions of locally anisotropic gauge and spinor theories can be found
in Refs. \cite{14vgauge,14vncf,14vmon1,14vmon2,14vnonc,14vspinors}). The action with $B$%
--field \ strength in (\ref{act5}) is defined as follows
\begin{equation*}
\widetilde{H}=\delta B-\frac{\kappa }{\sqrt{2}}\omega _{CS}\left( A\right) ,
\end{equation*}%
for
\begin{equation*}
\omega _{CS}\left( A\right) =tr\left( A\wedge \delta A+\frac{2}{3}%
g_{F}A\wedge A\wedge A\right) .
\end{equation*}%
Such constructions conclude in a theory with $S_{IIB}+S_{YM}+$ fermionic
terms with anholonomies and $\mathcal{N}=1$ supersymmetry.

Finally, we emphasize that the actions for supersymmetric anholonomic models
can considered in the framework of (super) geometric formulation of
supergravities in $n+m=10$ and $11$ dimensions on superbundles provided with
N--connection structure \cite{14vsuper,14vstr2,14vmon1}.

\subsection{Superstring effective actions and anisotropic toroidal
co\-mpactifica\-ti\-ons}

The supergravity actions presented in the previous subsection can be
included in different supersymmetric string theories which emphasize
anisotropic effects if spacetimes provided with N--connection structure are
considered. In this subsection we analyze a model with toroidal
compactification when the background is locally anisotropic. In order to
obtain four--dimensional (4D) theories, the simplest way is to make use of
the Kaluza--Klein idea: to propose a model when some of the dimensions are
curled--up into a compact manifold, the rest of dimensions living only for
non--compact manifold. Our aim is to show that in result of toroidal
compactifications the resulting 4D theory could be locally anisotropic.

The action (\ref{act5}) can be obtained also as a 10 dimensional heterotic
string effective action (in the locally isotropic variant see, for instance,
Ref. \cite{14kir})
\begin{equation}
\left( \alpha ^{\prime }\right) ^{8}S_{10-n^{\prime }-m^{\prime }}=\int
\delta ^{10}u\sqrt{|g_{\alpha \beta }|}e^{-\Phi ^{\prime }}[\widehat{R}%
+S+(D\Phi ^{\prime })^{2}-\frac{1}{12}\widetilde{H}^{2}-\frac{1}{4}F_{\mu
\nu }^{\widehat{a}}F^{\widehat{a}\mu \nu }]+o\left( \alpha ^{\prime }\right)
,  \label{act6}
\end{equation}%
where we redefined $2\Phi \rightarrow \Phi ^{\prime },$ use the string
constant $\alpha ^{\prime }$ and consider the $\left( n^{\prime },m^{\prime
}\right) $ as the (holonomic, anholonomic) dimensions of the compactified
spacetime (as a particular case we can consider $n^{\prime }+m^{\prime }=4,$
or $n^{\prime }+m^{\prime }<10$ for any brane configurations. Let us use
parametrizations of indices and of vierbeinds: Greek indices $\alpha ,\beta
,...\mu ...$ run values for a 10 dimensional spacetime and split as $\alpha
=\left( \alpha ^{\prime },\widehat{\alpha }\right) ,\beta =\left( \beta
^{\prime },\widehat{\beta }\right) ,...$ when primed indices $\alpha
^{\prime },\beta ^{\prime },...\mu ^{\prime }...$ run values for
compactified spacetime and split into h- and v--components like $\alpha
^{\prime }=\left( i^{\prime },a^{\prime }\right) ,$ $\beta ^{\prime }=\left(
j^{\prime },b^{\prime }\right) ,...;$ the frame coefficients are split as
\begin{equation*}
e_{\mu }^{~\underline{\mu }}(u)=\left(
\begin{array}{cc}
e_{\alpha ^{\prime }}^{~\underline{\alpha ^{\prime }}}(u^{\beta ^{\prime }})
& A_{\alpha ^{\prime }}^{\widehat{\alpha }}(u^{\beta ^{\prime }})e_{\widehat{%
\alpha }}^{~\underline{\widehat{\alpha }}}(u^{\beta ^{\prime }}) \\
0 & e_{\widehat{\alpha }}^{~\underline{\widehat{\alpha }}}(u^{\beta ^{\prime
}})%
\end{array}%
\right)
\end{equation*}%
where $e_{\alpha ^{\prime }}^{~\underline{\alpha ^{\prime }}}(u^{\beta
^{\prime }}),$ in their turn, are taken in the form (\ref{vielbtr}),
\begin{equation}
e_{\alpha ^{\prime }}^{~\underline{\alpha ^{\prime }}}(u^{\beta ^{\prime
}})=\left(
\begin{array}{cc}
e_{i^{\prime }}^{~\underline{i^{\prime }}}(x^{j^{\prime }},y^{a^{\prime }})
& N_{i^{\prime }}^{a^{\prime }}(x^{j^{\prime }},y^{a^{\prime }})e_{a^{\prime
}}^{~\underline{a^{\prime }}}(x^{j^{\prime }},y^{a^{\prime }}) \\
0 & e_{a^{\prime }}^{~\underline{a^{\prime }}}(x^{j^{\prime }},y^{a^{\prime
}})%
\end{array}%
\right) .  \label{2frame8}
\end{equation}%
For the metric we have the recurrent ansatz%
\begin{equation*}
\underline{g}_{\alpha \beta }=\left[
\begin{array}{cc}
g_{\alpha ^{\prime }\beta ^{\prime }}(u^{\beta ^{\prime }})+N_{\alpha
^{\prime }}^{\widehat{\alpha }}(u^{\beta ^{\prime }})N_{\beta ^{\prime }}^{%
\widehat{\beta }}(u^{\beta ^{\prime }})h_{\widehat{\alpha }\widehat{\beta }%
}(u^{\beta ^{\prime }}) & h_{\widehat{\alpha }\widehat{\beta }}(u^{\beta
^{\prime }})N_{\alpha ^{\prime }}^{\widehat{\alpha }}(u^{\beta ^{\prime }})
\\
h_{\widehat{\alpha }\widehat{\beta }}(u^{\beta ^{\prime }})N_{\beta ^{\prime
}}^{\widehat{\beta }}(u^{\beta ^{\prime }}) & h_{\widehat{\alpha }\widehat{%
\beta }}(u^{\beta ^{\prime }})%
\end{array}%
\right] .
\end{equation*}%
where%
\begin{equation}
g_{\alpha ^{\prime }\beta ^{\prime }}=\left[
\begin{array}{cc}
g_{i^{\prime }j^{\prime }}(u^{\beta ^{\prime }})+N_{i^{\prime }}^{a^{\prime
}}(u^{\beta ^{\prime }})N_{j^{\prime }}^{b^{\prime }}(u^{\beta ^{\prime
}})h_{a^{\prime }b^{\prime }}(u^{\beta ^{\prime }}) & h_{a^{\prime
}b^{\prime }}(u^{\beta ^{\prime }})N_{i^{\prime }}^{a^{\prime }}(u^{\beta
^{\prime }}) \\
h_{a^{\prime }b^{\prime }}(u^{\beta ^{\prime }})N_{j^{\prime }}^{b^{\prime
}}(u^{\beta ^{\prime }}) & h_{a^{\prime }b^{\prime }}(u^{\beta ^{\prime }})%
\end{array}%
\right] .  \label{2metr8}
\end{equation}%
The part of action (\ref{act6}) containing the gravity and dilaton terms
becomes%
\begin{eqnarray}
\left( \alpha ^{\prime }\right) ^{n^{\prime }+m^{\prime }}S_{n^{\prime
}+m^{\prime }}^{heterotic} &=&\int \delta ^{n^{\prime }+m^{\prime }}u\sqrt{%
|g_{\alpha \beta }|}e^{-\phi }[\widehat{R}^{\prime }+S^{\prime }+(\delta
_{\mu ^{\prime }}\phi )(\delta ^{\mu ^{\prime }}\phi )  \label{rterm} \\
&&+\frac{1}{4}(\delta _{\mu ^{\prime }}h_{\widehat{\alpha }\widehat{\beta }%
})(\delta ^{\mu ^{\prime }}h^{\widehat{\alpha }\widehat{\beta }})-\frac{1}{4}%
h_{\widehat{\alpha }\widehat{\beta }}F_{\mu ^{\prime }\nu ^{\prime }}^{[A]%
\widehat{\alpha }}F^{[A]\widehat{\beta }\mu ^{\prime }\nu ^{\prime }}],
\notag
\end{eqnarray}%
where $\phi =\Phi ^{\prime }-\frac{1}{2}\log \left( \det |h_{\widehat{\alpha
}\widehat{\beta }}|\right) $ and $F_{\mu ^{\prime }\nu ^{\prime }}^{[A]%
\widehat{\alpha }}=\delta _{\mu ^{\prime }}A_{\nu ^{\prime }}^{\widehat{%
\alpha }}-\delta _{\nu ^{\prime }}A_{\mu ^{\prime }}^{\widehat{\alpha }}$
and the h- and v--components of the induced scalar curvature, respectively, $%
\widehat{R}^{\prime }$ and $S^{\prime }$ (see formula (\ref{4dscalar}) in
Appendix) are primed in order to point that these values are for the lower
dimensional space. The \ antisymmetric tensor part may be decomposed in the
form%
\begin{eqnarray}
-\frac{1}{12}\int \delta ^{10}u\sqrt{|g_{\alpha \beta }|}e^{-\Phi ^{\prime
}}H^{\mu \nu \rho }H_{\mu \nu \rho } &=&-\frac{1}{4}\int \delta ^{n^{\prime
}+m^{\prime }}u\sqrt{|g_{\alpha ^{\prime }\beta ^{\prime }}|}e^{-\phi }\times
\label{hterm} \\
&&[H^{\mu ^{\prime }\widehat{\alpha }\widehat{\beta }}H_{\mu ^{\prime }%
\widehat{\alpha }\widehat{\beta }}+H^{\mu ^{\prime }\nu ^{\prime }\widehat{%
\beta }}H_{\mu ^{\prime }\nu ^{\prime }\widehat{\beta }}+\frac{1}{3}H^{\mu
^{\prime }\nu ^{\prime }\rho ^{\prime }}H_{\mu ^{\prime }\nu ^{\prime }\rho
^{\prime }}],  \notag
\end{eqnarray}%
where, for instance,
\begin{equation*}
H_{\mu ^{\prime }\widehat{\alpha }\widehat{\beta }}=e_{\mu ^{\prime }}^{~%
\underline{\mu ^{\prime }}}e_{\underline{\mu ^{\prime }}}^{\mu }H_{\mu
\widehat{\alpha }\widehat{\beta }}
\end{equation*}%
and we have considered $H_{\widehat{\alpha }\widehat{\beta }\widehat{\gamma }%
}=0.$ In a similar manner we can decompose the action for gauge fields $%
\widehat{A}_{\mu }^{I}$ with index $I=1,...,32,$%
\begin{equation}
\int \delta ^{10}u\sqrt{|g_{\alpha \beta }|}e^{-\Phi ^{\prime
}}\sum\limits_{I=1}^{16}\widehat{F}^{I,\mu \nu }\widehat{F}_{\mu \nu
}^{I}=\int \delta ^{n^{\prime }+m^{\prime }}u\sqrt{|g_{\alpha ^{\prime
}\beta ^{\prime }}|}e^{-\phi }~\sum\limits_{I=1}^{16}[\widehat{F}^{I,\mu
^{\prime }\nu ^{\prime }}\widehat{F}_{\mu ^{\prime }\nu ^{\prime }}^{I}+2%
\widehat{F}^{I,\mu ^{\prime }\widehat{\nu }}\widehat{F}_{\mu ^{\prime }%
\widehat{\nu }}^{I}],  \label{fterm}
\end{equation}%
with
\begin{eqnarray*}
Y_{\widehat{\alpha }}^{I} &=&A_{\widehat{\alpha }}^{I},~A_{\alpha ^{\prime
}}^{I}=\widehat{A}_{\alpha ^{\prime }}^{I}-Y_{\widehat{\alpha }}^{I}A_{\mu
}^{\widehat{\alpha }},~ \\
\widehat{F}_{\mu ^{\prime }\nu ^{\prime }}^{I} &=&F_{\mu ^{\prime }\nu
^{\prime }}^{I}+Y_{\widehat{\alpha }}^{I}F_{\mu ^{\prime }\nu ^{\prime
}}^{[A]\widehat{\alpha }},~\widehat{F}_{\mu ^{\prime }\widehat{\nu }%
}^{I}=\delta _{\mu ^{\prime }}Y_{\widehat{\alpha }}^{I},~\widehat{F}_{\mu
^{\prime }\nu ^{\prime }}^{I}=\delta _{\mu ^{\prime }}A_{\nu ^{\prime
}}^{I}-\delta _{\nu ^{\prime }}A_{\mu ^{\prime }}^{I},
\end{eqnarray*}%
where the scalars $Y_{\widehat{\alpha }}^{I}$ coming from the
ten--dimensional vectors should be associated to a normal Higgs phenomenon
generating a mass matrix for the gauge fields. Thy are related to the fact
that a non--Abelian gauge field strength contains nonlinear terms not being
certain derivatives of potentials.

After a straightforward calculus of the actions' components (\ref{rterm}), (%
\ref{hterm}) and (\ref{fterm}) (for locally isotropic gauge theories and
strings, see a similar calculus, for instance, in Refs. \cite{14kir}), putting
everything together, we can write the $n^{\prime }+m^{\prime }$ dimensional
action including anholonomic interactions in the form%
\begin{eqnarray}
S_{n^{\prime }+m^{\prime }}^{heterotic} &=&\int \delta ^{n^{\prime
}+m^{\prime }}u\sqrt{|g_{\alpha ^{\prime }\beta ^{\prime }}|}e^{-\phi }[%
\widehat{R}^{\prime }+S^{\prime }+(\delta _{\mu ^{\prime }}\phi )(\delta
^{\mu ^{\prime }}\phi )-\frac{1}{12}H^{\mu ^{\prime }\nu ^{\prime }\rho
^{\prime }}H_{\mu ^{\prime }\nu ^{\prime }\rho ^{\prime }}  \notag \\
&&-\frac{1}{4}(M^{-1})_{\overline{I}\overline{J}}F_{\mu ^{\prime }\nu
^{\prime }}^{\overline{I}}F^{\overline{J}\mu ^{\prime }\nu ^{\prime }}+\frac{%
1}{8}Tr\left( \delta _{\mu ^{\prime }}M\delta ^{\mu ^{\prime }}M^{-1}\right)
],  \label{act7}
\end{eqnarray}%
where ${\overleftarrow{R}}^{\prime }{=}\widehat{R}^{\prime }+S^{\prime }$ is
the d--scalar curvature of type (\ref{4dscalar}) induced after toroidal
compactification, the $\left( 2p+16\right) \times \left( 2p+16\right) $
dimensional symmetric matrix $M$ has the structure%
\begin{equation*}
M=\left(
\begin{array}{ccc}
\underline{g}^{-1} & \underline{g}^{-1}C & \underline{g}^{-1}Y^{t} \\
C^{t}\underline{g}^{-1} & \underline{g}+C^{t}\underline{g}^{-1}C+Y^{t}Y &
C^{t}\underline{g}^{-1}Y^{t}+Y^{t} \\
Y\underline{g}^{-1} & Y\underline{g}^{-1}C+Y & I_{16}+Y\underline{g}%
^{-1}Y^{t}%
\end{array}%
\right)
\end{equation*}%
with the block sub--matrices
\begin{equation*}
\underline{g}=\left( \underline{g}_{\alpha \beta }\right) ,C=\left( C_{%
\widehat{\alpha }\widehat{\beta }}=B_{\widehat{\alpha }\widehat{\beta }}-%
\frac{1}{2}Y_{\widehat{\alpha }}^{I}Y_{\widehat{\beta }}^{I}\right)
,Y=\left( Y_{\widehat{\alpha }}^{I}\right) ,
\end{equation*}%
for which $I_{16}$ is the 16 dimensional unit matrix; for instance, $Y^{t}$
denotes the transposition of the matrix $Y.$ The dimension $p$ satisfies the
condition $n^{\prime }+m^{\prime }-p=16$ relevant to the heterotic string
describing $p$ left--moving bosons and $n^{\prime }+m^{\prime }$
right--moving ones with $m^{\prime }$ constrained degrees of freedom. To
have good modular properties $p-n^{\prime }-m^{\prime }$ should be a
multiple of eight. The indices $\overline{I},\overline{J}$ run values $%
1,2,...\left( 2p+16\right) .$ The action (\ref{act7}) describes a heterotic
string effective action with local anisotropies (contained in the values $%
\widehat{R}^{\prime },S^{\prime }$ and $\delta _{\mu ^{\prime
}})$ induced by the fact that the dynamics of the right--moving
bosons are subjected to certain constraints. The induced metric
$g_{\alpha ^{\prime }\beta ^{\prime }}$ is of type
(\ref{7dmetric}) given with respect  to an N--elongated basis
(\ref{7ddif}) (in this case,  primed), $\delta _{\mu ^{\prime
}}=\partial _{\mu ^{\prime }}+N_{\mu ^{\prime }}.$ For
$N_{i^{\prime }}^{a^{\prime }}\rightarrow 0$ and $m^{\prime },$
i. e. for a subclass of effective backgrounds with block
$n^{\prime }\times n^{\prime }\oplus m^{\prime }\times m^{\prime
}$ metrics $g_{\alpha ^{\prime }\beta ^{\prime }},$ the action
(\ref{act7}) transforms in the well known isotropic form (see,
for instance, formula (C22), from the Appendix C in Ref.
\cite{14kir}, from which
following the 'geometric d--covariant rule' we could write down directly (%
\ref{act7}); this is a more formal approach which hides the physical meaning
and anholonomic character of the components (\ref{rterm}), (\ref{hterm}) and
(\ref{fterm})).

\subsection{4D NS--NS anholonomic field equations}

As a matter of principle, compactifications of all type in (super) string
theory can be performed in such ways as to include anholonomic frame effects
as in the previous subsection. The simplest way to define anisotropic
generalizations or such models is to apply the 'geometric d--covariant rule'
when the tensors, spinors and connections are changed into theirs
corresponding N--distinguished analogous. As an example, we write down here
the anholonomic variant of the toroidally compactified (from ten to four
dimensions) NS--NS action (we write in brief NS instead of Neveu--Schwarz) %
\cite{14cosm},%
\begin{equation}
S=\int \delta ^{4}u\sqrt{|g_{\alpha ^{\prime }\beta ^{\prime }}|}e^{-\varphi
}[\widehat{R}^{\prime }+S^{\prime }+(\delta _{\mu ^{\prime }}\phi )(\delta
^{\mu ^{\prime }}\phi )-\frac{1}{2}(\delta _{\mu ^{\prime }}\beta )(\delta
^{\mu ^{\prime }}\beta )-\frac{1}{2}e^{2\varphi }(\delta _{\mu ^{\prime
}}\sigma )(\delta ^{\mu ^{\prime }}\sigma )],  \label{act9}
\end{equation}%
for a d--metric parametrized as
\begin{equation*}
\delta s^{2}=-\epsilon \delta (x^{0^{\prime }})^{2}+g_{\underline{\alpha
^{\prime }}\underline{\beta ^{\prime }}}\delta u^{\underline{\alpha ^{\prime
}}}\delta u^{\underline{\beta ^{\prime }}}+e^{\beta /\sqrt{3}}\delta _{%
\widehat{\alpha }\widehat{\gamma }}\delta u^{\widehat{\alpha }}\delta u^{%
\widehat{\beta }},
\end{equation*}%
where, for instance, $u^{\alpha ^{\prime }}=(x^{0^{\prime }},u^{\underline{%
\alpha ^{\prime }}}),\underline{\alpha ^{\prime }}=1,2,3$ and $\widehat{%
\alpha },\widehat{\beta },...=4,5,...9$ are indices of extra dimension
coordinates, $\epsilon =\pm 1$ depending on signature (in usual string
theory one takes $x^{0^{\prime }}=t$ and $\epsilon =-1),$ the modulus field $%
\beta $ is normalized in such a way that it becomes minimally coupled to
gravity in the Einstein d--frame, $\sigma $ is a pseudo--scalar axion
d--field, related with the anti--symmetric strength,%
\begin{equation*}
H^{\alpha ^{\prime }\beta ^{\prime }\gamma ^{\prime }}(u^{\alpha ^{\prime
}})=\varepsilon ^{\alpha ^{\prime }\beta ^{\prime }\gamma ^{\prime }\tau
^{\prime }}e^{\varphi (u^{\alpha ^{\prime }})}D_{\tau ^{\prime }}\sigma
(u^{\alpha ^{\prime }}),
\end{equation*}%
$\varepsilon ^{\alpha ^{\prime }\beta ^{\prime }\gamma ^{\prime }\tau
^{\prime }}$ being completely antisymmetric and $\varphi (u^{\alpha ^{\prime
}})=\Phi ^{\prime }(u^{\alpha ^{\prime }})-\sqrt{3}\beta (u^{\alpha ^{\prime
}}),$ with $\Phi ^{\prime }(u^{\alpha ^{\prime }})$ taken as in (\ref{act6}).

We can derive certain locally anisotropic field equations from the action (%
\ref{act9}) by varying with respect to N--adapted frames for massless
excitations of $g_{\alpha ^{\prime }\beta ^{\prime }},B_{\alpha ^{\prime
}\beta ^{\prime }},\beta $ and $\varphi ,$ which are given by%
\begin{eqnarray}
2\left[ R_{\mu ^{\prime }\nu ^{\prime }}-\frac{1}{2}\left( \widehat{R}%
^{\prime }+S^{\prime }\right) g_{\mu ^{\prime }\nu ^{\prime }}\right] =\frac{%
1}{2}H_{\mu ^{\prime }\lambda ^{\prime }\tau ^{\prime }}H_{\nu ^{\prime
}}^{~~\lambda ^{\prime }\tau ^{\prime }}-H^{2}g_{\mu ^{\prime }\nu ^{\prime
}}+ &&  \label{eqfstr} \\
\left( \delta _{\mu ^{\prime }}^{\lambda ^{\prime }}\delta _{\nu ^{\prime
}}^{\tau ^{\prime }}-\frac{1}{2}g_{\mu ^{\prime }\nu ^{\prime }}g^{\lambda
^{\prime }\tau ^{\prime }}\right) D_{\lambda ^{\prime }}\beta D_{\tau
^{\prime }}\beta -g_{\mu ^{\prime }\nu ^{\prime }}(D\varphi )^{2}+2\left(
g_{\mu ^{\prime }\nu ^{\prime }}g^{\lambda ^{\prime }\tau ^{\prime }}-\delta
_{\mu ^{\prime }}^{\lambda ^{\prime }}\delta _{\nu ^{\prime }}^{\tau
^{\prime }}\right) D_{\lambda ^{\prime }}D_{\tau ^{\prime }}\varphi &=&0,
\notag \\
D_{\mu ^{\prime }}\left( e^{-\varphi }H^{\mu ^{\prime }\nu ^{\prime }\lambda
^{\prime }}\right) &=&0,  \notag \\
D_{\mu ^{\prime }}\left( e^{-\varphi }D^{\mu ^{\prime }}\beta \right) &=&0,
\notag \\
2D_{\mu ^{\prime }}D^{\mu ^{\prime }}\varphi =-\widehat{R}^{\prime
}-S^{\prime }+(D\varphi )^{2}+\frac{1}{2}(D\beta )^{2}+\frac{1}{12}H^{2}
&=&0,  \notag
\end{eqnarray}%
where $H^{2}=H_{\mu ^{\prime }\lambda ^{\prime }\tau ^{\prime }}H^{~\mu
^{\prime }~\lambda ^{\prime }\tau ^{\prime }}$ and, for instance, $(D\varphi
)^{2}=D_{\mu ^{\prime }}\varphi D^{\mu ^{\prime }}\varphi .$ We may select a
consistent solution of these field equations when the internal space is
static with $D_{\mu ^{\prime }}\beta =0.$

The equations (\ref{eqfstr}) can be decomposed in invariant h--
and v--components like the Einstein d--equations (\ref{3einsteq2})
(we omit a such trivial calculus). We recall \cite{14deligne} that
the NS--NS sector is common to both the heterotic \ and type II
string theories and is comprised of the dilaton, graviton and
antisymmetric two--form potential. The obtained equations
(\ref{eqfstr}) \ define respective anisotropic string corrections
to the anholonomic Einstein gravity.

\subsection{Distinguishing anholonomic Riemannian--Finsler \newline
(super\-) gra\-vities}

There are two classes of general anisotropies contained in supergravity and
superstring effective actions:

\begin{itemize}
\item Generic local anisotropies contained in the higher dimension (11, for
supergravity models, or 10, for superstring models) which can be also
induced in lower dimension after compactification (like it was considered
for actions (\ref{act4}), (\ref{act4b}), (\ref{act5}) and (\ref{act6})).

\item Local anisotropies which are in induced on the lower dimensional
spacetime (for instance, actions (\ref{act7}) and (\ref{act9}) and
respective field equations).
\end{itemize}

All types of general supergravity/superstring anisotropies may be in their
turn to be distinguished to be of ''pure'' $B$--field origin, of ''pure''
anholonomic frame origin with arbitrary $B$--field, or of a mixed type when
local anisotropies are both induced in a nonlinear form by both anholonomic
(super) vielbeins and $B$--field (like we considered in subsection \ref{efs}
for bosonic strings). In explicit form, a model of locally anisotropic
superstring corrected gravity is to be constructed following the type of
parametrizations we establish for the N--coefficients, d--metrics and
d--connections.

For instance, if we choose the frame ansatz (\ref{2frame8}) and corresponding
metric ansatz (\ref{2metr8}) with general coefficients $g_{i^{\prime
}j^{\prime }}(x^{j^{\prime }},y^{c^{\prime }}),h_{a^{\prime }b^{\prime
}}(x^{j^{\prime }},y^{a^{\prime }})$ and $N_{i^{\prime }}^{a^{\prime
}}(x^{j^{\prime }},y^{a^{\prime }})$ satisfying the effective field
equations (\ref{eqfstr}) (containing also the fields $H_{\mu ^{\prime
}\lambda ^{\prime }\tau ^{\prime }},\varphi $ and $\beta )$ we define an
anholonomic gravity model corrected by toroidally compactified (from ten to
four dimensions) NS--NS \ superstring model. In four and five dimensional
Einstein/ Kaluza--Klein gravities, there were constructed a number of
anisotropic black hole, wormhole, solitonic, spinor waive and Taub/NUT
metrics \cite{14vexsol,14vmethod,14vbel,14vsingl,14vsolsp}; in section
 \ref{1exsol} we
shall consider some generalizations to string gravity.

Another possibility is to impose the condition that $g_{i^{\prime }j^{\prime
}},h_{a^{\prime }b^{\prime }}$ and $N_{i^{\prime }}^{a^{\prime }}$ are of
Finsler type, $g_{i^{\prime }j^{\prime }}^{[F]}=h_{i^{\prime }j^{\prime
}}^{[F]}=\partial ^{2}F^{2}/2\partial y^{i}\partial y^{j}$ (\ref{fmetric})
and $N_{j}^{[F]i}(x,y)=\partial \left[ c_{lk}^{\iota }(x,y)y^{l}y^{k}\right]
/$ $4\partial y^{j}$ (\ref{2ncc}), with an effective d--metric (\ref{2dmetricf}%
). If a such set of metric/N--connection coefficients can \ found as a
solution of some string gravity equations, we may construct a lower
dimensional Finsler gravity model induced from string theory (it depends of
what kind of effective action, (\ref{act7}) or (\ref{act9}), we consider).
Instead of a Finsler gravity we may search for a Lagrange model of string
gravity if the d--metric coefficients are taken in the form (\ref{1mfl}).

We conclude this section by a remark that we may construct various type of
anho\-lonomic Riemannian and generalized Finsler/Lagrange string gravity
models, with aniso\-tropies in higher and/or lower dimensions by prescribing
corresponding parametrizations for $g_{ij},h_{ab}$ and $N_{i}^{a}$ (for
'higher' anisotropies) and $g_{i^{\prime }j^{\prime }},h_{a^{\prime
}b^{\prime }}$ and $N_{i^{\prime }}^{a^{\prime }}$ (for 'lower'
anisotropies). The anholonomic structures may be of mixed type, for
instance, in some dimensions being of Finsler configuration, in another ones
being with anholonomic Riemannian metric, in another one of Lagrange type
and different combinations and generalizations, see explicit examples in
Section \ref{1exsol}.

\section{ Noncommutative Anisotropic Field Interacti\-ons}

We define the noncommutative field theory in a new form when spacetimes and
configuration spaces are provided with some anholonomic frame and associated
N--connection structures. The equations of motions are derived from
functional integrals in a usual manner but considering N--elongated partial
derivatives and differentials.

\subsection{Basic definitions and conventions}

The basic concepts on noncommutative geometry are outlined here in a
somewhat pedestrian way by emphasizing anholonomic structures. More rigorous
approaches on mathematical aspects of noncommutative geometry may be found
in Refs. \ \cite{14nc,14majid,14qg,14landi}, physical versions are given in Refs. %
\cite{14dn,14strncg,14connes1,14sw} \ (the review \cite{14vncf} is a synthesis of
results on noncommutative geometry, N--connections and Finsler geometry,
Clifford structures and anholonomic gauge gravity \ based on monographs \ %
\cite{14landi,14vmon1,14vmon2,14ma}).

As a fundamental ingredient we use an associative, in general,
noncommutative algebra $\mathcal{A}$ with a product of some elements $a,b\in
\mathcal{A}$ denoted $ab=a\cdot b,$ or in the conotation to noncommutative
spaces, written as a ''star'' product $ab=a\star b.$ Every element $a\in
\mathcal{A}$ corresponds to a configuration of a classical complex scalar
field on a ''space'' $M,$ a topological manifold, which (in our approach)
can be enabled with a N--connection structure. This associated
noncommutative algebra generalize the algebra of complex valued functions $%
\mathcal{C}(M)$ on a manifold $M$ (for different theories we may consider
instead $M$ a tangent bundle $TM,$ or a vector bundle $E\left( M\right) ).$
We consider that all functions referring to the algebra $\mathcal{A},$ $\ $%
denoted as $\mathcal{A}\left( M\right) ,$ arising in  physical
considerations are of necessary smooth class (continuous, smooth, subjected to
certain bounded conditions etc.).

\subsubsection{Matrix algebras and noncommutativity}

As the most elementary examples of noncommutative algebras, which are
largely applied in quantum field theory and noncommutative geometry, one
considers the algebra $Mat_{k}(\C)$ of complex $k\times k$ matrices and the
algebra $Mat_{k}\left( \mathcal{C}(M)\right) $ of $k\times k$ matrices whose
matrix elements are elements of $\mathcal{C}(M).$ The last algebra may be
also defined as a tensor product,
\begin{equation*}
Mat_{k}\left( \mathcal{C}(M)\right) =Mat_{k}(\C)\otimes \mathcal{C}(M).
\end{equation*}%
The last construction is easy to be generalized for arbitrary noncommutative
algebra $\mathcal{A}$ as
\begin{equation*}
Mat_{k}\left( \mathcal{A}\right) =Mat_{k}(\C)\otimes \mathcal{A},
\end{equation*}
which is just the algebra of $k\times k$ matrices with elements in $\mathcal{%
A}.$ The algebra $Mat_{k}\left( \mathcal{A}\right) $ admits an authomorphism
group $GL(k,\C)$ with the action defined as $a\rightarrow \varsigma
^{-1}a\varsigma ,$ for $a\in \mathcal{A},\varsigma \in GL(k,\C).$ One
considers the subgroup $U\left( k\right) \subset GL(k,\C)$ which is
preserved by hermitian conjugations, $a\rightarrow a^{+},$ and reality
conditions, $a=a^{+}.$ To define the hermitian conjugation, for which the
hermitian matrices $a=a^{+}$ have real eigenvalues, it is considered that $%
\left( a^{+}\right) ^{+}=a$ and $\left( ca\right) ^{+}=c^{\ast }a^{+},$ for $%
c\in \C $ and $c^{\ast }$ being the complex conjugated element of $c,$ i. e.
it defined an antiholomorphic involution.

\subsubsection{Noncommutative Euclidean space $\R_{\protect\theta }^{k}$}

Another simple example of a noncommutative space is the 'noncommutative
Euclidean space' $\R_{\theta }^{k}$ defined by all complex linear
combinations of products of variables $x=\{x^{j}\}$ satisfying%
\begin{equation}
\lbrack x^{j},x^{l}]=x^{j}x^{l}-x^{l}x^{j}=i\theta ^{jl},  \label{nceucl}
\end{equation}%
where $i$ is the complex 'imaginary' unity and $\theta ^{jl}$ are real
constants treated as some noncommutative parameters or a ''Poison tensor''
by analogy to the Poison bracket in quantum mechanics where the commutator $%
\left[ ...\right] $ of hermitian operators is antihermitian. A set of
partial derivatives $\partial _{j}=\partial /\partial x^{i}$ on $\R_{\theta
}^{k}$ can be defined by postulating the relations
\begin{eqnarray}
\partial _{j}x^{n} &=&\delta _{j}^{n},  \notag \\
\lbrack \partial _{j},\partial _{n}] &=&-i\Xi _{jn}  \label{nceucder}
\end{eqnarray}%
where $\Xi _{jn}$ may be zero, but in general is non--trivial if we wont to
incorporate some additional magnetic fields or anholonomic relations. A
simplified noncommutative differential calculus can be constructed if $\Xi
_{jn}=-\left( \theta ^{-1}\right) _{jn}.$

The metric structure on $\R_{\theta }^{k}$ is stated by a constant symmetric
tensor $\eta _{nj}$ for which $\partial _{j}\eta _{nj}=0.$

Infinitesimal translations $x^{j}\rightarrow x^{j}+a^{j}$ on $\R_{\theta
}^{k}$ are defined as actions on functions $\varphi $ of type $%
\bigtriangleup \varphi =a^{j}\partial _{j}\varphi .$ Because the coordinates
are non--commuting there are formally defined inner derivations as%
\begin{equation}
\partial _{j}\varphi =\left[ -i\left( \theta ^{-1}\right) _{jn}x^{n},\varphi %
\right]  \label{inder}
\end{equation}%
which result in exponential global tanslations%
\begin{equation*}
\varphi \left( x^{j}+\epsilon ^{j}\right) =e^{-i\theta _{lj}\epsilon
^{l}x^{j}}\varphi \left( x^{j}\right) e^{i\theta _{lj}\epsilon ^{l}x^{j}}.
\end{equation*}

In order to understand the symmetries of the space $\R_{\theta }^{k}$ it is
better to write the metric and Poisson tensor in the forms%
\begin{eqnarray}
ds^{2} &=&\sum_{A=1}^{r}dz_{A}d\overline{z}_{A}+\sum_{B}dy_{B}^{2},
\label{strnce} \\
&=&dq_{A}^{2}+dp_{A}^{2}+dy_{B}^{2};  \notag \\
&&  \notag \\
\theta &=&\frac{1}{2}\sum_{A=1}\theta _{A}\partial _{z_{A}}\wedge \partial _{%
\overline{z}_{A}},~\theta _{A}>0,  \notag
\end{eqnarray}%
where $z_{A}=q_{A}+ip_{A}$ $\ $\ and $\ \overline{z}_{A}=q_{A}-ip_{A}$ are
some convenient complex coordinates for which there are satisfied the
commutation rules%
\begin{eqnarray*}
\left[ y_{A},y_{B}\right] &=&\left[ y_{B},q_{A}\right] =\left[ y_{B},p_{A}%
\right] =0, \\
\left[ q_{A},p_{B}\right] &=&i\theta _{A}\delta _{AB}.
\end{eqnarray*}%
Now, it is obvious that for fixed types of metric and Poisson structures (%
\ref{strnce}) there are two symmetry groups on $\R_{\theta }^{k},$ the group
of rotations, denoted $O\left( k\right) ,$ and the group of invariance of
the form $\theta ,$ denoted $Sp\left( 2r\right) .$

\subsubsection{The noncommutative derivative and integral}

In order to elaborate noncommutative field theories in terms of an
associative noncommutative algebra $\mathcal{A},$ additionally to the
derivatives $\partial _{j}$ we need an integral $\int Tr$ which following
the examples of noncommutative matrix spaces must contain also the ''trace''
operator. In this case we can not separate the notations of trace and
integral.

It should be noted here that the role of derivative $\partial _{j}$ can be
played by any sets of elements $d_{j}\in \mathcal{A}$ which some formal
derivatives as $\partial _{j}A=\left[ d_{j},A\right] ,$ for $A\in $ $%
\mathcal{A};$ derivations written in this form are called as inner
derivations while those which can not written in this form are referred to
as outer derivations.

The general derivation and integration operations are defined as some
general dual linear operators satisfying certain formal properties: 1) the
Leibnitz rule of the derivative, $\partial _{j}(AB)=\partial
_{j}(A)B+A(\partial _{j}B);$ 2) the integral of the trace of a total
derivative is zero, $\int Tr\partial _{j}A=0;$ 3) the integral of the trace
of a commutator is zero, $\int Tr[A,B]=0,$ for any $A,B\in $ $\mathcal{A}.$
For some particular classes of functions in some noncommutative models the
condition 2) and/or 3) may be violated, see details and discussion in Ref. %
\cite{14dn}.

Given a noncommutative space induced by some relations (\ref{nceucl}), the
algebra of functions on $\R^{k}$ is deformed on $\R_{\theta }^{k}$ such that%
\begin{eqnarray}
f\left( x\right) \star \varphi \left( x\right) &=&e^{\frac{i}{2}\theta ^{jk}%
\frac{\partial }{\partial \xi ^{j}}\frac{\partial }{\partial \zeta ^{k}}%
}f\left( x+\xi \right) \varphi \left( x+\zeta \right) _{|\xi =\varsigma =0}
\notag \\
&=&f\varphi +\frac{i}{2}\theta ^{jk}\partial _{j}f\partial _{k}\varphi
+o\left( \theta ^{2}\right) ,  \label{moyal}
\end{eqnarray}%
which define the Moyal bracket (product), or star product ($\star $%
--product), of functions which is associative compatible with integration in
the sense that for matrix valued functions $f$ and $\varphi $ that vanish
rapidly enough at infinity we can integrate by parts in the integrals
\begin{equation*}
\int Tr~f\star \varphi =\int Tr~\varphi \star f.
\end{equation*}

In a more rigorous operator form the star multiplication is defined by
considering a space $M_{\theta },$ locally covered by coordinate carts with
noncommutative coordinates (\ref{nceucl}), and choosing a linear map $S$
from $M_{\theta }$ to $\mathcal{C}(M),$ called the ''symbol'' of the
operator, when $\widehat{f}$ $\rightarrow S\left[ \widehat{f}\right] .$ This
way, the original operator multiplication is expressed in terms of the star
product of symbols as
\begin{equation*}
\widehat{f}\widehat{\varphi }=S^{-1}\left[ S\left[ \widehat{f}\right] \star S%
\left[ \widehat{\varphi }\right] \right] .
\end{equation*}%
It should be noted that there could be many valid definitions of $S,$
corresponding to different choices of operator ordering prescription for $%
S^{-1}.$ One writes, for simplicity, $\int Tr~f\star \varphi =\int
Tr~f\varphi $ in some special cases.

\subsection{Anholonomic frames and noncommutative spacetimes}

One may consider that noncommutative relations for coordinates and
partial derivatives (\ref{nceucl}) and (\ref{nceucder}) are
introduced by specific form of anholonomic relations (\ref{4anhol})
for some formal anholonomic frames of type (\ref{6dder}) and/or
(\ref{7ddif}) (see Appendix) when anholonomy coefficients are
complex and depend nonlinearly on frame coefficients. We shall
not consider in this work the method of complex nonlinear operator
anholonomic frames with associated nonlinear connection structure,
containing as particular cases various type of Finsler/Cartan and
Lagrange/Hamilton geometries in complexified form, which could
consist in a general complex geometric formalism for
noncommutative theories but we shall restrict our analysis to
noncommutative spaces for which the coordinates and partial
derivatives are distinguished by a N--connection structure into
certain holonomic and anholonomic subsets which generalize the
N--elongated \ commutative differential calculus (considered in
the previous Sections) to a variant of both $N$-- and $\theta
$--deformed one.

In order to emphasize the N--connection structure on respective spaces we
shall write $M_{\theta }^{N},TM_{\theta }^{N},E_{\theta }^{M}\left(
M_{\theta }\right) ,$ $\mathcal{C}(M^{N}),\mathcal{A}^{N}$ and $\mathcal{A}%
\left( M^{N}\right) .$ For a space $M^{N}$ provided with N--connection
structure, the matrix algebras considered in the previous subsection may be
denoted $Mat_{k}\left( \mathcal{C}(M^{N})\right) $ and $Mat_{k}\left(
\mathcal{A}^{N}\right) .$

\subsubsection{Noncommutative anholonomic derivatives}

\label{torus}We introduce splitting of indices, $\alpha =\left( i,a\right) ,$
$\beta =\left( j,b\right) ,...,$ and coordinates, $u^{\alpha }=\left(
x^{i},y^{a}\right) ,...,$ into 'horizontal' and 'vertical' components for a
space $M_{\theta }$ (being in general a manifold, tangeng/vector bundle, or
their duals, or higher order models \cite{14miron,14vspinors,14vstr2,14vmon1,14vmon2}.
\ The derivatives $\partial _{i}$ satisfying the conditions (\ref{nceucder})
must be changed into some N--elongated ones if both anholnomy and
noncommutative structures are introduced into consideration.

In explicit form, the anholonomic analogous of (\ref{nceucl}) is stated by a
set of coordinates $u^{\alpha }=\left( x^{i},y^{a}\right) $ satisfying the
relations
\begin{equation}
\lbrack u^{\alpha },u^{\beta }]=i\Theta ^{\alpha \beta },  \label{nceucln}
\end{equation}%
with $\Theta ^{\alpha \beta }=\left( \Theta ^{ij},\Theta ^{ab}\right) $
parametrized as to have a noncommutative structure locally adapted to the
N--connection, and the analogouses of (\ref{nceucder}) redefined for
operators (\ref{6dder}) as
\begin{eqnarray}
\delta _{\alpha }u^{\beta } &=&\delta _{\alpha }^{\beta },\mbox{
for }\delta _{\alpha }=\left( \delta _{i}=\partial
_{i}-N_{i}^{a}\partial _{a}, \partial _{b} \right) ,
\notag \\
\lbrack \delta _{\alpha },\delta _{\beta }] &=&-i\Xi _{\alpha \beta },
\label{nceucdern}
\end{eqnarray}%
where $\Xi _{\alpha \beta }=-\left( \Theta ^{-1}\right) _{\alpha \beta }$
for a simplified N--elongated noncommutative differential calculus. We
emphasize that if the vielbein transforms of type (\ref{vielbtr}) and frames
of type (\ref{dder1a}) and (\ref{ddif1a}) are considered, the values $\Theta
^{\alpha \beta }$ and $\Xi _{\alpha \beta }$ could be some complex functions
depending on variables $u^{\beta }$ including also the anholonomy
contributions of\ $N_{i}^{a}.$ In particular cases, they my constructed by
some anholonomic frame transforms from some constant real tensors.

An anholonomic noncommutative Euclidean space $\R_{N,\theta }^{n+m}$ is
defined as a usual one of dimension $k=n+m$ for which a N--connection
structure is prescribed by coefficients $N_{i}^{a}\left( x,y\right) $ which
states an N--elongated differential calculus. The d--metric $\eta _{\alpha
\beta }=\left( \eta _{ij},\eta _{ab}\right) $ and Poisson d--tensor $\Theta
^{\alpha \beta }=\left( \Theta ^{ij},\Theta ^{ab}\right) $ are introduced
via vielbein transforms (\ref{vielbtr}) depending on N--coefficients of the
corresponding constant values contained in (\ref{nceucl}) and (\ref{strnce}%
). As a matter of principle such noncommutative spaces are already curved.

The interior derivative (\ref{inder}) is to be extended on $\R_{N,\theta
}^{n+m}$ as%
\begin{equation*}
\delta _{\alpha }\varphi =\left[ -i\left( \Theta ^{-1}\right) _{\alpha \beta
}u^{\beta },\varphi \right] .
\end{equation*}%
In a similar form, by introducing operators $\delta _{\alpha }$ instead of $%
\partial _{\alpha },$ we can generalize the Moyal product (\ref{moyal}) for
anisotropic spaces:
\begin{eqnarray*}
f\left( x\right) \star \varphi \left( x\right) &=&e^{\frac{i}{2}\Theta
^{\alpha \beta }\frac{\delta }{\partial \xi ^{\alpha }}\frac{\delta }{%
\partial \zeta ^{\beta }}}f\left( x+\xi \right) \varphi \left( x+\zeta
\right) _{|\xi =\varsigma =0} \\
&=&f\varphi +\frac{i}{2}\Theta ^{\alpha \beta }\delta _{\alpha }f\delta
_{\beta }\varphi +o\left( \theta ^{2}\right) .
\end{eqnarray*}

\bigskip For elaborating of perturbation and scattering theory, the more
useful basis is the plane wave basis, which for anholonomic noncommutative
Euclidean spaces, consists of eigenfunctions of the derivatives%
\begin{equation*}
\delta _{\alpha }e^{ipu}=ip_{\alpha }e^{ipu},pu=p_{\alpha }u^{\alpha }.
\end{equation*}%
In this basis, the integral can be defined as
\begin{equation*}
^{N}\int Tr~e^{ipu}=\delta _{p,0}
\end{equation*}%
where the symbol $\int Tr$ is enabled with the left upper index $N$ in order
to emphasize that integration is to be performed on a N--deformed space (we
shall briefly call this as ''N--integration'') and the delta function may be
interpreted as usually (its value at zero represents the volume of physical
space, in our case, N--deformed). There is a specific multiplication low
with respect to the plane wave basis: for instance, by operator reordering,
\begin{equation*}
e^{ipu}\cdot e^{ip^{\prime }u}=e^{-\frac{1}{2}\Theta ^{\alpha \beta
}p_{\alpha }p_{\beta }^{\prime }}~e^{i\left( p+p^{\prime }\right) u},
\end{equation*}%
when $\Theta ^{\alpha \beta }p_{\alpha }p_{\beta }^{\prime }$ may be written
as $p\times p^{\prime }\equiv \Theta ^{\alpha \beta }p_{\alpha }p_{\beta
}^{\prime }=p\times _{\Theta }p^{\prime }.$ There is another example of
multiplication, when the N--elongated partial derivative is involved,
\begin{equation*}
e^{ipu}\cdot f\left( u\right) \cdot e^{-ipu}=e^{-\Theta ^{\alpha \beta
}p_{\alpha }\delta _{\beta }}f\left( u\right) =f\left( u^{\beta }-\Theta
^{\alpha \beta }p_{\alpha }\right) ,
\end{equation*}%
which shows that multiplication by a plane wave in anholonomic
noncommutative Euclidean space translates and N--deform a general function
by $u^{\beta }\rightarrow u^{\beta }-\Theta ^{\alpha \beta }p_{\alpha }.$
This exhibits both the nonlocality and anholonomy of the theory and
preserves the principles that large momenta lead to large nonlocality which
can be also locally anisotropic.

\subsubsection{Noncommutative anholonomic torus}

Let us define the concept of noncommutative anholonomic torus, $\mathbf{T}%
_{N,\theta }^{n+m},$ i. e. the algebra of functions on a noncommutative
torus with some splitting of coordinates into holonomic and anholonomic
ones. We note that a function $f$ on a anholonomic torus $\mathbf{T}%
_{N}^{n+m}$ with N--decomposition is a function on $\R$$_{N}^{n+m}$ which
satisfies a periodicity condition, $f\left( u^{\alpha }\right) =f\left(
u^{\alpha }+2\pi z^{\alpha }\right) $ for d--vectors $z^{\alpha }$ with
integer coordinates. Then the noncommutative extension is to define $\mathbf{%
T}_{N,\theta }^{n+m}$ as the algebra of all sums of products of arbitrary
integer powers of the set of distinguished $n+m$ variables $U_{\alpha
}=\left( U_{i},U_{j}\right) $ satisfying%
\begin{equation}
U_{\alpha }U_{\beta }=e^{-i\Theta ^{\alpha \beta }}U_{\beta }U_{\alpha }.
\label{nctorus}
\end{equation}%
The variables $U_{\alpha }$ are taken instead of $e^{iu^{\alpha }}$ for
plane waves and the derivation of a Weyl algebra from (\ref{nceucln}) is
possible if we take
\begin{eqnarray*}
\lbrack \delta _{\alpha }U_{\beta }] &=&i\delta _{\alpha \beta }U_{\beta },
\\
^{N}\int Tr~U_{1}^{z_{1}}...U_{n+m}^{z_{n+m}} &=&\delta _{\overrightarrow{z}%
,0}.
\end{eqnarray*}%
In addition to the usual topological aspects for nontrivial values of $N$%
--connection there is much more to say in dependence of the fact what type
of topology is induced by the $N$--connection curvature. We omit such
consideration in this paper. The introduced in this subsection formulas and
definitions transform into usual ones from noncommutative geometry if $%
N,m\rightarrow 0.$

\subsection{Anisotropic field theories and anholonomic symmetries}

In a formal sense, every field theory, commutative or noncommutative, can be
anholonomically transformed by changing partial derivatives into
N--elongated ones and redefining the integrating measure in corresponding
Lagrangians. We shall apply this rule to noncommutative scalar, gauge and
Dirac fields and make them to be locally anisotropic and to investigate
their anholonomic symmetries.

\subsubsection{Locally anisotropic matrix scalar field theory}

A generic matrix locally anisotropic matrix scalar field theory with a
hermitian matrix valued field $\phi \left( u\right) =\phi ^{+}\left(
u\right) $ and anholonomically N--deformed Euclidean action%
\begin{equation*}
S=~^{N}\int \delta ^{n+m}u\sqrt{|g_{\alpha \beta }|}\left[ \frac{1}{2}%
g^{\alpha \beta }Tr~\delta _{\alpha }\phi ~\delta _{\beta }\phi +V\left(
\phi \right) \right]
\end{equation*}%
where $V\left( \phi \right) $ is polynomial in variable $\phi ,g_{\alpha
\beta }$ is a d--metric of type (\ref{7dmetric}) and $\delta _{\alpha }$ are
N--elongated partial derivatives (\ref{6dder}). It is easy to check that if
we replace the matrix algebra by a general associative noncommutative
algebra $\mathcal{A},$ the standard procedure of derivation of motion
equations, classical symmetries from Noether's theorem and related physical
considerations go through but with N--elongated partial derivatives and
N--integration: The field equations are
\begin{equation*}
g^{\alpha \beta }\delta _{\alpha }~\delta _{\beta }\phi =\frac{\partial
V\left( \phi \right) }{\partial \phi }
\end{equation*}%
and the conservation laws
\begin{equation*}
\delta _{\alpha }J^{\alpha }=0
\end{equation*}%
for the current $J^{\alpha }$ is associated to a symmetry $\bigtriangleup
\phi \left( \epsilon ,\phi \right) $ determined by the N--adapted
variational procedure, $\bigtriangleup S=~^{N}\int Tr~J^{\alpha }\delta
_{\alpha }\epsilon .$ We emphasize that these equations are obtained
according the prescription that we at the first stage perform a usual
variational calculus then we change the usual derivatives and differentials
into N--elongated ones. If we treat the N--connection as an object which
generates and associated linear connection with corresponding curvature we
have to introduce into the motion equations and conservation laws necessary
d--covariant objects curvature/torsion terms.

We may define the momentum operator
\begin{equation*}
P_{\alpha }=-i\left( \Theta ^{-1}\right) _{\alpha \beta }~^{N}\int
Tr~u^{\beta }T^{0},
\end{equation*}%
which follows from the anholonomic transform of the restricted \
stress--energy tensor' constructed from the Noether procedure with
symmetries $\bigtriangleup \phi =i[\phi ,\epsilon ]$ resulting in
\begin{equation*}
T^{\alpha }=ig^{\alpha \beta }[\phi ,\delta _{\beta }\phi ].
\end{equation*}%
We chosen the simplest possibility to define for noncommutative scalar
fields certain energy--momentum values and their anholonomic deformations.
In general, in noncommutatie field theory one introduced more conventional
stress--energy tensors \cite{14abou}.

\subsubsection{Locally anisotropic noncommutative gauge fields}

\label{ncgt}Some models of locally anisotropic Yang--Mills and gauge gravity
noncommutative theories are analyzed in Refs. \cite{14vnonc,14vncf}. Here we say
only the basic facts about such theories with possible supersymmetry but not
concerning points of gauge gravity.

\paragraph{Anholonomic Yang--Mills actions and MSYM model}

{\qquad}

A gauge field is introduced as a one form $A_{\alpha }$ having each
component taking values in $\mathcal{A}$ and satisfying $A_{\alpha
}=A_{\alpha }^{+}$ and curvature (equivalently, field strength)
\begin{equation*}
F_{\alpha \beta }=\delta _{\alpha }A_{\beta }-\delta _{\beta }A_{\alpha }+i
\left[ A_{\alpha }A_{\beta }\right]
\end{equation*}%
with gauge locally anisotropic transformation laws,
\begin{equation}
\bigtriangleup F_{\alpha \beta }=i\left[ F_{\alpha \beta },\epsilon \right] %
\mbox{ for }\bigtriangleup A_{\alpha }=\delta _{\alpha }\epsilon +i\left[
A_{\alpha },\epsilon \right] .  \label{gauget}
\end{equation}%
Now we can introduce the noncommutative locally anisotropic Yang--Mills
action%
\begin{equation*}
S=-\frac{1}{4g_{YM}}^{N}\int Tr~F^{2}
\end{equation*}%
which describes the N--anholonomic dynamics of the gauge field $A_{\alpha }.$
Coupling to matter field can be introduced in a standard way by using
N--elongated partial derivatives $\delta _{\alpha },$%
\begin{equation*}
\bigtriangledown _{\alpha }\varphi =\delta _{\alpha }\varphi +i\left[
A_{\alpha },\varphi \right] .
\end{equation*}%
Here we note that by using $Mat_{Z}\left( \mathcal{A}\right) $ we can
construct both noncommutative and anisotropic analog of $U\left( Z\right) $
gauge theory, or, by introducing supervariables adapted to N--connections %
\cite{14vstr2} and locally anisotropic spinors \cite{14vspinors,14vmon2}, we can
generate supersymmetric Yang--Mills theories. For instance, the maximally
supersymmetric Yang--Mills (MSYM) Lagrangian in ten dimensions with $%
\mathcal{N}=4$ can be deduced in anisotropic form, by corresponding
dimensional reductions and anholonomic constraints, as%
\begin{equation*}
S=~^{N}\int \delta ^{10}u~Tr~\left( F_{\alpha \beta }^{2}+i\overline{\chi }%
\overrightarrow{\bigtriangledown }\chi \right)
\end{equation*}%
where $\chi $ is a 16 component adjoint Majorana--Weyl fermion and the
spinor d--covariant derivative operator $\overrightarrow{\bigtriangledown }$
is written by using N--anholonomic frames.

\paragraph{The emergence of locally anisotropic spacetime}

{\qquad}

It is well known that spacetime translations may arise from a gauge group
transforms in noncommutative gauge theory (see, for instance, Refs. \cite{14dn}%
). If the same procedure is reconsidered for N--elongated partial
derivatives and distinguished noncommutative parameters, we can write
\begin{equation*}
\delta A_{\alpha }=v^{\beta }\delta _{\beta }A_{\alpha }
\end{equation*}
as a gauge transform (\ref{gauget}) when the parameter $\epsilon $ is
expressed as%
\begin{equation*}
\epsilon =v^{\alpha }\left( \Theta ^{-1}\right) _{\alpha \beta }u^{\beta
}=v^{i}\left( \Theta ^{-1}\right) _{jk}x^{k}+v^{a}\left( \Theta ^{-1}\right)
_{ab}x^{b},
\end{equation*}%
which generates
\begin{equation*}
\bigtriangleup A_{\alpha }=v^{\beta }\delta _{\beta }A_{\alpha }+v^{\beta
}\left( \Theta ^{-1}\right) _{\alpha \beta }.
\end{equation*}%
This way the spacetime anholonomy is induced by a noncommutative gauge
anisotropy. For another type of functions $\epsilon (u),$ we may generate
another spacetime locally anisot\-rop\-ic transforms. For instance, we can
generate a Poisson bracket $\{\varphi ,\epsilon \}$ with N--elongat\-ed
derivatives,%
\begin{equation*}
\bigtriangleup \varphi =i\left[ \varphi ,\epsilon \right] =\Theta ^{\alpha
\beta }\delta _{\alpha }\varphi \delta _{\beta }\epsilon +o\left( \delta
_{\alpha }^{2}\varphi \delta _{\beta }^{2}\epsilon \right) \rightarrow
\{\varphi ,\epsilon \}
\end{equation*}%
which proves that at leading order the locally anisotropic gauge transforms
preserve the locally anisotropic noncommutative structure of parameter $%
\Theta ^{\alpha \beta }.$

Now, we demonstrate that the Yang--Mills action may be rewritten as a
''matrix model'' action even the spacetime background is N--deformed. This
is another side of unification of noncommutative spacetime and gauge field
with anholonomically deformed symmetries. We can absorb a inner derivation
into a vector potential by associating the covariant operator $%
\bigtriangledown _{\alpha }=\delta _{\alpha }+iA_{\alpha }$ to connection
operators in $\R$$_{N,\theta }^{n+m},$%
\begin{equation*}
\bigtriangledown _{\alpha }\varphi \rightarrow \left[ C_{\alpha },\varphi %
\right]
\end{equation*}%
for%
\begin{equation}
C_{\alpha }=\left( -i\Theta ^{-1}\right) _{\alpha \beta }u^{\beta
}+iA_{\alpha }.  \label{coper}
\end{equation}%
As in usual noncommutative gauge theory we introduce the ''covariant
coordinates'' but distinguished by the N--connection,%
\begin{equation*}
Y^{\alpha }=u^{\alpha }+\Theta ^{\alpha \beta }A_{\beta }\left( u\right) .
\end{equation*}%
For invertible $\Theta ^{\alpha \beta },$ one considers another notation, $%
Y^{\alpha }=i\Theta ^{\alpha \beta }C_{\beta }.$ Such transforms allow to
express $F_{\alpha \beta }=i\left[ \bigtriangledown _{\alpha
},\bigtriangledown _{\beta }\right] $ as
\begin{equation*}
F_{\alpha \beta }=i\left[ C_{\alpha },C_{\beta }\right] -\left( \Theta
^{-1}\right) _{\alpha \beta }
\end{equation*}%
for which the Yang--Mills action transform into a matrix relation,%
\begin{eqnarray}
S &=&~^{N}Tr\sum_{\alpha ,\beta }\left( \acute{\imath}\left[ C_{\alpha
},C_{\beta }\right] -\left( \Theta ^{-1}\right) _{\alpha \beta }\right) ^{2}
\label{actymmatr} \\
&=&~^{N}Tr\{\left[ \acute{\imath}\left[ C_{k},C_{j}\right] -\left( \Theta
^{-1}\right) _{kj}\right] \left[ \acute{\imath}\left[ C^{k},C^{j}\right]
-\left( \Theta ^{-1}\right) ^{kj}\right]   \notag \\
&&+\left[ \acute{\imath}\left[ C_{a},C_{b}\right] -\left( \Theta
^{-1}\right) _{ab}\right] \left[ \acute{\imath}\left[ C^{a},C^{b}\right]
-\left( \Theta ^{-1}\right) ^{ab}\right] \}  \notag
\end{eqnarray}%
where we emphasize the N--distinguished components.

\paragraph{The noncommutative Dirac d--operator}

{\qquad}

If we consider multiplications $a\cdot \psi $ with $a\in \mathcal{A}$ on a
Dirac spinor $\psi ,$ we can have two different physics depending on the
orders of such multiplications we consider, $a\psi $ or $\psi a.$ In order
to avoid infinite spectral densities, in the locally isotropic
noncommutative gauge theory, one writes the Dirac operator as
\begin{equation*}
\overrightarrow{\bigtriangledown }\psi =\gamma ^{i}\left( \overrightarrow{%
\bigtriangledown }_{i}\psi -\psi \partial _{i}\right) =0.
\end{equation*}%
In the locally anisotropic case we have to introduce N--elongated partial
derivatives,%
\begin{eqnarray*}
\overrightarrow{\bigtriangledown }\psi &=&\gamma ^{\alpha }\left(
\overrightarrow{\bigtriangledown }_{\alpha }\psi -\psi \delta _{\alpha
}\right) \\
&=&\gamma ^{i}\left( \overrightarrow{\bigtriangledown }_{i}\psi -\psi \delta
_{i}\right) +\gamma ^{a}\left( \overrightarrow{\bigtriangledown }_{a}\psi
-\psi \delta _{a}\right) =0
\end{eqnarray*}%
and use a d--covariant spinor calculus \cite{14vspinors,14vmon2}.

\paragraph{The N--adapted stress--energy tensor}

{\qquad}

The action (\ref{actymmatr}) produces a stress--energy d--tensor%
\begin{equation*}
T_{\alpha \beta }\left( p\right) =\sum_{\gamma }\int_{0}^{1}ds~~^{N}\int
Tr~e^{isp_{\tau }Y^{\tau }}\left[ C_{\alpha },C_{\gamma }\right]
~e^{i(1-s)p_{\tau }Y^{\tau }}\left[ C_{\beta },C_{\gamma }\right]
\end{equation*}%
as a Noether current derived by the variation $C_{\alpha }\rightarrow
C_{\alpha }+a_{\alpha }\left( p\right) e^{isp_{\tau }Y^{\tau }}.$ This
d--tensor has a property of conservation,
\begin{equation*}
p_{\tau }\Theta ^{\tau \lambda }T_{\lambda \beta }\left( p\right) =0
\end{equation*}%
for the solutions of field equations and seem to be a more natural object in
string theory, which admits an anholonomic generalizations by
''distinguishing of indices''.

\paragraph{The anholonomic Seiberg--Witten map}

{\qquad}

There are two different types of gauge theories: commutative and
noncommutative ones. They my be related by the so--called Seiberg--Witten
map \cite{14sw} which explicitly transforms a noncommutative vector potential
to a conventional Yang--Mills vector potential. This map can be generalized
in gauge gravity and for locally anisotropic gravity \cite{14vnonc,14vncf}. Here
we define the Seiberg--Witten map for locally anisotropic gauge fields with
N--elongated partial derivatives. \

The idea is that if there exists a standard, but locally anisotropic,
Yang--Mills potential $A_{\alpha }$ with gauge transformation laws
parametrized by the parameter $\epsilon $ like in (\ref{gauget}), a
noncommutative gauge potential $\widehat{A}_{\alpha }\left( A_{\alpha
}\right) $ with gauge transformation parameter $\widehat{\epsilon }\left(
A,\epsilon \right) ,$ when
\begin{equation*}
\widehat{\bigtriangleup }_{\widehat{\epsilon }}\widehat{A}_{\alpha }=\delta
_{\alpha }\widehat{\epsilon }+i\left( \widehat{A}_{\alpha }\star \widehat{%
\epsilon }-\widehat{\epsilon }\star \widehat{A}_{\alpha }\right) ,
\end{equation*}%
should satisfy the equation%
\begin{equation}
\widehat{A}\left( A\right) +\widehat{\bigtriangleup }_{\widehat{\epsilon }}%
\widehat{A}\left( A\right) =\widehat{A}\left( A+\bigtriangleup _{\epsilon
}A\right) ,  \label{swe}
\end{equation}%
where, for simplicity, the indices were omitted. This is the Seiberg--Witten
equation which, in our case, contains N--adapted operators $\delta _{\alpha
} $ (\ref{6dder}) and d--vector gauge potentials, respectively, $\widehat{A}%
_{\alpha }=\left( \widehat{A}_{i},\widehat{A}_{a}\right) $ and $A_{\alpha
}=\left( A_{i},A_{a}\right) .$ To first order in $\Theta ^{\alpha \beta
}=\bigtriangleup \Theta ^{\alpha \beta },$ the equation (\ref{swe}) can be
solved in a usual way, by related respectively the potentials and
transformation parameters,%
\begin{eqnarray*}
\widehat{A}_{\alpha }\left( A_{\alpha }\right) -A_{\alpha } &=&-\frac{1}{4}%
\bigtriangleup \Theta ^{\beta \lambda }\left[ A_{\beta }\left( \delta
_{\lambda }A_{\alpha }+F_{\lambda \alpha }\right) +\left( \delta _{\lambda
}A_{\alpha }+F_{\lambda \alpha }\right) A_{\beta }\right] +o(\bigtriangleup
\Theta ^{2}), \\
\widehat{\epsilon }\left( A,\epsilon \right) -\epsilon &=&\frac{1}{4}%
\bigtriangleup \Theta ^{\beta \lambda }\left( \delta _{\beta }\epsilon
~A_{\lambda }+A_{\lambda }\delta _{\beta }\epsilon \right) +o(\bigtriangleup
\Theta ^{2}),
\end{eqnarray*}%
from which we can also find a first order relation for the field strength,%
\begin{eqnarray*}
\widehat{F}_{\lambda \alpha }-F_{\lambda \alpha } &=&\frac{1}{2}%
\bigtriangleup \Theta ^{\beta \tau }\left( F_{\lambda \beta }F_{\alpha \tau
}+F_{\alpha \tau }F_{\lambda \beta }\right) \\
&&-A_{\beta }\left( \bigtriangledown _{\tau }F_{\lambda \alpha }+\delta
_{\tau }F_{\lambda \alpha }\right) -\left( \bigtriangledown _{\tau
}F_{\lambda \alpha }+\delta _{\tau }F_{\lambda \alpha }\right) A_{\beta
}+o(\bigtriangleup \Theta ^{2}).
\end{eqnarray*}%
By a recurrent procedure the solution of (\ref{swe}) can be constructed in
all orders \ of $\bigtriangleup \Theta ^{\alpha \beta }$ as in the locally
isotropic case (see details on recent supersymmetric generalizations in
Refs. \cite{14liu} which can be transformed at least in a formal form into
certain anisotropic analogs following the d--covariant geometric rule.

\section[String Anholonoic Noncommutativity]
{Anholonomy and Noncommutativity:\newline Relations to String/ M--Theory}

The aim of this Section is to discuss how both noncommutative and locally
anisotropic field theories arise from string theory and M--theory. The first
use of noncommutative geometry in string theory was suggested by E. Witten
(see Refs. \cite{14strncg,14sw} for details and developments). Noncommutativity
is natural in open string theory: interactions of open strings with two ends
contains formal similarities to matrix multiplication which explicitly
results in noncommutative structures. In other turn, matrix noncommutativity
is contained in off--diagonal metrics and anholonomic vielbeins with
associated N--connection and anholonomic relations (see (\ref{4anhol}) and
related details in Appendix) which are used in order to develop locally
anisotropic geometries and field theories. We emphasize that the constructed
exact solutions with off--diagonal metrics in general relativity \ and extra
dimension gravity together with the existence of a string field framework
strongly suggest that noncommutative locally anisotropic structures have a
deep underlying significance in such theories \cite%
{14vexsol,14vbel,14vsingl,14vsingl1,14vnonc,14vncf,14vstring,14vstr2}.

\subsection{Noncommutativity and anholonomy in string theory}

In this subsection, we will analyze strings in curved spacetimes with
constant coefficients $\{g_{ij},h_{ab}\}$ of d--metric (\ref{7dmetric}) (the
coefficients $N_{i}^{a}\left( x^{k},y^{a}\right) $ are not constant and the
off--diagonal metric (\ref{6ansatz}) has a non--trivial curvature tensor).
With respect to N--adapted frames (\ref{6dder}) and (\ref{7ddif}) the string
propagation is like in constant Neveu--Schwarz constant $B$--field and with $%
Dp$--branes. We work under the conditions of string and brane
theory which results in noncommutative geometry \cite{14strncg} but
the background under consideration here is an anholonomic one.
The $B$--field is a like constant magnetic field which is
polarized by the N--connection structure. The rank
of the matrix $B_{\alpha \beta }$ is denoted $k=n+m=11\leq p+1,$ where $%
p\geq 10$ is a constant. For a target space, defined with respect to
anholonomic frames, we will assume that $B_{0\beta }=0$ with $"0"$ the time
direction (for a Euclidean signature, this condition is not necessary). We
can similarly consider another dimensions than 11, or to suppose that some
dimensions are compactified. We can pick some torus like coordinates, in
general anholonomic, by certain conditions, $u^{\alpha }\sim $ $u^{\alpha
}+2\pi k^{\alpha }.$ For simplicity, we parametrize $B_{\alpha \beta
}=const\neq 0$ for $\alpha ,\beta =1,...,k$ and $g_{\alpha \beta }=0$ for $%
\alpha =1,...,r,\beta \neq 1,...,k=n+m$ with a further distinguishing of
indices

There are two possibilities of writing out the worldsheet action,%
\begin{eqnarray}
S &=&\frac{1}{4\pi \alpha ^{\prime }}\int_{\Sigma }\delta \underline{\mu }%
_{g}\left( g_{\underline{\alpha }\underline{\beta }}\partial _{A}u^{%
\underline{\alpha }}\partial ^{A}u^{\underline{\beta }}-2\pi \alpha ^{\prime
}iB_{\underline{\alpha }\underline{\beta }}\varepsilon ^{AB}\partial _{A}u^{%
\underline{\alpha }}\partial _{B}u^{\underline{\beta }}\right)  \label{act10}
\\
&=&\frac{1}{4\pi \alpha ^{\prime }}\int_{\Sigma }\delta \underline{\mu }%
_{g}g_{\underline{\alpha }\underline{\beta }}\partial _{A}u^{\underline{%
\alpha }}\partial ^{A}u^{\underline{\beta }}-\frac{i}{2}\int_{\partial
\Sigma }\delta \underline{\mu }_{g}B_{\underline{\alpha }\underline{\beta }%
}~u^{\underline{\alpha }}\partial _{\tan }u^{\underline{\beta }};  \notag
\end{eqnarray}
\begin{eqnarray}
&=&\frac{1}{4\pi \alpha ^{\prime }}\int_{\Sigma }\delta \mu
_{g}(g_{ij}\partial _{A}x^{i}\partial ^{A}x^{j}+h_{ab}\partial
_{A}y^{a}\partial ^{A}y^{b}  \notag \\
&&-2\pi \alpha ^{\prime }iB_{ij}\varepsilon ^{AB}\partial _{A}x^{i}\partial
_{B}x^{j}-2\pi \alpha ^{\prime }iB_{ab}\varepsilon ^{AB}\partial
_{A}y^{a}\partial _{B}y^{b})  \notag \\
&=&\frac{1}{4\pi \alpha ^{\prime }}\int_{\Sigma }\delta \mu _{g}\left(
g_{ij}\partial _{A}x^{i}\partial ^{A}x^{j}+h_{ab}\partial _{A}y^{a}\partial
^{A}y^{b}\right)  \notag \\
&&-\frac{i}{2}\int_{\partial \Sigma }\delta \mu _{g}B_{ij}~x^{i}\partial
_{\tan }x^{j}-\frac{i}{2}\int_{\partial \Sigma }\delta \mu
_{g}B_{ab}~y^{a}\partial _{\tan }y^{b},  \notag
\end{eqnarray}%
where the first variant is written by using metric ansatz $g_{\underline{%
\alpha }\underline{\beta }}$ (\ref{6ansatz}) but the second variant is just
the term $S_{g_{N},B}$ from action (\ref{act1a}) with d--metric (\ref%
{7dmetric}) and different boundary conditions and $\partial _{\tan }$ is the
tangential derivative along the worldwheet boundary $\partial \Sigma .$ We
emphasize that the values $g_{ij},h_{ab}$ and $B_{ij},B_{ab},$ given with
respect to N--adapted frames are constant, but the off--diagonal $g_{%
\underline{\alpha }\underline{\beta }}$ and $B_{\underline{\alpha }%
\underline{\beta }},$ in coordinate base, are some functions on $\left(
x,y\right) .$ The worldsheet $\Sigma $ is taken to be with Euclidean
signature (for a Lorentzian wolrdsheet the complex $i$ should be omitted
multiplying $B).$

The equation of motion of string in anholonomic constant background define
respective anholonomic, N--adapted boundary conditions. For coordinated $%
\alpha $ along the $Dp$--branes they are
\begin{eqnarray}
g_{\alpha \beta }\partial _{norm}u^{\beta }+2\pi i\alpha ^{\prime }B_{\alpha
\beta }\partial _{\tan }u^{\beta } &=&  \label{boundc} \\
g_{ij}\partial _{norm}x^{j}+h_{ab}\partial _{norm}y^{b}+2\pi i\alpha
^{\prime }B_{ij}\partial _{B}x^{j}-2\pi \alpha ^{\prime }iB_{ab}\partial
_{\tan }y^{b}|_{\partial \Sigma } &=&0,  \notag
\end{eqnarray}%
where $\partial _{norm}$ is a normal derivative to $\partial \Sigma .$ By
transforms of type $g_{\underline{\alpha }\underline{\beta }}=e_{\
\underline{\alpha }}^{\alpha }(u)e_{\ \underline{\beta }}^{\beta
}(u)g_{\alpha \beta }$ and $B_{\underline{\alpha }\underline{\beta }}=e_{\
\underline{\alpha }}^{\alpha }(u)e_{\ \underline{\beta }}^{\beta
}(u)B_{\alpha \beta }$ we can remove these boundary conditions into a
holonomic off--diagonal form which is more difficult to investigate. With
respect to N--adapted frames (with non--underlined indices) the analysis is
very similar to the case constant values of the metric and $B$--field. For $%
B=0,$ the boundary conditions (\ref{boundc}) are Neumann ones. If $B$ has
the rank $r=p$ and $B\rightarrow \infty $ (equivalently, $g_{\alpha \beta
}\rightarrow 0$ along the spactial directions of the brane, the boundary
conditions become of Dirichlet type). The effect of all such type conditions
and their possible interpolations can be investigated as in the usual open
string theory with constant $B$--field but, in this subsection, with respect
to N--adapted frames.

For instance, we can suppose that $\Sigma $ is a disc,
conformally and anholonomically mapped to the upper half plane
with complex variables $z$ and $\overline{z}$ and ${Im}\ z\geq 0.$
The propagator with such boundary conditions is the same as in
\cite{14fradkin} with coordinates redefined to  anholonomic frames,%
\begin{eqnarray*}
<x^{i}(z)x^{j}(z^{\prime })> &=&-\alpha ^{\prime }[g^{ij}\log \frac{%
|z-z^{\prime }|}{|z-\overline{z}^{\prime }|}+H^{ij}\log |z-\overline{z}%
^{\prime }|^{2} \\
&&+\frac{1}{2\pi \alpha ^{\prime }}\Theta ^{ij}\log \frac{|z-\overline{z}%
^{\prime }|}{|\overline{z}-z^{\prime }|}+Q^{ij}], \\
<y^{a}(z)y^{b}(z^{\prime })> &=&-\alpha ^{\prime }[h^{ab}\log \frac{%
|z-z^{\prime }|}{|z-\overline{z}^{\prime }|}+H^{ab}\log |z-\overline{z}%
^{\prime }|^{2} \\
&&+\frac{1}{2\pi \alpha ^{\prime }}\Theta ^{ab}\log \frac{|z-\overline{z}%
^{\prime }|}{|\overline{z}-z^{\prime }|}+Q^{ab}],
\end{eqnarray*}%
where the coefficients are correspondingly computed,%
\begin{eqnarray}
H_{ij} &=&g_{ij}-(2\pi \alpha ^{\prime })^{2}\left( Bg^{-1}B\right)
_{ij},~H_{ab}=h_{ab}-(2\pi \alpha ^{\prime })^{2}\left( Bg^{-1}B\right)
_{ab},  \label{constants} \\
&&  \notag
\end{eqnarray}%
\begin{eqnarray*}
H^{ij} &=&\left( \frac{1}{g+2\pi \alpha ^{\prime }B}\right)
_{[sym]}^{ij}=\left( \frac{1}{g+2\pi \alpha ^{\prime }B}g\frac{1}{g-2\pi
\alpha ^{\prime }B}\right) ^{ij}, \\
H^{ab} &=&\left( \frac{1}{h+2\pi \alpha ^{\prime }B}\right)
_{[sym]}^{ij}=\left( \frac{1}{h+2\pi \alpha ^{\prime }B}h\frac{1}{h-2\pi
\alpha ^{\prime }B}\right) ^{ij},
\end{eqnarray*}%
\begin{eqnarray*}
\Theta ^{ij} &=&2\pi \alpha ^{\prime }\left( \frac{1}{g+2\pi \alpha ^{\prime
}B}\right) _{[antisym]}^{ij}=-(2\pi \alpha ^{\prime })^{2}\left( \frac{1}{%
g+2\pi \alpha ^{\prime }B}g\frac{1}{g-2\pi \alpha ^{\prime }B}\right) ^{ij},
\\
\Theta ^{ab} &=&2\pi \alpha ^{\prime }\left( \frac{1}{g+2\pi \alpha ^{\prime
}B}\right) _{[antisym]}^{ab}=-(2\pi \alpha ^{\prime })^{2}\left( \frac{1}{%
g+2\pi \alpha ^{\prime }B}g\frac{1}{g-2\pi \alpha ^{\prime }B}\right) ^{ab},
\end{eqnarray*}%
with $[sym]$ and $[antisym]$ prescribing, respectively, the symmetric and
antisymmetric parts of a matrix and constants $Q^{ij}$ and $Q^{ab}$ (in
general, depending on $B,$ but not on $z$ or $z^{\prime })$ do to not play
an essential role which allows to set them to a convenient value. The last
two terms are signed--valued (if the branch cut of the logarithm is taken in
lower half plane) and the rest ones are manifestly sign--valued.

Restricting our considerations to the open string vertex operators and
interactions with real $z=\tau $ and $z=\tau ^{\prime },$ evaluating at
boundary points of $\Sigma $ for a convenient value of $D^{\alpha \beta },$
the propagator (in non--distinguished form) becomes%
\begin{equation*}
<u^{\alpha }(\tau )u^{\beta }(\tau ^{\prime })>=-\alpha ^{\prime }H^{\alpha
\beta }\log \left( \tau -\tau ^{\prime }\right) ^{2}+\frac{i}{2}\Theta
^{\alpha \beta }\epsilon \left( \tau -\tau ^{\prime }\right)
\end{equation*}%
for $\epsilon \left( \tau -\tau ^{\prime }\right) $ being 1 for $\tau >\tau
^{\prime }$ and -1 for $\tau <\tau ^{\prime }.$ The d--tensor $H_{\alpha
\beta }$ $\ $defines the effective metric seen by the open string subjected
to some anholonomic constraints being constant with respect to N--adapted
frames. Working as in conformal field theory, one can compute commutators of
operators from the short distance behavior of operator products (by
interpreting time ordering as operator ordering with time $\tau )$ and find
that the coordinate commutator
\begin{equation*}
\left[ u^{\alpha }(\tau ),u^{\beta }(\tau )\right] =i\Theta ^{\alpha \beta }
\end{equation*}%
which is just the relation (\ref{nceucln}) for noncommutative coordinates
with constant noncommutativity parameter $\Theta ^{\alpha \beta }$
distinguished by a N--connection structure.

In a similar manner we can introduce gauge fields and consider worldsheet
supersymmetry together with noncommutative relations with respect to
N--adapted frames. This results in locally anisotropic modifications of the
results from \cite{14strncg} via anholonomic frame transforms and
distinguished tensor and noncommutative calculus (we omit here the details
of such calculations).

We emphasize that even the values $H^{\alpha \beta }$ and $\Theta ^{\alpha
\beta }$ (\ref{constants}) are constant with respect to N--adapted frames
the anholonomic noncommutative string configurations are characterized by
locally anisotropic values $H^{\underline{\alpha }\underline{\beta }}$ and $%
\theta ^{\underline{\alpha }\underline{\beta }}$ which are defined with
respect to coordinate frames as
\begin{equation*}
H^{\underline{\alpha }\underline{\beta }}=e_{\alpha }^{~\underline{\alpha }%
}(u)e_{\beta }^{~\underline{\beta }}(u)H^{\alpha \beta }\mbox{ and }\theta ^{%
\underline{\alpha }\underline{\beta }}=e_{\alpha }^{~\underline{\alpha }%
}(u)e_{\beta }^{~\underline{\beta }}(u)\theta ^{\alpha \beta }
\end{equation*}%
with $e_{\alpha }^{~\underline{\alpha }}(u)$ (\ref{vielbtr}) defined by $%
N_{i}^{a}$ as in (\ref{viel1}), i. e.%
\begin{equation*}
e_{i}^{~\underline{i}}=\delta _{i}^{~\underline{i}},~e_{i}^{~\underline{a}%
}=-N_{i}^{\underline{a}}(u),~e_{a}^{~\underline{a}}=\delta _{a}^{\underline{a%
}},~e_{a}^{~\underline{i}}=0.
\end{equation*}

Now, we make use of he standard relation between world--sheet correlation
function of vertex operators, the S--matrix for string scattering and
effective actions which can reproduce this low energy string physics \cite%
{14deligne} but generalizing them for anholonomic structures. We consider that
operators in the bulk of the world--sheet correspond to closed strings,
while operators on the boundary correspond to open strings and thus fields
which propagate on the world volume of a D--brane. The basic idea is that
each local world--sheet operator $V_{s}\left( z\right) $ corresponds to an
interaction with a spacetime field $\varphi _{s}\left( z\right) $ which
results in the effective Lagrangian
\begin{equation*}
\int \delta ^{p+1}u\sqrt{|\det g_{\alpha \beta }|}~^{N}Tr~\varphi
_{1}\varphi _{2}...\varphi _{s}
\end{equation*}%
which is computed by integrating on $z_{s}$ following the prescribed order
for the correlation function%
\begin{equation*}
\left\langle \int dz_{1}V_{1}\left( z_{1}\right) \int dz_{2}V_{2}\left(
z_{2}\right) ...\int dz_{s}V_{s}\left( z_{s}\right) \right\rangle
\end{equation*}%
on a world--sheet $\Sigma $ with disk topology, with operators $V_{s}$ as
successive points $z_{s}$ on the boundary $\partial \Sigma .$ The
integrating measure is constructed from N--elongated values and coefficients
of d--metric. In the leading limit of the S--matrix with vertex operators
only for the massless fields we reproduce a locally anisotropic variant of
the MSYM effective action which describes the physics of a D--brane with
arbitrarily large but anisotropically and slowly varying field strength,
\begin{equation}
S_{BNI}^{[anh]}=\frac{1}{g_{s}l_{s}\left( 2\pi l_{s}\right) ^{p}}\int \delta
^{p+1}u\sqrt{|\det (g_{\alpha \beta }+2\pi l_{s}^{2}(B+F))|}  \label{act12}
\end{equation}%
where $g_{s}$ is the string coupling, the constant $l_{s}$ is the usual one
from D--brane theory and $g_{\alpha \beta }$ is the induced d--metric on the
brane world--volume. The action (\ref{act12}) is just the
Nambu--Born--Infeld (NBI) action \cite{14fradkin} but defined for d--metrics
and d--tensor fields with coefficients computed with respect to N--adapted
frames.

\subsection{Noncommutative anisotropic structures in M(atrix) theory}

For an introduction to M--theory, we refer to \cite{14polch,14deligne}.
Throughout this subsection we consider M--theory as to be not completely
defined but with a well--defined quantum gravity theory with the low energy
spectrum of the 11 dimensional supergravity theory \cite{14cjs}, containing
solitonic ''branes'', the 2--brane, or supermembrane, and five--branes and
that from M--theory there exists connections to the superstring theories.
Our claim is that in the low energy limits the noncommutative structures
are, in general, locally anisotropic.

The simplest way to derive noncommutativity from M--theory is to start with
a matrix model action such in subsection \ref{ncgt} and by introducing
operators of type $C_{\alpha }$ $\ $(\ref{coper}) and actions (\ref%
{actymmatr}). For instance, we can consider the action for maximally
supersymmetric quantum mechanics, i. e. a trivial case with $p=0$ of MSYM,
when
\begin{equation}
S=\int \delta t~~^{N}Tr\sum_{\alpha =1}^{9}\left( D_{t}X\right)
^{2}-\sum_{\alpha <\beta }\left[ X^{\alpha },X^{\beta }\right] ^{2}+\chi
^{+}\left( D_{t}+\Gamma _{\alpha }X^{\alpha }\right) \chi ,  \label{act11}
\end{equation}%
where $D_{t}=\delta /\partial t+iA_{0}$ with d--derivative
(\ref{6dder}) with varying $A_{0}$ which introduces constraints in
physical states because of restriction of unitary symmetry. This
action is written in anholonomic variables and generalizes the
approach of entering the M--theory as a regularized form of the
actions for the supermembranes \cite{14wit}. In this interpretation
the the compact eleventh dimension does not disappear and the
M--theory is to be considered as to be anisotropically
compactified on a light--like circle.

In order to understand how anisotropic torus compactifications may be
performed (see subsection \ref{torus}) we use the general theory of
D--branes on quotient spaces \cite{14taylor}. We consider $U_{\alpha }=\gamma
\left( \beta _{\alpha }\right) $ for a set of generators of $\Z$$^{n+m}$
with $\mathcal{A}=Mat_{n+m}\left( \C\right) $ which satisfy the equations%
\begin{equation*}
U_{\alpha }^{-1}X^{\beta }U_{\alpha }=X^{\beta }+\delta _{\alpha }^{\beta
}2\pi R_{\alpha }
\end{equation*}%
having solutions of type
\begin{equation*}
X_{\beta }=-i\delta /\partial \sigma ^{\beta }+A_{\beta }
\end{equation*}
for $A_{\beta }$ commuting with $U_{\alpha }$ and indices distinguished by a
N--connection structure as $\alpha =\left( i,a\right) ,\beta =\left(
j,b\right) .$ For such variables the action (\ref{act11}) leads to a locally
anisotropic MSYM on $T^{n+m}\times \R.$ Of course, this construction admits
a natural generalization for variables $U_{\alpha }$ satisfying relations (%
\ref{nceucln}) for noncommutative locally anisotropic tori which leads to
noncommutative anholonomic gauge theories \cite{14vnonc,14vncf}. In original
form this type of noncommutativity was introduced in M--theory (without
anisotropies) in Ref. \cite{14connes1}.

The anisotropic noncommutativity in M--theory can related to string model
via nontrivial components $C_{\alpha \beta -}$ of a three--form potential
(''-'' denotes the compact light--like direction). This potential has as \ a
background value if the M(atrix) theory is treated as M--theory on a
light--like circle as in usual isotropic models. In the IIA string
interpretation of $C_{\alpha \beta -}$ as a Neveu--Schwarz $B$--field which
minimally coupled to the string world--sheet, we obtain the action (\ref%
{act10}) compactified on a $\R\times T^{n+m}$ spacetime where torus has
constant d--metric and $B$--field coefficients.

\section{Anisotropic Gravity on Noncommutative D--Bra\-nes}

We develop a model of locally anisotropic \ gravity on noncommutative
D--branes (see Refs. \cite{14ardalan} for a locally isotropic variant). We
investigate what kind of deformations of the low energy effective action of
closed strings are induced in the presence of constant background
antisymmetric field (or it anholonomic transforms) and/or in the presence of
generic off--diagonal metric and associated nonlinear connection terms. It
should be noted that there were proposed and studied different models of
nocommutative deformations of gravity \cite{14ncg}, which were not derived
from string theory but introduced ''ad hoc''. Anholonomic and/or gauge
transforms in noncommutative gravity were considered in Refs. \cite%
{14vncf,14vnonc}. In this Section, we illustrate how such gravity models with
generic anisotropy and noncommutativity can be embedded in D--brane physics.

We can compute the tree level bosonic string scattering amplitude of two
massless closed string off a noncommutative D--brane \ with locally
anisotropic contributions by considering boundary conditions and correlators
stated with respect to anholonomic frames. By using the 'geometric
d--covariant rule' of changing the tensors, spinors and connections into
theirs corresponding N--distinguished d--objects we derive the locally
anisotropic variant of effective actions in a straightforword manner.

For instance, the action which describes this amplitude to order of the
string constant $(\alpha ^{\prime })^{0}$ is just the so--called DBI and
Einstein--Hilbert action. With respect to the Einstein N--emphasized frame
the DBI action is
\begin{equation}
S_{D-brane}^{[0]}=-T_{p}\int \delta ^{p+1}u~e^{-\Phi }\sqrt{\left| \det
\left( e^{-\gamma \Phi }g_{\alpha \beta }+\mathcal{B}_{\alpha \beta
}+f_{\alpha \beta }\right) \right| }  \label{act13}
\end{equation}%
where $g_{\alpha \beta }$ is the induced metric on the D--brane, $\mathcal{B}%
_{\alpha \beta }=B_{\alpha \beta }-2\kappa b_{\alpha \beta }$ is the pull
back of the antisymmetric d--field $B$ being constant \ with respect to
N--adapted frames along D--brane, $f_{\alpha \beta }$ is the gauge d--field
strength and $\gamma =-4/\left( n+m-2\right) $ and the constant $T_{p}=%
\mathcal{C}(\alpha ^{\prime })^{2}/C\kappa ^{2}$ is taken as in Ref. \cite%
{14ardalan} for usual D--brane theory (this allow to obtain in a limit the
Einstein--Hilber action in the bulk). There are used such parametrizations
of indices:
\begin{equation*}
\mu ^{\prime },\nu ^{\prime },...=0,...,25;\mu ^{\prime }=\left( \mu ,\hat{%
\mu}\right) ;~\hat{\mu},\hat{\nu}...=p+1,...,25;\hat{\mu}=\left( \hat{\imath}%
,\hat{a}\right)
\end{equation*}%
where $i$ takes $n$--dimensional 'horizontal' values and $a$ takes $m$%
--dimensional 'vertical' being used for a D--brane localized at $%
u^{p+1},...u^{25}$ with the boundary conditions given with respect to a
N--adapted frame,%
\begin{equation*}
g_{\alpha \beta }\left( \partial -\overline{\partial }\right) U^{\alpha
}+B_{\alpha \beta }\left( \partial +\overline{\partial }\right) U_{\mid z=%
\overline{z}}^{\alpha }=0,
\end{equation*}%
which should be distinguished in h- and v--components, and the two point
correlator of string anholonomic coordinates $U^{\alpha ^{\prime }}\left( z,%
\overline{z}\right) $ on the D--brane is%
\begin{eqnarray*}
&<&U_{\hat{\mu}}^{\alpha ^{\prime }}U_{\hat{\nu}}^{\beta ^{\prime }}>~=-%
\frac{\alpha ^{\prime }}{2}\{g^{\alpha ^{\prime }\beta ^{\prime }}\log \left[
\left( z_{\hat{\mu}}-z_{\hat{\nu}}\right) \left( \overline{z}_{\hat{\mu}}-%
\overline{z}_{\hat{\nu}}\right) \right] \\
&&+D^{\alpha ^{\prime }\beta ^{\prime }}\log \left( z_{\hat{\mu}}-\overline{z%
}_{\hat{\nu}}\right) +D^{\beta ^{\prime }\alpha ^{\prime }}\log \left(
\overline{z}_{\hat{\mu}}-z_{\hat{\nu}}\right) \}
\end{eqnarray*}%
where
\begin{equation*}
D^{\beta \alpha }=2\left( \frac{1}{\eta +B}\right) ^{\alpha \beta }-\eta
^{\alpha \beta },~D^{\hat{\mu}\hat{\nu}}=-\delta ^{\hat{\mu}\hat{\nu}%
},~D_{~\alpha ^{\prime }}^{\beta ^{\prime }}D^{\nu ^{\prime }\alpha ^{\prime
}}=\eta ^{\beta ^{\prime }\alpha ^{\prime }}
\end{equation*}%
for constant $\eta ^{\alpha ^{\prime }\beta ^{\prime }\text{ }}$ given with
respect to N--adapted frames.

The scattering amplitude of two closed strings off a D--brane is computed as
the integral
\begin{equation}
A=g_{c}^{2}~e^{-\lambda }\int d^{2}z_{\underline{1}}~d^{2}z_{\underline{2}%
}~<V\left( z_{\underline{1}},\overline{z}_{\underline{1}}\right) V\left( z_{%
\underline{2}},\overline{z}_{\underline{2}}\right) >,  \label{dbranea}
\end{equation}%
for $g_{c}$ being the closed string coupling constant, $\lambda $ being the
Euler number of the world sheet and the vertex operators for the massless
closed strings with the momenta $k_{\underline{i}\mu ^{\prime }}=\left( k_{%
\underline{i}i^{\prime }},k_{\underline{i}a^{\prime }}\right) $ $\ $\ and
polarizations $\epsilon _{\mu ^{\prime }\nu ^{\prime }}$ (satisfying the
conditions $\epsilon _{\mu ^{\prime }\nu ^{\prime }}k_{\underline{i}}^{\mu
^{\prime }}=\epsilon _{\mu ^{\prime }\nu ^{\prime }}k_{\underline{i}}^{\nu
^{\prime }}=0$ and $k_{\underline{i}\mu ^{\prime }}k_{\underline{i}}^{\mu
^{\prime }}=0$ taken as
\begin{equation*}
V\left( z_{i},\overline{z}_{i}\right) =\epsilon _{\mu ^{\prime }\nu ^{\prime
}}~~D_{~\alpha ^{\prime }}^{\nu ^{\prime }}:\partial X^{\mu ^{\prime
}}\left( z_{\underline{i}}\right) \exp \left[ ik_{\underline{i}}X\left( z_{%
\underline{i}}\right) \right] :~:\overline{\partial }X^{\alpha ^{\prime
}}\left( \overline{z}_{\underline{i}}\right) \exp \left[ ik_{\underline{i}%
\beta ^{\prime }}D_{~\tau ^{\prime }}^{\beta ^{\prime }}X^{\tau ^{\prime
}}\left( z_{\underline{i}}\right) \right] :.
\end{equation*}%
Calculation of such calculation functions can be performed as in usual
string theory with that difference that the tensors and derivatives are
distinguished by N--connections.

Decomposing the metric $g_{\alpha \beta }$ as
\begin{equation*}
g_{\alpha \beta }=\eta _{\alpha \beta }+2\kappa \chi _{\alpha \beta }
\end{equation*}%
where $\eta _{\alpha \beta }$ is constant (Minkowski metric but with respect
to N--adapted frames) and $\chi _{\alpha \beta }$ could be of (pseudo)
Riemannian or Finsler like type. Action (\ref{act13}) can be written to the
first order of $\chi ,$
\begin{equation}
S_{D-brane}^{[0]}=-\kappa T_{p}c\int \delta ^{p+1}u~\chi _{\alpha \beta
}Q^{\alpha \beta },  \label{act13a}
\end{equation}%
where
\begin{equation}
Q^{\alpha \beta }=\frac{1}{2}\left( \eta ^{\alpha \beta }+D^{\alpha \beta
}\right)  \label{qfield}
\end{equation}%
and $c=\sqrt{|\det \left( \eta _{\alpha \beta }+B_{\alpha \beta }\right) |},$
which exhibits a source for locally anisotropic gravity on D--brane,
\begin{equation*}
T_{\chi }^{\alpha \beta }=-\frac{1}{2}T_{p}\kappa C\left( \eta
^{\alpha \beta }+D_{(S)}^{\beta \alpha }\right) ,
\end{equation*}%
for $D_{(S)}^{\beta \alpha }$ denoting symmetrization of the matrix $%
D^{\beta \alpha }.$ This way we reproduce the same action as in superstring
theory \cite{14gar} but in a manner when anholonomic effects and anisotropic
scattering can be included.

Next order terms on $\alpha ^{\prime }$ in the string amplitude are included
by the term%
\begin{eqnarray*}
S_{bulk}^{[1]} &=&\frac{\alpha ^{\prime }}{8\kappa ^{2}}\int
\delta
^{26}u^{\prime }~e^{\gamma \Phi }\sqrt{|g_{\mu ^{\prime }\nu ^{\prime }}|}%
[R_{h^{\prime }i^{\prime }j^{\prime }k^{\prime }}R^{h^{\prime }i^{\prime
}j^{\prime }k^{\prime }}+R_{a^{\prime }b^{\prime }j^{\prime }k^{\prime
}}R^{a^{\prime }b^{\prime }j^{\prime }k^{\prime }}+P_{j^{\prime }i^{\prime
}k^{\prime }a^{\prime }}P^{j^{\prime }i^{\prime }k^{\prime }a^{\prime }} \\
&&+P_{c^{\prime }d^{\prime }k^{\prime }a^{\prime }}P^{c^{\prime }d^{\prime
}k^{\prime }a^{\prime }}+S_{j^{\prime }i^{\prime }b^{\prime }c^{\prime
}}S^{j^{\prime }i^{\prime }b^{\prime }c^{\prime }}+S_{d^{\prime }e^{\prime
}b^{\prime }c^{\prime }}S^{d^{\prime }e^{\prime }b^{\prime }c^{\prime }} \\
&&-4\left( R_{i^{\prime }j^{\prime }}R^{\iota ^{\prime }j^{\prime
}}+R_{i^{\prime }a^{\prime }}R^{i^{\prime }a^{\prime }}+P_{a^{\prime
}i^{\prime }}P^{a^{\prime }i^{\prime }}+R_{a^{\prime }b^{\prime
}}R^{a^{\prime }b^{\prime }}\right) +(g^{i^{\prime }j^{\prime }}R_{i^{\prime
}j^{\prime }}+h^{a^{\prime }b^{\prime }}S_{a^{\prime }b^{\prime }})^{2}]
\end{eqnarray*}%
where the indices are split as $\mu ^{\prime }=\left( i^{\prime },a^{\prime
}\right) $ and we use respectively the d--curvatures (\ref{3dcurvatures}),
Ricci d--tensors (\ref{6dricci}) and d--scalars (\ref{4dscalar}). Splitting of
''primed' indices reduces to splitting of D--brane values.

The DBI action on D--brane (\ref{dbranea}) is defined with a gauge field
strength
\begin{equation*}
f_{\alpha \beta }=\delta _{\alpha }a_{\beta }-\delta _{\beta }a_{\alpha }
\end{equation*}%
and with the induced metric
\begin{equation*}
g_{\alpha \beta }=\delta _{\alpha }X^{\mu ^{\prime }}\delta _{\beta }X_{\mu
^{\prime }}
\end{equation*}%
expanded around the flat space in the static gauge $U^{\mu }=u^{\mu },$%
\begin{equation*}
g_{\mu \nu }=\eta _{\mu \nu }+2\kappa \chi _{\mu \nu }+2\kappa \left( \chi _{%
\hat{\mu}\mu }\delta _{\nu }U^{\hat{\mu}}+\chi _{\hat{\mu}\nu }\delta _{\mu
}U^{\hat{\mu}}\right) +\delta _{\mu }U^{\hat{\mu}}\delta _{\nu }U_{\hat{\mu}%
}+2\kappa \chi _{\hat{\mu}\hat{\nu}}\delta _{\mu }U^{\hat{\mu}}\delta _{\nu
}U^{\hat{\nu}}.
\end{equation*}%
In order to describe D--brane locally anisotropic processes in the first
order in $\alpha ^{\prime }$ we need to add a new term to the DBI\ as follow,%
\begin{equation}
S^{1}=-\frac{\alpha ^{\prime }T_{p}}{2}\int \delta ^{p+1}u\sqrt{|\det q_{\mu
\nu }|}\{~R_{\alpha \beta \gamma \tau }q^{\alpha \tau }-\left( \Psi _{\alpha
\gamma }^{\hat{\mu}}\Psi _{\hat{\mu}\beta \tau }-\Psi _{\alpha \tau }^{\hat{%
\mu}}\Psi _{\hat{\mu}\beta \gamma }\right) \tilde{q}^{\alpha \tau }\}\tilde{q%
}^{\beta \gamma }  \label{act14}
\end{equation}%
where $q_{\mu \nu }=\eta _{\mu \nu }+B_{\mu \nu }+f_{\mu \nu },$ $q^{\mu \nu
}$ is the inverse of $q_{\mu \nu },\tilde{q}_{\mu \nu }=g_{\mu \nu }+B_{\mu
\nu }+f_{\mu \nu },$ $\tilde{q}^{\mu \nu }$ is the inverse of $\tilde{q}%
_{\mu \nu },$ the curvature d--tensor $~R_{\alpha \beta \gamma \tau }$ is
constructed from the induced d--metric by using the canonical d--connection
(see (\ref{3dcurvatures}) and (\ref{6dcon})) and
\begin{equation*}
\Psi _{\alpha \beta }^{\hat{\mu}}=\kappa \left( -\delta ^{\hat{\mu}}\chi
_{\alpha \beta }+\delta _{\alpha }\chi _{~\beta }^{\hat{\mu}}+\delta _{\beta
}\chi _{~\alpha }^{\hat{\mu}}\right) +\delta _{\alpha }\delta _{\beta }U^{%
\hat{\mu}}.
\end{equation*}

The action (\ref{act14}) can be related to the Einstein--Hilbert action on
the D--brane if the the $B$--field is turned off. To see this we consider
the field $Q^{\alpha \beta }=$ $\eta ^{\alpha \beta }$ (\ref{qfield}) which
reduces (up to some total d--derivatives, which by the momentum conservation
relation have no effects in scattering amplitudes, and ignoring gauge fields
because they do not any contraction with gravitons because of antisymmetry
of $f_{\alpha \beta }$) to
\begin{equation}
S_{D-brane}^{[1]}=-\frac{\alpha ^{\prime }T_{p}}{2}\int \delta ^{p+1}u\sqrt{%
|\det g_{\mu \nu }|}\left( \widehat{R}+S+\Psi _{~\alpha }^{\hat{\mu}~\alpha
}\Psi _{\hat{\mu}\beta }^{\quad \beta }-\Psi _{\alpha \beta }^{\hat{\mu}%
}\Psi _{\hat{\mu}}^{\quad \alpha \beta }\right)  \label{act15}
\end{equation}%
were $\widehat{R}$ and $S$ are computed as d--scalar objects (\ref{4dscalar})
and by following the relation at $\mathcal{O}(\chi ^{2}),$%
\begin{equation*}
\sqrt{|\det \eta _{\mu \nu }|}R_{\alpha \beta \gamma \tau }\eta ^{\alpha
\tau }g^{\beta \gamma }=\sqrt{|\det g_{\mu \nu }|}R_{\alpha \beta \gamma
\tau }g^{\alpha \gamma }g^{\beta \tau }+\mbox{total d--derivatives}.
\end{equation*}%
The action (\ref{act15}) transforms into the Einstein--Hilbert action as it
was proven for the locally isotropic D--brane theory \cite{14corley} for
vanishing N--connections and trivial vertical (anisotropic) dimensions.

In conclusion, it has been shown in this Section that the D--brane dynamics
can be transformed into a locally anisotropic one, which in low energy
limits contains different models of generalized Lagrange/\ Finsler or
anholonomic Riemannian spacetimes, by introducing corresponding anholonomic
frames with associated N--connection structures and d--metric fields (like (%
\ref{2ncc}) and (\ref{1mfl}) and (\ref{2dmetricf})).

\section[Anisotropic and Noncommutative Solutions]
{Exact Solutions: Noncommutative and/ or Locally Anisotropic
Structures}

\label{2exsol}

In the previous sections we demonstrated that locally anisotropic
noncommutative geometric structures are hidden in string/ M--theory. Our aim
here is to construct and analyze four classes of exact solutions in string
gravity with effective metrics possessing generic off--diagonal terms which
for associated anholonomic frames and N--connections can be extended to
commutative or noncommutative string configurations.

\subsection{Black ellipsoids from string gravity}

A simple string gravity model with antisymmetric two form potential field $%
H^{\alpha ^{\prime }\beta ^{\prime }\gamma ^{\prime }},$ for constant
dilaton $\phi ,$ and static internal space, $\beta ,$ is to be found for the
NS--NS sector which is common to both the heterotic and type II string
theories \cite{14lidsey}. The equations (\ref{eqfstr}) reduce to%
\begin{eqnarray}
R_{\mu ^{\prime }\nu ^{\prime }} &=&\frac{1}{4}H_{\mu ^{\prime }\lambda
^{\prime }\tau ^{\prime }}H_{\nu ^{\prime }}^{~~\lambda ^{\prime }\tau
^{\prime }},  \label{eq16} \\
D_{\mu ^{\prime }}H^{\mu ^{\prime }~\lambda ^{\prime }\tau ^{\prime }} &=&0,
\notag
\end{eqnarray}%
for
\begin{equation*}
H_{\mu ^{\prime }\nu ^{\prime }\lambda ^{\prime }}=\delta _{\mu ^{\prime
}}B_{\nu ^{\prime }\lambda ^{\prime }}+\delta _{\lambda ^{\prime }}B_{\mu
^{\prime }\nu ^{\prime }}+\delta _{\nu ^{\prime }}B_{\lambda ^{\prime }\mu
^{\prime }}.
\end{equation*}%
If $H_{\mu ^{\prime }\nu ^{\prime }\lambda ^{\prime }}=\sqrt{|g_{\mu
^{\prime }\nu ^{\prime }}|}\epsilon _{\mu ^{\prime }\nu ^{\prime }\lambda
^{\prime }},$ we obtain the vacuum equations for the gravity with
cosmological constant $\lambda $,%
\begin{equation}
R_{\mu ^{\prime }\nu ^{\prime }}=\lambda g_{\mu ^{\prime }\nu ^{\prime }},
\label{2eq17}
\end{equation}%
for $\lambda =1/4$ where $R_{\mu ^{\prime }\nu ^{\prime }}$ is
the Ricci d--tensor (\ref{6dricci}), with ''primed'' indices
emphasizing that the geometry is induced after a topological
compactification.

For an ansatz of type
\begin{eqnarray}
\delta s^{2} &=&g_{1}(dx^{1})^{2}+g_{2}(dx^{2})^{2}+h_{3}\left( x^{i^{\prime
}},y^{3}\right) (\delta y^{3})^{2}+h_{4}\left( x^{i^{\prime }},y^{3}\right)
(\delta y^{4})^{2},  \label{2ansatz18} \\
\delta y^{3} &=&dy^{3}+w_{i^{\prime }}\left( x^{k^{\prime }},y^{3}\right)
dx^{i^{\prime }},\quad \delta y^{4}=dy^{4}+n_{i^{\prime }}\left(
x^{k^{\prime }},y^{3}\right) dx^{i^{\prime }},  \notag
\end{eqnarray}%
for the d--metric (\ref{7dmetric}) the Einstein equations (\ref{2eq17}) are
written (see \cite{14vmethod,14vexsol} for details on computation)
\begin{eqnarray}
R_{1}^{1}=R_{2}^{2}=-\frac{1}{2g_{1}g_{2}}[g_{2}^{\bullet \bullet }-\frac{%
g_{1}^{\bullet }g_{2}^{\bullet }}{2g_{1}}-\frac{(g_{2}^{\bullet })^{2}}{%
2g_{2}}+g_{1}^{^{\prime \prime }}-\frac{g_{1}^{^{\prime }}g_{2}^{^{\prime }}%
}{2g_{2}}-\frac{(g_{1}^{^{\prime }})^{2}}{2g_{1}}] &=&\lambda ,
\label{7ricci1a} \\
R_{3}^{3}=R_{4}^{4}=-\frac{\beta }{2h_{3}h_{4}} &=&\lambda ,  \label{6ricci2a}
\\
R_{3i^{\prime }}=-w_{i^{\prime }}\frac{\beta }{2h_{4}}-\frac{\alpha
_{i^{\prime }}}{2h_{4}} &=&0,  \label{5ricci3a} \\
R_{4i^{\prime }}=-\frac{h_{4}}{2h_{3}}\left[ n_{i^{\prime }}^{\ast \ast
}+\gamma n_{i^{\prime }}^{\ast }\right] &=&0,  \label{5ricci4a}
\end{eqnarray}%
where the indices take values $i^{\prime },k^{\prime }=1,2$ and $a^{\prime
},b^{\prime }=3,4.$ The coefficients of equations (\ref{7ricci1a}) - (\ref%
{5ricci4a}) are given by
\begin{equation}
\alpha _{i}=\partial _{i}{h_{4}^{\ast }}-h_{4}^{\ast }\partial _{i}\ln \sqrt{%
|h_{3}h_{4}|},\qquad \beta =h_{4}^{\ast \ast }-h_{4}^{\ast }[\ln \sqrt{%
|h_{3}h_{4}|}]^{\ast },\qquad \gamma =\frac{3h_{4}^{\ast }}{2h_{4}}-\frac{%
h_{3}^{\ast }}{h_{3}}.  \label{6abc}
\end{equation}%
The various partial derivatives are denoted as $a^{\bullet }=\partial
a/\partial x^{1},a^{^{\prime }}=\partial a/\partial x^{2},a^{\ast }=\partial
a/\partial y^{3}.$ This system of equations (\ref{7ricci1a})--(\ref{5ricci4a})
can be solved by choosing one of the ansatz functions (\textit{e.g.} $%
g_{1}\left( x^{i}\right) $ or $g_{2}\left( x^{i}\right) )$ and one of the
ansatz functions (\textit{e.g.} $h_{3}\left( x^{i},y^{3}\right) $ or $%
h_{4}\left( x^{i},y^{3}\right) )$ to take some arbitrary, but
physically interesting form. Then the other ansatz functions can
be analytically determined up to an integration in terms of this
choice. In this way we can generate a lost of different
solutions, but we impose the condition that the initial,
arbitrary choice of the ansatz functions is ``physically
interesting'' which means that one wants to make this original
choice so that the generated final solution yield a well behaved
metric.

In references \cite{14vbel} it is proved that for
\begin{eqnarray}
g_{1} &=&-1,\quad g_{2}=r^{2}\left( \xi \right) q\left( \xi \right) ,
\label{data10} \\
h_{3} &=&-\eta _{3}\left( \xi ,\varphi \right) r^{2}\left( \xi \right) \sin
^{2}\theta ,\quad  \notag \\
h_{4} &=&\eta _{4}\left( \xi ,\varphi \right) h_{4[0]}\left( \xi \right) =1-%
\frac{2\mu }{r}+\varepsilon \frac{\Phi _{4}\left( \xi ,\varphi \right) }{%
2\mu ^{2}},  \notag
\end{eqnarray}%
with coordinates $x^{1}=\xi =\int dr\sqrt{1-2m/r+\varepsilon /r^{2}}%
,x^{2}=\theta ,y^{3}=\varphi ,y^{4}=t$ (the $(r,\theta ,\varphi )$ being
usual radial coordinates), the ansatz (\ref{2ansatz18}) is a vacuum solution
with $\lambda =0$ of the equations (\ref{2eq17}) which defines a black
ellipsoid with mass $\mu ,$ eccentricity $\varepsilon $ and gravitational
polarizations $q\left( \xi \right) ,\eta _{3}\left( \xi ,\varphi \right) $
and $\Phi _{4}\left( \xi ,\varphi \right) .$ Such black holes are certain
deformations of the Schwarzschild metrics to static configurations with
ellipsoidal horizons which is possible if generic off--diagonal metrics and
anholonomic frames are considered. In this subsection we show that the data (%
\ref{data10}) can be extended as to generate exact black
ellipsoid solutions with nontrivial cosmological constant $\lambda
=1/4$ which can be imbedded in string theory.

At the first \ step, we find a class of solutions with $g_{1}=-1$ and $\quad
g_{2}=g_{2}\left( \xi \right) $ solving the equation (\ref{7ricci1a}), which
under such parametrizations transforms to
\begin{equation*}
g_{2}^{\bullet \bullet }-\frac{(g_{2}^{\bullet })^{2}}{2g_{2}}=2g_{2}\lambda
.
\end{equation*}%
With respect to the variable $Z=(g_{2})^{2}$ this equation is written as
\begin{equation*}
Z^{\bullet \bullet }+2\lambda Z=0
\end{equation*}%
which can be integrated in explicit form, $Z=Z_{[0]}\sin \left( \sqrt{%
2\lambda }\xi +\xi _{\lbrack 0]}\right) ,$ for some constants $Z_{[0]}$ and $%
\xi _{\lbrack 0]}$ which means that
\begin{equation}
g_{2}=-Z_{[0]}^{2}\sin ^{2}\left( \sqrt{2\lambda }\xi +\xi _{\lbrack
0]}\right)  \label{2aux2}
\end{equation}%
parametrize a class of solution of (\ref{7ricci1a}) for the signature $\left(
-,-,-,+\right) .$ For $\lambda \rightarrow 0$ we can approximate $%
g_{2}=r^{2}\left( \xi \right) q\left( \xi \right) =-\xi ^{2}$ and $%
Z_{[0]}^{2}=1$ which has compatibility with the data
(\ref{data10}). The solution (\ref{2aux2}) with cosmological
constant (of string or non--string origin) induces oscillations
in the ''horozontal'' part of the d--metric.

The next step is to solve the equation (\ref{6ricci2a}),%
\begin{equation*}
h_{4}^{\ast \ast }-h_{4}^{\ast }[\ln \sqrt{|h_{3}h_{4}|}]^{\ast }=-2\lambda
h_{3}h_{4}.
\end{equation*}%
For $\lambda =0$ a class of solution is given by any $\widehat{h}_{3}$ and $%
\widehat{h}_{4}$ related as
\begin{equation*}
\widehat{h}_{3}=\eta _{0}\left[ \left( \sqrt{|\hat{h}_{4}|}\right) ^{\ast }%
\right]^2
\end{equation*}%
for a constant $\eta _{0}$ chosen to be negative in order to generate the
signature $\left( -,-,-,+\right) .$ For non--trivial $\lambda ,$ we may
search the solution as
\begin{equation}
h_{3}=\widehat{h}_{3}\left( \xi ,\varphi \right) ~q_{3}\left( \xi ,\varphi
\right) \mbox{ and }h_{4}=\widehat{h}_{4}\left( \xi ,\varphi \right) ,
\label{sol15}
\end{equation}%
which solves (\ref{6ricci2a}) if $q_{3}=1$ for $\lambda =0$ and
\begin{equation*}
q_{3}=\frac{1}{4\lambda }\left[ \int \frac{\hat{h}_{3}\hat{h}_{4}}{\hat{h}%
_{4}^{\ast }}d\varphi \right] ^{-1}\mbox{ for }\lambda \neq 0.
\end{equation*}

Now it is easy to write down the solutions of equations (\ref{5ricci3a})
(being a linear equation for $w_{i^{\prime }})$ and (\ref{5ricci4a}) (after
two integrations of $n_{i^{\prime }}$ on $\varphi ),$%
\begin{equation}
w_{i^{\prime }}=\varepsilon \widehat{w}_{i^{\prime }}=-\alpha _{i^{\prime
}}/\beta ,  \label{aux3}
\end{equation}%
were $\alpha _{i^{\prime }}$ and $\beta $ are computed by putting (\ref%
{sol15}) $\ $into corresponding values from (\ref{6abc}) (we chose the
initial conditions as $w_{i^{\prime }}\rightarrow 0$ for $\varepsilon
\rightarrow 0)$ and
\begin{equation*}
n_{1}=\varepsilon \widehat{n}_{1}\left( \xi ,\varphi \right)
\end{equation*}%
where
\begin{eqnarray}
\widehat{n}_{1}\left( \xi ,\varphi \right) &=&n_{1[1]}\left( \xi \right)
+n_{1[2]}\left( \xi \right) \int d\varphi \ \eta _{3}\left( \xi ,\varphi
\right) /\left( \sqrt{|\eta _{4}\left( \xi ,\varphi \right) |}\right)
^{3},\eta _{4}^{\ast }\neq 0;  \label{auxf4} \\
&=&n_{1[1]}\left( \xi \right) +n_{1[2]}\left( \xi \right) \int d\varphi \
\eta _{3}\left( \xi ,\varphi \right) ,\eta _{4}^{\ast }=0;  \notag \\
&=&n_{1[1]}\left( \xi \right) +n_{1[2]}\left( \xi \right) \int d\varphi
/\left( \sqrt{|\eta _{4}\left( \xi ,\varphi \right) |}\right) ^{3},\eta
_{3}^{\ast }=0;  \notag
\end{eqnarray}%
with the functions $n_{k[1,2]}\left( \xi \right) $ to be stated by boundary
conditions.

We conclude that the set of data $g_{1}=-1,$ with non--trivial $g_{2}\left(
\xi \right) ,h_{3},h_{4},w_{i^{\prime }},n_{1}$ stated respectively by (\ref%
{2aux2}), (\ref{sol15}), (\ref{aux3}), (\ref{auxf4}) define a black ellipsoid
solution with explicit dependence on cosmological constant $\lambda ,$ i. e.
a d--metric (\ref{2ansatz18}), which can be induced from string theory for $%
\lambda =1/4.$ The stability of such string static black ellipsoids can be
proven exactly as it was done in Refs. \cite{14vbel} for the vanishing
cosmological constant.

\subsection{2D Finsler structures in string theory}

There are some constructions which prove that two dimensional (2D) Finsler
structures can be embedded into the Einstein's theory of gravity \cite{14dv}.
Here we analyze the conditions when such Finsler configurations can be
generated from string theory. The aim is to include a 2D Finsler metric (\ref%
{fmetric}) into a d--metric (\ref{7dmetric}) being an exact solution of the
string corrected Einstein's equations (\ref{2eq17}).

If
\begin{equation*}
h_{a^{\prime }b^{\prime }}=\frac{1}{2}\frac{\partial ^{2}F^{2}(x^{i^{\prime
}},y^{c^{\prime }})}{\partial y^{a^{\prime }}\partial y^{b^{\prime }}}
\end{equation*}%
for $i^{\prime },j^{\prime },...=1,2$ and $a^{\prime },b^{\prime },...=3,4$
and following the homogeneity conditions for Finsler metric, we can write
\begin{equation*}
F\left( x^{i^{\prime }},y^{3},y^{4}\right) =y^{3}f\left( x^{i^{\prime
}},s\right)
\end{equation*}%
for $s=y^{4}/y^{3}$ with $f\left( x^{i^{\prime }},s\right) =F\left(
x^{i^{\prime }},1,s\right) ,$ that
\begin{eqnarray}
h_{33} &=&\frac{s^{2}}{2}(f^{2})^{\ast \ast }-s(f^{2})^{\ast }+f^{2},
\label{hcoeff} \\
h_{34} &=&-\frac{s^{2}}{2}(f^{2})^{\ast \ast }+\frac{1}{2}(f^{2})^{\ast },
\notag \\
h_{44} &=&\frac{1}{2}(f^{2})^{\ast \ast },  \notag
\end{eqnarray}%
in this subsection we denote $a^{\ast }=\partial a/\partial s.$ The
condition of vanishing of the off--diagonal term $h_{34}$ gives us the
trivial case, when $f^{2}\simeq s^{2}...+...s+...,$ i. e. Riemannian 2D
metrics, so we can not include some general Finsler coefficients (\ref%
{hcoeff}) directly into a diagonal d--metric ansatz (\ref{2ansatz18}). There
is also another problem related with the Cartan's N--connection (\ref{2ncc})
being computed directly from the coefficients (\ref{hcoeff}) generated by a
function $f^{2}:$ all such values substituted into the equations (\ref%
{6ricci2a}) - (\ref{5ricci4a}) result in systems of nonlinear equations
containing the 6th and higher derivatives of $f$ on $s$ which is very
difficult to deal with.

We can include 2D Finsler structures in the Einstein and string gravity via
additional 2D anholnomic frame transforms,%
\begin{equation*}
h_{ab}=e_{a}^{a^{\prime }}\left( x^{i^{\prime }},s\right) ~e_{b}^{b^{\prime
}}\left( x^{i^{\prime }},s\right) ~h_{a^{\prime }b^{\prime }}\left(
x^{i^{\prime }},s\right)
\end{equation*}%
where $h_{a^{\prime }b^{\prime }}$ are induced by a Finsler metric $f^{2}$
as in (\ref{hcoeff}) and $h_{ab}$ may be diagonal, $h_{ab}=diag[h_{a}].$ We
also should consider an embedding of the Cartan's N--connection into a more
general N--connection, $N_{b^{\prime }}^{a^{\prime }}\subset N_{i^{\prime
}}^{a^{\prime }},$ via transforms $N_{i^{\prime }}^{a^{\prime }}=\hat{e}%
_{i^{\prime }}^{b^{\prime }}\left( x^{i^{\prime }},s\right) N_{b^{\prime
}}^{a^{\prime }}$ where $\hat{e}_{i^{\prime }}^{b^{\prime }}\left(
x^{i^{\prime }},s\right) $ are some additional frame transforms in the
off--diagonal sector of the ansatz (\ref{6ansatz}). Such way generated
metrics,
\begin{eqnarray*}
\delta s^{2} &=&g_{i^{\prime }}(dx^{i^{\prime }})^{2}+e_{a}^{a^{\prime
}}~e_{a}^{b^{\prime }}h_{a^{\prime }b^{\prime }}(\delta y^{a})^{2}, \\
\delta y^{a} &=&dy^{a}+\hat{e}_{i^{\prime }}^{b^{\prime }}N_{b^{\prime
}}^{a^{\prime }}dx^{i^{\prime }}\quad
\end{eqnarray*}%
may be constrained by the condition to be an exact solution of the Einstein
equations with (or not) certain string corrections. As a matter of
principle, any string black ellipsoid \ configuration (of the type examined
in the previous subsection) can be related to a 2D Finsler configuration for
corresponding coefficients $e_{a}^{a^{\prime }}$ and $\hat{e}_{i^{\prime
}}^{b^{\prime }}.$ An explicit form of anisotropic configuration is to be
stated by corresponding boundary conditions and the type of anholonomic
transforms. Finally, we note that instead of a 2D Finsler metric (\ref%
{fmetric}) we can use a 2D Lagrange metric (\ref{1mfl}).

\subsection{Moving soliton--black hole string configurations}

In this subsection, we consider that the primed coordinates are 5D ones
obtained after a string compactification background for the NS--NS sector
being common to both the heterotic and type II string theories. The $%
u^{\alpha ^{\prime }}=(x^{i^{\prime }},y^{a^{\prime }})$ are split into
coordinates $x^{i},$ with indices $i^{\prime },j^{\prime },k^{\prime
}...=1,2,3,$ and coordinates $y^{a^{\prime }},$ with indices $a^{\prime
},b^{\prime },c^{\prime },...=4,5.$ Explicitly the coordinates are of the
form
\begin{equation*}
x^{i^{\prime }}=(x^{1}=\chi ,\quad x^{2}=\phi =\ln \widehat{\rho },\quad
x^{3}=\theta )\quad \mbox{and }\quad y^{a^{\prime }}=\left( y^{4}=v,\qquad
y^{5}=p\right) ,
\end{equation*}%
where $\chi $ is the 5th extra--dimensional coordinate and $\widehat{\rho }$
will be related with the 4D Schwarzschild coordinate. We analyze a metric
interval written as
\begin{equation}
ds^{2}=\Omega ^{2}(x^{i^{\prime }},v)\hat{{g}}_{\alpha ^{\prime }\beta
^{\prime }}\left( x^{i^{\prime }},v\right) du^{\alpha ^{\prime }}du^{\beta
^{\prime }},  \label{6cmetric}
\end{equation}%
were the coefficients $\hat{{g}}_{\alpha ^{\prime }\beta ^{\prime }}$ are
parametrized by the ansatz {\scriptsize
\begin{equation}
\left[
\begin{array}{ccccc}
g_{1}+(w_{1}^{\ 2}+\zeta _{1}^{\ 2})h_{4}+n_{1}^{\ 2}h_{5} &
(w_{1}w_{2}+\zeta _{1}\zeta _{2})h_{4}+n_{1}n_{2}h_{5} & (w_{1}w_{3}+\zeta
_{1}\zeta _{3})h_{4}+n_{1}n_{3}h_{5} & (w_{1}+\zeta _{1})h_{4} & n_{1}h_{5}
\\
(w_{1}w_{2}+\zeta _{1}\zeta _{2})h_{4}+n_{1}n_{2}h_{5} & g_{2}+(w_{2}^{\
2}+\zeta _{2}^{\ 2})h_{4}+n_{2}^{\ 2}h_{5} & (w_{2}w_{3}++\zeta _{2}\zeta
_{3})h_{4}+n_{2}n_{3}h_{5} & (w_{2}+\zeta _{2})h_{4} & n_{2}h_{5} \\
(w_{1}w_{3}+\zeta _{1}\zeta _{3})h_{4}+n_{1}n_{3}h_{5} & (w_{2}w_{3}++\zeta
_{2}\zeta _{3})h_{4}+n_{2}n_{3}h_{5} & g_{3}+(w_{3}^{\ 2}+\zeta _{3}^{\
2})h_{4}+n_{3}^{\ 2}h_{5} & (w_{3}+\zeta _{3})h_{4} & n_{3}h_{5} \\
(w_{1}+\zeta _{1})h_{4} & (w_{2}+\zeta _{2})h_{4} & (w_{3}+\zeta _{3})h_{4}
& h_{4} & 0 \\
n_{1}h_{5} & n_{2}h_{5} & n_{3}h_{5} & 0 & h_{5}%
\end{array}%
\right]  \label{6ansatzc}
\end{equation}%
} The metric coefficients are necessary class smooth functions of the form:
\begin{eqnarray}
g_{1} &=&\pm 1,\qquad g_{2,3}=g_{2,3}(x^{2},x^{3}),\qquad
h_{4,5}=h_{4,5}(x^{i^{\prime }},v)=\eta
_{4,5}(x^{i},v)h_{4,5[0]}(x^{k^{\prime }}),  \notag \\
w_{i^{\prime }} &=&w_{i^{\prime }}(x^{k^{\prime }},v),\qquad n_{i^{\prime
}}=n_{i^{\prime }}(x^{k^{\prime }},v),\qquad \zeta _{i^{\prime }}=\zeta
_{i^{\prime }}(x^{k^{\prime }},v),\qquad \Omega =\Omega (x^{i^{\prime }},v).
\label{par1}
\end{eqnarray}%
The quadratic line element (\ref{6cmetric}) with metric coefficients (\ref%
{6ansatzc}) can be diagonalized by anholonmic transforms,
\begin{equation}
\delta s^{2}=\Omega ^{2}(x^{i^{\prime
}},v)[g_{1}(dx^{1})^{2}+g_{2}(dx^{2})^{2}+g_{3}(dx^{3})^{2}+h_{4}(\hat{{%
\delta }}v)^{2}+h_{5}(\delta p)^{2}],  \label{4cdmetric}
\end{equation}%
with respect to the anholonomic co--frame $\left( dx^{i^{\prime }},\hat{{%
\delta }}v,\delta p\right) ,$ where
\begin{equation}
\hat{\delta}v=dv+(w_{i^{\prime }}+\zeta _{i^{\prime }})dx^{i^{\prime
}}+\zeta _{5}\delta p\qquad \mbox{
and }\qquad \delta p=dp+n_{i^{\prime }}dx^{i^{\prime }}  \label{2ddif2}
\end{equation}%
which is dual to the frame $\left( \hat{{\delta }}_{i^{\prime }},\partial
_{4},\hat{{\partial }}_{5}\right) ,$ where
\begin{equation}
\hat{{\delta }}_{i^{\prime }}=\partial _{i^{\prime }}-(w_{i^{\prime }}+\zeta
_{i^{\prime }})\partial _{4}+n_{i^{\prime }}\partial _{5},\qquad \hat{{%
\partial }}_{5}=\partial _{5}-\zeta _{5}\partial _{4}.  \label{2dder2}
\end{equation}%
The simplest way to compute the nontrivial coefficients of the Ricci tensor
for the (\ref{4cdmetric}) is to do this with respect to anholonomic bases ( %
\ref{2ddif2}) and (\ref{2dder2}) (see details in \cite{14vmethod,14vsingl}), which
reduces the 5D vacuum Einstein equations to the following system (in this
paper containing a non--trivial cosmological constant):
\begin{eqnarray}
\frac{1}{2}R_{1}^{1}=R_{2}^{2}=R_{3}^{3}=-\frac{1}{2g_{2}g_{3}}%
[g_{3}^{\bullet \bullet }-\frac{g_{2}^{\bullet }g_{3}^{\bullet }}{2g_{2}}-%
\frac{(g_{3}^{\bullet })^{2}}{2g_{3}}+g_{2}^{^{\prime \prime }}-\frac{%
g_{2}^{^{\prime }}g_{3}^{^{\prime }}}{2g_{3}}-\frac{(g_{2}^{^{\prime }})^{2}%
}{2g_{2}}] &=&\lambda ,  \label{ricci7a} \\
R_{4}^{4}=R_{5}^{5}=-\frac{\beta }{2h_{4}h_{5}} &=&\lambda ,  \label{ricci8a}
\\
R_{4i^{\prime }}=-w_{i^{\prime }}\frac{\beta }{2h_{5}}-\frac{\alpha
_{i^{\prime }}}{2h_{5}} &=&0,  \label{ricci9a} \\
R_{5i^{\prime }}=-\frac{h_{5}}{2h_{4}}\left[ n_{i^{\prime }}^{\ast \ast
}+\gamma n_{i^{\prime }}^{\ast }\right] &=&0,  \label{ricci10a}
\end{eqnarray}%
with the conditions that
\begin{equation}
\Omega ^{q_{1}/q_{2}}=h_{4}~(q_{1}\mbox{ and }q_{2}\mbox{ are
integers}),  \label{5confq}
\end{equation}%
and $\zeta _{i}$ satisfies the equations
\begin{equation}
\partial _{i^{\prime }}\Omega -(w_{i^{\prime }}+\zeta _{i^{\prime }})\Omega
^{\ast }=0,  \label{6confeq}
\end{equation}%
The coefficients of equations (\ref{ricci7a}) - (\ref{ricci10a}) are given
by
\begin{equation}
\alpha _{i^{\prime }}=\partial _{i}{h_{5}^{\ast }}-h_{5}^{\ast }\partial
_{i^{\prime }}\ln \sqrt{|h_{4}h_{5}|},\qquad \beta =h_{5}^{\ast \ast
}-h_{5}^{\ast }[\ln \sqrt{|h_{4}h_{5}|}]^{\ast },\qquad \gamma =\frac{%
3h_{5}^{\ast }}{2h_{5}}-\frac{h_{4}^{\ast }}{h_{4}}.  \label{abc1}
\end{equation}%
The various partial derivatives are denoted as $a^{\bullet }=\partial
a/\partial x^{2},a^{^{\prime }}=\partial a/\partial x^{3},a^{\ast }=\partial
a/\partial v.$

The system of equations (\ref{ricci7a})--(\ref{ricci10a}), (\ref{5confq}) and
(\ref{6confeq}) can be solved by choosing one of the ansatz functions (%
\textit{e.g.} $h_{4}\left( x^{i^{\prime }},v\right) $ or $h_{5}\left(
x^{i^{\prime }},v\right) )$ to take some arbitrary, but physically
interesting form. Then the other ansatz functions can be analytically
determined up to an integration in terms of this choice. In this way one can
generate many solutions, but the requirement that the initial, arbitrary
choice of the ansatz functions be ``physically interesting'' means that one
wants to make this original choice so that the final solution generated in
this way yield a well behaved solution. To satisfy this requirement we start
from well known solutions of Einstein's equations and then use the above
procedure to deform this solutions in a number of ways as to include it in a
string theory. In the simplest case we derive 5D locally anisotropic string
gravity solutions connected to the the Schwarzschild solution in \textit{%
isotropic spherical coordinates} \cite{14ll} given by the quadratic line
interval%
\begin{equation}
ds^{2}=\left( \frac{\widehat{\rho }-1}{\widehat{\rho }+1}\right)
^{2}dt^{2}-\rho _{g}^{2}\left( \frac{\widehat{\rho }+1}{\widehat{\rho }}%
\right) ^{4}\left( d\widehat{\rho }^{2}+\widehat{\rho }^{2}d\theta ^{2}+%
\widehat{\rho }^{2}\sin ^{2}\theta d\varphi ^{2}\right) .  \label{4schw}
\end{equation}%
We identify the coordinate $\widehat{\rho }$ with the re--scaled isotropic
radial coordinate, $\widehat{\rho }=\rho /\rho _{g},$ with $\rho
_{g}=r_{g}/4 $; $\rho $ is connected with the usual radial coordinate $r$ by
$r=\rho \left( 1+r_{g}/4\rho \right) ^{2}$; $r_{g}=2G_{[4]}m_{0}/c^{2}$ is
the 4D Schwarzschild radius of a point particle of mass $m_{0}$; $%
G_{[4]}=1/M_{P[4]}^{2}$ is the 4D Newton constant expressed via the Planck
mass $M_{P[4]}$ (in general, we may consider that $M_{P[4]}$ may be an
effective 4D mass scale which arises from a more fundamental scale of the
full, higher dimensional spacetime); we set $c=1.$

The metric (\ref{4schw}) is a vacuum static solution of 4D Einstein equations
with spherical symmetry describing the gravitational field of a point
particle of mass $m_{0}.$ It has a singularity for $r=0$ and a spherical
horizon at $r=r_{g},$ or at $\widehat{\rho }=1$ in the re--scaled isotropic
coordinates. This solution is parametrized by a diagonal metric given with
respect to holonomic coordinate frames. This spherically symmetric solution
can be deformed in various interesting ways using the anholonomic frames
method.

Vacuum gravitational 2D solitons in 4D Einstein vacuum gravity were
originally investigated by Belinski and Zakharov \cite{14belinski}. In Refs. %
\cite{14vsolsp} 3D solitonic configurations were constructed on anisotropic
Taub-NUT backgrounds. Here we show that 3D solitonic/black hole
configurations can be embedded into the 5D locally anisotropic string
gravity.

\subsubsection{3D solitonic deformations in string gravity}

The simplest way to construct a solitonic deformation of the off--diagonal
metric in equation (\ref{6ansatzc}) is to take one of the ``polarization''
factors $\eta _{4}$, $\eta _{5}$ from (\ref{par1}) or the ansatz function $%
n_{i^{\prime }}$ as a solitonic solution of some particular non-linear
equation. The rest of the ansatz functions can then be found by carrying out
the integrations of equations (\ref{ricci7a})-- (\ref{6confeq}).

As an example of this procedure we suggest to take $\eta _{5}(r,\theta ,\chi
)$ as a soliton solution of the Kadomtsev--Petviashvili (KdP) equation or
(2+1) sine-Gordon (SG) equation (Refs. \cite{14kad} contain the original
results, basic references and methods for handling such non-linear equations
with solitonic solutions). In the KdP case $\eta _{5}(v,\theta ,\chi )$
satisfies the following equation
\begin{equation}
\eta _{5}^{\ast \ast }+\epsilon \left( \dot{\eta}_{5}-6\eta _{5}\eta
_{5}^{\prime }+\eta _{5}^{\prime \prime \prime }\right) ^{\prime }=0,\qquad
\epsilon =\pm 1,  \label{kdp}
\end{equation}%
while in the most general SG case $\eta _{5}(v,\chi )$ satisfies
\begin{equation}
\pm \eta _{5}^{\ast \ast }\mp \ddot{\eta}_{5}=\sin (\eta _{5}).
\label{sineq}
\end{equation}%
For simplicity, we can also consider less general versions of the SG
equation where $\eta _{5}$ depends on only one (\textit{e.g.} $v$ and $x_{1}$%
) variable. We use the notation $\eta _{5}=\eta _{5}^{KP}$ or $\eta
_{5}=\eta _{5}^{SG}$ ($h_{5}=h_{5}^{KP}$ or $h_{5}=h_{5}^{SG}$) depending on
if ($\eta _{5}$ ) ($h_{5}$) satisfies equation (\ref{kdp}), or (\ref{sineq})
respectively.

For a stated solitonic form for $h_{5}=h_{5}^{KP,SG},$ $h_{4}$ can be found
from
\begin{equation}
h_{4}=h_{4}^{KP,SG}=h_{[0]}^{2}\left[ \left( \sqrt{|h_{5}^{KP,SG}(x^{i^{%
\prime }},v)|}\right) ^{\ast }\right] ^{2}  \label{p1b}
\end{equation}%
where $h_{[0]}$ is a constant. By direct substitution it can be shown that
equation (\ref{p1b}) solves equation (\ref{ricci8a}) with $\beta $ given by
( \ref{abc1}) when $h_{5}^{\ast }\neq 0$ but $\lambda =0.$ If $h_{5}^{\ast
}=0,$ then $\hat{h}_{4}$ is an arbitrary function $\hat{h}_{4}(x^{i^{\prime
}},v)$. In either case we will denote the ansatz function determined in this
way as $\hat{h}_{4}^{KP,SG}$ although it does not necessarily share the
solitonic character of $\hat{h}_{5}$. Substituting the values $\hat{h}%
_{4}^{KP,SG}$ and $\hat{h}_{5}^{KP,SG}$ into $\gamma $ from equation (\ref%
{6abc}) gives, after two $v$ integrations of equation (\ref{5ricci4a}), the
ansatz functions $n_{i^{\prime }}=n_{i^{\prime }}^{KP,SG}(v,\theta ,\chi ).$
Solutions with $\lambda \neq 0$ can be generated similarly as in (\ref{sol15}%
) by redefining
\begin{equation*}
h_{4}=\widehat{h}_{4}\left( x^{i^{\prime }},v\right) ~q_{4}\left(
x^{i^{\prime }},v\right) \mbox{ and }h_{5}=\widehat{h}_{5}\left(
x^{i^{\prime }},v\right) ,
\end{equation*}%
which solves (\ref{ricci8a}) if $q_{4}=1$ for $\lambda =0$ and
\begin{equation}
q_{4}=\frac{1}{4\lambda }\left[ \int \frac{\hat{h}_{5}\hat{h}_{4}}{\hat{h}%
_{5}^{\ast }}dv\right] ^{-1}\mbox{ for }\lambda \neq 0.  \label{quf}
\end{equation}

Here, for simplicity, we \ may set $g_{2}=-1$ but
\begin{equation}
g_{3}=-Z_{[0]}^{2}\sin ^{2}\left( \sqrt{2\lambda }x^{3}+\xi
_{\lbrack 0]}\right) ,\  Z_{[0]}, \xi _{[0]}=const, \label{aux5}
\end{equation}%
parametrize a class of solution of (\ref{ricci7a}) for the signature $\left(
-,-,-,-,+\right) $ like we constructed the solution (\ref{2aux2}). In ref. %
\cite{14vsolsp,14vsingl} it was shown how to generate solutions using 2D
solitonic configurations for $g_{2}$ or $g_{3}.$

The main conclusion to be formulated here is that the ansatz (\ref{6ansatzc}%
), when treated with anholonomic frames, has a degree of freedom that allows
one to pick one of the ansatz functions ($\eta _{4}$ , $\eta _{5}$ , or $%
n_{i^{\prime }}$) to satisfy some 3D solitonic equation. Then in terms of
this choice all the other ansatz functions can be generated up to carrying
out some explicit integrations and differentiations. In this way it is
possible to build exact solutions of the 5D string gravity equations with a
solitonic character.

\subsubsection{Solitonically propagating string black hole backgrounds}

\label{solitonical}

The Schwarzschild solution is given in terms of the parameterization in (\ref%
{6ansatzc}) by
\begin{eqnarray}
g_{1} &=&\pm 1,\qquad g_{2}=g_{3}=-1,\qquad h_{4}=h_{4[0]}(x^{i^{\prime
}}),\qquad h_{5}=h_{5[0]}(x^{i^{\prime }}),  \notag \\
w_{i^{\prime }} &=&0,\qquad n_{i^{\prime }}=0,\qquad \zeta _{i^{\prime
}}=0,\qquad \Omega =\Omega _{\lbrack 0]}(x^{i^{\prime }}),  \notag
\end{eqnarray}%
with
\begin{equation}
h_{4[0]}(x^{i})=\frac{b(\phi )}{a(\phi )},\qquad h_{5[0]}(x^{i^{\prime
}})=-\sin ^{2}\theta ,\qquad \Omega _{\lbrack 0]}^{2}(x^{i^{\prime
}})=a(\phi )  \label{3aux1}
\end{equation}%
or alternatively, for another class of solutions,
\begin{equation}
h_{4[0]}(x^{i^{\prime }})=-\sin ^{2}\theta ,\qquad h_{5[0]}(x^{i^{\prime }})=%
\frac{b(\phi )}{a(\phi )},  \label{aux2a}
\end{equation}%
were
\begin{equation}
a(\phi )=\rho _{g}^{2}\frac{\left( e^{\phi }+1\right) ^{4}}{e^{2\phi }}%
\qquad \mbox{ and }\qquad b(\phi )=\frac{\left( e^{\phi }-1\right) ^{2}}{%
\left( e^{\phi }+1\right) ^{2}},  \label{ab}
\end{equation}%
Putting this together gives
\begin{equation}
ds^{2}=\pm d\chi ^{2}-a(\phi )\left( d\lambda ^{2}+d\theta ^{2}+\sin
^{2}\theta d\varphi ^{2}\right) +b\left( \phi \right) dt^{2}  \label{schw5}
\end{equation}%
which represents a trivial embedding of the 4D Schwarzschild metric (\ref%
{4schw}) into the 5D spacetime. We now want to deform anisotropically the
coefficients of (\ref{schw5}) in the following way
\begin{eqnarray*}
h_{4[0]}(x^{i^{\prime }}) &\rightarrow &h_{4}(x^{i^{\prime }},v)=\eta
_{4}\left( x^{i^{\prime }},v\right) h_{4[0]}(x^{i^{\prime }}), \\
~h_{5[0]}(x^{i^{\prime }}) &\rightarrow &h_{5}(x^{i^{\prime }},v)=\eta
_{5}\left( x^{i^{\prime }},v\right) h_{5[0]}(x^{i^{\prime }}), \\
\Omega _{\lbrack 0]}^{2}(x^{i^{\prime }}) &\rightarrow &\Omega
^{2}(x^{i^{\prime }},v)=\Omega _{\lbrack 0]}^{2}(x^{i^{\prime }})\Omega
_{\lbrack 1]}^{2}(x^{i^{\prime }},v).
\end{eqnarray*}%
The factors $\eta _{i^{\prime }}$ and $\Omega _{\lbrack 1]}^{2}$ can be
interpreted as re-scaling or ''renormalizing'' the original ansatz
functions. These gravitational ``polarization'' factors -- $\eta _{4,5}$ and
$\Omega _{\lbrack 1]}^{2}$ -- generate non--trivial values for $w_{i^{\prime
}}(x^{i^{\prime }},v),n_{i^{\prime }}(x^{i^{\prime }},v)$ and $\zeta
_{i^{\prime }}(x^{i^{\prime }},v),$ via the vacuum equations (\ref{ricci7a}%
)-- (\ref{6confeq}). We shall also consider more general nonlinear
polarizations which can not be expresses as $h\sim $ $\eta h_{[0]}$ and show
how the coefficients $a(\phi )$ and $b(\phi )$ of the Schwarzschild metric
can be polarized by choosing the original, arbitrary ansatz function to be
some 3D soliton configuration.

The horizon is defined by the vanishing of the coefficient $b\left( \phi
\right) $ from equation (\ref{ab}). This occurs when $e^{\phi }=1$. In order
to create a solitonically propagating black hole we define the function $%
\tau =\phi -\tau _{0}\left( \chi ,v\right) $, and let $\tau _{0}\left( \chi
,v\right) $ be a soliton solution of either the 3D KdP equation (\ref{kdp}),
or the SG equation (\ref{sineq}). This redefines $b\left( \phi \right) $ as
\begin{equation*}
b\left( \phi \right) \rightarrow B\left( x^{i^{\prime }},v\right) =\frac{%
e^{\tau }-1}{e^{\phi }+1}.
\end{equation*}%
A class of 5D string gravity metrics can be constructed by parametrizing $%
h_{4}=\eta _{4}\left( x^{i^{\prime }},v\right) $ $h_{4[0]}(x^{i^{\prime }})$
and $h_{5}=B\left( x^{i^{\prime }},v\right) /a\left( \phi \right) $, or
inversely, $h_{4}=B\left( x^{i^{\prime }},v\right) /a\left( \phi \right) $
and $h_{5}=\eta _{5}\left( x^{i^{\prime }},v\right)$\\ $ h_{5[0]}(x^{i^{\prime
}}).$ The polarization $\eta _{4}\left( x^{i^{\prime }},v\right) $ \ (or \ $%
\eta _{5}\left( x^{i^{\prime }},v\right) )$ is determined from equation (\ref%
{p1b}) with the factor $q_{4}$ (\ref{quf}) \ included in $h^{2},$
\begin{equation*}
|\eta _{4}\left( x^{i^{\prime }},v\right) h_{4(0)}(x^{i^{\prime }})|=h^{2}
\left[ \left( \sqrt{\left| \frac{B\left( x^{i^{\prime }},v\right) }{a\left(
\phi \right) }\right| }\right) ^{\ast }\right] ^{2}
\end{equation*}%
or
\begin{equation*}
\left| \frac{B\left( x^{i},v\right) }{a\left( \phi \right) }\right|
=h^{2}h_{5(0)}(x^{i^{\prime }})\left[ \left( \sqrt{|\eta _{5}\left(
x^{i^{\prime }},v\right) |}\right) ^{\ast }\right] ^{2}.
\end{equation*}%
The last step in constructing of the form for these solitonically
propagating black hole solutions is to use $h_{4}$ and $h_{5}$ in equation (%
\ref{5ricci4a}) to determine $n_{k^{\prime }}$
\begin{eqnarray}
n_{k^{\prime }} &=&n_{k^{\prime }[1]}(x^{i^{\prime }})+n_{k^{\prime
}[2]}(x^{i^{\prime }})\int \frac{h_{4}}{(\sqrt{|h_{5}|})^{3}}dv,\qquad
h_{5}^{\ast }\neq 0;  \label{nnn1} \\
&=&n_{k^{\prime }[1]}(x^{i^{\prime }})+n_{k^{\prime }[2]}(x^{i^{\prime
}})\int h_{4}dv,\qquad h_{5}^{\ast }=0;  \notag \\
&=&n_{k^{\prime }[1]}(x^{i^{\prime }})+n_{k^{\prime }[2]}(x^{i^{\prime
}})\int \frac{1}{(\sqrt{|h_{5}|})^{3}}dv,\qquad h_{4}^{\ast }=0,  \notag
\end{eqnarray}%
where $n_{k[1,2]}\left( x^{i^{\prime }}\right) $ are set by boundary
conditions.

The simplest version of the above class of solutions are the so--called $t$%
--solutions (depending on $t$--variable), defined by a pair of ansatz
functions, $\left[ B\left( x^{i^{\prime }},t\right) ,h_{5(0)}\right] ,$ with
$h_{5}^{\ast }=0$ and $B\left( x^{i^{\prime }},t\right) $ being a 3D
solitonic configuration. Such solutions have a spherical horizon when $%
h_{4}=0,$ \textit{i.e.} when $\tau =0$. This solution describes a
propagating black hole horizon. The propagation occurs via a 3D solitonic
wave form depending on the time coordinate, $t$, and on the 5$^{th}$
coordinate $\chi $. The form of the ansatz functions for this solution (both
with trivial and non-trivial conformal factors) is
\begin{eqnarray}
\mbox{$t$--solutions} &:&(x^{1}=\chi ,\qquad x^{2}=\phi ,\qquad x^{3}=\theta
,\qquad y^{4}=v=t,\qquad y^{5}=p=\varphi ),  \notag \\
g_{1} &=&\pm 1,g_{2}=-1,g_{3}=-Z_{[0]}^{2}\sin ^{2}\left( \sqrt{2\lambda }%
x^{3}+\xi _{\lbrack 0]}\right) ,\tau =\phi -\tau _{0}\left( \chi ,t\right) ,
\notag \\
h_{4} &=&B/a(\phi ),h_{5}=h_{5(0)}(x^{i^{\prime }})=-\sin ^{2}\theta ,\omega
=\eta _{5}=1,B\left( x^{i^{\prime }},t\right) =\frac{e^{\tau }-1}{e^{\phi }+1%
},  \notag \\
w_{i^{\prime }} &=&\zeta _{i^{\prime }}=0,\qquad n_{k^{\prime }}\left(
x^{i^{\prime }},t\right) =n_{k^{\prime }[1]}(x^{i^{\prime }})+n_{k^{\prime
}[2]}(x^{i^{\prime }})\int B\left( x^{i^{\prime }},t\right) dt,
\label{sol6t}
\end{eqnarray}%
where $q_{4}$ is chosen to preserve the condition $w_{i^{\prime }}=\zeta
_{i^{\prime }}=0.$

As a simple example of the above solutions we take $\tau _{0}$ to satisfy
the SG equation $\partial _{\chi \chi }\tau _{0}-\partial _{tt}\tau
_{0}=\sin (\tau _{0})$. This has the standard propagating kink solution
\begin{equation*}
\tau _{0}(\chi ,t)=4\tan ^{-1}\left[ \pm \gamma (\chi -Vt)\right]
\end{equation*}%
where $\gamma =(1-V^{2})^{-1/2}$ and $V$ is the velocity at which the kink
moves into the extra dimension $\chi $. To obtain the simplest form of this
solution we also take $n_{k^{\prime }[1]}(x^{i^{\prime }})=n_{k^{\prime
}[2]}(x^{i^{\prime }})=0.$ This example can be easily extended to solutions
with a non-trivial conformal factor $\Omega $ that gives an exponentially
suppressing factor, $\exp [-2k|\chi |],$ see details in Ref. \cite{14vsingl}.
In this manner one has an effective 4D black hole which propagates from the
3D brane into the non-compact, but exponentially suppressed extra dimension,
$\chi $.

The solution constructed in this subsection describes propagating 4D
Schwarzschild black holes in a bulk 5D spacetime obtained from string
theory. The propagation arises from requiring that certain of the ansatz
functions take a 3D soliton form. In the simplest version of these
propagating solutions the parameters of the ansatz functions are constant,
and the horizons are spherical. It can be also shown that such propagating
solutions could be formed with a polarization of the parameters and/or
deformation of the horizons, see the non--string case in \cite{14vsingl}.

\subsection{Noncommutative anisotropic wormholes and strings}

Let us construct and analyze an exact 5D solution of the string gravity
which can also considered as a noncommutative structure in string theory. \
The d--metric ansatz is taken in the form%
\begin{eqnarray}
\delta s^{2} &=&g_{1}(dx^{1})^{2}+g_{2}(dx^{2})^{2}+g_{3}(dx^{3})^{2}+h_{4}({%
\delta }y^{4})^{2}+h_{5}(\delta y^{5})^{2},  \notag \\
{\delta }y^{4} &=&{d}y^{4}+w_{k^{\prime }}\left( x^{i^{\prime }},v\right)
dx^{k^{\prime }},{\delta }y^{5}={d}y^{5}+n_{k^{\prime }}\left( x^{i^{\prime
}},v\right) dx^{k^{\prime }};i^{\prime },k^{\prime }=1,2,3,  \label{ans20}
\end{eqnarray}%
where
\begin{eqnarray}
g_{1} &=&1,\quad g_{2}=g_{2}(r),\quad g_{3}=-a(r),  \label{anz6a} \\
h_{4} &=&\hat{h}_{4}=\widehat{\eta }_{4}\left( r,\theta ,\varphi \right)
h_{4[0]}(r),\quad h_{5}=\hat{h}_{5}=\widehat{\eta }_{5}\left( r,\theta
,\varphi \right) h_{5[0]}(r,\theta )  \notag
\end{eqnarray}%
for the parametrization of coordinate of type
\begin{equation}
x^{1}=t,x^{2}=r,x^{3}=\theta ,y^{4}=v=\varphi ,y^{5}=p=\chi  \label{coord5}
\end{equation}%
where $t$ is the time coordinate, $\left( r,\theta ,\varphi \right) $ are
spherical coordinates, $\chi $ is the 5th coordinate; $\varphi $ is the
anholonomic coordinate; for this ansatz there is not considered the
dependence of d--metric coefficients on the second anholonomic coordinate $%
\chi .$ The data%
\begin{eqnarray}
g_{1} &=&1,~\hat{g}_{2}=-1,~g_{3}=-a(r),  \label{data6a} \\
~h_{4[0]}(r) &=&-r_{0}^{2}e^{2\psi (r)},~\eta _{4}=1/\kappa _{r}^{2}\left(
r,\theta ,\varphi \right) ,~h_{5[0]}=-a\left( r\right) \sin ^{2}\theta
,~\eta _{5}=1,  \notag \\
w_{1} &=&\widehat{w}_{1}=\omega \left( r\right) ,~w_{2}=\widehat{w}%
_{2}=0,w_{3}=~\widehat{w}_{3}=n\cos \theta /\kappa _{n}^{2}\left( r,\theta
,\varphi \right) ,  \notag \\
n_{1} &=&\widehat{n}_{1}=0,~n_{2,3}=\widehat{n}_{2,3}=n_{2,3[1]}\left(
r,\theta \right) \int \ln |\kappa _{r}^{2}\left( r,\theta ,\varphi \right)
|d\varphi  \notag
\end{eqnarray}%
for some constants $r_{0}$ $\ $\ and $n$ and arbitrary functions $a(r),\psi
(r)$ and arbitrary vacuum gravitational polarizations $\kappa _{r}\left(
r,\theta ,\varphi \right) $ and $\kappa _{n}\left( r,\theta ,\varphi \right)
$ define an exact vacuum 5D solution of Kaluza--Klein gravity \cite{14vsingl1}
describing a locally anisotropic wormhole with elliptic gravitational vacuum
polarization of charges,
\begin{equation*}
\frac{q_{0}^{2}}{4a\left( 0\right) \kappa _{r}^{2}}+\frac{Q_{0}^{2}}{%
4a\left( 0\right) \kappa _{n}^{2}}=1,
\end{equation*}%
where $q_{0}=2\sqrt{a\left( 0\right) }\sin \alpha _{0}$ and $Q_{0}=2\sqrt{%
a\left( 0\right) }\cos \alpha _{0}$ are respectively the electric and
magnetic charges and $2\sqrt{a\left( 0\right) }\kappa _{r}$ and $2\sqrt{%
a\left( 0\right) }\kappa _{n}$ are ellipse's axes.

The first aim in this subsection is to prove that following the ansatz (\ref%
{ans20}) we can construct locally anisotropic wormhole metrics in string
gravity as solutions of the system of equations (\ref{ricci7a}) - (\ref%
{ricci10a}) with redefined coordinates as in (\ref{coord5}). Having the
vacuum data (\ref{data6a}) we may generalize the solution for a nontrivial
cosmological constant following the method presented in subsection \ref%
{solitonical}, when the new solutions are represented
\begin{equation}
h_{4}=\widehat{h}_{4}\left( x^{i^{\prime }},v\right) ~q_{4}\left(
x^{i^{\prime }},v\right) \mbox{ and }h_{5}=\widehat{h}_{5}\left(
x^{i^{\prime }},v\right) ,  \label{shift3}
\end{equation}%
with $\widehat{h}_{4,5}$ taken as in (\ref{anz6a}) which solves (\ref%
{ricci8a}) if $q_{4}=1$ for $\lambda =0$ and
\begin{equation*}
q_{4}=\frac{1}{4\lambda }\left[ \int \frac{\hat{h}_{5}\left( r,\theta
,\varphi \right) \hat{h}_{4}\left( r,\theta ,\varphi \right) }{\hat{h}%
_{5}^{\ast }\left( r,\theta ,\varphi \right) }d\varphi \right] ^{-1}%
\mbox{
for }\lambda \neq 0.
\end{equation*}%
This $q_{4}$ can be considered as an additional polarization to $\eta _{4}$
induced by the cosmological constant $\lambda .$ We state $g_{2}=-1$ but
\begin{equation*}
g_{3}=-\sin ^{2}\left( \sqrt{2\lambda }\theta +\xi _{\lbrack 0]}\right) ,
\end{equation*}%
which give of solution of (\ref{ricci7a}) with signature $\left(
+,-,-,-,-\right) $ which is different from the solution (\ref{2aux2}). A
non--trivial $q_{4}$ results in modification of coefficients (\ref{abc1}),
\begin{eqnarray*}
\alpha _{i^{\prime }} &=&\hat{\alpha}_{i^{\prime }}+\alpha _{i^{\prime
}}^{[q]},~\beta =\hat{\beta}+\beta ^{\lbrack q]},~\gamma =\hat{\gamma}%
+\gamma ^{\lbrack q]}, \\
\hat{\alpha}_{i^{\prime }} &=&\partial _{i}{\hat{h}_{5}^{\ast }}-\hat{h}%
_{5}^{\ast }\partial _{i^{\prime }}\ln \sqrt{|\hat{h}_{4}\hat{h}_{5}|}%
,\qquad \hat{\beta}=\hat{h}_{5}^{\ast \ast }-\hat{h}_{5}^{\ast }[\ln \sqrt{|%
\hat{h}_{4}\hat{h}_{5}|}]^{\ast },\qquad \hat{\gamma}=\frac{3\hat{h}%
_{5}^{\ast }}{2\hat{h}_{5}}-\frac{\hat{h}_{4}^{\ast }}{\hat{h}_{4}} \\
\alpha _{i^{\prime }}^{[q]} &=&-h_{5}^{\ast }\partial _{i^{\prime }}\ln
\sqrt{|q_{4}|},\qquad \beta ^{\lbrack q]}=-h_{5}^{\ast }[\ln \sqrt{|q_{4}|}%
]^{\ast },\qquad \gamma ^{\lbrack q]}=-\frac{q_{4}^{\ast }}{q_{4}},
\end{eqnarray*}%
which following formulas (\ref{ricci9a}) and (\ref{ricci10a}) result in
additional terms to the N--connec\-ti\-on coefficients, i. e.
\begin{equation}
w_{i^{\prime }}=\widehat{w}_{i^{\prime }}+w_{i^{\prime }}^{[q]}~\mbox{ and }%
n_{i^{\prime }}=\widehat{n}_{i^{\prime }}+n_{i^{\prime }}^{[q]},
\label{ncon05}
\end{equation}%
with $w_{i^{\prime }}^{[q]}$ and $n_{i^{\prime }}^{[q]}$ computed by using
respectively $\alpha _{i^{\prime }}^{[q]},\beta ^{\lbrack q]}$ and $\gamma
^{\lbrack q]}.$

The N--connection coefficients (\ref{ncon05}) can be transformed partially
into a $B$--field with $\{B_{i^{\prime }j^{\prime }},B_{b^{\prime }j^{\prime
}}\}$ defined by integrating the conditions (\ref{2aux02}), i. e.%
\begin{equation}
B_{i^{\prime }j^{\prime }}=B_{i^{\prime }j^{\prime }[0]}\left( x^{k^{\prime
}}\right) +\int h_{4}\delta _{\lbrack i^{\prime }}w_{j^{\prime }]}d\varphi
,~B_{4j^{\prime }}=B_{4j^{\prime }[0]}\left( x^{k^{\prime }}\right) +\int
h_{4}w_{j^{\prime }}^{\ast }d\varphi ,  \label{last}
\end{equation}%
for some arbitrary functions $B_{i^{\prime }j^{\prime }[0]}\left(
x^{k^{\prime }}\right) $ and $B_{4j^{\prime }[0]}\left( x^{k^{\prime
}}\right) .$ The string background corrections are presented via nontrivial $%
w_{i^{\prime }}^{[q]}$ induced by $\lambda =1/4.$ The formulas (\ref{last})
consist the second aim of this subsection: to illustrate how a a $B$--field
inducing noncommutativity may be related with a N--connection inducing
local anisotropy. This is an explicit example of locally anisotropic
noncommutative configuration contained in string theory. For the considered
class of wormhole solutions the coefficients $n_{i^{\prime }}$ do not
contribute into the noncommutative configuration, but, in general, following
(\ref{2aux01}), they can be also related to noncommutativity.

\section{Comments and Questions}

In this paper, we have developed the method of anholonomic frames
 and associated nonlinear connections from a viewpoint of
application in noncommutative geometry and string theory. We note
in this retrospect that several futures connecting Finsler like
generalizations of gravity and gauge theories, which in the past
were considered ad hoc and sophisticated, actually have a very
natural physical and geometric interpretation in the
noncommutative and D--brane picture in string/M--theory. Such
locally anisotropic and/ or noncommutative configurations are
hidden even in general relativity and its various Kaluza--Klein
like and supergravity extension. To emphasize them we have to
consider off--diagonal metrics which can be diagonalized in
result of certain anholonomic frame transforms which induce also
nonlinear connection structures in the curved spacetime, in
general, with noncompactified extra dimensions.

On general grounds, it could be said the the appearance of noncommutative
and Finsler like geometry when considering $B$--fields, off--diagonal
metrics and anholonomic frames (all parametrized, in general, by
noncommutative matrices) is a natural thing. Such implementations in the
presence of D--branes and matrix approaches to M--theory were proven here to
have explicit realizations and supported by six background constructions
elaborated in this paper:

First, both the local anisotropy and noncommutativity can be derived from
considering string propagation in general manifolds and bundles and in
various low energy string limits. This way the anholonomic Einstein and
Finsler generalized gravity models are generated from string theory.

Second, the anholonomic constructions with associated nonlinear connection
geometry can be explicitly modelled on superbundles which results in
superstring effective actions with anholonomic (super) field equations which
can be related to various superstring and supergravity theories.

Third, noncommutative geometries and associated differential calculi can be
distinguished in  anholonomic geometric form which allows  formulation of
locally anisotropic field theories with anholonomic symmetries.

Forth, anholonomy and noncommutativity can be related to string/M--theory \
following consequently the matrix algebra and geometry and/or associated to
nonlinear connections noncommutative covariant differential calculi.

Fifth, different models of locally anisotropic gravity with explicit limits
to string and Einstein gravity can be realized on noncommutative D--branes.

Sixth, the anholonomic frame method is a very powerful one in
constructing and investigating  new classes of exact solutions in
string and gravity theories; such solutions contain generic
noncommutativity and/or  local anisotropy and can be parametrized
as to describe  locally anisotropic black hole configurations,
Finsler like structures, anisoropic solitonic and moving string
black hole metrics, or noncommutative and anisotropic wormhole
structures which may be derived in Einstein gravity and/or its
Kaluza--Klein and (super) string generalizations.

The obtained in this paper results have a recent confirmation in
Ref. \cite{14risi} where the spacetime noncommutativity is obtained
in string theory with a constant off--diagonal metric background
when an appropriate form is present and one of the spatial
direction has Dirichlet boundary conditions. We note that in Refs.
\cite{14vmethod,14vbel,14vsolsp,14vsingl,14vsingl1} we constructed exact
solutions in the Einstein and extra dimension gravity with
off--diagonal metrics which were diagonalized by anholonomic
transforms to effective spacetimes with noncommutative structure
induced by anholonomic frames. Those results were extended to
noncommutative geometry and gauge gravity theories, in general,
containing local anisotropy, in Refs. \cite{14vnonc,14vncf}. The low
energy string spacetime with noncommutativity  constructed in
subsection 7.4 of this work is parametrized by an off--diagonal
metric which is a very general (non--constant) pseudo--Riemannian
one defining an exact solution in string gravity.

 \vskip6pt

Finally, our work raises a number of other interesting questions:

\begin{enumerate}
\item What kind of anholonomic quantum noncommutative structures are hidden
in string theory and gravity; how such constructions are to be
modelled by modern geometric methods.

\item How, in general, to relate the commutative and noncommutative gauge
models of (super) gravity with local anisotropy directly to string/M--theory.

\item What kind of quantum structure is more naturally associated to string
gravity and how to develop such anisotropic generalizations.

\item To formulate a nonlinear connection theory in quantum bundles and
relate it to various Finsler like quantum generalizations.

\item What kind of Clifford structures are more natural for developing a
unified geometric approach to anholonomic noncommutative and quantum
geometry following in various perturbative limits and non--perturbative
sectors of string/M--theory and when a such geometry is to be associated to
D--brane configurations.

\item To construct new classes of exact solutions with generic anisotropy
and noncommutativity and analyze theirs physical meaning and possible
applications.
\end{enumerate}

We hope to address some of these questions in future works.

\subsection*{Acknowledgements}

~~The author is grateful to S. Majid for collaboration and
discussions and  J.\ P. S. Lemos for hospitality and support.

%\newpage

%\appendix

\section{Appendix:\newline Anholonomic Frames and N--Connections}

We outline the basic definitions and formulas on anholonomic frames and
associated nonlinear connection (N--connection) structures on vector bundles %
\cite{14ma} and (pseu\-do) Riemannian manifolds \cite{14vexsol,14vmethod}.
  The
Einstein equations are written in mixed holonom\-lic-\--anholonomic
variables. We state the conditions when locally anisotropic
structures (Finsler like and another type ones) can be modelled in
general relativity and its extra dimension generalizations. This
Abstract contains the necessary formulas in coordinate form taken
from a
 geometric paper under preparation together with a co-author.

\subsection{The N--connection geometry}

The concept of N--connection came from Finsler geometry (as a set of
coefficients it is present in the works of E. Cartan \cite{14cartan}, then it
was elaborated in a more explicit fashion by A. Kawaguchi \cite{14kaw}). The
global definition of N--connections in commutative spaces is due to W.
Barthel \ \cite{14barthel}. The geometry of N--connections was developed in
details for vector, covector and higher order bundles \cite{14ma,14miron,14bejancu}%
, spinor bundles \cite{14vspinors,14vmon2} and superspaces and superstrings \cite%
{14vsuper,14vmon1,14vstring} with recent applications in modern anisotropic
kinetics and theormodynamics \cite{14vankin} and elaboration of new methods of
constructing exact off--diagonal solutions of the Einstein equations \cite%
{14vexsol,14vmethod}. The concept of N--connection can be extended in a similar
manner from commutative to noncommutative spaces if a differential calculus
is fixed on a noncommutative vector (or covector) bundle or another type of
quantum manifolds \ \cite{14vncf}.

\subsubsection{N--connections in vector bundles \ and (pseudo) Riemannian
spaces}

Let us consider a vector bundle $\xi =\left( E,\mu ,M\right) $ with typical
fibre $\R$$^{m}$ and the map%
\begin{equation*}
\mu ^{T}:TE\rightarrow TM
\end{equation*}%
being the differential of the map $\mu :E\rightarrow M.$ The map $\mu ^{T}$
is a fibre--preserving morphism of the tangent bundle $\left( TE,\tau
_{E},E\right) $ to $E$ and of tangent bundle $\left( TM,\tau ,M\right) $ to $%
M.$ The kernel of the morphism $\mu ^{T}$ is a vector subbundle of the
vector bundle $\left( TE,\tau _{E},E\right) .$ This kernel is denoted $%
\left( VE,\tau _{V},E\right) $ and called the vertical subbundle over $E.$
By
\begin{equation*}
i:VE\rightarrow TE
\end{equation*}%
it is denoted the inclusion mapping \ when the local coordinates of a point $%
u\in E$ are written $u^{\alpha }=\left( x^{i},y^{a}\right) ,$
where the values of indices are $i,j,k,...=1,2,...,n$ and
$a,b,c,...=1,2,...,m.$

A vector $X_{u}\in TE,$ tangent in the point $u\in E,$ is locally
represented
\begin{equation*}
\left( x,y,X,\widetilde{X}\right) =\left( x^{i},y^{a},X^{i},X^{a}\right) ,
\end{equation*}%
where $\left( X^{i}\right) \in $$\R$$^{n}$ and $\left( X^{a}\right) \in $$\R$%
$^{m}$ are defined by the equality
\begin{equation*}
X_{u}=X^{i}\partial _{i}+X^{a}\partial _{a}
\end{equation*}
[$\partial _{\alpha }=\left( \partial _{i},\partial _{a}\right) $ are usual
partial derivatives on respective coordinates $x^{i}$ and $y^{a}$]. For
instance, $\mu ^{T}\left( x,y,X,\widetilde{X}\right) =\left( x,X\right) $
and the submanifold $VE$ contains elements of type $\left( x,y,0,\widetilde{X%
}\right) $ and the local fibers of the vertical subbundle are isomorphic to $%
\R$$^{m}.$ Having $\mu ^{T}\left( \partial _{a}\right) =0,$ one comes out
that $\partial _{a}$ is a local basis of the vertical distribution $%
u\rightarrow V_{u}E$ on $E,$ which is an integrable distribution.

A nonlinear connection (in brief, N--connection) in the vector bundle $\xi
=\left( E,\mu ,M\right) $ is the splitting on the left of the exact sequence
\begin{equation*}
0\rightarrow VE\rightarrow TE/VE\rightarrow 0,
\end{equation*}%
i. e. a morphism of vector bundles $N:TE\rightarrow VE$ such that $C\circ i$
is the identity on $VE.$

The kernel of the morphism $N$ is a vector subbundle of $\left( TE,\tau
_{E},E\right) ,$ it is called the horizontal subbundle and denoted by $%
\left( HE,\tau _{H},E\right) .$ Every vector bundle $\left( TE,\tau
_{E},E\right) $ provided with a N--connection structure is Whitney sum of
the vertical and horizontal subbundles, i. e.
\begin{equation}
TE=HE\oplus VE.  \label{4wihit}
\end{equation}
It is proven that for every vector bundle $\xi =\left( E,\mu ,M\right) $
over a compact manifold $M$ there exists a nonlinear connection \cite{14ma}.

Locally a N--connection $N$ is parametrized by a set of coefficients\newline
$\left\{ N_{i}^{a}(u^{\alpha })=N_{i}^{a}(x^{j},y^{b})\right\} $ which
transform as
\begin{equation*}
N_{i^{\prime }}^{a^{\prime }}\frac{\partial x^{i^{\prime }}}{\partial x^{i}}%
=M_{a}^{a^{\prime }}N_{i}^{a}-\frac{\partial M_{a}^{a^{\prime }}}{\partial
x^{i}}y^{a}
\end{equation*}%
under coordinate transforms on the vector bundle $\xi =\left( E,\mu
,M\right) ,$%
\begin{equation*}
x^{i^{\prime }}=x^{i^{\prime }}\left( x^{i}\right) \mbox{ and
}y^{a^{\prime }}=M_{a}^{a^{\prime }}(x)y^{a}.
\end{equation*}

The well known class of linear connections consists a particular
parametization of the coefficients $N_{i}^{a}$ when
\begin{equation*}
N_{i}^{a}(x^{j},y^{b})=\Gamma _{bi}^{a}(x^{j})y^{b}
\end{equation*}%
are linear on variables $y^{b}.$

If a N--connection structure is associated to local frame (basis, vielbein)
on $\xi ,$ the operators of local partial derivatives $\partial _{\alpha
}=\left( \partial _{i},\partial _{a}\right) $ and differentials $d^{\alpha
}=du^{\alpha }=( d^{i}=dx^{i},$\\ $d^{a}=dy^{a}) $ should be elongated
as to adapt the local basis (and dual basis) structure to the Whitney
decomposition of the vector bundle into vertical and horizontal subbundles, (%
\ref{4wihit}):%
\begin{eqnarray}
\partial _{\alpha } &=&\left( \partial _{i},\partial _{a}\right) \rightarrow
\delta _{\alpha }=\left( \delta _{i}=\partial _{i}-N_{i}^{b}\partial
_{b},\partial _{a}\right) ,  \label{7dder} \\
d^{\alpha } &=&\left( d^{i},d^{a}\right) \rightarrow \delta ^{\alpha
}=\left( d^{i},\delta ^{a}=d^{a}+N_{i}^{b}d^{i}\right) .  \label{8ddif}
\end{eqnarray}%
The transforms can be considered as some particular case of frame transforms
of type
\begin{equation*}
\partial _{\alpha }\rightarrow \delta _{\alpha }=e_{\alpha }^{\beta
}\partial _{\beta }\mbox{ and }d^{\alpha }\rightarrow \delta ^{\alpha
}=(e^{-1})_{\beta }^{\alpha }\delta ^{\beta },
\end{equation*}%
$e_{\alpha }^{\beta }(e^{-1})_{\beta }^{\gamma }=\delta _{\alpha }^{\gamma
}, $ when the vielbein coefficients $e_{\alpha }^{\beta }$ are constructed
by using the Kronecker symbols $\delta _{a}^{b},\delta _{j}^{i}$ and $%
N_{i}^{b}. $

The bases $\delta _{\alpha }$ and $\delta ^{\alpha }$ satisfy, in general,
some anholonomy conditions, for instance,
\begin{equation}
\delta _{\alpha }\delta _{\beta }-\delta _{\beta }\delta _{\alpha
}=W_{\alpha \beta }^{\gamma }\delta _{\gamma },  \label{5anhol}
\end{equation}%
where $W_{\alpha \beta }^{\gamma }$ are called the anholonomy coefficients.
\ An explicit calculus of commutators of operators (\ref{7dder}) shows that
there are the non--trivial values:%
\begin{equation}
W_{ij}^{a}=R_{ij}^{a}=\delta _{i}N_{j}^{a}-\delta
_{j}N_{i}^{a},~W_{ai}^{b}=-W_{ia}^{b}=-\partial _{a}N_{i}^{b}.
\label{anholncoef}
\end{equation}

Tensor fields on a vector bundle $\xi =\left( E,\mu ,M\right) $ provided
with N--connection structure $N$ \ (we subject such spaces with the index $%
N, $ $\xi _{N})$ may be decomposed in N--adapted form with respect to the
bases $\delta _{\alpha }$ and $\delta ^{\alpha },$ and their tensor
products. For instance, for a tensor of rang (1,1) $T=\{T_{\alpha }^{~\beta
}=\left( T_{i}^{~j},T_{i}^{~a},T_{b}^{~j},T_{a}^{~b}\right) \}$ we have
\begin{equation}
T=T_{\alpha }^{~\beta }\delta ^{\alpha }\otimes \delta _{\beta
}=T_{i}^{~j}d^{i}\otimes \delta _{i}+T_{i}^{~a}d^{i}\otimes \partial
_{a}+T_{b}^{~j}\delta ^{b}\otimes \delta _{j}+T_{a}^{~b}\delta ^{a}\otimes
\partial _{b}.  \label{2dten}
\end{equation}

Every N--connection with coefficients $N_{i}^{b}$ $\ $generates also a
linear connection on $\xi _{N}$ \ as $\Gamma _{\alpha \beta }^{(N)\gamma
}=\{N_{bi}^{a}=\partial N_{i}^{a}(x,y)/\partial y^{b}\}$ which defines a
covariant derivative
\begin{equation*}
D_{\alpha }^{(N)}A^{\beta }=\delta _{\alpha }A^{\beta }+\Gamma _{\alpha
\gamma }^{(N)\beta }A^{\gamma }.
\end{equation*}

Another important characteristic of a N--connection is its curvature $\Omega
=\{\Omega _{ij}^{a}\}$ with the coefficients
\begin{equation}
\Omega _{ij}^{a}=\delta _{j}N_{i}^{a}-\delta _{i}N_{j}^{a}=\partial
_{j}N_{i}^{a}-\partial _{i}N_{j}^{a}+N_{i}^{b}N_{bj}^{a}-N_{j}^{b}N_{bi}^{a}.
\label{5ncurv}
\end{equation}

In general, on a vector bundle we may consider arbitrary linear connections
and metric structures adapted to the N--connection decomposition into
vertical and horizontal subbundles (one says that such objects are
distinguished by the N--connection, in brief, d--objects, like the d-tensor (%
\ref{2dten}), d--connection, d--metric:

\begin{itemize}
\item The coefficients of linear d--connections $\Gamma =\{\Gamma _{\alpha
\gamma }^{\beta }=\left( L_{jk}^{i},L_{bk}^{a},C_{jc}^{i},C_{ac}^{b}\right)
\}$ are defined for an arbitrary covariant derivative $D$ on $\xi $ being
adapted to the $N$--connection structure as $D_{\delta _{\alpha }}(\delta
_{\beta })=\Gamma _{\beta \alpha }^{\gamma }\delta _{\gamma }$ with the
coefficients being invariant under horizontal and vertical decomposition
\begin{equation*}
\quad D_{\delta _{i}}(\delta _{j})=L_{ji}^{k}\delta _{k},~D_{\delta
_{i}}(\partial _{a})=L_{ai}^{b}\partial _{b},~D_{\partial _{c}}(\delta
_{j})=C_{jc}^{k}\delta _{k},~~D_{\partial _{c}}(\partial
_{a})=C_{ac}^{b}\partial _{b}.
\end{equation*}%
The operator of covariant differentiation $D$ splits into the horizontal
covariant derivative $D^{[h]},$ stated by the coefficients $\left(
L_{jk}^{i},L_{bk}^{a}\right) ,$ for instance, and the operator of vertical
covariant derivative $D^{[v]},$ stated by the coefficients $\left(
C_{jc}^{i},C_{ac}^{b}\right) .$ For instance, for $A=A^{i}\delta
_{i}+A^{a}\partial _{a}=A_{i}\partial ^{i}+A_{a}\delta ^{a}$ one holds the
d--covariant derivation rules,%
\begin{eqnarray*}
D_{i}^{[h]}A^{k} &=&\delta
_{i}A^{k}+L_{ij}^{k}A^{j},~D_{i}^{[h]}A^{b}=\delta _{i}A^{b}+L_{ic}^{b}A^{c},
\\
D_{i}^{[h]}A_{k} &=&\delta _{i}A_{k}-L_{ik}^{j}A_{j},D_{i}^{[h]}A_{b}=\delta
_{i}A_{b}-L_{ib}^{c}A_{c},~ \\
D_{a}^{[v]}A^{k} &=&\partial
_{a}A^{k}+C_{aj}^{k}A^{j},~D_{a}^{[v]}A^{b}=\partial
_{a}A^{b}+C_{ac}^{b}A^{c}, \\
D_{a}^{[v]}A_{k} &=&\partial
_{a}A_{k}-C_{ak}^{j}A_{j},D_{a}^{[v]}A_{b}=\partial
_{a}A_{b}-C_{ab}^{c}A_{c}.
\end{eqnarray*}

\item The d--metric structure $G=g_{\alpha \beta }\delta ^{a}\otimes \delta
^{b}$ which has the invariant decomposition as $g_{\alpha \beta }=\left(
g_{ij},g_{ab}\right) $ following from%
\begin{equation}
G=g_{ij}(x,y)d^{i}\otimes d^{j}+g_{ab}(x,y)\delta ^{a}\otimes \delta ^{b}.
\label{9dmetric}
\end{equation}
\end{itemize}

We may impose the condition that a d--metric $g_{\alpha \beta }$ and a
d--connection $\Gamma _{\alpha \gamma }^{\beta }$ are compatible, i. e.
there are satisfied the conditions
\begin{equation}
D_{\gamma }g_{\alpha \beta }=0.  \label{2metrcond}
\end{equation}

With respect to the anholonomic frames (\ref{7dder}) and (\ref{8ddif}), there
is a linear connection, called the canonical distinguished linear
connection, which is similar to the metric connection introduced by the
Christoffel symbols in the case of holonomic bases, i. e. being constructed
only from the metric components and satisfying the metricity conditions (\ref%
{2metrcond}). It is parametrized by the coefficients,\ $\Gamma _{\ \beta
\gamma }^{\alpha }=\left( L_{\ jk}^{i},L_{\ bk}^{a},C_{\ jc}^{i},C_{\
bc}^{a}\right) $ where
\begin{eqnarray}
L_{\ jk}^{i} &=&\frac{1}{2}g^{in}\left( \delta _{k}g_{nj}+\delta
_{j}g_{nk}-\delta _{n}g_{jk}\right) ,  \label{7dcon} \\
L_{\ bk}^{a} &=&\partial _{b}N_{k}^{a}+\frac{1}{2}h^{ac}\left( \delta
_{k}h_{bc}-h_{dc}\partial _{b}N_{k}^{d}-h_{db}\partial _{c}N_{k}^{d}\right) ,
\notag \\
C_{\ jc}^{i} &=&\frac{1}{2}g^{ik}\partial _{c}g_{jk},\ C_{\ bc}^{a}=\frac{1}{%
2}h^{ad}\left( \partial _{c}h_{db}+\partial _{b}h_{dc}-\partial
_{d}h_{bc}\right) .  \notag
\end{eqnarray}%
Instead of this connection one can consider on $\xi $ another types of
linear connections which are/or not adapted to the N--connection structure
(see examples in \cite{14ma}).

\subsubsection{D--torsions and d--curvatures:}

The anholonomic coefficients $W_{\ \alpha \beta }^{\gamma }$ and
N--elongated derivatives give nontrivial coefficients for the torsion
tensor, $T(\delta _{\gamma },\delta _{\beta })=T_{\ \beta \gamma }^{\alpha
}\delta _{\alpha },$ where
\begin{equation}
T_{\ \beta \gamma }^{\alpha }=\Gamma _{\ \beta \gamma }^{\alpha }-\Gamma _{\
\gamma \beta }^{\alpha }+W_{\ \beta \gamma }^{\alpha },  \label{5torsion}
\end{equation}%
and for the curvature tensor, $R(\delta _{\tau },\delta _{\gamma })\delta
_{\beta }=R_{\beta \ \gamma \tau }^{\ \alpha }\delta _{\alpha },$ where
\begin{equation}
R_{\beta \ \gamma \tau }^{\ \alpha }=\delta _{\tau }\Gamma _{\ \beta \gamma
}^{\alpha }-\delta _{\gamma }\Gamma _{\ \beta \tau }^{\alpha }+\Gamma _{\
\beta \gamma }^{\varphi }\Gamma _{\ \varphi \tau }^{\alpha }-\Gamma _{\
\beta \tau }^{\varphi }\Gamma _{\ \varphi \gamma }^{\alpha }+\Gamma _{\
\beta \varphi }^{\alpha }W_{\ \gamma \tau }^{\varphi }.  \label{4curvature}
\end{equation}%
We emphasize that the torsion tensor on (pseudo) Riemannian spacetimes is
induced by an\-ho\-lonomic frames, whereas its components vanish with respect to
holonomic frames. All tensors are distinguished (d) by the N--connection
structure into irreducible (hori\-zont\-al--vertical) h--v--components, and are
called d--tensors. For instance, the torsion, d--tensor has the following
irreducible, nonvanishing, h--v--components,\\ $T_{\ \beta \gamma }^{\alpha
}=\{T_{\ jk}^{i},C_{\ ja}^{i},S_{\ bc}^{a},T_{\ ij}^{a},T_{\ bi}^{a}\},$
where
\begin{eqnarray}
T_{.jk}^{i} &=&T_{jk}^{i}=L_{jk}^{i}-L_{kj}^{i},\quad
T_{ja}^{i}=C_{.ja}^{i},\quad T_{aj}^{i}=-C_{ja}^{i},  \notag \\
T_{.ja}^{i} &=&0,\quad T_{.bc}^{a}=S_{.bc}^{a}=C_{bc}^{a}-C_{cb}^{a},
\label{4dtors} \\
T_{.ij}^{a} &=&-\Omega _{ij}^{a},\quad T_{.bi}^{a}=\partial
_{b}N_{i}^{a}-L_{.bi}^{a},\quad T_{.ib}^{a}=-T_{.bi}^{a}  \notag
\end{eqnarray}%
(the d--torsion is computed by substituting the h--v--compo\-nents of the
canonical d--connection (\ref{7dcon}) and anholonomy coefficients (\ref{5anhol}%
) into the formula for the torsion coefficients (\ref{5torsion})).

We emphasize that with respect to anholonomic frames the torsion is not zero
even for symmetric connections with $\Gamma _{\ \beta \gamma }^{\alpha
}=\Gamma _{\ \gamma \beta }^{\alpha }$ because the anholonomy coefficients $%
W_{\ \beta \gamma }^{\alpha }$ are contained in the formulas for the torsion
coefficients (\ref{5torsion}). By straightforward computations we can prove
that for nontrivial N--connection curvatures, $\Omega _{ij}^{a}\neq 0,$ even
the Levi--Civita connection for the metric (\ref{9dmetric}) contains
nonvanishing torsion coefficients. Of course, the torsion vanishes if the
Levi--Civita connection is defined as the usual Christoffel symbols with
respect to the coordinate frames, $\left( \partial _{i},\partial _{a}\right)
$ and $\left( d^{i},\partial ^{a}\right) ;$ in this case the d--metric (\ref%
{9dmetric}) is redefined into, in general, off--diagonal metric containing
products of $N_{i}^{a}$ and $h_{ab}.$

The curvature d--tensor has the following irreducible, non-vanishing,
h--v--compon\-ents\ $R_{\beta \ \gamma \tau }^{\ \alpha
}=%
\{R_{h.jk}^{.i},R_{b.jk}^{.a},P_{j.ka}^{.i},P_{b.ka}^{.c},S_{j.bc}^{.i},S_{b.cd}^{.a}\},
$\ where
\newpage
\begin{eqnarray}
R_{h.jk}^{.i} &=&\delta _{k}L_{.hj}^{i}-\delta
_{j}L_{.hk}^{i}+L_{.hj}^{m}L_{mk}^{i}-L_{.hk}^{m}L_{mj}^{i}-C_{.ha}^{i}%
\Omega _{.jk}^{a},  \label{4dcurvatures} \\
R_{b.jk}^{.a} &=&\delta _{k}L_{.bj}^{a}-\delta
_{j}L_{.bk}^{a}+L_{.bj}^{c}L_{.ck}^{a}-L_{.bk}^{c}L_{.cj}^{a}-C_{.bc}^{a}%
\Omega _{.jk}^{c},  \notag \\
P_{j.ka}^{.i} &=&\partial _{a}L_{.jk}^{i}+C_{.jb}^{i}T_{.ka}^{b}-(\delta
_{k}C_{.ja}^{i}+L_{.lk}^{i}C_{.ja}^{l}-L_{.jk}^{l}C_{.la}^{i}-L_{.ak}^{c}C_{.jc}^{i}),
\notag \\
P_{b.ka}^{.c} &=&\partial _{a}L_{.bk}^{c}+C_{.bd}^{c}T_{.ka}^{d}-(\delta
_{k}C_{.ba}^{c}+L_{.dk}^{c\
}C_{.ba}^{d}-L_{.bk}^{d}C_{.da}^{c}-L_{.ak}^{d}C_{.bd}^{c}),  \notag \\
S_{j.bc}^{.i} &=&\partial _{c}C_{.jb}^{i}-\partial
_{b}C_{.jc}^{i}+C_{.jb}^{h}C_{.hc}^{i}-C_{.jc}^{h}C_{hb}^{i},  \notag \\
S_{b.cd}^{.a} &=&\partial _{d}C_{.bc}^{a}-\partial
_{c}C_{.bd}^{a}+C_{.bc}^{e}C_{.ed}^{a}-C_{.bd}^{e}C_{.ec}^{a}  \notag
\end{eqnarray}%
(the d--curvature components are computed in a similar fashion by using the
formula for curvature coefficients (\ref{4curvature})).

\subsubsection{Einstein equations in d--variables}

In this subsection we write and analyze the Einstein equations on spaces
provided with anholonomic frame structures and associated N--connections.

The Ricci tensor $R_{\beta \gamma }=R_{\beta ~\gamma \alpha }^{~\alpha }$
has the d--components
\begin{equation}
R_{ij}=R_{i.jk}^{.k},\quad
R_{ia}=-^{2}P_{ia}=-P_{i.ka}^{.k},R_{ai}=^{1}P_{ai}=P_{a.ib}^{.b},\quad
R_{ab}=S_{a.bc}^{.c}.  \label{7dricci}
\end{equation}%
In general, since $^{1}P_{ai}\neq ~^{2}P_{ia}$, the Ricci d-tensor is
non-symmetric (we emphasize that this could be with respect to anholonomic
frames of reference because the N--connection and its curvature
coefficients, $N_{i}^{a}$ and $\Omega _{.jk}^{a},$ as well the anholonomy
coefficients $W_{\ \beta \gamma }^{\alpha }$ and d--torsions $T_{\ \beta
\gamma }^{\alpha }$ are contained in the formulas for d--curvatures (\ref%
{4curvature})). The scalar curvature of the metric d--connection, $%
\overleftarrow{R}=g^{\beta \gamma }R_{\beta \gamma },$ is computed
\begin{equation}
{\overleftarrow{R}}=G^{\alpha \beta }R_{\alpha \beta }=\widehat{R}+S,
\label{5dscalar}
\end{equation}%
where $\widehat{R}=g^{ij}R_{ij}$ and $S=h^{ab}S_{ab}.$

By substituting (\ref{7dricci}) and (\ref{5dscalar}) into the Einstein
equations
\begin{equation}
R_{\alpha \beta }-\frac{1}{2}g_{\alpha \beta }R=\kappa \Upsilon _{\alpha
\beta },  \label{35einstein}
\end{equation}%
where $\kappa $ and $\Upsilon _{\alpha \beta }$ are respectively the
coupling constant and the energy--momentum tensor we obtain the
h-v-decomposition by N--connection of the Einstein equations
\begin{eqnarray}
R_{ij}-\frac{1}{2}\left( \widehat{R}+S\right) g_{ij} &=&\kappa \Upsilon
_{ij},  \label{4einsteq2} \\
S_{ab}-\frac{1}{2}\left( \widehat{R}+S\right) h_{ab} &=&\kappa \Upsilon
_{ab},  \notag \\
^{1}P_{ai}=\kappa \Upsilon _{ai},\ ^{2}P_{ia} &=&\kappa \Upsilon _{ia}.
\notag
\end{eqnarray}%
The definition of matter sources with respect to anholonomic frames is
considered in Refs. \cite{14vspinors,14vmon1,14ma}.

The vacuum locally anisotropic gravitational field equations, in invariant
h-- v--components, are written
\begin{eqnarray}
R_{ij} &=&0,S_{ab}=0,^{1}P_{ai}=0,\ ^{2}P_{ia}=0.  \label{3einsteq3} \\
&&  \notag
\end{eqnarray}

We emphasize that general linear connections in vector bundles and even in
the (pseudo) Riemannian spacetimes have non--trivial torsion components if
off--diagonal metrics and anholomomic frames are introduced into
consideration. This is a ''pure'' anholonomic frame effect: the torsion
vanishes for the Levi--Civita connection stated with respect to a coordinate
frame, but even this metric connection contains some torsion coefficients if
it is defined with respect to anholonomic frames (this follows from the $w$%
--terms in (\ref{12lcsym})). For the (pseudo) Riemannian spaces we conclude
that the Einstein theory transforms into an effective Einstein--Cartan
theory with anholonomically induced torsion if the general relativity is
formulated with respect to general frame bases (both holonomic and
anholonomic).

The N--connection geometry can be similarly formulated for a tangent bundle $%
TM$ of a manifold $M$ (which is used in Finsler and Lagrange geometry \cite%
{14ma}), on cotangent bundle $T^{\ast }M$ and higher order bundles (higher
order Lagrange and Hamilton geometry \cite{14miron}) as well in the geometry
of locally anisotropic superspaces \cite{14vsuper}, superstrings \cite{14vstr2},
anisotropic spinor \cite{14vspinors} and gauge \ \cite{14vgauge} theories or
even on (pseudo) Riemannian spaces provided with anholonomic frame
structures \cite{14vmon2}.

\subsection{Anholonomic Frames in Commutative Gravity}

We introduce the concepts of generalized Lagrange and Finsler geometry and
outline the conditions when such structures can be modelled on a Riemannian
space by using anholnomic frames.

Different classes of commutative anisotropic spacetimes are
modelled by corresponding parametriztions of some compatible (or
even non--compatible) N--connection, d--connection and d--metric
structures on (pseudo) Riemannian spaces, tangent (or cotangent)
bundles, vector (or covector) bundles and their higher order
generalizations in their usual manifold,
supersymmetric, spinor, gauge like or another type approaches (see Refs. %
\cite{14vexsol,14miron,14ma,14bejancu,14vspinors,14vgauge,14vmon1,14vmon2}).

\subsubsection{Anholonomic structures on Riemannian spaces}

\bigskip We note that  the N--connection structure may be defined not
only in vector bundles but also on (pseudo) Riemannian spaces \cite{14vexsol}.
\ In this case the N--connection is an object completely defined by
anholonomic frames, when the coefficients of frame transforms, $e_{\alpha
}^{\beta }\left( u^{\gamma }\right) ,$ are parametrized explicitly via
certain values $\left( N_{i}^{a},\delta _{i}^{j},\delta _{b}^{a}\right) ,$
where $\delta _{i}^{j}$ $\ $and $\delta _{b}^{a}$ are the Kronecker symbols.
By straightforward calculations we can compute that the coefficients of the
Levi--Civita metric connection
\begin{equation*}
\Gamma _{\alpha \beta \gamma }^{\bigtriangledown }=g\left( \delta _{\alpha
},\bigtriangledown _{\gamma }\delta _{\beta }\right) =g_{\alpha \tau }\Gamma
_{\beta \gamma }^{\bigtriangledown \tau },\,
\end{equation*}%
associated to a covariant derivative operator $\bigtriangledown ,$
satisfying the metricity condition\\ $\bigtriangledown _{\gamma }g_{\alpha
\beta }=0$ for $g_{\alpha \beta }=\left( g_{ij},h_{ab}\right) $ and
\begin{equation}
\Gamma _{\alpha \beta \gamma }^{\bigtriangledown }=\frac{1}{2}\left[ \delta
_{\beta }g_{\alpha \gamma }+\delta _{\gamma }g_{\beta \alpha }-\delta
_{\alpha }g_{\gamma \beta }+g_{\alpha \tau }W_{\gamma \beta }^{\tau
}+g_{\beta \tau }W_{\alpha \gamma }^{\tau }-g_{\gamma \tau }W_{\beta \alpha
}^{\tau }\right] ,  \label{3lcsym}
\end{equation}%
are given with respect to the anholonomic basis (\ref{8ddif}) by the
coefficients
\begin{equation}
\Gamma _{\beta \gamma }^{\bigtriangledown \tau }=\left( L_{\ jk}^{i},L_{\
bk}^{a},C_{\ jc}^{i}+\frac{1}{2}g^{ik}\Omega _{jk}^{a}h_{ca},C_{\
bc}^{a}\right)  \label{3lccon}
\end{equation}%
when $L_{\ jk}^{i},L_{\ bk}^{a},C_{\ jc}^{i},C_{\ bc}^{a}$ and $\Omega
_{jk}^{a}$ are respectively computed by the formulas (\ref{7dcon}) and (\ref%
{5ncurv}). A specific property of off--diagonal metrics is that they can
define different classes of linear connections which satisfy the metricity
conditions for a given metric, or inversely, there is a certain class of
metrics which satisfy the metricity conditions for a given linear
connection. \ This result was originally obtained by A. Kawaguchi \cite{14kaw}
(Details can be found in Ref. \cite{14ma}, see Theorems 5.4 and 5.5 in Chapter
III, formulated for vector bundles; here we note that similar proofs hold
also on manifolds enabled with anholonomic frames associated to a
N--connection structure).

With respect to anholonomic frames, we can not distinguish the Levi--Civita
connection as the unique one being both metric and torsionless. For
instance, both linear connections (\ref{7dcon}) and (\ref{3lccon}) contain
anholonomically induced torsion coefficients, are compatible with the same
metric and transform into the usual Levi--Civita coefficients for vanishing
N--connection and ''pure'' holonomic coordinates. This means that to an
off--diagonal metric in general relativity one may be associated different
covariant differential calculi, all being compatible with the same metric
structure (like in the non--commutative geometry, which is not a surprising
\ fact because the anolonomic frames satisfy by definition some
non--commutative relations (\ref{5anhol})). In such cases we have to select a
particular type of connection following some physical or geometrical
arguments, or to impose some conditions when there is a single compatible
linear connection constructed only from the metric and N--coefficients. We
note that if $\Omega _{jk}^{a}=0$ the connections (\ref{7dcon}) and (\ref%
{3lccon}) coincide, i. e. $\Gamma _{\ \beta \gamma }^{\alpha }=\Gamma _{\beta
\gamma }^{\bigtriangledown \alpha }.$

If an anholonomic (equivalently, anisotropic) frame structure is defined on
a (pseu\-do) Riemannian space of dimension $(n+m)$ space, the space is
called to be an anholonom\-ic (pseudo) Riemannian one (denoted as $\
V^{(n+m)}).$ By fixing an anholonomic frame basis and co--basis with
associated N--connection $N_{i}^{a}(x,y),$ respectively, as (\ref{7dder}) and
(\ref{8ddif}), one splits the local coordinates $u^{\alpha }=(x^{i},y^{a})$
into two classes: the fist class consists from $n$ holonomic coordinates, $%
x^{i},$ and the second class consists from $m$ anholonomic coordinates, $%
y^{a}.$ The d--metric (\ref{9dmetric}) on $V^{(n+m)}$,
\begin{equation}
G^{[R]}=g_{ij}(x,y)dx^{i}\otimes dx^{j}+h_{ab}(x,y)\delta y^{a}\otimes
\delta y^{b}  \label{2dmetrr}
\end{equation}%
written with respect to a usual coordinate basis $du^{\alpha }=\left(
dx^{i},dy^{a}\right) ,$%
\begin{equation*}
ds^{2}=\underline{g}_{\alpha \beta }\left( x,y\right) du^{\alpha }du^{\beta }
\end{equation*}%
is a generic off--diagonal \ Riemannian metric parametrized as%
\begin{equation}
\underline{g}_{\alpha \beta }=\left[
\begin{array}{cc}
g_{ij}+N_{i}^{a}N_{j}^{b}g_{ab} & h_{ab}N_{i}^{a} \\
h_{ab}N_{j}^{b} & h_{ab}%
\end{array}%
\right] .  \label{odm}
\end{equation}%
Such type of metrics were largely investigated in the Kaluza--Klein gravity %
\cite{14salam}, but also in the Einstein gravity \cite{14vexsol}. An
off--diagonal metric (\ref{odm}) can be reduced to a block $\left( n\times
n\right) \oplus \left( m\times m\right) $ form $\left( g_{ij},g_{ab}\right)
, $ and even effectively diagonalized in result of a superposition of
anholonomic N--transforms. It can be defined as an exact solution of the
Einstein equations. With respect to anholonomic frames, in general, the
Levi--Civita connection obtains a torsion component (\ref{3lcsym}). Every
class of off--diagonal metrics can be anholonomically equivalent to another
ones for which it is not possible to a select the Levi--Civita metric defied
as the unique torsionless and metric compatible linear connection. \ The
conclusion is that if anholonomic frames of reference, which authomatically
induce the torsion via anholonomy coefficients, are considered on a
Riemannian space we have to postulate explicitly what type of linear
connection (adapted both to the anholonomic frame and metric structure) is
chosen in order to construct a Riemannian geometry and corresponding
physical models. For instance, we may postulate the connection (\ref{3lccon})
or the d--connection (\ref{7dcon}). Both these connections are metric
compatible and transform into the usual Christoffel symbols if the
N--connection vanishes, i. e. the local frames became holonomic. But, in
general, anholonomic frames and off--diagonal Riemannian metrics are
connected with anisotropic configurations which allow, in principle, to
model even Finsler like structures in (pseudo) Riemannian spaces \cite%
{14vankin,14vexsol}.

\subsubsection{Finsler geometry and its almost Kahlerian model}

The modern approaches to Finsler geometry are outlined in Refs. \cite%
{14finsler,14ma,14miron,14bejancu,14vmon1,14vmon2}. Here we emphasize that a Finsler
metric can be defined on a tangent bundle $TM$ with local coordinates $%
u^{\alpha }=(x^{i},y^{a}\rightarrow y^{i})$ of dimension $2n,$ with a
d--metric (\ref{9dmetric}) for which the Finsler metric, i. e. the quadratic
form
\begin{equation}
g_{ij}^{[F]}=h_{ab}=\frac{1}{2}\frac{\partial ^{2}F^{2}}{\partial
y^{i}\partial y^{j}}  \label{fmetric}
\end{equation}%
is positive definite, is defined in this way: \ 1) A Finsler metric on a
real manifold $M$ is a function $F:TM\rightarrow \R$ which on $\widetilde{TM}%
=TM\backslash \{0\}$ is of class $C^{\infty }$ and $F$ is only continuous on
the image of the null cross--sections in the tangent bundle to $M.$ 2) $%
F\left( x,\chi y\right) =\chi F\left( x,y\right) $ for every $\R_{+}^{\ast
}. $ 3) The restriction of $F$ to $\widetilde{TM}$ is a positive function.
4) $rank\left[ g_{ij}^{[F]}(x,y)\right] =n.$

The Finsler metric $F(x,y)$ and the quadratic form $g_{ij}^{[F]}$ can be
used to define the Christoffel symbols (not those from the usual Riemannian
geometry)%
\begin{equation*}
c_{jk}^{\iota }(x,y)=\frac{1}{2}g^{ih}\left( \partial
_{j}g_{hk}^{[F]}+\partial _{k}g_{jh}^{[F]}-\partial _{h}g_{jk}^{[F]}\right) ,
\end{equation*}%
where $\partial _{j}=\partial /\partial x^{j},$ $\ $which allows us to
define the Cartan nonlinear connection as
\begin{equation}
N_{j}^{[F]i}(x,y)=\frac{1}{4}\frac{\partial }{\partial y^{j}}\left[
c_{lk}^{\iota }(x,y)y^{l}y^{k}\right]  \label{3ncc}
\end{equation}%
where we may not distinguish the v- and h- indices taking on $TM$ \ the same
values.

In Finsler geometry there were investigated different classes of remarkable
Finsler linear connections introduced by Cartan, Berwald, Matsumoto and
other ones (see details in Refs. \cite{14finsler,14ma,14bejancu}). Here we note
that we can introduce $g_{ij}^{[F]}=g_{ab}$ and $N_{j}^{i}(x,y)$ in (\ref%
{9dmetric}) and construct a d--connection via formulas (\ref{7dcon}).

A usual Finsler space $F^{n}=\left( M,F\left( x,y\right) \right) $ is
completely defined by its fundamental tensor $g_{ij}^{[F]}(x,y)$ and Cartan
nonlinear connection $N_{j}^{i}(x,y)$ and its chosen d--connection
structure. But the N--connection allows us to define an almost complex
structure $I$ on $TM$ as follows%
\begin{equation*}
I\left( \delta _{i}\right) =-\partial /\partial y^{i}\mbox{ and
}I\left( \partial /\partial y^{i}\right) =\delta _{i}
\end{equation*}%
for which $I^{2}=-1.$

The pair $\left( G^{[F]},I\right) $ consisting from a Riemannian metric on $%
TM,$%
\begin{equation}
G^{[F]}=g_{ij}^{[F]}(x,y)dx^{i}\otimes dx^{j}+g_{ij}^{[F]}(x,y)\delta
y^{i}\otimes \delta y^{j}  \label{3dmetricf}
\end{equation}%
and the almost complex structure $I$ defines an almost Hermitian structure
on $\widetilde{TM}$ associated to a 2--form%
\begin{equation*}
\theta =g_{ij}^{[F]}(x,y)\delta y^{i}\wedge dx^{j}.
\end{equation*}%
This model of Finsler geometry is called almost Hermitian and denoted $%
H^{2n} $ and it is proven \cite{14ma} that is almost Kahlerian, i. e. the form
$\theta $ is closed. The almost Kahlerian space $K^{2n}=\left( \widetilde{TM}%
,G^{[F]},I\right) $ is also called the almost Kahlerian model of the Finsler
space $F^{n}.$

On Finsler (and their almost Kahlerian models) spaces one distinguishes the
almost Kahler linear connection of Finsler type, $D^{[I]}$ on $\widetilde{TM}
$ with the property that this covariant derivation preserves by parallelism
the vertical distribution and is compatible with the almost Kahler structure
$\left( G^{[F]},I\right) ,$ i.e.
\begin{equation*}
D_{X}^{[I]}G^{[F]}=0\mbox{ and }D_{X}^{[I]}I=0
\end{equation*}%
for \ every d--vector field on $\widetilde{TM}.$ This d--connection is
defined by the data
\begin{equation*}
\Gamma =\left( L_{jk}^{i},L_{bk}^{a}=0,C_{ja}^{i}=0,C_{bc}^{a}\rightarrow
C_{jk}^{i}\right)
\end{equation*}%
with $L_{jk}^{i}$ and $C_{jk}^{i}$ computed as in the formulas (\ref{7dcon})
by using $g_{ij}^{[F]}$ and $N_{j}^{i}$ from (\ref{3ncc}).

We emphasize that a Finsler space $F^{n}$ with a d--metric (\ref{3dmetricf})
and Cartan's N--connection structure (\ref{3ncc}), or the corresponding
almost Hermitian (Kahler) model $H^{2n},$ can be equivalently modelled on a
Riemannian space of dimension $2n$ provided with an off--diagonal Riemannian
metric (\ref{odm}). From this viewpoint a Finsler geometry is a
corresponding Riemannian geometry with a respective off--diagonal metric
(or, equivalently, with an anholonomic frame structure with associated
N--connection) and a corresponding prescription for the type of linear
connection chosen to be compatible with the metric and N--connection
structures.

\subsubsection{Lagrange and generalized Lagrange geometry}

Lagrange spaces were introduced in order to geometrize the fundamental
concepts in mechanics \cite{14kern} and investigated in Refs. \cite{14ma} (see %
\cite{14vspinors,14vgauge,14vsuper,14vstr2,14vmon1,14vmon2} for their spinor, gauge and
supersymmetric generalizations).

A Lagrange space $L^{n}=\left( M,L\left( x,y\right) \right) $ is defined as
a pair which consists of a real, smooth $n$--dimensional manifold $M$ and
regular Lagrangian $L:TM\rightarrow \R.$ Similarly as for Finsler spaces one
introduces the symmetric d--tensor field
\begin{equation}
g_{ij}^{[L]}=h_{ab}=\frac{1}{2}\frac{\partial ^{2}L}{\partial y^{i}\partial
y^{j}}.  \label{2mfl}
\end{equation}%
So, the Lagrangian $L(x,y)$ is like the square of the fundamental Finsler
metric, $F^{2}(x,y),$ but not subjected to any homogeneity conditions.

In the rest me can introduce similar concepts of almost Hermitian
(Kahlerian) models of Lagrange spaces as for the Finsler spaces, by using
the similar definitions and formulas as in the previous subsection, but
changing $g_{ij}^{[F]}\rightarrow g_{ij}^{[L]}.$

R. Miron introduced the concept of generalized Lagrange space, GL--space
(see details in \cite{14ma}) and a corresponding N--connection geometry on $TM$
when the fundamental metric function $g_{ij}=g_{ij}\left( x,y\right) $ is a
general one, not obligatory defined as a second derivative from a Lagrangian
as in (\ref{2mfl}). The corresponding almost Hermitian (Kahlerian) models of
GL--spaces were investigated and applied in order to elaborate
generalizations of gravity and gauge theories \cite{14ma,14vgauge}.

Finally, a few remarks on definition of gravity models with generic local
anisotropy on anholonomic Riemannian, Finsler or (generalized) Lagrange
spaces and vector bundles. So, by choosing a d-metric (\ref{9dmetric}) (in
particular cases (\ref{2dmetrr}), or (\ref{3dmetricf}) with $g_{ij}^{[F]},$ or
$g_{ij}^{[L]})$ we may compute the coefficients of, for instance,
d--connection (\ref{7dcon}), d--torsion (\ref{4dtors}) and (\ref{4dcurvatures})
and even to write down the explicit form of Einstein equations (\ref%
{4einsteq2}) which define such geometries. For instance, in a series of works %
\cite{14vankin,14vexsol,14vmon2} we found explicit solutions when Finsler like and
another type anisotropic configurations are modelled in anisotropic kinetic
theory and irreversible thermodynamics and even in Einstein or
low/extra--dimension gravity as exact solutions of the vacuum (\ref{4einsteq2}%
) and nonvacuum (\ref{3einsteq3}) Einstein equations. From the viewpoint of
the geometry of anholonomic frames is not much difference between the usual
Riemannian geometry and its Finsler like generalizations. The explicit form
and parametrizations of coefficients of metric, linear connections,
torsions, curvatures and Einstein equations in all types of mentioned
geometric models depends on the type of anholomic frame relations and
compatibility metric conditions between the associated N--connection
structure and linear connections we fixed. Such structures can be
correspondingly picked up from a noncommutative functional model, for
instance, from some almost Hermitian structures over projective modules
and/or generalized to some noncommutative configurations \cite{14vncf}.

%%%%%%%%%%%%%%%%%%%%%%%%%%%%%%%%%%%%%%%%%%%%%%%%%%%%%%%%%%%%%%%%%%%%%%%%%%%%%

\chapter[Noncommutative Clifford--Finsler  Gravity]
{Nonholonomic  Clifford Structures and Noncommutative
 Riemann--Finsler  Geometry}

{\bf Abstract}
\footnote{\copyright\
 S. Vacaru, Nonholonomic Clifford Structures and
  Noncommutative Riemann-Finsler Geometry,
 math.DG/0408121 }

We survey the geometry of Lagrange and Finsler spaces and discuss the
issues related to the definition of curvature of
 nonholonomic manifolds enabled with  nonlinear
connection structure. It is proved that  any commutative
Riemannian geometry (in general, any Riemann--Cartan space)
 defined by  a generic off--diagonal metric structure (with an
 additional  affine connection possessing  nontrivial torsion)
   is  equivalent  to a generalized
 Lagrange, or Finsler, geometry   modelled on
 nonholonomic manifolds.
This  results in  the problem of constructing
noncommutative  geometries with local anisotropy,
in particular, related to geometrization of classical and quantum
mechanical and field theories, even if we restrict our considerations only to
commutative and noncommutative Riemannian spaces.  We  elaborate a
geometric  approach  to the Clifford modules adapted to nonlinear connections,
 to the  theory of spinors and the Dirac operators  on nonholonomic
 spaces   and consider  possible  generalizations to
 noncommutative geometry. We argue that any  commutative
 Riemann--Finsler  geometry and  generalizations
 my be derived from  noncommutative geometry by applying certain
 methods elaborated for Riemannian spaces but extended to
 nonholonomic frame transforms and  manifolds provided with nonlinear
 connection structure.

\vskip0.2cm

AMS Subject Classification:\

 46L87, 51P05, 53B20, 53B40,  70G45, 83C65

\vskip0.2cm

\textbf{Keywords:}\  Noncommutative geometry, Lagrange and Finsler geometry,
nonlinear connection,   nonholonomic manifolds, Clifford modules,
spinors, Dirac operator, off--diagonal metrics and gravity.

\section{ Introduction}

The goal of this work is to provide a better understanding of the
relationship between the theory of nonholonomic manifolds with
associated nonlinear connection structure, locally anisotropic spin
configurations and  Dirac operators on such manifolds and
noncommutative Riemann--Finsler and Lagrange geometry. The latter approach is
based on geometrical modelling of mechanical and classical field
theories (defined, for simplicity, by regular Lagrangians in analytic
mechanics and Finsler like anisotropic structures)
  and  gravitational, gauge and spinor field interactions in
low energy limits of string theory.  This allows to apply the
Serre--Swan theorem and think of vector bundles as projective modules,
which, for our purposes, are provided with nonlinear connection (in
brief, N--connection) structure and  can be defined as a
nonintegrabele (nonholonomic) distribution into conventional
horizontal  and vertical submodules. We relay on the theory of
 Clifford and spinor structures adapted to N--connections which
  results in locally anisotropic (Finsler like, or more general ones
  defined by more general nonholonomic frame structures) Dirac
  operators. In the former item, it is the machinery of noncommutative
  geometry to derive distance formulas and to consider noncommutative
  extensions of Riemann--Finsler and Lagrange geometry and related off--diagonal
  metrics in gravity theories.

In \cite{15vnc} it was proposed that an equivalent reformulation of the
general relativity theory as a gauge model with nonlinear realizations
of the affine, Poincare and/or de Sitter groups allows a standard
extension of gravity theories in the language of noncommutative gauge
fields. The approach was developed in \cite{15vncg} as an attempt to
generalize the A. Connes' noncommutative geometry \cite{15connes1} to spaces with
generic local anisotropy. The nonlinear connection formalism was
elaborated for projective module spaces and the Dirac operator
associated to metrics in Finsler geometry and some generalizations
\cite{15vsp1,15vsph} (such as Sasaki type lifts of metrics to the tangent
bundles and vector bundle analogs) were considered as certain
examples of noncommutative Finsler geometry. The constructions were
synthesized and revised in connection to ideas about appearance of
both noncommutative and Finsler geometry in string theory with
nonvanishing B--field and/or anholonomic (super) frame structures
 \cite{15sw,15con2,15ard,15abd,15sah,15vncgs,15vstr0,15vstr1} and
  in supergravity and gauge
 gravity  \cite{15card,15gars,15vanc1,15vanc2,15dup}. In particular, one has
 considered hidden noncommutative and Finsler like structures in
 general relativity and extra dimension gravity
 \cite{15pan,15vs1,15v0,15dv,15v3}.

In this work, we confine ourselves to the classical aspects of
Lagrange--Finsler geometry (sprays, nonlinear connections, metric and
linear connection structures and almost complex structure derived from
from a Lagrange or Finsler fundamental form) in order  to
generalize the  doctrine of the "spectral action" and the theory of
 distance in noncommutative geometry which is an extension of the
previous results \cite{15connes1}. For a complete information on modern
 noncommutative geometry and physics,  we refer the reader to
\cite{15landi,15madore,15kastler2,15bondia,15douglas,15konechny},
 see a historical sketch in  Ref. \cite{15kastler2} as well the aspects
  related to quantum group theory \cite{15manin,15maj2,15kassel}
  (here we note that the first  quantum group Finsler
structure was considered in \cite{15vargqugr}).  The theory of Dirac
operators and related Clifford analysis is a subject of various
investigations in modern mathematics and mathematical physics
\cite{15martinmircea,15martinmircea1,15mitrea,15nistor,15fauser,15castro1,15castro2,15castro3,15castro4,15schroeder,15carey}
(see also a relation to Finsler geometry \cite{15vargcl} and an
off--diagonal "non" Kaluza--Klein compactified ansatz, but without
N--connection counstructions \cite{15castro2a}). \footnote{The theory of
N--connections should not be confused with nonlinear gauge theories
and nonlinear realizations of gauge groups.}
 For an exposition spelling out all the
details of proofs and important concepts preliminary undertaken on
  the subjects elaborated in our works, we refer to  proofs and
quotations in Refs.
\cite{15ma1,15ma2,15v1,15v2,15v3,15vncg,15vncgs,15bondia,15rennie1,15lord,15mart2}.

This paper consists of two heterogeneous parts:

 The first (commutative) contains an overview of the Lagrange and Finsler
geometry and the off--diagonal metric and nonholonomic frame geometry in
gravity theories. In Section 2, we formulate the N--connection geometry
for arbitrary manifolds with tangent bundles admitting splitting into
conventional horizontal and vertical subspaces. We illustrate how regular
Lagrangians induce natural semispray, N--connection, metric and almost
complex structures on tangent bundles and discuss the relation between
Lagrange and Finsler geometry and theirs generalizations.
 We formulate six most important Results \ref{1r1}--\ref{r6}
 demonstrating that the geometrization of Lagrange mechanics and the
 geometric models in gravity theories with generic off--diagonal metrics and
 nonholonomic frame structures are rigorously described by certain
 generalized Finsler geometries, which can be modelled equivalently
 both on Riemannian manifold and Riemann--Cartan nonholonomic
 manifolds. This give rise to the Conclusion \ref{ic} stating that a
 rigorous geometric study of nonholonmic frame and related metric,
 linear connection and spin structures in both commutative and
 noncommutative Riemann geometries requests the elaboration of
 noncommutative Lagrange--Finsler geometry.  Then, in
Section 3,  we consider  the theory of linear connections on N--anholonomic
manifolds (i. e. on manifolds with nonholonomic structure defined by
N--connections). We construct in explicit form the curvature
tensor of such spaces and define the Einstein equations for N--adapted
linear connection and metric structures.

The second (noncommutative)  part starts with Section 4 where we
define noncommutative N--anholonomic spaces. We consider the example
of noncommutative gauge theories adapted to the N--connection
structure.  Section 5 is devoted to
the geometry of nonholonomic Clifford--Lagrange structures. We define
 the Clifford--Lagrange modules and Clifford N--anholonomic bundles
 being induced by the Lagrange quadratic form and adapted to the
 corresponding N--connection. Then we prove the {\bf Main Result 1,}
 of this work,   ( Theorem \ref{mr1}), stating that any regular
 Lagrangian and/or N--con\-nec\-ti\-on structure define naturally
 the fundamental geometric objects  and structures
 (such as the Clifford--Lagrange module and Clifford d--modules, the
 Lagrange spin structure  and d--spinors) for the corresponding Lagrange
 spin manifold  and/or  N--anholo\-nom\-ic spinor (d--spinor)
 manifold. We conclude that the Lagrange mechanics and
  off--diagonal gravitational interactions (in general, with nontrivial
  torsion and nonholonomic constraints) can be adequately
  geometrized as certain Lagrange spin (N--anholonomic) manifolds.

 In Section 6, we link up the theory of Dirac operators to nonholonomic structures
and spectral triples. We prove that there is a canonical spin
d--connection on the N--anholonomic manifolds generalizing that
induced by the Levi--Civita to the naturally ones induced by regular
Lagrangians and off--diagonal metrics. We define the Dirac d--operator
and the Dirac--Lagrange operator and formulate the {\bf Main Result 2}
(Theorem \ref{mr2}) arguing   that such N--adapted operators can be induced
canonically by almost Hermitian spin operators.  The concept of
distinguished spin triple is introduced in order to  adapt the
constructions to the N--connection structure. Finally, the {\bf Main
  Result 3,} Theorem \ref{mr2}, is devoted to the definition, main
  properties  and  computation  of  distance in noncommutative spaces
 defined by  N--anholonomic spin varieties. In these lecture notes, we
  only sketch in brief the ideas
  of proofs of the Main  Results:\ the details will be published in our
  further works.

\section[Finser Geometry and Nonholonomy]
{Lagrange--Finsler Geometry\newline and Nonholono\-mic Manifolds}

This section presents some basic facts from the geometry of
nonholonomic manifolds  provided with nonlinear connection structure
 \cite{15vsp1,15nhm3,15nhm1,15nhm4,15mironnh1,15ver}.
 The constructions and methods are inspired from the Lagrange--Finsler
 geometry and generalizations
 \cite{15fin,15car1,15rund,15ma1,15asa1,15mats,15bej,15ma2,15vstr,15vstr1,15bcs,15mhss,15sh}
and gravity models on metric--affine
spaces provided with generic off--diagonal metric, nonholonomic frame and affine
 connection structures \cite{15vex1,15vt,15vs1,15dv,15v1,15v2}
(such spaces, in general, possess nontrivial torsion and
 nonmetricity).

\subsection{Preliminaries: Lagrange--Finsler metrics}\label{sslfm}

Let us consider a nondegenerate bilinear symmetric form $q(u,v)$ on a $n$--dimensional
 real vector space
 $V^n.$ With respect to a basis $\{e_i\}^n_{i=1}$ for  $V^n,$ we express
\begin{equation*}
q(u,v) \doteq q_{ij}u^i v^j
\end{equation*}%
for any vectors $u=u^i e_i,\ v= v^i e_i \in V^n$ and $q_{ij}$ being
  a nodegenerate symmetric matrix (the Einstein's
 convention on summing on repeating indices is adopted). This gives rise to
 the Euclidean inner product
 \begin{equation*}
u \rfloor v \doteq q_{E}(u, v),
 \end{equation*}
if  $q_{ij}$ is positive definite, and  to the Euclidean norm
 \begin{equation*}
{|{\ \cdot \ } |}  \doteq \sqrt{q_{E}(u, u)}
 \end{equation*}
defining an Euclidean space $ (V^n, {|{\ \cdot \ } |}).$ Every
 Euclidean space
 is linearly isometric to the standard
 Euclidean space $\R ^n =  (R^n, {|{\ \cdot \ } |})$ if $q_{ij}=diag
 [1,1,...,1]$
 with standard Euclidean norm,
${|{\ y \ } |}  \doteq \sqrt{\sum_{i=1}^n |y^i|^2},$ for any
 $y=(y^i)\in R^n,$ where  $ R^n$ denotes the n--dimensional
 canonical real vector space.

There are also  different types of  quadratic forms/norms then the Euclidean
one:
\begin{definition} \label{dlagf}
A Lagrange  fundamental form $q_{L}(u, v)$\
on vector space  $V^n$ is defined by a
Lagrange  functional $L:\ V^n \to \R ,$ with
\begin{equation}
q_{L(y)}(u,v)\doteq \frac{1}{2} \frac{\partial ^{2}}{\partial
  s\partial t}{[L(y+su+tu)]}_{\mid s=t=0} \label{lagrf}
\end{equation}
which is a ${\it C}^\infty$--function on $V^n \backslash \{0\}$ and
 nondegenerate for any  nonzero vector $y \in V^n$ and  real
parameters $s$ and $t.$
\end{definition}
Having taken a  basis $\{e_i\}^n_{i=1}$ for  $V^n,$ we
transform $L=L(y^i e_i)$ in a function of $(y^i)\in R^n.$

The Lagrange norm is ${|{\ \cdot \ } |}_L  \doteq \sqrt{q_{L}(u, u)}.$

\begin{definition} \label{dlargf} A Minkowski space is a pair $(V^n, F)$
where the   Minkowski functional $F$ is a positively homogeneous of degree two
  Lagrange functional with the fundamental form (\ref{lagrf})
  defined for $L=F^2$ satisfying $F(\lambda y)= \lambda  F(y)$ for any
  $\lambda > 0$ and $y \in V^n.$
\end{definition}

The Minkowski norm is defined by  ${|{\ \cdot \ } |}_F  \doteq
\sqrt{q_{F}(u, u)}.$

\begin{definition}
The Lagrange (or Minkowski) metric fundamental function is defined
\begin{equation}
g_{ij}=\frac{1}{2} \frac{\partial ^{2} L }{\partial
  y^i \partial y^j}(y) \label{lagm}
\end{equation}
(or \begin{equation}
g_{ij}=\frac{1}{2} \frac{\partial ^{2} F^2 }{\partial
  y^i \partial y^j} (y) \ ). \label{finm}
\end{equation}
\end{definition}

\begin{remark}\label{rem1} If $L$ is a Lagrange functional on $R^n$ (it could be
also any functional of class  ${\it C}^\infty )$ with local coordinates
$(y^2,y^3,...,y^n),$ it also defines a singular  Minkowski functional
\begin{equation}
F(y)=\lbrack y^1 L(\frac{y^2}{y^1},...,\frac{y^n}{y^1}) \rbrack ^2
\end{equation}
which is of class  ${\it C}^\infty )$ on $R^n \backslash \lbrace y^1=0
 \rbrace$.
\end{remark}

The Remark \ref{rem1} states that the Lagrange functionals are not
essentially more general than the Minkowski functionals
\cite{15sh}. Nevertheless, we must introduce more general functionals
if we extend our considerations in relativistic optics, string models
of gravity and the theory of locally anisotropic stochastic and/or
kinetic processes \cite{15ma2,15vex1,15vt,15vs1,15dv,15vkin}.

Let us consider a base manifold $M,\ dim M =n,$ and its tangent bundle
$(TM, \pi , M)$ with natural surjective projection $\pi : TM \to M.$
From now on, all manifolds and geometric objects are supposed to be of
class $\mathcal{C}^{\infty }.$ We write $\widetilde{TM}= TM \backslash
 \lbrace 0 \rbrace $  where $ \lbrace 0 \rbrace $ means the null
 section of the map $\pi .$

A differentiable Lagrangian $L(x,y)$ is defined by a map
 $L:(x,y)\in TM\rightarrow L(x,y)\in \R$ of class $%
\mathcal{C}^{\infty }$ on $\widetilde{TM}$ and continuous on the
null section $0:M\rightarrow TM$ of $\pi .$ For any point $x \in
M,$ the restriction $L_x \doteq L_{\mid T_xM}$ is a Lagrange
functional on $T_xM$ (see Definition \ref{dlagf}). For simplicity,
in this work we shall consider only  regular Lagrangians  with
nondegenerated Hessians,
\begin{equation}
\ ^{(L)}g_{ij}(x,y)=\frac{1}{2}\frac{\partial ^{2}L(x,y)}{\partial
y^{i}\partial y^{j}}  \label{lqf}
\end{equation}%
when $rank\left| g_{ij}\right| =n$ on $\widetilde{TM},$ which is a
Lagrange fundamental quadratic form (\ref{lagm}) on  $T_xM.$ In our further
considerations, we shall write $M_{(L)}$ if would be necessary to
emphasize that the manifold $M$ is provided in any its points with a
quadratic form (\ref{lqf}).

\begin{definition}
A Lagrange space is a pair $L^{n}=\left[ M,L(x,y)\right] $ with the
metric form  $\
^{(L)}g_{ij}(x,y)$ being of constant signature over $\widetilde{TM}.$
\end{definition}

\begin{definition}
A Finsler space is a pair $F^{n}=\left[ M,F(x,y)\right] $ where
$F_{\mid x}(y)$ defines a Minkowski space with metric fundamental
function of type (\ref{finm}).
\end{definition}

The notion of Lagrange space was introduced by J. Kern \cite{15kern} and
elaborated in details by the R. Miron's school on Finsler and Lagrange
geometry, see Refs. \cite{15ma1,15ma2}, as a natural extension of Finsler
geometry \cite{15fin,15car1,15rund,15asa1,15mats,15bej,15bcs,15sh}
 (see also Refs.\cite{15vstr,15vstr1}, on Lagrange--Finsler supergeometry, and
Refs. \cite{15vnc,15vncg,15vncgs}, on some examples of noncommutative
locally anisotropic gravity and string theory).

\subsection{Nonlinear connection geometry}
We consider two important examples when the nonlinear connection (in
brief, N--connection) is
naturally defined in Lagrange mechanics and in gravity theories with
generic off--diagonal metrics and nonholonomic frames.

\subsubsection{Geometrization of mechanics: some important results}
\label{ssmr}
The Lagrange mechanics was geometrized by using methods of Finsler
geometry \cite{15ma1,15ma2} on tangent bundles enabled with a
corresponding nonholonomic structure (nonintegrable distribution)
 defining a canonical N--connection.\footnote{We cite a recent review
   \cite{15ml} on alternative  approaches to geometric mechanics and
   geometry of classical fields related to investigation of  the
   geometric properties of
   Euler--Lagrange equations for various type of nonholonomic,
   singular or higher order systems. In the approach developed by
   R. Miron's school \cite{15ma1,15ma2,15mhss}, the nonlinear connection and
 fundamental   geometric structures are derived in general form  from
 the  Lagrangian  and/or Hamiltonian: the basic  geometric
 constructions are not related to the  particular properties of
 certain  systems of partial  differential equations,
 symmetries and constraints of mechanical  and field models.}
 By straightforward calculations, one proved the results:
\begin{result} \label{2r1}
The Euler--Lagrange equations%
\begin{equation}
\frac{d}{d\tau }\left( \frac{\partial L}{\partial y^{i}}\right) -\frac{%
\partial L}{\partial x^{i}}=0 \label{eleq}
\end{equation}%
where $y^{i}=\frac{dx^{i}}{d\tau }$ for $x^{i}(\tau )$ depending on
parameter $\tau ,$ are equivalent to the ''nonlinear'' geodesic equations
\begin{equation}
\frac{d^{2}x^{i}}{d\tau ^{2}}+2G^{i}(x^{k},\frac{dx^{j}}{d\tau })=0
\label{ngeq}
\end{equation}%
where
\begin{equation}
2G^{i}(x,y)=\frac{1}{2}\ ^{(L)}g^{ij}\left( \frac{\partial ^{2}L}{\partial
y^{i}\partial x^{k}}y^{k}-\frac{\partial L}{\partial x^{i}}\right)
\label{gcoeff}
\end{equation}%
with $^{(L)}g^{ij}$ being inverse to (\ref{lqf}).
\end{result}

\begin{result} \label{r2}
The coefficients $G^{i}(x,y)$ from (\ref{gcoeff}) define the solutions
of both type of equations (\ref{eleq}) and (\ref{ngeq})  as
 paths of the canonical semispray%
\begin{equation*}
S=y^{i}\frac{\partial }{\partial x^{i}}-2G^{i}(x,y)\frac{\partial }{\partial
y^{i}}
\end{equation*}%
and a canonical N--connection structure  on $\widetilde{TM},$
\begin{equation}
^{(L)}N_{j}^{i}=\frac{\partial G^{i}(x,y)}{\partial y^{i}},  \label{cncl}
\end{equation}%
induced by the fundamental Lagrange function $L(x,y)$ (see Section
\ref{snc} on exact definitions and main properties).
\end{result}

\begin{result} \label{resws}
The coefficients $\ ^{(L)}N_{j}^{i}$ defined by a Lagrange (Finsler)
fundamental function induce a global splitting on $TTM,$  a Whitney
sum,
\begin{equation*}
TTM=hTM \oplus vTM
\end{equation*}
as a nonintegrable distribution (nonholonomic, or equivalently,
anholonomic structure) into
horizontal (h) and vertical (v) subspaces parametrized
locally by frames (vielbeins)
 $\mathbf{e}_{\nu }=(e_{i},e_{a}),$
where
\begin{equation}
e_{i}=\frac{\partial }{\partial x^{i}}-N_{i}^{a}(u)\frac{\partial }{\partial
y^{a}}\mbox{ and }e_{a}=\frac{\partial }{\partial y^{a}},  \label{dder}
\end{equation}%
and the dual frames (coframes) $\mathbf{\vartheta }^{\mu
}=(\vartheta ^{i},\vartheta ^{a}),$ where
\begin{equation}
\vartheta ^{i}=dx^{i}\mbox{ and }\vartheta ^{a}=dy^{a}+N_{i}^{a}(u)dx^{i}.
\label{ddif}
\end{equation}%
\end{result}

The vielbeins (\ref{dder}) and (\ref{ddif}) are called N--adapted (co)
frames. We omitted the label $(L)$ and used vertical indices
$a,b,c,...$  for the N--connection coefficients in order to be able to
use the formulas for arbitrary N--connections). We also note that we
shall use 'boldfaced' symbols for the geometric objects and spaces
adapted/ enabled to N--connection structure. For instance, we shall
write in brief $\mathbf{e}=(e,\ ^{\star }e)$ and
$\mathbf{\vartheta }=(\vartheta,\ ^{\star }\vartheta )$, respectively, for
\begin{equation*}
\mathbf{e}_{\nu }=(e_{i},\ ^{\star }e_{k})=(e_{i},\  e_{a})
\mbox{ and }
\mathbf{\vartheta }^{\mu}=(\vartheta ^{i},\ ^{\star }\vartheta ^{k})=
(\vartheta ^{i}, \vartheta ^{a}).
\end{equation*}
The vielbeins (\ref{dder}) satisfy the nonholonomy relations
\begin{equation}
\lbrack \mathbf{e}_{\alpha },\mathbf{e}_{\beta }]=\mathbf{e}_{\alpha }%
\mathbf{e}_{\beta }-\mathbf{e}_{\beta }\mathbf{e}_{\alpha }=W_{\alpha \beta
}^{\gamma }\mathbf{e}_{\gamma }  \label{anhrel}
\end{equation}%
with (antisymmetric) nontrivial anholonomy coefficients $W_{ia}^{b}=\partial
_{a}N_{i}^{b}$ and $W_{ji}^{a}=\Omega _{ij}^{a}$ where
\begin{equation}
\Omega _{ij}^{a}=\delta _{\lbrack j}N_{i]}^{a}=\frac{\partial N_{i}^{a}}{%
\partial x^{j}}-\frac{\partial N_{j}^{a}}{\partial x^{i}}+N_{i}^{b}\frac{%
\partial N_{j}^{a}}{\partial y^{b}}-N_{j}^{b}\frac{\partial N_{i}^{a}}{%
\partial y^{b}}.  \label{ncurv}
\end{equation}

 In order to preserve a relation with our previous denotations \cite%
{15vex1,15vsp1,15vstr1}, we note that $\mathbf{e}_{\nu
}=(e_{i},e_{a})$ and $\mathbf{\vartheta }^{\mu }=(\vartheta ^{i},\vartheta
^{a})$ are, respectively, the former $\delta _{\nu }=\delta /\partial u^{\nu
}=(\delta _{i},\partial _{a})$ and $\delta ^{\mu }=\delta u^{\mu
}=(dx^{i},\delta y^{a})$ which emphasize that the  operators (\ref{dder}) and (%
\ref{ddif}) define, correspondingly, certain 'N--elongated' partial
derivatives and differentials which are more convenient for calculations on
spaces provided with nonholonomic structure.

\begin{result} \label{r4}
 On $\widetilde{TM},$ there is a canonical metric structure
$\ ^{(L)}\mathbf{g}= [g,\ ^{\star}g],$
\begin{equation}
\ ^{(L)}\mathbf{g}=\ ^{(L)}g_{ij}(x,y)\ \vartheta ^{i}\otimes \vartheta
^{j}+\ ^{(L)}g_{ij}(x,y)\ ^{\star }\vartheta ^{i}\otimes \ ^{\star }\vartheta
^{j}  \label{slm}
\end{equation}%
constructed as a Sasaki type lift from $M.$\footnote{In
  Refs. \cite{15sh,15mhss}, it was suggested to use lifts with h- and
  v--components of type
 $\ ^{(L)}\mathbf{g}=(g_{ij}, g_{ij}a/\parallel y \parallel)$
  where $a=const$ and $\parallel y \parallel = g_{ij}y^i y^j $ in order
  to elaborate more physical extensions of the general relativity to
  the tangent bundles of manifolds. In another turn, such
  modifications are not necessary if we model Lagrange--Finsler
  structures by exact solutions with generic off--diagonal metrics in
  Einstein and/or gravity \cite{15vex1,15vt,15vs1,15dv,15v2,15v3,15vkin}. For
  simplicity, in this work, we  consider only lifts of metrics of
  type (\ref{slm}).}
\end{result}
We note that a complete geometrical model of Lagrange mechanics or a
well defined Finsler geometry can be elaborated only by additional assumptions
about a linear connection structure, which can be adapted, or not, to
a defined N--connection (see Section \ref{sdlc}).

\begin{result} \label{r5}
 The canonical N--connection (\ref{cncl})
induces naturally an almost complex structure $\mathbf{F}:\chi (\widetilde{TM%
})\rightarrow \chi (\widetilde{TM}),$ where $\chi $ denotes the module of
vector fields on $\widetilde{TM},$%
\begin{equation*}
\mathbf{F}(e_{i})=\ ^{\star }e_{i}\mbox{ and }\mathbf{F}(\ ^{\star
}e_{i})=-e_{i},
\end{equation*}%
when
\begin{equation}
\mathbf{F}=\ ^{\star }e_{i}\otimes \vartheta ^{i}-e_{i}\otimes \ ^{\star
}\vartheta ^{i}  \label{acs1}
\end{equation}%
satisfies the condition $\mathbf{F\rfloor \ F=-I,}$ i. e. $F_{\ \ \beta
}^{\alpha }F_{\ \ \gamma }^{\beta }=-\delta _{\gamma }^{\alpha },$ where $%
\delta _{\gamma }^{\alpha }$ is the Kronecker symbol and ''$\mathbf{\rfloor }
$'' denotes the interior product.
\end{result}
The last result is important for elaborating  an approach to geometric
quantization of mechanical systems modelled on nonholonomic manifolds
\cite{15esv}  as well for definition of almost complex structures
derived from the real N--connection geometry related to nonholonomic
(anisotropic)  Clifford structures and  spinors in commutative
 \cite{15vsp1,15vsph,15vst,15vsp2,15vp} and noncommutative spaces
  \cite{15vnc,15vncg,15vncgs}.

\subsubsection{N--connections in  gravity theories}

For nonholonomic geometric models of gravity and string theories, one
 does not consider  the
bundle $\widetilde{TM}$ but a general manifold $\mathbf{V},\
dim\mathbf{V}=n+m,$ which is a (pseudo) Riemannian space or a certain
generalization with possible torsion and nonmetricity fields.
 A metric structure is defined on $\mathbf{V},$ with the coefficients
 stated with respect to a local coordinate basis $du^{\alpha }=\left(
dx^{i},dy^{a}\right) ,$ \footnote{the indices run correspondingly the values
 $i,j,k,...=1,2,...,n$ and $a,b,c,...=n+1,n+2,...,n+m.$}
\begin{equation*}
\mathbf{g}=\underline{g}_{\alpha \beta }(u)du^{\alpha }\otimes du^{\beta }
\end{equation*}%
where
\begin{equation}
\underline{g}_{\alpha \beta }=\left[
\begin{array}{cc}
g_{ij}+N_{i}^{a}N_{j}^{b}h_{ab} & N_{j}^{e}h_{ae} \\
N_{i}^{e}h_{be} & h_{ab}%
\end{array}%
\right] .  \label{ansatz}
\end{equation}

A metric, for instance, parametrized in the form (\ref{ansatz}), is generic
off--diagonal if it can not be diagonalized by any coordinate transforms.
Performing a frame transform with the coefficients
\begin{eqnarray}
\mathbf{e}_{\alpha }^{\ \underline{\alpha }}(u) &=&\left[
\begin{array}{cc}
e_{i}^{\ \underline{i}}(u) & N_{i}^{b}(u)e_{b}^{\ \underline{a}}(u) \\
0 & e_{a}^{\ \underline{a}}(u)%
\end{array}%
\right] ,  \label{3vt1} \\
\mathbf{e}_{\ \underline{\beta }}^{\beta }(u) &=&\left[
\begin{array}{cc}
e_{\ \underline{i}}^{i\ }(u) & -N_{k}^{b}(u)e_{\ \underline{i}}^{k\ }(u) \\
0 & e_{\ \underline{a}}^{a\ }(u)%
\end{array}%
\right] ,  \label{3vt2}
\end{eqnarray}%
we write equivalently the metric in the form
\begin{equation}
\mathbf{g}=\mathbf{g}_{\alpha \beta }\left( u\right) \mathbf{\vartheta }%
^{\alpha }\otimes \mathbf{\vartheta }^{\beta }=g_{ij}\left( u\right)
\vartheta ^{i}\otimes \vartheta ^{j}+h_{ab}\left( u\right) \ ^{\star
}\vartheta ^{a}\otimes \ ^{\star }\vartheta ^{b},  \label{metr}
\end{equation}%
where $g_{ij}\doteqdot \mathbf{g}\left( e_{i},e_{j}\right) $ and $%
h_{ab}\doteqdot \mathbf{g}\left( e_{a},e_{b}\right) $ \ and
\begin{equation*}
\mathbf{e}_{\alpha }=\mathbf{e}_{\alpha }^{\ \underline{\alpha }}\partial _{%
\underline{\alpha }}\mbox{ and }\mathbf{\vartheta }_{\ }^{\beta }=\mathbf{e}%
_{\ \underline{\beta }}^{\beta }du^{\underline{\beta }}.
\end{equation*}%
are vielbeins of type (\ref{dder}) and (\ref{ddif}) defined for arbitrary $%
N_{i}^{b}(u).$ We can consider a special class of manifolds provided with a
global splitting into conventional ''horizontal'' and ''vertical'' subspaces
(\ref{whitney}) induced by the ''off--diagonal'' terms $N_{i}^{b}(u)$ and
prescribed type of nonholonomic frame structure.

If the manifold $\mathbf{V}$ is (pseudo) Riemannian, there is a unique
linear connection (the Levi--Civita connection) $\nabla ,$ which is metric, $%
\nabla \mathbf{g=0,}$ and torsionless, $\ ^{\nabla }T=0.$ Nevertheless, the
connection $\nabla $ is not adapted to the nonintegrable distribution
induced by $N_{i}^{b}(u).$ In this case,
for instance, in order to construct exact solutions parametrized by generic
off--diagonal metrics, or for investigating nonholonomic frame structures in
gravity models with nontrivial torsion, it is more convenient to work with
more general classes of linear connections which are N--adapted but contain
nontrivial torsion coefficients because of nontrivial nonholonomy
coefficients $W_{\alpha \beta }^{\gamma }$ (\ref{anhrel}).

For a splitting of a (pseudo) Riemannian--Cartan space of dimension $(n+m)$
(under certain constraints, we can consider (pseudo) Riemannian
configurations), the Lagrange and Finsler type geometries were modelled by
N--anholonomic structures as exact solutions of gravitational field
equations \cite{15vex1,15vt,15vs1,15dv}, see also Refs. \cite{15v2,15v3} for exact
solutions with nonmetricity.  One holds \cite{15v1} the
\begin{result} \label{r6}
 The  geometry of any  Riemannian space of dimension $n+m$  where
 $n,m \geq 2$  (we can consider $n,m=1$ as special degenerated cases),
 provided with  off--diagonal metric structure of type (\ref{ansatz})
 can be equivalently modelled, by vielbein transforms of type
 (\ref{3vt1}) and (\ref{3vt2}) as a geometry of nonholonomic manifold
 enabled with N--connection structure $N_{i}^{b}(u)$ and 'more
 diagonalized' metric (\ref{metr}).
\end{result}
For particular cases, we present the
\begin{remark}
For certain special conditions
 when $n=m,$ $N_{i}^{b}=\ ^{(L)}N_{i}^{b}$ (\ref{cncl}) and the metric
(\ref{metr}) is of type (\ref{slm}), a such Riemann space of even
 dimension is 'nonholonomically' equivalent to a Lagrange space (for
 the corresponding homogeneity conditions, see Definition \ref{dlargf},
 one obtains the equivalence to a Finsler space).
\end{remark}
Roughly speaking, by prescribing corresponding nonholonomic frame
structures, we can model a Lagrange, or Finsler, geometry on
a Riemannian manifold and, inversely, a Riemannian geometry is 'not
only a Riemannian one' but also could be a generalized Finsler
one. It is possible to define similar constructions for the
 (pseudo) Riemannian spaces.
This is a quite surprising result if to compare it with the
"superficial" interpretation of the Finsler geometry as a nonlinear
extension, 'more sophisticate' on the tangent bundle,
 of the Riemannian geometry.

It is known the fact that the first example of Finsler geometry was
considered in 1854 in the famous B. Riemann's hability thesis (see
historical  details and discussion
in Refs. \cite{15sh,15bcs,15ma2,15v1}) who, for simplicity, restricted his
considerations only to the curvatures defined by quadratic
forms on hypersurfaces.  Sure,  for B. Riemann, it  was unknown  the fact that if we
consider general (nonholonomic) frames with  associated nonlinear
connections (the E. Cartan's geometry, see Refs. in \cite{15car1})
 and off--diagonal metrics, the Finsler geometry may be derived naturally
even from quadratic metric forms being adapted to the N--connection structure.

 More rigorous geometric constructions involving the Cartan--Miron metric
 connections and, respectively,    the Berwald and Chern--Rund
 nonmetric  connections in Finsler geometry and generalizations, see
 more details in subsection \ref{sdlc}, result in equivalence theorems
  to certain types of  Riemann--Cartan nonholonomic
 manifolds (with nontirvial N--connection and torsion) and
 metric--affine nonholonomic manifolds (with additional nontrivial
 nonmetricity structures) \cite{15v1}.

This Result \ref{r6} give rise to an important:
\begin{conclusion} \label{ic} To study  generalized Finsler spinor and
  noncommutative geometries is necessary
  even if we restrict our considerations only to
  (non) commutative Riemannian geometries.
\end{conclusion}

For simplicity, in this work we
 restrict our considerations only to certain Riemannian commutative and
 noncommutative geometries when the N--connection and torsion are
 defined by corresponding nonholonomic frames.

\subsection{N--anholonomic manifolds}
\label{snc}

Now we shall demonstrate how general N--connection structures define a
certain class of nonholonomic geometries. In this case, it is
convenient to work on a general manifold $\mathbf{V,}$ $\dim
\mathbf{V=}n+m,$ with global splitting, instead of the tangent bundle
$\widetilde{TM}.$ The constructions will contain  those from
geometric mechanics and gravity theories, as certain particular
cases.

Let $\mathbf{V}$ be a $(n+m)$--dimensional manifold. It is supposed that in
any point $u\in \mathbf{V}$ there is a local distribution (splitting)
 $\mathbf{V}_{u}=M_{u}\oplus V_{u},$
 where $M$ is a $n-$dimensional subspace and $V$ is
a $m$--dimensional subspace. The local coordinates (in
general, abstract ones both for holonomic and nonholonomic variables)
may be written in the
form $u=(x,y),$ or $u^{\alpha }=\left( x^{i},y^{a}\right).$  We denote by $\pi
^{\top }:T\mathbf{V}\rightarrow TM$ the differential of a map $\pi
:V^{n+m}\rightarrow V^{n}$ defined by fiber preserving morphisms of the
tangent bundles $T\mathbf{V}$ and $TM.$ The kernel of $\pi ^{\top }$ is just
the vertical subspace $v\mathbf{V}$ with a related inclusion mapping $i:v%
\mathbf{V}\rightarrow T\mathbf{V}.$

\begin{definition} \label{dncam}
A nonlinear connection (N--connection) $\mathbf{N}$ on a manifold $\mathbf{V}
$ is defined by the splitting on the left of an exact sequence
\begin{equation*}
0\rightarrow v\mathbf{V}\overset{i}{\rightarrow} T\mathbf{V}\rightarrow T%
\mathbf{V}/v\mathbf{V}\rightarrow 0,
\end{equation*}%
i. e. by a morphism of submanifolds $\mathbf{N:\ \ }T\mathbf{V}\rightarrow v%
\mathbf{V}$ such that $\mathbf{N\circ i}$ is the unity in $v\mathbf{V}.$
\end{definition}

In an equivalent form, we can say that a N--connection is defined by a
splitting to subspaces with a Whitney sum of conventional h--subspace,
$\left( h\mathbf{V}\right) ,$ and v--subspace, $\left( v%
\mathbf{V}\right) ,$
\begin{equation}
T\mathbf{V}=h\mathbf{V}\oplus v\mathbf{V}  \label{whitney}
\end{equation}%
where $h\mathbf{V}$ is isomorphic to $M.$ This  generalizes
the splitting considered in Result \ref{resws}.

Locally, a N--connection is defined by its coefficients $N_{i}^{a}(u),$%
\begin{equation}
\mathbf{N}=N_{i}^{a}(u)dx^{i}\otimes \frac{\partial }{\partial y^{a}}.
\label{nconcoef}
\end{equation}%
The well known class of linear connections consists a particular subclass
with the coefficients being linear on $y^{a},$ i. e. $N_{i}^{a}(u)=\Gamma
_{bj}^{a}(x)y^{b}.$

Any N--connection also defines a N--connection curvature
\begin{equation*}
\mathbf{\Omega }=\frac{1}{2}\Omega _{ij}^{a}d^{i}\wedge d^{j}\otimes
\partial _{a},
\end{equation*}%
with N--connection curvature coefficients given by formula (\ref{anhrel}). %

\begin{definition} \label{dnam}
A manifold \ $\mathbf{V}$ is called N--anholonomic if on the tangent space $T%
\mathbf{V}$ it is defined a local (nonintegrable) distribution (\ref{whitney}%
), i. e. $T\mathbf{V}$ is enabled with a N--connection (\ref{nconcoef})
inducing a vielbein structure (\ref{dder}) satisfying the nonholonomy
relations  (\ref{anhrel}) (such N--connections and
associated vielbeins may be general ones or any derived from a
Lagrange/ Finsler fundamental function).
\end{definition}

We note that the  boldfaced symbols are used for the spaces and
geometric objects provided/adapted to a N--connection structure. For
instance, a vector field $\mathbf{X}\in T\mathbf{V}$ \ is expressed $\mathbf{%
X}=(X \equiv\ ^{-}X,\ ^{\star }X),$ or $\mathbf{X}=X^{\alpha }\mathbf{e}_{\alpha
}=X^{i}e_{i}+X^{a}e_{a},$ where
$X=\  ^{-}X = X^{i}e_{i}$ and $^{\star }X=X^{a}e_{a}$
state, respectively, the irreducible (adapted to the N--connection
structure) h-- and v--components of the vector (which
following Refs. \cite{15ma1,15ma2} is called a distinguished vectors, in brief,
d--vector). In a similar fashion, the geometric objects on $\mathbf{V}$ like
tensors, spinors, connections, ... are respectively defined and called
 d--tensors, d--spinors, d--connections if they are adapted to the N--connection
splitting.\footnote{In order to emphasize h-- and v--splitting of any
d--objects $\mathbf{Y}, \mathbf{g},$ ... we shall write the
irreducible components as $\mathbf{Y} = (\ ^{-}Y,\ ^\star Y),$\
 $\mathbf{g} = (\ ^{-}g,\ ^\star g)$ but we shall omit "$-$" or
"$\star$" if the simplified denotations will not result in ambiguities.}

\begin{definition} \label{ddms}
  A d--metric structure on N--anholonomic manifold
  $\mathbf{V}$ is defined by a symmetric d--tensor field of type
 $\mathbf{g}=[g,\ ^{\star}h]$ (\ref{metr}).
\end{definition}

 For any fixed values of
  coordinates $u=(x,y) \in \mathbf{V}$  a d--metric it defines a symmetric
 quadratic d--metric form,
\begin{equation} \label{qdmf}
 \mathbf{q}(\mathbf{x},\mathbf{y}) \doteq g_{ij} x^i x^j + h_{ab} y^a
 y^b,
\end{equation}
where the $n+m$--splitting is defined by the N--connection structure
and  $\mathbf{x} = x^i e_i + x^a e_a,\ \mathbf{y}=y^i e_i + y^a e_a
\in V^{n+m}.$

Any d--metric is parametrized by a generic off--diagonal matrix
(\ref{ansatz}) if the coefficients are redefined with respect to a
 local coordinate basis (for corresponding parametri\-zations of the
 the data $[g,h,N]$ such ansatz model a  geometry of mechanics, or a
 Finsler like structure, in a  Riemann--Cartan--Weyl space provided
 with N--connection structure  \cite{15v1,15v2};
 for certain constraints, there are possible models derived as exact
solutions in Einsten gravity and noncommutative generalizations
 \cite{15vex1,15dv,15v3}).

\begin{remark} There is a  special case when
 $\dim \mathbf{V=}n+n,\ h_{ab}\rightarrow g_{ij}$ and
$N_{i}^{a}\rightarrow N_{\ i}^{j}$ in (\ref{metr}), which models
locally,  on $\mathbf{V,}$
 a tangent bundle  structure. We denote a
such space by $\widetilde{\mathbf{V}}_{(n,n)}.$ If the N--connection
and d--metric coefficients are just the canonical ones for the Lagrange
(Finsler) geometry (see, respectively, formulas (\ref{cncl}) and
(\ref{slm}) ),
 we model such locally anisotropic structures not
on a tangent bundle $TM$ but on a N--anholonomic manifold of dimension $2n.$
\end{remark}

We present some historical remarks on
 N--con\-nec\-ti\-ons and related subjects:\ The geometrical aspects of
the N--connection formalism has been studied since the first papers of E.
Cartan \cite{15car1}\ and A. Kawaguchi \cite{15ak1,15ak2}\ (who used it in
component form for Finsler geometry). Then one should be mentioned the so
called Ehressman connection \cite{15eh}) and the work of W. Barthel \cite{15wb}
where the global definition of N--connection was given. In monographs \cite%
{15ma1,15ma2,15mhss},  the N--connection formalism was elaborated in details
and applied to the geometry of generalized Finsler--Lagrange and
 Cartan--Hamilton spaces, see also the approaches \cite{15ml1,15ml2,15fe}.It
 should be  noted that the works related to  nonholonomic geometry
and N--connections have appeared many times in a rather dispersive
way when different schools of authors from geometry, mechanics and
physics have worked many times not having relation with another.
We cite only some our recent results with explicit applications in
modern mathematical physics and particle and string theories:
N--connection structures were modelled on Clifford and spinor
bundles \cite{15vsp1,15vsph,15vsp2,15vst},
 on superbundles and in some directions of (super) string theory %
\cite{15vstr,15vstr1}, as well in noncommutative geometry and gravity
 \cite{15vnc,15vncg,15vncgs}.
 The idea to apply the N--connections formalism as a new geometric method
of constructing exact solutions in gravity theories was suggested in Refs. %
\cite{15vex1,15vkin} and developed in a number of works, see for instance, Ref. %
\cite{15vt,15vs1,15dv}).

\section{Curvature of N--anholonomic Manifolds}

The geometry of nonholonomic manifolds has a long time history of yet
unfinished elaboration:\ For instance, in the review \cite{15ver} it is
stated that it is probably impossible to construct an analog of the
Riemannian tensor for the general nonholonomic manifold.  In a more
recent review \cite{15mironnh1}, it is emphasized that in the
past there were proposed well defined Riemannian tensors for a number
of spaces provided with nonholonomic distributions, like Finsler and
Lagrange spaces and various type of theirs higher order
generalizations, i. e. for nonholonomic manifolds possessing
corresponding  N--connection structures. As some examples of first
such  investigations, we cite the works  \cite{15nhm1,15nhm3,15nhm4}.
 In this section  we shall construct in explicit form the curvature tensor for
 the N--anholonomic manifolds.

\subsection{Distinguished connections }
\label{sdlc}

On N--anholonomic manifolds, the geometric
 constructions can  be adapted to the N--connection structure:

\begin{definition} \label{ddc}
A distinguished connection (d--connection) $\mathbf{D}$ on a manifold $%
\mathbf{V}$ is a linear connection conserving under parallelism the Whitney
sum (\ref{whitney}) defining a general N--connection.
\end{definition}

The N--adapted components $\mathbf{\Gamma }_{\beta \gamma }^{\alpha }$ of a
d-connection $\mathbf{D}_{\alpha }=(\delta _{\alpha }\rfloor \mathbf{D})$
are defined by the equations $
\mathbf{D}_{\alpha }\delta _{\beta }=\mathbf{\Gamma }_{\ \alpha \beta
}^{\gamma }\delta _{\gamma },$
or
\begin{equation}
\mathbf{\Gamma }_{\ \alpha \beta }^{\gamma }\left( u\right) =\left( \mathbf{D%
}_{\alpha }\delta _{\beta }\right) \rfloor \delta ^{\gamma }.  \label{dcon1}
\end{equation}%
In its turn, this defines a N--adapted splitting into h-- and v--covariant
derivatives, $\mathbf{D}=D+\ ^{\star }D,$ where $D_{k}=\left(
L_{jk}^{i},L_{bk\;}^{a}\right) $ and $\ ^{\star }D_{c}=\left(
C_{jk}^{i},C_{bc}^{a}\right) $ are introduced as corresponding h- and
v--parametrizations of (\ref{dcon1}),%
\begin{equation*}
L_{jk}^{i}=\left( \mathbf{D}_{k}e_{j}\right) \rfloor \vartheta ^{i},\quad
L_{bk}^{a}=\left( \mathbf{D}_{k}e_{b}\right) \rfloor \vartheta
^{a},~C_{jc}^{i}=\left( \mathbf{D}_{c}e_{j}\right) \rfloor \vartheta
^{i},\quad C_{bc}^{a}=\left( \mathbf{D}_{c}e_{b}\right) \rfloor \vartheta
^{a}.
\end{equation*}%
The components $\mathbf{\Gamma }_{\ \alpha \beta }^{\gamma }=\left(
L_{jk}^{i},L_{bk}^{a},C_{jc}^{i},C_{bc}^{a}\right) $ completely define a
d--connection $\mathbf{D}$ on a N--anholonomic manifold $\mathbf{V}.$

The simplest way to perform computations with d--connections is to use
N--adapted differential forms like $\mathbf{\Gamma }_{\beta }^{\alpha }=%
\mathbf{\Gamma }_{\beta \gamma }^{\alpha }\mathbf{\vartheta }^{\gamma }$
with the coefficients defined with respect to N--elongate bases (\ref{ddif})
 and (\ref{dder}).

The torsion  of  d--connection $\mathbf{D}$ is defined by the
usual formula%
\begin{equation*}
\mathbf{T}(\mathbf{X},\mathbf{Y})\doteqdot \mathbf{D}_{X}\mathbf{D}%
_{Y}-\mathbf{D}_{Y}\mathbf{D}_{X} -[\mathbf{X},\mathbf{Y}].
\end{equation*}%

\begin{theorem}
The torsion $\mathbf{T}^{\alpha }\doteqdot \mathbf{D\vartheta }^{\alpha }=d%
\mathbf{\vartheta }^{\alpha }+\Gamma _{\beta }^{\alpha }\wedge \mathbf{%
\vartheta }^{\beta }$ of a d--connection has the irreducible h- v--
components (d--torsions)with N--adapted coefficients
\begin{eqnarray}
T_{\ jk}^{i} &=&L_{\ [jk]}^{i},\ T_{\ ja}^{i}=-T_{\ aj}^{i}=C_{\ ja}^{i},\
T_{\ ji}^{a}=\Omega _{\ ji}^{a},\   \notag \\
T_{\ bi}^{a} &=&T_{\ ib}^{a}=\frac{\partial N_{i}^{a}}{\partial y^{b}}-L_{\
bi}^{a},\ T_{\ bc}^{a}=C_{\ [bc]}^{a}.  \label{dtors}
\end{eqnarray}
\end{theorem}

\begin{proof}
By a straightforward calculation we can verify the formulas.
\end{proof}

The Levi--Civita linear connection $\nabla =\{^{\nabla }\mathbf{\Gamma }%
_{\beta \gamma }^{\alpha }\},$ with vanishing both torsion and
nonmetricity, is not adapted to the global splitting
(\ref{whitney}). One holds:

\begin{proposition} \label{pcnc}
There is a preferred, canonical d--connection structure,
$\widehat{\mathbf{D}},$ on N--anholonomic manifold $\mathbf{V}$ constructed only
from the metric and N--con\-nec\-ti\-on coefficients $%
[g_{ij},h_{ab},N_{i}^{a}]$ and satisfying the metricity conditions
 $\widehat{\mathbf{D}%
}\mathbf{g}=0$ and $\widehat{T}_{\ jk}^{i}=0$ and $\widehat{T}_{\ bc}^{a}=0.$
\end{proposition}

\begin{proof}
By straightforward calculations with respect to the N--adapted bases (\ref%
{ddif}) and (\ref{dder}), we can verify that the connection
\begin{equation}
\widehat{\mathbf{\Gamma }}_{\beta \gamma }^{\alpha }=\ ^{\nabla }\mathbf{%
\Gamma }_{\beta \gamma }^{\alpha }+\ \widehat{\mathbf{P}}_{\beta \gamma
}^{\alpha }  \label{cdc}
\end{equation}%
with the deformation d--tensor
\begin{equation*}
\widehat{\mathbf{P}}_{\beta \gamma }^{\alpha }=(P_{jk}^{i}=0,P_{bk}^{a}=%
\frac{\partial N_{k}^{a}}{\partial y^{b}},P_{jc}^{i}=-\frac{1}{2}%
g^{ik}\Omega _{\ kj}^{a}h_{ca},P_{bc}^{a}=0)
\end{equation*}%
satisfies the conditions of this Proposition. It should be noted that, in
general, the components $\widehat{T}_{\ ja}^{i},\ \widehat{T}_{\ ji}^{a}$
and $\widehat{T}_{\ bi}^{a}$ are not zero. This is an anholonomic frame (or,
equivalently, off--diagonal metric) effect.
\end{proof}

With respect to the N--adapted frames, the coefficients\newline
$\widehat{\mathbf{\Gamma }}_{\ \alpha \beta }^{\gamma }=\left( \widehat{L}%
_{jk}^{i},\widehat{L}_{bk}^{a},\widehat{C}_{jc}^{i},\widehat{C}%
_{bc}^{a}\right) $ are computed:
\begin{eqnarray}
\widehat{L}_{jk}^{i} &=&\frac{1}{2}g^{ir}\left( \frac{\delta g_{jr}}{%
\partial x^{k}}+\frac{\delta g_{kr}}{\partial x^{j}}-\frac{\delta g_{jk}}{%
\partial x^{r}}\right) ,  \label{3candcon} \\
\widehat{L}_{bk}^{a} &=&\frac{\partial N_{k}^{a}}{\partial y^{b}}+\frac{1}{2}%
h^{ac}\left( \frac{\delta h_{bc}}{\partial x^{k}}-\frac{\partial N_{k}^{d}}{%
\partial y^{b}}h_{dc}-\frac{\partial N_{k}^{d}}{\partial y^{c}}h_{db}\right)
,  \notag \\
\widehat{C}_{jc}^{i} &=&\frac{1}{2}g^{ik}\frac{\partial g_{jk}}{\partial
y^{c}},  \notag \\
\widehat{C}_{bc}^{a} &=&\frac{1}{2}h^{ad}\left( \frac{\partial h_{bd}}{%
\partial y^{c}}+\frac{\partial h_{cd}}{\partial y^{b}}-\frac{\partial h_{bc}%
}{\partial y^{d}}\right) .  \notag
\end{eqnarray}%
The d--connection (\ref{3candcon}) defines the 'most minimal'    N--adapted
 extension of the Levi--Civita connection in order to preserve the
 metricity condition and to have zero torsions on the h-- and
 v--subspaces (the rest of nonzero torsion coefficients are defined by
 the condition of compatibility with the N--connection splitting).

\begin{remark}
The canonical d--connection $\widehat{\mathbf{D}}$ (\ref{3candcon}) for a
local modelling of a $\widetilde{TM}$ space on $\widetilde{\mathbf{V}}%
_{(n,n)}$ is defined by the coefficients
 $\widehat{\mathbf{\Gamma }}_{\ \alpha \beta
}^{\gamma }=(\widehat{L}_{jk}^{i},\widehat{C}_{jk}^{i})\ $ with%
\begin{equation}
\widehat{L}_{jk}^{i}=\frac{1}{2}g^{ir}\left( \frac{\delta g_{jr}}{\partial
x^{k}}+\frac{\delta g_{kr}}{\partial x^{j}}-\frac{\delta g_{jk}}{\partial
x^{r}}\right) , \widehat{C}_{jk}^{i}=\frac{1}{2}g^{ir}\left( \frac{\partial
g_{jr}}{\partial y^{k}}+\frac{\partial g_{kr}}{\partial y^{j}}-\frac{%
\partial g_{jk}}{\partial y^{r}}\right)  \label{candcon1}
\end{equation}%
 computed with respect to N--adapted bases (\ref{dder}) and
 (\ref{ddif}) when  $\widehat{L}_{jk}^{i}$ and $\widehat{C}_{jk}^{i}$ define
respectively the canonical h-- and v--covariant derivations.
\end{remark}

Various models of Finsler geometry and generalizations were elaborated
by using different types of d--connections which satisfy, or not, the
compatibility conditions with a fixed d--metric structure (for
instance, with  a Sasaki type one). Let us consider the main examples:
\begin{example} The Cartan's d--connection \cite{15car1}  with
  the coefficients (\ref{candcon1})  was defined by
   some generalized Christoffel symbols with the aim to have a 'minimal'
  torsion and to preserve the metricity condition. This approach was
  developed for generalized Lagrange spaces and on vector bundles
  provided with N--connection structure \cite{15ma1,15ma2}  by introducing the
  canonical d--connection (\ref{3candcon}). The direction emphasized
  metric compatible and N--adapted geometric constructions.
\end{example}

An alternative class of Finsler geometries is concluded in monographs
\cite{15bcs,15sh}:
\begin{example} \label{ecrdc} It was the idea of C. C. Chern
 \cite{15chern} (latter also
proposed by H. Rund \cite{15rund}) to consider a d--connection
 $\ ^{[Chern]}{\mathbf{\Gamma }}_{\ \alpha \beta
}^{\gamma }=(\widehat{L}_{jk}^{i},{C}_{jk}^{i}= 0)\ $ and to work not
on a tangent bundle $TM$ but to try to 'keep maximally' the constructions
 on the base manifold $M.$ The Chern d--connection, as well
the Berwald d--connection   $\ ^{[Berwald]}{\mathbf{\Gamma }}_{\ \alpha \beta
}^{\gamma }=({L}_{jk}^{i}=\frac{\partial {N}_{k}^{i}}{\partial y^j}
,{C}_{jk}^{i}= 0)\ $ \cite{15bw}, are not subjected to the metricity
conditions.
\end{example}
We note that the constructions mentioned in the last example
 define certain  'nonmetric geometries'
(a Finsler modification of the Riemann--Cartan--Weyl
spaces). For the Chern's connection, the torsion vanishes but there is
a nontrivial nonmetricity. A detailed study and classification  of Finsler--affine spaces
with general  nontrivial N--connection, torsion and nonmetricity was
recently performed in Refs. \cite{15v1,15v2,15v3}.
Here we also note that we may consider any  linear connection can be
generated by deformations of type
\begin{equation}
\mathbf{\Gamma }_{\beta \gamma }^{\alpha
}=\widehat{\mathbf{\Gamma }}_{\beta \gamma }^{\alpha }+\mathbf{P}_{\beta
\gamma }^{\alpha }. \label{2defcon}
\end{equation}
This  splits all geometric objects into canonical and
post-canonical pieces which results in N--adapted geometric
constructions.

In order to define spinors on generalized Lagrange and
Finsler spaces \cite{15vsp1,15vsph,15vst,15vsp2} the canonical d--connection
and Cartan's d--connection were used because the metric compatibility
allows the simplest definition of Clifford structures locally adapted
to the N--connection. This is also the simplest way to define
the Dirac operator for generalized Finsler spaces and to extend the
constructions to noncommutative Finsler geometry
\cite{15vnc,15vncg,15vncgs}. The geometric constructions with general metric
compatible affine connection (with torsion) are preferred in
modern gravity and string theories. Nevertheless, the geometrical and
physical models with generic nonmetricity also present certain
interest \cite{15hehl,15v1,15v2,15v3} (see also \cite{15maj} where nonmetricity
is considered to be important in quantum group co gravity). In such
cases, we can use deformations of connection (\ref{2defcon}) in order
to 'deform', for instance, the spinorial geometric constructions
defined by the canonical d--connection and to transform them into
certain 'nonmetric' configurations.

\subsection{ Curvature of  d--connections}

The curvature of a  d--connection $\mathbf{D}$ on an N--anholonomic
manifold is defined by the usual formula%
\begin{equation*}
\mathbf{R}(\mathbf{X},\mathbf{Y})\mathbf{Z}\doteqdot \mathbf{D}_{X}\mathbf{D}%
_{Y}\mathbf{Z}-\mathbf{D}_{Y}\mathbf{D}_{X}\mathbf{Z-D}_{[X,X]}\mathbf{Z.}
\end{equation*}%

By straightforward calculations we prove:
\begin{theorem}
The curvature $\mathcal{R}_{\ \beta }^{\alpha }\doteqdot \mathbf{D\Gamma }%
_{\beta }^{\alpha }=d\mathbf{\Gamma }_{\beta }^{\alpha }-\mathbf{\Gamma }%
_{\beta }^{\gamma }\wedge \mathbf{\Gamma }_{\gamma }^{\alpha }$ of a
d--connection $\mathcal{D} \doteq \mathbf{\Gamma }_{\gamma }^{\alpha }$ has the irreducible h-
v-- components (d--curvatures) of $\mathbf{R}_{\ \beta \gamma \delta
}^{\alpha }$,%
\begin{eqnarray}
R_{\ hjk}^{i} &=&e_{k}L_{\ hj}^{i}-e_{j}L_{\ hk}^{i}+L_{\ hj}^{m}L_{\
mk}^{i}-L_{\ hk}^{m}L_{\ mj}^{i}-C_{\ ha}^{i}\Omega _{\ kj}^{a},  \notag \\
R_{\ bjk}^{a} &=&e_{k}L_{\ bj}^{a}-e_{j}L_{\ bk}^{a}+L_{\ bj}^{c}L_{\
ck}^{a}-L_{\ bk}^{c}L_{\ cj}^{a}-C_{\ bc}^{a}\Omega _{\ kj}^{c},  \notag \\
R_{\ jka}^{i} &=&e_{a}L_{\ jk}^{i}-D_{k}C_{\ ja}^{i}+C_{\ jb}^{i}T_{\
ka}^{b},  \label{3dcurv} \\
R_{\ bka}^{c} &=&e_{a}L_{\ bk}^{c}-D_{k}C_{\ ba}^{c}+C_{\ bd}^{c}T_{\
ka}^{c},  \notag \\
R_{\ jbc}^{i} &=&e_{c}C_{\ jb}^{i}-e_{b}C_{\ jc}^{i}+C_{\ jb}^{h}C_{\
hc}^{i}-C_{\ jc}^{h}C_{\ hb}^{i},  \notag \\
R_{\ bcd}^{a} &=&e_{d}C_{\ bc}^{a}-e_{c}C_{\ bd}^{a}+C_{\ bc}^{e}C_{\
ed}^{a}-C_{\ bd}^{e}C_{\ ec}^{a}.  \notag
\end{eqnarray}
\end{theorem}

\begin{remark}
For an N--anholonomic manifold $\widetilde{\mathbf{V}}_{(n,n)}$ provided
with N--sym\-plet\-ic canonical d--connection $\widehat{\mathbf{\Gamma }}_{\
\alpha \beta }^{\tau },$ see (\ref{candcon1}), the d--curvatures (\ref{3dcurv}%
) reduces to three irreducible components
\begin{eqnarray}
R_{\ hjk}^{i} &=&e_{k}L_{\ hj}^{i}-e_{j}L_{\ hk}^{i}+L_{\ hj}^{m}L_{\
mk}^{i}-L_{\ hk}^{m}L_{\ mj}^{i}-C_{\ ha}^{i}\Omega _{\ kj}^{a},  \notag \\
R_{\ jka}^{i} &=&e_{a}L_{\ jk}^{i}-D_{k}C_{\ ja}^{i}+C_{\ jb}^{i}T_{\
ka}^{b},  \label{dcurv1} \\
R_{\ bcd}^{a} &=&e_{d}C_{\ bc}^{a}-e_{c}C_{\ bd}^{a}+C_{\ bc}^{e}C_{\
ed}^{a}-C_{\ bd}^{e}C_{\ ec}^{a}  \notag
\end{eqnarray}%
where all indices $i,j,k...$ and $a,b,..$ run the same values but label the
components with respect to different h-- or v--frames.
\end{remark}

Contracting respectively the components of (\ref{3dcurv}) and (\ref{dcurv1})
we prove:

\begin{corollary}
The Ricci d--tensor $\mathbf{R}_{\alpha \beta }\doteqdot \mathbf{R}_{\
\alpha \beta \tau }^{\tau }$ has the irreducible h- v--components%
\begin{equation}
R_{ij}\doteqdot R_{\ ijk}^{k},\ \ R_{ia}\doteqdot -R_{\ ika}^{k},\
R_{ai}\doteqdot R_{\ aib}^{b},\ R_{ab}\doteqdot R_{\ abc}^{c},
\label{dricci}
\end{equation}%
for a general N--holonomic manifold $\mathbf{V,}$ and
\begin{equation}
R_{ij}\doteqdot R_{\ ijk}^{k},\ \ R_{ia}\doteqdot -R_{\ ika}^{k},\ \
R_{ab}\doteqdot R_{\ abc}^{c},  \label{dricci1}
\end{equation}%
for an N--anholonomic manifold $\widetilde{\mathbf{V}}_{(n,n)}.$
\end{corollary}

\begin{corollary}
The scalar curvature of a d--connection is
\begin{equation} \label{sdccurv}
\overleftarrow{\mathbf{R}} \doteqdot \mathbf{g}^{\alpha \beta }\mathbf{R}%
_{\alpha \beta }=g^{ij}R_{ij}+h^{ab}R_{ab},
\end{equation}
defined by the "pure" h-- and v--components of (\ref{dricci1}).
\end{corollary}

\begin{corollary}
The Einstein d--tensor is computed $\mathbf{G}_{\alpha \beta }=\mathbf{R}%
_{\alpha \beta }-\frac{1}{2}\mathbf{g}_{\alpha \beta }\overleftarrow{\mathbf{%
R}}.$
\end{corollary}

For physical applications, the Riemann, Ricci and Einstein d--tensors
can be computed for the canonical d--connection. We can
redefine the constructions for arbitrary d--connections by using the
corresponding deformation tensors like in (\ref{2defcon}), for instance,
\begin{equation}
\mathcal{R}_{\ \beta }^{\alpha }=\widehat{\mathcal{R}}_{\ \beta }^{\alpha }+%
\mathbf{D}\mathcal{P}_{\ \beta }^{\alpha }+\mathcal{P}_{\ \gamma }^{\alpha
}\wedge \mathcal{P}_{\ \beta }^{\gamma }  \label{deformcurv}
\end{equation}%
for $\mathcal{P}_{\beta }^{\alpha }=\mathbf{P}_{\beta \gamma }^{\alpha }%
\mathbf{\vartheta }^{\gamma }.$
A set of examples of such deformations are analyzed in
Refs. \cite{15v1,15v2,15v3}.

\section{Noncommutative N--Anholonomic Spaces}

In this section, we outline how the analogs of basic objects in
commutative geometry of N--anholonomic manifolds, such as
vector/tangent bundles, N-- and d--connections can be defined in
noncommutative geometry \cite{15vncg,15vncgs}. We note that the
 A. Connes' functional analytic approach \cite{15connes1} to the
noncommutative topology and geometry is based on the theory of
noncommutative $C^{\ast }$--algebras. Any commutative $C^{\ast }$--algebra
can be realized as the $C^{\ast }$--algebra of complex valued functions over
locally compact Hausdorff space. A noncommutative $C^{\ast }$--algebra can
be thought of as the algebra of continuous functions on some 'noncommutative
space' (see main definitions and results in Refs.
 \cite{15connes1,15bondia,15landi,15madore}).

The starting idea of noncommutative geometry is to derive the
geometric properties of ``commutative'' spaces from their algebras of
functions  characterized by involutive algebras of operators by
dropping the condition of commutativity (see the Gelfand and Naimark
theorem \cite{15gelfand}). A space topology is defined by the algebra of
commutative continuous  functions, but the geometric constructions request a
differentiable structure. Usually, one considers a differentiable and
compact manifold $M,\ dim N=n$ (there is an open problem how to
include in noncommutative geometry spaces with indefinite metric signature like
pseudo--Euclidean and pseudo--Riemannian ones). In order to construct
models of commutative and noncommutative differential geometries it is
more or less obvious that the class of algebras of smooth functions,
 $\mathcal{C}\doteq C^{\infty}(M)$ is more appropriate. If $M$ is a
 smooth manifold, it is possible to reconstruct this manifold with its
 smooth structure and the attached objects (differential forms,
 etc...) by starting from  $\mathcal{C}$ considered as an abstract
 (commutative) unity $*$--algebra with involution. As a set $M$ can
 be identified with the set of characters of $\mathcal{C},$ but its
 differential structure is connected with the abundance of derivations
 of $\mathcal{C}$ which identify with the smooth vector fields on $M.$
There are two standard constructions: 1) when the vector fields are
 considered to be the derivations of  $\mathcal{C}$ (into itself) or
 2) one considers a generalization of the calculus of differential
 forms which is the Kahler differential calculus (see, details in
 Lectures \cite{15dubois}). The noncommutative versions of differential
 geometry may be elaborated if the algebra of smooth complex functions
 on a smooth manifold is replaced by a noncommutative associative
 unity complex $*$--algebra $\mathcal{A}.$

The geometry of commutative gauge and gravity theories is derived
from the notions of connections (linear and nonlinear ones),
 metrics and frames of references on manifolds and
vector bundle spaces. The possibility of extending such theories to some
noncommutative models is based on the Serre--Swan theorem \cite{15swan}
stating that there is a complete equivalence between the category of
(smooth) vector bundles over a smooth compact space (with bundle maps) and
the category of porjective modules of finite type over commutative algebras
and module morphisms. So, the space $\Gamma \left(
E\right) $ of smooth sections of a vector bundle $E$ over a compact space is
a projective module of finite type over the algebra $C\left( M\right) $ of
smooth functions over $M$ and any finite projective $C\left( M\right) $%
--module can be realized as the module of sections of some vector bundle
over $M.$ This construction may be extended if a noncommutative algebra $%
\mathcal{A}$ is taken as the starting ingredient:\ the noncommutative
analogue of vector bundles are projective modules of finite type over $%
\mathcal{A}$. This way one developed a theory of linear connections which
culminates in the definition of Yang--Mills type actions or, by some much
more general settings, one reproduced Lagrangians for the Standard model
with its Higgs sector or different type of gravity and Kaluza--Klein models
(see, for instance, Refs \cite{15connes1,15madore}).

\subsection{Modules as bundles}

A vector space $\mathcal{E}$ over the complex number field $\C$ can be
defined also as a right module of an algebra $\mathcal{A}$ over $\C$ \ which
carries a right representation of $\mathcal{A},$ when for every map of
elements $\mathcal{E}$ $\times \mathcal{A}\ni \left( \eta ,a\right)
\rightarrow \eta a\in \mathcal{E}$ one hold the properties
\begin{equation*}
\lambda (ab)=(\lambda a)b,~\lambda (a+b)=\lambda a+\lambda b,~(\lambda +\mu
)a=\lambda a+\mu a
\end{equation*}%
for every $\lambda ,\mu \in \mathcal{E}$ and $a,b\in \mathcal{A}.$

Having two $\mathcal{A}$--modules $\mathcal{E}$ and $\mathcal{F},$ a
morphism of $\mathcal{E}$ into $\mathcal{F}$ is \ any linear map $\rho :%
\mathcal{E}$ $\rightarrow $ $\mathcal{F}$  which is also $\mathcal{A}$%
--linear, i. e. $\rho (\eta a)=\rho (\eta )a$ for every $\eta \in \mathcal{E}
$ and $a\in \mathcal{A}.$

We can define in a similar (dual) manner the left modules and theirs
morphisms which are distinct from the right ones for noncommutative algebras
$\mathcal{A}.$ A bimodule over an algebra $\mathcal{A}$ is a vector space $%
\mathcal{E}$ which carries both a left and right module
structures. The bimodule structure is important for modelling of real
geometries starting from complex structures. We may
define the opposite algebra $\mathcal{A}^{o}$ with elements $a^{o}$ being in
bijective correspondence with the elements \ $a\in \mathcal{A}$ while the
multiplication is given by $\mathcal{\,}a^{o}b^{o}=\left( ba\right) ^{o}.$A
right (respectively, left) $\mathcal{A}$--module $\mathcal{E}$ is connected
to a left (respectively right) $\mathcal{A}^{o}$--module via relations $%
a^{o}\eta =\eta a^{o}$ (respectively, $a\eta =\eta a).$
One introduces the enveloping algebra $\mathcal{A}^{\varepsilon }=\mathcal{A}%
\otimes _{\C}\mathcal{A}^{o};$ any $\mathcal{A}$--bimodule $\mathcal{E}$ can
be regarded as a right [left] $\mathcal{A}^{\varepsilon }$--module by
setting $\eta \left( a\otimes b^{o}\right) =b\eta a$ $\quad \left[ \left(
a\otimes b^{o}\right) \eta =a\eta b\right] .$

For a (for instance, right) module $\mathcal{E}$ , we may introduce a family
of elements $\left( e_{t}\right) _{t\in T}$ parametrized by any (finite or
infinite) directed set $T$ for which any element $\eta \in \mathcal{E}$ is
expressed as a combination (in general, in \ more than one manner) $\eta
=\sum\nolimits_{t\in T}e_{t}a_{t}$ with $a_{t}\in \mathcal{A}$ and only a
finite number of non vanishing terms in the sum. A family $\left(
e_{t}\right) _{t\in T}$ is free if it consists from linearly independent
elements and defines a basis if any element $\eta \in \mathcal{E}$ can be
written as a unique combination (sum). One says a module to be free if it
admits a basis. The module $\mathcal{E}$ is said to be of finite type if \
it is finitely generated, i. e. it admits a generating family of finite
cardinality.

Let us consider the module $\mathcal{A}^{K}\doteqdot \C^{K}\otimes _{\C}%
\mathcal{A}.$ The elements of this module can be thought as $K$--dimensional
vectors with entries in $\mathcal{A}$ and written uniquely as a linear
combination $\eta =\sum\nolimits_{t=1}^{K}e_{t}a_{t}$ were the basis $e_{t}$
identified with the canonical basis of $\C^{K}.$ This is a free and finite
type module. In general, we can have bases of different cardinality.
However, if a module $\mathcal{E}$ \ is of finite type there is always an
integer $K$ and a module surjection $\rho :\mathcal{A}^{K}\rightarrow
\mathcal{E}$ with a base being a image of a free basis, $\epsilon _{j}=\rho
(e_{j});j=1,2,...,K.$

We say that a right $\mathcal{A}$--module $\mathcal{E}$ is projective if for
every surjective module morphism $\rho :\mathcal{M}$ $\rightarrow $ $%
\mathcal{N}$ splits, i. e. there exists a module morphism \ $s:\mathcal{E}$ $%
\rightarrow $ $\mathcal{M}$ such that $\rho \circ s=id_{\mathcal{E}}.$ There
are different definitions of porjective modules (see Ref. \cite{15landi} on
properties of such modules). Here we note the property that if a $\mathcal{A}
$--module $\mathcal{E}$ is projective, there exists a free module $\mathcal{F%
}$ and a module $\mathcal{E}^{\prime }$ (being a priory projective) such
that $\mathcal{F}=\mathcal{E}\oplus \mathcal{E}^{\prime }.$

For the right $\mathcal{A}$--module $\mathcal{E}$ being projective and of
finite type with surjection $\rho :\mathcal{A}^{K}\rightarrow \mathcal{E}$
and following the projective property we can find a lift $\widetilde{\lambda
}:\mathcal{E}$ $\rightarrow $ $\mathcal{A}^{K}$ such that $\rho \circ
\widetilde{\lambda }=id_{\mathcal{E}}.$ There is a proof of the property
that the module $\mathcal{E}$ is projective of finite type over $\mathcal{A}$
if and only if there exists an idempotent $p\in End_{\mathcal{A}}\mathcal{A}%
^{K}=M_{K}(\mathcal{A}),$ $p^{2}=p,$ the $M_{K}(\mathcal{A})$ denoting the
algebra of $K\times K$ matrices with entry in $\mathcal{A},$ such that $%
\mathcal{E}=p\mathcal{A}^{K}.$ We may associate the elements of $\mathcal{E}$
to $K$--dimensional column vectors whose elements are in $\mathcal{A},$ the
collection \ of which are invariant under the map $p,$ $\mathcal{E}$ $=\{\xi
=(\xi _{1},...,\xi _{K});\xi _{j}\in \mathcal{A},~p\xi =\xi \}.$ For
simplicity, we shall use the term finite projective to mean projective of
finite type.

\subsection{Nonlinear connections in projective modules}

The nonlinear connection (N--connection) for noncommutative spaces
can be defined similarly to commutative spaces by considering instead of
usual vector bundles theirs noncommutative analogs defined as finite
projective modules over noncommutative algebras \cite{15vncg}.
 The explicit constructions depend on the type of differential
 calculus we use for definition of tangent structures and theirs
 maps. In this subsection, we shall consider such projective modules
  provided with N--connection which define noncommutative analogous
  both of vector bundles and of N--anholonomic manifolds (see
  Definition \ref{dnam}).

In general, one can be defined several differential calculi over a
given algebra $\mathcal{A}$
 (for a more detailed discussion within the context of
noncommutative geometry, see Refs. \cite{15connes1,15madore}).\ For
simplicity, in this work we consider that a differential calculus on
$\mathcal{A}$ is fixed,  which means that we choose a (graded) algebra
 $\Omega ^{\ast }(\mathcal{A})=\cup _{p}\Omega ^{p}(\mathcal{A})$
  giving a differential structure to $\mathcal{A}.$ The elements
 of  $\Omega ^{p}(\mathcal{A})$ are called $p$--forms. There is a linear map $d$
which takes $p$--forms into $(p+1)$--forms and which satisfies a graded
Leibniz rule as well the condition $d^{2}=0.$
By definition $\Omega ^{0}(\mathcal{A})=\mathcal{A}.$

The differential $df$ of a real or complex variable on a
N--anholonomic manifold  $\mathbf{V}$
\begin{eqnarray*}
df &=&\delta _{i}f~dx^{i}+\partial _{a}f~\delta y^{a}, \\
\delta _{i}f~ &=&\partial _{i}f-N_{i}^{a}\partial _{a}f~,~\delta
y^{a}=dy^{a}+N_{i}^{a}dx^{i},
\end{eqnarray*}%
where the N--elongated derivatives and differentials are defined
respectively  by formulas (\ref{dder}) and (\ref{ddif}),
in the noncommutative case is replaced by a distinguished commutator
(d--commutator)%
\begin{equation*}
\overline{d}f=\left[ F,f\right] =\left[ F^{[h]},f\right] +\left[ F^{[v]},f%
\right]
\end{equation*}%
where the operator $F^{[h]}$ $\ (F^{[v]})$ acts on the horizontal
(vertical) projective submodule and this operator is  defined by a
fixed differential calculus $\Omega ^{\ast }(\mathcal{A}^{[h]})$
($\Omega ^{\ast }(\mathcal{A}^{[v]}))$ on the so--called horizontal
(vertical)  $\mathcal{A}^{[h]}$ ($\mathcal{A}^{[v]})$ algebras. We
conclude that in order to elaborated noncommutative versions of
N--anholonomic manifolds we need couples of 'horizontal' and 'vertical'
operators which reflects the nonholonomic splitting given by the
N--connection structure.

Let us consider instead of a N--anholonomic manifold $\mathbf{V}$
 an $\mathcal{A}$--module $\mathcal{E}$ being projective and of finite
 type.  For a fixed differential calculus on $\mathcal{E}$ we define
 the  tangent structures $T\mathcal{E}.$
\begin{definition} \label{dncamnc}
A nonlinear connection (N--connection) $\mathbf{N}$ on  an $\mathcal{A}$%
--module $\mathcal{E}$  is defined by the splitting on the left of an
 exact sequence of finite projective $\mathcal{A}$--moduli
\begin{equation*}
0\rightarrow v\mathcal{E}\overset{i}{\rightarrow} T\mathcal{E}\rightarrow T%
\mathcal{E}/v\mathcal{E}\rightarrow 0,
\end{equation*}%
i. e. by a morphism of submanifolds $\mathbf{N:\ }T\mathcal{E}\rightarrow v%
\mathcal{E}$ such that $\mathbf{N\circ i}$ is the unity in $v\mathcal{E}.$
\end{definition}
In an equivalent form, we can say that a N--connection is defined by a
splitting to projective submodules with a Whitney sum of conventional
 h--submodule, $\left( h\mathcal{E}\right) ,$ and v--submodule,
 $\left( v\mathcal{E}\right) ,$
\begin{equation}
T\mathcal{E}=h\mathcal{E}\oplus v\mathcal{E}.  \label{whitneync}
\end{equation}%
We note that locally $h\mathcal{E}$ is isomorphic to $TM$ where $M$ is a
 differential compact manifold of dimension $n.$

The Definition \ref{dncamnc} reconsiders  for noncommutative spaces the
Definition \ref{dncam}. In result, we may generalize the concept of
'commutative' N--anholonomic space:

\begin{definition}  A N--anholonomic noncommutative space $\mathcal{E}_N$ is an
 $\mathcal{A}$--mo\-du\-le $\mathcal{E}$  possessing a  tangent structure
 $T\mathcal{E}$   defined  by a Whitney sum of projective submodules
 (\ref{whitneync}).
\end{definition}
Such geometric  constructions  depend  on the type of fixed differential
calculus, i. e. on the procedure how the tangent spaces are
defined.

\begin{remark}
Locally always N--connections exist, but it is not obvious if they
could be glued together. In the classical case of vector bundles over
paracompact manifolds this is possible \cite{15ma1}. If there is an
appropriate partition of unity, a similar result can be proved for
finite projective modules.
 For certain applications,  it is more convenient to use the
Dirac operator already defined on N--anholonomic manifolds, see
Section \ref{sdonst}.
\end{remark}

In order to understand how the N--connection structure may be taken
 into account on noncommutative spaces but distinguished from the
 class of  linear  gauge fields,  we analyze   an example:

\subsection{Commutative and noncommutative gauge d--fields}

Let us consider a N--anholonomic manifold $\mathbf{V}$  and a
 vector bundle $\beta =\left( B,\pi ,\mathbf{V}\right) $ with
 $\pi :B\rightarrow \mathbf{V}$
with a typical $k$-dimensional vector fiber. In local coordinates a linear
connection (a gauge field) in $\beta $ is given by a collection of
differential operators%
\begin{equation*}
\nabla _{\alpha }=D_{\alpha }+B_{\alpha }(u),
\end{equation*}%
acting on $T\xi _{N}$ where
\begin{equation*}
D_{\alpha }=\delta _{\alpha }\pm \Gamma _{\cdot \alpha }^{\cdot }\mbox{ with
}D_{i}=\delta _{i}\pm \Gamma _{\cdot i}^{\cdot }\mbox{ and }D_{a}=\partial
_{a}\pm \Gamma _{\cdot a}^{\cdot }
\end{equation*}
is a d--connection in $\mathbf{V}$  ($\alpha =1,2,...,n+m),$ with the
operator  $\delta _{\alpha },$
 being  N--elongated as in (\ref{dder}), $u=(x,y)\in \xi _{N}$ and $%
B_{\alpha }$ are $k\times k$--matrix valued functions. For every vector
field
\begin{equation*}
X=X^{\alpha }(u)\delta _{\alpha }=X^{i}(u)\delta _{i}+X^{a}(u)\partial
_{a}\in T\mathbf{V}
\end{equation*}
we can consider the operator
\begin{equation}
X^{\alpha }(u)\nabla _{\alpha }(f\cdot s)=f\cdot \nabla
_{X}s+\delta _{X}f\cdot s  \label{2rul1c}
\end{equation}%
for any section $s\in \mathcal{B}$ \ and function
 $f\in C^{\infty }(\mathbf{V}),$ where%
\begin{equation*}
\delta _{X}f=X^{\alpha }\delta _{\alpha }~\mbox{ and
}\nabla _{fX}=f\nabla _{X}.
\end{equation*}%
In the simplest definition we assume that there is a Lie algebra $\mathcal{GL%
}B$ that acts on associative algebra $B$ by means of infinitesimal
automorphisms (derivations). This means that we have linear operators $%
\delta _{X}:B\rightarrow B$ which linearly depend on $X$ and satisfy%
\begin{equation*}
\delta _{X}(a\cdot b)=(\delta _{X}a)\cdot b+a\cdot (\delta _{X}b)
\end{equation*}%
for any $a,b\in B.$ The mapping $X\rightarrow \delta _{X}$ is a Lie algebra
homomorphism, i. e. $\delta _{\lbrack X,Y]}=[\delta _{X},\delta _{Y}].$

Now we consider respectively instead of commutative spaces
 $\mathbf{V}$  and $\beta $
the finite \ projective $\mathcal{A}$--module $\mathcal{E}_{N},$ provided
with N--connection structure, and the finite projective $\mathcal{B}$%
--module $\mathcal{E}_{\beta }.$

A d--connection $\nabla _{X}$ on $\mathcal{E}_{\beta }$ is
by definition a set of linear d--operators, adapted to the N--connection
structure, depending linearly on $X$ and satisfying the Leibniz rule%
\begin{equation}
\nabla _{X}(b\cdot e)=b\cdot \nabla  _{X}(e)+\delta
_{X}b\cdot e  \label{2rul1n}
\end{equation}%
for any $e\in \mathcal{E}_{\beta }$ and $b\in \mathcal{B}.$ The rule
(\ref{2rul1n}) is a noncommutative generalization of (\ref{2rul1c}). We emphasize
that both operators $\nabla _{X}$ and $\delta _{X}$ are
distinguished by the N--connection structure and that the difference of two
such linear d--operators, $\nabla _{X}-\nabla _{X}^{\prime }$
commutes with action of $B$ on $\mathcal{E}_{\beta },$ which
is an endomorphism of $\mathcal{E}_{\beta }.$ Hence, if we fix some fiducial
connection $\nabla _{X}^{\prime }$ (for instance, $%
\nabla _{X}^{\prime }=D_{X})$ on $\mathcal{E}_{\beta }$ an
arbitrary connection has the form
\begin{equation*}
\nabla _{X}=D_{X}+B_{X},
\end{equation*}%
where $B_{X}\in End_{B}\mathcal{E}_{\beta }$ depend linearly on $X.$

The curvature of connection $\nabla _{X}$ is a two--form $F_{XY}$
which values linear operator in $\mathcal{B}$ and measures a deviation of
mapping $X\rightarrow \nabla _{X}$ from being a Lie algebra
homomorphism,%
\begin{equation*}
F_{XY}=[\nabla _{X},\nabla _{Y}]-\nabla _{\lbrack X,Y]}.
\end{equation*}%
The usual curvature d--tensor is defined as
\begin{equation*}
F_{\alpha \beta }=\left[ \nabla _{\alpha },\nabla
_{\beta }\right] -\nabla _{\lbrack \alpha ,\beta ]}.
\end{equation*}

The simplest connection on a finite projective $\mathcal{B}$--module $%
\mathcal{E}_{\beta }$ is to be specified by a projector $P:\mathcal{B}%
^{k}\otimes \mathcal{B}^{k}$ when the d--operator $\delta _{X}$ acts
naturally on the free module $\mathcal{B}^{k}.$ The operator $%
\nabla _{X}^{LC}=P\cdot \delta _{X}\cdot P$ $\ $\ is called the
Levi--Civita operator and satisfy the condition
 $Tr[\nabla _{X}^{LC},\phi ]=0$ for any endomorphism
 $\phi \in End_{B}\mathcal{E}_{\beta}.$
From this identity, and from the fact that any two connections differ by
an endomorphism that $Tr[\nabla _{X},\phi ]=0$
for an arbitrary connection $\nabla _{X}$ and an arbitrary
endomorphism $\phi ,$ that instead of $\nabla _{X}^{LC}$ we may
consider equivalently the canonical d--connection, constructed only from
d-metric and N--connection coefficients.

\section{Nonholonomic Clifford--Lagrange Structu\-res}

The geometry of spinors on generalized Lagrange and Finsler spaces was
elaborated in Refs. \cite{15vsp1,15vsph,15vst,15vsp2}. It was applied for
definition of noncommutative extensions of the Finsler geometry
related to certain models of Einstein, gauge and string gravity
\cite{15vstr,15vncg,15vncgs,15vex1,15vt,15vp}. Recently, it is was proposed an
extended Clifford approach to relativity, strings and noncommutativity
based on the concept of "C--space'' \cite{15castro1,15castro2,15castro3,15castro4}.

 The aim of this section is to
 formulate the geometry of nonholonomic Clifford--Lagrange structures
 in a form adapted to generalizations for noncommutative spaces.

\subsection{Clifford d--module} \label{sscldm}
Let  $\mathbf{V}$ be a compact N--anholonomic manifold. We denote,
respectively, by  $T_x\mathbf{V}$ and $T^*_x\mathbf{V}$  the tangent
and cotangent spaces in a point $x\in\mathbf{V}.$ We  consider a
 complex vector bundle $\tau: E \rightarrow \mathbf{V}$
where, in general, both the base $\mathbf{V}$ and the total space $E$ may be
provided with N--connection structure, and denote by
$\Gamma ^{\infty}(E)$ (respectively, $\Gamma(E))$ the set of
differentiable (continuous) sections of $E.$ The symbols
$\chi (\mathbf{M})= \Gamma ^{\infty}(\mathbf{TM})$ and
 $\Omega ^1 (\mathbf{M}) \doteq \Gamma ^{\infty}\mathbf{(T^*M})$ are used
 respectively for the set of d--vectors and one d--forms on
$\mathbf{TM}.$

\subsubsection{Clifford--Lagrange  functionals}
In the simplest case, a generic nonholonomic Clifford structure
 can be associated  to a Lagrange metric on a $n$--dimensional real
  vector space $V^n$  provided with a Lagrange quadratic form
 $L(y)=q_L(y,y),$ see subsection \ref{sslfm}. We consider the exterior
 algebra $\wedge V^n$ defined by the identity element $\I$ and
 antisymmetric products $v_{[1]}\wedge ... \wedge v_{[k]}$ with
 $v_{[1]}, ...,  v_{[k]} \in V^n$ for $k \leq dim V^n$ where
$\I \wedge v = v,$\ $v_{[1]}\wedge v_{[2]}=-v_{[2]}\wedge v_{[1]}, ...$
\begin{definition}\label{dclalg} The Clifford--Lagrange (or Clifford--Minkowski)
 algebra is a $\wedge V^n$ algebra provided with a product
 \begin{equation}
 uv+vu = 2 ^{(L)}g(u,v)\ \I \label{clpl}
 \end{equation}
\begin{equation}
 \mbox{(or }uv+vu = 2 ^{(F)}g(u,v)\ \I \ ) \label{clpm}
\end{equation}
for any $u,v \in V^n$ and $\  ^{(L)}g(u,v)$ (or $\  ^{(F)}g(u,v))$
defined by formulas (\ref{lagm}) (or(\ref{finm})).
\end{definition}

For simplicity, hereafter we shall prefer to  write down  the formulas
for the Lagrange configurations instead of   dubbing of similar
formulas  for the Finsler configurations.

We can introduce the complex Clifford--Lagrange algebra $\C
l_{(L)}(V^n)$ structure by using the complex unity ``i'',
 $V_{\C} \doteq V^n + i V^n,$ enabled with complex metric
\begin{equation*}
   ^{(L)}g_{\C}(u,v+iw) \doteq \    ^{(L)}g(u,v) + i\    ^{(L)} g(u,w),
\end{equation*}
which results in certain isomorphisms of matrix algebras (see, for
instance, \cite{15bondia}),
\begin{eqnarray*}
\C l (\R ^{2m} )&\simeq & M_{2^m}(\C ), \\
 \C l (\R ^{2m+1}) &\simeq & M_{2^m}(\C ) \oplus  M_{2^m}(\C ) .
\end{eqnarray*}
We omitted the label $(L)$ because such isomorphisms hold true for any
quadratic forms.

The Clifford--Lagrange algebra possesses usual properties:
\begin{enumerate}
\item On $\C l_{(L)}(V^n)$, it is linearly defined the involution "*",
\begin{equation*} {(\lambda v_{[1]} ...  v_{[k]})}^{*} =
 \overline{\lambda} v_{[1]} ...  v_{[k]},\ \forall \lambda \in \C .
\end{equation*}
\item There is a $\Z _2$ graduation,
\begin{equation*}
\C l_{(L)}(V^n)= \C l_{(L)}^+(V^n) \oplus \C l_{(L)}^{-}(V^n)
\end{equation*}
with $\chi _{(L)}(a) = \pm 1$ for $a \in  \C l_{(L)}^{\pm}(V^n),$ where
$\C l_{(L)}^+(V^n),$ or  $\C l_{(L)}^{-}(V^n),$ are defined by
products of an odd, or even, number of vectors.
\item For positive definite forms $q_L(u,v)$, one defines the
  chirality of\\  $\C l_{(L)}(V^n),$
\begin{equation*}
\gamma _{(L)}= (-i)^n´ e_1 e_2... e_n,\ \gamma ^2 =\gamma ^* \gamma = \I
\end{equation*}
where $\lbrace e_i \rbrace ^n_{i=1}$ is an orthonormal basis of $V^n$
and $n=2n',$ or $=2n'+1.$
\end{enumerate}

In a more general case, a  nonholonomic Clifford structure
  is defined by quadratic d--metric form
  $\mathbf{q}(\mathbf{x},\mathbf{y})$ (\ref{qdmf}) on
 a $n+m$--dimensional real d--vector space $V^{n+m}$  with the
  $(n+m)$--splitting defined by the N--connection structure.
\begin{definition} \label{dcdalg} The Clifford d--algebra  is a
 $\wedge V^{n+m}$ algebra provided with a product
 \begin{equation}
 \mathbf{u}\mathbf{v} + \mathbf{v} \mathbf{u} =
 2 \mathbf{g} (\mathbf{u}, \mathbf{v})\ \I \label{cdalg}
 \end{equation}
or, equivalently, distinguished into h-- and v--products
\begin{equation*}
 uv+vu = 2 g(u,v)\ \I \
\end{equation*}
and
\begin{equation*}
 \ ^\star u\ ^\star v + \ ^\star v \ ^\star u
= 2\ ^\star h(\ ^\star u,\ ^\star v)\ \I \
\end{equation*}
for any $\mathbf{u}=(u,\ ^\star u),\ \mathbf{v} = (v,\ ^\star v) \in
 V^{n+m}.$
\end{definition}
Such Clifford d--algebras have similar properties on the
h-- and v--compo\-nents  as the
Clifford--Lagrange algebras. We may define a standard complexification but it
should be emphasized that for $n=m$ the N--connection (in
particular, the canonical Lagrange N--connection) induces naturally
an almost complex structure (\ref{acs1}) which gives the possibility
to define almost complex Clifford d--algebras
 (see details in \cite{15vsp1,15vsp2}).

\subsubsection{Clifford--Lagrange  and Clifford  N--anholonomic
  structures}

A metric  on a manifold $M$ is defined by sections of the
tangent bundle $TM$ provided with a bilinear symmetric form on
continuous sections $\Gamma (TM).$  In Lagrange geometry, the metric
structure is of type $ ^{(L)}g_{ij}(x,y)$ (\ref{lagm}) which allows us to
define Clifford--Lagrange algebras $\C l_{(L)}(T_x M),$ in any point
$x\in TM,$
\begin{equation*}
\gamma _i \gamma _j + \gamma _j \gamma _i = 2\  ^{(L)}g_{ij}\ \I.
\end{equation*}
For any point $x\in M$ and fixed $y=y_0,$
 one exists a standard complexification,
 $T_x M^{\C} \doteq T_x M + i T_x M,$ which can be used for definition
 of the 'involution' operator on sections of $T_x M^{\C},$
\begin{equation*}
 \sigma _1 \sigma _2 (x) \doteq \sigma _2 (x) \sigma _1 (x), \
 \sigma ^* (x) \doteq \sigma (x) ^*, \forall x \in M,
\end{equation*}
where  "*" denotes the involution on every $\C l_{(L)}(T_x M).$ The
norm is defined by using the Lagrange norm, see Definition
\ref{dlagf},
\begin{equation*}
 \parallel \sigma \parallel _L \doteq sup_{x\in M}\ \lbrace
\mid \sigma (x) \mid _L \rbrace,
\end{equation*}
which defines a $C^*_{L}$--algebra instead of the usual $C^*$--algebra
 of $\C l(T_x M).$ Such constructions can be also performed on the
 cotangent space $T_x M,$  or for any  vector bundle $E$ on $M$
 enabled with a symmetric bilinear form of class $C^\infty$ on
 $\Gamma ^\infty (E) \times \Gamma ^\infty (E).$

For Lagrange spaces modelled on $\widetilde{TM},$ there is a natural
almost complex structure $\mathbf{F}$\ (\ref{acs1}) induced by the
canonical N--connection $  ^{(L)}N,$  see the Results \ref{r2}, \ref{r4} and
\ref{r5}, which allows also to construct an almost Kahler model of
Lagrange geometry, see details in Refs. \cite{15ma1,15ma2}, and to define
an Clifford--Kahler d--algebra $\C l_{(KL)}(T_x M)$ \cite{15vsp1},
for $y=y_0,$ being provided with the norm
\begin{equation*}
\parallel \sigma \parallel _{KL} \doteq sup_{x\in M}\ \lbrace
\mid \sigma (x) \mid _{KL} \rbrace,
\end{equation*}
which on $T_x M$ is defined by projecting on $x$ the d--metric
 $\ ^{(L)}\mathbf{g}$ (\ref{slm}).

In order to model Clifford--Lagrange structures on $\widetilde{TM}$
and $\widetilde{T^*M}$ it is necessary to consider d--metrics
induced by Lagrangians:

\begin{definition} A Clifford--Lagrange space on a manifold $M$
  enabled with  a fundamental metric $\ ^{(L)}g_{ij}(x,y)$
 (\ref{lqf})  and canonical  N--connection $\ ^{(L)}N_{j}^{i}$
 (\ref{cncl}) inducing a Sasaki type d--metric $\ ^{(L)}\mathbf{g}$
(\ref{slm}) is  defined as a Clifford bundle
$\C l_{(L)}(M)\doteq \C l_{(L)}(T^*M).$
\end{definition}

For a general N--anholonomic manifold $\mathbf{V}$ of dimension $n+m$
provided with a general  d--metric structure $\mathbf{g}$ (\ref{metr})
 (for instance, in a gravitational model, or constructed by conformal
 transforms and  imbedding into higher dimensions of a Lagrange (or Finsler)
 d--metrics), we introduce
\begin{definition} A Clifford N--anholonomic bundle on $\mathbf{V}$ is
  defined as\\ $\C l_{(N)}(\mathbf{V})\doteq \C l_{(N)}(T^*\mathbf{V}).$
\end{definition}

Let us consider a complex vector bundle $\pi : \mathbf{E} \to M$
provided with N--connection structure which can be defined by a
corresponding exact chain of subbundles, or nonintegrable
distributions, like for real vector bundles, see \cite{15ma1,15ma2} and
 subsection \ref{snc}.   Denoting by
$V^m_{\C}$ the typical fiber (a complex vector space), we can define
the usual Clifford map
\begin{equation*}
c:\   \C l (T^*M) \rightarrow End(V^m_{\C})
\end{equation*}
via (by convention, left) action on sections
 $c(\sigma) \sigma ^1(x) \doteq c(\sigma (x)) \sigma ^1 (x).$
\begin{definition} \label{defcdma} The Clifford d--module (distinguished by a
  N--connection)  of a N--anholonomic vector bundle $\mathbf{E}$ is
  defined by the $C(M)$--module $\Gamma (\mathbf{E})$ of continuous
  sections in $\mathbf{E},$
 \begin{equation*}
 c:\ \Gamma(\C l (M)) \rightarrow End(\Gamma (\mathbf{E})).
 \end{equation*}
\end{definition}

In an alternative case,  one considers a complex
 vector  bundle $\pi:\ E \to  \mathbf{V}$   on an
N--anholonomic space $\mathbf{V}$ when the N--connection structure is
 given for the base manifold.
\begin{definition} \label{defcdmb} The Clifford d--module
  of a  vector bundle ${E}$ is
  defined by the $C(\mathbf{V})$--module $\Gamma ({E})$ of continuous
  sections in ${E},$
 \begin{equation*}
 c:\ \Gamma(\C l_{(N)} (\mathbf{V})) \rightarrow End(\Gamma ({E})).
 \end{equation*}
\end{definition}

A Clifford d--module with both N--anholonomic total space
$\mathbf{E}$ and base space $\mathbf{V}$ with corresponding
N--connections (in general, two independent ones, but the
N--connection in the distinguished complex vector bundle must be
adapted to the N--connection on the base) has to be defined by an
"interference" of Definitions \ref{defcdma} and \ref{defcdmb}.

\subsection{N--anholonomic spin structures}
Usually, the spinor bundle on a manifold $M,\ dim M=n,$ is constructed
on the tangent bundle by substituting the group $SO(n)$ by its
universal covering $Spin(n).$ If a Lagrange fundamental quadratic form
$\ ^{(L)}g_{ij}(x,y)$ (\ref{lqf}) is defined on $T_x,M$  we can
consider
Lagrange--spinor spaces in every point $x \in M.$  The constructions
can be completed on $\widetilde{TM}$ by using the Sasaki type metric
$\ ^{(L)}\mathbf{g}$ (\ref{slm}) being similar for any type of
N--connection and d--metric structure on $TM.$ On general
N--anholonomic manifolds $\mathbf{V}, dim \mathbf{V}=n+m,$  the
distinguished  spinor space (in brief, d--spinor
space) is to be  derived from the d--metric (\ref{metr}) and adapted
to the N--connection structure. In this case, the group $SO(n+m)$ is
not only substituted by $Spin(n+m)$ but with respect to N--adapted
frames (\ref{dder}) and (\ref{ddif}) one defines irreducible
decompositions to $Spin(n)\oplus Spin(m).$

\subsubsection{Lagrange spin groups}

Let us consider a vector space $V^n$ provided with Clifford--Lagrange
structures as in subsection \ref{sscldm}. We denote a such space as
$V^n_{(L)}$ in order to emphasize that  its tangent space
 is provided with a Lagrange
type quadratic form $\ ^{(L)}g.$ In a similar form, we shall write
 $\C l_{(L)} (V^n)  \equiv \C l (V^n_{(L)})$ if this will be more
 convenient.  A vector $u \in V^n_{(L)}$ has a unity length
 on the Lagrange quadratic form if $\ ^{(L)}g(u,u) = 1,$ or
 $u^2=\I,$ as an element of corresponding Clifford algebra,
 which follows from (\ref{clpl}). We define an endomorphism of $V^n:$
\begin{equation*}
\phi _{(L)}^{u}\doteq \chi _{(L)} (u) v u^{-1}=-uvu = \left( uv - 2\
^{(L)}g(u,v)\right) u = u - 2\ ^{(L)}g(u,v)u
\end{equation*}
where $\chi _{(L)}$ is the $\Z _2$ graduation defined by $\ ^{(L)}g.$
 By multiplication,
\begin{equation*}
\phi _{(L)}^{u_1 u_2}(v)\doteq u_2^{-1}  u_1^{-1} v u_1 u_2 =
 \phi _{(L)}^{u_2} \circ \phi _{(L)}^{u_1} (v),
\end{equation*}
which defines the subgroup $SO(V^n_{(L)}) \subset O(V^n_{(L)}).$ Now
we can define \cite{15vsp1,15vsp2}
\begin{definition} The space of complex Lagrange spins is defined by the
 subgroup   $Spin^c_{(L)}(n)\equiv Spin^c (V^n_{(L)})
 \subset \C l (V^n_{(L)}),$
 determined  by the products of pairs of
  vectors $w \in V_{(L)}^{\C}$ when $w\doteq \lambda u$ where
  $\lambda$ is a complex number of module 1 and $u$ is of unity length
  in $V^n_{(L)}.$
\end{definition}

We note that $ker \phi _{(L)} \cong U(1).$ We can define a
homomorphism $\nu _{(L)}$ with values in $U(1),$
\begin{equation*}
\nu _{(L)}(w)= w_{2k}...w_1 w_1...w_{2k} = \lambda _1 ... \lambda _{2k},
\end{equation*}
where $w =  w_1...w_{2k} \in Spin^c (V^n_{(L)})$
and $\lambda _i = w_i^2 \in U(1).$
\begin{definition}
The group of real Lagrange spins  $Spin^c_{(L)}(n) \equiv Spin
(V^n_{(L)})$ is defined by  $ker\  \nu _{(L)}.$
\end{definition}
The complex conjugation on $ \C l (V^n_{(L)})$  is usually defined as
$\overline{\lambda v} \doteq \overline{\lambda} v$ for $\lambda \in
\C,\ v\in V^n_{(L)}.$ So, any element $w \in Spin (V^n_{(L)})$
satisfies the conditions $\overline{w^*}w = w^* w = \I$  and
$\overline{w}=w.$ If we take $V^n_{(L)}=\R ^n$ provided with a (pseudo)
Euclidean quadratic form instead of the Lagrange norm, we obtain the
usual spin--group constructions from the (pseudo) Euclidean geometry.

\subsubsection{Lagrange spinors and d--spinors:\ Main Result 1}

A usual spinor is a section of a vector bundle $S$ on a manifold $M$
when  an irreducible representation of the group $Spin (M)
\doteq Spin (T^*_x M)$ is defined on the typical fiber. The set of
sections $\Gamma (S)$ is a irreducible Clifford module. If the base
manifold of type $M_{(L)},$ or  is a general N--anholonomic manifold
$\mathbf{V},$ we have to define the spinors on such spaces as to be
adapted to the respective N--connection structure.

In the case when the base space is of even dimension (the geometric
constructions in in this subsection will be considered for even
dimensions both for the base and typical fiber spaces), one should consider the
so--called Morita equivalence (see details in \cite{15bondia,15mart2}
 for a such equivalence between $C(M)$ and $\Gamma (\C l(M))$). One
 says that two algebras $\mathcal{A}$ and $\mathcal{B}$ are
 Morita--equivalent if
\begin{equation*}
\mathcal{E} \otimes _\mathcal{A} \mathcal{F} \simeq \mathcal{B}
 \mbox{ and }
\mathcal{F} \otimes _\mathcal{B} \mathcal{F} \simeq \mathcal{A},
\end{equation*}
respectively, for  $\mathcal{B}$-- and $\mathcal{A}$--bimodules and
 $\mathcal{B-A}$--bimodule $\mathcal{E}$ and
 $\mathcal{A-B}$--bimodule $\mathcal{F}.$
If we study algebras through theirs representations, we also have to
 consider various algebras related by the Morita equivalence.
\begin{definition}
  A Lagrange spinor  bundle $S_{(L)}$ on a manifold $M,\ dim
 M=n,$ is a complex vector bundle with both  defined action of the spin group
$Spin (V^n_{(L)})$ on the typical fiber  and an irreducible
 reprezentation of the group
$ Spin_{(L)} (M) \equiv Spin (M_{(L)}) \doteq Spin (T_x^*
 M_{(L)}).$ The set of sections $\Gamma (S_{(L)})$ defines an irreducible
 Clifford--Lagrange module.
\end{definition}

The so--called "d--spinors" have been introduced for the spaces
provided with N--connection structure \cite{15vsp1,15vsph,15vstr1}:
\begin{definition} \label{ddsp}
 A distinguished spinor (d--spinor) bundle $\mathbf{S}\doteq (S,\
  ^{\star} S)$ on an
  N--anho\-lo\-nom\-ic manifold $\mathbf{V},$ $\ dim \mathbf{V}=n+m,$ is
  a complex vector bundle with a defined action of the spin
  d--group  $Spin\ \mathbf{V} \doteq Spin(V^n) \oplus  Spin(V^m) $ with
  the splitting adapted to the N--connection structure which results
  in an irreducible  representation
 $Spin (\mathbf{V}) \doteq Spin (T^*\mathbf{V}).$ The set of sections
 $\Gamma (\mathbf{S}) = \Gamma\ ({S}) \oplus
 \Gamma (\ ^{\star} {S}) $ is an irreducible  Clifford d--module.
\end{definition}

The fact that $C(\mathbf{V})$ and $\Gamma
(\C l (\mathbf{V}))$ are Morita equivalent can be analyzed by applying
in N--adapted form, both on the base and fiber spaces, the
consequences of the Plymen's theorem (see Theorem 9.3 in
Ref. \cite{15bondia}). This is connected with the possibility to
distinguish the $Spin (n)$ (or, correspondingly $Spin (M_{(L)}),$\
 $Spin(V^n) \oplus  Spin(V^m) )$ an antilinear bijection $J:\ S\ \to \ S$
 (or $J:\ S_{(L)}\ \to \ S_{(L)}$ and $J:\ \mathbf{S}\ \to \
 \mathbf{S})$ with the properties:
\begin{eqnarray}
J(\psi f) &=& (J\psi )f \mbox{ for } f \in  C(M) (\mbox{ or }
C(M_{(L)}),\  C(\mathbf{V})); \nonumber \\
J(a\psi ) &=& \chi (a)J\psi, \mbox{ for } a\in \Gamma ^\infty (\C
l(M))  (\mbox{ or } \Gamma ^\infty (\C l(M_{(L)})),\
 \Gamma ^\infty (\C l(\mathbf{V}));   \nonumber \\
(J\phi | J\psi ) &=& (\psi | \phi) \mbox{ for } \phi , \psi \in S
(\mbox{ or } S_{(L)}, \mathbf{S} ).
 \label{jeq}
\end{eqnarray}

\begin{definition}
The spin structure on a manifold $M$ (respectively, on
 $M_{(L)},$ or on N--anholonomic manifold $\mathbf{V}$) with even dimensions
for the corresponding base and typical fiber spaces is defined by a
bimodule $S$ (respectively, $M_{(L)},$ or $\mathbf{V}$) obeying the
Morita equivalence $C(M)- \Gamma (\C l(M))$ (respectively,
 $C(M_{(L)})- \Gamma (\C l(M_{(L)})),$ or
$C(\mathbf{V})- \Gamma (\C l(\mathbf{V})))$   by a corresponding
bijections (\ref{jeq}) and a fixed orientation on $M$ (respectively,
on $M_{(L)}$ or $\mathbf{V}$).
\end{definition}

In brief, we may call $M$ ($M_{(L)},$ or $\mathbf{V}$) as a spin manifold
(Lagrange spin manifold, or N--anholonomic spin manifold). If any of
the base or typical fiber spaces is of odd dimension, we may perform
similar constructions by considering $\C l^+$ instead of $\C l.$

The considerations presented in this Section consists the proof of the
first main Result of this paper (let us conventionally say that it is
the 7th one after the Results \ref{2r1}--\ref{r6}:\
\begin{theorem} \label{mr1} {\bf (Main Result 1)}
Any regular Lagrangian and/or
N--con\-nec\-ti\-on structure define naturally the fundamental geometric objects
 and structures  (such as the Clifford--Lagrange module and Clifford d--modules,
 the Lagrange spin structure  and d--spinors) for the corresponding Lagrange
 spin manifold  and/or  N--anholo\-nom\-ic spinor (d--spinor) manifold.
\end{theorem}
We note that similar results were obtained in
Refs. \cite{15vsp1,15vsph,15vst,15vsp2} for the standard Finsler and Lagrange
geometries and theirs higher order generalizations. In a more
restricted form,  the idea  of  Theorem \ref{mr1}
can be found in Ref. \cite{15vncg}, where the first models of noncommutative
Finsler geometry and related gravity  were considered (in a more rough
form, for instance, with constructions not reflecting the Morita equivalence).

Finally, in this Section, we can make the
\begin{conclusion} Any regular Lagrange and/or N--connection structure
  (the second one being any admissible N--connection in Lagrange--Finsler geometry
  and their generalizations, or induced by any generic off--diagonal and/ or
   nonholonomic frame structure) define certain, corresponding,
  Clifford--La\-gran\-ge module and/or Clifford d--module and  related
  Lagrange spinor and/or d--spinor structures.
\end{conclusion}
It is a bit surprizing
  that a Lagrangian may define  not only the fundamental geometric objects of
 a  nonholonomic Lagrange space but also the structure of
 a naturally associated Lagrange spin manifold. The Lagrange mechanics and
  off--diagonal gravitational interactions (in general, with nontrivial
  torsion and nonholonomic constraints) can be completely
 geometrized on Lagrange spin (N--anholonomic) manifolds.

\section{The Dirac Operator, Nonholonomy, and \newline Spec\-tral
  Trip\-les} \label{sdonst}
The Dirac operator for a certain class of  (non) commutative Finsler
spaces provided with compatible metric structure  was  introduced in
Ref. \cite{15vncg} following previous constructions for  the Dirac equations
on locally anisotropic spaces \cite{15vsp1,15vsph,15vstr1,15vst,15vsp2}.
The aim of this Section is to elucidate the possibility of  definition of Dirac
operators for general N--anholonomic manifolds and
Lagrange--Finsler spaces. It should be noted that such geometric
constructions depend on the type of linear connections which are used
  for the complete definition of the  Dirac operator. They are metric
 compatible and N--adapted if the canonical d--connection
  is used, see Proposition \ref{pcnc} (we can also use
 any its deformation which results in a  metric compatible
 d--connection). The constructions  can be  more sophisticate and
 nonmetric (with some  geometric  objects not  completely
 defined  on the  tangent spaces) if the Chern, or the
  Berwald d--connection, is considered, see  Example \ref{ecrdc}.

\subsection{N--anholonomic Dirac operators}
We introduce the basic definitions and formulas with respect to
N--adapted frames of type (\ref{dder}) and (\ref{ddif}).
 Then we shall present the main  results in a global form.

\subsubsection{Noholonomic vielbeins and spin d--connections}
Let us consider a Hilbert space of finite dimension. For a local dual
coordinate basis $e^{\underline i} \doteq dx^{\underline i}$
 on a manifold  $M,\ dim M =n,$\
  we may respectively introduce certain classes of
    orthonormalized  vielbeins and the
  N--adapted vielbeins, \footnote{(depending both on the base coordinates $x
  \doteq\ x^i$   and some "fiber" coordinates $y\doteq y^a,$ the
  status of $y^a$ depends on what kind of models we shall consider:\
  elongated on $TM,$ for a Lagrange space, for a vector bundle, or on
  a N--anholonomic manifold)}
\begin{equation}
  e^{\hat i} \doteq  e^{\hat i}_{\ \underline i} (x,y)\ e^{\underline i}
 \mbox{ and }
   e^{i} \doteq e^{i}_{\ \underline i}(x,y)\ e^{\underline i}, \label{hatbvb}
\end{equation}
where
\begin{equation*}
 g^{\underline i \underline j}(x,y)\ e^{\hat i}_{\ \underline i} (x,y)
 e^{\hat j}_{\ \underline j} (x,y)
 = \delta ^{\hat i \hat j} \mbox{ and }
 g^{\underline i \underline j}(x,y)\ e^{i}_{\ \underline i} (x,y)
 e^{j}_{\ \underline j} (x,y)
 = g ^{ij}(x,y).
\end{equation*}
We define the the algebra of Dirac's gamma matrices (in brief,
h--gamma matrices defined by  self--adjoints
matrices $M_k(\C)$ where $k=2^{n/2}$ is the dimension of the
irreducible representation of $\C l(M)$ for even dimensions, or of $\C
l(M)^+$ for odd dimensions) from the relation
\begin{equation}
 \gamma ^{\hat i} \gamma ^{\hat j} +  \gamma ^{\hat j}\gamma ^{\hat i}
 = 2 \delta ^{\hat i \hat j}\ \I. \label{grelflat}
\end{equation}
We can consider the action of $dx^i\in \C l (M)$ on a spinor $\psi \in
S$ via representations
\begin{equation}
\ ^{-}c(dx^{\hat i}) \doteq \gamma ^{\hat i} \mbox{ and }
 \ ^{-}c(dx^i)\psi \doteq \gamma ^i
 \psi \equiv e^i_{\ \hat i}\ \gamma ^{\hat i} \psi. \label{gamfibb}
\end{equation}

For any type of spaces $T_xM, TM, \mathbf{V}$ possessing a local (in
any point) or global fibered structure and, in general, enabled with
a N--connection structure, we can introduce similar definitions of the
gamma matrices following algebraic relations and metric structures on
fiber subspaces,
\begin{equation}
  e^{\hat a} \doteq  e^{\hat a}_{\ \underline a} (x,y)\ e^{\underline a}
 \mbox{ and }
   e^{a} \doteq e^{a}_{\ \underline a}(x,y)\ e^{\underline a}, \label{hatbvf}
\end{equation}
where
\begin{equation*}
 g^{\underline a \underline b}(x,y)\ e^{\hat a}_{\ \underline a} (x,y)
 e^{\hat b}_{\ \underline b} (x,y)
 = \delta ^{\hat a \hat b} \mbox{ and }
 g^{\underline a \underline b}(x,y)\ e^{a}_{\ \underline a} (x,y)
 e^{b}_{\ \underline b} (x,y)
 = h^{ab}(x,y).
\end{equation*}
Similarly, we define the algebra of Dirac's matrices related to
typical fibers (in brief, v--gamma matrices defined by self--adjoints
matrices $M_k'(\C)$ where $k'=2^{m/2}$ is the dimension of the
irreducible representation of $\C l(F)$ for even dimensions, or of $\C
l(F)^+$ for odd dimensions, of the typical fiber) from the relation
\begin{equation}
 \gamma ^{\hat a} \gamma ^{\hat b} +  \gamma ^{\hat b}\gamma ^{\hat a}
 = 2 \delta ^{\hat a \hat b}\ \I. \label{grelflatf}
\end{equation}
The action of $dy^a\in \C l (F)$ on a spinor $\ ^\star
\psi \in \ ^{\star}S$ is considered via representations
\begin{equation}
\ ^{\star}c(dy^{\hat a}) \doteq \gamma ^{\hat a} \mbox{ and }
 \ ^{\star}c(dy^a)\ ^\star \psi \doteq \gamma ^a
 \ ^\star \psi \equiv e^a_{\ \hat a}\ \gamma ^{\hat a}
 \ ^\star \psi. \label{gamfibf}
\end{equation}

We note that additionally to formulas (\ref{gamfibb}) and
(\ref{gamfibf}) we may write respectively
\begin{equation*}
 c(dx^{\underline i})\psi \doteq \gamma ^{\underline i}
 \psi \equiv e^{\underline i}_{\ \hat i}\ \gamma ^{\hat i} \psi
\mbox{ and }
 c(dy^{\underline a})\ ^\star \psi \doteq \gamma ^{\underline a}
 \ ^\star \psi \equiv e^{\underline a}_{\ \hat a}\ \gamma ^{\hat a}
 \ ^\star \psi
\end{equation*}
but such operators are not adapted to the N--connection structure.

A more general gamma matrix calculus with distinguished gamma
matrices (in brief, d--gamma matrices\footnote{in our previous works
  \cite{15vsp1,15vsph,15vstr1,15vst,15vsp2} we wrote $\sigma$ instead of
  $\gamma$}) can be elaborated for N--anholonomic manifolds
$\mathbf{V}$ provided with d--metric structure $\mathbf{g}=[g,
^{\star}g]$  and for
 d--spinors $\breve{\psi} \doteq (\psi,\ ^{\star}\psi)
\in \mathbf{S} \doteq (S,\ ^{\star}S),$ see the corresponding
 Definitions \ref{dnam}, \ref{ddms} and \ref{ddsp}. Firstly, we should write
 in a unified form, related to a d--metric (\ref{metr}), the formulas
 (\ref{hatbvb}) and (\ref{hatbvf}),
 \begin{equation}
  e^{\hat \alpha} \doteq  e^{\hat \alpha}_{\ \underline a}
 (u)\ e^{\underline \alpha}
 \mbox{ and }
   e^{\alpha} \doteq e^{\alpha}_{\ \underline \alpha}(u)\
 e^{\underline \alpha}, \label{hatbvd}
\end{equation}
where
\begin{equation*}
 g^{\underline{\alpha} \underline{\beta}}
(u)\ e^{\hat \alpha}_{\ \underline \alpha} (u)
 e^{\hat \beta}_{\ \underline \beta} (u)
 = \delta ^{\hat \alpha \hat \beta} \mbox{ and }
 g^{\underline \alpha \underline \beta}(u)\ e^{\alpha}_{\ \underline
   \alpha}  (u)  e^{\beta}_{\ \underline \beta} (u)
 = g^{\alpha \beta}(u).
\end{equation*}
The second step, is to consider d--gamma matrix relations (unifying
(\ref{grelflat}) and (\ref{grelflatf}))
\begin{equation}
 \gamma ^{\hat \alpha} \gamma ^{\hat \beta} +  \gamma ^{\hat \beta}
\gamma ^{\hat \alpha}
 = 2 \delta ^{\hat \alpha \hat \beta}\ \I,  \label{grelflatd}
\end{equation}
 with the action of $du^\alpha \in \C l (\mathbf{V})$ on a d--spinor
 $\breve{\psi} \in \ \mathbf{S}$ resulting in distinguished irreducible
representations (unifying (\ref{gamfibb}) and (\ref{gamfibf}))
\begin{equation}
\mathbf{c}(du^{\hat \alpha}) \doteq \gamma ^{\hat \alpha} \mbox{ and }
 \mathbf{c}=(du^\alpha)\  \breve{\psi} \doteq \gamma ^\alpha
 \  \breve{\psi} \equiv e^\alpha_{\ \hat \alpha}\ \gamma ^{\hat \alpha}
 \  \breve{\psi} \label{gamfibd}
\end{equation}
which allows to write
\begin{equation}
 \gamma ^{\alpha}(u) \gamma ^{\beta}(u) +
 \gamma ^{\beta} (u) \gamma ^{\alpha} (u)
 = 2 g ^{\alpha \beta}(u)\ \I.  \label{grelnam}
\end{equation}
In the canonical representation we can write in irreducible form
$\breve \gamma \doteq \gamma \oplus\ ^\star \gamma$ and
 $\breve \psi \doteq \psi \oplus\ ^\star \psi,$  for instance, by
 using block type of h-- and v--matrices, or, writing alternatively
 as couples of gamma and/or h-- and v--spinor objects written
in N--adapted form,
\begin{equation} \label{crgs}
\gamma ^{\alpha} \doteq (\gamma ^i, \gamma ^a) \mbox{ and }
 \breve \psi \doteq (\psi,\ ^\star \psi).
\end{equation}
The decomposition (\ref{grelnam}) holds with respect to a N--adapted
vielbein (\ref{dder}). We also note that for a spinor calculus, the
indices of spinor objects should be treated as abstract spinorial ones
 possessing certain reducible, or irreducible, properties depending on
 the space dimension (see details in Refs.
 \cite{15vsp1,15vsph,15vstr1,15vst,15vsp2}). For simplicity, we shall consider
 that spinors like $\breve \psi, \psi,\ ^\star \psi$ and all type of
 gamma objects can be enabled with corresponding spinor indices
 running certain values which are different from the usual coordinate
 space indices. In a "rough" but brief form we can use the same indices $i,j,
 ...,a,b...,\alpha, \beta,...$ both for d--spinor and d--tensor
 objects.

The spin connection $\nabla ^S$  for the Riemannian manifolds is
induced by the Levi--Civita connection $\ ^\nabla \Gamma ,$
\begin{equation} \label{sclcc}
 \nabla ^S  \doteq d - \frac{1}{4}\ ^\nabla \Gamma ^i_{\ j k}
\gamma _i \gamma ^j\ dx^k.
\end{equation}
On N--anholonomic spaces, it is possible to define spin connections
which are N--adapted by replacing the Levi--Civita connection by any
d--connection (see Definition \ref{ddc}).
\begin{definition}
The canonical spin d--connection is defined by the canonical
d--connection (\ref{cdc}) as
\begin{equation} \label{csdc}
 \widehat{\nabla}^{\mathbf{S}} \doteq \delta
 - \frac{1}{4}\ \widehat{\mathbf{\Gamma}}^\alpha_{\ \beta \mu}
\gamma _\alpha \gamma ^\beta \delta u^\mu,
\end{equation}
where the absolute differential $\delta$ acts in  N--adapted form
resulting in 1--forms decomposed with respect to N--elongated
differentials like $\delta u^\mu = (dx^i, \delta y^a)$ (\ref{ddif}).
\end{definition}
We note that the canonical spin d--connection
$\widehat{\nabla}^{\mathbf{S}}$ is metric compatible and
contains nontrivial d--torsion coefficients
 induced by the N--anholonomy relations (see
the formulas (\ref{dtors}) proved for arbitrary d--connection).
 It is possible to introduce more general spin d--connections
 ${\mathbf{D}}^{\mathbf{S}}$ by using the same formula (\ref{csdc})
 but for arbitrary metric compatible d--connection
${\mathbf{\Gamma}}^\alpha_{\ \beta \mu}.$

In a particular case, we
 can define, for instance, the canonical spin d--connections for a
local modelling of a $\widetilde{TM}$ space on $\widetilde{\mathbf{V}}%
_{(n,n)}$ with the canonical d--connection
 $\widehat{\mathbf{\Gamma }}_{\ \alpha \beta
}^{\gamma }=(\widehat{L}_{jk}^{i},\widehat{C}_{jk}^{i}),$ see formulas
 (\ref{candcon1}). This allows us to prove (by considering d--connection
 and d--metric structure defined by the fundamental Lagrange, or Finsler,
 functions, we put formulas (\ref{cncl}) and (\ref{slm})  into
 (\ref{candcon1})):
 \begin{proposition} \label{pcslc} On Lagrange   spaces, there is a
 canonical spin d--connec\-ti\-on (the canonical spin--Lagrange connection),
 \begin{equation} \label{fcslc}
 \widehat{\nabla}^{(SL)} \doteq \delta
 - \frac{1}{4}\ ^{(L)}{\mathbf{\Gamma}}^\alpha_{\ \beta \mu}
\gamma _\alpha \gamma ^\beta \delta u^\mu,
 \end{equation}
 where $\delta u^\mu = (dx^i, \delta y^k = dy^k +\ ^{(L)} N^k_{\ i}\ dx^i).$
 \end{proposition}
 We emphasize that even regular Lagrangians of classical mechanics
 without spin particles induce in a canonical (but nonholonomic) form
 certain classes of  spin d--connections like (\ref{fcslc}).

For the spaces provided with generic
 off--diagonal metric structure (\ref{ansatz}) (in particular, for
 such  Riemannian manifolds) resulting in  equivalent
 N--anho\-lo\-nom\-ic manifolds, it is possible to
prove a result being similar to
 Proposition \ref{pcslc}:
\begin{remark} \
There is a canonical spin d--connection (\ref{csdc})
 induced by the off--diagonal metric
coefficients with nontrivial $N^a_i$ and associated nonholonomic
 frames in gravity theories.
\end{remark}

The N--connection structure  also states a global h-- and v--splitting
of spin d--connecti\-on operators, for instance,
\begin{equation} \label{cslc}
 \widehat{\nabla}^{(SL)} \doteq \delta
 - \frac{1}{4}\ ^{(L)}{\widehat{L}}^i_{\ jk }
\gamma _i \gamma ^j dx^k
 - \frac{1}{4}\ ^{(L)}{\widehat{C}}^a_{\ bc}
\gamma _a \gamma ^b \delta y^c.
 \end{equation}
So, any spin d--connection is a  d--operator with
conventional splitting of action like ${\nabla}^{(\mathbf{S})} \equiv
 ({\ ^{-}{\nabla}}^{(\mathbf{S})},{\ ^\star
   {\nabla}}^{(\mathbf{S})}),$  or  ${\nabla}^{(SL)} \equiv
 ({\ ^{-}{\nabla}}^{(SL)},{\ ^\star
   {\nabla}}^{(SL)}).$ For instance, for
$\widehat{\nabla}^{(SL)} \equiv  ({\ ^{-}\widehat{\nabla}}^{(SL)},{\ ^\star
   \widehat{\nabla}}^{(SL)}),$   the operators
 $\ ^{-}\widehat{\nabla}^{(SL)}$ and
$\ ^\star \widehat{\nabla}^{(SL)}$ act respectively
  on a h--spinor $\psi$ as
\begin{equation} \label{hdslop}
{\ ^{-}\widehat{\nabla}}^{(SL)} \psi \doteq  dx^i \  \frac{\delta \psi}{dx^i}
 -  dx^k  \frac{1}{4}\ ^{(L)}{\widehat{L}}^i_{\ jk }
\gamma _i \gamma ^j \ \psi
\end{equation}
and
\begin{equation*}
{\ ^\star \widehat{\nabla}}^{(SL)}
\psi \doteq  \delta y^a \  \frac{\partial \psi}{dy^a}
 - \delta y^c \  \frac{1}{4}\ ^{(L)}{\widehat{C}}^a_{\ bc}
\gamma _a \gamma ^b \ \psi
\end{equation*}
being defined by the canonical d--connection (\ref{candcon1}).
\begin{remark} \label{rscfs} We can consider that the h--operator (\ref{hdslop})
  defines a spin generalization of the Chern's d--connection
 $\ ^{[Chern]}{\mathbf{\Gamma }}_{\ \alpha \beta}^{\gamma }=
(\widehat{L}_{jk}^{i},{C}_{jk}^{i}= 0),\ $ see Example \ref{ecrdc},
 which may be introduced as a minimal extension, with Finsler structure,
 of the spin connection defined by the Levi--Civita connection
(\ref{sclcc})  preserving the torsionless condition. This is an
 example of nonmetric spin connection operator because
$\ ^{[Chern]}{\mathbf{\Gamma }}_{\ \alpha \beta}^{\gamma }$ does not
 satisfy the condition of  metric compatibility.
\end{remark}
We can define  spin Chern--Finsler  structures, considered in
 the Remark \ref{rscfs},  for  any point
 of an N--anholonomic  manifold. There are necessary
 some additional assumptions in order to completely define such  structures
 (for instance, on the tangent bundle).  We can say that this is
 a  deformed nonholonomic spin
 structure derived from a d--spinor one provided with the canonical
 spin d--connection by deforming the canonical  d--connection
  in a  manner that  the horizontal  torsion vanishes transforming
 into a nonmetricity d--tensor. The
 "nonspinor" aspects of such generalizations of the Riemann--Finsler
 spaces and gravity models with nontrivial nonmetricity are analyzed
 in Refs. \cite{15v2}.

\subsubsection{Dirac d--operators:\ Main Result 2}

We consider a vector bundle $\mathbf{E}$ on an N--anholonomic manifold
 $\mathbf{M}$ (with two compatible N--connections defined as h-- and
 v--splitting of $T\mathbf{E}$ and $T\mathbf{M}$)). A d--connection
\begin{equation*} \mathcal{D}:\ \Gamma ^\infty (\mathbf{E})
 \rightarrow \Gamma ^\infty (\mathbf{E}) \otimes \Omega ^1(\mathbf{M})
\end{equation*}
preserves by parallelism splitting of the tangent total and base
spaces and satisfy the Leibniz condition
\begin{equation*} \mathcal{D}(f \sigma ) = f (\mathcal{D} \sigma) +
  \delta f \otimes \sigma
\end{equation*}
for any $f\in C^\infty (\mathbf{M}),$ and
 $\sigma \in \Gamma ^\infty (\mathbf{E})$ and $\delta$ defining an
 N--adapted  exterior calculus by using N--elongated operators
 (\ref{dder}) and (\ref{ddif}) which emphasize d--forms instead of
 usual forms on $\mathbf{M},$ with the coefficients taking values in
$\mathbf{E}.$

The metricity and Leibniz conditions for $ \mathcal{D}$ are written
respectively
\begin{equation} \mathbf{g} (\mathcal{D}\mathbf{X}, \mathbf{Y})+
 \mathbf{g} (\mathbf{X}, \mathcal{D} \mathbf{Y})=
 \delta [\mathbf{g} (\mathbf{X}, \mathbf{Y})], \label{mc1}
\end{equation}
for any $\mathbf{X},\ \mathbf{Y} \in \chi(\mathbf{M}),$
and
\begin{equation} \mathcal{D} (\sigma \beta) \doteq
 \mathcal{D} (\sigma) \beta +  \sigma \mathcal{D}(\beta), \label{lc1}
\end{equation}
for any $\sigma, \beta \in \Gamma ^\infty (\mathbf{E}).$

For local computations, we may define the corresponding coefficients of
the geometric d--objects and write
\begin{equation*}
\mathcal{D} \sigma _{\acute \beta}
 \doteq {\mathbf{\Gamma}}^{\acute \alpha}_{\ {\acute   \beta} \mu}\
\sigma _{\acute \alpha} \otimes \delta u^\mu =
{\mathbf{\Gamma}}^{\acute \alpha}_{\ {\acute   \beta} i}\
\sigma _{\acute \alpha} \otimes dx^i +
{\mathbf{\Gamma}}^{\acute \alpha}_{\ {\acute   \beta} a}\
\sigma _{\acute \alpha} \otimes \delta y^a,
\end{equation*}
where fiber "acute" indices, in their turn, may  split
${\acute \alpha}\doteq ({\acute i}, {\acute a} )$ if any N--connection
structure is defined on $T\mathbf{E}.$ For some particular
constructions of particular interest, we can take $\mathbf{E} =
T^*\mathbf{V}, = T^*V_{(L)}$ and/or any Clifford d--algebra
$\mathbf{E} = \C l(\mathbf{V}), \C l(V_{(L)}),...$ with a
corresponding treating of "acute" indices to of d--tensor and/or
d--spinor type as well  when the d--operator  $\mathcal{D}$ transforms into
respective d--connection $\mathbf{D}$ and spin d--connections
 $\widehat{\nabla}^{\mathbf{S}}$\  (\ref{csdc}),
   $\widehat{\nabla}^{(SL)}$\ (\ref{fcslc}).... All such, adapted to the
   N--connections,  computations are similar for both N--anholonomic
 (co) vector and spinor  bundles.

The respective actions of the Clifford d--algebra and Clifford--Lagrange algebra
(see Definitions \ref{dcdalg} and \ref{dcdalg}) can be transformed into
 maps  $\Gamma ^\infty (\mathbf{S}) \otimes \Gamma ^(\C
l(\mathbf{V}))$ and $\Gamma ^\infty (S_{(L)}) \otimes \Gamma ^(\C
l(V_{(L)}))$ to $\Gamma ^\infty (\mathbf{S})$ and, respectively,
 $\Gamma ^\infty (S_{(L)})$ by considering maps of type
 (\ref{gamfibb}) and (\ref{gamfibd})
 \begin{equation*}
 \widehat{\mathbf{c}}(\breve{\psi} \otimes \mathbf{a}) \doteq
 \mathbf{c}(\mathbf{a}) \breve{\psi}
 \mbox{\ and\ }
 \widehat{c}({\psi} \otimes {a}) \doteq {c}({a}){\psi}.
 \end{equation*}
\begin{definition} \label{dddo} The Dirac d--operator (Dirac--Lagrange
 operator) on  a spin N--anholonomic manifold $(\mathbf{V},\mathbf{S}, J)$ (on a
 Lagrange spin manifold\\ $(M_{(L)}, S_{(L)}, J))$  is defined
\begin{eqnarray}
 \D &\doteq &  -i\ (\widehat{\mathbf{c}} \circ \nabla ^{\mathbf{S}})
\label{ddo} \\
 & = & \left( \ ^{-}\D =  -i\ (\ ^{-}\widehat{{c}}
 \circ \ ^{-}\nabla ^{\mathbf{S}})
,\ ^\star \D  = -i\ (\ ^{\star}\widehat{{c}}
 \circ \ ^{\star}\nabla ^{\mathbf{S}} ) \right) \nonumber  \\
 (\ _{(L)}\D  &\doteq &
 -i\ (\widehat{c} \circ \nabla ^{(SL)}) \ ) \label{dlo} \\
 & = & \left( \ _{(L)}{^{-}\D
= -i (\ ^{-}\widehat{c} \circ \ ^{-}\nabla ^{(SL)}}),
\ _{(L)}{^\star \D} =  -i (\ ^{\star} \widehat{c}
 \circ  \ ^{\star}\nabla ^{(SL)}  ) \right)\ ). \nonumber
\end{eqnarray}
Such N--adapted Dirac d--operators are called  canonical  and denoted
 $\widehat{\D} = ( \ ^{-}\widehat{\D},\
^\star\widehat{\D}\ )$\ ( $ _{(L)}\widehat{\D} = ( \
_{(L)}{^{-}\widehat{\D}},\  _{(L)}{^{\star}\widehat{\D}}\ )$\ )  if they
are defined for the canonical d--connection (\ref{3candcon})\ (\
(\ref{candcon1})) and respective spin d--connection (\ref{csdc})
(\ (\ref{fcslc})).
\end{definition}

Now we can formulate the
 \begin{theorem} \label{mr2} {\bf  (Main Result 2)} Let $(\mathbf{V},\mathbf{S},
   J)$ (\ $(M_{(L)}, S_{(L)}, J)$ be a spin N--anho\-lo\-no\-mic manifold ( spin
 Lagrange space). There is the canonical Dirac d--operator
 (Dirac--Lagrange operator) defined by the almost Hermitian  spin
  d--operator
\begin{equation*}
\widehat{\nabla}^{\mathbf{S}}:\ \Gamma ^\infty (\mathbf{S})\rightarrow
 \Gamma ^\infty (\mathbf{S}) \otimes \Omega ^1 (\mathbf{V})
\end{equation*}
 (spin Lagrange operator
\begin{equation*}
\widehat{\nabla}^{(SL)}:\ \Gamma ^\infty ({S_{(L)}})\rightarrow
 \Gamma ^\infty (S_{(L)}) \otimes \Omega ^1 (M_{(L)}) \ )
\end{equation*}
 commuting with $J$ (\ref{jeq}) and satisfying the  conditions
\begin{equation} \label{scmdcond}
 (\widehat{\nabla}^{\mathbf{S}} \breve{\psi}\ |\ \breve{\phi} )\  +
 ( \breve{\psi}\ |\ \widehat{\nabla}^{\mathbf{S}} \breve{\phi} )\ =
 \delta ( \breve{\psi}\ |\  \breve{\phi} )\
\end{equation}
and
\begin{equation*}
\widehat{\nabla}^{\mathbf{S}}( \mathbf{c}(\mathbf{a}) \breve{\psi})\
 = \mathbf{c} (\widehat{\mathbf{D}}\mathbf{a}) \breve{\psi}
 + \mathbf{c} (\mathbf{a})\widehat{\nabla}^{\mathbf{S}} \breve{\psi}
\end{equation*}
for $\mathbf{a} \in \C l(\mathbf{V})$ and $\breve{\psi}\in \Gamma
^\infty (\mathbf{S})$
\begin{equation} \label{scmlcond}
 (\ (\widehat{\nabla}^{(SL)} \breve{\psi}\ |\ \breve{\phi} )\  +
 ( \breve{\psi}\ |\ \widehat{\nabla}^{(SL)} \breve{\phi} )\ =
 \delta ( \breve{\psi}\ |\  \breve{\phi} )\
\end{equation}
and
\begin{equation*}
\widehat{\nabla}^{(SL)}( \mathbf{c}(\mathbf{a}) \breve{\psi})\
 = \mathbf{c} (\widehat{\mathbf{D}}\mathbf{a}) \breve{\psi}
 + \mathbf{c} (\mathbf{a})\widehat{\nabla}^{(SL)} \breve{\psi}
\end{equation*}
for $\mathbf{a} \in \C l(M_{(L)})$ and $\breve{\psi}\in \Gamma
^\infty (S_{(L)}$ )\
 determined by the metricity (\ref{mc1}) and Leibnitz (\ref{lc1})
conditions.
 \end{theorem}
\begin{proof} We sketch the main ideas of  such Proofs. There two
  ways:

 The first one  is similar to
  that given in Ref. \cite{15bondia}, Theorem 9.8, for the Levi--Civita
  connection, see similar considerations in \cite{15schroeder}.
In our case, we have to extend the constructions for
  d--metrics and canonical d--connections by applying N--elongated
  operators for differentials and partial derivatives. The
  formulas have to be distinguished into h-- and
  v--irreducible components. We are going to present the related technical details
  in our further publications.

In other turn, the second way, is to argue  a such proof is
 a straightforward consequence of the Result \ref{r6} stating that any
  Riemannian manifold can be modelled as a N--anholonomic manifold
 induced by the generic off--diagonal metric structure. If the results
 from  \cite{15bondia} hold true for any Riemannian space,
 the formulas may be rewritten with respect to any local frame
 system, as well with respect to (\ref{dder}) and
 (\ref{ddif}). Nevertheless, on N--anholonomic manifolds the
 canonical d--connection is not just the Levi--Civita connection but
 a deformation of type (\ref{cdc}):\ we must verify that such
 deformations results in  N--adapted constructions satisfying the metricity
 and Leibnitz conditions. The existence of such configurations was
 proven from the properties of the canonical d--connection completely
 defined from the d--metric and N--connection coefficients. The main
 difference from the case of the Levi--Civita configuration is that we
 have a nontrivial torsion induced by the frame nonholonomy. But it is not
 a problem to define the  Dirac operator with nontrivial torsion if
 the metricity conditions are satisfied. $\Box$
\end{proof}

The canonical Dirac d--operator has very similar
  properties for spin N--anholonomic manifolds and spin Lagrange
  spaces. Nevertheless, theirs geometric and physical meaning may be
  completely different and that why we have written the corresponding
  formulas with different labels and emphasized the existing
  differences. With respect to the {\bf Main Result 2}, one holds
 three important remarks:
\begin{remark}  The first type of canonical Dirac d--operators may be
  associated to Riemannian--Cartan (in particular, Riemann)
 off--diagonal metric and nonholonomic frame   structures and the
  second type of canonical Dirac--Lagrange operators are completely
  induced by a regular Lagrange mechanics.
  In both cases, such d--operators are
  completely determined by the coefficients of the corresponding
  Sasaki type d--metric and the N--connection structure.
\end{remark}
\begin{remark} The conditions of the Theorem \ref{mr2} may be revised
  for any d--connection and induced spin d--connection satisfying the
  metricity condition. But, for such cases, the corresponding Dirac
   d--operators are
  not  completely defined by the d--metric and N--connection
  structures. We can prescribe certain type of torsions of
  d--connections and, via such 'noncanonical' Dirac operators, we are able
  to define noncommutative geometries with prescribed d--torsions.
\end{remark}
\begin{remark} The properties (\ref{scmdcond}) and  (\ref{scmlcond})
  hold if and only if the metricity conditions are satisfied
  (\ref{mc1}). So, for the Chern or Berwald type d--connections which
  are nonmetric (see Example \ref{ecrdc}  and Remark \ref{rscfs} ),
 the conditions of Theorem \ref{mr2} do not hold.
\end{remark}
 It is a more   sophisticate problem to find applications in physics for
  such nonmetric
  constructions \footnote{See Refs. \cite{15hehl} and \cite{15v1,15v2,15v3}
  for details on elaborated geometrical and physical models being, respectively,
   locally   isotropic and locally anisotropic.}
  but they define positively some examples of nonmetric
  d--spinor and noncommutative structures minimally deformed from the
  Riemannian (non) commutative  geometry to certain Finsler type (non)
  commutative geometries.

\subsection{Distinguished spectral triples}
 The geometric information of a spin manifold (in particular, the
 metric) is contained in the Dirac operator. For nonholonomic
 manifolds, the canonical Dirac d--operator has h-- and v--irreducible parts
 related to  off--diagonal metric terms and nonholonomic frames with
 associated structure. In a more special case, the canonical
 Dirac--Lagrange operator is defined by a regular Lagrangian. So, such
 Driac d--operators contain more information than the usual,
 holonomic, ones.

For simplicity, hereafter, we shall formulate the results for the
general N--anholonomic spaces, by omitting the explicit formulas and
 proofs  for Lagrange and  Finsler spaces, which can be derived by
 imposing certain  conditions  that the
 N--connection,  d--connection and d--metric are just those
  defined canonically by a  Lagrangian. We shall only present the
  Main Result and some  important Remarks concerning
 Lagrange   mechanics and/or Finsler  structures.
\begin{proposition} \label{pdohv}
If $\widehat{\D} = ( \ ^{-}\widehat{\D},\ ^\star\widehat{\D}\ )$\ is
the canonical Dirac  d--operator then
\begin{eqnarray*}
\left[\widehat{\D},\ f \right] & = &  i \mathbf{c} (\delta f),
\mbox{ equivalently,} \\
    \left[\ ^{-}\widehat{\D},\  f \right]  +
\left[\  ^\star\widehat{\D} ,\ f \right] & = &
 i \ ^{-}c ( dx^i \frac{\delta  f}{\partial x^i})
 + i \ ^{\star}c( \delta y^a \frac{\partial f}{\partial  y^a}),
\end{eqnarray*}
for all $f \in C^\infty (\mathbf{V}).$
\end{proposition}
\begin{proof} It is a straightforward computation following from
  Definition \ref{dddo}.
\end{proof}

The canonical Dirac d--operator and its irreversible h-- and
v--components have all the properties of the usual Dirac operators
(for instance, they are self--adjoint but unbounded). It is possible
to define a scalar product on $\Gamma ^\infty (\mathbf{S})$,
\begin{equation} \label{scprod}
 <\breve{\psi}, \breve{\phi}>
\doteq \int_\mathbf{V}(\breve{\psi} | \breve{\phi}) |\nu _\mathbf{g}|
\end{equation}
where $$\nu _\mathbf{g} = \sqrt{det g}\ \sqrt{det h}\ dx^1...dx^n\
 dy^{n+1}...dy^{n+m}$$ is the volume d--form on the N--anholonomic
 manifold $\mathbf{V}.$

We denote by
\begin{equation} \label{dhs}
 \mathcal{H}_{N} \doteq L_2 (\mathbf{V}, \mathbf{S}) = \left[
  \ ^{-}\mathcal{H} = L_2  (\mathbf{V},\ ^{-}S),\ ^\star \mathcal{H} =
  L_2 (\mathbf{V},\ ^\star S)\right]
\end{equation}
the Hilbert d--space obtained by completing $\Gamma ^\infty
(\mathbf{S})$ with the norm defined by the scalar product
(\ref{scprod}).

Similarly to the holonomic spaces, by using formulas (\ref{ddo}) and
 (\ref{csdc}), one may prove that there is a self--adjoint unitary
 endomorphism $\Gamma ^{[cr]}$ of $ \mathcal{H}_{N},$ called
 "chirality", being a ${\Z}_2$ graduation of $ \mathcal{H}_{N},$
 \footnote{We use the  label\ $[cr]$\ in order to avoid
 misunderstanding  with the
 symbol $\Gamma$ used for the connections. }
 which satisfies the condition
 \begin{equation}
\widehat{\D}\ \Gamma ^{[cr]} = - \Gamma ^{[cr]}\ \widehat{\D}
\label{chrc}.
 \end{equation}
We note that the condition (\ref{chrc}) may be written also for
the irreducible components $\ ^{-}\widehat{\D}$ and $\ ^{\star}\widehat{\D}.$
\begin{definition}
A distinguished canonical spectral triple (canonical spectral d--triple)
 $(\mathcal{A}, \mathcal{H}_{N},\ \widehat{\D}  )$ for an algebra $\mathcal{A}$
 is defined by a Hilbert d--space $\mathcal{H}_{N},$ a representation
 of $\mathcal{A}$ in the algebra $\mathcal{B}(\mathcal{H})$ of d--operators
 bounded on $\mathcal{H}_{N},$ and by a self--adjoint d--operator
 $\widehat{\D},$ of compact resolution,\footnote{An operator $D$ is of
 compact resolution if  for any $\lambda \in sp(D)$ the operator
 $(D-\lambda \I)^{-1}$ is compact, see details in
 \cite{15mart2,15bondia}.}
 such that
 $[\widehat{\D},a] \in \mathcal{B}(\mathcal{H})$ for any $a\in \mathcal{A}.$
\end{definition}

Roughly speaking, every canonical spectral d--triple is defined by two
usual spectral triples which in our case corresponds to certain h--
and v--irreducible components induced by the corresponding h-- and
v--components of the Dirac d--operator. For such spectral h(v)--triples
we can define the notion of $KR^n$--cycle ($KR^m$--cycle) and consider
respective Hochschild complexes.  We note that in order to define a
noncommutative geometry  the h-- and v-- components of a
 canonical spectral d--triples must satisfy some well defined
 Conditions \cite{15connes1,15bondia} (Conditions 1 - 7, enumerated in
 \cite{15mart2}, section II.4) which states:\ 1) the spectral dimension, being of
 order $1/(n+m)$ for a Dirac d--operator, and of order $1/n$ (or $1/m)$
 for its h-- (or v)--components; 2) regularity; 3) finiteness; 4)
 reality; 5) representation of 1st order; 6) orientability; 7)
 Poincar\'{e} duality. Such conditions can be satisfied by any Dirac
 operators and canonical Dirac d--operators (in the second case we have
 to work with d--objects). \footnote{ We omit in this paper the
   details on axiomatics and related proofs for  such
   considerations:\ we shall present details and proofs in our further
   works. Roughly speaking, we are in right to do this because the
   canonical d--connection and the Sasaki type d--metric for
   N--anholonomic spaces satisfy the bulk of properties of the metric
   and connection on the Riemannian space but "slightly"
   nonholonomically modified). }
\begin{definition}\label{dncdg}
A spectral d--triple satisfying the mentioned seven
  Conditions for his h-- and v--irreversible components is a real one
  which defines a (d--spinor) N--anholono\-mic noncommutative geometry
  defined  by the data
$(\mathcal{A}, \mathcal{H}_{N},\ \widehat{\D},\ J,\  \Gamma ^{[cr]}\ )$
 and derived for the  Dirac d--operator (\ref{ddo}).
\end{definition}
For a particular case, when the N--distinguished structures are of Lagrange
(Finsler) type, we can consider:
\begin{definition} \label{dnclg} A spectral d--triple satisfying the
  mentioned  seven
  Conditions for his h-- and v--irreversible components is a real one
  which defines a Lagrange, or Finsler, (spinor)
  noncommutative geometry  defined by the data
 $(\mathcal{A}, \mathcal{H}_{(SL)},\ _{(L)}\widehat{\D},\ J,\ \Gamma ^{[cr]}\ )$
 and derived for the  Dirac d--operator (\ref{dlo}).
\end{definition}
In Ref. \cite{15vncg}, we used the concept of d--algebra
$\mathcal{A}_{N} \doteq (\ ^{-}\mathcal{A},\ ^\star \mathcal{A})$ which we
 introduced  as a "couple" of algebras for respective h-- and
 v--irreducible decomposition of constructions defined by the
 N--connection. This is possible if
$\mathcal{A}_{N} \doteq \ ^{-}\mathcal{A}\oplus\  ^\star \mathcal{A}),$ but
we can consider arbitrary noncommutative associative algebras
$\mathcal{A}$ if  the splitting is defined by the Dirac d--operator.

\subsection{Distance in d--spinor spaces: Main Result 3}

We can select N--anholonomic and Lagrange commutative geometries from
the corresponding Definitions \ref{dncdg} and \ref{dnclg} if we put
respectively $\mathcal{A}\doteq C^\infty (\mathbf{V})$ and
 $\mathcal{A}\doteq C^\infty (V_{(L)})$ and consider real spectral
 d--triples. One holds:
\begin{theorem} \label{mr3} {\bf  (Main Result 3)} Let
 $(\mathcal{A}, \mathcal{H}_{N},\ \widehat{\D},\ J,\  \Gamma ^{[cr]}\
  )$ \\
  (or $(\mathcal{A}, \mathcal{H}_{(SL)},\ _{(L)}\widehat{\D},\ J,\
  \Gamma ^{[cr]}\ )$) defines a noncommutative geometry being
  irreducible for $\mathcal{A}\doteq C^\infty (\mathbf{V})$ (or
 $\mathcal{A}\doteq C^\infty (V_{(L)})),$ where $\mathbf{V}$ (or
  $V_{(L)}$) is a compact, connected and  oriented  manifold without
  boundaries, of   spectral dimension $dim\ \mathbf{V}=n+m$
 (or $dim\ V_{(L)} =n+n$ ). In this case, there are
  satisfied the conditions:
\begin{enumerate}
\item There is a unique d--metric  $\mathbf{g}(\widehat{\D} ) = (g,\ ^\star g)$
 of type ((\ref{metr})) on  $\mathbf{V}$\ (or  of type (\ref{slm}) on
 $V_{(L)})$ with the "nonlinear" geodesic distance defined by
\begin{equation} \label{fngdd}
 d(u_1,u_2) = \sup _{f\in C(\mathbf{V})} \left\{ f(u_1,u_2) / \parallel [\D ,
 f ]\parallel \leq 1 \right\}
\end{equation}
 (we have to consider $f\in C(V_{(L)})$ and $\ _{(L)}\widehat{\D}$ if we compute
 $d(u_1,u_2)$ for Lagrange  configurations).
\item The N--anholonomic manifold $\mathbf{V}$\ (or Lagrange space
  $V_{(L)})$ is a spin N--anholonmic space (or a spin Lagrange
  manifold) for which the operators $\D ^\prime$ satisfying
  $\mathbf{g}(\D ^\prime) = \mathbf{g}(\widehat{\D})$ define an union of affine
  spaces identified by the d--spinor structures on $\mathbf{V}$ (we
  should consider the operators  $\ _{(L)}\D ^\prime$ satisfying
  $\ ^{(L)}\mathbf{g}(\ _{(L)}\D ^\prime) =\   ^{(L)}\mathbf{g}(\
  _{(L)}\widehat{\D})$ for the space   $V_{(L)})$).\
\item The functional $S(\D ) \doteq \int |\D |^{-n-m+2}$
    defines a   quadratic d--form with $(n+m)$--splitting
 for every affine spaces which is minimal for
 $\widehat{\D} = \overleftarrow{\D}$ as the Dirac d--operator
  corresponding to the d--spin structure with the minimum proportional
  to the Einstein--Hilbert action constructed for the canonical
  d--connection with the d--scalar curvature $
  \overleftarrow{\mathbf{R}}$ (\ref{sdccurv}),
\footnote{The  integral for the  usual Dirac operator related to
  the Levi--Civita     connection  $D\ $ is computed:\
 $\int |D|^{-n+2} \doteq \frac{1}{2^{[n/2]}{\Omega _n}} Wres
    |D|^{-n+2},$
 where $\Omega _n$ is the integral of the  volume on the sphere
  $S^n$ and $Wres$ is the Wodzicki residu, see  details in Theorem 7.5
  \cite{15bondia}. On N--anholonomic manifolds, we may consider
   similar definitions and computations but
    applying N--elongated partial derivatives and differentials. }
\begin{equation*}
  S(\overleftarrow{\D}) = - \frac{n+m-2}{24}\  \int _{\mathbf{V}}\
 \overleftarrow{\mathbf{R}}\ \sqrt{g}\ \sqrt{h}\ dx^1...dx^n\
\delta y^{n+1}... \delta y^{n+k}.
\end{equation*}
\end{enumerate}
\end{theorem}
\begin{proof} In this work, we sketch only the idea and the key points of
 a such Proof.  The Theorem is a generalization for N--anholonomic
 spaces of a similar one, formulated in Ref. \cite{15connes1}, with a
 detailed proof presented in  \cite{15bondia}, which seems to be a
 final  almost generally  accepted result. There are
 also alternative considerations,  with useful details, in
 Refs. \cite{15rennie1,15lord}.
 For the Dirac d--operators,  we have to start with the
 Proposition \ref{pdohv} and then to repeat all constructions from
 \cite{15connes1,15bondia}, both on h-- and v--subspaces, in N--adapted
 form.

 The  existence of a canonical d--connection structure which is metric
 compatible and  constructed from the coefficients of the
 d--metric and  N--connecti\-on  structure is a crucial result allowing
 the formulation and proof of the Main Results 1-3 of this
 work. Roughly speaking, if the commutative Riemannian geometry can be
 extracted from a noncommutative geometry, we can also generate (in a similar, but
 technically more sophisticate form)  Finsler like geometries and
 generalizations. To do this, we have to consider the
 corresponding parametrizations of the nonholonomic frame structure,
 off--diagonal metrics and deformations of the linear connection
 structure, all constructions being adapted to the N--connection
 splitting. If a fixed d--connection satisfies the metricity conditions, the
 resulting Lagrange--Finsler geometry belongs to a  class
 of nonholonomic Riemann--Cartan geometries, which (in their turns) are
   equivalents, related by nonholonomic maps, of  Riemannian spaces, see
 \cite{15v1,15v3}. However, it is not yet clear how
 to perform a such general proof for nonmetric d--connections (of
 Berwald or Chern type).  We shall present the technical details of such
 considerations in our further works.

Finally, we emphasize that for the Main Result 3 there is the
possibility to elaborate an alternative proof (like for the Main
Result 2) by verifying that the basic formulas proved for the
Riemannian geometry hold true on N--anholonomic manifolds by a
corresponding substitution of the N--elongated differential and
partial derivatives operators acting on canonical d--connections and
d--metrics. All such constructions are elaborated in N--adapted form
by preserving the respective h- and v--irreducible decompositions. $\Box$
\end{proof}

Finally, we can formulate three important conclusions:
\begin{conclusion} The formula (\ref{fngdd})  defines the
  distance in a manner as to be satisfied all necessary properties (finitenes,
  positivity conditions, ...) discussed in details in
  Ref. \cite{15bondia}. It allows to generalize the constructions for
  discrete spaces with anisotropies and to consider anisotropic
  fluctuations of noncommutative geometries \cite{15mart2,15mart3}
   (of Finsler type, and more
  general ones, we omit such constructions in this work).
 For the nonholonomic configurations we have to work
  with canonical d--connection and d--metric structures.
\end{conclusion}
 Following the N--connection formalism originally elaborated
  in the framework of  Finsler geometry, we may state:
\begin{conclusion} In the particular case of the canonical N--connection,
  d--con\-nec\-ti\-on and d--metrics defined by a regular Lagrangian, it is
  possible  a noncommutative geometrization of Lagrange mechanics
  related to  corresponding  classes of noncommutative Lagrange--Finsler geometry.
\end{conclusion}
Such geometric methods have a number of applications in modern gravity:
\begin{conclusion} By anholonomic frame transforms, we can generate
  noncommutative Riemann--Cartan and Lagrange--Finsler spaces, in
  particular exact solutions of the Einstein equations with
  noncommutative variables \footnote{see
  examples in Refs. \cite{15v0,15vs1,15v2,15v3,15vncgs}}, by considering
  N--anholonomic deformations of the Dirac operator.
\end{conclusion}

\subsection*{Acknowledgment}

The  work  summarizes the  results communicated in  a series of  Lectures and
Seminars:

\begin{enumerate}

\item S. Vacaru, A Survey of (Non) Commutative Lagrange and Finsler
Geometry, lecture 2: Noncommutative Lagrange and Finsler
Geometry.  Inst. Sup. Tecnico, Dep. Math., Lisboa,
Portugal, May 19, 2004 (host:  P. Almeida).

 \item S. Vacaru, A Survey of (Non) Commutative Lagrange and Finsler
 Geometry, lecture 1:  Commutative Lagrange and Finsler Spaces and
 Spinors. Inst. Sup. Tecnico, Dep. Math., Lisboa,
 Portugal, May 19, 2004 (host:  P. Almeida).

 \item S. Vacaru, Geometric Models in Mechanics and Field Theory,
 lecture at the Dep. Mathematics, University of Cantabria, Santander,
Spain, March 16, 2004 (host:
 F. Etayo).

 \item S. Vacaru, Noncommutative Symmetries Associated to the Lagrange and
 Hamilton Geometry,   seminar at the Instituto de Matematicas y
 Fisica Fundamental,  Consejo Superior de Investigationes, Ministerio
de Ciencia y Tecnologia, Madrid, Spain, March 10, 2004 (host: M. de Leon).

 \item S. Vacaru, Commutative and Noncommutative Gauge Models,   lecture
at the Department of Experimental Sciences, University of Hu\-el\-va,
Spain, March 5, 2004 (host:  M.E. Gomez).

\item S. Vacaru, Noncommutative Finsler Geometry, Gauge Fields and Gravity,
  seminar at the Dep. Theoretical Physics, University of Zaragoza, Spain,
 November 19, 2003 (host: L. Boya).

\end{enumerate}

The author is  grateful to the Organizes and hosts  for financial support and
collaboration. He also thanks  R. Picken for help  and  P. Martinetti
  for  useful  discussions.

%%%%%%%%%%%%%%%%%%%%%%%%%%%%%%%%%%%%%%%%%%%%%%%%%%%%%%%%%%%%%%%%%%%%%%%%%%%%%

\end{document}